\documentclass{JHEP3}
\usepackage{amsmath,amssymb,amsfonts}
\usepackage{epsfig}
\usepackage{lscape}
\usepackage{multirow}
\usepackage{cite}

\newcommand\fullversion[1]{}%{#1}

\makeatletter

\@addtoreset{equation}{section}

\newcommand\Ref[1]     {Ref.\,\cite{#1}}
\newcommand\Refs[1]    {Refs.\,\cite{#1}}
\newcommand\eqn[1]     {Eq.\,(\ref{#1})}
\newcommand\eqns[2]    {Eqs.\,(\ref{#1}) and~(\ref{#2})}
\newcommand\eqnss[2]   {Eqs.\,(\ref{#1})--(\ref{#2})}
\newcommand\fig[1]     {Fig.\,{\ref{#1}}}
\newcommand\sect[1]    {Sect.\,{\ref{#1}}}
\newcommand\sects[2]   {Sects.\,\ref{#1} and~\ref{#2}}
\newcommand\appx[1]    {Appendix~\ref{#1}}
\newcommand\tab[1]     {Table~\ref{#1}}
\newcommand\nn         {\nonumber}

\def\beq{\begin{equation}}
\def\eeq{\end{equation}}
\def\bsp#1\esp{\begin{split}#1\end{split}}
\def\bal#1\eal{\begin{align}#1\end{align}}
\def\beeq{\begin{eqnarray}}
\def\eeeq{\end{eqnarray}}
\newcommand\bom[1]     {{\mbox{\boldmath $#1$}}}

\newcommand{\CF}       {C_{\mathrm{F}}}
\newcommand{\CA}       {C_{\mathrm{A}}}
\newcommand{\TR}       {T_{\mathrm{R}}}

\newcommand{\Nf}       {n_{\mathrm{f}}}
\newcommand{\Ns}       {n_{\mathrm{s}}}

\newcommand\qb         {{\bar q}}

\newcommand\Oe[1]      {\ensuremath{\mathrm O(\ep^{#1})}}
\newcommand{\ep}       {\epsilon}
\newcommand{\eps}      {\varepsilon}

\newcommand\ldot       {\!\cdot\!}

\newcommand{\rd}       {{\mathrm{d}}}
\newcommand{\PS}[1]    {\rd\phi_{#1}}

\newcommand{\cI}       {{\cal I}}
\newcommand{\cJ}       {{\cal J}}
\newcommand{\cK}       {{\cal K}}

\newcommand{\cA}       {{\cal A}}
\newcommand{\cB}       {{\cal B}}
\newcommand{\cII}[1] {{\cal I}\kern-4pt *\kern-4pt{\cal I}_{#1}}
\newcommand{\cIJ}     {{\cal I}\kern-4pt *\kern-4pt{\cal J}}
\newcommand{\cJJ}[1] {{\cal J}\kern-4pt *\kern-4pt{\cal J}_{#1}}
\newcommand{\cJI}     {{\cal J}\kern-4pt *\kern-4pt{\cal I}}
\newcommand{\cKJ}[1] {{\cal K}\kern-4pt *\kern-4pt{\cal J}_{#1}}
\newcommand{\cKI}     {{\cal K}\kern-4pt *\kern-4pt{\cal I}}

\newcommand{\mom}[1]   {\{p\}^{#1}}
\newcommand{\momt}[1]   {\{\ti{p}\}^{#1}}

\newcommand{\cmap}[1]   {\stackrel{{\rm C}_{#1}}{\longrightarrow}}
\newcommand{\smap}[1]   {\stackrel{{\rm S}_{#1}}{\longrightarrow}}

\newcommand{\bSCS}[1]  {\bom{\mathrm C}\kern-2pt\bom{\mathrm S}_{#1}}

\newcommand{\cSCS}[2]  {{\cal C}\kern-2pt{\cal S}_{#1}^{#2}}

\newcommand{\Y}[2]     {Y_{\ti{#1}\ti{#2},Q}}

\newcommand{\ti}[1]    {\tilde{#1}}
\newcommand{\wti}[1]   {\widetilde{\,#1\,}}

\renewcommand\S        {{\scriptscriptstyle\rm S\!.}}
\newcommand\NS         {{\scriptscriptstyle\rm N\!.S\!.}}

\newcommand{\ie}[0]{\emph{i.e. }}
\newcommand{\eg}[0]{\emph{e.g. }}
\newcommand{\kap}{\kappa}
\newcommand{\tht}{\vartheta}
\newcommand{\ph}{\varphi}
\newcommand{\alp}{\alpha}
\newcommand{\hpls}{\emph{HPL's} }
\newcommand{\dhpls}{\emph{2dHPL's} }
\newcommand{\ocal}{\begin{cal}O\end{cal} }
\newcommand{\kint}{\begin{cal}K\end{cal} }
\newcommand{\ciint}{\begin{cal}I\end{cal} }
\newcommand{\aint}{\begin{cal}A\end{cal} }
\newcommand{\bint}{\begin{cal}B\end{cal} }
\newcommand{\ao}{\alpha_0}
\newcommand{\yo}{y_0}
\newcommand{\hypgeo}{\phantom{}_2F_1}
\newcommand{\cji}     {({j}\ast{i})}
\newcommand{\cki}     {({k}\ast{i})}

\newenvironment{respr}[0]{\sloppy\begin{flushleft}\hspace*{0.75cm}\(}{\)\end{flushleft}\fussy}

\newcommand{\brp}{\begin{respr}}
\newcommand{\erp}{\end{respr}}

\title{
Analytic integration of real-virtual counterterms
in NNLO jet cross sections I}

\author{Ugo Aglietti\\
Dipartimento di Fisica,
Universit\`a di Roma ``La Sapienza'',\\
and INFN, Sezione di Roma, Italy\\
E-mail: \email{Ugo.Aglietti@roma1.infn.it}}

\author{Vittorio Del Duca\footnote{On leave of absence from INFN, Sezione di Torino.}\\
INFN, Laboratori Nazionali di Frascati,\\
Via E. Fermi 40, I-00044Frascati, Italy\\
E-mail: \email{Vittorio.DelDuca@lnf.infn.it}}

\author{Claude Duhr\\
Institut de Physique Th\'eorique and
Centre for Particle Physics and Phenomenology (CP3)\\
Universit\'e Catholique de Louvain\\
Chemin du Cyclotron 2,
B-1348 Louvain-la-Neuve, Belgium\\
E-mail: \email{claude.duhr@uclouvain.be}}

\author{G\'abor Somogyi \\
Institute for Theoretical Physics, University of
Z\"urich\\ Winterthurerstrasse 190, CH-8057 Z\"urich, Switzerland\\
E-mail: \email{sgabi@physik.unizh.ch}}

\author{Zolt\'an Tr\'ocs\'anyi\\
University of Debrecen and Institute of Nuclear Research of the 
Hungarian Academy of Sciences, H-4001 Debrecen P.O.Box 51, Hungary\\
E-mail: \email{Zoltan.Trocsanyi@cern.ch}}

\abstract{
We present analytic evaluations of some integrals needed to give
explicitly the integrated
real-virtual integrated counterterms, based on a recently
proposed subtraction scheme for next-to-next-to-leading order
(NNLO) jet cross sections.
After an algebraic reduction of the integrals, integration-by-parts
identities are used for the reduction to master integrals and for
the computation of the master integrals themselves by means of
differential equations. The results are written in terms of
one- and two-dimensional harmonic polylogarithms, once an extension
of the standard basis is made.
We expect that the techniques described here will be useful in computing
other integrals emerging in calculations in perturbative quantum field
theories.
}

\preprint{arXiv:yymm.nnnn [hep-ph]\\
CP3-08-21\\
ZU-TH 10/08}

\begin{document}

%%%
%%% Introduction
%%%

\section{Introduction}
\label{sec:intro}

LHC physics demands calculating physical observables beyond leading
order (LO) accuracy, by including the virtual and real corrections that
appear at higher orders.  However, the evaluation of phase space
integrals beyond LO is not straightforward because it involves infrared
singularities that have to be consistently treated before any numerical
computation may be performed.  At next-to-leading order (NLO), infrared
divergences can be handled using a \emph{subtraction scheme} exploiting
the fact that the structure of the kinematical singularities of QCD
matrix elements is universal and independent of the hard process.  This
allows us to build process-independent counterterms which regularize
the one-loop (or virtual) corrections and real phase space integrals
simultaneously~\cite{Catani:1996vz}.

In recent years a lot of effort has been devoted to the extension of
the subtraction method to the computation of the radiative corrections
at the next-to-next-to-leading order
(NNLO)~\cite{GehrmannDeRidder:2004tv, GehrmannDeRidder:2004xe, %
Weinzierl:2003fx, Weinzierl:2003ra, Frixione:2004is, %
Gehrmann-DeRidder:2005hi, GehrmannDeRidder:2005aw, %
Gehrmann-DeRidder:2005cm, Weinzierl:2006ij, Somogyi:2005xz}.
In particular, in Ref.~\cite{Somogyi:2006da, Somogyi:2006db}, a
subtraction scheme was defined for computing NNLO corrections to QCD
jet cross sections to processes without coloured partons in the initial
state and an arbitrary number of massless particles (coloured or
colourless) in the final state.  This scheme however is of practical
utility only after the universal counterterms for the regularization of
the real emissions are integrated over the phase space of the unresolved
particles.  The integrated counterterms can be computed once and for all
and their knowledge is necessary to regularize the infrared divergences
appearing in virtual corrections.  That is indeed the task of this work:
we analytically evaluate some of the integrals needed for giving
explicitly the counterterms appearing in the
scheme~\cite{Somogyi:2006da, Somogyi:2006db}.  The method is an
adaptation of a technique developed in the last two decades to
compute multi-loop Feynman diagrams~\cite{Kotikov:1990kg, Kotikov:1991hm,
Kotikov:1991pm, Caffo:1998du, Caffo:1998yd, Remiddi:1997ny}.
To our knowledge this is the first time that these techniques are applied
to integrals of the type
\begin{equation}
F(z) \, = \,
\int_0^{1} \int_0^{\alpha_0} \rd x \, \rd y\,
{ x^{k_1 \ep}  \, (1-x)^{k_2 \ep} \, y^{k_3 \ep} \, (1-y)^{k_4 \ep}
\, (1 - x y z)^{k_5 \ep} }
\, f(x,y,z)\,,
\label{example}
\end{equation}
where
\begin{equation}
f(x,y,z) \, = \,
\frac{1}{ x^{n_1} } \, \frac{1}{(1-x)^{n_2}} \, \frac{1}{y^{n_3}} \,
\frac{1}{(1-y)^{n_4}} \, \frac{1}{(1-x y z)^{n_5}}\,,
\label{examplef}
\end{equation}
with $n_i$ being non-negative integers and $0<\alpha_0 \leq 1$.

An alternative method for computing the $\ep$-expansion of the
integrals is iterated sector decomposition. This approach allows one to
express the expansion coefficients of all functions we consider as
finite, multidimensional integrals. Integrating these representations
numerically, we obtain the expansion coefficients for any fixed value
of the arguments.  Every integral in this paper was computed
numerically as well, with this alternative method for selected values of
the parameters. We found that in all cases the analytical and numerical
results agreed up to the uncertainty associated with the numerical
integration.

The outline of the paper is the following.
In \sect{sec:Method} we outline the steps of our method.
In \sect{sec:Subtractions} we define the integrals of the subtraction
terms
that we will consider in the paper.
Our analytic results will be presented in terms of one- and
two-dimensional harmonic polylogarithms. We summarize those properties
of these functions that are important for our computations in
\sects{sec:1dhpls}{sec:2dhpls}, respectively. %\cite{Remiddi:1999ew}.
In \sects{sec:soft01}{sec:softcol01} we calculate analytically the
integrals needed for integrating the soft-type counterterms as a series
expansion in the dimensional regularization parameter $\ep$.
In \sect{sec:coll01} we calculate some of the integrlas needed for
integrating the collinear counterterms.
In \sects{sec:softR0}{sec:softcollR0} we calculate two sets of
convoluted integrated counterterms, which can be obtained from a
successive integration of the results obtained in \sect{sec:coll01}.
\sect{sec:numerical} briefly discusses the numerical calculation of the
integrated subtraction terms and the merits of both the analytical and
the numerical approaches.
Finally in \sect{conclusion} we present the conclusions of this work and
we
discuss possible developments concerning more complicated classes of
integrals.
\appx{app:APfcns} contains the spin-averaged splitting function at tree
level and at one-loop, which are needed for the evaluation of the
counterterms.
There are further appendices containing the (often rather lengthy)
expressions of the integrated counterterms.

\section{The method}
\label{sec:Method}

Our method of computing the integrals involves the following steps:

\paragraph{Algebraic reduction of the integrand by means of partial
fractioning.}
For each class of integrals, we perform a partial fractioning of the
integrand in order to obtain a set of independent integrals. For
example, for the integrand in \eqn{examplef} with $n_1=n_2=n_3=n_4=n_5=1$ 
one can perform partial
fractioning with respect to the integration variable $x$ first, so that 
\begin{equation}
\frac{1}{ x \, (1-x) \, (1 - x y z) }
\, = \,
\frac{1}{x} \, + \, \frac{1}{ 1 - y z} \, \frac{1}{1-x} \, - \, \frac{y^2
z^2}{1- y z} \, \frac{1}{ 1 - x y z } \, .
\end{equation}
Note the appearance of the new denominator $1 - y z$, not originally
present in the integrand and coming from $x$ partial fractioning\footnote{
By increasing the number of variables, the number of additional
denominators grows very fast.
}.
One then performs partial fractioning with respect to $y$, by
considering the denominator $1 - x y z$ as a constant: that is because
the latter was already involved in the $x$ partial fractioning and, to
avoid an infinite loop, it cannot be subjected to any further
transformation. For example:
\begin{eqnarray}
\frac{1}{ y \, (1 - y) \, (1 - y z) \, (1 - x y z) }
&=&
- \, \frac{ z^2 }{ 1 - z } \, \frac{1}{ (1 - y z) \, (1-x y z) }
\, + \, \frac{1}{y \, (1 - x y z)} \, +
\nonumber\\
&& + \, \frac{1}{ 1 - z } \, \frac{1}{(1-y) \, (1-x y z)} \, .
\end{eqnarray}
After this final partial fractioning over $y$, the original integrand
$f$, depending on five denominators, is transformed into a combination
of terms having at most two denominators, out of which at most one
depends on $x$.%
\footnote{
Performing first the partial fractioning in $y$ and then in $x$ results
in a different basis of independent amplitudes.
}

\paragraph{Reduction to master integrals by means of integration-by-parts
identities.}
We then write integration-by-parts identities (ibps) for the chosen set
of independent amplitudes.
If the upper limits in the $x$ or $y$ integrals in \eqn{example} differ
from one, $\alpha_0 < 1$, surface terms have to be taken into account.
That is to be contrasted with the case of loop calculations, in which
surface terms
always vanish.
By solving the ibps with the standard Laporta algorithm, complete
reduction to master integrals
is accomplished. 

\paragraph{Analytic evaluation of the Master Integrals.}
After having identified for each class of integrals a set of master
integrals,
we write the corresponding system of differential equations.
The $\ep$-expansion of the master integrals is obtained by solving
such systems expanded in powers of $\ep$.
A natural basis consists of one- and two-dimensional harmonic
polylogarithms~\cite{Remiddi:1999ew,Gehrmann:2000zt}; for representing
some master integrals,
an extension of the standard basis functions has proved to be necessary.

%%%
%%% Integrals of the subtraction terms
%%%

\section{Integrals needed for the integrated subtraction terms}
\label{sec:Subtractions}

The subtraction method developed in \Refs{Somogyi:2006da,Somogyi:2006db}
relies on the universal soft and collinear factorization properties of
QCD squared matrix elements. Although the necessary factorization
formulae for NNLO computations have been known for almost a decade, the
explicit definition of a subtraction scheme has been hampered for
several reasons. Firstly, the various factorization formulae overlap in
a rather complicated way beyond NLO accuracy and these overlaps have to
be disentangled in order to avoid multiple subtractions.  
At NNLO accuracy this was first achieved in \Ref{Somogyi:2005xz}.
A general and simple solution to this problem was subsequently given in 
\Ref{Nagy:2007mn}, where a method was described to obtain pure-soft 
factorization at any order in perturbation theory leading to soft-singular 
factors without collinear singularities.

Secondly, the factorization formulae are valid only in the strict soft
and collinear limits and have to be extended to the whole phase space. A
method that works at any order in perturbation theory requires a
mapping of the original $n$ momenta
$\mom{}_n = \{p_1,\dots, p_n\}$ to $m$ momenta
$\momt{}_m = \{\ti p_1,\dots, \ti p_m\}$
($m$ is the number of hard partons and $n-m$ is the number of unresolved 
ones)
that preserves momentum conservation.  Such a mapping leads to an exact
factorization of the original $n$-particle phase space of total
momentum $Q$,
\beq
\PS{n}(p_1,\ldots,p_n;Q) =
\prod_{i=1}^{n}\frac{\rd^d p_i}{(2\pi)^{d-1}}\,\delta_+(p_i^2)\,
(2\pi)^d \delta^{(d)}\left(Q-\sum_{i=1}^{n}p_i\right)\,,
\label{eq:PSn}
\eeq
in the form
\beq
\PS{n}(\mom{}_n;Q)=\PS{m}(\momt{}_{m};Q)
\: [\rd p_{n-m;m}(\mom{}_{n-m};Q)]\,.
\label{eq:PSfact}
\eeq
In the context of computing QCD corrections, this sort of exact
phase-space factorization was first introduced in \Ref{Catani:1996vz},
where only three of the original momenta $\mom{}$ -- that of the emitter
$p_i^\mu$, the spectator $p_k^\mu$ and the emitted particle $p_j^\mu$
-- were mapped to two momenta, $\ti p_{ij}^\mu$ and $\ti p_k^\mu$, the rest
of the phase space was left unchanged. This sort of mapping requires
that both $i$ and $k$ be hard partons, which is always satisfied
in a computation at NLO accuracy because only one parton is unresolved.
However, in a computation beyond NLO the spectator momentum may also
become unresolved unless this is explicitly avoided by using colour-ordered
subamplitudes \cite{Gehrmann-DeRidder:2005hi,GehrmannDeRidder:2005aw}. In order
to take into account the colour degrees of freedom explicitly, as well
as define a phase space mapping valid at any order in perturbation theory,
in \Ref{Somogyi:2006cz}, two types of `democratic' phase-space mappings
were introduced. In this paper we are concerned with the integrals of the
singly-unresolved counterterms, therefore, in the rest of the paper we
deal with the case when $m=1$. Symbolically, the mapping
\beq
\mom{}_n\cmap{ir}\momt{(ir)}_{n-1} =
\{\ti{p}_1,\ldots,\ti{p}_{ir},\ldots,\ti{p}_n\}\,,
\label{eq:cmap}
\eeq
used for collinear subtractions, denotes a mapping where the momenta
$p_i^\mu$ and $p_r^\mu$ are replaced by a single momentum
$\ti{p}_{ir}^\mu$ and all other momenta are rescaled, while
for soft-type subtractions,
\beq
\mom{}_n\smap{r}\momt{(r)}_{n-1} = \{\ti{p}_1,\ldots,\ti{p}_n\}
\label{eq:smap}
\eeq
denotes a mapping such that the momentum $p_r^\mu$, that may become
soft, is missing from the set, and all other momenta are rescaled and
transformed by a proper Lorentz transformation.  These mappings are
defined such that the recoil due to the emission of the unresolved
partons is taken by all hard partons.  In both cases the
factorized phase-space measure can be written in the form of a
convolution.

%
% Definition of the collinear integrals
%

\subsection{Definition of the collinear integrals}
\label{sec:collintegrals}

\begin{figure}
\begin{center}
\includegraphics{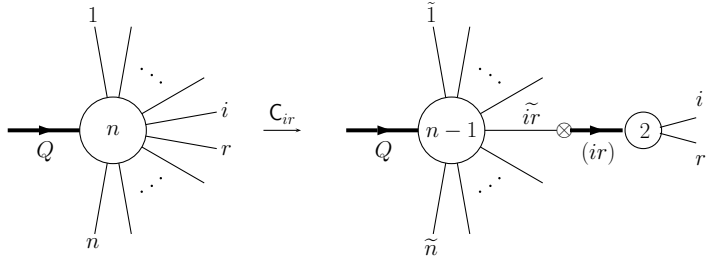}
\end{center}
\caption{Graphical representation of the collinear momentum mapping
and the implied phase space factorization.}
\label{fig:PSCir}
\end{figure}

In the case of collinear mapping the factorized phase-space measure can
be written as
\beq
[\rd p_{1;n-1}^{(ir)}(p_{r},\ti{p}_{ir};Q)] =
\int_{0}^{1}\!\rd\alpha\,(1-\alpha)^{2(n-2)(1-\ep)-1}
\,\frac{s_{\wti{ir}Q}}{2\pi}
\,\PS{2}(p_i,p_r; p_{(ir)})
\,,
\label{eq:dp1Cir}
\eeq
where $s_{\wti{ir}Q} = 2 \ti{p}_{ir} \ldot Q$ and 
$p_{(ir)}^\mu = (1-\alpha) \ti{p}_{ir}^\mu + \alpha Q^\mu$.
The collinear momentum mapping and the implied factorization of the
phase-space measure are represented graphically in \fig{fig:PSCir}.
The picture on the left shows the $n$-particle phase space
$\PS{n}(\mom{};Q)$, where in the circle we have indicated the
number of momenta. The picture on the right corresponds to
\eqn{eq:PSfact} (with $m = 1$) and \eqn{eq:dp1Cir}:
the two circles represent the $(n-1)$-particle phase space
$\PS{n-1}(\momt{(ir)};Q)$ and the two-particle phase
space $\PS{2}(p_i,p_r;p_{(ir)})$ respectively, while the symbol $\otimes$
stands for the convolution over $\alpha$, as precisely defined in
\eqn{eq:dp1Cir}.

Writing the factorized phase space in the form of \eqn{eq:dp1Cir} has
some advantages:
\begin{itemize}
\itemsep -2pt
\item It makes the symmetry property of the factorized phase space
under the permutation of the factorized momenta manifest. For instance,
for any function $f$,
\beq
\int [\rd p_{1;n-1}^{(ir)}(p_r,\ti{p}_{ir};Q)]\, f(p_i,p_r) =
\int [\rd p_{1;n-1}^{(ir)}(p_r,\ti{p}_{ir};Q)]\, f(p_r,p_i)
\,,
\label{eq:PS2identity}
\eeq
which can be used to reduce the number of independent integrals.
\item It exhibits the $n$-dependence of the factorized phase space
explicitly. This allows for including $n$-dependent factors of
$(1-\alpha)^{2d_0-2(n-2)(1-\ep)}\Theta(\alpha_0-\alpha)$ 
(with $d_0\big|_{\ep=0} \geq 2$)
in the subtraction terms such that the integrated counterterms will be
$n$-independent (for details see \Ref{Somogyi:2008XxYy}). 
\item \eqn{eq:dp1Cir} generalizes very straightforwardly for more
complicated factorizations. (The formula for the general case when
phase-spaces of $N$ groups of $r_1,r_2,\ldots,r_N$ partons are
factorized simultaneously can be given explicitly.)
\end{itemize}

To write the factorized two-particle phase-space measure we
introduce the variable $v$,
\beq
v = \frac{z_{r}-z_r^{(-)}}{z_r^{(+)}-z_r^{(-)}}\,.
\label{eq:defvri}
\eeq
In \eqn{eq:defvri} $z_{r}$ is the momentum fraction of parton $r$ in
the Altarelli--Parisi splitting function that describes the $f_{(ir)} \to
f_i + f_r$ collinear splittings ($f$ denotes the flavour of the partons).
This momentum fraction takes values between
\beq
z_r^{(-)} =
\frac{\alpha}{2\alpha+x-\alpha x}
\label{eq:zr+zr-}
\eeq
and $z_r^{(+)} = 1 - z_r^{(-)}$ ($x = s_{\wti{ir}Q}/Q^2$).
Using the variables $s_{ir}=2p_i\ldot p_r$, and $v$ the
two-particle phase-space measure reads
\beeq
\PS{2}(p_i,p_r; p_{(ir)}) &&=
\frac{s_{ir}^{-\ep}}{8\pi}\,S_\ep
\,\rd s_{ir}\,\rd v
\,\delta\!\left(s_{ir}-Q^2 \alpha\big(\alpha
+(1-\alpha) x\big)\right)
\nn\\[2mm]
&&\times
\,[v\,(1-v)]^{-\ep}
\,\Theta(1 - v)\Theta(v)\,,
\label{eq:dPS2}
\eeeq
where
\beq
S_\ep = \frac{(4\pi)^\ep}{\Gamma(1-\ep)}
\,.
\eeq
The integration of the collinear subtractions over the unresolved phase
space involves the integrals \cite{Somogyi:2008XxYy}
\beq
\frac{(4\pi)^2}{S_\ep}(Q^2)^{(1+\kappa)\ep}
\int_{0}^{\alpha_{0}}\!\rd\alpha\, (1-\alpha)^{2d_{0}-1}
\frac{s_{\wti{ir}Q}}{2\pi}\PS{2}(p_{i},p_{r};p_{(ir)})
\frac{1}{s_{ir}^{1+\kappa \ep}}P_{f_i f_r}^{(\kappa)}(z_i,z_r;\ep)
\,,\qquad \kappa=0,\,1\,,
\label{eq:collintegrals}
\eeq
where $\alpha_0 \in (0,1]$ while $P_{f_i f_r}^{(0)}$ and $P_{f_i f_r}^{(1)}$ 
denote the average of
the tree-level and one-loop splitting kernels over the spin states of
the parent parton (Altarelli--Parisi splitting functions),
respectively.  These spin-averaged splitting kernels depend, in
general, on $z_i$ and $z_r$, with the constraint
\beq
z_i + z_r = 1\,,
\label{eq:sumzizr}
\eeq
and are listed in \appx{app:APfcns}. Inspecting the actual form of the
Altarelli--Parisi splitting functions and using the symmetry property
of the factorized phase space under the interchange $i\leftrightarrow r$,
we find that (\ref{eq:collintegrals}) can be expressed as a linear
combination of the integrals
\beq
\frac{(4\pi)^2}{S_\ep}(Q^2)^{(1+\kappa)\ep}
\int_0^{\alpha_0}\!\rd \alpha_{ir}\,(1-\alpha_{ir})^{2d_0-1}
\,\frac{s_{\wti{ir}Q}}{2\pi}
\int\!\PS{2}(p_i,p_r; p_{(ir)})
\, \frac{ z_r^{k+\delta\ep}}{s_{ir}^{1+\kappa\ep}}
\, g_I^{(\pm)}(z_r)
\,,
\label{eq:I1k}
\eeq
for $k=-1,0,1,2$, $\kappa = 0,1$ and the values of $\delta$ and
functions $g_I^{(\pm)}$ as given in \tab{tab:I1ints}.
%%%%
%T
%%%%

Using \eqnss{eq:defvri}{eq:dPS2} and $z_{r}$ expressed with $v$,
\beq
z_{r} = \frac{\alpha+(1-\alpha)x v}
{2\alpha+(1-\alpha)x}\,,
\label{eq:yirzr_with_airvri}
\eeq
we can see that the integrals in \eqn{eq:I1k} take the form
\beeq
&&
\cI(x;\ep,\alpha_0,d_0;\kappa,k,\delta,g_I^{(\pm)}) =
x
\int_0^{\alpha_0}\! \rd \alpha\, \alpha^{-1-(1+\kappa)\ep}
\,(1-\alpha)^{2d_0-1}\,[\alpha+(1-\alpha)x]^{-1-(1+\kappa)\ep}
\nn\\[2mm]&&\qquad%\qquad\qquad
\times
\int_0^1\! \rd v
[v\,(1-v)]^{-\ep}
\left(\frac{\alpha+(1-\alpha)xv}{2\alpha+(1-\alpha)x}\right)^{k+\delta\ep}
\,g_I^{(\pm)}\left(\frac{\alpha+(1-\alpha)xv}{2\alpha+(1-\alpha)x}\right)
\,.
\label{eq:Iint}
\eeeq
We compute the integrals corresponding to the first two rows of
\tab{tab:I1ints} in \sect{sec:coll01}.

\TABLE{
\begin{tabular}{|c|c|c|}
\hline
\hline
$\delta$ & Function & $g_I^{(\pm)}(z)$ \\
\hline
 & & \\[-4mm]
$0$ & $g_A$ & $1$ \\[2mm]
$\mp 1$ & $g_B^{(\pm)}$ & $(1-z)^{\pm\ep}$ \\[2mm]
$0$ & $g_C^{(\pm)}$ &
$(1-z)^{\pm\ep}{}_2F_1(\pm\ep,\pm\ep,1\pm\ep,z)$\\[2mm]
$\pm 1$ & $g_D^{(\pm)}$ & ${}_2F_1(\pm\ep,\pm\ep,1\pm\ep,1-z)$ \\[2mm]
\hline
\hline
\end{tabular}
\label{tab:I1ints}
\caption{The values of $\delta$ and $g_I^{(\pm)}(z_r)$ at which
\eqn{eq:I1k} needs to be evaluated.}
}

%
% Definition of the soft-type integrals
%

\subsection{Definition of the soft-type integrals}
\label{sec:softintegrals}

\begin{figure}
\begin{center}
\includegraphics{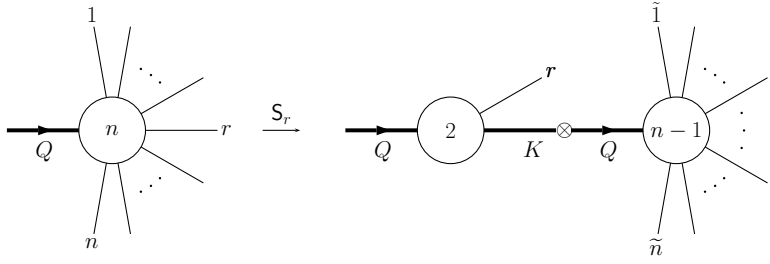}
\end{center}
\caption{Graphical representation of the soft momentum mapping
and the implied phase space factorization.}
\label{fig:PSSr}
\end{figure}

In the case of soft mapping the factorized phase-space measure can
be written as
\beq
\bsp &
[\rd p_{1;n-1}^{(r)}(p_r;Q)] =
\\ &
\int_{0}^{1}\rd y(1-y)^{(n-2)(1-\ep)-1}
\frac{Q^{2}}{2\pi}\PS{2}(p_{r},K;Q)
\label{eq:dp1Sr}
\esp
\eeq
where the timelike momentum $K$ is massive with $K^{2}=(1-y)Q^{2}$.
We show the soft momentum mapping and the implied phase space
factorization in \fig{fig:PSSr}.
The picture on the left shows again the $n$-particle phase space
$\PS{n}(\mom{};Q)$, while the picture on the right corresponds to
\eqn{eq:PSfact} (with $m = 1$) and \eqn{eq:dp1Sr}:
the two circles represent the two-particle phase
space $\PS{2}(p_r,K;Q)$ and the $(n-1)$-particle phase space
$\PS{n-1}(\momt{(r)};Q)$ respectively. The symbol $\otimes$
stands for the convolution over $y$ as defined in \eqn{eq:dp1Sr}.

The soft and soft-collinear subtraction terms involve the integral of the
eikonal factor and its collinear limit over the factorized phase space of
\eqn{eq:dp1Sr} \cite{Somogyi:2008XxYy}, namely the integrals
\beeq
-\frac{(4\pi)^2}{S_\ep}(Q^2)^{(1+\kappa)\ep}
\int_{0}^{y_{0}}\!\rd y\,(1-y)^{d'_{0}-1}
\frac{Q^{2}}{2\pi}\PS{2}(p_{r},K;Q)
\left(\frac{s_{ik}}{s_{ir}s_{kr}}\right)^{1+\kappa\ep}\,,
\qquad \kappa=0,1\,,
\label{eq:ISik0}
\\[2mm]
\frac{(4\pi)^2}{S_\ep}(Q^2)^{(1+\kappa)\ep}
\int_{0}^{y_{0}}\!\rd y\,(1-y)^{d'_{0}-1}
\frac{Q^{2}}{2\pi}\PS{2}(p_{r},K;Q)
2 \left(\frac{1}{s_{ir}}\frac{z_i}{z_r}\right)^{1+\kappa\ep}\,,
\qquad \kappa=0,1\,.
\label{eq:ICS0}
\eeeq
Here again, we included harmless factors of 
$(1-y)^{d'_0-(n-2)(1-\ep)}\Theta(y_0-y)$ 
(with $d'_0|_{\ep = 0} \geq 2$) in
the subtraction terms to make their integrals independent of $n$.
The computation of these integrals is fairly straightforward using energy
and angle variables.

In order to write the factorized phase-space measure, we choose a frame
in which
\beq
Q^\mu = \sqrt{s}(1,\ldots)\,,
\qquad
\ti{p}_i^\mu = \ti{E}_i(1,\ldots,1)\,,
\qquad
\ti{p}_k^\mu = \ti{E}_k(1,\ldots,\sin\chi,\cos\chi)\,,
\label{eq:frame_1}
\eeq
and
\beq
p_r^\mu = 
E_r(1,\mbox{..`angles'..},\sin\vartheta\sin\varphi,\sin\vartheta\cos\varphi,\cos\vartheta)\,.
\label{eq:frame_2}
\eeq
In \eqn{eq:frame_1} the dots stand for vanishing components, while the
notation `angles' in \eqn{eq:frame_2} denotes the dependence of $p_r$ 
on the $d-3$ angular variables that can be trivially integrated.
Then in terms of the scaled energy-like variable
\beq
\eps_r = \frac{2p_r\ldot Q}{Q^2} = \frac{2E_r}{\sqrt{s}}
\label{eq:yrQ}
\eeq
and the angular variables $\vartheta$ and $\varphi$ the two-particle
phase space reads
\beq
\bsp
\PS{2}(p_{r},K;Q) &=
\frac{(Q^{2})^{-\ep}}{16\pi^{2}}S_{\ep}
\frac{\Gamma^{2}(1-\ep)}{\Gamma(1-2\ep)}
\,\rd\eps_r\,\eps_r^{1-2\ep}\delta(y-\eps_r)\\
&\times
\rd(\cos\vartheta)\,\rd(\cos\varphi)
(\sin\vartheta)^{-2\ep}(\sin\varphi)^{-1-2\ep}
\,,
\label{eq:PS2Sr}
\esp
\eeq
where $y \in (0,1]$ and the cosines of both angles run from $-1$ to
$+1$.

To write the integrands in these variables, we observe that the precise
definitions of $\ti{p}_i$ and $\ti{p}_k$ as given in
\Ref{Somogyi:2006da} imply 
\beq
s_{ik} = (1-\eps_r) s_{\ti{i}\ti{k}}\,,
\qquad
s_{ir} = s_{\ti{i}r}\,,
\qquad
s_{kr} = s_{\ti{k}r}\,,
\label{eq:siktilde}
\eeq
and
\beq
s_{iQ} = (1-\eps_r) s_{\ti{i}Q} + s_{\ti{i}r}\,.
\label{eq:piQtilde}
\eeq
{}From Eqs.~(\ref{eq:frame_1}), (\ref{eq:frame_2}), (\ref{eq:siktilde})
and (\ref{eq:piQtilde}) we find
\beq
\frac{s_{ik}}{s_{ir}s_{kr}} =
(1-\eps_r) \frac{s_{\ti{i}\ti{k}}}{s_{\ti{i}r}s_{\ti{k}r}} =
\frac{4 \Y{i}{k}}{Q^2}\frac{(1-\eps_r)}{\eps_r^2}
\frac{1}{(1-\cos\vartheta)(1-\cos\chi\cos\vartheta-\sin\chi\sin\vartheta\cos\varphi)}
\,,
\label{eq:Sikrtilde}
\eeq
and
\beq
\frac{1}{s_{ir}}\frac{z_i}{z_r} =
\frac{1}{s_{\ti{i}r}}\frac{(1-\eps_r) s_{\ti{i}Q}+s_{\ti{i}r}}{s_{rQ}} =
\frac{1}{Q^2}\frac{1}{\eps_r}
\left[1+\frac{2(1-\eps_r)}{\eps_r(1-\cos\vartheta)}\right]\,.
\label{eq:csinttilde}
\eeq

Using Eqs.~(\ref{eq:PS2Sr}), (\ref{eq:Sikrtilde}) and
(\ref{eq:csinttilde}) we see that the integral of the soft subtraction
term in \eqn{eq:ISik0} may be written as
\beq
\bsp
\cJ(\Y{i}{k};\ep,y_0,d'_0;\kappa) &=
-(4Y_{\ti{i}\ti{k},Q})^{1+\kappa\ep}
\frac{\Gamma^2(1-\ep)}{2\pi\Gamma(1-2\ep)}
\Omega^{(1+\kappa\ep,1+\kappa\ep)}(\cos\chi)\\
&\times
\int_0^{y_0}\!\rd y\,
y^{-1-2(1+\kappa)\ep}(1-y)^{d'_0+\kappa\ep}\,,
\esp
\label{eq:Jint}
\eeq
where $\Omega^{(i,k)}(\cos\chi)$ denotes the angular integral
\beq
\bsp
\Omega^{(i,k)}(\cos\chi) &=
\int_{-1}^1\!\rd(\cos\vartheta)\;(\sin\vartheta)^{-2\ep}
\int_{-1}^1\!\rd(\cos\varphi)\;(\sin\varphi)^{-1-2\ep}\\
&\times
(1-\cos\vartheta)^{-i}
(1-\cos\chi\cos\vartheta-\sin\chi\sin\vartheta\cos\varphi)^{-k}
\,.
\esp
\label{eq:Omegajl}
\eeq
Furthermore, from \eqn{eq:frame_1} it is easy to see that
\beq
\cos\chi = 1-2\,\Y{i}{k}
\equiv 1 - \frac{2 Q^2 s_{\ti{i}\ti{k}}}{s_{\ti{i}Q}s_{\ti{k}Q}}
\,.
\label{eq:coschi}
\eeq
We compute the soft integrals $\cJ(X,\ep;y_0,d'_0;\kappa)$ in
\sect{sec:soft01}.

The soft-collinear subtraction term in \eqn{eq:ICS0} leads to the
integral
\beq
\bsp
\cK(\ep,y_0,d'_0;\kappa) &=
\,2\int_0^{y_0}\! \rd y\,y^{-(2+\kappa)\ep}
(1-y)^{d'_0-1}\int_{-1}^1\! \rd (\cos\vartheta)\,(\sin\vartheta)^{-2\ep}
\label{eq:Kint}
\\ &\times
\left[1+\frac{2(1-y)}{y(1-\cos\vartheta)}\right]^{1+\kappa\ep}
\frac{\Gamma^2(1-\ep)}{2\pi\Gamma(1-2\ep)}
\int_{-1}^{1}\! \rd (\cos\varphi)\,(\sin\varphi)^{-1-2\ep}
\,,
\esp
\eeq
which we compute in \sect{sec:softcol01}.

%
% Iterated integrals
%

\subsection{Iterated integrals}
\label{sec:iterated}

In an NNLO computation, iterations of the above integrals also appear.
In this paper we compute also two of those. The first one is the
integration of a soft integral with a collinear one in its argument,
\beq
\bsp
\cJI(\Y{i}{k};\ep,\alpha_{0},d_{0},y_{0},d'_{0};k) &=
- 4\Y{i}{k}
\,\frac{\Gamma^{2}(1-\ep)}{2\pi\Gamma(1-2\ep)}\,\Omega^{(1,1)}(\cos\chi)
\\[1mm] &\times
\int_{0}^{y_{0}}\rd y\,
y^{-1-2\ep}(1-y)^{d'_{0}}
\,\cI(y;\ep,\alpha_0,d_0;0,k,0,1)
\,,
\esp
\label{eq:JIint}
\eeq
which we need for $k = -1,0,1,2$. Details of the computation are given in
\sect{sec:softR0}. The second case is when the collinear integral appears in
the argument of a soft-collinear one,
\beq
\bsp
\cKI(\ep,\alpha_{0},d_{0},y_{0},d'_{0};k) &=
2\,\frac{\Gamma^{2}(1-\ep)}{2\pi\Gamma(1-2\ep)}
\,\int_{-1}^{1}\rd(\cos\vartheta)\,(\sin\vartheta)^{-2\ep}
\\[1mm]
&\times
\int_{-1}^{1}\rd(\cos\varphi)\,(\sin\varphi)^{-1-2\ep}
\,\int_{0}^{y_{0}}\rd y\,y^{-1-2\ep}\,(1-y)^{d'_{0}-1}
\\[1mm]
&\qquad\times
\frac{2-y(1+\cos\vartheta)}{1-\cos\vartheta}
\,\cI(y;\ep,\alpha_0,d_0;0,k,0,1)
\,,
\label{eq:KIint}
\esp
\eeq
needed again for $k = -1,0,1,2$. Details of the computation are given in
\sect{sec:softcollR0}.

%%%
%%% HPLS
%%%

\section{One-dimensional harmonic polylogarithms}
\label{sec:1dhpls}

As anticipated in the introduction, it is convenient to represent the integrals depending on
a single variable $x$ in terms of a general class of special functions called harmonic polylogarithms
(\emph{HPL's}) introduced in \Ref{Remiddi:1999ew}.
The \hpls  of weight one, i.e. depending on one index $w=-1,0,1$, are defined as:
\beq
H(-1;x) \, \equiv \, \log (1+x) \, ; ~~~
H(0;x) \, \equiv \, \log (x) \, ; ~~~
H(1;x) \, \equiv \,  - \, \log (1-x) \, .
\eeq
These functions are then just logarithms of linear functions of $x$.
The \hpls of higher weight are defined recursively by the relation
\beq
\label{eq:hpldef}
H(a,\vec{w}; x) \, \equiv \, \int_0^x f(a;x') \, H(\vec{w}; x') \, \rd x' 
~~~ {\rm for} ~ a \ne 0 ~ {\rm and } ~ \vec{w} \, \ne \, \vec{0}_n \, ,
\eeq
i.e. in the case in which not all the indices are zero.
The left-most index takes the values $a=-1,0,1$ and $\vec{w}$ is an 
$n-$dimensional vector with components $w_i = - 1,0,1$. 
We call $n$ the weight of the \hpls, so the above relation allows one
to increase the weight $w = n \to n+1$. 
The basis  functions $f(a;x)$ are given by
\beq
\label{basfun}
f(-1;x) \, \equiv \, \frac{1}{1+x} \, ; ~~~
f(0;x) \, \equiv \, \frac{1}{x} \, ; ~~~
f(1;x) \, \equiv \,  \frac{1}{1-x} \, .
\eeq
In the  case in which all indices are zero, one defines instead,
\beq
H(\vec{0}_n; x) \, \equiv \, \frac{1}{n!} \log^n (x) \, .
\eeq
The \hpls  introduced above fulfill many interesting relations, one of the most important ones
being that of generating a `shuffle algebra',
\beq
H(\vec w_1;x) \, H(\vec w_2;x) \, = \, \sum_{\vec w=\vec w_1\uplus\vec w_2} H(\vec w;x) \, ,
\eeq
where $\vec w_1\uplus\vec w_2$ denotes the merging of the two weight vectors $\vec w_1$ and $\vec w_2$, 
\ie all possible concatenations of $\vec w_1$ and $\vec w_2$ in which relative orderings of $\vec w_1$ 
and $\vec w_2$ are preserved.

The basis of \hpls  can be extended by adding some new basis functions to the set in \eqn{basfun}; 
for our computation we have to introduce the function
\beq
f(2;x) \, \equiv \, \frac{1}{x-2} \, .
\eeq
The \hpls  can be evaluated numerically in a fast and accurate way; there are various packages available 
for this purpose ~\cite{Gehrmann:2001pz, Maitre:2005uu, Maitre:2007kp}.

%%%
%%% 2dHPLS
%%%

\section{Two-dimensional harmonic polylogarithms}
\label{sec:2dhpls}

To represent integrals depending on two arguments, an extension of the \hpls  to functions of two
variables proves to be convenient \cite{Gehrmann:2000zt}.
Since a harmonic polylogarithm is basically a repeated integration on {\it one} variable, a second
independent variable is introduced as a parameter entering the basis functions: 
$f(i; x) \, \to \, f(i,\alpha; x)$.
%Roughly speaking, in addition to the discrete index $i$, we have now a 
We may say that in addition to the discrete index $i$, we have now a 
continuous index $\alpha$.
In \Ref{Gehrmann:2000zt} the following basis functions were originally introduced:
\beq
f(c_i(\alp);x) \, = \, \frac{1}{x-c_i(\alp)} \, ,
\eeq
where
\beq
c_1(\alp) \, = \, 1 - \alp \quad \textrm{or} 
\quad 
c_2(\alp) \, = \, - \alp \, .
\eeq
Let us remark that the above extension keeps most of the properties of the
one-dimensional \emph{HPL's}.
In this work we have to introduce the following new basis functions, which are slightly
more complicated than the ones above,
\beq
f(c_1(\alpha);x) =\frac{1}{x-c_1(\alpha)} ~~~
f(c_2(\alpha);x) =\frac{1}{x-c_2(\alpha)} \, ,
\eeq
with
\beq
c_1(\alpha) \, = \, \frac{ \alpha }{ \alpha - 1} \, ,
\qquad 
c_2(\alpha)\, = \, \frac{ 2 \alpha }{ \alpha - 1} \, .
\eeq
The explicit definition of the two-dimensional harmonic polylogarithms (\dhpls) reads:
\beq
H(c_i(\alpha), \vec{w}(\alpha); x)
\, \equiv \,
\int_0^x f(c_i(\alpha); x') \,  H(\vec{w}(\alpha); x') \, \rd x' \, .
\eeq
In general, the \dhpls  have complicated analyticity properties, with imaginary parts
coming from integrating over the zeroes of the basis functions.
Our computation does not involve such complications because we can always assume
$0 \le x\,,\alp \le 1$. That implies that $c_k(\alp) < 0$ for any $k$: the denominators 
are never singular and the \dhpls  are real.
The numerical evaluation of our \dhpls  can be achieved by extending the algorithm described 
and implemented in \Ref{Gehrmann:2001jv}.

%
% Special values
%

\subsection{Special values}

For some special values of the argument, the \dhpls  reduce to ordinary one-dimensional \emph{HPL's}.
It is easy to see that for $\alpha=0$ and $\alpha=1$ we have
\beq
f(c_k(\alpha=0);x) \, = \, f(0;x) ,
\quad
\lim_{\alpha\rightarrow 1} f(c_k( \alpha);x)=0.
\eeq
From this it follows that
\beq
\bsp
H(\ldots,c_i(\alpha=0),\ldots;x) &= H(\ldots,0,\ldots;x),\\
\lim_{\alpha\rightarrow 1}H(\ldots,c_i(\alpha),\ldots;x) &= 0.
\esp
\eeq
Similarly, for $x=1$, the \dhpls  reduce to combinations of one-dimensional \hpls  in $\alp$. 
This reduction can be performed using an extension of the algorithm presented in~\cite{Gehrmann:2000zt}. 
We first write the \dhpls  in $x=1$ as the integral of the derivative with respect to $\alp$, 
\beq
H(\vec w(\alpha);1)=H(\vec w(\alpha=1);1) + \int_1^\alpha\,\rd\alpha'\,\frac{\partial}{\partial \alpha'}\,H(\vec w(\alpha');1).
\eeq
In the case where $\vec w$ only contains objects of the type $c_i$, we have $H(\vec w(\alpha=1);x)=0$. Thus,
\beq
\label{eq:interchangealgorithm}
H(\vec w(\alpha);1)=\int_1^\alpha\,\rd\alpha'\,\frac{\partial}{\partial \alpha'}\,H(\vec w(\alpha');1).
\eeq
The derivative is then carried out on the integral representation of $H(\vec w(\alpha');1)$, 
and integrating back gives the desired reduction of $ H(\vec w(\alpha);1)$ to one-dimensional \hpls  in $\alp$, \eg
\beq
\bsp
H(c_1(\alp);1) & = - \, H(0;\alp) \, ,
\\
H(c_2(\alp);1) & = H(-1;\alp)-H(0;\alp)-\ln 2.
\esp
\eeq

%
% Interchange of arguments
%

\subsection{Interchange of arguments}

The basis of \dhpls  introduced above selects $x$ as the explicit (integration) variable and $\alp$ as a parameter, 
but an alternative representation involving a repeated integration over $\alpha$ of (different) basis functions 
depending on $x$ as an external parameter is also possible.
Therefore, we have to deal with the typical problem of analytic computations: multiple representations of the
same function. It is well known that a complete analytic control requires the absence of `hidden zeroes' in the formulae.
That means that one has to know all the transformation properties (identities) of the functions introduced in order to have 
a single representative out of each class of identical objects.
In \Ref{Gehrmann:2000zt} an algorithm was presented which allows one to interchange the roles of the two variables.
The algorithm is basically the same as the one presented for the special values at $x=1$: 
let us just replace everywhere $x=1$ by $x$ in \eqn{eq:interchangealgorithm}. 
Then we have to introduce the following set of basis functions for the \dhpls,
\beq
f( d_k(x); \alpha ) \, = \, \frac{1}{ \alpha + d_k(x) } \, ,
\eeq
where
\beq
d_k(x) \, = \, \frac{x}{x - k} \, .
\eeq
All the properties defined at the beginning of this section can be easily extended to this new class of denominators.
One finds for example:
\beq
\bsp
H(c_1(\alp); x) &= H(0;x) \, - \, H(0;\alp) \, + \, H(d_1(x);\alp) \, ,
\\
H(c_2(\alp);x) &= H(0;x)\, - \, H(0;\alp)\, - \, \ln 2 \, + \, H(d_2(x); \alp) \, .
\esp
\eeq

%%%
%%% The soft Integral J
%%%

\section{The soft integral $\cJ$}
\label{sec:soft01}

In this section we present the analytic calculation of the soft
integral defined in \eqn{eq:Jint} for $\kappa = 0,1$ and $d'_0 = D'_0 +
d'_0\ep$, with $D'_0 \geq 2$ being an integer. The angular integral
$\Omega^{(i,k)}(\cos\chi)$ was evaluated in \Ref{vanNeerven:1985xr}.
The integration over $y$ leads to a hypergeometric function, and for the
complete soft integral~(\ref{eq:Jint}) we obtain the analytic expression
\beq
\bsp
\label{eq:Jinthypgeo}
\cJ(Y,\ep;y_0,d'_0;\kappa) &=
-Y^{-(1+\kappa)\ep}\,y_0^{-2(1+\kap)\ep}\,\frac{1}{(1+\kap)^2\ep^2}\frac{\Gamma^2(1-(1+\kap)\ep)}{\Gamma(1-2(1+\kap)\ep)}\\
&\times \hypgeo(-d'_0-\kap\ep, -2(1+\kap)\ep, 1-2(1+\kap)\ep,y_0)\\
&\times\hypgeo(-(1+\kap)\ep,-(1+\kap)\ep,1-\ep,1-Y)\,,
\esp
\eeq
\ie, we only need to find the $\ep$-expansion of an integral of the form
\beq
\label{eq:hypgeoint}
f(x,\ep;n_1,n_2,n_3,r_1,r_2,r_3) = \int_0^1\rd
t\,t^{-n_1-r_1\ep}\,(1-t)^{-n_2-r_2\ep}\,(1-x t)^{-n_3-r_3\ep}.
\eeq
which can be obtained using the {\sc HypExp} {\em Mathematica} package
\cite{Huber:2005yg}. Nevertheless, we compute the expansion to show our
procedure. 
The first hypergeometric function on the right hand side of
\eqn{eq:Jinthypgeo} is of the specific form $\hypgeo(a,b,1+b;x)$, whose
expansion reduces to the expansion of the incomplete beta
function $B_x$, which is a simple case to illustrate the steps of our
procedure. It involves the integrals
\beq
\bsp
\label{eq:hypgeobeta}
\beta(x,\ep;n_1,n_3,r_1,r_3) =
f(x,\ep;n_1,0,n_3,r_1,0,r_3) &=
\int_0^1\rd t\,t^{-n_1-r_1\ep}\,(1-x t)^{-n_3-r_3\ep}
\\
&= x^{-1+n_1+r_1\ep}\,B_x(1-n_1-r_1\ep,1-n_3-r_3\ep).
\esp
\eeq
The class of independent integrals can be easily obtained using partial
fractioning in $x$.  However, when writing down the
integration-by-parts identities for the independent integrals, we have
to take into account a surface term coming from the fact that the
denominator in $(1-x t')$ does not vanish for $t'=1$,
\beq
\int_0^1\,\rd t'\,\frac{\partial}{\partial
t'}\,\Big(t'^{-n_1-r_1\ep}\,(1-xt')^{-n_3-r_3\ep}\Big)=(1-x)^{-n_3-r_3\ep}.
\eeq
Solving the inhomogeneous linear system we find a single master
integral
\beq
\beta^{(1)}(x,\ep)=\beta(x,\ep;0,0,r_1,r_3),
\eeq
which fulfills the differential equation
\beq
\frac{\partial}{\partial x}\beta^{(1)}=
\frac{r_1\ep-1}{x}\,\beta^{(1)}+\frac{(1-x)^{-r_3\ep}}{x}\,,
\eeq
with initial condition
\beq
\beta^{(1)}(x=0;\ep)=\int_0^1\,\rd
t'\,t'^{-r_1\ep}=\frac{1}{1-r_1\ep}=\sum_{k=0}^\infty r_1^k\ep^k\,.
\eeq
Solving this differential equation, we obtain the expansion of the
incomplete beta function in terms of \hpls and thus the expansion of
hypergeometric functions of the form $\hypgeo(a,b,1+b;x)$.

Turning to the general case, we note that if we want to calculate the
integral~(\ref{eq:hypgeoint}) using the integration-by-parts
identities, we must require $r_1\cdot r_2\cdot r_3\neq 0$, because the
integration-by-parts identities can exhibit poles in $r_i=0$.  It is
also useful to notice that not all of the integrals are independent,
but only those where just one of the indices $n_1$, $n_2$, $n_3$ is nonzero 
and where $n_2, n_3\ge0$. In fact, all other integrals can be
reduced to one of this class using partial fractioning, \eg
\beq
f(x,\ep;1,-1,1,r_1,r_2,r_3) =
f(x,\ep;1,0,0,r_1,r_2,r_3) - (1-x) f(x,\ep;0,0,1,r_1,r_2,r_3)\,.
\eeq 

If $r_1\cdot r_2\cdot r_3\neq 0$, we can write immediately the
integration-by-parts identities for the independent integrals for $f$
obtained by partial fractioning,
\beq
\int_0^1\rd t\,\frac{\partial}{\partial
t}\left(t^{-n_1-r_1\ep}\,(1-t)^{-n_2-r_2\ep}\,(1-x t)^{-n_3-r_3\ep}\right) = 0.
\eeq
Solving the integration-by-parts identities we find that $f$ has two
master integrals,
\beq
f^{(1)}(x,\ep) = f(x,\ep;0,0,0,r_1,r_2,r_3),\quad
f^{(2)}(x,\ep) = f(x,\ep;0,0,1,r_1,r_2,r_3).
\eeq
The master integrals fulfill the following differential equations
\beq
\bsp
\frac{\partial}{\partial x}f^{(1)} & = \frac{\ep r_3}{x}f^{(2)}-\frac{\ep r_3}{x}f^{(1)},\\
\frac{\partial}{\partial x}f^{(2)} & = f^{(1)} \left(\frac{-\ep r_1-\ep
r_2-\ep r_3+1}{x}+\frac{\ep r_1+\
\ep r_2+\ep r_3-1}{x-1}\right)+\\
&f^{(2)} \left(\frac{-\ep r_2-\ep \
r_3}{x-1}+\frac{\ep r_1+\ep r_2+\ep r_3-1}{x}\right),
\esp
\eeq
with initial condition
\beq
f^{(1)}(x=0,\ep)=f^{(2)}(x=0,\ep) = B(1-r_1\ep,1-r_2\ep).
\eeq
Solving this set of linear differential equations we can write down the
$\ep$-expansion of the hypergeometric function in terms of \hpls  in $x$.

The solution for the integral $\cJ$ can be easily obtained by using the
expansion of the hypergeometric function we just obtained. The results
for $\kap=0,1$ and $D_0'=3$ can be found in \appx{app:JIntegrals}.

\begin{figure}[t]
\includegraphics[scale=0.75]{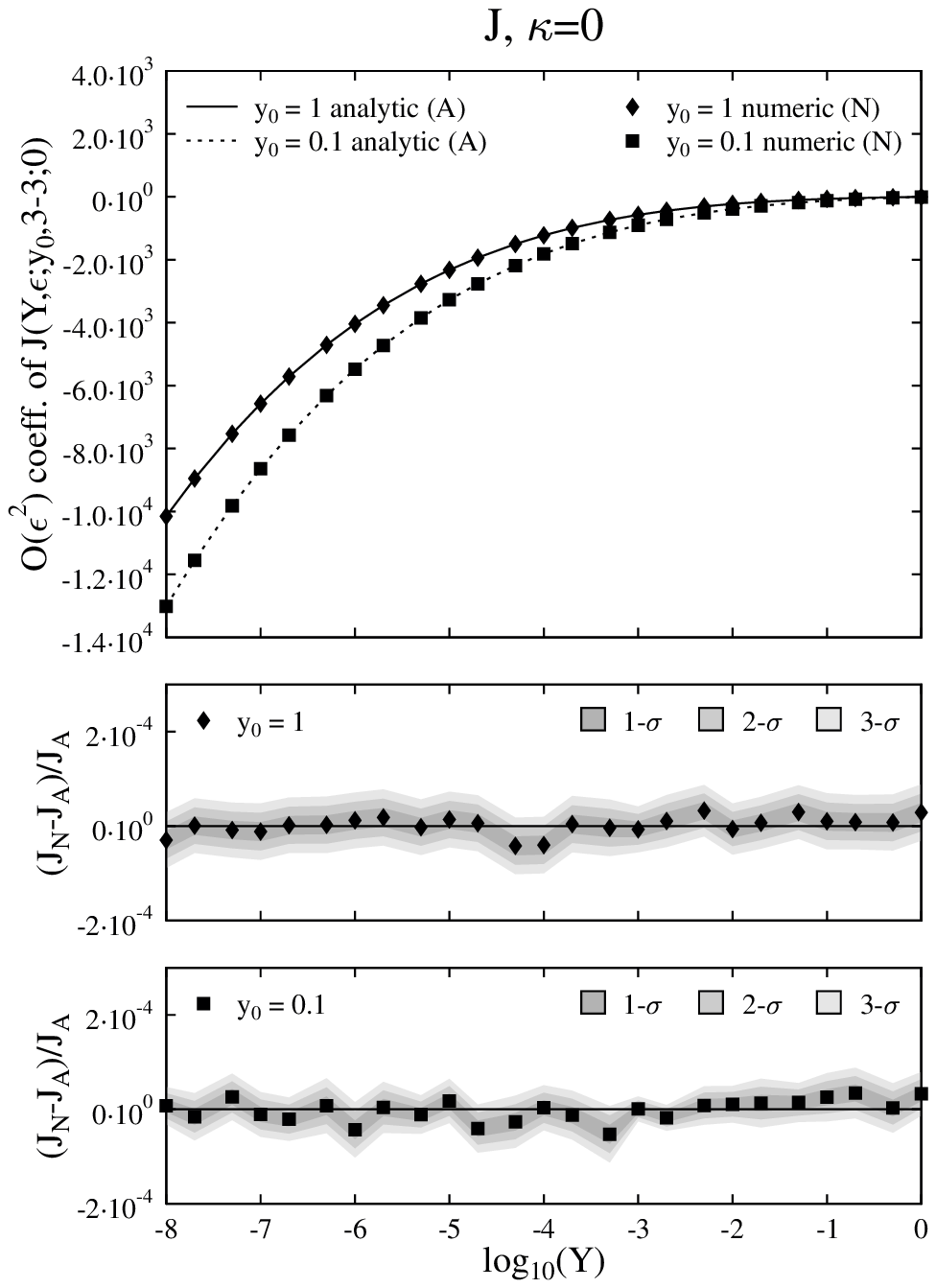}
\includegraphics[scale=0.75]{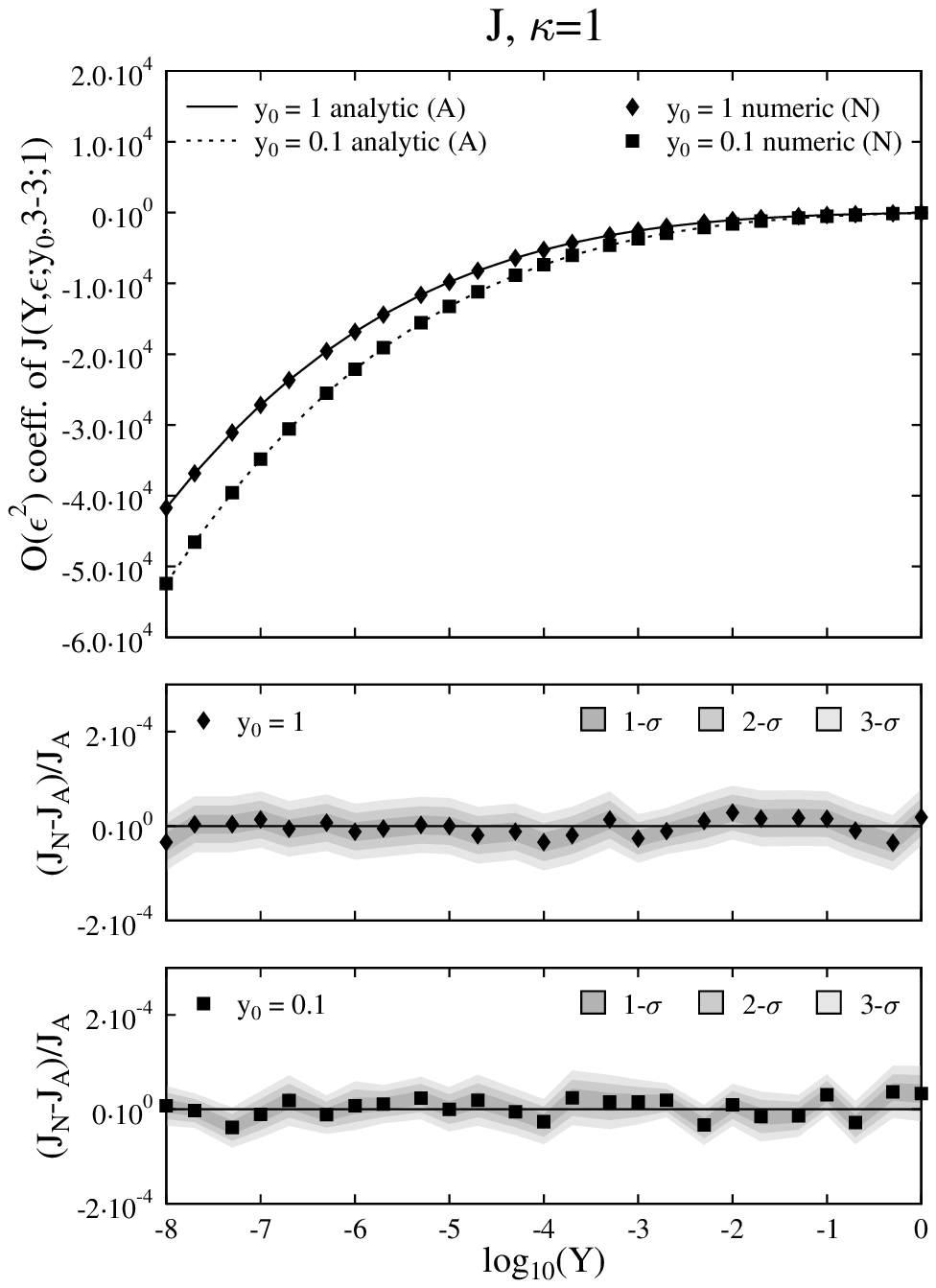}
\vspace{-2em}
\caption{\label{fig:Jfigs}
Representative results for the $\cJ$ integral. The plots show
the coefficient of the $\Oe{2}$ term in 
$\cJ(Y,\ep;y_0,3-3\ep;\kappa)$ for $\kappa=0$ (left figure) and $\kappa=1$
(right figure) with $y_0=0.1,1$.}
\end{figure}

As representative examples, in \fig{fig:Jfigs} we compare the analytic
and numeric results for the $\ep^2$ coefficient in the expansion of
$\cJ(Y,\ep;y_0,3-3\ep;\kappa)$ for $\kappa=0,1$ and $y_0=0.1,1$. The
agreement between the two computations is seen to be excellent for the
whole $Y$-range. We find a similar agreement for other (lower-order, thus
simpler) expansion coefficients and/or other values of the parameters.

% Figs for J integral

%\clearpage

%%%
%%% The soft-collinear Integral K
%%%

\section{The soft-collinear integral $\cK$}
\label{sec:softcol01}

In this section we calculate analytically the soft-collinear integral
defined in \eqn{eq:Kint} for $\kap =0,1$ and $d'_0=D'_0+d'_1\ep$,
$D'_0$ being an integer. The $\ph$ integral is trivial to perform and we find
\beq
\label{eq:phiint}
\frac{\Gamma^2(1-\ep)}{2\pi\Gamma(1-2\ep)}\,\int_{-1}^1\rd(\cos
\ph)\,(\sin\ph)^{-1-2\ep} = 2^{-1+2\ep}.
\eeq
Putting $\cos\tht = 2\xi -1$, we are left with the integral
\beq
\kint(\ep;y_0,d'_0;\kappa)  =  2\int_0^{y_0}\rd
y\,\int_0^1\rd\xi\,\,y^{-1-2(1+\kap)\ep}(1-y)^{d'_0-1}\xi^{-\ep}(1-\xi)^{-1-(1+\kap)\ep}(1-y\xi)^{1+\kap\ep}.
\label{eq:kint}
\eeq

%
% Analytic results for kappa=0
%

\subsection{Analytic result for $\kap = 0$}
For $\kap = 0$, the integral decouples into a product of two one-dimensional integrals and we get
\beq
\bsp
\kint(\ep;&y_0,d'_0;0) =2
\,B_{\yo}(-2\ep,d_0')\,B(1-\ep,-\ep)-2\,B_{\yo}(1-2\ep,d_0')\,B(2-\ep,-\ep),
\esp
\eeq
Using the expansion of the incomplete $B$-function, carried out in
\sect{sec:soft01}, we can immediately write down the expansion of $\cK$
for $\kap=0$. The result for $D_0'=3$ can be found in \appx{app:KIntegrals}.

%
% Analytic results for kappa=1
%

\subsection{Analytic result for $\kap = 1$}
The integral~(\ref{eq:Kint}) for $\kap=1$ reads
\beq
\kint(\ep;y_0,d'_0;1)  =  2\int_0^{y_0}\rd
y\,\int_0^1\rd\xi\,\,y^{-1-4\ep}(1-y)^{d'_0-1}\xi^{-\ep}(1-\xi)^{-1-2\ep}(1-y\xi)^{1+\ep}.
\eeq
The analytic solution for this integral cannot be obtained in a
straightforward way, due to the presence of the factor $(1-y\xi)^{\ep}$
that couples the two integrals. Therefore, we rewrite the integral in
the form 
\beq
\bsp
\kint(\ep;y_0,d'_0;1) = 2\,y_0^{-4\ep}\,K(\ep;y_0,d'_1;1,1-D'_0,0,1,-1),
\esp
\eeq
where
\beq
\bsp
K(\ep;y_0&,d'_1;n_1,n_2,n_3,n_4,n_5) \\
&= \int_0^{1}\rd
y\,\int_0^1\rd\xi\,\,y^{-n_1-4\ep}(1-y_0y)^{-n_2-d'_1\ep}\xi^{-n_3-\ep}(1-\xi)^{-n_4-2\ep}(1-y_0y\xi)^{-n_5+\ep}.
\esp
\eeq
We now calculate the integral $K$ using the Laporta algorithm. 
The independent integrals can be obtained by partial fractioning in $y$ and $\xi$, using the prescription that denominators depending on both integration variables are only partial fractioned in $\xi$, \eg
\beq
\bsp
\frac{1}{\xi(1-\yo y\xi)}&\rightarrow\frac{1}{\xi}+\frac{\yo y}{1-\yo y \xi},\\
\frac{1}{y(1-\yo y \xi)}&\rightarrow\frac{1}{y(1-\yo y \xi)}.
\esp
\eeq
When writing down the integration-by-parts identities for the independent integrals, we have to take into account a surface term coming from the fact that the denominator in $(1-\yo y)$ does not vanish in $y=1$,
\beq
\bsp
\label{eq:ibpK}
\int_0^{1}\rd& y\,\int_0^1\rd\xi\,\frac{\partial}{\partial
\xi}\,\left(y^{-n_1-4\ep}(1-y_0y)^{-n_2-d'_1\ep}\xi^{-n_3-\ep}(1-\xi)^{-n_4-2\ep}(1-y_0y\xi)^{-n_5+\ep}\right) \\ &= 0\\
\int_0^{1}\rd& y\,\int_0^1\rd\xi\,\frac{\partial}{\partial
y}\,\left(y^{-n_1-4\ep}(1-y_0y)^{-n_2-d'_1\ep}\xi^{-n_3-\ep}(1-\xi)^{-n_4-2\ep}(1-y_0y\xi)^{-n_5+\ep}\right)\\
&=(1-y_0)^{-n_3-d'_1\ep}\,K_S(\ep;y_0,d'_1;n_3,n_4,n_5),
\esp
\eeq
with
\beq
K_S(\ep;y_0,d'_1;n_3,n_4,n_5) = \int_0^1\rd
\xi\,\xi^{-n_3-\ep}(1-\xi)^{-n_4-2\ep}(1-y_0\xi)^{-n_5+\ep}.
\eeq
$K_S$ is just a hypergeometric function,
\beq
\bsp
K_S(\ep;y_0&,d'_1;n_3,n_4,n_5)\\
&=B(1-n_3-\ep,1-n_4-\ep)\,\hypgeo(1-n_3-\ep,n_5-2\ep,2-n_2-n_4-3\ep;y_0),
\esp
\eeq
and can thus be calculated using the technique presented in \sect{sec:soft01}.

Knowing the series expansion for the surface term $K_S$, we can solve the integration-by-parts identities for the $K$ integrals, \eqn{eq:ibpK}. We find the following two master integrals,
\beq
\bsp
K^{(1)}(\ep;y_0,d'_1) & =  K(\ep;y_0,d'_1;0,0,0,0,0),\\
K^{(2)}(\ep;y_0,d'_1) & = K(\ep;y_0,d'_1;-1,0,0,0,0),
\esp
\eeq
fulfilling the following differential equations,
\beq
\bsp
\frac{\partial}{\partial y_0} K^{(1)}  &
=\frac{4\ep-1}{y_0}\,K^{(1)}+\frac{(1-y_0)^{d_1\ep}}{y_0}\,f^{(1)},\\
\frac{\partial}{\partial y_0} K^{(2)}  &
=2\,\frac{2\ep-1}{y_0}\,K^{(2)}+\frac{(1-y_0)^{d_1\ep}}{y_0}\,f^{(1)},
\esp
\eeq
where $f^{(1)}$ denotes the master integral of the hypergeometric function calculated in \sect{sec:soft01} and where the initial conditions are given by
\beq
\bsp
K^{(1)}(\ep;y_0=0,d'_1)&=B(1-4\ep,1)\,B(1-\ep,1-2\ep),\\
K^{(2)}(\ep;y_0=0,d'_1)&=B(2-4\ep,1)\,B(1-\ep,1-2\ep).
\esp
\eeq
Plugging in the series expansion of $f^{(1)}$, and expanding
$(1-y_0)^{d'_1\ep}$ into a power series in $\ep$, we can solve for the
$K^{(1)}$ and $K^{(2)}$ as a power series in $\ep$ whose coefficients are written in terms of \hpls  in $y_0$.

Knowing the series expansions of $K^{(1)}$ and $K^{(2)}$, we can obtain
the integral $\kint(\ep;y_0,d'_0;1)$ for any fixed integer $D'_0$. In \appx{app:KIntegrals} we give the explicit result for $D'_0=3$.

%%%
%%% The collinear integrals I
%%%

\section{The collinear integrals $\cI$}
\label{sec:coll01}

In this section, we calculate the collinear integrals defined in
\eqn{eq:Iint} for $g_{I}=g_{A}$ and $g_{I}=g_{B}$ analytically.

%
% Collinear A-type k>=0
%

\subsection{The $\cA$-type collinear integrals for $k\ge0$}
\label{subsec:Aint}
The collinear integral for $g_I=g_A$ requires the evaluation of an integral of the form
\beq
\bsp
\label{eq:Aint}
\aint(x,\ep;\ao,d_0;\kap,k)  & = \frac{1}{x}\,\cI(x,\ep;\ao,d_0;\kap,k,0,g_A)\\
&=\int_0^{\ao} \rd\alp\,\int_0^1\rd
v\,\alp^{-1-(1+\kap)\ep}(1-\alp)^{2d_0-1}[\alp+(1-\alp)x]^{-1-(1+\kap)\ep}\\
& \quad\times \,v^{-\ep}(1-v)^{-\ep}\left(\frac{\alp+(1-\alp)xv}{2\alp+(1-\alp)x}\right)^k,
\esp
\eeq
where $k=-1,0,1,2$, $\kap = 0,1$ and $d_0=D_0+d_1\ep$ with $D_0$ an integer. For $k\ge0$ this two-dimensional integral decouples into the product of two one-dimensional integrals, out of which one is straightforward,
\begin{align}
\label{eq:Ap}
\aint(x,\ep&;\ao,d_0;\kap,k)  =\sum_{j=0}^{k}\binom{k}{j}\,x^{j}\,B(1+j-\ep,1-\ep)\\
&\times\int_0^{\ao}\rd\alp\,\alp^{k-j-1-(1+\kap)\ep}(1-\alp)^{j+2d_0-1}[\alp+(1-\alp)x]^{-1-(1+\kap)\ep}[2\alp+(1-\alp)x]^{-k}.\nonumber
\end{align}
We will therefore treat separately the cases $k\ge0$ and $k<0$.

For $k\ge0$ the calculation of the $\aint$ integrals reduces to the calculation of a one-dimensional integral of the form
\beq
\bsp
A_+(x,&\ep;\ao,d_1;\kap;n_1,n_2,n_3,n_4)  \\
& = \int_0^{\ao}\rd\alp\,\alp^{-n_1-(1+\kap)\ep}(1-\alp)^{-n_2+2d_1\ep}[\alp+(1-\alp)x]^{-n_3-(1+\kap)\ep}[2\alp+(1-\alp)x]^{-n_4},
\esp
\eeq
$n_i$ being integers. The integration-by-parts identities, including a surface term for the independent integrals, are
\beq
\bsp
\int_0^{\ao}\rd\alp\,&\frac{\partial}{\partial\alp}\left(\alp^{-n_1-(1+\kap)\ep}(1-\alp)^{-n_2+2d_1\ep}[\alp+(1-\alp)x]^{-n_3-(1+\kap)\ep}[2\alp+(1-\alp)x]^{-n_4}\right)\\
& = \ao^{-n_1-(1+\kap)\ep}(1-\ao)^{-n_2+2d_1\ep}[\ao+(1-\ao)x]^{-n_3-(1+\kap)\ep}[2\ao+(1-\ao)x]^{-n_4}.
\esp
\eeq
Using the Laporta algorithm we find three master integrals for $A_+$,
\beq
\bsp
A_+^{(1)}(x,\ep;\ao,d_1;\kap) & =  A_+(x,\ep;\ao,d_1;\kap;0,0,0,0) = \int_0^{\ao}\,\rd\mu_\ep(\alp;x),\\
A_+^{(2)}(x,\ep;\ao,d_1;\kap) & =  A_+(x,\ep;\ao,d_1;\kap;-1,0,0,0)= \int_0^{\ao}\,\rd\mu_\ep(\alp;x)\,\alp,\\
A_+^{(3)}(x,\ep;\ao,d_1;\kap) & =  A_+(x,\ep;\ao,d_1;\kap;0,0,0,1)= \int_0^{\ao}\,\frac{\rd\mu_\ep(\alp;x)}{2\alp+(1-\alp)x}.
\esp
\eeq
where
\beq
\bsp
\rd\mu_\ep&(\alp, x)=\rd\alp\,\alp^{-(1+\kappa)\ep}\,(1-\alp)^{2d_1\ep}\,
(\alp+(1-\alp)\;x)^{-(1+\kappa)\ep},\\
&=\rd\alp+\ep \,\rd\alp\big(2 d_1 \ln (1-\alpha)-(1+\kappa)  \ln \alpha 
  -(1+\kappa)  \ln (\alpha+ x -\alpha x)\big)+\begin{cal}O\end{cal}(\ep^2),\\
  &=\rd\alp+\ep \,\rd\alp\big(-(1+\kappa) H(0;\alpha)- (1+\kappa)
  H(0;x)-2 d_1 H(1;\alpha)-(\kappa
  +1) H(d_1(x);\alpha)\big)\\
  &\phantom{=}+\begin{cal}O\end{cal}(\ep^2).
\esp
\eeq
where we used the $d$-representation of the two-dimensional \hpls   defined in \sect{sec:2dhpls},
\beq
\bsp
H(d_1(x);\alp) & =  \ln\left(1+\frac{1-x}{x}\alp\right),\\
H(d_1(x),d_1(x);\alp) & =   \frac{1}{2}\ln^2\left(1+\frac{1-x}{x}\alp\right),
\\
&etc.  
\esp
\eeq
Notice that all three master integrals are finite for $\ep=0$. This allows us to expand the integrand into a power series in $\ep$ and integrate order by order in $\ep$, using the defining property of the \emph{HPL's}, \eqn{eq:hpldef}. We obtain in this way the series expansion of the master integrals as a power series in $\ep$ whose coefficients are written in terms of the $d$-representation of the two-dimensional \hpls. We can then switch back to the $c$-representation using the algorithm described in \sect{sec:2dhpls}.

Having a representation of the master integrals, we can immediately write down the solutions for $\aint(x,\ep;\ao,d_0;\kap,k)$ for $k\ge0$ and fixed $D_0$ using 
\eqn{eq:Ap}. In \appx{app:AIntegrals} we give as an example the series expansions up to order $\ep^2$ for $D_0=3$.

%
% Collinear A-type k=-1
%

\subsection{The $\cA$-type collinear integrals for $k=-1$}
For $k=-1$, the integral~(\ref{eq:Aint}) does not decouple, so we have to use the Laporta algorithm to calculate the full two-dimensional integral. However, for $k=-1$, we can get rid of the denominator in $(2\alp+(1-\alp)x)$ in the integrand. So we only have to deal with an integral of the form
\beq
\bsp
A_-(x,\ep&;\ao,d_1;\kap;n_1,n_2,n_3,n_4,n_5,n_6)=  \\
& = \int_0^{\ao}\rd\alp\,\int_0^1\rd v\,\alp^{-n_1-(1+\kap)\ep}(1-\alp)^{-n_2+2d_1\ep}[\alp+(1-\alp)x]^{-n_3-(1+\kap)\ep}\\
&\times v^{-n_4-\ep}(1-v)^{-n_5-\ep}[\alp+(1-\alp)xv]^{-n_6},
\esp
\eeq
$n_i$ being integers.

We write down the integration-by-parts identities for $A_-$ including a
surface term for $\alp$,
\beq
\bsp
\label{eq:IBPAm1}
\int_0^{\ao}&\rd\alp\,\int_0^1\rd v\,\frac{\partial}{\partial v}
\Big(\alp^{-n_1-(1+\kap)\ep}(1-\alp)^{-n_2+2d_1\ep}
[\alp+(1-\alp)x]^{-n_3-(1+\kap)\ep}
\\ &\times
v^{-n_4-\ep}(1-v)^{-n_5-\ep}[\alp+(1-\alp)xv]^{-n_6}\Big) =0\,,
\\
\int_0^{\ao}&\rd\alp\,\int_0^1\rd v\,\frac{\partial}{\partial\alp}
\Big(\alp^{-n_1-(1+\kap)\ep}(1-\alp)^{-n_2+2d_1\ep}
[\alp+(1-\alp)x]^{-n_3-(1+\kap)\ep}
\\ &\times 
v^{-n_4-\ep}(1-v)^{-n_5-\ep}[\alp+(1-\alp)xv]^{-n_6}\Big)
\\
& = \ao^{-n_1-(1+\kap)\ep}(1-\ao)^{-n_2+2d_1\ep}
[\ao+(1-\ao)x]^{-n_3-(1+\kap)\ep}\,A_{-,S}(x,\ep;\ao,d_1;n_4,n_5,n_6)\,,
\esp
\eeq
with
\beq
\bsp
A_{-,S}(x&,\ep;\ao,d_1;n_4,n_5,n_6) = \int_0^1\rd v\,v^{-n_4-\ep}(1-v)^{-n_5-\ep}[\ao+(1-\ao)xv]^{-n_6}\\
&=\alpha_0^{-n_6}B(1-n_4-\ep,1-n_5-\ep)\,\hypgeo\left(1-n_4-\ep,n_6,2-n_4-n_5-2\ep;\frac{\ao-1}{\ao}x\right).
\esp
\eeq
As in the case of $\kint$ we are going to evaluate this surface term using the Laporta algorithm, especially to get rid of the strange argument the hypergeometric function depends on, and to get an expression for $A_{-,S}$ in terms of two-dimensional \hpls  in $\ao$ and $x$.

% Evaluation of the surface term A_{-,S}

\paragraph{Evaluation of the surface term $A_{-,S}$.}
Because the $v$ integration is over the whole range $[0,1]$, we do not have to take into account a surface term in the integration-by-parts identities for $A_{-,S}$,
\beq
\int_0^1\rd v\,\frac{\partial}{\partial v}\left(v^{-n_4-\ep}(1-v)^{-n_5-\ep}[\ao+(1-\ao)xv]^{-n_6}\right) = 0.
\eeq
Using the Laporta algorithm we see that $A_{-,S}$ has two master integrals,
\beq
\bsp
A_{-,S}^{(1)}(x,\ep;\ao,d_1) & =  A_{-,S}(x,\ep;\ao,d_1;0,0,0),\\
A_{-,S}^{(2)}(x,\ep;\ao,d_1) & =  A_{-,S}(x,\ep;\ao,d_1;0,0,1).
\esp
\eeq
$A_{-,S}^{(i)}(x,\ep;\ao,d_1)$, $i=1,2$, are functions of the two
variables $x$ and $\ao$ defined on the square $[0,1]\times[0,1]$, so in
principle we should write down a set of partial differential equations
for the evolution of both $\ao$ and $x$. However, it is easy to see
that in $x=0$ we have
\beq
\bsp
\label{eq:initcondAS}
A_{-,S}^{(1)}(x=0,\ep;\ao,d_1) & =  B(1-\ep,1-\ep),\\
A_{-,S}^{(2)}(x=0,\ep;\ao,d_1) & =  \frac{1}{\ao} B(1-\ep,1-\ep),
\esp
\eeq
for arbitrary $\ao$. So we are in the special situation where we know the solutions on the line $\{x=0\}\times[0,1]$, and so we only need to consider the evolution for the $x$ variable. In other words, we consider $A_{-,S}^{(i)}$ as a function of $x$ only, keeping $\ao$ as a parameter.

The differential equations for the evolution in the $x$ variable read
\beq
\bsp
\frac{\partial}{\partial x}A_{-,S}^{(1)} & =  0,\\
\frac{\partial}{\partial x}A_{-,S}^{(2)}& =  
A_{-,S}^{(1)} \left(\frac{1-2 \ep}{\ao x}+\frac{(\ao-1) (2
  \ep-1)}{\ao (\ao x- x-\ao)}\right) +A_{-,S}^{(2)}
  \left(\frac{2
  \ep-1}{x}-\frac{(\ao-1) \ep}{\ao x-
  x-\ao}\right),
\esp
\eeq
and the initial condition for this system is given by \eqn{eq:initcondAS}. As the system is already triangular, we can immdiately solve for $A_{-,S}^{(1)}$ and $A_{-,S}^{(2)}$. Notice in particular that the denominator in $(\ao +x -x \ao)$ will give rise to two-dimensional \hpls  of the form $H(c_1(\ao);x)$, \emph{etc.}

% Evaluation of A_-

\paragraph{Evaluation of $A_-$.}
Having an expression for the $\ep$-expansion of the surface term, we can solve the integration-by-parts identities for $A_-$, \eqn{eq:IBPAm1}. We find four master integrals,
\beq
\bsp
A_-^{(1)}(x,\ep;\ao,d_1;\kap)&=A_-(x,\ep;\ao,d_1;\kap;0,0,0,0,0,0),\\
A_-^{(2)}(x,\ep;\ao,d_1;\kap)&=A_-(x,\ep;\ao,d_1;\kap;-1,0,0,0,0,0),\\
A_-^{(3)}(x,\ep;\ao,d_1;\kap)&=A_-(x,\ep;\ao,d_1;\kap;-1,0,0,0,0,1),\\
A_-^{(4)}(x,\ep;\ao,d_1;\kap)&=A_-(x,\ep;\ao,d_1;\kap;-2,0,0,0,0,1).
\esp
\eeq
It is easy to see that all of the master integrals are finite for $\ep=0$.

As in the case of the surface terms, we are only interested in the $x$ evolution, because the master integrals are known for $x=0$ for any value of $\ao$,
\beq
\bsp
\label{eq:initcondAm1}
A_-^{(1)}(x=0,\ep;\ao,d_1;\kap)&=B_{\ao}(1-2(1+\kap)\ep,1+2d_1\ep)\,B(1-\ep,1-\ep)\,,\\
A_-^{(2)}(x=0,\ep;\ao,d_1;\kap)&=B_{\ao}(2-2(1+\kap)\ep,1+2d_1\ep)\,B(1-\ep,1-\ep)\,,
\esp
\eeq
and
\beq
\bsp
A_-^{(3)}(x=0,\ep;\ao,d_1;\kap)&=A_-^{(1)}(x=0,\ep;\ao,d_1;\kap)\,,
\\
A_-^{(4)}(x=0,\ep;\ao,d_1;\kap)&=A_-^{(2)}(x=0,\ep;\ao,d_1;\kap)\,.
\esp
\eeq

The master integrals $A_-^{(1)}$ and $A_-^{(2)}$ form a subtopology, \ie the differential equations for these two master integrals close under themselves:
\beq
\bsp
\frac{\partial}{\partial x}A_-^{(1)} =& 
\frac{1-2(1+\kap) \ep}{x}\,A_-^{(1)} -\frac{2 (d_1
  \ep-(1+\kap)\ep+1)}{x}\,A_-^{(2)}\\
  &
-\frac{(1-\ao )^{1+2 d_1 \ep} 
  (-x \ao +\ao +x)^{-(1+\kap)\ep} \ao^{1-(1+\kap)\ep}}{x}\,A_{-,S}^{(1)},\\
\frac{\partial}{\partial x}A_-^{(2)} =&  
 \frac{1-(1+\kap)\ep}{x-1}\,A_-^{(1)}
 +\frac{-2 d_1
  \ep+(1+\kap)\ep-2}{x-1}\,A_-^{(2)}\\
  &
  -\frac{(1-\ao )^{1+2 d_1 \ep} 
  (-x \ao +\ao +x)^{-(1+\kap)\ep}
\ao^{1-(1+\kap)\ep}}{x-1}\,A_{-,S}^{(1)}.
\esp
\eeq
The two equations can be triangularized by the change of variable
\beq
\bsp
\tilde A_-^{(1)} & =  A_-^{(1)} - 2A_-^{(2)}, \\
\tilde A_-^{(2)} & =  A_-^{(2)}.
\esp
\eeq
The equations for the subtopology now take the triangularized form
\beq
\bsp
\label{eq:Am1Subtopo}
\frac{\partial}{\partial x}\tilde A_{-}^{(1)} =&
\left(\frac{2 \ep-2}{x-1}+\frac{1-2\ep}{x}\right)\,\tilde A_{-}^{(1)}
  +
  \left(\frac{4 d_1+2 }{x-1}-\frac{2 d_1 +2}{x}\right)\,\ep\,\tilde A_{-}^{(2)}\\
  &  +(1-\ao)^{2 d_1
  \ep} (-x \ao+\ao+x)^{-\ep}
  \ao^{1-\ep}
  \left(\frac{2-2\ao}{x-1}+\frac{
  \ao-1}{x}\right)\, A_{-,S}^{(1)},\\
\frac{\partial}{\partial x}\tilde A_{-}^{(2)} =&  
  \frac{1-\ep}{x-1}\tilde A_{-}^{(1)} -\frac{2 d_1+1
  }{x-1}\,\ep\,
  \tilde A_{-}^{(2)}
  +(1-\ao)^{2 d_1
  \ep} \ao^{1-\ep} (-x
  \ao+\ao+x)^{-\ep}
\frac{ a-1}{x-1}\, A_{-,S}^{(1)} .
 \esp
 \eeq
 The initial condition for $\tilde A_{-}^{(2)}$ can be obtained from 
\eqn{eq:initcondAm1}. For $\tilde A_{-}^{(1)}$ however, \eqn{eq:initcondAm1} gives only trivial information. Furthermore, the solution of the differential equation has in general a pole in $x=1$, but it is easy to convince oneself that $\tilde A_{-}^{(1)}$ is finite in $x=1$, which serves as the initial condition.

We can now solve for the remaining two master integrals. The differential equations for $A_{-}^{(3)}$ and $A_{-}^{(4)}$ read
\beq
\bsp
\label{eq:EqMIAm13}
  \frac{\partial}{\partial x}A_-^{(3)}=&
  \frac{1-2 (1+\kap)\ep}{x} \,A_-^{(3)}
  -\frac{2 (d_1
  \ep-(1+\kap)\ep+1)}{x}\,A_-^{(4)}\\
  &
  -\frac{(1-\ao )^{2 d_1 \ep} 
  (-x \ao +\ao +x)^{-(1+\kap)\ep}
\ao^{2-(1+\kap)\ep}}{x}\,A_{-,S}^{(2)},\\
  \frac{\partial}{\partial x}A_-^{(4)}=& 
  \left(\frac{1-2 \ep}{x}+\frac{2
  \ep-1}{x-1}\right)\,A_-^{(2)}
  -\frac{
  (2+\kap)\ep-2}{x-1}\,A_-^{(3)}\\
   &
   +
  \left(\frac{2 \ep-1}{x}-\frac{
  (2d_1 + \kap)\ep)}{x-1}\right)\,A_-^{(4)}\\
  &
  -\frac{(1-\ao )^{1+2 d_1 \ep} 
  (-x \ao +\ao +x)^{-(1+\kap)\ep}
\ao^{2-(1+\kap)\ep}}{x-1}\,A_{-,S}^{(2)}.
\esp
\eeq
These equations can be brought into a triangularized form via the change of variable
\beq
\bsp
\tilde A_-^{(3)} & =  A_-^{(3)} - A_-^{(4)}, \\
\tilde A_-^{(4)} & =  A_-^{(4)},
\esp
\eeq
and \eqn{eq:EqMIAm13} now reads
\beq
\bsp
\label{eq:eqG2z}
\frac{\partial}{\partial x}\tilde A_-^{(3)} =&
\left(\frac{1-2 \ep}{x}+\frac{2
  (\ep-1)}{x-1}\right)\,\tilde A_-^{(3)}
+ \left(\frac{2
  (d_1+1) \ep}{x-1}-\frac{2 (d_1+1)
  \ep}{x}\right)\,\tilde A_-^{(4)}\\
&-(1-\ao)^{1+2 d_1
  \ep}  \ao^{2-\ep} (-x
  \ao+\ao+x)^{-\ep}\left(\frac{1}{x}-\frac{1}{x-1}\right) A_{-,S}^{(2)}\\ &+ \left(\frac{1-2
  \ep}{x-1}+\frac{2
  \ep-1}{x}\right)\,A_-^{(2)}
 ,\\
\frac{\partial}{\partial x}\tilde A_-^{(4)} =& 
\frac{2-2 \ep}{x-1}\,\tilde A_-^{(3)}
 +
  \left(\frac{2 \ep-1}{x}-\frac{2 (d_1
  \ep+\ep)}{x-1}\right)\,\tilde A_-^{(4)}\\
  &   +\left(\frac{1-2
  \ep}{x}+\frac{2 \ep-1}{x-1}\right)\,A_-^{(2)}
-\frac{(1-\ao)^{1+2 d_1 \ep}  (-x
  \ao+\ao+x)^{-\ep} 
  \ao^{2-\ep}}{x-1}A_{-,S}^{(2)}.
\esp
\eeq
The initial condition for $A_-^{(3)}$ and $A_-^{(4)}$ can again be obtained from 
\eqn{eq:initcondAm1} and requiring $A_-^{(3)}$ to be finite in $x=1$. 

Having the analytic expressions for the master integrals, we can now easily obtain the solutions for $\aint$ for $k=-1$ for a fixed value of $D_0$. The results for $D_0=3$ can be found in \appx{app:AIntegrals}.

\begin{figure}[t]
\includegraphics[scale=0.75]{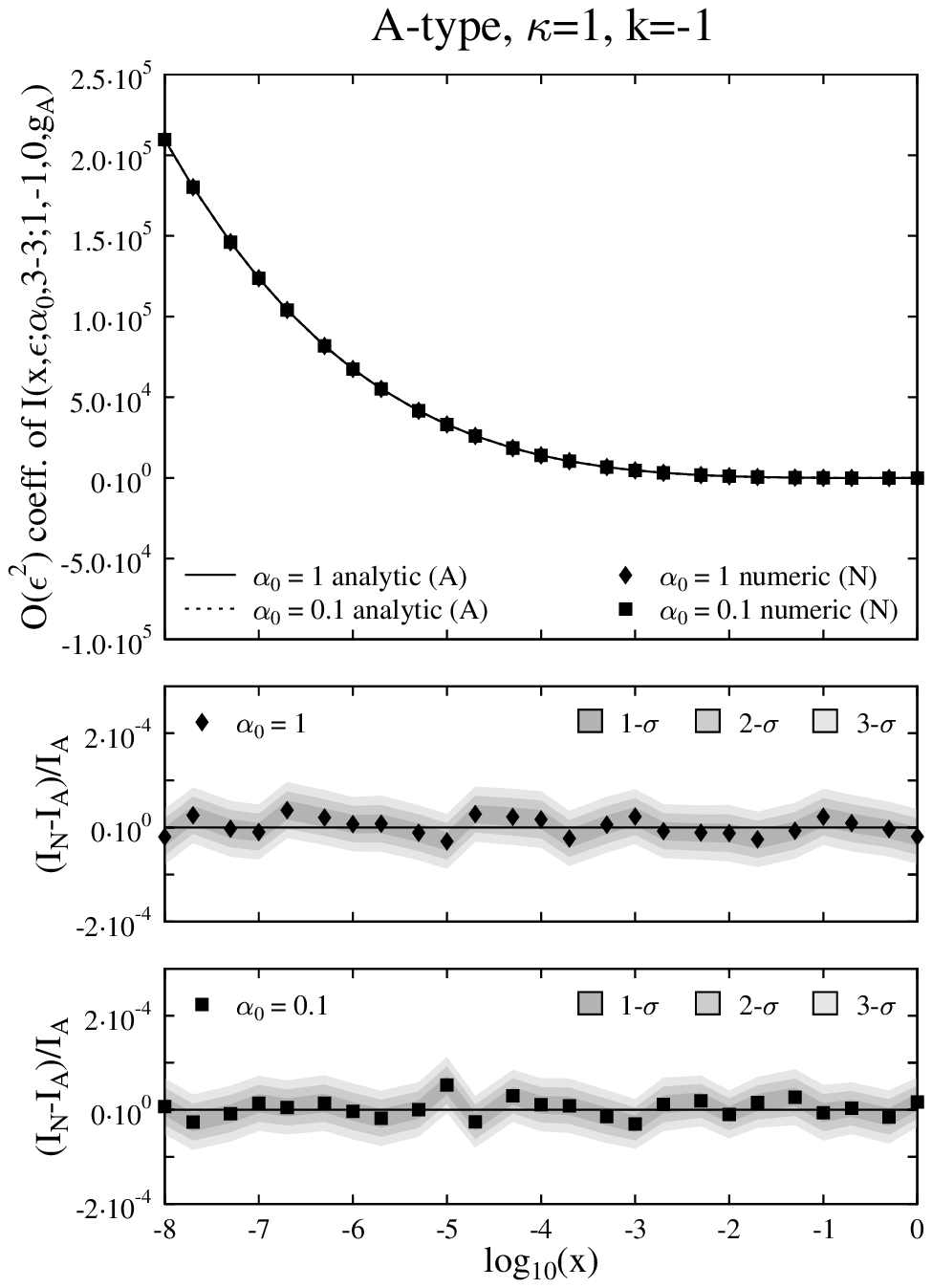}
\includegraphics[scale=0.75]{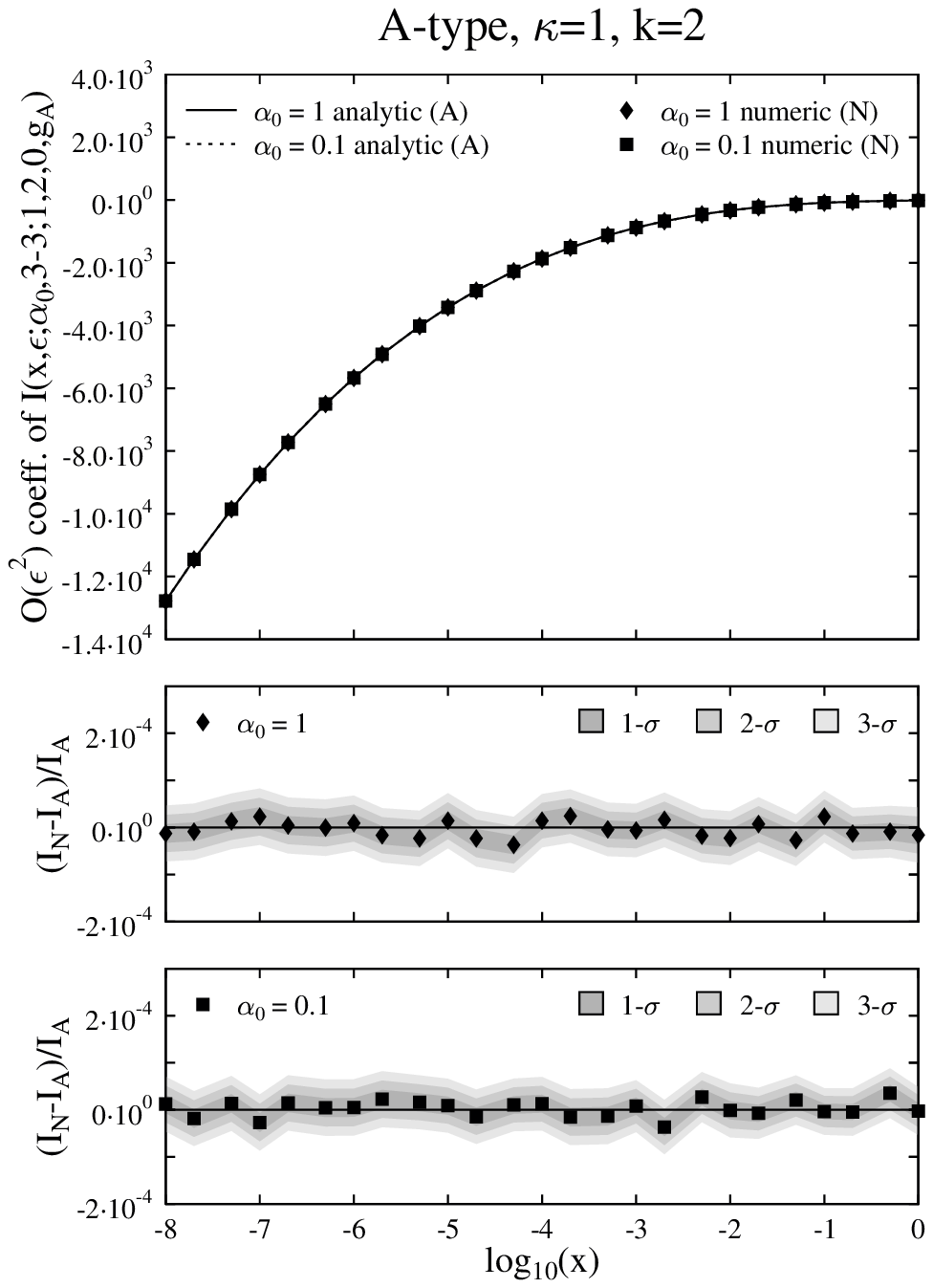}
\vspace{-2em}
\caption{\label{fig:Afigs}
Representative results for the ${\cal A}$-type integrals. The plots show
the coefficient of the $\Oe{2}$ term in 
$\cI(x,\ep;\alpha_0,3-3\ep;1,k,0,g_A)$ for $k=-1$ (left figure) and $k=2$ (right
figure) with $\alpha_0=0.1,1$.}
\end{figure}

In \fig{fig:Afigs} we compare the analytic and numeric results for the $\ep^2$ coefficient in the expansion of $\cI(x,\ep;\alpha_0,3-3\ep;1,k,0,g_A)$ for $k=-1,2$ and $\alpha_0=0.1,1$ as representative examples. The dependence on $\alpha_0$ is not visible on the plots. The agreement between the two computations is excellent for the whole $x$-range. We find a similar agreement for other (lower-order, thus simpler) expansion coefficients and/or other values of the parameters.

% Figs for A-type integrals

%
% Collinear B-type
%

\subsection{The $\cB$-type collinear integrals}

The $\cB$-type collinear integrals require the evaluation of an integral of the form
\beq
\bsp
\label{eq:Bint}
\bint(x,\ep;&\ao,d_0;\delta,k) = \frac{1}{x}\,\ciint(x,\ep;\ao,d_0;1,k,\delta,g_B)\\
&=\int_0^{\ao} \rd\alp\,\int_0^1\rd v\,\alp^{-1-2\ep}(1-\alp)^{2d_0-1}[\alp+(1-\alp)x]^{-1-2\ep}[2\alp+(1-\alp)x]^{-k}\\
& \times \,v^{-\ep}(1-v)^{-\ep}[\alp+(1-\alp)xv]^{k+\delta\ep}[\alp+(1-\alp)(1-v)x]^{-\delta\ep},
\esp
\eeq
 where $k=-1,0,1,2$, $\delta = \pm1$ and $d_0=D_0+d_1\ep$ (as before $D_0$ is an integer). Unlike the $\aint$-type integrals, the $\bint$-type integrals do not decouple for $k\ge0$, due to the appearance of the $\ep$ pieces in the exponents, so we have to consider the denominators altogether, and have to deal with an integral of the form
\beq
\bsp
B(x,&\ep;\ao,d_1;\delta;n_1,n_2,n_3,n_4,n_5,n_6,n_7,n_8)=\\
&=\int_0^{\ao} \rd\alp\,\int_0^1\rd v\,\alp^{-n_1-2\ep}(1-\alp)^{-n_2+2d_1}[\alp+(1-\alp)x]^{-n_3-2\ep}[2\alp+(1-\alp)x]^{-n_4}\\
& \quad\times \,v^{-n_5-\ep}(1-v)^{-n_6-\ep}[\alp+(1-\alp)xv]^{-n_7+\delta\ep}[\alp+(1-\alp)(1-v)x]^{-n_8-\delta\ep}.
\esp
\eeq

We use again the Laporta algorithm, and write down the integration-by-parts identities for $B$,
\beq
\bsp
\int_0^{\ao} \rd\alp\,\int_0^1\rd v\,\frac{\partial}{\partial v}  &\left(\alp^{-n_1-2\ep}(1-\alp)^{-n_2+2d_1}[\alp+(1-\alp)x]^{-n_3-2\ep}[2\alp+(1-\alp)x]^{-n_4}\right.\\
& \left. \times \,v^{-n_5-\ep}(1-v)^{-n_6-\ep}[\alp+(1-\alp)xv]^{-n_7+\delta\ep}[\alp+(1-\alp)(1-v)x]^{-n_8-\delta\ep}\right) \nonumber\\
=0,&\\
\int_0^{\ao} \rd\alp\,\int_0^1\rd v\,\frac{\partial}{\partial \alp}  &\left(\alp^{-n_1-2\ep}(1-\alp)^{-n_2+2d_1}[\alp+(1-\alp)x]^{-n_3-2\ep}[2\alp+(1-\alp)x]^{-n_4}\right.\\
& \left.\times \,v^{-n_5-\ep}(1-v)^{-n_6-\ep}[\alp+(1-\alp)xv]^{-n_7+\delta\ep}[\alp+(1-\alp)(1-v)x]^{-n_8-\delta\ep}\right) \nonumber\\
=\ao^{-n_1-2\ep}&(1-\ao)^{-n_2+2d_1}[\ao+(1-\ao)x]^{-n_3-2\ep}[2\ao+(1-\ao)x]^{-n_4}\\
&\times B_S(x,\ep;\ao,d_1;\delta,k;n_5,n_6,n_7,n_8),
\esp
\eeq
where the surface term is given by
\beq
\bsp
B_S(x,&\ep;\ao,d_1;\delta;n_5,n_6,n_7,n_8)=\\
&=\int_0^1\rd v\,v^{-n_5-\ep}(1-v)^{-n_6-\ep}[\ao+(1-\ao)xv]^{-n_7+\delta\ep}[\ao+(1-\ao)(1-v)x]^{-n_8-\delta\ep}.
\esp
\eeq

% Evaluation of the surface term B_S

\paragraph{Evaluation of the surface term $B_S$.}
The surface term $B_S$ is no longer a hypergeometric function as it was the case for the $\kint$ and $\aint$-type integrals. It can nevertheless be easily calculated using the Laporta algorithm. The integration-by-parts identities for $B_S$ read
\beq
\int_0^1\rd v\,\frac{\partial}{\partial v}\left(v^{-n_5-\ep}(1-v)^{-n_6-\ep}[\ao+(1-\ao)xv]^{-n_7+\delta\ep}[\ao+(1-\ao)(1-v)x]^{-n_8-\delta\ep}\right)=0.
\eeq
We find three master integrals for $B_S$,
\beq
\bsp
B_S^{(1)}(x,\ep;\ao,d_1;\delta) & =  B_S(x,\ep;\ao,d_1;\delta;0,0,0,0),\\
B_S^{(2)}(x,\ep;\ao,d_1;\delta) & =  B_S(x,\ep;\ao,d_1;\delta;-1,0,0,0),\\
B_S^{(3)}(x,\ep;\ao,d_1;\delta) & =  B_S(x,\ep;\ao,d_1;\delta;0,0,1,0),
\esp
\eeq
fulfilling the differential equations
\beq
\bsp
\frac{\partial}{\partial x} B_S^{(1)}  = &     
  \,B_S^{(1)} \Bigg(\frac{2
  \left(\ep \alpha _0-\alpha
  _0-\ep+1\right)}{\alpha _0 \left(\left(\alpha
  _0-1\right) x-2 \alpha _0\right)}+\frac{-2 \alpha _0
  \ep^2+2 \ep^2+3 \alpha _0 \delta  \ep-3
  \delta  \ep-\alpha _0+1}{\alpha _0
  \left(\left(\alpha _0-1\right) x-\alpha _0\right)
  (\ep \delta -1)}\\
&  +\frac{-2 \delta  \ep^2+2
  \ep^2-\delta  \ep+2 \ep-1}{\alpha _0 x
  (\ep \delta -1)}\Bigg)
  \\
  &+B_S^{(2)} \Bigg(\frac{4 \left(\alpha _0-1\right)
  (\ep-1)}{\alpha _0 \left(\left(\alpha _0-1\right)
  x-\alpha _0\right)}-\frac{4 \left(\alpha _0-1\right)
  (\ep-1)}{\alpha _0 \left(\left(\alpha _0-1\right)
  x-2 \alpha _0\right)}\Bigg)\\
&+B_S^{(3)}
  \Bigg(\frac{\left(\alpha _0-1\right) \left(\delta 
  \ep^2-\ep^2-\ep+1\right)}{\left(x
  \alpha _0-\alpha _0-x\right) (\ep \delta
  -1)}+\frac{2 \delta  \ep^2-2 \ep^2+\delta 
  \ep-2 \ep+1}{x (\ep \delta
  -1)}\\
  &-\frac{\left(\alpha _0-1\right) \left(2 \delta 
  \ep^2-2 \ep^2+\delta  \ep-2
  \ep+1\right)}{\left(\left(\alpha _0-1\right)
  x-\alpha _0\right) (\ep \delta -1)}\Bigg),\\
 \frac{\partial}{\partial x} B_S^{(2)}  = & \,-B_S^{(1)}\frac{ (-\delta 
  \ep+\ep-1)}{x}+B_S^{(2)}  \frac{2 (\ep-1)}{x}-B_S^{(3)}\frac{\alpha _0
  \ep \delta }{x},\\
   \frac{\partial}{\partial x} B_S^{(3)}  =
& \,B_S^{(1)} \Bigg(-\frac{\delta  \left(2 \alpha
  _0 \ep^2-2 \ep^2-2 \alpha _0 \ep-2
  \alpha _0 \delta  \ep+2 \delta  \ep+2
  \ep+2 \alpha _0 \delta -2 \delta \right)}{\alpha _0
  \left(\left(\alpha _0-1\right) x-2 \alpha _0\right)
  (1-\ep \delta )}\\
  &+\frac{2 \delta  \ep^2-2
  \ep^2+\delta  \ep-2 \ep+1}{\alpha _0 x
  (1-\ep \delta )}-\frac{\delta  \left(-2 \alpha _0
  \delta  \ep^2+2 \delta  \ep^2+3 \alpha _0
  \ep-3 \ep-\alpha _0 \delta +\delta
  \right)}{\alpha _0 \left(\left(\alpha _0-1\right)
  x-\alpha _0\right) (1-\ep \delta )}\Bigg)\\
  &+B_S^{(2)} \Bigg(\frac{2 \left(\alpha _0-1\right)
  (\ep-1) (2 \ep-2 \delta ) \delta }{\alpha _0
  \left(\left(\alpha _0-1\right) x-2 \alpha _0\right)
  (1-\ep \delta )}-\frac{2 \left(\alpha _0-1\right)
  (\ep-1) (2 \ep-2 \delta ) \delta }{\alpha _0
  \left(\left(\alpha _0-1\right) x-\alpha _0\right)
  (1-\ep \delta )}\Bigg)\\
&+B_S^{(3)} \Bigg(\frac{-2
  \delta  \ep^2+2 \ep^2-\delta  \ep+2
  \ep-1}{x (1-\ep \delta )}-\frac{\left(\alpha
  _0-1\right) \left(-2 \delta  \ep^2+2
  \ep^2-\delta  \ep+2
  \ep-1\right)}{\left(\left(\alpha _0-1\right)
  x-\alpha _0\right) (1-\ep \delta
  )}\\
  &+\frac{\left(\alpha _0-1\right) \left(-\delta 
  \ep^2+\ep^2+\ep-1\right)}{\left(x
  \alpha _0-\alpha _0-x\right) (1-\ep \delta
  )}\Bigg).
  \esp
  \eeq
The initial conditions for the differential equations are
\beq
\bsp
B_S^{(1)}(x=0,\ep;\ao,d_1;\delta) & =  B(1-\ep,1-\ep),\\
B_S^{(2)}(x=0,\ep;\ao,d_1;\delta) & =  B(2-\ep,1-\ep),\\
 B_S^{(3)}(x=0,\ep;\ao,d_1;\delta) & =  \frac{1}{\ao}B(1-\ep,1-\ep),\\
 \esp
 \eeq
 The system can be triangularized by the change of variable
 \beq
\tilde B_S^{(1)} = B_S^{(1)}-2B_S^{(2)},\quad\tilde B_S^{(2)} = B_S^{(2)},\quad\tilde B_S^{(3)} = B_S^{(3)},\\
\eeq
and then solved in the usual way.

% Evaluation of the B integral

\paragraph{Evaluation of the $B$ integral.}
Solving the integration-by-parts identities for the $B$ integrals, we find nine master integrals
\beq
\bsp
B^{(1)}(x,\ep;\ao,d_1;\delta) & =  B(x,\ep;\ao,d_1;\delta;0,0,0,0,0,0,0,0),\\
B^{(2)}(x,\ep;\ao,d_1;\delta) & =  B(x,\ep;\ao,d_1;\delta;0,0,0,0,0,0,0,1),\\
B^{(3)}(x,\ep;\ao,d_1;\delta) & =  B(x,\ep;\ao,d_1;\delta;0,0,0,0,0,0,1,0),\\
B^{(4)}(x,\ep;\ao,d_1;\delta) & =  B(x,\ep;\ao,d_1;\delta;0,0,0,1,0,0,0,0),\\
B^{(5)}(x,\ep;\ao,d_1;\delta) & =  B(x,\ep;\ao,d_1;\delta;0,0,1,0,0,0,0,0),\\
B^{(6)}(x,\ep;\ao,d_1;\delta) & =  B(x,\ep;\ao,d_1;\delta;-1,0,0,0,0,0,1,0),\\
B^{(7)}(x,\ep;\ao,d_1;\delta) & =  B(x,\ep;\ao,d_1;\delta;0,0,0,1,0,0,1,0),\\
B^{(8)}(x,\ep;\ao,d_1;\delta) & =  B(x,\ep;\ao,d_1;\delta;0,0,1,0,0,0,1,0),\\
B^{(9)}(x,\ep;\ao,d_1;\delta) & =  B(x,\ep;\ao,d_1;\delta;0,0,1,0,0,0,0,1),
\esp
\eeq
The master integrals $B^{(i)}$, $i\neq 4, 7$, form a subtopology, \ie the
differential equations for these master integrals close under themselves.
Furthermore the differential equations for $B^{(1)}, B^{(3)}, B^{(5)}$
and  $B^{(6)}$ have a triangular structure in $\ep$, \ie all other master
integrals are suppressed by a power of $\ep$.
For $\delta=+1$, the corresponding differential equations are given by
\beq
\bsp
%B1
\frac{\partial}{\partial x}B^{(1)}
&=\frac{2 (\varepsilon -1) B_S^{(2)} (\ao-1)^2}{(2 d_1 \varepsilon -4 \
\varepsilon +1) (x \ao-\ao-x)}-\frac{2 \varepsilon  \
B^{(1)}}{x-1}
+\Bigg(\frac{4 \varepsilon ^2}{(2 d_1 \varepsilon -4 \varepsilon \
+1) (x-1)}-\\
&\frac{\varepsilon  (8 \varepsilon -1)}{(2 d_1 \varepsilon -4 \
\varepsilon +1) x}\Bigg) B^{(2)}
+\Bigg(\frac{2 \varepsilon ^2}{(2 d_1 \
\varepsilon -4 \varepsilon +1) (x-1)}-\frac{\varepsilon  (4 \varepsilon -1)}{(2 \
d_1 \varepsilon -4 \varepsilon +1) x}\Bigg) B^{(3)}+\\
&\Bigg(\frac{2 \varepsilon \
}{x-1}+\frac{2 (2 \varepsilon -1) \varepsilon }{(2 d_1 \varepsilon -4 \varepsilon \
+1) x}\Bigg) B^{(5)}-\frac{2 \varepsilon  B^{(6)}}{x}-\frac{2 \varepsilon ^2 \
B^{(8)}}{(2 d_1 \varepsilon -4 \varepsilon +1) (x-1)}-\\
&\frac{4 \varepsilon ^2 \
B^{(9)}}{(2 d_1 \varepsilon -4 \varepsilon +1) (x-1)}+\Bigg(-\frac{2 \
\varepsilon  (\ao-1)^2}{(2 d_1 \varepsilon -4 \varepsilon +1) ((\ao-1) \
x-\ao)}+\\
&\frac{(\ao-1)^2}{(2 d_1 \varepsilon -4 \varepsilon +1) (x \
\ao-\ao-x)}+\frac{2 \varepsilon  (\ao-1)}{(2 d_1 \varepsilon -4 \varepsilon \
+1) x}\Bigg) B_S^{(1)}\,,\\
%B3
\frac{\partial}{\partial x}B^{(3)} &=-\frac{4 \ep  B^{(3)}}{x}+\frac{(-2 d_1 \ep +4 \ep -1) \
B^{(6)}}{x}+\frac{(\ao-1) \ao B_S^{(3)}}{x}\,,\\
%B5
\frac{\partial}{\partial x}B^{(5)} &=\Bigg(\frac{-2 d_1 \ep +4 \ep -1}{x}+\frac{2 d_1 \ep -4 \ep \
+1}{x-1}\Bigg) B^{(1)}+\Bigg(\frac{-2 d_1 \ep +4 \ep -1}{x-1}-\frac{4 \
\ep }{x}\Bigg) B^{(5)}+\\
&\Bigg(\frac{\ao-1}{x}-\frac{(\ao-1)^2}{(\ao-1) \
x-\ao}\Bigg) B_S^{(1)}\,,\\
%\esp
%\eeq
%\beq
%\bsp
%B6
\frac{\partial}{\partial x}B^{(6)} &=\nonumber
\Bigg(\frac{1-2 \ep }{x}+\frac{2 \ep -1}{x-1}\Bigg) \
B^{(1)}+\Bigg(\frac{8 \ep ^2}{(2 d_1 \ep -4 \ep +1) \
(x-1)}-\frac{\ep }{(2 d_1 \ep -4 \ep +1) (x-2)}-\\
&\frac{(8 \ep -1) \
\ep }{(2 d_1 \ep -4 \ep +1) x}\Bigg) B^{(2)}+\Bigg(-\frac{2 \ep \
}{(x-1)^2}+\frac{2 d_1 \ep -5 \ep +1}{(2 d_1 \ep -4 \ep +1) \
(x-2)}-\\
&\frac{2(2 d_1 \ep ^2-6 \ep ^2+\ep)}{(2 d_1 \ep \
-4 \ep +1) (x-1)}+\frac{\ep -4 \ep ^2}{(2 d_1 \ep -4 \ep +1) \
x}\Bigg) B^{(3)}+\Bigg(\frac{1-2 \ep }{x-1}+\\
&\frac{4 d_1 \ep ^2-12 \
\ep ^2-2 d_1 \ep +8 \ep -1}{(2 d_1 \ep -4 \ep +1) (x-2)}+\frac{2 \
\ep  (2 \ep -1)}{(2 d_1 \ep -4 \ep +1) x}\Bigg) \
B^{(5)}+\Bigg(-\frac{2 \ep }{x-1}+\\
&\frac{-2 d_1 \ep +4 \ep -1}{x-2}-\
\frac{1}{x}\Bigg) B^{(6)}+\Bigg(\frac{2 \ep }{(x-1)^2}-\frac{4(2 \
d_1 \ep ^2-5 \ep ^2+\ep)}{(2 d_1 \ep -4 \ep +1) \
(x-2)}+\\
&\frac{4(2 d_1 \ep ^2-5 \ep ^2+\ep)}{(2 d_1 \ep \
-4 \ep +1) (x-1)}\Bigg) B^{(8)}+\\
&\Bigg(\frac{8 \ep ^2}{(2 d_1 \ep -4 \
\ep +1) (x-2)}-\frac{8 \ep ^2}{(2 d_1 \ep -4 \ep +1) (x-1)}\Bigg) \
B^{(9)}+%
\esp
\eeq
\beq\bsp
&\Bigg(\frac{4 (\ao-1)^3 (\ep -1)}{(\ao-2) (2 d_1 \ep -4 \ep \
+1) ((\ao-1) x-\ao)}-\\
&\frac{4 (\ao-1)^2 (\ep -1)}{(\ao-2) (2 d_1 \ep \
-4 \ep +1) (x-2)}\Bigg) B_S^{(2)}+\\
&\Bigg(-\frac{2 (2 \ep -1) \
(\ao-1)^3}{(\ao-2) (2 d_1 \ep -4 \ep +1) ((\ao-1) x-\ao)}+\frac{2 \
\ep  (\ao-1)}{(2 d_1 \ep -4 \ep +1) x}+\\
&\frac{2(\ep  \
\ao^2-\ao^2-\ep  \ao+2 \ao-1)}{(\ao-2) (2 d_1 \ep -4 \ep +1) \
(x-2)}\Bigg) B_S^{(1)}+\frac{(\ao-1) \ao B_S^{(3)}}{x-2},
\esp
\eeq
whereas for $\delta=-1$, the differential equations are 
\beq
\bsp
%B1
\frac{\partial}{\partial x}B^{(1)} &=
\frac{2 (\ep -1) B_S^{(2)} (\ao-1)^2}{(2 d_1 \ep -4 \ep +1) (x \
\ao-\ao-x)}-\frac{2 \ep  B^{(1)}}{x-1}+\Bigg(\frac{\ep  (4 \ep \
-1)}{(2 d_1 \ep -4 \ep +1) x}-\\
&\frac{2 \ep ^2}{(2 d_1 \ep -4 \ep \
+1) (x-1)}\Bigg) B^{(2)}+\Bigg(\frac{\ep  (8 \ep -1)}{(2 d_1 \ep -4 \
\ep +1) x}-\frac{4 \ep ^2}{(2 d_1 \ep -4 \ep +1) (x-1)}\Bigg) \\
&
B^{(3)}+\Bigg(\frac{2 \ep }{x-1}-\frac{2 \ep  (2 \ep -1)}{(2 d_1 \
\ep -4 \ep +1) x}\Bigg) B^{(5)}+\frac{2 \ep  B^{(6)}}{x}+\frac{4 \
\ep ^2 B^{(8)}}{(2 d_1 \ep -4 \ep +1) (x-1)}+\\
&\frac{2 \ep ^2 \
B^{(9)}}{(2 d_1 \ep -4 \ep +1) (x-1)}+\Bigg(\frac{2 \ep  \
(\ao-1)^2}{(2 d_1 \ep -4 \ep +1) ((\ao-1) x-\ao)}-\\
&\frac{(2 \ep -1) \
(\ao-1)^2}{(2 d_1 \ep -4 \ep +1) (x \ao-\ao-x)}-\frac{2 \ep  \
(\ao-1)}{(2 d_1 \ep -4 \ep +1) x}\Bigg) B_S^{(1)}\,,\\
%\esp
%\eeq
%\beq
%\bsp
%B3
\frac{\partial}{\partial x}B^{(3)} &=-\frac{4 \ep  B^{(3)}}{x}+\frac{(-2 d_1 \ep +4 \ep -1) \
B^{(6)}}{x}+\frac{(\ao-1) \ao B_S^{(3)}}{x}\,,
\\
%B5
\frac{\partial}{\partial x}B^{(5)} &=\Bigg(\frac{-2 d_1 \ep +4 \ep -1}{x}+\frac{2 d_1 \ep -4 \ep \
+1}{x-1}\Bigg) B^{(1)}+\Bigg(\frac{-2 d_1 \ep +4 \ep -1}{x-1}-\frac{4 \
\ep }{x}\Bigg) B^{(5)}+\\
&\Bigg(\frac{\ao-1}{x}-\frac{(\ao-1)^2}{(\ao-1) \
x-\ao}\Bigg) B_S^{(1)}\,,\\
%\esp
%\eeq
%\beq
%\bsp
%B6
\frac{\partial}{\partial x}B^{(6)} &=\Bigg(\frac{1-2 \ep }{x}+\frac{2 \ep
-1}{x-1}\Bigg) \
B^{(1)}+\Bigg(-\frac{4 \ep ^2}{(2 d_1 \ep -4 \ep +1) \
(x-1)}+\\
&\frac{\ep }{(2 d_1 \ep -4 \ep +1) (x-2)}+\frac{(4 \ep -1) \
\ep }{(2 d_1 \ep -4 \ep +1) x}\Bigg) B^{(2)}+\Bigg(-\frac{4 \ep \
}{(x-1)^2}+\\
&\frac{2 d_1 \ep -3 \ep +1}{(2 d_1 \ep -4 \ep +1) \
(x-2)}-\frac{4(2 d_1 \ep ^2-2 \ep ^2+\ep)}{(2 d_1 \ep \
-4 \ep +1) (x-1)}+\frac{8 \ep ^2-\ep }{(2 d_1 \ep -4 \ep +1) \
x}\Bigg) B^{(3)}+\\
&\Bigg(\frac{1-2 \ep }{x-1}+\frac{4 d_1 \ep ^2-4 \ep \
^2-2 d_1 \ep +4 \ep -1}{(2 d_1 \ep -4 \ep +1) (x-2)}-\frac{2 \ep \
(2 \ep -1)}{(2 d_1 \ep -4 \ep +1) x}\Bigg) B^{(5)}+\Bigg(-\frac{4 \
\ep }{x-1}+%
\esp
\eeq
\beq\bsp
&\frac{-2 d_1 \ep +4 \ep -1}{x-2}+\frac{4 \ep \
-1}{x}\Bigg) B^{(6)}+\Bigg(\frac{4 \ep }{(x-1)^2}-\frac{8(2 d_1 \
\ep ^2-3 \ep ^2+\ep)}{(2 d_1 \ep -4 \ep +1) (x-2)}+\\
&\frac{8 \
(2 d_1 \ep ^2-3 \ep ^2+\ep)}{(2 d_1 \ep -4 \ep +1) \
(x-1)}\Bigg) B^{(8)}+\Bigg(\frac{4 \ep ^2}{(2 d_1 \ep -4 \ep +1) \
(x-1)}-\\
&\frac{4 \ep ^2}{(2 d_1 \ep -4 \ep +1) (x-2)}\Bigg) \
B^{(9)}+\Bigg(\frac{4 (\ao-1)^3 (\ep -1)}{(\ao-2) (2 d_1 \ep -4 \ep \
+1) ((\ao-1) x-\ao)}-\\
&\frac{4 (\ao-1)^2 (\ep -1)}{(\ao-2) (2 d_1 \ep \
-4 \ep +1) (x-2)}\Bigg) B_S^{(2)}+\\
&\Bigg(\frac{2 (\ao-1)^3}{(\ao-2) (2 \
d_1 \ep -4 \ep +1) ((\ao-1) x-\ao)}-\frac{2 \ep  (\ao-1)}{(2 d_1 \
\ep -4 \ep +1) x}+\\
&\frac{2(\ep  \ao^2-\ao^2-3 \ep  \ao+2 \
\ao+2 \ep -1)}{(\ao-2) (2 d_1 \ep -4 \ep +1) (x-2)}\Bigg) \
B_S^{(1)}+\frac{(\ao-1) \ao B_S^{(3)}}{x-2}.\nonumber
\esp
\eeq
The differential equations for $B^{(2)}, B^{(8)}$ and $B^{(9)}$ read, for $\delta=+1$,
\beq
\bsp
%B2
\frac{\partial}{\partial x}B^{(2)} &=\Bigg(\frac{4 \ep -1}{x}-\frac{4 \ep
}{x-1}\Bigg) \
B^{(2)}+\Bigg(\frac{4 \ep -1}{x}-\frac{2 \ep }{x-1}\Bigg) \
B^{(3)}-\frac{2 (2 \ep -1) B^{(5)}}{x}+\\
&\frac{(2 d_1 \ep -4 \ep +1) \
B^{(6)}}{x}+\frac{2 \ep  B^{(8)}}{x-1}+\frac{4 \ep  \
B^{(9)}}{x-1}-\frac{(\ao-1) \ao B_S^{(3)}}{x},\\
%B8
\frac{\partial}{\partial x}B^{(8)} &=\Bigg(\frac{2 (d_1-2) \ep
}{x-1}-\frac{2 (d_1-2) \ep }{x}\Bigg) \
B^{(3)}+\Bigg(\frac{-4 \ep -1}{x}-\frac{2 (d_1 \ep -2 \ep \
)}{x-1}\Bigg) B^{(8)}+\\
&\Bigg(\frac{\ao-1}{x}-\frac{(\ao-1)^2}{(\ao-1) x-\
\ao}\Bigg) B_S^{(3)},\\
%B9
\frac{\partial}{\partial x}B^{(9)} &=-\frac{2 (\ep -1) B_S^{(2)} (\ao-1)^2}{\ep  (x \
\ao-\ao-x)^2}+\Bigg(\frac{2 (d_1-2) \ep }{x-1}-\frac{2 (d_1-2) \ep \
}{x}\Bigg) B^{(2)}+\\
&\Bigg(\frac{-4 \ep -1}{x}-\frac{2 (d_1 \ep -2 \ep \
)}{x-1}\Bigg) B^{(9)}+\Bigg(-\frac{2 (\ao-1)^2}{\ao (x \
\ao-\ao-x)}+\frac{2 (\ao-1)}{\ao x}+
\\&\frac{2 \ep  \ao^2-\ao^2-4 \ep  \
\ao+2 \ao+2 \ep -1}{\ep  (x \ao-\ao-x)^2}\Bigg) \
B_S^{(1)}+\Bigg(\frac{(\ao-1)^2}{(\ao-1) x-\ao}+\frac{1-\ao}{x}\Bigg) \
B_S^{(3)},
\esp
\eeq
and for $\delta =-1$
\beq
\bsp
%B2
\frac{\partial}{\partial x}B^{(2)} &=\Bigg(-\frac{2 \ep
}{x-1}-\frac{1}{x}\Bigg) B^{(2)}+\Bigg(\frac{8 \ep \
-1}{x}-\frac{4 \ep }{x-1}\Bigg) B^{(3)}-\frac{2 (2 \ep -1) \
B^{(5)}}{x}+\\
&\frac{(2 d_1 \ep -4 \ep +1) B^{(6)}}{x}+\frac{4 \ep  \
B^{(8)}}{x-1}+\frac{2 \ep  B^{(9)}}{x-1}-\frac{(\ao-1) \ao \
B_S^{(3)}}{x},
\\
%B8
\frac{\partial}{\partial x}B^{(8)} &=\Bigg(\frac{2 (d_1-2) \ep
}{x-1}-\frac{2 (d_1-2) \ep }{x}\Bigg) \
B^{(3)}+\Bigg(\frac{-4 \ep -1}{x}-\frac{2 (d_1 \ep -2 \ep \
)}{x-1}\Bigg) B^{(8)}+\\
&\Bigg(\frac{\ao-1}{x}-\frac{(\ao-1)^2}{(\ao-1) x-\
\ao}\Bigg) B_S^{(3)},
\\
%B9
\frac{\partial}{\partial x}B^{(9)} &=\frac{2 (\ep -1) B_S^{(2)} (\ao-1)^2}{\ep  (x \
\ao-\ao-x)^2}+\Bigg(\frac{2 (d_1-2) \ep }{x-1}-\frac{2 (d_1-2) \ep \
}{x}\Bigg) B^{(2)}+\\
&\Bigg(\frac{-4 \ep -1}{x}-\frac{2 (d_1 \ep -2 \ep \
)}{x-1}\Bigg) B^{(9)}+\Bigg(-\frac{2 (\ao-1)^2}{\ao (x \
\ao-\ao-x)}+\frac{(\ao-1)^2}{\ep  (x \ao-\ao-x)^2}+\\
&\frac{2 (\ao-1)}{\
\ao x}\Bigg) B_S^{(1)}+\Bigg(\frac{(\ao-1)^2}{(\ao-1) \
x-\ao}+\frac{1-\ao}{x}\Bigg) B_S^{(3)}.
\esp
\eeq
Knowing the solutions for the subtopology, we can solve for the remaining two master integrals $B^{(4)}$ and $B^{(7)}$. They fulfill the following differential equations, for $\delta=+1$,
\beq
\bsp
%B4
\frac{\partial}{\partial x}B^{(4)} &=\Bigg(\frac{-2 d_1 \ep +4 \ep -1}{2 x}+\frac{2 d_1 \ep -4 \ep \
+1}{2 (x-2)}\Bigg) B^{(1)}+\Bigg(\frac{-2 d_1 \ep +4 \ep \
-1}{x-2}-\frac{4 \ep }{x}\Bigg) B^{(4)}+\\
&\Bigg(\frac{\ao-1}{2 \
x}-\frac{(\ao-1)^2}{2 ((\ao-1) x-2 \ao)}\Bigg) B_S^{(1)},
\\
%B7
\frac{\partial}{\partial x}B^{(7)} &=\Bigg(\frac{(d_1-2) \ep
}{x-2}-\frac{(d_1-2) \ep }{x}\Bigg) \
B^{(3)}+\Bigg(\frac{-4 \ep -1}{x}-\frac{2 (d_1 \ep -2 \ep \
)}{x-2}\Bigg) B^{(7)}+\\
&\Bigg(\frac{\ao-1}{2 x}-\frac{(\ao-1)^2}{2 \
((\ao-1) x-2 \ao)}\Bigg) B_S^{(3)},
\esp
\eeq
whereas for $\delta=-1$ the differential equations read
\beq
\bsp
%B4
\frac{\partial}{\partial x}B^{(4)} &=\Bigg(\frac{-2 d_1 \ep +4 \ep -1}{2 x}+\frac{2 d_1 \ep -4 \ep \
+1}{2 (x-2)}\Bigg) B^{(1)}+\Bigg(\frac{-2 d_1 \ep +4 \ep \
-1}{x-2}-\frac{4 \ep }{x}\Bigg) B^{(4)}+\\
&\Bigg(\frac{\ao-1}{2 \
x}-\frac{(\ao-1)^2}{2 ((\ao-1) x-2 \ao)}\Bigg) B_S^{(1)},
\\
%B7
\frac{\partial}{\partial x}B^{(7)} &=\Bigg(\frac{(d_1-2) \ep
}{x-2}-\frac{(d_1-2) \ep }{x}\Bigg) \
B^{(3)}+\Bigg(\frac{-4 \ep -1}{x}-\frac{2 (d_1 \ep -2 \ep \
)}{x-2}\Bigg) B^{(7)}+\\
&\Bigg(\frac{\ao-1}{2 x}-\frac{(\ao-1)^2}{2 \
((\ao-1) x-2 \ao)}\Bigg) B_S^{(3)}.
\esp
\eeq
The initial conditions are the following.
At $x=0$, we have
\beq
B^{(1)}(x=0,\ep;\ao,d_1;\delta) = B^{(6)}(x=0,\ep;\ao,d_1;\delta) = B_{\ao}(1-4\ep,1+2d_1\ep)\,B(1-\ep,1-\ep).
\eeq
At $x=1$, we have
\beq
\bsp
B^{(5)}(x=1,\ep;\ao,d_1;\delta) &= B^{(1)}(x=1,\ep;\ao,d_1;\delta),\\
B^{(8)}(x=1,\ep;\ao,d_1;\delta) &= B^{(2)}(x=1,\ep;\ao,d_1;\delta),\\
B^{(9)}(x=1,\ep;\ao,d_1;\delta) &= B^{(3)}(x=1,\ep;\ao,d_1;\delta).
\esp
\eeq
At $x=2$, we have
\beq
\bsp
B^{(4)}(x=2,\ep;\ao,d_1;\delta) &= \frac{1}{2}\,B^{(1)}(x=2,\ep;\ao,d_1;\delta),\\
B^{(7)}(x=2,\ep;\ao,d_1;\delta) &= \frac{1}{2}\,B^{(3)}(x=2,\ep;\ao,d_1;\delta).
\esp
\eeq
It is easy to check that $B^{(1)}$ is finite at $x=0$ and $x=2$.
The integration constants of $B^{(2)}$ and $B^{(3)}$ can then be fixed in
an implicit way by requiring the residues of the general solution for
$B^{(1)}$ to vanish at $x=0$ and $x=2$.

Having the analytic expression for the master integrals, 
we can calculate the $\bint$-type integrals for a fixed integer value of $D_0$. 
We give the explicit results for $D_0=3$ in \appx{app:BIntegrals}.

\begin{figure}[t]
\includegraphics[scale=0.75]{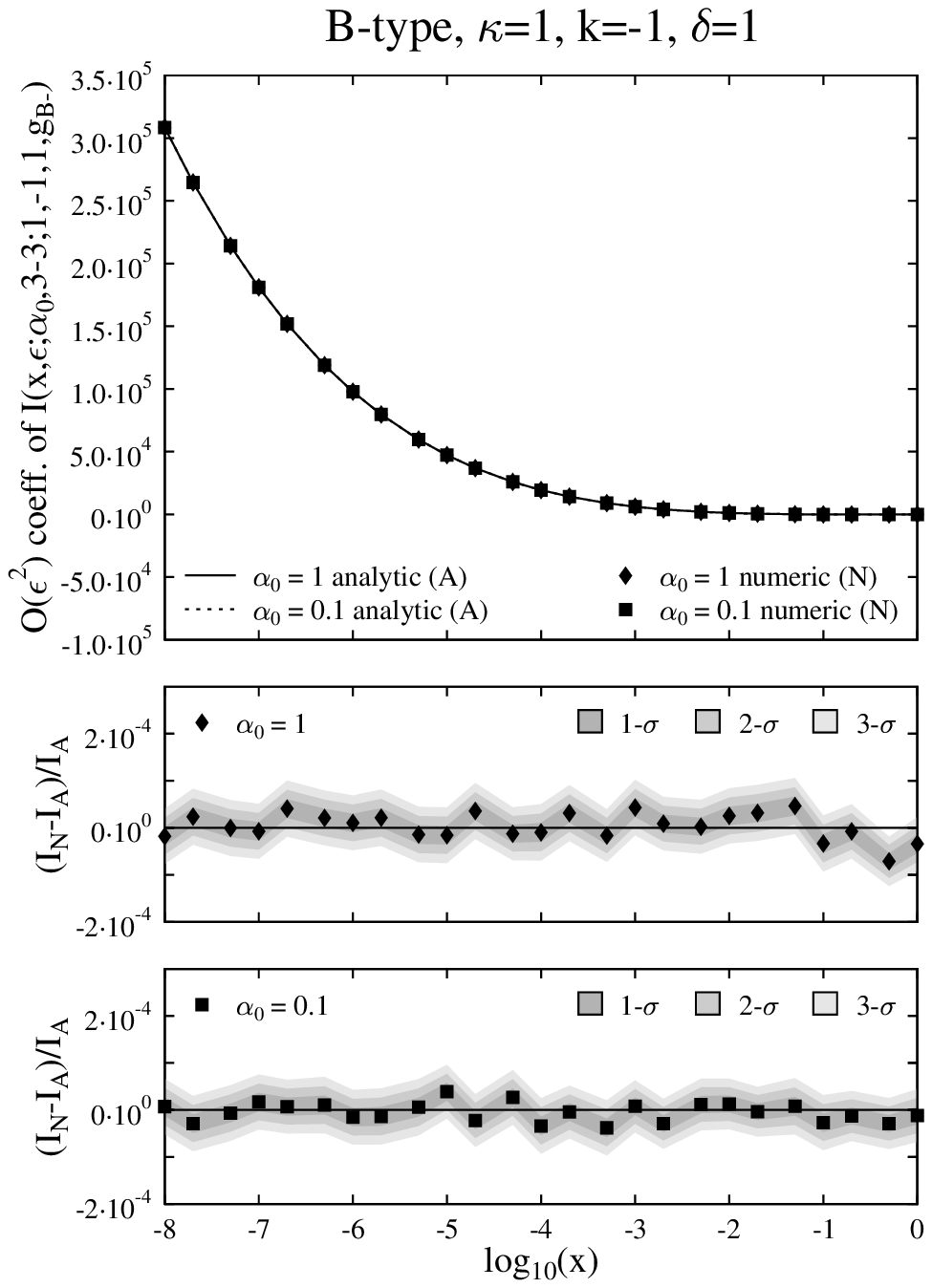}
\includegraphics[scale=0.75]{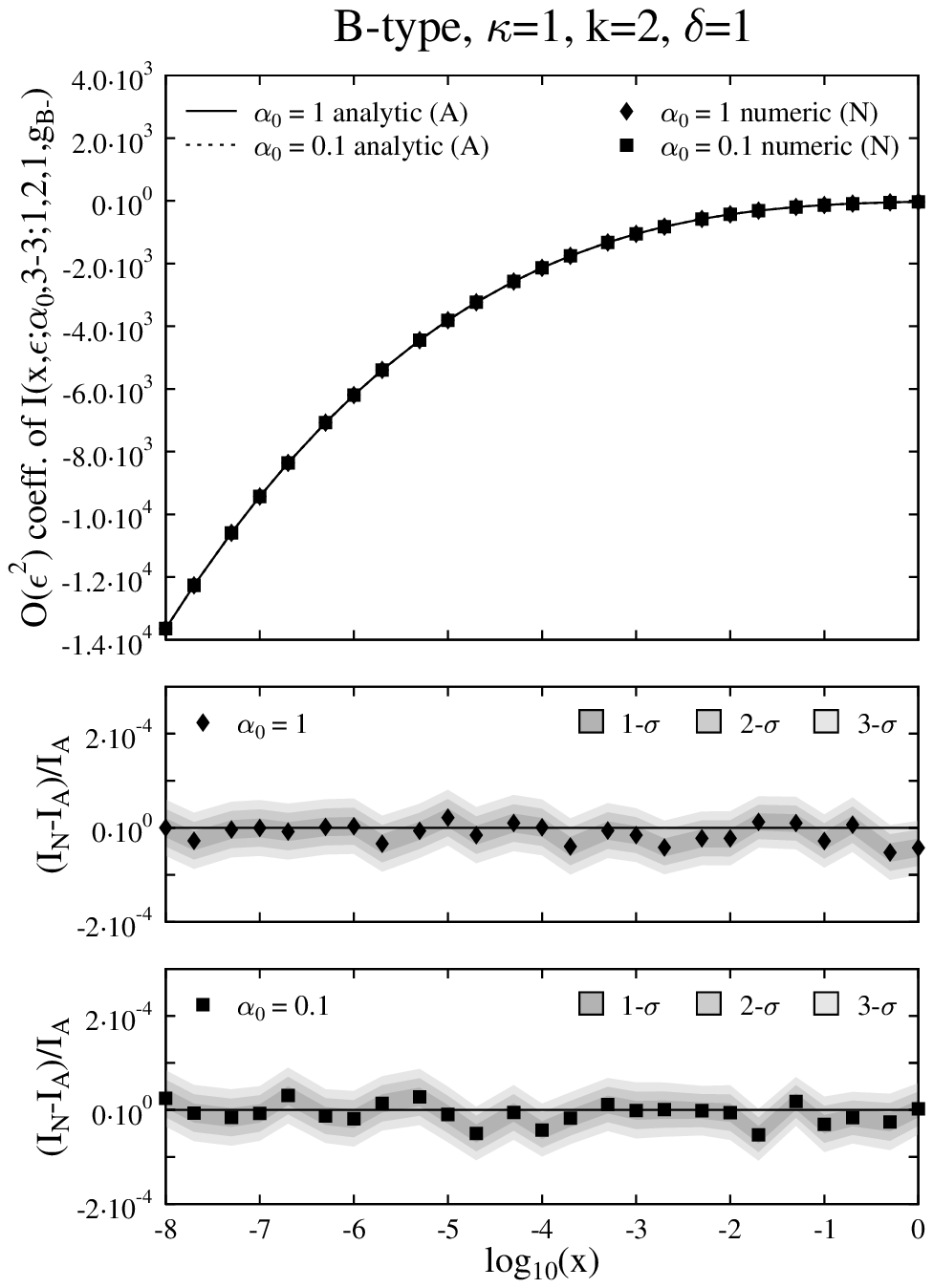}
\vspace{-2em}
\caption{\label{fig:Bfigs}
Representative results for the ${\cal B}$-type integrals. The plots show
the coefficient of the $\Oe{2}$ term in 
$\cI(x,\ep;\alpha_0,3-3\ep;1,k,1,g_{B_-})$ for $k=-1$ (left figure) and $k=2$ (right
figure) with $\alpha_0=0.1,1$.}
\end{figure}

In \fig{fig:Bfigs} we show some representative results of comparing the analytic and numeric computations for the $\ep^2$ coefficient in the expansion of $\cI(x,\ep;\alpha_0,3-3\ep;1,k,1,g_{B_-})$ for $k=-1,2$ and $\alpha_0=0.1,1$. The dependence on $\alpha_0$ is not visible on the plots. The two sets of results are in  excellent agreement for the whole $x$-range. For other (lower-order, thus simpler) expansion coefficients and/or other values of the parameters, we find similar agreement.

% Figs for B-type integrals

%%%
%%% The soft R\times(0)-type J*I
%%%

\section{The soft $R\times(0)$-type $\cJI$ integrals}
\label{sec:softR0}

In this section we calculate the integral defined in \eqn{eq:JIint}.
Substituting the result for the angular integral $\Omega^{(1,1)}$,
%obtained from \eqn{eq:Omegaik},
we can rewrite \eqn{eq:JIint} as 
\beq
\bsp
\label{eq:JIInt}
\cJI(Y,\ep;y_{0},d'_{0},\alpha_{0},d_{0};k) &=-Y\,B(-\ep,-\ep)\,\hypgeo(1,1,1-\ep;1-Y)\\
&\times
\int_{0}^{y_{0}}\rd y\,y^{-1-2\ep}(1-y)^{d'_{0}}\,\cI(y;\ep,\alpha_0,d_0;0,k,0,g_{A}).
\esp
\eeq
The hypergeometric function can be easily evaluated using the technique
described in \sect{sec:soft01}. The evaluation of the $y$ integral order
by order in $\ep$ is a little bit more cumbersome because the integrand
has two kinds of singularities,
\begin{enumerate}
\item The pole in $y=0$.
\item The integral $\cI$ is order by order logarithmically divergent for $y\sim0$, as can be easily seen from the $\ep$-expansion given in \appx{app:AIntegrals}. 
\end{enumerate}
The pole in $y=0$ can easily be factorized by performing the integration
by parts in $y$. The logarithmic singularities in $\cI$ however are more
subtle. We have to resum all these singularities before expanding the
integral. We find that we can write\footnote{We checked this assumption
explicitly on the $\ep$-expansion of $\cI$ given in 
\appx{app:AIntegrals}.}
\beq
\label{eq:Ifunc}
\cI(y;\ep,\alpha_0,d_0;0,k,0,g_{A}) = y^{-2\ep}\,I(y;\ep,\alpha_0,d_0;0,k,0,g_{A}),
\eeq
where $I$ is a function that is order by order finite in $y=0$. \eqn{eq:JIInt} can now be written as
\beq
\bsp
\cJI(Y,\ep;y_{0},d'_{0},\alpha_{0},d_{0};k) &=-Y\,B(-\ep,-\ep)\,\hypgeo(1,1,1-\ep;1-Y)\\
&\times\left\{-\frac{1}{4\ep}\,\yo^{-4\ep}(1-\yo)^{d'_{0}}\,I(\yo;\ep,\alpha_0,d_0;0,k,0,g_{A})\right.\\
&+\left.\frac{1}{4\ep}\,\int_{0}^{y_{0}}\rd y\,y^{-4\ep}\frac{\partial}{\partial y}\left[(1-y)^{d'_{0}}\,I(y;\ep,\alpha_0,d_0;0,k,0,g_{A})\right]\right\}.
\esp
\eeq
As $I$ does not have logarithmic divergences, the derivative does not
produce any poles, and so the integral is uniformly convergent.
We can thus just expand the integrand into
a power series in $\ep$ and integrate order by order, using the
definition of the \hpls\footnote{Notice that the rational part of $I$
gives a non-vanishing contribution to the lower integration limit that
has to be subtracted.}, \eqn{eq:hpldef}. The result for $D_0=D'_0=3$ is
given in 
\appx{app:JIIntegrals}.

\begin{figure}[t]
\includegraphics[scale=0.75]{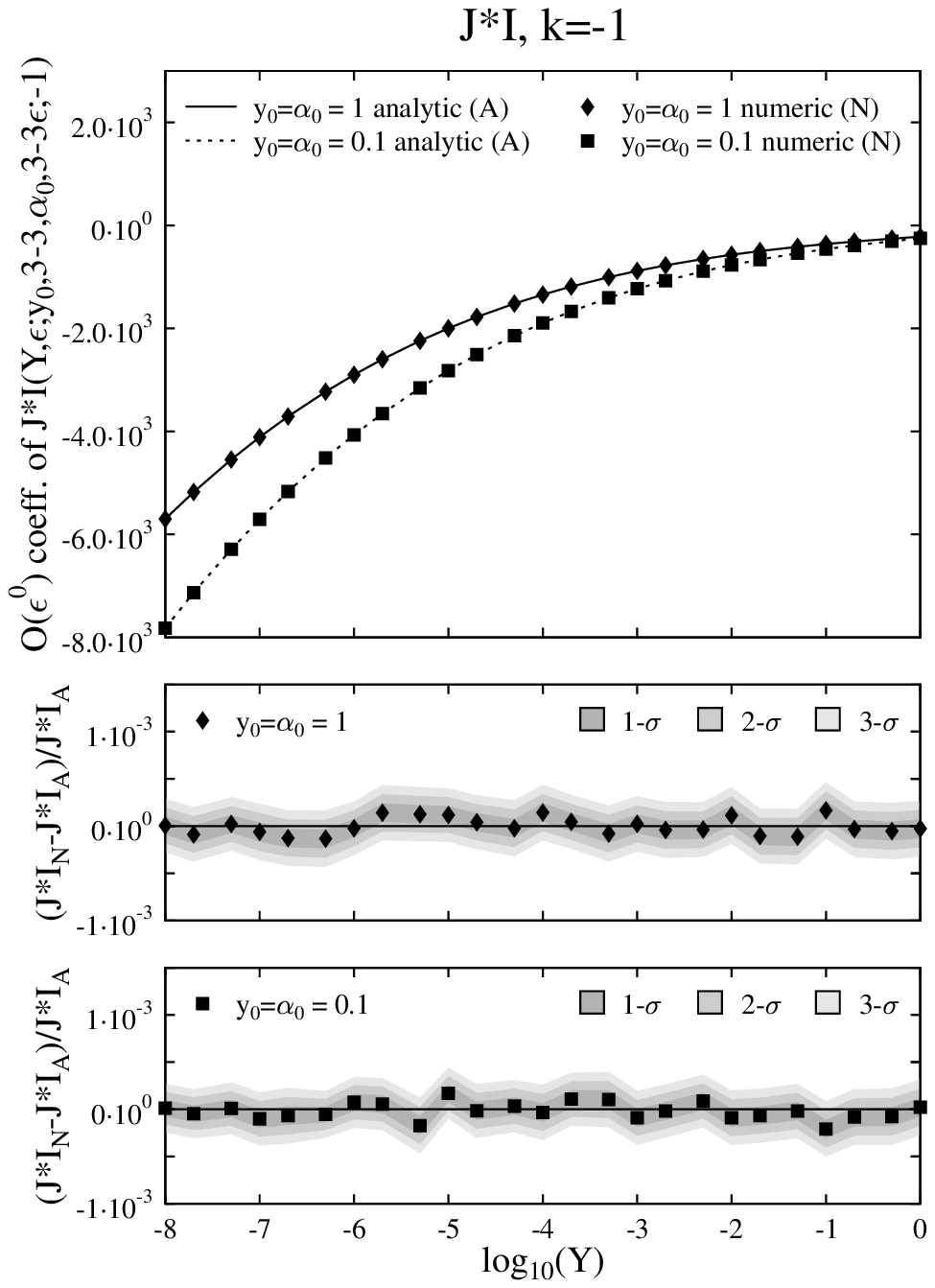}
\includegraphics[scale=0.75]{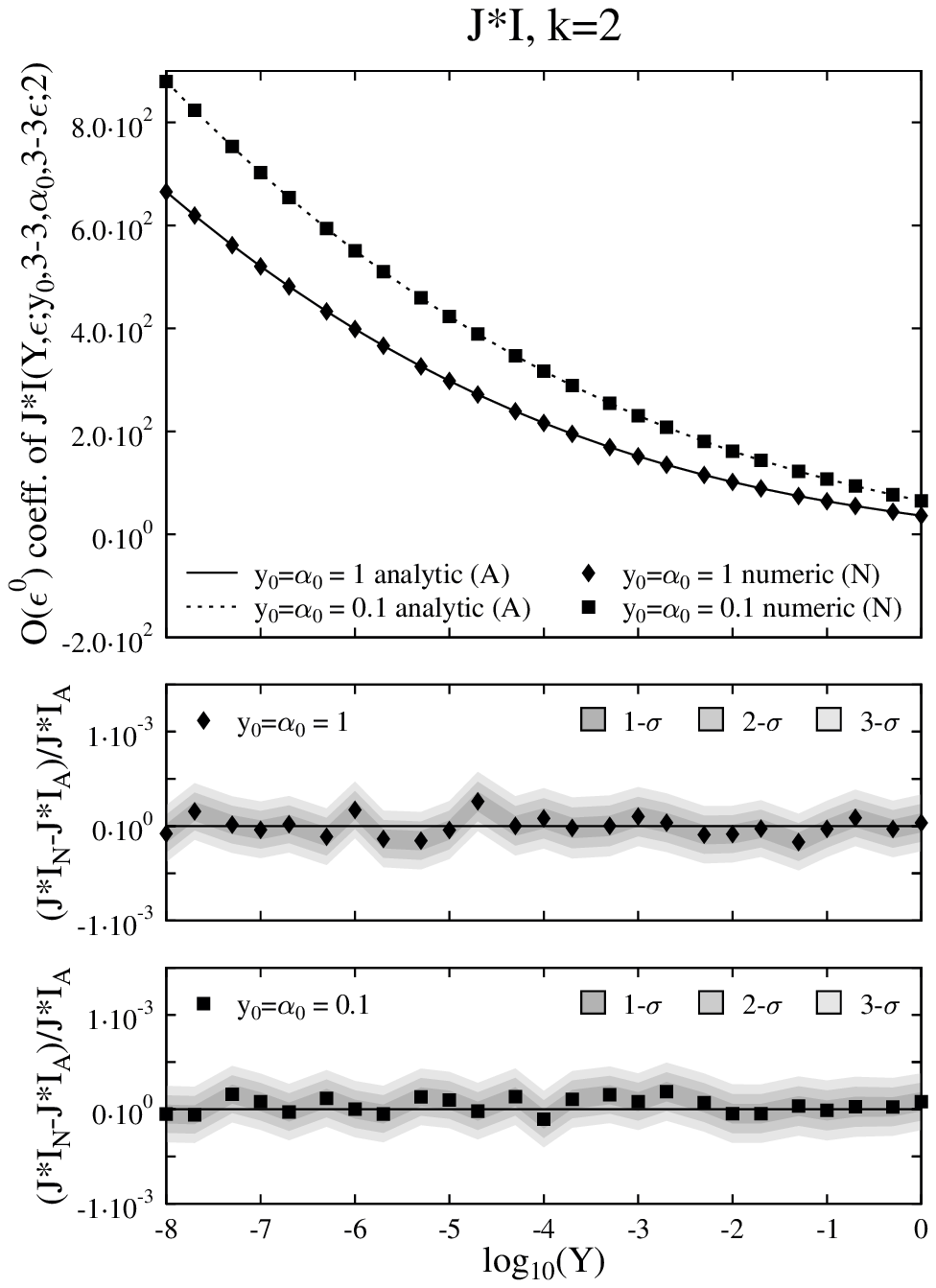}
\vspace{-2em}
\caption{\label{fig:JIfigs}
Representative results for the $\cJI$ integrals. The plots show
the coefficient of the $\Oe{0}$ term in 
$\cJI(Y,\ep;y_0,3-3\ep,\alpha_0,3-3\ep;k)$ for $k=-1$ (left figure) and $k=2$ (right figure) with $y_0=\alpha_0=0.1,1$.}
\end{figure}

As representative examples, in \fig{fig:JIfigs} we compare the analytic
and numeric results for the $\ep^0$ coefficient in the expansion of
$\cJI(Y,\ep;y_0,3-3\ep,\alpha_0,3-3\ep;k)$ for $k=-1,2$ and
$y_0=\alpha_0=0.1,1$. The two computations  agree very well over the
whole $Y$-range. Other (lower-order, thus simpler) expansion coefficients
and/or other values of the parameters show similar agreement.

% Figs for J*I integrals

%%%
%%% The soft-collinear, R\times(0)-type K*I
%%%

\section{The soft-collinear $R\times(0)$-type $\cKI$ integrals}
\label{sec:softcollR0}
In this section we calculate the integral defined in \eqn{eq:KIint}.
The $\varphi$ integral is given in \eqn{eq:phiint}. Putting $\cos \vartheta = 2\xi-1$, the integral can be rewritten as
\beq
\bsp
\cKI(\ep&;y_{0},d'_{0},\alpha_{0},d_{0};k) =\\
&2\,B(1-\ep,-\ep)\,\int_{0}^{y_{0}}\rd y\,y^{-1-4\ep}(1-y)^{d'_{0}}\,I(y;\ep,\alpha_0,d_0;0,k,0,g_{A})\\
&+2\,B(1-\ep,1-\ep)\,\,\int_{0}^{y_{0}}\rd y\,y^{-4\ep}(1-y)^{d'_{0}-1}\,I(y;\ep,\alpha_0,d_0;0,k,0,g_{A}),
\esp\eeq
where $I$ was defined in \eqn{eq:Ifunc}. The first integral is exactly
the same as in \sect{sec:softR0}.  The second integral
is uniformly convergent, so we can just
expand under the integration sign, and integrate order by order. 
The result for $D_0=D'_0=3$ is given in \appx{app:KIIntegrals}.

%%%
%%% Numerical evaluation of integrated subtraction terms
%%%

\section{Numerical evaluation of integrated subtraction terms}
\label{sec:numerical}

Let us briefly discuss the numerical evaluation of the integrals which were
analytically 
computed in the previous sections. First of all, if the singular integrals in the
chosen integration variables are non-overlapping and furthermore occur in a
single point in the integration region (which can always be mapped to the origin)
then we can isolate the poles using standard residuum subtraction. Consider as
an example the $\kint(\ep;y_0,d'_0;\kappa)$ integral of \eqn{eq:kint}:
\beq
\kint(\ep;y_0,d'_0;\kappa) =
2\int_0^{y_0}\rd y\,\int_0^1\rd\xi\,\,y^{-1-2(1+\kap)\ep}(1-y)^{d'_0-1}\xi^{-\ep}(1-\xi)^{-1-(1+\kap)\ep}(1-y\xi)^{1+\kap\ep}.
\eeq
We see that the singularities come only from $y\to 0$ and $\xi\to 1$
and there are no overlapping singularities%
\footnote{Overlapping singularities would be signaled by the presence
of `composite' denominators, i.e.\ denominators which vanish only when
{\it both} $y\to 0$ and $\xi\to 1$, but not otherwise.}.
After remapping the singularity at $\xi\to 1$ to the origin by setting
$\xi\to 1-\xi$, we can easily extract the poles using residuum
subtraction. The finite integrals that are left over are
straightforward to evaluate numerically.  

In general, we encounter integrals which contain overlapping divergences. 
A typical example occurs in the ${\cal A}$-type collinear integral for $k=-1$:
\beeq
{\cal A}(x,\ep;\alpha_0,d_0;\kappa,-1) &=& 
\int_0^{\alpha_0} \rd\alpha\,\int_0^1\rd v\,\alpha^{-1-(1+\kappa)\ep}
(1-\alpha)^{2d_0-1}[\alpha-(1-\alpha)x]^{-1-(1+\kappa)\ep}
\nn\\&&\times
v^{-\ep}(1-v)^{-\ep}\,\frac{2\alpha+(1-\alpha)x}{\alpha+(1-\alpha)xv}\,.
\label{eq:Anum}
\eeeq
The singularities are at $\alpha\to 0$ and $v\to 0$, but the presence of the `composite' denominator $\alpha+(1-\alpha)xv$ (which vanishes only when 
both $\alpha\to 0$ and $v\to 0$) does not allow the use of simple residuum subtraction to isolate the poles. Instead one has to disentangle the 
overlapping singularities first, which we achieve by using sector decomposition \cite{Heinrich:2008si}.
This consists of the following steps
\begin{itemize}
\item We first transform the integral so that the range of integration is the unit
square. This is easily achieved by setting $\alpha \to \alpha_0\alpha$.
\item Then we split the integral into two `sectors' by inserting
$1=[\Theta(\alpha-v)+\Theta(v-\alpha)]$ in the integrand.
\item Next, we transform the variables in each sector such that the integration 
region is remapped to the unit square. When $\alpha\ge v$, we use $v\to \alpha v$, while for $v\ge\alpha$ we need $\alpha\to v\alpha$.
\item Notice that in each sector either $\alpha$ or $v$ now factorizes from the
composite denominator and the remainder is finite at $\alpha=0,v=0$.
\item We can now apply residuum subtraction in each sector to extract the $\ep$ poles.
\end{itemize}
As before, the finite integrals left over are straightforward to compute numerically.

We have written a {\em Mathematica} package for the extraction of poles
using these techniques. The program produces {\sc Fortran} codes that
may immediately be used in numerical integration programs. To produce
the numerical results, we used the Monte Carlo integrator {\sc Vegas}
\cite{Lepage:1980dq}. The program {\sc SectorDecomposition} of 
\Ref{Bogner:2007cr} was used to check our implementation.  

The main advantage of the numerical approach based on sector
decomposition is its ability to handle a very wide class of integrals
in a unified way.  For example, the appearance of a new type of
denominator in the integrals does not require any changes in the
implementation of the algorithm. The drawbacks of the method stem from
the fact that it does not give the expansion coefficients of the
integrals directly. Rather it produces (usually very cumbersome) integral
representations of them.  Thus the evaluation of (any of) the integrated
subtraction terms at any one point requires performing a numerical
integration, which is time consuming and has an intrinsic numerical
uncertainty. It might then be objected that this gives a slow and
inaccurate%
\footnote{It certainly does not seem practical to obtain e.g.\ eight
significant digits with this technique.}
evaluation of the integrals. There is weight behind this objection,
however let us note two things: first of all, the integrated subtraction
terms can be computed once and for all. (At least for given values of
$\alpha_0$ and $y_0$. Changing these values requires the recomputation of
all numerical integrals.)
The final results are smooth functions and can be given e.g.\ in the
form of interpolating tables. Second, in an actual computation, one
expects that for any observable the relative uncertainty associated with
the phase space integrations would be much greater than the relative
uncertainty of the integrated subtraction terms. Therefore it can be
argued that numerical results alone could be enough to produce physical
higher-order computations.

Nevertheless, having analytical results is very useful for many purposes. First of all, in a higher-order computation, the $\ep$ poles of the integrated subtraction terms need to cancel the $\ep$ poles coming from the loop matrix elements in the virtual corrections. The correct cancellation of these poles can only be demonstrated rigorously once the pole structure of the integrated subtraction terms is exhibited analytically. With numerical results, one can only cancel the poles to whatever precision the numerical integrations were carried to. Second, in terms of speed and precision of evaluation, analytical results are very fast and very accurate compared to numerical ones. In fact, the analytical computation is only limited in this regard by the capability to compute \dhpls in a fast and accurate way. Building a routine which accomplishes this task is not too difficult and has already been done in other cases \cite{Gehrmann:2001jv}. 
It is feasible to construct an implementation that evaluates a \emph{2dHPL} with a relative accuracy of ${\cal O}(10^{-15})$ in less than 1 millisecond on a standard PC. 
In the present context, the evaluation of the \dhpls is further simplified by the fact that the arguments are always between zero and one, and the functions are all real in this range. There is then no need to worry about logarithms developing imaginary parts that have to be treated consistently. Finally, analytical results are `user-friendly' in the sense that their use is not tied to any specific implementation or code. This is a particularly relevant concern when setting up a universal subtraction algorithm.

%%%
%%% Conclusions
%%%

\section{Conclusions}
\label{conclusion}

In this work we have analytically evaluated some of the integrals needed
for computing the integrated real-virtual counterterms that appear in
the subtraction scheme for computing NNLO jet cross sections proposed
in \Refs{Somogyi:2006da, Somogyi:2006db}.  Such integrals have to be
computed once and for all and their knowledge is necessary in order to
make the subtraction scheme an effective tool.  Our method is an
adaptation of the current technique used to compute multi-loop Feynman
diagrams: after an algebraic reduction to a class of independent
amplitudes, integration-by-parts identities are generated and solved
with the Laporta algorithm to achieve reduction to master integrals. 
The latter are computed with the differential equation method and are
expressed in terms of one- and two-dimensional harmonic polylogarithms;
the $\ep$-expansion has been performed up to the required order in
$\ep$.  The numerical evaluation of harmonic polylogarithms has been
treated in many works, where it has been shown that it can be fast and
accurate \cite{Gehrmann:2001pz, Gehrmann:2001jv}; there is not any
specific problem in our case either.  A check of all our analytic
results has been made by means of a direct numerical calculation of the
integrals, typically with an accuracy $\lesssim\mathcal{O}(10^{-4})$.
Specific properties of the present calculation are:
\begin{enumerate}
\item
the partial fractioning in many variables of the integrands, which
requires in general the introduction of new denominators;
\item
the occurrence of surface terms in integration-by-parts identities,
consisting of integrals of lower dimensionality than the original ones;
\item
a non-trivial basis extension for two-dimensional harmonic
polylogarithms, together with corresponding consistency relations in
order to have complete analytic control over the results.
\end{enumerate}
Our method can in principle be applied to the analytic evaluation of
classes of more complicated real-virtual integrated counter-terms, such
as the $C$ and $D$-type integrals of \sect{sec:coll01}, even though the
solution of the ibps in the latter case can be rather lengthy and the
explicit expressions of the integrals can become rather cumbersome. 
For more complicated integrals, it is probably convenient to modify the
algorithm used in this work in order to avoid the generation of
additional denominators through the multiple partial fractioning.  A
preliminary study shows for example that our algorithm produces a lot
of new denominators in the case of 3-dimensional integrals. For
example, in reducing the integral over $x$ and $y$ considered in the
introduction, one should not subject the ``overlapping denominator''
$1/(1-x y)$ to any partial fractioning; this way one ends up with
3-denominator integrals without any additional denominator.

%%%
%%% Acknowledgments
%%%

\acknowledgments{C.D. is a research fellow of the \emph{Fonds National
de la Recherche Scientifque}, Belgium.  This work was partly supported
by MIUR under contract 2006020509$_0$04, by the EC Marie-Curie Research
Training Network ``Tools and Precision Calculations for Physics
Discoveries at Colliders'' under contract MRTN-CT-2006-035505, by the
Hungarian Scientific Research Fund grand OTKA K-60432 and by the Swiss
National Science Foundation (SNF) under contract 200020-117602. We
thank the Galileo Galilei Institute for Theoretical Physics and the ``Laboratori Nazionali di Frascati'' for the
hospitality and the INFN for partial support during the completion of
this work.
}

%%%
%%% The bibliography
%%%

\newpage

\bibliography{RVIntegralsBib}

%%%
%%% appendices
%%%

\newpage
\appendix

%%%
%%% Spin-averaged splitting kernels
%%%

\section{Spin-averaged splitting kernels}
\label{app:APfcns}

In this Appendix we recall the explicit expressions for the 
spin-averaged splitting kernels that enter \eqn{eq:collintegrals}.

The azimuthally averaged Altarelli--Parisi splitting kernels read
\beeq
P^{(0)}_{g_ig_r}(z_i,z_r;\ep) 
&=&
2\CA \left[\frac{1}{z_i}+\frac{1}{z_r}-2+z_i z_r\right]\,,
\label{eq:P0gg}
\\
P^{(0)}_{q_i\qb_r}(z_i,z_r;\ep) 
&=&
\TR \left[1-\frac{2}{1-\ep}z_i z_r\right]\,,
\label{eq:P0qq}
\\
P^{(0)}_{q_ig_r}(z_i,z_r;\ep) 
&=&
\CF \left[\frac{2}{z_r}-2 + (1-\ep)z_r\right]\,,
\label{eq:P0qg}
\eeeq
while their one-loop generalizations are
\beq
P^{(1)}_{f_i f_r}(z_i,z_r;\ep) 
=
r_{\S,\rm{ren}}^{f_i f_r}(z_i,z_r;\ep)
P^{(0)}_{f_i f_r}(z_i,z_r;\ep) 
+\left\{
\begin{array}{lr}
\displaystyle{
2\CA\, r_\NS^{gg}\,\frac{1-2\ep z_i z_r}{1-\ep}}\,,
& \mbox{if}\quad f_if_r = gg\,,
\\ 0\,, & \mbox{if}\quad f_if_r = q\qb\,, \\
\displaystyle{
\CF\, r_\NS^{qg}\,(1-\ep z_r)}\,, 
& \mbox{if}\quad f_if_r = qg\,.
\end{array}
\right.
\label{eq:P1av}
\eeq
The renormalized $r^{f_i f_r}_{\S,\rm{ren}}(z_i,z_r;\ep)$ functions 
that appear above are expressed in terms of the corresponding 
unrenormalized ones as
\beq
r^{f_i f_r}_{\S,\rm ren}(z_i,z_r;\ep)=
r^{f_i f_r}_\S(z_i,z_r;\ep)
- \frac{\beta_0}{2\ep}\,\frac{S_\ep}{(4\pi)^2 c_\Gamma}
\,\left[\left(\frac{\mu^2}{s_{ir}}\right)^{\ep}\cos(\pi\ep)\right]^{-1}
\,,
\label{rSUV}
\eeq
where 
the unrenormalized $r^{f_i f_r}_{\S}(z_i,z_r;\ep)$ factors may be
written in the following form
\beeq
r_{\S}^{gg}(z_i,z_r;\ep) &=& 
\frac{\CA}{\ep^2}\bigg[-\frac{\pi\ep}{\sin(\pi\ep)} 
\left(\frac{z_i}{z_r}\right)^\ep + z_i^\ep {}_2F_1(\ep,\ep,1+\ep,z_r)
\nn\\&&\qquad \quad
- z_i^{-\ep} {}_2F_1(-\ep,-\ep,1-\ep,z_r)\bigg]\,,
\label{rSgg}
\\
r_{\S}^{q\qb}(z_i,z_r;\ep) &=& 
\frac{1}{\ep^2}(\CA-2\CF)+
\frac{\CA}{\ep^2}\bigg[-\frac{\pi\ep}{\sin(\pi\ep)} 
\left(\frac{z_i}{z_r}\right)^\ep + z_i^\ep {}_2F_1(\ep,\ep,1+\ep,z_r)
\nn\\&&\qquad\qquad\qquad\qquad\quad\;\;
-\,\frac{\pi\ep}{\sin(\pi\ep)} 
\left(\frac{z_r}{z_i}\right)^\ep + z_r^\ep {}_2F_1(\ep,\ep,1+\ep,z_i)
\bigg]
\nn\\&&\qquad\quad
+\,\frac{1}{1-2\ep}\left[\frac{\beta_0-3\CF}{\ep}+\CA-2\CF
+\frac{\CA+4\TR(\Nf-\Ns)}{3(3-2\ep)}\right]\,,
\label{rSqq}
\\
r_{\S}^{qg}(z_i,z_r;\ep) &=& 
-\frac{1}{\ep^2}\bigg[2(\CA-\CF) + \CA \frac{\pi\ep}{\sin(\pi\ep)} 
\left(\frac{z_i}{z_r}\right)^\ep - \CA z_i^\ep {}_2F_1(\ep,\ep,1+\ep,z_r)
\nn\\&&\qquad \quad
-(\CA-2\CF)z_i^{-\ep} {}_2F_1(-\ep,-\ep,1-\ep,z_r)\bigg]\,.
\label{rSqg}
\eeeq
The $r^{f_i f_r}_\NS$ non-singular factors are
\beq
r_\NS^{g g} =
\frac{\CA(1-\ep)-2\TR (\Nf-\Ns)}{(1-2\ep)(2-2\ep)(3-2\ep)}\,,
\qquad
r_\NS^{q g} =
\frac{\CA-\CF}{1-2\ep}\,.
\label{rNSqg}
\eeq
For QCD, $\Ns=0$.
Finally $\beta_0$ in \eqns{rSUV}{rSqq} is given by
\beq
\beta_0 = \frac{11}{3}\CA-\frac{4}{3}\TR\Nf-\frac{2}{3}\TR\Ns\,.
\eeq

%%%
%%% Results
%%%

\newpage

%%%
%%% The J integrals
%%%

\section{The $\cJ$ integrals}
\label{app:JIntegrals}
%
% This file contains the TeX output produced by Mathematica for the integral K, for arbitrary kap  and D0 = 3 +d'1 ep
%
The $\eps$ expansion for this integral reads
\beq
\cJ(Y,\eps;y_0,3+d'_1\eps;\kappa)=\frac{1}{\eps^2}i_{-2}^{(\kap)}+\frac{1}{\eps}i_{-1}^{(\kap)}+i_0+\eps i_1^{(\kap)}+\eps^2i_2^{(\kap)} +\ocal\left(\eps^3\right),
\eeq
where
%1/ep piece
\brp
i_{-2}^{(\kap)}=-\frac{1}{(\kappa +1)^2},
\erp
\brp
i_{-1}^{(\kap)}=-\frac{2 \yo^3}{3 (\kappa +1)}+\frac{3 \yo^2}{\kappa +1}-\frac{6 \
\yo}{\kappa +1}+\frac{2 H(0;\yo)}{\kappa +1}+\frac{H(0;Y)}{\kappa +1},
\erp
% ep^0
\brp
i_0^{(\kap)}=\frac{2 d_1' \yo^3}{9 (3 \kappa +1)}+\frac{2 d_1' \kappa  \yo^3}{9 (3 \
\kappa +1)}-\frac{8 \kappa  \yo^3}{9 (3 \kappa +1)}-\frac{4 \yo^3}{9 \
(3 \kappa +1)}-\frac{7 d_1' \yo^2}{6 (3 \kappa +1)}-\frac{7 d_1' \
\kappa  \yo^2}{6 (3 \kappa +1)}+\frac{20 \kappa  \yo^2}{3 (3 \kappa \
+1)}+\frac{3 \yo^2}{3 \kappa +1}+\frac{11 d_1' \yo}{3 (3 \kappa +1)}+\
\frac{11 d_1' \kappa  \yo}{3 (3 \kappa +1)}-\frac{86 \kappa  \yo}{3 \
(3 \kappa +1)}-\frac{12 \yo}{3 \kappa +1}+\Big(\frac{4 \yo^3}{3}-6 \
\yo^2+12 \yo\Big) H(0;\yo)+\Big(\frac{2 \yo^3}{3}-3 \yo^2+6 \yo-2 \
H(0;\yo)\Big) H(0;Y)+\Big(\frac{2 d_1' \yo^3}{3 (3 \kappa \
+1)}+\frac{2 d_1' \kappa  \yo^3}{3 (3 \kappa +1)}+\frac{4 \kappa  \
\yo^3}{3 (3 \kappa +1)}-\frac{3 d_1' \yo^2}{3 \kappa +1}-\frac{3 d_1' \
\kappa  \yo^2}{3 \kappa +1}-\frac{6 \kappa  \yo^2}{3 \kappa \
+1}+\frac{6 d_1' \yo}{3 \kappa +1}+\frac{6 d_1' \kappa  \yo}{3 \kappa \
+1}+\frac{12 \kappa  \yo}{3 \kappa +1}-\frac{11 d_1'}{3 (3 \kappa \
+1)}-\frac{11 d_1' \kappa }{3 (3 \kappa +1)}-\frac{22 \kappa }{3 (3 \
\kappa +1)}\Big) H(1;\yo)-4 H(0,0;\yo)-H(0,0;Y)+\Big(-\frac{2 \kappa  \
d_1'}{(\kappa +1)^2}-\frac{2 d_1'}{(\kappa +1)^2}-\frac{4 \kappa \
}{(\kappa +1)^2}\Big) H(0,1;\yo)-H(1,0;Y),
\erp
% ep^1
\brp
i_1^{(\kap)}=-\frac{2 d_1'^2 \yo^3}{27 (3 \kappa +1)}+\frac{8 d_1' \yo^3}{27 (3 \
\kappa +1)}-\frac{2 d_1'^2 \kappa  \yo^3}{27 (3 \kappa +1)}+\frac{16 \
d_1' \kappa  \yo^3}{27 (3 \kappa +1)}-\frac{28 \kappa  \yo^3}{27 (3 \
\kappa +1)}-\frac{8 \yo^3}{27 (3 \kappa +1)}+\frac{17 d_1'^2 \
\yo^2}{36 (3 \kappa +1)}-\frac{22 d_1' \yo^2}{9 (3 \kappa \
+1)}+\frac{17 d_1'^2 \kappa  \yo^2}{36 (3 \kappa +1)}-\frac{49 d_1' \
\kappa  \yo^2}{9 (3 \kappa +1)}+\frac{73 \kappa  \yo^2}{6 (3 \kappa \
+1)}+\frac{3 \yo^2}{3 \kappa +1}-\frac{49 d_1'^2 \yo}{18 (3 \kappa \
+1)}+\frac{151 d_1' \yo}{9 (3 \kappa +1)}-\frac{49 d_1'^2 \kappa  \
\yo}{18 (3 \kappa +1)}+\frac{355 d_1' \kappa  \yo}{9 (3 \kappa \
+1)}-\frac{319 \kappa  \yo}{3 (3 \kappa +1)}-\frac{24 \yo}{3 \kappa \
+1}+\Big(-\frac{4 d_1' \yo^3}{9 (3 \kappa +1)}-\frac{4 d_1' \kappa  \
\yo^3}{3 (3 \kappa +1)}+\frac{40 \kappa  \yo^3}{9 (3 \kappa \
+1)}+\frac{8 \yo^3}{9 (3 \kappa +1)}+\frac{7 d_1' \yo^2}{3 (3 \kappa \
+1)}+\frac{7 d_1' \kappa  \yo^2}{3 \kappa +1}-\frac{98 \kappa  \
\yo^2}{3 (3 \kappa +1)}-\frac{6 \yo^2}{3 \kappa +1}-\frac{22 d_1' \
\yo}{3 (3 \kappa +1)}-\frac{22 d_1' \kappa  \yo}{3 \kappa \
+1}+\frac{416 \kappa  \yo}{3 (3 \kappa +1)}+\frac{24 \yo}{3 \kappa \
+1}\Big) H(0;\yo)+\Big(-\frac{2 d_1'^2 \yo^3}{9 (3 \kappa \
+1)}+\frac{4 d_1' \yo^3}{9 (3 \kappa +1)}-\frac{2 d_1'^2 \kappa  \
\yo^3}{9 (3 \kappa +1)}+\frac{4 d_1' \kappa  \yo^3}{9 (3 \kappa +1)}+\
\frac{4 \kappa  \yo^3}{3 (3 \kappa +1)}+\frac{7 d_1'^2 \yo^2}{6 (3 \
\kappa +1)}-\frac{3 d_1' \yo^2}{3 \kappa +1}+\frac{7 d_1'^2 \kappa  \
\yo^2}{6 (3 \kappa +1)}-\frac{13 d_1' \kappa  \yo^2}{3 (3 \kappa \
+1)}-\frac{29 \kappa  \yo^2}{3 (3 \kappa +1)}-\frac{11 d_1'^2 \yo}{3 \
(3 \kappa +1)}+\frac{12 d_1' \yo}{3 \kappa +1}-\frac{11 d_1'^2 \kappa \
 \yo}{3 (3 \kappa +1)}+\frac{64 d_1' \kappa  \yo}{3 (3 \kappa \
+1)}+\frac{122 \kappa  \yo}{3 (3 \kappa +1)}+\frac{49 d_1'^2}{18 (3 \
\kappa +1)}-\frac{85 d_1'}{9 (3 \kappa +1)}+\frac{49 d_1'^2 \kappa \
}{18 (3 \kappa +1)}-\frac{157 d_1' \kappa }{9 (3 \kappa +1)}-\frac{97 \
\kappa }{3 (3 \kappa +1)}\Big) H(1;\yo)+\frac{1}{6} \pi ^2 \
H(1;Y)+\Big(-\frac{56 \kappa  \yo^3}{3 (3 \kappa +1)}-\frac{8 \
\yo^3}{3 (3 \kappa +1)}+\frac{84 \kappa  \yo^2}{3 \kappa +1}+\frac{12 \
\yo^2}{3 \kappa +1}-\frac{168 \kappa  \yo}{3 \kappa +1}-\frac{24 \
\yo}{3 \kappa +1}\Big) H(0,0;\yo)+\Big(-\frac{14 \kappa  \yo^3}{3 (3 \
\kappa +1)}-\frac{2 \yo^3}{3 (3 \kappa +1)}+\frac{21 \kappa  \yo^2}{3 \
\kappa +1}+\frac{3 \yo^2}{3 \kappa +1}-\frac{42 \kappa  \yo}{3 \kappa \
+1}-\frac{6 \yo}{3 \kappa +1}+\Big(\frac{14 \kappa }{(\kappa \
+1)^2}+\frac{2}{(\kappa +1)^2}\Big) H(0;\yo)\Big) \
H(0,0;Y)+\Big(-\frac{4 d_1' \yo^3}{3 (3 \kappa +1)}-\frac{4 d_1' \
\kappa  \yo^3}{3 \kappa +1}-\frac{16 \kappa  \yo^3}{3 (3 \kappa +1)}+\
\frac{6 d_1' \yo^2}{3 \kappa +1}+\frac{18 d_1' \kappa  \yo^2}{3 \
\kappa +1}+\frac{24 \kappa  \yo^2}{3 \kappa +1}-\frac{12 d_1' \yo}{3 \
\kappa +1}-\frac{36 d_1' \kappa  \yo}{3 \kappa +1}-\frac{48 \kappa  \
\yo}{3 \kappa +1}\Big) H(0,1;\yo)+H(0;Y) \Big(-\frac{2 d_1' \yo^3}{9 \
(3 \kappa +1)}-\frac{2 d_1' \kappa  \yo^3}{3 (3 \kappa +1)}+\frac{20 \
\kappa  \yo^3}{9 (3 \kappa +1)}+\frac{4 \yo^3}{9 (3 \kappa \
+1)}+\frac{7 d_1' \yo^2}{6 (3 \kappa +1)}+\frac{7 d_1' \kappa  \
\yo^2}{2 (3 \kappa +1)}-\frac{49 \kappa  \yo^2}{3 (3 \kappa \
+1)}-\frac{3 \yo^2}{3 \kappa +1}-\frac{11 d_1' \yo}{3 (3 \kappa +1)}-\
\frac{11 d_1' \kappa  \yo}{3 \kappa +1}+\frac{208 \kappa  \yo}{3 (3 \
\kappa +1)}+\frac{12 \yo}{3 \kappa +1}+\Big(-\frac{28 \kappa  \
\yo^3}{3 (3 \kappa +1)}-\frac{4 \yo^3}{3 (3 \kappa +1)}+\frac{42 \
\kappa  \yo^2}{3 \kappa +1}+\frac{6 \yo^2}{3 \kappa +1}-\frac{84 \
\kappa  \yo}{3 \kappa +1}-\frac{12 \yo}{3 \kappa +1}\Big) \
H(0;\yo)+\Big(-\frac{2 d_1' \yo^3}{3 (3 \kappa +1)}-\frac{2 d_1' \
\kappa  \yo^3}{3 \kappa +1}-\frac{8 \kappa  \yo^3}{3 (3 \kappa \
+1)}+\frac{3 d_1' \yo^2}{3 \kappa +1}+\frac{9 d_1' \kappa  \yo^2}{3 \
\kappa +1}+\frac{12 \kappa  \yo^2}{3 \kappa +1}-\frac{6 d_1' \yo}{3 \
\kappa +1}-\frac{18 d_1' \kappa  \yo}{3 \kappa +1}-\frac{24 \kappa  \
\yo}{3 \kappa +1}+\frac{11 d_1'}{3 (3 \kappa +1)}+\frac{11 d_1' \
\kappa }{3 \kappa +1}+\frac{44 \kappa }{3 (3 \kappa +1)}\Big) \
H(1;\yo)+\Big(\frac{28 \kappa }{(\kappa +1)^2}+\frac{4}{(\kappa \
+1)^2}\Big) H(0,0;\yo)+\Big(\frac{6 \kappa  d_1'}{(\kappa \
+1)^2}+\frac{2 d_1'}{(\kappa +1)^2}+\frac{8 \kappa }{(\kappa \
+1)^2}\Big) H(0,1;\yo)\Big)+\Big(-\frac{4 d_1' \yo^3}{3 (3 \kappa \
+1)}-\frac{4 d_1' \kappa  \yo^3}{3 \kappa +1}-\frac{16 \kappa  \
\yo^3}{3 (3 \kappa +1)}+\frac{6 d_1' \yo^2}{3 \kappa +1}+\frac{18 \
d_1' \kappa  \yo^2}{3 \kappa +1}+\frac{24 \kappa  \yo^2}{3 \kappa \
+1}-\frac{12 d_1' \yo}{3 \kappa +1}-\frac{36 d_1' \kappa  \yo}{3 \
\kappa +1}-\frac{48 \kappa  \yo}{3 \kappa +1}+\frac{22 d_1'}{3 (3 \
\kappa +1)}+\frac{22 d_1' \kappa }{3 \kappa +1}+\frac{88 \kappa }{3 \
(3 \kappa +1)}\Big) H(1,0;\yo)+\Big(-\frac{14 \kappa  \yo^3}{3 (3 \
\kappa +1)}-\frac{2 \yo^3}{3 (3 \kappa +1)}+\frac{21 \kappa  \yo^2}{3 \
\kappa +1}+\frac{3 \yo^2}{3 \kappa +1}-\frac{42 \kappa  \yo}{3 \kappa \
+1}-\frac{6 \yo}{3 \kappa +1}+\Big(\frac{14 \kappa }{(\kappa \
+1)^2}+\frac{2}{(\kappa +1)^2}\Big) H(0;\yo)\Big) \
H(1,0;Y)+\Big(-\frac{2 d_1'^2 \yo^3}{3 (3 \kappa +1)}-\frac{2 d_1'^2 \
\kappa  \yo^3}{3 (3 \kappa +1)}-\frac{8 d_1' \kappa  \yo^3}{3 (3 \
\kappa +1)}-\frac{4 \kappa  \yo^3}{3 (3 \kappa +1)}+\frac{3 d_1'^2 \
\yo^2}{3 \kappa +1}+\frac{3 d_1'^2 \kappa  \yo^2}{3 \kappa \
+1}+\frac{12 d_1' \kappa  \yo^2}{3 \kappa +1}+\frac{6 \kappa  \
\yo^2}{3 \kappa +1}-\frac{6 d_1'^2 \yo}{3 \kappa +1}-\frac{6 d_1'^2 \
\kappa  \yo}{3 \kappa +1}-\frac{24 d_1' \kappa  \yo}{3 \kappa \
+1}-\frac{12 \kappa  \yo}{3 \kappa +1}+\frac{11 d_1'^2}{3 (3 \kappa \
+1)}+\frac{11 d_1'^2 \kappa }{3 (3 \kappa +1)}+\frac{44 d_1' \kappa \
}{3 (3 \kappa +1)}+\frac{22 \kappa }{3 (3 \kappa +1)}\Big) \
H(1,1;\yo)+\Big(\frac{56 \kappa }{(\kappa +1)^2}+\frac{8}{(\kappa \
+1)^2}\Big) H(0,0,0;\yo)+\Big(\frac{7 \kappa }{(\kappa \
+1)^2}+\frac{1}{(\kappa +1)^2}\Big) H(0,0,0;Y)+\Big(\frac{12 \kappa  \
d_1'}{(\kappa +1)^2}+\frac{4 d_1'}{(\kappa +1)^2}+\frac{16 \kappa }{(\
\kappa +1)^2}\Big) H(0,0,1;\yo)+\Big(\frac{12 \kappa  d_1'}{(\kappa \
+1)^2}+\frac{4 d_1'}{(\kappa +1)^2}+\frac{16 \kappa }{(\kappa \
+1)^2}\Big) H(0,1,0;\yo)+\Big(\frac{7 \kappa }{(\kappa \
+1)^2}+\frac{1}{(\kappa +1)^2}\Big) H(0,1,0;Y)+\Big(\frac{2 \kappa  \
d_1'^2}{(\kappa +1)^2}+\frac{2 d_1'^2}{(\kappa +1)^2}+\frac{8 \kappa  \
d_1'}{(\kappa +1)^2}+\frac{4 \kappa }{(\kappa +1)^2}\Big) \
H(0,1,1;\yo)+H(1,0,0;Y)+H(1,1,0;Y)+\frac{22 \kappa  \zeta_3}{3 \kappa \
+1}+\frac{2 \zeta_3}{3 \kappa +1},
\erp
%ep^2
\brp
i_2^{(\kap)}=\frac{2 \kappa  \yo^3 d_1'^3}{81 (3 \kappa +1)}+\frac{2 \yo^3 \
d_1'^3}{81 (3 \kappa +1)}-\frac{43 \kappa  \yo^2 d_1'^3}{216 (3 \
\kappa +1)}-\frac{43 \yo^2 d_1'^3}{216 (3 \kappa +1)}+\frac{251 \
\kappa  \yo d_1'^3}{108 (3 \kappa +1)}+\frac{251 \yo d_1'^3}{108 (3 \
\kappa +1)}-\frac{8 \kappa  \yo^3 d_1'^2}{27 (3 \kappa +1)}-\frac{4 \
\yo^3 d_1'^2}{27 (3 \kappa +1)}+\frac{31 \kappa  \yo^2 d_1'^2}{9 (3 \
\kappa +1)}+\frac{167 \yo^2 d_1'^2}{108 (3 \kappa +1)}-\frac{833 \
\kappa  \yo d_1'^2}{18 (3 \kappa +1)}-\frac{542 \yo d_1'^2}{27 (3 \
\kappa +1)}+\frac{28 \kappa  \yo^3 d_1'}{27 (3 \kappa +1)}+\frac{8 \
\yo^3 d_1'}{27 (3 \kappa +1)}-\frac{1663 \kappa  \yo^2 d_1'}{108 (3 \
\kappa +1)}-\frac{205 \yo^2 d_1'}{54 (3 \kappa +1)}+\frac{12815 \
\kappa  \yo d_1'}{54 (3 \kappa +1)}+\frac{1481 \yo d_1'}{27 (3 \kappa \
+1)}-\frac{92 \kappa  \yo^3}{81 (3 \kappa +1)}-\frac{16 \yo^3}{81 (3 \
\kappa +1)}+\frac{245 \kappa  \yo^2}{12 (3 \kappa +1)}+\frac{3 \
\yo^2}{3 \kappa +1}-\frac{2141 \kappa  \yo}{6 (3 \kappa +1)}-\frac{48 \
\yo}{3 \kappa +1}+\Big(\frac{2 \kappa  \yo^3 d_1'^3}{27 (3 \kappa \
+1)}+\frac{2 \yo^3 d_1'^3}{27 (3 \kappa +1)}-\frac{17 \kappa  \yo^2 \
d_1'^3}{36 (3 \kappa +1)}-\frac{17 \yo^2 d_1'^3}{36 (3 \kappa \
+1)}+\frac{49 \kappa  \yo d_1'^3}{18 (3 \kappa +1)}+\frac{49 \yo \
d_1'^3}{18 (3 \kappa +1)}-\frac{251 \kappa  d_1'^3}{108 (3 \kappa \
+1)}-\frac{251 d_1'^3}{108 (3 \kappa +1)}-\frac{4 \kappa  \yo^3 \
d_1'^2}{9 (3 \kappa +1)}-\frac{8 \yo^3 d_1'^2}{27 (3 \kappa \
+1)}+\frac{9 \kappa  \yo^2 d_1'^2}{2 (3 \kappa +1)}+\frac{22 \yo^2 \
d_1'^2}{9 (3 \kappa +1)}-\frac{34 \kappa  \yo d_1'^2}{3 \kappa \
+1}-\frac{151 \yo d_1'^2}{9 (3 \kappa +1)}+\frac{539 \kappa  \
d_1'^2}{18 (3 \kappa +1)}+\frac{395 d_1'^2}{27 (3 \kappa +1)}+\frac{4 \
\kappa  \yo^3 d_1'}{27 (3 \kappa +1)}+\frac{8 \yo^3 d_1'}{27 (3 \
\kappa +1)}-\frac{77 \kappa  \yo^2 d_1'}{18 (3 \kappa +1)}-\frac{3 \
\yo^2 d_1'}{3 \kappa +1}+\frac{451 \kappa  \yo d_1'}{9 (3 \kappa \
+1)}+\frac{24 \yo d_1'}{3 \kappa +1}-\frac{2483 \kappa  d_1'}{54 (3 \
\kappa +1)}-\frac{575 d_1'}{27 (3 \kappa +1)}+\frac{4 \kappa  \
\yo^3}{3 (3 \kappa +1)}-\frac{91 \kappa  \yo^2}{6 (3 \kappa \
+1)}+\frac{391 \kappa  \yo}{3 (3 \kappa +1)}-\frac{233 \kappa }{2 (3 \
\kappa +1)}\Big) H(1;\yo)+\Big(\frac{8 d_1' \yo^3}{9 (3 \kappa \
+1)}+\frac{56 d_1' \kappa  \yo^3}{9 (3 \kappa +1)}-\frac{176 \kappa  \
\yo^3}{9 (3 \kappa +1)}-\frac{16 \yo^3}{9 (3 \kappa +1)}-\frac{14 \
d_1' \yo^2}{3 (3 \kappa +1)}-\frac{98 d_1' \kappa  \yo^2}{3 (3 \kappa \
+1)}+\frac{428 \kappa  \yo^2}{3 (3 \kappa +1)}+\frac{12 \yo^2}{3 \
\kappa +1}+\frac{44 d_1' \yo}{3 (3 \kappa +1)}+\frac{308 d_1' \kappa  \
\yo}{3 (3 \kappa +1)}-\frac{1808 \kappa  \yo}{3 (3 \kappa \
+1)}-\frac{48 \yo}{3 \kappa +1}\Big) H(0,0;\yo)+\Big(\frac{4 d_1'^2 \
\yo^3}{9 (3 \kappa +1)}-\frac{8 d_1' \yo^3}{9 (3 \kappa +1)}+\frac{4 \
d_1'^2 \kappa  \yo^3}{3 (3 \kappa +1)}-\frac{8 d_1' \kappa  \yo^3}{3 \
(3 \kappa +1)}-\frac{16 \kappa  \yo^3}{3 (3 \kappa +1)}-\frac{7 \
d_1'^2 \yo^2}{3 (3 \kappa +1)}+\frac{6 d_1' \yo^2}{3 \kappa \
+1}-\frac{7 d_1'^2 \kappa  \yo^2}{3 \kappa +1}+\frac{70 d_1' \kappa  \
\yo^2}{3 (3 \kappa +1)}+\frac{116 \kappa  \yo^2}{3 (3 \kappa \
+1)}+\frac{22 d_1'^2 \yo}{3 (3 \kappa +1)}-\frac{24 d_1' \yo}{3 \
\kappa +1}+\frac{22 d_1'^2 \kappa  \yo}{3 \kappa +1}-\frac{328 d_1' \
\kappa  \yo}{3 (3 \kappa +1)}-\frac{488 \kappa  \yo}{3 (3 \kappa +1)}\
\Big) H(0,1;\yo)+H(0,0;Y) \Big(\frac{2 d_1' \yo^3}{9 (3 \kappa \
+1)}+\frac{14 d_1' \kappa  \yo^3}{9 (3 \kappa +1)}-\frac{44 \kappa  \
\yo^3}{9 (3 \kappa +1)}-\frac{4 \yo^3}{9 (3 \kappa +1)}-\frac{7 d_1' \
\yo^2}{6 (3 \kappa +1)}-\frac{49 d_1' \kappa  \yo^2}{6 (3 \kappa \
+1)}+\frac{107 \kappa  \yo^2}{3 (3 \kappa +1)}+\frac{3 \yo^2}{3 \
\kappa +1}+\frac{11 d_1' \yo}{3 (3 \kappa +1)}+\frac{77 d_1' \kappa  \
\yo}{3 (3 \kappa +1)}-\frac{452 \kappa  \yo}{3 (3 \kappa \
+1)}-\frac{12 \yo}{3 \kappa +1}+\Big(\frac{20 \kappa  \yo^3}{3 \kappa \
+1}+\frac{4 \yo^3}{3 (3 \kappa +1)}-\frac{90 \kappa  \yo^2}{3 \kappa \
+1}-\frac{6 \yo^2}{3 \kappa +1}+\frac{180 \kappa  \yo}{3 \kappa \
+1}+\frac{12 \yo}{3 \kappa +1}\Big) H(0;\yo)+\Big(\frac{2 d_1' \
\yo^3}{3 (3 \kappa +1)}+\frac{14 d_1' \kappa  \yo^3}{3 (3 \kappa \
+1)}+\frac{16 \kappa  \yo^3}{3 (3 \kappa +1)}-\frac{3 d_1' \yo^2}{3 \
\kappa +1}-\frac{21 d_1' \kappa  \yo^2}{3 \kappa +1}-\frac{24 \kappa  \
\yo^2}{3 \kappa +1}+\frac{6 d_1' \yo}{3 \kappa +1}+\frac{42 d_1' \
\kappa  \yo}{3 \kappa +1}+\frac{48 \kappa  \yo}{3 \kappa +1}-\frac{11 \
d_1'}{3 (3 \kappa +1)}-\frac{77 d_1' \kappa }{3 (3 \kappa \
+1)}-\frac{88 \kappa }{3 (3 \kappa +1)}\Big) H(1;\yo)+\Big(-\frac{60 \
\kappa }{(\kappa +1)^2}-\frac{4}{(\kappa +1)^2}\Big) \
H(0,0;\yo)+\Big(-\frac{14 \kappa  d_1'}{(\kappa +1)^2}-\frac{2 \
d_1'}{(\kappa +1)^2}-\frac{16 \kappa }{(\kappa +1)^2}\Big) H(0,1;\yo)\
\Big)+\Big(-\frac{7 \pi ^2 \kappa }{6 (3 \kappa +1)}-\frac{\pi ^2}{6 \
(3 \kappa +1)}\Big) H(0,1;Y)+\Big(\frac{4 d_1'^2 \yo^3}{9 (3 \kappa \
+1)}-\frac{8 d_1' \yo^3}{9 (3 \kappa +1)}+\frac{4 d_1'^2 \kappa  \
\yo^3}{3 (3 \kappa +1)}-\frac{8 d_1' \kappa  \yo^3}{3 (3 \kappa +1)}-\
\frac{16 \kappa  \yo^3}{3 (3 \kappa +1)}-\frac{7 d_1'^2 \yo^2}{3 (3 \
\kappa +1)}+\frac{6 d_1' \yo^2}{3 \kappa +1}-\frac{7 d_1'^2 \kappa  \
\yo^2}{3 \kappa +1}+\frac{70 d_1' \kappa  \yo^2}{3 (3 \kappa \
+1)}+\frac{116 \kappa  \yo^2}{3 (3 \kappa +1)}+\frac{22 d_1'^2 \yo}{3 \
(3 \kappa +1)}-\frac{24 d_1' \yo}{3 \kappa +1}+\frac{22 d_1'^2 \kappa \
 \yo}{3 \kappa +1}-\frac{328 d_1' \kappa  \yo}{3 (3 \kappa \
+1)}-\frac{488 \kappa  \yo}{3 (3 \kappa +1)}-\frac{49 d_1'^2}{9 (3 \
\kappa +1)}+\frac{170 d_1'}{9 (3 \kappa +1)}-\frac{49 d_1'^2 \kappa \
}{3 (3 \kappa +1)}+\frac{266 d_1' \kappa }{3 (3 \kappa +1)}+\frac{388 \
\kappa }{3 (3 \kappa +1)}\Big) H(1,0;\yo)+\Big(\frac{2 d_1' \yo^3}{9 \
(3 \kappa +1)}+\frac{14 d_1' \kappa  \yo^3}{9 (3 \kappa +1)}-\frac{44 \
\kappa  \yo^3}{9 (3 \kappa +1)}-\frac{4 \yo^3}{9 (3 \kappa \
+1)}-\frac{7 d_1' \yo^2}{6 (3 \kappa +1)}-\frac{49 d_1' \kappa  \
\yo^2}{6 (3 \kappa +1)}+\frac{107 \kappa  \yo^2}{3 (3 \kappa \
+1)}+\frac{3 \yo^2}{3 \kappa +1}+\frac{11 d_1' \yo}{3 (3 \kappa +1)}+\
\frac{77 d_1' \kappa  \yo}{3 (3 \kappa +1)}-\frac{452 \kappa  \yo}{3 \
(3 \kappa +1)}-\frac{12 \yo}{3 \kappa +1}+\Big(\frac{20 \kappa  \
\yo^3}{3 \kappa +1}+\frac{4 \yo^3}{3 (3 \kappa +1)}-\frac{90 \kappa  \
\yo^2}{3 \kappa +1}-\frac{6 \yo^2}{3 \kappa +1}+\frac{180 \kappa  \
\yo}{3 \kappa +1}+\frac{12 \yo}{3 \kappa +1}\Big) \
H(0;\yo)+\Big(\frac{2 d_1' \yo^3}{3 (3 \kappa +1)}+\frac{14 d_1' \
\kappa  \yo^3}{3 (3 \kappa +1)}+\frac{16 \kappa  \yo^3}{3 (3 \kappa \
+1)}-\frac{3 d_1' \yo^2}{3 \kappa +1}-\frac{21 d_1' \kappa  \yo^2}{3 \
\kappa +1}-\frac{24 \kappa  \yo^2}{3 \kappa +1}+\frac{6 d_1' \yo}{3 \
\kappa +1}+\frac{42 d_1' \kappa  \yo}{3 \kappa +1}+\frac{48 \kappa  \
\yo}{3 \kappa +1}-\frac{11 d_1'}{3 (3 \kappa +1)}-\frac{77 d_1' \
\kappa }{3 (3 \kappa +1)}-\frac{88 \kappa }{3 (3 \kappa +1)}\Big) \
H(1;\yo)+\Big(-\frac{60 \kappa }{(\kappa +1)^2}-\frac{4}{(\kappa \
+1)^2}\Big) H(0,0;\yo)+\Big(-\frac{14 \kappa  d_1'}{(\kappa \
+1)^2}-\frac{2 d_1'}{(\kappa +1)^2}-\frac{16 \kappa }{(\kappa \
+1)^2}\Big) H(0,1;\yo)+\frac{2 \kappa  \pi ^2}{3 (3 \kappa +1)}\Big) \
H(1,0;Y)+\Big(\frac{2 \kappa  \yo^3 d_1'^3}{9 (3 \kappa +1)}+\frac{2 \
\yo^3 d_1'^3}{9 (3 \kappa +1)}-\frac{7 \kappa  \yo^2 d_1'^3}{6 (3 \
\kappa +1)}-\frac{7 \yo^2 d_1'^3}{6 (3 \kappa +1)}+\frac{11 \kappa  \
\yo d_1'^3}{3 (3 \kappa +1)}+\frac{11 \yo d_1'^3}{3 (3 \kappa \
+1)}-\frac{49 \kappa  d_1'^3}{18 (3 \kappa +1)}-\frac{49 d_1'^3}{18 \
(3 \kappa +1)}-\frac{4 \yo^3 d_1'^2}{9 (3 \kappa +1)}+\frac{2 \kappa  \
\yo^2 d_1'^2}{3 \kappa +1}+\frac{3 \yo^2 d_1'^2}{3 \kappa \
+1}-\frac{14 \kappa  \yo d_1'^2}{3 \kappa +1}-\frac{12 \yo d_1'^2}{3 \
\kappa +1}+\frac{12 \kappa  d_1'^2}{3 \kappa +1}+\frac{85 d_1'^2}{9 \
(3 \kappa +1)}-\frac{20 \kappa  \yo^3 d_1'}{9 (3 \kappa +1)}+\frac{17 \
\kappa  \yo^2 d_1'}{3 \kappa +1}-\frac{74 \kappa  \yo d_1'}{3 \kappa \
+1}+\frac{533 \kappa  d_1'}{9 (3 \kappa +1)}-\frac{4 \kappa  \yo^3}{3 \
(3 \kappa +1)}+\frac{29 \kappa  \yo^2}{3 (3 \kappa +1)}-\frac{122 \
\kappa  \yo}{3 (3 \kappa +1)}+\frac{97 \kappa }{3 (3 \kappa +1)}\Big) \
H(1,1;\yo)-\frac{1}{6} \pi ^2 H(1,1;Y)+\Big(\frac{80 \kappa  \yo^3}{3 \
\kappa +1}+\frac{16 \yo^3}{3 (3 \kappa +1)}-\frac{360 \kappa  \
\yo^2}{3 \kappa +1}-\frac{24 \yo^2}{3 \kappa +1}+\frac{720 \kappa  \
\yo}{3 \kappa +1}+\frac{48 \yo}{3 \kappa +1}\Big) \
H(0,0,0;\yo)+\Big(\frac{10 \kappa  \yo^3}{3 \kappa +1}+\frac{2 \
\yo^3}{3 (3 \kappa +1)}-\frac{45 \kappa  \yo^2}{3 \kappa +1}-\frac{3 \
\yo^2}{3 \kappa +1}+\frac{90 \kappa  \yo}{3 \kappa +1}+\frac{6 \yo}{3 \
\kappa +1}+\Big(-\frac{30 \kappa }{(\kappa +1)^2}-\frac{2}{(\kappa \
+1)^2}\Big) H(0;\yo)\Big) H(0,0,0;Y)+\Big(\frac{8 d_1' \yo^3}{3 (3 \
\kappa +1)}+\frac{56 d_1' \kappa  \yo^3}{3 (3 \kappa +1)}+\frac{64 \
\kappa  \yo^3}{3 (3 \kappa +1)}-\frac{12 d_1' \yo^2}{3 \kappa \
+1}-\frac{84 d_1' \kappa  \yo^2}{3 \kappa +1}-\frac{96 \kappa  \
\yo^2}{3 \kappa +1}+\frac{24 d_1' \yo}{3 \kappa +1}+\frac{168 d_1' \
\kappa  \yo}{3 \kappa +1}+\frac{192 \kappa  \yo}{3 \kappa +1}\Big) \
H(0,0,1;\yo)+\Big(\frac{8 d_1' \yo^3}{3 (3 \kappa +1)}+\frac{56 d_1' \
\kappa  \yo^3}{3 (3 \kappa +1)}+\frac{64 \kappa  \yo^3}{3 (3 \kappa \
+1)}-\frac{12 d_1' \yo^2}{3 \kappa +1}-\frac{84 d_1' \kappa  \yo^2}{3 \
\kappa +1}-\frac{96 \kappa  \yo^2}{3 \kappa +1}+\frac{24 d_1' \yo}{3 \
\kappa +1}+\frac{168 d_1' \kappa  \yo}{3 \kappa +1}+\frac{192 \kappa  \
\yo}{3 \kappa +1}\Big) H(0,1,0;\yo)+\Big(\frac{10 \kappa  \yo^3}{3 \
\kappa +1}+\frac{2 \yo^3}{3 (3 \kappa +1)}-\frac{45 \kappa  \yo^2}{3 \
\kappa +1}-\frac{3 \yo^2}{3 \kappa +1}+\frac{90 \kappa  \yo}{3 \kappa \
+1}+\frac{6 \yo}{3 \kappa +1}+\Big(-\frac{30 \kappa }{(\kappa +1)^2}-\
\frac{2}{(\kappa +1)^2}\Big) H(0;\yo)\Big) H(0,1,0;Y)+\Big(\frac{4 \
d_1'^2 \yo^3}{3 (3 \kappa +1)}+\frac{4 d_1'^2 \kappa  \yo^3}{3 \kappa \
+1}+\frac{32 d_1' \kappa  \yo^3}{3 (3 \kappa +1)}+\frac{16 \kappa  \
\yo^3}{3 (3 \kappa +1)}-\frac{6 d_1'^2 \yo^2}{3 \kappa +1}-\frac{18 \
d_1'^2 \kappa  \yo^2}{3 \kappa +1}-\frac{48 d_1' \kappa  \yo^2}{3 \
\kappa +1}-\frac{24 \kappa  \yo^2}{3 \kappa +1}+\frac{12 d_1'^2 \
\yo}{3 \kappa +1}+\frac{36 d_1'^2 \kappa  \yo}{3 \kappa +1}+\frac{96 \
d_1' \kappa  \yo}{3 \kappa +1}+\frac{48 \kappa  \yo}{3 \kappa \
+1}\Big) H(0,1,1;\yo)+\Big(\frac{8 d_1' \yo^3}{3 (3 \kappa \
+1)}+\frac{56 d_1' \kappa  \yo^3}{3 (3 \kappa +1)}+\frac{64 \kappa  \
\yo^3}{3 (3 \kappa +1)}-\frac{12 d_1' \yo^2}{3 \kappa +1}-\frac{84 \
d_1' \kappa  \yo^2}{3 \kappa +1}-\frac{96 \kappa  \yo^2}{3 \kappa \
+1}+\frac{24 d_1' \yo}{3 \kappa +1}+\frac{168 d_1' \kappa  \yo}{3 \
\kappa +1}+\frac{192 \kappa  \yo}{3 \kappa +1}-\frac{44 d_1'}{3 (3 \
\kappa +1)}-\frac{308 d_1' \kappa }{3 (3 \kappa +1)}-\frac{352 \kappa \
}{3 (3 \kappa +1)}\Big) H(1,0,0;\yo)+\Big(\frac{14 \kappa  \yo^3}{3 \
(3 \kappa +1)}+\frac{2 \yo^3}{3 (3 \kappa +1)}-\frac{21 \kappa  \
\yo^2}{3 \kappa +1}-\frac{3 \yo^2}{3 \kappa +1}+\frac{42 \kappa  \
\yo}{3 \kappa +1}+\frac{6 \yo}{3 \kappa +1}+\Big(-\frac{14 \kappa }{(\
\kappa +1)^2}-\frac{2}{(\kappa +1)^2}\Big) H(0;\yo)\Big) \
H(1,0,0;Y)+\Big(\frac{4 d_1'^2 \yo^3}{3 (3 \kappa +1)}+\frac{4 d_1'^2 \
\kappa  \yo^3}{3 \kappa +1}+\frac{32 d_1' \kappa  \yo^3}{3 (3 \kappa \
+1)}+\frac{16 \kappa  \yo^3}{3 (3 \kappa +1)}-\frac{6 d_1'^2 \yo^2}{3 \
\kappa +1}-\frac{18 d_1'^2 \kappa  \yo^2}{3 \kappa +1}-\frac{48 d_1' \
\kappa  \yo^2}{3 \kappa +1}-\frac{24 \kappa  \yo^2}{3 \kappa \
+1}+\frac{12 d_1'^2 \yo}{3 \kappa +1}+\frac{36 d_1'^2 \kappa  \yo}{3 \
\kappa +1}+\frac{96 d_1' \kappa  \yo}{3 \kappa +1}+\frac{48 \kappa  \
\yo}{3 \kappa +1}-\frac{22 d_1'^2}{3 (3 \kappa +1)}-\frac{22 d_1'^2 \
\kappa }{3 \kappa +1}-\frac{176 d_1' \kappa }{3 (3 \kappa \
+1)}-\frac{88 \kappa }{3 (3 \kappa +1)}\Big) \
H(1,0,1;\yo)+\Big(\frac{4 d_1'^2 \yo^3}{3 (3 \kappa +1)}+\frac{4 \
d_1'^2 \kappa  \yo^3}{3 \kappa +1}+\frac{32 d_1' \kappa  \yo^3}{3 (3 \
\kappa +1)}+\frac{16 \kappa  \yo^3}{3 (3 \kappa +1)}-\frac{6 d_1'^2 \
\yo^2}{3 \kappa +1}-\frac{18 d_1'^2 \kappa  \yo^2}{3 \kappa \
+1}-\frac{48 d_1' \kappa  \yo^2}{3 \kappa +1}-\frac{24 \kappa  \
\yo^2}{3 \kappa +1}+\frac{12 d_1'^2 \yo}{3 \kappa +1}+\frac{36 d_1'^2 \
\kappa  \yo}{3 \kappa +1}+\frac{96 d_1' \kappa  \yo}{3 \kappa \
+1}+\frac{48 \kappa  \yo}{3 \kappa +1}-\frac{22 d_1'^2}{3 (3 \kappa \
+1)}-\frac{22 d_1'^2 \kappa }{3 \kappa +1}-\frac{176 d_1' \kappa }{3 \
(3 \kappa +1)}-\frac{88 \kappa }{3 (3 \kappa +1)}\Big) \
H(1,1,0;\yo)+\Big(\frac{14 \kappa  \yo^3}{3 (3 \kappa +1)}+\frac{2 \
\yo^3}{3 (3 \kappa +1)}-\frac{21 \kappa  \yo^2}{3 \kappa +1}-\frac{3 \
\yo^2}{3 \kappa +1}+\frac{42 \kappa  \yo}{3 \kappa +1}+\frac{6 \yo}{3 \
\kappa +1}+\Big(-\frac{14 \kappa }{(\kappa +1)^2}-\frac{2}{(\kappa \
+1)^2}\Big) H(0;\yo)\Big) H(1,1,0;Y)+\Big(\frac{2 \kappa  \yo^3 \
d_1'^3}{3 (3 \kappa +1)}+\frac{2 \yo^3 d_1'^3}{3 (3 \kappa \
+1)}-\frac{3 \kappa  \yo^2 d_1'^3}{3 \kappa +1}-\frac{3 \yo^2 \
d_1'^3}{3 \kappa +1}+\frac{6 \kappa  \yo d_1'^3}{3 \kappa +1}+\frac{6 \
\yo d_1'^3}{3 \kappa +1}-\frac{11 \kappa  d_1'^3}{3 (3 \kappa \
+1)}-\frac{11 d_1'^3}{3 (3 \kappa +1)}+\frac{4 \kappa  \yo^3 \
d_1'^2}{3 \kappa +1}-\frac{18 \kappa  \yo^2 d_1'^2}{3 \kappa \
+1}+\frac{36 \kappa  \yo d_1'^2}{3 \kappa +1}-\frac{22 \kappa  \
d_1'^2}{3 \kappa +1}+\frac{4 \kappa  \yo^3 d_1'}{3 \kappa \
+1}-\frac{18 \kappa  \yo^2 d_1'}{3 \kappa +1}+\frac{36 \kappa  \yo \
d_1'}{3 \kappa +1}-\frac{22 \kappa  d_1'}{3 \kappa +1}+\frac{4 \kappa \
 \yo^3}{3 (3 \kappa +1)}-\frac{6 \kappa  \yo^2}{3 \kappa +1}+\frac{12 \
\kappa  \yo}{3 \kappa +1}-\frac{22 \kappa }{3 (3 \kappa +1)}\Big) \
H(1,1,1;\yo)+\Big(-\frac{240 \kappa }{(\kappa \
+1)^2}-\frac{16}{(\kappa +1)^2}\Big) H(0,0,0,0;\yo)+\Big(-\frac{15 \
\kappa }{(\kappa +1)^2}-\frac{1}{(\kappa +1)^2}\Big) \
H(0,0,0,0;Y)+\Big(-\frac{56 \kappa  d_1'}{(\kappa +1)^2}-\frac{8 \
d_1'}{(\kappa +1)^2}-\frac{64 \kappa }{(\kappa +1)^2}\Big) H(0,0,0,1;\
\yo)+\Big(-\frac{56 \kappa  d_1'}{(\kappa +1)^2}-\frac{8 \
d_1'}{(\kappa +1)^2}-\frac{64 \kappa }{(\kappa +1)^2}\Big) H(0,0,1,0;\
\yo)+\Big(-\frac{15 \kappa }{(\kappa +1)^2}-\frac{1}{(\kappa \
+1)^2}\Big) H(0,0,1,0;Y)+\Big(-\frac{12 \kappa  d_1'^2}{(\kappa \
+1)^2}-\frac{4 d_1'^2}{(\kappa +1)^2}-\frac{32 \kappa  d_1'}{(\kappa \
+1)^2}-\frac{16 \kappa }{(\kappa +1)^2}\Big) \
H(0,0,1,1;\yo)+\Big(-\frac{56 \kappa  d_1'}{(\kappa +1)^2}-\frac{8 \
d_1'}{(\kappa +1)^2}-\frac{64 \kappa }{(\kappa +1)^2}\Big) H(0,1,0,0;\
\yo)+\Big(-\frac{7 \kappa }{(\kappa +1)^2}-\frac{1}{(\kappa \
+1)^2}\Big) H(0,1,0,0;Y)+\Big(-\frac{12 \kappa  d_1'^2}{(\kappa \
+1)^2}-\frac{4 d_1'^2}{(\kappa +1)^2}-\frac{32 \kappa  d_1'}{(\kappa \
+1)^2}-\frac{16 \kappa }{(\kappa +1)^2}\Big) \
H(0,1,0,1;\yo)+\Big(-\frac{12 \kappa  d_1'^2}{(\kappa +1)^2}-\frac{4 \
d_1'^2}{(\kappa +1)^2}-\frac{32 \kappa  d_1'}{(\kappa +1)^2}-\frac{16 \
\kappa }{(\kappa +1)^2}\Big) H(0,1,1,0;\yo)+\Big(-\frac{7 \kappa \
}{(\kappa +1)^2}-\frac{1}{(\kappa +1)^2}\Big) \
H(0,1,1,0;Y)+\Big(-\frac{2 \kappa  d_1'^3}{(\kappa +1)^2}-\frac{2 \
d_1'^3}{(\kappa +1)^2}-\frac{12 \kappa  d_1'^2}{(\kappa \
+1)^2}-\frac{12 \kappa  d_1'}{(\kappa +1)^2}-\frac{4 \kappa }{(\kappa \
+1)^2}\Big) H(0,1,1,1;\yo)+\Big(-\frac{11 \kappa }{(\kappa \
+1)^2}-\frac{1}{(\kappa +1)^2}\Big) H(1,0,0,0;Y)+\Big(-\frac{11 \
\kappa }{(\kappa +1)^2}-\frac{1}{(\kappa +1)^2}\Big) \
H(1,0,1,0;Y)-H(1,1,0,0;Y)-H(1,1,1,0;Y)+H(0;\yo) \Big(\frac{4 d_1'^2 \
\yo^3}{27 (3 \kappa +1)}-\frac{16 d_1' \yo^3}{27 (3 \kappa \
+1)}+\frac{4 d_1'^2 \kappa  \yo^3}{9 (3 \kappa +1)}-\frac{80 d_1' \
\kappa  \yo^3}{27 (3 \kappa +1)}+\frac{128 \kappa  \yo^3}{27 (3 \
\kappa +1)}+\frac{16 \yo^3}{27 (3 \kappa +1)}-\frac{17 d_1'^2 \
\yo^2}{18 (3 \kappa +1)}+\frac{44 d_1' \yo^2}{9 (3 \kappa \
+1)}-\frac{17 d_1'^2 \kappa  \yo^2}{6 (3 \kappa +1)}+\frac{80 d_1' \
\kappa  \yo^2}{3 (3 \kappa +1)}-\frac{164 \kappa  \yo^2}{3 (3 \kappa \
+1)}-\frac{6 \yo^2}{3 \kappa +1}+\frac{49 d_1'^2 \yo}{9 (3 \kappa \
+1)}-\frac{302 d_1' \yo}{9 (3 \kappa +1)}+\frac{49 d_1'^2 \kappa  \
\yo}{3 (3 \kappa +1)}-\frac{574 d_1' \kappa  \yo}{3 (3 \kappa \
+1)}+\frac{1420 \kappa  \yo}{3 (3 \kappa +1)}+\frac{48 \yo}{3 \kappa \
+1}-\frac{92 \kappa  \zeta_3}{3 \kappa +1}-\frac{4 \zeta_3}{3 \kappa \
+1}\Big)+H(0;Y) \Big(\frac{2 d_1'^2 \yo^3}{27 (3 \kappa +1)}-\frac{8 \
d_1' \yo^3}{27 (3 \kappa +1)}+\frac{2 d_1'^2 \kappa  \yo^3}{9 (3 \
\kappa +1)}-\frac{40 d_1' \kappa  \yo^3}{27 (3 \kappa +1)}+\frac{64 \
\kappa  \yo^3}{27 (3 \kappa +1)}+\frac{8 \yo^3}{27 (3 \kappa \
+1)}-\frac{17 d_1'^2 \yo^2}{36 (3 \kappa +1)}+\frac{22 d_1' \yo^2}{9 \
(3 \kappa +1)}-\frac{17 d_1'^2 \kappa  \yo^2}{12 (3 \kappa \
+1)}+\frac{40 d_1' \kappa  \yo^2}{3 (3 \kappa +1)}-\frac{82 \kappa  \
\yo^2}{3 (3 \kappa +1)}-\frac{3 \yo^2}{3 \kappa +1}+\frac{49 d_1'^2 \
\yo}{18 (3 \kappa +1)}-\frac{151 d_1' \yo}{9 (3 \kappa +1)}+\frac{49 \
d_1'^2 \kappa  \yo}{6 (3 \kappa +1)}-\frac{287 d_1' \kappa  \yo}{3 (3 \
\kappa +1)}+\frac{710 \kappa  \yo}{3 (3 \kappa +1)}+\frac{24 \yo}{3 \
\kappa +1}+\Big(\frac{4 d_1' \yo^3}{9 (3 \kappa +1)}+\frac{28 d_1' \
\kappa  \yo^3}{9 (3 \kappa +1)}-\frac{88 \kappa  \yo^3}{9 (3 \kappa \
+1)}-\frac{8 \yo^3}{9 (3 \kappa +1)}-\frac{7 d_1' \yo^2}{3 (3 \kappa \
+1)}-\frac{49 d_1' \kappa  \yo^2}{3 (3 \kappa +1)}+\frac{214 \kappa  \
\yo^2}{3 (3 \kappa +1)}+\frac{6 \yo^2}{3 \kappa +1}+\frac{22 d_1' \
\yo}{3 (3 \kappa +1)}+\frac{154 d_1' \kappa  \yo}{3 (3 \kappa \
+1)}-\frac{904 \kappa  \yo}{3 (3 \kappa +1)}-\frac{24 \yo}{3 \kappa \
+1}\Big) H(0;\yo)+\Big(\frac{2 d_1'^2 \yo^3}{9 (3 \kappa +1)}-\frac{4 \
d_1' \yo^3}{9 (3 \kappa +1)}+\frac{2 d_1'^2 \kappa  \yo^3}{3 (3 \
\kappa +1)}-\frac{4 d_1' \kappa  \yo^3}{3 (3 \kappa +1)}-\frac{8 \
\kappa  \yo^3}{3 (3 \kappa +1)}-\frac{7 d_1'^2 \yo^2}{6 (3 \kappa \
+1)}+\frac{3 d_1' \yo^2}{3 \kappa +1}-\frac{7 d_1'^2 \kappa  \yo^2}{2 \
(3 \kappa +1)}+\frac{35 d_1' \kappa  \yo^2}{3 (3 \kappa +1)}+\frac{58 \
\kappa  \yo^2}{3 (3 \kappa +1)}+\frac{11 d_1'^2 \yo}{3 (3 \kappa \
+1)}-\frac{12 d_1' \yo}{3 \kappa +1}+\frac{11 d_1'^2 \kappa  \yo}{3 \
\kappa +1}-\frac{164 d_1' \kappa  \yo}{3 (3 \kappa +1)}-\frac{244 \
\kappa  \yo}{3 (3 \kappa +1)}-\frac{49 d_1'^2}{18 (3 \kappa \
+1)}+\frac{85 d_1'}{9 (3 \kappa +1)}-\frac{49 d_1'^2 \kappa }{6 (3 \
\kappa +1)}+\frac{133 d_1' \kappa }{3 (3 \kappa +1)}+\frac{194 \kappa \
}{3 (3 \kappa +1)}\Big) H(1;\yo)+\Big(\frac{40 \kappa  \yo^3}{3 \
\kappa +1}+\frac{8 \yo^3}{3 (3 \kappa +1)}-\frac{180 \kappa  \yo^2}{3 \
\kappa +1}-\frac{12 \yo^2}{3 \kappa +1}+\frac{360 \kappa  \yo}{3 \
\kappa +1}+\frac{24 \yo}{3 \kappa +1}\Big) H(0,0;\yo)+\Big(\frac{4 \
d_1' \yo^3}{3 (3 \kappa +1)}+\frac{28 d_1' \kappa  \yo^3}{3 (3 \kappa \
+1)}+\frac{32 \kappa  \yo^3}{3 (3 \kappa +1)}-\frac{6 d_1' \yo^2}{3 \
\kappa +1}-\frac{42 d_1' \kappa  \yo^2}{3 \kappa +1}-\frac{48 \kappa  \
\yo^2}{3 \kappa +1}+\frac{12 d_1' \yo}{3 \kappa +1}+\frac{84 d_1' \
\kappa  \yo}{3 \kappa +1}+\frac{96 \kappa  \yo}{3 \kappa +1}\Big) \
H(0,1;\yo)+\Big(\frac{4 d_1' \yo^3}{3 (3 \kappa +1)}+\frac{28 d_1' \
\kappa  \yo^3}{3 (3 \kappa +1)}+\frac{32 \kappa  \yo^3}{3 (3 \kappa \
+1)}-\frac{6 d_1' \yo^2}{3 \kappa +1}-\frac{42 d_1' \kappa  \yo^2}{3 \
\kappa +1}-\frac{48 \kappa  \yo^2}{3 \kappa +1}+\frac{12 d_1' \yo}{3 \
\kappa +1}+\frac{84 d_1' \kappa  \yo}{3 \kappa +1}+\frac{96 \kappa  \
\yo}{3 \kappa +1}-\frac{22 d_1'}{3 (3 \kappa +1)}-\frac{154 d_1' \
\kappa }{3 (3 \kappa +1)}-\frac{176 \kappa }{3 (3 \kappa +1)}\Big) \
H(1,0;\yo)+\Big(\frac{2 d_1'^2 \yo^3}{3 (3 \kappa +1)}+\frac{2 d_1'^2 \
\kappa  \yo^3}{3 \kappa +1}+\frac{16 d_1' \kappa  \yo^3}{3 (3 \kappa \
+1)}+\frac{8 \kappa  \yo^3}{3 (3 \kappa +1)}-\frac{3 d_1'^2 \yo^2}{3 \
\kappa +1}-\frac{9 d_1'^2 \kappa  \yo^2}{3 \kappa +1}-\frac{24 d_1' \
\kappa  \yo^2}{3 \kappa +1}-\frac{12 \kappa  \yo^2}{3 \kappa \
+1}+\frac{6 d_1'^2 \yo}{3 \kappa +1}+\frac{18 d_1'^2 \kappa  \yo}{3 \
\kappa +1}+\frac{48 d_1' \kappa  \yo}{3 \kappa +1}+\frac{24 \kappa  \
\yo}{3 \kappa +1}-\frac{11 d_1'^2}{3 (3 \kappa +1)}-\frac{11 d_1'^2 \
\kappa }{3 \kappa +1}-\frac{88 d_1' \kappa }{3 (3 \kappa \
+1)}-\frac{44 \kappa }{3 (3 \kappa +1)}\Big) \
H(1,1;\yo)+\Big(-\frac{120 \kappa }{(\kappa +1)^2}-\frac{8}{(\kappa \
+1)^2}\Big) H(0,0,0;\yo)+\Big(-\frac{28 \kappa  d_1'}{(\kappa +1)^2}-\
\frac{4 d_1'}{(\kappa +1)^2}-\frac{32 \kappa }{(\kappa +1)^2}\Big) \
H(0,0,1;\yo)+\Big(-\frac{28 \kappa  d_1'}{(\kappa +1)^2}-\frac{4 \
d_1'}{(\kappa +1)^2}-\frac{32 \kappa }{(\kappa +1)^2}\Big) \
H(0,1,0;\yo)+\Big(-\frac{6 \kappa  d_1'^2}{(\kappa +1)^2}-\frac{2 \
d_1'^2}{(\kappa +1)^2}-\frac{16 \kappa  d_1'}{(\kappa +1)^2}-\frac{8 \
\kappa }{(\kappa +1)^2}\Big) H(0,1,1;\yo)-\frac{46 \kappa  \zeta_3}{3 \
\kappa +1}-\frac{2 \zeta_3}{3 \kappa +1}\Big)+H(1;Y) \Big(\frac{7 \
\kappa  \pi ^2 \yo^3}{9 (3 \kappa +1)}+\frac{\pi ^2 \yo^3}{9 (3 \
\kappa +1)}-\frac{7 \kappa  \pi ^2 \yo^2}{2 (3 \kappa +1)}-\frac{\pi \
^2 \yo^2}{2 (3 \kappa +1)}+\frac{7 \kappa  \pi ^2 \yo}{3 \kappa \
+1}+\frac{\pi ^2 \yo}{3 \kappa +1}+\Big(-\frac{7 \pi ^2 \kappa }{3 (3 \
\kappa +1)}-\frac{\pi ^2}{3 (3 \kappa +1)}\Big) H(0;\yo)-\frac{16 \
\kappa  \zeta_3}{3 \kappa +1}\Big)+\frac{92 \kappa  \yo^3 \zeta_3}{3 \
(3 \kappa +1)}+\frac{4 \yo^3 \zeta_3}{3 (3 \kappa +1)}-\frac{138 \
\kappa  \yo^2 \zeta_3}{3 \kappa +1}-\frac{6 \yo^2 \zeta_3}{3 \kappa \
+1}+\frac{276 \kappa  \yo \zeta_3}{3 \kappa +1}+\frac{12 \yo \
\zeta_3}{3 \kappa +1}+\frac{11 \kappa  \pi ^4}{30 (3 \kappa \
+1)}+\frac{\pi ^4}{30 (3 \kappa +1)}.
\erp

%%%
%%% The K integrals
%%%

\section{The $\cK$ integrals}
\label{app:KIntegrals}

%
% The K integral for kappa=0
%

\subsection{The $\cK$ integral for $\kap =0$}
%
% This file contains the TeX output produced by Mathematica for the integral K, for arbitrary kap = 0 and D0 = 3 +d'1 ep
%
The $\eps$ expansion for this integral reads
\beq
\kint(\eps;y_0,3+d'_1\eps;0)=\frac{1}{\eps^2}k_{-2}^{(0)}+\frac{1}{\eps}k_{-1}^{(0)}+k_0^{(0)}+\eps k_1^{(0)}+\eps^2k_2^{(0)} +\ocal\left(\eps^3\right),
\eeq
where
%1/ep piece
\brp
k_{-2}^{(0)}=1,
\erp
\brp
k_{-1}^{(0)}=\frac{2 \yo^3}{3}-3 \yo^2+6 \yo-2 H(0;\yo),
\erp
% ep^0
\brp
k_0^{(0)}=-\frac{2 d_1' x^3}{9}+\frac{10 x^3}{9}+\frac{7 d_1' x^2}{6}-5 \
x^2-\frac{11 d_1' x}{3}+14 x+\Big(-\frac{4 x^3}{3}+6 x^2-12 x\Big) \
H(0;x)+\Big(-\frac{2 d_1' x^3}{3}+3 d_1' x^2-6 d_1' x+\frac{11 \
d_1'}{3}\Big) H(1;x)+4 H(0,0;x)+2 d_1' H(0,1;x)-\frac{\pi ^2}{6},
\erp
% ep^1
\brp
k_{1}^{(0)}=\frac{2 d_1'^2 x^3}{27}-\frac{14 d_1' x^3}{27}-\frac{\pi ^2 \
x^3}{9}+\frac{56 x^3}{27}-\frac{17 d_1'^2 x^2}{36}+\frac{28 d_1' \
x^2}{9}+\frac{\pi ^2 x^2}{2}-9 x^2+\frac{49 d_1'^2 x}{18}-\frac{157 \
d_1' x}{9}-\pi ^2 x+32 x+\Big(\frac{4 d_1' x^3}{9}-\frac{20 \
x^3}{9}-\frac{7 d_1' x^2}{3}+10 x^2+\frac{22 d_1' x}{3}-28 \
x+\frac{\pi ^2}{3}\Big) H(0;x)+\Big(\frac{2 d_1'^2 x^3}{9}-\frac{10 \
d_1' x^3}{9}-\frac{7 d_1'^2 x^2}{6}+5 d_1' x^2+\frac{11 d_1'^2 \
x}{3}-14 d_1' x-\frac{49 d_1'^2}{18}+\frac{91 d_1'}{9}\Big) \
H(1;x)+\Big(\frac{8 x^3}{3}-12 x^2+24 x\Big) H(0,0;x)+\Big(\frac{4 \
d_1' x^3}{3}-6 d_1' x^2+12 d_1' x\Big) H(0,1;x)+\Big(\frac{4 d_1' \
x^3}{3}-6 d_1' x^2+12 d_1' x-\frac{22 d_1'}{3}\Big) \
H(1,0;x)+\Big(\frac{2 d_1'^2 x^3}{3}-3 d_1'^2 x^2+6 d_1'^2 x-\frac{11 \
d_1'^2}{3}\Big) H(1,1;x)-8 H(0,0,0;x)-4 d_1' H(0,0,1;x)-4 d_1' \
H(0,1,0;x)-2 d_1'^2 H(0,1,1;x)-2 \zeta_3,
\erp
%ep^2
\brp
k_2^{(0)}=-\frac{2}{81} x^3 d_1'^3+\frac{43 x^2 d_1'^3}{216}-\frac{251 x \
d_1'^3}{108}+2 H(0,1,1,1;x) d_1'^3+\frac{2 x^3 d_1'^2}{9}-\frac{191 \
x^2 d_1'^2}{108}+\frac{548 x d_1'^2}{27}+4 H(0,0,1,1;x) d_1'^2+4 \
H(0,1,0,1;x) d_1'^2+4 H(0,1,1,0;x) d_1'^2+\frac{1}{27} \pi ^2 x^3 \
d_1'-\frac{28 x^3 d_1'}{27}-\frac{7}{36} \pi ^2 x^2 d_1'+\frac{355 \
x^2 d_1'}{54}+\frac{11}{18} \pi ^2 x d_1'-\frac{1619 x d_1'}{27}+8 \
H(0,0,0,1;x) d_1'+8 H(0,0,1,0;x) d_1'+8 H(0,1,0,0;x) d_1'-\frac{5 \pi \
^2 x^3}{27}+\frac{328 x^3}{81}+\frac{5 \pi ^2 x^2}{6}-17 x^2-\frac{7 \
\pi ^2 x}{3}+72 x+\Big(-\frac{2}{27} x^3 d_1'^3+\frac{17 x^2 \
d_1'^3}{36}-\frac{49 x d_1'^3}{18}+\frac{251 d_1'^3}{108}+\frac{14 \
x^3 d_1'^2}{27}-\frac{28 x^2 d_1'^2}{9}+\frac{157 x \
d_1'^2}{9}-\frac{401 d_1'^2}{27}+\frac{1}{9} \pi ^2 x^3 d_1'-\frac{56 \
x^3 d_1'}{27}-\frac{1}{2} \pi ^2 x^2 d_1'+9 x^2 d_1'+\pi ^2 x d_1'-32 \
x d_1'-\frac{11 \pi ^2 d_1'}{18}+\frac{677 d_1'}{27}\Big) \
H(1;x)+\Big(-\frac{8 d_1' x^3}{9}+\frac{40 x^3}{9}+\frac{14 d_1' \
x^2}{3}-20 x^2-\frac{44 d_1' x}{3}+56 x-\frac{2 \pi ^2}{3}\Big) \
H(0,0;x)+\Big(-\frac{4}{9} d_1'^2 x^3+\frac{20 d_1' x^3}{9}+\frac{7 \
d_1'^2 x^2}{3}-10 d_1' x^2-\frac{22 d_1'^2 x}{3}+28 d_1' x-\frac{d_1' \
\pi ^2}{3}\Big) H(0,1;x)+\Big(-\frac{4}{9} d_1'^2 x^3+\frac{20 d_1' \
x^3}{9}+\frac{7 d_1'^2 x^2}{3}-10 d_1' x^2-\frac{22 d_1'^2 x}{3}+28 \
d_1' x+\frac{49 d_1'^2}{9}-\frac{182 d_1'}{9}\Big) \
H(1,0;x)+\Big(-\frac{2}{9} x^3 d_1'^3+\frac{7 x^2 d_1'^3}{6}-\frac{11 \
x d_1'^3}{3}+\frac{49 d_1'^3}{18}+\frac{10 x^3 d_1'^2}{9}-5 x^2 \
d_1'^2+14 x d_1'^2-\frac{91 d_1'^2}{9}\Big) H(1,1;x)+\Big(-\frac{16 \
x^3}{3}+24 x^2-48 x\Big) H(0,0,0;x)+\Big(-\frac{8 d_1' x^3}{3}+12 \
d_1' x^2-24 d_1' x\Big) H(0,0,1;x)+\Big(-\frac{8 d_1' x^3}{3}+12 d_1' \
x^2-24 d_1' x\Big) H(0,1,0;x)+\Big(-\frac{4}{3} d_1'^2 x^3+6 d_1'^2 \
x^2-12 d_1'^2 x\Big) H(0,1,1;x)+\Big(-\frac{8 d_1' x^3}{3}+12 d_1' \
x^2-24 d_1' x+\frac{44 d_1'}{3}\Big) H(1,0,0;x)+\Big(-\frac{4}{3} \
d_1'^2 x^3+6 d_1'^2 x^2-12 d_1'^2 x+\frac{22 d_1'^2}{3}\Big) \
H(1,0,1;x)+\Big(-\frac{4}{3} d_1'^2 x^3+6 d_1'^2 x^2-12 d_1'^2 \
x+\frac{22 d_1'^2}{3}\Big) H(1,1,0;x)+\Big(-\frac{2}{3} x^3 d_1'^3+3 \
x^2 d_1'^3-6 x d_1'^3+\frac{11 d_1'^3}{3}\Big) H(1,1,1;x)+16 \
H(0,0,0,0;x)+H(0;x) \Big(-\frac{4}{27} d_1'^2 x^3+\frac{28 d_1' \
x^3}{27}+\frac{2 \pi ^2 x^3}{9}-\frac{112 x^3}{27}+\frac{17 d_1'^2 \
x^2}{18}-\frac{56 d_1' x^2}{9}-\pi ^2 x^2+18 x^2-\frac{49 d_1'^2 \
x}{9}+\frac{314 d_1' x}{9}+2 \pi ^2 x-64 x+4 \zeta_3\Big)-\frac{4}{3} \
x^3 \zeta_3+6 x^2 \zeta_3-12 x \zeta_3-\frac{\pi ^4}{40}.
\erp

%
% The K integral for kappa=1
%

\subsection{The $\cK$ integral for $\kap =1$}
%
% This file contains the TeX output produced by Mathematica for the integral K, for arbitrary kap = 1 and D0 = 3 +d'1 ep
%
The $\eps$ expansion for this integral reads
\beq
\kint(\eps;y_0,3+d'_1\eps;)=\frac{1}{\eps^2}k_{-2}^{(1)}+\frac{1}{\eps}k_{-1}^{(1)}+k_0^{(1)}+\eps k_1^{(1)}+\eps^2k_2^{(1)} +\ocal\left(\eps^3\right),
\eeq
where
%1/ep piece
\brp
k_{-2}^{(1)}=\frac{1}{4},
\erp
\brp
k_{-1}^{(1)}=\frac{\yo^3}{3}-\frac{3 \yo^2}{2}+3 \yo-H(0;\yo),
\erp
% ep^0
\brp
k_0^{(1)}=-\frac{d_1' x^3}{9}+x^3+\frac{7 d_1' x^2}{12}-\frac{53 \
x^2}{12}-\frac{11 d_1' x}{6}+\frac{73 x}{6}+\Big(-\frac{4 x^3}{3}+6 \
x^2-12 x\Big) H(0;x)+\Big(-\frac{d_1' x^3}{3}-\frac{x^3}{3}+\frac{3 \
d_1' x^2}{2}+\frac{3 x^2}{2}-3 d_1' x-3 x+\frac{11 \
d_1'}{6}+\frac{11}{6}\Big) H(1;x)+4 H(0,0;x)+(d_1'+1) H(0,1;x)-\frac{\
\pi ^2}{12},
\erp
% ep^1
\brp
k_1^{(1)}=\frac{d_1'^2 x^3}{27}-\frac{4 d_1' x^3}{9}-\frac{\pi ^2 \
x^3}{9}+\frac{7 x^3}{3}-\frac{17 d_1'^2 x^2}{72}+\frac{95 d_1' \
x^2}{36}+\frac{\pi ^2 x^2}{2}-\frac{259 x^2}{24}+\frac{49 d_1'^2 \
x}{36}-\frac{265 d_1' x}{18}-\pi ^2 x+\frac{515 x}{12}+\Big(\frac{4 \
d_1' x^3}{9}-4 x^3-\frac{7 d_1' x^2}{3}+\frac{53 x^2}{3}+\frac{22 \
d_1' x}{3}-\frac{146 x}{3}+\frac{\pi ^2}{3}\Big) \
H(0;x)+\Big(\frac{d_1'^2 x^3}{9}-\frac{8 d_1' x^3}{9}-x^3-\frac{7 \
d_1'^2 x^2}{12}+\frac{23 d_1' x^2}{6}+\frac{61 x^2}{12}+\frac{11 \
d_1'^2 x}{6}-\frac{31 d_1' x}{3}-\frac{83 x}{6}-\frac{49 d_1'^2}{36}+\
\frac{133 d_1'}{18}+\frac{39}{4}\Big) H(1;x)+\Big(\frac{16 x^3}{3}-24 \
x^2+48 x\Big) H(0,0;x)+\Big(\frac{4 d_1' x^3}{3}+\frac{2 x^3}{3}-6 \
d_1' x^2-3 x^2+12 d_1' x+6 x+2\Big) H(0,1;x)+\Big(\frac{4 d_1' \
x^3}{3}+\frac{4 x^3}{3}-6 d_1' x^2-6 x^2+12 d_1' x+12 x-\frac{22 \
d_1'}{3}-\frac{22}{3}\Big) H(1,0;x)+\Big(\frac{d_1'^2 x^3}{3}+\frac{2 \
d_1' x^3}{3}-\frac{x^3}{3}-\frac{3 d_1'^2 x^2}{2}-3 d_1' x^2+\frac{3 \
x^2}{2}+3 d_1'^2 x+6 d_1' x-3 x-\frac{11 d_1'^2}{6}-\frac{11 \
d_1'}{3}+\frac{11}{6}\Big) H(1,1;x)-16 H(0,0,0;x)+(-4 d_1'-2) \
H(0,0,1;x)+(-4 d_1'-4) H(0,1,0;x)+\Big(-d_1'^2-2 d_1'+1\Big) \
H(0,1,1;x)-\frac{3 \zeta_3}{2},
\erp
%ep^2
\brp
k_2^{(1)}=-\frac{1}{81} x^3 d_1'^3+\frac{43 x^2 d_1'^3}{432}-\frac{251 x \
d_1'^3}{216}+\frac{5 x^3 d_1'^2}{27}-\frac{635 x^2 \
d_1'^2}{432}+\frac{3631 x d_1'^2}{216}+\frac{1}{27} \pi ^2 x^3 \
d_1'-\frac{11 x^3 d_1'}{9}-\frac{7}{36} \pi ^2 x^2 d_1'+\frac{1195 \
x^2 d_1'}{144}+\frac{11}{18} \pi ^2 x d_1'-\frac{5831 x \
d_1'}{72}-\frac{\pi ^2 x^3}{3}+5 x^3+\frac{53 \pi ^2 \
x^2}{36}-\frac{1169 x^2}{48}-\frac{73 \pi ^2 x}{18}+\frac{1139 x}{8}+\
\Big(-\frac{1}{27} x^3 d_1'^3+\frac{17 x^2 d_1'^3}{72}-\frac{49 x \
d_1'^3}{36}+\frac{251 d_1'^3}{216}+\frac{11 x^3 d_1'^2}{27}-\frac{173 \
x^2 d_1'^2}{72}+\frac{481 x d_1'^2}{36}-\frac{2455 \
d_1'^2}{216}+\frac{1}{9} \pi ^2 x^3 d_1'-\frac{17 x^3 \
d_1'}{9}-\frac{1}{2} \pi ^2 x^2 d_1'+\frac{571 x^2 d_1'}{72}+\pi ^2 x \
d_1'-\frac{1013 x d_1'}{36}-\frac{11 \pi ^2 d_1'}{18}+\frac{1591 \
d_1'}{72}+\frac{\pi ^2 x^3}{9}-\frac{7 x^3}{3}-\frac{\pi ^2 \
x^2}{2}+\frac{307 x^2}{24}+\pi ^2 x-\frac{191 x}{4}-\frac{11 \pi \
^2}{18}+\frac{895}{24}\Big) H(1;x)+\Big(-\frac{16 d_1' x^3}{9}+16 \
x^3+\frac{28 d_1' x^2}{3}-\frac{212 x^2}{3}-\frac{88 d_1' \
x}{3}+\frac{584 x}{3}-\frac{4 \pi ^2}{3}\Big) \
H(0,0;x)+\Big(-\frac{4}{9} d_1'^2 x^3+\frac{34 d_1' x^3}{9}+2 \
x^3+\frac{7 d_1'^2 x^2}{3}-\frac{33 d_1' x^2}{2}-\frac{19 \
x^2}{2}-\frac{22 d_1'^2 x}{3}+45 d_1' x+26 x-\frac{d_1' \pi \
^2}{3}-\frac{\pi ^2}{3}+4\Big) H(0,1;x)+\Big(-\frac{4}{9} d_1'^2 x^3+\
\frac{32 d_1' x^3}{9}+4 x^3+\frac{7 d_1'^2 x^2}{3}-\frac{46 d_1' \
x^2}{3}-\frac{61 x^2}{3}-\frac{22 d_1'^2 x}{3}+\frac{124 d_1' \
x}{3}+\frac{166 x}{3}+\frac{49 d_1'^2}{9}-\frac{266 d_1'}{9}-39\Big) \
H(1,0;x)+\Big(-\frac{1}{9} x^3 d_1'^3+\frac{7 x^2 \
d_1'^3}{12}-\frac{11 x d_1'^3}{6}+\frac{49 d_1'^3}{36}+\frac{7 x^3 \
d_1'^2}{9}-\frac{13 x^2 d_1'^2}{4}+\frac{17 x d_1'^2}{2}-\frac{217 \
d_1'^2}{36}+\frac{19 x^3 d_1'}{9}-\frac{43 x^2 d_1'}{4}+\frac{59 x \
d_1'}{2}-\frac{751 d_1'}{36}-x^3+\frac{61 x^2}{12}-\frac{83 \
x}{6}+\frac{39}{4}\Big) H(1,1;x)+\Big(-\frac{64 x^3}{3}+96 x^2-192 \
x\Big) H(0,0,0;x)+\Big(-\frac{16 d_1' x^3}{3}-2 x^3+24 d_1' x^2+9 \
x^2-48 d_1' x-18 x-2\Big) H(0,0,1;x)+\Big(-\frac{16 d_1' \
x^3}{3}-\frac{8 x^3}{3}+24 d_1' x^2+12 x^2-48 d_1' x-24 x-8\Big) \
H(0,1,0;x)+\Big(-\frac{4}{3} d_1'^2 x^3-\frac{4 d_1' x^3}{3}+\frac{2 \
x^3}{3}+6 d_1'^2 x^2+6 d_1' x^2-3 x^2-12 d_1'^2 x-12 d_1' x+6 x-4 \
d_1'+2\Big) H(0,1,1;x)+\Big(-\frac{16 d_1' x^3}{3}-\frac{16 \
x^3}{3}+24 d_1' x^2+24 x^2-48 d_1' x-48 x+\frac{88 \
d_1'}{3}+\frac{88}{3}\Big) H(1,0,0;x)+\Big(-\frac{4}{3} d_1'^2 x^3-2 \
d_1' x^3+6 d_1'^2 x^2+9 d_1' x^2-12 d_1'^2 x-18 d_1' x+\frac{22 \
d_1'^2}{3}+11 d_1'\Big) H(1,0,1;x)+\Big(-\frac{4}{3} d_1'^2 \
x^3-\frac{8 d_1' x^3}{3}+\frac{4 x^3}{3}+6 d_1'^2 x^2+12 d_1' x^2-6 \
x^2-12 d_1'^2 x-24 d_1' x+12 x+\frac{22 d_1'^2}{3}+\frac{44 d_1'}{3}-\
\frac{22}{3}\Big) H(1,1,0;x)+\Big(-\frac{1}{3} x^3 d_1'^3+\frac{3 x^2 \
d_1'^3}{2}-3 x d_1'^3+\frac{11 d_1'^3}{6}-x^3 d_1'^2+\frac{9 x^2 \
d_1'^2}{2}-9 x d_1'^2+\frac{11 d_1'^2}{2}+x^3 d_1'-\frac{9 x^2 \
d_1'}{2}+9 x d_1'-\frac{11 d_1'}{2}-\frac{x^3}{3}+\frac{3 x^2}{2}-3 \
x+\frac{11}{6}\Big) H(1,1,1;x)+64 H(0,0,0,0;x)+(16 d_1'+6) \
H(0,0,0,1;x)+(16 d_1'+8) H(0,0,1,0;x)+\Big(4 d_1'^2+4 d_1'-2\Big) \
H(0,0,1,1;x)+(16 d_1'+16) H(0,1,0,0;x)+\Big(4 d_1'^2+6 d_1'\Big) \
H(0,1,0,1;x)+\Big(4 d_1'^2+8 d_1'-4\Big) H(0,1,1,0;x)+\Big(d_1'^3+3 \
d_1'^2-3 d_1'+1\Big) H(0,1,1,1;x)+H(0;x) \Big(-\frac{4}{27} d_1'^2 \
x^3+\frac{16 d_1' x^3}{9}+\frac{4 \pi ^2 x^3}{9}-\frac{28 \
x^3}{3}+\frac{17 d_1'^2 x^2}{18}-\frac{95 d_1' x^2}{9}-2 \pi ^2 \
x^2+\frac{259 x^2}{6}-\frac{49 d_1'^2 x}{9}+\frac{530 d_1' x}{9}+4 \
\pi ^2 x-\frac{515 x}{3}+6 \zeta_3\Big)-2 x^3 \zeta_3+9 x^2 \
\zeta_3-18 x \zeta_3-\frac{11 \pi ^4}{360}.
\erp

%%%
%%% The collinear A-type
%%%

\section{The $\cA$-type collinear integrals}
\label{app:AIntegrals}

%
% The A integral for k=0
%

\subsection{The $\cA$ integral for $k=0$ and arbitrary $\kappa$}
%
% This file contains the TeX output produced by Mathematica for the integral A0, for arbitrary kappa
%
The $\eps$ expansion for this integral reads
\beq
\bsp
\begin{cal}I\end{cal}(x,\eps;\ao,3+d_1\eps;\kappa,0,0,g_A) &= x\,\aint(\eps,x;3+d_1\eps;\kap,2)\\
&=\frac{1}{\eps}a_{-1}^{(\kap,0)}+a_0^{(\kap,0)}+\eps a_1^{(\kap,0)}+\eps^2a_2^{(\kap,0)} +\ocal\left(\eps^3\right),
\esp
\eeq
where
%1/ep piece
\brp
a_{-1}^{(\kap,0)}=-\frac{1}{(\kappa +1)},
\erp
% ep^0
\brp
a_0^{(\kap,0)} = 
\frac{\ao^4}{4 (x-1)}+\frac{\kappa  \ao^4}{4 (\kappa \
+1)}+\frac{\ao^4}{4 (\kappa +1)}-\frac{\ao^3}{x-1}-\frac{4 \kappa  \
\ao^3}{3 (\kappa +1)}-\frac{4 \ao^3}{3 (\kappa +1)}+\frac{\ao^3}{3 \
(x-1)^2}+\frac{3 \ao^2}{2 (x-1)}+\frac{3 \kappa  \ao^2}{\kappa \
+1}+\frac{3 \ao^2}{\kappa +1}-\frac{\ao^2}{(x-1)^2}+\frac{\ao^2}{2 \
(x-1)^3}-\frac{\ao}{x-1}-\frac{4 \kappa  \ao}{\kappa +1}-\frac{4 \
\ao}{\kappa \
+1}+\frac{\ao}{(x-1)^2}-\frac{\ao}{(x-1)^3}+\frac{\ao}{(x-1)^4}+\Big(\
1+\frac{1}{(x-1)^5}\Big) H(0;\ao)+\Big(1-\frac{1}{(x-1)^5}\Big) \
H(0;x)+\frac{H(c_1(\ao);x)}{(x-1)^5}-\frac{2}{\kappa +1},
\erp
% ep^1
\brp
a_1^{(\kap,0)} =-\frac{d_1 \ao^4}{8 (\kappa +1)}-\frac{d_1 \kappa  \ao^4}{8 (\kappa \
+1)}-\frac{d_1 \kappa  \ao^4}{8 (x-1) (\kappa +1)}+\frac{7 \kappa  \
\ao^4}{8 (x-1) (\kappa +1)}+\frac{7 \kappa  \ao^4}{8 (\kappa \
+1)}-\frac{d_1 \ao^4}{8 (x-1) (\kappa +1)}+\frac{5 \ao^4}{8 (x-1) \
(\kappa +1)}+\frac{5 \ao^4}{8 (\kappa +1)}+\frac{13 d_1 \ao^3}{18 \
(\kappa +1)}+\frac{13 d_1 \kappa  \ao^3}{18 (\kappa +1)}+\frac{d_1 \
\kappa  \ao^3}{2 (x-1) (\kappa +1)}-\frac{7 \kappa  \ao^3}{2 (x-1) \
(\kappa +1)}-\frac{2 d_1 \kappa  \ao^3}{9 (x-1)^2 (\kappa \
+1)}+\frac{19 \kappa  \ao^3}{12 (x-1)^2 (\kappa +1)}-\frac{61 \kappa  \
\ao^3}{12 (\kappa +1)}+\frac{d_1 \ao^3}{2 (x-1) (\kappa +1)}-\frac{5 \
\ao^3}{2 (x-1) (\kappa +1)}-\frac{2 d_1 \ao^3}{9 (x-1)^2 (\kappa \
+1)}+\frac{35 \ao^3}{36 (x-1)^2 (\kappa +1)}-\frac{125 \ao^3}{36 \
(\kappa +1)}-\frac{23 d_1 \ao^2}{12 (\kappa +1)}-\frac{23 d_1 \kappa  \
\ao^2}{12 (\kappa +1)}-\frac{3 d_1 \kappa  \ao^2}{4 (x-1) (\kappa \
+1)}+\frac{41 \kappa  \ao^2}{8 (x-1) (\kappa +1)}+\frac{2 d_1 \kappa  \
\ao^2}{3 (x-1)^2 (\kappa +1)}-\frac{39 \kappa  \ao^2}{8 (x-1)^2 \
(\kappa +1)}-\frac{d_1 \kappa  \ao^2}{2 (x-1)^3 (\kappa +1)}+\frac{27 \
\kappa  \ao^2}{8 (x-1)^3 (\kappa +1)}+\frac{107 \kappa  \ao^2}{8 \
(\kappa +1)}-\frac{3 d_1 \ao^2}{4 (x-1) (\kappa +1)}+\frac{89 \
\ao^2}{24 (x-1) (\kappa +1)}+\frac{2 d_1 \ao^2}{3 (x-1)^2 (\kappa \
+1)}-\frac{71 \ao^2}{24 (x-1)^2 (\kappa +1)}-\frac{d_1 \ao^2}{2 \
(x-1)^3 (\kappa +1)}+\frac{43 \ao^2}{24 (x-1)^3 (\kappa \
+1)}+\frac{203 \ao^2}{24 (\kappa +1)}+\frac{25 d_1 \ao}{6 (\kappa \
+1)}+\frac{25 d_1 \kappa  \ao}{6 (\kappa +1)}+\frac{d_1 \kappa  \
\ao}{2 (x-1) (\kappa +1)}-\frac{5 \kappa  \ao}{2 (x-1) (\kappa \
+1)}-\frac{2 d_1 \kappa  \ao}{3 (x-1)^2 (\kappa +1)}+\frac{5 \kappa  \
\ao}{(x-1)^2 (\kappa +1)}+\frac{d_1 \kappa  \ao}{(x-1)^3 (\kappa \
+1)}-\frac{15 \kappa  \ao}{2 (x-1)^3 (\kappa +1)}-\frac{2 d_1 \kappa  \
\ao}{(x-1)^4 (\kappa +1)}+\frac{45 \kappa  \ao}{4 (x-1)^4 (\kappa \
+1)}-\frac{105 \kappa  \ao}{4 (\kappa +1)}+\frac{d_1 \ao}{2 (x-1) \
(\kappa +1)}-\frac{13 \ao}{6 (x-1) (\kappa +1)}-\frac{2 d_1 \ao}{3 \
(x-1)^2 (\kappa +1)}+\frac{3 \ao}{(x-1)^2 (\kappa +1)}+\frac{d_1 \
\ao}{(x-1)^3 (\kappa +1)}-\frac{23 \ao}{6 (x-1)^3 (\kappa \
+1)}-\frac{2 d_1 \ao}{(x-1)^4 (\kappa +1)}+\frac{61 \ao}{12 (x-1)^4 (\
\kappa +1)}-\frac{169 \ao}{12 (\kappa +1)}+\Big(-\frac{\kappa  \
\ao^4}{2 (x-1)}-\frac{\kappa  \ao^4}{2}-\frac{\ao^4}{2 \
(x-1)}-\frac{\ao^4}{2}+\frac{2 \kappa  \ao^3}{x-1}-\frac{2 \kappa  \
\ao^3}{3 (x-1)^2}+\frac{8 \kappa  \ao^3}{3}+\frac{2 \
\ao^3}{x-1}-\frac{2 \ao^3}{3 (x-1)^2}+\frac{8 \ao^3}{3}-\frac{3 \
\kappa  \ao^2}{x-1}+\frac{2 \kappa  \ao^2}{(x-1)^2}-\frac{\kappa  \
\ao^2}{(x-1)^3}-6 \kappa  \ao^2-\frac{3 \ao^2}{x-1}+\frac{2 \
\ao^2}{(x-1)^2}-\frac{\ao^2}{(x-1)^3}-6 \ao^2+\frac{2 \kappa  \
\ao}{x-1}-\frac{2 \kappa  \ao}{(x-1)^2}+\frac{2 \kappa  \
\ao}{(x-1)^3}-\frac{2 \kappa  \ao}{(x-1)^4}+8 \kappa  \ao+\frac{2 \
\ao}{x-1}-\frac{2 \ao}{(x-1)^2}+\frac{2 \ao}{(x-1)^3}-\frac{2 \
\ao}{(x-1)^4}+8 \ao-\frac{5 \kappa }{4 (x-1)}+\frac{5 \kappa }{6 \
(x-1)^2}-\frac{5 \kappa }{6 (x-1)^3}+\frac{5 \kappa }{4 \
(x-1)^4}+\frac{25 \kappa }{12 (x-1)^5}-\frac{25 \kappa \
}{12}-\frac{5}{4 (x-1)}+\frac{5}{6 (x-1)^2}-\frac{5}{6 \
(x-1)^3}+\frac{5}{4 (x-1)^4}+\frac{49}{12 (x-1)^5}-\frac{1}{12}\Big) \
H(0;\ao)+\Big(\frac{5 \kappa }{4 (x-1)}-\frac{5 \kappa }{6 \
(x-1)^2}+\frac{5 \kappa }{6 (x-1)^3}-\frac{5 \kappa }{4 \
(x-1)^4}-\frac{25 \kappa }{12 (x-1)^5}+\frac{25 \kappa \
}{12}+\frac{5}{4 (x-1)}-\frac{5}{6 (x-1)^2}+\frac{5}{6 \
(x-1)^3}-\frac{5}{4 (x-1)^4}-\frac{49}{12 (x-1)^5}+\frac{49}{12}\Big) \
H(0;x)+\Big(-\frac{d_1 \ao^4}{2}-\frac{d_1 \ao^4}{2 (x-1)}+\frac{8 \
d_1 \ao^3}{3}+\frac{2 d_1 \ao^3}{x-1}-\frac{2 d_1 \ao^3}{3 (x-1)^2}-6 \
d_1 \ao^2-\frac{3 d_1 \ao^2}{x-1}+\frac{2 d_1 \
\ao^2}{(x-1)^2}-\frac{d_1 \ao^2}{(x-1)^3}+8 d_1 \ao+\frac{2 d_1 \
\ao}{x-1}-\frac{2 d_1 \ao}{(x-1)^2}+\frac{2 d_1 \ao}{(x-1)^3}-\frac{2 \
d_1 \ao}{(x-1)^4}-\frac{25 d_1}{6}-\frac{d_1}{2 (x-1)}+\frac{2 d_1}{3 \
(x-1)^2}-\frac{d_1}{(x-1)^3}+\frac{2 d_1}{(x-1)^4}\Big) \
H(1;\ao)+\Big(\frac{2 d_1}{(x-1)^5}-\frac{\kappa }{(x-1)^5}+\kappa \
-\frac{1}{(x-1)^5}+1\Big) H(0;\ao) H(1;x)+\Big(-\frac{\kappa  \
\ao^4}{4 (x-1)}-\frac{\kappa  \ao^4}{4}-\frac{\ao^4}{4 \
(x-1)}-\frac{\ao^4}{4}+\frac{\kappa  \ao^3}{x-1}-\frac{\kappa  \
\ao^3}{3 (x-1)^2}+\frac{4 \kappa  \
\ao^3}{3}+\frac{\ao^3}{x-1}-\frac{\ao^3}{3 (x-1)^2}+\frac{4 \
\ao^3}{3}-\frac{3 \kappa  \ao^2}{2 (x-1)}+\frac{\kappa  \
\ao^2}{(x-1)^2}-\frac{\kappa  \ao^2}{2 (x-1)^3}-3 \kappa  \
\ao^2-\frac{3 \ao^2}{2 (x-1)}+\frac{\ao^2}{(x-1)^2}-\frac{\ao^2}{2 \
(x-1)^3}-3 \ao^2+\frac{\kappa  \ao}{x-1}-\frac{\kappa  \ao}{(x-1)^2}+\
\frac{\kappa  \ao}{(x-1)^3}-\frac{\kappa  \ao}{(x-1)^4}+4 \kappa  \
\ao+\frac{\ao}{x-1}-\frac{\ao}{(x-1)^2}+\frac{\ao}{(x-1)^3}-\frac{\ao}\
{(x-1)^4}+4 \ao-\frac{5 \kappa }{4 (x-1)}+\frac{5 \kappa }{6 \
(x-1)^2}-\frac{5 \kappa }{6 (x-1)^3}+\frac{5 \kappa }{4 \
(x-1)^4}+\frac{25 \kappa }{12 (x-1)^5}-\frac{25 \kappa \
}{12}+\Big(-\frac{2 \kappa }{(x-1)^5}-\frac{2}{(x-1)^5}\Big) \
H(0;\ao)-\frac{2 d_1 H(1;\ao)}{(x-1)^5}-\frac{5}{4 (x-1)}+\frac{5}{6 \
(x-1)^2}-\frac{5}{6 (x-1)^3}+\frac{5}{4 (x-1)^4}+\frac{49}{12 \
(x-1)^5}-\frac{25}{12}\Big) H(c_1(\ao);x)+\Big(-\frac{2 \kappa \
}{(x-1)^5}-2 \kappa -\frac{2}{(x-1)^5}-2\Big) H(0,0;\ao)+\Big(\frac{2 \
\kappa }{(x-1)^5}-2 \kappa +\frac{2}{(x-1)^5}-2\Big) \
H(0,0;x)+\Big(-\frac{2 d_1}{(x-1)^5}-2 d_1\Big) \
H(0,1;\ao)+\Big(-\frac{\kappa }{(x-1)^5}+\kappa \
-\frac{1}{(x-1)^5}+1\Big) H(0,c_1(\ao);x)+\Big(-\frac{2 \
d_1}{(x-1)^5}+\frac{\kappa }{(x-1)^5}-\kappa \
+\frac{1}{(x-1)^5}-1\Big) H(1,0;x)+\Big(\frac{2 \
d_1}{(x-1)^5}-\frac{\kappa }{(x-1)^5}+\kappa \
-\frac{1}{(x-1)^5}+1\Big) H(1,c_1(\ao);x)+\Big(-\frac{\kappa \
}{(x-1)^5}-\frac{1}{(x-1)^5}\Big) H(c_1(\ao),c_1(\ao);x)+\frac{\pi ^2 \
\kappa }{2 (x-1)^5 (\kappa +1)}-\frac{\pi ^2 \kappa }{2 (\kappa +1)}+\
\frac{\pi ^2}{6 (x-1)^5 (\kappa +1)}-\frac{4}{\kappa +1},
\erp
% ep^2
\brp
a_2^{(\kap,0)}=\frac{d_1^2 \ao^4}{16 (\kappa +1)}-\frac{3 d_1 \ao^4}{8 (\kappa +1)}+\
\frac{d_1^2 \kappa  \ao^4}{16 (\kappa +1)}-\frac{5 d_1 \kappa  \
\ao^4}{8 (\kappa +1)}+\frac{d_1^2 \kappa  \ao^4}{16 (x-1) (\kappa \
+1)}-\frac{5 d_1 \kappa  \ao^4}{8 (x-1) (\kappa +1)}-\frac{\pi ^2 \
\kappa  \ao^4}{24 (x-1) (\kappa +1)}+\frac{35 \kappa  \ao^4}{16 (x-1) \
(\kappa +1)}+\frac{35 \kappa  \ao^4}{16 (\kappa +1)}+\frac{d_1^2 \
\ao^4}{16 (x-1) (\kappa +1)}-\frac{3 d_1 \ao^4}{8 (x-1) (\kappa +1)}-\
\frac{\pi ^2 \ao^4}{24 (x-1) (\kappa +1)}+\frac{21 \ao^4}{16 (x-1) \
(\kappa +1)}+\frac{21 \ao^4}{16 (\kappa +1)}-\frac{\pi ^2 \ao^4}{24}-\
\frac{43 d_1^2 \ao^3}{108 (\kappa +1)}+\frac{505 d_1 \ao^3}{216 \
(\kappa +1)}-\frac{43 d_1^2 \kappa  \ao^3}{108 (\kappa +1)}+\frac{33 \
d_1 \kappa  \ao^3}{8 (\kappa +1)}-\frac{d_1^2 \kappa  \ao^3}{4 (x-1) \
(\kappa +1)}+\frac{5 d_1 \kappa  \ao^3}{2 (x-1) (\kappa \
+1)}+\frac{\pi ^2 \kappa  \ao^3}{6 (x-1) (\kappa +1)}-\frac{35 \kappa \
 \ao^3}{4 (x-1) (\kappa +1)}+\frac{4 d_1^2 \kappa  \ao^3}{27 (x-1)^2 \
(\kappa +1)}-\frac{13 d_1 \kappa  \ao^3}{8 (x-1)^2 (\kappa \
+1)}-\frac{\pi ^2 \kappa  \ao^3}{18 (x-1)^2 (\kappa +1)}+\frac{1055 \
\kappa  \ao^3}{216 (x-1)^2 (\kappa +1)}-\frac{2945 \kappa  \ao^3}{216 \
(\kappa +1)}-\frac{d_1^2 \ao^3}{4 (x-1) (\kappa +1)}+\frac{3 d_1 \
\ao^3}{2 (x-1) (\kappa +1)}+\frac{\pi ^2 \ao^3}{6 (x-1) (\kappa +1)}-\
\frac{21 \ao^3}{4 (x-1) (\kappa +1)}+\frac{4 d_1^2 \ao^3}{27 (x-1)^2 \
(\kappa +1)}-\frac{181 d_1 \ao^3}{216 (x-1)^2 (\kappa +1)}-\frac{\pi \
^2 \ao^3}{18 (x-1)^2 (\kappa +1)}+\frac{473 \ao^3}{216 (x-1)^2 \
(\kappa +1)}-\frac{1607 \ao^3}{216 (\kappa +1)}+\frac{2 \pi ^2 \
\ao^3}{9}+\frac{95 d_1^2 \ao^2}{72 (\kappa +1)}-\frac{347 d_1 \
\ao^2}{48 (\kappa +1)}+\frac{95 d_1^2 \kappa  \ao^2}{72 (\kappa +1)}-\
\frac{673 d_1 \kappa  \ao^2}{48 (\kappa +1)}+\frac{3 d_1^2 \kappa  \
\ao^2}{8 (x-1) (\kappa +1)}-\frac{167 d_1 \kappa  \ao^2}{48 (x-1) \
(\kappa +1)}-\frac{\pi ^2 \kappa  \ao^2}{4 (x-1) (\kappa \
+1)}+\frac{1721 \kappa  \ao^2}{144 (x-1) (\kappa +1)}-\frac{4 d_1^2 \
\kappa  \ao^2}{9 (x-1)^2 (\kappa +1)}+\frac{247 d_1 \kappa  \ao^2}{48 \
(x-1)^2 (\kappa +1)}+\frac{\pi ^2 \kappa  \ao^2}{6 (x-1)^2 (\kappa \
+1)}-\frac{2279 \kappa  \ao^2}{144 (x-1)^2 (\kappa +1)}+\frac{d_1^2 \
\kappa  \ao^2}{2 (x-1)^3 (\kappa +1)}-\frac{259 d_1 \kappa  \ao^2}{48 \
(x-1)^3 (\kappa +1)}-\frac{\pi ^2 \kappa  \ao^2}{12 (x-1)^3 (\kappa \
+1)}+\frac{1987 \kappa  \ao^2}{144 (x-1)^3 (\kappa +1)}+\frac{5987 \
\kappa  \ao^2}{144 (\kappa +1)}+\frac{3 d_1^2 \ao^2}{8 (x-1) (\kappa \
+1)}-\frac{311 d_1 \ao^2}{144 (x-1) (\kappa +1)}-\frac{\pi ^2 \
\ao^2}{4 (x-1) (\kappa +1)}+\frac{1103 \ao^2}{144 (x-1) (\kappa +1)}-\
\frac{4 d_1^2 \ao^2}{9 (x-1)^2 (\kappa +1)}+\frac{125 d_1 \ao^2}{48 \
(x-1)^2 (\kappa +1)}+\frac{\pi ^2 \ao^2}{6 (x-1)^2 (\kappa \
+1)}-\frac{977 \ao^2}{144 (x-1)^2 (\kappa +1)}+\frac{d_1^2 \ao^2}{2 \
(x-1)^3 (\kappa +1)}-\frac{355 d_1 \ao^2}{144 (x-1)^3 (\kappa \
+1)}-\frac{\pi ^2 \ao^2}{12 (x-1)^3 (\kappa +1)}+\frac{661 \ao^2}{144 \
(x-1)^3 (\kappa +1)}+\frac{2741 \ao^2}{144 (\kappa +1)}-\frac{\pi ^2 \
\ao^2}{2}-\frac{205 d_1^2 \ao}{36 (\kappa +1)}+\frac{575 d_1 \ao}{24 \
(\kappa +1)}-\frac{205 d_1^2 \kappa  \ao}{36 (\kappa +1)}+\frac{1325 \
d_1 \kappa  \ao}{24 (\kappa +1)}-\frac{d_1^2 \kappa  \ao}{4 (x-1) \
(\kappa +1)}-\frac{2 d_1 \kappa  \ao}{3 (x-1) (\kappa +1)}+\frac{\pi \
^2 \kappa  \ao}{6 (x-1) (\kappa +1)}+\frac{32 \kappa  \ao}{9 (x-1) \
(\kappa +1)}+\frac{4 d_1^2 \kappa  \ao}{9 (x-1)^2 (\kappa \
+1)}-\frac{65 d_1 \kappa  \ao}{12 (x-1)^2 (\kappa +1)}-\frac{\pi ^2 \
\kappa  \ao}{6 (x-1)^2 (\kappa +1)}+\frac{17 \kappa  \ao}{(x-1)^2 \
(\kappa +1)}-\frac{d_1^2 \kappa  \ao}{(x-1)^3 (\kappa +1)}+\frac{161 \
d_1 \kappa  \ao}{12 (x-1)^3 (\kappa +1)}+\frac{\pi ^2 \kappa  \ao}{6 \
(x-1)^3 (\kappa +1)}-\frac{338 \kappa  \ao}{9 (x-1)^3 (\kappa \
+1)}+\frac{4 d_1^2 \kappa  \ao}{(x-1)^4 (\kappa +1)}-\frac{889 d_1 \
\kappa  \ao}{24 (x-1)^4 (\kappa +1)}-\frac{\pi ^2 \kappa  \ao}{6 \
(x-1)^4 (\kappa +1)}+\frac{668 \kappa  \ao}{9 (x-1)^4 (\kappa \
+1)}-\frac{1127 \kappa  \ao}{9 (\kappa +1)}-\frac{d_1^2 \ao}{4 (x-1) \
(\kappa +1)}+\frac{4 d_1 \ao}{9 (x-1) (\kappa +1)}+\frac{\pi ^2 \
\ao}{6 (x-1) (\kappa +1)}-\frac{28 \ao}{9 (x-1) (\kappa +1)}+\frac{4 \
d_1^2 \ao}{9 (x-1)^2 (\kappa +1)}-\frac{97 d_1 \ao}{36 (x-1)^2 \
(\kappa +1)}-\frac{\pi ^2 \ao}{6 (x-1)^2 (\kappa +1)}+\frac{7 \
\ao}{(x-1)^2 (\kappa +1)}-\frac{d_1^2 \ao}{(x-1)^3 (\kappa \
+1)}+\frac{209 d_1 \ao}{36 (x-1)^3 (\kappa +1)}+\frac{\pi ^2 \ao}{6 \
(x-1)^3 (\kappa +1)}-\frac{98 \ao}{9 (x-1)^3 (\kappa +1)}+\frac{4 \
d_1^2 \ao}{(x-1)^4 (\kappa +1)}-\frac{1081 d_1 \ao}{72 (x-1)^4 \
(\kappa +1)}-\frac{\pi ^2 \ao}{6 (x-1)^4 (\kappa +1)}+\frac{158 \
\ao}{9 (x-1)^4 (\kappa +1)}-\frac{347 \ao}{9 (\kappa +1)}+\frac{2 \pi \
^2 \ao}{3}+\Big(\frac{d_1 \ao^4}{4}+\frac{1}{4} d_1 \kappa  \
\ao^4+\frac{d_1 \kappa  \ao^4}{4 (x-1)}-\frac{7 \kappa  \ao^4}{4 \
(x-1)}-\frac{7 \kappa  \ao^4}{4}+\frac{d_1 \ao^4}{4 (x-1)}-\frac{5 \
\ao^4}{4 (x-1)}-\frac{5 \ao^4}{4}-\frac{13 d_1 \ao^3}{9}-\frac{13}{9} \
d_1 \kappa  \ao^3-\frac{d_1 \kappa  \ao^3}{x-1}+\frac{7 \kappa  \
\ao^3}{x-1}+\frac{4 d_1 \kappa  \ao^3}{9 (x-1)^2}-\frac{19 \kappa  \
\ao^3}{6 (x-1)^2}+\frac{61 \kappa  \ao^3}{6}-\frac{d_1 \
\ao^3}{x-1}+\frac{5 \ao^3}{x-1}+\frac{4 d_1 \ao^3}{9 \
(x-1)^2}-\frac{35 \ao^3}{18 (x-1)^2}+\frac{125 \ao^3}{18}+\frac{23 \
d_1 \ao^2}{6}+\frac{23}{6} d_1 \kappa  \ao^2+\frac{3 d_1 \kappa  \
\ao^2}{2 (x-1)}-\frac{41 \kappa  \ao^2}{4 (x-1)}-\frac{4 d_1 \kappa  \
\ao^2}{3 (x-1)^2}+\frac{39 \kappa  \ao^2}{4 (x-1)^2}+\frac{d_1 \kappa \
 \ao^2}{(x-1)^3}-\frac{27 \kappa  \ao^2}{4 (x-1)^3}-\frac{107 \kappa  \
\ao^2}{4}+\frac{3 d_1 \ao^2}{2 (x-1)}-\frac{89 \ao^2}{12 \
(x-1)}-\frac{4 d_1 \ao^2}{3 (x-1)^2}+\frac{71 \ao^2}{12 \
(x-1)^2}+\frac{d_1 \ao^2}{(x-1)^3}-\frac{43 \ao^2}{12 \
(x-1)^3}-\frac{203 \ao^2}{12}-\frac{25 d_1 \ao}{3}-\frac{25 d_1 \
\kappa  \ao}{3}-\frac{d_1 \kappa  \ao}{x-1}+\frac{5 \kappa  \
\ao}{x-1}+\frac{4 d_1 \kappa  \ao}{3 (x-1)^2}-\frac{10 \kappa  \
\ao}{(x-1)^2}-\frac{2 d_1 \kappa  \ao}{(x-1)^3}+\frac{15 \kappa  \
\ao}{(x-1)^3}+\frac{4 d_1 \kappa  \ao}{(x-1)^4}-\frac{45 \kappa  \
\ao}{2 (x-1)^4}+\frac{105 \kappa  \ao}{2}-\frac{d_1 \
\ao}{x-1}+\frac{13 \ao}{3 (x-1)}+\frac{4 d_1 \ao}{3 (x-1)^2}-\frac{6 \
\ao}{(x-1)^2}-\frac{2 d_1 \ao}{(x-1)^3}+\frac{23 \ao}{3 \
(x-1)^3}+\frac{4 d_1 \ao}{(x-1)^4}-\frac{61 \ao}{6 (x-1)^4}+\frac{169 \
\ao}{6}+\frac{205 d_1}{72}+\frac{205 d_1 \kappa }{72}+\frac{17 d_1 \
\kappa }{8 (x-1)}-\frac{45 \kappa }{4 (x-1)}-\frac{13 d_1 \kappa }{18 \
(x-1)^2}+\frac{35 \kappa }{6 (x-1)^2}+\frac{13 d_1 \kappa }{18 \
(x-1)^3}-\frac{35 \kappa }{6 (x-1)^3}-\frac{17 d_1 \kappa }{8 \
(x-1)^4}+\frac{45 \kappa }{4 (x-1)^4}-\frac{205 d_1 \kappa }{72 \
(x-1)^5}+\frac{205 \kappa }{12 (x-1)^5}-\frac{205 \kappa \
}{12}+\frac{17 d_1}{8 (x-1)}-\frac{65}{12 (x-1)}-\frac{13 d_1}{18 \
(x-1)^2}+\frac{55}{18 (x-1)^2}+\frac{13 d_1}{18 (x-1)^3}-\frac{55}{18 \
(x-1)^3}-\frac{17 d_1}{8 (x-1)^4}+\frac{65}{12 (x-1)^4}-\frac{205 \
d_1}{72 (x-1)^5}-\frac{\pi ^2}{6 (x-1)^5}+\frac{449}{36 \
(x-1)^5}-\frac{\pi ^2}{6}-\frac{161}{36}\Big) H(0;\ao)+\Big(-\frac{17 \
\kappa  d_1}{8 (x-1)}+\frac{13 \kappa  d_1}{18 (x-1)^2}-\frac{13 \
\kappa  d_1}{18 (x-1)^3}+\frac{17 \kappa  d_1}{8 (x-1)^4}+\frac{205 \
\kappa  d_1}{72 (x-1)^5}-\frac{205 \kappa  d_1}{72}-\frac{17 d_1}{8 \
(x-1)}+\frac{13 d_1}{18 (x-1)^2}-\frac{13 d_1}{18 (x-1)^3}+\frac{17 \
d_1}{8 (x-1)^4}+\frac{205 d_1}{72 (x-1)^5}-\frac{205 \
d_1}{72}+\frac{45 \kappa }{4 (x-1)}-\frac{35 \kappa }{6 \
(x-1)^2}+\frac{35 \kappa }{6 (x-1)^3}-\frac{45 \kappa }{4 \
(x-1)^4}-\frac{\pi ^2 \kappa }{(x-1)^5}-\frac{205 \kappa }{12 \
(x-1)^5}+\pi ^2 \kappa +\frac{205 \kappa }{12}+\frac{65}{12 \
(x-1)}-\frac{55}{18 (x-1)^2}+\frac{55}{18 (x-1)^3}-\frac{65}{12 \
(x-1)^4}-\frac{\pi ^2}{6 (x-1)^5}-\frac{449}{36 (x-1)^5}+\frac{\pi \
^2}{6}+\frac{449}{36}\Big) H(0;x)+\Big(\frac{d_1^2 \ao^4}{4}-\frac{5 \
d_1 \ao^4}{4}-\frac{1}{4} d_1 \kappa  \ao^4-\frac{d_1 \kappa  \
\ao^4}{4 (x-1)}+\frac{d_1^2 \ao^4}{4 (x-1)}-\frac{5 d_1 \ao^4}{4 \
(x-1)}-\frac{13 d_1^2 \ao^3}{9}+\frac{125 d_1 \
\ao^3}{18}+\frac{29}{18} d_1 \kappa  \ao^3+\frac{d_1 \kappa  \
\ao^3}{x-1}-\frac{11 d_1 \kappa  \ao^3}{18 (x-1)^2}-\frac{d_1^2 \
\ao^3}{x-1}+\frac{5 d_1 \ao^3}{x-1}+\frac{4 d_1^2 \ao^3}{9 \
(x-1)^2}-\frac{35 d_1 \ao^3}{18 (x-1)^2}+\frac{23 d_1^2 \
\ao^2}{6}-\frac{203 d_1 \ao^2}{12}-\frac{59}{12} d_1 \kappa  \
\ao^2-\frac{17 d_1 \kappa  \ao^2}{12 (x-1)}+\frac{23 d_1 \kappa  \
\ao^2}{12 (x-1)^2}-\frac{19 d_1 \kappa  \ao^2}{12 (x-1)^3}+\frac{3 \
d_1^2 \ao^2}{2 (x-1)}-\frac{89 d_1 \ao^2}{12 (x-1)}-\frac{4 d_1^2 \
\ao^2}{3 (x-1)^2}+\frac{71 d_1 \ao^2}{12 (x-1)^2}+\frac{d_1^2 \
\ao^2}{(x-1)^3}-\frac{43 d_1 \ao^2}{12 (x-1)^3}-\frac{25 d_1^2 \
\ao}{3}+\frac{169 d_1 \ao}{6}+\frac{73 d_1 \kappa  \ao}{6}+\frac{d_1 \
\kappa  \ao}{3 (x-1)}-\frac{2 d_1 \kappa  \ao}{(x-1)^2}+\frac{11 d_1 \
\kappa  \ao}{3 (x-1)^3}-\frac{37 d_1 \kappa  \ao}{6 \
(x-1)^4}-\frac{d_1^2 \ao}{x-1}+\frac{13 d_1 \ao}{3 (x-1)}+\frac{4 \
d_1^2 \ao}{3 (x-1)^2}-\frac{6 d_1 \ao}{(x-1)^2}-\frac{2 d_1^2 \
\ao}{(x-1)^3}+\frac{23 d_1 \ao}{3 (x-1)^3}+\frac{4 d_1^2 \
\ao}{(x-1)^4}-\frac{61 d_1 \ao}{6 (x-1)^4}+\frac{205 \
d_1^2}{36}-\frac{305 d_1}{18}-\frac{155 d_1 \kappa }{18}+\frac{d_1 \
\kappa }{3 (x-1)}+\frac{25 d_1 \kappa }{36 (x-1)^2}-\frac{25 d_1 \
\kappa }{12 (x-1)^3}+\frac{37 d_1 \kappa }{6 (x-1)^4}+\frac{d_1^2}{4 \
(x-1)}-\frac{2 d_1}{3 (x-1)}-\frac{4 d_1^2}{9 (x-1)^2}+\frac{73 \
d_1}{36 (x-1)^2}+\frac{d_1^2}{(x-1)^3}-\frac{49 d_1}{12 \
(x-1)^3}-\frac{4 d_1^2}{(x-1)^4}+\frac{61 d_1}{6 (x-1)^4}\Big) \
H(1;\ao)+\Big(\frac{3 \kappa  \ao^4}{x-1}+3 \kappa  \
\ao^4+\frac{\ao^4}{x-1}+\ao^4-\frac{12 \kappa  \ao^3}{x-1}+\frac{4 \
\kappa  \ao^3}{(x-1)^2}-16 \kappa  \ao^3-\frac{4 \ao^3}{x-1}+\frac{4 \
\ao^3}{3 (x-1)^2}-\frac{16 \ao^3}{3}+\frac{18 \kappa  \
\ao^2}{x-1}-\frac{12 \kappa  \ao^2}{(x-1)^2}+\frac{6 \kappa  \
\ao^2}{(x-1)^3}+36 \kappa  \ao^2+\frac{6 \ao^2}{x-1}-\frac{4 \
\ao^2}{(x-1)^2}+\frac{2 \ao^2}{(x-1)^3}+12 \ao^2-\frac{12 \kappa  \
\ao}{x-1}+\frac{12 \kappa  \ao}{(x-1)^2}-\frac{12 \kappa  \
\ao}{(x-1)^3}+\frac{12 \kappa  \ao}{(x-1)^4}-48 \kappa  \ao-\frac{4 \
\ao}{x-1}+\frac{4 \ao}{(x-1)^2}-\frac{4 \ao}{(x-1)^3}+\frac{4 \
\ao}{(x-1)^4}-16 \ao+\frac{15 \kappa }{2 (x-1)}-\frac{5 \kappa \
}{(x-1)^2}+\frac{5 \kappa }{(x-1)^3}-\frac{15 \kappa }{2 \
(x-1)^4}-\frac{33 \kappa }{2 (x-1)^5}+\frac{17 \kappa }{2}+\frac{5}{2 \
(x-1)}-\frac{5}{3 (x-1)^2}+\frac{5}{3 (x-1)^3}-\frac{5}{2 \
(x-1)^4}-\frac{49}{6 (x-1)^5}+\frac{1}{6}\Big) \
H(0,0;\ao)+\Big(-\frac{15 \kappa }{2 (x-1)}+\frac{5 \kappa \
}{(x-1)^2}-\frac{5 \kappa }{(x-1)^3}+\frac{15 \kappa }{2 \
(x-1)^4}+\frac{33 \kappa }{2 (x-1)^5}-\frac{33 \kappa }{2}-\frac{5}{2 \
(x-1)}+\frac{5}{3 (x-1)^2}-\frac{5}{3 (x-1)^3}+\frac{5}{2 \
(x-1)^4}+\frac{49}{6 (x-1)^5}-\frac{49}{6}\Big) H(0,0;x)+\Big(d_1 \
\ao^4+d_1 \kappa  \ao^4+\frac{d_1 \kappa  \ao^4}{x-1}+\frac{d_1 \
\ao^4}{x-1}-\frac{16 d_1 \ao^3}{3}-\frac{16}{3} d_1 \kappa  \
\ao^3-\frac{4 d_1 \kappa  \ao^3}{x-1}+\frac{4 d_1 \kappa  \ao^3}{3 \
(x-1)^2}-\frac{4 d_1 \ao^3}{x-1}+\frac{4 d_1 \ao^3}{3 (x-1)^2}+12 d_1 \
\ao^2+12 d_1 \kappa  \ao^2+\frac{6 d_1 \kappa  \ao^2}{x-1}-\frac{4 \
d_1 \kappa  \ao^2}{(x-1)^2}+\frac{2 d_1 \kappa  \
\ao^2}{(x-1)^3}+\frac{6 d_1 \ao^2}{x-1}-\frac{4 d_1 \
\ao^2}{(x-1)^2}+\frac{2 d_1 \ao^2}{(x-1)^3}-16 d_1 \ao-16 d_1 \kappa  \
\ao-\frac{4 d_1 \kappa  \ao}{x-1}+\frac{4 d_1 \kappa  \
\ao}{(x-1)^2}-\frac{4 d_1 \kappa  \ao}{(x-1)^3}+\frac{4 d_1 \kappa  \
\ao}{(x-1)^4}-\frac{4 d_1 \ao}{x-1}+\frac{4 d_1 \ao}{(x-1)^2}-\frac{4 \
d_1 \ao}{(x-1)^3}+\frac{4 d_1 \ao}{(x-1)^4}+\frac{d_1}{6}+\frac{25 \
d_1 \kappa }{6}+\frac{5 d_1 \kappa }{2 (x-1)}-\frac{5 d_1 \kappa }{3 \
(x-1)^2}+\frac{5 d_1 \kappa }{3 (x-1)^3}-\frac{5 d_1 \kappa }{2 \
(x-1)^4}-\frac{25 d_1 \kappa }{6 (x-1)^5}+\frac{5 d_1}{2 \
(x-1)}-\frac{5 d_1}{3 (x-1)^2}+\frac{5 d_1}{3 (x-1)^3}-\frac{5 d_1}{2 \
(x-1)^4}-\frac{49 d_1}{6 (x-1)^5}\Big) H(0,1;\ao)+H(1;x) \
\Big(\frac{\pi ^2 \kappa  d_1}{3 (x-1)^5}+\frac{\pi ^2 d_1}{3 \
(x-1)^5}-\frac{\pi ^2 \kappa }{2 (x-1)^5}+\frac{\pi ^2 \kappa \
}{2}+\Big(-\frac{2 \kappa  d_1}{x-1}+\frac{\kappa  \
d_1}{(x-1)^2}-\frac{2 \kappa  d_1}{3 (x-1)^3}+\frac{\kappa  d_1}{2 \
(x-1)^4}+\frac{25 \kappa  d_1}{6 (x-1)^5}-\frac{2 \
d_1}{x-1}+\frac{d_1}{(x-1)^2}-\frac{2 d_1}{3 (x-1)^3}+\frac{d_1}{2 \
(x-1)^4}+\frac{49 d_1}{6 (x-1)^5}+\frac{15 \kappa }{4 (x-1)}-\frac{5 \
\kappa }{2 (x-1)^2}+\frac{5 \kappa }{2 (x-1)^3}-\frac{15 \kappa }{4 \
(x-1)^4}-\frac{33 \kappa }{4 (x-1)^5}+\frac{33 \kappa }{4}+\frac{5}{4 \
(x-1)}-\frac{5}{6 (x-1)^2}+\frac{5}{6 (x-1)^3}-\frac{5}{4 \
(x-1)^4}-\frac{49}{12 (x-1)^5}+\frac{49}{12}\Big) \
H(0;\ao)+\Big(-\frac{4 \kappa  d_1}{(x-1)^5}-\frac{4 \
d_1}{(x-1)^5}+\frac{6 \kappa }{(x-1)^5}-6 \kappa +\frac{2}{(x-1)^5}-2\
\Big) H(0,0;\ao)+\Big(-\frac{4 d_1^2}{(x-1)^5}+\frac{2 \kappa  \
d_1}{(x-1)^5}-2 \kappa  d_1+\frac{2 d_1}{(x-1)^5}-2 d_1\Big) \
H(0,1;\ao)-\frac{\pi ^2}{6 (x-1)^5}+\frac{\pi \
^2}{6}\Big)+\Big(-\frac{2 \kappa  d_1}{(x-1)^5}+2 \kappa  d_1-\frac{2 \
d_1}{(x-1)^5}+2 d_1+\frac{6 \kappa }{(x-1)^5}-6 \kappa \
+\frac{2}{(x-1)^5}-2\Big) H(0;\ao) H(0,1;x)+\Big(\frac{15 \kappa }{4 \
(x-1)}-\frac{5 \kappa }{2 (x-1)^2}+\frac{5 \kappa }{2 \
(x-1)^3}-\frac{15 \kappa }{4 (x-1)^4}-\frac{33 \kappa }{4 \
(x-1)^5}+\frac{33 \kappa }{4}+\Big(\frac{6 \kappa }{(x-1)^5}-6 \kappa \
+\frac{2}{(x-1)^5}-2\Big) H(0;\ao)+\Big(\frac{2 \kappa  \
d_1}{(x-1)^5}-2 \kappa  d_1+\frac{2 d_1}{(x-1)^5}-2 d_1\Big) \
H(1;\ao)+\frac{5}{4 (x-1)}-\frac{5}{6 (x-1)^2}+\frac{5}{6 \
(x-1)^3}-\frac{5}{4 (x-1)^4}-\frac{49}{12 (x-1)^5}+\frac{49}{12}\Big) \
H(0,c_1(\ao);x)+\Big(d_1 \ao^4+d_1 \kappa  \ao^4+\frac{d_1 \kappa  \
\ao^4}{x-1}+\frac{d_1 \ao^4}{x-1}-\frac{16 d_1 \ao^3}{3}-\frac{16}{3} \
d_1 \kappa  \ao^3-\frac{4 d_1 \kappa  \ao^3}{x-1}+\frac{4 d_1 \kappa  \
\ao^3}{3 (x-1)^2}-\frac{4 d_1 \ao^3}{x-1}+\frac{4 d_1 \ao^3}{3 \
(x-1)^2}+12 d_1 \ao^2+12 d_1 \kappa  \ao^2+\frac{6 d_1 \kappa  \
\ao^2}{x-1}-\frac{4 d_1 \kappa  \ao^2}{(x-1)^2}+\frac{2 d_1 \kappa  \
\ao^2}{(x-1)^3}+\frac{6 d_1 \ao^2}{x-1}-\frac{4 d_1 \
\ao^2}{(x-1)^2}+\frac{2 d_1 \ao^2}{(x-1)^3}-16 d_1 \ao-16 d_1 \kappa  \
\ao-\frac{4 d_1 \kappa  \ao}{x-1}+\frac{4 d_1 \kappa  \
\ao}{(x-1)^2}-\frac{4 d_1 \kappa  \ao}{(x-1)^3}+\frac{4 d_1 \kappa  \
\ao}{(x-1)^4}-\frac{4 d_1 \ao}{x-1}+\frac{4 d_1 \ao}{(x-1)^2}-\frac{4 \
d_1 \ao}{(x-1)^3}+\frac{4 d_1 \ao}{(x-1)^4}+\frac{25 d_1}{3}+\frac{25 \
d_1 \kappa }{3}+\frac{d_1 \kappa }{x-1}-\frac{4 d_1 \kappa }{3 \
(x-1)^2}+\frac{2 d_1 \kappa }{(x-1)^3}-\frac{4 d_1 \kappa }{(x-1)^4}+\
\frac{d_1}{x-1}-\frac{4 d_1}{3 (x-1)^2}+\frac{2 d_1}{(x-1)^3}-\frac{4 \
d_1}{(x-1)^4}\Big) H(1,0;\ao)+\Big(\frac{2 \kappa  \
d_1}{x-1}-\frac{\kappa  d_1}{(x-1)^2}+\frac{2 \kappa  d_1}{3 \
(x-1)^3}-\frac{\kappa  d_1}{2 (x-1)^4}-\frac{25 \kappa  d_1}{6 \
(x-1)^5}+\frac{2 d_1}{x-1}-\frac{d_1}{(x-1)^2}+\frac{2 d_1}{3 \
(x-1)^3}-\frac{d_1}{2 (x-1)^4}-\frac{49 d_1}{6 (x-1)^5}-\frac{15 \
\kappa }{4 (x-1)}+\frac{5 \kappa }{2 (x-1)^2}-\frac{5 \kappa }{2 \
(x-1)^3}+\frac{15 \kappa }{4 (x-1)^4}+\frac{33 \kappa }{4 \
(x-1)^5}-\frac{33 \kappa }{4}-\frac{5}{4 (x-1)}+\frac{5}{6 \
(x-1)^2}-\frac{5}{6 (x-1)^3}+\frac{5}{4 (x-1)^4}+\frac{49}{12 \
(x-1)^5}-\frac{49}{12}\Big) H(1,0;x)+\Big(d_1^2 \ao^4+\frac{d_1^2 \
\ao^4}{x-1}-\frac{16 d_1^2 \ao^3}{3}-\frac{4 d_1^2 \
\ao^3}{x-1}+\frac{4 d_1^2 \ao^3}{3 (x-1)^2}+12 d_1^2 \ao^2+\frac{6 \
d_1^2 \ao^2}{x-1}-\frac{4 d_1^2 \ao^2}{(x-1)^2}+\frac{2 d_1^2 \
\ao^2}{(x-1)^3}-16 d_1^2 \ao-\frac{4 d_1^2 \ao}{x-1}+\frac{4 d_1^2 \
\ao}{(x-1)^2}-\frac{4 d_1^2 \ao}{(x-1)^3}+\frac{4 d_1^2 \
\ao}{(x-1)^4}+\frac{25 d_1^2}{3}+\frac{d_1^2}{x-1}-\frac{4 d_1^2}{3 \
(x-1)^2}+\frac{2 d_1^2}{(x-1)^3}-\frac{4 d_1^2}{(x-1)^4}\Big) \
H(1,1;\ao)+H(c_1(\ao);x) \Big(\frac{d_1 \ao^4}{8}+\frac{1}{8} d_1 \
\kappa  \ao^4+\frac{d_1 \kappa  \ao^4}{8 (x-1)}-\frac{7 \kappa  \
\ao^4}{8 (x-1)}-\frac{7 \kappa  \ao^4}{8}+\frac{d_1 \ao^4}{8 \
(x-1)}-\frac{5 \ao^4}{8 (x-1)}-\frac{5 \ao^4}{8}-\frac{13 d_1 \
\ao^3}{18}-\frac{13}{18} d_1 \kappa  \ao^3-\frac{d_1 \kappa  \ao^3}{2 \
(x-1)}+\frac{7 \kappa  \ao^3}{2 (x-1)}+\frac{2 d_1 \kappa  \ao^3}{9 \
(x-1)^2}-\frac{19 \kappa  \ao^3}{12 (x-1)^2}+\frac{61 \kappa  \
\ao^3}{12}-\frac{d_1 \ao^3}{2 (x-1)}+\frac{5 \ao^3}{2 (x-1)}+\frac{2 \
d_1 \ao^3}{9 (x-1)^2}-\frac{35 \ao^3}{36 (x-1)^2}+\frac{125 \
\ao^3}{36}+\frac{23 d_1 \ao^2}{12}+\frac{23}{12} d_1 \kappa  \
\ao^2+\frac{3 d_1 \kappa  \ao^2}{4 (x-1)}-\frac{41 \kappa  \ao^2}{8 \
(x-1)}-\frac{2 d_1 \kappa  \ao^2}{3 (x-1)^2}+\frac{39 \kappa  \
\ao^2}{8 (x-1)^2}+\frac{d_1 \kappa  \ao^2}{2 (x-1)^3}-\frac{27 \kappa \
 \ao^2}{8 (x-1)^3}-\frac{107 \kappa  \ao^2}{8}+\frac{3 d_1 \ao^2}{4 \
(x-1)}-\frac{89 \ao^2}{24 (x-1)}-\frac{2 d_1 \ao^2}{3 \
(x-1)^2}+\frac{71 \ao^2}{24 (x-1)^2}+\frac{d_1 \ao^2}{2 \
(x-1)^3}-\frac{43 \ao^2}{24 (x-1)^3}-\frac{203 \ao^2}{24}-\frac{25 \
d_1 \ao}{6}-\frac{25 d_1 \kappa  \ao}{6}-\frac{d_1 \kappa  \ao}{2 \
(x-1)}+\frac{5 \kappa  \ao}{2 (x-1)}+\frac{2 d_1 \kappa  \ao}{3 \
(x-1)^2}-\frac{5 \kappa  \ao}{(x-1)^2}-\frac{d_1 \kappa  \
\ao}{(x-1)^3}+\frac{15 \kappa  \ao}{2 (x-1)^3}+\frac{2 d_1 \kappa  \
\ao}{(x-1)^4}-\frac{45 \kappa  \ao}{4 (x-1)^4}+\frac{105 \kappa  \
\ao}{4}-\frac{d_1 \ao}{2 (x-1)}+\frac{13 \ao}{6 (x-1)}+\frac{2 d_1 \
\ao}{3 (x-1)^2}-\frac{3 \ao}{(x-1)^2}-\frac{d_1 \
\ao}{(x-1)^3}+\frac{23 \ao}{6 (x-1)^3}+\frac{2 d_1 \
\ao}{(x-1)^4}-\frac{61 \ao}{12 (x-1)^4}+\frac{169 \ao}{12}+\frac{205 \
d_1}{72}+\frac{205 d_1 \kappa }{72}+\frac{17 d_1 \kappa }{8 \
(x-1)}-\frac{45 \kappa }{4 (x-1)}-\frac{13 d_1 \kappa }{18 \
(x-1)^2}+\frac{35 \kappa }{6 (x-1)^2}+\frac{13 d_1 \kappa }{18 \
(x-1)^3}-\frac{35 \kappa }{6 (x-1)^3}-\frac{17 d_1 \kappa }{8 \
(x-1)^4}+\frac{45 \kappa }{4 (x-1)^4}-\frac{205 d_1 \kappa }{72 \
(x-1)^5}+\frac{205 \kappa }{12 (x-1)^5}-\frac{205 \kappa \
}{12}+\Big(\frac{3 \kappa  \ao^4}{2 (x-1)}+\frac{3 \kappa  \ao^4}{2}+\
\frac{\ao^4}{2 (x-1)}+\frac{\ao^4}{2}-\frac{6 \kappa  \
\ao^3}{x-1}+\frac{2 \kappa  \ao^3}{(x-1)^2}-8 \kappa  \ao^3-\frac{2 \
\ao^3}{x-1}+\frac{2 \ao^3}{3 (x-1)^2}-\frac{8 \ao^3}{3}+\frac{9 \
\kappa  \ao^2}{x-1}-\frac{6 \kappa  \ao^2}{(x-1)^2}+\frac{3 \kappa  \
\ao^2}{(x-1)^3}+18 \kappa  \ao^2+\frac{3 \ao^2}{x-1}-\frac{2 \
\ao^2}{(x-1)^2}+\frac{\ao^2}{(x-1)^3}+6 \ao^2-\frac{6 \kappa  \
\ao}{x-1}+\frac{6 \kappa  \ao}{(x-1)^2}-\frac{6 \kappa  \
\ao}{(x-1)^3}+\frac{6 \kappa  \ao}{(x-1)^4}-24 \kappa  \ao-\frac{2 \
\ao}{x-1}+\frac{2 \ao}{(x-1)^2}-\frac{2 \ao}{(x-1)^3}+\frac{2 \
\ao}{(x-1)^4}-8 \ao+\frac{15 \kappa }{2 (x-1)}-\frac{5 \kappa \
}{(x-1)^2}+\frac{5 \kappa }{(x-1)^3}-\frac{15 \kappa }{2 \
(x-1)^4}-\frac{33 \kappa }{2 (x-1)^5}+\frac{25 \kappa }{2}+\frac{5}{2 \
(x-1)}-\frac{5}{3 (x-1)^2}+\frac{5}{3 (x-1)^3}-\frac{5}{2 \
(x-1)^4}-\frac{49}{6 (x-1)^5}+\frac{25}{6}\Big) \
H(0;\ao)+\Big(\frac{d_1 \ao^4}{2}+\frac{1}{2} d_1 \kappa  \
\ao^4+\frac{d_1 \kappa  \ao^4}{2 (x-1)}+\frac{d_1 \ao^4}{2 \
(x-1)}-\frac{8 d_1 \ao^3}{3}-\frac{8}{3} d_1 \kappa  \ao^3-\frac{2 \
d_1 \kappa  \ao^3}{x-1}+\frac{2 d_1 \kappa  \ao^3}{3 (x-1)^2}-\frac{2 \
d_1 \ao^3}{x-1}+\frac{2 d_1 \ao^3}{3 (x-1)^2}+6 d_1 \ao^2+6 d_1 \
\kappa  \ao^2+\frac{3 d_1 \kappa  \ao^2}{x-1}-\frac{2 d_1 \kappa  \
\ao^2}{(x-1)^2}+\frac{d_1 \kappa  \ao^2}{(x-1)^3}+\frac{3 d_1 \
\ao^2}{x-1}-\frac{2 d_1 \ao^2}{(x-1)^2}+\frac{d_1 \ao^2}{(x-1)^3}-8 \
d_1 \ao-8 d_1 \kappa  \ao-\frac{2 d_1 \kappa  \ao}{x-1}+\frac{2 d_1 \
\kappa  \ao}{(x-1)^2}-\frac{2 d_1 \kappa  \ao}{(x-1)^3}+\frac{2 d_1 \
\kappa  \ao}{(x-1)^4}-\frac{2 d_1 \ao}{x-1}+\frac{2 d_1 \
\ao}{(x-1)^2}-\frac{2 d_1 \ao}{(x-1)^3}+\frac{2 d_1 \
\ao}{(x-1)^4}+\frac{25 d_1}{6}+\frac{25 d_1 \kappa }{6}+\frac{5 d_1 \
\kappa }{2 (x-1)}-\frac{5 d_1 \kappa }{3 (x-1)^2}+\frac{5 d_1 \kappa \
}{3 (x-1)^3}-\frac{5 d_1 \kappa }{2 (x-1)^4}-\frac{25 d_1 \kappa }{6 \
(x-1)^5}+\frac{5 d_1}{2 (x-1)}-\frac{5 d_1}{3 (x-1)^2}+\frac{5 d_1}{3 \
(x-1)^3}-\frac{5 d_1}{2 (x-1)^4}-\frac{49 d_1}{6 (x-1)^5}\Big) \
H(1;\ao)+\Big(\frac{12 \kappa }{(x-1)^5}+\frac{4}{(x-1)^5}\Big) \
H(0,0;\ao)+\Big(\frac{4 \kappa  d_1}{(x-1)^5}+\frac{4 \
d_1}{(x-1)^5}\Big) H(0,1;\ao)+\Big(\frac{4 \kappa  \
d_1}{(x-1)^5}+\frac{4 d_1}{(x-1)^5}\Big) H(1,0;\ao)+\frac{4 d_1^2 \
H(1,1;\ao)}{(x-1)^5}+\frac{17 d_1}{8 (x-1)}-\frac{65}{12 \
(x-1)}-\frac{13 d_1}{18 (x-1)^2}+\frac{55}{18 (x-1)^2}+\frac{13 \
d_1}{18 (x-1)^3}-\frac{55}{18 (x-1)^3}-\frac{17 d_1}{8 \
(x-1)^4}+\frac{65}{12 (x-1)^4}-\frac{205 d_1}{72 (x-1)^5}-\frac{\pi \
^2}{6 (x-1)^5}+\frac{449}{36 \
(x-1)^5}-\frac{305}{36}\Big)+\Big(\frac{4 d_1^2}{(x-1)^5}-\frac{4 \
\kappa  d_1}{(x-1)^5}+2 \kappa  d_1-\frac{4 d_1}{(x-1)^5}+2 \
d_1+\frac{3 \kappa }{(x-1)^5}-3 \kappa +\frac{1}{(x-1)^5}-1\Big) H(0;\
\ao) H(1,1;x)+\Big(-\frac{2 \kappa  d_1}{x-1}+\frac{\kappa  \
d_1}{(x-1)^2}-\frac{2 \kappa  d_1}{3 (x-1)^3}+\frac{\kappa  d_1}{2 \
(x-1)^4}+\frac{25 \kappa  d_1}{6 (x-1)^5}-\frac{2 \
d_1}{x-1}+\frac{d_1}{(x-1)^2}-\frac{2 d_1}{3 (x-1)^3}+\frac{d_1}{2 \
(x-1)^4}+\frac{49 d_1}{6 (x-1)^5}+\frac{15 \kappa }{4 (x-1)}-\frac{5 \
\kappa }{2 (x-1)^2}+\frac{5 \kappa }{2 (x-1)^3}-\frac{15 \kappa }{4 \
(x-1)^4}-\frac{33 \kappa }{4 (x-1)^5}+\frac{33 \kappa \
}{4}+\Big(-\frac{4 \kappa  d_1}{(x-1)^5}-\frac{4 \
d_1}{(x-1)^5}+\frac{6 \kappa }{(x-1)^5}-6 \kappa +\frac{2}{(x-1)^5}-2\
\Big) H(0;\ao)+\Big(-\frac{4 d_1^2}{(x-1)^5}+\frac{2 \kappa  \
d_1}{(x-1)^5}-2 \kappa  d_1+\frac{2 d_1}{(x-1)^5}-2 d_1\Big) \
H(1;\ao)+\frac{5}{4 (x-1)}-\frac{5}{6 (x-1)^2}+\frac{5}{6 \
(x-1)^3}-\frac{5}{4 (x-1)^4}-\frac{49}{12 (x-1)^5}+\frac{49}{12}\Big) \
H(1,c_1(\ao);x)+\Big(\frac{3 \kappa  \ao^4}{4 (x-1)}+\frac{3 \kappa  \
\ao^4}{4}+\frac{\ao^4}{4 (x-1)}+\frac{\ao^4}{4}-\frac{3 \kappa  \
\ao^3}{x-1}+\frac{\kappa  \ao^3}{(x-1)^2}-4 \kappa  \
\ao^3-\frac{\ao^3}{x-1}+\frac{\ao^3}{3 (x-1)^2}-\frac{4 \
\ao^3}{3}+\frac{9 \kappa  \ao^2}{2 (x-1)}-\frac{3 \kappa  \
\ao^2}{(x-1)^2}+\frac{3 \kappa  \ao^2}{2 (x-1)^3}+9 \kappa  \
\ao^2+\frac{3 \ao^2}{2 (x-1)}-\frac{\ao^2}{(x-1)^2}+\frac{\ao^2}{2 \
(x-1)^3}+3 \ao^2-\frac{3 \kappa  \ao}{x-1}+\frac{3 \kappa  \
\ao}{(x-1)^2}-\frac{3 \kappa  \ao}{(x-1)^3}+\frac{3 \kappa  \
\ao}{(x-1)^4}-12 \kappa  \
\ao-\frac{\ao}{x-1}+\frac{\ao}{(x-1)^2}-\frac{\ao}{(x-1)^3}+\frac{\ao}\
{(x-1)^4}-4 \ao+\frac{15 \kappa }{4 (x-1)}-\frac{5 \kappa }{2 \
(x-1)^2}+\frac{5 \kappa }{2 (x-1)^3}-\frac{15 \kappa }{4 \
(x-1)^4}-\frac{33 \kappa }{4 (x-1)^5}+\frac{25 \kappa \
}{4}+\Big(\frac{6 \kappa }{(x-1)^5}+\frac{2}{(x-1)^5}\Big) \
H(0;\ao)+\Big(\frac{2 \kappa  d_1}{(x-1)^5}+\frac{2 \
d_1}{(x-1)^5}\Big) H(1;\ao)+\frac{5}{4 (x-1)}-\frac{5}{6 \
(x-1)^2}+\frac{5}{6 (x-1)^3}-\frac{5}{4 (x-1)^4}-\frac{49}{12 \
(x-1)^5}+\frac{25}{12}\Big) H(c_1(\ao),c_1(\ao);x)+\Big(\frac{12 \
\kappa }{(x-1)^5}+12 \kappa +\frac{4}{(x-1)^5}+4\Big) \
H(0,0,0;\ao)+\Big(-\frac{12 \kappa }{(x-1)^5}+12 \kappa \
-\frac{4}{(x-1)^5}+4\Big) H(0,0,0;x)+\Big(\frac{4 \kappa  \
d_1}{(x-1)^5}+4 \kappa  d_1+\frac{4 d_1}{(x-1)^5}+4 d_1\Big) H(0,0,1;\
\ao)+\Big(\frac{6 \kappa }{(x-1)^5}-6 \kappa \
+\frac{2}{(x-1)^5}-2\Big) H(0,0,c_1(\ao);x)+\Big(\frac{4 \kappa  \
d_1}{(x-1)^5}+4 \kappa  d_1+\frac{4 d_1}{(x-1)^5}+4 d_1\Big) H(0,1,0;\
\ao)+\Big(\frac{2 \kappa  d_1}{(x-1)^5}-2 \kappa  d_1+\frac{2 \
d_1}{(x-1)^5}-2 d_1-\frac{6 \kappa }{(x-1)^5}+6 \kappa \
-\frac{2}{(x-1)^5}+2\Big) H(0,1,0;x)+\Big(\frac{4 d_1^2}{(x-1)^5}+4 \
d_1^2\Big) H(0,1,1;\ao)+\Big(-\frac{2 \kappa  d_1}{(x-1)^5}+2 \kappa  \
d_1-\frac{2 d_1}{(x-1)^5}+2 d_1+\frac{6 \kappa }{(x-1)^5}-6 \kappa \
+\frac{2}{(x-1)^5}-2\Big) H(0,1,c_1(\ao);x)+\Big(\frac{3 \kappa \
}{(x-1)^5}-3 \kappa +\frac{1}{(x-1)^5}-1\Big) \
H(0,c_1(\ao),c_1(\ao);x)+\Big(\frac{4 \kappa  d_1}{(x-1)^5}+\frac{4 \
d_1}{(x-1)^5}-\frac{6 \kappa }{(x-1)^5}+6 \kappa -\frac{2}{(x-1)^5}+2\
\Big) H(1,0,0;x)+\Big(-\frac{2 \kappa  d_1}{(x-1)^5}-\frac{2 \
d_1}{(x-1)^5}+\frac{3 \kappa }{(x-1)^5}-3 \kappa +\frac{1}{(x-1)^5}-1\
\Big) H(1,0,c_1(\ao);x)+\Big(-\frac{4 d_1^2}{(x-1)^5}+\frac{4 \kappa  \
d_1}{(x-1)^5}-2 \kappa  d_1+\frac{4 d_1}{(x-1)^5}-2 d_1-\frac{3 \
\kappa }{(x-1)^5}+3 \kappa -\frac{1}{(x-1)^5}+1\Big) H(1,1,0;x)+\Big(\
\frac{4 d_1^2}{(x-1)^5}-\frac{4 \kappa  d_1}{(x-1)^5}+2 \kappa  \
d_1-\frac{4 d_1}{(x-1)^5}+2 d_1+\frac{3 \kappa }{(x-1)^5}-3 \kappa \
+\frac{1}{(x-1)^5}-1\Big) H(1,1,c_1(\ao);x)+\Big(-\frac{2 \kappa  \
d_1}{(x-1)^5}-\frac{2 d_1}{(x-1)^5}+\frac{3 \kappa }{(x-1)^5}-3 \
\kappa +\frac{1}{(x-1)^5}-1\Big) \
H(1,c_1(\ao),c_1(\ao);x)+\Big(\frac{3 \kappa \
}{(x-1)^5}+\frac{1}{(x-1)^5}\Big) \
H(c_1(\ao),c_1(\ao),c_1(\ao);x)-\frac{35 \pi ^2 \kappa }{24 (x-1) \
(\kappa +1)}+\frac{35 \pi ^2 \kappa }{36 (x-1)^2 (\kappa \
+1)}-\frac{35 \pi ^2 \kappa }{36 (x-1)^3 (\kappa +1)}+\frac{35 \pi ^2 \
\kappa }{24 (x-1)^4 (\kappa +1)}+\frac{247 \pi ^2 \kappa }{72 (x-1)^5 \
(\kappa +1)}-\frac{247 \pi ^2 \kappa }{72 (\kappa +1)}-\frac{5 \pi \
^2}{24 (x-1) (\kappa +1)}+\frac{5 \pi ^2}{36 (x-1)^2 (\kappa \
+1)}-\frac{5 \pi ^2}{36 (x-1)^3 (\kappa +1)}+\frac{5 \pi ^2}{24 \
(x-1)^4 (\kappa +1)}+\frac{49 \pi ^2}{72 (x-1)^5 (\kappa \
+1)}-\frac{25 \pi ^2}{72 (\kappa +1)}-\frac{8}{\kappa +1}-\frac{7 \
\kappa  \zeta_3}{(x-1)^5 (\kappa +1)}+\frac{7 \kappa  \zeta_3}{\kappa \
+1}-\frac{\zeta_3}{(x-1)^5 (\kappa +1)}+\frac{3 \zeta_3}{\kappa +1}.
\erp

%
% The A integral for k=1
%

\subsection{The $\cA$ integral for $k=1$ and arbitrary $\kappa$}
%
% This file contains the TeX output produced by Mathematica for the integral A1, for arbitrary kappa and d0 = 3 +d1 ep
%
The $\eps$ expansion for this integral reads
\beq
\bsp
\begin{cal}I\end{cal}(x,\eps;\ao,3+d_1\eps;\kappa,1,0,g_A) &= x\,\aint(\eps,x;3+d_1\eps;\kap,1)\\
&=\frac{1}{\eps}a_{-1}^{(\kap,1)}+a_0^{(\kap,1)}+\eps a_1^{(\kap,1)}+\eps^2 a_2^{(\kap,1)}+\ocal\left(\eps^3\right),
\esp
\eeq
where
%1/ep piece
\brp
a_{-1}^{(\kap,1)}=-\frac{1}{2 (\kappa +1)},
\erp
% ep^0
\brp
a_0^{(\kap,1)} = \frac{\ao^4}{8 (x-1)}+\frac{\kappa  \ao^4}{8 (\kappa \
+1)}+\frac{\ao^4}{8 (\kappa +1)}-\frac{\ao^3}{2 (x-1)}-\frac{2 \kappa \
 \ao^3}{3 (\kappa +1)}-\frac{2 \ao^3}{3 (\kappa +1)}+\frac{\ao^3}{6 \
(x-1)^2}+\frac{3 \ao^2}{4 (x-1)}+\frac{3 \kappa  \ao^2}{2 (\kappa \
+1)}+\frac{3 \ao^2}{2 (\kappa +1)}-\frac{\ao^2}{2 \
(x-1)^2}+\frac{\ao^2}{4 (x-1)^3}-\frac{\ao}{2 (x-1)}-\frac{2 \kappa  \
\ao}{\kappa +1}-\frac{2 \ao}{\kappa +1}+\frac{\ao}{2 \
(x-1)^2}-\frac{\ao}{2 (x-1)^3}+\frac{\ao}{2 \
(x-1)^4}+\Big(\frac{1}{2}+\frac{1}{2 (x-1)^5}\Big) \
H(0;\ao)+\Big(\frac{1}{2}-\frac{1}{2 (x-1)^5}\Big) \
H(0;x)+\frac{H(c_1(\ao);x)}{2 (x-1)^5}-\frac{1}{\kappa +1},
\erp% ep^1
\brp
a_1^{(\kap,1)} = -\frac{d_1 \ao^4}{16 (\kappa +1)}-\frac{d_1 \kappa  \ao^4}{16 (\kappa \
+1)}-\frac{d_1 \kappa  \ao^4}{16 (x-1) (\kappa +1)}+\frac{7 \kappa  \
\ao^4}{16 (x-1) (\kappa +1)}+\frac{7 \kappa  \ao^4}{16 (\kappa \
+1)}-\frac{d_1 \ao^4}{16 (x-1) (\kappa +1)}+\frac{5 \ao^4}{16 (x-1) (\
\kappa +1)}+\frac{5 \ao^4}{16 (\kappa +1)}+\frac{13 d_1 \ao^3}{36 \
(\kappa +1)}+\frac{13 d_1 \kappa  \ao^3}{36 (\kappa +1)}+\frac{d_1 \
\kappa  \ao^3}{4 (x-1) (\kappa +1)}-\frac{7 \kappa  \ao^3}{4 (x-1) \
(\kappa +1)}-\frac{d_1 \kappa  \ao^3}{9 (x-1)^2 (\kappa +1)}+\frac{19 \
\kappa  \ao^3}{24 (x-1)^2 (\kappa +1)}-\frac{61 \kappa  \ao^3}{24 \
(\kappa +1)}+\frac{d_1 \ao^3}{4 (x-1) (\kappa +1)}-\frac{5 \ao^3}{4 \
(x-1) (\kappa +1)}-\frac{d_1 \ao^3}{9 (x-1)^2 (\kappa +1)}+\frac{35 \
\ao^3}{72 (x-1)^2 (\kappa +1)}-\frac{125 \ao^3}{72 (\kappa \
+1)}-\frac{23 d_1 \ao^2}{24 (\kappa +1)}-\frac{23 d_1 \kappa  \
\ao^2}{24 (\kappa +1)}-\frac{3 d_1 \kappa  \ao^2}{8 (x-1) (\kappa \
+1)}+\frac{41 \kappa  \ao^2}{16 (x-1) (\kappa +1)}+\frac{d_1 \kappa  \
\ao^2}{3 (x-1)^2 (\kappa +1)}-\frac{39 \kappa  \ao^2}{16 (x-1)^2 \
(\kappa +1)}-\frac{d_1 \kappa  \ao^2}{4 (x-1)^3 (\kappa +1)}+\frac{27 \
\kappa  \ao^2}{16 (x-1)^3 (\kappa +1)}+\frac{107 \kappa  \ao^2}{16 \
(\kappa +1)}-\frac{3 d_1 \ao^2}{8 (x-1) (\kappa +1)}+\frac{89 \
\ao^2}{48 (x-1) (\kappa +1)}+\frac{d_1 \ao^2}{3 (x-1)^2 (\kappa +1)}-\
\frac{71 \ao^2}{48 (x-1)^2 (\kappa +1)}-\frac{d_1 \ao^2}{4 (x-1)^3 \
(\kappa +1)}+\frac{43 \ao^2}{48 (x-1)^3 (\kappa +1)}+\frac{203 \
\ao^2}{48 (\kappa +1)}+\frac{25 d_1 \ao}{12 (\kappa +1)}+\frac{25 d_1 \
\kappa  \ao}{12 (\kappa +1)}+\frac{d_1 \kappa  \ao}{4 (x-1) (\kappa \
+1)}-\frac{5 \kappa  \ao}{4 (x-1) (\kappa +1)}-\frac{d_1 \kappa  \
\ao}{3 (x-1)^2 (\kappa +1)}+\frac{5 \kappa  \ao}{2 (x-1)^2 (\kappa \
+1)}+\frac{d_1 \kappa  \ao}{2 (x-1)^3 (\kappa +1)}-\frac{15 \kappa  \
\ao}{4 (x-1)^3 (\kappa +1)}-\frac{d_1 \kappa  \ao}{(x-1)^4 (\kappa \
+1)}+\frac{45 \kappa  \ao}{8 (x-1)^4 (\kappa +1)}-\frac{105 \kappa  \
\ao}{8 (\kappa +1)}+\frac{d_1 \ao}{4 (x-1) (\kappa +1)}-\frac{13 \
\ao}{12 (x-1) (\kappa +1)}-\frac{d_1 \ao}{3 (x-1)^2 (\kappa \
+1)}+\frac{3 \ao}{2 (x-1)^2 (\kappa +1)}+\frac{d_1 \ao}{2 (x-1)^3 \
(\kappa +1)}-\frac{23 \ao}{12 (x-1)^3 (\kappa +1)}-\frac{d_1 \
\ao}{(x-1)^4 (\kappa +1)}+\frac{61 \ao}{24 (x-1)^4 (\kappa \
+1)}-\frac{169 \ao}{24 (\kappa +1)}+\Big(-\frac{\kappa  \ao^4}{4 \
(x-1)}-\frac{\kappa  \ao^4}{4}-\frac{\ao^4}{4 (x-1)}-\frac{\ao^4}{4}+\
\frac{\kappa  \ao^3}{x-1}-\frac{\kappa  \ao^3}{3 (x-1)^2}+\frac{4 \
\kappa  \ao^3}{3}+\frac{\ao^3}{x-1}-\frac{\ao^3}{3 (x-1)^2}+\frac{4 \
\ao^3}{3}-\frac{3 \kappa  \ao^2}{2 (x-1)}+\frac{\kappa  \
\ao^2}{(x-1)^2}-\frac{\kappa  \ao^2}{2 (x-1)^3}-3 \kappa  \
\ao^2-\frac{3 \ao^2}{2 (x-1)}+\frac{\ao^2}{(x-1)^2}-\frac{\ao^2}{2 \
(x-1)^3}-3 \ao^2+\frac{\kappa  \ao}{x-1}-\frac{\kappa  \ao}{(x-1)^2}+\
\frac{\kappa  \ao}{(x-1)^3}-\frac{\kappa  \ao}{(x-1)^4}+4 \kappa  \
\ao+\frac{\ao}{x-1}-\frac{\ao}{(x-1)^2}+\frac{\ao}{(x-1)^3}-\frac{\ao}\
{(x-1)^4}+4 \ao-\frac{5 \kappa }{8 (x-1)}+\frac{5 \kappa }{12 \
(x-1)^2}-\frac{5 \kappa }{12 (x-1)^3}+\frac{5 \kappa }{8 \
(x-1)^4}+\frac{25 \kappa }{24 (x-1)^5}-\frac{25 \kappa \
}{24}-\frac{5}{8 (x-1)}+\frac{5}{12 (x-1)^2}-\frac{5}{12 \
(x-1)^3}+\frac{5}{8 (x-1)^4}+\frac{49}{24 (x-1)^5}-\frac{1}{24}\Big) \
H(0;\ao)+\Big(\frac{5 \kappa }{8 (x-1)}-\frac{5 \kappa }{12 (x-1)^2}+\
\frac{5 \kappa }{12 (x-1)^3}-\frac{5 \kappa }{8 (x-1)^4}-\frac{25 \
\kappa }{24 (x-1)^5}+\frac{25 \kappa }{24}+\frac{5}{8 \
(x-1)}-\frac{5}{12 (x-1)^2}+\frac{5}{12 (x-1)^3}-\frac{5}{8 (x-1)^4}-\
\frac{49}{24 (x-1)^5}+\frac{49}{24}\Big) H(0;x)+\Big(-\frac{d_1 \
\ao^4}{4}-\frac{d_1 \ao^4}{4 (x-1)}+\frac{4 d_1 \ao^3}{3}+\frac{d_1 \
\ao^3}{x-1}-\frac{d_1 \ao^3}{3 (x-1)^2}-3 d_1 \ao^2-\frac{3 d_1 \
\ao^2}{2 (x-1)}+\frac{d_1 \ao^2}{(x-1)^2}-\frac{d_1 \ao^2}{2 \
(x-1)^3}+4 d_1 \ao+\frac{d_1 \ao}{x-1}-\frac{d_1 \
\ao}{(x-1)^2}+\frac{d_1 \ao}{(x-1)^3}-\frac{d_1 \
\ao}{(x-1)^4}-\frac{25 d_1}{12}-\frac{d_1}{4 (x-1)}+\frac{d_1}{3 \
(x-1)^2}-\frac{d_1}{2 (x-1)^3}+\frac{d_1}{(x-1)^4}\Big) \
H(1;\ao)+\Big(\frac{d_1}{(x-1)^5}-\frac{\kappa }{2 \
(x-1)^5}+\frac{\kappa }{2}-\frac{1}{2 (x-1)^5}+\frac{1}{2}\Big) \
H(0;\ao) H(1;x)+\Big(-\frac{\kappa  \ao^4}{8 (x-1)}-\frac{\kappa  \
\ao^4}{8}-\frac{\ao^4}{8 (x-1)}-\frac{\ao^4}{8}+\frac{\kappa  \
\ao^3}{2 (x-1)}-\frac{\kappa  \ao^3}{6 (x-1)^2}+\frac{2 \kappa  \
\ao^3}{3}+\frac{\ao^3}{2 (x-1)}-\frac{\ao^3}{6 (x-1)^2}+\frac{2 \
\ao^3}{3}-\frac{3 \kappa  \ao^2}{4 (x-1)}+\frac{\kappa  \ao^2}{2 \
(x-1)^2}-\frac{\kappa  \ao^2}{4 (x-1)^3}-\frac{3 \kappa  \
\ao^2}{2}-\frac{3 \ao^2}{4 (x-1)}+\frac{\ao^2}{2 \
(x-1)^2}-\frac{\ao^2}{4 (x-1)^3}-\frac{3 \ao^2}{2}+\frac{\kappa  \
\ao}{2 (x-1)}-\frac{\kappa  \ao}{2 (x-1)^2}+\frac{\kappa  \ao}{2 \
(x-1)^3}-\frac{\kappa  \ao}{2 (x-1)^4}+2 \kappa  \ao+\frac{\ao}{2 \
(x-1)}-\frac{\ao}{2 (x-1)^2}+\frac{\ao}{2 (x-1)^3}-\frac{\ao}{2 \
(x-1)^4}+2 \ao-\frac{5 \kappa }{8 (x-1)}+\frac{5 \kappa }{12 \
(x-1)^2}-\frac{5 \kappa }{12 (x-1)^3}+\frac{5 \kappa }{8 \
(x-1)^4}+\frac{25 \kappa }{24 (x-1)^5}-\frac{25 \kappa \
}{24}+\Big(-\frac{\kappa }{(x-1)^5}-\frac{1}{(x-1)^5}\Big) \
H(0;\ao)-\frac{d_1 H(1;\ao)}{(x-1)^5}-\frac{5}{8 (x-1)}+\frac{5}{12 \
(x-1)^2}-\frac{5}{12 (x-1)^3}+\frac{5}{8 (x-1)^4}+\frac{49}{24 \
(x-1)^5}-\frac{25}{24}\Big) H(c_1(\ao);x)+\Big(-\frac{\kappa \
}{(x-1)^5}-\kappa -\frac{1}{(x-1)^5}-1\Big) \
H(0,0;\ao)+\Big(\frac{\kappa }{(x-1)^5}-\kappa \
+\frac{1}{(x-1)^5}-1\Big) H(0,0;x)+\Big(-\frac{d_1}{(x-1)^5}-d_1\Big) \
H(0,1;\ao)+\Big(-\frac{\kappa }{2 (x-1)^5}+\frac{\kappa \
}{2}-\frac{1}{2 (x-1)^5}+\frac{1}{2}\Big) \
H(0,c_1(\ao);x)+\Big(-\frac{d_1}{(x-1)^5}+\frac{\kappa }{2 \
(x-1)^5}-\frac{\kappa }{2}+\frac{1}{2 (x-1)^5}-\frac{1}{2}\Big) \
H(1,0;x)+\Big(\frac{d_1}{(x-1)^5}-\frac{\kappa }{2 \
(x-1)^5}+\frac{\kappa }{2}-\frac{1}{2 (x-1)^5}+\frac{1}{2}\Big) \
H(1,c_1(\ao);x)+\Big(-\frac{\kappa }{2 (x-1)^5}-\frac{1}{2 \
(x-1)^5}\Big) H(c_1(\ao),c_1(\ao);x)+\frac{\pi ^2 \kappa }{4 (x-1)^5 \
(\kappa +1)}-\frac{\pi ^2 \kappa }{4 (\kappa +1)}+\frac{\pi ^2}{12 \
(x-1)^5 (\kappa +1)}-\frac{2}{\kappa +1},
\erp
% ep^2
\brp
a_2^{(\kap,1)} = \frac{d_1^2 \ao^4}{32 (\kappa +1)}-\frac{3 d_1 \ao^4}{16 (\kappa \
+1)}+\frac{d_1^2 \kappa  \ao^4}{32 (\kappa +1)}-\frac{5 d_1 \kappa  \
\ao^4}{16 (\kappa +1)}+\frac{d_1^2 \kappa  \ao^4}{32 (x-1) (\kappa \
+1)}-\frac{5 d_1 \kappa  \ao^4}{16 (x-1) (\kappa +1)}-\frac{\pi ^2 \
\kappa  \ao^4}{48 (x-1) (\kappa +1)}+\frac{35 \kappa  \ao^4}{32 (x-1) \
(\kappa +1)}+\frac{35 \kappa  \ao^4}{32 (\kappa +1)}+\frac{d_1^2 \
\ao^4}{32 (x-1) (\kappa +1)}-\frac{3 d_1 \ao^4}{16 (x-1) (\kappa \
+1)}-\frac{\pi ^2 \ao^4}{48 (x-1) (\kappa +1)}+\frac{21 \ao^4}{32 \
(x-1) (\kappa +1)}+\frac{21 \ao^4}{32 (\kappa +1)}-\frac{\pi ^2 \
\ao^4}{48}-\frac{43 d_1^2 \ao^3}{216 (\kappa +1)}+\frac{505 d_1 \
\ao^3}{432 (\kappa +1)}-\frac{43 d_1^2 \kappa  \ao^3}{216 (\kappa \
+1)}+\frac{33 d_1 \kappa  \ao^3}{16 (\kappa +1)}-\frac{d_1^2 \kappa  \
\ao^3}{8 (x-1) (\kappa +1)}+\frac{5 d_1 \kappa  \ao^3}{4 (x-1) \
(\kappa +1)}+\frac{\pi ^2 \kappa  \ao^3}{12 (x-1) (\kappa \
+1)}-\frac{35 \kappa  \ao^3}{8 (x-1) (\kappa +1)}+\frac{2 d_1^2 \
\kappa  \ao^3}{27 (x-1)^2 (\kappa +1)}-\frac{13 d_1 \kappa  \ao^3}{16 \
(x-1)^2 (\kappa +1)}-\frac{\pi ^2 \kappa  \ao^3}{36 (x-1)^2 (\kappa \
+1)}+\frac{1055 \kappa  \ao^3}{432 (x-1)^2 (\kappa +1)}-\frac{2945 \
\kappa  \ao^3}{432 (\kappa +1)}-\frac{d_1^2 \ao^3}{8 (x-1) (\kappa \
+1)}+\frac{3 d_1 \ao^3}{4 (x-1) (\kappa +1)}+\frac{\pi ^2 \ao^3}{12 \
(x-1) (\kappa +1)}-\frac{21 \ao^3}{8 (x-1) (\kappa +1)}+\frac{2 d_1^2 \
\ao^3}{27 (x-1)^2 (\kappa +1)}-\frac{181 d_1 \ao^3}{432 (x-1)^2 \
(\kappa +1)}-\frac{\pi ^2 \ao^3}{36 (x-1)^2 (\kappa +1)}+\frac{473 \
\ao^3}{432 (x-1)^2 (\kappa +1)}-\frac{1607 \ao^3}{432 (\kappa \
+1)}+\frac{\pi ^2 \ao^3}{9}+\frac{95 d_1^2 \ao^2}{144 (\kappa \
+1)}-\frac{347 d_1 \ao^2}{96 (\kappa +1)}+\frac{95 d_1^2 \kappa  \
\ao^2}{144 (\kappa +1)}-\frac{673 d_1 \kappa  \ao^2}{96 (\kappa +1)}+\
\frac{3 d_1^2 \kappa  \ao^2}{16 (x-1) (\kappa +1)}-\frac{167 d_1 \
\kappa  \ao^2}{96 (x-1) (\kappa +1)}-\frac{\pi ^2 \kappa  \ao^2}{8 \
(x-1) (\kappa +1)}+\frac{1721 \kappa  \ao^2}{288 (x-1) (\kappa \
+1)}-\frac{2 d_1^2 \kappa  \ao^2}{9 (x-1)^2 (\kappa +1)}+\frac{247 \
d_1 \kappa  \ao^2}{96 (x-1)^2 (\kappa +1)}+\frac{\pi ^2 \kappa  \
\ao^2}{12 (x-1)^2 (\kappa +1)}-\frac{2279 \kappa  \ao^2}{288 (x-1)^2 \
(\kappa +1)}+\frac{d_1^2 \kappa  \ao^2}{4 (x-1)^3 (\kappa \
+1)}-\frac{259 d_1 \kappa  \ao^2}{96 (x-1)^3 (\kappa +1)}-\frac{\pi \
^2 \kappa  \ao^2}{24 (x-1)^3 (\kappa +1)}+\frac{1987 \kappa  \
\ao^2}{288 (x-1)^3 (\kappa +1)}+\frac{5987 \kappa  \ao^2}{288 (\kappa \
+1)}+\frac{3 d_1^2 \ao^2}{16 (x-1) (\kappa +1)}-\frac{311 d_1 \
\ao^2}{288 (x-1) (\kappa +1)}-\frac{\pi ^2 \ao^2}{8 (x-1) (\kappa \
+1)}+\frac{1103 \ao^2}{288 (x-1) (\kappa +1)}-\frac{2 d_1^2 \ao^2}{9 \
(x-1)^2 (\kappa +1)}+\frac{125 d_1 \ao^2}{96 (x-1)^2 (\kappa \
+1)}+\frac{\pi ^2 \ao^2}{12 (x-1)^2 (\kappa +1)}-\frac{977 \ao^2}{288 \
(x-1)^2 (\kappa +1)}+\frac{d_1^2 \ao^2}{4 (x-1)^3 (\kappa \
+1)}-\frac{355 d_1 \ao^2}{288 (x-1)^3 (\kappa +1)}-\frac{\pi ^2 \
\ao^2}{24 (x-1)^3 (\kappa +1)}+\frac{661 \ao^2}{288 (x-1)^3 (\kappa \
+1)}+\frac{2741 \ao^2}{288 (\kappa +1)}-\frac{\pi ^2 \
\ao^2}{4}-\frac{205 d_1^2 \ao}{72 (\kappa +1)}+\frac{575 d_1 \ao}{48 \
(\kappa +1)}-\frac{205 d_1^2 \kappa  \ao}{72 (\kappa +1)}+\frac{1325 \
d_1 \kappa  \ao}{48 (\kappa +1)}-\frac{d_1^2 \kappa  \ao}{8 (x-1) \
(\kappa +1)}-\frac{d_1 \kappa  \ao}{3 (x-1) (\kappa +1)}+\frac{\pi ^2 \
\kappa  \ao}{12 (x-1) (\kappa +1)}+\frac{16 \kappa  \ao}{9 (x-1) \
(\kappa +1)}+\frac{2 d_1^2 \kappa  \ao}{9 (x-1)^2 (\kappa \
+1)}-\frac{65 d_1 \kappa  \ao}{24 (x-1)^2 (\kappa +1)}-\frac{\pi ^2 \
\kappa  \ao}{12 (x-1)^2 (\kappa +1)}+\frac{17 \kappa  \ao}{2 (x-1)^2 \
(\kappa +1)}-\frac{d_1^2 \kappa  \ao}{2 (x-1)^3 (\kappa \
+1)}+\frac{161 d_1 \kappa  \ao}{24 (x-1)^3 (\kappa +1)}+\frac{\pi ^2 \
\kappa  \ao}{12 (x-1)^3 (\kappa +1)}-\frac{169 \kappa  \ao}{9 (x-1)^3 \
(\kappa +1)}+\frac{2 d_1^2 \kappa  \ao}{(x-1)^4 (\kappa \
+1)}-\frac{889 d_1 \kappa  \ao}{48 (x-1)^4 (\kappa +1)}-\frac{\pi ^2 \
\kappa  \ao}{12 (x-1)^4 (\kappa +1)}+\frac{334 \kappa  \ao}{9 (x-1)^4 \
(\kappa +1)}-\frac{1127 \kappa  \ao}{18 (\kappa +1)}-\frac{d_1^2 \
\ao}{8 (x-1) (\kappa +1)}+\frac{2 d_1 \ao}{9 (x-1) (\kappa \
+1)}+\frac{\pi ^2 \ao}{12 (x-1) (\kappa +1)}-\frac{14 \ao}{9 (x-1) \
(\kappa +1)}+\frac{2 d_1^2 \ao}{9 (x-1)^2 (\kappa +1)}-\frac{97 d_1 \
\ao}{72 (x-1)^2 (\kappa +1)}-\frac{\pi ^2 \ao}{12 (x-1)^2 (\kappa \
+1)}+\frac{7 \ao}{2 (x-1)^2 (\kappa +1)}-\frac{d_1^2 \ao}{2 (x-1)^3 (\
\kappa +1)}+\frac{209 d_1 \ao}{72 (x-1)^3 (\kappa +1)}+\frac{\pi ^2 \
\ao}{12 (x-1)^3 (\kappa +1)}-\frac{49 \ao}{9 (x-1)^3 (\kappa \
+1)}+\frac{2 d_1^2 \ao}{(x-1)^4 (\kappa +1)}-\frac{1081 d_1 \ao}{144 \
(x-1)^4 (\kappa +1)}-\frac{\pi ^2 \ao}{12 (x-1)^4 (\kappa \
+1)}+\frac{79 \ao}{9 (x-1)^4 (\kappa +1)}-\frac{347 \ao}{18 (\kappa \
+1)}+\frac{\pi ^2 \ao}{3}+\Big(\frac{d_1 \ao^4}{8}+\frac{1}{8} d_1 \
\kappa  \ao^4+\frac{d_1 \kappa  \ao^4}{8 (x-1)}-\frac{7 \kappa  \
\ao^4}{8 (x-1)}-\frac{7 \kappa  \ao^4}{8}+\frac{d_1 \ao^4}{8 \
(x-1)}-\frac{5 \ao^4}{8 (x-1)}-\frac{5 \ao^4}{8}-\frac{13 d_1 \
\ao^3}{18}-\frac{13}{18} d_1 \kappa  \ao^3-\frac{d_1 \kappa  \ao^3}{2 \
(x-1)}+\frac{7 \kappa  \ao^3}{2 (x-1)}+\frac{2 d_1 \kappa  \ao^3}{9 \
(x-1)^2}-\frac{19 \kappa  \ao^3}{12 (x-1)^2}+\frac{61 \kappa  \
\ao^3}{12}-\frac{d_1 \ao^3}{2 (x-1)}+\frac{5 \ao^3}{2 (x-1)}+\frac{2 \
d_1 \ao^3}{9 (x-1)^2}-\frac{35 \ao^3}{36 (x-1)^2}+\frac{125 \
\ao^3}{36}+\frac{23 d_1 \ao^2}{12}+\frac{23}{12} d_1 \kappa  \
\ao^2+\frac{3 d_1 \kappa  \ao^2}{4 (x-1)}-\frac{41 \kappa  \ao^2}{8 \
(x-1)}-\frac{2 d_1 \kappa  \ao^2}{3 (x-1)^2}+\frac{39 \kappa  \
\ao^2}{8 (x-1)^2}+\frac{d_1 \kappa  \ao^2}{2 (x-1)^3}-\frac{27 \kappa \
 \ao^2}{8 (x-1)^3}-\frac{107 \kappa  \ao^2}{8}+\frac{3 d_1 \ao^2}{4 \
(x-1)}-\frac{89 \ao^2}{24 (x-1)}-\frac{2 d_1 \ao^2}{3 \
(x-1)^2}+\frac{71 \ao^2}{24 (x-1)^2}+\frac{d_1 \ao^2}{2 \
(x-1)^3}-\frac{43 \ao^2}{24 (x-1)^3}-\frac{203 \ao^2}{24}-\frac{25 \
d_1 \ao}{6}-\frac{25 d_1 \kappa  \ao}{6}-\frac{d_1 \kappa  \ao}{2 \
(x-1)}+\frac{5 \kappa  \ao}{2 (x-1)}+\frac{2 d_1 \kappa  \ao}{3 \
(x-1)^2}-\frac{5 \kappa  \ao}{(x-1)^2}-\frac{d_1 \kappa  \
\ao}{(x-1)^3}+\frac{15 \kappa  \ao}{2 (x-1)^3}+\frac{2 d_1 \kappa  \
\ao}{(x-1)^4}-\frac{45 \kappa  \ao}{4 (x-1)^4}+\frac{105 \kappa  \
\ao}{4}-\frac{d_1 \ao}{2 (x-1)}+\frac{13 \ao}{6 (x-1)}+\frac{2 d_1 \
\ao}{3 (x-1)^2}-\frac{3 \ao}{(x-1)^2}-\frac{d_1 \
\ao}{(x-1)^3}+\frac{23 \ao}{6 (x-1)^3}+\frac{2 d_1 \
\ao}{(x-1)^4}-\frac{61 \ao}{12 (x-1)^4}+\frac{169 \ao}{12}+\frac{205 \
d_1}{144}+\frac{205 d_1 \kappa }{144}+\frac{17 d_1 \kappa }{16 \
(x-1)}-\frac{45 \kappa }{8 (x-1)}-\frac{13 d_1 \kappa }{36 \
(x-1)^2}+\frac{35 \kappa }{12 (x-1)^2}+\frac{13 d_1 \kappa }{36 \
(x-1)^3}-\frac{35 \kappa }{12 (x-1)^3}-\frac{17 d_1 \kappa }{16 \
(x-1)^4}+\frac{45 \kappa }{8 (x-1)^4}-\frac{205 d_1 \kappa }{144 \
(x-1)^5}+\frac{205 \kappa }{24 (x-1)^5}-\frac{205 \kappa \
}{24}+\frac{17 d_1}{16 (x-1)}-\frac{65}{24 (x-1)}-\frac{13 d_1}{36 \
(x-1)^2}+\frac{55}{36 (x-1)^2}+\frac{13 d_1}{36 (x-1)^3}-\frac{55}{36 \
(x-1)^3}-\frac{17 d_1}{16 (x-1)^4}+\frac{65}{24 (x-1)^4}-\frac{205 \
d_1}{144 (x-1)^5}-\frac{\pi ^2}{12 (x-1)^5}+\frac{449}{72 \
(x-1)^5}-\frac{\pi ^2}{12}-\frac{161}{72}\Big) \
H(0;\ao)+\Big(-\frac{17 \kappa  d_1}{16 (x-1)}+\frac{13 \kappa  \
d_1}{36 (x-1)^2}-\frac{13 \kappa  d_1}{36 (x-1)^3}+\frac{17 \kappa  \
d_1}{16 (x-1)^4}+\frac{205 \kappa  d_1}{144 (x-1)^5}-\frac{205 \kappa \
 d_1}{144}-\frac{17 d_1}{16 (x-1)}+\frac{13 d_1}{36 (x-1)^2}-\frac{13 \
d_1}{36 (x-1)^3}+\frac{17 d_1}{16 (x-1)^4}+\frac{205 d_1}{144 \
(x-1)^5}-\frac{205 d_1}{144}+\frac{45 \kappa }{8 (x-1)}-\frac{35 \
\kappa }{12 (x-1)^2}+\frac{35 \kappa }{12 (x-1)^3}-\frac{45 \kappa \
}{8 (x-1)^4}-\frac{\pi ^2 \kappa }{2 (x-1)^5}-\frac{205 \kappa }{24 \
(x-1)^5}+\frac{\pi ^2 \kappa }{2}+\frac{205 \kappa }{24}+\frac{65}{24 \
(x-1)}-\frac{55}{36 (x-1)^2}+\frac{55}{36 (x-1)^3}-\frac{65}{24 \
(x-1)^4}-\frac{\pi ^2}{12 (x-1)^5}-\frac{449}{72 (x-1)^5}+\frac{\pi \
^2}{12}+\frac{449}{72}\Big) H(0;x)+\Big(\frac{d_1^2 \ao^4}{8}-\frac{5 \
d_1 \ao^4}{8}-\frac{1}{8} d_1 \kappa  \ao^4-\frac{d_1 \kappa  \
\ao^4}{8 (x-1)}+\frac{d_1^2 \ao^4}{8 (x-1)}-\frac{5 d_1 \ao^4}{8 \
(x-1)}-\frac{13 d_1^2 \ao^3}{18}+\frac{125 d_1 \
\ao^3}{36}+\frac{29}{36} d_1 \kappa  \ao^3+\frac{d_1 \kappa  \ao^3}{2 \
(x-1)}-\frac{11 d_1 \kappa  \ao^3}{36 (x-1)^2}-\frac{d_1^2 \ao^3}{2 \
(x-1)}+\frac{5 d_1 \ao^3}{2 (x-1)}+\frac{2 d_1^2 \ao^3}{9 \
(x-1)^2}-\frac{35 d_1 \ao^3}{36 (x-1)^2}+\frac{23 d_1^2 \
\ao^2}{12}-\frac{203 d_1 \ao^2}{24}-\frac{59}{24} d_1 \kappa  \
\ao^2-\frac{17 d_1 \kappa  \ao^2}{24 (x-1)}+\frac{23 d_1 \kappa  \
\ao^2}{24 (x-1)^2}-\frac{19 d_1 \kappa  \ao^2}{24 (x-1)^3}+\frac{3 \
d_1^2 \ao^2}{4 (x-1)}-\frac{89 d_1 \ao^2}{24 (x-1)}-\frac{2 d_1^2 \
\ao^2}{3 (x-1)^2}+\frac{71 d_1 \ao^2}{24 (x-1)^2}+\frac{d_1^2 \
\ao^2}{2 (x-1)^3}-\frac{43 d_1 \ao^2}{24 (x-1)^3}-\frac{25 d_1^2 \
\ao}{6}+\frac{169 d_1 \ao}{12}+\frac{73 d_1 \kappa  \
\ao}{12}+\frac{d_1 \kappa  \ao}{6 (x-1)}-\frac{d_1 \kappa  \
\ao}{(x-1)^2}+\frac{11 d_1 \kappa  \ao}{6 (x-1)^3}-\frac{37 d_1 \
\kappa  \ao}{12 (x-1)^4}-\frac{d_1^2 \ao}{2 (x-1)}+\frac{13 d_1 \
\ao}{6 (x-1)}+\frac{2 d_1^2 \ao}{3 (x-1)^2}-\frac{3 d_1 \
\ao}{(x-1)^2}-\frac{d_1^2 \ao}{(x-1)^3}+\frac{23 d_1 \ao}{6 (x-1)^3}+\
\frac{2 d_1^2 \ao}{(x-1)^4}-\frac{61 d_1 \ao}{12 (x-1)^4}+\frac{205 \
d_1^2}{72}-\frac{305 d_1}{36}-\frac{155 d_1 \kappa }{36}+\frac{d_1 \
\kappa }{6 (x-1)}+\frac{25 d_1 \kappa }{72 (x-1)^2}-\frac{25 d_1 \
\kappa }{24 (x-1)^3}+\frac{37 d_1 \kappa }{12 (x-1)^4}+\frac{d_1^2}{8 \
(x-1)}-\frac{d_1}{3 (x-1)}-\frac{2 d_1^2}{9 (x-1)^2}+\frac{73 d_1}{72 \
(x-1)^2}+\frac{d_1^2}{2 (x-1)^3}-\frac{49 d_1}{24 (x-1)^3}-\frac{2 \
d_1^2}{(x-1)^4}+\frac{61 d_1}{12 (x-1)^4}\Big) H(1;\ao)+\Big(\frac{3 \
\kappa  \ao^4}{2 (x-1)}+\frac{3 \kappa  \ao^4}{2}+\frac{\ao^4}{2 \
(x-1)}+\frac{\ao^4}{2}-\frac{6 \kappa  \ao^3}{x-1}+\frac{2 \kappa  \
\ao^3}{(x-1)^2}-8 \kappa  \ao^3-\frac{2 \ao^3}{x-1}+\frac{2 \ao^3}{3 \
(x-1)^2}-\frac{8 \ao^3}{3}+\frac{9 \kappa  \ao^2}{x-1}-\frac{6 \kappa \
 \ao^2}{(x-1)^2}+\frac{3 \kappa  \ao^2}{(x-1)^3}+18 \kappa  \
\ao^2+\frac{3 \ao^2}{x-1}-\frac{2 \
\ao^2}{(x-1)^2}+\frac{\ao^2}{(x-1)^3}+6 \ao^2-\frac{6 \kappa  \
\ao}{x-1}+\frac{6 \kappa  \ao}{(x-1)^2}-\frac{6 \kappa  \
\ao}{(x-1)^3}+\frac{6 \kappa  \ao}{(x-1)^4}-24 \kappa  \ao-\frac{2 \
\ao}{x-1}+\frac{2 \ao}{(x-1)^2}-\frac{2 \ao}{(x-1)^3}+\frac{2 \
\ao}{(x-1)^4}-8 \ao+\frac{15 \kappa }{4 (x-1)}-\frac{5 \kappa }{2 \
(x-1)^2}+\frac{5 \kappa }{2 (x-1)^3}-\frac{15 \kappa }{4 \
(x-1)^4}-\frac{33 \kappa }{4 (x-1)^5}+\frac{17 \kappa }{4}+\frac{5}{4 \
(x-1)}-\frac{5}{6 (x-1)^2}+\frac{5}{6 (x-1)^3}-\frac{5}{4 \
(x-1)^4}-\frac{49}{12 (x-1)^5}+\frac{1}{12}\Big) \
H(0,0;\ao)+\Big(-\frac{15 \kappa }{4 (x-1)}+\frac{5 \kappa }{2 \
(x-1)^2}-\frac{5 \kappa }{2 (x-1)^3}+\frac{15 \kappa }{4 \
(x-1)^4}+\frac{33 \kappa }{4 (x-1)^5}-\frac{33 \kappa }{4}-\frac{5}{4 \
(x-1)}+\frac{5}{6 (x-1)^2}-\frac{5}{6 (x-1)^3}+\frac{5}{4 \
(x-1)^4}+\frac{49}{12 (x-1)^5}-\frac{49}{12}\Big) \
H(0,0;x)+\Big(\frac{d_1 \ao^4}{2}+\frac{1}{2} d_1 \kappa  \
\ao^4+\frac{d_1 \kappa  \ao^4}{2 (x-1)}+\frac{d_1 \ao^4}{2 \
(x-1)}-\frac{8 d_1 \ao^3}{3}-\frac{8}{3} d_1 \kappa  \ao^3-\frac{2 \
d_1 \kappa  \ao^3}{x-1}+\frac{2 d_1 \kappa  \ao^3}{3 (x-1)^2}-\frac{2 \
d_1 \ao^3}{x-1}+\frac{2 d_1 \ao^3}{3 (x-1)^2}+6 d_1 \ao^2+6 d_1 \
\kappa  \ao^2+\frac{3 d_1 \kappa  \ao^2}{x-1}-\frac{2 d_1 \kappa  \
\ao^2}{(x-1)^2}+\frac{d_1 \kappa  \ao^2}{(x-1)^3}+\frac{3 d_1 \
\ao^2}{x-1}-\frac{2 d_1 \ao^2}{(x-1)^2}+\frac{d_1 \ao^2}{(x-1)^3}-8 \
d_1 \ao-8 d_1 \kappa  \ao-\frac{2 d_1 \kappa  \ao}{x-1}+\frac{2 d_1 \
\kappa  \ao}{(x-1)^2}-\frac{2 d_1 \kappa  \ao}{(x-1)^3}+\frac{2 d_1 \
\kappa  \ao}{(x-1)^4}-\frac{2 d_1 \ao}{x-1}+\frac{2 d_1 \
\ao}{(x-1)^2}-\frac{2 d_1 \ao}{(x-1)^3}+\frac{2 d_1 \
\ao}{(x-1)^4}+\frac{d_1}{12}+\frac{25 d_1 \kappa }{12}+\frac{5 d_1 \
\kappa }{4 (x-1)}-\frac{5 d_1 \kappa }{6 (x-1)^2}+\frac{5 d_1 \kappa \
}{6 (x-1)^3}-\frac{5 d_1 \kappa }{4 (x-1)^4}-\frac{25 d_1 \kappa }{12 \
(x-1)^5}+\frac{5 d_1}{4 (x-1)}-\frac{5 d_1}{6 (x-1)^2}+\frac{5 d_1}{6 \
(x-1)^3}-\frac{5 d_1}{4 (x-1)^4}-\frac{49 d_1}{12 (x-1)^5}\Big) \
H(0,1;\ao)+H(1;x) \Big(\frac{\pi ^2 \kappa  d_1}{6 (x-1)^5}+\frac{\pi \
^2 d_1}{6 (x-1)^5}-\frac{\pi ^2 \kappa }{4 (x-1)^5}+\frac{\pi ^2 \
\kappa }{4}+\Big(-\frac{\kappa  d_1}{x-1}+\frac{\kappa  d_1}{2 \
(x-1)^2}-\frac{\kappa  d_1}{3 (x-1)^3}+\frac{\kappa  d_1}{4 (x-1)^4}+\
\frac{25 \kappa  d_1}{12 (x-1)^5}-\frac{d_1}{x-1}+\frac{d_1}{2 \
(x-1)^2}-\frac{d_1}{3 (x-1)^3}+\frac{d_1}{4 (x-1)^4}+\frac{49 d_1}{12 \
(x-1)^5}+\frac{15 \kappa }{8 (x-1)}-\frac{5 \kappa }{4 \
(x-1)^2}+\frac{5 \kappa }{4 (x-1)^3}-\frac{15 \kappa }{8 \
(x-1)^4}-\frac{33 \kappa }{8 (x-1)^5}+\frac{33 \kappa }{8}+\frac{5}{8 \
(x-1)}-\frac{5}{12 (x-1)^2}+\frac{5}{12 (x-1)^3}-\frac{5}{8 (x-1)^4}-\
\frac{49}{24 (x-1)^5}+\frac{49}{24}\Big) H(0;\ao)+\Big(-\frac{2 \
\kappa  d_1}{(x-1)^5}-\frac{2 d_1}{(x-1)^5}+\frac{3 \kappa \
}{(x-1)^5}-3 \kappa +\frac{1}{(x-1)^5}-1\Big) \
H(0,0;\ao)+\Big(-\frac{2 d_1^2}{(x-1)^5}+\frac{\kappa  d_1}{(x-1)^5}-\
\kappa  d_1+\frac{d_1}{(x-1)^5}-d_1\Big) H(0,1;\ao)-\frac{\pi ^2}{12 \
(x-1)^5}+\frac{\pi ^2}{12}\Big)+\Big(-\frac{\kappa  \
d_1}{(x-1)^5}+\kappa  d_1-\frac{d_1}{(x-1)^5}+d_1+\frac{3 \kappa \
}{(x-1)^5}-3 \kappa +\frac{1}{(x-1)^5}-1\Big) H(0;\ao) H(0,1;x)+\Big(\
\frac{15 \kappa }{8 (x-1)}-\frac{5 \kappa }{4 (x-1)^2}+\frac{5 \kappa \
}{4 (x-1)^3}-\frac{15 \kappa }{8 (x-1)^4}-\frac{33 \kappa }{8 \
(x-1)^5}+\frac{33 \kappa }{8}+\Big(\frac{3 \kappa }{(x-1)^5}-3 \kappa \
+\frac{1}{(x-1)^5}-1\Big) H(0;\ao)+\Big(\frac{\kappa  \
d_1}{(x-1)^5}-\kappa  d_1+\frac{d_1}{(x-1)^5}-d_1\Big) \
H(1;\ao)+\frac{5}{8 (x-1)}-\frac{5}{12 (x-1)^2}+\frac{5}{12 (x-1)^3}-\
\frac{5}{8 (x-1)^4}-\frac{49}{24 (x-1)^5}+\frac{49}{24}\Big) H(0,c_1(\
\ao);x)+\Big(\frac{d_1 \ao^4}{2}+\frac{1}{2} d_1 \kappa  \
\ao^4+\frac{d_1 \kappa  \ao^4}{2 (x-1)}+\frac{d_1 \ao^4}{2 \
(x-1)}-\frac{8 d_1 \ao^3}{3}-\frac{8}{3} d_1 \kappa  \ao^3-\frac{2 \
d_1 \kappa  \ao^3}{x-1}+\frac{2 d_1 \kappa  \ao^3}{3 (x-1)^2}-\frac{2 \
d_1 \ao^3}{x-1}+\frac{2 d_1 \ao^3}{3 (x-1)^2}+6 d_1 \ao^2+6 d_1 \
\kappa  \ao^2+\frac{3 d_1 \kappa  \ao^2}{x-1}-\frac{2 d_1 \kappa  \
\ao^2}{(x-1)^2}+\frac{d_1 \kappa  \ao^2}{(x-1)^3}+\frac{3 d_1 \
\ao^2}{x-1}-\frac{2 d_1 \ao^2}{(x-1)^2}+\frac{d_1 \ao^2}{(x-1)^3}-8 \
d_1 \ao-8 d_1 \kappa  \ao-\frac{2 d_1 \kappa  \ao}{x-1}+\frac{2 d_1 \
\kappa  \ao}{(x-1)^2}-\frac{2 d_1 \kappa  \ao}{(x-1)^3}+\frac{2 d_1 \
\kappa  \ao}{(x-1)^4}-\frac{2 d_1 \ao}{x-1}+\frac{2 d_1 \
\ao}{(x-1)^2}-\frac{2 d_1 \ao}{(x-1)^3}+\frac{2 d_1 \
\ao}{(x-1)^4}+\frac{25 d_1}{6}+\frac{25 d_1 \kappa }{6}+\frac{d_1 \
\kappa }{2 (x-1)}-\frac{2 d_1 \kappa }{3 (x-1)^2}+\frac{d_1 \kappa \
}{(x-1)^3}-\frac{2 d_1 \kappa }{(x-1)^4}+\frac{d_1}{2 (x-1)}-\frac{2 \
d_1}{3 (x-1)^2}+\frac{d_1}{(x-1)^3}-\frac{2 d_1}{(x-1)^4}\Big) H(1,0;\
\ao)+\Big(\frac{\kappa  d_1}{x-1}-\frac{\kappa  d_1}{2 \
(x-1)^2}+\frac{\kappa  d_1}{3 (x-1)^3}-\frac{\kappa  d_1}{4 (x-1)^4}-\
\frac{25 \kappa  d_1}{12 (x-1)^5}+\frac{d_1}{x-1}-\frac{d_1}{2 \
(x-1)^2}+\frac{d_1}{3 (x-1)^3}-\frac{d_1}{4 (x-1)^4}-\frac{49 d_1}{12 \
(x-1)^5}-\frac{15 \kappa }{8 (x-1)}+\frac{5 \kappa }{4 \
(x-1)^2}-\frac{5 \kappa }{4 (x-1)^3}+\frac{15 \kappa }{8 \
(x-1)^4}+\frac{33 \kappa }{8 (x-1)^5}-\frac{33 \kappa }{8}-\frac{5}{8 \
(x-1)}+\frac{5}{12 (x-1)^2}-\frac{5}{12 (x-1)^3}+\frac{5}{8 (x-1)^4}+\
\frac{49}{24 (x-1)^5}-\frac{49}{24}\Big) H(1,0;x)+\Big(\frac{d_1^2 \
\ao^4}{2}+\frac{d_1^2 \ao^4}{2 (x-1)}-\frac{8 d_1^2 \ao^3}{3}-\frac{2 \
d_1^2 \ao^3}{x-1}+\frac{2 d_1^2 \ao^3}{3 (x-1)^2}+6 d_1^2 \
\ao^2+\frac{3 d_1^2 \ao^2}{x-1}-\frac{2 d_1^2 \
\ao^2}{(x-1)^2}+\frac{d_1^2 \ao^2}{(x-1)^3}-8 d_1^2 \ao-\frac{2 d_1^2 \
\ao}{x-1}+\frac{2 d_1^2 \ao}{(x-1)^2}-\frac{2 d_1^2 \
\ao}{(x-1)^3}+\frac{2 d_1^2 \ao}{(x-1)^4}+\frac{25 \
d_1^2}{6}+\frac{d_1^2}{2 (x-1)}-\frac{2 d_1^2}{3 \
(x-1)^2}+\frac{d_1^2}{(x-1)^3}-\frac{2 d_1^2}{(x-1)^4}\Big) \
H(1,1;\ao)+H(c_1(\ao);x) \Big(\frac{d_1 \ao^4}{16}+\frac{1}{16} d_1 \
\kappa  \ao^4+\frac{d_1 \kappa  \ao^4}{16 (x-1)}-\frac{7 \kappa  \
\ao^4}{16 (x-1)}-\frac{7 \kappa  \ao^4}{16}+\frac{d_1 \ao^4}{16 \
(x-1)}-\frac{5 \ao^4}{16 (x-1)}-\frac{5 \ao^4}{16}-\frac{13 d_1 \
\ao^3}{36}-\frac{13}{36} d_1 \kappa  \ao^3-\frac{d_1 \kappa  \ao^3}{4 \
(x-1)}+\frac{7 \kappa  \ao^3}{4 (x-1)}+\frac{d_1 \kappa  \ao^3}{9 \
(x-1)^2}-\frac{19 \kappa  \ao^3}{24 (x-1)^2}+\frac{61 \kappa  \
\ao^3}{24}-\frac{d_1 \ao^3}{4 (x-1)}+\frac{5 \ao^3}{4 \
(x-1)}+\frac{d_1 \ao^3}{9 (x-1)^2}-\frac{35 \ao^3}{72 \
(x-1)^2}+\frac{125 \ao^3}{72}+\frac{23 d_1 \ao^2}{24}+\frac{23}{24} \
d_1 \kappa  \ao^2+\frac{3 d_1 \kappa  \ao^2}{8 (x-1)}-\frac{41 \kappa \
 \ao^2}{16 (x-1)}-\frac{d_1 \kappa  \ao^2}{3 (x-1)^2}+\frac{39 \kappa \
 \ao^2}{16 (x-1)^2}+\frac{d_1 \kappa  \ao^2}{4 (x-1)^3}-\frac{27 \
\kappa  \ao^2}{16 (x-1)^3}-\frac{107 \kappa  \ao^2}{16}+\frac{3 d_1 \
\ao^2}{8 (x-1)}-\frac{89 \ao^2}{48 (x-1)}-\frac{d_1 \ao^2}{3 \
(x-1)^2}+\frac{71 \ao^2}{48 (x-1)^2}+\frac{d_1 \ao^2}{4 \
(x-1)^3}-\frac{43 \ao^2}{48 (x-1)^3}-\frac{203 \ao^2}{48}-\frac{25 \
d_1 \ao}{12}-\frac{25 d_1 \kappa  \ao}{12}-\frac{d_1 \kappa  \ao}{4 \
(x-1)}+\frac{5 \kappa  \ao}{4 (x-1)}+\frac{d_1 \kappa  \ao}{3 \
(x-1)^2}-\frac{5 \kappa  \ao}{2 (x-1)^2}-\frac{d_1 \kappa  \ao}{2 \
(x-1)^3}+\frac{15 \kappa  \ao}{4 (x-1)^3}+\frac{d_1 \kappa  \
\ao}{(x-1)^4}-\frac{45 \kappa  \ao}{8 (x-1)^4}+\frac{105 \kappa  \
\ao}{8}-\frac{d_1 \ao}{4 (x-1)}+\frac{13 \ao}{12 (x-1)}+\frac{d_1 \
\ao}{3 (x-1)^2}-\frac{3 \ao}{2 (x-1)^2}-\frac{d_1 \ao}{2 \
(x-1)^3}+\frac{23 \ao}{12 (x-1)^3}+\frac{d_1 \ao}{(x-1)^4}-\frac{61 \
\ao}{24 (x-1)^4}+\frac{169 \ao}{24}+\frac{205 d_1}{144}+\frac{205 d_1 \
\kappa }{144}+\frac{17 d_1 \kappa }{16 (x-1)}-\frac{45 \kappa }{8 \
(x-1)}-\frac{13 d_1 \kappa }{36 (x-1)^2}+\frac{35 \kappa }{12 \
(x-1)^2}+\frac{13 d_1 \kappa }{36 (x-1)^3}-\frac{35 \kappa }{12 \
(x-1)^3}-\frac{17 d_1 \kappa }{16 (x-1)^4}+\frac{45 \kappa }{8 \
(x-1)^4}-\frac{205 d_1 \kappa }{144 (x-1)^5}+\frac{205 \kappa }{24 \
(x-1)^5}-\frac{205 \kappa }{24}+\Big(\frac{3 \kappa  \ao^4}{4 (x-1)}+\
\frac{3 \kappa  \ao^4}{4}+\frac{\ao^4}{4 \
(x-1)}+\frac{\ao^4}{4}-\frac{3 \kappa  \ao^3}{x-1}+\frac{\kappa  \
\ao^3}{(x-1)^2}-4 \kappa  \ao^3-\frac{\ao^3}{x-1}+\frac{\ao^3}{3 \
(x-1)^2}-\frac{4 \ao^3}{3}+\frac{9 \kappa  \ao^2}{2 (x-1)}-\frac{3 \
\kappa  \ao^2}{(x-1)^2}+\frac{3 \kappa  \ao^2}{2 (x-1)^3}+9 \kappa  \
\ao^2+\frac{3 \ao^2}{2 (x-1)}-\frac{\ao^2}{(x-1)^2}+\frac{\ao^2}{2 \
(x-1)^3}+3 \ao^2-\frac{3 \kappa  \ao}{x-1}+\frac{3 \kappa  \
\ao}{(x-1)^2}-\frac{3 \kappa  \ao}{(x-1)^3}+\frac{3 \kappa  \
\ao}{(x-1)^4}-12 \kappa  \
\ao-\frac{\ao}{x-1}+\frac{\ao}{(x-1)^2}-\frac{\ao}{(x-1)^3}+\frac{\ao}\
{(x-1)^4}-4 \ao+\frac{15 \kappa }{4 (x-1)}-\frac{5 \kappa }{2 \
(x-1)^2}+\frac{5 \kappa }{2 (x-1)^3}-\frac{15 \kappa }{4 \
(x-1)^4}-\frac{33 \kappa }{4 (x-1)^5}+\frac{25 \kappa }{4}+\frac{5}{4 \
(x-1)}-\frac{5}{6 (x-1)^2}+\frac{5}{6 (x-1)^3}-\frac{5}{4 \
(x-1)^4}-\frac{49}{12 (x-1)^5}+\frac{25}{12}\Big) \
H(0;\ao)+\Big(\frac{d_1 \ao^4}{4}+\frac{1}{4} d_1 \kappa  \
\ao^4+\frac{d_1 \kappa  \ao^4}{4 (x-1)}+\frac{d_1 \ao^4}{4 \
(x-1)}-\frac{4 d_1 \ao^3}{3}-\frac{4}{3} d_1 \kappa  \ao^3-\frac{d_1 \
\kappa  \ao^3}{x-1}+\frac{d_1 \kappa  \ao^3}{3 (x-1)^2}-\frac{d_1 \
\ao^3}{x-1}+\frac{d_1 \ao^3}{3 (x-1)^2}+3 d_1 \ao^2+3 d_1 \kappa  \
\ao^2+\frac{3 d_1 \kappa  \ao^2}{2 (x-1)}-\frac{d_1 \kappa  \
\ao^2}{(x-1)^2}+\frac{d_1 \kappa  \ao^2}{2 (x-1)^3}+\frac{3 d_1 \
\ao^2}{2 (x-1)}-\frac{d_1 \ao^2}{(x-1)^2}+\frac{d_1 \ao^2}{2 \
(x-1)^3}-4 d_1 \ao-4 d_1 \kappa  \ao-\frac{d_1 \kappa  \
\ao}{x-1}+\frac{d_1 \kappa  \ao}{(x-1)^2}-\frac{d_1 \kappa  \
\ao}{(x-1)^3}+\frac{d_1 \kappa  \ao}{(x-1)^4}-\frac{d_1 \
\ao}{x-1}+\frac{d_1 \ao}{(x-1)^2}-\frac{d_1 \ao}{(x-1)^3}+\frac{d_1 \
\ao}{(x-1)^4}+\frac{25 d_1}{12}+\frac{25 d_1 \kappa }{12}+\frac{5 d_1 \
\kappa }{4 (x-1)}-\frac{5 d_1 \kappa }{6 (x-1)^2}+\frac{5 d_1 \kappa \
}{6 (x-1)^3}-\frac{5 d_1 \kappa }{4 (x-1)^4}-\frac{25 d_1 \kappa }{12 \
(x-1)^5}+\frac{5 d_1}{4 (x-1)}-\frac{5 d_1}{6 (x-1)^2}+\frac{5 d_1}{6 \
(x-1)^3}-\frac{5 d_1}{4 (x-1)^4}-\frac{49 d_1}{12 (x-1)^5}\Big) \
H(1;\ao)+\Big(\frac{6 \kappa }{(x-1)^5}+\frac{2}{(x-1)^5}\Big) H(0,0;\
\ao)+\Big(\frac{2 \kappa  d_1}{(x-1)^5}+\frac{2 d_1}{(x-1)^5}\Big) \
H(0,1;\ao)+\Big(\frac{2 \kappa  d_1}{(x-1)^5}+\frac{2 \
d_1}{(x-1)^5}\Big) H(1,0;\ao)+\frac{2 d_1^2 \
H(1,1;\ao)}{(x-1)^5}+\frac{17 d_1}{16 (x-1)}-\frac{65}{24 \
(x-1)}-\frac{13 d_1}{36 (x-1)^2}+\frac{55}{36 (x-1)^2}+\frac{13 \
d_1}{36 (x-1)^3}-\frac{55}{36 (x-1)^3}-\frac{17 d_1}{16 \
(x-1)^4}+\frac{65}{24 (x-1)^4}-\frac{205 d_1}{144 (x-1)^5}-\frac{\pi \
^2}{12 (x-1)^5}+\frac{449}{72 \
(x-1)^5}-\frac{305}{72}\Big)+\Big(\frac{2 d_1^2}{(x-1)^5}-\frac{2 \
\kappa  d_1}{(x-1)^5}+\kappa  d_1-\frac{2 d_1}{(x-1)^5}+d_1+\frac{3 \
\kappa }{2 (x-1)^5}-\frac{3 \kappa }{2}+\frac{1}{2 \
(x-1)^5}-\frac{1}{2}\Big) H(0;\ao) H(1,1;x)+\Big(-\frac{\kappa  \
d_1}{x-1}+\frac{\kappa  d_1}{2 (x-1)^2}-\frac{\kappa  d_1}{3 \
(x-1)^3}+\frac{\kappa  d_1}{4 (x-1)^4}+\frac{25 \kappa  d_1}{12 \
(x-1)^5}-\frac{d_1}{x-1}+\frac{d_1}{2 (x-1)^2}-\frac{d_1}{3 (x-1)^3}+\
\frac{d_1}{4 (x-1)^4}+\frac{49 d_1}{12 (x-1)^5}+\frac{15 \kappa }{8 \
(x-1)}-\frac{5 \kappa }{4 (x-1)^2}+\frac{5 \kappa }{4 \
(x-1)^3}-\frac{15 \kappa }{8 (x-1)^4}-\frac{33 \kappa }{8 \
(x-1)^5}+\frac{33 \kappa }{8}+\Big(-\frac{2 \kappa  \
d_1}{(x-1)^5}-\frac{2 d_1}{(x-1)^5}+\frac{3 \kappa }{(x-1)^5}-3 \
\kappa +\frac{1}{(x-1)^5}-1\Big) H(0;\ao)+\Big(-\frac{2 \
d_1^2}{(x-1)^5}+\frac{\kappa  d_1}{(x-1)^5}-\kappa  \
d_1+\frac{d_1}{(x-1)^5}-d_1\Big) H(1;\ao)+\frac{5}{8 \
(x-1)}-\frac{5}{12 (x-1)^2}+\frac{5}{12 (x-1)^3}-\frac{5}{8 (x-1)^4}-\
\frac{49}{24 (x-1)^5}+\frac{49}{24}\Big) H(1,c_1(\ao);x)+\Big(\frac{3 \
\kappa  \ao^4}{8 (x-1)}+\frac{3 \kappa  \ao^4}{8}+\frac{\ao^4}{8 \
(x-1)}+\frac{\ao^4}{8}-\frac{3 \kappa  \ao^3}{2 (x-1)}+\frac{\kappa  \
\ao^3}{2 (x-1)^2}-2 \kappa  \ao^3-\frac{\ao^3}{2 \
(x-1)}+\frac{\ao^3}{6 (x-1)^2}-\frac{2 \ao^3}{3}+\frac{9 \kappa  \
\ao^2}{4 (x-1)}-\frac{3 \kappa  \ao^2}{2 (x-1)^2}+\frac{3 \kappa  \
\ao^2}{4 (x-1)^3}+\frac{9 \kappa  \ao^2}{2}+\frac{3 \ao^2}{4 \
(x-1)}-\frac{\ao^2}{2 (x-1)^2}+\frac{\ao^2}{4 (x-1)^3}+\frac{3 \
\ao^2}{2}-\frac{3 \kappa  \ao}{2 (x-1)}+\frac{3 \kappa  \ao}{2 \
(x-1)^2}-\frac{3 \kappa  \ao}{2 (x-1)^3}+\frac{3 \kappa  \ao}{2 \
(x-1)^4}-6 \kappa  \ao-\frac{\ao}{2 (x-1)}+\frac{\ao}{2 \
(x-1)^2}-\frac{\ao}{2 (x-1)^3}+\frac{\ao}{2 (x-1)^4}-2 \ao+\frac{15 \
\kappa }{8 (x-1)}-\frac{5 \kappa }{4 (x-1)^2}+\frac{5 \kappa }{4 \
(x-1)^3}-\frac{15 \kappa }{8 (x-1)^4}-\frac{33 \kappa }{8 \
(x-1)^5}+\frac{25 \kappa }{8}+\Big(\frac{3 \kappa \
}{(x-1)^5}+\frac{1}{(x-1)^5}\Big) H(0;\ao)+\Big(\frac{\kappa  \
d_1}{(x-1)^5}+\frac{d_1}{(x-1)^5}\Big) H(1;\ao)+\frac{5}{8 \
(x-1)}-\frac{5}{12 (x-1)^2}+\frac{5}{12 (x-1)^3}-\frac{5}{8 (x-1)^4}-\
\frac{49}{24 (x-1)^5}+\frac{25}{24}\Big) H(c_1(\ao),c_1(\ao);x)+\Big(\
\frac{6 \kappa }{(x-1)^5}+6 \kappa +\frac{2}{(x-1)^5}+2\Big) H(0,0,0;\
\ao)+\Big(-\frac{6 \kappa }{(x-1)^5}+6 \kappa \
-\frac{2}{(x-1)^5}+2\Big) H(0,0,0;x)+\Big(\frac{2 \kappa  \
d_1}{(x-1)^5}+2 \kappa  d_1+\frac{2 d_1}{(x-1)^5}+2 d_1\Big) H(0,0,1;\
\ao)+\Big(\frac{3 \kappa }{(x-1)^5}-3 \kappa \
+\frac{1}{(x-1)^5}-1\Big) H(0,0,c_1(\ao);x)+\Big(\frac{2 \kappa  \
d_1}{(x-1)^5}+2 \kappa  d_1+\frac{2 d_1}{(x-1)^5}+2 d_1\Big) H(0,1,0;\
\ao)+\Big(\frac{\kappa  d_1}{(x-1)^5}-\kappa  \
d_1+\frac{d_1}{(x-1)^5}-d_1-\frac{3 \kappa }{(x-1)^5}+3 \kappa \
-\frac{1}{(x-1)^5}+1\Big) H(0,1,0;x)+\Big(\frac{2 d_1^2}{(x-1)^5}+2 \
d_1^2\Big) H(0,1,1;\ao)+\Big(-\frac{\kappa  d_1}{(x-1)^5}+\kappa  \
d_1-\frac{d_1}{(x-1)^5}+d_1+\frac{3 \kappa }{(x-1)^5}-3 \kappa \
+\frac{1}{(x-1)^5}-1\Big) H(0,1,c_1(\ao);x)+\Big(\frac{3 \kappa }{2 \
(x-1)^5}-\frac{3 \kappa }{2}+\frac{1}{2 (x-1)^5}-\frac{1}{2}\Big) \
H(0,c_1(\ao),c_1(\ao);x)+\Big(\frac{2 \kappa  d_1}{(x-1)^5}+\frac{2 \
d_1}{(x-1)^5}-\frac{3 \kappa }{(x-1)^5}+3 \kappa -\frac{1}{(x-1)^5}+1\
\Big) H(1,0,0;x)+\Big(-\frac{\kappa  \
d_1}{(x-1)^5}-\frac{d_1}{(x-1)^5}+\frac{3 \kappa }{2 (x-1)^5}-\frac{3 \
\kappa }{2}+\frac{1}{2 (x-1)^5}-\frac{1}{2}\Big) \
H(1,0,c_1(\ao);x)+\Big(-\frac{2 d_1^2}{(x-1)^5}+\frac{2 \kappa  \
d_1}{(x-1)^5}-\kappa  d_1+\frac{2 d_1}{(x-1)^5}-d_1-\frac{3 \kappa \
}{2 (x-1)^5}+\frac{3 \kappa }{2}-\frac{1}{2 (x-1)^5}+\frac{1}{2}\Big) \
H(1,1,0;x)+\Big(\frac{2 d_1^2}{(x-1)^5}-\frac{2 \kappa  \
d_1}{(x-1)^5}+\kappa  d_1-\frac{2 d_1}{(x-1)^5}+d_1+\frac{3 \kappa \
}{2 (x-1)^5}-\frac{3 \kappa }{2}+\frac{1}{2 (x-1)^5}-\frac{1}{2}\Big) \
H(1,1,c_1(\ao);x)+\Big(-\frac{\kappa  \
d_1}{(x-1)^5}-\frac{d_1}{(x-1)^5}+\frac{3 \kappa }{2 (x-1)^5}-\frac{3 \
\kappa }{2}+\frac{1}{2 (x-1)^5}-\frac{1}{2}\Big) \
H(1,c_1(\ao),c_1(\ao);x)+\Big(\frac{3 \kappa }{2 (x-1)^5}+\frac{1}{2 \
(x-1)^5}\Big) H(c_1(\ao),c_1(\ao),c_1(\ao);x)-\frac{35 \pi ^2 \kappa \
}{48 (x-1) (\kappa +1)}+\frac{35 \pi ^2 \kappa }{72 (x-1)^2 (\kappa \
+1)}-\frac{35 \pi ^2 \kappa }{72 (x-1)^3 (\kappa +1)}+\frac{35 \pi ^2 \
\kappa }{48 (x-1)^4 (\kappa +1)}+\frac{247 \pi ^2 \kappa }{144 \
(x-1)^5 (\kappa +1)}-\frac{247 \pi ^2 \kappa }{144 (\kappa \
+1)}-\frac{5 \pi ^2}{48 (x-1) (\kappa +1)}+\frac{5 \pi ^2}{72 (x-1)^2 \
(\kappa +1)}-\frac{5 \pi ^2}{72 (x-1)^3 (\kappa +1)}+\frac{5 \pi \
^2}{48 (x-1)^4 (\kappa +1)}+\frac{49 \pi ^2}{144 (x-1)^5 (\kappa \
+1)}-\frac{25 \pi ^2}{144 (\kappa +1)}-\frac{4}{\kappa +1}-\frac{7 \
\kappa  \zeta_3}{2 (x-1)^5 (\kappa +1)}+\frac{7 \kappa  \zeta_3}{2 \
(\kappa +1)}-\frac{\zeta_3}{2 (x-1)^5 (\kappa +1)}+\frac{3 \zeta_3}{2 \
(\kappa +1)}.
\erp

%
% The A integral for k=2
%

\subsection{The $\cA$ integral for $k=2$ and arbitrary $\kappa$}
%
% This file contains the TeX output produced by Mathematica for the integral A2, for arbitrary kappa
%
The $\eps$ expansion for this integral reads
\beq
\bsp
\begin{cal}I\end{cal}(x,\eps;\ao,3+d_1\eps;\kappa,2,0,g_A) &= x\,\aint(\eps,x;3+d_1\eps;\kap,2)\\
&=\frac{1}{\eps}a_{-1}^{(\kap,2)}+a_0^{(\kap,2)}+\eps a_1^{(\kap,2)}+\eps a_2^{(\kap,2)}+\ocal\left(\eps^3\right),
\esp
\eeq
where
%1/ep piece
\brp
a_{-1}^{(\kap,2)}=-\frac{1}{3(\kappa +1)},
\erp
% ep^0
\brp
a_0^{(\kap,2)} = -\frac{\ao^6}{3 (x \ao-2 \ao-x)}+\frac{\ao^5}{3 (x-2)}+\frac{5 \
\ao^5}{3 (x \ao-2 \ao-x)}-\frac{5 \ao^4}{4 (x-2)}+\frac{\ao^4}{12 \
(x-1)}-\frac{10 \ao^4}{3 (x \ao-2 \ao-x)}+\frac{\kappa  \ao^4}{12 \
(\kappa +1)}+\frac{\ao^4}{12 (\kappa +1)}+\frac{5 \ao^4}{6 \
(x-2)^2}+\frac{5 \ao^3}{3 (x-2)}-\frac{\ao^3}{3 (x-1)}+\frac{10 \
\ao^3}{3 (x \ao-2 \ao-x)}-\frac{4 \kappa  \ao^3}{9 (\kappa \
+1)}-\frac{4 \ao^3}{9 (\kappa +1)}-\frac{20 \ao^3}{9 \
(x-2)^2}+\frac{\ao^3}{9 (x-1)^2}+\frac{20 \ao^3}{9 (x-2)^3}-\frac{5 \
\ao^2}{6 (x-2)}+\frac{\ao^2}{2 (x-1)}-\frac{5 \ao^2}{3 (x \ao-2 \
\ao-x)}+\frac{\kappa  \ao^2}{\kappa +1}+\frac{\ao^2}{\kappa \
+1}+\frac{5 \ao^2}{3 (x-2)^2}-\frac{\ao^2}{3 (x-1)^2}-\frac{10 \
\ao^2}{3 (x-2)^3}+\frac{\ao^2}{6 (x-1)^3}+\frac{20 \ao^2}{3 (x-2)^4}-\
\frac{\ao}{3 (x-1)}+\frac{\ao}{3 (x \ao-2 \ao-x)}-\frac{4 \kappa  \
\ao}{3 (\kappa +1)}-\frac{4 \ao}{3 (\kappa +1)}+\frac{\ao}{3 \
(x-1)^2}-\frac{\ao}{3 (x-1)^3}+\frac{\ao}{3 (x-1)^4}+\frac{80 \ao}{3 \
(x-2)^5}+\Big(\frac{1}{3 (x-1)^5}+\frac{1}{3}+\frac{80}{3 \
(x-2)^5}+\frac{160}{3 (x-2)^6}\Big) H(0;\ao)+\Big(-\frac{1}{3 \
(x-1)^5}+\frac{1}{3}-\frac{80}{3 (x-2)^5}-\frac{160}{3 (x-2)^6}\Big) \
H(0;x)+\frac{H(c_1(\ao);x)}{3 (x-1)^5}+\Big(\frac{80}{3 \
(x-2)^5}+\frac{160}{3 (x-2)^6}\Big) H(c_2(\ao);x)-\frac{13}{18 \
(\kappa +1)}+\frac{80 \ln 2\, }{3 (x-2)^5}+\frac{160 \ln 2\, }{3 \
(x-2)^6},
\erp
% ep^1
\brp
a_1^{(\kap,2)} = \frac{1}{\ao(x-2)-x}\Big\{-\frac{1}{24} d_1 x \ao^5-\frac{d_1 \ao^5}{12 (x-2)}+\frac{d_1 \
\ao^5}{24 (x-1)}+\frac{2 x \ao^5}{9 (\kappa +1)}+\frac{11 x \kappa  \
\ao^5}{36 (\kappa +1)}+\frac{25 \kappa  \ao^5}{36 (x-2) (\kappa +1)}-\
\frac{11 \kappa  \ao^5}{36 (x-1) (\kappa +1)}+\frac{\kappa  \ao^5}{24 \
(\kappa +1)}+\frac{19 \ao^5}{36 (x-2) (\kappa +1)}-\frac{2 \ao^5}{9 \
(x-1) (\kappa +1)}+\frac{\ao^5}{24 (\kappa +1)}-\frac{7 d_1 \
\ao^4}{108}+\frac{61}{216} d_1 x \ao^4+\frac{11 d_1 \ao^4}{54 (x-2)}-\
\frac{43 d_1 \ao^4}{216 (x-1)}-\frac{157 x \ao^4}{108 (\kappa \
+1)}-\frac{56 x \kappa  \ao^4}{27 (\kappa +1)}-\frac{113 \kappa  \
\ao^4}{54 (x-2) (\kappa +1)}+\frac{149 \kappa  \ao^4}{108 (x-1) \
(\kappa +1)}+\frac{149 \kappa  \ao^4}{54 (x-2)^2 (\kappa \
+1)}-\frac{59 \kappa  \ao^4}{108 (x-1)^2 (\kappa +1)}+\frac{41 \kappa \
 \ao^4}{216 (\kappa +1)}-\frac{91 \ao^4}{54 (x-2) (\kappa \
+1)}+\frac{53 \ao^4}{54 (x-1) (\kappa +1)}+\frac{103 \ao^4}{54 \
(x-2)^2 (\kappa +1)}-\frac{37 \ao^4}{108 (x-1)^2 (\kappa \
+1)}+\frac{\ao^4}{216 (\kappa +1)}-\frac{23 d_1 \ao^4}{54 \
(x-2)^2}+\frac{2 d_1 \ao^4}{27 (x-1)^2}+\frac{4 d_1 \
\ao^3}{9}-\frac{95}{108} d_1 x \ao^3+\frac{d_1 \ao^3}{54 \
(x-2)}+\frac{41 d_1 \ao^3}{108 (x-1)}+\frac{911 x \ao^3}{216 (\kappa \
+1)}+\frac{1381 x \kappa  \ao^3}{216 (\kappa +1)}+\frac{25 \kappa  \
\ao^3}{27 (x-2) (\kappa +1)}-\frac{239 \kappa  \ao^3}{108 (x-1) \
(\kappa +1)}-\frac{104 \kappa  \ao^3}{27 (x-2)^2 (\kappa \
+1)}+\frac{467 \kappa  \ao^3}{216 (x-1)^2 (\kappa +1)}+\frac{388 \
\kappa  \ao^3}{27 (x-2)^3 (\kappa +1)}-\frac{83 \kappa  \ao^3}{72 \
(x-1)^3 (\kappa +1)}-\frac{83 \kappa  \ao^3}{36 (\kappa +1)}+\frac{35 \
\ao^3}{27 (x-2) (\kappa +1)}-\frac{175 \ao^3}{108 (x-1) (\kappa +1)}-\
\frac{88 \ao^3}{27 (x-2)^2 (\kappa +1)}+\frac{277 \ao^3}{216 (x-1)^2 \
(\kappa +1)}+\frac{236 \ao^3}{27 (x-2)^3 (\kappa +1)}-\frac{5 \
\ao^3}{8 (x-1)^3 (\kappa +1)}-\frac{35 \ao^3}{36 (\kappa +1)}+\frac{8 \
d_1 \ao^3}{27 (x-2)^2}-\frac{17 d_1 \ao^3}{54 (x-1)^2}-\frac{76 d_1 \
\ao^3}{27 (x-2)^3}+\frac{d_1 \ao^3}{6 (x-1)^3}-\frac{35 d_1 \
\ao^2}{18}+\frac{73}{36} d_1 x \ao^2-\frac{4 d_1 \ao^2}{9 \
(x-2)}-\frac{13 d_1 \ao^2}{36 (x-1)}-\frac{569 x \ao^2}{72 (\kappa \
+1)}-\frac{979 x \kappa  \ao^2}{72 (\kappa +1)}+\frac{16 \kappa  \
\ao^2}{3 (x-2) (\kappa +1)}-\frac{2 \kappa  \ao^2}{3 (x-1) (\kappa \
+1)}-\frac{10 \kappa  \ao^2}{(x-2)^2 (\kappa +1)}-\frac{67 \kappa  \
\ao^2}{24 (x-1)^2 (\kappa +1)}+\frac{80 \kappa  \ao^2}{3 (x-2)^3 \
(\kappa +1)}+\frac{119 \kappa  \ao^2}{24 (x-1)^3 (\kappa \
+1)}+\frac{1256 \kappa  \ao^2}{9 (x-2)^4 (\kappa +1)}-\frac{137 \
\kappa  \ao^2}{36 (x-1)^4 (\kappa +1)}+\frac{32 \kappa  \ao^2}{3 \
(\kappa +1)}+\frac{16 \ao^2}{9 (x-2) (\kappa +1)}+\frac{\ao^2}{2 \
(x-1) (\kappa +1)}-\frac{10 \ao^2}{3 (x-2)^2 (\kappa +1)}-\frac{119 \
\ao^2}{72 (x-1)^2 (\kappa +1)}+\frac{80 \ao^2}{9 (x-2)^3 (\kappa \
+1)}+\frac{19 \ao^2}{8 (x-1)^3 (\kappa +1)}+\frac{632 \ao^2}{9 \
(x-2)^4 (\kappa +1)}-\frac{7 \ao^2}{4 (x-1)^4 (\kappa +1)}+\frac{85 \
\ao^2}{18 (\kappa +1)}+\frac{5 d_1 \ao^2}{3 (x-2)^2}+\frac{d_1 \
\ao^2}{2 (x-1)^2}-\frac{80 d_1 \ao^2}{9 (x-2)^3}-\frac{5 d_1 \ao^2}{6 \
(x-1)^3}-\frac{104 d_1 \ao^2}{3 (x-2)^4}+\frac{2 d_1 \ao^2}{3 \
(x-1)^4}-\frac{d_1 \ao}{3 (\kappa +1)}-\frac{25 d_1 x \ao}{18 (\kappa \
+1)}+\frac{371 x \ao}{108 (\kappa +1)}-\frac{d_1 \kappa  \ao}{3 \
(\kappa +1)}-\frac{25 d_1 x \kappa  \ao}{18 (\kappa +1)}-\frac{\pi ^2 \
x \kappa  \ao}{6 (\kappa +1)}+\frac{323 x \kappa  \ao}{36 (\kappa \
+1)}+\frac{2 d_1 \kappa  \ao}{9 (x-2) (\kappa +1)}+\frac{4 \kappa  \
\ao}{3 (x-2) (\kappa +1)}+\frac{d_1 \kappa  \ao}{18 (x-1) (\kappa \
+1)}-\frac{11 \kappa  \ao}{4 (x-1) (\kappa +1)}-\frac{10 d_1 \kappa  \
\ao}{9 (x-2)^2 (\kappa +1)}-\frac{d_1 \kappa  \ao}{9 (x-1)^2 (\kappa \
+1)}+\frac{17 \kappa  \ao}{12 (x-1)^2 (\kappa +1)}+\frac{80 d_1 \
\kappa  \ao}{9 (x-2)^3 (\kappa +1)}-\frac{40 \kappa  \ao}{3 (x-2)^3 (\
\kappa +1)}+\frac{d_1 \kappa  \ao}{3 (x-1)^3 (\kappa +1)}-\frac{3 \
\kappa  \ao}{2 (x-1)^3 (\kappa +1)}+\frac{208 d_1 \kappa  \ao}{3 \
(x-2)^4 (\kappa +1)}+\frac{20 \pi ^2 \kappa  \ao}{3 (x-2)^4 (\kappa \
+1)}-\frac{2512 \kappa  \ao}{9 (x-2)^4 (\kappa +1)}+\frac{2 d_1 \
\kappa  \ao}{3 (x-1)^4 (\kappa +1)}+\frac{\pi ^2 \kappa  \ao}{6 \
(x-1)^4 (\kappa +1)}-\frac{137 \kappa  \ao}{36 (x-1)^4 (\kappa \
+1)}+\frac{256 d_1 \kappa  \ao}{3 (x-2)^5 (\kappa +1)}+\frac{40 \pi \
^2 \kappa  \ao}{3 (x-2)^5 (\kappa +1)}-\frac{3584 \kappa  \ao}{9 \
(x-2)^5 (\kappa +1)}-\frac{\pi ^2 \kappa  \ao}{6 (x-1)^5 (\kappa \
+1)}+\frac{\pi ^2 \kappa  \ao}{3 (\kappa +1)}+\frac{109 \kappa  \
\ao}{36 (\kappa +1)}+\frac{2 d_1 \ao}{9 (x-2) (\kappa +1)}+\frac{4 \
\ao}{9 (x-2) (\kappa +1)}+\frac{d_1 \ao}{18 (x-1) (\kappa \
+1)}-\frac{11 \ao}{12 (x-1) (\kappa +1)}-\frac{10 d_1 \ao}{9 (x-2)^2 \
(\kappa +1)}-\frac{d_1 \ao}{9 (x-1)^2 (\kappa +1)}+\frac{17 \ao}{36 \
(x-1)^2 (\kappa +1)}+\frac{80 d_1 \ao}{9 (x-2)^3 (\kappa \
+1)}-\frac{40 \ao}{9 (x-2)^3 (\kappa +1)}+\frac{d_1 \ao}{3 (x-1)^3 \
(\kappa +1)}-\frac{\ao}{2 (x-1)^3 (\kappa +1)}+\frac{208 d_1 \ao}{3 \
(x-2)^4 (\kappa +1)}+\frac{20 \pi ^2 \ao}{9 (x-2)^4 (\kappa \
+1)}-\frac{1264 \ao}{9 (x-2)^4 (\kappa +1)}+\frac{2 d_1 \ao}{3 \
(x-1)^4 (\kappa +1)}+\frac{\pi ^2 \ao}{18 (x-1)^4 (\kappa \
+1)}-\frac{7 \ao}{4 (x-1)^4 (\kappa +1)}+\frac{256 d_1 \ao}{3 (x-2)^5 \
(\kappa +1)}+\frac{40 \pi ^2 \ao}{9 (x-2)^5 (\kappa +1)}-\frac{2048 \
\ao}{9 (x-2)^5 (\kappa +1)}-\frac{\pi ^2 \ao}{18 (x-1)^5 (\kappa \
+1)}+\frac{545 \ao}{108 (\kappa +1)}+\frac{80 \kappa  \ln ^22\,  \
\ao}{(x-2)^4 (\kappa +1)}+\frac{160 \kappa  \ln ^22\,  \ao}{(x-2)^5 (\
\kappa +1)}+\frac{80 \ln ^22\,  \ao}{3 (x-2)^4 (\kappa +1)}+\frac{160 \
\ln ^22\,  \ao}{3 (x-2)^5 (\kappa +1)}+\frac{32 d_1 \kappa  \ln 2\,  \
\ao}{3 (x-2)^4 (\kappa +1)}+\frac{352 \kappa  \ln 2\,  \ao}{9 (x-2)^4 \
(\kappa +1)}+\frac{64 d_1 \kappa  \ln 2\,  \ao}{3 (x-2)^5 (\kappa \
+1)}+\frac{704 \kappa  \ln 2\,  \ao}{9 (x-2)^5 (\kappa +1)}+\frac{32 \
d_1 \ln 2\,  \ao}{3 (x-2)^4 (\kappa +1)}+\frac{544 \ln 2\,  \ao}{9 \
(x-2)^4 (\kappa +1)}+\frac{64 d_1 \ln 2\,  \ao}{3 (x-2)^5 (\kappa \
+1)}+\frac{1088 \ln 2\,  \ao}{9 (x-2)^5 (\kappa +1)}+\Big(-\frac{x \
\ao^5}{6}-\frac{1}{6} x \kappa  \ao^5-\frac{\kappa  \ao^5}{3 \
(x-2)}+\frac{\kappa  \ao^5}{6 (x-1)}-\frac{\ao^5}{3 \
(x-2)}+\frac{\ao^5}{6 (x-1)}+\frac{19 x \ao^4}{18}+\frac{19}{18} x \
\kappa  \ao^4+\frac{10 \kappa  \ao^4}{9 (x-2)}-\frac{13 \kappa  \
\ao^4}{18 (x-1)}-\frac{10 \kappa  \ao^4}{9 (x-2)^2}+\frac{2 \kappa  \
\ao^4}{9 (x-1)^2}-\frac{\kappa  \ao^4}{9}+\frac{10 \ao^4}{9 \
(x-2)}-\frac{13 \ao^4}{18 (x-1)}-\frac{10 \ao^4}{9 (x-2)^2}+\frac{2 \
\ao^4}{9 (x-1)^2}-\frac{\ao^4}{9}-\frac{26 x \ao^3}{9}-\frac{26}{9} x \
\kappa  \ao^3-\frac{10 \kappa  \ao^3}{9 (x-2)}+\frac{11 \kappa  \
\ao^3}{9 (x-1)}+\frac{20 \kappa  \ao^3}{9 (x-2)^2}-\frac{7 \kappa  \
\ao^3}{9 (x-1)^2}-\frac{40 \kappa  \ao^3}{9 (x-2)^3}+\frac{\kappa  \
\ao^3}{3 (x-1)^3}+\frac{2 \kappa  \ao^3}{3}-\frac{10 \ao^3}{9 (x-2)}+\
\frac{11 \ao^3}{9 (x-1)}+\frac{20 \ao^3}{9 (x-2)^2}-\frac{7 \ao^3}{9 \
(x-1)^2}-\frac{40 \ao^3}{9 (x-2)^3}+\frac{\ao^3}{3 (x-1)^3}+\frac{2 \
\ao^3}{3}+\frac{14 x \ao^2}{3}+\frac{14}{3} x \kappa  \
\ao^2-\frac{\kappa  \ao^2}{x-1}+\frac{\kappa  \
\ao^2}{(x-1)^2}-\frac{\kappa  \ao^2}{(x-1)^3}-\frac{80 \kappa  \
\ao^2}{3 (x-2)^4}+\frac{2 \kappa  \ao^2}{3 (x-1)^4}-2 \kappa  \
\ao^2-\frac{\ao^2}{x-1}+\frac{\ao^2}{(x-1)^2}-\frac{\ao^2}{(x-1)^3}-\frac{80 \ao^2}{3 (x-2)^4}+\frac{2 \ao^2}{3 (x-1)^4}-2 \ao^2-\frac{95 x \
\ao}{36}-\frac{121 x \kappa  \ao}{36}-\frac{40 \kappa  \ao}{9 (x-2)}+\
\frac{65 \kappa  \ao}{12 (x-1)}+\frac{40 \kappa  \ao}{9 \
(x-2)^2}-\frac{25 \kappa  \ao}{18 (x-1)^2}+\frac{35 \kappa  \ao}{36 \
(x-1)^3}+\frac{96 \kappa  \ao}{(x-2)^4}+\frac{31 \kappa  \ao}{36 \
(x-1)^4}+\frac{256 \kappa  \ao}{3 (x-2)^5}-\frac{25 \kappa  \ao}{36 \
(x-1)^5}-\frac{4 \kappa  \ao}{9}-\frac{40 \ao}{9 (x-2)}+\frac{65 \
\ao}{12 (x-1)}+\frac{40 \ao}{9 (x-2)^2}-\frac{25 \ao}{18 \
(x-1)^2}+\frac{35 \ao}{36 (x-1)^3}+\frac{32 d_1 \ao}{3 \
(x-2)^4}+\frac{1504 \ao}{9 (x-2)^4}+\frac{19 \ao}{12 \
(x-1)^4}+\frac{64 d_1 \ao}{3 (x-2)^5}+\frac{2048 \ao}{9 \
(x-2)^5}-\frac{17 \ao}{12 (x-1)^5}-\frac{17 \
\ao}{9}-\frac{x}{36}+\frac{25 x \kappa }{36}-\frac{32 \kappa }{9 \
(x-2)}+\frac{43 \kappa }{12 (x-1)}+\frac{40 \kappa }{9 \
(x-2)^2}-\frac{4 \kappa }{9 (x-1)^2}-\frac{80 \kappa }{9 \
(x-2)^3}-\frac{\kappa }{36 (x-1)^3}-\frac{128 \kappa }{3 \
(x-2)^4}-\frac{43 \kappa }{36 (x-1)^4}-\frac{64 \kappa \
}{(x-2)^5}-\frac{25 \kappa }{36 (x-1)^5}+\frac{128 \kappa }{3 \
(x-2)^6}+\frac{\kappa }{2}-\frac{32}{9 (x-2)}+\frac{43}{12 \
(x-1)}+\frac{40}{9 (x-2)^2}-\frac{4}{9 (x-1)^2}-\frac{80}{9 (x-2)^3}-\
\frac{1}{36 (x-1)^3}-\frac{32 d_1}{3 (x-2)^4}-\frac{1024}{9 (x-2)^4}-\
\frac{23}{12 (x-1)^4}-\frac{128 d_1}{3 (x-2)^5}-\frac{3136}{9 \
(x-2)^5}-\frac{17}{12 (x-1)^5}-\frac{128 d_1}{3 \
(x-2)^6}-\frac{2176}{9 (x-2)^6}+\frac{1}{2}\Big) \
H(0;\ao)+\Big(-\frac{1}{6} d_1 x \ao^5-\frac{d_1 \ao^5}{3 \
(x-2)}+\frac{d_1 \ao^5}{6 (x-1)}-\frac{d_1 \ao^4}{9}+\frac{19}{18} \
d_1 x \ao^4+\frac{10 d_1 \ao^4}{9 (x-2)}-\frac{13 d_1 \ao^4}{18 \
(x-1)}-\frac{10 d_1 \ao^4}{9 (x-2)^2}+\frac{2 d_1 \ao^4}{9 \
(x-1)^2}+\frac{2 d_1 \ao^3}{3}-\frac{26}{9} d_1 x \ao^3-\frac{10 d_1 \
\ao^3}{9 (x-2)}+\frac{11 d_1 \ao^3}{9 (x-1)}+\frac{20 d_1 \ao^3}{9 \
(x-2)^2}-\frac{7 d_1 \ao^3}{9 (x-1)^2}-\frac{40 d_1 \ao^3}{9 \
(x-2)^3}+\frac{d_1 \ao^3}{3 (x-1)^3}-2 d_1 \ao^2+\frac{14}{3} d_1 x \
\ao^2-\frac{d_1 \ao^2}{x-1}+\frac{d_1 \ao^2}{(x-1)^2}-\frac{d_1 \
\ao^2}{(x-1)^3}-\frac{80 d_1 \ao^2}{3 (x-2)^4}+\frac{2 d_1 \ao^2}{3 \
(x-1)^4}+\frac{10 d_1 \ao}{9}-\frac{73 d_1 x \ao}{18}+\frac{5 d_1 \
\ao}{9 (x-2)}+\frac{7 d_1 \ao}{18 (x-1)}-\frac{20 d_1 \ao}{9 \
(x-2)^2}-\frac{5 d_1 \ao}{9 (x-1)^2}+\frac{40 d_1 \ao}{3 \
(x-2)^3}+\frac{d_1 \ao}{(x-1)^3}+\frac{320 d_1 \ao}{3 \
(x-2)^4}+\frac{320 d_1 \ao}{3 (x-2)^5}+\frac{d_1}{3}+\frac{25 d_1 \
x}{18}-\frac{2 d_1}{9 (x-2)}-\frac{d_1}{18 (x-1)}+\frac{10 d_1}{9 \
(x-2)^2}+\frac{d_1}{9 (x-1)^2}-\frac{80 d_1}{9 (x-2)^3}-\frac{d_1}{3 \
(x-1)^3}-\frac{80 d_1}{(x-2)^4}-\frac{2 d_1}{3 (x-1)^4}-\frac{320 \
d_1}{3 (x-2)^5}\Big) H(1;\ao)+\Big(\frac{x \ao}{3}+\frac{x \kappa  \
\ao}{3}+\frac{80 \kappa  \ao}{3 (x-2)^4}-\frac{\kappa  \ao}{3 \
(x-1)^4}+\frac{160 \kappa  \ao}{3 (x-2)^5}+\frac{\kappa  \ao}{3 \
(x-1)^5}-\frac{2 \kappa  \ao}{3}+\frac{80 \ao}{3 (x-2)^4}+\frac{2 d_1 \
\ao}{3 (x-1)^4}-\frac{\ao}{3 (x-1)^4}+\frac{160 \ao}{3 \
(x-2)^5}-\frac{2 d_1 \ao}{3 (x-1)^5}+\frac{\ao}{3 (x-1)^5}-\frac{2 \
\ao}{3}-\frac{x}{3}-\frac{x \kappa }{3}-\frac{80 \kappa }{3 (x-2)^4}+\
\frac{\kappa }{3 (x-1)^4}-\frac{320 \kappa }{3 (x-2)^5}+\frac{\kappa \
}{3 (x-1)^5}-\frac{320 \kappa }{3 (x-2)^6}-\frac{80}{3 \
(x-2)^4}-\frac{2 d_1}{3 (x-1)^4}+\frac{1}{3 (x-1)^4}-\frac{320}{3 \
(x-2)^5}-\frac{2 d_1}{3 (x-1)^5}+\frac{1}{3 (x-1)^5}-\frac{320}{3 \
(x-2)^6}\Big) H(0;\ao) H(1;x)+\Big(-\frac{x \ao^5}{12}-\frac{1}{12} x \
\kappa  \ao^5-\frac{\kappa  \ao^5}{6 (x-2)}+\frac{\kappa  \ao^5}{12 \
(x-1)}-\frac{\ao^5}{6 (x-2)}+\frac{\ao^5}{12 (x-1)}+\frac{19 x \
\ao^4}{36}+\frac{19}{36} x \kappa  \ao^4+\frac{5 \kappa  \ao^4}{9 \
(x-2)}-\frac{13 \kappa  \ao^4}{36 (x-1)}-\frac{5 \kappa  \ao^4}{9 \
(x-2)^2}+\frac{\kappa  \ao^4}{9 (x-1)^2}-\frac{\kappa  \
\ao^4}{18}+\frac{5 \ao^4}{9 (x-2)}-\frac{13 \ao^4}{36 (x-1)}-\frac{5 \
\ao^4}{9 (x-2)^2}+\frac{\ao^4}{9 (x-1)^2}-\frac{\ao^4}{18}-\frac{13 x \
\ao^3}{9}-\frac{13}{9} x \kappa  \ao^3-\frac{5 \kappa  \ao^3}{9 \
(x-2)}+\frac{11 \kappa  \ao^3}{18 (x-1)}+\frac{10 \kappa  \ao^3}{9 \
(x-2)^2}-\frac{7 \kappa  \ao^3}{18 (x-1)^2}-\frac{20 \kappa  \ao^3}{9 \
(x-2)^3}+\frac{\kappa  \ao^3}{6 (x-1)^3}+\frac{\kappa  \
\ao^3}{3}-\frac{5 \ao^3}{9 (x-2)}+\frac{11 \ao^3}{18 (x-1)}+\frac{10 \
\ao^3}{9 (x-2)^2}-\frac{7 \ao^3}{18 (x-1)^2}-\frac{20 \ao^3}{9 \
(x-2)^3}+\frac{\ao^3}{6 (x-1)^3}+\frac{\ao^3}{3}+\frac{7 x \ao^2}{3}+\
\frac{7}{3} x \kappa  \ao^2-\frac{\kappa  \ao^2}{2 \
(x-1)}+\frac{\kappa  \ao^2}{2 (x-1)^2}-\frac{\kappa  \ao^2}{2 \
(x-1)^3}-\frac{40 \kappa  \ao^2}{3 (x-2)^4}+\frac{\kappa  \ao^2}{3 \
(x-1)^4}-\kappa  \ao^2-\frac{\ao^2}{2 (x-1)}+\frac{\ao^2}{2 (x-1)^2}-\
\frac{\ao^2}{2 (x-1)^3}-\frac{40 \ao^2}{3 (x-2)^4}+\frac{\ao^2}{3 \
(x-1)^4}-\ao^2-\frac{73 x \ao}{36}-\frac{73 x \kappa  \
\ao}{36}-\frac{40 \kappa  \ao}{9 (x-2)}+\frac{65 \kappa  \ao}{12 \
(x-1)}+\frac{40 \kappa  \ao}{9 (x-2)^2}-\frac{25 \kappa  \ao}{18 \
(x-1)^2}+\frac{35 \kappa  \ao}{36 (x-1)^3}+\frac{80 \kappa  \
\ao}{(x-2)^4}+\frac{19 \kappa  \ao}{36 (x-1)^4}+\frac{160 \kappa  \
\ao}{3 (x-2)^5}-\frac{25 \kappa  \ao}{36 (x-1)^5}+\frac{2 \kappa  \
\ao}{9}-\frac{40 \ao}{9 (x-2)}+\frac{65 \ao}{12 (x-1)}+\frac{40 \
\ao}{9 (x-2)^2}-\frac{25 \ao}{18 (x-1)^2}+\frac{35 \ao}{36 \
(x-1)^3}+\frac{80 \ao}{(x-2)^4}+\frac{5 \ao}{4 (x-1)^4}+\frac{160 \
\ao}{3 (x-2)^5}-\frac{17 \ao}{12 (x-1)^5}+\frac{2 \ao}{9}+\frac{25 \
x}{36}+\frac{25 x \kappa }{36}-\frac{32 \kappa }{9 (x-2)}+\frac{43 \
\kappa }{12 (x-1)}+\frac{40 \kappa }{9 (x-2)^2}-\frac{4 \kappa }{9 \
(x-1)^2}-\frac{80 \kappa }{9 (x-2)^3}-\frac{\kappa }{36 \
(x-1)^3}-\frac{160 \kappa }{3 (x-2)^4}-\frac{43 \kappa }{36 (x-1)^4}-\
\frac{320 \kappa }{3 (x-2)^5}-\frac{25 \kappa }{36 \
(x-1)^5}+\frac{\kappa }{2}+\Big(-\frac{2 \kappa  \ao}{3 \
(x-1)^4}+\frac{2 \kappa  \ao}{3 (x-1)^5}-\frac{2 \ao}{3 \
(x-1)^4}+\frac{2 \ao}{3 (x-1)^5}+\frac{2 \kappa }{3 (x-1)^4}+\frac{2 \
\kappa }{3 (x-1)^5}+\frac{2}{3 (x-1)^4}+\frac{2}{3 (x-1)^5}\Big) H(0;\
\ao)+\Big(-\frac{2 \ao d_1}{3 (x-1)^4}+\frac{2 d_1}{3 \
(x-1)^4}+\frac{2 \ao d_1}{3 (x-1)^5}+\frac{2 d_1}{3 (x-1)^5}\Big) \
H(1;\ao)-\frac{32}{9 (x-2)}+\frac{43}{12 (x-1)}+\frac{40}{9 (x-2)^2}-\
\frac{4}{9 (x-1)^2}-\frac{80}{9 (x-2)^3}-\frac{1}{36 \
(x-1)^3}-\frac{160}{3 (x-2)^4}-\frac{23}{12 (x-1)^4}-\frac{320}{3 \
(x-2)^5}-\frac{17}{12 (x-1)^5}+\frac{1}{2}\Big) \
H(c_1(\ao);x)+\Big(-\frac{32 \kappa  \ao}{3 (x-2)^4}-\frac{64 \kappa  \
\ao}{3 (x-2)^5}+\frac{32 d_1 \ao}{3 (x-2)^4}+\frac{544 \ao}{9 \
(x-2)^4}+\frac{64 d_1 \ao}{3 (x-2)^5}+\frac{1088 \ao}{9 \
(x-2)^5}+\frac{32 \kappa }{3 (x-2)^4}+\frac{128 \kappa }{3 \
(x-2)^5}+\frac{128 \kappa }{3 (x-2)^6}+\Big(-\frac{160 \kappa  \ao}{3 \
(x-2)^4}-\frac{320 \kappa  \ao}{3 (x-2)^5}-\frac{160 \ao}{3 (x-2)^4}-\
\frac{320 \ao}{3 (x-2)^5}+\frac{160 \kappa }{3 (x-2)^4}+\frac{640 \
\kappa }{3 (x-2)^5}+\frac{640 \kappa }{3 (x-2)^6}+\frac{160}{3 \
(x-2)^4}+\frac{640}{3 (x-2)^5}+\frac{640}{3 (x-2)^6}\Big) \
H(0;\ao)+\Big(-\frac{160 \ao d_1}{3 (x-2)^4}+\frac{160 d_1}{3 \
(x-2)^4}-\frac{320 \ao d_1}{3 (x-2)^5}+\frac{640 d_1}{3 \
(x-2)^5}+\frac{640 d_1}{3 (x-2)^6}\Big) H(1;\ao)-\frac{32 d_1}{3 \
(x-2)^4}-\frac{544}{9 (x-2)^4}-\frac{128 d_1}{3 \
(x-2)^5}-\frac{2176}{9 (x-2)^5}-\frac{128 d_1}{3 \
(x-2)^6}-\frac{2176}{9 (x-2)^6}\Big) H(c_2(\ao);x)+\Big(-\frac{2 x \
\ao}{3}-\frac{2 x \kappa  \ao}{3}-\frac{160 \kappa  \ao}{3 \
(x-2)^4}-\frac{2 \kappa  \ao}{3 (x-1)^4}-\frac{320 \kappa  \ao}{3 \
(x-2)^5}+\frac{2 \kappa  \ao}{3 (x-1)^5}+\frac{4 \kappa  \
\ao}{3}-\frac{160 \ao}{3 (x-2)^4}-\frac{2 \ao}{3 (x-1)^4}-\frac{320 \
\ao}{3 (x-2)^5}+\frac{2 \ao}{3 (x-1)^5}+\frac{4 \ao}{3}+\frac{2 \
x}{3}+\frac{2 x \kappa }{3}+\frac{160 \kappa }{3 (x-2)^4}+\frac{2 \
\kappa }{3 (x-1)^4}+\frac{640 \kappa }{3 (x-2)^5}+\frac{2 \kappa }{3 \
(x-1)^5}+\frac{640 \kappa }{3 (x-2)^6}+\frac{160}{3 \
(x-2)^4}+\frac{2}{3 (x-1)^4}+\frac{640}{3 (x-2)^5}+\frac{2}{3 \
(x-1)^5}+\frac{640}{3 (x-2)^6}\Big) H(0,0;\ao)+\Big(-\frac{2 x \
\ao}{3}-\frac{2 x \kappa  \ao}{3}+\frac{160 \kappa  \ao}{3 \
(x-2)^4}+\frac{2 \kappa  \ao}{3 (x-1)^4}+\frac{320 \kappa  \ao}{3 \
(x-2)^5}-\frac{2 \kappa  \ao}{3 (x-1)^5}+\frac{4 \kappa  \
\ao}{3}+\frac{160 \ao}{3 (x-2)^4}+\frac{2 \ao}{3 (x-1)^4}+\frac{320 \
\ao}{3 (x-2)^5}-\frac{2 \ao}{3 (x-1)^5}+\frac{4 \ao}{3}+\frac{2 \
x}{3}+\frac{2 x \kappa }{3}-\frac{160 \kappa }{3 (x-2)^4}-\frac{2 \
\kappa }{3 (x-1)^4}-\frac{640 \kappa }{3 (x-2)^5}-\frac{2 \kappa }{3 \
(x-1)^5}-\frac{640 \kappa }{3 (x-2)^6}-\frac{160}{3 \
(x-2)^4}-\frac{2}{3 (x-1)^4}-\frac{640}{3 (x-2)^5}-\frac{2}{3 \
(x-1)^5}-\frac{640}{3 (x-2)^6}\Big) H(0,0;x)+\Big(\frac{4 \ao \
d_1}{3}-\frac{2 \ao x d_1}{3}+\frac{2 x d_1}{3}-\frac{160 \ao d_1}{3 \
(x-2)^4}+\frac{160 d_1}{3 (x-2)^4}-\frac{2 \ao d_1}{3 \
(x-1)^4}+\frac{2 d_1}{3 (x-1)^4}-\frac{320 \ao d_1}{3 \
(x-2)^5}+\frac{640 d_1}{3 (x-2)^5}+\frac{2 \ao d_1}{3 \
(x-1)^5}+\frac{2 d_1}{3 (x-1)^5}+\frac{640 d_1}{3 (x-2)^6}\Big) \
H(0,1;\ao)+\Big(\frac{x \ao}{3}+\frac{x \kappa  \ao}{3}+\frac{80 \
\kappa  \ao}{3 (x-2)^4}-\frac{\kappa  \ao}{3 (x-1)^4}+\frac{160 \
\kappa  \ao}{3 (x-2)^5}+\frac{\kappa  \ao}{3 (x-1)^5}-\frac{2 \kappa  \
\ao}{3}+\frac{80 \ao}{3 (x-2)^4}-\frac{\ao}{3 (x-1)^4}+\frac{160 \
\ao}{3 (x-2)^5}+\frac{\ao}{3 (x-1)^5}-\frac{2 \
\ao}{3}-\frac{x}{3}-\frac{x \kappa }{3}-\frac{80 \kappa }{3 (x-2)^4}+\
\frac{\kappa }{3 (x-1)^4}-\frac{320 \kappa }{3 (x-2)^5}+\frac{\kappa \
}{3 (x-1)^5}-\frac{320 \kappa }{3 (x-2)^6}-\frac{80}{3 \
(x-2)^4}+\frac{1}{3 (x-1)^4}-\frac{320}{3 (x-2)^5}+\frac{1}{3 \
(x-1)^5}-\frac{320}{3 (x-2)^6}\Big) H(0,c_1(\ao);x)+\Big(-\frac{160 \
\kappa  \ao}{3 (x-2)^4}-\frac{320 \kappa  \ao}{3 (x-2)^5}-\frac{160 \
\ao}{3 (x-2)^4}-\frac{320 \ao}{3 (x-2)^5}+\frac{160 \kappa }{3 \
(x-2)^4}+\frac{640 \kappa }{3 (x-2)^5}+\frac{640 \kappa }{3 (x-2)^6}+\
\frac{160}{3 (x-2)^4}+\frac{640}{3 (x-2)^5}+\frac{640}{3 \
(x-2)^6}\Big) H(0,c_2(\ao);x)+\Big(-\frac{x \ao}{3}-\frac{x \kappa  \
\ao}{3}-\frac{80 \kappa  \ao}{3 (x-2)^4}+\frac{\kappa  \ao}{3 \
(x-1)^4}-\frac{160 \kappa  \ao}{3 (x-2)^5}-\frac{\kappa  \ao}{3 \
(x-1)^5}+\frac{2 \kappa  \ao}{3}-\frac{80 \ao}{3 (x-2)^4}-\frac{2 d_1 \
\ao}{3 (x-1)^4}+\frac{\ao}{3 (x-1)^4}-\frac{160 \ao}{3 \
(x-2)^5}+\frac{2 d_1 \ao}{3 (x-1)^5}-\frac{\ao}{3 (x-1)^5}+\frac{2 \
\ao}{3}+\frac{x}{3}+\frac{x \kappa }{3}+\frac{80 \kappa }{3 (x-2)^4}-\
\frac{\kappa }{3 (x-1)^4}+\frac{320 \kappa }{3 (x-2)^5}-\frac{\kappa \
}{3 (x-1)^5}+\frac{320 \kappa }{3 (x-2)^6}+\frac{80}{3 \
(x-2)^4}+\frac{2 d_1}{3 (x-1)^4}-\frac{1}{3 (x-1)^4}+\frac{320}{3 \
(x-2)^5}+\frac{2 d_1}{3 (x-1)^5}-\frac{1}{3 (x-1)^5}+\frac{320}{3 \
(x-2)^6}\Big) H(1,0;x)+\Big(\frac{x \ao}{3}+\frac{x \kappa  \
\ao}{3}+\frac{80 \kappa  \ao}{3 (x-2)^4}-\frac{\kappa  \ao}{3 \
(x-1)^4}+\frac{160 \kappa  \ao}{3 (x-2)^5}+\frac{\kappa  \ao}{3 \
(x-1)^5}-\frac{2 \kappa  \ao}{3}+\frac{80 \ao}{3 (x-2)^4}+\frac{2 d_1 \
\ao}{3 (x-1)^4}-\frac{\ao}{3 (x-1)^4}+\frac{160 \ao}{3 \
(x-2)^5}-\frac{2 d_1 \ao}{3 (x-1)^5}+\frac{\ao}{3 (x-1)^5}-\frac{2 \
\ao}{3}-\frac{x}{3}-\frac{x \kappa }{3}-\frac{80 \kappa }{3 (x-2)^4}+\
\frac{\kappa }{3 (x-1)^4}-\frac{320 \kappa }{3 (x-2)^5}+\frac{\kappa \
}{3 (x-1)^5}-\frac{320 \kappa }{3 (x-2)^6}-\frac{80}{3 \
(x-2)^4}-\frac{2 d_1}{3 (x-1)^4}+\frac{1}{3 (x-1)^4}-\frac{320}{3 \
(x-2)^5}-\frac{2 d_1}{3 (x-1)^5}+\frac{1}{3 (x-1)^5}-\frac{320}{3 \
(x-2)^6}\Big) H(1,c_1(\ao);x)+\Big(-\frac{160 \kappa  \ao}{3 \
(x-2)^4}-\frac{320 \kappa  \ao}{3 (x-2)^5}+\frac{160 d_1 \ao}{3 \
(x-2)^4}-\frac{160 \ao}{3 (x-2)^4}+\frac{320 d_1 \ao}{3 \
(x-2)^5}-\frac{320 \ao}{3 (x-2)^5}+\frac{160 \kappa }{3 \
(x-2)^4}+\frac{640 \kappa }{3 (x-2)^5}+\frac{640 \kappa }{3 (x-2)^6}-\
\frac{160 d_1}{3 (x-2)^4}+\frac{160}{3 (x-2)^4}-\frac{640 d_1}{3 \
(x-2)^5}+\frac{640}{3 (x-2)^5}-\frac{640 d_1}{3 (x-2)^6}+\frac{640}{3 \
(x-2)^6}\Big) H(2,0;x)+\Big(\frac{160 \kappa  \ao}{3 \
(x-2)^4}+\frac{320 \kappa  \ao}{3 (x-2)^5}-\frac{160 d_1 \ao}{3 \
(x-2)^4}+\frac{160 \ao}{3 (x-2)^4}-\frac{320 d_1 \ao}{3 \
(x-2)^5}+\frac{320 \ao}{3 (x-2)^5}-\frac{160 \kappa }{3 \
(x-2)^4}-\frac{640 \kappa }{3 (x-2)^5}-\frac{640 \kappa }{3 (x-2)^6}+\
\frac{160 d_1}{3 (x-2)^4}-\frac{160}{3 (x-2)^4}+\frac{640 d_1}{3 \
(x-2)^5}-\frac{640}{3 (x-2)^5}+\frac{640 d_1}{3 (x-2)^6}-\frac{640}{3 \
(x-2)^6}\Big) H(2,c_2(\ao);x)+\Big(-\frac{\kappa  \ao}{3 \
(x-1)^4}+\frac{\kappa  \ao}{3 (x-1)^5}-\frac{\ao}{3 \
(x-1)^4}+\frac{\ao}{3 (x-1)^5}+\frac{\kappa }{3 (x-1)^4}+\frac{\kappa \
}{3 (x-1)^5}+\frac{1}{3 (x-1)^4}+\frac{1}{3 (x-1)^5}\Big) \
H(c_1(\ao),c_1(\ao);x)+\Big(-\frac{80 \kappa  \ao}{3 \
(x-2)^4}-\frac{160 \kappa  \ao}{3 (x-2)^5}-\frac{80 \ao}{3 \
(x-2)^4}-\frac{160 \ao}{3 (x-2)^5}+\frac{80 \kappa }{3 \
(x-2)^4}+\frac{320 \kappa }{3 (x-2)^5}+\frac{320 \kappa }{3 (x-2)^6}+\
\frac{80}{3 (x-2)^4}+\frac{320}{3 (x-2)^5}+\frac{320}{3 (x-2)^6}\Big) \
H(c_2(\ao),c_1(\ao);x)+H(0;x) \Big(\frac{17 x \ao}{12}+\frac{25 x \
\kappa  \ao}{36}+\frac{40 \kappa  \ao}{9 (x-2)}-\frac{65 \kappa  \
\ao}{12 (x-1)}-\frac{40 \kappa  \ao}{9 (x-2)^2}+\frac{25 \kappa  \
\ao}{18 (x-1)^2}-\frac{35 \kappa  \ao}{36 (x-1)^3}-\frac{128 \kappa  \
\ao}{3 (x-2)^4}-\frac{7 \kappa  \ao}{36 (x-1)^4}+\frac{64 \kappa  \
\ao}{3 (x-2)^5}+\frac{25 \kappa  \ao}{36 (x-1)^5}-\frac{8 \kappa  \
\ao}{9}+\frac{40 \ao}{9 (x-2)}-\frac{65 \ao}{12 (x-1)}-\frac{40 \
\ao}{9 (x-2)^2}+\frac{25 \ao}{18 (x-1)^2}-\frac{35 \ao}{36 \
(x-1)^3}-\frac{32 d_1 \ao}{3 (x-2)^4}-\frac{1024 \ao}{9 \
(x-2)^4}-\frac{11 \ao}{12 (x-1)^4}-\frac{64 d_1 \ao}{3 \
(x-2)^5}-\frac{1088 \ao}{9 (x-2)^5}+\frac{17 \ao}{12 \
(x-1)^5}-\frac{160 \kappa  \ln 2\,  \ao}{3 (x-2)^4}-\frac{320 \kappa  \
\ln 2\,  \ao}{3 (x-2)^5}-\frac{160 \ln 2\,  \ao}{3 (x-2)^4}-\frac{320 \
\ln 2\,  \ao}{3 (x-2)^5}-\frac{7 \ao}{3}-\frac{17 x}{12}-\frac{25 x \
\kappa }{36}+\frac{32 \kappa }{9 (x-2)}-\frac{43 \kappa }{12 \
(x-1)}-\frac{40 \kappa }{9 (x-2)^2}+\frac{4 \kappa }{9 \
(x-1)^2}+\frac{80 \kappa }{9 (x-2)^3}+\frac{\kappa }{36 \
(x-1)^3}+\frac{128 \kappa }{3 (x-2)^4}+\frac{43 \kappa }{36 (x-1)^4}+\
\frac{64 \kappa }{(x-2)^5}+\frac{25 \kappa }{36 (x-1)^5}-\frac{128 \
\kappa }{3 (x-2)^6}-\frac{\kappa }{2}+\frac{32}{9 (x-2)}-\frac{43}{12 \
(x-1)}-\frac{40}{9 (x-2)^2}+\frac{4}{9 (x-1)^2}+\frac{80}{9 (x-2)^3}+\
\frac{1}{36 (x-1)^3}+\frac{32 d_1}{3 (x-2)^4}+\frac{1024}{9 (x-2)^4}+\
\frac{23}{12 (x-1)^4}+\frac{128 d_1}{3 (x-2)^5}+\frac{3136}{9 \
(x-2)^5}+\frac{17}{12 (x-1)^5}+\frac{128 d_1}{3 \
(x-2)^6}+\frac{2176}{9 (x-2)^6}+\frac{160 \kappa  \ln 2\, }{3 \
(x-2)^4}+\frac{640 \kappa  \ln 2\, }{3 (x-2)^5}+\frac{640 \kappa  \ln \
2\, }{3 (x-2)^6}+\frac{160 \ln 2\, }{3 (x-2)^4}+\frac{640 \ln 2\, }{3 \
(x-2)^5}+\frac{640 \ln 2\, }{3 (x-2)^6}-\frac{1}{2}\Big)+H(2;x) \Big(\
\frac{160 \kappa  \ln 2\,  \ao}{3 (x-2)^4}+\frac{320 \kappa  \ln 2\,  \
\ao}{3 (x-2)^5}-\frac{160 d_1 \ln 2\,  \ao}{3 (x-2)^4}+\frac{160 \ln \
2\,  \ao}{3 (x-2)^4}-\frac{320 d_1 \ln 2\,  \ao}{3 (x-2)^5}+\frac{320 \
\ln 2\,  \ao}{3 (x-2)^5}+\Big(\frac{160 \kappa  \ao}{3 \
(x-2)^4}+\frac{320 \kappa  \ao}{3 (x-2)^5}-\frac{160 d_1 \ao}{3 \
(x-2)^4}+\frac{160 \ao}{3 (x-2)^4}-\frac{320 d_1 \ao}{3 \
(x-2)^5}+\frac{320 \ao}{3 (x-2)^5}-\frac{160 \kappa }{3 \
(x-2)^4}-\frac{640 \kappa }{3 (x-2)^5}-\frac{640 \kappa }{3 (x-2)^6}+\
\frac{160 d_1}{3 (x-2)^4}-\frac{160}{3 (x-2)^4}+\frac{640 d_1}{3 \
(x-2)^5}-\frac{640}{3 (x-2)^5}+\frac{640 d_1}{3 (x-2)^6}-\frac{640}{3 \
(x-2)^6}\Big) H(0;\ao)-\frac{160 \kappa  \ln 2\, }{3 \
(x-2)^4}-\frac{640 \kappa  \ln 2\, }{3 (x-2)^5}-\frac{640 \kappa  \ln \
2\, }{3 (x-2)^6}+\frac{160 d_1 \ln 2\, }{3 (x-2)^4}-\frac{160 \ln 2\, \
}{3 (x-2)^4}+\frac{640 d_1 \ln 2\, }{3 (x-2)^5}-\frac{640 \ln 2\, }{3 \
(x-2)^5}+\frac{640 d_1 \ln 2\, }{3 (x-2)^6}-\frac{640 \ln 2\, }{3 \
(x-2)^6}\Big)+\frac{40 x}{27 (\kappa +1)}+\frac{\pi ^2 x \kappa }{6 (\
\kappa +1)}-\frac{20 \pi ^2 \kappa }{3 (x-2)^4 (\kappa +1)}-\frac{\pi \
^2 \kappa }{6 (x-1)^4 (\kappa +1)}-\frac{80 \pi ^2 \kappa }{3 (x-2)^5 \
(\kappa +1)}-\frac{\pi ^2 \kappa }{6 (x-1)^5 (\kappa +1)}-\frac{80 \
\pi ^2 \kappa }{3 (x-2)^6 (\kappa +1)}-\frac{20 \pi ^2}{9 (x-2)^4 \
(\kappa +1)}-\frac{\pi ^2}{18 (x-1)^4 (\kappa +1)}-\frac{80 \pi ^2}{9 \
(x-2)^5 (\kappa +1)}-\frac{\pi ^2}{18 (x-1)^5 (\kappa +1)}-\frac{80 \
\pi ^2}{9 (x-2)^6 (\kappa +1)}-\frac{80 \kappa  \ln ^22\, }{(x-2)^4 (\
\kappa +1)}-\frac{320 \kappa  \ln ^22\, }{(x-2)^5 (\kappa \
+1)}-\frac{320 \kappa  \ln ^22\, }{(x-2)^6 (\kappa +1)}-\frac{80 \ln \
^22\, }{3 (x-2)^4 (\kappa +1)}-\frac{320 \ln ^22\, }{3 (x-2)^5 \
(\kappa +1)}-\frac{320 \ln ^22\, }{3 (x-2)^6 (\kappa +1)}-\frac{32 \
d_1 \kappa  \ln 2\, }{3 (x-2)^4 (\kappa +1)}-\frac{352 \kappa  \ln \
2\, }{9 (x-2)^4 (\kappa +1)}-\frac{128 d_1 \kappa  \ln 2\, }{3 \
(x-2)^5 (\kappa +1)}-\frac{1408 \kappa  \ln 2\, }{9 (x-2)^5 (\kappa \
+1)}-\frac{128 d_1 \kappa  \ln 2\, }{3 (x-2)^6 (\kappa \
+1)}-\frac{1408 \kappa  \ln 2\, }{9 (x-2)^6 (\kappa +1)}-\frac{32 d_1 \
\ln 2\, }{3 (x-2)^4 (\kappa +1)}-\frac{544 \ln 2\, }{9 (x-2)^4 \
(\kappa +1)}-\frac{128 d_1 \ln 2\, }{3 (x-2)^5 (\kappa \
+1)}-\frac{2176 \ln 2\, }{9 (x-2)^5 (\kappa +1)}-\frac{128 d_1 \ln \
2\, }{3 (x-2)^6 (\kappa +1)}-\frac{2176 \ln 2\, }{9 (x-2)^6 (\kappa \
+1)}
\Big\},
\erp
% ep^2
\brp
a_2^{(\kap,2)}= \frac{1}{\ao(x-2)-x}\Big\{\frac{1}{48} d_1^2 x \ao^5-\frac{1}{72} \pi ^2 x \ao^5+\frac{d_1^2 \
\ao^5}{24 (x-2)}-\frac{\pi ^2 \ao^5}{36 (x-2)}-\frac{d_1^2 \ao^5}{48 \
(x-1)}+\frac{\pi ^2 \ao^5}{72 (x-1)}-\frac{d_1 \ao^5}{48 (\kappa \
+1)}-\frac{19 d_1 x \ao^5}{144 (\kappa +1)}+\frac{13 x \ao^5}{27 \
(\kappa +1)}-\frac{d_1 \kappa  \ao^5}{48 (\kappa +1)}-\frac{31 d_1 x \
\kappa  \ao^5}{144 (\kappa +1)}+\frac{85 x \kappa  \ao^5}{108 (\kappa \
+1)}-\frac{17 d_1 \kappa  \ao^5}{36 (x-2) (\kappa +1)}+\frac{415 \
\kappa  \ao^5}{216 (x-2) (\kappa +1)}+\frac{31 d_1 \kappa  \ao^5}{144 \
(x-1) (\kappa +1)}-\frac{85 \kappa  \ao^5}{108 (x-1) (\kappa \
+1)}+\frac{25 \kappa  \ao^5}{144 (\kappa +1)}-\frac{11 d_1 \ao^5}{36 \
(x-2) (\kappa +1)}+\frac{265 \ao^5}{216 (x-2) (\kappa +1)}+\frac{19 \
d_1 \ao^5}{144 (x-1) (\kappa +1)}-\frac{13 \ao^5}{27 (x-1) (\kappa \
+1)}+\frac{19 \ao^5}{144 (\kappa +1)}+\frac{37 d_1^2 \
\ao^4}{648}-\frac{199 d_1^2 x \ao^4}{1296}+\frac{19}{216} \pi ^2 x \
\ao^4-\frac{17 d_1^2 \ao^4}{324 (x-2)}+\frac{5 \pi ^2 \ao^4}{54 \
(x-2)}+\frac{145 d_1^2 \ao^4}{1296 (x-1)}-\frac{13 \pi ^2 \ao^4}{216 \
(x-1)}-\frac{79 d_1 \ao^4}{432 (\kappa +1)}+\frac{137 d_1 x \
\ao^4}{144 (\kappa +1)}-\frac{173 x \ao^4}{54 (\kappa +1)}-\frac{617 \
d_1 \kappa  \ao^4}{1296 (\kappa +1)}+\frac{2113 d_1 x \kappa  \
\ao^4}{1296 (\kappa +1)}-\frac{1835 x \kappa  \ao^4}{324 (\kappa \
+1)}+\frac{139 d_1 \kappa  \ao^4}{162 (x-2) (\kappa +1)}-\frac{1687 \
\kappa  \ao^4}{324 (x-2) (\kappa +1)}-\frac{1429 d_1 \kappa  \
\ao^4}{1296 (x-1) (\kappa +1)}+\frac{2359 \kappa  \ao^4}{648 (x-1) \
(\kappa +1)}-\frac{487 d_1 \kappa  \ao^4}{162 (x-2)^2 (\kappa \
+1)}+\frac{2851 \kappa  \ao^4}{324 (x-2)^2 (\kappa +1)}+\frac{359 d_1 \
\kappa  \ao^4}{648 (x-1)^2 (\kappa +1)}-\frac{140 \kappa  \ao^4}{81 \
(x-1)^2 (\kappa +1)}+\frac{733 \kappa  \ao^4}{1296 (\kappa \
+1)}+\frac{35 d_1 \ao^4}{54 (x-2) (\kappa +1)}-\frac{137 \ao^4}{36 \
(x-2) (\kappa +1)}-\frac{283 d_1 \ao^4}{432 (x-1) (\kappa \
+1)}+\frac{461 \ao^4}{216 (x-1) (\kappa +1)}-\frac{95 d_1 \ao^4}{54 \
(x-2)^2 (\kappa +1)}+\frac{503 \ao^4}{108 (x-2)^2 (\kappa \
+1)}+\frac{7 d_1 \ao^4}{24 (x-1)^2 (\kappa +1)}-\frac{43 \ao^4}{54 \
(x-1)^2 (\kappa +1)}-\frac{59 \ao^4}{432 (\kappa +1)}+\frac{101 d_1^2 \
\ao^4}{324 (x-2)^2}-\frac{5 \pi ^2 \ao^4}{54 (x-2)^2}-\frac{4 d_1^2 \
\ao^4}{81 (x-1)^2}+\frac{\pi ^2 \ao^4}{54 (x-1)^2}-\frac{\pi ^2 \
\ao^4}{108}-\frac{25 d_1^2 \ao^3}{54}+\frac{371}{648} d_1^2 x \
\ao^3-\frac{13}{54} \pi ^2 x \ao^3-\frac{67 d_1^2 \ao^3}{324 \
(x-2)}-\frac{5 \pi ^2 \ao^3}{54 (x-2)}-\frac{155 d_1^2 \ao^3}{648 \
(x-1)}+\frac{11 \pi ^2 \ao^3}{108 (x-1)}+\frac{437 d_1 \ao^3}{216 \
(\kappa +1)}-\frac{1441 d_1 x \ao^3}{432 (\kappa +1)}+\frac{521 x \
\ao^3}{54 (\kappa +1)}+\frac{985 d_1 \kappa  \ao^3}{216 (\kappa +1)}-\
\frac{8029 d_1 x \kappa  \ao^3}{1296 (\kappa +1)}+\frac{6347 x \kappa \
 \ao^3}{324 (\kappa +1)}+\frac{769 d_1 \kappa  \ao^3}{324 (x-2) \
(\kappa +1)}-\frac{182 \kappa  \ao^3}{81 (x-2) (\kappa +1)}+\frac{139 \
d_1 \kappa  \ao^3}{81 (x-1) (\kappa +1)}-\frac{3085 \kappa  \
\ao^3}{648 (x-1) (\kappa +1)}-\frac{110 d_1 \kappa  \ao^3}{81 (x-2)^2 \
(\kappa +1)}-\frac{509 \kappa  \ao^3}{81 (x-2)^2 (\kappa \
+1)}-\frac{3677 d_1 \kappa  \ao^3}{1296 (x-1)^2 (\kappa \
+1)}+\frac{10099 \kappa  \ao^3}{1296 (x-1)^2 (\kappa +1)}-\frac{2168 \
d_1 \kappa  \ao^3}{81 (x-2)^3 (\kappa +1)}+\frac{4684 \kappa  \
\ao^3}{81 (x-2)^3 (\kappa +1)}+\frac{263 d_1 \kappa  \ao^3}{144 \
(x-1)^3 (\kappa +1)}-\frac{173 \kappa  \ao^3}{36 (x-1)^3 (\kappa \
+1)}-\frac{4183 \kappa  \ao^3}{432 (\kappa +1)}+\frac{89 d_1 \
\ao^3}{108 (x-2) (\kappa +1)}+\frac{58 \ao^3}{27 (x-2) (\kappa \
+1)}+\frac{121 d_1 \ao^3}{108 (x-1) (\kappa +1)}-\frac{239 \ao^3}{72 \
(x-1) (\kappa +1)}+\frac{2 d_1 \ao^3}{27 (x-2)^2 (\kappa \
+1)}-\frac{185 \ao^3}{27 (x-2)^2 (\kappa +1)}-\frac{605 d_1 \
\ao^3}{432 (x-1)^2 (\kappa +1)}+\frac{1351 \ao^3}{432 (x-1)^2 (\kappa \
+1)}-\frac{376 d_1 \ao^3}{27 (x-2)^3 (\kappa +1)}+\frac{212 \ao^3}{9 \
(x-2)^3 (\kappa +1)}+\frac{367 d_1 \ao^3}{432 (x-1)^3 (\kappa \
+1)}-\frac{89 \ao^3}{54 (x-1)^3 (\kappa +1)}-\frac{1021 \ao^3}{432 \
(\kappa +1)}+\frac{29 d_1^2 \ao^3}{81 (x-2)^2}+\frac{5 \pi ^2 \
\ao^3}{27 (x-2)^2}+\frac{43 d_1^2 \ao^3}{162 (x-1)^2}-\frac{7 \pi ^2 \
\ao^3}{108 (x-1)^2}+\frac{260 d_1^2 \ao^3}{81 (x-2)^3}-\frac{10 \pi \
^2 \ao^3}{27 (x-2)^3}-\frac{d_1^2 \ao^3}{6 (x-1)^3}+\frac{\pi ^2 \
\ao^3}{36 (x-1)^3}+\frac{\pi ^2 \ao^3}{18}+\frac{365 d_1^2 \
\ao^2}{108}-\frac{505}{216} d_1^2 x \ao^2+\frac{7}{18} \pi ^2 x \
\ao^2+\frac{14 d_1^2 \ao^2}{27 (x-2)}+\frac{55 d_1^2 \ao^2}{216 \
(x-1)}-\frac{\pi ^2 \ao^2}{12 (x-1)}-\frac{1445 d_1 \ao^2}{108 \
(\kappa +1)}+\frac{4637 d_1 x \ao^2}{432 (\kappa +1)}-\frac{497 x \
\ao^2}{24 (\kappa +1)}-\frac{1735 d_1 \kappa  \ao^2}{54 (\kappa +1)}+\
\frac{10115 d_1 x \kappa  \ao^2}{432 (\kappa +1)}-\frac{12541 x \
\kappa  \ao^2}{216 (\kappa +1)}-\frac{458 d_1 \kappa  \ao^2}{27 (x-2) \
(\kappa +1)}+\frac{1280 \kappa  \ao^2}{27 (x-2) (\kappa \
+1)}+\frac{1619 d_1 \kappa  \ao^2}{216 (x-1) (\kappa +1)}-\frac{575 \
\kappa  \ao^2}{24 (x-1) (\kappa +1)}+\frac{737 d_1 \kappa  \ao^2}{18 \
(x-2)^2 (\kappa +1)}-\frac{821 \kappa  \ao^2}{9 (x-2)^2 (\kappa +1)}+\
\frac{569 d_1 \kappa  \ao^2}{144 (x-1)^2 (\kappa +1)}-\frac{3731 \
\kappa  \ao^2}{432 (x-1)^2 (\kappa +1)}-\frac{5344 d_1 \kappa  \
\ao^2}{27 (x-2)^3 (\kappa +1)}+\frac{7408 \kappa  \ao^2}{27 (x-2)^3 (\
\kappa +1)}-\frac{2093 d_1 \kappa  \ao^2}{144 (x-1)^3 (\kappa \
+1)}+\frac{743 \kappa  \ao^2}{24 (x-1)^3 (\kappa +1)}-\frac{5008 d_1 \
\kappa  \ao^2}{9 (x-2)^4 (\kappa +1)}+\frac{24184 \kappa  \ao^2}{27 \
(x-2)^4 (\kappa +1)}+\frac{299 d_1 \kappa  \ao^2}{24 (x-1)^4 (\kappa \
+1)}-\frac{1829 \kappa  \ao^2}{72 (x-1)^4 (\kappa +1)}+\frac{28453 \
\kappa  \ao^2}{432 (\kappa +1)}-\frac{58 d_1 \ao^2}{9 (x-2) (\kappa \
+1)}+\frac{256 \ao^2}{27 (x-2) (\kappa +1)}+\frac{427 d_1 \ao^2}{216 \
(x-1) (\kappa +1)}-\frac{61 \ao^2}{24 (x-1) (\kappa +1)}+\frac{299 \
d_1 \ao^2}{18 (x-2)^2 (\kappa +1)}-\frac{163 \ao^2}{9 (x-2)^2 (\kappa \
+1)}+\frac{881 d_1 \ao^2}{432 (x-1)^2 (\kappa +1)}-\frac{1621 \
\ao^2}{432 (x-1)^2 (\kappa +1)}-\frac{736 d_1 \ao^2}{9 (x-2)^3 \
(\kappa +1)}+\frac{1424 \ao^2}{27 (x-2)^3 (\kappa +1)}-\frac{871 d_1 \
\ao^2}{144 (x-1)^3 (\kappa +1)}+\frac{569 \ao^2}{72 (x-1)^3 (\kappa \
+1)}-\frac{2224 d_1 \ao^2}{9 (x-2)^4 (\kappa +1)}+\frac{6664 \
\ao^2}{27 (x-2)^4 (\kappa +1)}+\frac{1105 d_1 \ao^2}{216 (x-1)^4 \
(\kappa +1)}-\frac{1333 \ao^2}{216 (x-1)^4 (\kappa +1)}+\frac{6791 \
\ao^2}{432 (\kappa +1)}-\frac{53 d_1^2 \ao^2}{18 (x-2)^2}-\frac{d_1^2 \
\ao^2}{2 (x-1)^2}+\frac{\pi ^2 \ao^2}{12 (x-1)^2}+\frac{784 d_1^2 \
\ao^2}{27 (x-2)^3}+\frac{3 d_1^2 \ao^2}{2 (x-1)^3}-\frac{\pi ^2 \
\ao^2}{12 (x-1)^3}+\frac{232 d_1^2 \ao^2}{3 (x-2)^4}-\frac{20 \pi ^2 \
\ao^2}{9 (x-2)^4}-\frac{4 d_1^2 \ao^2}{3 (x-1)^4}+\frac{\pi ^2 \
\ao^2}{18 (x-1)^4}-\frac{\pi ^2 \ao^2}{6}+\frac{d_1^2 \ao}{6 (\kappa \
+1)}-\frac{263 d_1 \ao}{216 (\kappa +1)}+\frac{205 d_1^2 x \ao}{108 (\
\kappa +1)}-\frac{1775 d_1 x \ao}{216 (\kappa +1)}-\frac{73 \pi ^2 x \
\ao}{216 (\kappa +1)}+\frac{6995 x \ao}{648 (\kappa +1)}+\frac{d_1^2 \
\kappa  \ao}{6 (\kappa +1)}-\frac{49 d_1 \kappa  \ao}{24 (\kappa \
+1)}+\frac{205 d_1^2 x \kappa  \ao}{108 (\kappa +1)}-\frac{4025 d_1 x \
\kappa  \ao}{216 (\kappa +1)}-\frac{301 \pi ^2 x \kappa  \ao}{216 \
(\kappa +1)}+\frac{3121 x \kappa  \ao}{72 (\kappa +1)}-\frac{7 d_1^2 \
\kappa  \ao}{27 (x-2) (\kappa +1)}-\frac{86 d_1 \kappa  \ao}{27 (x-2) \
(\kappa +1)}-\frac{140 \pi ^2 \kappa  \ao}{27 (x-2) (\kappa \
+1)}+\frac{320 \kappa  \ao}{27 (x-2) (\kappa +1)}-\frac{7 d_1^2 \
\kappa  \ao}{108 (x-1) (\kappa +1)}+\frac{1787 d_1 \kappa  \ao}{216 \
(x-1) (\kappa +1)}+\frac{455 \pi ^2 \kappa  \ao}{72 (x-1) (\kappa \
+1)}-\frac{865 \kappa  \ao}{36 (x-1) (\kappa +1)}+\frac{53 d_1^2 \
\kappa  \ao}{27 (x-2)^2 (\kappa +1)}-\frac{239 d_1 \kappa  \ao}{27 \
(x-2)^2 (\kappa +1)}+\frac{140 \pi ^2 \kappa  \ao}{27 (x-2)^2 (\kappa \
+1)}+\frac{5 d_1^2 \kappa  \ao}{27 (x-1)^2 (\kappa +1)}-\frac{889 d_1 \
\kappa  \ao}{216 (x-1)^2 (\kappa +1)}-\frac{175 \pi ^2 \kappa  \
\ao}{108 (x-1)^2 (\kappa +1)}+\frac{1315 \kappa  \ao}{108 (x-1)^2 \
(\kappa +1)}-\frac{784 d_1^2 \kappa  \ao}{27 (x-2)^3 (\kappa \
+1)}+\frac{4168 d_1 \kappa  \ao}{27 (x-2)^3 (\kappa +1)}-\frac{3704 \
\kappa  \ao}{27 (x-2)^3 (\kappa +1)}-\frac{d_1^2 \kappa  \ao}{(x-1)^3 \
(\kappa +1)}+\frac{77 d_1 \kappa  \ao}{9 (x-1)^3 (\kappa \
+1)}+\frac{245 \pi ^2 \kappa  \ao}{216 (x-1)^3 (\kappa +1)}-\frac{361 \
\kappa  \ao}{24 (x-1)^3 (\kappa +1)}-\frac{464 d_1^2 \kappa  \ao}{3 \
(x-2)^4 (\kappa +1)}+\frac{10016 d_1 \kappa  \ao}{9 (x-2)^4 (\kappa \
+1)}+\frac{8 d_1 \pi ^2 \kappa  \ao}{3 (x-2)^4 (\kappa +1)}+\frac{704 \
\pi ^2 \kappa  \ao}{9 (x-2)^4 (\kappa +1)}-\frac{48368 \kappa  \
\ao}{27 (x-2)^4 (\kappa +1)}-\frac{4 d_1^2 \kappa  \ao}{3 (x-1)^4 \
(\kappa +1)}+\frac{299 d_1 \kappa  \ao}{24 (x-1)^4 (\kappa \
+1)}+\frac{139 \pi ^2 \kappa  \ao}{216 (x-1)^4 (\kappa \
+1)}-\frac{1829 \kappa  \ao}{72 (x-1)^4 (\kappa +1)}-\frac{512 d_1^2 \
\kappa  \ao}{3 (x-2)^5 (\kappa +1)}+\frac{11264 d_1 \kappa  \ao}{9 \
(x-2)^5 (\kappa +1)}+\frac{16 d_1 \pi ^2 \kappa  \ao}{3 (x-2)^5 \
(\kappa +1)}+\frac{32 \pi ^2 \kappa  \ao}{(x-2)^5 (\kappa \
+1)}-\frac{59008 \kappa  \ao}{27 (x-2)^5 (\kappa +1)}-\frac{253 \pi \
^2 \kappa  \ao}{216 (x-1)^5 (\kappa +1)}+\frac{89 \pi ^2 \kappa  \
\ao}{54 (\kappa +1)}+\frac{307 \kappa  \ao}{36 (\kappa +1)}-\frac{7 \
d_1^2 \ao}{27 (x-2) (\kappa +1)}-\frac{2 d_1 \ao}{3 (x-2) (\kappa \
+1)}-\frac{20 \pi ^2 \ao}{27 (x-2) (\kappa +1)}+\frac{64 \ao}{27 \
(x-2) (\kappa +1)}-\frac{7 d_1^2 \ao}{108 (x-1) (\kappa \
+1)}+\frac{613 d_1 \ao}{216 (x-1) (\kappa +1)}+\frac{65 \pi ^2 \
\ao}{72 (x-1) (\kappa +1)}-\frac{19 \ao}{4 (x-1) (\kappa \
+1)}+\frac{53 d_1^2 \ao}{27 (x-2)^2 (\kappa +1)}-\frac{133 d_1 \
\ao}{27 (x-2)^2 (\kappa +1)}+\frac{20 \pi ^2 \ao}{27 (x-2)^2 (\kappa \
+1)}+\frac{5 d_1^2 \ao}{27 (x-1)^2 (\kappa +1)}-\frac{331 d_1 \
\ao}{216 (x-1)^2 (\kappa +1)}-\frac{25 \pi ^2 \ao}{108 (x-1)^2 \
(\kappa +1)}+\frac{257 \ao}{108 (x-1)^2 (\kappa +1)}-\frac{784 d_1^2 \
\ao}{27 (x-2)^3 (\kappa +1)}+\frac{1816 d_1 \ao}{27 (x-2)^3 (\kappa \
+1)}-\frac{712 \ao}{27 (x-2)^3 (\kappa +1)}-\frac{d_1^2 \ao}{(x-1)^3 \
(\kappa +1)}+\frac{10 d_1 \ao}{3 (x-1)^3 (\kappa +1)}+\frac{35 \pi ^2 \
\ao}{216 (x-1)^3 (\kappa +1)}-\frac{67 \ao}{24 (x-1)^3 (\kappa \
+1)}-\frac{464 d_1^2 \ao}{3 (x-2)^4 (\kappa +1)}+\frac{4448 d_1 \
\ao}{9 (x-2)^4 (\kappa +1)}+\frac{8 d_1 \pi ^2 \ao}{9 (x-2)^4 (\kappa \
+1)}+\frac{496 \pi ^2 \ao}{27 (x-2)^4 (\kappa +1)}-\frac{13328 \
\ao}{27 (x-2)^4 (\kappa +1)}-\frac{4 d_1^2 \ao}{3 (x-1)^4 (\kappa \
+1)}+\frac{1105 d_1 \ao}{216 (x-1)^4 (\kappa +1)}+\frac{5 \pi ^2 \
\ao}{24 (x-1)^4 (\kappa +1)}-\frac{1333 \ao}{216 (x-1)^4 (\kappa \
+1)}-\frac{512 d_1^2 \ao}{3 (x-2)^5 (\kappa +1)}+\frac{5120 d_1 \
\ao}{9 (x-2)^5 (\kappa +1)}+\frac{16 d_1 \pi ^2 \ao}{9 (x-2)^5 \
(\kappa +1)}+\frac{512 \pi ^2 \ao}{27 (x-2)^5 (\kappa \
+1)}-\frac{19072 \ao}{27 (x-2)^5 (\kappa +1)}-\frac{17 \pi ^2 \ao}{72 \
(x-1)^5 (\kappa +1)}+\frac{\pi ^2 \ao}{27 (\kappa +1)}+\frac{3505 \
\ao}{324 (\kappa +1)}+\frac{x \zeta_3 \ao}{\kappa +1}+\frac{7 x \
\kappa  \zeta_3 \ao}{3 (\kappa +1)}-\frac{7 \kappa  \zeta_3 \ao}{3 \
(x-1)^4 (\kappa +1)}+\frac{7 \kappa  \zeta_3 \ao}{3 (x-1)^5 (\kappa \
+1)}-\frac{14 \kappa  \zeta_3 \ao}{3 (\kappa +1)}-\frac{\zeta_3 \
\ao}{3 (x-1)^4 (\kappa +1)}+\frac{\zeta_3 \ao}{3 (x-1)^5 (\kappa \
+1)}-\frac{2 \zeta_3 \ao}{\kappa +1}+\frac{1120 \kappa  \ln ^32\,  \
\ao}{9 (x-2)^4 (\kappa +1)}+\frac{2240 \kappa  \ln ^32\,  \ao}{9 \
(x-2)^5 (\kappa +1)}+\frac{160 \ln ^32\,  \ao}{9 (x-2)^4 (\kappa \
+1)}+\frac{320 \ln ^32\,  \ao}{9 (x-2)^5 (\kappa +1)}+\frac{32 d_1 \
\kappa  \ln ^22\,  \ao}{(x-2)^4 (\kappa +1)}+\frac{416 \kappa  \ln \
^22\,  \ao}{3 (x-2)^4 (\kappa +1)}+\frac{64 d_1 \kappa  \ln ^22\,  \
\ao}{(x-2)^5 (\kappa +1)}+\frac{832 \kappa  \ln ^22\,  \ao}{3 (x-2)^5 \
(\kappa +1)}+\frac{32 d_1 \ln ^22\,  \ao}{3 (x-2)^4 (\kappa \
+1)}+\frac{544 \ln ^22\,  \ao}{9 (x-2)^4 (\kappa +1)}+\frac{64 d_1 \
\ln ^22\,  \ao}{3 (x-2)^5 (\kappa +1)}+\frac{1088 \ln ^22\,  \ao}{9 \
(x-2)^5 (\kappa +1)}+\frac{256 d_1 \kappa  \ln 2\,  \ao}{9 (x-2)^4 \
(\kappa +1)}+\frac{80 \pi ^2 \kappa  \ln 2\,  \ao}{3 (x-2)^4 (\kappa \
+1)}+\frac{1856 \kappa  \ln 2\,  \ao}{27 (x-2)^4 (\kappa \
+1)}+\frac{512 d_1 \kappa  \ln 2\,  \ao}{9 (x-2)^5 (\kappa \
+1)}+\frac{160 \pi ^2 \kappa  \ln 2\,  \ao}{3 (x-2)^5 (\kappa \
+1)}+\frac{3712 \kappa  \ln 2\,  \ao}{27 (x-2)^5 (\kappa \
+1)}+\frac{256 d_1 \ln 2\,  \ao}{9 (x-2)^4 (\kappa +1)}+\frac{3392 \
\ln 2\,  \ao}{27 (x-2)^4 (\kappa +1)}+\frac{512 d_1 \ln 2\,  \ao}{9 \
(x-2)^5 (\kappa +1)}+\frac{6784 \ln 2\,  \ao}{27 (x-2)^5 (\kappa \
+1)}+\Big(\frac{1}{12} d_1 x \ao^5-\frac{4 x \ao^5}{9}+\frac{1}{12} \
d_1 x \kappa  \ao^5-\frac{11}{18} x \kappa  \ao^5+\frac{d_1 \kappa  \
\ao^5}{6 (x-2)}-\frac{25 \kappa  \ao^5}{18 (x-2)}-\frac{d_1 \kappa  \
\ao^5}{12 (x-1)}+\frac{11 \kappa  \ao^5}{18 (x-1)}-\frac{\kappa  \
\ao^5}{12}+\frac{d_1 \ao^5}{6 (x-2)}-\frac{19 \ao^5}{18 \
(x-2)}-\frac{d_1 \ao^5}{12 (x-1)}+\frac{4 \ao^5}{9 \
(x-1)}-\frac{\ao^5}{12}+\frac{7 d_1 \ao^4}{54}-\frac{61}{108} d_1 x \
\ao^4+\frac{157 x \ao^4}{54}+\frac{7}{54} d_1 \kappa  \
\ao^4-\frac{61}{108} d_1 x \kappa  \ao^4+\frac{112}{27} x \kappa  \
\ao^4-\frac{11 d_1 \kappa  \ao^4}{27 (x-2)}+\frac{113 \kappa  \
\ao^4}{27 (x-2)}+\frac{43 d_1 \kappa  \ao^4}{108 (x-1)}-\frac{149 \
\kappa  \ao^4}{54 (x-1)}+\frac{23 d_1 \kappa  \ao^4}{27 \
(x-2)^2}-\frac{149 \kappa  \ao^4}{27 (x-2)^2}-\frac{4 d_1 \kappa  \
\ao^4}{27 (x-1)^2}+\frac{59 \kappa  \ao^4}{54 (x-1)^2}-\frac{41 \
\kappa  \ao^4}{108}-\frac{11 d_1 \ao^4}{27 (x-2)}+\frac{91 \ao^4}{27 \
(x-2)}+\frac{43 d_1 \ao^4}{108 (x-1)}-\frac{53 \ao^4}{27 \
(x-1)}+\frac{23 d_1 \ao^4}{27 (x-2)^2}-\frac{103 \ao^4}{27 \
(x-2)^2}-\frac{4 d_1 \ao^4}{27 (x-1)^2}+\frac{37 \ao^4}{54 \
(x-1)^2}-\frac{\ao^4}{108}-\frac{8 d_1 \ao^3}{9}+\frac{95}{54} d_1 x \
\ao^3-\frac{911 x \ao^3}{108}-\frac{8}{9} d_1 \kappa  \
\ao^3+\frac{95}{54} d_1 x \kappa  \ao^3-\frac{1381}{108} x \kappa  \
\ao^3-\frac{d_1 \kappa  \ao^3}{27 (x-2)}-\frac{50 \kappa  \ao^3}{27 \
(x-2)}-\frac{41 d_1 \kappa  \ao^3}{54 (x-1)}+\frac{239 \kappa  \
\ao^3}{54 (x-1)}-\frac{16 d_1 \kappa  \ao^3}{27 (x-2)^2}+\frac{208 \
\kappa  \ao^3}{27 (x-2)^2}+\frac{17 d_1 \kappa  \ao^3}{27 \
(x-1)^2}-\frac{467 \kappa  \ao^3}{108 (x-1)^2}+\frac{152 d_1 \kappa  \
\ao^3}{27 (x-2)^3}-\frac{776 \kappa  \ao^3}{27 (x-2)^3}-\frac{d_1 \
\kappa  \ao^3}{3 (x-1)^3}+\frac{83 \kappa  \ao^3}{36 \
(x-1)^3}+\frac{83 \kappa  \ao^3}{18}-\frac{d_1 \ao^3}{27 \
(x-2)}-\frac{70 \ao^3}{27 (x-2)}-\frac{41 d_1 \ao^3}{54 \
(x-1)}+\frac{175 \ao^3}{54 (x-1)}-\frac{16 d_1 \ao^3}{27 \
(x-2)^2}+\frac{176 \ao^3}{27 (x-2)^2}+\frac{17 d_1 \ao^3}{27 \
(x-1)^2}-\frac{277 \ao^3}{108 (x-1)^2}+\frac{152 d_1 \ao^3}{27 \
(x-2)^3}-\frac{472 \ao^3}{27 (x-2)^3}-\frac{d_1 \ao^3}{3 \
(x-1)^3}+\frac{5 \ao^3}{4 (x-1)^3}+\frac{35 \ao^3}{18}+\frac{35 d_1 \
\ao^2}{9}-\frac{73}{18} d_1 x \ao^2+\frac{569 x \
\ao^2}{36}+\frac{35}{9} d_1 \kappa  \ao^2-\frac{73}{18} d_1 x \kappa  \
\ao^2+\frac{979}{36} x \kappa  \ao^2+\frac{8 d_1 \kappa  \ao^2}{9 \
(x-2)}-\frac{32 \kappa  \ao^2}{3 (x-2)}+\frac{13 d_1 \kappa  \
\ao^2}{18 (x-1)}+\frac{4 \kappa  \ao^2}{3 (x-1)}-\frac{10 d_1 \kappa  \
\ao^2}{3 (x-2)^2}+\frac{20 \kappa  \ao^2}{(x-2)^2}-\frac{d_1 \kappa  \
\ao^2}{(x-1)^2}+\frac{67 \kappa  \ao^2}{12 (x-1)^2}+\frac{160 d_1 \
\kappa  \ao^2}{9 (x-2)^3}-\frac{160 \kappa  \ao^2}{3 (x-2)^3}+\frac{5 \
d_1 \kappa  \ao^2}{3 (x-1)^3}-\frac{119 \kappa  \ao^2}{12 \
(x-1)^3}+\frac{208 d_1 \kappa  \ao^2}{3 (x-2)^4}-\frac{2512 \kappa  \
\ao^2}{9 (x-2)^4}-\frac{4 d_1 \kappa  \ao^2}{3 (x-1)^4}+\frac{137 \
\kappa  \ao^2}{18 (x-1)^4}-\frac{64 \kappa  \ao^2}{3}+\frac{8 d_1 \
\ao^2}{9 (x-2)}-\frac{32 \ao^2}{9 (x-2)}+\frac{13 d_1 \ao^2}{18 \
(x-1)}-\frac{\ao^2}{x-1}-\frac{10 d_1 \ao^2}{3 (x-2)^2}+\frac{20 \
\ao^2}{3 (x-2)^2}-\frac{d_1 \ao^2}{(x-1)^2}+\frac{119 \ao^2}{36 \
(x-1)^2}+\frac{160 d_1 \ao^2}{9 (x-2)^3}-\frac{160 \ao^2}{9 (x-2)^3}+\
\frac{5 d_1 \ao^2}{3 (x-1)^3}-\frac{19 \ao^2}{4 (x-1)^3}+\frac{208 \
d_1 \ao^2}{3 (x-2)^4}-\frac{1264 \ao^2}{9 (x-2)^4}-\frac{4 d_1 \
\ao^2}{3 (x-1)^4}+\frac{7 \ao^2}{2 (x-1)^4}-\frac{85 \
\ao^2}{9}-\frac{13 d_1 \ao}{27}+\frac{805 d_1 x \
\ao}{216}-\frac{1}{18} \pi ^2 x \ao-\frac{271 x \ao}{24}-\frac{13 d_1 \
\kappa  \ao}{27}+\frac{805}{216} d_1 x \kappa  \ao-\frac{5131 x \
\kappa  \ao}{216}+\frac{380 d_1 \kappa  \ao}{27 (x-2)}-\frac{1208 \
\kappa  \ao}{27 (x-2)}-\frac{379 d_1 \kappa  \ao}{24 \
(x-1)}+\frac{4009 \kappa  \ao}{72 (x-1)}-\frac{332 d_1 \kappa  \
\ao}{27 (x-2)^2}+\frac{1136 \kappa  \ao}{27 (x-2)^2}+\frac{205 d_1 \
\kappa  \ao}{108 (x-1)^2}-\frac{713 \kappa  \ao}{54 \
(x-1)^2}-\frac{160 d_1 \kappa  \ao}{9 (x-2)^3}+\frac{80 \kappa  \
\ao}{3 (x-2)^3}-\frac{395 d_1 \kappa  \ao}{216 (x-1)^3}+\frac{2399 \
\kappa  \ao}{216 (x-1)^3}-\frac{224 d_1 \kappa  \
\ao}{(x-2)^4}+\frac{928 \kappa  \ao}{(x-2)^4}-\frac{331 d_1 \kappa  \
\ao}{216 (x-1)^4}+\frac{475 \kappa  \ao}{54 (x-1)^4}-\frac{512 d_1 \
\kappa  \ao}{3 (x-2)^5}+\frac{6656 \kappa  \ao}{9 (x-2)^5}+\frac{205 \
d_1 \kappa  \ao}{216 (x-1)^5}-\frac{1255 \kappa  \ao}{216 \
(x-1)^5}+\frac{203 \kappa  \ao}{216}+\frac{380 d_1 \ao}{27 \
(x-2)}-\frac{616 \ao}{27 (x-2)}-\frac{379 d_1 \ao}{24 \
(x-1)}+\frac{671 \ao}{24 (x-1)}-\frac{332 d_1 \ao}{27 \
(x-2)^2}+\frac{592 \ao}{27 (x-2)^2}+\frac{205 d_1 \ao}{108 \
(x-1)^2}-\frac{361 \ao}{54 (x-1)^2}-\frac{160 d_1 \ao}{9 \
(x-2)^3}+\frac{80 \ao}{9 (x-2)^3}-\frac{395 d_1 \ao}{216 \
(x-1)^3}+\frac{1123 \ao}{216 (x-1)^3}-\frac{1760 d_1 \ao}{9 (x-2)^4}-\
\frac{40 \pi ^2 \ao}{9 (x-2)^4}+\frac{17120 \ao}{27 \
(x-2)^4}-\frac{331 d_1 \ao}{216 (x-1)^4}-\frac{\pi ^2 \ao}{18 \
(x-1)^4}+\frac{152 \ao}{27 (x-1)^4}-\frac{1024 d_1 \ao}{9 \
(x-2)^5}-\frac{80 \pi ^2 \ao}{9 (x-2)^5}+\frac{19072 \ao}{27 \
(x-2)^5}+\frac{205 d_1 \ao}{216 (x-1)^5}+\frac{\pi ^2 \ao}{18 \
(x-1)^5}-\frac{955 \ao}{216 (x-1)^5}+\frac{\pi ^2 \ao}{9}-\frac{85 \
\ao}{24}-\frac{3 d_1}{4}-\frac{205 d_1 x}{216}+\frac{\pi ^2 \
x}{18}+\frac{35 x}{24}-\frac{3 d_1 \kappa }{4}-\frac{205 d_1 x \kappa \
}{216}+\frac{1255 x \kappa }{216}+\frac{346 d_1 \kappa }{27 \
(x-2)}-\frac{952 \kappa }{27 (x-2)}-\frac{947 d_1 \kappa }{72 (x-1)}+\
\frac{2621 \kappa }{72 (x-1)}-\frac{392 d_1 \kappa }{27 \
(x-2)^2}+\frac{1136 \kappa }{27 (x-2)^2}+\frac{23 d_1 \kappa }{27 \
(x-1)^2}-\frac{92 \kappa }{27 (x-1)^2}+\frac{784 d_1 \kappa }{27 \
(x-2)^3}-\frac{2272 \kappa }{27 (x-2)^3}+\frac{73 d_1 \kappa }{216 \
(x-1)^3}-\frac{247 \kappa }{216 (x-1)^3}+\frac{256 d_1 \kappa }{3 \
(x-2)^4}-\frac{3328 \kappa }{9 (x-2)^4}+\frac{367 d_1 \kappa }{216 \
(x-1)^4}-\frac{1127 \kappa }{108 (x-1)^4}+\frac{512 d_1 \kappa }{3 \
(x-2)^5}-\frac{2048 \kappa }{3 (x-2)^5}+\frac{205 d_1 \kappa }{216 \
(x-1)^5}-\frac{1255 \kappa }{216 (x-1)^5}+\frac{1024 \kappa }{9 \
(x-2)^6}+\frac{37 \kappa }{8}+\frac{346 d_1}{27 (x-2)}-\frac{488}{27 \
(x-2)}-\frac{947 d_1}{72 (x-1)}+\frac{443}{24 (x-1)}-\frac{392 \
d_1}{27 (x-2)^2}+\frac{592}{27 (x-2)^2}+\frac{23 d_1}{27 \
(x-1)^2}-\frac{52}{27 (x-1)^2}+\frac{784 d_1}{27 \
(x-2)^3}-\frac{1184}{27 (x-2)^3}+\frac{73 d_1}{216 \
(x-1)^3}-\frac{83}{216 (x-1)^3}+\frac{512 d_1}{9 (x-2)^4}+\frac{40 \
\pi ^2}{9 (x-2)^4}-\frac{9536}{27 (x-2)^4}+\frac{367 d_1}{216 \
(x-1)^4}+\frac{\pi ^2}{18 (x-1)^4}-\frac{725}{108 (x-1)^4}+\frac{512 \
d_1}{9 (x-2)^5}+\frac{160 \pi ^2}{9 (x-2)^5}-\frac{25856}{27 \
(x-2)^5}+\frac{205 d_1}{216 (x-1)^5}+\frac{\pi ^2}{18 \
(x-1)^5}-\frac{955}{216 (x-1)^5}-\frac{1024 d_1}{9 (x-2)^6}+\frac{160 \
\pi ^2}{9 (x-2)^6}-\frac{13568}{27 (x-2)^6}+\frac{55}{24}\Big) \
H(0;\ao)+\Big(-\frac{d_1 \ao^5}{12}+\frac{1}{12} d_1^2 x \
\ao^5-\frac{4}{9} d_1 x \ao^5-\frac{1}{12} d_1 x \kappa  \
\ao^5-\frac{d_1 \kappa  \ao^5}{6 (x-2)}+\frac{d_1 \kappa  \ao^5}{12 \
(x-1)}+\frac{d_1^2 \ao^5}{6 (x-2)}-\frac{19 d_1 \ao^5}{18 \
(x-2)}-\frac{d_1^2 \ao^5}{12 (x-1)}+\frac{4 d_1 \ao^5}{9 \
(x-1)}+\frac{7 d_1^2 \ao^4}{54}-\frac{d_1 \ao^4}{108}-\frac{61}{108} \
d_1^2 x \ao^4+\frac{157}{54} d_1 x \ao^4-\frac{5}{27} d_1 \kappa  \
\ao^4+\frac{67}{108} d_1 x \kappa  \ao^4+\frac{11 d_1 \kappa  \
\ao^4}{27 (x-2)}-\frac{43 d_1 \kappa  \ao^4}{108 (x-1)}-\frac{23 d_1 \
\kappa  \ao^4}{27 (x-2)^2}+\frac{11 d_1 \kappa  \ao^4}{54 \
(x-1)^2}-\frac{11 d_1^2 \ao^4}{27 (x-2)}+\frac{91 d_1 \ao^4}{27 \
(x-2)}+\frac{43 d_1^2 \ao^4}{108 (x-1)}-\frac{53 d_1 \ao^4}{27 \
(x-1)}+\frac{23 d_1^2 \ao^4}{27 (x-2)^2}-\frac{103 d_1 \ao^4}{27 \
(x-2)^2}-\frac{4 d_1^2 \ao^4}{27 (x-1)^2}+\frac{37 d_1 \ao^4}{54 \
(x-1)^2}-\frac{8 d_1^2 \ao^3}{9}+\frac{35 d_1 \
\ao^3}{18}+\frac{95}{54} d_1^2 x \ao^3-\frac{911}{108} d_1 x \
\ao^3+\frac{4}{3} d_1 \kappa  \ao^3-\frac{235}{108} d_1 x \kappa  \
\ao^3+\frac{10 d_1 \kappa  \ao^3}{27 (x-2)}+\frac{16 d_1 \kappa  \
\ao^3}{27 (x-1)}+\frac{16 d_1 \kappa  \ao^3}{27 (x-2)^2}-\frac{95 d_1 \
\kappa  \ao^3}{108 (x-1)^2}-\frac{152 d_1 \kappa  \ao^3}{27 (x-2)^3}+\
\frac{19 d_1 \kappa  \ao^3}{36 (x-1)^3}-\frac{d_1^2 \ao^3}{27 (x-2)}-\
\frac{70 d_1 \ao^3}{27 (x-2)}-\frac{41 d_1^2 \ao^3}{54 \
(x-1)}+\frac{175 d_1 \ao^3}{54 (x-1)}-\frac{16 d_1^2 \ao^3}{27 \
(x-2)^2}+\frac{176 d_1 \ao^3}{27 (x-2)^2}+\frac{17 d_1^2 \ao^3}{27 \
(x-1)^2}-\frac{277 d_1 \ao^3}{108 (x-1)^2}+\frac{152 d_1^2 \ao^3}{27 \
(x-2)^3}-\frac{472 d_1 \ao^3}{27 (x-2)^3}-\frac{d_1^2 \ao^3}{3 \
(x-1)^3}+\frac{5 d_1 \ao^3}{4 (x-1)^3}+\frac{35 d_1^2 \
\ao^2}{9}-\frac{85 d_1 \ao^2}{9}-\frac{73}{18} d_1^2 x \
\ao^2+\frac{569}{36} d_1 x \ao^2-\frac{107}{18} d_1 \kappa  \
\ao^2+\frac{205}{36} d_1 x \kappa  \ao^2-\frac{32 d_1 \kappa  \
\ao^2}{9 (x-2)}+\frac{7 d_1 \kappa  \ao^2}{6 (x-1)}+\frac{20 d_1 \
\kappa  \ao^2}{3 (x-2)^2}+\frac{41 d_1 \kappa  \ao^2}{36 \
(x-1)^2}-\frac{160 d_1 \kappa  \ao^2}{9 (x-2)^3}-\frac{31 d_1 \kappa  \
\ao^2}{12 (x-1)^3}-\frac{208 d_1 \kappa  \ao^2}{3 (x-2)^4}+\frac{37 \
d_1 \kappa  \ao^2}{18 (x-1)^4}+\frac{8 d_1^2 \ao^2}{9 (x-2)}-\frac{32 \
d_1 \ao^2}{9 (x-2)}+\frac{13 d_1^2 \ao^2}{18 (x-1)}-\frac{d_1 \
\ao^2}{x-1}-\frac{10 d_1^2 \ao^2}{3 (x-2)^2}+\frac{20 d_1 \ao^2}{3 \
(x-2)^2}-\frac{d_1^2 \ao^2}{(x-1)^2}+\frac{119 d_1 \ao^2}{36 \
(x-1)^2}+\frac{160 d_1^2 \ao^2}{9 (x-2)^3}-\frac{160 d_1 \ao^2}{9 \
(x-2)^3}+\frac{5 d_1^2 \ao^2}{3 (x-1)^3}-\frac{19 d_1 \ao^2}{4 \
(x-1)^3}+\frac{208 d_1^2 \ao^2}{3 (x-2)^4}-\frac{1264 d_1 \ao^2}{9 \
(x-2)^4}-\frac{4 d_1^2 \ao^2}{3 (x-1)^4}+\frac{7 d_1 \ao^2}{2 \
(x-1)^4}-\frac{80 d_1^2 \ao}{27}+\frac{703 d_1 \
\ao}{108}+\frac{505}{108} d_1^2 x \ao-\frac{1697 d_1 x \
\ao}{108}+\frac{122 d_1 \kappa  \ao}{27}-\frac{187}{27} d_1 x \kappa  \
\ao+\frac{125 d_1 \kappa  \ao}{54 (x-2)}-\frac{5 d_1 \kappa  \ao}{108 \
(x-1)}-\frac{196 d_1 \kappa  \ao}{27 (x-2)^2}-\frac{133 d_1 \kappa  \
\ao}{108 (x-1)^2}+\frac{392 d_1 \kappa  \ao}{9 (x-2)^3}+\frac{43 d_1 \
\kappa  \ao}{12 (x-1)^3}+\frac{224 d_1 \kappa  \
\ao}{(x-2)^4}+\frac{512 d_1 \kappa  \ao}{3 (x-2)^5}-\frac{47 d_1^2 \
\ao}{54 (x-2)}+\frac{205 d_1 \ao}{54 (x-2)}-\frac{37 d_1^2 \ao}{108 \
(x-1)}+\frac{43 d_1 \ao}{54 (x-1)}+\frac{136 d_1^2 \ao}{27 \
(x-2)^2}-\frac{356 d_1 \ao}{27 (x-2)^2}+\frac{19 d_1^2 \ao}{27 \
(x-1)^2}-\frac{263 d_1 \ao}{108 (x-1)^2}-\frac{472 d_1^2 \ao}{9 \
(x-2)^3}+\frac{712 d_1 \ao}{9 (x-2)^3}-\frac{7 d_1^2 \ao}{3 (x-1)^3}+\
\frac{23 d_1 \ao}{4 (x-1)^3}-\frac{224 d_1^2 \ao}{(x-2)^4}+\frac{4576 \
d_1 \ao}{9 (x-2)^4}-\frac{512 d_1^2 \ao}{3 (x-2)^5}+\frac{4096 d_1 \
\ao}{9 (x-2)^5}-\frac{d_1^2}{6}+\frac{13 d_1}{12}-\frac{205 d_1^2 \
x}{108}+\frac{635 d_1 x}{108}+\frac{5 d_1 \kappa }{18}+\frac{155 d_1 \
x \kappa }{54}+\frac{17 d_1 \kappa }{27 (x-2)}-\frac{151 d_1 \kappa \
}{108 (x-1)}+\frac{23 d_1 \kappa }{27 (x-2)^2}+\frac{83 d_1 \kappa \
}{108 (x-1)^2}-\frac{544 d_1 \kappa }{27 (x-2)^3}-\frac{55 d_1 \kappa \
}{36 (x-1)^3}-\frac{464 d_1 \kappa }{3 (x-2)^4}-\frac{37 d_1 \kappa \
}{18 (x-1)^4}-\frac{512 d_1 \kappa }{3 (x-2)^5}+\frac{7 d_1^2}{27 \
(x-2)}+\frac{d_1}{27 (x-2)}+\frac{7 d_1^2}{108 (x-1)}-\frac{41 \
d_1}{27 (x-1)}-\frac{53 d_1^2}{27 (x-2)^2}+\frac{103 d_1}{27 \
(x-2)^2}-\frac{5 d_1^2}{27 (x-1)^2}+\frac{109 d_1}{108 \
(x-1)^2}+\frac{784 d_1^2}{27 (x-2)^3}-\frac{1184 d_1}{27 \
(x-2)^3}+\frac{d_1^2}{(x-1)^3}-\frac{9 d_1}{4 (x-1)^3}+\frac{464 \
d_1^2}{3 (x-2)^4}-\frac{368 d_1}{(x-2)^4}+\frac{4 d_1^2}{3 \
(x-1)^4}-\frac{7 d_1}{2 (x-1)^4}+\frac{512 d_1^2}{3 \
(x-2)^5}-\frac{4096 d_1}{9 (x-2)^5}\Big) H(1;\ao)+\Big(\frac{x \
\ao^5}{3}+x \kappa  \ao^5+\frac{2 \kappa  \ao^5}{x-2}-\frac{\kappa  \
\ao^5}{x-1}+\frac{2 \ao^5}{3 (x-2)}-\frac{\ao^5}{3 (x-1)}-\frac{19 x \
\ao^4}{9}-\frac{19}{3} x \kappa  \ao^4-\frac{20 \kappa  \ao^4}{3 \
(x-2)}+\frac{13 \kappa  \ao^4}{3 (x-1)}+\frac{20 \kappa  \ao^4}{3 \
(x-2)^2}-\frac{4 \kappa  \ao^4}{3 (x-1)^2}+\frac{2 \kappa  \ao^4}{3}-\
\frac{20 \ao^4}{9 (x-2)}+\frac{13 \ao^4}{9 (x-1)}+\frac{20 \ao^4}{9 \
(x-2)^2}-\frac{4 \ao^4}{9 (x-1)^2}+\frac{2 \ao^4}{9}+\frac{52 x \
\ao^3}{9}+\frac{52}{3} x \kappa  \ao^3+\frac{20 \kappa  \ao^3}{3 \
(x-2)}-\frac{22 \kappa  \ao^3}{3 (x-1)}-\frac{40 \kappa  \ao^3}{3 \
(x-2)^2}+\frac{14 \kappa  \ao^3}{3 (x-1)^2}+\frac{80 \kappa  \ao^3}{3 \
(x-2)^3}-\frac{2 \kappa  \ao^3}{(x-1)^3}-4 \kappa  \ao^3+\frac{20 \
\ao^3}{9 (x-2)}-\frac{22 \ao^3}{9 (x-1)}-\frac{40 \ao^3}{9 \
(x-2)^2}+\frac{14 \ao^3}{9 (x-1)^2}+\frac{80 \ao^3}{9 \
(x-2)^3}-\frac{2 \ao^3}{3 (x-1)^3}-\frac{4 \ao^3}{3}-\frac{28 x \
\ao^2}{3}-28 x \kappa  \ao^2+\frac{6 \kappa  \ao^2}{x-1}-\frac{6 \
\kappa  \ao^2}{(x-1)^2}+\frac{6 \kappa  \ao^2}{(x-1)^3}+\frac{160 \
\kappa  \ao^2}{(x-2)^4}-\frac{4 \kappa  \ao^2}{(x-1)^4}+12 \kappa  \
\ao^2+\frac{2 \ao^2}{x-1}-\frac{2 \ao^2}{(x-1)^2}+\frac{2 \
\ao^2}{(x-1)^3}+\frac{160 \ao^2}{3 (x-2)^4}-\frac{4 \ao^2}{3 \
(x-1)^4}+4 \ao^2+\frac{95 x \ao}{18}+\frac{337 x \kappa  \
\ao}{18}+\frac{80 \kappa  \ao}{3 (x-2)}-\frac{65 \kappa  \ao}{2 \
(x-1)}-\frac{80 \kappa  \ao}{3 (x-2)^2}+\frac{25 \kappa  \ao}{3 \
(x-1)^2}-\frac{35 \kappa  \ao}{6 (x-1)^3}-\frac{64 d_1 \kappa  \ao}{3 \
(x-2)^4}-\frac{6464 \kappa  \ao}{9 (x-2)^4}-\frac{119 \kappa  \ao}{18 \
(x-1)^4}-\frac{128 d_1 \kappa  \ao}{3 (x-2)^5}-\frac{7168 \kappa  \
\ao}{9 (x-2)^5}+\frac{101 \kappa  \ao}{18 (x-1)^5}+\frac{50 \kappa  \
\ao}{9}+\frac{80 \ao}{9 (x-2)}-\frac{65 \ao}{6 (x-1)}-\frac{80 \ao}{9 \
(x-2)^2}+\frac{25 \ao}{9 (x-1)^2}-\frac{35 \ao}{18 (x-1)^3}-\frac{64 \
d_1 \ao}{3 (x-2)^4}-\frac{3008 \ao}{9 (x-2)^4}-\frac{19 \ao}{6 \
(x-1)^4}-\frac{128 d_1 \ao}{3 (x-2)^5}-\frac{4096 \ao}{9 \
(x-2)^5}+\frac{17 \ao}{6 (x-1)^5}+\frac{34 \
\ao}{9}+\frac{x}{18}-\frac{49 x \kappa }{18}+\frac{64 \kappa }{3 \
(x-2)}-\frac{43 \kappa }{2 (x-1)}-\frac{80 \kappa }{3 \
(x-2)^2}+\frac{8 \kappa }{3 (x-1)^2}+\frac{160 \kappa }{3 \
(x-2)^3}+\frac{\kappa }{6 (x-1)^3}+\frac{64 d_1 \kappa }{3 \
(x-2)^4}+\frac{3584 \kappa }{9 (x-2)^4}+\frac{155 \kappa }{18 \
(x-1)^4}+\frac{256 d_1 \kappa }{3 (x-2)^5}+\frac{8576 \kappa }{9 \
(x-2)^5}+\frac{101 \kappa }{18 (x-1)^5}+\frac{256 d_1 \kappa }{3 \
(x-2)^6}+\frac{2816 \kappa }{9 (x-2)^6}-3 \kappa +\frac{64}{9 (x-2)}-\
\frac{43}{6 (x-1)}-\frac{80}{9 (x-2)^2}+\frac{8}{9 \
(x-1)^2}+\frac{160}{9 (x-2)^3}+\frac{1}{18 (x-1)^3}+\frac{64 d_1}{3 \
(x-2)^4}+\frac{2048}{9 (x-2)^4}+\frac{23}{6 (x-1)^4}+\frac{256 d_1}{3 \
(x-2)^5}+\frac{6272}{9 (x-2)^5}+\frac{17}{6 (x-1)^5}+\frac{256 d_1}{3 \
(x-2)^6}+\frac{4352}{9 (x-2)^6}-1\Big) H(0,0;\ao)+\Big(\frac{1}{3} \
d_1 x \ao^5+\frac{1}{3} d_1 x \kappa  \ao^5+\frac{2 d_1 \kappa  \
\ao^5}{3 (x-2)}-\frac{d_1 \kappa  \ao^5}{3 (x-1)}+\frac{2 d_1 \
\ao^5}{3 (x-2)}-\frac{d_1 \ao^5}{3 (x-1)}+\frac{2 d_1 \
\ao^4}{9}-\frac{19}{9} d_1 x \ao^4+\frac{2}{9} d_1 \kappa  \
\ao^4-\frac{19}{9} d_1 x \kappa  \ao^4-\frac{20 d_1 \kappa  \ao^4}{9 \
(x-2)}+\frac{13 d_1 \kappa  \ao^4}{9 (x-1)}+\frac{20 d_1 \kappa  \
\ao^4}{9 (x-2)^2}-\frac{4 d_1 \kappa  \ao^4}{9 (x-1)^2}-\frac{20 d_1 \
\ao^4}{9 (x-2)}+\frac{13 d_1 \ao^4}{9 (x-1)}+\frac{20 d_1 \ao^4}{9 \
(x-2)^2}-\frac{4 d_1 \ao^4}{9 (x-1)^2}-\frac{4 d_1 \
\ao^3}{3}+\frac{52}{9} d_1 x \ao^3-\frac{4}{3} d_1 \kappa  \
\ao^3+\frac{52}{9} d_1 x \kappa  \ao^3+\frac{20 d_1 \kappa  \ao^3}{9 \
(x-2)}-\frac{22 d_1 \kappa  \ao^3}{9 (x-1)}-\frac{40 d_1 \kappa  \
\ao^3}{9 (x-2)^2}+\frac{14 d_1 \kappa  \ao^3}{9 (x-1)^2}+\frac{80 d_1 \
\kappa  \ao^3}{9 (x-2)^3}-\frac{2 d_1 \kappa  \ao^3}{3 \
(x-1)^3}+\frac{20 d_1 \ao^3}{9 (x-2)}-\frac{22 d_1 \ao^3}{9 \
(x-1)}-\frac{40 d_1 \ao^3}{9 (x-2)^2}+\frac{14 d_1 \ao^3}{9 (x-1)^2}+\
\frac{80 d_1 \ao^3}{9 (x-2)^3}-\frac{2 d_1 \ao^3}{3 (x-1)^3}+4 d_1 \
\ao^2-\frac{28}{3} d_1 x \ao^2+4 d_1 \kappa  \ao^2-\frac{28}{3} d_1 x \
\kappa  \ao^2+\frac{2 d_1 \kappa  \ao^2}{x-1}-\frac{2 d_1 \kappa  \
\ao^2}{(x-1)^2}+\frac{2 d_1 \kappa  \ao^2}{(x-1)^3}+\frac{160 d_1 \
\kappa  \ao^2}{3 (x-2)^4}-\frac{4 d_1 \kappa  \ao^2}{3 \
(x-1)^4}+\frac{2 d_1 \ao^2}{x-1}-\frac{2 d_1 \ao^2}{(x-1)^2}+\frac{2 \
d_1 \ao^2}{(x-1)^3}+\frac{160 d_1 \ao^2}{3 (x-2)^4}-\frac{4 d_1 \
\ao^2}{3 (x-1)^4}+\frac{34 d_1 \ao}{9}+\frac{95 d_1 x \
\ao}{18}+\frac{8 d_1 \kappa  \ao}{9}+\frac{121}{18} d_1 x \kappa  \
\ao+\frac{80 d_1 \kappa  \ao}{9 (x-2)}-\frac{65 d_1 \kappa  \ao}{6 \
(x-1)}-\frac{80 d_1 \kappa  \ao}{9 (x-2)^2}+\frac{25 d_1 \kappa  \
\ao}{9 (x-1)^2}-\frac{35 d_1 \kappa  \ao}{18 (x-1)^3}-\frac{192 d_1 \
\kappa  \ao}{(x-2)^4}-\frac{31 d_1 \kappa  \ao}{18 (x-1)^4}-\frac{512 \
d_1 \kappa  \ao}{3 (x-2)^5}+\frac{25 d_1 \kappa  \ao}{18 \
(x-1)^5}+\frac{80 d_1 \ao}{9 (x-2)}-\frac{65 d_1 \ao}{6 \
(x-1)}-\frac{80 d_1 \ao}{9 (x-2)^2}+\frac{25 d_1 \ao}{9 \
(x-1)^2}-\frac{35 d_1 \ao}{18 (x-1)^3}-\frac{64 d_1^2 \ao}{3 \
(x-2)^4}-\frac{3008 d_1 \ao}{9 (x-2)^4}-\frac{19 d_1 \ao}{6 (x-1)^4}-\
\frac{128 d_1^2 \ao}{3 (x-2)^5}-\frac{4096 d_1 \ao}{9 \
(x-2)^5}+\frac{17 d_1 \ao}{6 (x-1)^5}-d_1+\frac{d_1 x}{18}-d_1 \kappa \
-\frac{25 d_1 x \kappa }{18}+\frac{64 d_1 \kappa }{9 (x-2)}-\frac{43 \
d_1 \kappa }{6 (x-1)}-\frac{80 d_1 \kappa }{9 (x-2)^2}+\frac{8 d_1 \
\kappa }{9 (x-1)^2}+\frac{160 d_1 \kappa }{9 (x-2)^3}+\frac{d_1 \
\kappa }{18 (x-1)^3}+\frac{256 d_1 \kappa }{3 (x-2)^4}+\frac{43 d_1 \
\kappa }{18 (x-1)^4}+\frac{128 d_1 \kappa }{(x-2)^5}+\frac{25 d_1 \
\kappa }{18 (x-1)^5}-\frac{256 d_1 \kappa }{3 (x-2)^6}+\frac{64 \
d_1}{9 (x-2)}-\frac{43 d_1}{6 (x-1)}-\frac{80 d_1}{9 (x-2)^2}+\frac{8 \
d_1}{9 (x-1)^2}+\frac{160 d_1}{9 (x-2)^3}+\frac{d_1}{18 \
(x-1)^3}+\frac{64 d_1^2}{3 (x-2)^4}+\frac{2048 d_1}{9 \
(x-2)^4}+\frac{23 d_1}{6 (x-1)^4}+\frac{256 d_1^2}{3 \
(x-2)^5}+\frac{6272 d_1}{9 (x-2)^5}+\frac{17 d_1}{6 \
(x-1)^5}+\frac{256 d_1^2}{3 (x-2)^6}+\frac{4352 d_1}{9 (x-2)^6}\Big) \
H(0,1;\ao)+H(1;x) \Big(\frac{1}{18} \pi ^2 x \ao+\frac{1}{6} \pi ^2 x \
\kappa  \ao+\frac{40 \pi ^2 \kappa  \ao}{3 (x-2)^4}+\frac{d_1 \pi ^2 \
\kappa  \ao}{9 (x-1)^4}-\frac{\pi ^2 \kappa  \ao}{6 (x-1)^4}+\frac{80 \
\pi ^2 \kappa  \ao}{3 (x-2)^5}-\frac{d_1 \pi ^2 \kappa  \ao}{9 \
(x-1)^5}+\frac{\pi ^2 \kappa  \ao}{6 (x-1)^5}-\frac{1}{3} \pi ^2 \
\kappa  \ao+\frac{40 \pi ^2 \ao}{9 (x-2)^4}+\frac{d_1 \pi ^2 \ao}{9 \
(x-1)^4}-\frac{\pi ^2 \ao}{18 (x-1)^4}+\frac{80 \pi ^2 \ao}{9 \
(x-2)^5}-\frac{d_1 \pi ^2 \ao}{9 (x-1)^5}+\frac{\pi ^2 \ao}{18 \
(x-1)^5}-\frac{\pi ^2 \ao}{9}-\frac{\pi ^2 x}{18}-\frac{1}{6} \pi ^2 \
x \kappa -\frac{40 \pi ^2 \kappa }{3 (x-2)^4}-\frac{d_1 \pi ^2 \kappa \
}{9 (x-1)^4}+\frac{\pi ^2 \kappa }{6 (x-1)^4}-\frac{160 \pi ^2 \kappa \
}{3 (x-2)^5}-\frac{d_1 \pi ^2 \kappa }{9 (x-1)^5}+\frac{\pi ^2 \kappa \
}{6 (x-1)^5}-\frac{160 \pi ^2 \kappa }{3 (x-2)^6}+\Big(-\frac{2 d_1 \
\ao}{3}+\frac{17 x \ao}{12}-\frac{2 d_1 \kappa  \ao}{3}+\frac{101 x \
\kappa  \ao}{36}-\frac{85 d_1 \kappa  \ao}{9 (x-2)}+\frac{40 \kappa  \
\ao}{3 (x-2)}+\frac{94 d_1 \kappa  \ao}{9 (x-1)}-\frac{65 \kappa  \
\ao}{4 (x-1)}+\frac{100 d_1 \kappa  \ao}{9 (x-2)^2}-\frac{40 \kappa  \
\ao}{3 (x-2)^2}-\frac{20 d_1 \kappa  \ao}{9 (x-1)^2}+\frac{25 \kappa  \
\ao}{6 (x-1)^2}-\frac{40 d_1 \kappa  \ao}{3 (x-2)^3}+\frac{17 d_1 \
\kappa  \ao}{18 (x-1)^3}-\frac{35 \kappa  \ao}{12 (x-1)^3}+\frac{64 \
d_1 \kappa  \ao}{(x-2)^4}-\frac{1088 \kappa  \ao}{9 (x-2)^4}+\frac{19 \
d_1 \kappa  \ao}{18 (x-1)^4}-\frac{47 \kappa  \ao}{36 \
(x-1)^4}+\frac{64 d_1 \kappa  \ao}{3 (x-2)^5}+\frac{704 \kappa  \
\ao}{9 (x-2)^5}-\frac{25 d_1 \kappa  \ao}{18 (x-1)^5}+\frac{101 \
\kappa  \ao}{36 (x-1)^5}-\frac{37 \kappa  \ao}{9}-\frac{85 d_1 \ao}{9 \
(x-2)}+\frac{40 \ao}{9 (x-2)}+\frac{94 d_1 \ao}{9 (x-1)}-\frac{65 \
\ao}{12 (x-1)}+\frac{100 d_1 \ao}{9 (x-2)^2}-\frac{40 \ao}{9 \
(x-2)^2}-\frac{20 d_1 \ao}{9 (x-1)^2}+\frac{25 \ao}{18 \
(x-1)^2}-\frac{40 d_1 \ao}{3 (x-2)^3}+\frac{17 d_1 \ao}{18 \
(x-1)^3}-\frac{35 \ao}{36 (x-1)^3}+\frac{64 d_1 \
\ao}{(x-2)^4}+\frac{64 \ao}{9 (x-2)^4}+\frac{5 d_1 \ao}{2 \
(x-1)^4}-\frac{11 \ao}{12 (x-1)^4}+\frac{64 d_1 \ao}{3 \
(x-2)^5}+\frac{1088 \ao}{9 (x-2)^5}-\frac{17 d_1 \ao}{6 \
(x-1)^5}+\frac{17 \ao}{12 (x-1)^5}-\frac{7 \ao}{3}+\frac{2 \
d_1}{3}-\frac{17 x}{12}+\frac{2 d_1 \kappa }{3}-\frac{101 x \kappa \
}{36}-\frac{62 d_1 \kappa }{9 (x-2)}+\frac{32 \kappa }{3 \
(x-2)}+\frac{65 d_1 \kappa }{9 (x-1)}-\frac{43 \kappa }{4 \
(x-1)}+\frac{70 d_1 \kappa }{9 (x-2)^2}-\frac{40 \kappa }{3 (x-2)^2}-\
\frac{d_1 \kappa }{(x-1)^2}+\frac{4 \kappa }{3 (x-1)^2}-\frac{80 d_1 \
\kappa }{9 (x-2)^3}+\frac{80 \kappa }{3 (x-2)^3}+\frac{5 d_1 \kappa \
}{18 (x-1)^3}+\frac{\kappa }{12 (x-1)^3}-\frac{112 d_1 \kappa }{3 \
(x-2)^4}+\frac{1088 \kappa }{9 (x-2)^4}-\frac{31 d_1 \kappa }{18 \
(x-1)^4}+\frac{155 \kappa }{36 (x-1)^4}-\frac{448 d_1 \kappa }{3 \
(x-2)^5}+\frac{1472 \kappa }{9 (x-2)^5}-\frac{25 d_1 \kappa }{18 \
(x-1)^5}+\frac{101 \kappa }{36 (x-1)^5}-\frac{128 d_1 \kappa }{3 \
(x-2)^6}-\frac{1408 \kappa }{9 (x-2)^6}-\frac{3 \kappa }{2}-\frac{62 \
d_1}{9 (x-2)}+\frac{32}{9 (x-2)}+\frac{65 d_1}{9 (x-1)}-\frac{43}{12 \
(x-1)}+\frac{70 d_1}{9 (x-2)^2}-\frac{40}{9 \
(x-2)^2}-\frac{d_1}{(x-1)^2}+\frac{4}{9 (x-1)^2}-\frac{80 d_1}{9 \
(x-2)^3}+\frac{80}{9 (x-2)^3}+\frac{5 d_1}{18 (x-1)^3}+\frac{1}{36 \
(x-1)^3}-\frac{112 d_1}{3 (x-2)^4}-\frac{64}{9 (x-2)^4}-\frac{19 \
d_1}{6 (x-1)^4}+\frac{23}{12 (x-1)^4}-\frac{448 d_1}{3 \
(x-2)^5}-\frac{1216}{9 (x-2)^5}-\frac{17 d_1}{6 (x-1)^5}+\frac{17}{12 \
(x-1)^5}-\frac{128 d_1}{3 (x-2)^6}-\frac{2176}{9 (x-2)^6}-\frac{1}{2}\
\Big) H(0;\ao)+\Big(-\frac{2 x \ao}{3}-2 x \kappa  \ao-\frac{160 \
\kappa  \ao}{(x-2)^4}-\frac{4 d_1 \kappa  \ao}{3 (x-1)^4}+\frac{2 \
\kappa  \ao}{(x-1)^4}-\frac{320 \kappa  \ao}{(x-2)^5}+\frac{4 d_1 \
\kappa  \ao}{3 (x-1)^5}-\frac{2 \kappa  \ao}{(x-1)^5}+4 \kappa  \
\ao-\frac{160 \ao}{3 (x-2)^4}-\frac{4 d_1 \ao}{3 (x-1)^4}+\frac{2 \
\ao}{3 (x-1)^4}-\frac{320 \ao}{3 (x-2)^5}+\frac{4 d_1 \ao}{3 \
(x-1)^5}-\frac{2 \ao}{3 (x-1)^5}+\frac{4 \ao}{3}+\frac{2 x}{3}+2 x \
\kappa +\frac{160 \kappa }{(x-2)^4}+\frac{4 d_1 \kappa }{3 \
(x-1)^4}-\frac{2 \kappa }{(x-1)^4}+\frac{640 \kappa \
}{(x-2)^5}+\frac{4 d_1 \kappa }{3 (x-1)^5}-\frac{2 \kappa }{(x-1)^5}+\
\frac{640 \kappa }{(x-2)^6}+\frac{160}{3 (x-2)^4}+\frac{4 d_1}{3 \
(x-1)^4}-\frac{2}{3 (x-1)^4}+\frac{640}{3 (x-2)^5}+\frac{4 d_1}{3 \
(x-1)^5}-\frac{2}{3 (x-1)^5}+\frac{640}{3 (x-2)^6}\Big) \
H(0,0;\ao)+\Big(-\frac{4 \ao d_1^2}{3 (x-1)^4}+\frac{4 d_1^2}{3 \
(x-1)^4}+\frac{4 \ao d_1^2}{3 (x-1)^5}+\frac{4 d_1^2}{3 \
(x-1)^5}+\frac{4 \ao d_1}{3}-\frac{2 \ao x d_1}{3}+\frac{2 x d_1}{3}+\
\frac{4 \ao \kappa  d_1}{3}-\frac{2}{3} \ao x \kappa  d_1+\frac{2 x \
\kappa  d_1}{3}-\frac{160 \ao \kappa  d_1}{3 (x-2)^4}+\frac{160 \
\kappa  d_1}{3 (x-2)^4}+\frac{2 \ao \kappa  d_1}{3 (x-1)^4}-\frac{2 \
\kappa  d_1}{3 (x-1)^4}-\frac{320 \ao \kappa  d_1}{3 \
(x-2)^5}+\frac{640 \kappa  d_1}{3 (x-2)^5}-\frac{2 \ao \kappa  d_1}{3 \
(x-1)^5}-\frac{2 \kappa  d_1}{3 (x-1)^5}+\frac{640 \kappa  d_1}{3 \
(x-2)^6}-\frac{160 \ao d_1}{3 (x-2)^4}+\frac{160 d_1}{3 \
(x-2)^4}+\frac{2 \ao d_1}{3 (x-1)^4}-\frac{2 d_1}{3 \
(x-1)^4}-\frac{320 \ao d_1}{3 (x-2)^5}+\frac{640 d_1}{3 \
(x-2)^5}-\frac{2 \ao d_1}{3 (x-1)^5}-\frac{2 d_1}{3 \
(x-1)^5}+\frac{640 d_1}{3 (x-2)^6}\Big) H(0,1;\ao)-\frac{40 \pi ^2}{9 \
(x-2)^4}-\frac{d_1 \pi ^2}{9 (x-1)^4}+\frac{\pi ^2}{18 \
(x-1)^4}-\frac{160 \pi ^2}{9 (x-2)^5}-\frac{d_1 \pi ^2}{9 \
(x-1)^5}+\frac{\pi ^2}{18 (x-1)^5}-\frac{160 \pi ^2}{9 (x-2)^6}\Big)+\
\Big(-\frac{4 d_1 \ao}{3}+\frac{2 d_1 x \ao}{3}-\frac{2 x \
\ao}{3}-\frac{4 d_1 \kappa  \ao}{3}+\frac{2}{3} d_1 x \kappa  \ao-2 x \
\kappa  \ao+\frac{160 d_1 \kappa  \ao}{3 (x-2)^4}-\frac{160 \kappa  \
\ao}{(x-2)^4}-\frac{2 d_1 \kappa  \ao}{3 (x-1)^4}+\frac{2 \kappa  \
\ao}{(x-1)^4}+\frac{320 d_1 \kappa  \ao}{3 (x-2)^5}-\frac{320 \kappa  \
\ao}{(x-2)^5}+\frac{2 d_1 \kappa  \ao}{3 (x-1)^5}-\frac{2 \kappa  \
\ao}{(x-1)^5}+4 \kappa  \ao+\frac{160 d_1 \ao}{3 (x-2)^4}-\frac{160 \
\ao}{3 (x-2)^4}-\frac{2 d_1 \ao}{3 (x-1)^4}+\frac{2 \ao}{3 \
(x-1)^4}+\frac{320 d_1 \ao}{3 (x-2)^5}-\frac{320 \ao}{3 \
(x-2)^5}+\frac{2 d_1 \ao}{3 (x-1)^5}-\frac{2 \ao}{3 (x-1)^5}+\frac{4 \
\ao}{3}-\frac{2 d_1 x}{3}+\frac{2 x}{3}-\frac{2 d_1 x \kappa }{3}+2 x \
\kappa -\frac{160 d_1 \kappa }{3 (x-2)^4}+\frac{160 \kappa \
}{(x-2)^4}+\frac{2 d_1 \kappa }{3 (x-1)^4}-\frac{2 \kappa }{(x-1)^4}-\
\frac{640 d_1 \kappa }{3 (x-2)^5}+\frac{640 \kappa }{(x-2)^5}+\frac{2 \
d_1 \kappa }{3 (x-1)^5}-\frac{2 \kappa }{(x-1)^5}-\frac{640 d_1 \
\kappa }{3 (x-2)^6}+\frac{640 \kappa }{(x-2)^6}-\frac{160 d_1}{3 \
(x-2)^4}+\frac{160}{3 (x-2)^4}+\frac{2 d_1}{3 (x-1)^4}-\frac{2}{3 \
(x-1)^4}-\frac{640 d_1}{3 (x-2)^5}+\frac{640}{3 (x-2)^5}+\frac{2 \
d_1}{3 (x-1)^5}-\frac{2}{3 (x-1)^5}-\frac{640 d_1}{3 \
(x-2)^6}+\frac{640}{3 (x-2)^6}\Big) H(0;\ao) H(0,1;x)+\Big(\frac{17 x \
\ao}{12}+\frac{101 x \kappa  \ao}{36}+\frac{40 \kappa  \ao}{3 (x-2)}-\
\frac{65 \kappa  \ao}{4 (x-1)}-\frac{40 \kappa  \ao}{3 \
(x-2)^2}+\frac{25 \kappa  \ao}{6 (x-1)^2}-\frac{35 \kappa  \ao}{12 \
(x-1)^3}+\frac{32 d_1 \kappa  \ao}{3 (x-2)^4}-\frac{1088 \kappa  \
\ao}{9 (x-2)^4}-\frac{47 \kappa  \ao}{36 (x-1)^4}+\frac{64 d_1 \kappa \
 \ao}{3 (x-2)^5}+\frac{704 \kappa  \ao}{9 (x-2)^5}+\frac{101 \kappa  \
\ao}{36 (x-1)^5}-\frac{37 \kappa  \ao}{9}+\frac{40 \ao}{9 \
(x-2)}-\frac{65 \ao}{12 (x-1)}-\frac{40 \ao}{9 (x-2)^2}+\frac{25 \
\ao}{18 (x-1)^2}-\frac{35 \ao}{36 (x-1)^3}+\frac{32 d_1 \ao}{3 \
(x-2)^4}+\frac{64 \ao}{9 (x-2)^4}-\frac{11 \ao}{12 (x-1)^4}+\frac{64 \
d_1 \ao}{3 (x-2)^5}+\frac{1088 \ao}{9 (x-2)^5}+\frac{17 \ao}{12 \
(x-1)^5}-\frac{7 \ao}{3}-\frac{17 x}{12}-\frac{101 x \kappa \
}{36}+\frac{32 \kappa }{3 (x-2)}-\frac{43 \kappa }{4 (x-1)}-\frac{40 \
\kappa }{3 (x-2)^2}+\frac{4 \kappa }{3 (x-1)^2}+\frac{80 \kappa }{3 \
(x-2)^3}+\frac{\kappa }{12 (x-1)^3}-\frac{32 d_1 \kappa }{3 (x-2)^4}+\
\frac{1088 \kappa }{9 (x-2)^4}+\frac{155 \kappa }{36 \
(x-1)^4}-\frac{128 d_1 \kappa }{3 (x-2)^5}+\frac{1472 \kappa }{9 \
(x-2)^5}+\frac{101 \kappa }{36 (x-1)^5}-\frac{128 d_1 \kappa }{3 \
(x-2)^6}-\frac{1408 \kappa }{9 (x-2)^6}-\frac{3 \kappa \
}{2}+\Big(-\frac{2 x \ao}{3}-2 x \kappa  \ao-\frac{160 \kappa  \
\ao}{(x-2)^4}+\frac{2 \kappa  \ao}{(x-1)^4}-\frac{320 \kappa  \
\ao}{(x-2)^5}-\frac{2 \kappa  \ao}{(x-1)^5}+4 \kappa  \ao-\frac{160 \
\ao}{3 (x-2)^4}+\frac{2 \ao}{3 (x-1)^4}-\frac{320 \ao}{3 \
(x-2)^5}-\frac{2 \ao}{3 (x-1)^5}+\frac{4 \ao}{3}+\frac{2 x}{3}+2 x \
\kappa +\frac{160 \kappa }{(x-2)^4}-\frac{2 \kappa \
}{(x-1)^4}+\frac{640 \kappa }{(x-2)^5}-\frac{2 \kappa \
}{(x-1)^5}+\frac{640 \kappa }{(x-2)^6}+\frac{160}{3 \
(x-2)^4}-\frac{2}{3 (x-1)^4}+\frac{640}{3 (x-2)^5}-\frac{2}{3 \
(x-1)^5}+\frac{640}{3 (x-2)^6}\Big) H(0;\ao)+\Big(\frac{4 \ao \
d_1}{3}-\frac{2 \ao x d_1}{3}+\frac{2 x d_1}{3}+\frac{4 \ao \kappa  \
d_1}{3}-\frac{2}{3} \ao x \kappa  d_1+\frac{2 x \kappa  \
d_1}{3}-\frac{160 \ao \kappa  d_1}{3 (x-2)^4}+\frac{160 \kappa  \
d_1}{3 (x-2)^4}+\frac{2 \ao \kappa  d_1}{3 (x-1)^4}-\frac{2 \kappa  \
d_1}{3 (x-1)^4}-\frac{320 \ao \kappa  d_1}{3 (x-2)^5}+\frac{640 \
\kappa  d_1}{3 (x-2)^5}-\frac{2 \ao \kappa  d_1}{3 (x-1)^5}-\frac{2 \
\kappa  d_1}{3 (x-1)^5}+\frac{640 \kappa  d_1}{3 (x-2)^6}-\frac{160 \
\ao d_1}{3 (x-2)^4}+\frac{160 d_1}{3 (x-2)^4}+\frac{2 \ao d_1}{3 \
(x-1)^4}-\frac{2 d_1}{3 (x-1)^4}-\frac{320 \ao d_1}{3 \
(x-2)^5}+\frac{640 d_1}{3 (x-2)^5}-\frac{2 \ao d_1}{3 \
(x-1)^5}-\frac{2 d_1}{3 (x-1)^5}+\frac{640 d_1}{3 (x-2)^6}\Big) \
H(1;\ao)+\frac{32}{9 (x-2)}-\frac{43}{12 (x-1)}-\frac{40}{9 (x-2)^2}+\
\frac{4}{9 (x-1)^2}+\frac{80}{9 (x-2)^3}+\frac{1}{36 \
(x-1)^3}-\frac{32 d_1}{3 (x-2)^4}-\frac{64}{9 (x-2)^4}+\frac{23}{12 \
(x-1)^4}-\frac{128 d_1}{3 (x-2)^5}-\frac{1216}{9 \
(x-2)^5}+\frac{17}{12 (x-1)^5}-\frac{128 d_1}{3 \
(x-2)^6}-\frac{2176}{9 (x-2)^6}-\frac{1}{2}\Big) \
H(0,c_1(\ao);x)+\Big(-\frac{64 d_1 \kappa  \ao}{3 (x-2)^4}-\frac{704 \
\kappa  \ao}{9 (x-2)^4}-\frac{128 d_1 \kappa  \ao}{3 \
(x-2)^5}-\frac{1408 \kappa  \ao}{9 (x-2)^5}-\frac{64 d_1 \ao}{3 \
(x-2)^4}-\frac{1088 \ao}{9 (x-2)^4}-\frac{128 d_1 \ao}{3 \
(x-2)^5}-\frac{2176 \ao}{9 (x-2)^5}+\frac{64 d_1 \kappa }{3 (x-2)^4}+\
\frac{704 \kappa }{9 (x-2)^4}+\frac{256 d_1 \kappa }{3 \
(x-2)^5}+\frac{2816 \kappa }{9 (x-2)^5}+\frac{256 d_1 \kappa }{3 \
(x-2)^6}+\frac{2816 \kappa }{9 (x-2)^6}+\Big(\frac{320 \kappa  \
\ao}{(x-2)^4}+\frac{640 \kappa  \ao}{(x-2)^5}+\frac{320 \ao}{3 \
(x-2)^4}+\frac{640 \ao}{3 (x-2)^5}-\frac{320 \kappa \
}{(x-2)^4}-\frac{1280 \kappa }{(x-2)^5}-\frac{1280 \kappa }{(x-2)^6}-\
\frac{320}{3 (x-2)^4}-\frac{1280}{3 (x-2)^5}-\frac{1280}{3 \
(x-2)^6}\Big) H(0;\ao)+\Big(\frac{320 \ao \kappa  d_1}{3 \
(x-2)^4}-\frac{320 \kappa  d_1}{3 (x-2)^4}+\frac{640 \ao \kappa  \
d_1}{3 (x-2)^5}-\frac{1280 \kappa  d_1}{3 (x-2)^5}-\frac{1280 \kappa  \
d_1}{3 (x-2)^6}+\frac{320 \ao d_1}{3 (x-2)^4}-\frac{320 d_1}{3 \
(x-2)^4}+\frac{640 \ao d_1}{3 (x-2)^5}-\frac{1280 d_1}{3 \
(x-2)^5}-\frac{1280 d_1}{3 (x-2)^6}\Big) H(1;\ao)+\frac{64 d_1}{3 \
(x-2)^4}+\frac{1088}{9 (x-2)^4}+\frac{256 d_1}{3 \
(x-2)^5}+\frac{4352}{9 (x-2)^5}+\frac{256 d_1}{3 \
(x-2)^6}+\frac{4352}{9 (x-2)^6}\Big) H(0,c_2(\ao);x)+\Big(\frac{1}{3} \
d_1 x \ao^5+\frac{1}{3} d_1 x \kappa  \ao^5+\frac{2 d_1 \kappa  \
\ao^5}{3 (x-2)}-\frac{d_1 \kappa  \ao^5}{3 (x-1)}+\frac{2 d_1 \
\ao^5}{3 (x-2)}-\frac{d_1 \ao^5}{3 (x-1)}+\frac{2 d_1 \
\ao^4}{9}-\frac{19}{9} d_1 x \ao^4+\frac{2}{9} d_1 \kappa  \
\ao^4-\frac{19}{9} d_1 x \kappa  \ao^4-\frac{20 d_1 \kappa  \ao^4}{9 \
(x-2)}+\frac{13 d_1 \kappa  \ao^4}{9 (x-1)}+\frac{20 d_1 \kappa  \
\ao^4}{9 (x-2)^2}-\frac{4 d_1 \kappa  \ao^4}{9 (x-1)^2}-\frac{20 d_1 \
\ao^4}{9 (x-2)}+\frac{13 d_1 \ao^4}{9 (x-1)}+\frac{20 d_1 \ao^4}{9 \
(x-2)^2}-\frac{4 d_1 \ao^4}{9 (x-1)^2}-\frac{4 d_1 \
\ao^3}{3}+\frac{52}{9} d_1 x \ao^3-\frac{4}{3} d_1 \kappa  \
\ao^3+\frac{52}{9} d_1 x \kappa  \ao^3+\frac{20 d_1 \kappa  \ao^3}{9 \
(x-2)}-\frac{22 d_1 \kappa  \ao^3}{9 (x-1)}-\frac{40 d_1 \kappa  \
\ao^3}{9 (x-2)^2}+\frac{14 d_1 \kappa  \ao^3}{9 (x-1)^2}+\frac{80 d_1 \
\kappa  \ao^3}{9 (x-2)^3}-\frac{2 d_1 \kappa  \ao^3}{3 \
(x-1)^3}+\frac{20 d_1 \ao^3}{9 (x-2)}-\frac{22 d_1 \ao^3}{9 \
(x-1)}-\frac{40 d_1 \ao^3}{9 (x-2)^2}+\frac{14 d_1 \ao^3}{9 (x-1)^2}+\
\frac{80 d_1 \ao^3}{9 (x-2)^3}-\frac{2 d_1 \ao^3}{3 (x-1)^3}+4 d_1 \
\ao^2-\frac{28}{3} d_1 x \ao^2+4 d_1 \kappa  \ao^2-\frac{28}{3} d_1 x \
\kappa  \ao^2+\frac{2 d_1 \kappa  \ao^2}{x-1}-\frac{2 d_1 \kappa  \
\ao^2}{(x-1)^2}+\frac{2 d_1 \kappa  \ao^2}{(x-1)^3}+\frac{160 d_1 \
\kappa  \ao^2}{3 (x-2)^4}-\frac{4 d_1 \kappa  \ao^2}{3 \
(x-1)^4}+\frac{2 d_1 \ao^2}{x-1}-\frac{2 d_1 \ao^2}{(x-1)^2}+\frac{2 \
d_1 \ao^2}{(x-1)^3}+\frac{160 d_1 \ao^2}{3 (x-2)^4}-\frac{4 d_1 \
\ao^2}{3 (x-1)^4}-\frac{20 d_1 \ao}{9}+\frac{73 d_1 x \
\ao}{9}-\frac{20 d_1 \kappa  \ao}{9}+\frac{73}{9} d_1 x \kappa  \
\ao-\frac{10 d_1 \kappa  \ao}{9 (x-2)}-\frac{7 d_1 \kappa  \ao}{9 \
(x-1)}+\frac{40 d_1 \kappa  \ao}{9 (x-2)^2}+\frac{10 d_1 \kappa  \
\ao}{9 (x-1)^2}-\frac{80 d_1 \kappa  \ao}{3 (x-2)^3}-\frac{2 d_1 \
\kappa  \ao}{(x-1)^3}-\frac{640 d_1 \kappa  \ao}{3 (x-2)^4}-\frac{640 \
d_1 \kappa  \ao}{3 (x-2)^5}-\frac{10 d_1 \ao}{9 (x-2)}-\frac{7 d_1 \
\ao}{9 (x-1)}+\frac{40 d_1 \ao}{9 (x-2)^2}+\frac{10 d_1 \ao}{9 \
(x-1)^2}-\frac{80 d_1 \ao}{3 (x-2)^3}-\frac{2 d_1 \
\ao}{(x-1)^3}-\frac{640 d_1 \ao}{3 (x-2)^4}-\frac{640 d_1 \ao}{3 \
(x-2)^5}-\frac{2 d_1}{3}-\frac{25 d_1 x}{9}-\frac{2 d_1 \kappa \
}{3}-\frac{25 d_1 x \kappa }{9}+\frac{4 d_1 \kappa }{9 \
(x-2)}+\frac{d_1 \kappa }{9 (x-1)}-\frac{20 d_1 \kappa }{9 \
(x-2)^2}-\frac{2 d_1 \kappa }{9 (x-1)^2}+\frac{160 d_1 \kappa }{9 \
(x-2)^3}+\frac{2 d_1 \kappa }{3 (x-1)^3}+\frac{160 d_1 \kappa \
}{(x-2)^4}+\frac{4 d_1 \kappa }{3 (x-1)^4}+\frac{640 d_1 \kappa }{3 \
(x-2)^5}+\frac{4 d_1}{9 (x-2)}+\frac{d_1}{9 (x-1)}-\frac{20 d_1}{9 \
(x-2)^2}-\frac{2 d_1}{9 (x-1)^2}+\frac{160 d_1}{9 (x-2)^3}+\frac{2 \
d_1}{3 (x-1)^3}+\frac{160 d_1}{(x-2)^4}+\frac{4 d_1}{3 \
(x-1)^4}+\frac{640 d_1}{3 (x-2)^5}\Big) H(1,0;\ao)+\Big(\frac{2 d_1 \
\ao}{3}-\frac{17 x \ao}{12}+\frac{2 d_1 \kappa  \ao}{3}-\frac{101 x \
\kappa  \ao}{36}+\frac{85 d_1 \kappa  \ao}{9 (x-2)}-\frac{40 \kappa  \
\ao}{3 (x-2)}-\frac{94 d_1 \kappa  \ao}{9 (x-1)}+\frac{65 \kappa  \
\ao}{4 (x-1)}-\frac{100 d_1 \kappa  \ao}{9 (x-2)^2}+\frac{40 \kappa  \
\ao}{3 (x-2)^2}+\frac{20 d_1 \kappa  \ao}{9 (x-1)^2}-\frac{25 \kappa  \
\ao}{6 (x-1)^2}+\frac{40 d_1 \kappa  \ao}{3 (x-2)^3}-\frac{17 d_1 \
\kappa  \ao}{18 (x-1)^3}+\frac{35 \kappa  \ao}{12 (x-1)^3}-\frac{64 \
d_1 \kappa  \ao}{(x-2)^4}+\frac{1088 \kappa  \ao}{9 (x-2)^4}-\frac{19 \
d_1 \kappa  \ao}{18 (x-1)^4}+\frac{47 \kappa  \ao}{36 \
(x-1)^4}-\frac{64 d_1 \kappa  \ao}{3 (x-2)^5}-\frac{704 \kappa  \
\ao}{9 (x-2)^5}+\frac{25 d_1 \kappa  \ao}{18 (x-1)^5}-\frac{101 \
\kappa  \ao}{36 (x-1)^5}+\frac{37 \kappa  \ao}{9}+\frac{85 d_1 \ao}{9 \
(x-2)}-\frac{40 \ao}{9 (x-2)}-\frac{94 d_1 \ao}{9 (x-1)}+\frac{65 \
\ao}{12 (x-1)}-\frac{100 d_1 \ao}{9 (x-2)^2}+\frac{40 \ao}{9 \
(x-2)^2}+\frac{20 d_1 \ao}{9 (x-1)^2}-\frac{25 \ao}{18 \
(x-1)^2}+\frac{40 d_1 \ao}{3 (x-2)^3}-\frac{17 d_1 \ao}{18 \
(x-1)^3}+\frac{35 \ao}{36 (x-1)^3}-\frac{64 d_1 \
\ao}{(x-2)^4}-\frac{64 \ao}{9 (x-2)^4}-\frac{5 d_1 \ao}{2 \
(x-1)^4}+\frac{11 \ao}{12 (x-1)^4}-\frac{64 d_1 \ao}{3 \
(x-2)^5}-\frac{1088 \ao}{9 (x-2)^5}+\frac{17 d_1 \ao}{6 \
(x-1)^5}-\frac{17 \ao}{12 (x-1)^5}+\frac{7 \ao}{3}-\frac{2 \
d_1}{3}+\frac{17 x}{12}-\frac{2 d_1 \kappa }{3}+\frac{101 x \kappa \
}{36}+\frac{62 d_1 \kappa }{9 (x-2)}-\frac{32 \kappa }{3 \
(x-2)}-\frac{65 d_1 \kappa }{9 (x-1)}+\frac{43 \kappa }{4 \
(x-1)}-\frac{70 d_1 \kappa }{9 (x-2)^2}+\frac{40 \kappa }{3 (x-2)^2}+\
\frac{d_1 \kappa }{(x-1)^2}-\frac{4 \kappa }{3 (x-1)^2}+\frac{80 d_1 \
\kappa }{9 (x-2)^3}-\frac{80 \kappa }{3 (x-2)^3}-\frac{5 d_1 \kappa \
}{18 (x-1)^3}-\frac{\kappa }{12 (x-1)^3}+\frac{112 d_1 \kappa }{3 \
(x-2)^4}-\frac{1088 \kappa }{9 (x-2)^4}+\frac{31 d_1 \kappa }{18 \
(x-1)^4}-\frac{155 \kappa }{36 (x-1)^4}+\frac{448 d_1 \kappa }{3 \
(x-2)^5}-\frac{1472 \kappa }{9 (x-2)^5}+\frac{25 d_1 \kappa }{18 \
(x-1)^5}-\frac{101 \kappa }{36 (x-1)^5}+\frac{128 d_1 \kappa }{3 \
(x-2)^6}+\frac{1408 \kappa }{9 (x-2)^6}+\frac{3 \kappa }{2}+\frac{62 \
d_1}{9 (x-2)}-\frac{32}{9 (x-2)}-\frac{65 d_1}{9 (x-1)}+\frac{43}{12 \
(x-1)}-\frac{70 d_1}{9 (x-2)^2}+\frac{40}{9 \
(x-2)^2}+\frac{d_1}{(x-1)^2}-\frac{4}{9 (x-1)^2}+\frac{80 d_1}{9 \
(x-2)^3}-\frac{80}{9 (x-2)^3}-\frac{5 d_1}{18 (x-1)^3}-\frac{1}{36 \
(x-1)^3}+\frac{112 d_1}{3 (x-2)^4}+\frac{64}{9 (x-2)^4}+\frac{19 \
d_1}{6 (x-1)^4}-\frac{23}{12 (x-1)^4}+\frac{448 d_1}{3 \
(x-2)^5}+\frac{1216}{9 (x-2)^5}+\frac{17 d_1}{6 (x-1)^5}-\frac{17}{12 \
(x-1)^5}+\frac{128 d_1}{3 (x-2)^6}+\frac{2176}{9 (x-2)^6}+\frac{1}{2}\
\Big) H(1,0;x)+\Big(\frac{1}{3} d_1^2 x \ao^5+\frac{2 d_1^2 \ao^5}{3 \
(x-2)}-\frac{d_1^2 \ao^5}{3 (x-1)}+\frac{2 d_1^2 \
\ao^4}{9}-\frac{19}{9} d_1^2 x \ao^4-\frac{20 d_1^2 \ao^4}{9 \
(x-2)}+\frac{13 d_1^2 \ao^4}{9 (x-1)}+\frac{20 d_1^2 \ao^4}{9 \
(x-2)^2}-\frac{4 d_1^2 \ao^4}{9 (x-1)^2}-\frac{4 d_1^2 \
\ao^3}{3}+\frac{52}{9} d_1^2 x \ao^3+\frac{20 d_1^2 \ao^3}{9 \
(x-2)}-\frac{22 d_1^2 \ao^3}{9 (x-1)}-\frac{40 d_1^2 \ao^3}{9 \
(x-2)^2}+\frac{14 d_1^2 \ao^3}{9 (x-1)^2}+\frac{80 d_1^2 \ao^3}{9 \
(x-2)^3}-\frac{2 d_1^2 \ao^3}{3 (x-1)^3}+4 d_1^2 \ao^2-\frac{28}{3} \
d_1^2 x \ao^2+\frac{2 d_1^2 \ao^2}{x-1}-\frac{2 d_1^2 \
\ao^2}{(x-1)^2}+\frac{2 d_1^2 \ao^2}{(x-1)^3}+\frac{160 d_1^2 \
\ao^2}{3 (x-2)^4}-\frac{4 d_1^2 \ao^2}{3 (x-1)^4}-\frac{20 d_1^2 \
\ao}{9}+\frac{73}{9} d_1^2 x \ao-\frac{10 d_1^2 \ao}{9 (x-2)}-\frac{7 \
d_1^2 \ao}{9 (x-1)}+\frac{40 d_1^2 \ao}{9 (x-2)^2}+\frac{10 d_1^2 \
\ao}{9 (x-1)^2}-\frac{80 d_1^2 \ao}{3 (x-2)^3}-\frac{2 d_1^2 \
\ao}{(x-1)^3}-\frac{640 d_1^2 \ao}{3 (x-2)^4}-\frac{640 d_1^2 \ao}{3 \
(x-2)^5}-\frac{2 d_1^2}{3}-\frac{25 d_1^2 x}{9}+\frac{4 d_1^2}{9 \
(x-2)}+\frac{d_1^2}{9 (x-1)}-\frac{20 d_1^2}{9 (x-2)^2}-\frac{2 \
d_1^2}{9 (x-1)^2}+\frac{160 d_1^2}{9 (x-2)^3}+\frac{2 d_1^2}{3 \
(x-1)^3}+\frac{160 d_1^2}{(x-2)^4}+\frac{4 d_1^2}{3 \
(x-1)^4}+\frac{640 d_1^2}{3 (x-2)^5}\Big) H(1,1;\ao)+H(c_2(\ao);x) \
\Big(-\frac{256 \kappa  \ao}{9 (x-2)^4}-\frac{512 \kappa  \ao}{9 \
(x-2)^5}+\frac{256 d_1 \ao}{9 (x-2)^4}-\frac{40 \pi ^2 \ao}{9 \
(x-2)^4}+\frac{3392 \ao}{27 (x-2)^4}+\frac{512 d_1 \ao}{9 \
(x-2)^5}-\frac{80 \pi ^2 \ao}{9 (x-2)^5}+\frac{6784 \ao}{27 (x-2)^5}+\
\frac{256 \kappa }{9 (x-2)^4}+\frac{1024 \kappa }{9 \
(x-2)^5}+\frac{1024 \kappa }{9 (x-2)^6}+\Big(-\frac{64 d_1 \kappa  \
\ao}{3 (x-2)^4}-\frac{704 \kappa  \ao}{9 (x-2)^4}-\frac{128 d_1 \
\kappa  \ao}{3 (x-2)^5}-\frac{1408 \kappa  \ao}{9 (x-2)^5}-\frac{64 \
d_1 \ao}{3 (x-2)^4}-\frac{1088 \ao}{9 (x-2)^4}-\frac{128 d_1 \ao}{3 \
(x-2)^5}-\frac{2176 \ao}{9 (x-2)^5}+\frac{64 d_1 \kappa }{3 (x-2)^4}+\
\frac{704 \kappa }{9 (x-2)^4}+\frac{256 d_1 \kappa }{3 \
(x-2)^5}+\frac{2816 \kappa }{9 (x-2)^5}+\frac{256 d_1 \kappa }{3 \
(x-2)^6}+\frac{2816 \kappa }{9 (x-2)^6}+\frac{64 d_1}{3 \
(x-2)^4}+\frac{1088}{9 (x-2)^4}+\frac{256 d_1}{3 \
(x-2)^5}+\frac{4352}{9 (x-2)^5}+\frac{256 d_1}{3 \
(x-2)^6}+\frac{4352}{9 (x-2)^6}\Big) H(0;\ao)+\Big(-\frac{64 \ao \
d_1^2}{3 (x-2)^4}+\frac{64 d_1^2}{3 (x-2)^4}-\frac{128 \ao d_1^2}{3 \
(x-2)^5}+\frac{256 d_1^2}{3 (x-2)^5}+\frac{256 d_1^2}{3 \
(x-2)^6}+\frac{64 \ao \kappa  d_1}{3 (x-2)^4}-\frac{64 \kappa  d_1}{3 \
(x-2)^4}+\frac{128 \ao \kappa  d_1}{3 (x-2)^5}-\frac{256 \kappa  \
d_1}{3 (x-2)^5}-\frac{256 \kappa  d_1}{3 (x-2)^6}-\frac{1088 \ao \
d_1}{9 (x-2)^4}+\frac{1088 d_1}{9 (x-2)^4}-\frac{2176 \ao d_1}{9 \
(x-2)^5}+\frac{4352 d_1}{9 (x-2)^5}+\frac{4352 d_1}{9 (x-2)^6}\Big) \
H(1;\ao)+\Big(\frac{320 \kappa  \ao}{(x-2)^4}+\frac{640 \kappa  \
\ao}{(x-2)^5}+\frac{320 \ao}{3 (x-2)^4}+\frac{640 \ao}{3 \
(x-2)^5}-\frac{320 \kappa }{(x-2)^4}-\frac{1280 \kappa \
}{(x-2)^5}-\frac{1280 \kappa }{(x-2)^6}-\frac{320}{3 \
(x-2)^4}-\frac{1280}{3 (x-2)^5}-\frac{1280}{3 (x-2)^6}\Big) \
H(0,0;\ao)+\Big(\frac{320 \ao \kappa  d_1}{3 (x-2)^4}-\frac{320 \
\kappa  d_1}{3 (x-2)^4}+\frac{640 \ao \kappa  d_1}{3 \
(x-2)^5}-\frac{1280 \kappa  d_1}{3 (x-2)^5}-\frac{1280 \kappa  d_1}{3 \
(x-2)^6}+\frac{320 \ao d_1}{3 (x-2)^4}-\frac{320 d_1}{3 \
(x-2)^4}+\frac{640 \ao d_1}{3 (x-2)^5}-\frac{1280 d_1}{3 \
(x-2)^5}-\frac{1280 d_1}{3 (x-2)^6}\Big) H(0,1;\ao)+\Big(\frac{320 \
\ao \kappa  d_1}{3 (x-2)^4}-\frac{320 \kappa  d_1}{3 \
(x-2)^4}+\frac{640 \ao \kappa  d_1}{3 (x-2)^5}-\frac{1280 \kappa  \
d_1}{3 (x-2)^5}-\frac{1280 \kappa  d_1}{3 (x-2)^6}+\frac{320 \ao \
d_1}{3 (x-2)^4}-\frac{320 d_1}{3 (x-2)^4}+\frac{640 \ao d_1}{3 \
(x-2)^5}-\frac{1280 d_1}{3 (x-2)^5}-\frac{1280 d_1}{3 (x-2)^6}\Big) \
H(1,0;\ao)+\Big(\frac{320 \ao d_1^2}{3 (x-2)^4}-\frac{320 d_1^2}{3 \
(x-2)^4}+\frac{640 \ao d_1^2}{3 (x-2)^5}-\frac{1280 d_1^2}{3 \
(x-2)^5}-\frac{1280 d_1^2}{3 (x-2)^6}\Big) H(1,1;\ao)-\frac{256 \
d_1}{9 (x-2)^4}+\frac{40 \pi ^2}{9 (x-2)^4}-\frac{3392}{27 \
(x-2)^4}-\frac{1024 d_1}{9 (x-2)^5}+\frac{160 \pi ^2}{9 \
(x-2)^5}-\frac{13568}{27 (x-2)^5}-\frac{1024 d_1}{9 \
(x-2)^6}+\frac{160 \pi ^2}{9 (x-2)^6}-\frac{13568}{27 \
(x-2)^6}\Big)+H(c_1(\ao);x) \Big(\frac{1}{24} d_1 x \ao^5-\frac{2 x \
\ao^5}{9}+\frac{1}{24} d_1 x \kappa  \ao^5-\frac{11}{36} x \kappa  \
\ao^5+\frac{d_1 \kappa  \ao^5}{12 (x-2)}-\frac{25 \kappa  \ao^5}{36 \
(x-2)}-\frac{d_1 \kappa  \ao^5}{24 (x-1)}+\frac{11 \kappa  \ao^5}{36 \
(x-1)}-\frac{\kappa  \ao^5}{24}+\frac{d_1 \ao^5}{12 (x-2)}-\frac{19 \
\ao^5}{36 (x-2)}-\frac{d_1 \ao^5}{24 (x-1)}+\frac{2 \ao^5}{9 \
(x-1)}-\frac{\ao^5}{24}+\frac{7 d_1 \ao^4}{108}-\frac{61}{216} d_1 x \
\ao^4+\frac{157 x \ao^4}{108}+\frac{7}{108} d_1 \kappa  \
\ao^4-\frac{61}{216} d_1 x \kappa  \ao^4+\frac{56}{27} x \kappa  \
\ao^4-\frac{11 d_1 \kappa  \ao^4}{54 (x-2)}+\frac{113 \kappa  \
\ao^4}{54 (x-2)}+\frac{43 d_1 \kappa  \ao^4}{216 (x-1)}-\frac{149 \
\kappa  \ao^4}{108 (x-1)}+\frac{23 d_1 \kappa  \ao^4}{54 \
(x-2)^2}-\frac{149 \kappa  \ao^4}{54 (x-2)^2}-\frac{2 d_1 \kappa  \
\ao^4}{27 (x-1)^2}+\frac{59 \kappa  \ao^4}{108 (x-1)^2}-\frac{41 \
\kappa  \ao^4}{216}-\frac{11 d_1 \ao^4}{54 (x-2)}+\frac{91 \ao^4}{54 \
(x-2)}+\frac{43 d_1 \ao^4}{216 (x-1)}-\frac{53 \ao^4}{54 \
(x-1)}+\frac{23 d_1 \ao^4}{54 (x-2)^2}-\frac{103 \ao^4}{54 \
(x-2)^2}-\frac{2 d_1 \ao^4}{27 (x-1)^2}+\frac{37 \ao^4}{108 (x-1)^2}-\
\frac{\ao^4}{216}-\frac{4 d_1 \ao^3}{9}+\frac{95}{108} d_1 x \
\ao^3-\frac{911 x \ao^3}{216}-\frac{4}{9} d_1 \kappa  \
\ao^3+\frac{95}{108} d_1 x \kappa  \ao^3-\frac{1381}{216} x \kappa  \
\ao^3-\frac{d_1 \kappa  \ao^3}{54 (x-2)}-\frac{25 \kappa  \ao^3}{27 \
(x-2)}-\frac{41 d_1 \kappa  \ao^3}{108 (x-1)}+\frac{239 \kappa  \
\ao^3}{108 (x-1)}-\frac{8 d_1 \kappa  \ao^3}{27 (x-2)^2}+\frac{104 \
\kappa  \ao^3}{27 (x-2)^2}+\frac{17 d_1 \kappa  \ao^3}{54 \
(x-1)^2}-\frac{467 \kappa  \ao^3}{216 (x-1)^2}+\frac{76 d_1 \kappa  \
\ao^3}{27 (x-2)^3}-\frac{388 \kappa  \ao^3}{27 (x-2)^3}-\frac{d_1 \
\kappa  \ao^3}{6 (x-1)^3}+\frac{83 \kappa  \ao^3}{72 \
(x-1)^3}+\frac{83 \kappa  \ao^3}{36}-\frac{d_1 \ao^3}{54 \
(x-2)}-\frac{35 \ao^3}{27 (x-2)}-\frac{41 d_1 \ao^3}{108 \
(x-1)}+\frac{175 \ao^3}{108 (x-1)}-\frac{8 d_1 \ao^3}{27 \
(x-2)^2}+\frac{88 \ao^3}{27 (x-2)^2}+\frac{17 d_1 \ao^3}{54 (x-1)^2}-\
\frac{277 \ao^3}{216 (x-1)^2}+\frac{76 d_1 \ao^3}{27 \
(x-2)^3}-\frac{236 \ao^3}{27 (x-2)^3}-\frac{d_1 \ao^3}{6 \
(x-1)^3}+\frac{5 \ao^3}{8 (x-1)^3}+\frac{35 \ao^3}{36}+\frac{35 d_1 \
\ao^2}{18}-\frac{73}{36} d_1 x \ao^2+\frac{569 x \
\ao^2}{72}+\frac{35}{18} d_1 \kappa  \ao^2-\frac{73}{36} d_1 x \kappa \
 \ao^2+\frac{979}{72} x \kappa  \ao^2+\frac{4 d_1 \kappa  \ao^2}{9 \
(x-2)}-\frac{16 \kappa  \ao^2}{3 (x-2)}+\frac{13 d_1 \kappa  \
\ao^2}{36 (x-1)}+\frac{2 \kappa  \ao^2}{3 (x-1)}-\frac{5 d_1 \kappa  \
\ao^2}{3 (x-2)^2}+\frac{10 \kappa  \ao^2}{(x-2)^2}-\frac{d_1 \kappa  \
\ao^2}{2 (x-1)^2}+\frac{67 \kappa  \ao^2}{24 (x-1)^2}+\frac{80 d_1 \
\kappa  \ao^2}{9 (x-2)^3}-\frac{80 \kappa  \ao^2}{3 (x-2)^3}+\frac{5 \
d_1 \kappa  \ao^2}{6 (x-1)^3}-\frac{119 \kappa  \ao^2}{24 \
(x-1)^3}+\frac{104 d_1 \kappa  \ao^2}{3 (x-2)^4}-\frac{1256 \kappa  \
\ao^2}{9 (x-2)^4}-\frac{2 d_1 \kappa  \ao^2}{3 (x-1)^4}+\frac{137 \
\kappa  \ao^2}{36 (x-1)^4}-\frac{32 \kappa  \ao^2}{3}+\frac{4 d_1 \
\ao^2}{9 (x-2)}-\frac{16 \ao^2}{9 (x-2)}+\frac{13 d_1 \ao^2}{36 \
(x-1)}-\frac{\ao^2}{2 (x-1)}-\frac{5 d_1 \ao^2}{3 (x-2)^2}+\frac{10 \
\ao^2}{3 (x-2)^2}-\frac{d_1 \ao^2}{2 (x-1)^2}+\frac{119 \ao^2}{72 \
(x-1)^2}+\frac{80 d_1 \ao^2}{9 (x-2)^3}-\frac{80 \ao^2}{9 \
(x-2)^3}+\frac{5 d_1 \ao^2}{6 (x-1)^3}-\frac{19 \ao^2}{8 \
(x-1)^3}+\frac{104 d_1 \ao^2}{3 (x-2)^4}-\frac{632 \ao^2}{9 (x-2)^4}-\
\frac{2 d_1 \ao^2}{3 (x-1)^4}+\frac{7 \ao^2}{4 (x-1)^4}-\frac{85 \
\ao^2}{18}-\frac{22 d_1 \ao}{27}+\frac{505 d_1 x \ao}{216}-\frac{1697 \
x \ao}{216}-\frac{22 d_1 \kappa  \ao}{27}+\frac{505}{216} d_1 x \
\kappa  \ao-\frac{3193 x \kappa  \ao}{216}+\frac{386 d_1 \kappa  \
\ao}{27 (x-2)}-\frac{1172 \kappa  \ao}{27 (x-2)}-\frac{1133 d_1 \
\kappa  \ao}{72 (x-1)}+\frac{3811 \kappa  \ao}{72 (x-1)}-\frac{362 \
d_1 \kappa  \ao}{27 (x-2)^2}+\frac{1136 \kappa  \ao}{27 \
(x-2)^2}+\frac{193 d_1 \kappa  \ao}{108 (x-1)^2}-\frac{1273 \kappa  \
\ao}{108 (x-1)^2}-\frac{80 d_1 \kappa  \ao}{9 (x-2)^3}+\frac{40 \
\kappa  \ao}{3 (x-2)^3}-\frac{323 d_1 \kappa  \ao}{216 \
(x-1)^3}+\frac{2075 \kappa  \ao}{216 (x-1)^3}-\frac{464 d_1 \kappa  \
\ao}{3 (x-2)^4}+\frac{2032 \kappa  \ao}{3 (x-2)^4}-\frac{187 d_1 \
\kappa  \ao}{216 (x-1)^4}+\frac{539 \kappa  \ao}{108 \
(x-1)^4}-\frac{256 d_1 \kappa  \ao}{3 (x-2)^5}+\frac{3584 \kappa  \
\ao}{9 (x-2)^5}+\frac{205 d_1 \kappa  \ao}{216 (x-1)^5}-\frac{1255 \
\kappa  \ao}{216 (x-1)^5}+\frac{857 \kappa  \ao}{216}+\frac{386 d_1 \
\ao}{27 (x-2)}-\frac{604 \ao}{27 (x-2)}-\frac{1133 d_1 \ao}{72 \
(x-1)}+\frac{649 \ao}{24 (x-1)}-\frac{362 d_1 \ao}{27 \
(x-2)^2}+\frac{592 \ao}{27 (x-2)^2}+\frac{193 d_1 \ao}{108 \
(x-1)^2}-\frac{671 \ao}{108 (x-1)^2}-\frac{80 d_1 \ao}{9 \
(x-2)^3}+\frac{40 \ao}{9 (x-2)^3}-\frac{323 d_1 \ao}{216 \
(x-1)^3}+\frac{1015 \ao}{216 (x-1)^3}-\frac{464 d_1 \ao}{3 \
(x-2)^4}+\frac{368 \ao}{(x-2)^4}-\frac{187 d_1 \ao}{216 \
(x-1)^4}-\frac{\pi ^2 \ao}{18 (x-1)^4}+\frac{419 \ao}{108 \
(x-1)^4}-\frac{256 d_1 \ao}{3 (x-2)^5}+\frac{2048 \ao}{9 \
(x-2)^5}+\frac{205 d_1 \ao}{216 (x-1)^5}+\frac{\pi ^2 \ao}{18 \
(x-1)^5}-\frac{955 \ao}{216 (x-1)^5}+\frac{325 \ao}{216}-\frac{3 \
d_1}{4}-\frac{205 d_1 x}{216}+\frac{635 x}{216}-\frac{3 d_1 \kappa \
}{4}-\frac{205 d_1 x \kappa }{216}+\frac{1255 x \kappa \
}{216}+\frac{346 d_1 \kappa }{27 (x-2)}-\frac{952 \kappa }{27 (x-2)}-\
\frac{947 d_1 \kappa }{72 (x-1)}+\frac{2621 \kappa }{72 \
(x-1)}-\frac{392 d_1 \kappa }{27 (x-2)^2}+\frac{1136 \kappa }{27 \
(x-2)^2}+\frac{23 d_1 \kappa }{27 (x-1)^2}-\frac{92 \kappa }{27 \
(x-1)^2}+\frac{784 d_1 \kappa }{27 (x-2)^3}-\frac{2272 \kappa }{27 \
(x-2)^3}+\frac{73 d_1 \kappa }{216 (x-1)^3}-\frac{247 \kappa }{216 \
(x-1)^3}+\frac{256 d_1 \kappa }{3 (x-2)^4}-\frac{3584 \kappa }{9 \
(x-2)^4}+\frac{367 d_1 \kappa }{216 (x-1)^4}-\frac{1127 \kappa }{108 \
(x-1)^4}+\frac{512 d_1 \kappa }{3 (x-2)^5}-\frac{7168 \kappa }{9 \
(x-2)^5}+\frac{205 d_1 \kappa }{216 (x-1)^5}-\frac{1255 \kappa }{216 \
(x-1)^5}+\frac{37 \kappa }{8}+\Big(\frac{x \ao^5}{6}+\frac{1}{2} x \
\kappa  \ao^5+\frac{\kappa  \ao^5}{x-2}-\frac{\kappa  \ao^5}{2 \
(x-1)}+\frac{\ao^5}{3 (x-2)}-\frac{\ao^5}{6 (x-1)}-\frac{19 x \
\ao^4}{18}-\frac{19}{6} x \kappa  \ao^4-\frac{10 \kappa  \ao^4}{3 \
(x-2)}+\frac{13 \kappa  \ao^4}{6 (x-1)}+\frac{10 \kappa  \ao^4}{3 \
(x-2)^2}-\frac{2 \kappa  \ao^4}{3 (x-1)^2}+\frac{\kappa  \
\ao^4}{3}-\frac{10 \ao^4}{9 (x-2)}+\frac{13 \ao^4}{18 (x-1)}+\frac{10 \
\ao^4}{9 (x-2)^2}-\frac{2 \ao^4}{9 (x-1)^2}+\frac{\ao^4}{9}+\frac{26 \
x \ao^3}{9}+\frac{26}{3} x \kappa  \ao^3+\frac{10 \kappa  \ao^3}{3 \
(x-2)}-\frac{11 \kappa  \ao^3}{3 (x-1)}-\frac{20 \kappa  \ao^3}{3 \
(x-2)^2}+\frac{7 \kappa  \ao^3}{3 (x-1)^2}+\frac{40 \kappa  \ao^3}{3 \
(x-2)^3}-\frac{\kappa  \ao^3}{(x-1)^3}-2 \kappa  \ao^3+\frac{10 \
\ao^3}{9 (x-2)}-\frac{11 \ao^3}{9 (x-1)}-\frac{20 \ao^3}{9 \
(x-2)^2}+\frac{7 \ao^3}{9 (x-1)^2}+\frac{40 \ao^3}{9 \
(x-2)^3}-\frac{\ao^3}{3 (x-1)^3}-\frac{2 \ao^3}{3}-\frac{14 x \
\ao^2}{3}-14 x \kappa  \ao^2+\frac{3 \kappa  \ao^2}{x-1}-\frac{3 \
\kappa  \ao^2}{(x-1)^2}+\frac{3 \kappa  \ao^2}{(x-1)^3}+\frac{80 \
\kappa  \ao^2}{(x-2)^4}-\frac{2 \kappa  \ao^2}{(x-1)^4}+6 \kappa  \
\ao^2+\frac{\ao^2}{x-1}-\frac{\ao^2}{(x-1)^2}+\frac{\ao^2}{(x-1)^3}+\frac{80 \ao^2}{3 (x-2)^4}-\frac{2 \ao^2}{3 (x-1)^4}+2 \ao^2+\frac{73 x \
\ao}{18}+\frac{73 x \kappa  \ao}{6}+\frac{80 \kappa  \ao}{3 \
(x-2)}-\frac{65 \kappa  \ao}{2 (x-1)}-\frac{80 \kappa  \ao}{3 \
(x-2)^2}+\frac{25 \kappa  \ao}{3 (x-1)^2}-\frac{35 \kappa  \ao}{6 \
(x-1)^3}-\frac{480 \kappa  \ao}{(x-2)^4}-\frac{83 \kappa  \ao}{18 \
(x-1)^4}-\frac{320 \kappa  \ao}{(x-2)^5}+\frac{101 \kappa  \ao}{18 \
(x-1)^5}-\frac{4 \kappa  \ao}{3}+\frac{80 \ao}{9 (x-2)}-\frac{65 \
\ao}{6 (x-1)}-\frac{80 \ao}{9 (x-2)^2}+\frac{25 \ao}{9 \
(x-1)^2}-\frac{35 \ao}{18 (x-1)^3}-\frac{160 \ao}{(x-2)^4}-\frac{5 \
\ao}{2 (x-1)^4}-\frac{320 \ao}{3 (x-2)^5}+\frac{17 \ao}{6 \
(x-1)^5}-\frac{4 \ao}{9}-\frac{25 x}{18}-\frac{25 x \kappa \
}{6}+\frac{64 \kappa }{3 (x-2)}-\frac{43 \kappa }{2 (x-1)}-\frac{80 \
\kappa }{3 (x-2)^2}+\frac{8 \kappa }{3 (x-1)^2}+\frac{160 \kappa }{3 \
(x-2)^3}+\frac{\kappa }{6 (x-1)^3}+\frac{320 \kappa \
}{(x-2)^4}+\frac{155 \kappa }{18 (x-1)^4}+\frac{640 \kappa \
}{(x-2)^5}+\frac{101 \kappa }{18 (x-1)^5}-3 \kappa +\frac{64}{9 \
(x-2)}-\frac{43}{6 (x-1)}-\frac{80}{9 (x-2)^2}+\frac{8}{9 \
(x-1)^2}+\frac{160}{9 (x-2)^3}+\frac{1}{18 (x-1)^3}+\frac{320}{3 \
(x-2)^4}+\frac{23}{6 (x-1)^4}+\frac{640}{3 (x-2)^5}+\frac{17}{6 \
(x-1)^5}-1\Big) H(0;\ao)+\Big(\frac{1}{6} d_1 x \ao^5+\frac{1}{6} d_1 \
x \kappa  \ao^5+\frac{d_1 \kappa  \ao^5}{3 (x-2)}-\frac{d_1 \kappa  \
\ao^5}{6 (x-1)}+\frac{d_1 \ao^5}{3 (x-2)}-\frac{d_1 \ao^5}{6 \
(x-1)}+\frac{d_1 \ao^4}{9}-\frac{19}{18} d_1 x \ao^4+\frac{1}{9} d_1 \
\kappa  \ao^4-\frac{19}{18} d_1 x \kappa  \ao^4-\frac{10 d_1 \kappa  \
\ao^4}{9 (x-2)}+\frac{13 d_1 \kappa  \ao^4}{18 (x-1)}+\frac{10 d_1 \
\kappa  \ao^4}{9 (x-2)^2}-\frac{2 d_1 \kappa  \ao^4}{9 \
(x-1)^2}-\frac{10 d_1 \ao^4}{9 (x-2)}+\frac{13 d_1 \ao^4}{18 \
(x-1)}+\frac{10 d_1 \ao^4}{9 (x-2)^2}-\frac{2 d_1 \ao^4}{9 \
(x-1)^2}-\frac{2 d_1 \ao^3}{3}+\frac{26}{9} d_1 x \ao^3-\frac{2}{3} \
d_1 \kappa  \ao^3+\frac{26}{9} d_1 x \kappa  \ao^3+\frac{10 d_1 \
\kappa  \ao^3}{9 (x-2)}-\frac{11 d_1 \kappa  \ao^3}{9 (x-1)}-\frac{20 \
d_1 \kappa  \ao^3}{9 (x-2)^2}+\frac{7 d_1 \kappa  \ao^3}{9 \
(x-1)^2}+\frac{40 d_1 \kappa  \ao^3}{9 (x-2)^3}-\frac{d_1 \kappa  \
\ao^3}{3 (x-1)^3}+\frac{10 d_1 \ao^3}{9 (x-2)}-\frac{11 d_1 \ao^3}{9 \
(x-1)}-\frac{20 d_1 \ao^3}{9 (x-2)^2}+\frac{7 d_1 \ao^3}{9 \
(x-1)^2}+\frac{40 d_1 \ao^3}{9 (x-2)^3}-\frac{d_1 \ao^3}{3 (x-1)^3}+2 \
d_1 \ao^2-\frac{14}{3} d_1 x \ao^2+2 d_1 \kappa  \ao^2-\frac{14}{3} \
d_1 x \kappa  \ao^2+\frac{d_1 \kappa  \ao^2}{x-1}-\frac{d_1 \kappa  \
\ao^2}{(x-1)^2}+\frac{d_1 \kappa  \ao^2}{(x-1)^3}+\frac{80 d_1 \kappa \
 \ao^2}{3 (x-2)^4}-\frac{2 d_1 \kappa  \ao^2}{3 (x-1)^4}+\frac{d_1 \
\ao^2}{x-1}-\frac{d_1 \ao^2}{(x-1)^2}+\frac{d_1 \
\ao^2}{(x-1)^3}+\frac{80 d_1 \ao^2}{3 (x-2)^4}-\frac{2 d_1 \ao^2}{3 \
(x-1)^4}-\frac{4 d_1 \ao}{9}+\frac{73 d_1 x \ao}{18}-\frac{4 d_1 \
\kappa  \ao}{9}+\frac{73}{18} d_1 x \kappa  \ao+\frac{80 d_1 \kappa  \
\ao}{9 (x-2)}-\frac{65 d_1 \kappa  \ao}{6 (x-1)}-\frac{80 d_1 \kappa  \
\ao}{9 (x-2)^2}+\frac{25 d_1 \kappa  \ao}{9 (x-1)^2}-\frac{35 d_1 \
\kappa  \ao}{18 (x-1)^3}-\frac{160 d_1 \kappa  \ao}{(x-2)^4}-\frac{19 \
d_1 \kappa  \ao}{18 (x-1)^4}-\frac{320 d_1 \kappa  \ao}{3 \
(x-2)^5}+\frac{25 d_1 \kappa  \ao}{18 (x-1)^5}+\frac{80 d_1 \ao}{9 \
(x-2)}-\frac{65 d_1 \ao}{6 (x-1)}-\frac{80 d_1 \ao}{9 \
(x-2)^2}+\frac{25 d_1 \ao}{9 (x-1)^2}-\frac{35 d_1 \ao}{18 \
(x-1)^3}-\frac{160 d_1 \ao}{(x-2)^4}-\frac{5 d_1 \ao}{2 \
(x-1)^4}-\frac{320 d_1 \ao}{3 (x-2)^5}+\frac{17 d_1 \ao}{6 \
(x-1)^5}-d_1-\frac{25 d_1 x}{18}-d_1 \kappa -\frac{25 d_1 x \kappa \
}{18}+\frac{64 d_1 \kappa }{9 (x-2)}-\frac{43 d_1 \kappa }{6 \
(x-1)}-\frac{80 d_1 \kappa }{9 (x-2)^2}+\frac{8 d_1 \kappa }{9 \
(x-1)^2}+\frac{160 d_1 \kappa }{9 (x-2)^3}+\frac{d_1 \kappa }{18 \
(x-1)^3}+\frac{320 d_1 \kappa }{3 (x-2)^4}+\frac{43 d_1 \kappa }{18 \
(x-1)^4}+\frac{640 d_1 \kappa }{3 (x-2)^5}+\frac{25 d_1 \kappa }{18 \
(x-1)^5}+\frac{64 d_1}{9 (x-2)}-\frac{43 d_1}{6 (x-1)}-\frac{80 \
d_1}{9 (x-2)^2}+\frac{8 d_1}{9 (x-1)^2}+\frac{160 d_1}{9 \
(x-2)^3}+\frac{d_1}{18 (x-1)^3}+\frac{320 d_1}{3 (x-2)^4}+\frac{23 \
d_1}{6 (x-1)^4}+\frac{640 d_1}{3 (x-2)^5}+\frac{17 d_1}{6 \
(x-1)^5}\Big) H(1;\ao)+\Big(\frac{4 \kappa  \ao}{(x-1)^4}-\frac{4 \
\kappa  \ao}{(x-1)^5}+\frac{4 \ao}{3 (x-1)^4}-\frac{4 \ao}{3 \
(x-1)^5}-\frac{4 \kappa }{(x-1)^4}-\frac{4 \kappa \
}{(x-1)^5}-\frac{4}{3 (x-1)^4}-\frac{4}{3 (x-1)^5}\Big) \
H(0,0;\ao)+\Big(\frac{4 \ao \kappa  d_1}{3 (x-1)^4}-\frac{4 \kappa  \
d_1}{3 (x-1)^4}-\frac{4 \ao \kappa  d_1}{3 (x-1)^5}-\frac{4 \kappa  \
d_1}{3 (x-1)^5}+\frac{4 \ao d_1}{3 (x-1)^4}-\frac{4 d_1}{3 \
(x-1)^4}-\frac{4 \ao d_1}{3 (x-1)^5}-\frac{4 d_1}{3 (x-1)^5}\Big) \
H(0,1;\ao)+\Big(\frac{4 \ao \kappa  d_1}{3 (x-1)^4}-\frac{4 \kappa  \
d_1}{3 (x-1)^4}-\frac{4 \ao \kappa  d_1}{3 (x-1)^5}-\frac{4 \kappa  \
d_1}{3 (x-1)^5}+\frac{4 \ao d_1}{3 (x-1)^4}-\frac{4 d_1}{3 \
(x-1)^4}-\frac{4 \ao d_1}{3 (x-1)^5}-\frac{4 d_1}{3 (x-1)^5}\Big) \
H(1,0;\ao)+\Big(\frac{4 \ao d_1^2}{3 (x-1)^4}-\frac{4 d_1^2}{3 \
(x-1)^4}-\frac{4 \ao d_1^2}{3 (x-1)^5}-\frac{4 d_1^2}{3 (x-1)^5}\Big) \
H(1,1;\ao)+\frac{346 d_1}{27 (x-2)}-\frac{488}{27 (x-2)}-\frac{947 \
d_1}{72 (x-1)}+\frac{443}{24 (x-1)}-\frac{392 d_1}{27 \
(x-2)^2}+\frac{592}{27 (x-2)^2}+\frac{23 d_1}{27 \
(x-1)^2}-\frac{52}{27 (x-1)^2}+\frac{784 d_1}{27 \
(x-2)^3}-\frac{1184}{27 (x-2)^3}+\frac{73 d_1}{216 \
(x-1)^3}-\frac{83}{216 (x-1)^3}+\frac{256 d_1}{3 \
(x-2)^4}-\frac{2048}{9 (x-2)^4}+\frac{367 d_1}{216 (x-1)^4}+\frac{\pi \
^2}{18 (x-1)^4}-\frac{725}{108 (x-1)^4}+\frac{512 d_1}{3 \
(x-2)^5}-\frac{4096}{9 (x-2)^5}+\frac{205 d_1}{216 (x-1)^5}+\frac{\pi \
^2}{18 (x-1)^5}-\frac{955}{216 \
(x-1)^5}+\frac{55}{24}\Big)+\Big(\frac{4 \ao d_1^2}{3 \
(x-1)^4}-\frac{4 d_1^2}{3 (x-1)^4}-\frac{4 \ao d_1^2}{3 \
(x-1)^5}-\frac{4 d_1^2}{3 (x-1)^5}-\frac{4 \ao d_1}{3}+\frac{2 \ao x \
d_1}{3}-\frac{2 x d_1}{3}-\frac{4 \ao \kappa  d_1}{3}+\frac{2}{3} \ao \
x \kappa  d_1-\frac{2 x \kappa  d_1}{3}+\frac{160 \ao \kappa  d_1}{3 \
(x-2)^4}-\frac{160 \kappa  d_1}{3 (x-2)^4}-\frac{4 \ao \kappa  d_1}{3 \
(x-1)^4}+\frac{4 \kappa  d_1}{3 (x-1)^4}+\frac{320 \ao \kappa  d_1}{3 \
(x-2)^5}-\frac{640 \kappa  d_1}{3 (x-2)^5}+\frac{4 \ao \kappa  d_1}{3 \
(x-1)^5}+\frac{4 \kappa  d_1}{3 (x-1)^5}-\frac{640 \kappa  d_1}{3 \
(x-2)^6}+\frac{160 \ao d_1}{3 (x-2)^4}-\frac{160 d_1}{3 \
(x-2)^4}-\frac{4 \ao d_1}{3 (x-1)^4}+\frac{4 d_1}{3 \
(x-1)^4}+\frac{320 \ao d_1}{3 (x-2)^5}-\frac{640 d_1}{3 \
(x-2)^5}+\frac{4 \ao d_1}{3 (x-1)^5}+\frac{4 d_1}{3 \
(x-1)^5}-\frac{640 d_1}{3 (x-2)^6}+\frac{2 \ao}{3}-\frac{\ao \
x}{3}+\frac{x}{3}+2 \ao \kappa -\ao x \kappa +x \kappa -\frac{80 \ao \
\kappa }{(x-2)^4}+\frac{80 \kappa }{(x-2)^4}+\frac{\ao \kappa \
}{(x-1)^4}-\frac{\kappa }{(x-1)^4}-\frac{160 \ao \kappa \
}{(x-2)^5}+\frac{320 \kappa }{(x-2)^5}-\frac{\ao \kappa \
}{(x-1)^5}-\frac{\kappa }{(x-1)^5}+\frac{320 \kappa \
}{(x-2)^6}-\frac{80 \ao}{3 (x-2)^4}+\frac{80}{3 (x-2)^4}+\frac{\ao}{3 \
(x-1)^4}-\frac{1}{3 (x-1)^4}-\frac{160 \ao}{3 (x-2)^5}+\frac{320}{3 \
(x-2)^5}-\frac{\ao}{3 (x-1)^5}-\frac{1}{3 (x-1)^5}+\frac{320}{3 \
(x-2)^6}\Big) H(0;\ao) H(1,1;x)+\Big(-\frac{2 d_1 \ao}{3}+\frac{17 x \
\ao}{12}-\frac{2 d_1 \kappa  \ao}{3}+\frac{101 x \kappa  \
\ao}{36}-\frac{85 d_1 \kappa  \ao}{9 (x-2)}+\frac{40 \kappa  \ao}{3 \
(x-2)}+\frac{94 d_1 \kappa  \ao}{9 (x-1)}-\frac{65 \kappa  \ao}{4 \
(x-1)}+\frac{100 d_1 \kappa  \ao}{9 (x-2)^2}-\frac{40 \kappa  \ao}{3 \
(x-2)^2}-\frac{20 d_1 \kappa  \ao}{9 (x-1)^2}+\frac{25 \kappa  \ao}{6 \
(x-1)^2}-\frac{40 d_1 \kappa  \ao}{3 (x-2)^3}+\frac{17 d_1 \kappa  \
\ao}{18 (x-1)^3}-\frac{35 \kappa  \ao}{12 (x-1)^3}+\frac{64 d_1 \
\kappa  \ao}{(x-2)^4}-\frac{1088 \kappa  \ao}{9 (x-2)^4}+\frac{19 d_1 \
\kappa  \ao}{18 (x-1)^4}-\frac{47 \kappa  \ao}{36 (x-1)^4}+\frac{64 \
d_1 \kappa  \ao}{3 (x-2)^5}+\frac{704 \kappa  \ao}{9 \
(x-2)^5}-\frac{25 d_1 \kappa  \ao}{18 (x-1)^5}+\frac{101 \kappa  \
\ao}{36 (x-1)^5}-\frac{37 \kappa  \ao}{9}-\frac{85 d_1 \ao}{9 (x-2)}+\
\frac{40 \ao}{9 (x-2)}+\frac{94 d_1 \ao}{9 (x-1)}-\frac{65 \ao}{12 \
(x-1)}+\frac{100 d_1 \ao}{9 (x-2)^2}-\frac{40 \ao}{9 \
(x-2)^2}-\frac{20 d_1 \ao}{9 (x-1)^2}+\frac{25 \ao}{18 \
(x-1)^2}-\frac{40 d_1 \ao}{3 (x-2)^3}+\frac{17 d_1 \ao}{18 \
(x-1)^3}-\frac{35 \ao}{36 (x-1)^3}+\frac{64 d_1 \
\ao}{(x-2)^4}+\frac{64 \ao}{9 (x-2)^4}+\frac{5 d_1 \ao}{2 \
(x-1)^4}-\frac{11 \ao}{12 (x-1)^4}+\frac{64 d_1 \ao}{3 \
(x-2)^5}+\frac{1088 \ao}{9 (x-2)^5}-\frac{17 d_1 \ao}{6 \
(x-1)^5}+\frac{17 \ao}{12 (x-1)^5}-\frac{7 \ao}{3}+\frac{2 \
d_1}{3}-\frac{17 x}{12}+\frac{2 d_1 \kappa }{3}-\frac{101 x \kappa \
}{36}-\frac{62 d_1 \kappa }{9 (x-2)}+\frac{32 \kappa }{3 \
(x-2)}+\frac{65 d_1 \kappa }{9 (x-1)}-\frac{43 \kappa }{4 \
(x-1)}+\frac{70 d_1 \kappa }{9 (x-2)^2}-\frac{40 \kappa }{3 (x-2)^2}-\
\frac{d_1 \kappa }{(x-1)^2}+\frac{4 \kappa }{3 (x-1)^2}-\frac{80 d_1 \
\kappa }{9 (x-2)^3}+\frac{80 \kappa }{3 (x-2)^3}+\frac{5 d_1 \kappa \
}{18 (x-1)^3}+\frac{\kappa }{12 (x-1)^3}-\frac{112 d_1 \kappa }{3 \
(x-2)^4}+\frac{1088 \kappa }{9 (x-2)^4}-\frac{31 d_1 \kappa }{18 \
(x-1)^4}+\frac{155 \kappa }{36 (x-1)^4}-\frac{448 d_1 \kappa }{3 \
(x-2)^5}+\frac{1472 \kappa }{9 (x-2)^5}-\frac{25 d_1 \kappa }{18 \
(x-1)^5}+\frac{101 \kappa }{36 (x-1)^5}-\frac{128 d_1 \kappa }{3 \
(x-2)^6}-\frac{1408 \kappa }{9 (x-2)^6}-\frac{3 \kappa \
}{2}+\Big(-\frac{2 x \ao}{3}-2 x \kappa  \ao-\frac{160 \kappa  \
\ao}{(x-2)^4}-\frac{4 d_1 \kappa  \ao}{3 (x-1)^4}+\frac{2 \kappa  \
\ao}{(x-1)^4}-\frac{320 \kappa  \ao}{(x-2)^5}+\frac{4 d_1 \kappa  \
\ao}{3 (x-1)^5}-\frac{2 \kappa  \ao}{(x-1)^5}+4 \kappa  \ao-\frac{160 \
\ao}{3 (x-2)^4}-\frac{4 d_1 \ao}{3 (x-1)^4}+\frac{2 \ao}{3 \
(x-1)^4}-\frac{320 \ao}{3 (x-2)^5}+\frac{4 d_1 \ao}{3 \
(x-1)^5}-\frac{2 \ao}{3 (x-1)^5}+\frac{4 \ao}{3}+\frac{2 x}{3}+2 x \
\kappa +\frac{160 \kappa }{(x-2)^4}+\frac{4 d_1 \kappa }{3 \
(x-1)^4}-\frac{2 \kappa }{(x-1)^4}+\frac{640 \kappa \
}{(x-2)^5}+\frac{4 d_1 \kappa }{3 (x-1)^5}-\frac{2 \kappa }{(x-1)^5}+\
\frac{640 \kappa }{(x-2)^6}+\frac{160}{3 (x-2)^4}+\frac{4 d_1}{3 \
(x-1)^4}-\frac{2}{3 (x-1)^4}+\frac{640}{3 (x-2)^5}+\frac{4 d_1}{3 \
(x-1)^5}-\frac{2}{3 (x-1)^5}+\frac{640}{3 (x-2)^6}\Big) \
H(0;\ao)+\Big(-\frac{4 \ao d_1^2}{3 (x-1)^4}+\frac{4 d_1^2}{3 \
(x-1)^4}+\frac{4 \ao d_1^2}{3 (x-1)^5}+\frac{4 d_1^2}{3 \
(x-1)^5}+\frac{4 \ao d_1}{3}-\frac{2 \ao x d_1}{3}+\frac{2 x d_1}{3}+\
\frac{4 \ao \kappa  d_1}{3}-\frac{2}{3} \ao x \kappa  d_1+\frac{2 x \
\kappa  d_1}{3}-\frac{160 \ao \kappa  d_1}{3 (x-2)^4}+\frac{160 \
\kappa  d_1}{3 (x-2)^4}+\frac{2 \ao \kappa  d_1}{3 (x-1)^4}-\frac{2 \
\kappa  d_1}{3 (x-1)^4}-\frac{320 \ao \kappa  d_1}{3 \
(x-2)^5}+\frac{640 \kappa  d_1}{3 (x-2)^5}-\frac{2 \ao \kappa  d_1}{3 \
(x-1)^5}-\frac{2 \kappa  d_1}{3 (x-1)^5}+\frac{640 \kappa  d_1}{3 \
(x-2)^6}-\frac{160 \ao d_1}{3 (x-2)^4}+\frac{160 d_1}{3 \
(x-2)^4}+\frac{2 \ao d_1}{3 (x-1)^4}-\frac{2 d_1}{3 \
(x-1)^4}-\frac{320 \ao d_1}{3 (x-2)^5}+\frac{640 d_1}{3 \
(x-2)^5}-\frac{2 \ao d_1}{3 (x-1)^5}-\frac{2 d_1}{3 \
(x-1)^5}+\frac{640 d_1}{3 (x-2)^6}\Big) H(1;\ao)-\frac{62 d_1}{9 \
(x-2)}+\frac{32}{9 (x-2)}+\frac{65 d_1}{9 (x-1)}-\frac{43}{12 (x-1)}+\
\frac{70 d_1}{9 (x-2)^2}-\frac{40}{9 \
(x-2)^2}-\frac{d_1}{(x-1)^2}+\frac{4}{9 (x-1)^2}-\frac{80 d_1}{9 \
(x-2)^3}+\frac{80}{9 (x-2)^3}+\frac{5 d_1}{18 (x-1)^3}+\frac{1}{36 \
(x-1)^3}-\frac{112 d_1}{3 (x-2)^4}-\frac{64}{9 (x-2)^4}-\frac{19 \
d_1}{6 (x-1)^4}+\frac{23}{12 (x-1)^4}-\frac{448 d_1}{3 \
(x-2)^5}-\frac{1216}{9 (x-2)^5}-\frac{17 d_1}{6 (x-1)^5}+\frac{17}{12 \
(x-1)^5}-\frac{128 d_1}{3 (x-2)^6}-\frac{2176}{9 (x-2)^6}-\frac{1}{2}\
\Big) H(1,c_1(\ao);x)+\Big(-\frac{160 d_1 \kappa  \ao}{3 \
(x-2)^4}+\frac{160 \kappa  \ao}{(x-2)^4}-\frac{320 d_1 \kappa  \ao}{3 \
(x-2)^5}+\frac{320 \kappa  \ao}{(x-2)^5}-\frac{160 d_1 \ao}{3 \
(x-2)^4}+\frac{160 \ao}{3 (x-2)^4}-\frac{320 d_1 \ao}{3 \
(x-2)^5}+\frac{320 \ao}{3 (x-2)^5}+\frac{160 d_1 \kappa }{3 (x-2)^4}-\
\frac{160 \kappa }{(x-2)^4}+\frac{640 d_1 \kappa }{3 \
(x-2)^5}-\frac{640 \kappa }{(x-2)^5}+\frac{640 d_1 \kappa }{3 \
(x-2)^6}-\frac{640 \kappa }{(x-2)^6}+\frac{160 d_1}{3 \
(x-2)^4}-\frac{160}{3 (x-2)^4}+\frac{640 d_1}{3 (x-2)^5}-\frac{640}{3 \
(x-2)^5}+\frac{640 d_1}{3 (x-2)^6}-\frac{640}{3 (x-2)^6}\Big) \
H(0;\ao) H(2,1;x)+\Big(-\frac{64 \ao d_1^2}{3 (x-2)^4}+\frac{64 \
d_1^2}{3 (x-2)^4}-\frac{128 \ao d_1^2}{3 (x-2)^5}+\frac{256 d_1^2}{3 \
(x-2)^5}+\frac{256 d_1^2}{3 (x-2)^6}+\frac{128 \ao \kappa  d_1}{3 \
(x-2)^4}-\frac{128 \kappa  d_1}{3 (x-2)^4}+\frac{256 \ao \kappa  \
d_1}{3 (x-2)^5}-\frac{512 \kappa  d_1}{3 (x-2)^5}-\frac{512 \kappa  \
d_1}{3 (x-2)^6}-\frac{896 \ao d_1}{9 (x-2)^4}+\frac{896 d_1}{9 \
(x-2)^4}-\frac{1792 \ao d_1}{9 (x-2)^5}+\frac{3584 d_1}{9 \
(x-2)^5}+\frac{3584 d_1}{9 (x-2)^6}+\frac{704 \ao \kappa }{9 \
(x-2)^4}-\frac{704 \kappa }{9 (x-2)^4}+\frac{1408 \ao \kappa }{9 \
(x-2)^5}-\frac{2816 \kappa }{9 (x-2)^5}-\frac{2816 \kappa }{9 \
(x-2)^6}+\Big(\frac{320 d_1 \kappa  \ao}{3 (x-2)^4}-\frac{320 \kappa  \
\ao}{(x-2)^4}+\frac{640 d_1 \kappa  \ao}{3 (x-2)^5}-\frac{640 \kappa  \
\ao}{(x-2)^5}+\frac{320 d_1 \ao}{3 (x-2)^4}-\frac{320 \ao}{3 \
(x-2)^4}+\frac{640 d_1 \ao}{3 (x-2)^5}-\frac{640 \ao}{3 \
(x-2)^5}-\frac{320 d_1 \kappa }{3 (x-2)^4}+\frac{320 \kappa \
}{(x-2)^4}-\frac{1280 d_1 \kappa }{3 (x-2)^5}+\frac{1280 \kappa \
}{(x-2)^5}-\frac{1280 d_1 \kappa }{3 (x-2)^6}+\frac{1280 \kappa \
}{(x-2)^6}-\frac{320 d_1}{3 (x-2)^4}+\frac{320}{3 (x-2)^4}-\frac{1280 \
d_1}{3 (x-2)^5}+\frac{1280}{3 (x-2)^5}-\frac{1280 d_1}{3 \
(x-2)^6}+\frac{1280}{3 (x-2)^6}\Big) H(0;\ao)+\Big(\frac{320 \ao \
d_1^2}{3 (x-2)^4}-\frac{320 d_1^2}{3 (x-2)^4}+\frac{640 \ao d_1^2}{3 \
(x-2)^5}-\frac{1280 d_1^2}{3 (x-2)^5}-\frac{1280 d_1^2}{3 \
(x-2)^6}-\frac{320 \ao \kappa  d_1}{3 (x-2)^4}+\frac{320 \kappa  \
d_1}{3 (x-2)^4}-\frac{640 \ao \kappa  d_1}{3 (x-2)^5}+\frac{1280 \
\kappa  d_1}{3 (x-2)^5}+\frac{1280 \kappa  d_1}{3 (x-2)^6}-\frac{320 \
\ao d_1}{3 (x-2)^4}+\frac{320 d_1}{3 (x-2)^4}-\frac{640 \ao d_1}{3 \
(x-2)^5}+\frac{1280 d_1}{3 (x-2)^5}+\frac{1280 d_1}{3 (x-2)^6}\Big) \
H(1;\ao)+\frac{1088 \ao}{9 (x-2)^4}-\frac{1088}{9 (x-2)^4}+\frac{2176 \
\ao}{9 (x-2)^5}-\frac{4352}{9 (x-2)^5}-\frac{4352}{9 (x-2)^6}\Big) \
H(2,c_2(\ao);x)+\Big(\frac{x \ao^5}{12}+\frac{1}{4} x \kappa  \
\ao^5+\frac{\kappa  \ao^5}{2 (x-2)}-\frac{\kappa  \ao^5}{4 \
(x-1)}+\frac{\ao^5}{6 (x-2)}-\frac{\ao^5}{12 (x-1)}-\frac{19 x \
\ao^4}{36}-\frac{19}{12} x \kappa  \ao^4-\frac{5 \kappa  \ao^4}{3 \
(x-2)}+\frac{13 \kappa  \ao^4}{12 (x-1)}+\frac{5 \kappa  \ao^4}{3 \
(x-2)^2}-\frac{\kappa  \ao^4}{3 (x-1)^2}+\frac{\kappa  \
\ao^4}{6}-\frac{5 \ao^4}{9 (x-2)}+\frac{13 \ao^4}{36 (x-1)}+\frac{5 \
\ao^4}{9 (x-2)^2}-\frac{\ao^4}{9 (x-1)^2}+\frac{\ao^4}{18}+\frac{13 x \
\ao^3}{9}+\frac{13}{3} x \kappa  \ao^3+\frac{5 \kappa  \ao^3}{3 \
(x-2)}-\frac{11 \kappa  \ao^3}{6 (x-1)}-\frac{10 \kappa  \ao^3}{3 \
(x-2)^2}+\frac{7 \kappa  \ao^3}{6 (x-1)^2}+\frac{20 \kappa  \ao^3}{3 \
(x-2)^3}-\frac{\kappa  \ao^3}{2 (x-1)^3}-\kappa  \ao^3+\frac{5 \
\ao^3}{9 (x-2)}-\frac{11 \ao^3}{18 (x-1)}-\frac{10 \ao^3}{9 (x-2)^2}+\
\frac{7 \ao^3}{18 (x-1)^2}+\frac{20 \ao^3}{9 (x-2)^3}-\frac{\ao^3}{6 \
(x-1)^3}-\frac{\ao^3}{3}-\frac{7 x \ao^2}{3}-7 x \kappa  \
\ao^2+\frac{3 \kappa  \ao^2}{2 (x-1)}-\frac{3 \kappa  \ao^2}{2 \
(x-1)^2}+\frac{3 \kappa  \ao^2}{2 (x-1)^3}+\frac{40 \kappa  \
\ao^2}{(x-2)^4}-\frac{\kappa  \ao^2}{(x-1)^4}+3 \kappa  \
\ao^2+\frac{\ao^2}{2 (x-1)}-\frac{\ao^2}{2 (x-1)^2}+\frac{\ao^2}{2 \
(x-1)^3}+\frac{40 \ao^2}{3 (x-2)^4}-\frac{\ao^2}{3 \
(x-1)^4}+\ao^2+\frac{73 x \ao}{36}+\frac{73 x \kappa  \
\ao}{12}+\frac{40 \kappa  \ao}{3 (x-2)}-\frac{65 \kappa  \ao}{4 \
(x-1)}-\frac{40 \kappa  \ao}{3 (x-2)^2}+\frac{25 \kappa  \ao}{6 \
(x-1)^2}-\frac{35 \kappa  \ao}{12 (x-1)^3}-\frac{240 \kappa  \
\ao}{(x-2)^4}-\frac{83 \kappa  \ao}{36 (x-1)^4}-\frac{160 \kappa  \
\ao}{(x-2)^5}+\frac{101 \kappa  \ao}{36 (x-1)^5}-\frac{2 \kappa  \
\ao}{3}+\frac{40 \ao}{9 (x-2)}-\frac{65 \ao}{12 (x-1)}-\frac{40 \
\ao}{9 (x-2)^2}+\frac{25 \ao}{18 (x-1)^2}-\frac{35 \ao}{36 \
(x-1)^3}-\frac{80 \ao}{(x-2)^4}-\frac{5 \ao}{4 (x-1)^4}-\frac{160 \
\ao}{3 (x-2)^5}+\frac{17 \ao}{12 (x-1)^5}-\frac{2 \ao}{9}-\frac{25 \
x}{36}-\frac{25 x \kappa }{12}+\frac{32 \kappa }{3 (x-2)}-\frac{43 \
\kappa }{4 (x-1)}-\frac{40 \kappa }{3 (x-2)^2}+\frac{4 \kappa }{3 \
(x-1)^2}+\frac{80 \kappa }{3 (x-2)^3}+\frac{\kappa }{12 \
(x-1)^3}+\frac{160 \kappa }{(x-2)^4}+\frac{155 \kappa }{36 \
(x-1)^4}+\frac{320 \kappa }{(x-2)^5}+\frac{101 \kappa }{36 \
(x-1)^5}-\frac{3 \kappa }{2}+\Big(\frac{2 \kappa  \
\ao}{(x-1)^4}-\frac{2 \kappa  \ao}{(x-1)^5}+\frac{2 \ao}{3 \
(x-1)^4}-\frac{2 \ao}{3 (x-1)^5}-\frac{2 \kappa }{(x-1)^4}-\frac{2 \
\kappa }{(x-1)^5}-\frac{2}{3 (x-1)^4}-\frac{2}{3 (x-1)^5}\Big) \
H(0;\ao)+\Big(\frac{2 \ao \kappa  d_1}{3 (x-1)^4}-\frac{2 \kappa  \
d_1}{3 (x-1)^4}-\frac{2 \ao \kappa  d_1}{3 (x-1)^5}-\frac{2 \kappa  \
d_1}{3 (x-1)^5}+\frac{2 \ao d_1}{3 (x-1)^4}-\frac{2 d_1}{3 \
(x-1)^4}-\frac{2 \ao d_1}{3 (x-1)^5}-\frac{2 d_1}{3 (x-1)^5}\Big) \
H(1;\ao)+\frac{32}{9 (x-2)}-\frac{43}{12 (x-1)}-\frac{40}{9 (x-2)^2}+\
\frac{4}{9 (x-1)^2}+\frac{80}{9 (x-2)^3}+\frac{1}{36 \
(x-1)^3}+\frac{160}{3 (x-2)^4}+\frac{23}{12 (x-1)^4}+\frac{320}{3 \
(x-2)^5}+\frac{17}{12 (x-1)^5}-\frac{1}{2}\Big) \
H(c_1(\ao),c_1(\ao);x)+\Big(-\frac{32 d_1 \kappa  \ao}{3 \
(x-2)^4}-\frac{352 \kappa  \ao}{9 (x-2)^4}-\frac{64 d_1 \kappa  \
\ao}{3 (x-2)^5}-\frac{704 \kappa  \ao}{9 (x-2)^5}-\frac{32 d_1 \ao}{3 \
(x-2)^4}-\frac{544 \ao}{9 (x-2)^4}-\frac{64 d_1 \ao}{3 \
(x-2)^5}-\frac{1088 \ao}{9 (x-2)^5}+\frac{32 d_1 \kappa }{3 (x-2)^4}+\
\frac{352 \kappa }{9 (x-2)^4}+\frac{128 d_1 \kappa }{3 \
(x-2)^5}+\frac{1408 \kappa }{9 (x-2)^5}+\frac{128 d_1 \kappa }{3 \
(x-2)^6}+\frac{1408 \kappa }{9 (x-2)^6}+\Big(\frac{160 \kappa  \
\ao}{(x-2)^4}+\frac{320 \kappa  \ao}{(x-2)^5}+\frac{160 \ao}{3 \
(x-2)^4}+\frac{320 \ao}{3 (x-2)^5}-\frac{160 \kappa \
}{(x-2)^4}-\frac{640 \kappa }{(x-2)^5}-\frac{640 \kappa \
}{(x-2)^6}-\frac{160}{3 (x-2)^4}-\frac{640}{3 (x-2)^5}-\frac{640}{3 \
(x-2)^6}\Big) H(0;\ao)+\Big(\frac{160 \ao \kappa  d_1}{3 \
(x-2)^4}-\frac{160 \kappa  d_1}{3 (x-2)^4}+\frac{320 \ao \kappa  \
d_1}{3 (x-2)^5}-\frac{640 \kappa  d_1}{3 (x-2)^5}-\frac{640 \kappa  \
d_1}{3 (x-2)^6}+\frac{160 \ao d_1}{3 (x-2)^4}-\frac{160 d_1}{3 \
(x-2)^4}+\frac{320 \ao d_1}{3 (x-2)^5}-\frac{640 d_1}{3 \
(x-2)^5}-\frac{640 d_1}{3 (x-2)^6}\Big) H(1;\ao)+\frac{32 d_1}{3 \
(x-2)^4}+\frac{544}{9 (x-2)^4}+\frac{128 d_1}{3 \
(x-2)^5}+\frac{2176}{9 (x-2)^5}+\frac{128 d_1}{3 \
(x-2)^6}+\frac{2176}{9 (x-2)^6}\Big) \
H(c_2(\ao),c_1(\ao);x)+\Big(\frac{4 x \ao}{3}+4 x \kappa  \
\ao+\frac{320 \kappa  \ao}{(x-2)^4}+\frac{4 \kappa  \
\ao}{(x-1)^4}+\frac{640 \kappa  \ao}{(x-2)^5}-\frac{4 \kappa  \
\ao}{(x-1)^5}-8 \kappa  \ao+\frac{320 \ao}{3 (x-2)^4}+\frac{4 \ao}{3 \
(x-1)^4}+\frac{640 \ao}{3 (x-2)^5}-\frac{4 \ao}{3 (x-1)^5}-\frac{8 \
\ao}{3}-\frac{4 x}{3}-4 x \kappa -\frac{320 \kappa }{(x-2)^4}-\frac{4 \
\kappa }{(x-1)^4}-\frac{1280 \kappa }{(x-2)^5}-\frac{4 \kappa \
}{(x-1)^5}-\frac{1280 \kappa }{(x-2)^6}-\frac{320}{3 \
(x-2)^4}-\frac{4}{3 (x-1)^4}-\frac{1280}{3 (x-2)^5}-\frac{4}{3 \
(x-1)^5}-\frac{1280}{3 (x-2)^6}\Big) H(0,0,0;\ao)+\Big(\frac{4 x \
\ao}{3}+4 x \kappa  \ao-\frac{320 \kappa  \ao}{(x-2)^4}-\frac{4 \
\kappa  \ao}{(x-1)^4}-\frac{640 \kappa  \ao}{(x-2)^5}+\frac{4 \kappa  \
\ao}{(x-1)^5}-8 \kappa  \ao-\frac{320 \ao}{3 (x-2)^4}-\frac{4 \ao}{3 \
(x-1)^4}-\frac{640 \ao}{3 (x-2)^5}+\frac{4 \ao}{3 (x-1)^5}-\frac{8 \
\ao}{3}-\frac{4 x}{3}-4 x \kappa +\frac{320 \kappa }{(x-2)^4}+\frac{4 \
\kappa }{(x-1)^4}+\frac{1280 \kappa }{(x-2)^5}+\frac{4 \kappa \
}{(x-1)^5}+\frac{1280 \kappa }{(x-2)^6}+\frac{320}{3 \
(x-2)^4}+\frac{4}{3 (x-1)^4}+\frac{1280}{3 (x-2)^5}+\frac{4}{3 \
(x-1)^5}+\frac{1280}{3 (x-2)^6}\Big) H(0,0,0;x)+\Big(-\frac{8 \ao \
d_1}{3}+\frac{4 \ao x d_1}{3}-\frac{4 x d_1}{3}-\frac{8 \ao \kappa  \
d_1}{3}+\frac{4}{3} \ao x \kappa  d_1-\frac{4 x \kappa  \
d_1}{3}+\frac{320 \ao \kappa  d_1}{3 (x-2)^4}-\frac{320 \kappa  \
d_1}{3 (x-2)^4}+\frac{4 \ao \kappa  d_1}{3 (x-1)^4}-\frac{4 \kappa  \
d_1}{3 (x-1)^4}+\frac{640 \ao \kappa  d_1}{3 (x-2)^5}-\frac{1280 \
\kappa  d_1}{3 (x-2)^5}-\frac{4 \ao \kappa  d_1}{3 (x-1)^5}-\frac{4 \
\kappa  d_1}{3 (x-1)^5}-\frac{1280 \kappa  d_1}{3 (x-2)^6}+\frac{320 \
\ao d_1}{3 (x-2)^4}-\frac{320 d_1}{3 (x-2)^4}+\frac{4 \ao d_1}{3 \
(x-1)^4}-\frac{4 d_1}{3 (x-1)^4}+\frac{640 \ao d_1}{3 \
(x-2)^5}-\frac{1280 d_1}{3 (x-2)^5}-\frac{4 \ao d_1}{3 \
(x-1)^5}-\frac{4 d_1}{3 (x-1)^5}-\frac{1280 d_1}{3 (x-2)^6}\Big) \
H(0,0,1;\ao)+\Big(-\frac{2 x \ao}{3}-2 x \kappa  \ao-\frac{160 \kappa \
 \ao}{(x-2)^4}+\frac{2 \kappa  \ao}{(x-1)^4}-\frac{320 \kappa  \
\ao}{(x-2)^5}-\frac{2 \kappa  \ao}{(x-1)^5}+4 \kappa  \ao-\frac{160 \
\ao}{3 (x-2)^4}+\frac{2 \ao}{3 (x-1)^4}-\frac{320 \ao}{3 \
(x-2)^5}-\frac{2 \ao}{3 (x-1)^5}+\frac{4 \ao}{3}+\frac{2 x}{3}+2 x \
\kappa +\frac{160 \kappa }{(x-2)^4}-\frac{2 \kappa \
}{(x-1)^4}+\frac{640 \kappa }{(x-2)^5}-\frac{2 \kappa \
}{(x-1)^5}+\frac{640 \kappa }{(x-2)^6}+\frac{160}{3 \
(x-2)^4}-\frac{2}{3 (x-1)^4}+\frac{640}{3 (x-2)^5}-\frac{2}{3 \
(x-1)^5}+\frac{640}{3 (x-2)^6}\Big) H(0,0,c_1(\ao);x)+\Big(\frac{320 \
\kappa  \ao}{(x-2)^4}+\frac{640 \kappa  \ao}{(x-2)^5}+\frac{320 \
\ao}{3 (x-2)^4}+\frac{640 \ao}{3 (x-2)^5}-\frac{320 \kappa \
}{(x-2)^4}-\frac{1280 \kappa }{(x-2)^5}-\frac{1280 \kappa }{(x-2)^6}-\
\frac{320}{3 (x-2)^4}-\frac{1280}{3 (x-2)^5}-\frac{1280}{3 \
(x-2)^6}\Big) H(0,0,c_2(\ao);x)+\Big(-\frac{8 \ao d_1}{3}+\frac{4 \ao \
x d_1}{3}-\frac{4 x d_1}{3}-\frac{8 \ao \kappa  d_1}{3}+\frac{4}{3} \
\ao x \kappa  d_1-\frac{4 x \kappa  d_1}{3}+\frac{320 \ao \kappa  \
d_1}{3 (x-2)^4}-\frac{320 \kappa  d_1}{3 (x-2)^4}+\frac{4 \ao \kappa  \
d_1}{3 (x-1)^4}-\frac{4 \kappa  d_1}{3 (x-1)^4}+\frac{640 \ao \kappa  \
d_1}{3 (x-2)^5}-\frac{1280 \kappa  d_1}{3 (x-2)^5}-\frac{4 \ao \kappa \
 d_1}{3 (x-1)^5}-\frac{4 \kappa  d_1}{3 (x-1)^5}-\frac{1280 \kappa  \
d_1}{3 (x-2)^6}+\frac{320 \ao d_1}{3 (x-2)^4}-\frac{320 d_1}{3 \
(x-2)^4}+\frac{4 \ao d_1}{3 (x-1)^4}-\frac{4 d_1}{3 \
(x-1)^4}+\frac{640 \ao d_1}{3 (x-2)^5}-\frac{1280 d_1}{3 \
(x-2)^5}-\frac{4 \ao d_1}{3 (x-1)^5}-\frac{4 d_1}{3 \
(x-1)^5}-\frac{1280 d_1}{3 (x-2)^6}\Big) H(0,1,0;\ao)+\Big(\frac{4 \
d_1 \ao}{3}-\frac{2 d_1 x \ao}{3}+\frac{2 x \ao}{3}+\frac{4 d_1 \
\kappa  \ao}{3}-\frac{2}{3} d_1 x \kappa  \ao+2 x \kappa  \
\ao-\frac{160 d_1 \kappa  \ao}{3 (x-2)^4}+\frac{160 \kappa  \
\ao}{(x-2)^4}+\frac{2 d_1 \kappa  \ao}{3 (x-1)^4}-\frac{2 \kappa  \
\ao}{(x-1)^4}-\frac{320 d_1 \kappa  \ao}{3 (x-2)^5}+\frac{320 \kappa  \
\ao}{(x-2)^5}-\frac{2 d_1 \kappa  \ao}{3 (x-1)^5}+\frac{2 \kappa  \
\ao}{(x-1)^5}-4 \kappa  \ao-\frac{160 d_1 \ao}{3 (x-2)^4}+\frac{160 \
\ao}{3 (x-2)^4}+\frac{2 d_1 \ao}{3 (x-1)^4}-\frac{2 \ao}{3 \
(x-1)^4}-\frac{320 d_1 \ao}{3 (x-2)^5}+\frac{320 \ao}{3 \
(x-2)^5}-\frac{2 d_1 \ao}{3 (x-1)^5}+\frac{2 \ao}{3 (x-1)^5}-\frac{4 \
\ao}{3}+\frac{2 d_1 x}{3}-\frac{2 x}{3}+\frac{2 d_1 x \kappa }{3}-2 x \
\kappa +\frac{160 d_1 \kappa }{3 (x-2)^4}-\frac{160 \kappa \
}{(x-2)^4}-\frac{2 d_1 \kappa }{3 (x-1)^4}+\frac{2 \kappa }{(x-1)^4}+\
\frac{640 d_1 \kappa }{3 (x-2)^5}-\frac{640 \kappa }{(x-2)^5}-\frac{2 \
d_1 \kappa }{3 (x-1)^5}+\frac{2 \kappa }{(x-1)^5}+\frac{640 d_1 \
\kappa }{3 (x-2)^6}-\frac{640 \kappa }{(x-2)^6}+\frac{160 d_1}{3 \
(x-2)^4}-\frac{160}{3 (x-2)^4}-\frac{2 d_1}{3 (x-1)^4}+\frac{2}{3 \
(x-1)^4}+\frac{640 d_1}{3 (x-2)^5}-\frac{640}{3 (x-2)^5}-\frac{2 \
d_1}{3 (x-1)^5}+\frac{2}{3 (x-1)^5}+\frac{640 d_1}{3 \
(x-2)^6}-\frac{640}{3 (x-2)^6}\Big) H(0,1,0;x)+\Big(-\frac{8 \ao \
d_1^2}{3}+\frac{4}{3} \ao x d_1^2-\frac{4 x d_1^2}{3}+\frac{320 \ao \
d_1^2}{3 (x-2)^4}-\frac{320 d_1^2}{3 (x-2)^4}+\frac{4 \ao d_1^2}{3 \
(x-1)^4}-\frac{4 d_1^2}{3 (x-1)^4}+\frac{640 \ao d_1^2}{3 \
(x-2)^5}-\frac{1280 d_1^2}{3 (x-2)^5}-\frac{4 \ao d_1^2}{3 \
(x-1)^5}-\frac{4 d_1^2}{3 (x-1)^5}-\frac{1280 d_1^2}{3 (x-2)^6}\Big) \
H(0,1,1;\ao)+\Big(-\frac{4 d_1 \ao}{3}+\frac{2 d_1 x \ao}{3}-\frac{2 \
x \ao}{3}-\frac{4 d_1 \kappa  \ao}{3}+\frac{2}{3} d_1 x \kappa  \ao-2 \
x \kappa  \ao+\frac{160 d_1 \kappa  \ao}{3 (x-2)^4}-\frac{160 \kappa  \
\ao}{(x-2)^4}-\frac{2 d_1 \kappa  \ao}{3 (x-1)^4}+\frac{2 \kappa  \
\ao}{(x-1)^4}+\frac{320 d_1 \kappa  \ao}{3 (x-2)^5}-\frac{320 \kappa  \
\ao}{(x-2)^5}+\frac{2 d_1 \kappa  \ao}{3 (x-1)^5}-\frac{2 \kappa  \
\ao}{(x-1)^5}+4 \kappa  \ao+\frac{160 d_1 \ao}{3 (x-2)^4}-\frac{160 \
\ao}{3 (x-2)^4}-\frac{2 d_1 \ao}{3 (x-1)^4}+\frac{2 \ao}{3 \
(x-1)^4}+\frac{320 d_1 \ao}{3 (x-2)^5}-\frac{320 \ao}{3 \
(x-2)^5}+\frac{2 d_1 \ao}{3 (x-1)^5}-\frac{2 \ao}{3 (x-1)^5}+\frac{4 \
\ao}{3}-\frac{2 d_1 x}{3}+\frac{2 x}{3}-\frac{2 d_1 x \kappa }{3}+2 x \
\kappa -\frac{160 d_1 \kappa }{3 (x-2)^4}+\frac{160 \kappa \
}{(x-2)^4}+\frac{2 d_1 \kappa }{3 (x-1)^4}-\frac{2 \kappa }{(x-1)^4}-\
\frac{640 d_1 \kappa }{3 (x-2)^5}+\frac{640 \kappa }{(x-2)^5}+\frac{2 \
d_1 \kappa }{3 (x-1)^5}-\frac{2 \kappa }{(x-1)^5}-\frac{640 d_1 \
\kappa }{3 (x-2)^6}+\frac{640 \kappa }{(x-2)^6}-\frac{160 d_1}{3 \
(x-2)^4}+\frac{160}{3 (x-2)^4}+\frac{2 d_1}{3 (x-1)^4}-\frac{2}{3 \
(x-1)^4}-\frac{640 d_1}{3 (x-2)^5}+\frac{640}{3 (x-2)^5}+\frac{2 \
d_1}{3 (x-1)^5}-\frac{2}{3 (x-1)^5}-\frac{640 d_1}{3 \
(x-2)^6}+\frac{640}{3 (x-2)^6}\Big) H(0,1,c_1(\ao);x)+\Big(-\frac{320 \
d_1 \kappa  \ao}{3 (x-2)^4}+\frac{320 \kappa  \ao}{(x-2)^4}-\frac{640 \
d_1 \kappa  \ao}{3 (x-2)^5}+\frac{640 \kappa  \ao}{(x-2)^5}-\frac{320 \
d_1 \ao}{3 (x-2)^4}+\frac{320 \ao}{3 (x-2)^4}-\frac{640 d_1 \ao}{3 \
(x-2)^5}+\frac{640 \ao}{3 (x-2)^5}+\frac{320 d_1 \kappa }{3 (x-2)^4}-\
\frac{320 \kappa }{(x-2)^4}+\frac{1280 d_1 \kappa }{3 \
(x-2)^5}-\frac{1280 \kappa }{(x-2)^5}+\frac{1280 d_1 \kappa }{3 \
(x-2)^6}-\frac{1280 \kappa }{(x-2)^6}+\frac{320 d_1}{3 \
(x-2)^4}-\frac{320}{3 (x-2)^4}+\frac{1280 d_1}{3 \
(x-2)^5}-\frac{1280}{3 (x-2)^5}+\frac{1280 d_1}{3 \
(x-2)^6}-\frac{1280}{3 (x-2)^6}\Big) H(0,2,0;x)+\Big(\frac{320 d_1 \
\kappa  \ao}{3 (x-2)^4}-\frac{320 \kappa  \ao}{(x-2)^4}+\frac{640 d_1 \
\kappa  \ao}{3 (x-2)^5}-\frac{640 \kappa  \ao}{(x-2)^5}+\frac{320 d_1 \
\ao}{3 (x-2)^4}-\frac{320 \ao}{3 (x-2)^4}+\frac{640 d_1 \ao}{3 \
(x-2)^5}-\frac{640 \ao}{3 (x-2)^5}-\frac{320 d_1 \kappa }{3 (x-2)^4}+\
\frac{320 \kappa }{(x-2)^4}-\frac{1280 d_1 \kappa }{3 \
(x-2)^5}+\frac{1280 \kappa }{(x-2)^5}-\frac{1280 d_1 \kappa }{3 \
(x-2)^6}+\frac{1280 \kappa }{(x-2)^6}-\frac{320 d_1}{3 \
(x-2)^4}+\frac{320}{3 (x-2)^4}-\frac{1280 d_1}{3 \
(x-2)^5}+\frac{1280}{3 (x-2)^5}-\frac{1280 d_1}{3 \
(x-2)^6}+\frac{1280}{3 (x-2)^6}\Big) H(0,2,c_2(\ao);x)+\Big(-\frac{x \
\ao}{3}-x \kappa  \ao-\frac{80 \kappa  \ao}{(x-2)^4}+\frac{\kappa  \
\ao}{(x-1)^4}-\frac{160 \kappa  \ao}{(x-2)^5}-\frac{\kappa  \
\ao}{(x-1)^5}+2 \kappa  \ao-\frac{80 \ao}{3 (x-2)^4}+\frac{\ao}{3 \
(x-1)^4}-\frac{160 \ao}{3 (x-2)^5}-\frac{\ao}{3 (x-1)^5}+\frac{2 \
\ao}{3}+\frac{x}{3}+x \kappa +\frac{80 \kappa }{(x-2)^4}-\frac{\kappa \
}{(x-1)^4}+\frac{320 \kappa }{(x-2)^5}-\frac{\kappa \
}{(x-1)^5}+\frac{320 \kappa }{(x-2)^6}+\frac{80}{3 \
(x-2)^4}-\frac{1}{3 (x-1)^4}+\frac{320}{3 (x-2)^5}-\frac{1}{3 \
(x-1)^5}+\frac{320}{3 (x-2)^6}\Big) \
H(0,c_1(\ao),c_1(\ao);x)+\Big(\frac{160 \kappa  \
\ao}{(x-2)^4}+\frac{320 \kappa  \ao}{(x-2)^5}+\frac{160 \ao}{3 \
(x-2)^4}+\frac{320 \ao}{3 (x-2)^5}-\frac{160 \kappa \
}{(x-2)^4}-\frac{640 \kappa }{(x-2)^5}-\frac{640 \kappa \
}{(x-2)^6}-\frac{160}{3 (x-2)^4}-\frac{640}{3 (x-2)^5}-\frac{640}{3 \
(x-2)^6}\Big) H(0,c_2(\ao),c_1(\ao);x)+\Big(\frac{2 x \ao}{3}+2 x \
\kappa  \ao+\frac{160 \kappa  \ao}{(x-2)^4}+\frac{4 d_1 \kappa  \
\ao}{3 (x-1)^4}-\frac{2 \kappa  \ao}{(x-1)^4}+\frac{320 \kappa  \
\ao}{(x-2)^5}-\frac{4 d_1 \kappa  \ao}{3 (x-1)^5}+\frac{2 \kappa  \
\ao}{(x-1)^5}-4 \kappa  \ao+\frac{160 \ao}{3 (x-2)^4}+\frac{4 d_1 \
\ao}{3 (x-1)^4}-\frac{2 \ao}{3 (x-1)^4}+\frac{320 \ao}{3 \
(x-2)^5}-\frac{4 d_1 \ao}{3 (x-1)^5}+\frac{2 \ao}{3 (x-1)^5}-\frac{4 \
\ao}{3}-\frac{2 x}{3}-2 x \kappa -\frac{160 \kappa }{(x-2)^4}-\frac{4 \
d_1 \kappa }{3 (x-1)^4}+\frac{2 \kappa }{(x-1)^4}-\frac{640 \kappa \
}{(x-2)^5}-\frac{4 d_1 \kappa }{3 (x-1)^5}+\frac{2 \kappa }{(x-1)^5}-\
\frac{640 \kappa }{(x-2)^6}-\frac{160}{3 (x-2)^4}-\frac{4 d_1}{3 \
(x-1)^4}+\frac{2}{3 (x-1)^4}-\frac{640}{3 (x-2)^5}-\frac{4 d_1}{3 \
(x-1)^5}+\frac{2}{3 (x-1)^5}-\frac{640}{3 (x-2)^6}\Big) \
H(1,0,0;x)+\Big(-\frac{x \ao}{3}-x \kappa  \ao-\frac{80 \kappa  \
\ao}{(x-2)^4}-\frac{2 d_1 \kappa  \ao}{3 (x-1)^4}+\frac{\kappa  \
\ao}{(x-1)^4}-\frac{160 \kappa  \ao}{(x-2)^5}+\frac{2 d_1 \kappa  \
\ao}{3 (x-1)^5}-\frac{\kappa  \ao}{(x-1)^5}+2 \kappa  \ao-\frac{80 \
\ao}{3 (x-2)^4}-\frac{2 d_1 \ao}{3 (x-1)^4}+\frac{\ao}{3 \
(x-1)^4}-\frac{160 \ao}{3 (x-2)^5}+\frac{2 d_1 \ao}{3 (x-1)^5}-\frac{\
\ao}{3 (x-1)^5}+\frac{2 \ao}{3}+\frac{x}{3}+x \kappa +\frac{80 \kappa \
}{(x-2)^4}+\frac{2 d_1 \kappa }{3 (x-1)^4}-\frac{\kappa \
}{(x-1)^4}+\frac{320 \kappa }{(x-2)^5}+\frac{2 d_1 \kappa }{3 \
(x-1)^5}-\frac{\kappa }{(x-1)^5}+\frac{320 \kappa \
}{(x-2)^6}+\frac{80}{3 (x-2)^4}+\frac{2 d_1}{3 (x-1)^4}-\frac{1}{3 \
(x-1)^4}+\frac{320}{3 (x-2)^5}+\frac{2 d_1}{3 (x-1)^5}-\frac{1}{3 \
(x-1)^5}+\frac{320}{3 (x-2)^6}\Big) H(1,0,c_1(\ao);x)+\Big(-\frac{4 \
\ao d_1^2}{3 (x-1)^4}+\frac{4 d_1^2}{3 (x-1)^4}+\frac{4 \ao d_1^2}{3 \
(x-1)^5}+\frac{4 d_1^2}{3 (x-1)^5}+\frac{4 \ao d_1}{3}-\frac{2 \ao x \
d_1}{3}+\frac{2 x d_1}{3}+\frac{4 \ao \kappa  d_1}{3}-\frac{2}{3} \ao \
x \kappa  d_1+\frac{2 x \kappa  d_1}{3}-\frac{160 \ao \kappa  d_1}{3 \
(x-2)^4}+\frac{160 \kappa  d_1}{3 (x-2)^4}+\frac{4 \ao \kappa  d_1}{3 \
(x-1)^4}-\frac{4 \kappa  d_1}{3 (x-1)^4}-\frac{320 \ao \kappa  d_1}{3 \
(x-2)^5}+\frac{640 \kappa  d_1}{3 (x-2)^5}-\frac{4 \ao \kappa  d_1}{3 \
(x-1)^5}-\frac{4 \kappa  d_1}{3 (x-1)^5}+\frac{640 \kappa  d_1}{3 \
(x-2)^6}-\frac{160 \ao d_1}{3 (x-2)^4}+\frac{160 d_1}{3 \
(x-2)^4}+\frac{4 \ao d_1}{3 (x-1)^4}-\frac{4 d_1}{3 \
(x-1)^4}-\frac{320 \ao d_1}{3 (x-2)^5}+\frac{640 d_1}{3 \
(x-2)^5}-\frac{4 \ao d_1}{3 (x-1)^5}-\frac{4 d_1}{3 \
(x-1)^5}+\frac{640 d_1}{3 (x-2)^6}-\frac{2 \ao}{3}+\frac{\ao \
x}{3}-\frac{x}{3}-2 \ao \kappa +\ao x \kappa -x \kappa +\frac{80 \ao \
\kappa }{(x-2)^4}-\frac{80 \kappa }{(x-2)^4}-\frac{\ao \kappa \
}{(x-1)^4}+\frac{\kappa }{(x-1)^4}+\frac{160 \ao \kappa \
}{(x-2)^5}-\frac{320 \kappa }{(x-2)^5}+\frac{\ao \kappa \
}{(x-1)^5}+\frac{\kappa }{(x-1)^5}-\frac{320 \kappa \
}{(x-2)^6}+\frac{80 \ao}{3 (x-2)^4}-\frac{80}{3 (x-2)^4}-\frac{\ao}{3 \
(x-1)^4}+\frac{1}{3 (x-1)^4}+\frac{160 \ao}{3 (x-2)^5}-\frac{320}{3 \
(x-2)^5}+\frac{\ao}{3 (x-1)^5}+\frac{1}{3 (x-1)^5}-\frac{320}{3 \
(x-2)^6}\Big) H(1,1,0;x)+\Big(\frac{4 \ao d_1^2}{3 (x-1)^4}-\frac{4 \
d_1^2}{3 (x-1)^4}-\frac{4 \ao d_1^2}{3 (x-1)^5}-\frac{4 d_1^2}{3 \
(x-1)^5}-\frac{4 \ao d_1}{3}+\frac{2 \ao x d_1}{3}-\frac{2 x d_1}{3}-\
\frac{4 \ao \kappa  d_1}{3}+\frac{2}{3} \ao x \kappa  d_1-\frac{2 x \
\kappa  d_1}{3}+\frac{160 \ao \kappa  d_1}{3 (x-2)^4}-\frac{160 \
\kappa  d_1}{3 (x-2)^4}-\frac{4 \ao \kappa  d_1}{3 (x-1)^4}+\frac{4 \
\kappa  d_1}{3 (x-1)^4}+\frac{320 \ao \kappa  d_1}{3 \
(x-2)^5}-\frac{640 \kappa  d_1}{3 (x-2)^5}+\frac{4 \ao \kappa  d_1}{3 \
(x-1)^5}+\frac{4 \kappa  d_1}{3 (x-1)^5}-\frac{640 \kappa  d_1}{3 \
(x-2)^6}+\frac{160 \ao d_1}{3 (x-2)^4}-\frac{160 d_1}{3 \
(x-2)^4}-\frac{4 \ao d_1}{3 (x-1)^4}+\frac{4 d_1}{3 \
(x-1)^4}+\frac{320 \ao d_1}{3 (x-2)^5}-\frac{640 d_1}{3 \
(x-2)^5}+\frac{4 \ao d_1}{3 (x-1)^5}+\frac{4 d_1}{3 \
(x-1)^5}-\frac{640 d_1}{3 (x-2)^6}+\frac{2 \ao}{3}-\frac{\ao \
x}{3}+\frac{x}{3}+2 \ao \kappa -\ao x \kappa +x \kappa -\frac{80 \ao \
\kappa }{(x-2)^4}+\frac{80 \kappa }{(x-2)^4}+\frac{\ao \kappa \
}{(x-1)^4}-\frac{\kappa }{(x-1)^4}-\frac{160 \ao \kappa \
}{(x-2)^5}+\frac{320 \kappa }{(x-2)^5}-\frac{\ao \kappa \
}{(x-1)^5}-\frac{\kappa }{(x-1)^5}+\frac{320 \kappa \
}{(x-2)^6}-\frac{80 \ao}{3 (x-2)^4}+\frac{80}{3 (x-2)^4}+\frac{\ao}{3 \
(x-1)^4}-\frac{1}{3 (x-1)^4}-\frac{160 \ao}{3 (x-2)^5}+\frac{320}{3 \
(x-2)^5}-\frac{\ao}{3 (x-1)^5}-\frac{1}{3 (x-1)^5}+\frac{320}{3 \
(x-2)^6}\Big) H(1,1,c_1(\ao);x)+\Big(-\frac{x \ao}{3}-x \kappa  \
\ao-\frac{80 \kappa  \ao}{(x-2)^4}-\frac{2 d_1 \kappa  \ao}{3 \
(x-1)^4}+\frac{\kappa  \ao}{(x-1)^4}-\frac{160 \kappa  \ao}{(x-2)^5}+\
\frac{2 d_1 \kappa  \ao}{3 (x-1)^5}-\frac{\kappa  \ao}{(x-1)^5}+2 \
\kappa  \ao-\frac{80 \ao}{3 (x-2)^4}-\frac{2 d_1 \ao}{3 \
(x-1)^4}+\frac{\ao}{3 (x-1)^4}-\frac{160 \ao}{3 (x-2)^5}+\frac{2 d_1 \
\ao}{3 (x-1)^5}-\frac{\ao}{3 (x-1)^5}+\frac{2 \ao}{3}+\frac{x}{3}+x \
\kappa +\frac{80 \kappa }{(x-2)^4}+\frac{2 d_1 \kappa }{3 \
(x-1)^4}-\frac{\kappa }{(x-1)^4}+\frac{320 \kappa }{(x-2)^5}+\frac{2 \
d_1 \kappa }{3 (x-1)^5}-\frac{\kappa }{(x-1)^5}+\frac{320 \kappa \
}{(x-2)^6}+\frac{80}{3 (x-2)^4}+\frac{2 d_1}{3 (x-1)^4}-\frac{1}{3 \
(x-1)^4}+\frac{320}{3 (x-2)^5}+\frac{2 d_1}{3 (x-1)^5}-\frac{1}{3 \
(x-1)^5}+\frac{320}{3 (x-2)^6}\Big) \
H(1,c_1(\ao),c_1(\ao);x)+\Big(-\frac{320 d_1 \kappa  \ao}{3 (x-2)^4}+\
\frac{320 \kappa  \ao}{(x-2)^4}-\frac{640 d_1 \kappa  \ao}{3 \
(x-2)^5}+\frac{640 \kappa  \ao}{(x-2)^5}-\frac{320 d_1 \ao}{3 \
(x-2)^4}+\frac{320 \ao}{3 (x-2)^4}-\frac{640 d_1 \ao}{3 \
(x-2)^5}+\frac{640 \ao}{3 (x-2)^5}+\frac{320 d_1 \kappa }{3 (x-2)^4}-\
\frac{320 \kappa }{(x-2)^4}+\frac{1280 d_1 \kappa }{3 \
(x-2)^5}-\frac{1280 \kappa }{(x-2)^5}+\frac{1280 d_1 \kappa }{3 \
(x-2)^6}-\frac{1280 \kappa }{(x-2)^6}+\frac{320 d_1}{3 \
(x-2)^4}-\frac{320}{3 (x-2)^4}+\frac{1280 d_1}{3 \
(x-2)^5}-\frac{1280}{3 (x-2)^5}+\frac{1280 d_1}{3 \
(x-2)^6}-\frac{1280}{3 (x-2)^6}\Big) H(2,0,0;x)+\Big(-\frac{160 d_1 \
\kappa  \ao}{3 (x-2)^4}+\frac{160 \kappa  \ao}{(x-2)^4}-\frac{320 d_1 \
\kappa  \ao}{3 (x-2)^5}+\frac{320 \kappa  \ao}{(x-2)^5}-\frac{160 d_1 \
\ao}{3 (x-2)^4}+\frac{160 \ao}{3 (x-2)^4}-\frac{320 d_1 \ao}{3 \
(x-2)^5}+\frac{320 \ao}{3 (x-2)^5}+\frac{160 d_1 \kappa }{3 (x-2)^4}-\
\frac{160 \kappa }{(x-2)^4}+\frac{640 d_1 \kappa }{3 \
(x-2)^5}-\frac{640 \kappa }{(x-2)^5}+\frac{640 d_1 \kappa }{3 \
(x-2)^6}-\frac{640 \kappa }{(x-2)^6}+\frac{160 d_1}{3 \
(x-2)^4}-\frac{160}{3 (x-2)^4}+\frac{640 d_1}{3 (x-2)^5}-\frac{640}{3 \
(x-2)^5}+\frac{640 d_1}{3 (x-2)^6}-\frac{640}{3 (x-2)^6}\Big) \
H(2,0,c_1(\ao);x)+\Big(\frac{320 d_1 \kappa  \ao}{3 \
(x-2)^4}-\frac{320 \kappa  \ao}{(x-2)^4}+\frac{640 d_1 \kappa  \ao}{3 \
(x-2)^5}-\frac{640 \kappa  \ao}{(x-2)^5}+\frac{320 d_1 \ao}{3 \
(x-2)^4}-\frac{320 \ao}{3 (x-2)^4}+\frac{640 d_1 \ao}{3 \
(x-2)^5}-\frac{640 \ao}{3 (x-2)^5}-\frac{320 d_1 \kappa }{3 (x-2)^4}+\
\frac{320 \kappa }{(x-2)^4}-\frac{1280 d_1 \kappa }{3 \
(x-2)^5}+\frac{1280 \kappa }{(x-2)^5}-\frac{1280 d_1 \kappa }{3 \
(x-2)^6}+\frac{1280 \kappa }{(x-2)^6}-\frac{320 d_1}{3 \
(x-2)^4}+\frac{320}{3 (x-2)^4}-\frac{1280 d_1}{3 \
(x-2)^5}+\frac{1280}{3 (x-2)^5}-\frac{1280 d_1}{3 \
(x-2)^6}+\frac{1280}{3 (x-2)^6}\Big) H(2,0,c_2(\ao);x)+\Big(\frac{160 \
d_1 \kappa  \ao}{3 (x-2)^4}-\frac{160 \kappa  \ao}{(x-2)^4}+\frac{320 \
d_1 \kappa  \ao}{3 (x-2)^5}-\frac{320 \kappa  \ao}{(x-2)^5}+\frac{160 \
d_1 \ao}{3 (x-2)^4}-\frac{160 \ao}{3 (x-2)^4}+\frac{320 d_1 \ao}{3 \
(x-2)^5}-\frac{320 \ao}{3 (x-2)^5}-\frac{160 d_1 \kappa }{3 (x-2)^4}+\
\frac{160 \kappa }{(x-2)^4}-\frac{640 d_1 \kappa }{3 \
(x-2)^5}+\frac{640 \kappa }{(x-2)^5}-\frac{640 d_1 \kappa }{3 \
(x-2)^6}+\frac{640 \kappa }{(x-2)^6}-\frac{160 d_1}{3 \
(x-2)^4}+\frac{160}{3 (x-2)^4}-\frac{640 d_1}{3 (x-2)^5}+\frac{640}{3 \
(x-2)^5}-\frac{640 d_1}{3 (x-2)^6}+\frac{640}{3 (x-2)^6}\Big) \
H(2,1,0;x)+\Big(-\frac{160 d_1 \kappa  \ao}{3 (x-2)^4}+\frac{160 \
\kappa  \ao}{(x-2)^4}-\frac{320 d_1 \kappa  \ao}{3 (x-2)^5}+\frac{320 \
\kappa  \ao}{(x-2)^5}-\frac{160 d_1 \ao}{3 (x-2)^4}+\frac{160 \ao}{3 \
(x-2)^4}-\frac{320 d_1 \ao}{3 (x-2)^5}+\frac{320 \ao}{3 \
(x-2)^5}+\frac{160 d_1 \kappa }{3 (x-2)^4}-\frac{160 \kappa \
}{(x-2)^4}+\frac{640 d_1 \kappa }{3 (x-2)^5}-\frac{640 \kappa \
}{(x-2)^5}+\frac{640 d_1 \kappa }{3 (x-2)^6}-\frac{640 \kappa \
}{(x-2)^6}+\frac{160 d_1}{3 (x-2)^4}-\frac{160}{3 (x-2)^4}+\frac{640 \
d_1}{3 (x-2)^5}-\frac{640}{3 (x-2)^5}+\frac{640 d_1}{3 \
(x-2)^6}-\frac{640}{3 (x-2)^6}\Big) H(2,1,c_1(\ao);x)+\Big(-\frac{320 \
\ao d_1^2}{3 (x-2)^4}+\frac{320 d_1^2}{3 (x-2)^4}-\frac{640 \ao \
d_1^2}{3 (x-2)^5}+\frac{1280 d_1^2}{3 (x-2)^5}+\frac{1280 d_1^2}{3 \
(x-2)^6}+\frac{640 \ao \kappa  d_1}{3 (x-2)^4}-\frac{640 \kappa  \
d_1}{3 (x-2)^4}+\frac{1280 \ao \kappa  d_1}{3 (x-2)^5}-\frac{2560 \
\kappa  d_1}{3 (x-2)^5}-\frac{2560 \kappa  d_1}{3 (x-2)^6}+\frac{640 \
\ao d_1}{3 (x-2)^4}-\frac{640 d_1}{3 (x-2)^4}+\frac{1280 \ao d_1}{3 \
(x-2)^5}-\frac{2560 d_1}{3 (x-2)^5}-\frac{2560 d_1}{3 \
(x-2)^6}-\frac{320 \ao \kappa }{(x-2)^4}+\frac{320 \kappa }{(x-2)^4}-\
\frac{640 \ao \kappa }{(x-2)^5}+\frac{1280 \kappa \
}{(x-2)^5}+\frac{1280 \kappa }{(x-2)^6}-\frac{320 \ao}{3 \
(x-2)^4}+\frac{320}{3 (x-2)^4}-\frac{640 \ao}{3 \
(x-2)^5}+\frac{1280}{3 (x-2)^5}+\frac{1280}{3 (x-2)^6}\Big) \
H(2,2,0;x)+\Big(\frac{320 \ao d_1^2}{3 (x-2)^4}-\frac{320 d_1^2}{3 \
(x-2)^4}+\frac{640 \ao d_1^2}{3 (x-2)^5}-\frac{1280 d_1^2}{3 \
(x-2)^5}-\frac{1280 d_1^2}{3 (x-2)^6}-\frac{640 \ao \kappa  d_1}{3 \
(x-2)^4}+\frac{640 \kappa  d_1}{3 (x-2)^4}-\frac{1280 \ao \kappa  \
d_1}{3 (x-2)^5}+\frac{2560 \kappa  d_1}{3 (x-2)^5}+\frac{2560 \kappa  \
d_1}{3 (x-2)^6}-\frac{640 \ao d_1}{3 (x-2)^4}+\frac{640 d_1}{3 \
(x-2)^4}-\frac{1280 \ao d_1}{3 (x-2)^5}+\frac{2560 d_1}{3 \
(x-2)^5}+\frac{2560 d_1}{3 (x-2)^6}+\frac{320 \ao \kappa \
}{(x-2)^4}-\frac{320 \kappa }{(x-2)^4}+\frac{640 \ao \kappa \
}{(x-2)^5}-\frac{1280 \kappa }{(x-2)^5}-\frac{1280 \kappa }{(x-2)^6}+\
\frac{320 \ao}{3 (x-2)^4}-\frac{320}{3 (x-2)^4}+\frac{640 \ao}{3 \
(x-2)^5}-\frac{1280}{3 (x-2)^5}-\frac{1280}{3 (x-2)^6}\Big) \
H(2,2,c_2(\ao);x)+\Big(\frac{160 d_1 \kappa  \ao}{3 \
(x-2)^4}-\frac{160 \kappa  \ao}{(x-2)^4}+\frac{320 d_1 \kappa  \ao}{3 \
(x-2)^5}-\frac{320 \kappa  \ao}{(x-2)^5}+\frac{160 d_1 \ao}{3 \
(x-2)^4}-\frac{160 \ao}{3 (x-2)^4}+\frac{320 d_1 \ao}{3 \
(x-2)^5}-\frac{320 \ao}{3 (x-2)^5}-\frac{160 d_1 \kappa }{3 (x-2)^4}+\
\frac{160 \kappa }{(x-2)^4}-\frac{640 d_1 \kappa }{3 \
(x-2)^5}+\frac{640 \kappa }{(x-2)^5}-\frac{640 d_1 \kappa }{3 \
(x-2)^6}+\frac{640 \kappa }{(x-2)^6}-\frac{160 d_1}{3 \
(x-2)^4}+\frac{160}{3 (x-2)^4}-\frac{640 d_1}{3 (x-2)^5}+\frac{640}{3 \
(x-2)^5}-\frac{640 d_1}{3 (x-2)^6}+\frac{640}{3 (x-2)^6}\Big) \
H(2,c_2(\ao),c_1(\ao);x)+\Big(\frac{\kappa  \
\ao}{(x-1)^4}-\frac{\kappa  \ao}{(x-1)^5}+\frac{\ao}{3 \
(x-1)^4}-\frac{\ao}{3 (x-1)^5}-\frac{\kappa }{(x-1)^4}-\frac{\kappa \
}{(x-1)^5}-\frac{1}{3 (x-1)^4}-\frac{1}{3 (x-1)^5}\Big) \
H(c_1(\ao),c_1(\ao),c_1(\ao);x)+\Big(\frac{80 \kappa  \
\ao}{(x-2)^4}+\frac{160 \kappa  \ao}{(x-2)^5}+\frac{80 \ao}{3 \
(x-2)^4}+\frac{160 \ao}{3 (x-2)^5}-\frac{80 \kappa \
}{(x-2)^4}-\frac{320 \kappa }{(x-2)^5}-\frac{320 \kappa \
}{(x-2)^6}-\frac{80}{3 (x-2)^4}-\frac{320}{3 (x-2)^5}-\frac{320}{3 \
(x-2)^6}\Big) H(c_2(\ao),c_1(\ao),c_1(\ao);x)+H(0,0;x) \Big(-\frac{17 \
x \ao}{6}-\frac{101 x \kappa  \ao}{18}-\frac{80 \kappa  \ao}{3 \
(x-2)}+\frac{65 \kappa  \ao}{2 (x-1)}+\frac{80 \kappa  \ao}{3 \
(x-2)^2}-\frac{25 \kappa  \ao}{3 (x-1)^2}+\frac{35 \kappa  \ao}{6 \
(x-1)^3}+\frac{64 d_1 \kappa  \ao}{3 (x-2)^4}+\frac{3584 \kappa  \
\ao}{9 (x-2)^4}+\frac{47 \kappa  \ao}{18 (x-1)^4}+\frac{128 d_1 \
\kappa  \ao}{3 (x-2)^5}+\frac{1408 \kappa  \ao}{9 (x-2)^5}-\frac{101 \
\kappa  \ao}{18 (x-1)^5}+\frac{74 \kappa  \ao}{9}-\frac{80 \ao}{9 \
(x-2)}+\frac{65 \ao}{6 (x-1)}+\frac{80 \ao}{9 (x-2)^2}-\frac{25 \
\ao}{9 (x-1)^2}+\frac{35 \ao}{18 (x-1)^3}+\frac{64 d_1 \ao}{3 \
(x-2)^4}+\frac{2048 \ao}{9 (x-2)^4}+\frac{11 \ao}{6 \
(x-1)^4}+\frac{128 d_1 \ao}{3 (x-2)^5}+\frac{2176 \ao}{9 \
(x-2)^5}-\frac{17 \ao}{6 (x-1)^5}+\frac{320 \kappa  \ln 2\,  \
\ao}{(x-2)^4}+\frac{640 \kappa  \ln 2\,  \ao}{(x-2)^5}+\frac{320 \ln \
2\,  \ao}{3 (x-2)^4}+\frac{640 \ln 2\,  \ao}{3 (x-2)^5}+\frac{14 \
\ao}{3}+\frac{17 x}{6}+\frac{101 x \kappa }{18}-\frac{64 \kappa }{3 \
(x-2)}+\frac{43 \kappa }{2 (x-1)}+\frac{80 \kappa }{3 \
(x-2)^2}-\frac{8 \kappa }{3 (x-1)^2}-\frac{160 \kappa }{3 \
(x-2)^3}-\frac{\kappa }{6 (x-1)^3}-\frac{64 d_1 \kappa }{3 \
(x-2)^4}-\frac{3584 \kappa }{9 (x-2)^4}-\frac{155 \kappa }{18 \
(x-1)^4}-\frac{256 d_1 \kappa }{3 (x-2)^5}-\frac{8576 \kappa }{9 \
(x-2)^5}-\frac{101 \kappa }{18 (x-1)^5}-\frac{256 d_1 \kappa }{3 \
(x-2)^6}-\frac{2816 \kappa }{9 (x-2)^6}+3 \kappa -\frac{64}{9 (x-2)}+\
\frac{43}{6 (x-1)}+\frac{80}{9 (x-2)^2}-\frac{8}{9 \
(x-1)^2}-\frac{160}{9 (x-2)^3}-\frac{1}{18 (x-1)^3}-\frac{64 d_1}{3 \
(x-2)^4}-\frac{2048}{9 (x-2)^4}-\frac{23}{6 (x-1)^4}-\frac{256 d_1}{3 \
(x-2)^5}-\frac{6272}{9 (x-2)^5}-\frac{17}{6 (x-1)^5}-\frac{256 d_1}{3 \
(x-2)^6}-\frac{4352}{9 (x-2)^6}-\frac{320 \kappa  \ln 2\, }{(x-2)^4}-\
\frac{1280 \kappa  \ln 2\, }{(x-2)^5}-\frac{1280 \kappa  \ln 2\, \
}{(x-2)^6}-\frac{320 \ln 2\, }{3 (x-2)^4}-\frac{1280 \ln 2\, }{3 \
(x-2)^5}-\frac{1280 \ln 2\, }{3 (x-2)^6}+1\Big)+H(2,2;x) \
\Big(\frac{320 \ao \ln 2\,  d_1^2}{3 (x-2)^4}-\frac{320 \ln 2\,  \
d_1^2}{3 (x-2)^4}+\frac{640 \ao \ln 2\,  d_1^2}{3 (x-2)^5}-\frac{1280 \
\ln 2\,  d_1^2}{3 (x-2)^5}-\frac{1280 \ln 2\,  d_1^2}{3 \
(x-2)^6}-\frac{640 \ao \kappa  \ln 2\,  d_1}{3 (x-2)^4}+\frac{640 \
\kappa  \ln 2\,  d_1}{3 (x-2)^4}-\frac{1280 \ao \kappa  \ln 2\,  \
d_1}{3 (x-2)^5}+\frac{2560 \kappa  \ln 2\,  d_1}{3 \
(x-2)^5}+\frac{2560 \kappa  \ln 2\,  d_1}{3 (x-2)^6}-\frac{640 \ao \
\ln 2\,  d_1}{3 (x-2)^4}+\frac{640 \ln 2\,  d_1}{3 \
(x-2)^4}-\frac{1280 \ao \ln 2\,  d_1}{3 (x-2)^5}+\frac{2560 \ln 2\,  \
d_1}{3 (x-2)^5}+\frac{2560 \ln 2\,  d_1}{3 (x-2)^6}+\Big(\frac{320 \
\ao d_1^2}{3 (x-2)^4}-\frac{320 d_1^2}{3 (x-2)^4}+\frac{640 \ao \
d_1^2}{3 (x-2)^5}-\frac{1280 d_1^2}{3 (x-2)^5}-\frac{1280 d_1^2}{3 \
(x-2)^6}-\frac{640 \ao \kappa  d_1}{3 (x-2)^4}+\frac{640 \kappa  \
d_1}{3 (x-2)^4}-\frac{1280 \ao \kappa  d_1}{3 (x-2)^5}+\frac{2560 \
\kappa  d_1}{3 (x-2)^5}+\frac{2560 \kappa  d_1}{3 (x-2)^6}-\frac{640 \
\ao d_1}{3 (x-2)^4}+\frac{640 d_1}{3 (x-2)^4}-\frac{1280 \ao d_1}{3 \
(x-2)^5}+\frac{2560 d_1}{3 (x-2)^5}+\frac{2560 d_1}{3 \
(x-2)^6}+\frac{320 \ao \kappa }{(x-2)^4}-\frac{320 \kappa }{(x-2)^4}+\
\frac{640 \ao \kappa }{(x-2)^5}-\frac{1280 \kappa \
}{(x-2)^5}-\frac{1280 \kappa }{(x-2)^6}+\frac{320 \ao}{3 \
(x-2)^4}-\frac{320}{3 (x-2)^4}+\frac{640 \ao}{3 \
(x-2)^5}-\frac{1280}{3 (x-2)^5}-\frac{1280}{3 (x-2)^6}\Big) H(0;\ao)+\
\frac{320 \ao \kappa  \ln 2\, }{(x-2)^4}-\frac{320 \kappa  \ln 2\, \
}{(x-2)^4}+\frac{640 \ao \kappa  \ln 2\, }{(x-2)^5}-\frac{1280 \kappa \
 \ln 2\, }{(x-2)^5}-\frac{1280 \kappa  \ln 2\, }{(x-2)^6}+\frac{320 \
\ao \ln 2\, }{3 (x-2)^4}-\frac{320 \ln 2\, }{3 (x-2)^4}+\frac{640 \ao \
\ln 2\, }{3 (x-2)^5}-\frac{1280 \ln 2\, }{3 (x-2)^5}-\frac{1280 \ln 2\
\, }{3 (x-2)^6}\Big)+H(2,0;x) \Big(\frac{64 \ao d_1^2}{3 \
(x-2)^4}-\frac{64 d_1^2}{3 (x-2)^4}+\frac{128 \ao d_1^2}{3 \
(x-2)^5}-\frac{256 d_1^2}{3 (x-2)^5}-\frac{256 d_1^2}{3 \
(x-2)^6}-\frac{128 \ao \kappa  d_1}{3 (x-2)^4}+\frac{128 \kappa  \
d_1}{3 (x-2)^4}-\frac{256 \ao \kappa  d_1}{3 (x-2)^5}+\frac{512 \
\kappa  d_1}{3 (x-2)^5}+\frac{512 \kappa  d_1}{3 (x-2)^6}+\frac{896 \
\ao d_1}{9 (x-2)^4}-\frac{896 d_1}{9 (x-2)^4}+\frac{1792 \ao d_1}{9 \
(x-2)^5}-\frac{3584 d_1}{9 (x-2)^5}-\frac{3584 d_1}{9 \
(x-2)^6}+\frac{320 \ao \kappa  \ln 2\,  d_1}{3 (x-2)^4}-\frac{320 \
\kappa  \ln 2\,  d_1}{3 (x-2)^4}+\frac{640 \ao \kappa  \ln 2\,  \
d_1}{3 (x-2)^5}-\frac{1280 \kappa  \ln 2\,  d_1}{3 \
(x-2)^5}-\frac{1280 \kappa  \ln 2\,  d_1}{3 (x-2)^6}+\frac{320 \ao \
\ln 2\,  d_1}{3 (x-2)^4}-\frac{320 \ln 2\,  d_1}{3 (x-2)^4}+\frac{640 \
\ao \ln 2\,  d_1}{3 (x-2)^5}-\frac{1280 \ln 2\,  d_1}{3 \
(x-2)^5}-\frac{1280 \ln 2\,  d_1}{3 (x-2)^6}-\frac{704 \ao \kappa }{9 \
(x-2)^4}+\frac{704 \kappa }{9 (x-2)^4}-\frac{1408 \ao \kappa }{9 \
(x-2)^5}+\frac{2816 \kappa }{9 (x-2)^5}+\frac{2816 \kappa }{9 \
(x-2)^6}-\frac{1088 \ao}{9 (x-2)^4}+\frac{1088}{9 (x-2)^4}-\frac{2176 \
\ao}{9 (x-2)^5}+\frac{4352}{9 (x-2)^5}+\frac{4352}{9 \
(x-2)^6}-\frac{320 \ao \kappa  \ln 2\, }{(x-2)^4}+\frac{320 \kappa  \
\ln 2\, }{(x-2)^4}-\frac{640 \ao \kappa  \ln 2\, \
}{(x-2)^5}+\frac{1280 \kappa  \ln 2\, }{(x-2)^5}+\frac{1280 \kappa  \
\ln 2\, }{(x-2)^6}-\frac{320 \ao \ln 2\, }{3 (x-2)^4}+\frac{320 \ln 2\
\, }{3 (x-2)^4}-\frac{640 \ao \ln 2\, }{3 (x-2)^5}+\frac{1280 \ln 2\, \
}{3 (x-2)^5}+\frac{1280 \ln 2\, }{3 (x-2)^6}\Big)+H(0,2;x) \
\Big(\frac{320 d_1 \kappa  \ln 2\,  \ao}{3 (x-2)^4}-\frac{320 \kappa  \
\ln 2\,  \ao}{(x-2)^4}+\frac{640 d_1 \kappa  \ln 2\,  \ao}{3 \
(x-2)^5}-\frac{640 \kappa  \ln 2\,  \ao}{(x-2)^5}+\frac{320 d_1 \ln 2\
\,  \ao}{3 (x-2)^4}-\frac{320 \ln 2\,  \ao}{3 (x-2)^4}+\frac{640 d_1 \
\ln 2\,  \ao}{3 (x-2)^5}-\frac{640 \ln 2\,  \ao}{3 \
(x-2)^5}+\Big(\frac{320 d_1 \kappa  \ao}{3 (x-2)^4}-\frac{320 \kappa  \
\ao}{(x-2)^4}+\frac{640 d_1 \kappa  \ao}{3 (x-2)^5}-\frac{640 \kappa  \
\ao}{(x-2)^5}+\frac{320 d_1 \ao}{3 (x-2)^4}-\frac{320 \ao}{3 \
(x-2)^4}+\frac{640 d_1 \ao}{3 (x-2)^5}-\frac{640 \ao}{3 \
(x-2)^5}-\frac{320 d_1 \kappa }{3 (x-2)^4}+\frac{320 \kappa \
}{(x-2)^4}-\frac{1280 d_1 \kappa }{3 (x-2)^5}+\frac{1280 \kappa \
}{(x-2)^5}-\frac{1280 d_1 \kappa }{3 (x-2)^6}+\frac{1280 \kappa \
}{(x-2)^6}-\frac{320 d_1}{3 (x-2)^4}+\frac{320}{3 (x-2)^4}-\frac{1280 \
d_1}{3 (x-2)^5}+\frac{1280}{3 (x-2)^5}-\frac{1280 d_1}{3 \
(x-2)^6}+\frac{1280}{3 (x-2)^6}\Big) H(0;\ao)-\frac{320 d_1 \kappa  \
\ln 2\, }{3 (x-2)^4}+\frac{320 \kappa  \ln 2\, }{(x-2)^4}-\frac{1280 \
d_1 \kappa  \ln 2\, }{3 (x-2)^5}+\frac{1280 \kappa  \ln 2\, \
}{(x-2)^5}-\frac{1280 d_1 \kappa  \ln 2\, }{3 (x-2)^6}+\frac{1280 \
\kappa  \ln 2\, }{(x-2)^6}-\frac{320 d_1 \ln 2\, }{3 \
(x-2)^4}+\frac{320 \ln 2\, }{3 (x-2)^4}-\frac{1280 d_1 \ln 2\, }{3 \
(x-2)^5}+\frac{1280 \ln 2\, }{3 (x-2)^5}-\frac{1280 d_1 \ln 2\, }{3 \
(x-2)^6}+\frac{1280 \ln 2\, }{3 (x-2)^6}\Big)+H(0;x) \Big(\frac{31 \
d_1 \ao}{27}-\frac{205 d_1 x \ao}{216}+\frac{1}{18} \pi ^2 x \
\ao+\frac{955 x \ao}{216}+\frac{31 d_1 \kappa  \
\ao}{27}-\frac{205}{216} d_1 x \kappa  \ao+\frac{1}{3} \pi ^2 x \
\kappa  \ao+\frac{1255 x \kappa  \ao}{216}-\frac{392 d_1 \kappa  \
\ao}{27 (x-2)}+\frac{1136 \kappa  \ao}{27 (x-2)}+\frac{1129 d_1 \
\kappa  \ao}{72 (x-1)}-\frac{3613 \kappa  \ao}{72 (x-1)}+\frac{392 \
d_1 \kappa  \ao}{27 (x-2)^2}-\frac{1136 \kappa  \ao}{27 \
(x-2)^2}-\frac{181 d_1 \kappa  \ao}{108 (x-1)^2}+\frac{280 \kappa  \
\ao}{27 (x-1)^2}+\frac{251 d_1 \kappa  \ao}{216 (x-1)^3}-\frac{1751 \
\kappa  \ao}{216 (x-1)^3}+\frac{256 d_1 \kappa  \ao}{3 \
(x-2)^4}-\frac{40 \pi ^2 \kappa  \ao}{3 (x-2)^4}-\frac{3328 \kappa  \
\ao}{9 (x-2)^4}+\frac{43 d_1 \kappa  \ao}{216 (x-1)^4}-\frac{\pi ^2 \
\kappa  \ao}{3 (x-1)^4}-\frac{32 \kappa  \ao}{27 (x-1)^4}-\frac{80 \
\pi ^2 \kappa  \ao}{3 (x-2)^5}+\frac{512 \kappa  \ao}{9 \
(x-2)^5}-\frac{205 d_1 \kappa  \ao}{216 (x-1)^5}+\frac{\pi ^2 \kappa  \
\ao}{3 (x-1)^5}+\frac{1255 \kappa  \ao}{216 (x-1)^5}-\frac{2}{3} \pi \
^2 \kappa  \ao-\frac{1511 \kappa  \ao}{216}-\frac{392 d_1 \ao}{27 \
(x-2)}+\frac{592 \ao}{27 (x-2)}+\frac{1129 d_1 \ao}{72 \
(x-1)}-\frac{209 \ao}{8 (x-1)}+\frac{392 d_1 \ao}{27 \
(x-2)^2}-\frac{592 \ao}{27 (x-2)^2}-\frac{181 d_1 \ao}{108 \
(x-1)^2}+\frac{155 \ao}{27 (x-1)^2}+\frac{251 d_1 \ao}{216 \
(x-1)^3}-\frac{907 \ao}{216 (x-1)^3}+\frac{512 d_1 \ao}{9 \
(x-2)^4}-\frac{9536 \ao}{27 (x-2)^4}+\frac{43 d_1 \ao}{216 \
(x-1)^4}-\frac{\pi ^2 \ao}{18 (x-1)^4}-\frac{115 \ao}{54 \
(x-1)^4}-\frac{512 d_1 \ao}{9 (x-2)^5}-\frac{6784 \ao}{27 \
(x-2)^5}-\frac{205 d_1 \ao}{216 (x-1)^5}+\frac{\pi ^2 \ao}{18 \
(x-1)^5}+\frac{955 \ao}{216 (x-1)^5}-\frac{160 \kappa  \ln ^22\,  \
\ao}{(x-2)^4}-\frac{320 \kappa  \ln ^22\,  \ao}{(x-2)^5}-\frac{160 \
\ln ^22\,  \ao}{3 (x-2)^4}-\frac{320 \ln ^22\,  \ao}{3 \
(x-2)^5}-\frac{64 d_1 \kappa  \ln 2\,  \ao}{3 (x-2)^4}-\frac{704 \
\kappa  \ln 2\,  \ao}{9 (x-2)^4}-\frac{128 d_1 \kappa  \ln 2\,  \
\ao}{3 (x-2)^5}-\frac{1408 \kappa  \ln 2\,  \ao}{9 (x-2)^5}-\frac{64 \
d_1 \ln 2\,  \ao}{3 (x-2)^4}-\frac{1088 \ln 2\,  \ao}{9 \
(x-2)^4}-\frac{128 d_1 \ln 2\,  \ao}{3 (x-2)^5}-\frac{2176 \ln 2\,  \
\ao}{9 (x-2)^5}-\frac{\pi ^2 \ao}{9}-\frac{1415 \ao}{216}+\frac{3 \
d_1}{4}+\frac{205 d_1 x}{216}-\frac{\pi ^2 x}{18}-\frac{955 \
x}{216}+\frac{3 d_1 \kappa }{4}+\frac{205 d_1 x \kappa \
}{216}-\frac{1}{3} \pi ^2 x \kappa -\frac{1255 x \kappa \
}{216}-\frac{346 d_1 \kappa }{27 (x-2)}+\frac{952 \kappa }{27 (x-2)}+\
\frac{947 d_1 \kappa }{72 (x-1)}-\frac{2621 \kappa }{72 \
(x-1)}+\frac{392 d_1 \kappa }{27 (x-2)^2}-\frac{1136 \kappa }{27 \
(x-2)^2}-\frac{23 d_1 \kappa }{27 (x-1)^2}+\frac{92 \kappa }{27 \
(x-1)^2}-\frac{784 d_1 \kappa }{27 (x-2)^3}+\frac{2272 \kappa }{27 \
(x-2)^3}-\frac{73 d_1 \kappa }{216 (x-1)^3}+\frac{247 \kappa }{216 \
(x-1)^3}-\frac{256 d_1 \kappa }{3 (x-2)^4}+\frac{40 \pi ^2 \kappa }{3 \
(x-2)^4}+\frac{3328 \kappa }{9 (x-2)^4}-\frac{367 d_1 \kappa }{216 \
(x-1)^4}+\frac{\pi ^2 \kappa }{3 (x-1)^4}+\frac{1127 \kappa }{108 \
(x-1)^4}-\frac{512 d_1 \kappa }{3 (x-2)^5}+\frac{160 \pi ^2 \kappa \
}{3 (x-2)^5}+\frac{2048 \kappa }{3 (x-2)^5}-\frac{205 d_1 \kappa \
}{216 (x-1)^5}+\frac{\pi ^2 \kappa }{3 (x-1)^5}+\frac{1255 \kappa \
}{216 (x-1)^5}+\frac{160 \pi ^2 \kappa }{3 (x-2)^6}-\frac{1024 \kappa \
}{9 (x-2)^6}-\frac{37 \kappa }{8}-\frac{346 d_1}{27 \
(x-2)}+\frac{488}{27 (x-2)}+\frac{947 d_1}{72 (x-1)}-\frac{443}{24 \
(x-1)}+\frac{392 d_1}{27 (x-2)^2}-\frac{592}{27 (x-2)^2}-\frac{23 \
d_1}{27 (x-1)^2}+\frac{52}{27 (x-1)^2}-\frac{784 d_1}{27 \
(x-2)^3}+\frac{1184}{27 (x-2)^3}-\frac{73 d_1}{216 \
(x-1)^3}+\frac{83}{216 (x-1)^3}-\frac{512 d_1}{9 \
(x-2)^4}+\frac{9536}{27 (x-2)^4}-\frac{367 d_1}{216 \
(x-1)^4}+\frac{\pi ^2}{18 (x-1)^4}+\frac{725}{108 (x-1)^4}-\frac{512 \
d_1}{9 (x-2)^5}+\frac{25856}{27 (x-2)^5}-\frac{205 d_1}{216 (x-1)^5}+\
\frac{\pi ^2}{18 (x-1)^5}+\frac{955}{216 (x-1)^5}+\frac{1024 d_1}{9 \
(x-2)^6}+\frac{13568}{27 (x-2)^6}+\frac{160 \kappa  \ln ^22\, \
}{(x-2)^4}+\frac{640 \kappa  \ln ^22\, }{(x-2)^5}+\frac{640 \kappa  \
\ln ^22\, }{(x-2)^6}+\frac{160 \ln ^22\, }{3 (x-2)^4}+\frac{640 \ln \
^22\, }{3 (x-2)^5}+\frac{640 \ln ^22\, }{3 (x-2)^6}+\frac{64 d_1 \
\kappa  \ln 2\, }{3 (x-2)^4}+\frac{704 \kappa  \ln 2\, }{9 \
(x-2)^4}+\frac{256 d_1 \kappa  \ln 2\, }{3 (x-2)^5}+\frac{2816 \kappa \
 \ln 2\, }{9 (x-2)^5}+\frac{256 d_1 \kappa  \ln 2\, }{3 \
(x-2)^6}+\frac{2816 \kappa  \ln 2\, }{9 (x-2)^6}+\frac{64 d_1 \ln 2\, \
}{3 (x-2)^4}+\frac{1088 \ln 2\, }{9 (x-2)^4}+\frac{256 d_1 \ln 2\, \
}{3 (x-2)^5}+\frac{4352 \ln 2\, }{9 (x-2)^5}+\frac{256 d_1 \ln 2\, \
}{3 (x-2)^6}+\frac{4352 \ln 2\, }{9 \
(x-2)^6}-\frac{55}{24}\Big)+H(2;x) \Big(-\frac{64 \ao \ln 2\,  \
d_1^2}{3 (x-2)^4}+\frac{64 \ln 2\,  d_1^2}{3 (x-2)^4}-\frac{128 \ao \
\ln 2\,  d_1^2}{3 (x-2)^5}+\frac{256 \ln 2\,  d_1^2}{3 \
(x-2)^5}+\frac{256 \ln 2\,  d_1^2}{3 (x-2)^6}-\frac{40 \ao \pi ^2 \
\kappa  d_1}{9 (x-2)^4}+\frac{40 \pi ^2 \kappa  d_1}{9 \
(x-2)^4}-\frac{80 \ao \pi ^2 \kappa  d_1}{9 (x-2)^5}+\frac{160 \pi ^2 \
\kappa  d_1}{9 (x-2)^5}+\frac{160 \pi ^2 \kappa  d_1}{9 \
(x-2)^6}-\frac{40 \ao \pi ^2 d_1}{9 (x-2)^4}+\frac{40 \pi ^2 d_1}{9 \
(x-2)^4}-\frac{80 \ao \pi ^2 d_1}{9 (x-2)^5}+\frac{160 \pi ^2 d_1}{9 \
(x-2)^5}+\frac{160 \pi ^2 d_1}{9 (x-2)^6}-\frac{160 \ao \kappa  \ln \
^22\,  d_1}{3 (x-2)^4}+\frac{160 \kappa  \ln ^22\,  d_1}{3 \
(x-2)^4}-\frac{320 \ao \kappa  \ln ^22\,  d_1}{3 (x-2)^5}+\frac{640 \
\kappa  \ln ^22\,  d_1}{3 (x-2)^5}+\frac{640 \kappa  \ln ^22\,  \
d_1}{3 (x-2)^6}-\frac{160 \ao \ln ^22\,  d_1}{3 (x-2)^4}+\frac{160 \
\ln ^22\,  d_1}{3 (x-2)^4}-\frac{320 \ao \ln ^22\,  d_1}{3 \
(x-2)^5}+\frac{640 \ln ^22\,  d_1}{3 (x-2)^5}+\frac{640 \ln ^22\,  \
d_1}{3 (x-2)^6}+\frac{128 \ao \kappa  \ln 2\,  d_1}{3 \
(x-2)^4}-\frac{128 \kappa  \ln 2\,  d_1}{3 (x-2)^4}+\frac{256 \ao \
\kappa  \ln 2\,  d_1}{3 (x-2)^5}-\frac{512 \kappa  \ln 2\,  d_1}{3 \
(x-2)^5}-\frac{512 \kappa  \ln 2\,  d_1}{3 (x-2)^6}-\frac{896 \ao \ln \
2\,  d_1}{9 (x-2)^4}+\frac{896 \ln 2\,  d_1}{9 (x-2)^4}-\frac{1792 \
\ao \ln 2\,  d_1}{9 (x-2)^5}+\frac{3584 \ln 2\,  d_1}{9 \
(x-2)^5}+\frac{3584 \ln 2\,  d_1}{9 (x-2)^6}+\frac{40 \ao \pi ^2 \
\kappa }{3 (x-2)^4}-\frac{40 \pi ^2 \kappa }{3 (x-2)^4}+\frac{80 \ao \
\pi ^2 \kappa }{3 (x-2)^5}-\frac{160 \pi ^2 \kappa }{3 \
(x-2)^5}-\frac{160 \pi ^2 \kappa }{3 (x-2)^6}+\Big(-\frac{64 \ao \
d_1^2}{3 (x-2)^4}+\frac{64 d_1^2}{3 (x-2)^4}-\frac{128 \ao d_1^2}{3 \
(x-2)^5}+\frac{256 d_1^2}{3 (x-2)^5}+\frac{256 d_1^2}{3 \
(x-2)^6}+\frac{128 \ao \kappa  d_1}{3 (x-2)^4}-\frac{128 \kappa  \
d_1}{3 (x-2)^4}+\frac{256 \ao \kappa  d_1}{3 (x-2)^5}-\frac{512 \
\kappa  d_1}{3 (x-2)^5}-\frac{512 \kappa  d_1}{3 (x-2)^6}-\frac{896 \
\ao d_1}{9 (x-2)^4}+\frac{896 d_1}{9 (x-2)^4}-\frac{1792 \ao d_1}{9 \
(x-2)^5}+\frac{3584 d_1}{9 (x-2)^5}+\frac{3584 d_1}{9 \
(x-2)^6}+\frac{704 \ao \kappa }{9 (x-2)^4}-\frac{704 \kappa }{9 \
(x-2)^4}+\frac{1408 \ao \kappa }{9 (x-2)^5}-\frac{2816 \kappa }{9 \
(x-2)^5}-\frac{2816 \kappa }{9 (x-2)^6}+\frac{1088 \ao}{9 \
(x-2)^4}-\frac{1088}{9 (x-2)^4}+\frac{2176 \ao}{9 \
(x-2)^5}-\frac{4352}{9 (x-2)^5}-\frac{4352}{9 (x-2)^6}\Big) H(0;\ao)+\
\Big(\frac{320 d_1 \kappa  \ao}{3 (x-2)^4}-\frac{320 \kappa  \
\ao}{(x-2)^4}+\frac{640 d_1 \kappa  \ao}{3 (x-2)^5}-\frac{640 \kappa  \
\ao}{(x-2)^5}+\frac{320 d_1 \ao}{3 (x-2)^4}-\frac{320 \ao}{3 \
(x-2)^4}+\frac{640 d_1 \ao}{3 (x-2)^5}-\frac{640 \ao}{3 \
(x-2)^5}-\frac{320 d_1 \kappa }{3 (x-2)^4}+\frac{320 \kappa \
}{(x-2)^4}-\frac{1280 d_1 \kappa }{3 (x-2)^5}+\frac{1280 \kappa \
}{(x-2)^5}-\frac{1280 d_1 \kappa }{3 (x-2)^6}+\frac{1280 \kappa \
}{(x-2)^6}-\frac{320 d_1}{3 (x-2)^4}+\frac{320}{3 (x-2)^4}-\frac{1280 \
d_1}{3 (x-2)^5}+\frac{1280}{3 (x-2)^5}-\frac{1280 d_1}{3 \
(x-2)^6}+\frac{1280}{3 (x-2)^6}\Big) H(0,0;\ao)+\Big(\frac{320 \ao \
d_1^2}{3 (x-2)^4}-\frac{320 d_1^2}{3 (x-2)^4}+\frac{640 \ao d_1^2}{3 \
(x-2)^5}-\frac{1280 d_1^2}{3 (x-2)^5}-\frac{1280 d_1^2}{3 \
(x-2)^6}-\frac{320 \ao \kappa  d_1}{3 (x-2)^4}+\frac{320 \kappa  \
d_1}{3 (x-2)^4}-\frac{640 \ao \kappa  d_1}{3 (x-2)^5}+\frac{1280 \
\kappa  d_1}{3 (x-2)^5}+\frac{1280 \kappa  d_1}{3 (x-2)^6}-\frac{320 \
\ao d_1}{3 (x-2)^4}+\frac{320 d_1}{3 (x-2)^4}-\frac{640 \ao d_1}{3 \
(x-2)^5}+\frac{1280 d_1}{3 (x-2)^5}+\frac{1280 d_1}{3 (x-2)^6}\Big) \
H(0,1;\ao)+\frac{40 \ao \pi ^2}{9 (x-2)^4}-\frac{40 \pi ^2}{9 \
(x-2)^4}+\frac{80 \ao \pi ^2}{9 (x-2)^5}-\frac{160 \pi ^2}{9 \
(x-2)^5}-\frac{160 \pi ^2}{9 (x-2)^6}+\frac{160 \ao \kappa  \ln ^22\, \
}{(x-2)^4}-\frac{160 \kappa  \ln ^22\, }{(x-2)^4}+\frac{320 \ao \
\kappa  \ln ^22\, }{(x-2)^5}-\frac{640 \kappa  \ln ^22\, \
}{(x-2)^5}-\frac{640 \kappa  \ln ^22\, }{(x-2)^6}+\frac{160 \ao \ln \
^22\, }{3 (x-2)^4}-\frac{160 \ln ^22\, }{3 (x-2)^4}+\frac{320 \ao \ln \
^22\, }{3 (x-2)^5}-\frac{640 \ln ^22\, }{3 (x-2)^5}-\frac{640 \ln ^22\
\, }{3 (x-2)^6}+\frac{704 \ao \kappa  \ln 2\, }{9 (x-2)^4}-\frac{704 \
\kappa  \ln 2\, }{9 (x-2)^4}+\frac{1408 \ao \kappa  \ln 2\, }{9 \
(x-2)^5}-\frac{2816 \kappa  \ln 2\, }{9 (x-2)^5}-\frac{2816 \kappa  \
\ln 2\, }{9 (x-2)^6}+\frac{1088 \ao \ln 2\, }{9 (x-2)^4}-\frac{1088 \
\ln 2\, }{9 (x-2)^4}+\frac{2176 \ao \ln 2\, }{9 (x-2)^5}-\frac{4352 \
\ln 2\, }{9 (x-2)^5}-\frac{4352 \ln 2\, }{9 (x-2)^6}\Big)+\frac{25 \
\pi ^2 x}{216 (\kappa +1)}+\frac{242 x}{81 (\kappa +1)}+\frac{253 \pi \
^2 x \kappa }{216 (\kappa +1)}-\frac{112 \pi ^2 \kappa }{27 (x-2) \
(\kappa +1)}+\frac{301 \pi ^2 \kappa }{72 (x-1) (\kappa \
+1)}+\frac{140 \pi ^2 \kappa }{27 (x-2)^2 (\kappa +1)}-\frac{14 \pi \
^2 \kappa }{27 (x-1)^2 (\kappa +1)}-\frac{280 \pi ^2 \kappa }{27 \
(x-2)^3 (\kappa +1)}-\frac{7 \pi ^2 \kappa }{216 (x-1)^3 (\kappa \
+1)}-\frac{8 d_1 \pi ^2 \kappa }{3 (x-2)^4 (\kappa +1)}-\frac{664 \pi \
^2 \kappa }{9 (x-2)^4 (\kappa +1)}-\frac{379 \pi ^2 \kappa }{216 \
(x-1)^4 (\kappa +1)}-\frac{32 d_1 \pi ^2 \kappa }{3 (x-2)^5 (\kappa \
+1)}-\frac{512 \pi ^2 \kappa }{3 (x-2)^5 (\kappa +1)}-\frac{253 \pi \
^2 \kappa }{216 (x-1)^5 (\kappa +1)}-\frac{32 d_1 \pi ^2 \kappa }{3 \
(x-2)^6 (\kappa +1)}-\frac{416 \pi ^2 \kappa }{9 (x-2)^6 (\kappa \
+1)}+\frac{7 \pi ^2 \kappa }{12 (\kappa +1)}-\frac{16 \pi ^2}{27 \
(x-2) (\kappa +1)}+\frac{43 \pi ^2}{72 (x-1) (\kappa +1)}+\frac{20 \
\pi ^2}{27 (x-2)^2 (\kappa +1)}-\frac{2 \pi ^2}{27 (x-1)^2 (\kappa \
+1)}-\frac{40 \pi ^2}{27 (x-2)^3 (\kappa +1)}-\frac{\pi ^2}{216 \
(x-1)^3 (\kappa +1)}-\frac{8 d_1 \pi ^2}{9 (x-2)^4 (\kappa \
+1)}-\frac{376 \pi ^2}{27 (x-2)^4 (\kappa +1)}-\frac{23 \pi ^2}{72 \
(x-1)^4 (\kappa +1)}-\frac{32 d_1 \pi ^2}{9 (x-2)^5 (\kappa \
+1)}-\frac{1024 \pi ^2}{27 (x-2)^5 (\kappa +1)}-\frac{17 \pi ^2}{72 \
(x-1)^5 (\kappa +1)}-\frac{32 d_1 \pi ^2}{9 (x-2)^6 (\kappa \
+1)}-\frac{544 \pi ^2}{27 (x-2)^6 (\kappa +1)}+\frac{\pi ^2}{12 \
(\kappa +1)}-\frac{x \zeta_3}{\kappa +1}-\frac{7 x \kappa  \zeta_3}{3 \
(\kappa +1)}+\frac{7 \kappa  \zeta_3}{3 (x-1)^4 (\kappa +1)}+\frac{7 \
\kappa  \zeta_3}{3 (x-1)^5 (\kappa +1)}+\frac{\zeta_3}{3 (x-1)^4 \
(\kappa +1)}+\frac{\zeta_3}{3 (x-1)^5 (\kappa +1)}-\frac{1120 \kappa  \
\ln ^32\, }{9 (x-2)^4 (\kappa +1)}-\frac{4480 \kappa  \ln ^32\, }{9 \
(x-2)^5 (\kappa +1)}-\frac{4480 \kappa  \ln ^32\, }{9 (x-2)^6 (\kappa \
+1)}-\frac{160 \ln ^32\, }{9 (x-2)^4 (\kappa +1)}-\frac{640 \ln ^32\, \
}{9 (x-2)^5 (\kappa +1)}-\frac{640 \ln ^32\, }{9 (x-2)^6 (\kappa \
+1)}-\frac{32 d_1 \kappa  \ln ^22\, }{(x-2)^4 (\kappa +1)}-\frac{416 \
\kappa  \ln ^22\, }{3 (x-2)^4 (\kappa +1)}-\frac{128 d_1 \kappa  \ln \
^22\, }{(x-2)^5 (\kappa +1)}-\frac{1664 \kappa  \ln ^22\, }{3 (x-2)^5 \
(\kappa +1)}-\frac{128 d_1 \kappa  \ln ^22\, }{(x-2)^6 (\kappa \
+1)}-\frac{1664 \kappa  \ln ^22\, }{3 (x-2)^6 (\kappa +1)}-\frac{32 \
d_1 \ln ^22\, }{3 (x-2)^4 (\kappa +1)}-\frac{544 \ln ^22\, }{9 \
(x-2)^4 (\kappa +1)}-\frac{128 d_1 \ln ^22\, }{3 (x-2)^5 (\kappa \
+1)}-\frac{2176 \ln ^22\, }{9 (x-2)^5 (\kappa +1)}-\frac{128 d_1 \ln \
^22\, }{3 (x-2)^6 (\kappa +1)}-\frac{2176 \ln ^22\, }{9 (x-2)^6 \
(\kappa +1)}-\frac{256 d_1 \kappa  \ln 2\, }{9 (x-2)^4 (\kappa \
+1)}-\frac{80 \pi ^2 \kappa  \ln 2\, }{3 (x-2)^4 (\kappa \
+1)}-\frac{1856 \kappa  \ln 2\, }{27 (x-2)^4 (\kappa +1)}-\frac{1024 \
d_1 \kappa  \ln 2\, }{9 (x-2)^5 (\kappa +1)}-\frac{320 \pi ^2 \kappa  \
\ln 2\, }{3 (x-2)^5 (\kappa +1)}-\frac{7424 \kappa  \ln 2\, }{27 \
(x-2)^5 (\kappa +1)}-\frac{1024 d_1 \kappa  \ln 2\, }{9 (x-2)^6 \
(\kappa +1)}-\frac{320 \pi ^2 \kappa  \ln 2\, }{3 (x-2)^6 (\kappa \
+1)}-\frac{7424 \kappa  \ln 2\, }{27 (x-2)^6 (\kappa +1)}-\frac{256 \
d_1 \ln 2\, }{9 (x-2)^4 (\kappa +1)}-\frac{3392 \ln 2\, }{27 (x-2)^4 \
(\kappa +1)}-\frac{1024 d_1 \ln 2\, }{9 (x-2)^5 (\kappa \
+1)}-\frac{13568 \ln 2\, }{27 (x-2)^5 (\kappa +1)}-\frac{1024 d_1 \ln \
2\, }{9 (x-2)^6 (\kappa +1)}-\frac{13568 \ln 2\, }{27 (x-2)^6 (\kappa \
+1)}
\Big\}.
\erp

%
% The A integral for k=-1 and kappa=0
%

\subsection{The $\cA$ integral for $k=-1$ and $\kappa=0$}
%
% This file contains the TeX output produced by Mathematica for the integral Am1, for kappa = 0
%
The $\eps$ expansion for this integral reads
\beq
\bsp
\begin{cal}I\end{cal}(x,\eps;\ao,3+d_1\eps;0,2,0,g_A) &= x\,\aint(\eps,x;3+d_1\eps;0,2)\\
&=\frac{1}{\eps^2}a_{-2}^{(0,-1)}+\frac{1}{\eps}a_{-1}^{(0,-1)}+a_0^{(0,-1)}+\eps a_1^{(0,-1)}+\eps^2 a_2^{(0,-1)} \ocal\left(\eps^3\right),
\esp
\eeq
where
%1/ep piece
\brp
a_{-2}^{(0,-1)}=\frac{1}{2},
\erp
\brp
a_{-1}^{(0,-1)} = -H(0;x),
\erp
% ep^0
\brp
a_0^{(0,-1)} = \frac{\ao^3}{12 (x-1)^2}-\frac{\ao^3}{12}+\frac{\ao^2}{24 \
(x-1)}-\frac{5 \ao^2}{24 (x-1)^2}+\frac{7 \ao^2}{24 (x-1)^3}+\frac{13 \
\ao^2}{24}-\frac{\ao}{3 (x-1)}+\frac{\ao}{6 (x-1)^2}-\frac{\ao}{3 \
(x-1)^3}+\frac{13 \ao}{12 (x-1)^4}-\frac{23 \
\ao}{12}+\Big(\frac{25}{12}+\frac{3}{4 (x-1)}-\frac{1}{6 \
(x-1)^2}-\frac{1}{6 (x-1)^3}+\frac{3}{4 (x-1)^4}+\frac{25}{12 \
(x-1)^5}\Big) H(0;\ao)+\Big(-\frac{25}{12}-\frac{3}{4 \
(x-1)}+\frac{1}{6 (x-1)^2}+\frac{1}{6 (x-1)^3}-\frac{3}{4 \
(x-1)^4}-\frac{25}{12 (x-1)^5}\Big) \
H(0;x)+\Big(\frac{1}{(x-1)^5}-1\Big) H(0;\ao) \
H(1;x)+\Big(-\frac{\ao^4}{4 (x-1)}+\frac{\ao^4}{4}+\frac{\ao^3}{x-1}-\
\frac{\ao^3}{3 (x-1)^2}-\frac{4 \ao^3}{3}-\frac{3 \ao^2}{2 \
(x-1)}+\frac{\ao^2}{(x-1)^2}-\frac{\ao^2}{2 (x-1)^3}+3 \
\ao^2+\frac{\ao}{x-1}-\frac{\ao}{(x-1)^2}+\frac{\ao}{(x-1)^3}-\frac{\ao}{(x-1)^4}-4 \ao+\frac{3}{4 (x-1)}-\frac{1}{6 (x-1)^2}-\frac{1}{6 \
(x-1)^3}+\frac{3}{4 (x-1)^4}+\frac{25}{12 (x-1)^5}+\frac{25}{12}\Big) \
H(c_1(\ao);x)+2 H(0,0;x)+\Big(\frac{1}{(x-1)^5}-1\Big) \
H(0,c_1(\ao);x)+\Big(1-\frac{1}{(x-1)^5}\Big) \
H(1,0;x)+\Big(\frac{1}{(x-1)^5}-1\Big) \
H(1,c_1(\ao);x)-\frac{H(c_1(\ao),c_1(\ao);x)}{(x-1)^5}-\frac{\pi^2}{6 (x-1)^5}-\frac{\pi ^2}{12},
\erp
% ep^0
\brp
a_1^{(0,-1)} = \frac{7 d_1 \ao^3}{72}-\frac{7 d_1 \ao^3}{72 (x-1)^2}+\frac{7 \
\ao^3}{72 (x-1)^2}-\frac{7 \ao^3}{72}-\frac{109 d_1 \
\ao^2}{144}-\frac{13 d_1 \ao^2}{144 (x-1)}-\frac{5 \ao^2}{144 (x-1)}+\
\frac{29 d_1 \ao^2}{144 (x-1)^2}-\frac{47 \ao^2}{144 \
(x-1)^2}-\frac{67 d_1 \ao^2}{144 (x-1)^3}+\frac{85 \ao^2}{144 \
(x-1)^3}+\frac{127 \ao^2}{144}+\frac{305 d_1 \ao}{72}+\frac{19 d_1 \
\ao}{18 (x-1)}-\frac{2 \ao}{9 (x-1)}-\frac{d_1 \ao}{9 \
(x-1)^2}+\frac{13 \ao}{36 (x-1)^2}+\frac{d_1 \ao}{18 (x-1)^3}-\frac{8 \
\ao}{9 (x-1)^3}-\frac{217 d_1 \ao}{72 (x-1)^4}+\frac{149 \ao}{36 \
(x-1)^4}-\frac{101 \ao}{18}+\Big(-\frac{\ao^3}{6 \
(x-1)^2}+\frac{\ao^3}{6}-\frac{\ao^2}{12 (x-1)}+\frac{5 \ao^2}{12 \
(x-1)^2}-\frac{7 \ao^2}{12 (x-1)^3}-\frac{13 \ao^2}{12}+\frac{2 \
\ao}{3 (x-1)}-\frac{\ao}{3 (x-1)^2}+\frac{2 \ao}{3 (x-1)^3}-\frac{13 \
\ao}{6 (x-1)^4}+\frac{23 \ao}{6}-\frac{205 d_1}{72}-\frac{15 d_1}{8 \
(x-1)}+\frac{1}{4 (x-1)}+\frac{5 d_1}{18 (x-1)^2}-\frac{7}{36 \
(x-1)^2}+\frac{5 d_1}{18 (x-1)^3}-\frac{13}{36 (x-1)^3}-\frac{15 \
d_1}{8 (x-1)^4}+\frac{3}{(x-1)^4}-\frac{205 d_1}{72 \
(x-1)^5}+\frac{65}{9 (x-1)^5}+\frac{155}{36}\Big) \
H(0;\ao)+\Big(\frac{15 d_1}{8 (x-1)}-\frac{5 d_1}{18 (x-1)^2}-\frac{5 \
d_1}{18 (x-1)^3}+\frac{15 d_1}{8 (x-1)^4}+\frac{205 d_1}{72 (x-1)^5}+\
\frac{205 d_1}{72}-\frac{1}{4 (x-1)}+\frac{7}{36 \
(x-1)^2}+\frac{13}{36 (x-1)^3}-\frac{3}{(x-1)^4}+\frac{\pi ^2}{3 \
(x-1)^5}-\frac{65}{9 (x-1)^5}+\frac{\pi ^2}{6}-\frac{155}{36}\Big) \
H(0;x)+\Big(\frac{d_1 \ao^3}{6}-\frac{d_1 \ao^3}{6 (x-1)^2}-\frac{13 \
d_1 \ao^2}{12}-\frac{d_1 \ao^2}{12 (x-1)}+\frac{5 d_1 \ao^2}{12 \
(x-1)^2}-\frac{7 d_1 \ao^2}{12 (x-1)^3}+\frac{23 d_1 \ao}{6}+\frac{2 \
d_1 \ao}{3 (x-1)}-\frac{d_1 \ao}{3 (x-1)^2}+\frac{2 d_1 \ao}{3 \
(x-1)^3}-\frac{13 d_1 \ao}{6 (x-1)^4}-\frac{35 d_1}{12}-\frac{7 \
d_1}{12 (x-1)}+\frac{d_1}{12 (x-1)^2}-\frac{d_1}{12 (x-1)^3}+\frac{13 \
d_1}{6 (x-1)^4}\Big) H(1;\ao)+\Big(-\frac{d_1 \ao^4}{8}+\frac{d_1 \
\ao^4}{8 (x-1)}-\frac{\ao^4}{8 (x-1)}+\frac{\ao^4}{8}+\frac{13 d_1 \
\ao^3}{18}-\frac{d_1 \ao^3}{2 (x-1)}+\frac{5 \ao^3}{6 (x-1)}+\frac{2 \
d_1 \ao^3}{9 (x-1)^2}-\frac{17 \ao^3}{36 (x-1)^2}-\frac{29 \
\ao^3}{36}-\frac{23 d_1 \ao^2}{12}+\frac{3 d_1 \ao^2}{4 \
(x-1)}-\frac{19 \ao^2}{8 (x-1)}-\frac{2 d_1 \ao^2}{3 \
(x-1)^2}+\frac{47 \ao^2}{24 (x-1)^2}+\frac{d_1 \ao^2}{2 \
(x-1)^3}-\frac{11 \ao^2}{8 (x-1)^3}+\frac{59 \ao^2}{24}+\frac{25 d_1 \
\ao}{6}-\frac{d_1 \ao}{2 (x-1)}+\frac{9 \ao}{2 (x-1)}+\frac{2 d_1 \
\ao}{3 (x-1)^2}-\frac{23 \ao}{6 (x-1)^2}-\frac{d_1 \
\ao}{(x-1)^3}+\frac{4 \ao}{(x-1)^3}+\frac{2 d_1 \
\ao}{(x-1)^4}-\frac{21 \ao}{4 (x-1)^4}-\frac{73 \ao}{12}-\frac{205 \
d_1}{72}+\Big(\frac{\ao^4}{2 (x-1)}-\frac{\ao^4}{2}-\frac{2 \
\ao^3}{x-1}+\frac{2 \ao^3}{3 (x-1)^2}+\frac{8 \ao^3}{3}+\frac{3 \
\ao^2}{x-1}-\frac{2 \ao^2}{(x-1)^2}+\frac{\ao^2}{(x-1)^3}-6 \
\ao^2-\frac{2 \ao}{x-1}+\frac{2 \ao}{(x-1)^2}-\frac{2 \
\ao}{(x-1)^3}+\frac{2 \ao}{(x-1)^4}+8 \ao-\frac{3}{2 \
(x-1)}+\frac{1}{3 (x-1)^2}+\frac{1}{3 (x-1)^3}-\frac{3}{2 \
(x-1)^4}-\frac{25}{6 (x-1)^5}-\frac{25}{6}\Big) \
H(0;\ao)+\Big(-\frac{d_1 \ao^4}{2}+\frac{d_1 \ao^4}{2 (x-1)}+\frac{8 \
d_1 \ao^3}{3}-\frac{2 d_1 \ao^3}{x-1}+\frac{2 d_1 \ao^3}{3 (x-1)^2}-6 \
d_1 \ao^2+\frac{3 d_1 \ao^2}{x-1}-\frac{2 d_1 \
\ao^2}{(x-1)^2}+\frac{d_1 \ao^2}{(x-1)^3}+8 d_1 \ao-\frac{2 d_1 \
\ao}{x-1}+\frac{2 d_1 \ao}{(x-1)^2}-\frac{2 d_1 \ao}{(x-1)^3}+\frac{2 \
d_1 \ao}{(x-1)^4}-\frac{25 d_1}{6}-\frac{3 d_1}{2 (x-1)}+\frac{d_1}{3 \
(x-1)^2}+\frac{d_1}{3 (x-1)^3}-\frac{3 d_1}{2 (x-1)^4}-\frac{25 \
d_1}{6 (x-1)^5}\Big) H(1;\ao)-\frac{15 d_1}{8 (x-1)}+\frac{1}{4 \
(x-1)}+\frac{5 d_1}{18 (x-1)^2}-\frac{7}{36 (x-1)^2}+\frac{5 d_1}{18 \
(x-1)^3}-\frac{13}{36 (x-1)^3}-\frac{15 d_1}{8 \
(x-1)^4}+\frac{3}{(x-1)^4}-\frac{205 d_1}{72 (x-1)^5}+\frac{65}{9 \
(x-1)^5}+\frac{155}{36}\Big) \
H(c_1(\ao);x)+\Big(-\frac{25}{6}-\frac{3}{2 (x-1)}+\frac{1}{3 \
(x-1)^2}+\frac{1}{3 (x-1)^3}-\frac{3}{2 (x-1)^4}-\frac{25}{6 (x-1)^5}\
\Big) H(0,0;\ao)+\Big(\frac{25}{6}+\frac{3}{2 (x-1)}-\frac{1}{3 \
(x-1)^2}-\frac{1}{3 (x-1)^3}+\frac{3}{2 (x-1)^4}+\frac{25}{6 (x-1)^5}\
\Big) H(0,0;x)+\Big(-\frac{3 d_1}{2 (x-1)}+\frac{d_1}{3 \
(x-1)^2}+\frac{d_1}{3 (x-1)^3}-\frac{3 d_1}{2 (x-1)^4}-\frac{25 \
d_1}{6 (x-1)^5}-\frac{25 d_1}{6}\Big) H(0,1;\ao)+H(1;x) \
\Big(-\frac{\pi ^2 d_1}{3 (x-1)^5}+\Big(\frac{2 \
d_1}{x-1}-\frac{d_1}{(x-1)^2}+\frac{2 d_1}{3 (x-1)^3}-\frac{d_1}{2 \
(x-1)^4}+\frac{25 d_1}{6 (x-1)^5}+\frac{5}{4 (x-1)}-\frac{5}{6 \
(x-1)^2}+\frac{5}{6 (x-1)^3}-\frac{5}{4 (x-1)^4}+\frac{25}{12 \
(x-1)^5}-\frac{25}{12}\Big) H(0;\ao)+\Big(2-\frac{2}{(x-1)^5}\Big) \
H(0,0;\ao)+\Big(2 d_1-\frac{2 d_1}{(x-1)^5}\Big) H(0,1;\ao)+\frac{\pi^2}{6 (x-1)^5}+\frac{\pi ^2}{6}\Big)+\Big(\frac{2 d_1}{(x-1)^5}-2 d_1\
\Big) H(0;\ao) H(0,1;x)+\Big(-\frac{\ao^4}{2 \
(x-1)}+\frac{\ao^4}{2}+\frac{2 \ao^3}{x-1}-\frac{2 \ao^3}{3 (x-1)^2}-\
\frac{8 \ao^3}{3}-\frac{3 \ao^2}{x-1}+\frac{2 \
\ao^2}{(x-1)^2}-\frac{\ao^2}{(x-1)^3}+6 \ao^2+\frac{2 \
\ao}{x-1}-\frac{2 \ao}{(x-1)^2}+\frac{2 \ao}{(x-1)^3}-\frac{2 \
\ao}{(x-1)^4}-8 \ao+\Big(2-\frac{2}{(x-1)^5}\Big) H(0;\ao)+\Big(2 \
d_1-\frac{2 d_1}{(x-1)^5}\Big) H(1;\ao)+\frac{3}{4 (x-1)}-\frac{1}{6 \
(x-1)^2}-\frac{1}{6 (x-1)^3}+\frac{3}{4 (x-1)^4}+\frac{25}{12 \
(x-1)^5}+\frac{25}{12}\Big) H(0,c_1(\ao);x)+\Big(-\frac{2 \
d_1}{x-1}+\frac{d_1}{(x-1)^2}-\frac{2 d_1}{3 (x-1)^3}+\frac{d_1}{2 \
(x-1)^4}-\frac{25 d_1}{6 (x-1)^5}-\frac{5}{4 (x-1)}+\frac{5}{6 \
(x-1)^2}-\frac{5}{6 (x-1)^3}+\frac{5}{4 (x-1)^4}-\frac{25}{12 \
(x-1)^5}+\frac{25}{12}\Big) H(1,0;x)+\Big(\frac{4 d_1}{(x-1)^5}-2 \
d_1-\frac{1}{(x-1)^5}-1\Big) H(0;\ao) H(1,1;x)+\Big(\frac{2 \
d_1}{x-1}-\frac{d_1}{(x-1)^2}+\frac{2 d_1}{3 (x-1)^3}-\frac{d_1}{2 \
(x-1)^4}+\frac{25 d_1}{6 (x-1)^5}+\Big(2-\frac{2}{(x-1)^5}\Big) \
H(0;\ao)+\Big(2 d_1-\frac{2 d_1}{(x-1)^5}\Big) H(1;\ao)+\frac{5}{4 \
(x-1)}-\frac{5}{6 (x-1)^2}+\frac{5}{6 (x-1)^3}-\frac{5}{4 \
(x-1)^4}+\frac{25}{12 (x-1)^5}-\frac{25}{12}\Big) \
H(1,c_1(\ao);x)+\Big(\frac{3 \ao^4}{4 (x-1)}-\frac{3 \
\ao^4}{4}-\frac{3 \ao^3}{x-1}+\frac{\ao^3}{(x-1)^2}+4 \ao^3+\frac{9 \
\ao^2}{2 (x-1)}-\frac{3 \ao^2}{(x-1)^2}+\frac{3 \ao^2}{2 (x-1)^3}-9 \
\ao^2-\frac{3 \ao}{x-1}+\frac{3 \ao}{(x-1)^2}-\frac{3 \
\ao}{(x-1)^3}+\frac{3 \ao}{(x-1)^4}+12 \ao+\frac{2 \
H(0;\ao)}{(x-1)^5}+\frac{2 d_1 H(1;\ao)}{(x-1)^5}-\frac{9}{4 \
(x-1)}+\frac{1}{2 (x-1)^2}+\frac{1}{2 (x-1)^3}-\frac{9}{4 \
(x-1)^4}-\frac{25}{4 (x-1)^5}-\frac{25}{4}\Big) \
H(c_1(\ao),c_1(\ao);x)-4 H(0,0,0;x)+\Big(2 d_1-\frac{2 \
d_1}{(x-1)^5}\Big) H(0,1,0;x)+\Big(\frac{2 d_1}{(x-1)^5}-2 d_1\Big) \
H(0,1,c_1(\ao);x)+\Big(3-\frac{1}{(x-1)^5}\Big) \
H(0,c_1(\ao),c_1(\ao);x)+\Big(\frac{2}{(x-1)^5}-2\Big) \
H(1,0,0;x)+\Big(\frac{2 d_1}{(x-1)^5}-\frac{1}{(x-1)^5}-1\Big) \
H(1,0,c_1(\ao);x)+\Big(-\frac{4 d_1}{(x-1)^5}+2 \
d_1+\frac{1}{(x-1)^5}+1\Big) H(1,1,0;x)+\Big(\frac{4 d_1}{(x-1)^5}-2 \
d_1-\frac{1}{(x-1)^5}-1\Big) H(1,1,c_1(\ao);x)+\Big(-\frac{2 \
d_1}{(x-1)^5}-\frac{1}{(x-1)^5}+3\Big) \
H(1,c_1(\ao),c_1(\ao);x)-\frac{2 \
H(c_1(\ao),0,c_1(\ao);x)}{(x-1)^5}+\frac{3 \
H(c_1(\ao),c_1(\ao),c_1(\ao);x)}{(x-1)^5}-\frac{\pi ^2}{8 \
(x-1)}+\frac{\pi ^2}{36 (x-1)^2}+\frac{\pi ^2}{36 (x-1)^3}-\frac{\pi^2}{8 (x-1)^4}-\frac{25 \pi ^2}{72 (x-1)^5}-\frac{3 \
\zeta_3}{(x-1)^5}-5 \zeta_3-\frac{25 \pi ^2}{72},
\erp
% ep^2
\brp
a_2^{(0,-1)} = -\frac{37}{432} d_1^2 \ao^3+\frac{37 d_1 \ao^3}{216}+\frac{37 d_1^2 \
\ao^3}{432 (x-1)^2}-\frac{37 d_1 \ao^3}{216 (x-1)^2}-\frac{\pi ^2 \
\ao^3}{72 (x-1)^2}+\frac{37 \ao^3}{432 (x-1)^2}+\frac{\pi ^2 \
\ao^3}{72}-\frac{37 \ao^3}{432}+\frac{715 d_1^2 \ao^2}{864}-\frac{52 \
d_1 \ao^2}{27}+\frac{115 d_1^2 \ao^2}{864 (x-1)}+\frac{d_1 \ao^2}{216 \
(x-1)}-\frac{\pi ^2 \ao^2}{144 (x-1)}-\frac{119 \ao^2}{864 \
(x-1)}-\frac{107 d_1^2 \ao^2}{864 (x-1)^2}+\frac{14 d_1 \ao^2}{27 \
(x-1)^2}+\frac{5 \pi ^2 \ao^2}{144 (x-1)^2}-\frac{341 \ao^2}{864 \
(x-1)^2}+\frac{493 d_1^2 \ao^2}{864 (x-1)^3}-\frac{305 d_1 \ao^2}{216 \
(x-1)^3}-\frac{7 \pi ^2 \ao^2}{144 (x-1)^3}+\frac{727 \ao^2}{864 \
(x-1)^3}-\frac{13 \pi ^2 \ao^2}{144}+\frac{949 \ao^2}{864}-\frac{3515 \
d_1^2 \ao}{432}+\frac{8965 d_1 \ao}{432}-\frac{265 d_1^2 \ao}{108 \
(x-1)}+\frac{263 d_1 \ao}{108 (x-1)}+\frac{\pi ^2 \ao}{18 \
(x-1)}+\frac{65 \ao}{108 (x-1)}-\frac{d_1^2 \ao}{108 \
(x-1)^2}-\frac{113 d_1 \ao}{216 (x-1)^2}-\frac{\pi ^2 \ao}{36 \
(x-1)^2}+\frac{115 \ao}{216 (x-1)^2}+\frac{113 d_1^2 \ao}{108 \
(x-1)^3}+\frac{41 d_1 \ao}{108 (x-1)^3}+\frac{\pi ^2 \ao}{18 \
(x-1)^3}-\frac{217 \ao}{108 (x-1)^3}+\frac{2911 d_1^2 \ao}{432 \
(x-1)^4}-\frac{7523 d_1 \ao}{432 (x-1)^4}-\frac{13 \pi ^2 \ao}{72 \
(x-1)^4}+\frac{2369 \ao}{216 (x-1)^4}+\frac{23 \pi ^2 \
\ao}{72}-\frac{697 \ao}{54}+\Big(-\frac{7 d_1 \ao^3}{36}+\frac{7 d_1 \
\ao^3}{36 (x-1)^2}-\frac{7 \ao^3}{36 (x-1)^2}+\frac{7 \
\ao^3}{36}+\frac{109 d_1 \ao^2}{72}+\frac{13 d_1 \ao^2}{72 \
(x-1)}+\frac{5 \ao^2}{72 (x-1)}-\frac{29 d_1 \ao^2}{72 \
(x-1)^2}+\frac{47 \ao^2}{72 (x-1)^2}+\frac{67 d_1 \ao^2}{72 (x-1)^3}-\
\frac{85 \ao^2}{72 (x-1)^3}-\frac{127 \ao^2}{72}-\frac{305 d_1 \
\ao}{36}-\frac{19 d_1 \ao}{9 (x-1)}+\frac{4 \ao}{9 (x-1)}+\frac{2 d_1 \
\ao}{9 (x-1)^2}-\frac{13 \ao}{18 (x-1)^2}-\frac{d_1 \ao}{9 \
(x-1)^3}+\frac{16 \ao}{9 (x-1)^3}+\frac{217 d_1 \ao}{36 \
(x-1)^4}-\frac{149 \ao}{18 (x-1)^4}+\frac{101 \ao}{9}+\frac{2035 \
d_1^2}{432}-\frac{5615 d_1}{432}+\frac{63 d_1^2}{16 (x-1)}-\frac{209 \
d_1}{48 (x-1)}-\frac{\pi ^2}{8 (x-1)}-\frac{2}{x-1}-\frac{19 \
d_1^2}{54 (x-1)^2}+\frac{347 d_1}{216 (x-1)^2}+\frac{\pi ^2}{36 \
(x-1)^2}-\frac{19}{216 (x-1)^2}-\frac{19 d_1^2}{54 (x-1)^3}+\frac{173 \
d_1}{216 (x-1)^3}+\frac{\pi ^2}{36 (x-1)^3}-\frac{205}{216 \
(x-1)^3}+\frac{63 d_1^2}{16 (x-1)^4}-\frac{197 d_1}{16 \
(x-1)^4}-\frac{\pi ^2}{8 (x-1)^4}+\frac{163}{24 (x-1)^4}+\frac{2035 \
d_1^2}{432 (x-1)^5}-\frac{8705 d_1}{432 (x-1)^5}-\frac{25 \pi ^2}{72 \
(x-1)^5}+\frac{3965}{216 (x-1)^5}-\frac{25 \pi ^2}{72}+\frac{235}{27}\
\Big) H(0;\ao)+\Big(-\frac{7}{36} d_1^2 \ao^3+\frac{7 d_1 \ao^3}{36}+\
\frac{7 d_1^2 \ao^3}{36 (x-1)^2}-\frac{7 d_1 \ao^3}{36 \
(x-1)^2}+\frac{109 d_1^2 \ao^2}{72}-\frac{127 d_1 \ao^2}{72}+\frac{13 \
d_1^2 \ao^2}{72 (x-1)}+\frac{5 d_1 \ao^2}{72 (x-1)}-\frac{29 d_1^2 \
\ao^2}{72 (x-1)^2}+\frac{47 d_1 \ao^2}{72 (x-1)^2}+\frac{67 d_1^2 \
\ao^2}{72 (x-1)^3}-\frac{85 d_1 \ao^2}{72 (x-1)^3}-\frac{305 d_1^2 \
\ao}{36}+\frac{101 d_1 \ao}{9}-\frac{19 d_1^2 \ao}{9 (x-1)}+\frac{4 \
d_1 \ao}{9 (x-1)}+\frac{2 d_1^2 \ao}{9 (x-1)^2}-\frac{13 d_1 \ao}{18 \
(x-1)^2}-\frac{d_1^2 \ao}{9 (x-1)^3}+\frac{16 d_1 \ao}{9 \
(x-1)^3}+\frac{217 d_1^2 \ao}{36 (x-1)^4}-\frac{149 d_1 \ao}{18 \
(x-1)^4}+\frac{515 d_1^2}{72}-\frac{695 d_1}{72}+\frac{139 d_1^2}{72 \
(x-1)}-\frac{37 d_1}{72 (x-1)}-\frac{d_1^2}{72 (x-1)^2}+\frac{19 \
d_1}{72 (x-1)^2}-\frac{59 d_1^2}{72 (x-1)^3}-\frac{43 d_1}{72 \
(x-1)^3}-\frac{217 d_1^2}{36 (x-1)^4}+\frac{149 d_1}{18 (x-1)^4}\Big) \
H(1;\ao)+\Big(\frac{\ao^3}{3 (x-1)^2}-\frac{\ao^3}{3}+\frac{\ao^2}{6 \
(x-1)}-\frac{5 \ao^2}{6 (x-1)^2}+\frac{7 \ao^2}{6 (x-1)^3}+\frac{13 \
\ao^2}{6}-\frac{4 \ao}{3 (x-1)}+\frac{2 \ao}{3 (x-1)^2}-\frac{4 \
\ao}{3 (x-1)^3}+\frac{13 \ao}{3 (x-1)^4}-\frac{23 \ao}{3}+\frac{205 \
d_1}{36}+\frac{15 d_1}{4 (x-1)}-\frac{1}{2 (x-1)}-\frac{5 d_1}{9 \
(x-1)^2}+\frac{7}{18 (x-1)^2}-\frac{5 d_1}{9 (x-1)^3}+\frac{13}{18 \
(x-1)^3}+\frac{15 d_1}{4 (x-1)^4}-\frac{6}{(x-1)^4}+\frac{205 d_1}{36 \
(x-1)^5}-\frac{130}{9 (x-1)^5}-\frac{155}{18}\Big) \
H(0,0;\ao)+\Big(-\frac{15 d_1}{4 (x-1)}+\frac{5 d_1}{9 \
(x-1)^2}+\frac{5 d_1}{9 (x-1)^3}-\frac{15 d_1}{4 (x-1)^4}-\frac{205 \
d_1}{36 (x-1)^5}-\frac{205 d_1}{36}+\frac{1}{2 (x-1)}-\frac{7}{18 \
(x-1)^2}-\frac{13}{18 (x-1)^3}+\frac{6}{(x-1)^4}-\frac{2 \pi ^2}{3 \
(x-1)^5}+\frac{130}{9 (x-1)^5}-\frac{\pi ^2}{3}+\frac{155}{18}\Big) \
H(0,0;x)+\Big(-\frac{d_1 \ao^3}{3}+\frac{d_1 \ao^3}{3 \
(x-1)^2}+\frac{13 d_1 \ao^2}{6}+\frac{d_1 \ao^2}{6 (x-1)}-\frac{5 d_1 \
\ao^2}{6 (x-1)^2}+\frac{7 d_1 \ao^2}{6 (x-1)^3}-\frac{23 d_1 \ao}{3}-\
\frac{4 d_1 \ao}{3 (x-1)}+\frac{2 d_1 \ao}{3 (x-1)^2}-\frac{4 d_1 \
\ao}{3 (x-1)^3}+\frac{13 d_1 \ao}{3 (x-1)^4}+\frac{205 \
d_1^2}{36}-\frac{155 d_1}{18}+\frac{15 d_1^2}{4 (x-1)}-\frac{d_1}{2 \
(x-1)}-\frac{5 d_1^2}{9 (x-1)^2}+\frac{7 d_1}{18 (x-1)^2}-\frac{5 \
d_1^2}{9 (x-1)^3}+\frac{13 d_1}{18 (x-1)^3}+\frac{15 d_1^2}{4 \
(x-1)^4}-\frac{6 d_1}{(x-1)^4}+\frac{205 d_1^2}{36 (x-1)^5}-\frac{130 \
d_1}{9 (x-1)^5}\Big) H(0,1;\ao)+\Big(\frac{\pi ^2 d_1}{3 \
(x-1)^5}+\frac{\pi ^2 d_1}{3}+\Big(\frac{5 d_1}{2 (x-1)}-\frac{5 \
d_1}{3 (x-1)^2}+\frac{5 d_1}{3 (x-1)^3}-\frac{5 d_1}{2 \
(x-1)^4}+\frac{25 d_1}{6 (x-1)^5}-\frac{25 d_1}{6}+\frac{5}{2 (x-1)}-\
\frac{5}{3 (x-1)^2}+\frac{5}{3 (x-1)^3}-\frac{5}{2 \
(x-1)^4}+\frac{25}{6 (x-1)^5}-\frac{25}{6}\Big) H(0;\ao)+\Big(4 \
d_1-\frac{4 d_1}{(x-1)^5}\Big) H(0,0;\ao)+\Big(4 d_1^2-\frac{4 \
d_1^2}{(x-1)^5}\Big) H(0,1;\ao)-\frac{2 \pi ^2}{3 (x-1)^5}\Big) \
H(0,1;x)+\Big(-\frac{d_1 \ao^3}{3}+\frac{d_1 \ao^3}{3 \
(x-1)^2}+\frac{13 d_1 \ao^2}{6}+\frac{d_1 \ao^2}{6 (x-1)}-\frac{5 d_1 \
\ao^2}{6 (x-1)^2}+\frac{7 d_1 \ao^2}{6 (x-1)^3}-\frac{23 d_1 \ao}{3}-\
\frac{4 d_1 \ao}{3 (x-1)}+\frac{2 d_1 \ao}{3 (x-1)^2}-\frac{4 d_1 \
\ao}{3 (x-1)^3}+\frac{13 d_1 \ao}{3 (x-1)^4}+\frac{35 d_1}{6}+\frac{7 \
d_1}{6 (x-1)}-\frac{d_1}{6 (x-1)^2}+\frac{d_1}{6 (x-1)^3}-\frac{13 \
d_1}{3 (x-1)^4}\Big) H(1,0;\ao)+\Big(\frac{4 \
d_1^2}{x-1}-\frac{d_1^2}{(x-1)^2}+\frac{4 d_1^2}{9 \
(x-1)^3}-\frac{d_1^2}{4 (x-1)^4}+\frac{205 d_1^2}{36 \
(x-1)^5}-\frac{97 d_1}{24 (x-1)}+\frac{157 d_1}{36 (x-1)^2}-\frac{137 \
d_1}{36 (x-1)^3}+\frac{19 d_1}{8 (x-1)^4}+\frac{2 \pi ^2 d_1}{3 \
(x-1)^5}-\frac{835 d_1}{72 (x-1)^5}-\frac{205 d_1}{72}-\frac{97}{12 \
(x-1)}+\frac{179}{36 (x-1)^2}-\frac{179}{36 (x-1)^3}+\frac{97}{12 \
(x-1)^4}-\frac{\pi ^2}{6 (x-1)^5}-\frac{155}{36 (x-1)^5}-\frac{\pi \
^2}{2}+\frac{155}{36}\Big) H(1,0;x)+\Big(-\frac{1}{3} d_1^2 \
\ao^3+\frac{d_1^2 \ao^3}{3 (x-1)^2}+\frac{13 d_1^2 \
\ao^2}{6}+\frac{d_1^2 \ao^2}{6 (x-1)}-\frac{5 d_1^2 \ao^2}{6 \
(x-1)^2}+\frac{7 d_1^2 \ao^2}{6 (x-1)^3}-\frac{23 d_1^2 \
\ao}{3}-\frac{4 d_1^2 \ao}{3 (x-1)}+\frac{2 d_1^2 \ao}{3 \
(x-1)^2}-\frac{4 d_1^2 \ao}{3 (x-1)^3}+\frac{13 d_1^2 \ao}{3 \
(x-1)^4}+\frac{35 d_1^2}{6}+\frac{7 d_1^2}{6 (x-1)}-\frac{d_1^2}{6 \
(x-1)^2}+\frac{d_1^2}{6 (x-1)^3}-\frac{13 d_1^2}{3 (x-1)^4}\Big) \
H(1,1;\ao)+H(0,c_1(\ao);x) \Big(-\frac{d_1 \ao^4}{4}+\frac{d_1 \
\ao^4}{4 (x-1)}-\frac{\ao^4}{4 (x-1)}+\frac{\ao^4}{4}+\frac{13 d_1 \
\ao^3}{9}-\frac{d_1 \ao^3}{x-1}+\frac{5 \ao^3}{3 (x-1)}+\frac{4 d_1 \
\ao^3}{9 (x-1)^2}-\frac{7 \ao^3}{9 (x-1)^2}-\frac{16 \
\ao^3}{9}-\frac{23 d_1 \ao^2}{6}+\frac{3 d_1 \ao^2}{2 (x-1)}-\frac{14 \
\ao^2}{3 (x-1)}-\frac{4 d_1 \ao^2}{3 (x-1)^2}+\frac{7 \ao^2}{2 \
(x-1)^2}+\frac{d_1 \ao^2}{(x-1)^3}-\frac{13 \ao^2}{6 (x-1)^3}+6 \
\ao^2+\frac{25 d_1 \ao}{3}-\frac{d_1 \ao}{x-1}+\frac{25 \ao}{3 \
(x-1)}+\frac{4 d_1 \ao}{3 (x-1)^2}-\frac{22 \ao}{3 (x-1)^2}-\frac{2 \
d_1 \ao}{(x-1)^3}+\frac{22 \ao}{3 (x-1)^3}+\frac{4 d_1 \ao}{(x-1)^4}-\
\frac{25 \ao}{3 (x-1)^4}-16 \ao-\frac{205 \
d_1}{72}+\Big(\frac{\ao^4}{x-1}-\ao^4-\frac{4 \ao^3}{x-1}+\frac{4 \
\ao^3}{3 (x-1)^2}+\frac{16 \ao^3}{3}+\frac{6 \ao^2}{x-1}-\frac{4 \
\ao^2}{(x-1)^2}+\frac{2 \ao^2}{(x-1)^3}-12 \ao^2-\frac{4 \
\ao}{x-1}+\frac{4 \ao}{(x-1)^2}-\frac{4 \ao}{(x-1)^3}+\frac{4 \
\ao}{(x-1)^4}+16 \ao-\frac{3}{2 (x-1)}+\frac{1}{3 (x-1)^2}+\frac{1}{3 \
(x-1)^3}-\frac{3}{2 (x-1)^4}-\frac{25}{6 (x-1)^5}-\frac{25}{6}\Big) \
H(0;\ao)+\Big(-d_1 \ao^4+\frac{d_1 \ao^4}{x-1}+\frac{16 d_1 \
\ao^3}{3}-\frac{4 d_1 \ao^3}{x-1}+\frac{4 d_1 \ao^3}{3 (x-1)^2}-12 \
d_1 \ao^2+\frac{6 d_1 \ao^2}{x-1}-\frac{4 d_1 \ao^2}{(x-1)^2}+\frac{2 \
d_1 \ao^2}{(x-1)^3}+16 d_1 \ao-\frac{4 d_1 \ao}{x-1}+\frac{4 d_1 \
\ao}{(x-1)^2}-\frac{4 d_1 \ao}{(x-1)^3}+\frac{4 d_1 \
\ao}{(x-1)^4}-\frac{25 d_1}{6}-\frac{3 d_1}{2 (x-1)}+\frac{d_1}{3 \
(x-1)^2}+\frac{d_1}{3 (x-1)^3}-\frac{3 d_1}{2 (x-1)^4}-\frac{25 \
d_1}{6 (x-1)^5}\Big) H(1;\ao)+\Big(\frac{4}{(x-1)^5}-4\Big) \
H(0,0;\ao)+\Big(\frac{4 d_1}{(x-1)^5}-4 d_1\Big) \
H(0,1;\ao)+\Big(\frac{4 d_1}{(x-1)^5}-4 d_1\Big) \
H(1,0;\ao)+\Big(\frac{4 d_1^2}{(x-1)^5}-4 d_1^2\Big) \
H(1,1;\ao)-\frac{15 d_1}{8 (x-1)}+\frac{3}{x-1}+\frac{5 d_1}{18 \
(x-1)^2}-\frac{13}{36 (x-1)^2}+\frac{5 d_1}{18 (x-1)^3}-\frac{7}{36 \
(x-1)^3}-\frac{15 d_1}{8 (x-1)^4}+\frac{1}{4 (x-1)^4}-\frac{205 \
d_1}{72 (x-1)^5}-\frac{\pi ^2}{6 (x-1)^5}+\frac{155}{36 \
(x-1)^5}+\frac{\pi ^2}{6}+\frac{65}{9}\Big)+H(c_1(\ao);x) \
\Big(\frac{d_1^2 \ao^4}{16}-\frac{d_1 \ao^4}{8}-\frac{d_1^2 \ao^4}{16 \
(x-1)}+\frac{d_1 \ao^4}{8 (x-1)}+\frac{\pi ^2 \ao^4}{24 (x-1)}-\frac{\
\ao^4}{16 (x-1)}-\frac{\pi ^2 \ao^4}{24}+\frac{\ao^4}{16}-\frac{43 \
d_1^2 \ao^3}{108}+\frac{193 d_1 \ao^3}{216}+\frac{d_1^2 \ao^3}{4 \
(x-1)}-\frac{8 d_1 \ao^3}{9 (x-1)}-\frac{\pi ^2 \ao^3}{6 \
(x-1)}+\frac{23 \ao^3}{36 (x-1)}-\frac{4 d_1^2 \ao^3}{27 \
(x-1)^2}+\frac{127 d_1 \ao^3}{216 (x-1)^2}+\frac{\pi ^2 \ao^3}{18 \
(x-1)^2}-\frac{95 \ao^3}{216 (x-1)^2}+\frac{2 \pi ^2 \
\ao^3}{9}-\frac{107 \ao^3}{216}+\frac{95 d_1^2 \ao^2}{72}-\frac{163 \
d_1 \ao^2}{48}-\frac{3 d_1^2 \ao^2}{8 (x-1)}+\frac{47 d_1 \ao^2}{16 \
(x-1)}+\frac{\pi ^2 \ao^2}{4 (x-1)}-\frac{47 \ao^2}{16 (x-1)}+\frac{4 \
d_1^2 \ao^2}{9 (x-1)^2}-\frac{125 d_1 \ao^2}{48 (x-1)^2}-\frac{\pi ^2 \
\ao^2}{6 (x-1)^2}+\frac{401 \ao^2}{144 (x-1)^2}-\frac{d_1^2 \ao^2}{2 \
(x-1)^3}+\frac{115 d_1 \ao^2}{48 (x-1)^3}+\frac{\pi ^2 \ao^2}{12 \
(x-1)^3}-\frac{35 \ao^2}{16 (x-1)^3}-\frac{\pi ^2 \ao^2}{2}+\frac{305 \
\ao^2}{144}-\frac{205 d_1^2 \ao}{36}+\frac{125 d_1 \
\ao}{8}+\frac{d_1^2 \ao}{4 (x-1)}-\frac{53 d_1 \ao}{6 \
(x-1)}-\frac{\pi ^2 \ao}{6 (x-1)}+\frac{47 \ao}{4 (x-1)}-\frac{4 \
d_1^2 \ao}{9 (x-1)^2}+\frac{7 d_1 \ao}{(x-1)^2}+\frac{\pi ^2 \ao}{6 \
(x-1)^2}-\frac{389 \ao}{36 (x-1)^2}+\frac{d_1^2 \
\ao}{(x-1)^3}-\frac{49 d_1 \ao}{6 (x-1)^3}-\frac{\pi ^2 \ao}{6 \
(x-1)^3}+\frac{34 \ao}{3 (x-1)^3}-\frac{4 d_1^2 \
\ao}{(x-1)^4}+\frac{409 d_1 \ao}{24 (x-1)^4}+\frac{\pi ^2 \ao}{6 \
(x-1)^4}-\frac{47 \ao}{3 (x-1)^4}+\frac{2 \pi ^2 \ao}{3}-\frac{187 \
\ao}{18}+\frac{2035 d_1^2}{432}-\frac{5615 d_1}{432}+\Big(\frac{d_1 \
\ao^4}{4}-\frac{d_1 \ao^4}{4 (x-1)}+\frac{\ao^4}{4 \
(x-1)}-\frac{\ao^4}{4}-\frac{13 d_1 \ao^3}{9}+\frac{d_1 \
\ao^3}{x-1}-\frac{5 \ao^3}{3 (x-1)}-\frac{4 d_1 \ao^3}{9 \
(x-1)^2}+\frac{17 \ao^3}{18 (x-1)^2}+\frac{29 \ao^3}{18}+\frac{23 d_1 \
\ao^2}{6}-\frac{3 d_1 \ao^2}{2 (x-1)}+\frac{19 \ao^2}{4 \
(x-1)}+\frac{4 d_1 \ao^2}{3 (x-1)^2}-\frac{47 \ao^2}{12 \
(x-1)^2}-\frac{d_1 \ao^2}{(x-1)^3}+\frac{11 \ao^2}{4 \
(x-1)^3}-\frac{59 \ao^2}{12}-\frac{25 d_1 \ao}{3}+\frac{d_1 \
\ao}{x-1}-\frac{9 \ao}{x-1}-\frac{4 d_1 \ao}{3 (x-1)^2}+\frac{23 \
\ao}{3 (x-1)^2}+\frac{2 d_1 \ao}{(x-1)^3}-\frac{8 \
\ao}{(x-1)^3}-\frac{4 d_1 \ao}{(x-1)^4}+\frac{21 \ao}{2 \
(x-1)^4}+\frac{73 \ao}{6}+\frac{205 d_1}{36}+\frac{15 d_1}{4 \
(x-1)}-\frac{1}{2 (x-1)}-\frac{5 d_1}{9 (x-1)^2}+\frac{7}{18 \
(x-1)^2}-\frac{5 d_1}{9 (x-1)^3}+\frac{13}{18 (x-1)^3}+\frac{15 \
d_1}{4 (x-1)^4}-\frac{6}{(x-1)^4}+\frac{205 d_1}{36 \
(x-1)^5}-\frac{130}{9 (x-1)^5}-\frac{155}{18}\Big) \
H(0;\ao)+\Big(\frac{d_1^2 \ao^4}{4}-\frac{d_1 \ao^4}{4}-\frac{d_1^2 \
\ao^4}{4 (x-1)}+\frac{d_1 \ao^4}{4 (x-1)}-\frac{13 d_1^2 \
\ao^3}{9}+\frac{29 d_1 \ao^3}{18}+\frac{d_1^2 \ao^3}{x-1}-\frac{5 d_1 \
\ao^3}{3 (x-1)}-\frac{4 d_1^2 \ao^3}{9 (x-1)^2}+\frac{17 d_1 \
\ao^3}{18 (x-1)^2}+\frac{23 d_1^2 \ao^2}{6}-\frac{59 d_1 \
\ao^2}{12}-\frac{3 d_1^2 \ao^2}{2 (x-1)}+\frac{19 d_1 \ao^2}{4 \
(x-1)}+\frac{4 d_1^2 \ao^2}{3 (x-1)^2}-\frac{47 d_1 \ao^2}{12 \
(x-1)^2}-\frac{d_1^2 \ao^2}{(x-1)^3}+\frac{11 d_1 \ao^2}{4 \
(x-1)^3}-\frac{25 d_1^2 \ao}{3}+\frac{73 d_1 \ao}{6}+\frac{d_1^2 \
\ao}{x-1}-\frac{9 d_1 \ao}{x-1}-\frac{4 d_1^2 \ao}{3 \
(x-1)^2}+\frac{23 d_1 \ao}{3 (x-1)^2}+\frac{2 d_1^2 \
\ao}{(x-1)^3}-\frac{8 d_1 \ao}{(x-1)^3}-\frac{4 d_1^2 \
\ao}{(x-1)^4}+\frac{21 d_1 \ao}{2 (x-1)^4}+\frac{205 \
d_1^2}{36}-\frac{155 d_1}{18}+\frac{15 d_1^2}{4 (x-1)}-\frac{d_1}{2 \
(x-1)}-\frac{5 d_1^2}{9 (x-1)^2}+\frac{7 d_1}{18 (x-1)^2}-\frac{5 \
d_1^2}{9 (x-1)^3}+\frac{13 d_1}{18 (x-1)^3}+\frac{15 d_1^2}{4 \
(x-1)^4}-\frac{6 d_1}{(x-1)^4}+\frac{205 d_1^2}{36 (x-1)^5}-\frac{130 \
d_1}{9 (x-1)^5}\Big) H(1;\ao)+\Big(-\frac{\ao^4}{x-1}+\ao^4+\frac{4 \
\ao^3}{x-1}-\frac{4 \ao^3}{3 (x-1)^2}-\frac{16 \ao^3}{3}-\frac{6 \
\ao^2}{x-1}+\frac{4 \ao^2}{(x-1)^2}-\frac{2 \ao^2}{(x-1)^3}+12 \ao^2+\
\frac{4 \ao}{x-1}-\frac{4 \ao}{(x-1)^2}+\frac{4 \ao}{(x-1)^3}-\frac{4 \
\ao}{(x-1)^4}-16 \ao+\frac{3}{x-1}-\frac{2}{3 (x-1)^2}-\frac{2}{3 \
(x-1)^3}+\frac{3}{(x-1)^4}+\frac{25}{3 (x-1)^5}+\frac{25}{3}\Big) \
H(0,0;\ao)+\Big(d_1 \ao^4-\frac{d_1 \ao^4}{x-1}-\frac{16 d_1 \
\ao^3}{3}+\frac{4 d_1 \ao^3}{x-1}-\frac{4 d_1 \ao^3}{3 (x-1)^2}+12 \
d_1 \ao^2-\frac{6 d_1 \ao^2}{x-1}+\frac{4 d_1 \ao^2}{(x-1)^2}-\frac{2 \
d_1 \ao^2}{(x-1)^3}-16 d_1 \ao+\frac{4 d_1 \ao}{x-1}-\frac{4 d_1 \
\ao}{(x-1)^2}+\frac{4 d_1 \ao}{(x-1)^3}-\frac{4 d_1 \
\ao}{(x-1)^4}+\frac{25 d_1}{3}+\frac{3 d_1}{x-1}-\frac{2 d_1}{3 \
(x-1)^2}-\frac{2 d_1}{3 (x-1)^3}+\frac{3 d_1}{(x-1)^4}+\frac{25 \
d_1}{3 (x-1)^5}\Big) H(0,1;\ao)+\Big(d_1 \ao^4-\frac{d_1 \ao^4}{x-1}-\
\frac{16 d_1 \ao^3}{3}+\frac{4 d_1 \ao^3}{x-1}-\frac{4 d_1 \ao^3}{3 \
(x-1)^2}+12 d_1 \ao^2-\frac{6 d_1 \ao^2}{x-1}+\frac{4 d_1 \
\ao^2}{(x-1)^2}-\frac{2 d_1 \ao^2}{(x-1)^3}-16 d_1 \ao+\frac{4 d_1 \
\ao}{x-1}-\frac{4 d_1 \ao}{(x-1)^2}+\frac{4 d_1 \ao}{(x-1)^3}-\frac{4 \
d_1 \ao}{(x-1)^4}+\frac{25 d_1}{3}+\frac{3 d_1}{x-1}-\frac{2 d_1}{3 \
(x-1)^2}-\frac{2 d_1}{3 (x-1)^3}+\frac{3 d_1}{(x-1)^4}+\frac{25 \
d_1}{3 (x-1)^5}\Big) H(1,0;\ao)+\Big(d_1^2 \ao^4-\frac{d_1^2 \
\ao^4}{x-1}-\frac{16 d_1^2 \ao^3}{3}+\frac{4 d_1^2 \
\ao^3}{x-1}-\frac{4 d_1^2 \ao^3}{3 (x-1)^2}+12 d_1^2 \ao^2-\frac{6 \
d_1^2 \ao^2}{x-1}+\frac{4 d_1^2 \ao^2}{(x-1)^2}-\frac{2 d_1^2 \
\ao^2}{(x-1)^3}-16 d_1^2 \ao+\frac{4 d_1^2 \ao}{x-1}-\frac{4 d_1^2 \
\ao}{(x-1)^2}+\frac{4 d_1^2 \ao}{(x-1)^3}-\frac{4 d_1^2 \
\ao}{(x-1)^4}+\frac{25 d_1^2}{3}+\frac{3 d_1^2}{x-1}-\frac{2 d_1^2}{3 \
(x-1)^2}-\frac{2 d_1^2}{3 (x-1)^3}+\frac{3 d_1^2}{(x-1)^4}+\frac{25 \
d_1^2}{3 (x-1)^5}\Big) H(1,1;\ao)+\frac{63 d_1^2}{16 (x-1)}-\frac{209 \
d_1}{48 (x-1)}-\frac{\pi ^2}{8 (x-1)}-\frac{2}{x-1}-\frac{19 \
d_1^2}{54 (x-1)^2}+\frac{347 d_1}{216 (x-1)^2}+\frac{\pi ^2}{36 \
(x-1)^2}-\frac{19}{216 (x-1)^2}-\frac{19 d_1^2}{54 (x-1)^3}+\frac{173 \
d_1}{216 (x-1)^3}+\frac{\pi ^2}{36 (x-1)^3}-\frac{205}{216 \
(x-1)^3}+\frac{63 d_1^2}{16 (x-1)^4}-\frac{197 d_1}{16 \
(x-1)^4}-\frac{\pi ^2}{8 (x-1)^4}+\frac{163}{24 (x-1)^4}+\frac{2035 \
d_1^2}{432 (x-1)^5}-\frac{8705 d_1}{432 (x-1)^5}-\frac{25 \pi ^2}{72 \
(x-1)^5}+\frac{3965}{216 (x-1)^5}-\frac{25 \pi ^2}{72}+\frac{235}{27}\
\Big)+\Big(-\frac{2 \pi ^2 d_1^2}{3 (x-1)^5}+\frac{2 \pi ^2 d_1}{3 \
(x-1)^5}+\frac{\pi ^2 d_1}{3}+\Big(\frac{4 d_1^2}{x-1}-\frac{2 \
d_1^2}{(x-1)^2}+\frac{4 d_1^2}{3 \
(x-1)^3}-\frac{d_1^2}{(x-1)^4}+\frac{25 d_1^2}{3 (x-1)^5}+\frac{9 \
d_1}{2 (x-1)}-\frac{8 d_1}{3 (x-1)^2}+\frac{7 d_1}{3 (x-1)^3}-\frac{3 \
d_1}{(x-1)^4}+\frac{25 d_1}{3 (x-1)^5}-\frac{25 d_1}{6}-\frac{1}{4 \
(x-1)}-\frac{1}{2 (x-1)^2}+\frac{7}{6 (x-1)^3}-\frac{11}{4 \
(x-1)^4}-\frac{25}{12 (x-1)^5}-\frac{25}{4}\Big) \
H(0;\ao)+\Big(-\frac{8 d_1}{(x-1)^5}+4 d_1+\frac{2}{(x-1)^5}+2\Big) \
H(0,0;\ao)+\Big(-\frac{8 d_1^2}{(x-1)^5}+4 d_1^2+\frac{2 \
d_1}{(x-1)^5}+2 d_1\Big) H(0,1;\ao)-\frac{\pi ^2}{6 \
(x-1)^5}-\frac{\pi ^2}{6}\Big) H(1,1;x)+\Big(-\frac{4 \
d_1^2}{x-1}+\frac{d_1^2}{(x-1)^2}-\frac{4 d_1^2}{9 \
(x-1)^3}+\frac{d_1^2}{4 (x-1)^4}-\frac{205 d_1^2}{36 \
(x-1)^5}+\frac{97 d_1}{24 (x-1)}-\frac{157 d_1}{36 (x-1)^2}+\frac{137 \
d_1}{36 (x-1)^3}-\frac{19 d_1}{8 (x-1)^4}+\frac{835 d_1}{72 (x-1)^5}+\
\frac{205 d_1}{72}+\Big(-\frac{4 d_1}{x-1}+\frac{2 \
d_1}{(x-1)^2}-\frac{4 d_1}{3 (x-1)^3}+\frac{d_1}{(x-1)^4}-\frac{25 \
d_1}{3 (x-1)^5}-\frac{5}{2 (x-1)}+\frac{5}{3 (x-1)^2}-\frac{5}{3 \
(x-1)^3}+\frac{5}{2 (x-1)^4}-\frac{25}{6 (x-1)^5}+\frac{25}{6}\Big) \
H(0;\ao)+\Big(-\frac{4 d_1^2}{x-1}+\frac{2 d_1^2}{(x-1)^2}-\frac{4 \
d_1^2}{3 (x-1)^3}+\frac{d_1^2}{(x-1)^4}-\frac{25 d_1^2}{3 \
(x-1)^5}-\frac{5 d_1}{2 (x-1)}+\frac{5 d_1}{3 (x-1)^2}-\frac{5 d_1}{3 \
(x-1)^3}+\frac{5 d_1}{2 (x-1)^4}-\frac{25 d_1}{6 (x-1)^5}+\frac{25 \
d_1}{6}\Big) H(1;\ao)+\Big(\frac{4}{(x-1)^5}-4\Big) \
H(0,0;\ao)+\Big(\frac{4 d_1}{(x-1)^5}-4 d_1\Big) \
H(0,1;\ao)+\Big(\frac{4 d_1}{(x-1)^5}-4 d_1\Big) \
H(1,0;\ao)+\Big(\frac{4 d_1^2}{(x-1)^5}-4 d_1^2\Big) \
H(1,1;\ao)+\frac{97}{12 (x-1)}-\frac{179}{36 (x-1)^2}+\frac{179}{36 \
(x-1)^3}-\frac{97}{12 (x-1)^4}-\frac{\pi ^2}{6 (x-1)^5}+\frac{155}{36 \
(x-1)^5}+\frac{\pi ^2}{6}-\frac{155}{36}\Big) \
H(1,c_1(\ao);x)+\Big(\frac{3 d_1 \ao^4}{8}-\frac{3 d_1 \ao^4}{8 \
(x-1)}+\frac{3 \ao^4}{8 (x-1)}-\frac{3 \ao^4}{8}-\frac{13 d_1 \
\ao^3}{6}+\frac{3 d_1 \ao^3}{2 (x-1)}-\frac{5 \ao^3}{2 (x-1)}-\frac{2 \
d_1 \ao^3}{3 (x-1)^2}+\frac{5 \ao^3}{4 (x-1)^2}+\frac{31 \
\ao^3}{12}+\frac{23 d_1 \ao^2}{4}-\frac{9 d_1 \ao^2}{4 \
(x-1)}+\frac{169 \ao^2}{24 (x-1)}+\frac{2 d_1 \
\ao^2}{(x-1)^2}-\frac{131 \ao^2}{24 (x-1)^2}-\frac{3 d_1 \ao^2}{2 \
(x-1)^3}+\frac{85 \ao^2}{24 (x-1)^3}-\frac{203 \ao^2}{24}-\frac{25 \
d_1 \ao}{2}+\frac{3 d_1 \ao}{2 (x-1)}-\frac{77 \ao}{6 (x-1)}-\frac{2 \
d_1 \ao}{(x-1)^2}+\frac{67 \ao}{6 (x-1)^2}+\frac{3 d_1 \ao}{(x-1)^3}-\
\frac{34 \ao}{3 (x-1)^3}-\frac{6 d_1 \ao}{(x-1)^4}+\frac{163 \ao}{12 \
(x-1)^4}+\frac{265 \ao}{12}+\frac{205 d_1}{24}+\Big(-\frac{3 \ao^4}{2 \
(x-1)}+\frac{3 \ao^4}{2}+\frac{6 \ao^3}{x-1}-\frac{2 \
\ao^3}{(x-1)^2}-8 \ao^3-\frac{9 \ao^2}{x-1}+\frac{6 \
\ao^2}{(x-1)^2}-\frac{3 \ao^2}{(x-1)^3}+18 \ao^2+\frac{6 \
\ao}{x-1}-\frac{6 \ao}{(x-1)^2}+\frac{6 \ao}{(x-1)^3}-\frac{6 \
\ao}{(x-1)^4}-24 \ao+\frac{9}{2 \
(x-1)}-\frac{1}{(x-1)^2}-\frac{1}{(x-1)^3}+\frac{9}{2 \
(x-1)^4}+\frac{25}{2 (x-1)^5}+\frac{25}{2}\Big) H(0;\ao)+\Big(\frac{3 \
d_1 \ao^4}{2}-\frac{3 d_1 \ao^4}{2 (x-1)}-8 d_1 \ao^3+\frac{6 d_1 \
\ao^3}{x-1}-\frac{2 d_1 \ao^3}{(x-1)^2}+18 d_1 \ao^2-\frac{9 d_1 \
\ao^2}{x-1}+\frac{6 d_1 \ao^2}{(x-1)^2}-\frac{3 d_1 \
\ao^2}{(x-1)^3}-24 d_1 \ao+\frac{6 d_1 \ao}{x-1}-\frac{6 d_1 \
\ao}{(x-1)^2}+\frac{6 d_1 \ao}{(x-1)^3}-\frac{6 d_1 \
\ao}{(x-1)^4}+\frac{25 d_1}{2}+\frac{9 d_1}{2 \
(x-1)}-\frac{d_1}{(x-1)^2}-\frac{d_1}{(x-1)^3}+\frac{9 d_1}{2 \
(x-1)^4}+\frac{25 d_1}{2 (x-1)^5}\Big) H(1;\ao)-\frac{4 \
H(0,0;\ao)}{(x-1)^5}-\frac{4 d_1 H(0,1;\ao)}{(x-1)^5}-\frac{4 d_1 \
H(1,0;\ao)}{(x-1)^5}-\frac{4 d_1^2 H(1,1;\ao)}{(x-1)^5}+\frac{45 \
d_1}{8 (x-1)}-\frac{7}{2 (x-1)}-\frac{5 d_1}{6 (x-1)^2}+\frac{3}{4 \
(x-1)^2}-\frac{5 d_1}{6 (x-1)^3}+\frac{11}{12 (x-1)^3}+\frac{45 \
d_1}{8 (x-1)^4}-\frac{25}{4 (x-1)^4}+\frac{205 d_1}{24 \
(x-1)^5}+\frac{\pi ^2}{6 (x-1)^5}-\frac{75}{4 \
(x-1)^5}-\frac{95}{6}\Big) \
H(c_1(\ao),c_1(\ao);x)+\Big(\frac{25}{3}+\frac{3}{x-1}-\frac{2}{3 \
(x-1)^2}-\frac{2}{3 (x-1)^3}+\frac{3}{(x-1)^4}+\frac{25}{3 \
(x-1)^5}\Big) \
H(0,0,0;\ao)+\Big(-\frac{25}{3}-\frac{3}{x-1}+\frac{2}{3 \
(x-1)^2}+\frac{2}{3 (x-1)^3}-\frac{3}{(x-1)^4}-\frac{25}{3 \
(x-1)^5}\Big) H(0,0,0;x)+\Big(\frac{3 d_1}{x-1}-\frac{2 d_1}{3 \
(x-1)^2}-\frac{2 d_1}{3 (x-1)^3}+\frac{3 d_1}{(x-1)^4}+\frac{25 \
d_1}{3 (x-1)^5}+\frac{25 d_1}{3}\Big) \
H(0,0,1;\ao)+\Big(\frac{4}{(x-1)^5}-4\Big) H(0;\ao) H(0,0,1;x)+\Big(-\
\frac{\ao^4}{x-1}+\ao^4+\frac{4 \ao^3}{x-1}-\frac{4 \ao^3}{3 \
(x-1)^2}-\frac{16 \ao^3}{3}-\frac{6 \ao^2}{x-1}+\frac{4 \
\ao^2}{(x-1)^2}-\frac{2 \ao^2}{(x-1)^3}+12 \ao^2+\frac{4 \
\ao}{x-1}-\frac{4 \ao}{(x-1)^2}+\frac{4 \ao}{(x-1)^3}-\frac{4 \
\ao}{(x-1)^4}-16 \ao+\frac{3}{2 (x-1)}-\frac{1}{3 (x-1)^2}-\frac{1}{3 \
(x-1)^3}+\frac{3}{2 (x-1)^4}+\frac{25}{6 (x-1)^5}+\frac{25}{6}\Big) \
H(0,0,c_1(\ao);x)+\Big(\frac{3 d_1}{x-1}-\frac{2 d_1}{3 \
(x-1)^2}-\frac{2 d_1}{3 (x-1)^3}+\frac{3 d_1}{(x-1)^4}+\frac{25 \
d_1}{3 (x-1)^5}+\frac{25 d_1}{3}\Big) H(0,1,0;\ao)+\Big(-\frac{5 \
d_1}{2 (x-1)}+\frac{5 d_1}{3 (x-1)^2}-\frac{5 d_1}{3 (x-1)^3}+\frac{5 \
d_1}{2 (x-1)^4}-\frac{25 d_1}{6 (x-1)^5}+\frac{25 d_1}{6}-\frac{5}{2 \
(x-1)}+\frac{5}{3 (x-1)^2}-\frac{5}{3 (x-1)^3}+\frac{5}{2 \
(x-1)^4}-\frac{25}{6 (x-1)^5}+\frac{25}{6}\Big) \
H(0,1,0;x)+\Big(\frac{3 d_1^2}{x-1}-\frac{2 d_1^2}{3 (x-1)^2}-\frac{2 \
d_1^2}{3 (x-1)^3}+\frac{3 d_1^2}{(x-1)^4}+\frac{25 d_1^2}{3 (x-1)^5}+\
\frac{25 d_1^2}{3}\Big) H(0,1,1;\ao)+\Big(\frac{4 d_1^2}{(x-1)^5}-4 \
d_1^2-\frac{2 d_1}{(x-1)^5}-2 d_1+\frac{4}{(x-1)^5}\Big) H(0;\ao) \
H(0,1,1;x)+\Big(\frac{5 d_1}{2 (x-1)}-\frac{5 d_1}{3 (x-1)^2}+\frac{5 \
d_1}{3 (x-1)^3}-\frac{5 d_1}{2 (x-1)^4}+\frac{25 d_1}{6 \
(x-1)^5}-\frac{25 d_1}{6}+\Big(4 d_1-\frac{4 d_1}{(x-1)^5}\Big) \
H(0;\ao)+\Big(4 d_1^2-\frac{4 d_1^2}{(x-1)^5}\Big) \
H(1;\ao)+\frac{5}{2 (x-1)}-\frac{5}{3 (x-1)^2}+\frac{5}{3 \
(x-1)^3}-\frac{5}{2 (x-1)^4}+\frac{25}{6 (x-1)^5}-\frac{25}{6}\Big) \
H(0,1,c_1(\ao);x)+\Big(\frac{3 \ao^4}{2 (x-1)}-\frac{3 \
\ao^4}{2}-\frac{6 \ao^3}{x-1}+\frac{2 \ao^3}{(x-1)^2}+8 \ao^3+\frac{9 \
\ao^2}{x-1}-\frac{6 \ao^2}{(x-1)^2}+\frac{3 \ao^2}{(x-1)^3}-18 \ao^2-\
\frac{6 \ao}{x-1}+\frac{6 \ao}{(x-1)^2}-\frac{6 \ao}{(x-1)^3}+\frac{6 \
\ao}{(x-1)^4}+24 \ao+\Big(\frac{2}{(x-1)^5}-6\Big) \
H(0;\ao)+\Big(\frac{2 d_1}{(x-1)^5}-6 d_1\Big) H(1;\ao)-\frac{3}{4 \
(x-1)}+\frac{1}{6 (x-1)^2}+\frac{1}{6 (x-1)^3}-\frac{3}{4 \
(x-1)^4}-\frac{25}{12 (x-1)^5}-\frac{25}{12}\Big) \
H(0,c_1(\ao),c_1(\ao);x)+\Big(\frac{4 d_1}{x-1}-\frac{2 \
d_1}{(x-1)^2}+\frac{4 d_1}{3 (x-1)^3}-\frac{d_1}{(x-1)^4}+\frac{25 \
d_1}{3 (x-1)^5}+\frac{5}{2 (x-1)}-\frac{5}{3 (x-1)^2}+\frac{5}{3 \
(x-1)^3}-\frac{5}{2 (x-1)^4}+\frac{25}{6 (x-1)^5}-\frac{25}{6}\Big) \
H(1,0,0;x)+\Big(\frac{4 d_1^2}{(x-1)^5}-\frac{2 d_1}{(x-1)^5}-2 \
d_1+\frac{2}{(x-1)^5}-2\Big) H(0;\ao) H(1,0,1;x)+\Big(\frac{2 \
d_1}{x-1}-\frac{d_1}{(x-1)^2}+\frac{2 d_1}{3 (x-1)^3}-\frac{d_1}{2 \
(x-1)^4}+\frac{25 d_1}{6 (x-1)^5}+\Big(-\frac{4 \
d_1}{(x-1)^5}+\frac{2}{(x-1)^5}+2\Big) H(0;\ao)+\Big(-\frac{4 \
d_1^2}{(x-1)^5}+\frac{2 d_1}{(x-1)^5}+2 d_1\Big) H(1;\ao)-\frac{1}{4 \
(x-1)}-\frac{1}{2 (x-1)^2}+\frac{7}{6 (x-1)^3}-\frac{11}{4 \
(x-1)^4}-\frac{25}{12 (x-1)^5}-\frac{25}{4}\Big) \
H(1,0,c_1(\ao);x)+\Big(-\frac{4 d_1^2}{x-1}+\frac{2 \
d_1^2}{(x-1)^2}-\frac{4 d_1^2}{3 \
(x-1)^3}+\frac{d_1^2}{(x-1)^4}-\frac{25 d_1^2}{3 (x-1)^5}-\frac{9 \
d_1}{2 (x-1)}+\frac{8 d_1}{3 (x-1)^2}-\frac{7 d_1}{3 (x-1)^3}+\frac{3 \
d_1}{(x-1)^4}-\frac{25 d_1}{3 (x-1)^5}+\frac{25 d_1}{6}+\frac{1}{4 \
(x-1)}+\frac{1}{2 (x-1)^2}-\frac{7}{6 (x-1)^3}+\frac{11}{4 \
(x-1)^4}+\frac{25}{12 (x-1)^5}+\frac{25}{4}\Big) \
H(1,1,0;x)+\Big(\frac{12 d_1^2}{(x-1)^5}-4 d_1^2-\frac{6 \
d_1}{(x-1)^5}-4 d_1+\frac{1}{(x-1)^5}+1\Big) H(0;\ao) \
H(1,1,1;x)+\Big(\frac{4 d_1^2}{x-1}-\frac{2 d_1^2}{(x-1)^2}+\frac{4 \
d_1^2}{3 (x-1)^3}-\frac{d_1^2}{(x-1)^4}+\frac{25 d_1^2}{3 \
(x-1)^5}+\frac{9 d_1}{2 (x-1)}-\frac{8 d_1}{3 (x-1)^2}+\frac{7 d_1}{3 \
(x-1)^3}-\frac{3 d_1}{(x-1)^4}+\frac{25 d_1}{3 (x-1)^5}-\frac{25 \
d_1}{6}+\Big(-\frac{8 d_1}{(x-1)^5}+4 d_1+\frac{2}{(x-1)^5}+2\Big) \
H(0;\ao)+\Big(-\frac{8 d_1^2}{(x-1)^5}+4 d_1^2+\frac{2 \
d_1}{(x-1)^5}+2 d_1\Big) H(1;\ao)-\frac{1}{4 (x-1)}-\frac{1}{2 \
(x-1)^2}+\frac{7}{6 (x-1)^3}-\frac{11}{4 (x-1)^4}-\frac{25}{12 \
(x-1)^5}-\frac{25}{4}\Big) H(1,1,c_1(\ao);x)+\Big(-\frac{6 d_1}{x-1}+\
\frac{3 d_1}{(x-1)^2}-\frac{2 d_1}{(x-1)^3}+\frac{3 d_1}{2 \
(x-1)^4}-\frac{25 d_1}{2 (x-1)^5}+\Big(\frac{4 \
d_1}{(x-1)^5}+\frac{2}{(x-1)^5}-6\Big) H(0;\ao)+\Big(\frac{4 \
d_1^2}{(x-1)^5}+\frac{2 d_1}{(x-1)^5}-6 d_1\Big) H(1;\ao)-\frac{9}{4 \
(x-1)}+\frac{13}{6 (x-1)^2}-\frac{17}{6 (x-1)^3}+\frac{21}{4 \
(x-1)^4}-\frac{25}{12 (x-1)^5}+\frac{125}{12}\Big) \
H(1,c_1(\ao),c_1(\ao);x)+\Big(\frac{\ao^4}{x-1}-\ao^4-\frac{4 \
\ao^3}{x-1}+\frac{4 \ao^3}{3 (x-1)^2}+\frac{16 \ao^3}{3}+\frac{6 \
\ao^2}{x-1}-\frac{4 \ao^2}{(x-1)^2}+\frac{2 \ao^2}{(x-1)^3}-12 \ao^2-\
\frac{4 \ao}{x-1}+\frac{4 \ao}{(x-1)^2}-\frac{4 \ao}{(x-1)^3}+\frac{4 \
\ao}{(x-1)^4}+16 \ao+\frac{4 H(0;\ao)}{(x-1)^5}+\frac{4 d_1 \
H(1;\ao)}{(x-1)^5}-\frac{3}{x-1}+\frac{2}{3 (x-1)^2}+\frac{2}{3 \
(x-1)^3}-\frac{3}{(x-1)^4}-\frac{25}{3 (x-1)^5}-\frac{25}{3}\Big) \
H(c_1(\ao),0,c_1(\ao);x)+\Big(-\frac{7 \ao^4}{4 (x-1)}+\frac{7 \
\ao^4}{4}+\frac{7 \ao^3}{x-1}-\frac{7 \ao^3}{3 (x-1)^2}-\frac{28 \
\ao^3}{3}-\frac{21 \ao^2}{2 (x-1)}+\frac{7 \ao^2}{(x-1)^2}-\frac{7 \
\ao^2}{2 (x-1)^3}+21 \ao^2+\frac{7 \ao}{x-1}-\frac{7 \
\ao}{(x-1)^2}+\frac{7 \ao}{(x-1)^3}-\frac{7 \ao}{(x-1)^4}-28 \
\ao-\frac{6 H(0;\ao)}{(x-1)^5}-\frac{6 d_1 \
H(1;\ao)}{(x-1)^5}+\frac{21}{4 (x-1)}-\frac{7}{6 (x-1)^2}-\frac{7}{6 \
(x-1)^3}+\frac{21}{4 (x-1)^4}+\frac{175}{12 \
(x-1)^5}+\frac{175}{12}\Big) H(c_1(\ao),c_1(\ao),c_1(\ao);x)+8 \
H(0,0,0,0;x)+\Big(\frac{4}{(x-1)^5}-4\Big) \
H(0,0,0,c_1(\ao);x)+\Big(4-\frac{4}{(x-1)^5}\Big) \
H(0,0,1,0;x)+\Big(\frac{4}{(x-1)^5}-4\Big) \
H(0,0,1,c_1(\ao);x)-\frac{4 \
H(0,0,c_1(\ao),c_1(\ao);x)}{(x-1)^5}+\Big(\frac{4 d_1}{(x-1)^5}-4 d_1\
\Big) H(0,1,0,0;x)+\Big(-\frac{2 d_1}{(x-1)^5}-2 \
d_1+\frac{4}{(x-1)^5}\Big) H(0,1,0,c_1(\ao);x)+\Big(-\frac{4 \
d_1^2}{(x-1)^5}+4 d_1^2+\frac{2 d_1}{(x-1)^5}+2 d_1-\frac{4}{(x-1)^5}\
\Big) H(0,1,1,0;x)+\Big(\frac{4 d_1^2}{(x-1)^5}-4 d_1^2-\frac{2 \
d_1}{(x-1)^5}-2 d_1+\frac{4}{(x-1)^5}\Big) H(0,1,1,c_1(\ao);x)+\Big(-\
\frac{2 d_1}{(x-1)^5}+6 d_1-\frac{4}{(x-1)^5}\Big) \
H(0,1,c_1(\ao),c_1(\ao);x)+4 \
H(0,c_1(\ao),0,c_1(\ao);x)+\Big(\frac{1}{(x-1)^5}-7\Big) \
H(0,c_1(\ao),c_1(\ao),c_1(\ao);x)+\Big(4-\frac{4}{(x-1)^5}\Big) \
H(1,0,0,0;x)+\Big(\frac{2}{(x-1)^5}-2\Big) H(1,0,0,c_1(\ao);x)+\Big(-\
\frac{4 d_1^2}{(x-1)^5}+\frac{2 d_1}{(x-1)^5}+2 \
d_1-\frac{2}{(x-1)^5}+2\Big) H(1,0,1,0;x)+\Big(\frac{4 \
d_1^2}{(x-1)^5}-\frac{2 d_1}{(x-1)^5}-2 d_1+\frac{2}{(x-1)^5}-2\Big) \
H(1,0,1,c_1(\ao);x)+\Big(-\frac{2 \
d_1}{(x-1)^5}+\frac{1}{(x-1)^5}+1\Big) \
H(1,0,c_1(\ao),c_1(\ao);x)+\Big(\frac{8 d_1}{(x-1)^5}-4 \
d_1-\frac{2}{(x-1)^5}-2\Big) H(1,1,0,0;x)+\Big(\frac{4 \
d_1^2}{(x-1)^5}-\frac{4 d_1}{(x-1)^5}-2 d_1+\frac{1}{(x-1)^5}+1\Big) \
H(1,1,0,c_1(\ao);x)+\Big(-\frac{12 d_1^2}{(x-1)^5}+4 d_1^2+\frac{6 \
d_1}{(x-1)^5}+4 d_1-\frac{1}{(x-1)^5}-1\Big) \
H(1,1,1,0;x)+\Big(\frac{12 d_1^2}{(x-1)^5}-4 d_1^2-\frac{6 \
d_1}{(x-1)^5}-4 d_1+\frac{1}{(x-1)^5}+1\Big) \
H(1,1,1,c_1(\ao);x)+\Big(-\frac{4 d_1^2}{(x-1)^5}-\frac{4 \
d_1}{(x-1)^5}+6 d_1+\frac{1}{(x-1)^5}+1\Big) \
H(1,1,c_1(\ao),c_1(\ao);x)+\Big(4-\frac{4 d_1}{(x-1)^5}\Big) H(1,c_1(\
\ao),0,c_1(\ao);x)+\Big(\frac{6 \
d_1}{(x-1)^5}+\frac{1}{(x-1)^5}-7\Big) \
H(1,c_1(\ao),c_1(\ao),c_1(\ao);x)-\frac{4 \
H(c_1(\ao),0,0,c_1(\ao);x)}{(x-1)^5}+\frac{6 \
H(c_1(\ao),0,c_1(\ao),c_1(\ao);x)}{(x-1)^5}+\frac{4 \
H(c_1(\ao),c_1(\ao),0,c_1(\ao);x)}{(x-1)^5}-\frac{7 \
H(c_1(\ao),c_1(\ao),c_1(\ao),c_1(\ao);x)}{(x-1)^5}+H(0;x) \
\Big(-\frac{63 d_1^2}{16 (x-1)}+\frac{19 d_1^2}{54 (x-1)^2}+\frac{19 \
d_1^2}{54 (x-1)^3}-\frac{63 d_1^2}{16 (x-1)^4}-\frac{2035 d_1^2}{432 \
(x-1)^5}-\frac{2035 d_1^2}{432}+\frac{209 d_1}{48 (x-1)}-\frac{347 \
d_1}{216 (x-1)^2}-\frac{173 d_1}{216 (x-1)^3}+\frac{197 d_1}{16 \
(x-1)^4}+\frac{8705 d_1}{432 (x-1)^5}+\frac{5615 d_1}{432}+\frac{3 \
\pi ^2}{8 (x-1)}+\frac{2}{x-1}-\frac{\pi ^2}{12 \
(x-1)^2}+\frac{19}{216 (x-1)^2}-\frac{\pi ^2}{12 \
(x-1)^3}+\frac{205}{216 (x-1)^3}+\frac{3 \pi ^2}{8 \
(x-1)^4}-\frac{163}{24 (x-1)^4}+\frac{25 \pi ^2}{24 \
(x-1)^5}-\frac{3965}{216 (x-1)^5}+\frac{6 \zeta_3}{(x-1)^5}+10 \
\zeta_3+\frac{25 \pi ^2}{24}-\frac{235}{27}\Big)+H(1;x) \
\Big(-\frac{\pi ^2 d_1}{3 (x-1)}+\frac{\pi ^2 d_1}{6 \
(x-1)^2}-\frac{\pi ^2 d_1}{9 (x-1)^3}+\frac{\pi ^2 d_1}{12 \
(x-1)^4}-\frac{25 \pi ^2 d_1}{36 (x-1)^5}-\frac{6 \zeta_3 \
d_1}{(x-1)^5}+\Big(-\frac{4 d_1^2}{x-1}+\frac{d_1^2}{(x-1)^2}-\frac{4 \
d_1^2}{9 (x-1)^3}+\frac{d_1^2}{4 (x-1)^4}-\frac{205 d_1^2}{36 \
(x-1)^5}+\frac{97 d_1}{24 (x-1)}-\frac{157 d_1}{36 (x-1)^2}+\frac{137 \
d_1}{36 (x-1)^3}-\frac{19 d_1}{8 (x-1)^4}+\frac{835 d_1}{72 (x-1)^5}+\
\frac{205 d_1}{72}+\frac{97}{12 (x-1)}-\frac{179}{36 \
(x-1)^2}+\frac{179}{36 (x-1)^3}-\frac{97}{12 (x-1)^4}-\frac{\pi ^2}{6 \
(x-1)^5}+\frac{155}{36 (x-1)^5}+\frac{\pi ^2}{6}-\frac{155}{36}\Big) \
H(0;\ao)+\Big(-\frac{4 d_1}{x-1}+\frac{2 d_1}{(x-1)^2}-\frac{4 d_1}{3 \
(x-1)^3}+\frac{d_1}{(x-1)^4}-\frac{25 d_1}{3 (x-1)^5}-\frac{5}{2 \
(x-1)}+\frac{5}{3 (x-1)^2}-\frac{5}{3 (x-1)^3}+\frac{5}{2 \
(x-1)^4}-\frac{25}{6 (x-1)^5}+\frac{25}{6}\Big) \
H(0,0;\ao)+\Big(-\frac{4 d_1^2}{x-1}+\frac{2 d_1^2}{(x-1)^2}-\frac{4 \
d_1^2}{3 (x-1)^3}+\frac{d_1^2}{(x-1)^4}-\frac{25 d_1^2}{3 \
(x-1)^5}-\frac{5 d_1}{2 (x-1)}+\frac{5 d_1}{3 (x-1)^2}-\frac{5 d_1}{3 \
(x-1)^3}+\frac{5 d_1}{2 (x-1)^4}-\frac{25 d_1}{6 (x-1)^5}+\frac{25 \
d_1}{6}\Big) H(0,1;\ao)+\Big(\frac{4}{(x-1)^5}-4\Big) \
H(0,0,0;\ao)+\Big(\frac{4 d_1}{(x-1)^5}-4 d_1\Big) H(0,0,1;\ao)+\Big(\
\frac{4 d_1}{(x-1)^5}-4 d_1\Big) H(0,1,0;\ao)+\Big(\frac{4 \
d_1^2}{(x-1)^5}-4 d_1^2\Big) H(0,1,1;\ao)+\frac{\pi ^2}{24 \
(x-1)}+\frac{\pi ^2}{12 (x-1)^2}-\frac{7 \pi ^2}{36 (x-1)^3}+\frac{11 \
\pi^2}{24 (x-1)^4}+\frac{25 \pi ^2}{72 \
(x-1)^5}-\frac{\zeta_3}{(x-1)^5}+7 \zeta_3+\frac{25 \pi ^2}{24}\Big)+\
\frac{5 d_1 \pi ^2}{16 (x-1)}-\frac{\pi ^2}{2 (x-1)}-\frac{5 d_1 \pi^2}{108 (x-1)^2}+\frac{13 \pi ^2}{216 (x-1)^2}-\frac{5 d_1 \pi^2}{108 (x-1)^3}+\frac{7 \pi ^2}{216 (x-1)^3}+\frac{5 d_1 \pi ^2}{16 \
(x-1)^4}-\frac{\pi ^2}{24 (x-1)^4}-\frac{23 \pi ^4}{180 \
(x-1)^5}+\frac{205 d_1 \pi ^2}{432 (x-1)^5}-\frac{155 \pi ^2}{216 \
(x-1)^5}-\frac{21 \zeta_3}{4 (x-1)}+\frac{7 \zeta_3}{6 \
(x-1)^2}+\frac{7 \zeta_3}{6 (x-1)^3}-\frac{21 \zeta_3}{4 \
(x-1)^4}-\frac{175 \zeta_3}{12 (x-1)^5}-\frac{175 \
\zeta_3}{12}-\frac{17 \pi ^4}{144}+\frac{205 d_1 \pi^2}{432}-\frac{65 \pi^2}{54}.
\erp  

%
% The A integral for k=-1 and kappa=1
%

\subsection{The $\cA$ integral for $k=-1$ and $\kappa=1$}
%
% This file contains the TeX output produced by Mathematica for the integral Am1, for kappa = 0
%
The $\eps$ expansion for this integral reads
\beq
\bsp
\begin{cal}I\end{cal}(x,\eps;\ao,3+d_1\eps;1,2,0,g_A) &= x\,\aint(\eps,x;3+d_1\eps;1,2)\\
&=\frac{1}{\eps^2}a_{-2}^{(1,-1)}+\frac{1}{\eps}a_{-1}^{(1,-1)}+a_0^{(1,-1)}+\eps a_1^{(1,-1)}+\eps^2 a_2^{(1,-1)} +\ocal\left(\eps^3\right),
\esp
\eeq
where
%1/ep piece
\brp
a_{-2}^{(1,-1)}=\frac{1}{6},
\erp
\brp
a_{-1}^{(1,-1)}=-\frac{2}{3}H(0;x),
\erp
% ep^0
\brp
a_0^{(1,-1)}=\frac{\ao^3}{12 (x-1)^2}-\frac{\ao^3}{12}+\frac{\ao^2}{24 \
(x-1)}-\frac{5 \ao^2}{24 (x-1)^2}+\frac{7 \ao^2}{24 (x-1)^3}+\frac{13 \
\ao^2}{24}-\frac{\ao}{3 (x-1)}+\frac{\ao}{6 (x-1)^2}-\frac{\ao}{3 \
(x-1)^3}+\frac{13 \ao}{12 (x-1)^4}-\frac{23 \
\ao}{12}+\Big(\frac{25}{12}+\frac{3}{4 (x-1)}-\frac{1}{6 \
(x-1)^2}-\frac{1}{6 (x-1)^3}+\frac{3}{4 (x-1)^4}+\frac{25}{12 \
(x-1)^5}\Big) H(0;\ao)+\Big(-\frac{25}{12}-\frac{3}{4 \
(x-1)}+\frac{1}{6 (x-1)^2}+\frac{1}{6 (x-1)^3}-\frac{3}{4 \
(x-1)^4}-\frac{25}{12 (x-1)^5}\Big) \
H(0;x)+\Big(\frac{1}{(x-1)^5}-1\Big) H(0;\ao) \
H(1;x)+\Big(-\frac{\ao^4}{4 (x-1)}+\frac{\ao^4}{4}+\frac{\ao^3}{x-1}-\
\frac{\ao^3}{3 (x-1)^2}-\frac{4 \ao^3}{3}-\frac{3 \ao^2}{2 \
(x-1)}+\frac{\ao^2}{(x-1)^2}-\frac{\ao^2}{2 (x-1)^3}+3 \
\ao^2+\frac{\ao}{x-1}-\frac{\ao}{(x-1)^2}+\frac{\ao}{(x-1)^3}-\frac{\ao}{(x-1)^4}-4 \ao+\frac{3}{4 (x-1)}-\frac{1}{6 (x-1)^2}-\frac{1}{6 \
(x-1)^3}+\frac{3}{4 (x-1)^4}+\frac{25}{12 (x-1)^5}+\frac{25}{12}\Big) \
H(c_1(\ao);x)+\frac{8}{3} H(0,0;x)+\Big(\frac{1}{(x-1)^5}-1\Big) \
H(0,c_1(\ao);x)+\Big(1-\frac{1}{(x-1)^5}\Big) \
H(1,0;x)+\Big(\frac{1}{(x-1)^5}-1\Big) \
H(1,c_1(\ao);x)-\frac{H(c_1(\ao),c_1(\ao);x)}{(x-1)^5}-\frac{\pi^2}{6 (x-1)^5}+\frac{\pi ^2}{36},
\erp
% ep^1
\brp
a_1^{(1,-1)} = \frac{7 d_1 \ao^3}{72}-\frac{7 d_1 \ao^3}{72 (x-1)^2}+\frac{7 \
\ao^3}{36 (x-1)^2}-\frac{7 \ao^3}{36}-\frac{109 d_1 \
\ao^2}{144}-\frac{13 d_1 \ao^2}{144 (x-1)}-\frac{\ao^2}{36 \
(x-1)}+\frac{29 d_1 \ao^2}{144 (x-1)^2}-\frac{11 \ao^2}{18 \
(x-1)^2}-\frac{67 d_1 \ao^2}{144 (x-1)^3}+\frac{41 \ao^2}{36 \
(x-1)^3}+\frac{31 \ao^2}{18}+\frac{305 d_1 \ao}{72}+\frac{19 d_1 \
\ao}{18 (x-1)}-\frac{25 \ao}{36 (x-1)}-\frac{d_1 \ao}{9 \
(x-1)^2}+\frac{23 \ao}{36 (x-1)^2}+\frac{d_1 \ao}{18 \
(x-1)^3}-\frac{55 \ao}{36 (x-1)^3}-\frac{217 d_1 \ao}{72 \
(x-1)^4}+\frac{569 \ao}{72 (x-1)^4}-\frac{775 \
\ao}{72}+\Big(-\frac{\ao^3}{3 (x-1)^2}+\frac{\ao^3}{3}-\frac{\ao^2}{6 \
(x-1)}+\frac{5 \ao^2}{6 (x-1)^2}-\frac{7 \ao^2}{6 (x-1)^3}-\frac{13 \
\ao^2}{6}+\frac{4 \ao}{3 (x-1)}-\frac{2 \ao}{3 (x-1)^2}+\frac{4 \
\ao}{3 (x-1)^3}-\frac{13 \ao}{3 (x-1)^4}+\frac{23 \ao}{3}-\frac{205 \
d_1}{72}-\frac{15 d_1}{8 (x-1)}+\frac{5}{8 (x-1)}+\frac{5 d_1}{18 \
(x-1)^2}-\frac{7}{18 (x-1)^2}+\frac{5 d_1}{18 (x-1)^3}-\frac{13}{18 \
(x-1)^3}-\frac{15 d_1}{8 (x-1)^4}+\frac{49}{8 (x-1)^4}-\frac{205 \
d_1}{72 (x-1)^5}+\frac{935}{72 (x-1)^5}+\frac{515}{72}\Big) H(0;\ao)+\
\Big(\frac{15 d_1}{8 (x-1)}-\frac{5 d_1}{18 (x-1)^2}-\frac{5 d_1}{18 \
(x-1)^3}+\frac{15 d_1}{8 (x-1)^4}+\frac{205 d_1}{72 \
(x-1)^5}+\frac{205 d_1}{72}-\frac{5}{8 (x-1)}+\frac{7}{18 \
(x-1)^2}+\frac{13}{18 (x-1)^3}-\frac{49}{8 (x-1)^4}+\frac{2 \pi ^2}{3 \
(x-1)^5}-\frac{935}{72 (x-1)^5}-\frac{\pi ^2}{9}-\frac{515}{72}\Big) \
H(0;x)+\Big(\frac{d_1 \ao^3}{6}-\frac{d_1 \ao^3}{6 (x-1)^2}-\frac{13 \
d_1 \ao^2}{12}-\frac{d_1 \ao^2}{12 (x-1)}+\frac{5 d_1 \ao^2}{12 \
(x-1)^2}-\frac{7 d_1 \ao^2}{12 (x-1)^3}+\frac{23 d_1 \ao}{6}+\frac{2 \
d_1 \ao}{3 (x-1)}-\frac{d_1 \ao}{3 (x-1)^2}+\frac{2 d_1 \ao}{3 \
(x-1)^3}-\frac{13 d_1 \ao}{6 (x-1)^4}-\frac{35 d_1}{12}-\frac{7 \
d_1}{12 (x-1)}+\frac{d_1}{12 (x-1)^2}-\frac{d_1}{12 (x-1)^3}+\frac{13 \
d_1}{6 (x-1)^4}\Big) H(1;\ao)+\Big(-\frac{d_1 \ao^4}{8}+\frac{d_1 \
\ao^4}{8 (x-1)}-\frac{\ao^4}{4 (x-1)}+\frac{\ao^4}{4}+\frac{13 d_1 \
\ao^3}{18}-\frac{d_1 \ao^3}{2 (x-1)}+\frac{3 \ao^3}{2 (x-1)}+\frac{2 \
d_1 \ao^3}{9 (x-1)^2}-\frac{31 \ao^3}{36 (x-1)^2}-\frac{55 \
\ao^3}{36}-\frac{23 d_1 \ao^2}{12}+\frac{3 d_1 \ao^2}{4 \
(x-1)}-\frac{95 \ao^2}{24 (x-1)}-\frac{2 d_1 \ao^2}{3 \
(x-1)^2}+\frac{27 \ao^2}{8 (x-1)^2}+\frac{d_1 \ao^2}{2 \
(x-1)^3}-\frac{59 \ao^2}{24 (x-1)^3}+\frac{35 \ao^2}{8}+\frac{25 d_1 \
\ao}{6}-\frac{d_1 \ao}{2 (x-1)}+\frac{43 \ao}{6 (x-1)}+\frac{2 d_1 \
\ao}{3 (x-1)^2}-\frac{37 \ao}{6 (x-1)^2}-\frac{d_1 \
\ao}{(x-1)^3}+\frac{20 \ao}{3 (x-1)^3}+\frac{2 d_1 \
\ao}{(x-1)^4}-\frac{113 \ao}{12 (x-1)^4}-\frac{41 \ao}{4}-\frac{205 \
d_1}{72}+\Big(\frac{\ao^4}{x-1}-\ao^4-\frac{4 \ao^3}{x-1}+\frac{4 \
\ao^3}{3 (x-1)^2}+\frac{16 \ao^3}{3}+\frac{6 \ao^2}{x-1}-\frac{4 \
\ao^2}{(x-1)^2}+\frac{2 \ao^2}{(x-1)^3}-12 \ao^2-\frac{4 \
\ao}{x-1}+\frac{4 \ao}{(x-1)^2}-\frac{4 \ao}{(x-1)^3}+\frac{4 \
\ao}{(x-1)^4}+16 \ao-\frac{3}{x-1}+\frac{2}{3 (x-1)^2}+\frac{2}{3 \
(x-1)^3}-\frac{3}{(x-1)^4}-\frac{25}{3 (x-1)^5}-\frac{25}{3}\Big) \
H(0;\ao)+\Big(-\frac{d_1 \ao^4}{2}+\frac{d_1 \ao^4}{2 (x-1)}+\frac{8 \
d_1 \ao^3}{3}-\frac{2 d_1 \ao^3}{x-1}+\frac{2 d_1 \ao^3}{3 (x-1)^2}-6 \
d_1 \ao^2+\frac{3 d_1 \ao^2}{x-1}-\frac{2 d_1 \
\ao^2}{(x-1)^2}+\frac{d_1 \ao^2}{(x-1)^3}+8 d_1 \ao-\frac{2 d_1 \
\ao}{x-1}+\frac{2 d_1 \ao}{(x-1)^2}-\frac{2 d_1 \ao}{(x-1)^3}+\frac{2 \
d_1 \ao}{(x-1)^4}-\frac{25 d_1}{6}-\frac{3 d_1}{2 (x-1)}+\frac{d_1}{3 \
(x-1)^2}+\frac{d_1}{3 (x-1)^3}-\frac{3 d_1}{2 (x-1)^4}-\frac{25 \
d_1}{6 (x-1)^5}\Big) H(1;\ao)-\frac{15 d_1}{8 (x-1)}+\frac{5}{8 \
(x-1)}+\frac{5 d_1}{18 (x-1)^2}-\frac{7}{18 (x-1)^2}+\frac{5 d_1}{18 \
(x-1)^3}-\frac{13}{18 (x-1)^3}-\frac{15 d_1}{8 (x-1)^4}+\frac{49}{8 \
(x-1)^4}-\frac{205 d_1}{72 (x-1)^5}+\frac{935}{72 \
(x-1)^5}+\frac{515}{72}\Big) \
H(c_1(\ao);x)+\Big(-\frac{25}{3}-\frac{3}{x-1}+\frac{2}{3 \
(x-1)^2}+\frac{2}{3 (x-1)^3}-\frac{3}{(x-1)^4}-\frac{25}{3 \
(x-1)^5}\Big) H(0,0;\ao)+\Big(\frac{25}{3}+\frac{3}{x-1}-\frac{2}{3 \
(x-1)^2}-\frac{2}{3 (x-1)^3}+\frac{3}{(x-1)^4}+\frac{25}{3 \
(x-1)^5}\Big) H(0,0;x)+\Big(-\frac{3 d_1}{2 (x-1)}+\frac{d_1}{3 \
(x-1)^2}+\frac{d_1}{3 (x-1)^3}-\frac{3 d_1}{2 (x-1)^4}-\frac{25 \
d_1}{6 (x-1)^5}-\frac{25 d_1}{6}\Big) H(0,1;\ao)+H(1;x) \
\Big(-\frac{\pi ^2 d_1}{3 (x-1)^5}+\Big(\frac{2 \
d_1}{x-1}-\frac{d_1}{(x-1)^2}+\frac{2 d_1}{3 (x-1)^3}-\frac{d_1}{2 \
(x-1)^4}+\frac{25 d_1}{6 (x-1)^5}+\frac{5}{4 (x-1)}-\frac{5}{6 \
(x-1)^2}+\frac{5}{6 (x-1)^3}-\frac{5}{4 (x-1)^4}+\frac{25}{12 \
(x-1)^5}-\frac{25}{12}\Big) H(0;\ao)+\Big(4-\frac{4}{(x-1)^5}\Big) \
H(0,0;\ao)+\Big(2 d_1-\frac{2 d_1}{(x-1)^5}\Big) H(0,1;\ao)+\frac{\pi ^2}{2 (x-1)^5}+\frac{\pi ^2}{6}\Big)+\Big(\frac{2 d_1}{(x-1)^5}-2 \
d_1-\frac{2}{(x-1)^5}+2\Big) H(0;\ao) H(0,1;x)+\Big(-\frac{\ao^4}{2 \
(x-1)}+\frac{\ao^4}{2}+\frac{2 \ao^3}{x-1}-\frac{2 \ao^3}{3 (x-1)^2}-\
\frac{8 \ao^3}{3}-\frac{3 \ao^2}{x-1}+\frac{2 \
\ao^2}{(x-1)^2}-\frac{\ao^2}{(x-1)^3}+6 \ao^2+\frac{2 \
\ao}{x-1}-\frac{2 \ao}{(x-1)^2}+\frac{2 \ao}{(x-1)^3}-\frac{2 \
\ao}{(x-1)^4}-8 \ao+\Big(4-\frac{4}{(x-1)^5}\Big) H(0;\ao)+\Big(2 \
d_1-\frac{2 d_1}{(x-1)^5}\Big) H(1;\ao)+\frac{3}{4 (x-1)}-\frac{1}{6 \
(x-1)^2}-\frac{1}{6 (x-1)^3}+\frac{3}{4 (x-1)^4}+\frac{25}{12 \
(x-1)^5}+\frac{25}{12}\Big) H(0,c_1(\ao);x)+\Big(-\frac{2 \
d_1}{x-1}+\frac{d_1}{(x-1)^2}-\frac{2 d_1}{3 (x-1)^3}+\frac{d_1}{2 \
(x-1)^4}-\frac{25 d_1}{6 (x-1)^5}-\frac{5}{4 (x-1)}+\frac{5}{6 \
(x-1)^2}-\frac{5}{6 (x-1)^3}+\frac{5}{4 (x-1)^4}-\frac{25}{12 \
(x-1)^5}+\frac{25}{12}\Big) H(1,0;x)+\Big(\frac{4 d_1}{(x-1)^5}-2 \
d_1-\frac{3}{(x-1)^5}-1\Big) H(0;\ao) H(1,1;x)+\Big(\frac{2 \
d_1}{x-1}-\frac{d_1}{(x-1)^2}+\frac{2 d_1}{3 (x-1)^3}-\frac{d_1}{2 \
(x-1)^4}+\frac{25 d_1}{6 (x-1)^5}+\Big(4-\frac{4}{(x-1)^5}\Big) \
H(0;\ao)+\Big(2 d_1-\frac{2 d_1}{(x-1)^5}\Big) H(1;\ao)+\frac{5}{4 \
(x-1)}-\frac{5}{6 (x-1)^2}+\frac{5}{6 (x-1)^3}-\frac{5}{4 \
(x-1)^4}+\frac{25}{12 (x-1)^5}-\frac{25}{12}\Big) \
H(1,c_1(\ao);x)+\Big(\frac{5 \ao^4}{4 (x-1)}-\frac{5 \
\ao^4}{4}-\frac{5 \ao^3}{x-1}+\frac{5 \ao^3}{3 (x-1)^2}+\frac{20 \
\ao^3}{3}+\frac{15 \ao^2}{2 (x-1)}-\frac{5 \ao^2}{(x-1)^2}+\frac{5 \
\ao^2}{2 (x-1)^3}-15 \ao^2-\frac{5 \ao}{x-1}+\frac{5 \
\ao}{(x-1)^2}-\frac{5 \ao}{(x-1)^3}+\frac{5 \ao}{(x-1)^4}+20 \
\ao+\frac{4 H(0;\ao)}{(x-1)^5}+\frac{2 d_1 \
H(1;\ao)}{(x-1)^5}-\frac{15}{4 (x-1)}+\frac{5}{6 (x-1)^2}+\frac{5}{6 \
(x-1)^3}-\frac{15}{4 (x-1)^4}-\frac{125}{12 \
(x-1)^5}-\frac{125}{12}\Big) H(c_1(\ao),c_1(\ao);x)-\frac{32}{3} \
H(0,0,0;x)+\Big(2-\frac{2}{(x-1)^5}\Big) \
H(0,0,c_1(\ao);x)+\Big(-\frac{2 d_1}{(x-1)^5}+2 \
d_1+\frac{2}{(x-1)^5}-2\Big) H(0,1,0;x)+\Big(\frac{2 d_1}{(x-1)^5}-2 \
d_1-\frac{2}{(x-1)^5}+2\Big) \
H(0,1,c_1(\ao);x)+\Big(5-\frac{1}{(x-1)^5}\Big) \
H(0,c_1(\ao),c_1(\ao);x)+\Big(\frac{4}{(x-1)^5}-4\Big) \
H(1,0,0;x)+\Big(\frac{2 d_1}{(x-1)^5}-\frac{3}{(x-1)^5}-1\Big) \
H(1,0,c_1(\ao);x)+\Big(-\frac{4 d_1}{(x-1)^5}+2 \
d_1+\frac{3}{(x-1)^5}+1\Big) H(1,1,0;x)+\Big(\frac{4 d_1}{(x-1)^5}-2 \
d_1-\frac{3}{(x-1)^5}-1\Big) H(1,1,c_1(\ao);x)+\Big(-\frac{2 \
d_1}{(x-1)^5}-\frac{1}{(x-1)^5}+5\Big) \
H(1,c_1(\ao),c_1(\ao);x)-\frac{2 \
H(c_1(\ao),0,c_1(\ao);x)}{(x-1)^5}+\frac{5 \
H(c_1(\ao),c_1(\ao),c_1(\ao);x)}{(x-1)^5}-\frac{\pi ^2}{8 \
(x-1)}+\frac{\pi ^2}{36 (x-1)^2}+\frac{\pi ^2}{36 (x-1)^3}-\frac{\pi^2}{8 (x-1)^4}-\frac{25 \pi ^2}{72 (x-1)^5}-\frac{3 \
\zeta_3}{(x-1)^5}-6 \zeta_3-\frac{25 \pi ^2}{72},
\erp
% ep^2
\brp
a_2^{(1,-1)} = -\frac{37}{432} d_1^2 \ao^3+\frac{37 d_1 \ao^3}{108}+\frac{37 d_1^2 \
\ao^3}{432 (x-1)^2}-\frac{37 d_1 \ao^3}{108 (x-1)^2}-\frac{\pi ^2 \
\ao^3}{72 (x-1)^2}+\frac{37 \ao^3}{108 (x-1)^2}+\frac{\pi ^2 \
\ao^3}{72}-\frac{37 \ao^3}{108}+\frac{715 d_1^2 \
\ao^2}{864}-\frac{1625 d_1 \ao^2}{432}+\frac{115 d_1^2 \ao^2}{864 \
(x-1)}-\frac{35 d_1 \ao^2}{432 (x-1)}-\frac{\pi ^2 \ao^2}{144 (x-1)}-\
\frac{10 \ao^2}{27 (x-1)}-\frac{107 d_1^2 \ao^2}{864 \
(x-1)^2}+\frac{409 d_1 \ao^2}{432 (x-1)^2}+\frac{5 \pi ^2 \ao^2}{144 \
(x-1)^2}-\frac{151 \ao^2}{108 (x-1)^2}+\frac{493 d_1^2 \ao^2}{864 \
(x-1)^3}-\frac{1181 d_1 \ao^2}{432 (x-1)^3}-\frac{7 \pi ^2 \ao^2}{144 \
(x-1)^3}+\frac{86 \ao^2}{27 (x-1)^3}-\frac{13 \pi ^2 \
\ao^2}{144}+\frac{455 \ao^2}{108}-\frac{3515 d_1^2 \
\ao}{432}+\frac{17285 d_1 \ao}{432}-\frac{265 d_1^2 \ao}{108 \
(x-1)}+\frac{1205 d_1 \ao}{216 (x-1)}+\frac{\pi ^2 \ao}{18 \
(x-1)}+\frac{13 \ao}{54 (x-1)}-\frac{d_1^2 \ao}{108 \
(x-1)^2}-\frac{187 d_1 \ao}{216 (x-1)^2}-\frac{\pi ^2 \ao}{36 \
(x-1)^2}+\frac{191 \ao}{108 (x-1)^2}+\frac{113 d_1^2 \ao}{108 \
(x-1)^3}+\frac{11 d_1 \ao}{216 (x-1)^3}+\frac{\pi ^2 \ao}{18 \
(x-1)^3}-\frac{317 \ao}{54 (x-1)^3}+\frac{2911 d_1^2 \ao}{432 \
(x-1)^4}-\frac{14479 d_1 \ao}{432 (x-1)^4}-\frac{13 \pi ^2 \ao}{72 \
(x-1)^4}+\frac{2207 \ao}{54 (x-1)^4}+\frac{23 \pi ^2 \
\ao}{72}-\frac{5213 \ao}{108}+\Big(-\frac{7 d_1 \ao^3}{18}+\frac{7 \
d_1 \ao^3}{18 (x-1)^2}-\frac{7 \ao^3}{9 (x-1)^2}+\frac{7 \
\ao^3}{9}+\frac{109 d_1 \ao^2}{36}+\frac{13 d_1 \ao^2}{36 \
(x-1)}+\frac{\ao^2}{9 (x-1)}-\frac{29 d_1 \ao^2}{36 (x-1)^2}+\frac{22 \
\ao^2}{9 (x-1)^2}+\frac{67 d_1 \ao^2}{36 (x-1)^3}-\frac{41 \ao^2}{9 \
(x-1)^3}-\frac{62 \ao^2}{9}-\frac{305 d_1 \ao}{18}-\frac{38 d_1 \
\ao}{9 (x-1)}+\frac{25 \ao}{9 (x-1)}+\frac{4 d_1 \ao}{9 \
(x-1)^2}-\frac{23 \ao}{9 (x-1)^2}-\frac{2 d_1 \ao}{9 \
(x-1)^3}+\frac{55 \ao}{9 (x-1)^3}+\frac{217 d_1 \ao}{18 \
(x-1)^4}-\frac{569 \ao}{18 (x-1)^4}+\frac{775 \ao}{18}+\frac{2035 \
d_1^2}{432}-\frac{9685 d_1}{432}+\frac{63 d_1^2}{16 (x-1)}-\frac{407 \
d_1}{48 (x-1)}-\frac{\pi ^2}{8 (x-1)}-\frac{37}{8 (x-1)}-\frac{19 \
d_1^2}{54 (x-1)^2}+\frac{80 d_1}{27 (x-1)^2}+\frac{\pi ^2}{36 \
(x-1)^2}-\frac{101}{108 (x-1)^2}-\frac{19 d_1^2}{54 (x-1)^3}+\frac{73 \
d_1}{54 (x-1)^3}+\frac{\pi ^2}{36 (x-1)^3}-\frac{365}{108 \
(x-1)^3}+\frac{63 d_1^2}{16 (x-1)^4}-\frac{1171 d_1}{48 \
(x-1)^4}-\frac{\pi ^2}{8 (x-1)^4}+\frac{239}{8 (x-1)^4}+\frac{2035 \
d_1^2}{432 (x-1)^5}-\frac{15865 d_1}{432 (x-1)^5}-\frac{25 \pi ^2}{72 \
(x-1)^5}+\frac{13505}{216 (x-1)^5}-\frac{25 \pi \
^2}{72}+\frac{5525}{216}\Big) H(0;\ao)+\Big(-\frac{7}{36} d_1^2 \
\ao^3+\frac{7 d_1 \ao^3}{18}+\frac{7 d_1^2 \ao^3}{36 (x-1)^2}-\frac{7 \
d_1 \ao^3}{18 (x-1)^2}+\frac{109 d_1^2 \ao^2}{72}-\frac{31 d_1 \
\ao^2}{9}+\frac{13 d_1^2 \ao^2}{72 (x-1)}+\frac{d_1 \ao^2}{18 (x-1)}-\
\frac{29 d_1^2 \ao^2}{72 (x-1)^2}+\frac{11 d_1 \ao^2}{9 \
(x-1)^2}+\frac{67 d_1^2 \ao^2}{72 (x-1)^3}-\frac{41 d_1 \ao^2}{18 \
(x-1)^3}-\frac{305 d_1^2 \ao}{36}+\frac{775 d_1 \ao}{36}-\frac{19 \
d_1^2 \ao}{9 (x-1)}+\frac{25 d_1 \ao}{18 (x-1)}+\frac{2 d_1^2 \ao}{9 \
(x-1)^2}-\frac{23 d_1 \ao}{18 (x-1)^2}-\frac{d_1^2 \ao}{9 \
(x-1)^3}+\frac{55 d_1 \ao}{18 (x-1)^3}+\frac{217 d_1^2 \ao}{36 \
(x-1)^4}-\frac{569 d_1 \ao}{36 (x-1)^4}+\frac{515 \
d_1^2}{72}-\frac{665 d_1}{36}+\frac{139 d_1^2}{72 (x-1)}-\frac{13 \
d_1}{9 (x-1)}-\frac{d_1^2}{72 (x-1)^2}+\frac{4 d_1}{9 \
(x-1)^2}-\frac{59 d_1^2}{72 (x-1)^3}-\frac{7 d_1}{9 \
(x-1)^3}-\frac{217 d_1^2}{36 (x-1)^4}+\frac{569 d_1}{36 (x-1)^4}\Big) \
H(1;\ao)+\Big(\frac{4 \ao^3}{3 (x-1)^2}-\frac{4 \ao^3}{3}+\frac{2 \
\ao^2}{3 (x-1)}-\frac{10 \ao^2}{3 (x-1)^2}+\frac{14 \ao^2}{3 \
(x-1)^3}+\frac{26 \ao^2}{3}-\frac{16 \ao}{3 (x-1)}+\frac{8 \ao}{3 \
(x-1)^2}-\frac{16 \ao}{3 (x-1)^3}+\frac{52 \ao}{3 (x-1)^4}-\frac{92 \
\ao}{3}+\frac{205 d_1}{18}+\frac{15 d_1}{2 (x-1)}-\frac{5}{2 \
(x-1)}-\frac{10 d_1}{9 (x-1)^2}+\frac{14}{9 (x-1)^2}-\frac{10 d_1}{9 \
(x-1)^3}+\frac{26}{9 (x-1)^3}+\frac{15 d_1}{2 (x-1)^4}-\frac{49}{2 \
(x-1)^4}+\frac{205 d_1}{18 (x-1)^5}-\frac{935}{18 \
(x-1)^5}-\frac{515}{18}\Big) H(0,0;\ao)+\Big(-\frac{15 d_1}{2 (x-1)}+\
\frac{10 d_1}{9 (x-1)^2}+\frac{10 d_1}{9 (x-1)^3}-\frac{15 d_1}{2 \
(x-1)^4}-\frac{205 d_1}{18 (x-1)^5}-\frac{205 d_1}{18}+\frac{5}{2 \
(x-1)}-\frac{14}{9 (x-1)^2}-\frac{26}{9 (x-1)^3}+\frac{49}{2 \
(x-1)^4}-\frac{8 \pi ^2}{3 (x-1)^5}+\frac{935}{18 (x-1)^5}+\frac{4 \
\pi ^2}{9}+\frac{515}{18}\Big) H(0,0;x)+\Big(-\frac{2 d_1 \
\ao^3}{3}+\frac{2 d_1 \ao^3}{3 (x-1)^2}+\frac{13 d_1 \
\ao^2}{3}+\frac{d_1 \ao^2}{3 (x-1)}-\frac{5 d_1 \ao^2}{3 \
(x-1)^2}+\frac{7 d_1 \ao^2}{3 (x-1)^3}-\frac{46 d_1 \ao}{3}-\frac{8 \
d_1 \ao}{3 (x-1)}+\frac{4 d_1 \ao}{3 (x-1)^2}-\frac{8 d_1 \ao}{3 \
(x-1)^3}+\frac{26 d_1 \ao}{3 (x-1)^4}+\frac{205 d_1^2}{36}-\frac{515 \
d_1}{36}+\frac{15 d_1^2}{4 (x-1)}-\frac{5 d_1}{4 (x-1)}-\frac{5 \
d_1^2}{9 (x-1)^2}+\frac{7 d_1}{9 (x-1)^2}-\frac{5 d_1^2}{9 \
(x-1)^3}+\frac{13 d_1}{9 (x-1)^3}+\frac{15 d_1^2}{4 (x-1)^4}-\frac{49 \
d_1}{4 (x-1)^4}+\frac{205 d_1^2}{36 (x-1)^5}-\frac{935 d_1}{36 \
(x-1)^5}\Big) H(0,1;\ao)+\Big(\frac{\pi ^2 d_1}{(x-1)^5}+\frac{\pi ^2 \
d_1}{3}+\Big(\frac{5 d_1}{2 (x-1)}-\frac{5 d_1}{3 (x-1)^2}+\frac{5 \
d_1}{3 (x-1)^3}-\frac{5 d_1}{2 (x-1)^4}+\frac{25 d_1}{6 \
(x-1)^5}-\frac{25 d_1}{6}+\frac{5}{2 (x-1)}-\frac{5}{3 \
(x-1)^2}+\frac{5}{3 (x-1)^3}-\frac{5}{2 (x-1)^4}+\frac{25}{6 \
(x-1)^5}-\frac{25}{6}\Big) H(0;\ao)+\Big(-\frac{8 d_1}{(x-1)^5}+8 \
d_1+\frac{8}{(x-1)^5}-8\Big) H(0,0;\ao)+\Big(-\frac{4 \
d_1^2}{(x-1)^5}+4 d_1^2+\frac{4 d_1}{(x-1)^5}-4 d_1\Big) \
H(0,1;\ao)-\frac{7 \pi ^2}{3 (x-1)^5}-\frac{\pi ^2}{3}\Big) H(0,1;x)+\
\Big(-\frac{2 d_1 \ao^3}{3}+\frac{2 d_1 \ao^3}{3 (x-1)^2}+\frac{13 \
d_1 \ao^2}{3}+\frac{d_1 \ao^2}{3 (x-1)}-\frac{5 d_1 \ao^2}{3 \
(x-1)^2}+\frac{7 d_1 \ao^2}{3 (x-1)^3}-\frac{46 d_1 \ao}{3}-\frac{8 \
d_1 \ao}{3 (x-1)}+\frac{4 d_1 \ao}{3 (x-1)^2}-\frac{8 d_1 \ao}{3 \
(x-1)^3}+\frac{26 d_1 \ao}{3 (x-1)^4}+\frac{35 d_1}{3}+\frac{7 d_1}{3 \
(x-1)}-\frac{d_1}{3 (x-1)^2}+\frac{d_1}{3 (x-1)^3}-\frac{26 d_1}{3 \
(x-1)^4}\Big) H(1,0;\ao)+\Big(\frac{4 \
d_1^2}{x-1}-\frac{d_1^2}{(x-1)^2}+\frac{4 d_1^2}{9 \
(x-1)^3}-\frac{d_1^2}{4 (x-1)^4}+\frac{205 d_1^2}{36 \
(x-1)^5}-\frac{193 d_1}{24 (x-1)}+\frac{265 d_1}{36 (x-1)^2}-\frac{25 \
d_1}{4 (x-1)^3}+\frac{107 d_1}{24 (x-1)^4}+\frac{4 \pi ^2 d_1}{3 \
(x-1)^5}-\frac{185 d_1}{8 (x-1)^5}-\frac{205 d_1}{72}-\frac{133}{8 \
(x-1)}+\frac{289}{36 (x-1)^2}-\frac{289}{36 (x-1)^3}+\frac{133}{8 \
(x-1)^4}-\frac{11 \pi ^2}{6 (x-1)^5}-\frac{305}{72 (x-1)^5}-\frac{5 \
\pi ^2}{6}+\frac{305}{72}\Big) H(1,0;x)+\Big(-\frac{1}{3} d_1^2 \
\ao^3+\frac{d_1^2 \ao^3}{3 (x-1)^2}+\frac{13 d_1^2 \
\ao^2}{6}+\frac{d_1^2 \ao^2}{6 (x-1)}-\frac{5 d_1^2 \ao^2}{6 \
(x-1)^2}+\frac{7 d_1^2 \ao^2}{6 (x-1)^3}-\frac{23 d_1^2 \
\ao}{3}-\frac{4 d_1^2 \ao}{3 (x-1)}+\frac{2 d_1^2 \ao}{3 \
(x-1)^2}-\frac{4 d_1^2 \ao}{3 (x-1)^3}+\frac{13 d_1^2 \ao}{3 \
(x-1)^4}+\frac{35 d_1^2}{6}+\frac{7 d_1^2}{6 (x-1)}-\frac{d_1^2}{6 \
(x-1)^2}+\frac{d_1^2}{6 (x-1)^3}-\frac{13 d_1^2}{3 (x-1)^4}\Big) \
H(1,1;\ao)+H(0,c_1(\ao);x) \Big(-\frac{d_1 \ao^4}{4}+\frac{d_1 \
\ao^4}{4 (x-1)}-\frac{\ao^4}{2 (x-1)}+\frac{\ao^4}{2}+\frac{13 d_1 \
\ao^3}{9}-\frac{d_1 \ao^3}{x-1}+\frac{3 \ao^3}{x-1}+\frac{4 d_1 \
\ao^3}{9 (x-1)^2}-\frac{25 \ao^3}{18 (x-1)^2}-\frac{61 \
\ao^3}{18}-\frac{23 d_1 \ao^2}{6}+\frac{3 d_1 \ao^2}{2 \
(x-1)}-\frac{31 \ao^2}{4 (x-1)}-\frac{4 d_1 \ao^2}{3 \
(x-1)^2}+\frac{71 \ao^2}{12 (x-1)^2}+\frac{d_1 \
\ao^2}{(x-1)^3}-\frac{15 \ao^2}{4 (x-1)^3}+\frac{131 \
\ao^2}{12}+\frac{25 d_1 \ao}{3}-\frac{d_1 \ao}{x-1}+\frac{13 \
\ao}{x-1}+\frac{4 d_1 \ao}{3 (x-1)^2}-\frac{35 \ao}{3 \
(x-1)^2}-\frac{2 d_1 \ao}{(x-1)^3}+\frac{12 \ao}{(x-1)^3}+\frac{4 d_1 \
\ao}{(x-1)^4}-\frac{29 \ao}{2 (x-1)^4}-\frac{169 \ao}{6}-\frac{205 \
d_1}{72}+\Big(\frac{2 \ao^4}{x-1}-2 \ao^4-\frac{8 \ao^3}{x-1}+\frac{8 \
\ao^3}{3 (x-1)^2}+\frac{32 \ao^3}{3}+\frac{12 \ao^2}{x-1}-\frac{8 \
\ao^2}{(x-1)^2}+\frac{4 \ao^2}{(x-1)^3}-24 \ao^2-\frac{8 \
\ao}{x-1}+\frac{8 \ao}{(x-1)^2}-\frac{8 \ao}{(x-1)^3}+\frac{8 \
\ao}{(x-1)^4}+32 \ao-\frac{3}{x-1}+\frac{2}{3 (x-1)^2}+\frac{2}{3 \
(x-1)^3}-\frac{3}{(x-1)^4}-\frac{25}{3 (x-1)^5}-\frac{25}{3}\Big) \
H(0;\ao)+\Big(-d_1 \ao^4+\frac{d_1 \ao^4}{x-1}+\frac{16 d_1 \
\ao^3}{3}-\frac{4 d_1 \ao^3}{x-1}+\frac{4 d_1 \ao^3}{3 (x-1)^2}-12 \
d_1 \ao^2+\frac{6 d_1 \ao^2}{x-1}-\frac{4 d_1 \ao^2}{(x-1)^2}+\frac{2 \
d_1 \ao^2}{(x-1)^3}+16 d_1 \ao-\frac{4 d_1 \ao}{x-1}+\frac{4 d_1 \
\ao}{(x-1)^2}-\frac{4 d_1 \ao}{(x-1)^3}+\frac{4 d_1 \
\ao}{(x-1)^4}-\frac{25 d_1}{6}-\frac{3 d_1}{2 (x-1)}+\frac{d_1}{3 \
(x-1)^2}+\frac{d_1}{3 (x-1)^3}-\frac{3 d_1}{2 (x-1)^4}-\frac{25 \
d_1}{6 (x-1)^5}\Big) H(1;\ao)+\Big(\frac{16}{(x-1)^5}-16\Big) \
H(0,0;\ao)+\Big(\frac{8 d_1}{(x-1)^5}-8 d_1\Big) \
H(0,1;\ao)+\Big(\frac{8 d_1}{(x-1)^5}-8 d_1\Big) \
H(1,0;\ao)+\Big(\frac{4 d_1^2}{(x-1)^5}-4 d_1^2\Big) \
H(1,1;\ao)-\frac{15 d_1}{8 (x-1)}+\frac{71}{8 (x-1)}+\frac{5 d_1}{18 \
(x-1)^2}-\frac{8}{9 (x-1)^2}+\frac{5 d_1}{18 (x-1)^3}-\frac{2}{9 \
(x-1)^3}-\frac{15 d_1}{8 (x-1)^4}-\frac{17}{8 (x-1)^4}-\frac{205 \
d_1}{72 (x-1)^5}-\frac{\pi ^2}{6 (x-1)^5}+\frac{305}{72 \
(x-1)^5}+\frac{\pi ^2}{6}+\frac{1145}{72}\Big)+H(c_1(\ao);x) \
\Big(\frac{d_1^2 \ao^4}{16}-\frac{d_1 \ao^4}{4}-\frac{d_1^2 \ao^4}{16 \
(x-1)}+\frac{d_1 \ao^4}{4 (x-1)}+\frac{\pi ^2 \ao^4}{24 (x-1)}-\frac{\
\ao^4}{4 (x-1)}-\frac{\pi ^2 \ao^4}{24}+\frac{\ao^4}{4}-\frac{43 \
d_1^2 \ao^3}{108}+\frac{365 d_1 \ao^3}{216}+\frac{d_1^2 \ao^3}{4 \
(x-1)}-\frac{19 d_1 \ao^3}{12 (x-1)}-\frac{\pi ^2 \ao^3}{6 \
(x-1)}+\frac{13 \ao^3}{6 (x-1)}-\frac{4 d_1^2 \ao^3}{27 \
(x-1)^2}+\frac{233 d_1 \ao^3}{216 (x-1)^2}+\frac{\pi ^2 \ao^3}{18 \
(x-1)^2}-\frac{169 \ao^3}{108 (x-1)^2}+\frac{2 \pi ^2 \
\ao^3}{9}-\frac{193 \ao^3}{108}+\frac{95 d_1^2 \ao^2}{72}-\frac{869 \
d_1 \ao^2}{144}-\frac{3 d_1^2 \ao^2}{8 (x-1)}+\frac{695 d_1 \
\ao^2}{144 (x-1)}+\frac{\pi ^2 \ao^2}{4 (x-1)}-\frac{319 \ao^2}{36 \
(x-1)}+\frac{4 d_1^2 \ao^2}{9 (x-1)^2}-\frac{641 d_1 \ao^2}{144 \
(x-1)^2}-\frac{\pi ^2 \ao^2}{6 (x-1)^2}+\frac{26 \ao^2}{3 \
(x-1)^2}-\frac{d_1^2 \ao^2}{2 (x-1)^3}+\frac{623 d_1 \ao^2}{144 \
(x-1)^3}+\frac{\pi ^2 \ao^2}{12 (x-1)^3}-\frac{67 \ao^2}{9 \
(x-1)^3}-\frac{\pi ^2 \ao^2}{2}+\frac{27 \ao^2}{4}-\frac{205 d_1^2 \
\ao}{36}+\frac{1945 d_1 \ao}{72}+\frac{d_1^2 \ao}{4 (x-1)}-\frac{505 \
d_1 \ao}{36 (x-1)}-\frac{\pi ^2 \ao}{6 (x-1)}+\frac{299 \ao}{9 \
(x-1)}-\frac{4 d_1^2 \ao}{9 (x-1)^2}+\frac{11 d_1 \
\ao}{(x-1)^2}+\frac{\pi ^2 \ao}{6 (x-1)^2}-\frac{541 \ao}{18 \
(x-1)^2}+\frac{d_1^2 \ao}{(x-1)^3}-\frac{239 d_1 \ao}{18 \
(x-1)^3}-\frac{\pi ^2 \ao}{6 (x-1)^3}+\frac{296 \ao}{9 \
(x-1)^3}-\frac{4 d_1^2 \ao}{(x-1)^4}+\frac{2237 d_1 \ao}{72 (x-1)^4}+\
\frac{\pi ^2 \ao}{6 (x-1)^4}-\frac{3835 \ao}{72 (x-1)^4}+\frac{2 \pi \
^2 \ao}{3}-\frac{739 \ao}{24}+\frac{2035 d_1^2}{432}-\frac{9685 \
d_1}{432}+\Big(\frac{d_1 \ao^4}{2}-\frac{d_1 \ao^4}{2 \
(x-1)}+\frac{\ao^4}{x-1}-\ao^4-\frac{26 d_1 \ao^3}{9}+\frac{2 d_1 \
\ao^3}{x-1}-\frac{6 \ao^3}{x-1}-\frac{8 d_1 \ao^3}{9 \
(x-1)^2}+\frac{31 \ao^3}{9 (x-1)^2}+\frac{55 \ao^3}{9}+\frac{23 d_1 \
\ao^2}{3}-\frac{3 d_1 \ao^2}{x-1}+\frac{95 \ao^2}{6 (x-1)}+\frac{8 \
d_1 \ao^2}{3 (x-1)^2}-\frac{27 \ao^2}{2 (x-1)^2}-\frac{2 d_1 \
\ao^2}{(x-1)^3}+\frac{59 \ao^2}{6 (x-1)^3}-\frac{35 \
\ao^2}{2}-\frac{50 d_1 \ao}{3}+\frac{2 d_1 \ao}{x-1}-\frac{86 \ao}{3 \
(x-1)}-\frac{8 d_1 \ao}{3 (x-1)^2}+\frac{74 \ao}{3 (x-1)^2}+\frac{4 \
d_1 \ao}{(x-1)^3}-\frac{80 \ao}{3 (x-1)^3}-\frac{8 d_1 \ao}{(x-1)^4}+\
\frac{113 \ao}{3 (x-1)^4}+41 \ao+\frac{205 d_1}{18}+\frac{15 d_1}{2 \
(x-1)}-\frac{5}{2 (x-1)}-\frac{10 d_1}{9 (x-1)^2}+\frac{14}{9 \
(x-1)^2}-\frac{10 d_1}{9 (x-1)^3}+\frac{26}{9 (x-1)^3}+\frac{15 \
d_1}{2 (x-1)^4}-\frac{49}{2 (x-1)^4}+\frac{205 d_1}{18 \
(x-1)^5}-\frac{935}{18 (x-1)^5}-\frac{515}{18}\Big) \
H(0;\ao)+\Big(\frac{d_1^2 \ao^4}{4}-\frac{d_1 \ao^4}{2}-\frac{d_1^2 \
\ao^4}{4 (x-1)}+\frac{d_1 \ao^4}{2 (x-1)}-\frac{13 d_1^2 \
\ao^3}{9}+\frac{55 d_1 \ao^3}{18}+\frac{d_1^2 \ao^3}{x-1}-\frac{3 d_1 \
\ao^3}{x-1}-\frac{4 d_1^2 \ao^3}{9 (x-1)^2}+\frac{31 d_1 \ao^3}{18 \
(x-1)^2}+\frac{23 d_1^2 \ao^2}{6}-\frac{35 d_1 \ao^2}{4}-\frac{3 \
d_1^2 \ao^2}{2 (x-1)}+\frac{95 d_1 \ao^2}{12 (x-1)}+\frac{4 d_1^2 \
\ao^2}{3 (x-1)^2}-\frac{27 d_1 \ao^2}{4 (x-1)^2}-\frac{d_1^2 \
\ao^2}{(x-1)^3}+\frac{59 d_1 \ao^2}{12 (x-1)^3}-\frac{25 d_1^2 \
\ao}{3}+\frac{41 d_1 \ao}{2}+\frac{d_1^2 \ao}{x-1}-\frac{43 d_1 \
\ao}{3 (x-1)}-\frac{4 d_1^2 \ao}{3 (x-1)^2}+\frac{37 d_1 \ao}{3 \
(x-1)^2}+\frac{2 d_1^2 \ao}{(x-1)^3}-\frac{40 d_1 \ao}{3 \
(x-1)^3}-\frac{4 d_1^2 \ao}{(x-1)^4}+\frac{113 d_1 \ao}{6 \
(x-1)^4}+\frac{205 d_1^2}{36}-\frac{515 d_1}{36}+\frac{15 d_1^2}{4 \
(x-1)}-\frac{5 d_1}{4 (x-1)}-\frac{5 d_1^2}{9 (x-1)^2}+\frac{7 d_1}{9 \
(x-1)^2}-\frac{5 d_1^2}{9 (x-1)^3}+\frac{13 d_1}{9 (x-1)^3}+\frac{15 \
d_1^2}{4 (x-1)^4}-\frac{49 d_1}{4 (x-1)^4}+\frac{205 d_1^2}{36 \
(x-1)^5}-\frac{935 d_1}{36 (x-1)^5}\Big) H(1;\ao)+\Big(-\frac{4 \
\ao^4}{x-1}+4 \ao^4+\frac{16 \ao^3}{x-1}-\frac{16 \ao^3}{3 \
(x-1)^2}-\frac{64 \ao^3}{3}-\frac{24 \ao^2}{x-1}+\frac{16 \
\ao^2}{(x-1)^2}-\frac{8 \ao^2}{(x-1)^3}+48 \ao^2+\frac{16 \
\ao}{x-1}-\frac{16 \ao}{(x-1)^2}+\frac{16 \ao}{(x-1)^3}-\frac{16 \
\ao}{(x-1)^4}-64 \ao+\frac{12}{x-1}-\frac{8}{3 (x-1)^2}-\frac{8}{3 \
(x-1)^3}+\frac{12}{(x-1)^4}+\frac{100}{3 (x-1)^5}+\frac{100}{3}\Big) \
H(0,0;\ao)+\Big(2 d_1 \ao^4-\frac{2 d_1 \ao^4}{x-1}-\frac{32 d_1 \
\ao^3}{3}+\frac{8 d_1 \ao^3}{x-1}-\frac{8 d_1 \ao^3}{3 (x-1)^2}+24 \
d_1 \ao^2-\frac{12 d_1 \ao^2}{x-1}+\frac{8 d_1 \
\ao^2}{(x-1)^2}-\frac{4 d_1 \ao^2}{(x-1)^3}-32 d_1 \ao+\frac{8 d_1 \
\ao}{x-1}-\frac{8 d_1 \ao}{(x-1)^2}+\frac{8 d_1 \ao}{(x-1)^3}-\frac{8 \
d_1 \ao}{(x-1)^4}+\frac{50 d_1}{3}+\frac{6 d_1}{x-1}-\frac{4 d_1}{3 \
(x-1)^2}-\frac{4 d_1}{3 (x-1)^3}+\frac{6 d_1}{(x-1)^4}+\frac{50 \
d_1}{3 (x-1)^5}\Big) H(0,1;\ao)+\Big(2 d_1 \ao^4-\frac{2 d_1 \
\ao^4}{x-1}-\frac{32 d_1 \ao^3}{3}+\frac{8 d_1 \ao^3}{x-1}-\frac{8 \
d_1 \ao^3}{3 (x-1)^2}+24 d_1 \ao^2-\frac{12 d_1 \ao^2}{x-1}+\frac{8 \
d_1 \ao^2}{(x-1)^2}-\frac{4 d_1 \ao^2}{(x-1)^3}-32 d_1 \ao+\frac{8 \
d_1 \ao}{x-1}-\frac{8 d_1 \ao}{(x-1)^2}+\frac{8 d_1 \
\ao}{(x-1)^3}-\frac{8 d_1 \ao}{(x-1)^4}+\frac{50 d_1}{3}+\frac{6 \
d_1}{x-1}-\frac{4 d_1}{3 (x-1)^2}-\frac{4 d_1}{3 (x-1)^3}+\frac{6 \
d_1}{(x-1)^4}+\frac{50 d_1}{3 (x-1)^5}\Big) H(1,0;\ao)+\Big(d_1^2 \
\ao^4-\frac{d_1^2 \ao^4}{x-1}-\frac{16 d_1^2 \ao^3}{3}+\frac{4 d_1^2 \
\ao^3}{x-1}-\frac{4 d_1^2 \ao^3}{3 (x-1)^2}+12 d_1^2 \ao^2-\frac{6 \
d_1^2 \ao^2}{x-1}+\frac{4 d_1^2 \ao^2}{(x-1)^2}-\frac{2 d_1^2 \
\ao^2}{(x-1)^3}-16 d_1^2 \ao+\frac{4 d_1^2 \ao}{x-1}-\frac{4 d_1^2 \
\ao}{(x-1)^2}+\frac{4 d_1^2 \ao}{(x-1)^3}-\frac{4 d_1^2 \
\ao}{(x-1)^4}+\frac{25 d_1^2}{3}+\frac{3 d_1^2}{x-1}-\frac{2 d_1^2}{3 \
(x-1)^2}-\frac{2 d_1^2}{3 (x-1)^3}+\frac{3 d_1^2}{(x-1)^4}+\frac{25 \
d_1^2}{3 (x-1)^5}\Big) H(1,1;\ao)+\frac{63 d_1^2}{16 (x-1)}-\frac{407 \
d_1}{48 (x-1)}-\frac{\pi ^2}{8 (x-1)}-\frac{37}{8 (x-1)}-\frac{19 \
d_1^2}{54 (x-1)^2}+\frac{80 d_1}{27 (x-1)^2}+\frac{\pi ^2}{36 \
(x-1)^2}-\frac{101}{108 (x-1)^2}-\frac{19 d_1^2}{54 (x-1)^3}+\frac{73 \
d_1}{54 (x-1)^3}+\frac{\pi ^2}{36 (x-1)^3}-\frac{365}{108 \
(x-1)^3}+\frac{63 d_1^2}{16 (x-1)^4}-\frac{1171 d_1}{48 \
(x-1)^4}-\frac{\pi ^2}{8 (x-1)^4}+\frac{239}{8 (x-1)^4}+\frac{2035 \
d_1^2}{432 (x-1)^5}-\frac{15865 d_1}{432 (x-1)^5}-\frac{25 \pi ^2}{72 \
(x-1)^5}+\frac{13505}{216 (x-1)^5}-\frac{25 \pi \
^2}{72}+\frac{5525}{216}\Big)+\Big(-\frac{2 \pi ^2 d_1^2}{3 (x-1)^5}+\
\frac{2 \pi ^2 d_1}{(x-1)^5}+\frac{\pi ^2 d_1}{3}+\Big(\frac{4 \
d_1^2}{x-1}-\frac{2 d_1^2}{(x-1)^2}+\frac{4 d_1^2}{3 \
(x-1)^3}-\frac{d_1^2}{(x-1)^4}+\frac{25 d_1^2}{3 (x-1)^5}+\frac{9 \
d_1}{2 (x-1)}-\frac{8 d_1}{3 (x-1)^2}+\frac{7 d_1}{3 (x-1)^3}-\frac{3 \
d_1}{(x-1)^4}+\frac{25 d_1}{3 (x-1)^5}-\frac{25 d_1}{6}-\frac{13}{4 \
(x-1)}+\frac{1}{6 (x-1)^2}+\frac{11}{6 (x-1)^3}-\frac{23}{4 (x-1)^4}-\
\frac{125}{12 (x-1)^5}-\frac{175}{12}\Big) H(0;\ao)+\Big(-\frac{16 \
d_1}{(x-1)^5}+8 d_1+\frac{12}{(x-1)^5}+4\Big) \
H(0,0;\ao)+\Big(-\frac{8 d_1^2}{(x-1)^5}+4 d_1^2+\frac{6 \
d_1}{(x-1)^5}+2 d_1\Big) H(0,1;\ao)-\frac{7 \pi ^2}{6 \
(x-1)^5}-\frac{5 \pi ^2}{6}\Big) H(1,1;x)+\Big(-\frac{4 \
d_1^2}{x-1}+\frac{d_1^2}{(x-1)^2}-\frac{4 d_1^2}{9 \
(x-1)^3}+\frac{d_1^2}{4 (x-1)^4}-\frac{205 d_1^2}{36 \
(x-1)^5}+\frac{193 d_1}{24 (x-1)}-\frac{265 d_1}{36 (x-1)^2}+\frac{25 \
d_1}{4 (x-1)^3}-\frac{107 d_1}{24 (x-1)^4}+\frac{185 d_1}{8 (x-1)^5}+\
\frac{205 d_1}{72}+\Big(-\frac{8 d_1}{x-1}+\frac{4 \
d_1}{(x-1)^2}-\frac{8 d_1}{3 (x-1)^3}+\frac{2 d_1}{(x-1)^4}-\frac{50 \
d_1}{3 (x-1)^5}-\frac{5}{x-1}+\frac{10}{3 (x-1)^2}-\frac{10}{3 \
(x-1)^3}+\frac{5}{(x-1)^4}-\frac{25}{3 (x-1)^5}+\frac{25}{3}\Big) \
H(0;\ao)+\Big(-\frac{4 d_1^2}{x-1}+\frac{2 d_1^2}{(x-1)^2}-\frac{4 \
d_1^2}{3 (x-1)^3}+\frac{d_1^2}{(x-1)^4}-\frac{25 d_1^2}{3 \
(x-1)^5}-\frac{5 d_1}{2 (x-1)}+\frac{5 d_1}{3 (x-1)^2}-\frac{5 d_1}{3 \
(x-1)^3}+\frac{5 d_1}{2 (x-1)^4}-\frac{25 d_1}{6 (x-1)^5}+\frac{25 \
d_1}{6}\Big) H(1;\ao)+\Big(\frac{16}{(x-1)^5}-16\Big) \
H(0,0;\ao)+\Big(\frac{8 d_1}{(x-1)^5}-8 d_1\Big) \
H(0,1;\ao)+\Big(\frac{8 d_1}{(x-1)^5}-8 d_1\Big) \
H(1,0;\ao)+\Big(\frac{4 d_1^2}{(x-1)^5}-4 d_1^2\Big) \
H(1,1;\ao)+\frac{133}{8 (x-1)}-\frac{289}{36 (x-1)^2}+\frac{289}{36 \
(x-1)^3}-\frac{133}{8 (x-1)^4}-\frac{\pi ^2}{6 (x-1)^5}+\frac{305}{72 \
(x-1)^5}+\frac{\pi ^2}{6}-\frac{305}{72}\Big) \
H(1,c_1(\ao);x)+\Big(\frac{5 d_1 \ao^4}{8}-\frac{5 d_1 \ao^4}{8 \
(x-1)}+\frac{5 \ao^4}{4 (x-1)}-\frac{5 \ao^4}{4}-\frac{65 d_1 \
\ao^3}{18}+\frac{5 d_1 \ao^3}{2 (x-1)}-\frac{15 \ao^3}{2 \
(x-1)}-\frac{10 d_1 \ao^3}{9 (x-1)^2}+\frac{137 \ao^3}{36 \
(x-1)^2}+\frac{293 \ao^3}{36}+\frac{115 d_1 \ao^2}{12}-\frac{15 d_1 \
\ao^2}{4 (x-1)}+\frac{469 \ao^2}{24 (x-1)}+\frac{10 d_1 \ao^2}{3 \
(x-1)^2}-\frac{125 \ao^2}{8 (x-1)^2}-\frac{5 d_1 \ao^2}{2 \
(x-1)^3}+\frac{253 \ao^2}{24 (x-1)^3}-\frac{201 \ao^2}{8}-\frac{125 \
d_1 \ao}{6}+\frac{5 d_1 \ao}{2 (x-1)}-\frac{203 \ao}{6 \
(x-1)}-\frac{10 d_1 \ao}{3 (x-1)^2}+\frac{179 \ao}{6 (x-1)^2}+\frac{5 \
d_1 \ao}{(x-1)^3}-\frac{94 \ao}{3 (x-1)^3}-\frac{10 d_1 \
\ao}{(x-1)^4}+\frac{487 \ao}{12 (x-1)^4}+\frac{251 \ao}{4}+\frac{1025 \
d_1}{72}+\Big(-\frac{5 \ao^4}{x-1}+5 \ao^4+\frac{20 \
\ao^3}{x-1}-\frac{20 \ao^3}{3 (x-1)^2}-\frac{80 \ao^3}{3}-\frac{30 \
\ao^2}{x-1}+\frac{20 \ao^2}{(x-1)^2}-\frac{10 \ao^2}{(x-1)^3}+60 \
\ao^2+\frac{20 \ao}{x-1}-\frac{20 \ao}{(x-1)^2}+\frac{20 \
\ao}{(x-1)^3}-\frac{20 \ao}{(x-1)^4}-80 \
\ao+\frac{15}{x-1}-\frac{10}{3 (x-1)^2}-\frac{10}{3 \
(x-1)^3}+\frac{15}{(x-1)^4}+\frac{125}{3 (x-1)^5}+\frac{125}{3}\Big) \
H(0;\ao)+\Big(\frac{5 d_1 \ao^4}{2}-\frac{5 d_1 \ao^4}{2 \
(x-1)}-\frac{40 d_1 \ao^3}{3}+\frac{10 d_1 \ao^3}{x-1}-\frac{10 d_1 \
\ao^3}{3 (x-1)^2}+30 d_1 \ao^2-\frac{15 d_1 \ao^2}{x-1}+\frac{10 d_1 \
\ao^2}{(x-1)^2}-\frac{5 d_1 \ao^2}{(x-1)^3}-40 d_1 \ao+\frac{10 d_1 \
\ao}{x-1}-\frac{10 d_1 \ao}{(x-1)^2}+\frac{10 d_1 \
\ao}{(x-1)^3}-\frac{10 d_1 \ao}{(x-1)^4}+\frac{125 d_1}{6}+\frac{15 \
d_1}{2 (x-1)}-\frac{5 d_1}{3 (x-1)^2}-\frac{5 d_1}{3 \
(x-1)^3}+\frac{15 d_1}{2 (x-1)^4}+\frac{125 d_1}{6 (x-1)^5}\Big) H(1;\
\ao)-\frac{16 H(0,0;\ao)}{(x-1)^5}-\frac{8 d_1 \
H(0,1;\ao)}{(x-1)^5}-\frac{8 d_1 H(1,0;\ao)}{(x-1)^5}-\frac{4 d_1^2 \
H(1,1;\ao)}{(x-1)^5}+\frac{75 d_1}{8 (x-1)}-\frac{91}{8 \
(x-1)}-\frac{25 d_1}{18 (x-1)^2}+\frac{22}{9 (x-1)^2}-\frac{25 \
d_1}{18 (x-1)^3}+\frac{28}{9 (x-1)^3}+\frac{75 d_1}{8 \
(x-1)^4}-\frac{179}{8 (x-1)^4}+\frac{1025 d_1}{72 (x-1)^5}+\frac{\pi \
^2}{6 (x-1)^5}-\frac{4045}{72 (x-1)^5}-\frac{3205}{72}\Big) \
H(c_1(\ao),c_1(\ao);x)+\Big(\frac{100}{3}+\frac{12}{x-1}-\frac{8}{3 \
(x-1)^2}-\frac{8}{3 (x-1)^3}+\frac{12}{(x-1)^4}+\frac{100}{3 (x-1)^5}\
\Big) H(0,0,0;\ao)+\Big(-\frac{100}{3}-\frac{12}{x-1}+\frac{8}{3 \
(x-1)^2}+\frac{8}{3 (x-1)^3}-\frac{12}{(x-1)^4}-\frac{100}{3 (x-1)^5}\
\Big) H(0,0,0;x)+\Big(\frac{6 d_1}{x-1}-\frac{4 d_1}{3 \
(x-1)^2}-\frac{4 d_1}{3 (x-1)^3}+\frac{6 d_1}{(x-1)^4}+\frac{50 \
d_1}{3 (x-1)^5}+\frac{50 d_1}{3}\Big) H(0,0,1;\ao)+\Big(-\frac{4 \
d_1}{(x-1)^5}+4 d_1+\frac{12}{(x-1)^5}-12\Big) H(0;\ao) \
H(0,0,1;x)+\Big(-\frac{\ao^4}{x-1}+\ao^4+\frac{4 \ao^3}{x-1}-\frac{4 \
\ao^3}{3 (x-1)^2}-\frac{16 \ao^3}{3}-\frac{6 \ao^2}{x-1}+\frac{4 \
\ao^2}{(x-1)^2}-\frac{2 \ao^2}{(x-1)^3}+12 \ao^2+\frac{4 \
\ao}{x-1}-\frac{4 \ao}{(x-1)^2}+\frac{4 \ao}{(x-1)^3}-\frac{4 \
\ao}{(x-1)^4}-16 \ao+\Big(\frac{8}{(x-1)^5}-8\Big) \
H(0;\ao)+\Big(\frac{4 d_1}{(x-1)^5}-4 d_1\Big) H(1;\ao)+\frac{3}{2 \
(x-1)}-\frac{1}{3 (x-1)^2}-\frac{1}{3 (x-1)^3}+\frac{3}{2 \
(x-1)^4}+\frac{25}{6 (x-1)^5}+\frac{25}{6}\Big) \
H(0,0,c_1(\ao);x)+\Big(\frac{6 d_1}{x-1}-\frac{4 d_1}{3 \
(x-1)^2}-\frac{4 d_1}{3 (x-1)^3}+\frac{6 d_1}{(x-1)^4}+\frac{50 \
d_1}{3 (x-1)^5}+\frac{50 d_1}{3}\Big) H(0,1,0;\ao)+\Big(-\frac{5 \
d_1}{2 (x-1)}+\frac{5 d_1}{3 (x-1)^2}-\frac{5 d_1}{3 (x-1)^3}+\frac{5 \
d_1}{2 (x-1)^4}-\frac{25 d_1}{6 (x-1)^5}+\frac{25 d_1}{6}-\frac{5}{2 \
(x-1)}+\frac{5}{3 (x-1)^2}-\frac{5}{3 (x-1)^3}+\frac{5}{2 \
(x-1)^4}-\frac{25}{6 (x-1)^5}+\frac{25}{6}\Big) \
H(0,1,0;x)+\Big(\frac{3 d_1^2}{x-1}-\frac{2 d_1^2}{3 (x-1)^2}-\frac{2 \
d_1^2}{3 (x-1)^3}+\frac{3 d_1^2}{(x-1)^4}+\frac{25 d_1^2}{3 (x-1)^5}+\
\frac{25 d_1^2}{3}\Big) H(0,1,1;\ao)+\Big(\frac{4 d_1^2}{(x-1)^5}-4 \
d_1^2-\frac{10 d_1}{(x-1)^5}+2 d_1+\frac{14}{(x-1)^5}+2\Big) H(0;\ao) \
H(0,1,1;x)+\Big(\frac{5 d_1}{2 (x-1)}-\frac{5 d_1}{3 (x-1)^2}+\frac{5 \
d_1}{3 (x-1)^3}-\frac{5 d_1}{2 (x-1)^4}+\frac{25 d_1}{6 \
(x-1)^5}-\frac{25 d_1}{6}+\Big(-\frac{8 d_1}{(x-1)^5}+8 \
d_1+\frac{8}{(x-1)^5}-8\Big) H(0;\ao)+\Big(-\frac{4 d_1^2}{(x-1)^5}+4 \
d_1^2+\frac{4 d_1}{(x-1)^5}-4 d_1\Big) H(1;\ao)+\frac{5}{2 \
(x-1)}-\frac{5}{3 (x-1)^2}+\frac{5}{3 (x-1)^3}-\frac{5}{2 \
(x-1)^4}+\frac{25}{6 (x-1)^5}-\frac{25}{6}\Big) \
H(0,1,c_1(\ao);x)+\Big(\frac{5 \ao^4}{2 (x-1)}-\frac{5 \
\ao^4}{2}-\frac{10 \ao^3}{x-1}+\frac{10 \ao^3}{3 (x-1)^2}+\frac{40 \
\ao^3}{3}+\frac{15 \ao^2}{x-1}-\frac{10 \ao^2}{(x-1)^2}+\frac{5 \
\ao^2}{(x-1)^3}-30 \ao^2-\frac{10 \ao}{x-1}+\frac{10 \
\ao}{(x-1)^2}-\frac{10 \ao}{(x-1)^3}+\frac{10 \ao}{(x-1)^4}+40 \
\ao+\Big(\frac{4}{(x-1)^5}-20\Big) H(0;\ao)+\Big(\frac{2 \
d_1}{(x-1)^5}-10 d_1\Big) H(1;\ao)+\frac{3}{4 (x-1)}-\frac{1}{6 \
(x-1)^2}-\frac{1}{6 (x-1)^3}+\frac{3}{4 (x-1)^4}+\frac{25}{12 \
(x-1)^5}+\frac{25}{12}\Big) H(0,c_1(\ao),c_1(\ao);x)+\Big(\frac{8 \
d_1}{x-1}-\frac{4 d_1}{(x-1)^2}+\frac{8 d_1}{3 (x-1)^3}-\frac{2 \
d_1}{(x-1)^4}+\frac{50 d_1}{3 (x-1)^5}+\frac{5}{x-1}-\frac{10}{3 \
(x-1)^2}+\frac{10}{3 (x-1)^3}-\frac{5}{(x-1)^4}+\frac{25}{3 (x-1)^5}-\
\frac{25}{3}\Big) H(1,0,0;x)+\Big(\frac{4 d_1^2}{(x-1)^5}-\frac{10 \
d_1}{(x-1)^5}-2 d_1+\frac{10}{(x-1)^5}-2\Big) H(0;\ao) \
H(1,0,1;x)+\Big(\frac{2 d_1}{x-1}-\frac{d_1}{(x-1)^2}+\frac{2 d_1}{3 \
(x-1)^3}-\frac{d_1}{2 (x-1)^4}+\frac{25 d_1}{6 (x-1)^5}+\Big(-\frac{8 \
d_1}{(x-1)^5}+\frac{12}{(x-1)^5}+4\Big) H(0;\ao)+\Big(-\frac{4 \
d_1^2}{(x-1)^5}+\frac{6 d_1}{(x-1)^5}+2 d_1\Big) H(1;\ao)-\frac{13}{4 \
(x-1)}+\frac{1}{6 (x-1)^2}+\frac{11}{6 (x-1)^3}-\frac{23}{4 (x-1)^4}-\
\frac{125}{12 (x-1)^5}-\frac{175}{12}\Big) \
H(1,0,c_1(\ao);x)+\Big(-\frac{4 d_1^2}{x-1}+\frac{2 \
d_1^2}{(x-1)^2}-\frac{4 d_1^2}{3 \
(x-1)^3}+\frac{d_1^2}{(x-1)^4}-\frac{25 d_1^2}{3 (x-1)^5}-\frac{9 \
d_1}{2 (x-1)}+\frac{8 d_1}{3 (x-1)^2}-\frac{7 d_1}{3 (x-1)^3}+\frac{3 \
d_1}{(x-1)^4}-\frac{25 d_1}{3 (x-1)^5}+\frac{25 d_1}{6}+\frac{13}{4 \
(x-1)}-\frac{1}{6 (x-1)^2}-\frac{11}{6 (x-1)^3}+\frac{23}{4 (x-1)^4}+\
\frac{125}{12 (x-1)^5}+\frac{175}{12}\Big) H(1,1,0;x)+\Big(\frac{12 \
d_1^2}{(x-1)^5}-4 d_1^2-\frac{18 d_1}{(x-1)^5}-4 \
d_1+\frac{7}{(x-1)^5}+5\Big) H(0;\ao) H(1,1,1;x)+\Big(\frac{4 \
d_1^2}{x-1}-\frac{2 d_1^2}{(x-1)^2}+\frac{4 d_1^2}{3 \
(x-1)^3}-\frac{d_1^2}{(x-1)^4}+\frac{25 d_1^2}{3 (x-1)^5}+\frac{9 \
d_1}{2 (x-1)}-\frac{8 d_1}{3 (x-1)^2}+\frac{7 d_1}{3 (x-1)^3}-\frac{3 \
d_1}{(x-1)^4}+\frac{25 d_1}{3 (x-1)^5}-\frac{25 \
d_1}{6}+\Big(-\frac{16 d_1}{(x-1)^5}+8 d_1+\frac{12}{(x-1)^5}+4\Big) \
H(0;\ao)+\Big(-\frac{8 d_1^2}{(x-1)^5}+4 d_1^2+\frac{6 \
d_1}{(x-1)^5}+2 d_1\Big) H(1;\ao)-\frac{13}{4 (x-1)}+\frac{1}{6 \
(x-1)^2}+\frac{11}{6 (x-1)^3}-\frac{23}{4 (x-1)^4}-\frac{125}{12 \
(x-1)^5}-\frac{175}{12}\Big) H(1,1,c_1(\ao);x)+\Big(-\frac{10 \
d_1}{x-1}+\frac{5 d_1}{(x-1)^2}-\frac{10 d_1}{3 (x-1)^3}+\frac{5 \
d_1}{2 (x-1)^4}-\frac{125 d_1}{6 (x-1)^5}+\Big(\frac{8 d_1}{(x-1)^5}+\
\frac{4}{(x-1)^5}-20\Big) H(0;\ao)+\Big(\frac{4 \
d_1^2}{(x-1)^5}+\frac{2 d_1}{(x-1)^5}-10 d_1\Big) H(1;\ao)-\frac{7}{4 \
(x-1)}+\frac{19}{6 (x-1)^2}-\frac{31}{6 (x-1)^3}+\frac{43}{4 \
(x-1)^4}+\frac{25}{12 (x-1)^5}+\frac{275}{12}\Big) \
H(1,c_1(\ao),c_1(\ao);x)+\Big(\frac{3 \ao^4}{2 (x-1)}-\frac{3 \
\ao^4}{2}-\frac{6 \ao^3}{x-1}+\frac{2 \ao^3}{(x-1)^2}+8 \ao^3+\frac{9 \
\ao^2}{x-1}-\frac{6 \ao^2}{(x-1)^2}+\frac{3 \ao^2}{(x-1)^3}-18 \ao^2-\
\frac{6 \ao}{x-1}+\frac{6 \ao}{(x-1)^2}-\frac{6 \ao}{(x-1)^3}+\frac{6 \
\ao}{(x-1)^4}+24 \ao+\frac{8 H(0;\ao)}{(x-1)^5}+\frac{4 d_1 \
H(1;\ao)}{(x-1)^5}-\frac{9}{2 \
(x-1)}+\frac{1}{(x-1)^2}+\frac{1}{(x-1)^3}-\frac{9}{2 \
(x-1)^4}-\frac{25}{2 (x-1)^5}-\frac{25}{2}\Big) \
H(c_1(\ao),0,c_1(\ao);x)+\Big(-\frac{19 \ao^4}{4 (x-1)}+\frac{19 \
\ao^4}{4}+\frac{19 \ao^3}{x-1}-\frac{19 \ao^3}{3 (x-1)^2}-\frac{76 \
\ao^3}{3}-\frac{57 \ao^2}{2 (x-1)}+\frac{19 \ao^2}{(x-1)^2}-\frac{19 \
\ao^2}{2 (x-1)^3}+57 \ao^2+\frac{19 \ao}{x-1}-\frac{19 \ao}{(x-1)^2}+\
\frac{19 \ao}{(x-1)^3}-\frac{19 \ao}{(x-1)^4}-76 \ao-\frac{20 \
H(0;\ao)}{(x-1)^5}-\frac{10 d_1 H(1;\ao)}{(x-1)^5}+\frac{57}{4 \
(x-1)}-\frac{19}{6 (x-1)^2}-\frac{19}{6 (x-1)^3}+\frac{57}{4 \
(x-1)^4}+\frac{475}{12 (x-1)^5}+\frac{475}{12}\Big) \
H(c_1(\ao),c_1(\ao),c_1(\ao);x)+\frac{128}{3} \
H(0,0,0,0;x)+\Big(\frac{12}{(x-1)^5}-12\Big) \
H(0,0,0,c_1(\ao);x)+\Big(\frac{4 d_1}{(x-1)^5}-4 \
d_1-\frac{12}{(x-1)^5}+12\Big) H(0,0,1,0;x)+\Big(-\frac{4 \
d_1}{(x-1)^5}+4 d_1+\frac{12}{(x-1)^5}-12\Big) \
H(0,0,1,c_1(\ao);x)+\Big(-10-\frac{6}{(x-1)^5}\Big) \
H(0,0,c_1(\ao),c_1(\ao);x)+\Big(\frac{8 d_1}{(x-1)^5}-8 \
d_1-\frac{8}{(x-1)^5}+8\Big) H(0,1,0,0;x)+\Big(-\frac{6 \
d_1}{(x-1)^5}-2 d_1+\frac{14}{(x-1)^5}+2\Big) \
H(0,1,0,c_1(\ao);x)+\Big(-\frac{4 d_1^2}{(x-1)^5}+4 d_1^2+\frac{10 \
d_1}{(x-1)^5}-2 d_1-\frac{14}{(x-1)^5}-2\Big) \
H(0,1,1,0;x)+\Big(\frac{4 d_1^2}{(x-1)^5}-4 d_1^2-\frac{10 \
d_1}{(x-1)^5}+2 d_1+\frac{14}{(x-1)^5}+2\Big) \
H(0,1,1,c_1(\ao);x)+\Big(-\frac{2 d_1}{(x-1)^5}+10 \
d_1-\frac{6}{(x-1)^5}-10\Big) \
H(0,1,c_1(\ao),c_1(\ao);x)+\Big(6+\frac{2}{(x-1)^5}\Big) \
H(0,c_1(\ao),0,c_1(\ao);x)+\Big(-19-\frac{1}{(x-1)^5}\Big) \
H(0,c_1(\ao),c_1(\ao),c_1(\ao);x)+\Big(16-\frac{16}{(x-1)^5}\Big) \
H(1,0,0,0;x)+\Big(-\frac{4 d_1}{(x-1)^5}+\frac{10}{(x-1)^5}-2\Big) \
H(1,0,0,c_1(\ao);x)+\Big(-\frac{4 d_1^2}{(x-1)^5}+\frac{10 \
d_1}{(x-1)^5}+2 d_1-\frac{10}{(x-1)^5}+2\Big) \
H(1,0,1,0;x)+\Big(\frac{4 d_1^2}{(x-1)^5}-\frac{10 d_1}{(x-1)^5}-2 \
d_1+\frac{10}{(x-1)^5}-2\Big) H(1,0,1,c_1(\ao);x)+\Big(-\frac{2 \
d_1}{(x-1)^5}+\frac{5}{(x-1)^5}-1\Big) \
H(1,0,c_1(\ao),c_1(\ao);x)+\Big(\frac{16 d_1}{(x-1)^5}-8 \
d_1-\frac{12}{(x-1)^5}-4\Big) H(1,1,0,0;x)+\Big(\frac{4 \
d_1^2}{(x-1)^5}-\frac{12 d_1}{(x-1)^5}-2 d_1+\frac{7}{(x-1)^5}+5\Big) \
H(1,1,0,c_1(\ao);x)+\Big(-\frac{12 d_1^2}{(x-1)^5}+4 d_1^2+\frac{18 \
d_1}{(x-1)^5}+4 d_1-\frac{7}{(x-1)^5}-5\Big) \
H(1,1,1,0;x)+\Big(\frac{12 d_1^2}{(x-1)^5}-4 d_1^2-\frac{18 \
d_1}{(x-1)^5}-4 d_1+\frac{7}{(x-1)^5}+5\Big) \
H(1,1,1,c_1(\ao);x)+\Big(-\frac{4 d_1^2}{(x-1)^5}-\frac{4 \
d_1}{(x-1)^5}+10 d_1+\frac{5}{(x-1)^5}-1\Big) \
H(1,1,c_1(\ao),c_1(\ao);x)+\Big(-\frac{4 \
d_1}{(x-1)^5}+\frac{2}{(x-1)^5}+6\Big) \
H(1,c_1(\ao),0,c_1(\ao);x)+\Big(\frac{10 \
d_1}{(x-1)^5}-\frac{1}{(x-1)^5}-19\Big) \
H(1,c_1(\ao),c_1(\ao),c_1(\ao);x)-\frac{4 \
H(c_1(\ao),0,0,c_1(\ao);x)}{(x-1)^5}+\frac{10 \
H(c_1(\ao),0,c_1(\ao),c_1(\ao);x)}{(x-1)^5}+\frac{6 \
H(c_1(\ao),c_1(\ao),0,c_1(\ao);x)}{(x-1)^5}-\frac{19 \
H(c_1(\ao),c_1(\ao),c_1(\ao),c_1(\ao);x)}{(x-1)^5}+H(0;x) \
\Big(-\frac{63 d_1^2}{16 (x-1)}+\frac{19 d_1^2}{54 (x-1)^2}+\frac{19 \
d_1^2}{54 (x-1)^3}-\frac{63 d_1^2}{16 (x-1)^4}-\frac{2035 d_1^2}{432 \
(x-1)^5}-\frac{2035 d_1^2}{432}+\frac{407 d_1}{48 (x-1)}-\frac{80 \
d_1}{27 (x-1)^2}-\frac{73 d_1}{54 (x-1)^3}+\frac{1171 d_1}{48 \
(x-1)^4}+\frac{15865 d_1}{432 (x-1)^5}+\frac{9685 d_1}{432}+\frac{5 \
\pi ^2}{8 (x-1)}+\frac{37}{8 (x-1)}-\frac{5 \pi ^2}{36 \
(x-1)^2}+\frac{101}{108 (x-1)^2}-\frac{5 \pi ^2}{36 \
(x-1)^3}+\frac{365}{108 (x-1)^3}+\frac{5 \pi ^2}{8 \
(x-1)^4}-\frac{239}{8 (x-1)^4}+\frac{125 \pi ^2}{72 \
(x-1)^5}-\frac{13505}{216 (x-1)^5}+\frac{12 \zeta_3}{(x-1)^5}+24 \
\zeta_3+\frac{125 \pi ^2}{72}-\frac{5525}{216}\Big)+H(1;x) \
\Big(-\frac{\pi ^2 d_1}{3 (x-1)}+\frac{\pi ^2 d_1}{6 \
(x-1)^2}-\frac{\pi ^2 d_1}{9 (x-1)^3}+\frac{\pi ^2 d_1}{12 \
(x-1)^4}-\frac{25 \pi ^2 d_1}{36 (x-1)^5}-\frac{6 \zeta_3 \
d_1}{(x-1)^5}+\Big(-\frac{4 d_1^2}{x-1}+\frac{d_1^2}{(x-1)^2}-\frac{4 \
d_1^2}{9 (x-1)^3}+\frac{d_1^2}{4 (x-1)^4}-\frac{205 d_1^2}{36 \
(x-1)^5}+\frac{193 d_1}{24 (x-1)}-\frac{265 d_1}{36 (x-1)^2}+\frac{25 \
d_1}{4 (x-1)^3}-\frac{107 d_1}{24 (x-1)^4}+\frac{185 d_1}{8 (x-1)^5}+\
\frac{205 d_1}{72}+\frac{133}{8 (x-1)}-\frac{289}{36 \
(x-1)^2}+\frac{289}{36 (x-1)^3}-\frac{133}{8 (x-1)^4}-\frac{\pi ^2}{6 \
(x-1)^5}+\frac{305}{72 (x-1)^5}+\frac{\pi ^2}{6}-\frac{305}{72}\Big) \
H(0;\ao)+\Big(-\frac{8 d_1}{x-1}+\frac{4 d_1}{(x-1)^2}-\frac{8 d_1}{3 \
(x-1)^3}+\frac{2 d_1}{(x-1)^4}-\frac{50 d_1}{3 \
(x-1)^5}-\frac{5}{x-1}+\frac{10}{3 (x-1)^2}-\frac{10}{3 \
(x-1)^3}+\frac{5}{(x-1)^4}-\frac{25}{3 (x-1)^5}+\frac{25}{3}\Big) \
H(0,0;\ao)+\Big(-\frac{4 d_1^2}{x-1}+\frac{2 d_1^2}{(x-1)^2}-\frac{4 \
d_1^2}{3 (x-1)^3}+\frac{d_1^2}{(x-1)^4}-\frac{25 d_1^2}{3 \
(x-1)^5}-\frac{5 d_1}{2 (x-1)}+\frac{5 d_1}{3 (x-1)^2}-\frac{5 d_1}{3 \
(x-1)^3}+\frac{5 d_1}{2 (x-1)^4}-\frac{25 d_1}{6 (x-1)^5}+\frac{25 \
d_1}{6}\Big) H(0,1;\ao)+\Big(\frac{16}{(x-1)^5}-16\Big) H(0,0,0;\ao)+\
\Big(\frac{8 d_1}{(x-1)^5}-8 d_1\Big) H(0,0,1;\ao)+\Big(\frac{8 \
d_1}{(x-1)^5}-8 d_1\Big) H(0,1,0;\ao)+\Big(\frac{4 d_1^2}{(x-1)^5}-4 \
d_1^2\Big) H(0,1,1;\ao)+\frac{13 \pi ^2}{24 (x-1)}-\frac{\pi ^2}{36 \
(x-1)^2}-\frac{11 \pi ^2}{36 (x-1)^3}+\frac{23 \pi ^2}{24 \
(x-1)^4}+\frac{125 \pi ^2}{72 (x-1)^5}-\frac{\zeta_3}{(x-1)^5}+13 \
\zeta_3+\frac{175 \pi ^2}{72}\Big)+\frac{5 d_1 \pi ^2}{16 \
(x-1)}-\frac{71 \pi ^2}{48 (x-1)}-\frac{5 d_1 \pi ^2}{108 \
(x-1)^2}+\frac{4 \pi ^2}{27 (x-1)^2}-\frac{5 d_1 \pi ^2}{108 \
(x-1)^3}+\frac{\pi ^2}{27 (x-1)^3}+\frac{5 d_1 \pi ^2}{16 \
(x-1)^4}+\frac{17 \pi ^2}{48 (x-1)^4}-\frac{61 \pi ^4}{180 \
(x-1)^5}+\frac{205 d_1 \pi ^2}{432 (x-1)^5}-\frac{305 \pi ^2}{432 \
(x-1)^5}-\frac{39 \zeta_3}{4 (x-1)}+\frac{13 \zeta_3}{6 \
(x-1)^2}+\frac{13 \zeta_3}{6 (x-1)^3}-\frac{39 \zeta_3}{4 \
(x-1)^4}-\frac{325 \zeta_3}{12 (x-1)^5}-\frac{325 \
\zeta_3}{12}-\frac{49 \pi ^4}{432}+\frac{205 d_1 \pi \
^2}{432}-\frac{1145 \pi ^2}{432}.
\erp  

%%%
%%% Collinear B-type
%%%

\section{The $\cB$-type collinear integrals}
\label{app:BIntegrals}

%
% The B integral for k=0 and delta=-1
%

\subsection{The $\cB$ integral for $k=0$ and $\delta=-1$}
%
% This file contains the TeX output produced by Mathematica for the integral B0, for delta = -1
%
The $\eps$ expansion for this integral reads
\beq
\bsp
\begin{cal}I\end{cal}(x,\eps;\ao,3+d_1\eps;1,0,-1,g_B) &= x\,\bint(\eps,x;3+d_1\eps;-1,0)\\
&=\frac{1}{\eps}b_{-1}^{(-1,0)}+b_0^{(-1,0)}+\eps b_1^{(-1,0)}+\eps^2b_2^{(-1,0)} +\ocal\left(\eps^3\right),
\esp
\eeq
where
%1/ep piece
\brp
b_{-1}^{(-1,0)}=-\frac{1}{2},
\erp
% ep^0
\brp
b_0^{(-1,0)}=\frac{\ao^4}{4 \
(x-1)}+\frac{\ao^4}{4}-\frac{\ao^3}{x-1}+\frac{\ao^3}{3 \
(x-1)^2}-\frac{4 \ao^3}{3}+\frac{3 \ao^2}{2 \
(x-1)}-\frac{\ao^2}{(x-1)^2}+\frac{\ao^2}{2 (x-1)^3}+3 \
\ao^2-\frac{\ao}{x-1}+\frac{\ao}{(x-1)^2}-\frac{\ao}{(x-1)^3}+\frac{\ao}{(x-1)^4}-4 \ao+\Big(1+\frac{1}{(x-1)^5}\Big) \
H(0;\ao)+\Big(1-\frac{1}{(x-1)^5}\Big) \
H(0;x)+\frac{H(c_1(\ao);x)}{(x-1)^5}-1,
\erp
% ep^1
\brp
b_1^{(-1,0)}=-\frac{d_1 \ao^4}{8}-\frac{d_1 \ao^4}{8 (x-1)}+\frac{3 \ao^4}{4 \
(x-1)}+\frac{3 \ao^4}{4}+\frac{13 d_1 \ao^3}{18}+\frac{d_1 \ao^3}{2 \
(x-1)}-\frac{3 \ao^3}{x-1}-\frac{2 d_1 \ao^3}{9 (x-1)^2}+\frac{23 \
\ao^3}{18 (x-1)^2}-\frac{77 \ao^3}{18}-\frac{23 d_1 \
\ao^2}{12}-\frac{3 d_1 \ao^2}{4 (x-1)}+\frac{53 \ao^2}{12 \
(x-1)}+\frac{2 d_1 \ao^2}{3 (x-1)^2}-\frac{47 \ao^2}{12 \
(x-1)^2}-\frac{d_1 \ao^2}{2 (x-1)^3}+\frac{31 \ao^2}{12 \
(x-1)^3}+\frac{131 \ao^2}{12}+\frac{25 d_1 \ao}{6}+\frac{d_1 \ao}{2 \
(x-1)}-\frac{7 \ao}{3 (x-1)}-\frac{2 d_1 \ao}{3 (x-1)^2}+\frac{4 \
\ao}{(x-1)^2}+\frac{d_1 \ao}{(x-1)^3}-\frac{17 \ao}{3 \
(x-1)^3}-\frac{2 d_1 \ao}{(x-1)^4}+\frac{49 \ao}{6 (x-1)^4}-\frac{121 \
\ao}{6}+\Big(-\frac{\ao^4}{x-1}-\ao^4+\frac{4 \ao^3}{x-1}-\frac{4 \
\ao^3}{3 (x-1)^2}+\frac{16 \ao^3}{3}-\frac{6 \ao^2}{x-1}+\frac{4 \
\ao^2}{(x-1)^2}-\frac{2 \ao^2}{(x-1)^3}-12 \ao^2+\frac{4 \
\ao}{x-1}-\frac{4 \ao}{(x-1)^2}+\frac{4 \ao}{(x-1)^3}-\frac{4 \
\ao}{(x-1)^4}+16 \ao-\frac{5}{2 (x-1)}+\frac{5}{3 (x-1)^2}-\frac{5}{3 \
(x-1)^3}+\frac{5}{2 (x-1)^4}+\frac{37}{6 (x-1)^5}-\frac{13}{6}\Big) \
H(0;\ao)+\Big(\frac{37}{6}+\frac{5}{2 (x-1)}-\frac{5}{3 \
(x-1)^2}+\frac{5}{3 (x-1)^3}-\frac{5}{2 (x-1)^4}-\frac{37}{6 (x-1)^5}\
\Big) H(0;x)+\Big(-\frac{d_1 \ao^4}{2}-\frac{d_1 \ao^4}{2 \
(x-1)}+\frac{8 d_1 \ao^3}{3}+\frac{2 d_1 \ao^3}{x-1}-\frac{2 d_1 \
\ao^3}{3 (x-1)^2}-6 d_1 \ao^2-\frac{3 d_1 \ao^2}{x-1}+\frac{2 d_1 \
\ao^2}{(x-1)^2}-\frac{d_1 \ao^2}{(x-1)^3}+8 d_1 \ao+\frac{2 d_1 \
\ao}{x-1}-\frac{2 d_1 \ao}{(x-1)^2}+\frac{2 d_1 \ao}{(x-1)^3}-\frac{2 \
d_1 \ao}{(x-1)^4}-\frac{25 d_1}{6}-\frac{d_1}{2 (x-1)}+\frac{2 d_1}{3 \
(x-1)^2}-\frac{d_1}{(x-1)^3}+\frac{2 d_1}{(x-1)^4}\Big) \
H(1;\ao)+\Big(\frac{2 d_1}{(x-1)^5}-\frac{2}{(x-1)^5}+2\Big) H(0;\ao) \
H(1;x)+\Big(-\frac{\ao^4}{2 (x-1)}-\frac{\ao^4}{2}+\frac{2 \
\ao^3}{x-1}-\frac{2 \ao^3}{3 (x-1)^2}+\frac{8 \ao^3}{3}-\frac{3 \
\ao^2}{x-1}+\frac{2 \ao^2}{(x-1)^2}-\frac{\ao^2}{(x-1)^3}-6 \
\ao^2+\frac{2 \ao}{x-1}-\frac{2 \ao}{(x-1)^2}+\frac{2 \
\ao}{(x-1)^3}-\frac{2 \ao}{(x-1)^4}+8 \ao-\frac{4 H(0;\ao)}{(x-1)^5}-\
\frac{2 d_1 H(1;\ao)}{(x-1)^5}-\frac{5}{2 (x-1)}+\frac{5}{3 (x-1)^2}-\
\frac{5}{3 (x-1)^3}+\frac{5}{2 (x-1)^4}+\frac{37}{6 \
(x-1)^5}-\frac{25}{6}\Big) \
H(c_1(\ao);x)+\Big(-4-\frac{4}{(x-1)^5}\Big) \
H(0,0;\ao)+\Big(\frac{4}{(x-1)^5}-4\Big) H(0,0;x)+\Big(-\frac{2 \
d_1}{(x-1)^5}-2 d_1\Big) H(0,1;\ao)+\Big(2-\frac{2}{(x-1)^5}\Big) \
H(0,c_1(\ao);x)+\Big(-\frac{2 d_1}{(x-1)^5}+\frac{2}{(x-1)^5}-2\Big) \
H(1,0;x)+\Big(\frac{2 d_1}{(x-1)^5}-\frac{2}{(x-1)^5}+2\Big) H(1,c_1(\
\ao);x)-\frac{2 H(c_1(\ao),c_1(\ao);x)}{(x-1)^5}+\frac{\pi ^2}{3 \
(x-1)^5}-\frac{\pi ^2}{3}-2,
\erp
% ep^2
\brp
b_2^{(-1,0)}=\frac{d_1^2 \ao^4}{16}-\frac{d_1 \ao^4}{2}+\frac{d_1^2 \ao^4}{16 \
(x-1)}-\frac{d_1 \ao^4}{2 (x-1)}-\frac{\pi ^2 \ao^4}{24 \
(x-1)}+\frac{7 \ao^4}{4 (x-1)}-\frac{\pi ^2 \ao^4}{24}+\frac{7 \
\ao^4}{4}-\frac{43 d_1^2 \ao^3}{108}+\frac{349 d_1 \
\ao^3}{108}-\frac{d_1^2 \ao^3}{4 (x-1)}+\frac{2 d_1 \
\ao^3}{x-1}+\frac{\pi ^2 \ao^3}{6 (x-1)}-\frac{7 \ao^3}{x-1}+\frac{4 \
d_1^2 \ao^3}{27 (x-1)^2}-\frac{133 d_1 \ao^3}{108 (x-1)^2}-\frac{\pi \
^2 \ao^3}{18 (x-1)^2}+\frac{191 \ao^3}{54 (x-1)^2}+\frac{2 \pi ^2 \
\ao^3}{9}-\frac{569 \ao^3}{54}+\frac{95 d_1^2 \ao^2}{72}-\frac{85 d_1 \
\ao^2}{8}+\frac{3 d_1^2 \ao^2}{8 (x-1)}-\frac{203 d_1 \ao^2}{72 \
(x-1)}-\frac{\pi ^2 \ao^2}{4 (x-1)}+\frac{353 \ao^2}{36 \
(x-1)}-\frac{4 d_1^2 \ao^2}{9 (x-1)^2}+\frac{31 d_1 \ao^2}{8 \
(x-1)^2}+\frac{\pi ^2 \ao^2}{6 (x-1)^2}-\frac{407 \ao^2}{36 (x-1)^2}+\
\frac{d_1^2 \ao^2}{2 (x-1)^3}-\frac{283 d_1 \ao^2}{72 (x-1)^3}-\frac{\
\pi ^2 \ao^2}{12 (x-1)^3}+\frac{331 \ao^2}{36 (x-1)^3}-\frac{\pi ^2 \
\ao^2}{2}+\frac{1091 \ao^2}{36}-\frac{205 d_1^2 \ao}{36}+\frac{475 \
d_1 \ao}{12}+\frac{2 \ao}{3 (x-2)}-\frac{d_1^2 \ao}{4 \
(x-1)}-\frac{d_1 \ao}{9 (x-1)}+\frac{\pi ^2 \ao}{6 (x-1)}-\frac{13 \
\ao}{36 (x-1)}+\frac{4 d_1^2 \ao}{9 (x-1)^2}-\frac{73 d_1 \ao}{18 \
(x-1)^2}-\frac{\pi ^2 \ao}{6 (x-1)^2}+\frac{35 \ao}{3 \
(x-1)^2}-\frac{d_1^2 \ao}{(x-1)^3}+\frac{173 d_1 \ao}{18 \
(x-1)^3}+\frac{\pi ^2 \ao}{6 (x-1)^3}-\frac{875 \ao}{36 \
(x-1)^3}+\frac{4 d_1^2 \ao}{(x-1)^4}-\frac{937 d_1 \ao}{36 \
(x-1)^4}-\frac{\pi ^2 \ao}{6 (x-1)^4}+\frac{413 \ao}{9 \
(x-1)^4}+\frac{2 \pi ^2 \ao}{3}-\frac{737 \ao}{9}+\Big(\frac{d_1 \
\ao^4}{2}+\frac{d_1 \ao^4}{2 (x-1)}-\frac{3 \ao^4}{x-1}-3 \
\ao^4-\frac{26 d_1 \ao^3}{9}-\frac{2 d_1 \ao^3}{x-1}+\frac{12 \
\ao^3}{x-1}+\frac{8 d_1 \ao^3}{9 (x-1)^2}-\frac{46 \ao^3}{9 (x-1)^2}+\
\frac{154 \ao^3}{9}+\frac{23 d_1 \ao^2}{3}+\frac{3 d_1 \
\ao^2}{x-1}-\frac{53 \ao^2}{3 (x-1)}-\frac{8 d_1 \ao^2}{3 \
(x-1)^2}+\frac{47 \ao^2}{3 (x-1)^2}+\frac{2 d_1 \
\ao^2}{(x-1)^3}-\frac{31 \ao^2}{3 (x-1)^3}-\frac{131 \
\ao^2}{3}-\frac{50 d_1 \ao}{3}-\frac{2 d_1 \ao}{x-1}+\frac{28 \ao}{3 \
(x-1)}+\frac{8 d_1 \ao}{3 (x-1)^2}-\frac{16 \ao}{(x-1)^2}-\frac{4 d_1 \
\ao}{(x-1)^3}+\frac{68 \ao}{3 (x-1)^3}+\frac{8 d_1 \
\ao}{(x-1)^4}-\frac{98 \ao}{3 (x-1)^4}+\frac{242 \ao}{3}+\frac{205 \
d_1}{36}-\frac{4}{x-2}+\frac{17 d_1}{4 (x-1)}-\frac{53}{4 \
(x-1)}+\frac{8}{3 (x-2)^2}-\frac{13 d_1}{9 (x-1)^2}+\frac{317}{36 \
(x-1)^2}+\frac{13 d_1}{9 (x-1)^3}-\frac{371}{36 (x-1)^3}-\frac{17 \
d_1}{4 (x-1)^4}+\frac{193}{12 (x-1)^4}-\frac{205 d_1}{36 \
(x-1)^5}-\frac{\pi ^2}{6 (x-1)^5}+\frac{266}{9 (x-1)^5}-\frac{\pi \
^2}{6}-\frac{194}{9}\Big) H(0;\ao)+\Big(-\frac{17 d_1}{4 \
(x-1)}+\frac{13 d_1}{9 (x-1)^2}-\frac{13 d_1}{9 (x-1)^3}+\frac{17 \
d_1}{4 (x-1)^4}+\frac{205 d_1}{36 (x-1)^5}-\frac{205 \
d_1}{36}+\frac{4}{x-2}+\frac{53}{4 (x-1)}-\frac{8}{3 \
(x-2)^2}-\frac{317}{36 (x-1)^2}+\frac{371}{36 (x-1)^3}-\frac{193}{12 \
(x-1)^4}-\frac{7 \pi ^2}{6 (x-1)^5}-\frac{266}{9 (x-1)^5}+\frac{3 \pi \
^2}{2}+\frac{266}{9}\Big) H(0;x)+\Big(\frac{d_1^2 \ao^4}{4}-\frac{3 \
d_1 \ao^4}{2}+\frac{d_1^2 \ao^4}{4 (x-1)}-\frac{3 d_1 \ao^4}{2 \
(x-1)}-\frac{13 d_1^2 \ao^3}{9}+\frac{77 d_1 \ao^3}{9}-\frac{d_1^2 \
\ao^3}{x-1}+\frac{6 d_1 \ao^3}{x-1}+\frac{4 d_1^2 \ao^3}{9 \
(x-1)^2}-\frac{23 d_1 \ao^3}{9 (x-1)^2}+\frac{23 d_1^2 \
\ao^2}{6}-\frac{131 d_1 \ao^2}{6}+\frac{3 d_1^2 \ao^2}{2 \
(x-1)}-\frac{53 d_1 \ao^2}{6 (x-1)}-\frac{4 d_1^2 \ao^2}{3 \
(x-1)^2}+\frac{47 d_1 \ao^2}{6 (x-1)^2}+\frac{d_1^2 \
\ao^2}{(x-1)^3}-\frac{31 d_1 \ao^2}{6 (x-1)^3}-\frac{25 d_1^2 \
\ao}{3}+\frac{121 d_1 \ao}{3}-\frac{d_1^2 \ao}{x-1}+\frac{14 d_1 \
\ao}{3 (x-1)}+\frac{4 d_1^2 \ao}{3 (x-1)^2}-\frac{8 d_1 \
\ao}{(x-1)^2}-\frac{2 d_1^2 \ao}{(x-1)^3}+\frac{34 d_1 \ao}{3 \
(x-1)^3}+\frac{4 d_1^2 \ao}{(x-1)^4}-\frac{49 d_1 \ao}{3 \
(x-1)^4}+\frac{205 d_1^2}{36}-\frac{230 d_1}{9}+\frac{d_1^2}{4 \
(x-1)}-\frac{d_1}{3 (x-1)}-\frac{4 d_1^2}{9 (x-1)^2}+\frac{49 d_1}{18 \
(x-1)^2}+\frac{d_1^2}{(x-1)^3}-\frac{37 d_1}{6 (x-1)^3}-\frac{4 \
d_1^2}{(x-1)^4}+\frac{49 d_1}{3 (x-1)^4}\Big) H(1;\ao)+\Big(\frac{\pi \
^2}{2 (x-1)^5}-\frac{\pi ^2}{2}\Big) H(2;x)+\Big(\frac{4 \
\ao^4}{x-1}+4 \ao^4-\frac{16 \ao^3}{x-1}+\frac{16 \ao^3}{3 \
(x-1)^2}-\frac{64 \ao^3}{3}+\frac{24 \ao^2}{x-1}-\frac{16 \
\ao^2}{(x-1)^2}+\frac{8 \ao^2}{(x-1)^3}+48 \ao^2-\frac{16 \
\ao}{x-1}+\frac{16 \ao}{(x-1)^2}-\frac{16 \ao}{(x-1)^3}+\frac{16 \
\ao}{(x-1)^4}-64 \ao+\frac{10}{x-1}-\frac{20}{3 (x-1)^2}+\frac{20}{3 \
(x-1)^3}-\frac{10}{(x-1)^4}-\frac{74}{3 (x-1)^5}+\frac{26}{3}\Big) \
H(0,0;\ao)+\Big(-\frac{74}{3}-\frac{10}{x-1}+\frac{20}{3 \
(x-1)^2}-\frac{20}{3 (x-1)^3}+\frac{10}{(x-1)^4}+\frac{74}{3 (x-1)^5}\
\Big) H(0,0;x)+\Big(2 d_1 \ao^4+\frac{2 d_1 \ao^4}{x-1}-\frac{32 d_1 \
\ao^3}{3}-\frac{8 d_1 \ao^3}{x-1}+\frac{8 d_1 \ao^3}{3 (x-1)^2}+24 \
d_1 \ao^2+\frac{12 d_1 \ao^2}{x-1}-\frac{8 d_1 \
\ao^2}{(x-1)^2}+\frac{4 d_1 \ao^2}{(x-1)^3}-32 d_1 \ao-\frac{8 d_1 \
\ao}{x-1}+\frac{8 d_1 \ao}{(x-1)^2}-\frac{8 d_1 \ao}{(x-1)^3}+\frac{8 \
d_1 \ao}{(x-1)^4}+\frac{13 d_1}{3}+\frac{5 d_1}{x-1}-\frac{10 d_1}{3 \
(x-1)^2}+\frac{10 d_1}{3 (x-1)^3}-\frac{5 d_1}{(x-1)^4}-\frac{37 \
d_1}{3 (x-1)^5}\Big) H(0,1;\ao)+H(1;x) \Big(\frac{2 \pi ^2 d_1}{3 \
(x-1)^5}+\Big(-\frac{4 d_1}{x-1}+\frac{2 d_1}{(x-1)^2}-\frac{4 d_1}{3 \
(x-1)^3}+\frac{d_1}{(x-1)^4}+\frac{37 d_1}{3 \
(x-1)^5}+\frac{4}{x-2}+\frac{5}{2 (x-1)}-\frac{8}{3 \
(x-2)^2}-\frac{3}{(x-1)^2}+\frac{8}{3 (x-2)^3}+\frac{5}{3 \
(x-1)^3}-\frac{13}{2 (x-1)^4}-\frac{37}{3 (x-1)^5}+\frac{37}{3}\Big) \
H(0;\ao)+\Big(-\frac{8 d_1}{(x-1)^5}+\frac{8}{(x-1)^5}-8\Big) \
H(0,0;\ao)+\Big(-\frac{4 d_1^2}{(x-1)^5}+\frac{4 d_1}{(x-1)^5}-4 \
d_1\Big) H(0,1;\ao)-\frac{2 \pi ^2}{3 (x-1)^5}+\frac{\pi ^2}{3}\Big)+\
\Big(-\frac{4 d_1}{(x-1)^5}+4 d_1+\frac{8}{(x-1)^5}-8\Big) H(0;\ao) \
H(0,1;x)+\Big(\Big(\frac{8}{(x-1)^5}-8\Big) H(0;\ao)+\Big(\frac{4 \
d_1}{(x-1)^5}-4 d_1\Big) H(1;\ao)+\frac{4}{x-2}+\frac{5}{2 \
(x-1)}-\frac{8}{3 (x-2)^2}-\frac{3}{(x-1)^2}+\frac{8}{3 \
(x-2)^3}+\frac{5}{3 (x-1)^3}-\frac{13}{2 (x-1)^4}-\frac{37}{3 \
(x-1)^5}+\frac{37}{3}\Big) H(0,c_1(\ao);x)+\Big(2 d_1 \ao^4+\frac{2 \
d_1 \ao^4}{x-1}-\frac{32 d_1 \ao^3}{3}-\frac{8 d_1 \
\ao^3}{x-1}+\frac{8 d_1 \ao^3}{3 (x-1)^2}+24 d_1 \ao^2+\frac{12 d_1 \
\ao^2}{x-1}-\frac{8 d_1 \ao^2}{(x-1)^2}+\frac{4 d_1 \
\ao^2}{(x-1)^3}-32 d_1 \ao-\frac{8 d_1 \ao}{x-1}+\frac{8 d_1 \
\ao}{(x-1)^2}-\frac{8 d_1 \ao}{(x-1)^3}+\frac{8 d_1 \
\ao}{(x-1)^4}+\frac{50 d_1}{3}+\frac{2 d_1}{x-1}-\frac{8 d_1}{3 \
(x-1)^2}+\frac{4 d_1}{(x-1)^3}-\frac{8 d_1}{(x-1)^4}\Big) H(1,0;\ao)+\
\Big(\frac{4 d_1}{x-1}-\frac{2 d_1}{(x-1)^2}+\frac{4 d_1}{3 (x-1)^3}-\
\frac{d_1}{(x-1)^4}-\frac{37 d_1}{3 (x-1)^5}-\frac{4}{x-2}-\frac{5}{2 \
(x-1)}+\frac{8}{3 (x-2)^2}+\frac{3}{(x-1)^2}-\frac{8}{3 \
(x-2)^3}-\frac{5}{3 (x-1)^3}+\frac{13}{2 (x-1)^4}+\frac{37}{3 \
(x-1)^5}-\frac{37}{3}\Big) H(1,0;x)+\Big(d_1^2 \ao^4+\frac{d_1^2 \
\ao^4}{x-1}-\frac{16 d_1^2 \ao^3}{3}-\frac{4 d_1^2 \
\ao^3}{x-1}+\frac{4 d_1^2 \ao^3}{3 (x-1)^2}+12 d_1^2 \ao^2+\frac{6 \
d_1^2 \ao^2}{x-1}-\frac{4 d_1^2 \ao^2}{(x-1)^2}+\frac{2 d_1^2 \
\ao^2}{(x-1)^3}-16 d_1^2 \ao-\frac{4 d_1^2 \ao}{x-1}+\frac{4 d_1^2 \
\ao}{(x-1)^2}-\frac{4 d_1^2 \ao}{(x-1)^3}+\frac{4 d_1^2 \
\ao}{(x-1)^4}+\frac{25 d_1^2}{3}+\frac{d_1^2}{x-1}-\frac{4 d_1^2}{3 \
(x-1)^2}+\frac{2 d_1^2}{(x-1)^3}-\frac{4 d_1^2}{(x-1)^4}\Big) \
H(1,1;\ao)+H(c_1(\ao);x) \Big(\frac{d_1 \ao^4}{4}+\frac{d_1 \ao^4}{4 \
(x-1)}-\frac{3 \ao^4}{2 (x-1)}-\frac{3 \ao^4}{2}-\frac{13 d_1 \
\ao^3}{9}-\frac{d_1 \ao^3}{x-1}+\frac{6 \ao^3}{x-1}+\frac{4 d_1 \
\ao^3}{9 (x-1)^2}-\frac{23 \ao^3}{9 (x-1)^2}+\frac{77 \
\ao^3}{9}+\frac{23 d_1 \ao^2}{6}-\frac{\ao^2}{3 (x-2)}+\frac{3 d_1 \
\ao^2}{2 (x-1)}-\frac{103 \ao^2}{12 (x-1)}-\frac{4 d_1 \ao^2}{3 \
(x-1)^2}+\frac{95 \ao^2}{12 (x-1)^2}+\frac{d_1 \
\ao^2}{(x-1)^3}-\frac{31 \ao^2}{6 (x-1)^3}-\frac{131 \
\ao^2}{6}-\frac{25 d_1 \ao}{3}+\frac{2 \ao}{x-2}-\frac{d_1 \ao}{x-1}+\
\frac{10 \ao}{3 (x-1)}-\frac{4 \ao}{3 (x-2)^2}+\frac{4 d_1 \ao}{3 \
(x-1)^2}-\frac{15 \ao}{2 (x-1)^2}-\frac{2 d_1 \ao}{(x-1)^3}+\frac{71 \
\ao}{6 (x-1)^3}+\frac{4 d_1 \ao}{(x-1)^4}-\frac{49 \ao}{3 \
(x-1)^4}+\frac{121 \ao}{3}+\frac{205 d_1}{36}+\Big(\frac{2 \
\ao^4}{x-1}+2 \ao^4-\frac{8 \ao^3}{x-1}+\frac{8 \ao^3}{3 \
(x-1)^2}-\frac{32 \ao^3}{3}+\frac{12 \ao^2}{x-1}-\frac{8 \
\ao^2}{(x-1)^2}+\frac{4 \ao^2}{(x-1)^3}+24 \ao^2-\frac{8 \
\ao}{x-1}+\frac{8 \ao}{(x-1)^2}-\frac{8 \ao}{(x-1)^3}+\frac{8 \
\ao}{(x-1)^4}-32 \ao+\frac{10}{x-1}-\frac{20}{3 (x-1)^2}+\frac{20}{3 \
(x-1)^3}-\frac{10}{(x-1)^4}-\frac{74}{3 (x-1)^5}+\frac{50}{3}\Big) \
H(0;\ao)+\Big(d_1 \ao^4+\frac{d_1 \ao^4}{x-1}-\frac{16 d_1 \ao^3}{3}-\
\frac{4 d_1 \ao^3}{x-1}+\frac{4 d_1 \ao^3}{3 (x-1)^2}+12 d_1 \
\ao^2+\frac{6 d_1 \ao^2}{x-1}-\frac{4 d_1 \ao^2}{(x-1)^2}+\frac{2 d_1 \
\ao^2}{(x-1)^3}-16 d_1 \ao-\frac{4 d_1 \ao}{x-1}+\frac{4 d_1 \
\ao}{(x-1)^2}-\frac{4 d_1 \ao}{(x-1)^3}+\frac{4 d_1 \
\ao}{(x-1)^4}+\frac{25 d_1}{3}+\frac{5 d_1}{x-1}-\frac{10 d_1}{3 \
(x-1)^2}+\frac{10 d_1}{3 (x-1)^3}-\frac{5 d_1}{(x-1)^4}-\frac{37 \
d_1}{3 (x-1)^5}\Big) H(1;\ao)+\frac{16 H(0,0;\ao)}{(x-1)^5}+\frac{8 \
d_1 H(0,1;\ao)}{(x-1)^5}+\frac{8 d_1 H(1,0;\ao)}{(x-1)^5}+\frac{4 \
d_1^2 H(1,1;\ao)}{(x-1)^5}-\frac{4}{x-2}+\frac{17 d_1}{4 \
(x-1)}-\frac{53}{4 (x-1)}+\frac{8}{3 (x-2)^2}-\frac{13 d_1}{9 \
(x-1)^2}+\frac{317}{36 (x-1)^2}+\frac{13 d_1}{9 \
(x-1)^3}-\frac{371}{36 (x-1)^3}-\frac{17 d_1}{4 \
(x-1)^4}+\frac{193}{12 (x-1)^4}-\frac{205 d_1}{36 (x-1)^5}-\frac{\pi \
^2}{6 (x-1)^5}+\frac{266}{9 (x-1)^5}-\frac{230}{9}\Big)+\Big(\frac{4 \
d_1^2}{(x-1)^5}-\frac{8 d_1}{(x-1)^5}+4 d_1+\frac{4}{(x-1)^5}-2\Big) \
H(0;\ao) H(1,1;x)+\Big(-\frac{4 d_1}{x-1}+\frac{2 \
d_1}{(x-1)^2}-\frac{4 d_1}{3 (x-1)^3}+\frac{d_1}{(x-1)^4}+\frac{37 \
d_1}{3 (x-1)^5}+\Big(-\frac{8 d_1}{(x-1)^5}+\frac{8}{(x-1)^5}-8\Big) \
H(0;\ao)+\Big(-\frac{4 d_1^2}{(x-1)^5}+\frac{4 d_1}{(x-1)^5}-4 \
d_1\Big) H(1;\ao)+\frac{4}{x-2}+\frac{5}{2 (x-1)}-\frac{8}{3 \
(x-2)^2}-\frac{3}{(x-1)^2}+\frac{8}{3 (x-2)^3}+\frac{5}{3 \
(x-1)^3}-\frac{13}{2 (x-1)^4}-\frac{37}{3 (x-1)^5}+\frac{37}{3}\Big) \
H(1,c_1(\ao);x)+\Big(2-\frac{2}{(x-1)^5}\Big) H(0;\ao) H(2,1;x)+\Big(\
\frac{\ao^4}{x-1}+\frac{3 \ao^4}{2}-\frac{4 \ao^3}{x-1}+\frac{4 \
\ao^3}{3 (x-1)^2}-8 \ao^3+\frac{6 \ao^2}{x-1}-\frac{4 \
\ao^2}{(x-1)^2}+\frac{2 \ao^2}{(x-1)^3}+18 \ao^2-\frac{4 \
\ao}{x-1}+\frac{4 \ao}{(x-1)^2}-\frac{4 \ao}{(x-1)^3}+\frac{4 \
\ao}{(x-1)^4}-24 \ao+\frac{8 H(0;\ao)}{(x-1)^5}+\frac{4 d_1 \
H(1;\ao)}{(x-1)^5}+\frac{7}{x-1}-\frac{13}{3 \
(x-1)^2}+\frac{4}{(x-1)^3}-\frac{11}{2 (x-1)^4}-\frac{37}{3 (x-1)^5}+\
\frac{25}{2}\Big) H(c_1(\ao),c_1(\ao);x)+\Big(\frac{\ao^4}{2 \
(x-1)}-\frac{\ao^4}{2}-\frac{2 \ao^3}{x-1}+\frac{2 \ao^3}{3 (x-1)^2}+\
\frac{8 \ao^3}{3}+\frac{3 \ao^2}{x-1}-\frac{2 \
\ao^2}{(x-1)^2}+\frac{\ao^2}{(x-1)^3}-6 \ao^2-\frac{2 \
\ao}{x-1}+\frac{2 \ao}{(x-1)^2}-\frac{2 \ao}{(x-1)^3}+\frac{2 \
\ao}{(x-1)^4}+8 \ao-\frac{4}{x-2}+\frac{1}{2 (x-1)}+\frac{8}{3 \
(x-2)^2}+\frac{2}{3 (x-1)^2}-\frac{8}{3 \
(x-2)^3}+\frac{1}{(x-1)^3}+\frac{2}{(x-1)^4}-\frac{25}{6}\Big) H(c_2(\
\ao),c_1(\ao);x)+\Big(16+\frac{16}{(x-1)^5}\Big) \
H(0,0,0;\ao)+\Big(16-\frac{16}{(x-1)^5}\Big) H(0,0,0;x)+\Big(\frac{8 \
d_1}{(x-1)^5}+8 d_1\Big) H(0,0,1;\ao)+\Big(\frac{8}{(x-1)^5}-8\Big) \
H(0,0,c_1(\ao);x)+\Big(\frac{8 d_1}{(x-1)^5}+8 d_1\Big) H(0,1,0;\ao)+\
\Big(\frac{4 d_1}{(x-1)^5}-4 d_1-\frac{8}{(x-1)^5}+8\Big) H(0,1,0;x)+\
\Big(\frac{4 d_1^2}{(x-1)^5}+4 d_1^2\Big) H(0,1,1;\ao)+\Big(-\frac{4 \
d_1}{(x-1)^5}+4 d_1+\frac{8}{(x-1)^5}-8\Big) \
H(0,1,c_1(\ao);x)+\Big(\frac{4}{(x-1)^5}-6\Big) \
H(0,c_1(\ao),c_1(\ao);x)+\Big(2-\frac{2}{(x-1)^5}\Big) \
H(0,c_2(\ao),c_1(\ao);x)+\Big(\frac{8 \
d_1}{(x-1)^5}-\frac{8}{(x-1)^5}+8\Big) H(1,0,0;x)+\Big(-\frac{4 \
d_1}{(x-1)^5}+\frac{4}{(x-1)^5}-2\Big) \
H(1,0,c_1(\ao);x)+\Big(-\frac{4 d_1^2}{(x-1)^5}+\frac{8 \
d_1}{(x-1)^5}-4 d_1-\frac{4}{(x-1)^5}+2\Big) H(1,1,0;x)+\Big(\frac{4 \
d_1^2}{(x-1)^5}-\frac{8 d_1}{(x-1)^5}+4 d_1+\frac{4}{(x-1)^5}-2\Big) \
H(1,1,c_1(\ao);x)+\Big(-\frac{4 \
d_1}{(x-1)^5}+\frac{4}{(x-1)^5}-6\Big) \
H(1,c_1(\ao),c_1(\ao);x)+\Big(2-\frac{2}{(x-1)^5}\Big) \
H(2,0,c_1(\ao);x)+\Big(\frac{2}{(x-1)^5}-2\Big) \
H(2,1,0;x)+\Big(2-\frac{2}{(x-1)^5}\Big) \
H(2,1,c_1(\ao);x)+\Big(\frac{2}{(x-1)^5}-2\Big) \
H(2,c_2(\ao),c_1(\ao);x)+\frac{4 \
H(c_1(\ao),c_1(\ao),c_1(\ao);x)}{(x-1)^5}+\frac{2 \
H(c_1(\ao),c_2(\ao),c_1(\ao);x)}{(x-1)^5}-\frac{\pi ^2}{x-2}-\frac{3 \
\pi ^2}{8 (x-1)}+\frac{2 \pi ^2}{3 (x-2)^2}+\frac{5 \pi ^2}{9 \
(x-1)^2}-\frac{2 \pi ^2}{3 (x-2)^3}-\frac{7 \pi ^2}{36 \
(x-1)^3}+\frac{5 \pi ^2}{4 (x-1)^4}+\frac{37 \pi ^2}{18 \
(x-1)^5}-\frac{21 \zeta_3}{4 (x-1)^5}+\frac{17 \zeta_3}{4}+\frac{\pi \
^2 \ln 2\, }{2 (x-1)^5}-\frac{1}{2} \pi ^2 \ln 2\, -\frac{173 \pi \
^2}{72}-4.
\erp

%
% The B integral for k=1 and delta=-1
%

\subsection{The $\cB$ integral for $k=1$ and $\delta=-1$}
%
% This file contains the TeX output produced by Mathematica for the integral B1, for delta = -1
%
The $\eps$ expansion for this integral reads
\beq
\bsp
\begin{cal}I\end{cal}(x,\eps;\ao,3+d_1\eps;1,1,-1,g_B) &= x\,\bint(\eps,x;3+d_1\eps;-1,1)\\
&=\frac{1}{\eps}b_{-1}^{(-1,1)}+b_0^{(-1,1)}+\eps b_1^{(-1,1)}+\eps^2b_2^{(-1,1)} +\ocal\left(\eps^3\right),
\esp
\eeq
where
%1/ep piece
\brp
b_{-1}^{(-1,1)}=-\frac{1}{4},
\erp
% ep^0
\brp
b_0^{(-1,1)}=\frac{\ao^4}{8 (x-1)}+\frac{\ao^4}{8}-\frac{\ao^3}{2 \
(x-1)}+\frac{\ao^3}{6 (x-1)^2}-\frac{2 \ao^3}{3}+\frac{3 \ao^2}{4 \
(x-1)}-\frac{\ao^2}{2 (x-1)^2}+\frac{\ao^2}{4 (x-1)^3}+\frac{3 \
\ao^2}{2}-\frac{\ao}{2 (x-1)}+\frac{\ao}{2 (x-1)^2}-\frac{\ao}{2 \
(x-1)^3}+\frac{\ao}{2 (x-1)^4}-2 \ao+\Big(\frac{1}{2}+\frac{1}{2 \
(x-1)^5}\Big) H(0;\ao)+\Big(\frac{1}{2}-\frac{1}{2 (x-1)^5}\Big) \
H(0;x)+\frac{H(c_1(\ao);x)}{2 (x-1)^5}-\frac{1}{4},
\erp
% ep^1
\brp
b_1^{(-1,1)}=-\frac{d_1 \ao^4}{16}-\frac{d_1 \ao^4}{16 (x-1)}+\frac{\ao^4}{4 \
(x-1)}+\frac{\ao^4}{4}+\frac{13 d_1 \ao^3}{36}+\frac{\ao^3}{6 (x-2)}+\
\frac{d_1 \ao^3}{4 (x-1)}-\frac{13 \ao^3}{12 (x-1)}-\frac{d_1 \
\ao^3}{9 (x-1)^2}+\frac{17 \ao^3}{36 (x-1)^2}-\frac{53 \
\ao^3}{36}-\frac{23 d_1 \ao^2}{24}-\frac{5 \ao^2}{6 (x-2)}-\frac{3 \
d_1 \ao^2}{8 (x-1)}+\frac{23 \ao^2}{12 (x-1)}+\frac{2 \ao^2}{3 \
(x-2)^2}+\frac{d_1 \ao^2}{3 (x-1)^2}-\frac{19 \ao^2}{12 \
(x-1)^2}-\frac{d_1 \ao^2}{4 (x-1)^3}+\frac{25 \ao^2}{24 \
(x-1)^3}+\frac{95 \ao^2}{24}+\frac{25 d_1 \ao}{12}+\frac{7 \ao}{3 \
(x-2)}+\frac{d_1 \ao}{4 (x-1)}-\frac{9 \ao}{4 (x-1)}-\frac{10 \ao}{3 \
(x-2)^2}-\frac{d_1 \ao}{3 (x-1)^2}+\frac{13 \ao}{6 (x-1)^2}+\frac{4 \
\ao}{(x-2)^3}+\frac{d_1 \ao}{2 (x-1)^3}-\frac{31 \ao}{12 \
(x-1)^3}-\frac{d_1 \ao}{(x-1)^4}+\frac{43 \ao}{12 (x-1)^4}-\frac{97 \
\ao}{12}+\Big(-\frac{\ao^4}{2 (x-1)}-\frac{\ao^4}{2}+\frac{2 \
\ao^3}{x-1}-\frac{2 \ao^3}{3 (x-1)^2}+\frac{8 \ao^3}{3}-\frac{3 \
\ao^2}{x-1}+\frac{2 \ao^2}{(x-1)^2}-\frac{\ao^2}{(x-1)^3}-6 \
\ao^2+\frac{2 \ao}{x-1}-\frac{2 \ao}{(x-1)^2}+\frac{2 \
\ao}{(x-1)^3}-\frac{2 \ao}{(x-1)^4}+8 \ao-\frac{4}{x-2}+\frac{5}{2 \
(x-1)}+\frac{16}{3 (x-2)^2}-\frac{1}{12 \
(x-1)^2}-\frac{8}{(x-2)^3}-\frac{5}{12 \
(x-1)^3}+\frac{16}{(x-2)^4}+\frac{1}{(x-1)^4}+\frac{31}{12 \
(x-1)^5}-\frac{19}{12}\Big) H(0;\ao)+\Big(-\frac{5}{2 \
(x-1)}+\frac{1}{12 (x-1)^2}+\frac{5}{12 \
(x-1)^3}-\frac{1}{(x-1)^4}-\frac{31}{12 \
(x-1)^5}+\frac{31}{12}+\frac{4}{x-2}-\frac{16}{3 \
(x-2)^2}+\frac{8}{(x-2)^3}-\frac{16}{(x-2)^4}\Big) \
H(0;x)+\Big(-\frac{d_1 \ao^4}{4}-\frac{d_1 \ao^4}{4 (x-1)}+\frac{4 \
d_1 \ao^3}{3}+\frac{d_1 \ao^3}{x-1}-\frac{d_1 \ao^3}{3 (x-1)^2}-3 d_1 \
\ao^2-\frac{3 d_1 \ao^2}{2 (x-1)}+\frac{d_1 \ao^2}{(x-1)^2}-\frac{d_1 \
\ao^2}{2 (x-1)^3}+4 d_1 \ao+\frac{d_1 \ao}{x-1}-\frac{d_1 \
\ao}{(x-1)^2}+\frac{d_1 \ao}{(x-1)^3}-\frac{d_1 \
\ao}{(x-1)^4}-\frac{25 d_1}{12}-\frac{d_1}{4 (x-1)}+\frac{d_1}{3 \
(x-1)^2}-\frac{d_1}{2 (x-1)^3}+\frac{d_1}{(x-1)^4}\Big) \
H(1;\ao)+\Big(\frac{d_1}{(x-1)^5}+\frac{16}{(x-2)^5}-\frac{1}{(x-1)^5}\
+1\Big) H(0;\ao) H(1;x)+\Big(-\frac{\ao^4}{4 (x-2)}-\frac{\ao^4}{4 \
(x-1)}-\frac{\ao^4}{4}+\frac{\ao^3}{x-2}+\frac{\ao^3}{x-1}-\frac{2 \
\ao^3}{3 (x-2)^2}-\frac{\ao^3}{3 (x-1)^2}+\frac{4 \ao^3}{3}-\frac{3 \
\ao^2}{2 (x-2)}-\frac{3 \ao^2}{2 (x-1)}+\frac{2 \
\ao^2}{(x-2)^2}+\frac{\ao^2}{(x-1)^2}-\frac{2 \
\ao^2}{(x-2)^3}-\frac{\ao^2}{2 (x-1)^3}-3 \
\ao^2+\frac{\ao}{x-2}+\frac{\ao}{x-1}-\frac{2 \
\ao}{(x-2)^2}-\frac{\ao}{(x-1)^2}+\frac{4 \
\ao}{(x-2)^3}+\frac{\ao}{(x-1)^3}-\frac{8 \
\ao}{(x-2)^4}-\frac{\ao}{(x-1)^4}+4 \ao-\frac{2 \
H(0;\ao)}{(x-1)^5}-\frac{d_1 \
H(1;\ao)}{(x-1)^5}-\frac{4}{x-2}+\frac{5}{2 (x-1)}+\frac{16}{3 \
(x-2)^2}-\frac{1}{12 (x-1)^2}-\frac{8}{(x-2)^3}-\frac{5}{12 (x-1)^3}+\
\frac{16}{(x-2)^4}+\frac{1}{(x-1)^4}+\frac{31}{12 \
(x-1)^5}-\frac{25}{12}\Big) \
H(c_1(\ao);x)+\Big(-2-\frac{2}{(x-1)^5}\Big) \
H(0,0;\ao)+\Big(\frac{2}{(x-1)^5}-2\Big) \
H(0,0;x)+\Big(-\frac{d_1}{(x-1)^5}-d_1\Big) \
H(0,1;\ao)+\Big(-\frac{1}{(x-1)^5}+1+\frac{16}{(x-2)^5}\Big) H(0,c_1(\
\ao);x)+\Big(-\frac{d_1}{(x-1)^5}-\frac{16}{(x-2)^5}+\frac{1}{(x-1)^5}\
-1\Big) H(1,0;x)+\Big(\frac{d_1}{(x-1)^5}+\frac{16}{(x-2)^5}-\frac{1}{\
(x-1)^5}+1\Big) \
H(1,c_1(\ao);x)-\frac{H(c_1(\ao),c_1(\ao);x)}{(x-1)^5}-\frac{16 \
H(c_2(\ao),c_1(\ao);x)}{(x-2)^5}-\frac{4 \pi ^2}{(x-2)^5}+\frac{\pi \
^2}{6 (x-1)^5}-\frac{\pi ^2}{6}-\frac{1}{4},
\erp
% ep^2
\brp
b_2^{(-1,1)}=\frac{d_1^2 \ao^4}{32}-\frac{3 d_1 \ao^4}{16}+\frac{d_1^2 \ao^4}{32 \
(x-1)}-\frac{3 d_1 \ao^4}{16 (x-1)}-\frac{\pi ^2 \ao^4}{48 \
(x-1)}+\frac{3 \ao^4}{8 (x-1)}-\frac{\pi ^2 \ao^4}{48}+\frac{3 \
\ao^4}{8}-\frac{43 d_1^2 \ao^3}{216}+\frac{271 d_1 \
\ao^3}{216}-\frac{7 d_1 \ao^3}{36 (x-2)}+\frac{5 \ao^3}{9 \
(x-2)}-\frac{d_1^2 \ao^3}{8 (x-1)}+\frac{61 d_1 \ao^3}{72 \
(x-1)}+\frac{\pi ^2 \ao^3}{12 (x-1)}-\frac{16 \ao^3}{9 (x-1)}+\frac{2 \
d_1^2 \ao^3}{27 (x-1)^2}-\frac{109 d_1 \ao^3}{216 (x-1)^2}-\frac{\pi \
^2 \ao^3}{36 (x-1)^2}+\frac{26 \ao^3}{27 (x-1)^2}+\frac{\pi ^2 \
\ao^3}{9}-\frac{133 \ao^3}{54}+\frac{95 d_1^2 \ao^2}{144}-\frac{209 \
d_1 \ao^2}{48}+\frac{41 d_1 \ao^2}{36 (x-2)}-\frac{28 \ao^2}{9 \
(x-2)}+\frac{3 d_1^2 \ao^2}{16 (x-1)}-\frac{61 d_1 \ao^2}{36 \
(x-1)}-\frac{\pi ^2 \ao^2}{8 (x-1)}+\frac{253 \ao^2}{72 \
(x-1)}-\frac{10 d_1 \ao^2}{9 (x-2)^2}+\frac{26 \ao^2}{9 \
(x-2)^2}-\frac{2 d_1^2 \ao^2}{9 (x-1)^2}+\frac{43 d_1 \ao^2}{24 \
(x-1)^2}+\frac{\pi ^2 \ao^2}{12 (x-1)^2}-\frac{281 \ao^2}{72 \
(x-1)^2}+\frac{d_1^2 \ao^2}{4 (x-1)^3}-\frac{247 d_1 \ao^2}{144 \
(x-1)^3}-\frac{\pi ^2 \ao^2}{24 (x-1)^3}+\frac{55 \ao^2}{18 (x-1)^3}-\
\frac{\pi ^2 \ao^2}{4}+\frac{295 \ao^2}{36}-\frac{205 d_1^2 \ao}{72}+\
\frac{425 d_1 \ao}{24}-\frac{97 d_1 \ao}{18 (x-2)}+\frac{221 \ao}{18 \
(x-2)}-\frac{d_1^2 \ao}{8 (x-1)}+\frac{91 d_1 \ao}{24 \
(x-1)}+\frac{\pi ^2 \ao}{12 (x-1)}-\frac{17 \ao}{3 (x-1)}+\frac{74 \
d_1 \ao}{9 (x-2)^2}-\frac{178 \ao}{9 (x-2)^2}+\frac{2 d_1^2 \ao}{9 \
(x-1)^2}-\frac{29 d_1 \ao}{9 (x-1)^2}-\frac{\pi ^2 \ao}{12 \
(x-1)^2}+\frac{145 \ao}{18 (x-1)^2}-\frac{12 d_1 \
\ao}{(x-2)^3}+\frac{28 \ao}{(x-2)^3}-\frac{d_1^2 \ao}{2 \
(x-1)^3}+\frac{355 d_1 \ao}{72 (x-1)^3}+\frac{\pi ^2 \ao}{12 \
(x-1)^3}-\frac{409 \ao}{36 (x-1)^3}+\frac{2 d_1^2 \
\ao}{(x-1)^4}-\frac{865 d_1 \ao}{72 (x-1)^4}-\frac{\pi ^2 \ao}{12 \
(x-1)^4}+\frac{661 \ao}{36 (x-1)^4}+\frac{\pi ^2 \ao}{3}-\frac{1039 \
\ao}{36}+\Big(\frac{d_1 \ao^4}{4}+\frac{d_1 \ao^4}{4 \
(x-1)}-\frac{\ao^4}{x-1}-\ao^4-\frac{13 d_1 \ao^3}{9}-\frac{2 \
\ao^3}{3 (x-2)}-\frac{d_1 \ao^3}{x-1}+\frac{13 \ao^3}{3 \
(x-1)}+\frac{4 d_1 \ao^3}{9 (x-1)^2}-\frac{17 \ao^3}{9 \
(x-1)^2}+\frac{53 \ao^3}{9}+\frac{23 d_1 \ao^2}{6}+\frac{10 \ao^2}{3 \
(x-2)}+\frac{3 d_1 \ao^2}{2 (x-1)}-\frac{23 \ao^2}{3 (x-1)}-\frac{8 \
\ao^2}{3 (x-2)^2}-\frac{4 d_1 \ao^2}{3 (x-1)^2}+\frac{19 \ao^2}{3 \
(x-1)^2}+\frac{d_1 \ao^2}{(x-1)^3}-\frac{25 \ao^2}{6 \
(x-1)^3}-\frac{95 \ao^2}{6}-\frac{25 d_1 \ao}{3}-\frac{28 \ao}{3 \
(x-2)}-\frac{d_1 \ao}{x-1}+\frac{9 \ao}{x-1}+\frac{40 \ao}{3 \
(x-2)^2}+\frac{4 d_1 \ao}{3 (x-1)^2}-\frac{26 \ao}{3 \
(x-1)^2}-\frac{16 \ao}{(x-2)^3}-\frac{2 d_1 \ao}{(x-1)^3}+\frac{31 \
\ao}{3 (x-1)^3}+\frac{4 d_1 \ao}{(x-1)^4}-\frac{43 \ao}{3 \
(x-1)^4}+\frac{97 \ao}{3}+\frac{205 d_1}{72}+\frac{34 d_1}{3 \
(x-2)}-\frac{8}{x-2}-\frac{109 d_1}{12 (x-1)}-\frac{1}{3 \
(x-1)}-\frac{116 d_1}{9 (x-2)^2}+\frac{112}{9 (x-2)^2}+\frac{37 \
d_1}{72 (x-1)^2}+\frac{2}{9 (x-1)^2}+\frac{16 \
d_1}{(x-2)^3}-\frac{24}{(x-2)^3}+\frac{29 d_1}{72 \
(x-1)^3}-\frac{31}{18 (x-1)^3}-\frac{32 \
d_1}{(x-2)^4}+\frac{80}{(x-2)^4}-\frac{2 \
d_1}{(x-1)^4}+\frac{4}{(x-1)^4}-\frac{205 d_1}{72 (x-1)^5}-\frac{\pi \
^2}{12 (x-1)^5}+\frac{403}{36 (x-1)^5}-\frac{\pi \
^2}{12}-\frac{367}{36}\Big) H(0;\ao)+\Big(-\frac{34 d_1}{3 \
(x-2)}+\frac{109 d_1}{12 (x-1)}+\frac{116 d_1}{9 (x-2)^2}-\frac{37 \
d_1}{72 (x-1)^2}-\frac{16 d_1}{(x-2)^3}-\frac{29 d_1}{72 \
(x-1)^3}+\frac{32 d_1}{(x-2)^4}+\frac{2 d_1}{(x-1)^4}+\frac{205 \
d_1}{72 (x-1)^5}-\frac{205 d_1}{72}+\frac{8}{x-2}+\frac{1}{3 \
(x-1)}-\frac{112}{9 (x-2)^2}-\frac{2}{9 \
(x-1)^2}+\frac{24}{(x-2)^3}+\frac{31}{18 (x-1)^3}-\frac{80}{(x-2)^4}-\
\frac{4}{(x-1)^4}+\frac{16 \pi ^2}{(x-2)^5}-\frac{7 \pi ^2}{12 \
(x-1)^5}-\frac{403}{36 (x-1)^5}+\frac{3 \pi \
^2}{4}+\frac{403}{36}\Big) H(0;x)+\Big(\frac{d_1^2 \
\ao^4}{8}-\frac{d_1 \ao^4}{2}+\frac{d_1^2 \ao^4}{8 (x-1)}-\frac{d_1 \
\ao^4}{2 (x-1)}-\frac{13 d_1^2 \ao^3}{18}+\frac{53 d_1 \
\ao^3}{18}-\frac{d_1 \ao^3}{3 (x-2)}-\frac{d_1^2 \ao^3}{2 \
(x-1)}+\frac{13 d_1 \ao^3}{6 (x-1)}+\frac{2 d_1^2 \ao^3}{9 \
(x-1)^2}-\frac{17 d_1 \ao^3}{18 (x-1)^2}+\frac{23 d_1^2 \
\ao^2}{12}-\frac{95 d_1 \ao^2}{12}+\frac{5 d_1 \ao^2}{3 \
(x-2)}+\frac{3 d_1^2 \ao^2}{4 (x-1)}-\frac{23 d_1 \ao^2}{6 \
(x-1)}-\frac{4 d_1 \ao^2}{3 (x-2)^2}-\frac{2 d_1^2 \ao^2}{3 (x-1)^2}+\
\frac{19 d_1 \ao^2}{6 (x-1)^2}+\frac{d_1^2 \ao^2}{2 (x-1)^3}-\frac{25 \
d_1 \ao^2}{12 (x-1)^3}-\frac{25 d_1^2 \ao}{6}+\frac{97 d_1 \
\ao}{6}-\frac{14 d_1 \ao}{3 (x-2)}-\frac{d_1^2 \ao}{2 (x-1)}+\frac{9 \
d_1 \ao}{2 (x-1)}+\frac{20 d_1 \ao}{3 (x-2)^2}+\frac{2 d_1^2 \ao}{3 \
(x-1)^2}-\frac{13 d_1 \ao}{3 (x-1)^2}-\frac{8 d_1 \
\ao}{(x-2)^3}-\frac{d_1^2 \ao}{(x-1)^3}+\frac{31 d_1 \ao}{6 (x-1)^3}+\
\frac{2 d_1^2 \ao}{(x-1)^4}-\frac{43 d_1 \ao}{6 (x-1)^4}+\frac{205 \
d_1^2}{72}-\frac{385 d_1}{36}+\frac{10 d_1}{3 (x-2)}+\frac{d_1^2}{8 \
(x-1)}-\frac{7 d_1}{3 (x-1)}-\frac{16 d_1}{3 (x-2)^2}-\frac{2 \
d_1^2}{9 (x-1)^2}+\frac{19 d_1}{9 (x-1)^2}+\frac{8 \
d_1}{(x-2)^3}+\frac{d_1^2}{2 (x-1)^3}-\frac{37 d_1}{12 \
(x-1)^3}-\frac{2 d_1^2}{(x-1)^4}+\frac{43 d_1}{6 (x-1)^4}\Big) \
H(1;\ao)+\Big(\frac{8 \pi ^2 d_1}{(x-2)^5}-\frac{16 \pi ^2}{(x-2)^5}+\
\frac{\pi ^2}{4 (x-1)^5}-\frac{\pi ^2}{4}\Big) H(2;x)+\Big(\frac{2 \
\ao^4}{x-1}+2 \ao^4-\frac{8 \ao^3}{x-1}+\frac{8 \ao^3}{3 \
(x-1)^2}-\frac{32 \ao^3}{3}+\frac{12 \ao^2}{x-1}-\frac{8 \
\ao^2}{(x-1)^2}+\frac{4 \ao^2}{(x-1)^3}+24 \ao^2-\frac{8 \
\ao}{x-1}+\frac{8 \ao}{(x-1)^2}-\frac{8 \ao}{(x-1)^3}+\frac{8 \
\ao}{(x-1)^4}-32 \ao+\frac{16}{x-2}-\frac{10}{x-1}-\frac{64}{3 \
(x-2)^2}+\frac{1}{3 (x-1)^2}+\frac{32}{(x-2)^3}+\frac{5}{3 \
(x-1)^3}-\frac{64}{(x-2)^4}-\frac{4}{(x-1)^4}-\frac{31}{3 \
(x-1)^5}+\frac{19}{3}\Big) H(0,0;\ao)+\Big(\frac{10}{x-1}-\frac{1}{3 \
(x-1)^2}-\frac{5}{3 (x-1)^3}+\frac{4}{(x-1)^4}+\frac{31}{3 \
(x-1)^5}-\frac{31}{3}-\frac{16}{x-2}+\frac{64}{3 \
(x-2)^2}-\frac{32}{(x-2)^3}+\frac{64}{(x-2)^4}\Big) H(0,0;x)+\Big(d_1 \
\ao^4+\frac{d_1 \ao^4}{x-1}-\frac{16 d_1 \ao^3}{3}-\frac{4 d_1 \
\ao^3}{x-1}+\frac{4 d_1 \ao^3}{3 (x-1)^2}+12 d_1 \ao^2+\frac{6 d_1 \
\ao^2}{x-1}-\frac{4 d_1 \ao^2}{(x-1)^2}+\frac{2 d_1 \
\ao^2}{(x-1)^3}-16 d_1 \ao-\frac{4 d_1 \ao}{x-1}+\frac{4 d_1 \
\ao}{(x-1)^2}-\frac{4 d_1 \ao}{(x-1)^3}+\frac{4 d_1 \
\ao}{(x-1)^4}+\frac{19 d_1}{6}+\frac{8 d_1}{x-2}-\frac{5 \
d_1}{x-1}-\frac{32 d_1}{3 (x-2)^2}+\frac{d_1}{6 (x-1)^2}+\frac{16 \
d_1}{(x-2)^3}+\frac{5 d_1}{6 (x-1)^3}-\frac{32 d_1}{(x-2)^4}-\frac{2 \
d_1}{(x-1)^4}-\frac{31 d_1}{6 (x-1)^5}\Big) H(0,1;\ao)+H(1;x) \
\Big(\frac{\pi ^2 d_1}{3 (x-1)^5}+\Big(-\frac{15 d_1}{2 \
(x-2)}+\frac{11 d_1}{2 (x-1)}+\frac{28 d_1}{3 (x-2)^2}-\frac{5 d_1}{6 \
(x-1)^2}-\frac{12 d_1}{(x-2)^3}+\frac{d_1}{6 (x-1)^3}+\frac{16 \
d_1}{(x-2)^4}+\frac{31 d_1}{6 (x-1)^5}-\frac{3}{2 \
(x-2)}+\frac{5}{x-1}+\frac{8}{3 (x-2)^2}-\frac{29}{12 \
(x-1)^2}-\frac{8}{3 (x-2)^3}+\frac{5}{4 (x-1)^3}-\frac{7}{2 (x-1)^4}+\
\frac{16}{(x-2)^5}-\frac{31}{6 (x-1)^5}+\frac{31}{6}\Big) \
H(0;\ao)+\Big(-\frac{4 \
d_1}{(x-1)^5}-\frac{64}{(x-2)^5}+\frac{4}{(x-1)^5}-4\Big) H(0,0;\ao)+\
\Big(-\frac{2 d_1^2}{(x-1)^5}-\frac{32 d_1}{(x-2)^5}+\frac{2 \
d_1}{(x-1)^5}-2 d_1\Big) H(0,1;\ao)-\frac{8 \pi ^2}{3 (x-2)^5}-\frac{\
\pi ^2}{3 (x-1)^5}+\frac{\pi ^2}{6}\Big)+\Big(\frac{32 d_1}{(x-2)^5}-\
\frac{2 d_1}{(x-1)^5}+2 \
d_1-\frac{32}{(x-2)^5}+\frac{4}{(x-1)^5}-4\Big) H(0;\ao) \
H(0,1;x)+\Big(-\frac{\ao^4}{2 (x-2)}+\frac{2 \ao^3}{x-2}-\frac{4 \
\ao^3}{3 (x-2)^2}-\frac{3 \ao^2}{x-2}+\frac{4 \ao^2}{(x-2)^2}-\frac{4 \
\ao^2}{(x-2)^3}+\frac{2 \ao}{x-2}-\frac{4 \ao}{(x-2)^2}+\frac{8 \
\ao}{(x-2)^3}-\frac{16 \
\ao}{(x-2)^4}+\Big(\frac{4}{(x-1)^5}-4-\frac{64}{(x-2)^5}\Big) \
H(0;\ao)+\Big(-\frac{32 d_1}{(x-2)^5}+\frac{2 d_1}{(x-1)^5}-2 \
d_1\Big) H(1;\ao)-\frac{2}{x-2}+\frac{5}{x-1}+\frac{4}{(x-2)^2}-\frac{\
29}{12 (x-1)^2}-\frac{20}{3 (x-2)^3}+\frac{5}{4 \
(x-1)^3}+\frac{16}{(x-2)^4}-\frac{7}{2 \
(x-1)^4}+\frac{16}{(x-2)^5}-\frac{31}{6 (x-1)^5}+\frac{31}{6}\Big) \
H(0,c_1(\ao);x)+\Big(d_1 \ao^4+\frac{d_1 \ao^4}{x-1}-\frac{16 d_1 \
\ao^3}{3}-\frac{4 d_1 \ao^3}{x-1}+\frac{4 d_1 \ao^3}{3 (x-1)^2}+12 \
d_1 \ao^2+\frac{6 d_1 \ao^2}{x-1}-\frac{4 d_1 \ao^2}{(x-1)^2}+\frac{2 \
d_1 \ao^2}{(x-1)^3}-16 d_1 \ao-\frac{4 d_1 \ao}{x-1}+\frac{4 d_1 \
\ao}{(x-1)^2}-\frac{4 d_1 \ao}{(x-1)^3}+\frac{4 d_1 \
\ao}{(x-1)^4}+\frac{25 d_1}{3}+\frac{d_1}{x-1}-\frac{4 d_1}{3 \
(x-1)^2}+\frac{2 d_1}{(x-1)^3}-\frac{4 d_1}{(x-1)^4}\Big) H(1,0;\ao)+\
\Big(\frac{15 d_1}{2 (x-2)}-\frac{11 d_1}{2 (x-1)}-\frac{28 d_1}{3 \
(x-2)^2}+\frac{5 d_1}{6 (x-1)^2}+\frac{12 d_1}{(x-2)^3}-\frac{d_1}{6 \
(x-1)^3}-\frac{16 d_1}{(x-2)^4}-\frac{31 d_1}{6 (x-1)^5}+\frac{3}{2 \
(x-2)}-\frac{5}{x-1}-\frac{8}{3 (x-2)^2}+\frac{29}{12 \
(x-1)^2}+\frac{8}{3 (x-2)^3}-\frac{5}{4 (x-1)^3}+\frac{7}{2 (x-1)^4}-\
\frac{16}{(x-2)^5}+\frac{31}{6 (x-1)^5}-\frac{31}{6}\Big) \
H(1,0;x)+\Big(\frac{d_1^2 \ao^4}{2}+\frac{d_1^2 \ao^4}{2 \
(x-1)}-\frac{8 d_1^2 \ao^3}{3}-\frac{2 d_1^2 \ao^3}{x-1}+\frac{2 \
d_1^2 \ao^3}{3 (x-1)^2}+6 d_1^2 \ao^2+\frac{3 d_1^2 \
\ao^2}{x-1}-\frac{2 d_1^2 \ao^2}{(x-1)^2}+\frac{d_1^2 \
\ao^2}{(x-1)^3}-8 d_1^2 \ao-\frac{2 d_1^2 \ao}{x-1}+\frac{2 d_1^2 \
\ao}{(x-1)^2}-\frac{2 d_1^2 \ao}{(x-1)^3}+\frac{2 d_1^2 \
\ao}{(x-1)^4}+\frac{25 d_1^2}{6}+\frac{d_1^2}{2 (x-1)}-\frac{2 \
d_1^2}{3 (x-1)^2}+\frac{d_1^2}{(x-1)^3}-\frac{2 d_1^2}{(x-1)^4}\Big) \
H(1,1;\ao)+H(c_1(\ao);x) \Big(\frac{d_1 \ao^4}{8}+\frac{d_1 \ao^4}{8 \
(x-2)}-\frac{\ao^4}{2 (x-2)}+\frac{d_1 \ao^4}{8 (x-1)}-\frac{\ao^4}{2 \
(x-1)}-\frac{\ao^4}{2}-\frac{13 d_1 \ao^3}{18}-\frac{d_1 \ao^3}{2 \
(x-2)}+\frac{5 \ao^3}{3 (x-2)}-\frac{d_1 \ao^3}{2 (x-1)}+\frac{29 \
\ao^3}{12 (x-1)}+\frac{4 d_1 \ao^3}{9 (x-2)^2}-\frac{14 \ao^3}{9 \
(x-2)^2}+\frac{2 d_1 \ao^3}{9 (x-1)^2}-\frac{17 \ao^3}{18 \
(x-1)^2}+\frac{53 \ao^3}{18}+\frac{23 d_1 \ao^2}{12}+\frac{3 d_1 \
\ao^2}{4 (x-2)}-\frac{5 \ao^2}{4 (x-2)}+\frac{3 d_1 \ao^2}{4 \
(x-1)}-\frac{61 \ao^2}{12 (x-1)}-\frac{4 d_1 \ao^2}{3 \
(x-2)^2}+\frac{10 \ao^2}{3 (x-2)^2}-\frac{2 d_1 \ao^2}{3 \
(x-1)^2}+\frac{43 \ao^2}{12 (x-1)^2}+\frac{2 d_1 \
\ao^2}{(x-2)^3}-\frac{6 \ao^2}{(x-2)^3}+\frac{d_1 \ao^2}{2 \
(x-1)^3}-\frac{25 \ao^2}{12 (x-1)^3}-\frac{95 \ao^2}{12}-\frac{25 d_1 \
\ao}{6}-\frac{d_1 \ao}{2 (x-2)}-\frac{25 \ao}{6 (x-2)}-\frac{d_1 \
\ao}{2 (x-1)}+\frac{103 \ao}{12 (x-1)}+\frac{4 d_1 \ao}{3 \
(x-2)^2}+\frac{10 \ao}{3 (x-2)^2}+\frac{2 d_1 \ao}{3 \
(x-1)^2}-\frac{73 \ao}{12 (x-1)^2}-\frac{4 d_1 \ao}{(x-2)^3}+\frac{4 \
\ao}{(x-2)^3}-\frac{d_1 \ao}{(x-1)^3}+\frac{37 \ao}{6 \
(x-1)^3}+\frac{16 d_1 \ao}{(x-2)^4}-\frac{40 \ao}{(x-2)^4}+\frac{2 \
d_1 \ao}{(x-1)^4}-\frac{43 \ao}{6 (x-1)^4}+\frac{97 \ao}{6}+\frac{205 \
d_1}{72}+\Big(\frac{\ao^4}{x-2}+\frac{\ao^4}{x-1}+\ao^4-\frac{4 \
\ao^3}{x-2}-\frac{4 \ao^3}{x-1}+\frac{8 \ao^3}{3 (x-2)^2}+\frac{4 \
\ao^3}{3 (x-1)^2}-\frac{16 \ao^3}{3}+\frac{6 \ao^2}{x-2}+\frac{6 \
\ao^2}{x-1}-\frac{8 \ao^2}{(x-2)^2}-\frac{4 \ao^2}{(x-1)^2}+\frac{8 \
\ao^2}{(x-2)^3}+\frac{2 \ao^2}{(x-1)^3}+12 \ao^2-\frac{4 \
\ao}{x-2}-\frac{4 \ao}{x-1}+\frac{8 \ao}{(x-2)^2}+\frac{4 \
\ao}{(x-1)^2}-\frac{16 \ao}{(x-2)^3}-\frac{4 \ao}{(x-1)^3}+\frac{32 \
\ao}{(x-2)^4}+\frac{4 \ao}{(x-1)^4}-16 \
\ao+\frac{16}{x-2}-\frac{10}{x-1}-\frac{64}{3 (x-2)^2}+\frac{1}{3 \
(x-1)^2}+\frac{32}{(x-2)^3}+\frac{5}{3 \
(x-1)^3}-\frac{64}{(x-2)^4}-\frac{4}{(x-1)^4}-\frac{31}{3 \
(x-1)^5}+\frac{25}{3}\Big) H(0;\ao)+\Big(\frac{d_1 \
\ao^4}{2}+\frac{d_1 \ao^4}{2 (x-2)}+\frac{d_1 \ao^4}{2 (x-1)}-\frac{8 \
d_1 \ao^3}{3}-\frac{2 d_1 \ao^3}{x-2}-\frac{2 d_1 \ao^3}{x-1}+\frac{4 \
d_1 \ao^3}{3 (x-2)^2}+\frac{2 d_1 \ao^3}{3 (x-1)^2}+6 d_1 \
\ao^2+\frac{3 d_1 \ao^2}{x-2}+\frac{3 d_1 \ao^2}{x-1}-\frac{4 d_1 \
\ao^2}{(x-2)^2}-\frac{2 d_1 \ao^2}{(x-1)^2}+\frac{4 d_1 \
\ao^2}{(x-2)^3}+\frac{d_1 \ao^2}{(x-1)^3}-8 d_1 \ao-\frac{2 d_1 \
\ao}{x-2}-\frac{2 d_1 \ao}{x-1}+\frac{4 d_1 \ao}{(x-2)^2}+\frac{2 d_1 \
\ao}{(x-1)^2}-\frac{8 d_1 \ao}{(x-2)^3}-\frac{2 d_1 \
\ao}{(x-1)^3}+\frac{16 d_1 \ao}{(x-2)^4}+\frac{2 d_1 \
\ao}{(x-1)^4}+\frac{25 d_1}{6}+\frac{8 d_1}{x-2}-\frac{5 \
d_1}{x-1}-\frac{32 d_1}{3 (x-2)^2}+\frac{d_1}{6 (x-1)^2}+\frac{16 \
d_1}{(x-2)^3}+\frac{5 d_1}{6 (x-1)^3}-\frac{32 d_1}{(x-2)^4}-\frac{2 \
d_1}{(x-1)^4}-\frac{31 d_1}{6 (x-1)^5}\Big) H(1;\ao)+\frac{8 \
H(0,0;\ao)}{(x-1)^5}+\frac{4 d_1 H(0,1;\ao)}{(x-1)^5}+\frac{4 d_1 \
H(1,0;\ao)}{(x-1)^5}+\frac{2 d_1^2 H(1,1;\ao)}{(x-1)^5}+\frac{34 \
d_1}{3 (x-2)}-\frac{8}{x-2}-\frac{109 d_1}{12 (x-1)}-\frac{1}{3 \
(x-1)}-\frac{116 d_1}{9 (x-2)^2}+\frac{112}{9 (x-2)^2}+\frac{37 \
d_1}{72 (x-1)^2}+\frac{2}{9 (x-1)^2}+\frac{16 \
d_1}{(x-2)^3}-\frac{24}{(x-2)^3}+\frac{29 d_1}{72 \
(x-1)^3}-\frac{31}{18 (x-1)^3}-\frac{32 \
d_1}{(x-2)^4}+\frac{80}{(x-2)^4}-\frac{2 \
d_1}{(x-1)^4}+\frac{4}{(x-1)^4}-\frac{205 d_1}{72 (x-1)^5}-\frac{\pi \
^2}{12 (x-1)^5}+\frac{403}{36 \
(x-1)^5}-\frac{385}{36}\Big)+\Big(\frac{2 d_1^2}{(x-1)^5}+\frac{32 \
d_1}{(x-2)^5}-\frac{4 d_1}{(x-1)^5}+2 \
d_1+\frac{16}{(x-2)^5}+\frac{2}{(x-1)^5}-1\Big) H(0;\ao) \
H(1,1;x)+\Big(-\frac{15 d_1}{2 (x-2)}+\frac{11 d_1}{2 (x-1)}+\frac{28 \
d_1}{3 (x-2)^2}-\frac{5 d_1}{6 (x-1)^2}-\frac{12 \
d_1}{(x-2)^3}+\frac{d_1}{6 (x-1)^3}+\frac{16 d_1}{(x-2)^4}+\frac{31 \
d_1}{6 (x-1)^5}+\Big(-\frac{4 \
d_1}{(x-1)^5}-\frac{64}{(x-2)^5}+\frac{4}{(x-1)^5}-4\Big) \
H(0;\ao)+\Big(-\frac{2 d_1^2}{(x-1)^5}-\frac{32 d_1}{(x-2)^5}+\frac{2 \
d_1}{(x-1)^5}-2 d_1\Big) H(1;\ao)-\frac{3}{2 \
(x-2)}+\frac{5}{x-1}+\frac{8}{3 (x-2)^2}-\frac{29}{12 \
(x-1)^2}-\frac{8}{3 (x-2)^3}+\frac{5}{4 (x-1)^3}-\frac{7}{2 (x-1)^4}+\
\frac{16}{(x-2)^5}-\frac{31}{6 (x-1)^5}+\frac{31}{6}\Big) \
H(1,c_1(\ao);x)+\Big(-\frac{32 \
d_1}{(x-2)^5}+\frac{64}{(x-2)^5}-\frac{1}{(x-1)^5}+1\Big) H(0;\ao) \
H(2,1;x)+\Big(\frac{5 \ao^4}{4 (x-2)}+\frac{\ao^4}{2 (x-1)}+\frac{3 \
\ao^4}{4}-\frac{5 \ao^3}{x-2}-\frac{2 \ao^3}{x-1}+\frac{10 \ao^3}{3 \
(x-2)^2}+\frac{2 \ao^3}{3 (x-1)^2}-4 \ao^3+\frac{15 \ao^2}{2 \
(x-2)}+\frac{3 \ao^2}{x-1}-\frac{10 \ao^2}{(x-2)^2}-\frac{2 \
\ao^2}{(x-1)^2}+\frac{10 \ao^2}{(x-2)^3}+\frac{\ao^2}{(x-1)^3}+9 \
\ao^2-\frac{5 \ao}{x-2}-\frac{2 \ao}{x-1}+\frac{10 \
\ao}{(x-2)^2}+\frac{2 \ao}{(x-1)^2}-\frac{20 \ao}{(x-2)^3}-\frac{2 \
\ao}{(x-1)^3}+\frac{40 \ao}{(x-2)^4}+\frac{2 \ao}{(x-1)^4}-12 \
\ao+\frac{4 H(0;\ao)}{(x-1)^5}+\frac{2 d_1 \
H(1;\ao)}{(x-1)^5}+\frac{20}{x-2}-\frac{61}{4 (x-1)}-\frac{80}{3 \
(x-2)^2}+\frac{29}{12 (x-1)^2}+\frac{40}{(x-2)^3}-\frac{1}{12 \
(x-1)^3}-\frac{80}{(x-2)^4}-\frac{3}{2 (x-1)^4}-\frac{31}{6 (x-1)^5}+\
\frac{25}{4}\Big) H(c_1(\ao),c_1(\ao);x)+\Big(\frac{\ao^4}{4 \
(x-1)}-\frac{\ao^4}{4}-\frac{\ao^3}{x-1}+\frac{\ao^3}{3 \
(x-1)^2}+\frac{4 \ao^3}{3}+\frac{3 \ao^2}{2 \
(x-1)}-\frac{\ao^2}{(x-1)^2}+\frac{\ao^2}{2 (x-1)^3}-3 \
\ao^2-\frac{\ao}{x-1}+\frac{\ao}{(x-1)^2}-\frac{\ao}{(x-1)^3}+\frac{\ao}{(x-1)^4}+4 \ao+\frac{64 H(0;\ao)}{(x-2)^5}+\frac{32 d_1 \
H(1;\ao)}{(x-2)^5}-\frac{2}{x-2}+\frac{1}{4 (x-1)}+\frac{4}{3 \
(x-2)^2}+\frac{1}{3 (x-1)^2}-\frac{4}{3 (x-2)^3}+\frac{1}{2 (x-1)^3}+\
\frac{1}{(x-1)^4}-\frac{16}{(x-2)^5}-\frac{25}{12}\Big) \
H(c_2(\ao),c_1(\ao);x)+\Big(8+\frac{8}{(x-1)^5}\Big) \
H(0,0,0;\ao)+\Big(8-\frac{8}{(x-1)^5}\Big) H(0,0,0;x)+\Big(\frac{4 \
d_1}{(x-1)^5}+4 d_1\Big) \
H(0,0,1;\ao)+\Big(\frac{4}{(x-1)^5}-4-\frac{32}{(x-2)^5}\Big) \
H(0,0,c_1(\ao);x)+\Big(\frac{4 d_1}{(x-1)^5}+4 d_1\Big) H(0,1,0;\ao)+\
\Big(-\frac{32 d_1}{(x-2)^5}+\frac{2 d_1}{(x-1)^5}-2 \
d_1+\frac{32}{(x-2)^5}-\frac{4}{(x-1)^5}+4\Big) \
H(0,1,0;x)+\Big(\frac{2 d_1^2}{(x-1)^5}+2 d_1^2\Big) \
H(0,1,1;\ao)+\Big(\frac{32 d_1}{(x-2)^5}-\frac{2 d_1}{(x-1)^5}+2 d_1-\
\frac{32}{(x-2)^5}+\frac{4}{(x-1)^5}-4\Big) \
H(0,1,c_1(\ao);x)+\Big(\frac{2}{(x-1)^5}-3-\frac{80}{(x-2)^5}\Big) \
H(0,c_1(\ao),c_1(\ao);x)+\Big(-\frac{1}{(x-1)^5}+1+\frac{64}{(x-2)^5}\
\Big) H(0,c_2(\ao),c_1(\ao);x)+\Big(\frac{4 \
d_1}{(x-1)^5}+\frac{64}{(x-2)^5}-\frac{4}{(x-1)^5}+4\Big) H(1,0,0;x)+\
\Big(-\frac{2 \
d_1}{(x-1)^5}+\frac{16}{(x-2)^5}+\frac{2}{(x-1)^5}-1\Big) \
H(1,0,c_1(\ao);x)+\Big(-\frac{2 d_1^2}{(x-1)^5}-\frac{32 \
d_1}{(x-2)^5}+\frac{4 d_1}{(x-1)^5}-2 \
d_1-\frac{16}{(x-2)^5}-\frac{2}{(x-1)^5}+1\Big) \
H(1,1,0;x)+\Big(\frac{2 d_1^2}{(x-1)^5}+\frac{32 \
d_1}{(x-2)^5}-\frac{4 d_1}{(x-1)^5}+2 \
d_1+\frac{16}{(x-2)^5}+\frac{2}{(x-1)^5}-1\Big) \
H(1,1,c_1(\ao);x)+\Big(-\frac{2 \
d_1}{(x-1)^5}-\frac{80}{(x-2)^5}+\frac{2}{(x-1)^5}-3\Big) \
H(1,c_1(\ao),c_1(\ao);x)+\Big(-\frac{32 \
d_1}{(x-2)^5}+\frac{64}{(x-2)^5}-\frac{1}{(x-1)^5}+1\Big) \
H(2,0,c_1(\ao);x)+\Big(\frac{32 \
d_1}{(x-2)^5}-\frac{64}{(x-2)^5}+\frac{1}{(x-1)^5}-1\Big) H(2,1,0;x)+\
\Big(-\frac{32 \
d_1}{(x-2)^5}+\frac{64}{(x-2)^5}-\frac{1}{(x-1)^5}+1\Big) \
H(2,1,c_1(\ao);x)+\Big(\frac{32 \
d_1}{(x-2)^5}-\frac{64}{(x-2)^5}+\frac{1}{(x-1)^5}-1\Big) \
H(2,c_2(\ao),c_1(\ao);x)+\frac{2 \
H(c_1(\ao),c_1(\ao),c_1(\ao);x)}{(x-1)^5}+\frac{H(c_1(\ao),c_2(\ao),c_1(\ao);x)}{(x-1)^5}-\frac{32 \
H(c_2(\ao),0,c_1(\ao);x)}{(x-2)^5}+\frac{80 \
H(c_2(\ao),c_1(\ao),c_1(\ao);x)}{(x-2)^5}+\frac{\pi ^2}{6 \
(x-2)}-\frac{13 \pi ^2}{16 (x-1)}-\frac{5 \pi ^2}{9 (x-2)^2}+\frac{31 \
\pi ^2}{72 (x-1)^2}+\frac{\pi ^2}{(x-2)^3}-\frac{\pi ^2}{6 \
(x-1)^3}-\frac{8 \pi ^2}{3 (x-2)^4}+\frac{2 \pi ^2}{3 \
(x-1)^4}-\frac{4 \pi ^2}{(x-2)^5}+\frac{31 \pi ^2}{36 \
(x-1)^5}+\frac{28 \zeta_3}{(x-2)^5}-\frac{21 \zeta_3}{8 \
(x-1)^5}+\frac{17 \zeta_3}{8}-\frac{24 \pi ^2 \ln 2\, \
}{(x-2)^5}+\frac{\pi ^2 \ln 2\, }{4 (x-1)^5}-\frac{1}{4} \pi ^2 \ln 2\
\, -\frac{149 \pi ^2}{144}-\frac{1}{4}
\erp

%
% The B integral for k=2 and delta=-1
%

\subsection{The $\cB$ integral for $k=2$ and $\delta=-1$ and $d_1=-3$}
%
% This file contains the TeX output produced by Mathematica for the integral B1, for delta = -1
%
The $\eps$ expansion for this integral reads
\beq
\bsp
\begin{cal}I\end{cal}(x,\eps;\ao,3+d_1\eps;1,2,-1,g_B) &= x\,\bint(\eps,x;3+d_1\eps;-1,2)\\
&=\frac{1}{\eps}b_{-1}^{(-1,2)}+b_0^{(-1,2)}+\eps b_1^{(-1,2)}+\eps^2b_2^{(-1,2)} +\ocal\left(\eps^3\right),
\esp
\eeq
where
%1/ep piece
\brp
b_{-1}^{(-1,2)}=-\frac{1}{6},
\erp
% ep^0
\brp
b_0^{(-1,2)}=-\frac{\ao^{10}}{3 (\ao+1)^4 (x \ao-2 \ao-x)}+\frac{\ao^9}{3
(\ao+1)^4 (x \ao-2 \ao-x)}+\frac{\ao^8}{12 (\ao+1)^4 (x-1)}+\frac{4 \
\ao^8}{3 (\ao+1)^4 (x \ao-2 \ao-x)}-\frac{4 \ao^7}{3 (\ao+1)^4 (x \
\ao-2 \ao-x)}-\frac{\ao^6}{3 (\ao+1)^4 (x-1)}-\frac{2 \
\ao^6}{(\ao+1)^4 (x \ao-2 \ao-x)}+\frac{\ao^6}{9 (\ao+1)^3 \
(x-1)^2}+\frac{\ao^5}{3 (x-2)}+\frac{2 \ao^5}{(\ao+1)^4 (x \ao-2 \
\ao-x)}-\frac{5 \ao^4}{4 (x-2)}+\frac{5 \ao^4}{12 (\ao+1)^4 \
(x-1)}+\frac{4 \ao^4}{3 (\ao+1)^4 (x \ao-2 \ao-x)}+\frac{5 \ao^4}{6 \
(x-2)^2}-\frac{\ao^4}{3 (\ao+1)^3 (x-1)^2}+\frac{\ao^4}{6 (\ao+1)^2 \
(x-1)^3}+\frac{\ao^4}{12}+\frac{5 \ao^3}{3 (x-2)}-\frac{\ao^3}{3 \
(\ao+1)^4 (x-1)}-\frac{4 \ao^3}{3 (\ao+1)^4 (x \ao-2 \ao-x)}-\frac{20 \
\ao^3}{9 (x-2)^2}+\frac{\ao^3}{9 (\ao+1)^3 (x-1)^2}+\frac{20 \ao^3}{9 \
(x-2)^3}-\frac{4 \ao^3}{9}-\frac{5 \ao^2}{6 (x-2)}-\frac{5 \ao^2}{6 (\
\ao+1)^4 (x-1)}-\frac{\ao^2}{3 (\ao+1)^4 (x \ao-2 \ao-x)}+\frac{5 \
\ao^2}{3 (x-2)^2}+\frac{2 \ao^2}{3 (\ao+1)^3 (x-1)^2}-\frac{10 \
\ao^2}{3 (x-2)^3}-\frac{\ao^2}{2 (\ao+1)^2 (x-1)^3}+\frac{20 \ao^2}{3 \
(x-2)^4}+\frac{\ao^2}{3 (\ao+1) (x-1)^4}+\ao^2-\frac{\ao}{3 (\ao+1)^4 \
(x-1)}+\frac{\ao}{3 (\ao+1)^4 (x \ao-2 \ao-x)}+\frac{\ao}{3 (\ao+1)^3 \
(x-1)^2}-\frac{\ao}{3 (\ao+1)^2 (x-1)^3}+\frac{\ao}{3 (\ao+1) \
(x-1)^4}+\frac{80 \ao}{3 (x-2)^5}-\frac{4 \ao}{3}+\Big(\frac{1}{3 \
(x-1)^5}+\frac{1}{3}+\frac{80}{3 (x-2)^5}+\frac{160}{3 (x-2)^6}\Big) \
H(0;\ao)+\Big(-\frac{1}{3 (x-1)^5}+\frac{1}{3}-\frac{80}{3 \
(x-2)^5}-\frac{160}{3 (x-2)^6}\Big) H(0;x)+\frac{H(c_1(\ao);x)}{3 \
(x-1)^5}+\Big(\frac{80}{3 (x-2)^5}+\frac{160}{3 (x-2)^6}\Big) \
H(c_2(\ao);x)+\frac{80 \ln 2\, }{3 (x-2)^5}+\frac{160 \ln 2\, }{3 \
(x-2)^6}-\frac{1}{9},
\erp
% ep^1
\brp
b_1^{(-1,2)}=\frac{1}{\ao x-x-2\ao}\Big\{\frac{19 x \ao^5}{72}+\frac{31 \ao^5}{36 (x-2)}-\frac{19 \ao^5}{72 \
(x-1)}+\frac{\ao^5}{6}-\frac{131 x \ao^4}{72}-\frac{5 \ao^4}{2 \
(x-2)}+\frac{95 \ao^4}{72 (x-1)}+\frac{65 \ao^4}{18 \
(x-2)^2}-\frac{\ao^4}{2 (x-1)^2}-\frac{\ao^4}{3}+\frac{52 x \
\ao^3}{9}+\frac{25 \ao^3}{18 (x-2)}-\frac{95 \ao^3}{36 \
(x-1)}-\frac{40 \ao^3}{9 (x-2)^2}+\frac{53 \ao^3}{24 \
(x-1)^2}+\frac{20 \ao^3}{(x-2)^3}-\frac{41 \ao^3}{36 \
(x-1)^3}-\frac{121 \ao^3}{72}-\frac{40 x \ao^2}{3}+\frac{23 \ao^2}{9 \
(x-2)}+\frac{13 \ao^2}{6 (x-1)}-\frac{9 \ao^2}{(x-2)^2}-\frac{271 \
\ao^2}{72 (x-1)^2}+\frac{400 \ao^2}{9 (x-2)^3}+\frac{17 \ao^2}{3 \
(x-1)^3}+\frac{1880 \ao^2}{9 (x-2)^4}-\frac{77 \ao^2}{18 \
(x-1)^4}+\frac{857 \ao^2}{72}-\frac{1}{9} \pi ^2 x \ao+\frac{244 x \
\ao}{27}-\frac{10 \ao}{9 (x-2)}-\frac{13 \ao}{12 (x-1)}+\frac{6 \
\ao}{(x-2)^2}+\frac{31 \ao}{36 (x-1)^2}-\frac{392 \ao}{9 \
(x-2)^3}-\frac{7 \ao}{4 (x-1)^3}+\frac{4 \pi ^2 \ao}{9 \
(x-2)^4}-\frac{3760 \ao}{9 (x-2)^4}+\frac{\pi ^2 \ao}{9 \
(x-1)^4}-\frac{77 \ao}{18 (x-1)^4}+\frac{80 \pi ^2 \ao}{9 \
(x-2)^5}-\frac{5120 \ao}{9 (x-2)^5}-\frac{\pi ^2 \ao}{9 \
(x-1)^5}+\frac{160 \ln ^22\,  \ao}{3 (x-2)^4}+\frac{320 \ln ^22\,  \
\ao}{3 (x-2)^5}+\frac{160 \ln 2\,  \ao}{9 (x-2)^4}+\frac{320 \ln 2\,  \
\ao}{9 (x-2)^5}+\frac{2 \pi ^2 \ao}{9}+\frac{265 \ao}{108}+\frac{\pi \
^2 x}{9}+\frac{2 x}{27}+\Big(-\frac{x \ao^5}{3}-\frac{2 \ao^5}{3 \
(x-2)}+\frac{\ao^5}{3 (x-1)}+\frac{19 x \ao^4}{9}+\frac{20 \ao^4}{9 \
(x-2)}-\frac{13 \ao^4}{9 (x-1)}-\frac{20 \ao^4}{9 (x-2)^2}+\frac{4 \
\ao^4}{9 (x-1)^2}-\frac{2 \ao^4}{9}-\frac{52 x \ao^3}{9}-\frac{20 \
\ao^3}{9 (x-2)}+\frac{22 \ao^3}{9 (x-1)}+\frac{40 \ao^3}{9 \
(x-2)^2}-\frac{14 \ao^3}{9 (x-1)^2}-\frac{80 \ao^3}{9 \
(x-2)^3}+\frac{2 \ao^3}{3 (x-1)^3}+\frac{4 \ao^3}{3}+\frac{28 x \
\ao^2}{3}-\frac{2 \ao^2}{x-1}+\frac{2 \ao^2}{(x-1)^2}-\frac{2 \
\ao^2}{(x-1)^3}-\frac{160 \ao^2}{3 (x-2)^4}+\frac{4 \ao^2}{3 \
(x-1)^4}-4 \ao^2-\frac{13 x \ao}{2}-\frac{32 \ao}{9 (x-2)}+\frac{37 \
\ao}{6 (x-1)}+\frac{8 \ao}{9 (x-2)^2}-\frac{13 \ao}{9 \
(x-1)^2}+\frac{16 \ao}{(x-2)^3}+\frac{23 \ao}{18 (x-1)^3}+\frac{2080 \
\ao}{9 (x-2)^4}+\frac{79 \ao}{36 (x-1)^4}+\frac{2240 \ao}{9 (x-2)^5}-\
\frac{29 \ao}{18 (x-1)^5}-\frac{19 \ao}{12}+\frac{7 x}{6}-\frac{40}{9 \
(x-2)}+\frac{13}{3 (x-1)}+\frac{56}{9 (x-2)^2}-\frac{7}{18 \
(x-1)^2}-\frac{160}{9 (x-2)^3}-\frac{2}{9 (x-1)^3}-\frac{1408}{9 \
(x-2)^4}-\frac{85}{36 (x-1)^4}-\frac{2560}{9 (x-2)^5}-\frac{29}{18 \
(x-1)^5}-\frac{640}{9 (x-2)^6}+\frac{5}{4}\Big) H(0;\ao)+\Big(\frac{x \
\ao^5}{2}+\frac{\ao^5}{x-2}-\frac{\ao^5}{2 (x-1)}-\frac{19 x \
\ao^4}{6}-\frac{10 \ao^4}{3 (x-2)}+\frac{13 \ao^4}{6 (x-1)}+\frac{10 \
\ao^4}{3 (x-2)^2}-\frac{2 \ao^4}{3 (x-1)^2}+\frac{\ao^4}{3}+\frac{26 \
x \ao^3}{3}+\frac{10 \ao^3}{3 (x-2)}-\frac{11 \ao^3}{3 \
(x-1)}-\frac{20 \ao^3}{3 (x-2)^2}+\frac{7 \ao^3}{3 (x-1)^2}+\frac{40 \
\ao^3}{3 (x-2)^3}-\frac{\ao^3}{(x-1)^3}-2 \ao^3-14 x \ao^2+\frac{3 \
\ao^2}{x-1}-\frac{3 \ao^2}{(x-1)^2}+\frac{3 \ao^2}{(x-1)^3}+\frac{80 \
\ao^2}{(x-2)^4}-\frac{2 \ao^2}{(x-1)^4}+6 \ao^2+\frac{73 x \
\ao}{6}-\frac{5 \ao}{3 (x-2)}-\frac{7 \ao}{6 (x-1)}+\frac{20 \ao}{3 \
(x-2)^2}+\frac{5 \ao}{3 (x-1)^2}-\frac{40 \ao}{(x-2)^3}-\frac{3 \
\ao}{(x-1)^3}-\frac{320 \ao}{(x-2)^4}-\frac{320 \
\ao}{(x-2)^5}-\frac{10 \ao}{3}-\frac{25 x}{6}+\frac{2}{3 \
(x-2)}+\frac{1}{6 (x-1)}-\frac{10}{3 (x-2)^2}-\frac{1}{3 \
(x-1)^2}+\frac{80}{3 \
(x-2)^3}+\frac{1}{(x-1)^3}+\frac{240}{(x-2)^4}+\frac{2}{(x-1)^4}+\frac{320}{(x-2)^5}-1\Big) H(1;\ao)+\Big(\frac{2 x \ao}{3}+\frac{208 \
\ao}{3 (x-2)^4}-\frac{8 \ao}{3 (x-1)^4}+\frac{320 \ao}{3 \
(x-2)^5}+\frac{8 \ao}{3 (x-1)^5}-\frac{4 \ao}{3}-\frac{2 \
x}{3}-\frac{208}{3 (x-2)^4}+\frac{8}{3 (x-1)^4}-\frac{736}{3 \
(x-2)^5}+\frac{8}{3 (x-1)^5}-\frac{640}{3 (x-2)^6}\Big) H(0;\ao) \
H(1;x)+\Big(-\frac{x \ao^5}{6}-\frac{\ao^5}{3 (x-2)}+\frac{\ao^5}{6 \
(x-1)}-\frac{\ao^5}{4}+\frac{19 x \ao^4}{18}+\frac{17 \ao^4}{18 \
(x-2)}-\frac{13 \ao^4}{18 (x-1)}-\frac{10 \ao^4}{9 (x-2)^2}+\frac{2 \
\ao^4}{9 (x-1)^2}+\frac{41 \ao^4}{36}-\frac{26 x \ao^3}{9}-\frac{4 \
\ao^3}{9 (x-2)}+\frac{11 \ao^3}{9 (x-1)}+\frac{14 \ao^3}{9 \
(x-2)^2}-\frac{7 \ao^3}{9 (x-1)^2}-\frac{40 \ao^3}{9 \
(x-2)^3}+\frac{\ao^3}{3 (x-1)^3}-\frac{11 \ao^3}{6}+\frac{14 x \
\ao^2}{3}-\frac{\ao^2}{x-2}-\frac{\ao^2}{x-1}+\frac{2 \
\ao^2}{(x-2)^2}+\frac{\ao^2}{(x-1)^2}-\frac{4 \
\ao^2}{(x-2)^3}-\frac{\ao^2}{(x-1)^3}-\frac{80 \ao^2}{3 \
(x-2)^4}+\frac{2 \ao^2}{3 (x-1)^4}+\frac{\ao^2}{2}-\frac{73 x \
\ao}{18}-\frac{32 \ao}{9 (x-2)}+\frac{37 \ao}{6 (x-1)}+\frac{8 \ao}{9 \
(x-2)^2}-\frac{13 \ao}{9 (x-1)^2}+\frac{16 \ao}{(x-2)^3}+\frac{23 \
\ao}{18 (x-1)^3}+\frac{176 \ao}{(x-2)^4}+\frac{55 \ao}{36 \
(x-1)^4}+\frac{320 \ao}{3 (x-2)^5}-\frac{29 \ao}{18 (x-1)^5}-\frac{29 \
\ao}{36}+\frac{25 x}{18}+\Big(-\frac{4 \ao}{3 (x-1)^4}+\frac{4 \ao}{3 \
(x-1)^5}+\frac{4}{3 (x-1)^4}+\frac{4}{3 (x-1)^5}\Big) \
H(0;\ao)+\Big(\frac{2 \ao}{(x-1)^4}-\frac{2 \
\ao}{(x-1)^5}-\frac{2}{(x-1)^4}-\frac{2}{(x-1)^5}\Big) \
H(1;\ao)-\frac{40}{9 (x-2)}+\frac{13}{3 (x-1)}+\frac{56}{9 \
(x-2)^2}-\frac{7}{18 (x-1)^2}-\frac{160}{9 (x-2)^3}-\frac{2}{9 \
(x-1)^3}-\frac{416}{3 (x-2)^4}-\frac{85}{36 (x-1)^4}-\frac{640}{3 \
(x-2)^5}-\frac{29}{18 (x-1)^5}+\frac{5}{4}\Big) \
H(c_1(\ao);x)+\Big(\frac{160 \ao}{9 (x-2)^4}+\frac{320 \ao}{9 \
(x-2)^5}+\Big(-\frac{320 \ao}{3 (x-2)^4}-\frac{640 \ao}{3 \
(x-2)^5}+\frac{320}{3 (x-2)^4}+\frac{1280}{3 (x-2)^5}+\frac{1280}{3 \
(x-2)^6}\Big) H(0;\ao)+\Big(\frac{160 \ao}{(x-2)^4}+\frac{320 \
\ao}{(x-2)^5}-\frac{160}{(x-2)^4}-\frac{640}{(x-2)^5}-\frac{640}{(x-2)\
^6}\Big) H(1;\ao)-\frac{160}{9 (x-2)^4}-\frac{640}{9 \
(x-2)^5}-\frac{640}{9 (x-2)^6}\Big) H(c_2(\ao);x)+\Big(-\frac{4 x \
\ao}{3}-\frac{320 \ao}{3 (x-2)^4}-\frac{4 \ao}{3 (x-1)^4}-\frac{640 \
\ao}{3 (x-2)^5}+\frac{4 \ao}{3 (x-1)^5}+\frac{8 \ao}{3}+\frac{4 \
x}{3}+\frac{320}{3 (x-2)^4}+\frac{4}{3 (x-1)^4}+\frac{1280}{3 \
(x-2)^5}+\frac{4}{3 (x-1)^5}+\frac{1280}{3 (x-2)^6}\Big) \
H(0,0;\ao)+\Big(-\frac{4 x \ao}{3}+\frac{320 \ao}{3 (x-2)^4}+\frac{4 \
\ao}{3 (x-1)^4}+\frac{640 \ao}{3 (x-2)^5}-\frac{4 \ao}{3 \
(x-1)^5}+\frac{8 \ao}{3}+\frac{4 x}{3}-\frac{320}{3 \
(x-2)^4}-\frac{4}{3 (x-1)^4}-\frac{1280}{3 (x-2)^5}-\frac{4}{3 \
(x-1)^5}-\frac{1280}{3 (x-2)^6}\Big) H(0,0;x)+\Big(2 x \ao+\frac{160 \
\ao}{(x-2)^4}+\frac{2 \ao}{(x-1)^4}+\frac{320 \ao}{(x-2)^5}-\frac{2 \
\ao}{(x-1)^5}-4 \ao-2 \
x-\frac{160}{(x-2)^4}-\frac{2}{(x-1)^4}-\frac{640}{(x-2)^5}-\frac{2}{(\
x-1)^5}-\frac{640}{(x-2)^6}\Big) H(0,1;\ao)+\Big(\frac{2 x \
\ao}{3}+\frac{208 \ao}{3 (x-2)^4}-\frac{2 \ao}{3 (x-1)^4}+\frac{320 \
\ao}{3 (x-2)^5}+\frac{2 \ao}{3 (x-1)^5}-\frac{4 \ao}{3}-\frac{2 \
x}{3}-\frac{208}{3 (x-2)^4}+\frac{2}{3 (x-1)^4}-\frac{736}{3 \
(x-2)^5}+\frac{2}{3 (x-1)^5}-\frac{640}{3 (x-2)^6}\Big) \
H(0,c_1(\ao);x)+\Big(-\frac{320 \ao}{3 (x-2)^4}-\frac{640 \ao}{3 \
(x-2)^5}+\frac{320}{3 (x-2)^4}+\frac{1280}{3 (x-2)^5}+\frac{1280}{3 \
(x-2)^6}\Big) H(0,c_2(\ao);x)+\Big(-\frac{2 x \ao}{3}-\frac{208 \
\ao}{3 (x-2)^4}+\frac{8 \ao}{3 (x-1)^4}-\frac{320 \ao}{3 \
(x-2)^5}-\frac{8 \ao}{3 (x-1)^5}+\frac{4 \ao}{3}+\frac{2 \
x}{3}+\frac{208}{3 (x-2)^4}-\frac{8}{3 (x-1)^4}+\frac{736}{3 \
(x-2)^5}-\frac{8}{3 (x-1)^5}+\frac{640}{3 (x-2)^6}\Big) \
H(1,0;x)+\Big(\frac{2 x \ao}{3}+\frac{208 \ao}{3 (x-2)^4}-\frac{8 \
\ao}{3 (x-1)^4}+\frac{320 \ao}{3 (x-2)^5}+\frac{8 \ao}{3 \
(x-1)^5}-\frac{4 \ao}{3}-\frac{2 x}{3}-\frac{208}{3 \
(x-2)^4}+\frac{8}{3 (x-1)^4}-\frac{736}{3 (x-2)^5}+\frac{8}{3 \
(x-1)^5}-\frac{640}{3 (x-2)^6}\Big) H(1,c_1(\ao);x)+\Big(-\frac{800 \
\ao}{3 (x-2)^4}-\frac{1600 \ao}{3 (x-2)^5}+\frac{800}{3 \
(x-2)^4}+\frac{3200}{3 (x-2)^5}+\frac{3200}{3 (x-2)^6}\Big) H(2,0;x)+\
\Big(\frac{800 \ao}{3 (x-2)^4}+\frac{1600 \ao}{3 \
(x-2)^5}-\frac{800}{3 (x-2)^4}-\frac{3200}{3 (x-2)^5}-\frac{3200}{3 \
(x-2)^6}\Big) H(2,c_2(\ao);x)+\Big(-\frac{2 \ao}{3 (x-1)^4}+\frac{2 \
\ao}{3 (x-1)^5}+\frac{2}{3 (x-1)^4}+\frac{2}{3 (x-1)^5}\Big) \
H(c_1(\ao),c_1(\ao);x)+\Big(-\frac{208 \ao}{3 (x-2)^4}-\frac{320 \
\ao}{3 (x-2)^5}+\frac{208}{3 (x-2)^4}+\frac{736}{3 \
(x-2)^5}+\frac{640}{3 (x-2)^6}\Big) H(c_2(\ao),c_1(\ao);x)+H(0;x) \
\Big(\frac{29 x \ao}{18}+\frac{32 \ao}{9 (x-2)}-\frac{37 \ao}{6 \
(x-1)}-\frac{8 \ao}{9 (x-2)^2}+\frac{13 \ao}{9 (x-1)^2}-\frac{16 \
\ao}{(x-2)^3}-\frac{23 \ao}{18 (x-1)^3}-\frac{1120 \ao}{9 \
(x-2)^4}-\frac{31 \ao}{36 (x-1)^4}-\frac{320 \ao}{9 (x-2)^5}+\frac{29 \
\ao}{18 (x-1)^5}-\frac{320 \ln 2\,  \ao}{3 (x-2)^4}-\frac{640 \ln 2\, \
 \ao}{3 (x-2)^5}-\frac{71 \ao}{36}-\frac{29 x}{18}+\frac{40}{9 \
(x-2)}-\frac{13}{3 (x-1)}-\frac{56}{9 (x-2)^2}+\frac{7}{18 \
(x-1)^2}+\frac{160}{9 (x-2)^3}+\frac{2}{9 (x-1)^3}+\frac{1408}{9 \
(x-2)^4}+\frac{85}{36 (x-1)^4}+\frac{2560}{9 (x-2)^5}+\frac{29}{18 \
(x-1)^5}+\frac{640}{9 (x-2)^6}+\frac{320 \ln 2\, }{3 \
(x-2)^4}+\frac{1280 \ln 2\, }{3 (x-2)^5}+\frac{1280 \ln 2\, }{3 \
(x-2)^6}-\frac{5}{4}\Big)+H(2;x) \Big(\frac{800 \ln 2\,  \ao}{3 \
(x-2)^4}+\frac{1600 \ln 2\,  \ao}{3 (x-2)^5}+\Big(\frac{800 \ao}{3 \
(x-2)^4}+\frac{1600 \ao}{3 (x-2)^5}-\frac{800}{3 \
(x-2)^4}-\frac{3200}{3 (x-2)^5}-\frac{3200}{3 (x-2)^6}\Big) H(0;\ao)-\
\frac{800 \ln 2\, }{3 (x-2)^4}-\frac{3200 \ln 2\, }{3 \
(x-2)^5}-\frac{3200 \ln 2\, }{3 (x-2)^6}\Big)-\frac{4 \pi ^2}{9 \
(x-2)^4}-\frac{\pi ^2}{9 (x-1)^4}-\frac{88 \pi ^2}{9 \
(x-2)^5}-\frac{\pi ^2}{9 (x-1)^5}-\frac{160 \pi ^2}{9 \
(x-2)^6}-\frac{160 \ln ^22\, }{3 (x-2)^4}-\frac{640 \ln ^22\, }{3 \
(x-2)^5}-\frac{640 \ln ^22\, }{3 (x-2)^6}-\frac{160 \ln 2\, }{9 \
(x-2)^4}-\frac{640 \ln 2\, }{9 (x-2)^5}-\frac{640 \ln 2\, }{9 (x-2)^6}\},
\erp
% ep^2
\brp
b_2^{(-1,2)}=\frac{1}{\ao x-x-2\ao}\Big\{-\frac{1}{72} \pi ^2 x \ao^5+\frac{301 x \ao^5}{432}-\frac{\pi ^2 \
\ao^5}{36 (x-2)}+\frac{673 \ao^5}{216 (x-2)}+\frac{\pi ^2 \ao^5}{72 \
(x-1)}-\frac{301 \ao^5}{432 (x-1)}+\frac{31 \ao^5}{36}+\frac{19}{216} \
\pi ^2 x \ao^4-\frac{6967 x \ao^4}{1296}+\frac{5 \pi ^2 \ao^4}{54 \
(x-2)}-\frac{2345 \ao^4}{324 (x-2)}-\frac{13 \pi ^2 \ao^4}{216 \
(x-1)}+\frac{5251 \ao^4}{1296 (x-1)}-\frac{5 \pi ^2 \ao^4}{54 \
(x-2)^2}+\frac{5405 \ao^4}{324 (x-2)^2}+\frac{\pi ^2 \ao^4}{54 \
(x-1)^2}-\frac{613 \ao^4}{324 (x-1)^2}-\frac{\pi ^2 \
\ao^4}{108}-\frac{299 \ao^4}{162}-\frac{13}{54} \pi ^2 x \
\ao^3+\frac{3413 x \ao^3}{162}-\frac{5 \pi ^2 \ao^3}{54 \
(x-2)}-\frac{895 \ao^3}{324 (x-2)}+\frac{11 \pi ^2 \ao^3}{108 (x-1)}-\
\frac{5807 \ao^3}{648 (x-1)}+\frac{5 \pi ^2 \ao^3}{27 \
(x-2)^2}-\frac{115 \ao^3}{81 (x-2)^2}-\frac{7 \pi ^2 \ao^3}{108 \
(x-1)^2}+\frac{14261 \ao^3}{1296 (x-1)^2}-\frac{10 \pi ^2 \ao^3}{27 \
(x-2)^3}+\frac{10580 \ao^3}{81 (x-2)^3}+\frac{\pi ^2 \ao^3}{36 \
(x-1)^3}-\frac{1411 \ao^3}{216 (x-1)^3}+\frac{\pi ^2 \
\ao^3}{18}-\frac{4837 \ao^3}{432}+\frac{7}{18} \pi ^2 x \
\ao^2-\frac{4571 x \ao^2}{54}+\frac{545 \ao^2}{18 (x-2)}-\frac{\pi ^2 \
\ao^2}{12 (x-1)}-\frac{121 \ao^2}{36 (x-1)}-\frac{2075 \ao^2}{18 \
(x-2)^2}+\frac{\pi ^2 \ao^2}{12 (x-1)^2}-\frac{3577 \ao^2}{144 \
(x-1)^2}+\frac{7600 \ao^2}{9 (x-2)^3}-\frac{\pi ^2 \ao^2}{12 \
(x-1)^3}+\frac{524 \ao^2}{9 (x-1)^3}-\frac{20 \pi ^2 \ao^2}{9 \
(x-2)^4}+\frac{66760 \ao^2}{27 (x-2)^4}+\frac{\pi ^2 \ao^2}{18 \
(x-1)^4}-\frac{5047 \ao^2}{108 (x-1)^4}-\frac{\pi ^2 \
\ao^2}{6}+\frac{17483 \ao^2}{144}-\frac{7}{8} \pi ^2 x \ao+\frac{5525 \
x \ao}{81}-\frac{26 \pi ^2 \ao}{9 (x-2)}-\frac{17 \ao}{9 \
(x-2)}+\frac{745 \pi ^2 \ao}{216 (x-1)}-\frac{467 \ao}{24 \
(x-1)}+\frac{10 \pi ^2 \ao}{3 (x-2)^2}+\frac{565 \ao}{9 \
(x-2)^2}-\frac{8 \pi ^2 \ao}{9 (x-1)^2}+\frac{91 \ao}{8 \
(x-1)^2}-\frac{8 \pi ^2 \ao}{3 (x-2)^3}-\frac{7240 \ao}{9 \
(x-2)^3}+\frac{79 \pi ^2 \ao}{108 (x-1)^3}-\frac{689 \ao}{24 \
(x-1)^3}+\frac{920 \pi ^2 \ao}{27 (x-2)^4}-\frac{133520 \ao}{27 \
(x-2)^4}+\frac{5 \pi ^2 \ao}{54 (x-1)^4}-\frac{5047 \ao}{108 \
(x-1)^4}+\frac{400 \pi ^2 \ao}{27 (x-2)^5}-\frac{154240 \ao}{27 \
(x-2)^5}-\frac{29 \pi ^2 \ao}{54 (x-1)^5}+\frac{17}{12} x \zeta_3 \
\ao+\frac{224 \zeta_3 \ao}{3 (x-2)^4}-\frac{7 \zeta_3 \ao}{4 \
(x-1)^4}+\frac{280 \zeta_3 \ao}{3 (x-2)^5}+\frac{7 \zeta_3 \ao}{4 \
(x-1)^5}-\frac{17}{6} \zeta_3 \ao+\frac{640 \ln ^32\,  \ao}{9 \
(x-2)^4}+\frac{1280 \ln ^32\,  \ao}{9 (x-2)^5}+\frac{320 \ln ^22\,  \
\ao}{9 (x-2)^4}+\frac{640 \ln ^22\,  \ao}{9 (x-2)^5}-\frac{1}{6} \pi \
^2 x \ln 2\,  \ao-\frac{32 \pi ^2 \ln 2\,  \ao}{3 (x-2)^4}+\frac{320 \
\ln 2\,  \ao}{27 (x-2)^4}+\frac{\pi ^2 \ln 2\,  \ao}{6 \
(x-1)^4}+\frac{80 \pi ^2 \ln 2\,  \ao}{3 (x-2)^5}+\frac{640 \ln 2\,  \
\ao}{27 (x-2)^5}-\frac{\pi ^2 \ln 2\,  \ao}{6 (x-1)^5}+\frac{1}{3} \
\pi ^2 \ln 2\,  \ao+\frac{53 \pi ^2 \ao}{72}+\frac{1669 \
\ao}{648}+\frac{47 \pi ^2 x}{72}+\frac{4 x}{81}+\Big(-\frac{19 x \
\ao^5}{18}-\frac{31 \ao^5}{9 (x-2)}+\frac{19 \ao^5}{18 (x-1)}-\frac{2 \
\ao^5}{3}+\frac{131 x \ao^4}{18}+\frac{10 \ao^4}{x-2}-\frac{95 \
\ao^4}{18 (x-1)}-\frac{130 \ao^4}{9 (x-2)^2}+\frac{2 \ao^4}{(x-1)^2}+\
\frac{4 \ao^4}{3}-\frac{208 x \ao^3}{9}-\frac{50 \ao^3}{9 \
(x-2)}+\frac{95 \ao^3}{9 (x-1)}+\frac{160 \ao^3}{9 (x-2)^2}-\frac{53 \
\ao^3}{6 (x-1)^2}-\frac{80 \ao^3}{(x-2)^3}+\frac{41 \ao^3}{9 \
(x-1)^3}+\frac{121 \ao^3}{18}+\frac{160 x \ao^2}{3}-\frac{92 \ao^2}{9 \
(x-2)}-\frac{26 \ao^2}{3 (x-1)}+\frac{36 \ao^2}{(x-2)^2}+\frac{271 \
\ao^2}{18 (x-1)^2}-\frac{1600 \ao^2}{9 (x-2)^3}-\frac{68 \ao^2}{3 \
(x-1)^3}-\frac{7520 \ao^2}{9 (x-2)^4}+\frac{154 \ao^2}{9 \
(x-1)^4}-\frac{857 \ao^2}{18}-\frac{1}{18} \pi ^2 x \ao-\frac{5255 x \
\ao}{108}-\frac{812 \ao}{9 (x-2)}+\frac{239 \ao}{2 (x-1)}+\frac{448 \
\ao}{9 (x-2)^2}-\frac{18 \ao}{(x-1)^2}+\frac{3248 \ao}{9 \
(x-2)^3}+\frac{164 \ao}{9 (x-1)^3}-\frac{40 \pi ^2 \ao}{9 \
(x-2)^4}+\frac{76160 \ao}{27 (x-2)^4}-\frac{\pi ^2 \ao}{18 \
(x-1)^4}+\frac{4877 \ao}{216 (x-1)^4}-\frac{80 \pi ^2 \ao}{9 \
(x-2)^5}+\frac{62080 \ao}{27 (x-2)^5}+\frac{\pi ^2 \ao}{18 \
(x-1)^5}-\frac{1351 \ao}{108 (x-1)^5}+\frac{\pi ^2 \ao}{9}+\frac{809 \
\ao}{216}+\frac{\pi ^2 x}{18}+\frac{1319 \
x}{108}-\frac{92}{x-2}+\frac{1133}{12 (x-1)}+\frac{1040}{9 \
(x-2)^2}-\frac{223}{36 (x-1)^2}-\frac{3008}{9 (x-2)^3}-\frac{103}{36 \
(x-1)^3}+\frac{40 \pi ^2}{9 (x-2)^4}-\frac{41120}{27 \
(x-2)^4}+\frac{\pi ^2}{18 (x-1)^4}-\frac{4223}{216 (x-1)^4}+\frac{160 \
\pi ^2}{9 (x-2)^5}-\frac{62720}{27 (x-2)^5}+\frac{\pi ^2}{18 \
(x-1)^5}-\frac{1351}{108 (x-1)^5}+\frac{160 \pi ^2}{9 \
(x-2)^6}-\frac{1280}{27 (x-2)^6}+\frac{275}{24}\Big) \
H(0;\ao)+\Big(\frac{19 x \ao^5}{12}+\frac{31 \ao^5}{6 (x-2)}-\frac{19 \
\ao^5}{12 (x-1)}+\ao^5-\frac{131 x \ao^4}{12}-\frac{15 \
\ao^4}{x-2}+\frac{95 \ao^4}{12 (x-1)}+\frac{65 \ao^4}{3 \
(x-2)^2}-\frac{3 \ao^4}{(x-1)^2}-2 \ao^4+\frac{104 x \
\ao^3}{3}+\frac{25 \ao^3}{3 (x-2)}-\frac{95 \ao^3}{6 (x-1)}-\frac{80 \
\ao^3}{3 (x-2)^2}+\frac{53 \ao^3}{4 (x-1)^2}+\frac{120 \
\ao^3}{(x-2)^3}-\frac{41 \ao^3}{6 (x-1)^3}-\frac{121 \ao^3}{12}-80 x \
\ao^2+\frac{46 \ao^2}{3 (x-2)}+\frac{13 \ao^2}{x-1}-\frac{54 \
\ao^2}{(x-2)^2}-\frac{271 \ao^2}{12 (x-1)^2}+\frac{800 \ao^2}{3 \
(x-2)^3}+\frac{34 \ao^2}{(x-1)^3}+\frac{3760 \ao^2}{3 \
(x-2)^4}-\frac{77 \ao^2}{3 (x-1)^4}+\frac{857 \ao^2}{12}+\frac{367 x \
\ao}{4}-\frac{109 \ao}{6 (x-2)}-\frac{103 \ao}{12 (x-1)}+\frac{296 \
\ao}{3 (x-2)^2}+\frac{205 \ao}{12 (x-1)^2}-\frac{888 \
\ao}{(x-2)^3}-\frac{89 \ao}{2 (x-1)^3}-\frac{12640 \ao}{3 \
(x-2)^4}-\frac{10240 \ao}{3 (x-2)^5}-\frac{233 \ao}{4}-\frac{445 \
x}{12}+\frac{13}{3 (x-2)}+\frac{61}{12 (x-1)}-\frac{119}{3 \
(x-2)^2}-\frac{19}{4 (x-1)^2}+\frac{1504}{3 (x-2)^3}+\frac{52}{3 \
(x-1)^3}+\frac{2960}{(x-2)^4}+\frac{77}{3 (x-1)^4}+\frac{10240}{3 \
(x-2)^5}-\frac{25}{12}\Big) H(1;\ao)+\Big(\frac{4 x \ao^5}{3}+\frac{8 \
\ao^5}{3 (x-2)}-\frac{4 \ao^5}{3 (x-1)}-\frac{76 x \ao^4}{9}-\frac{80 \
\ao^4}{9 (x-2)}+\frac{52 \ao^4}{9 (x-1)}+\frac{80 \ao^4}{9 \
(x-2)^2}-\frac{16 \ao^4}{9 (x-1)^2}+\frac{8 \ao^4}{9}+\frac{208 x \
\ao^3}{9}+\frac{80 \ao^3}{9 (x-2)}-\frac{88 \ao^3}{9 (x-1)}-\frac{160 \
\ao^3}{9 (x-2)^2}+\frac{56 \ao^3}{9 (x-1)^2}+\frac{320 \ao^3}{9 \
(x-2)^3}-\frac{8 \ao^3}{3 (x-1)^3}-\frac{16 \ao^3}{3}-\frac{112 x \
\ao^2}{3}+\frac{8 \ao^2}{x-1}-\frac{8 \ao^2}{(x-1)^2}+\frac{8 \
\ao^2}{(x-1)^3}+\frac{640 \ao^2}{3 (x-2)^4}-\frac{16 \ao^2}{3 \
(x-1)^4}+16 \ao^2+26 x \ao+\frac{128 \ao}{9 (x-2)}-\frac{74 \ao}{3 \
(x-1)}-\frac{32 \ao}{9 (x-2)^2}+\frac{52 \ao}{9 (x-1)^2}-\frac{64 \
\ao}{(x-2)^3}-\frac{46 \ao}{9 (x-1)^3}-\frac{8320 \ao}{9 \
(x-2)^4}-\frac{79 \ao}{9 (x-1)^4}-\frac{8960 \ao}{9 (x-2)^5}+\frac{58 \
\ao}{9 (x-1)^5}+\frac{19 \ao}{3}-\frac{14 x}{3}+\frac{160}{9 \
(x-2)}-\frac{52}{3 (x-1)}-\frac{224}{9 (x-2)^2}+\frac{14}{9 (x-1)^2}+\
\frac{640}{9 (x-2)^3}+\frac{8}{9 (x-1)^3}+\frac{5632}{9 \
(x-2)^4}+\frac{85}{9 (x-1)^4}+\frac{10240}{9 (x-2)^5}+\frac{58}{9 \
(x-1)^5}+\frac{2560}{9 (x-2)^6}-5\Big) H(0,0;\ao)+\Big(-2 x \
\ao^5-\frac{4 \ao^5}{x-2}+\frac{2 \ao^5}{x-1}+\frac{38 x \
\ao^4}{3}+\frac{40 \ao^4}{3 (x-2)}-\frac{26 \ao^4}{3 (x-1)}-\frac{40 \
\ao^4}{3 (x-2)^2}+\frac{8 \ao^4}{3 (x-1)^2}-\frac{4 \
\ao^4}{3}-\frac{104 x \ao^3}{3}-\frac{40 \ao^3}{3 (x-2)}+\frac{44 \
\ao^3}{3 (x-1)}+\frac{80 \ao^3}{3 (x-2)^2}-\frac{28 \ao^3}{3 \
(x-1)^2}-\frac{160 \ao^3}{3 (x-2)^3}+\frac{4 \ao^3}{(x-1)^3}+8 \
\ao^3+56 x \ao^2-\frac{12 \ao^2}{x-1}+\frac{12 \
\ao^2}{(x-1)^2}-\frac{12 \ao^2}{(x-1)^3}-\frac{320 \
\ao^2}{(x-2)^4}+\frac{8 \ao^2}{(x-1)^4}-24 \ao^2-39 x \ao-\frac{64 \
\ao}{3 (x-2)}+\frac{37 \ao}{x-1}+\frac{16 \ao}{3 (x-2)^2}-\frac{26 \
\ao}{3 (x-1)^2}+\frac{96 \ao}{(x-2)^3}+\frac{23 \ao}{3 \
(x-1)^3}+\frac{4160 \ao}{3 (x-2)^4}+\frac{79 \ao}{6 \
(x-1)^4}+\frac{4480 \ao}{3 (x-2)^5}-\frac{29 \ao}{3 (x-1)^5}-\frac{19 \
\ao}{2}+7 x-\frac{80}{3 (x-2)}+\frac{26}{x-1}+\frac{112}{3 \
(x-2)^2}-\frac{7}{3 (x-1)^2}-\frac{320}{3 (x-2)^3}-\frac{4}{3 \
(x-1)^3}-\frac{2816}{3 (x-2)^4}-\frac{85}{6 (x-1)^4}-\frac{5120}{3 \
(x-2)^5}-\frac{29}{3 (x-1)^5}-\frac{1280}{3 \
(x-2)^6}+\frac{15}{2}\Big) H(0,1;\ao)+H(1;x) \Big(\frac{1}{9} \pi ^2 \
x \ao+\frac{32 \pi ^2 \ao}{3 (x-2)^4}-\frac{8 \pi ^2 \ao}{9 (x-1)^4}+\
\frac{80 \pi ^2 \ao}{3 (x-2)^5}+\frac{8 \pi ^2 \ao}{9 \
(x-1)^5}-\frac{2 \pi ^2 \ao}{9}-\frac{\pi ^2 x}{9}+\Big(\frac{29 x \
\ao}{9}+\frac{406 \ao}{9 (x-2)}-\frac{166 \ao}{3 (x-1)}-\frac{424 \
\ao}{9 (x-2)^2}+\frac{193 \ao}{18 (x-1)^2}+\frac{32 \
\ao}{(x-2)^3}-\frac{53 \ao}{9 (x-1)^3}-\frac{4000 \ao}{9 \
(x-2)^4}-\frac{175 \ao}{18 (x-1)^4}+\frac{640 \ao}{9 \
(x-2)^5}+\frac{116 \ao}{9 (x-1)^5}+\frac{2 \ao}{9}-\frac{29 \
x}{9}+\frac{326}{9 (x-2)}-\frac{118}{3 (x-1)}-\frac{388}{9 \
(x-2)^2}+\frac{95}{18 (x-1)^2}+\frac{560}{9 (x-2)^3}+\frac{4}{9 \
(x-1)^3}+\frac{3424}{9 (x-2)^4}+\frac{289}{18 (x-1)^4}+\frac{7360}{9 \
(x-2)^5}+\frac{116}{9 (x-1)^5}-\frac{1280}{9 \
(x-2)^6}-\frac{20}{3}\Big) H(0;\ao)+\Big(-\frac{8 x \ao}{3}-\frac{832 \
\ao}{3 (x-2)^4}+\frac{32 \ao}{3 (x-1)^4}-\frac{1280 \ao}{3 \
(x-2)^5}-\frac{32 \ao}{3 (x-1)^5}+\frac{16 \ao}{3}+\frac{8 \
x}{3}+\frac{832}{3 (x-2)^4}-\frac{32}{3 (x-1)^4}+\frac{2944}{3 \
(x-2)^5}-\frac{32}{3 (x-1)^5}+\frac{2560}{3 (x-2)^6}\Big) H(0,0;\ao)+\
\Big(4 x \ao+\frac{416 \ao}{(x-2)^4}-\frac{16 \ao}{(x-1)^4}+\frac{640 \
\ao}{(x-2)^5}+\frac{16 \ao}{(x-1)^5}-8 \ao-4 \
x-\frac{416}{(x-2)^4}+\frac{16}{(x-1)^4}-\frac{1472}{(x-2)^5}+\frac{\
16}{(x-1)^5}-\frac{1280}{(x-2)^6}\Big) H(0,1;\ao)-\frac{32 \pi ^2}{3 \
(x-2)^4}+\frac{8 \pi ^2}{9 (x-1)^4}-\frac{48 \pi ^2}{(x-2)^5}+\frac{8 \
\pi ^2}{9 (x-1)^5}-\frac{160 \pi ^2}{3 (x-2)^6}\Big)+\Big(-\frac{20 x \
\ao}{3}-\frac{1984 \ao}{3 (x-2)^4}+\frac{20 \ao}{3 \
(x-1)^4}-\frac{3200 \ao}{3 (x-2)^5}-\frac{20 \ao}{3 (x-1)^5}+\frac{40 \
\ao}{3}+\frac{20 x}{3}+\frac{1984}{3 (x-2)^4}-\frac{20}{3 \
(x-1)^4}+\frac{7168}{3 (x-2)^5}-\frac{20}{3 (x-1)^5}+\frac{6400}{3 \
(x-2)^6}\Big) H(0;\ao) H(0,1;x)+\Big(-\frac{\ao^5}{2}-\frac{\ao^4}{3 \
(x-2)}+\frac{5 \ao^4}{2}+\frac{4 \ao^3}{3 (x-2)}-\frac{4 \ao^3}{3 \
(x-2)^2}-5 \ao^3-\frac{2 \ao^2}{x-2}+\frac{4 \ao^2}{(x-2)^2}-\frac{8 \
\ao^2}{(x-2)^3}+5 \ao^2+\frac{29 x \ao}{9}+\frac{160 \ao}{9 \
(x-2)}-\frac{62 \ao}{3 (x-1)}-\frac{184 \ao}{9 (x-2)^2}+\frac{97 \
\ao}{18 (x-1)^2}+\frac{16 \ao}{(x-2)^3}-\frac{38 \ao}{9 \
(x-1)^3}-\frac{832 \ao}{9 (x-2)^4}-\frac{5 \ao}{9 (x-1)^4}+\frac{640 \
\ao}{9 (x-2)^5}+\frac{29 \ao}{9 (x-1)^5}-\frac{113 \ao}{18}-\frac{29 \
x}{9}+\Big(-\frac{8 x \ao}{3}-\frac{832 \ao}{3 (x-2)^4}+\frac{8 \
\ao}{3 (x-1)^4}-\frac{1280 \ao}{3 (x-2)^5}-\frac{8 \ao}{3 \
(x-1)^5}+\frac{16 \ao}{3}+\frac{8 x}{3}+\frac{832}{3 \
(x-2)^4}-\frac{8}{3 (x-1)^4}+\frac{2944}{3 (x-2)^5}-\frac{8}{3 \
(x-1)^5}+\frac{2560}{3 (x-2)^6}\Big) H(0;\ao)+\Big(4 x \ao+\frac{416 \
\ao}{(x-2)^4}-\frac{4 \ao}{(x-1)^4}+\frac{640 \ao}{(x-2)^5}+\frac{4 \
\ao}{(x-1)^5}-8 \ao-4 \
x-\frac{416}{(x-2)^4}+\frac{4}{(x-1)^4}-\frac{1472}{(x-2)^5}+\frac{4}{\
(x-1)^5}-\frac{1280}{(x-2)^6}\Big) H(1;\ao)+\frac{104}{9 \
(x-2)}-\frac{13}{x-1}-\frac{136}{9 (x-2)^2}+\frac{41}{18 \
(x-1)^2}+\frac{224}{9 (x-2)^3}+\frac{10}{9 (x-1)^3}+\frac{832}{9 \
(x-2)^4}+\frac{53}{9 (x-1)^4}+\frac{1600}{9 (x-2)^5}+\frac{29}{9 \
(x-1)^5}-\frac{1280}{9 (x-2)^6}-\frac{13}{6}\Big) \
H(0,c_1(\ao);x)+\Big(-\frac{640 \ao}{9 (x-2)^4}-\frac{1280 \ao}{9 \
(x-2)^5}+\Big(\frac{1280 \ao}{3 (x-2)^4}+\frac{2560 \ao}{3 \
(x-2)^5}-\frac{1280}{3 (x-2)^4}-\frac{5120}{3 (x-2)^5}-\frac{5120}{3 \
(x-2)^6}\Big) H(0;\ao)+\Big(-\frac{640 \ao}{(x-2)^4}-\frac{1280 \
\ao}{(x-2)^5}+\frac{640}{(x-2)^4}+\frac{2560}{(x-2)^5}+\frac{2560}{(x-\
2)^6}\Big) H(1;\ao)+\frac{640}{9 (x-2)^4}+\frac{2560}{9 \
(x-2)^5}+\frac{2560}{9 (x-2)^6}\Big) H(0,c_2(\ao);x)+\Big(-2 x \ao^5-\
\frac{4 \ao^5}{x-2}+\frac{2 \ao^5}{x-1}+\frac{38 x \ao^4}{3}+\frac{40 \
\ao^4}{3 (x-2)}-\frac{26 \ao^4}{3 (x-1)}-\frac{40 \ao^4}{3 \
(x-2)^2}+\frac{8 \ao^4}{3 (x-1)^2}-\frac{4 \ao^4}{3}-\frac{104 x \
\ao^3}{3}-\frac{40 \ao^3}{3 (x-2)}+\frac{44 \ao^3}{3 (x-1)}+\frac{80 \
\ao^3}{3 (x-2)^2}-\frac{28 \ao^3}{3 (x-1)^2}-\frac{160 \ao^3}{3 \
(x-2)^3}+\frac{4 \ao^3}{(x-1)^3}+8 \ao^3+56 x \ao^2-\frac{12 \
\ao^2}{x-1}+\frac{12 \ao^2}{(x-1)^2}-\frac{12 \
\ao^2}{(x-1)^3}-\frac{320 \ao^2}{(x-2)^4}+\frac{8 \ao^2}{(x-1)^4}-24 \
\ao^2-\frac{146 x \ao}{3}+\frac{20 \ao}{3 (x-2)}+\frac{14 \ao}{3 \
(x-1)}-\frac{80 \ao}{3 (x-2)^2}-\frac{20 \ao}{3 (x-1)^2}+\frac{160 \
\ao}{(x-2)^3}+\frac{12 \ao}{(x-1)^3}+\frac{1280 \
\ao}{(x-2)^4}+\frac{1280 \ao}{(x-2)^5}+\frac{40 \ao}{3}+\frac{50 \
x}{3}-\frac{8}{3 (x-2)}-\frac{2}{3 (x-1)}+\frac{40}{3 \
(x-2)^2}+\frac{4}{3 (x-1)^2}-\frac{320}{3 (x-2)^3}-\frac{4}{(x-1)^3}-\
\frac{960}{(x-2)^4}-\frac{8}{(x-1)^4}-\frac{1280}{(x-2)^5}+4\Big) \
H(1,0;\ao)+\Big(-\frac{29 x \ao}{9}-\frac{406 \ao}{9 (x-2)}+\frac{166 \
\ao}{3 (x-1)}+\frac{424 \ao}{9 (x-2)^2}-\frac{193 \ao}{18 \
(x-1)^2}-\frac{32 \ao}{(x-2)^3}+\frac{53 \ao}{9 (x-1)^3}+\frac{4000 \
\ao}{9 (x-2)^4}+\frac{175 \ao}{18 (x-1)^4}-\frac{640 \ao}{9 (x-2)^5}-\
\frac{116 \ao}{9 (x-1)^5}-\frac{2 \ao}{9}+\frac{29 x}{9}-\frac{326}{9 \
(x-2)}+\frac{118}{3 (x-1)}+\frac{388}{9 (x-2)^2}-\frac{95}{18 \
(x-1)^2}-\frac{560}{9 (x-2)^3}-\frac{4}{9 (x-1)^3}-\frac{3424}{9 \
(x-2)^4}-\frac{289}{18 (x-1)^4}-\frac{7360}{9 (x-2)^5}-\frac{116}{9 \
(x-1)^5}+\frac{1280}{9 (x-2)^6}+\frac{20}{3}\Big) H(1,0;x)+\Big(3 x \
\ao^5+\frac{6 \ao^5}{x-2}-\frac{3 \ao^5}{x-1}-19 x \ao^4-\frac{20 \
\ao^4}{x-2}+\frac{13 \ao^4}{x-1}+\frac{20 \ao^4}{(x-2)^2}-\frac{4 \
\ao^4}{(x-1)^2}+2 \ao^4+52 x \ao^3+\frac{20 \ao^3}{x-2}-\frac{22 \
\ao^3}{x-1}-\frac{40 \ao^3}{(x-2)^2}+\frac{14 \
\ao^3}{(x-1)^2}+\frac{80 \ao^3}{(x-2)^3}-\frac{6 \ao^3}{(x-1)^3}-12 \
\ao^3-84 x \ao^2+\frac{18 \ao^2}{x-1}-\frac{18 \
\ao^2}{(x-1)^2}+\frac{18 \ao^2}{(x-1)^3}+\frac{480 \
\ao^2}{(x-2)^4}-\frac{12 \ao^2}{(x-1)^4}+36 \ao^2+73 x \ao-\frac{10 \
\ao}{x-2}-\frac{7 \ao}{x-1}+\frac{40 \ao}{(x-2)^2}+\frac{10 \
\ao}{(x-1)^2}-\frac{240 \ao}{(x-2)^3}-\frac{18 \
\ao}{(x-1)^3}-\frac{1920 \ao}{(x-2)^4}-\frac{1920 \ao}{(x-2)^5}-20 \
\ao-25 x+\frac{4}{x-2}+\frac{1}{x-1}-\frac{20}{(x-2)^2}-\frac{2}{(x-1)\
^2}+\frac{160}{(x-2)^3}+\frac{6}{(x-1)^3}+\frac{1440}{(x-2)^4}+\frac{\
12}{(x-1)^4}+\frac{1920}{(x-2)^5}-6\Big) H(1,1;\ao)+H(c_2(\ao);x) \
\Big(-\frac{40 \pi ^2 \ao}{9 (x-2)^4}+\frac{320 \ao}{27 \
(x-2)^4}-\frac{80 \pi ^2 \ao}{9 (x-2)^5}+\frac{640 \ao}{27 \
(x-2)^5}+\Big(-\frac{640 \ao}{9 (x-2)^4}-\frac{1280 \ao}{9 \
(x-2)^5}+\frac{640}{9 (x-2)^4}+\frac{2560}{9 (x-2)^5}+\frac{2560}{9 \
(x-2)^6}\Big) H(0;\ao)+\Big(\frac{320 \ao}{3 (x-2)^4}+\frac{640 \
\ao}{3 (x-2)^5}-\frac{320}{3 (x-2)^4}-\frac{1280}{3 \
(x-2)^5}-\frac{1280}{3 (x-2)^6}\Big) H(1;\ao)+\Big(\frac{1280 \ao}{3 \
(x-2)^4}+\frac{2560 \ao}{3 (x-2)^5}-\frac{1280}{3 \
(x-2)^4}-\frac{5120}{3 (x-2)^5}-\frac{5120}{3 (x-2)^6}\Big) \
H(0,0;\ao)+\Big(-\frac{640 \ao}{(x-2)^4}-\frac{1280 \
\ao}{(x-2)^5}+\frac{640}{(x-2)^4}+\frac{2560}{(x-2)^5}+\frac{2560}{(x-\
2)^6}\Big) H(0,1;\ao)+\Big(-\frac{640 \ao}{(x-2)^4}-\frac{1280 \
\ao}{(x-2)^5}+\frac{640}{(x-2)^4}+\frac{2560}{(x-2)^5}+\frac{2560}{(x-\
2)^6}\Big) H(1,0;\ao)+\Big(\frac{960 \ao}{(x-2)^4}+\frac{1920 \
\ao}{(x-2)^5}-\frac{960}{(x-2)^4}-\frac{3840}{(x-2)^5}-\frac{3840}{(x-\
2)^6}\Big) H(1,1;\ao)+\frac{40 \pi ^2}{9 (x-2)^4}-\frac{320}{27 \
(x-2)^4}+\frac{160 \pi ^2}{9 (x-2)^5}-\frac{1280}{27 \
(x-2)^5}+\frac{160 \pi ^2}{9 (x-2)^6}-\frac{1280}{27 \
(x-2)^6}\Big)+H(c_1(\ao);x) \Big(-\frac{19 x \ao^5}{36}-\frac{31 \
\ao^5}{18 (x-2)}+\frac{19 \ao^5}{36 (x-1)}-\frac{9 \
\ao^5}{8}+\frac{131 x \ao^4}{36}+\frac{137 \ao^4}{36 (x-2)}-\frac{103 \
\ao^4}{36 (x-1)}-\frac{65 \ao^4}{9 \
(x-2)^2}+\frac{\ao^4}{(x-1)^2}+\frac{349 \ao^4}{72}-\frac{104 x \
\ao^3}{9}+\frac{2 \ao^3}{x-2}+\frac{235 \ao^3}{36 (x-1)}+\frac{20 \
\ao^3}{9 (x-2)^2}-\frac{43 \ao^3}{9 (x-1)^2}-\frac{40 \
\ao^3}{(x-2)^3}+\frac{41 \ao^3}{18 (x-1)^3}-\frac{209 \
\ao^3}{36}+\frac{80 x \ao^2}{3}-\frac{103 \ao^2}{9 (x-2)}-\frac{307 \
\ao^2}{36 (x-1)}+\frac{38 \ao^2}{(x-2)^2}+\frac{86 \ao^2}{9 (x-1)^2}-\
\frac{1420 \ao^2}{9 (x-2)^3}-\frac{73 \ao^2}{6 (x-1)^3}-\frac{3760 \
\ao^2}{9 (x-2)^4}+\frac{77 \ao^2}{9 (x-1)^4}-\frac{455 \
\ao^2}{36}-\frac{367 x \ao}{12}-\frac{802 \ao}{9 (x-2)}+\frac{231 \
\ao}{2 (x-1)}+\frac{52 \ao}{(x-2)^2}-\frac{140 \ao}{9 \
(x-1)^2}+\frac{2864 \ao}{9 (x-2)^3}+\frac{125 \ao}{9 \
(x-1)^3}+\frac{2160 \ao}{(x-2)^4}-\frac{\pi ^2 \ao}{18 \
(x-1)^4}+\frac{3029 \ao}{216 (x-1)^4}+\frac{10240 \ao}{9 \
(x-2)^5}+\frac{\pi ^2 \ao}{18 (x-1)^5}-\frac{1351 \ao}{108 \
(x-1)^5}+\frac{235 \ao}{72}+\frac{445 x}{36}+\Big(\frac{2 x \
\ao^5}{3}+\frac{4 \ao^5}{3 (x-2)}-\frac{2 \ao^5}{3 \
(x-1)}+\ao^5-\frac{38 x \ao^4}{9}-\frac{34 \ao^4}{9 (x-2)}+\frac{26 \
\ao^4}{9 (x-1)}+\frac{40 \ao^4}{9 (x-2)^2}-\frac{8 \ao^4}{9 (x-1)^2}-\
\frac{41 \ao^4}{9}+\frac{104 x \ao^3}{9}+\frac{16 \ao^3}{9 \
(x-2)}-\frac{44 \ao^3}{9 (x-1)}-\frac{56 \ao^3}{9 (x-2)^2}+\frac{28 \
\ao^3}{9 (x-1)^2}+\frac{160 \ao^3}{9 (x-2)^3}-\frac{4 \ao^3}{3 \
(x-1)^3}+\frac{22 \ao^3}{3}-\frac{56 x \ao^2}{3}+\frac{4 \ao^2}{x-2}+\
\frac{4 \ao^2}{x-1}-\frac{8 \ao^2}{(x-2)^2}-\frac{4 \
\ao^2}{(x-1)^2}+\frac{16 \ao^2}{(x-2)^3}+\frac{4 \
\ao^2}{(x-1)^3}+\frac{320 \ao^2}{3 (x-2)^4}-\frac{8 \ao^2}{3 \
(x-1)^4}-2 \ao^2+\frac{146 x \ao}{9}+\frac{128 \ao}{9 (x-2)}-\frac{74 \
\ao}{3 (x-1)}-\frac{32 \ao}{9 (x-2)^2}+\frac{52 \ao}{9 \
(x-1)^2}-\frac{64 \ao}{(x-2)^3}-\frac{46 \ao}{9 (x-1)^3}-\frac{704 \
\ao}{(x-2)^4}-\frac{55 \ao}{9 (x-1)^4}-\frac{1280 \ao}{3 \
(x-2)^5}+\frac{58 \ao}{9 (x-1)^5}+\frac{29 \ao}{9}-\frac{50 \
x}{9}+\frac{160}{9 (x-2)}-\frac{52}{3 (x-1)}-\frac{224}{9 \
(x-2)^2}+\frac{14}{9 (x-1)^2}+\frac{640}{9 (x-2)^3}+\frac{8}{9 \
(x-1)^3}+\frac{1664}{3 (x-2)^4}+\frac{85}{9 (x-1)^4}+\frac{2560}{3 \
(x-2)^5}+\frac{58}{9 (x-1)^5}-5\Big) H(0;\ao)+\Big(-x \ao^5-\frac{2 \
\ao^5}{x-2}+\frac{\ao^5}{x-1}-\frac{3 \ao^5}{2}+\frac{19 x \ao^4}{3}+\
\frac{17 \ao^4}{3 (x-2)}-\frac{13 \ao^4}{3 (x-1)}-\frac{20 \ao^4}{3 \
(x-2)^2}+\frac{4 \ao^4}{3 (x-1)^2}+\frac{41 \ao^4}{6}-\frac{52 x \
\ao^3}{3}-\frac{8 \ao^3}{3 (x-2)}+\frac{22 \ao^3}{3 (x-1)}+\frac{28 \
\ao^3}{3 (x-2)^2}-\frac{14 \ao^3}{3 (x-1)^2}-\frac{80 \ao^3}{3 \
(x-2)^3}+\frac{2 \ao^3}{(x-1)^3}-11 \ao^3+28 x \ao^2-\frac{6 \
\ao^2}{x-2}-\frac{6 \ao^2}{x-1}+\frac{12 \ao^2}{(x-2)^2}+\frac{6 \
\ao^2}{(x-1)^2}-\frac{24 \ao^2}{(x-2)^3}-\frac{6 \
\ao^2}{(x-1)^3}-\frac{160 \ao^2}{(x-2)^4}+\frac{4 \ao^2}{(x-1)^4}+3 \
\ao^2-\frac{73 x \ao}{3}-\frac{64 \ao}{3 (x-2)}+\frac{37 \
\ao}{x-1}+\frac{16 \ao}{3 (x-2)^2}-\frac{26 \ao}{3 (x-1)^2}+\frac{96 \
\ao}{(x-2)^3}+\frac{23 \ao}{3 (x-1)^3}+\frac{1056 \
\ao}{(x-2)^4}+\frac{55 \ao}{6 (x-1)^4}+\frac{640 \
\ao}{(x-2)^5}-\frac{29 \ao}{3 (x-1)^5}-\frac{29 \ao}{6}+\frac{25 \
x}{3}-\frac{80}{3 (x-2)}+\frac{26}{x-1}+\frac{112}{3 \
(x-2)^2}-\frac{7}{3 (x-1)^2}-\frac{320}{3 (x-2)^3}-\frac{4}{3 \
(x-1)^3}-\frac{832}{(x-2)^4}-\frac{85}{6 \
(x-1)^4}-\frac{1280}{(x-2)^5}-\frac{29}{3 (x-1)^5}+\frac{15}{2}\Big) \
H(1;\ao)+\Big(\frac{16 \ao}{3 (x-1)^4}-\frac{16 \ao}{3 \
(x-1)^5}-\frac{16}{3 (x-1)^4}-\frac{16}{3 (x-1)^5}\Big) \
H(0,0;\ao)+\Big(-\frac{8 \ao}{(x-1)^4}+\frac{8 \
\ao}{(x-1)^5}+\frac{8}{(x-1)^4}+\frac{8}{(x-1)^5}\Big) \
H(0,1;\ao)+\Big(-\frac{8 \ao}{(x-1)^4}+\frac{8 \
\ao}{(x-1)^5}+\frac{8}{(x-1)^4}+\frac{8}{(x-1)^5}\Big) \
H(1,0;\ao)+\Big(\frac{12 \ao}{(x-1)^4}-\frac{12 \
\ao}{(x-1)^5}-\frac{12}{(x-1)^4}-\frac{12}{(x-1)^5}\Big) \
H(1,1;\ao)-\frac{92}{x-2}+\frac{1133}{12 (x-1)}+\frac{1040}{9 \
(x-2)^2}-\frac{223}{36 (x-1)^2}-\frac{3008}{9 (x-2)^3}-\frac{103}{36 \
(x-1)^3}-\frac{13600}{9 (x-2)^4}+\frac{\pi ^2}{18 \
(x-1)^4}-\frac{4223}{216 (x-1)^4}-\frac{20480}{9 (x-2)^5}+\frac{\pi \
^2}{18 (x-1)^5}-\frac{1351}{108 \
(x-1)^5}+\frac{275}{24}\Big)+\Big(-\frac{14 x \ao}{3}-\frac{480 \
\ao}{(x-2)^4}+\frac{64 \ao}{3 (x-1)^4}-\frac{800 \
\ao}{(x-2)^5}-\frac{64 \ao}{3 (x-1)^5}+\frac{28 \ao}{3}+\frac{14 \
x}{3}+\frac{480}{(x-2)^4}-\frac{64}{3 \
(x-1)^4}+\frac{1760}{(x-2)^5}-\frac{64}{3 \
(x-1)^5}+\frac{1600}{(x-2)^6}\Big) H(0;\ao) H(1,1;x)+\Big(\frac{29 x \
\ao}{9}+\frac{406 \ao}{9 (x-2)}-\frac{166 \ao}{3 (x-1)}-\frac{424 \
\ao}{9 (x-2)^2}+\frac{193 \ao}{18 (x-1)^2}+\frac{32 \
\ao}{(x-2)^3}-\frac{53 \ao}{9 (x-1)^3}-\frac{4000 \ao}{9 \
(x-2)^4}-\frac{175 \ao}{18 (x-1)^4}+\frac{640 \ao}{9 \
(x-2)^5}+\frac{116 \ao}{9 (x-1)^5}+\frac{2 \ao}{9}-\frac{29 \
x}{9}+\Big(-\frac{8 x \ao}{3}-\frac{832 \ao}{3 (x-2)^4}+\frac{32 \
\ao}{3 (x-1)^4}-\frac{1280 \ao}{3 (x-2)^5}-\frac{32 \ao}{3 \
(x-1)^5}+\frac{16 \ao}{3}+\frac{8 x}{3}+\frac{832}{3 \
(x-2)^4}-\frac{32}{3 (x-1)^4}+\frac{2944}{3 (x-2)^5}-\frac{32}{3 \
(x-1)^5}+\frac{2560}{3 (x-2)^6}\Big) H(0;\ao)+\Big(4 x \ao+\frac{416 \
\ao}{(x-2)^4}-\frac{16 \ao}{(x-1)^4}+\frac{640 \ao}{(x-2)^5}+\frac{16 \
\ao}{(x-1)^5}-8 \ao-4 \
x-\frac{416}{(x-2)^4}+\frac{16}{(x-1)^4}-\frac{1472}{(x-2)^5}+\frac{\
16}{(x-1)^5}-\frac{1280}{(x-2)^6}\Big) H(1;\ao)+\frac{326}{9 \
(x-2)}-\frac{118}{3 (x-1)}-\frac{388}{9 (x-2)^2}+\frac{95}{18 \
(x-1)^2}+\frac{560}{9 (x-2)^3}+\frac{4}{9 (x-1)^3}+\frac{3424}{9 \
(x-2)^4}+\frac{289}{18 (x-1)^4}+\frac{7360}{9 (x-2)^5}+\frac{116}{9 \
(x-1)^5}-\frac{1280}{9 (x-2)^6}-\frac{20}{3}\Big) \
H(1,c_1(\ao);x)+\Big(\frac{2 x \ao}{3}+\frac{2080 \ao}{3 \
(x-2)^4}-\frac{2 \ao}{3 (x-1)^4}+\frac{3200 \ao}{3 (x-2)^5}+\frac{2 \
\ao}{3 (x-1)^5}-\frac{4 \ao}{3}-\frac{2 x}{3}-\frac{2080}{3 (x-2)^4}+\
\frac{2}{3 (x-1)^4}-\frac{7360}{3 (x-2)^5}+\frac{2}{3 \
(x-1)^5}-\frac{6400}{3 (x-2)^6}\Big) H(0;\ao) \
H(2,1;x)+\Big(\frac{1600 \ao}{9 (x-2)^4}+\frac{3200 \ao}{9 \
(x-2)^5}+\Big(-\frac{3200 \ao}{3 (x-2)^4}-\frac{6400 \ao}{3 (x-2)^5}+\
\frac{3200}{3 (x-2)^4}+\frac{12800}{3 (x-2)^5}+\frac{12800}{3 \
(x-2)^6}\Big) H(0;\ao)+\Big(\frac{1600 \ao}{(x-2)^4}+\frac{3200 \
\ao}{(x-2)^5}-\frac{1600}{(x-2)^4}-\frac{6400}{(x-2)^5}-\frac{6400}{(\
x-2)^6}\Big) H(1;\ao)-\frac{1600}{9 (x-2)^4}-\frac{6400}{9 \
(x-2)^5}-\frac{6400}{9 (x-2)^6}\Big) H(2,c_2(\ao);x)+\Big(\frac{x \
\ao^5}{2}+\frac{5 \ao^5}{6 (x-2)}-\frac{\ao^5}{3 \
(x-1)}+\ao^5-\frac{19 x \ao^4}{6}-\frac{35 \ao^4}{18 (x-2)}+\frac{13 \
\ao^4}{9 (x-1)}+\frac{25 \ao^4}{9 (x-2)^2}-\frac{4 \ao^4}{9 (x-1)^2}-\
\frac{14 \ao^4}{3}+\frac{26 x \ao^3}{3}-\frac{5 \ao^3}{9 \
(x-2)}-\frac{22 \ao^3}{9 (x-1)}-\frac{20 \ao^3}{9 (x-2)^2}+\frac{14 \
\ao^3}{9 (x-1)^2}+\frac{100 \ao^3}{9 (x-2)^3}-\frac{2 \ao^3}{3 \
(x-1)^3}+8 \ao^3-14 x \ao^2+\frac{5 \ao^2}{x-2}+\frac{2 \
\ao^2}{x-1}-\frac{10 \ao^2}{(x-2)^2}-\frac{2 \ao^2}{(x-1)^2}+\frac{20 \
\ao^2}{(x-2)^3}+\frac{2 \ao^2}{(x-1)^3}+\frac{200 \ao^2}{3 \
(x-2)^4}-\frac{4 \ao^2}{3 (x-1)^4}-4 \ao^2+\frac{73 x \
\ao}{6}-\frac{40 \ao}{9 (x-2)}-\frac{73 \ao}{18 (x-1)}+\frac{160 \
\ao}{9 (x-2)^2}+\frac{5 \ao}{18 (x-1)^2}-\frac{80 \
\ao}{(x-2)^3}-\frac{11 \ao}{9 (x-1)^3}-\frac{480 \
\ao}{(x-2)^4}-\frac{32 \ao}{9 (x-1)^4}-\frac{800 \ao}{3 \
(x-2)^5}+\frac{29 \ao}{9 (x-1)^5}+\frac{11 \ao}{3}-\frac{25 \
x}{6}+\Big(\frac{8 \ao}{3 (x-1)^4}-\frac{8 \ao}{3 (x-1)^5}-\frac{8}{3 \
(x-1)^4}-\frac{8}{3 (x-1)^5}\Big) H(0;\ao)+\Big(-\frac{4 \
\ao}{(x-1)^4}+\frac{4 \
\ao}{(x-1)^5}+\frac{4}{(x-1)^4}+\frac{4}{(x-1)^5}\Big) \
H(1;\ao)+\frac{40}{9 (x-2)}-\frac{71}{18 (x-1)}-\frac{80}{9 (x-2)^2}-\
\frac{1}{6 (x-1)^2}+\frac{400}{9 (x-2)^3}+\frac{7}{9 \
(x-1)^3}+\frac{1280}{3 (x-2)^4}+\frac{38}{9 (x-1)^4}+\frac{1600}{3 \
(x-2)^5}+\frac{29}{9 (x-1)^5}-4\Big) \
H(c_1(\ao),c_1(\ao);x)+\Big(-\frac{x \ao^5}{6}-\frac{\ao^5}{6 (x-1)}+\
\frac{\ao^5}{2}+\frac{19 x \ao^4}{18}+\frac{13 \ao^4}{18 \
(x-1)}-\frac{2 \ao^4}{9 (x-1)^2}-\frac{47 \ao^4}{18}-\frac{26 x \
\ao^3}{9}-\frac{11 \ao^3}{9 (x-1)}+\frac{7 \ao^3}{9 \
(x-1)^2}-\frac{\ao^3}{3 (x-1)^3}+\frac{17 \ao^3}{3}+\frac{14 x \
\ao^2}{3}+\frac{\ao^2}{x-1}-\frac{\ao^2}{(x-1)^2}+\frac{\ao^2}{(x-1)^3}-\frac{2 \ao^2}{3 (x-1)^4}-7 \ao^2-\frac{73 x \ao}{18}+\frac{8 \
\ao}{9 (x-2)}+\frac{\ao}{18 (x-1)}-\frac{8 \ao}{9 \
(x-2)^2}+\frac{\ao}{9 (x-1)^2}+\frac{\ao}{3 (x-1)^3}-\frac{320 \ao}{9 \
(x-2)^4}-\frac{4 \ao}{3 (x-1)^4}-\frac{640 \ao}{9 (x-2)^5}+\frac{41 \
\ao}{18}+\frac{25 x}{18}+\Big(\frac{832 \ao}{3 (x-2)^4}+\frac{1280 \
\ao}{3 (x-2)^5}-\frac{832}{3 (x-2)^4}-\frac{2944}{3 \
(x-2)^5}-\frac{2560}{3 (x-2)^6}\Big) H(0;\ao)+\Big(-\frac{416 \
\ao}{(x-2)^4}-\frac{640 \
\ao}{(x-2)^5}+\frac{416}{(x-2)^4}+\frac{1472}{(x-2)^5}+\frac{1280}{(x-\
2)^6}\Big) H(1;\ao)+\frac{16}{9 (x-2)}-\frac{7}{18 (x-1)}-\frac{8}{9 \
(x-2)^2}-\frac{5}{9 (x-1)^2}+\frac{16}{9 \
(x-2)^3}-\frac{1}{(x-1)^3}+\frac{320}{9 (x-2)^4}-\frac{2}{3 (x-1)^4}+\
\frac{1280}{9 (x-2)^5}+\frac{1280}{9 (x-2)^6}+\frac{7}{6}\Big) H(c_2(\
\ao),c_1(\ao);x)+\Big(\frac{16 x \ao}{3}+\frac{1280 \ao}{3 \
(x-2)^4}+\frac{16 \ao}{3 (x-1)^4}+\frac{2560 \ao}{3 (x-2)^5}-\frac{16 \
\ao}{3 (x-1)^5}-\frac{32 \ao}{3}-\frac{16 x}{3}-\frac{1280}{3 \
(x-2)^4}-\frac{16}{3 (x-1)^4}-\frac{5120}{3 (x-2)^5}-\frac{16}{3 \
(x-1)^5}-\frac{5120}{3 (x-2)^6}\Big) H(0,0,0;\ao)+\Big(\frac{16 x \
\ao}{3}-\frac{1280 \ao}{3 (x-2)^4}-\frac{16 \ao}{3 \
(x-1)^4}-\frac{2560 \ao}{3 (x-2)^5}+\frac{16 \ao}{3 (x-1)^5}-\frac{32 \
\ao}{3}-\frac{16 x}{3}+\frac{1280}{3 (x-2)^4}+\frac{16}{3 \
(x-1)^4}+\frac{5120}{3 (x-2)^5}+\frac{16}{3 (x-1)^5}+\frac{5120}{3 \
(x-2)^6}\Big) H(0,0,0;x)+\Big(-8 x \ao-\frac{640 \
\ao}{(x-2)^4}-\frac{8 \ao}{(x-1)^4}-\frac{1280 \ao}{(x-2)^5}+\frac{8 \
\ao}{(x-1)^5}+16 \ao+8 \
x+\frac{640}{(x-2)^4}+\frac{8}{(x-1)^4}+\frac{2560}{(x-2)^5}+\frac{8}{\
(x-1)^5}+\frac{2560}{(x-2)^6}\Big) H(0,0,1;\ao)+\Big(-\frac{8 x \
\ao}{3}-\frac{736 \ao}{3 (x-2)^4}+\frac{8 \ao}{3 (x-1)^4}-\frac{1280 \
\ao}{3 (x-2)^5}-\frac{8 \ao}{3 (x-1)^5}+\frac{16 \ao}{3}+\frac{8 \
x}{3}+\frac{736}{3 (x-2)^4}-\frac{8}{3 (x-1)^4}+\frac{2752}{3 \
(x-2)^5}-\frac{8}{3 (x-1)^5}+\frac{2560}{3 (x-2)^6}\Big) \
H(0,0,c_1(\ao);x)+\Big(\frac{1280 \ao}{3 (x-2)^4}+\frac{2560 \ao}{3 \
(x-2)^5}-\frac{1280}{3 (x-2)^4}-\frac{5120}{3 (x-2)^5}-\frac{5120}{3 \
(x-2)^6}\Big) H(0,0,c_2(\ao);x)+\Big(-8 x \ao-\frac{640 \
\ao}{(x-2)^4}-\frac{8 \ao}{(x-1)^4}-\frac{1280 \ao}{(x-2)^5}+\frac{8 \
\ao}{(x-1)^5}+16 \ao+8 \
x+\frac{640}{(x-2)^4}+\frac{8}{(x-1)^4}+\frac{2560}{(x-2)^5}+\frac{8}{\
(x-1)^5}+\frac{2560}{(x-2)^6}\Big) H(0,1,0;\ao)+\Big(\frac{20 x \
\ao}{3}+\frac{1984 \ao}{3 (x-2)^4}-\frac{20 \ao}{3 \
(x-1)^4}+\frac{3200 \ao}{3 (x-2)^5}+\frac{20 \ao}{3 (x-1)^5}-\frac{40 \
\ao}{3}-\frac{20 x}{3}-\frac{1984}{3 (x-2)^4}+\frac{20}{3 \
(x-1)^4}-\frac{7168}{3 (x-2)^5}+\frac{20}{3 (x-1)^5}-\frac{6400}{3 \
(x-2)^6}\Big) H(0,1,0;x)+\Big(12 x \ao+\frac{960 \
\ao}{(x-2)^4}+\frac{12 \ao}{(x-1)^4}+\frac{1920 \
\ao}{(x-2)^5}-\frac{12 \ao}{(x-1)^5}-24 \ao-12 x-\frac{960}{(x-2)^4}-\
\frac{12}{(x-1)^4}-\frac{3840}{(x-2)^5}-\frac{12}{(x-1)^5}-\frac{3840}\
{(x-2)^6}\Big) H(0,1,1;\ao)+\Big(-\frac{20 x \ao}{3}-\frac{1984 \
\ao}{3 (x-2)^4}+\frac{20 \ao}{3 (x-1)^4}-\frac{3200 \ao}{3 \
(x-2)^5}-\frac{20 \ao}{3 (x-1)^5}+\frac{40 \ao}{3}+\frac{20 \
x}{3}+\frac{1984}{3 (x-2)^4}-\frac{20}{3 (x-1)^4}+\frac{7168}{3 \
(x-2)^5}-\frac{20}{3 (x-1)^5}+\frac{6400}{3 (x-2)^6}\Big) \
H(0,1,c_1(\ao);x)+\Big(\frac{3200 \ao}{3 (x-2)^4}+\frac{6400 \ao}{3 \
(x-2)^5}-\frac{3200}{3 (x-2)^4}-\frac{12800}{3 \
(x-2)^5}-\frac{12800}{3 (x-2)^6}\Big) H(0,2,0;x)+\Big(-\frac{3200 \
\ao}{3 (x-2)^4}-\frac{6400 \ao}{3 (x-2)^5}+\frac{3200}{3 \
(x-2)^4}+\frac{12800}{3 (x-2)^5}+\frac{12800}{3 (x-2)^6}\Big) \
H(0,2,c_2(\ao);x)+\Big(-2 x \ao-\frac{640 \ao}{3 (x-2)^4}+\frac{4 \
\ao}{3 (x-1)^4}-\frac{800 \ao}{3 (x-2)^5}-\frac{4 \ao}{3 (x-1)^5}+4 \
\ao+2 x+\frac{640}{3 (x-2)^4}-\frac{4}{3 (x-1)^4}+\frac{2080}{3 \
(x-2)^5}-\frac{4}{3 (x-1)^5}+\frac{1600}{3 (x-2)^6}\Big) \
H(0,c_1(\ao),c_1(\ao);x)+\Big(\frac{2 x \ao}{3}+\frac{832 \ao}{3 \
(x-2)^4}-\frac{2 \ao}{3 (x-1)^4}+\frac{1280 \ao}{3 (x-2)^5}+\frac{2 \
\ao}{3 (x-1)^5}-\frac{4 \ao}{3}-\frac{2 x}{3}-\frac{832}{3 \
(x-2)^4}+\frac{2}{3 (x-1)^4}-\frac{2944}{3 (x-2)^5}+\frac{2}{3 \
(x-1)^5}-\frac{2560}{3 (x-2)^6}\Big) \
H(0,c_2(\ao),c_1(\ao);x)+\Big(\frac{8 x \ao}{3}+\frac{832 \ao}{3 \
(x-2)^4}-\frac{32 \ao}{3 (x-1)^4}+\frac{1280 \ao}{3 (x-2)^5}+\frac{32 \
\ao}{3 (x-1)^5}-\frac{16 \ao}{3}-\frac{8 x}{3}-\frac{832}{3 (x-2)^4}+\
\frac{32}{3 (x-1)^4}-\frac{2944}{3 (x-2)^5}+\frac{32}{3 \
(x-1)^5}-\frac{2560}{3 (x-2)^6}\Big) H(1,0,0;x)+\Big(-\frac{2 x \
\ao}{3}-\frac{64 \ao}{(x-2)^4}+\frac{16 \ao}{3 (x-1)^4}-\frac{160 \
\ao}{(x-2)^5}-\frac{16 \ao}{3 (x-1)^5}+\frac{4 \ao}{3}+\frac{2 x}{3}+\
\frac{64}{(x-2)^4}-\frac{16}{3 \
(x-1)^4}+\frac{288}{(x-2)^5}-\frac{16}{3 (x-1)^5}+\frac{320}{(x-2)^6}\
\Big) H(1,0,c_1(\ao);x)+\Big(\frac{14 x \ao}{3}+\frac{480 \
\ao}{(x-2)^4}-\frac{64 \ao}{3 (x-1)^4}+\frac{800 \
\ao}{(x-2)^5}+\frac{64 \ao}{3 (x-1)^5}-\frac{28 \ao}{3}-\frac{14 \
x}{3}-\frac{480}{(x-2)^4}+\frac{64}{3 \
(x-1)^4}-\frac{1760}{(x-2)^5}+\frac{64}{3 \
(x-1)^5}-\frac{1600}{(x-2)^6}\Big) H(1,1,0;x)+\Big(-\frac{14 x \
\ao}{3}-\frac{480 \ao}{(x-2)^4}+\frac{64 \ao}{3 (x-1)^4}-\frac{800 \
\ao}{(x-2)^5}-\frac{64 \ao}{3 (x-1)^5}+\frac{28 \ao}{3}+\frac{14 \
x}{3}+\frac{480}{(x-2)^4}-\frac{64}{3 \
(x-1)^4}+\frac{1760}{(x-2)^5}-\frac{64}{3 \
(x-1)^5}+\frac{1600}{(x-2)^6}\Big) H(1,1,c_1(\ao);x)+\Big(-2 x \
\ao-\frac{640 \ao}{3 (x-2)^4}+\frac{16 \ao}{3 (x-1)^4}-\frac{800 \
\ao}{3 (x-2)^5}-\frac{16 \ao}{3 (x-1)^5}+4 \ao+2 x+\frac{640}{3 \
(x-2)^4}-\frac{16}{3 (x-1)^4}+\frac{2080}{3 (x-2)^5}-\frac{16}{3 \
(x-1)^5}+\frac{1600}{3 (x-2)^6}\Big) \
H(1,c_1(\ao),c_1(\ao);x)+\Big(\frac{3200 \ao}{3 (x-2)^4}+\frac{6400 \
\ao}{3 (x-2)^5}-\frac{3200}{3 (x-2)^4}-\frac{12800}{3 \
(x-2)^5}-\frac{12800}{3 (x-2)^6}\Big) H(2,0,0;x)+\Big(\frac{2 x \
\ao}{3}+\frac{2080 \ao}{3 (x-2)^4}-\frac{2 \ao}{3 (x-1)^4}+\frac{3200 \
\ao}{3 (x-2)^5}+\frac{2 \ao}{3 (x-1)^5}-\frac{4 \ao}{3}-\frac{2 \
x}{3}-\frac{2080}{3 (x-2)^4}+\frac{2}{3 (x-1)^4}-\frac{7360}{3 \
(x-2)^5}+\frac{2}{3 (x-1)^5}-\frac{6400}{3 (x-2)^6}\Big) \
H(2,0,c_1(\ao);x)+\Big(-\frac{3200 \ao}{3 (x-2)^4}-\frac{6400 \ao}{3 \
(x-2)^5}+\frac{3200}{3 (x-2)^4}+\frac{12800}{3 \
(x-2)^5}+\frac{12800}{3 (x-2)^6}\Big) H(2,0,c_2(\ao);x)+\Big(-\frac{2 \
x \ao}{3}-\frac{2080 \ao}{3 (x-2)^4}+\frac{2 \ao}{3 \
(x-1)^4}-\frac{3200 \ao}{3 (x-2)^5}-\frac{2 \ao}{3 (x-1)^5}+\frac{4 \
\ao}{3}+\frac{2 x}{3}+\frac{2080}{3 (x-2)^4}-\frac{2}{3 \
(x-1)^4}+\frac{7360}{3 (x-2)^5}-\frac{2}{3 (x-1)^5}+\frac{6400}{3 \
(x-2)^6}\Big) H(2,1,0;x)+\Big(\frac{2 x \ao}{3}+\frac{2080 \ao}{3 \
(x-2)^4}-\frac{2 \ao}{3 (x-1)^4}+\frac{3200 \ao}{3 (x-2)^5}+\frac{2 \
\ao}{3 (x-1)^5}-\frac{4 \ao}{3}-\frac{2 x}{3}-\frac{2080}{3 (x-2)^4}+\
\frac{2}{3 (x-1)^4}-\frac{7360}{3 (x-2)^5}+\frac{2}{3 \
(x-1)^5}-\frac{6400}{3 (x-2)^6}\Big) \
H(2,1,c_1(\ao);x)+\Big(-\frac{8000 \ao}{3 (x-2)^4}-\frac{16000 \ao}{3 \
(x-2)^5}+\frac{8000}{3 (x-2)^4}+\frac{32000}{3 \
(x-2)^5}+\frac{32000}{3 (x-2)^6}\Big) H(2,2,0;x)+\Big(\frac{8000 \
\ao}{3 (x-2)^4}+\frac{16000 \ao}{3 (x-2)^5}-\frac{8000}{3 \
(x-2)^4}-\frac{32000}{3 (x-2)^5}-\frac{32000}{3 (x-2)^6}\Big) \
H(2,2,c_2(\ao);x)+\Big(-\frac{2 x \ao}{3}-\frac{2080 \ao}{3 (x-2)^4}+\
\frac{2 \ao}{3 (x-1)^4}-\frac{3200 \ao}{3 (x-2)^5}-\frac{2 \ao}{3 \
(x-1)^5}+\frac{4 \ao}{3}+\frac{2 x}{3}+\frac{2080}{3 \
(x-2)^4}-\frac{2}{3 (x-1)^4}+\frac{7360}{3 (x-2)^5}-\frac{2}{3 \
(x-1)^5}+\frac{6400}{3 (x-2)^6}\Big) \
H(2,c_2(\ao),c_1(\ao);x)+\Big(\frac{4 \ao}{3 (x-1)^4}-\frac{4 \ao}{3 \
(x-1)^5}-\frac{4}{3 (x-1)^4}-\frac{4}{3 (x-1)^5}\Big) H(c_1(\ao),c_1(\
\ao),c_1(\ao);x)+\Big(\frac{2 \ao}{3 (x-1)^4}-\frac{2 \ao}{3 \
(x-1)^5}-\frac{2}{3 (x-1)^4}-\frac{2}{3 (x-1)^5}\Big) H(c_1(\ao),c_2(\
\ao),c_1(\ao);x)+\Big(-\frac{32 \
\ao}{(x-2)^4}+\frac{32}{(x-2)^4}+\frac{64}{(x-2)^5}\Big) \
H(c_2(\ao),0,c_1(\ao);x)+\Big(\frac{640 \ao}{3 (x-2)^4}+\frac{800 \
\ao}{3 (x-2)^5}-\frac{640}{3 (x-2)^4}-\frac{2080}{3 \
(x-2)^5}-\frac{1600}{3 (x-2)^6}\Big) \
H(c_2(\ao),c_1(\ao),c_1(\ao);x)+H(2,0;x) \Big(-\frac{1600 \ao}{9 \
(x-2)^4}-\frac{3200 \ao}{9 (x-2)^5}-\frac{3200 \ln 2\,  \ao}{3 \
(x-2)^4}-\frac{6400 \ln 2\,  \ao}{3 (x-2)^5}+\frac{1600}{9 \
(x-2)^4}+\frac{6400}{9 (x-2)^5}+\frac{6400}{9 (x-2)^6}+\frac{3200 \ln \
2\, }{3 (x-2)^4}+\frac{12800 \ln 2\, }{3 (x-2)^5}+\frac{12800 \ln 2\, \
}{3 (x-2)^6}\Big)+H(0,2;x) \Big(-\frac{3200 \ln 2\,  \ao}{3 (x-2)^4}-\
\frac{6400 \ln 2\,  \ao}{3 (x-2)^5}+\Big(-\frac{3200 \ao}{3 (x-2)^4}-\
\frac{6400 \ao}{3 (x-2)^5}+\frac{3200}{3 (x-2)^4}+\frac{12800}{3 \
(x-2)^5}+\frac{12800}{3 (x-2)^6}\Big) H(0;\ao)+\frac{3200 \ln 2\, }{3 \
(x-2)^4}+\frac{12800 \ln 2\, }{3 (x-2)^5}+\frac{12800 \ln 2\, }{3 \
(x-2)^6}\Big)+H(0,0;x) \Big(-\frac{58 x \ao}{9}-\frac{128 \ao}{9 \
(x-2)}+\frac{74 \ao}{3 (x-1)}+\frac{32 \ao}{9 (x-2)^2}-\frac{52 \
\ao}{9 (x-1)^2}+\frac{64 \ao}{(x-2)^3}+\frac{46 \ao}{9 \
(x-1)^3}+\frac{4480 \ao}{9 (x-2)^4}+\frac{31 \ao}{9 \
(x-1)^4}+\frac{1280 \ao}{9 (x-2)^5}-\frac{58 \ao}{9 \
(x-1)^5}+\frac{1280 \ln 2\,  \ao}{3 (x-2)^4}+\frac{2560 \ln 2\,  \
\ao}{3 (x-2)^5}+\frac{71 \ao}{9}+\frac{58 x}{9}-\frac{160}{9 \
(x-2)}+\frac{52}{3 (x-1)}+\frac{224}{9 (x-2)^2}-\frac{14}{9 (x-1)^2}-\
\frac{640}{9 (x-2)^3}-\frac{8}{9 (x-1)^3}-\frac{5632}{9 \
(x-2)^4}-\frac{85}{9 (x-1)^4}-\frac{10240}{9 (x-2)^5}-\frac{58}{9 \
(x-1)^5}-\frac{2560}{9 (x-2)^6}-\frac{1280 \ln 2\, }{3 \
(x-2)^4}-\frac{5120 \ln 2\, }{3 (x-2)^5}-\frac{5120 \ln 2\, }{3 \
(x-2)^6}+5\Big)+H(2,2;x) \Big(\frac{8000 \ln 2\,  \ao}{3 \
(x-2)^4}+\frac{16000 \ln 2\,  \ao}{3 (x-2)^5}+\Big(\frac{8000 \ao}{3 \
(x-2)^4}+\frac{16000 \ao}{3 (x-2)^5}-\frac{8000}{3 \
(x-2)^4}-\frac{32000}{3 (x-2)^5}-\frac{32000}{3 (x-2)^6}\Big) \
H(0;\ao)-\frac{8000 \ln 2\, }{3 (x-2)^4}-\frac{32000 \ln 2\, }{3 \
(x-2)^5}-\frac{32000 \ln 2\, }{3 (x-2)^6}\Big)+H(0;x) \
\Big(\frac{1}{2} \pi ^2 x \ao+\frac{1351 x \ao}{108}+\frac{284 \ao}{3 \
(x-2)}-\frac{691 \ao}{6 (x-1)}-\frac{664 \ao}{9 (x-2)^2}+\frac{131 \
\ao}{9 (x-1)^2}-\frac{560 \ao}{3 (x-2)^3}-\frac{101 \ao}{9 \
(x-1)^3}+\frac{8 \pi ^2 \ao}{3 (x-2)^4}-\frac{31040 \ao}{27 (x-2)^4}-\
\frac{7 \pi ^2 \ao}{18 (x-1)^4}-\frac{1181 \ao}{216 (x-1)^4}-\frac{80 \
\pi ^2 \ao}{3 (x-2)^5}-\frac{640 \ao}{27 (x-2)^5}+\frac{7 \pi ^2 \
\ao}{18 (x-1)^5}+\frac{1351 \ao}{108 (x-1)^5}-\frac{640 \ln ^22\,  \
\ao}{3 (x-2)^4}-\frac{1280 \ln ^22\,  \ao}{3 (x-2)^5}-\frac{640 \ln 2\
\,  \ao}{9 (x-2)^4}-\frac{1280 \ln 2\,  \ao}{9 (x-2)^5}-\pi ^2 \
\ao-\frac{2929 \ao}{216}-\frac{\pi ^2 x}{2}-\frac{1351 \
x}{108}+\frac{92}{x-2}-\frac{1133}{12 (x-1)}-\frac{1040}{9 \
(x-2)^2}+\frac{223}{36 (x-1)^2}+\frac{3008}{9 (x-2)^3}+\frac{103}{36 \
(x-1)^3}-\frac{8 \pi ^2}{3 (x-2)^4}+\frac{41120}{27 (x-2)^4}+\frac{7 \
\pi ^2}{18 (x-1)^4}+\frac{4223}{216 (x-1)^4}+\frac{64 \pi ^2}{3 \
(x-2)^5}+\frac{62720}{27 (x-2)^5}+\frac{7 \pi ^2}{18 \
(x-1)^5}+\frac{1351}{108 (x-1)^5}+\frac{160 \pi ^2}{3 \
(x-2)^6}+\frac{1280}{27 (x-2)^6}+\frac{640 \ln ^22\, }{3 \
(x-2)^4}+\frac{2560 \ln ^22\, }{3 (x-2)^5}+\frac{2560 \ln ^22\, }{3 \
(x-2)^6}+\frac{640 \ln 2\, }{9 (x-2)^4}+\frac{2560 \ln 2\, }{9 \
(x-2)^5}+\frac{2560 \ln 2\, }{9 (x-2)^6}-\frac{275}{24}\Big)+H(2;x) \
\Big(-\frac{1}{6} \pi ^2 x \ao+\frac{40 \pi ^2 \ao}{9 (x-2)^4}+\frac{\
\pi ^2 \ao}{6 (x-1)^4}+\frac{800 \pi ^2 \ao}{9 (x-2)^5}-\frac{\pi ^2 \
\ao}{6 (x-1)^5}+\frac{1600 \ln ^22\,  \ao}{3 (x-2)^4}+\frac{3200 \ln \
^22\,  \ao}{3 (x-2)^5}+\frac{1600 \ln 2\,  \ao}{9 (x-2)^4}+\frac{3200 \
\ln 2\,  \ao}{9 (x-2)^5}+\frac{\pi ^2 \ao}{3}+\frac{\pi ^2 \
x}{6}+\Big(\frac{1600 \ao}{9 (x-2)^4}+\frac{3200 \ao}{9 \
(x-2)^5}-\frac{1600}{9 (x-2)^4}-\frac{6400}{9 (x-2)^5}-\frac{6400}{9 \
(x-2)^6}\Big) H(0;\ao)+\Big(-\frac{3200 \ao}{3 (x-2)^4}-\frac{6400 \
\ao}{3 (x-2)^5}+\frac{3200}{3 (x-2)^4}+\frac{12800}{3 \
(x-2)^5}+\frac{12800}{3 (x-2)^6}\Big) H(0,0;\ao)+\Big(\frac{1600 \
\ao}{(x-2)^4}+\frac{3200 \
\ao}{(x-2)^5}-\frac{1600}{(x-2)^4}-\frac{6400}{(x-2)^5}-\frac{6400}{(\
x-2)^6}\Big) H(0,1;\ao)-\frac{40 \pi ^2}{9 (x-2)^4}-\frac{\pi ^2}{6 \
(x-1)^4}-\frac{880 \pi ^2}{9 (x-2)^5}-\frac{\pi ^2}{6 \
(x-1)^5}-\frac{1600 \pi ^2}{9 (x-2)^6}-\frac{1600 \ln ^22\, }{3 \
(x-2)^4}-\frac{6400 \ln ^22\, }{3 (x-2)^5}-\frac{6400 \ln ^22\, }{3 \
(x-2)^6}-\frac{1600 \ln 2\, }{9 (x-2)^4}-\frac{6400 \ln 2\, }{9 \
(x-2)^5}-\frac{6400 \ln 2\, }{9 (x-2)^6}\Big)-\frac{16 \pi ^2}{9 \
(x-2)}+\frac{461 \pi ^2}{216 (x-1)}+\frac{22 \pi ^2}{9 \
(x-2)^2}-\frac{23 \pi ^2}{54 (x-1)^2}-\frac{4 \pi \
^2}{(x-2)^3}-\frac{29 \pi ^2}{108 (x-1)^3}-\frac{656 \pi ^2}{27 \
(x-2)^4}-\frac{28 \pi ^2}{27 (x-1)^4}-\frac{1760 \pi ^2}{27 (x-2)^5}-\
\frac{29 \pi ^2}{54 (x-1)^5}-\frac{320 \pi ^2}{27 \
(x-2)^6}-\frac{17}{12} x \zeta_3-\frac{224 \zeta_3}{3 \
(x-2)^4}+\frac{7 \zeta_3}{4 (x-1)^4}-\frac{728 \zeta_3}{3 \
(x-2)^5}+\frac{7 \zeta_3}{4 (x-1)^5}-\frac{560 \zeta_3}{3 \
(x-2)^6}-\frac{640 \ln ^32\, }{9 (x-2)^4}-\frac{2560 \ln ^32\, }{9 \
(x-2)^5}-\frac{2560 \ln ^32\, }{9 (x-2)^6}-\frac{320 \ln ^22\, }{9 \
(x-2)^4}-\frac{1280 \ln ^22\, }{9 (x-2)^5}-\frac{1280 \ln ^22\, }{9 \
(x-2)^6}+\frac{1}{6} \pi ^2 x \ln 2\, +\frac{32 \pi ^2 \ln 2\, }{3 \
(x-2)^4}-\frac{320 \ln 2\, }{27 (x-2)^4}-\frac{\pi ^2 \ln 2\, }{6 \
(x-1)^4}-\frac{16 \pi ^2 \ln 2\, }{3 (x-2)^5}-\frac{1280 \ln 2\, }{27 \
(x-2)^5}-\frac{\pi ^2 \ln 2\, }{6 (x-1)^5}-\frac{160 \pi ^2 \ln 2\, \
}{3 (x-2)^6}-\frac{1280 \ln 2\, }{27 (x-2)^6}+\frac{11 \pi ^2}{24}\}.
\erp

%
% The B integral for k=-1 and delta=-1
%

\subsection{The $\cB$ integral for $k=-1$ and $\delta=-1$}
%
% This file contains the TeX output produced by Mathematica for the integral Bm1, for delta = -1
%
The $\eps$ expansion for this integral reads
\beq
\bsp
\begin{cal}I\end{cal}(x,\eps;\ao,3+d_1\eps&;1,-1,-1,g_B) = x\,\bint(\eps,x;3+d_1\eps;-1,-1)\\
&=\frac{1}{\eps^2}b_{-2}^{(-1,-1)}+\frac{1}{\eps}b_{-1}^{(-1,-1)}+b_0^{(-1,-1)}+\eps b_1^{(-1,-1)}+\eps^2b_2^{(-1,-1)} +\ocal\left(\eps^3\right),
\esp
\eeq
where
%1/ep piece
\brp
b_{-2}^{(-1,-1)}=\frac{1}{8},
\erp
\brp
b_{-1}^{(-1,-1)}=-\frac{1}{2} H(0;x),
\erp
% ep^0
\brp
b_0^{(-1,-1)}=\frac{\ao^3}{12 (x-1)^2}-\frac{\ao^3}{12}+\frac{\ao^2}{24 \
(x-1)}-\frac{5 \ao^2}{24 (x-1)^2}+\frac{7 \ao^2}{24 (x-1)^3}+\frac{13 \
\ao^2}{24}-\frac{\ao}{3 (x-1)}+\frac{\ao}{6 (x-1)^2}-\frac{\ao}{3 \
(x-1)^3}+\frac{13 \ao}{12 (x-1)^4}-\frac{23 \
\ao}{12}+\Big(\frac{25}{12}+\frac{3}{4 (x-1)}-\frac{1}{6 \
(x-1)^2}-\frac{1}{6 (x-1)^3}+\frac{3}{4 (x-1)^4}+\frac{25}{12 \
(x-1)^5}\Big) H(0;\ao)+\Big(-\frac{25}{12}-\frac{3}{4 \
(x-1)}+\frac{1}{6 (x-1)^2}+\frac{1}{6 (x-1)^3}-\frac{3}{4 \
(x-1)^4}-\frac{25}{12 (x-1)^5}\Big) \
H(0;x)+\Big(\frac{1}{(x-1)^5}-1\Big) H(0;\ao) \
H(1;x)+\Big(-\frac{\ao^4}{4 (x-1)}+\frac{\ao^4}{4}+\frac{\ao^3}{x-1}-\
\frac{\ao^3}{3 (x-1)^2}-\frac{4 \ao^3}{3}-\frac{3 \ao^2}{2 \
(x-1)}+\frac{\ao^2}{(x-1)^2}-\frac{\ao^2}{2 (x-1)^3}+3 \
\ao^2+\frac{\ao}{x-1}-\frac{\ao}{(x-1)^2}+\frac{\ao}{(x-1)^3}-\frac{\ao}{(x-1)^4}-4 \ao+\frac{3}{4 (x-1)}-\frac{1}{6 (x-1)^2}-\frac{1}{6 \
(x-1)^3}+\frac{3}{4 (x-1)^4}+\frac{25}{12 (x-1)^5}+\frac{25}{12}\Big) \
H(c_1(\ao);x)+2 H(0,0;x)+\Big(\frac{1}{(x-1)^5}-1\Big) \
H(0,c_1(\ao);x)+\Big(1-\frac{1}{(x-1)^5}\Big) \
H(1,0;x)+\Big(\frac{1}{(x-1)^5}-1\Big) \
H(1,c_1(\ao);x)-\frac{H(c_1(\ao),c_1(\ao);x)}{(x-1)^5}-\frac{\pi \
^2}{6 (x-1)^5}+\frac{\pi ^2}{8},
\erp
% ep^1
\brp
b_1^{(-1,-1)}=\frac{7 d_1 \ao^3}{72}-\frac{7 d_1 \ao^3}{72 (x-1)^2}+\frac{7 \
\ao^3}{36 (x-1)^2}-\frac{7 \ao^3}{36}-\frac{109 d_1 \
\ao^2}{144}-\frac{13 d_1 \ao^2}{144 (x-1)}+\frac{7 \ao^2}{72 \
(x-1)}+\frac{29 d_1 \ao^2}{144 (x-1)^2}-\frac{35 \ao^2}{72 \
(x-1)^2}-\frac{67 d_1 \ao^2}{144 (x-1)^3}+\frac{85 \ao^2}{72 \
(x-1)^3}+\frac{127 \ao^2}{72}+\frac{305 d_1 \ao}{72}-\frac{2 \ao}{3 \
(x-2)}+\frac{19 d_1 \ao}{18 (x-1)}-\frac{10 \ao}{9 (x-1)}-\frac{d_1 \
\ao}{9 (x-1)^2}+\frac{13 \ao}{18 (x-1)^2}+\frac{d_1 \ao}{18 (x-1)^3}-\
\frac{7 \ao}{9 (x-1)^3}-\frac{217 d_1 \ao}{72 (x-1)^4}+\frac{149 \
\ao}{18 (x-1)^4}-\frac{101 \ao}{9}+\Big(-\frac{\ao^3}{3 \
(x-1)^2}+\frac{\ao^3}{3}-\frac{\ao^2}{6 (x-1)}+\frac{5 \ao^2}{6 \
(x-1)^2}-\frac{7 \ao^2}{6 (x-1)^3}-\frac{13 \ao^2}{6}+\frac{4 \ao}{3 \
(x-1)}-\frac{2 \ao}{3 (x-1)^2}+\frac{4 \ao}{3 (x-1)^3}-\frac{13 \
\ao}{3 (x-1)^4}+\frac{23 \ao}{3}-\frac{205 \
d_1}{72}+\frac{4}{x-2}-\frac{15 d_1}{8 (x-1)}-\frac{1}{2 \
(x-1)}-\frac{8}{3 (x-2)^2}+\frac{5 d_1}{18 (x-1)^2}-\frac{13}{18 \
(x-1)^2}+\frac{5 d_1}{18 (x-1)^3}+\frac{5}{18 (x-1)^3}-\frac{15 \
d_1}{8 (x-1)^4}+\frac{9}{(x-1)^4}-\frac{205 d_1}{72 \
(x-1)^5}+\frac{130}{9 (x-1)^5}+\frac{155}{18}\Big) \
H(0;\ao)+\Big(\frac{15 d_1}{8 (x-1)}-\frac{5 d_1}{18 (x-1)^2}-\frac{5 \
d_1}{18 (x-1)^3}+\frac{15 d_1}{8 (x-1)^4}+\frac{205 d_1}{72 (x-1)^5}+\
\frac{205 d_1}{72}-\frac{4}{x-2}+\frac{1}{2 (x-1)}+\frac{8}{3 \
(x-2)^2}+\frac{13}{18 (x-1)^2}-\frac{5}{18 \
(x-1)^3}-\frac{9}{(x-1)^4}+\frac{2 \pi ^2}{3 (x-1)^5}-\frac{130}{9 \
(x-1)^5}-\frac{\pi ^2}{2}-\frac{155}{18}\Big) H(0;x)+\Big(\frac{d_1 \
\ao^3}{6}-\frac{d_1 \ao^3}{6 (x-1)^2}-\frac{13 d_1 \
\ao^2}{12}-\frac{d_1 \ao^2}{12 (x-1)}+\frac{5 d_1 \ao^2}{12 (x-1)^2}-\
\frac{7 d_1 \ao^2}{12 (x-1)^3}+\frac{23 d_1 \ao}{6}+\frac{2 d_1 \
\ao}{3 (x-1)}-\frac{d_1 \ao}{3 (x-1)^2}+\frac{2 d_1 \ao}{3 \
(x-1)^3}-\frac{13 d_1 \ao}{6 (x-1)^4}-\frac{35 d_1}{12}-\frac{7 \
d_1}{12 (x-1)}+\frac{d_1}{12 (x-1)^2}-\frac{d_1}{12 (x-1)^3}+\frac{13 \
d_1}{6 (x-1)^4}\Big) H(1;\ao)+\Big(\frac{\pi ^2}{2}-\frac{\pi ^2}{2 \
(x-1)^5}\Big) H(2;x)+\Big(-\frac{d_1 \ao^4}{8}+\frac{d_1 \ao^4}{8 \
(x-1)}-\frac{\ao^4}{4 (x-1)}+\frac{\ao^4}{4}+\frac{13 d_1 \ao^3}{18}-\
\frac{d_1 \ao^3}{2 (x-1)}+\frac{4 \ao^3}{3 (x-1)}+\frac{2 d_1 \
\ao^3}{9 (x-1)^2}-\frac{17 \ao^3}{18 (x-1)^2}-\frac{29 \
\ao^3}{18}-\frac{23 d_1 \ao^2}{12}+\frac{\ao^2}{3 (x-2)}+\frac{3 d_1 \
\ao^2}{4 (x-1)}-\frac{41 \ao^2}{12 (x-1)}-\frac{2 d_1 \ao^2}{3 \
(x-1)^2}+\frac{13 \ao^2}{4 (x-1)^2}+\frac{d_1 \ao^2}{2 \
(x-1)^3}-\frac{11 \ao^2}{4 (x-1)^3}+\frac{59 \ao^2}{12}+\frac{25 d_1 \
\ao}{6}-\frac{2 \ao}{x-2}-\frac{d_1 \ao}{2 (x-1)}+\frac{20 \ao}{3 \
(x-1)}+\frac{4 \ao}{3 (x-2)^2}+\frac{2 d_1 \ao}{3 (x-1)^2}-\frac{17 \
\ao}{3 (x-1)^2}-\frac{d_1 \ao}{(x-1)^3}+\frac{6 \ao}{(x-1)^3}+\frac{2 \
d_1 \ao}{(x-1)^4}-\frac{21 \ao}{2 (x-1)^4}-\frac{73 \ao}{6}-\frac{205 \
d_1}{72}+\Big(\frac{\ao^4}{x-1}-\ao^4-\frac{4 \ao^3}{x-1}+\frac{4 \
\ao^3}{3 (x-1)^2}+\frac{16 \ao^3}{3}+\frac{6 \ao^2}{x-1}-\frac{4 \
\ao^2}{(x-1)^2}+\frac{2 \ao^2}{(x-1)^3}-12 \ao^2-\frac{4 \
\ao}{x-1}+\frac{4 \ao}{(x-1)^2}-\frac{4 \ao}{(x-1)^3}+\frac{4 \
\ao}{(x-1)^4}+16 \ao-\frac{3}{x-1}+\frac{2}{3 (x-1)^2}+\frac{2}{3 \
(x-1)^3}-\frac{3}{(x-1)^4}-\frac{25}{3 (x-1)^5}-\frac{25}{3}\Big) \
H(0;\ao)+\Big(-\frac{d_1 \ao^4}{2}+\frac{d_1 \ao^4}{2 (x-1)}+\frac{8 \
d_1 \ao^3}{3}-\frac{2 d_1 \ao^3}{x-1}+\frac{2 d_1 \ao^3}{3 (x-1)^2}-6 \
d_1 \ao^2+\frac{3 d_1 \ao^2}{x-1}-\frac{2 d_1 \
\ao^2}{(x-1)^2}+\frac{d_1 \ao^2}{(x-1)^3}+8 d_1 \ao-\frac{2 d_1 \
\ao}{x-1}+\frac{2 d_1 \ao}{(x-1)^2}-\frac{2 d_1 \ao}{(x-1)^3}+\frac{2 \
d_1 \ao}{(x-1)^4}-\frac{25 d_1}{6}-\frac{3 d_1}{2 (x-1)}+\frac{d_1}{3 \
(x-1)^2}+\frac{d_1}{3 (x-1)^3}-\frac{3 d_1}{2 (x-1)^4}-\frac{25 \
d_1}{6 (x-1)^5}\Big) H(1;\ao)+\frac{4}{x-2}-\frac{15 d_1}{8 \
(x-1)}-\frac{1}{2 (x-1)}-\frac{8}{3 (x-2)^2}+\frac{5 d_1}{18 \
(x-1)^2}-\frac{13}{18 (x-1)^2}+\frac{5 d_1}{18 (x-1)^3}+\frac{5}{18 \
(x-1)^3}-\frac{15 d_1}{8 (x-1)^4}+\frac{9}{(x-1)^4}-\frac{205 d_1}{72 \
(x-1)^5}+\frac{130}{9 (x-1)^5}+\frac{155}{18}\Big) \
H(c_1(\ao);x)+\Big(-\frac{25}{3}-\frac{3}{x-1}+\frac{2}{3 \
(x-1)^2}+\frac{2}{3 (x-1)^3}-\frac{3}{(x-1)^4}-\frac{25}{3 \
(x-1)^5}\Big) H(0,0;\ao)+\Big(\frac{25}{3}+\frac{3}{x-1}-\frac{2}{3 \
(x-1)^2}-\frac{2}{3 (x-1)^3}+\frac{3}{(x-1)^4}+\frac{25}{3 \
(x-1)^5}\Big) H(0,0;x)+\Big(-\frac{3 d_1}{2 (x-1)}+\frac{d_1}{3 \
(x-1)^2}+\frac{d_1}{3 (x-1)^3}-\frac{3 d_1}{2 (x-1)^4}-\frac{25 \
d_1}{6 (x-1)^5}-\frac{25 d_1}{6}\Big) H(0,1;\ao)+H(1;x) \
\Big(-\frac{\pi ^2 d_1}{3 (x-1)^5}+\Big(\frac{2 \
d_1}{x-1}-\frac{d_1}{(x-1)^2}+\frac{2 d_1}{3 (x-1)^3}-\frac{d_1}{2 \
(x-1)^4}+\frac{25 d_1}{6 (x-1)^5}-\frac{4}{x-2}+\frac{5}{2 \
(x-1)}+\frac{8}{3 (x-2)^2}-\frac{1}{3 (x-1)^2}-\frac{8}{3 \
(x-2)^3}+\frac{5}{3 (x-1)^3}+\frac{3}{2 (x-1)^4}+\frac{25}{6 \
(x-1)^5}-\frac{25}{6}\Big) H(0;\ao)+\Big(4-\frac{4}{(x-1)^5}\Big) \
H(0,0;\ao)+\Big(2 d_1-\frac{2 d_1}{(x-1)^5}\Big) H(0,1;\ao)+\frac{\pi \
^2}{3 (x-1)^5}+\frac{\pi ^2}{3}\Big)+\Big(\frac{2 d_1}{(x-1)^5}-2 \
d_1-\frac{2}{(x-1)^5}+2\Big) H(0;\ao) H(0,1;x)+\Big(-\frac{\ao^4}{2 \
(x-1)}+\frac{\ao^4}{2}+\frac{2 \ao^3}{x-1}-\frac{2 \ao^3}{3 (x-1)^2}-\
\frac{8 \ao^3}{3}-\frac{3 \ao^2}{x-1}+\frac{2 \
\ao^2}{(x-1)^2}-\frac{\ao^2}{(x-1)^3}+6 \ao^2+\frac{2 \
\ao}{x-1}-\frac{2 \ao}{(x-1)^2}+\frac{2 \ao}{(x-1)^3}-\frac{2 \
\ao}{(x-1)^4}-8 \ao+\Big(4-\frac{4}{(x-1)^5}\Big) H(0;\ao)+\Big(2 \
d_1-\frac{2 d_1}{(x-1)^5}\Big) \
H(1;\ao)-\frac{4}{x-2}+\frac{2}{x-1}+\frac{8}{3 (x-2)^2}+\frac{1}{3 \
(x-1)^2}-\frac{8}{3 (x-2)^3}+\frac{2}{3 (x-1)^3}+\frac{7}{2 (x-1)^4}+\
\frac{25}{6 (x-1)^5}\Big) H(0,c_1(\ao);x)+\Big(-\frac{2 \
d_1}{x-1}+\frac{d_1}{(x-1)^2}-\frac{2 d_1}{3 (x-1)^3}+\frac{d_1}{2 \
(x-1)^4}-\frac{25 d_1}{6 (x-1)^5}+\frac{4}{x-2}-\frac{5}{2 \
(x-1)}-\frac{8}{3 (x-2)^2}+\frac{1}{3 (x-1)^2}+\frac{8}{3 \
(x-2)^3}-\frac{5}{3 (x-1)^3}-\frac{3}{2 (x-1)^4}-\frac{25}{6 \
(x-1)^5}+\frac{25}{6}\Big) H(1,0;x)+\Big(\frac{4 d_1}{(x-1)^5}-2 d_1-\
\frac{2}{(x-1)^5}-2\Big) H(0;\ao) H(1,1;x)+\Big(\frac{2 \
d_1}{x-1}-\frac{d_1}{(x-1)^2}+\frac{2 d_1}{3 (x-1)^3}-\frac{d_1}{2 \
(x-1)^4}+\frac{25 d_1}{6 (x-1)^5}+\Big(4-\frac{4}{(x-1)^5}\Big) \
H(0;\ao)+\Big(2 d_1-\frac{2 d_1}{(x-1)^5}\Big) \
H(1;\ao)-\frac{4}{x-2}+\frac{5}{2 (x-1)}+\frac{8}{3 \
(x-2)^2}-\frac{1}{3 (x-1)^2}-\frac{8}{3 (x-2)^3}+\frac{5}{3 (x-1)^3}+\
\frac{3}{2 (x-1)^4}+\frac{25}{6 (x-1)^5}-\frac{25}{6}\Big) \
H(1,c_1(\ao);x)+\Big(\frac{2}{(x-1)^5}-2\Big) H(0;\ao) H(2,1;x)+\Big(\
\frac{3 \ao^4}{2 (x-1)}-\frac{3 \ao^4}{2}-\frac{6 \ao^3}{x-1}+\frac{2 \
\ao^3}{(x-1)^2}+8 \ao^3+\frac{9 \ao^2}{x-1}-\frac{6 \
\ao^2}{(x-1)^2}+\frac{3 \ao^2}{(x-1)^3}-18 \ao^2-\frac{6 \
\ao}{x-1}+\frac{6 \ao}{(x-1)^2}-\frac{6 \ao}{(x-1)^3}+\frac{6 \
\ao}{(x-1)^4}+24 \ao+\frac{4 H(0;\ao)}{(x-1)^5}+\frac{2 d_1 \
H(1;\ao)}{(x-1)^5}-\frac{9}{2 \
(x-1)}+\frac{1}{(x-1)^2}+\frac{1}{(x-1)^3}-\frac{9}{2 \
(x-1)^4}-\frac{25}{2 (x-1)^5}-\frac{25}{2}\Big) \
H(c_1(\ao),c_1(\ao);x)+\Big(-\frac{\ao^4}{2 \
(x-1)}+\frac{\ao^4}{2}+\frac{2 \ao^3}{x-1}-\frac{2 \ao^3}{3 (x-1)^2}-\
\frac{8 \ao^3}{3}-\frac{3 \ao^2}{x-1}+\frac{2 \
\ao^2}{(x-1)^2}-\frac{\ao^2}{(x-1)^3}+6 \ao^2+\frac{2 \
\ao}{x-1}-\frac{2 \ao}{(x-1)^2}+\frac{2 \ao}{(x-1)^3}-\frac{2 \
\ao}{(x-1)^4}-8 \ao+\frac{4}{x-2}-\frac{1}{2 (x-1)}-\frac{8}{3 \
(x-2)^2}-\frac{2}{3 (x-1)^2}+\frac{8}{3 \
(x-2)^3}-\frac{1}{(x-1)^3}-\frac{2}{(x-1)^4}+\frac{25}{6}\Big) H(c_2(\
\ao),c_1(\ao);x)-8 H(0,0,0;x)+\Big(2-\frac{2}{(x-1)^5}\Big) \
H(0,0,c_1(\ao);x)+\Big(-\frac{2 d_1}{(x-1)^5}+2 \
d_1+\frac{2}{(x-1)^5}-2\Big) H(0,1,0;x)+\Big(\frac{2 d_1}{(x-1)^5}-2 \
d_1-\frac{2}{(x-1)^5}+2\Big) \
H(0,1,c_1(\ao);x)+\Big(6-\frac{2}{(x-1)^5}\Big) \
H(0,c_1(\ao),c_1(\ao);x)+\Big(\frac{2}{(x-1)^5}-2\Big) \
H(0,c_2(\ao),c_1(\ao);x)+\Big(\frac{4}{(x-1)^5}-4\Big) \
H(1,0,0;x)+\Big(\frac{2 d_1}{(x-1)^5}-\frac{2}{(x-1)^5}-2\Big) \
H(1,0,c_1(\ao);x)+\Big(-\frac{4 d_1}{(x-1)^5}+2 \
d_1+\frac{2}{(x-1)^5}+2\Big) H(1,1,0;x)+\Big(\frac{4 d_1}{(x-1)^5}-2 \
d_1-\frac{2}{(x-1)^5}-2\Big) H(1,1,c_1(\ao);x)+\Big(-\frac{2 \
d_1}{(x-1)^5}-\frac{2}{(x-1)^5}+6\Big) H(1,c_1(\ao),c_1(\ao);x)+\Big(\
\frac{2}{(x-1)^5}-2\Big) \
H(2,0,c_1(\ao);x)+\Big(2-\frac{2}{(x-1)^5}\Big) \
H(2,1,0;x)+\Big(\frac{2}{(x-1)^5}-2\Big) \
H(2,1,c_1(\ao);x)+\Big(2-\frac{2}{(x-1)^5}\Big) \
H(2,c_2(\ao),c_1(\ao);x)-\frac{2 \
H(c_1(\ao),0,c_1(\ao);x)}{(x-1)^5}+\frac{6 \
H(c_1(\ao),c_1(\ao),c_1(\ao);x)}{(x-1)^5}-\frac{2 \
H(c_1(\ao),c_2(\ao),c_1(\ao);x)}{(x-1)^5}+\frac{\pi ^2}{x-2}-\frac{3 \
\pi ^2}{8 (x-1)}-\frac{2 \pi ^2}{3 (x-2)^2}-\frac{\pi ^2}{9 (x-1)^2}+\
\frac{2 \pi ^2}{3 (x-2)^3}-\frac{7 \pi ^2}{36 (x-1)^3}-\frac{3 \pi \
^2}{4 (x-1)^4}-\frac{25 \pi ^2}{36 (x-1)^5}-\frac{3 \zeta_3}{4 \
(x-1)^5}-4 \zeta_3-\frac{\pi ^2 \ln 2\, }{2 (x-1)^5}+\frac{1}{2} \pi \
^2 \ln 2\, +\frac{25 \pi ^2}{72},
\erp
% ep^2
\brp
b_2^{(-1,-1)}=-\frac{37}{432} d_1^2 \ao^3+\frac{37 d_1 \ao^3}{108}+\frac{37 d_1^2 \
\ao^3}{432 (x-1)^2}-\frac{37 d_1 \ao^3}{108 (x-1)^2}-\frac{\pi ^2 \
\ao^3}{72 (x-1)^2}+\frac{37 \ao^3}{108 (x-1)^2}+\frac{\pi ^2 \
\ao^3}{72}-\frac{37 \ao^3}{108}+\frac{715 d_1^2 \ao^2}{864}-\frac{104 \
d_1 \ao^2}{27}+\frac{115 d_1^2 \ao^2}{864 (x-1)}-\frac{19 d_1 \
\ao^2}{54 (x-1)}-\frac{\pi ^2 \ao^2}{144 (x-1)}+\frac{37 \ao^2}{216 \
(x-1)}-\frac{107 d_1^2 \ao^2}{864 (x-1)^2}+\frac{73 d_1 \ao^2}{108 \
(x-1)^2}+\frac{5 \pi ^2 \ao^2}{144 (x-1)^2}-\frac{185 \ao^2}{216 \
(x-1)^2}+\frac{493 d_1^2 \ao^2}{864 (x-1)^3}-\frac{305 d_1 \ao^2}{108 \
(x-1)^3}-\frac{7 \pi ^2 \ao^2}{144 (x-1)^3}+\frac{727 \ao^2}{216 \
(x-1)^3}-\frac{13 \pi ^2 \ao^2}{144}+\frac{949 \ao^2}{216}-\frac{3515 \
d_1^2 \ao}{432}+\frac{8965 d_1 \ao}{216}+\frac{25 d_1 \ao}{9 \
(x-2)}-\frac{50 \ao}{9 (x-2)}-\frac{265 d_1^2 \ao}{108 \
(x-1)}+\frac{341 d_1 \ao}{54 (x-1)}+\frac{\pi ^2 \ao}{18 \
(x-1)}-\frac{107 \ao}{54 (x-1)}-\frac{d_1^2 \ao}{108 \
(x-1)^2}-\frac{185 d_1 \ao}{108 (x-1)^2}-\frac{\pi ^2 \ao}{36 \
(x-1)^2}+\frac{187 \ao}{54 (x-1)^2}+\frac{113 d_1^2 \ao}{108 \
(x-1)^3}-\frac{74 d_1 \ao}{27 (x-1)^3}+\frac{\pi ^2 \ao}{18 (x-1)^3}+\
\frac{25 \ao}{54 (x-1)^3}+\frac{2911 d_1^2 \ao}{432 \
(x-1)^4}-\frac{7523 d_1 \ao}{216 (x-1)^4}-\frac{13 \pi ^2 \ao}{72 \
(x-1)^4}+\frac{2369 \ao}{54 (x-1)^4}+\frac{23 \pi ^2 \
\ao}{72}-\frac{1394 \ao}{27}+\Big(-\frac{7 d_1 \ao^3}{18}+\frac{7 d_1 \
\ao^3}{18 (x-1)^2}-\frac{7 \ao^3}{9 (x-1)^2}+\frac{7 \
\ao^3}{9}+\frac{109 d_1 \ao^2}{36}+\frac{13 d_1 \ao^2}{36 \
(x-1)}-\frac{7 \ao^2}{18 (x-1)}-\frac{29 d_1 \ao^2}{36 \
(x-1)^2}+\frac{35 \ao^2}{18 (x-1)^2}+\frac{67 d_1 \ao^2}{36 (x-1)^3}-\
\frac{85 \ao^2}{18 (x-1)^3}-\frac{127 \ao^2}{18}-\frac{305 d_1 \
\ao}{18}+\frac{8 \ao}{3 (x-2)}-\frac{38 d_1 \ao}{9 (x-1)}+\frac{40 \
\ao}{9 (x-1)}+\frac{4 d_1 \ao}{9 (x-1)^2}-\frac{26 \ao}{9 \
(x-1)^2}-\frac{2 d_1 \ao}{9 (x-1)^3}+\frac{28 \ao}{9 \
(x-1)^3}+\frac{217 d_1 \ao}{18 (x-1)^4}-\frac{298 \ao}{9 \
(x-1)^4}+\frac{404 \ao}{9}+\frac{2035 d_1^2}{432}-\frac{5615 \
d_1}{216}-\frac{38 d_1}{3 (x-2)}+\frac{68}{3 (x-2)}+\frac{63 \
d_1^2}{16 (x-1)}-\frac{85 d_1}{24 (x-1)}-\frac{\pi ^2}{8 \
(x-1)}-\frac{37}{4 (x-1)}+\frac{76 d_1}{9 (x-2)^2}-\frac{152}{9 \
(x-2)^2}-\frac{19 d_1^2}{54 (x-1)^2}+\frac{317 d_1}{108 \
(x-1)^2}+\frac{\pi ^2}{36 (x-1)^2}+\frac{49}{108 (x-1)^2}-\frac{19 \
d_1^2}{54 (x-1)^3}-\frac{313 d_1}{108 (x-1)^3}+\frac{\pi ^2}{36 \
(x-1)^3}+\frac{949}{108 (x-1)^3}+\frac{63 d_1^2}{16 \
(x-1)^4}-\frac{257 d_1}{8 (x-1)^4}-\frac{\pi ^2}{8 \
(x-1)^4}+\frac{213}{4 (x-1)^4}+\frac{2035 d_1^2}{432 \
(x-1)^5}-\frac{8705 d_1}{216 (x-1)^5}-\frac{25 \pi ^2}{72 \
(x-1)^5}+\frac{3965}{54 (x-1)^5}-\frac{25 \pi \
^2}{72}+\frac{940}{27}\Big) H(0;\ao)+\Big(-\frac{7}{36} d_1^2 \
\ao^3+\frac{7 d_1 \ao^3}{18}+\frac{7 d_1^2 \ao^3}{36 (x-1)^2}-\frac{7 \
d_1 \ao^3}{18 (x-1)^2}+\frac{109 d_1^2 \ao^2}{72}-\frac{127 d_1 \
\ao^2}{36}+\frac{13 d_1^2 \ao^2}{72 (x-1)}-\frac{7 d_1 \ao^2}{36 \
(x-1)}-\frac{29 d_1^2 \ao^2}{72 (x-1)^2}+\frac{35 d_1 \ao^2}{36 \
(x-1)^2}+\frac{67 d_1^2 \ao^2}{72 (x-1)^3}-\frac{85 d_1 \ao^2}{36 \
(x-1)^3}-\frac{305 d_1^2 \ao}{36}+\frac{202 d_1 \ao}{9}+\frac{4 d_1 \
\ao}{3 (x-2)}-\frac{19 d_1^2 \ao}{9 (x-1)}+\frac{20 d_1 \ao}{9 \
(x-1)}+\frac{2 d_1^2 \ao}{9 (x-1)^2}-\frac{13 d_1 \ao}{9 \
(x-1)^2}-\frac{d_1^2 \ao}{9 (x-1)^3}+\frac{14 d_1 \ao}{9 \
(x-1)^3}+\frac{217 d_1^2 \ao}{36 (x-1)^4}-\frac{149 d_1 \ao}{9 \
(x-1)^4}+\frac{515 d_1^2}{72}-\frac{695 d_1}{36}-\frac{4 d_1}{3 \
(x-2)}+\frac{139 d_1^2}{72 (x-1)}-\frac{73 d_1}{36 \
(x-1)}-\frac{d_1^2}{72 (x-1)^2}+\frac{31 d_1}{36 (x-1)^2}-\frac{59 \
d_1^2}{72 (x-1)^3}+\frac{29 d_1}{36 (x-1)^3}-\frac{217 d_1^2}{36 \
(x-1)^4}+\frac{149 d_1}{9 (x-1)^4}\Big) H(1;\ao)+\Big(\frac{4 \
\ao^3}{3 (x-1)^2}-\frac{4 \ao^3}{3}+\frac{2 \ao^2}{3 (x-1)}-\frac{10 \
\ao^2}{3 (x-1)^2}+\frac{14 \ao^2}{3 (x-1)^3}+\frac{26 \
\ao^2}{3}-\frac{16 \ao}{3 (x-1)}+\frac{8 \ao}{3 (x-1)^2}-\frac{16 \
\ao}{3 (x-1)^3}+\frac{52 \ao}{3 (x-1)^4}-\frac{92 \ao}{3}+\frac{205 \
d_1}{18}-\frac{16}{x-2}+\frac{15 d_1}{2 \
(x-1)}+\frac{2}{x-1}+\frac{32}{3 (x-2)^2}-\frac{10 d_1}{9 \
(x-1)^2}+\frac{26}{9 (x-1)^2}-\frac{10 d_1}{9 (x-1)^3}-\frac{10}{9 \
(x-1)^3}+\frac{15 d_1}{2 (x-1)^4}-\frac{36}{(x-1)^4}+\frac{205 \
d_1}{18 (x-1)^5}-\frac{520}{9 (x-1)^5}-\frac{310}{9}\Big) H(0,0;\ao)+\
\Big(-\frac{15 d_1}{2 (x-1)}+\frac{10 d_1}{9 (x-1)^2}+\frac{10 d_1}{9 \
(x-1)^3}-\frac{15 d_1}{2 (x-1)^4}-\frac{205 d_1}{18 \
(x-1)^5}-\frac{205 d_1}{18}+\frac{16}{x-2}-\frac{2}{x-1}-\frac{32}{3 \
(x-2)^2}-\frac{26}{9 (x-1)^2}+\frac{10}{9 \
(x-1)^3}+\frac{36}{(x-1)^4}-\frac{8 \pi ^2}{3 (x-1)^5}+\frac{520}{9 \
(x-1)^5}+2 \pi ^2+\frac{310}{9}\Big) H(0,0;x)+\Big(-\frac{2 d_1 \
\ao^3}{3}+\frac{2 d_1 \ao^3}{3 (x-1)^2}+\frac{13 d_1 \
\ao^2}{3}+\frac{d_1 \ao^2}{3 (x-1)}-\frac{5 d_1 \ao^2}{3 \
(x-1)^2}+\frac{7 d_1 \ao^2}{3 (x-1)^3}-\frac{46 d_1 \ao}{3}-\frac{8 \
d_1 \ao}{3 (x-1)}+\frac{4 d_1 \ao}{3 (x-1)^2}-\frac{8 d_1 \ao}{3 \
(x-1)^3}+\frac{26 d_1 \ao}{3 (x-1)^4}+\frac{205 d_1^2}{36}-\frac{155 \
d_1}{9}-\frac{8 d_1}{x-2}+\frac{15 d_1^2}{4 \
(x-1)}+\frac{d_1}{x-1}+\frac{16 d_1}{3 (x-2)^2}-\frac{5 d_1^2}{9 \
(x-1)^2}+\frac{13 d_1}{9 (x-1)^2}-\frac{5 d_1^2}{9 (x-1)^3}-\frac{5 \
d_1}{9 (x-1)^3}+\frac{15 d_1^2}{4 (x-1)^4}-\frac{18 \
d_1}{(x-1)^4}+\frac{205 d_1^2}{36 (x-1)^5}-\frac{260 d_1}{9 \
(x-1)^5}\Big) H(0,1;\ao)+\Big(\frac{2 \pi ^2 d_1}{3 (x-1)^5}+\frac{2 \
\pi ^2 d_1}{3}+\Big(-\frac{8 d_1}{x-2}+\frac{5 d_1}{x-1}+\frac{16 \
d_1}{3 (x-2)^2}-\frac{2 d_1}{3 (x-1)^2}-\frac{16 d_1}{3 \
(x-2)^3}+\frac{10 d_1}{3 (x-1)^3}+\frac{3 d_1}{(x-1)^4}+\frac{25 \
d_1}{3 (x-1)^5}-\frac{25 d_1}{3}+\frac{8}{x-2}-\frac{16}{3 \
(x-2)^2}-\frac{8}{3 (x-1)^2}+\frac{16}{3 \
(x-2)^3}-\frac{8}{(x-1)^4}\Big) H(0;\ao)+\Big(-\frac{8 \
d_1}{(x-1)^5}+8 d_1+\frac{8}{(x-1)^5}-8\Big) H(0,0;\ao)+\Big(-\frac{4 \
d_1^2}{(x-1)^5}+4 d_1^2+\frac{4 d_1}{(x-1)^5}-4 d_1\Big) \
H(0,1;\ao)-\frac{2 \pi ^2}{(x-1)^5}-\frac{2 \pi ^2}{3}\Big) H(0,1;x)+\
\Big(-\frac{\pi ^2 d_1}{(x-1)^5}+\pi ^2 d_1+\frac{\pi \
^2}{(x-1)^5}-\pi ^2\Big) H(0,2;x)+\Big(-\frac{2 d_1 \ao^3}{3}+\frac{2 \
d_1 \ao^3}{3 (x-1)^2}+\frac{13 d_1 \ao^2}{3}+\frac{d_1 \ao^2}{3 \
(x-1)}-\frac{5 d_1 \ao^2}{3 (x-1)^2}+\frac{7 d_1 \ao^2}{3 \
(x-1)^3}-\frac{46 d_1 \ao}{3}-\frac{8 d_1 \ao}{3 (x-1)}+\frac{4 d_1 \
\ao}{3 (x-1)^2}-\frac{8 d_1 \ao}{3 (x-1)^3}+\frac{26 d_1 \ao}{3 \
(x-1)^4}+\frac{35 d_1}{3}+\frac{7 d_1}{3 (x-1)}-\frac{d_1}{3 \
(x-1)^2}+\frac{d_1}{3 (x-1)^3}-\frac{26 d_1}{3 (x-1)^4}\Big) \
H(1,0;\ao)+\Big(\frac{4 d_1^2}{x-1}-\frac{d_1^2}{(x-1)^2}+\frac{4 \
d_1^2}{9 (x-1)^3}-\frac{d_1^2}{4 (x-1)^4}+\frac{205 d_1^2}{36 \
(x-1)^5}-\frac{46 d_1}{3 (x-2)}-\frac{3 d_1}{4 (x-1)}+\frac{52 d_1}{9 \
(x-2)^2}+\frac{185 d_1}{18 (x-1)^2}-\frac{28 d_1}{9 (x-2)^3}-\frac{29 \
d_1}{18 (x-1)^3}+\frac{27 d_1}{4 (x-1)^4}+\frac{4 \pi ^2 d_1}{3 \
(x-1)^5}-\frac{835 d_1}{36 (x-1)^5}-\frac{205 d_1}{36}+\frac{80}{3 \
(x-2)}-\frac{70}{3 (x-1)}-\frac{56}{9 (x-2)^2}-\frac{13}{18 (x-1)^2}+\
\frac{56}{9 (x-2)^3}-\frac{355}{18 (x-1)^3}-\frac{10}{3 \
(x-1)^4}-\frac{7 \pi ^2}{6 (x-1)^5}-\frac{155}{9 (x-1)^5}-\frac{3 \pi \
^2}{2}+\frac{155}{9}\Big) H(1,0;x)+\Big(-\frac{1}{3} d_1^2 \
\ao^3+\frac{d_1^2 \ao^3}{3 (x-1)^2}+\frac{13 d_1^2 \
\ao^2}{6}+\frac{d_1^2 \ao^2}{6 (x-1)}-\frac{5 d_1^2 \ao^2}{6 \
(x-1)^2}+\frac{7 d_1^2 \ao^2}{6 (x-1)^3}-\frac{23 d_1^2 \
\ao}{3}-\frac{4 d_1^2 \ao}{3 (x-1)}+\frac{2 d_1^2 \ao}{3 \
(x-1)^2}-\frac{4 d_1^2 \ao}{3 (x-1)^3}+\frac{13 d_1^2 \ao}{3 \
(x-1)^4}+\frac{35 d_1^2}{6}+\frac{7 d_1^2}{6 (x-1)}-\frac{d_1^2}{6 \
(x-1)^2}+\frac{d_1^2}{6 (x-1)^3}-\frac{13 d_1^2}{3 (x-1)^4}\Big) \
H(1,1;\ao)+H(0,c_1(\ao);x) \Big(-\frac{d_1 \ao^4}{4}+\frac{d_1 \
\ao^4}{4 (x-1)}-\frac{\ao^4}{2 (x-1)}+\frac{\ao^4}{2}+\frac{13 d_1 \
\ao^3}{9}-\frac{d_1 \ao^3}{x-1}+\frac{8 \ao^3}{3 (x-1)}+\frac{4 d_1 \
\ao^3}{9 (x-1)^2}-\frac{14 \ao^3}{9 (x-1)^2}-\frac{32 \
\ao^3}{9}-\frac{23 d_1 \ao^2}{6}+\frac{2 \ao^2}{3 (x-2)}+\frac{3 d_1 \
\ao^2}{2 (x-1)}-\frac{20 \ao^2}{3 (x-1)}-\frac{4 d_1 \ao^2}{3 \
(x-1)^2}+\frac{17 \ao^2}{3 (x-1)^2}+\frac{d_1 \
\ao^2}{(x-1)^3}-\frac{13 \ao^2}{3 (x-1)^3}+12 \ao^2+\frac{25 d_1 \
\ao}{3}-\frac{4 \ao}{x-2}-\frac{d_1 \ao}{x-1}+\frac{12 \
\ao}{x-1}+\frac{8 \ao}{3 (x-2)^2}+\frac{4 d_1 \ao}{3 \
(x-1)^2}-\frac{32 \ao}{3 (x-1)^2}-\frac{2 d_1 \ao}{(x-1)^3}+\frac{32 \
\ao}{3 (x-1)^3}+\frac{4 d_1 \ao}{(x-1)^4}-\frac{50 \ao}{3 (x-1)^4}-32 \
\ao+\Big(\frac{2 \ao^4}{x-1}-2 \ao^4-\frac{8 \ao^3}{x-1}+\frac{8 \
\ao^3}{3 (x-1)^2}+\frac{32 \ao^3}{3}+\frac{12 \ao^2}{x-1}-\frac{8 \
\ao^2}{(x-1)^2}+\frac{4 \ao^2}{(x-1)^3}-24 \ao^2-\frac{8 \
\ao}{x-1}+\frac{8 \ao}{(x-1)^2}-\frac{8 \ao}{(x-1)^3}+\frac{8 \
\ao}{(x-1)^4}+32 \ao+\frac{16}{x-2}-\frac{8}{x-1}-\frac{32}{3 \
(x-2)^2}-\frac{4}{3 (x-1)^2}+\frac{32}{3 (x-2)^3}-\frac{8}{3 \
(x-1)^3}-\frac{14}{(x-1)^4}-\frac{50}{3 (x-1)^5}\Big) \
H(0;\ao)+\Big(-d_1 \ao^4+\frac{d_1 \ao^4}{x-1}+\frac{16 d_1 \
\ao^3}{3}-\frac{4 d_1 \ao^3}{x-1}+\frac{4 d_1 \ao^3}{3 (x-1)^2}-12 \
d_1 \ao^2+\frac{6 d_1 \ao^2}{x-1}-\frac{4 d_1 \ao^2}{(x-1)^2}+\frac{2 \
d_1 \ao^2}{(x-1)^3}+16 d_1 \ao-\frac{4 d_1 \ao}{x-1}+\frac{4 d_1 \
\ao}{(x-1)^2}-\frac{4 d_1 \ao}{(x-1)^3}+\frac{4 d_1 \
\ao}{(x-1)^4}+\frac{8 d_1}{x-2}-\frac{4 d_1}{x-1}-\frac{16 d_1}{3 \
(x-2)^2}-\frac{2 d_1}{3 (x-1)^2}+\frac{16 d_1}{3 (x-2)^3}-\frac{4 \
d_1}{3 (x-1)^3}-\frac{7 d_1}{(x-1)^4}-\frac{25 d_1}{3 (x-1)^5}\Big) \
H(1;\ao)+\Big(\frac{16}{(x-1)^5}-16\Big) H(0,0;\ao)+\Big(\frac{8 \
d_1}{(x-1)^5}-8 d_1\Big) H(0,1;\ao)+\Big(\frac{8 d_1}{(x-1)^5}-8 \
d_1\Big) H(1,0;\ao)+\Big(\frac{4 d_1^2}{(x-1)^5}-4 d_1^2\Big) \
H(1,1;\ao)+\frac{32 d_1}{3 (x-2)}-\frac{70}{3 (x-2)}-\frac{20 d_1}{3 \
(x-1)}+\frac{95}{6 (x-1)}-\frac{28 d_1}{9 (x-2)^2}+\frac{32}{9 \
(x-2)^2}-\frac{23 d_1}{9 (x-1)^2}+\frac{131}{18 (x-1)^2}+\frac{28 \
d_1}{9 (x-2)^3}-\frac{56}{9 (x-2)^3}-\frac{40 d_1}{9 \
(x-1)^3}+\frac{241}{18 (x-1)^3}-\frac{31 d_1}{4 \
(x-1)^4}+\frac{20}{(x-1)^4}-\frac{205 d_1}{36 (x-1)^5}-\frac{\pi \
^2}{6 (x-1)^5}+\frac{155}{9 (x-1)^5}+\frac{\pi \
^2}{6}+\frac{35}{6}\Big)+H(c_1(\ao);x) \Big(\frac{d_1^2 \
\ao^4}{16}-\frac{d_1 \ao^4}{4}-\frac{d_1^2 \ao^4}{16 (x-1)}+\frac{d_1 \
\ao^4}{4 (x-1)}+\frac{\pi ^2 \ao^4}{24 (x-1)}-\frac{\ao^4}{4 \
(x-1)}-\frac{\pi ^2 \ao^4}{24}+\frac{\ao^4}{4}-\frac{43 d_1^2 \
\ao^3}{108}+\frac{193 d_1 \ao^3}{108}+\frac{d_1^2 \ao^3}{4 \
(x-1)}-\frac{25 d_1 \ao^3}{18 (x-1)}-\frac{\pi ^2 \ao^3}{6 \
(x-1)}+\frac{16 \ao^3}{9 (x-1)}-\frac{4 d_1^2 \ao^3}{27 \
(x-1)^2}+\frac{127 d_1 \ao^3}{108 (x-1)^2}+\frac{\pi ^2 \ao^3}{18 \
(x-1)^2}-\frac{95 \ao^3}{54 (x-1)^2}+\frac{2 \pi ^2 \
\ao^3}{9}-\frac{107 \ao^3}{54}+\frac{95 d_1^2 \ao^2}{72}-\frac{163 \
d_1 \ao^2}{24}-\frac{13 d_1 \ao^2}{18 (x-2)}+\frac{13 \ao^2}{9 \
(x-2)}-\frac{3 d_1^2 \ao^2}{8 (x-1)}+\frac{311 d_1 \ao^2}{72 \
(x-1)}+\frac{\pi ^2 \ao^2}{4 (x-1)}-\frac{281 \ao^2}{36 \
(x-1)}+\frac{4 d_1^2 \ao^2}{9 (x-1)^2}-\frac{295 d_1 \ao^2}{72 \
(x-1)^2}-\frac{\pi ^2 \ao^2}{6 (x-1)^2}+\frac{89 \ao^2}{12 \
(x-1)^2}-\frac{d_1^2 \ao^2}{2 (x-1)^3}+\frac{115 d_1 \ao^2}{24 \
(x-1)^3}+\frac{\pi ^2 \ao^2}{12 (x-1)^3}-\frac{35 \ao^2}{4 \
(x-1)^3}-\frac{\pi ^2 \ao^2}{2}+\frac{305 \ao^2}{36}-\frac{205 d_1^2 \
\ao}{36}+\frac{125 d_1 \ao}{4}+\frac{17 d_1 \ao}{3 (x-2)}-\frac{10 \
\ao}{x-2}+\frac{d_1^2 \ao}{4 (x-1)}-\frac{251 d_1 \ao}{18 \
(x-1)}-\frac{\pi ^2 \ao}{6 (x-1)}+\frac{1115 \ao}{36 (x-1)}-\frac{38 \
d_1 \ao}{9 (x-2)^2}+\frac{76 \ao}{9 (x-2)^2}-\frac{4 d_1^2 \ao}{9 \
(x-1)^2}+\frac{32 d_1 \ao}{3 (x-1)^2}+\frac{\pi ^2 \ao}{6 \
(x-1)^2}-\frac{511 \ao}{18 (x-1)^2}+\frac{d_1^2 \
\ao}{(x-1)^3}-\frac{31 d_1 \ao}{3 (x-1)^3}-\frac{\pi ^2 \ao}{6 \
(x-1)^3}+\frac{95 \ao}{4 (x-1)^3}-\frac{4 d_1^2 \
\ao}{(x-1)^4}+\frac{409 d_1 \ao}{12 (x-1)^4}+\frac{\pi ^2 \ao}{6 \
(x-1)^4}-\frac{188 \ao}{3 (x-1)^4}+\frac{2 \pi ^2 \ao}{3}-\frac{374 \
\ao}{9}+\frac{2035 d_1^2}{432}-\frac{5615 d_1}{216}+\Big(\frac{d_1 \
\ao^4}{2}-\frac{d_1 \ao^4}{2 (x-1)}+\frac{\ao^4}{x-1}-\ao^4-\frac{26 \
d_1 \ao^3}{9}+\frac{2 d_1 \ao^3}{x-1}-\frac{16 \ao^3}{3 \
(x-1)}-\frac{8 d_1 \ao^3}{9 (x-1)^2}+\frac{34 \ao^3}{9 \
(x-1)^2}+\frac{58 \ao^3}{9}+\frac{23 d_1 \ao^2}{3}-\frac{4 \ao^2}{3 \
(x-2)}-\frac{3 d_1 \ao^2}{x-1}+\frac{41 \ao^2}{3 (x-1)}+\frac{8 d_1 \
\ao^2}{3 (x-1)^2}-\frac{13 \ao^2}{(x-1)^2}-\frac{2 d_1 \
\ao^2}{(x-1)^3}+\frac{11 \ao^2}{(x-1)^3}-\frac{59 \ao^2}{3}-\frac{50 \
d_1 \ao}{3}+\frac{8 \ao}{x-2}+\frac{2 d_1 \ao}{x-1}-\frac{80 \ao}{3 \
(x-1)}-\frac{16 \ao}{3 (x-2)^2}-\frac{8 d_1 \ao}{3 (x-1)^2}+\frac{68 \
\ao}{3 (x-1)^2}+\frac{4 d_1 \ao}{(x-1)^3}-\frac{24 \
\ao}{(x-1)^3}-\frac{8 d_1 \ao}{(x-1)^4}+\frac{42 \
\ao}{(x-1)^4}+\frac{146 \ao}{3}+\frac{205 \
d_1}{18}-\frac{16}{x-2}+\frac{15 d_1}{2 \
(x-1)}+\frac{2}{x-1}+\frac{32}{3 (x-2)^2}-\frac{10 d_1}{9 \
(x-1)^2}+\frac{26}{9 (x-1)^2}-\frac{10 d_1}{9 (x-1)^3}-\frac{10}{9 \
(x-1)^3}+\frac{15 d_1}{2 (x-1)^4}-\frac{36}{(x-1)^4}+\frac{205 \
d_1}{18 (x-1)^5}-\frac{520}{9 (x-1)^5}-\frac{310}{9}\Big) \
H(0;\ao)+\Big(\frac{d_1^2 \ao^4}{4}-\frac{d_1 \ao^4}{2}-\frac{d_1^2 \
\ao^4}{4 (x-1)}+\frac{d_1 \ao^4}{2 (x-1)}-\frac{13 d_1^2 \
\ao^3}{9}+\frac{29 d_1 \ao^3}{9}+\frac{d_1^2 \ao^3}{x-1}-\frac{8 d_1 \
\ao^3}{3 (x-1)}-\frac{4 d_1^2 \ao^3}{9 (x-1)^2}+\frac{17 d_1 \ao^3}{9 \
(x-1)^2}+\frac{23 d_1^2 \ao^2}{6}-\frac{59 d_1 \ao^2}{6}-\frac{2 d_1 \
\ao^2}{3 (x-2)}-\frac{3 d_1^2 \ao^2}{2 (x-1)}+\frac{41 d_1 \ao^2}{6 \
(x-1)}+\frac{4 d_1^2 \ao^2}{3 (x-1)^2}-\frac{13 d_1 \ao^2}{2 \
(x-1)^2}-\frac{d_1^2 \ao^2}{(x-1)^3}+\frac{11 d_1 \ao^2}{2 \
(x-1)^3}-\frac{25 d_1^2 \ao}{3}+\frac{73 d_1 \ao}{3}+\frac{4 d_1 \
\ao}{x-2}+\frac{d_1^2 \ao}{x-1}-\frac{40 d_1 \ao}{3 (x-1)}-\frac{8 \
d_1 \ao}{3 (x-2)^2}-\frac{4 d_1^2 \ao}{3 (x-1)^2}+\frac{34 d_1 \ao}{3 \
(x-1)^2}+\frac{2 d_1^2 \ao}{(x-1)^3}-\frac{12 d_1 \
\ao}{(x-1)^3}-\frac{4 d_1^2 \ao}{(x-1)^4}+\frac{21 d_1 \ao}{(x-1)^4}+\
\frac{205 d_1^2}{36}-\frac{155 d_1}{9}-\frac{8 d_1}{x-2}+\frac{15 \
d_1^2}{4 (x-1)}+\frac{d_1}{x-1}+\frac{16 d_1}{3 (x-2)^2}-\frac{5 \
d_1^2}{9 (x-1)^2}+\frac{13 d_1}{9 (x-1)^2}-\frac{5 d_1^2}{9 (x-1)^3}-\
\frac{5 d_1}{9 (x-1)^3}+\frac{15 d_1^2}{4 (x-1)^4}-\frac{18 \
d_1}{(x-1)^4}+\frac{205 d_1^2}{36 (x-1)^5}-\frac{260 d_1}{9 \
(x-1)^5}\Big) H(1;\ao)+\Big(-\frac{4 \ao^4}{x-1}+4 \ao^4+\frac{16 \
\ao^3}{x-1}-\frac{16 \ao^3}{3 (x-1)^2}-\frac{64 \ao^3}{3}-\frac{24 \
\ao^2}{x-1}+\frac{16 \ao^2}{(x-1)^2}-\frac{8 \ao^2}{(x-1)^3}+48 \
\ao^2+\frac{16 \ao}{x-1}-\frac{16 \ao}{(x-1)^2}+\frac{16 \
\ao}{(x-1)^3}-\frac{16 \ao}{(x-1)^4}-64 \ao+\frac{12}{x-1}-\frac{8}{3 \
(x-1)^2}-\frac{8}{3 (x-1)^3}+\frac{12}{(x-1)^4}+\frac{100}{3 \
(x-1)^5}+\frac{100}{3}\Big) H(0,0;\ao)+\Big(2 d_1 \ao^4-\frac{2 d_1 \
\ao^4}{x-1}-\frac{32 d_1 \ao^3}{3}+\frac{8 d_1 \ao^3}{x-1}-\frac{8 \
d_1 \ao^3}{3 (x-1)^2}+24 d_1 \ao^2-\frac{12 d_1 \ao^2}{x-1}+\frac{8 \
d_1 \ao^2}{(x-1)^2}-\frac{4 d_1 \ao^2}{(x-1)^3}-32 d_1 \ao+\frac{8 \
d_1 \ao}{x-1}-\frac{8 d_1 \ao}{(x-1)^2}+\frac{8 d_1 \
\ao}{(x-1)^3}-\frac{8 d_1 \ao}{(x-1)^4}+\frac{50 d_1}{3}+\frac{6 \
d_1}{x-1}-\frac{4 d_1}{3 (x-1)^2}-\frac{4 d_1}{3 (x-1)^3}+\frac{6 \
d_1}{(x-1)^4}+\frac{50 d_1}{3 (x-1)^5}\Big) H(0,1;\ao)+\Big(2 d_1 \
\ao^4-\frac{2 d_1 \ao^4}{x-1}-\frac{32 d_1 \ao^3}{3}+\frac{8 d_1 \
\ao^3}{x-1}-\frac{8 d_1 \ao^3}{3 (x-1)^2}+24 d_1 \ao^2-\frac{12 d_1 \
\ao^2}{x-1}+\frac{8 d_1 \ao^2}{(x-1)^2}-\frac{4 d_1 \
\ao^2}{(x-1)^3}-32 d_1 \ao+\frac{8 d_1 \ao}{x-1}-\frac{8 d_1 \
\ao}{(x-1)^2}+\frac{8 d_1 \ao}{(x-1)^3}-\frac{8 d_1 \
\ao}{(x-1)^4}+\frac{50 d_1}{3}+\frac{6 d_1}{x-1}-\frac{4 d_1}{3 \
(x-1)^2}-\frac{4 d_1}{3 (x-1)^3}+\frac{6 d_1}{(x-1)^4}+\frac{50 \
d_1}{3 (x-1)^5}\Big) H(1,0;\ao)+\Big(d_1^2 \ao^4-\frac{d_1^2 \
\ao^4}{x-1}-\frac{16 d_1^2 \ao^3}{3}+\frac{4 d_1^2 \
\ao^3}{x-1}-\frac{4 d_1^2 \ao^3}{3 (x-1)^2}+12 d_1^2 \ao^2-\frac{6 \
d_1^2 \ao^2}{x-1}+\frac{4 d_1^2 \ao^2}{(x-1)^2}-\frac{2 d_1^2 \
\ao^2}{(x-1)^3}-16 d_1^2 \ao+\frac{4 d_1^2 \ao}{x-1}-\frac{4 d_1^2 \
\ao}{(x-1)^2}+\frac{4 d_1^2 \ao}{(x-1)^3}-\frac{4 d_1^2 \
\ao}{(x-1)^4}+\frac{25 d_1^2}{3}+\frac{3 d_1^2}{x-1}-\frac{2 d_1^2}{3 \
(x-1)^2}-\frac{2 d_1^2}{3 (x-1)^3}+\frac{3 d_1^2}{(x-1)^4}+\frac{25 \
d_1^2}{3 (x-1)^5}\Big) H(1,1;\ao)-\frac{38 d_1}{3 (x-2)}+\frac{68}{3 \
(x-2)}+\frac{63 d_1^2}{16 (x-1)}-\frac{85 d_1}{24 (x-1)}-\frac{\pi \
^2}{8 (x-1)}-\frac{37}{4 (x-1)}+\frac{76 d_1}{9 (x-2)^2}-\frac{152}{9 \
(x-2)^2}-\frac{19 d_1^2}{54 (x-1)^2}+\frac{317 d_1}{108 \
(x-1)^2}+\frac{\pi ^2}{36 (x-1)^2}+\frac{49}{108 (x-1)^2}-\frac{19 \
d_1^2}{54 (x-1)^3}-\frac{313 d_1}{108 (x-1)^3}+\frac{\pi ^2}{36 \
(x-1)^3}+\frac{949}{108 (x-1)^3}+\frac{63 d_1^2}{16 \
(x-1)^4}-\frac{257 d_1}{8 (x-1)^4}-\frac{\pi ^2}{8 \
(x-1)^4}+\frac{213}{4 (x-1)^4}+\frac{2035 d_1^2}{432 \
(x-1)^5}-\frac{8705 d_1}{216 (x-1)^5}-\frac{25 \pi ^2}{72 \
(x-1)^5}+\frac{3965}{54 (x-1)^5}-\frac{25 \pi \
^2}{72}+\frac{940}{27}\Big)+\Big(-\frac{2 \pi ^2 d_1^2}{3 \
(x-1)^5}+\frac{4 \pi ^2 d_1}{3 (x-1)^5}+\frac{2 \pi ^2 \
d_1}{3}+\Big(\frac{4 d_1^2}{x-1}-\frac{2 d_1^2}{(x-1)^2}+\frac{4 \
d_1^2}{3 (x-1)^3}-\frac{d_1^2}{(x-1)^4}+\frac{25 d_1^2}{3 \
(x-1)^5}-\frac{8 d_1}{x-2}+\frac{9 d_1}{x-1}+\frac{16 d_1}{3 \
(x-2)^2}-\frac{8 d_1}{3 (x-1)^2}-\frac{16 d_1}{3 (x-2)^3}+\frac{14 \
d_1}{3 (x-1)^3}+\frac{2 d_1}{(x-1)^4}+\frac{50 d_1}{3 \
(x-1)^5}-\frac{25 d_1}{3}-\frac{4}{x-2}-\frac{3}{2 (x-1)}+\frac{8}{3 \
(x-2)^2}-\frac{8}{3 (x-2)^3}+\frac{11}{3 \
(x-1)^3}-\frac{5}{(x-1)^4}-\frac{25}{3 (x-1)^5}-\frac{125}{6}\Big) \
H(0;\ao)+\Big(-\frac{16 d_1}{(x-1)^5}+8 d_1+\frac{8}{(x-1)^5}+8\Big) \
H(0,0;\ao)+\Big(-\frac{8 d_1^2}{(x-1)^5}+4 d_1^2+\frac{4 \
d_1}{(x-1)^5}+4 d_1\Big) H(0,1;\ao)-\frac{2 \pi ^2}{3 \
(x-1)^5}-\frac{2 \pi ^2}{3}\Big) H(1,1;x)+\Big(-\frac{\pi ^2 \
d_1}{(x-1)^5}+\frac{\pi ^2}{2 (x-1)^5}+\frac{3 \pi ^2}{2}\Big) \
H(1,2;x)+\Big(-\frac{4 d_1^2}{x-1}+\frac{d_1^2}{(x-1)^2}-\frac{4 \
d_1^2}{9 (x-1)^3}+\frac{d_1^2}{4 (x-1)^4}-\frac{205 d_1^2}{36 \
(x-1)^5}+\frac{46 d_1}{3 (x-2)}+\frac{3 d_1}{4 (x-1)}-\frac{52 d_1}{9 \
(x-2)^2}-\frac{185 d_1}{18 (x-1)^2}+\frac{28 d_1}{9 (x-2)^3}+\frac{29 \
d_1}{18 (x-1)^3}-\frac{27 d_1}{4 (x-1)^4}+\frac{835 d_1}{36 (x-1)^5}+\
\frac{205 d_1}{36}+\Big(-\frac{8 d_1}{x-1}+\frac{4 \
d_1}{(x-1)^2}-\frac{8 d_1}{3 (x-1)^3}+\frac{2 d_1}{(x-1)^4}-\frac{50 \
d_1}{3 (x-1)^5}+\frac{16}{x-2}-\frac{10}{x-1}-\frac{32}{3 \
(x-2)^2}+\frac{4}{3 (x-1)^2}+\frac{32}{3 (x-2)^3}-\frac{20}{3 \
(x-1)^3}-\frac{6}{(x-1)^4}-\frac{50}{3 (x-1)^5}+\frac{50}{3}\Big) \
H(0;\ao)+\Big(-\frac{4 d_1^2}{x-1}+\frac{2 d_1^2}{(x-1)^2}-\frac{4 \
d_1^2}{3 (x-1)^3}+\frac{d_1^2}{(x-1)^4}-\frac{25 d_1^2}{3 \
(x-1)^5}+\frac{8 d_1}{x-2}-\frac{5 d_1}{x-1}-\frac{16 d_1}{3 \
(x-2)^2}+\frac{2 d_1}{3 (x-1)^2}+\frac{16 d_1}{3 (x-2)^3}-\frac{10 \
d_1}{3 (x-1)^3}-\frac{3 d_1}{(x-1)^4}-\frac{25 d_1}{3 \
(x-1)^5}+\frac{25 d_1}{3}\Big) \
H(1;\ao)+\Big(\frac{16}{(x-1)^5}-16\Big) H(0,0;\ao)+\Big(\frac{8 \
d_1}{(x-1)^5}-8 d_1\Big) H(0,1;\ao)+\Big(\frac{8 d_1}{(x-1)^5}-8 \
d_1\Big) H(1,0;\ao)+\Big(\frac{4 d_1^2}{(x-1)^5}-4 d_1^2\Big) \
H(1,1;\ao)-\frac{80}{3 (x-2)}+\frac{70}{3 (x-1)}+\frac{56}{9 \
(x-2)^2}+\frac{13}{18 (x-1)^2}-\frac{56}{9 (x-2)^3}+\frac{355}{18 \
(x-1)^3}+\frac{10}{3 (x-1)^4}-\frac{\pi ^2}{6 (x-1)^5}+\frac{155}{9 \
(x-1)^5}+\frac{\pi ^2}{6}-\frac{155}{9}\Big) \
H(1,c_1(\ao);x)+\Big(\frac{2 \pi ^2}{(x-1)^5}-2 \pi ^2\Big) H(2,0;x)+\
\Big(\Big(\frac{8 d_1}{x-2}-\frac{16 d_1}{3 (x-2)^2}-\frac{8 d_1}{3 \
(x-1)^2}+\frac{16 d_1}{3 (x-2)^3}-\frac{8 \
d_1}{(x-1)^4}-\frac{16}{x-2}+\frac{15}{2 (x-1)}+\frac{32}{3 (x-2)^2}+\
\frac{1}{3 (x-1)^2}-\frac{32}{3 \
(x-2)^3}+\frac{5}{(x-1)^3}+\frac{17}{2 (x-1)^4}+\frac{25}{2 (x-1)^5}-\
\frac{25}{2}\Big) H(0;\ao)+\Big(8-\frac{8}{(x-1)^5}\Big) \
H(0,0;\ao)+\Big(4 d_1-\frac{4 d_1}{(x-1)^5}\Big) H(0,1;\ao)-\frac{\pi \
^2}{3 (x-1)^5}+\frac{\pi ^2}{3}\Big) H(2,1;x)+\Big(\frac{\pi ^2 \
d_1}{(x-1)^5}-\pi ^2 d_1-\frac{2 \pi ^2}{(x-1)^5}+2 \pi ^2\Big) \
H(2,2;x)+\Big(\frac{3 d_1 \ao^4}{4}-\frac{3 d_1 \ao^4}{4 \
(x-1)}+\frac{3 \ao^4}{2 (x-1)}-\frac{3 \ao^4}{2}-\frac{13 d_1 \
\ao^3}{3}+\frac{3 d_1 \ao^3}{x-1}-\frac{49 \ao^3}{6 (x-1)}-\frac{4 \
d_1 \ao^3}{3 (x-1)^2}+\frac{5 \ao^3}{(x-1)^2}+\frac{61 \
\ao^3}{6}+\frac{23 d_1 \ao^2}{2}-\frac{5 \ao^2}{3 (x-2)}-\frac{9 d_1 \
\ao^2}{2 (x-1)}+\frac{83 \ao^2}{4 (x-1)}+\frac{4 d_1 \ao^2}{(x-1)^2}-\
\frac{109 \ao^2}{6 (x-1)^2}-\frac{3 d_1 \ao^2}{(x-1)^3}+\frac{85 \
\ao^2}{6 (x-1)^3}-\frac{131 \ao^2}{4}-25 d_1 \ao+\frac{10 \
\ao}{x-2}+\frac{3 d_1 \ao}{x-1}-\frac{229 \ao}{6 (x-1)}-\frac{20 \
\ao}{3 (x-2)^2}-\frac{4 d_1 \ao}{(x-1)^2}+\frac{100 \ao}{3 \
(x-1)^2}+\frac{6 d_1 \ao}{(x-1)^3}-\frac{103 \ao}{3 (x-1)^3}-\frac{12 \
d_1 \ao}{(x-1)^4}+\frac{163 \ao}{3 (x-1)^4}+\frac{169 \
\ao}{2}+\frac{205 d_1}{12}+\Big(-\frac{6 \ao^4}{x-1}+6 \ao^4+\frac{24 \
\ao^3}{x-1}-\frac{8 \ao^3}{(x-1)^2}-32 \ao^3-\frac{36 \
\ao^2}{x-1}+\frac{24 \ao^2}{(x-1)^2}-\frac{12 \ao^2}{(x-1)^3}+72 \
\ao^2+\frac{24 \ao}{x-1}-\frac{24 \ao}{(x-1)^2}+\frac{24 \
\ao}{(x-1)^3}-\frac{24 \ao}{(x-1)^4}-96 \
\ao+\frac{18}{x-1}-\frac{4}{(x-1)^2}-\frac{4}{(x-1)^3}+\frac{18}{(x-1)\
^4}+\frac{50}{(x-1)^5}+50\Big) H(0;\ao)+\Big(3 d_1 \ao^4-\frac{3 d_1 \
\ao^4}{x-1}-16 d_1 \ao^3+\frac{12 d_1 \ao^3}{x-1}-\frac{4 d_1 \
\ao^3}{(x-1)^2}+36 d_1 \ao^2-\frac{18 d_1 \ao^2}{x-1}+\frac{12 d_1 \
\ao^2}{(x-1)^2}-\frac{6 d_1 \ao^2}{(x-1)^3}-48 d_1 \ao+\frac{12 d_1 \
\ao}{x-1}-\frac{12 d_1 \ao}{(x-1)^2}+\frac{12 d_1 \
\ao}{(x-1)^3}-\frac{12 d_1 \ao}{(x-1)^4}+25 d_1+\frac{9 \
d_1}{x-1}-\frac{2 d_1}{(x-1)^2}-\frac{2 d_1}{(x-1)^3}+\frac{9 \
d_1}{(x-1)^4}+\frac{25 d_1}{(x-1)^5}\Big) H(1;\ao)-\frac{16 \
H(0,0;\ao)}{(x-1)^5}-\frac{8 d_1 H(0,1;\ao)}{(x-1)^5}-\frac{8 d_1 \
H(1,0;\ao)}{(x-1)^5}-\frac{4 d_1^2 \
H(1,1;\ao)}{(x-1)^5}-\frac{20}{x-2}+\frac{45 d_1}{4 \
(x-1)}-\frac{31}{4 (x-1)}+\frac{40}{3 (x-2)^2}-\frac{5 d_1}{3 \
(x-1)^2}+\frac{55}{12 (x-1)^2}-\frac{5 d_1}{3 (x-1)^3}-\frac{5}{4 \
(x-1)^3}+\frac{45 d_1}{4 (x-1)^4}-\frac{83}{2 (x-1)^4}+\frac{205 \
d_1}{12 (x-1)^5}+\frac{\pi ^2}{6 \
(x-1)^5}-\frac{75}{(x-1)^5}-\frac{725}{12}\Big) \
H(c_1(\ao),c_1(\ao);x)+\Big(-\frac{d_1 \ao^4}{4}+\frac{d_1 \ao^4}{4 \
(x-1)}-\frac{\ao^4}{2 (x-1)}+\frac{\ao^4}{2}+\frac{13 d_1 \
\ao^3}{9}-\frac{d_1 \ao^3}{x-1}+\frac{3 \ao^3}{x-1}+\frac{4 d_1 \
\ao^3}{9 (x-1)^2}-\frac{25 \ao^3}{18 (x-1)^2}-\frac{61 \
\ao^3}{18}-\frac{23 d_1 \ao^2}{6}+\frac{3 d_1 \ao^2}{2 \
(x-1)}-\frac{31 \ao^2}{4 (x-1)}-\frac{4 d_1 \ao^2}{3 \
(x-1)^2}+\frac{71 \ao^2}{12 (x-1)^2}+\frac{d_1 \
\ao^2}{(x-1)^3}-\frac{15 \ao^2}{4 (x-1)^3}+\frac{131 \
\ao^2}{12}+\frac{25 d_1 \ao}{3}-\frac{d_1 \ao}{x-1}+\frac{13 \
\ao}{x-1}+\frac{4 d_1 \ao}{3 (x-1)^2}-\frac{35 \ao}{3 \
(x-1)^2}-\frac{2 d_1 \ao}{(x-1)^3}+\frac{12 \ao}{(x-1)^3}+\frac{4 d_1 \
\ao}{(x-1)^4}-\frac{29 \ao}{2 (x-1)^4}-\frac{169 \ao}{6}-\frac{205 \
d_1}{36}+\Big(\frac{2 \ao^4}{x-1}-2 \ao^4-\frac{8 \ao^3}{x-1}+\frac{8 \
\ao^3}{3 (x-1)^2}+\frac{32 \ao^3}{3}+\frac{12 \ao^2}{x-1}-\frac{8 \
\ao^2}{(x-1)^2}+\frac{4 \ao^2}{(x-1)^3}-24 \ao^2-\frac{8 \
\ao}{x-1}+\frac{8 \ao}{(x-1)^2}-\frac{8 \ao}{(x-1)^3}+\frac{8 \
\ao}{(x-1)^4}+32 \ao-\frac{16}{x-2}+\frac{2}{x-1}+\frac{32}{3 \
(x-2)^2}+\frac{8}{3 (x-1)^2}-\frac{32}{3 \
(x-2)^3}+\frac{4}{(x-1)^3}+\frac{8}{(x-1)^4}-\frac{50}{3}\Big) \
H(0;\ao)+\Big(-d_1 \ao^4+\frac{d_1 \ao^4}{x-1}+\frac{16 d_1 \
\ao^3}{3}-\frac{4 d_1 \ao^3}{x-1}+\frac{4 d_1 \ao^3}{3 (x-1)^2}-12 \
d_1 \ao^2+\frac{6 d_1 \ao^2}{x-1}-\frac{4 d_1 \ao^2}{(x-1)^2}+\frac{2 \
d_1 \ao^2}{(x-1)^3}+16 d_1 \ao-\frac{4 d_1 \ao}{x-1}+\frac{4 d_1 \
\ao}{(x-1)^2}-\frac{4 d_1 \ao}{(x-1)^3}+\frac{4 d_1 \
\ao}{(x-1)^4}-\frac{25 d_1}{3}-\frac{8 \
d_1}{x-2}+\frac{d_1}{x-1}+\frac{16 d_1}{3 (x-2)^2}+\frac{4 d_1}{3 \
(x-1)^2}-\frac{16 d_1}{3 (x-2)^3}+\frac{2 d_1}{(x-1)^3}+\frac{4 \
d_1}{(x-1)^4}\Big) H(1;\ao)-\frac{32 d_1}{3 (x-2)}+\frac{82}{3 \
(x-2)}+\frac{35 d_1}{12 (x-1)}-\frac{73}{12 (x-1)}+\frac{28 d_1}{9 \
(x-2)^2}-\frac{56}{9 (x-2)^2}+\frac{28 d_1}{9 (x-1)^2}-\frac{323}{36 \
(x-1)^2}-\frac{28 d_1}{9 (x-2)^3}+\frac{56}{9 (x-2)^3}+\frac{5 \
d_1}{(x-1)^3}-\frac{53}{4 (x-1)^3}+\frac{4 d_1}{(x-1)^4}-\frac{29}{2 \
(x-1)^4}+\frac{725}{36}\Big) \
H(c_2(\ao),c_1(\ao);x)+\Big(\frac{100}{3}+\frac{12}{x-1}-\frac{8}{3 \
(x-1)^2}-\frac{8}{3 (x-1)^3}+\frac{12}{(x-1)^4}+\frac{100}{3 (x-1)^5}\
\Big) H(0,0,0;\ao)+\Big(-\frac{100}{3}-\frac{12}{x-1}+\frac{8}{3 \
(x-1)^2}+\frac{8}{3 (x-1)^3}-\frac{12}{(x-1)^4}-\frac{100}{3 (x-1)^5}\
\Big) H(0,0,0;x)+\Big(\frac{6 d_1}{x-1}-\frac{4 d_1}{3 \
(x-1)^2}-\frac{4 d_1}{3 (x-1)^3}+\frac{6 d_1}{(x-1)^4}+\frac{50 \
d_1}{3 (x-1)^5}+\frac{50 d_1}{3}\Big) H(0,0,1;\ao)+\Big(-\frac{4 \
d_1}{(x-1)^5}+4 d_1+\frac{12}{(x-1)^5}-12\Big) H(0;\ao) \
H(0,0,1;x)+\Big(-\frac{\ao^4}{x-1}+\ao^4+\frac{4 \ao^3}{x-1}-\frac{4 \
\ao^3}{3 (x-1)^2}-\frac{16 \ao^3}{3}-\frac{6 \ao^2}{x-1}+\frac{4 \
\ao^2}{(x-1)^2}-\frac{2 \ao^2}{(x-1)^3}+12 \ao^2+\frac{4 \
\ao}{x-1}-\frac{4 \ao}{(x-1)^2}+\frac{4 \ao}{(x-1)^3}-\frac{4 \
\ao}{(x-1)^4}-16 \ao+\Big(\frac{8}{(x-1)^5}-8\Big) \
H(0;\ao)+\Big(\frac{4 d_1}{(x-1)^5}-4 d_1\Big) \
H(1;\ao)+\frac{8}{x-2}-\frac{1}{x-1}-\frac{16}{3 (x-2)^2}-\frac{4}{3 \
(x-1)^2}+\frac{16}{3 \
(x-2)^3}-\frac{2}{(x-1)^3}-\frac{4}{(x-1)^4}+\frac{25}{3}\Big) \
H(0,0,c_1(\ao);x)+\Big(\frac{6 d_1}{x-1}-\frac{4 d_1}{3 \
(x-1)^2}-\frac{4 d_1}{3 (x-1)^3}+\frac{6 d_1}{(x-1)^4}+\frac{50 \
d_1}{3 (x-1)^5}+\frac{50 d_1}{3}\Big) H(0,1,0;\ao)+\Big(\frac{8 \
d_1}{x-2}-\frac{5 d_1}{x-1}-\frac{16 d_1}{3 (x-2)^2}+\frac{2 d_1}{3 \
(x-1)^2}+\frac{16 d_1}{3 (x-2)^3}-\frac{10 d_1}{3 (x-1)^3}-\frac{3 \
d_1}{(x-1)^4}-\frac{25 d_1}{3 (x-1)^5}+\frac{25 \
d_1}{3}-\frac{8}{x-2}+\frac{16}{3 (x-2)^2}+\frac{8}{3 \
(x-1)^2}-\frac{16}{3 (x-2)^3}+\frac{8}{(x-1)^4}\Big) H(0,1,0;x)+\Big(\
\frac{3 d_1^2}{x-1}-\frac{2 d_1^2}{3 (x-1)^2}-\frac{2 d_1^2}{3 \
(x-1)^3}+\frac{3 d_1^2}{(x-1)^4}+\frac{25 d_1^2}{3 (x-1)^5}+\frac{25 \
d_1^2}{3}\Big) H(0,1,1;\ao)+\Big(\frac{4 d_1^2}{(x-1)^5}-4 \
d_1^2-\frac{8 d_1}{(x-1)^5}+\frac{12}{(x-1)^5}+4\Big) H(0;\ao) \
H(0,1,1;x)+\Big(-\frac{8 d_1}{x-2}+\frac{5 d_1}{x-1}+\frac{16 d_1}{3 \
(x-2)^2}-\frac{2 d_1}{3 (x-1)^2}-\frac{16 d_1}{3 (x-2)^3}+\frac{10 \
d_1}{3 (x-1)^3}+\frac{3 d_1}{(x-1)^4}+\frac{25 d_1}{3 \
(x-1)^5}-\frac{25 d_1}{3}+\Big(-\frac{8 d_1}{(x-1)^5}+8 \
d_1+\frac{8}{(x-1)^5}-8\Big) H(0;\ao)+\Big(-\frac{4 d_1^2}{(x-1)^5}+4 \
d_1^2+\frac{4 d_1}{(x-1)^5}-4 d_1\Big) \
H(1;\ao)+\frac{8}{x-2}-\frac{16}{3 (x-2)^2}-\frac{8}{3 \
(x-1)^2}+\frac{16}{3 (x-2)^3}-\frac{8}{(x-1)^4}\Big) \
H(0,1,c_1(\ao);x)+\Big(\frac{4 d_1}{(x-1)^5}-4 \
d_1-\frac{4}{(x-1)^5}+4\Big) H(0;\ao) H(0,2,1;x)+\Big(\frac{3 \
\ao^4}{x-1}-3 \ao^4-\frac{12 \ao^3}{x-1}+\frac{4 \ao^3}{(x-1)^2}+16 \
\ao^3+\frac{18 \ao^2}{x-1}-\frac{12 \ao^2}{(x-1)^2}+\frac{6 \
\ao^2}{(x-1)^3}-36 \ao^2-\frac{12 \ao}{x-1}+\frac{12 \
\ao}{(x-1)^2}-\frac{12 \ao}{(x-1)^3}+\frac{12 \ao}{(x-1)^4}+48 \
\ao+\Big(\frac{8}{(x-1)^5}-24\Big) H(0;\ao)+\Big(\frac{4 \
d_1}{(x-1)^5}-12 d_1\Big) H(1;\ao)+\frac{20}{x-2}-\frac{11}{2 (x-1)}-\
\frac{40}{3 (x-2)^2}-\frac{8}{3 (x-1)^2}+\frac{40}{3 \
(x-2)^3}-\frac{13}{3 (x-1)^3}-\frac{13}{(x-1)^4}-\frac{25}{3 \
(x-1)^5}+\frac{25}{2}\Big) \
H(0,c_1(\ao),c_1(\ao);x)+\Big(-\frac{\ao^4}{x-1}+\ao^4+\frac{4 \
\ao^3}{x-1}-\frac{4 \ao^3}{3 (x-1)^2}-\frac{16 \ao^3}{3}-\frac{6 \
\ao^2}{x-1}+\frac{4 \ao^2}{(x-1)^2}-\frac{2 \ao^2}{(x-1)^3}+12 \ao^2+\
\frac{4 \ao}{x-1}-\frac{4 \ao}{(x-1)^2}+\frac{4 \ao}{(x-1)^3}-\frac{4 \
\ao}{(x-1)^4}-16 \ao+\Big(8-\frac{8}{(x-1)^5}\Big) H(0;\ao)+\Big(4 \
d_1-\frac{4 d_1}{(x-1)^5}\Big) H(1;\ao)-\frac{16}{x-2}+\frac{13}{2 \
(x-1)}+\frac{32}{3 (x-2)^2}+\frac{5}{3 (x-1)^2}-\frac{32}{3 (x-2)^3}+\
\frac{3}{(x-1)^3}+\frac{25}{2 (x-1)^4}+\frac{25}{2 \
(x-1)^5}-\frac{25}{6}\Big) H(0,c_2(\ao),c_1(\ao);x)+\Big(\frac{8 \
d_1}{x-1}-\frac{4 d_1}{(x-1)^2}+\frac{8 d_1}{3 (x-1)^3}-\frac{2 \
d_1}{(x-1)^4}+\frac{50 d_1}{3 (x-1)^5}-\frac{16}{x-2}+\frac{10}{x-1}+\
\frac{32}{3 (x-2)^2}-\frac{4}{3 (x-1)^2}-\frac{32}{3 \
(x-2)^3}+\frac{20}{3 (x-1)^3}+\frac{6}{(x-1)^4}+\frac{50}{3 (x-1)^5}-\
\frac{50}{3}\Big) H(1,0,0;x)+\Big(\frac{4 d_1^2}{(x-1)^5}-\frac{8 \
d_1}{(x-1)^5}-4 d_1+\frac{8}{(x-1)^5}\Big) H(0;\ao) \
H(1,0,1;x)+\Big(\frac{4 d_1}{x-1}-\frac{2 d_1}{(x-1)^2}+\frac{4 \
d_1}{3 (x-1)^3}-\frac{d_1}{(x-1)^4}+\frac{25 d_1}{3 \
(x-1)^5}+\Big(-\frac{8 d_1}{(x-1)^5}+\frac{8}{(x-1)^5}+8\Big) \
H(0;\ao)+\Big(-\frac{4 d_1^2}{(x-1)^5}+\frac{4 d_1}{(x-1)^5}+4 \
d_1\Big) H(1;\ao)-\frac{4}{x-2}-\frac{3}{2 (x-1)}+\frac{8}{3 \
(x-2)^2}-\frac{8}{3 (x-2)^3}+\frac{11}{3 \
(x-1)^3}-\frac{5}{(x-1)^4}-\frac{25}{3 (x-1)^5}-\frac{125}{6}\Big) \
H(1,0,c_1(\ao);x)+\Big(-\frac{4 d_1^2}{x-1}+\frac{2 \
d_1^2}{(x-1)^2}-\frac{4 d_1^2}{3 \
(x-1)^3}+\frac{d_1^2}{(x-1)^4}-\frac{25 d_1^2}{3 (x-1)^5}+\frac{8 \
d_1}{x-2}-\frac{9 d_1}{x-1}-\frac{16 d_1}{3 (x-2)^2}+\frac{8 d_1}{3 \
(x-1)^2}+\frac{16 d_1}{3 (x-2)^3}-\frac{14 d_1}{3 (x-1)^3}-\frac{2 \
d_1}{(x-1)^4}-\frac{50 d_1}{3 (x-1)^5}+\frac{25 \
d_1}{3}+\frac{4}{x-2}+\frac{3}{2 (x-1)}-\frac{8}{3 \
(x-2)^2}+\frac{8}{3 (x-2)^3}-\frac{11}{3 \
(x-1)^3}+\frac{5}{(x-1)^4}+\frac{25}{3 (x-1)^5}+\frac{125}{6}\Big) \
H(1,1,0;x)+\Big(\frac{12 d_1^2}{(x-1)^5}-4 d_1^2-\frac{12 \
d_1}{(x-1)^5}-8 d_1+\frac{4}{(x-1)^5}+4\Big) H(0;\ao) \
H(1,1,1;x)+\Big(\frac{4 d_1^2}{x-1}-\frac{2 d_1^2}{(x-1)^2}+\frac{4 \
d_1^2}{3 (x-1)^3}-\frac{d_1^2}{(x-1)^4}+\frac{25 d_1^2}{3 \
(x-1)^5}-\frac{8 d_1}{x-2}+\frac{9 d_1}{x-1}+\frac{16 d_1}{3 \
(x-2)^2}-\frac{8 d_1}{3 (x-1)^2}-\frac{16 d_1}{3 (x-2)^3}+\frac{14 \
d_1}{3 (x-1)^3}+\frac{2 d_1}{(x-1)^4}+\frac{50 d_1}{3 \
(x-1)^5}-\frac{25 d_1}{3}+\Big(-\frac{16 d_1}{(x-1)^5}+8 \
d_1+\frac{8}{(x-1)^5}+8\Big) H(0;\ao)+\Big(-\frac{8 d_1^2}{(x-1)^5}+4 \
d_1^2+\frac{4 d_1}{(x-1)^5}+4 d_1\Big) \
H(1;\ao)-\frac{4}{x-2}-\frac{3}{2 (x-1)}+\frac{8}{3 \
(x-2)^2}-\frac{8}{3 (x-2)^3}+\frac{11}{3 \
(x-1)^3}-\frac{5}{(x-1)^4}-\frac{25}{3 (x-1)^5}-\frac{125}{6}\Big) \
H(1,1,c_1(\ao);x)+\Big(\frac{4 d_1}{(x-1)^5}-\frac{2}{(x-1)^5}-6\Big) \
H(0;\ao) H(1,2,1;x)+\Big(-\frac{12 d_1}{x-1}+\frac{6 \
d_1}{(x-1)^2}-\frac{4 d_1}{(x-1)^3}+\frac{3 d_1}{(x-1)^4}-\frac{25 \
d_1}{(x-1)^5}+\Big(\frac{8 d_1}{(x-1)^5}+\frac{8}{(x-1)^5}-24\Big) \
H(0;\ao)+\Big(\frac{4 d_1^2}{(x-1)^5}+\frac{4 d_1}{(x-1)^5}-12 \
d_1\Big) H(1;\ao)+\frac{20}{x-2}-\frac{17}{2 (x-1)}-\frac{40}{3 \
(x-2)^2}+\frac{4}{3 (x-1)^2}+\frac{40}{3 (x-2)^3}-\frac{31}{3 \
(x-1)^3}-\frac{1}{(x-1)^4}-\frac{25}{3 (x-1)^5}+\frac{75}{2}\Big) \
H(1,c_1(\ao),c_1(\ao);x)+\Big(\frac{4 d_1}{(x-1)^5}-4 \
d_1-\frac{4}{(x-1)^5}+4\Big) H(0;\ao) H(2,0,1;x)+\Big(\frac{8 \
d_1}{x-2}-\frac{16 d_1}{3 (x-2)^2}-\frac{8 d_1}{3 (x-1)^2}+\frac{16 \
d_1}{3 (x-2)^3}-\frac{8 d_1}{(x-1)^4}+\Big(8-\frac{8}{(x-1)^5}\Big) \
H(0;\ao)+\Big(4 d_1-\frac{4 d_1}{(x-1)^5}\Big) \
H(1;\ao)-\frac{16}{x-2}+\frac{15}{2 (x-1)}+\frac{32}{3 \
(x-2)^2}+\frac{1}{3 (x-1)^2}-\frac{32}{3 \
(x-2)^3}+\frac{5}{(x-1)^3}+\frac{17}{2 (x-1)^4}+\frac{25}{2 (x-1)^5}-\
\frac{25}{2}\Big) H(2,0,c_1(\ao);x)+\Big(-\frac{8 d_1}{x-2}+\frac{16 \
d_1}{3 (x-2)^2}+\frac{8 d_1}{3 (x-1)^2}-\frac{16 d_1}{3 \
(x-2)^3}+\frac{8 d_1}{(x-1)^4}+\frac{16}{x-2}-\frac{15}{2 \
(x-1)}-\frac{32}{3 (x-2)^2}-\frac{1}{3 (x-1)^2}+\frac{32}{3 (x-2)^3}-\
\frac{5}{(x-1)^3}-\frac{17}{2 (x-1)^4}-\frac{25}{2 \
(x-1)^5}+\frac{25}{2}\Big) H(2,1,0;x)+\Big(\frac{4 d_1}{(x-1)^5}-4 \
d_1+\frac{2}{(x-1)^5}-2\Big) H(0;\ao) H(2,1,1;x)+\Big(\frac{8 \
d_1}{x-2}-\frac{16 d_1}{3 (x-2)^2}-\frac{8 d_1}{3 (x-1)^2}+\frac{16 \
d_1}{3 (x-2)^3}-\frac{8 d_1}{(x-1)^4}+\Big(8-\frac{8}{(x-1)^5}\Big) \
H(0;\ao)+\Big(4 d_1-\frac{4 d_1}{(x-1)^5}\Big) \
H(1;\ao)-\frac{16}{x-2}+\frac{15}{2 (x-1)}+\frac{32}{3 \
(x-2)^2}+\frac{1}{3 (x-1)^2}-\frac{32}{3 \
(x-2)^3}+\frac{5}{(x-1)^3}+\frac{17}{2 (x-1)^4}+\frac{25}{2 (x-1)^5}-\
\frac{25}{2}\Big) H(2,1,c_1(\ao);x)+\Big(-\frac{4 d_1}{(x-1)^5}+4 \
d_1+\frac{8}{(x-1)^5}-8\Big) H(0;\ao) H(2,2,1;x)+\Big(-\frac{8 \
d_1}{x-2}+\frac{16 d_1}{3 (x-2)^2}+\frac{8 d_1}{3 (x-1)^2}-\frac{16 \
d_1}{3 (x-2)^3}+\frac{8 d_1}{(x-1)^4}+\Big(\frac{8}{(x-1)^5}-8\Big) \
H(0;\ao)+\Big(\frac{4 d_1}{(x-1)^5}-4 d_1\Big) \
H(1;\ao)+\frac{16}{x-2}-\frac{15}{2 (x-1)}-\frac{32}{3 \
(x-2)^2}-\frac{1}{3 (x-1)^2}+\frac{32}{3 \
(x-2)^3}-\frac{5}{(x-1)^3}-\frac{17}{2 (x-1)^4}-\frac{25}{2 (x-1)^5}+\
\frac{25}{2}\Big) H(2,c_2(\ao),c_1(\ao);x)+\Big(\frac{2 \ao^4}{x-1}-2 \
\ao^4-\frac{8 \ao^3}{x-1}+\frac{8 \ao^3}{3 (x-1)^2}+\frac{32 \
\ao^3}{3}+\frac{12 \ao^2}{x-1}-\frac{8 \ao^2}{(x-1)^2}+\frac{4 \
\ao^2}{(x-1)^3}-24 \ao^2-\frac{8 \ao}{x-1}+\frac{8 \
\ao}{(x-1)^2}-\frac{8 \ao}{(x-1)^3}+\frac{8 \ao}{(x-1)^4}+32 \
\ao+\frac{8 H(0;\ao)}{(x-1)^5}+\frac{4 d_1 \
H(1;\ao)}{(x-1)^5}-\frac{6}{x-1}+\frac{4}{3 (x-1)^2}+\frac{4}{3 \
(x-1)^3}-\frac{6}{(x-1)^4}-\frac{50}{3 (x-1)^5}-\frac{50}{3}\Big) \
H(c_1(\ao),0,c_1(\ao);x)+\Big(-\frac{7 \ao^4}{x-1}+7 \ao^4+\frac{28 \
\ao^3}{x-1}-\frac{28 \ao^3}{3 (x-1)^2}-\frac{112 \ao^3}{3}-\frac{42 \
\ao^2}{x-1}+\frac{28 \ao^2}{(x-1)^2}-\frac{14 \ao^2}{(x-1)^3}+84 \
\ao^2+\frac{28 \ao}{x-1}-\frac{28 \ao}{(x-1)^2}+\frac{28 \
\ao}{(x-1)^3}-\frac{28 \ao}{(x-1)^4}-112 \ao-\frac{24 \
H(0;\ao)}{(x-1)^5}-\frac{12 d_1 \
H(1;\ao)}{(x-1)^5}+\frac{21}{x-1}-\frac{14}{3 (x-1)^2}-\frac{14}{3 \
(x-1)^3}+\frac{21}{(x-1)^4}+\frac{175}{3 (x-1)^5}+\frac{175}{3}\Big) \
H(c_1(\ao),c_1(\ao),c_1(\ao);x)+\Big(\frac{3 \ao^4}{2 (x-1)}-\frac{3 \
\ao^4}{2}-\frac{6 \ao^3}{x-1}+\frac{2 \ao^3}{(x-1)^2}+8 \ao^3+\frac{9 \
\ao^2}{x-1}-\frac{6 \ao^2}{(x-1)^2}+\frac{3 \ao^2}{(x-1)^3}-18 \ao^2-\
\frac{6 \ao}{x-1}+\frac{6 \ao}{(x-1)^2}-\frac{6 \ao}{(x-1)^3}+\frac{6 \
\ao}{(x-1)^4}+24 \ao+\frac{8 H(0;\ao)}{(x-1)^5}+\frac{4 d_1 \
H(1;\ao)}{(x-1)^5}-\frac{9}{2 \
(x-1)}+\frac{1}{(x-1)^2}+\frac{1}{(x-1)^3}-\frac{9}{2 \
(x-1)^4}-\frac{25}{2 (x-1)^5}-\frac{25}{2}\Big) \
H(c_1(\ao),c_2(\ao),c_1(\ao);x)+\Big(-\frac{\ao^4}{x-1}+\ao^4+\frac{4 \
\ao^3}{x-1}-\frac{4 \ao^3}{3 (x-1)^2}-\frac{16 \ao^3}{3}-\frac{6 \
\ao^2}{x-1}+\frac{4 \ao^2}{(x-1)^2}-\frac{2 \ao^2}{(x-1)^3}+12 \ao^2+\
\frac{4 \ao}{x-1}-\frac{4 \ao}{(x-1)^2}+\frac{4 \ao}{(x-1)^3}-\frac{4 \
\ao}{(x-1)^4}-16 \ao+\frac{8}{x-2}-\frac{1}{x-1}-\frac{16}{3 \
(x-2)^2}-\frac{4}{3 (x-1)^2}+\frac{16}{3 \
(x-2)^3}-\frac{2}{(x-1)^3}-\frac{4}{(x-1)^4}+\frac{25}{3}\Big) H(c_2(\
\ao),0,c_1(\ao);x)+\Big(\frac{5 \ao^4}{2 (x-1)}-\frac{5 \
\ao^4}{2}-\frac{10 \ao^3}{x-1}+\frac{10 \ao^3}{3 (x-1)^2}+\frac{40 \
\ao^3}{3}+\frac{15 \ao^2}{x-1}-\frac{10 \ao^2}{(x-1)^2}+\frac{5 \
\ao^2}{(x-1)^3}-30 \ao^2-\frac{10 \ao}{x-1}+\frac{10 \
\ao}{(x-1)^2}-\frac{10 \ao}{(x-1)^3}+\frac{10 \ao}{(x-1)^4}+40 \
\ao-\frac{20}{x-2}+\frac{5}{2 (x-1)}+\frac{40}{3 (x-2)^2}+\frac{10}{3 \
(x-1)^2}-\frac{40}{3 \
(x-2)^3}+\frac{5}{(x-1)^3}+\frac{10}{(x-1)^4}-\frac{125}{6}\Big) \
H(c_2(\ao),c_1(\ao),c_1(\ao);x)+32 \
H(0,0,0,0;x)+\Big(\frac{12}{(x-1)^5}-12\Big) \
H(0,0,0,c_1(\ao);x)+\Big(\frac{4 d_1}{(x-1)^5}-4 \
d_1-\frac{12}{(x-1)^5}+12\Big) H(0,0,1,0;x)+\Big(-\frac{4 \
d_1}{(x-1)^5}+4 d_1+\frac{12}{(x-1)^5}-12\Big) \
H(0,0,1,c_1(\ao);x)+\Big(-12-\frac{4}{(x-1)^5}\Big) \
H(0,0,c_1(\ao),c_1(\ao);x)+\Big(4-\frac{4}{(x-1)^5}\Big) \
H(0,0,c_2(\ao),c_1(\ao);x)+\Big(\frac{8 d_1}{(x-1)^5}-8 \
d_1-\frac{8}{(x-1)^5}+8\Big) H(0,1,0,0;x)+\Big(-\frac{4 \
d_1}{(x-1)^5}-4 d_1+\frac{12}{(x-1)^5}+4\Big) \
H(0,1,0,c_1(\ao);x)+\Big(-\frac{4 d_1^2}{(x-1)^5}+4 d_1^2+\frac{8 \
d_1}{(x-1)^5}-\frac{12}{(x-1)^5}-4\Big) H(0,1,1,0;x)+\Big(\frac{4 \
d_1^2}{(x-1)^5}-4 d_1^2-\frac{8 \
d_1}{(x-1)^5}+\frac{12}{(x-1)^5}+4\Big) \
H(0,1,1,c_1(\ao);x)+\Big(-\frac{4 d_1}{(x-1)^5}+12 \
d_1-\frac{4}{(x-1)^5}-12\Big) H(0,1,c_1(\ao),c_1(\ao);x)+\Big(\frac{4 \
d_1}{(x-1)^5}-4 d_1-\frac{4}{(x-1)^5}+4\Big) \
H(0,2,0,c_1(\ao);x)+\Big(-\frac{4 d_1}{(x-1)^5}+4 \
d_1+\frac{4}{(x-1)^5}-4\Big) H(0,2,1,0;x)+\Big(\frac{4 \
d_1}{(x-1)^5}-4 d_1-\frac{4}{(x-1)^5}+4\Big) \
H(0,2,1,c_1(\ao);x)+\Big(-\frac{4 d_1}{(x-1)^5}+4 \
d_1+\frac{4}{(x-1)^5}-4\Big) H(0,2,c_2(\ao),c_1(\ao);x)+8 \
H(0,c_1(\ao),0,c_1(\ao);x)+\Big(\frac{4}{(x-1)^5}-28\Big) \
H(0,c_1(\ao),c_1(\ao),c_1(\ao);x)+\Big(6+\frac{2}{(x-1)^5}\Big) \
H(0,c_1(\ao),c_2(\ao),c_1(\ao);x)+\Big(\frac{4}{(x-1)^5}-4\Big) \
H(0,c_2(\ao),0,c_1(\ao);x)+\Big(10-\frac{10}{(x-1)^5}\Big) \
H(0,c_2(\ao),c_1(\ao),c_1(\ao);x)+\Big(16-\frac{16}{(x-1)^5}\Big) \
H(1,0,0,0;x)+\Big(\frac{8}{(x-1)^5}-\frac{4 d_1}{(x-1)^5}\Big) \
H(1,0,0,c_1(\ao);x)+\Big(-\frac{4 d_1^2}{(x-1)^5}+\frac{8 \
d_1}{(x-1)^5}+4 d_1-\frac{8}{(x-1)^5}\Big) H(1,0,1,0;x)+\Big(\frac{4 \
d_1^2}{(x-1)^5}-\frac{8 d_1}{(x-1)^5}-4 d_1+\frac{8}{(x-1)^5}\Big) \
H(1,0,1,c_1(\ao);x)+\Big(-\frac{4 \
d_1}{(x-1)^5}+\frac{4}{(x-1)^5}+4\Big) \
H(1,0,c_1(\ao),c_1(\ao);x)+\Big(\frac{4 \
d_1}{(x-1)^5}-\frac{2}{(x-1)^5}-6\Big) \
H(1,0,c_2(\ao),c_1(\ao);x)+\Big(\frac{16 d_1}{(x-1)^5}-8 \
d_1-\frac{8}{(x-1)^5}-8\Big) H(1,1,0,0;x)+\Big(\frac{4 \
d_1^2}{(x-1)^5}-\frac{8 d_1}{(x-1)^5}-4 d_1+\frac{4}{(x-1)^5}+4\Big) \
H(1,1,0,c_1(\ao);x)+\Big(-\frac{12 d_1^2}{(x-1)^5}+4 d_1^2+\frac{12 \
d_1}{(x-1)^5}+8 d_1-\frac{4}{(x-1)^5}-4\Big) \
H(1,1,1,0;x)+\Big(\frac{12 d_1^2}{(x-1)^5}-4 d_1^2-\frac{12 \
d_1}{(x-1)^5}-8 d_1+\frac{4}{(x-1)^5}+4\Big) \
H(1,1,1,c_1(\ao);x)+\Big(-\frac{4 d_1^2}{(x-1)^5}-\frac{8 \
d_1}{(x-1)^5}+12 d_1+\frac{4}{(x-1)^5}+4\Big) \
H(1,1,c_1(\ao),c_1(\ao);x)+\Big(\frac{4 \
d_1}{(x-1)^5}-\frac{2}{(x-1)^5}-6\Big) \
H(1,2,0,c_1(\ao);x)+\Big(-\frac{4 \
d_1}{(x-1)^5}+\frac{2}{(x-1)^5}+6\Big) H(1,2,1,0;x)+\Big(\frac{4 \
d_1}{(x-1)^5}-\frac{2}{(x-1)^5}-6\Big) \
H(1,2,1,c_1(\ao);x)+\Big(-\frac{4 \
d_1}{(x-1)^5}+\frac{2}{(x-1)^5}+6\Big) \
H(1,2,c_2(\ao),c_1(\ao);x)+\Big(8-\frac{4 d_1}{(x-1)^5}\Big) H(1,c_1(\
\ao),0,c_1(\ao);x)+\Big(\frac{12 \
d_1}{(x-1)^5}+\frac{4}{(x-1)^5}-28\Big) \
H(1,c_1(\ao),c_1(\ao),c_1(\ao);x)+\Big(-\frac{4 \
d_1}{(x-1)^5}+\frac{2}{(x-1)^5}+6\Big) \
H(1,c_1(\ao),c_2(\ao),c_1(\ao);x)+\Big(4-\frac{4}{(x-1)^5}\Big) \
H(2,0,0,c_1(\ao);x)+\Big(-\frac{4 d_1}{(x-1)^5}+4 \
d_1+\frac{4}{(x-1)^5}-4\Big) H(2,0,1,0;x)+\Big(\frac{4 \
d_1}{(x-1)^5}-4 d_1-\frac{4}{(x-1)^5}+4\Big) \
H(2,0,1,c_1(\ao);x)+\Big(10-\frac{10}{(x-1)^5}\Big) \
H(2,0,c_1(\ao),c_1(\ao);x)+\Big(\frac{8}{(x-1)^5}-8\Big) \
H(2,0,c_2(\ao),c_1(\ao);x)+\Big(\frac{8}{(x-1)^5}-8\Big) \
H(2,1,0,0;x)+\Big(\frac{2}{(x-1)^5}-2\Big) H(2,1,0,c_1(\ao);x)+\Big(-\
\frac{4 d_1}{(x-1)^5}+4 d_1-\frac{2}{(x-1)^5}+2\Big) \
H(2,1,1,0;x)+\Big(\frac{4 d_1}{(x-1)^5}-4 \
d_1+\frac{2}{(x-1)^5}-2\Big) \
H(2,1,1,c_1(\ao);x)+\Big(10-\frac{10}{(x-1)^5}\Big) \
H(2,1,c_1(\ao),c_1(\ao);x)+\Big(-\frac{4 d_1}{(x-1)^5}+4 \
d_1+\frac{8}{(x-1)^5}-8\Big) H(2,2,0,c_1(\ao);x)+\Big(\frac{4 \
d_1}{(x-1)^5}-4 d_1-\frac{8}{(x-1)^5}+8\Big) \
H(2,2,1,0;x)+\Big(-\frac{4 d_1}{(x-1)^5}+4 \
d_1+\frac{8}{(x-1)^5}-8\Big) H(2,2,1,c_1(\ao);x)+\Big(\frac{4 \
d_1}{(x-1)^5}-4 d_1-\frac{8}{(x-1)^5}+8\Big) \
H(2,2,c_2(\ao),c_1(\ao);x)+\Big(4-\frac{4}{(x-1)^5}\Big) \
H(2,c_2(\ao),0,c_1(\ao);x)+\Big(\frac{10}{(x-1)^5}-10\Big) \
H(2,c_2(\ao),c_1(\ao),c_1(\ao);x)-\frac{4 \
H(c_1(\ao),0,0,c_1(\ao);x)}{(x-1)^5}+\frac{12 \
H(c_1(\ao),0,c_1(\ao),c_1(\ao);x)}{(x-1)^5}-\frac{4 H(c_1(\ao),0,c_2(\
\ao),c_1(\ao);x)}{(x-1)^5}+\frac{8 \
H(c_1(\ao),c_1(\ao),0,c_1(\ao);x)}{(x-1)^5}-\frac{28 \
H(c_1(\ao),c_1(\ao),c_1(\ao),c_1(\ao);x)}{(x-1)^5}+\frac{6 \
H(c_1(\ao),c_1(\ao),c_2(\ao),c_1(\ao);x)}{(x-1)^5}-\frac{4 \
H(c_1(\ao),c_2(\ao),0,c_1(\ao);x)}{(x-1)^5}+\frac{10 \
H(c_1(\ao),c_2(\ao),c_1(\ao),c_1(\ao);x)}{(x-1)^5}+H(0;x) \
\Big(-\frac{63 d_1^2}{16 (x-1)}+\frac{19 d_1^2}{54 (x-1)^2}+\frac{19 \
d_1^2}{54 (x-1)^3}-\frac{63 d_1^2}{16 (x-1)^4}-\frac{2035 d_1^2}{432 \
(x-1)^5}-\frac{2035 d_1^2}{432}+\frac{38 d_1}{3 (x-2)}+\frac{85 \
d_1}{24 (x-1)}-\frac{76 d_1}{9 (x-2)^2}-\frac{317 d_1}{108 \
(x-1)^2}+\frac{313 d_1}{108 (x-1)^3}+\frac{257 d_1}{8 \
(x-1)^4}+\frac{8705 d_1}{216 (x-1)^5}+\frac{5615 d_1}{216}-\frac{4 \
\pi ^2}{x-2}-\frac{68}{3 (x-2)}+\frac{13 \pi ^2}{8 (x-1)}+\frac{37}{4 \
(x-1)}+\frac{8 \pi ^2}{3 (x-2)^2}+\frac{152}{9 (x-2)^2}+\frac{5 \pi \
^2}{12 (x-1)^2}-\frac{49}{108 (x-1)^2}-\frac{8 \pi ^2}{3 \
(x-2)^3}+\frac{3 \pi ^2}{4 (x-1)^3}-\frac{949}{108 (x-1)^3}+\frac{25 \
\pi ^2}{8 (x-1)^4}-\frac{213}{4 (x-1)^4}+\frac{25 \pi ^2}{8 (x-1)^5}-\
\frac{3965}{54 (x-1)^5}+\frac{3 \zeta_3}{(x-1)^5}+16 \zeta_3+\frac{2 \
\pi ^2 \ln 2\, }{(x-1)^5}-2 \pi ^2 \ln 2\, -\frac{25 \pi \
^2}{24}-\frac{940}{27}\Big)+H(2;x) \Big(-\frac{2 \pi ^2 \
d_1}{x-2}+\frac{4 \pi ^2 d_1}{3 (x-2)^2}+\frac{2 \pi ^2 d_1}{3 \
(x-1)^2}-\frac{4 \pi ^2 d_1}{3 (x-2)^3}+\frac{2 \pi ^2 d_1}{(x-1)^4}+\
\frac{4 \pi ^2}{x-2}-\frac{15 \pi ^2}{8 (x-1)}-\frac{8 \pi ^2}{3 \
(x-2)^2}-\frac{\pi ^2}{12 (x-1)^2}+\frac{8 \pi ^2}{3 (x-2)^3}-\frac{5 \
\pi ^2}{4 (x-1)^3}-\frac{17 \pi ^2}{8 (x-1)^4}-\frac{25 \pi ^2}{8 \
(x-1)^5}+\frac{7 \zeta_3}{2 (x-1)^5}-\frac{7 \zeta_3}{2}-\frac{3 \pi \
^2 \ln 2\, }{(x-1)^5}+3 \pi ^2 \ln 2\, +\frac{25 \pi \
^2}{8}\Big)+H(1;x) \Big(-\frac{2 \pi ^2 d_1}{3 (x-1)}+\frac{\pi ^2 \
d_1}{3 (x-1)^2}-\frac{2 \pi ^2 d_1}{9 (x-1)^3}+\frac{\pi ^2 d_1}{6 \
(x-1)^4}-\frac{25 \pi ^2 d_1}{18 (x-1)^5}-\frac{3 \zeta_3 d_1}{2 \
(x-1)^5}-\frac{\pi ^2 \ln 2\,  d_1}{(x-1)^5}+\Big(-\frac{4 \
d_1^2}{x-1}+\frac{d_1^2}{(x-1)^2}-\frac{4 d_1^2}{9 \
(x-1)^3}+\frac{d_1^2}{4 (x-1)^4}-\frac{205 d_1^2}{36 \
(x-1)^5}+\frac{46 d_1}{3 (x-2)}+\frac{3 d_1}{4 (x-1)}-\frac{52 d_1}{9 \
(x-2)^2}-\frac{185 d_1}{18 (x-1)^2}+\frac{28 d_1}{9 (x-2)^3}+\frac{29 \
d_1}{18 (x-1)^3}-\frac{27 d_1}{4 (x-1)^4}+\frac{835 d_1}{36 (x-1)^5}+\
\frac{205 d_1}{36}-\frac{80}{3 (x-2)}+\frac{70}{3 (x-1)}+\frac{56}{9 \
(x-2)^2}+\frac{13}{18 (x-1)^2}-\frac{56}{9 (x-2)^3}+\frac{355}{18 \
(x-1)^3}+\frac{10}{3 (x-1)^4}-\frac{\pi ^2}{6 (x-1)^5}+\frac{155}{9 \
(x-1)^5}+\frac{\pi ^2}{6}-\frac{155}{9}\Big) H(0;\ao)+\Big(-\frac{8 \
d_1}{x-1}+\frac{4 d_1}{(x-1)^2}-\frac{8 d_1}{3 (x-1)^3}+\frac{2 \
d_1}{(x-1)^4}-\frac{50 d_1}{3 (x-1)^5}+\frac{16}{x-2}-\frac{10}{x-1}-\
\frac{32}{3 (x-2)^2}+\frac{4}{3 (x-1)^2}+\frac{32}{3 \
(x-2)^3}-\frac{20}{3 (x-1)^3}-\frac{6}{(x-1)^4}-\frac{50}{3 (x-1)^5}+\
\frac{50}{3}\Big) H(0,0;\ao)+\Big(-\frac{4 d_1^2}{x-1}+\frac{2 \
d_1^2}{(x-1)^2}-\frac{4 d_1^2}{3 \
(x-1)^3}+\frac{d_1^2}{(x-1)^4}-\frac{25 d_1^2}{3 (x-1)^5}+\frac{8 \
d_1}{x-2}-\frac{5 d_1}{x-1}-\frac{16 d_1}{3 (x-2)^2}+\frac{2 d_1}{3 \
(x-1)^2}+\frac{16 d_1}{3 (x-2)^3}-\frac{10 d_1}{3 (x-1)^3}-\frac{3 \
d_1}{(x-1)^4}-\frac{25 d_1}{3 (x-1)^5}+\frac{25 d_1}{3}\Big) \
H(0,1;\ao)+\Big(\frac{16}{(x-1)^5}-16\Big) H(0,0,0;\ao)+\Big(\frac{8 \
d_1}{(x-1)^5}-8 d_1\Big) H(0,0,1;\ao)+\Big(\frac{8 d_1}{(x-1)^5}-8 \
d_1\Big) H(0,1,0;\ao)+\Big(\frac{4 d_1^2}{(x-1)^5}-4 d_1^2\Big) \
H(0,1,1;\ao)+\frac{2 \pi ^2}{3 (x-2)}+\frac{\pi ^2}{4 (x-1)}-\frac{4 \
\pi ^2}{9 (x-2)^2}+\frac{4 \pi ^2}{9 (x-2)^3}-\frac{11 \pi ^2}{18 \
(x-1)^3}+\frac{5 \pi ^2}{6 (x-1)^4}+\frac{25 \pi ^2}{18 \
(x-1)^5}-\frac{21 \zeta_3}{4 (x-1)^5}+\frac{33 \zeta_3}{4}+\frac{\pi \
^2 \ln 2\, }{2 (x-1)^5}+\frac{3}{2} \pi ^2 \ln 2\, +\frac{125 \pi \
^2}{36}\Big)-\frac{8 d_1 \pi ^2}{3 (x-2)}+\frac{37 \pi ^2}{6 \
(x-2)}+\frac{65 d_1 \pi ^2}{48 (x-1)}-\frac{151 \pi ^2}{48 \
(x-1)}+\frac{7 d_1 \pi ^2}{9 (x-2)^2}-\frac{10 \pi ^2}{9 \
(x-2)^2}+\frac{37 d_1 \pi ^2}{54 (x-1)^2}-\frac{847 \pi ^2}{432 \
(x-1)^2}-\frac{7 d_1 \pi ^2}{9 (x-2)^3}+\frac{14 \pi ^2}{9 \
(x-2)^3}+\frac{125 d_1 \pi ^2}{108 (x-1)^3}-\frac{1441 \pi ^2}{432 \
(x-1)^3}+\frac{13 d_1 \pi ^2}{8 (x-1)^4}-\frac{109 \pi ^2}{24 \
(x-1)^4}-\frac{241 \pi ^4}{720 (x-1)^5}+\frac{205 d_1 \pi ^2}{216 \
(x-1)^5}-\frac{155 \pi ^2}{54 (x-1)^5}-\frac{7 \zeta_3}{x-2}-\frac{85 \
\zeta_3}{16 (x-1)}+\frac{14 \zeta_3}{3 (x-2)^2}+\frac{61 \zeta_3}{24 \
(x-1)^2}-\frac{14 \zeta_3}{3 (x-2)^3}+\frac{25 \zeta_3}{8 \
(x-1)^3}-\frac{43 \zeta_3}{16 (x-1)^4}-\frac{275 \zeta_3}{16 \
(x-1)^5}-\frac{1175 \zeta_3}{48}-\frac{4 \text{Li}_4\frac{1}{2}}{(x-1)^5}+4 \text{Li}_4\frac{1}{2}-\frac{\ln ^42\, }{6 (x-1)^5}+\frac{\ln \
^42\, }{6}-\frac{4 \pi ^2 \ln ^22\, }{3 (x-1)^5}+\frac{4}{3} \pi ^2 \
\ln ^22\, +\frac{6 \pi ^2 \ln 2\, }{x-2}-\frac{15 \pi ^2 \ln 2\, }{8 \
(x-1)}-\frac{4 \pi ^2 \ln 2\, }{(x-2)^2}-\frac{3 \pi ^2 \ln 2\, }{4 \
(x-1)^2}+\frac{4 \pi ^2 \ln 2\, }{(x-2)^3}-\frac{5 \pi ^2 \ln 2\, }{4 \
(x-1)^3}-\frac{33 \pi ^2 \ln 2\, }{8 (x-1)^4}-\frac{25 \pi ^2 \ln 2\, \
}{8 (x-1)^5}+\frac{25}{8} \pi ^2 \ln 2\, +\frac{\pi ^4}{30}-\frac{205 \
d_1 \pi ^2}{432}+\frac{305 \pi ^2}{432}.
\erp

%
% The B integral for k=0 and delta=1
%

\subsection{The $\cB$ integral for $k=0$ and $\delta=1$}
%
% This file contains the TeX output produced by Mathematica for the integral B0, for delta = -1
%
The $\eps$ expansion for this integral reads
\beq
\bsp
\begin{cal}I\end{cal}(x,\eps;\ao,3+d_1\eps;1,0,1,g_B) &= x\,\bint(\eps,x;3+d_1\eps;1,0)\\
&=\frac{1}{\eps}b_{-1}^{(1,0)}+b_0^{(1,0)}+\eps b_1^{(1,0)}+\eps^2b_2^{(1,0)} +\ocal\left(\eps^3\right),
\esp
\eeq
where
%1/ep piece
\brp
b_{-1}^{(1,0)}=-\frac{1}{2},
\erp
% ep^0
\brp
b_0^{(1,0)}=\frac{\ao^4}{4 \
(x-1)}+\frac{\ao^4}{4}-\frac{\ao^3}{x-1}+\frac{\ao^3}{3 \
(x-1)^2}-\frac{4 \ao^3}{3}+\frac{3 \ao^2}{2 \
(x-1)}-\frac{\ao^2}{(x-1)^2}+\frac{\ao^2}{2 (x-1)^3}+3 \
\ao^2-\frac{\ao}{x-1}+\frac{\ao}{(x-1)^2}-\frac{\ao}{(x-1)^3}+\frac{\ao}{(x-1)^4}-4 \ao+\Big(1+\frac{1}{(x-1)^5}\Big) \
H(0;\ao)+\Big(1-\frac{1}{(x-1)^5}\Big) \
H(0;x)+\frac{H(c_1(\ao);x)}{(x-1)^5}-1,
\erp
% ep^1
\brp
b_1^{(1,0)}=-\frac{d_1 \ao^4}{8}-\frac{d_1 \ao^4}{8 (x-1)}+\frac{3 \ao^4}{4 \
(x-1)}+\frac{3 \ao^4}{4}+\frac{13 d_1 \ao^3}{18}+\frac{d_1 \ao^3}{2 \
(x-1)}-\frac{3 \ao^3}{x-1}-\frac{2 d_1 \ao^3}{9 (x-1)^2}+\frac{23 \
\ao^3}{18 (x-1)^2}-\frac{77 \ao^3}{18}-\frac{23 d_1 \
\ao^2}{12}-\frac{3 d_1 \ao^2}{4 (x-1)}+\frac{53 \ao^2}{12 \
(x-1)}+\frac{2 d_1 \ao^2}{3 (x-1)^2}-\frac{47 \ao^2}{12 \
(x-1)^2}-\frac{d_1 \ao^2}{2 (x-1)^3}+\frac{31 \ao^2}{12 \
(x-1)^3}+\frac{131 \ao^2}{12}+\frac{25 d_1 \ao}{6}+\frac{d_1 \ao}{2 \
(x-1)}-\frac{7 \ao}{3 (x-1)}-\frac{2 d_1 \ao}{3 (x-1)^2}+\frac{4 \
\ao}{(x-1)^2}+\frac{d_1 \ao}{(x-1)^3}-\frac{17 \ao}{3 \
(x-1)^3}-\frac{2 d_1 \ao}{(x-1)^4}+\frac{49 \ao}{6 (x-1)^4}-\frac{121 \
\ao}{6}+\Big(-\frac{\ao^4}{x-1}-\ao^4+\frac{4 \ao^3}{x-1}-\frac{4 \
\ao^3}{3 (x-1)^2}+\frac{16 \ao^3}{3}-\frac{6 \ao^2}{x-1}+\frac{4 \
\ao^2}{(x-1)^2}-\frac{2 \ao^2}{(x-1)^3}-12 \ao^2+\frac{4 \
\ao}{x-1}-\frac{4 \ao}{(x-1)^2}+\frac{4 \ao}{(x-1)^3}-\frac{4 \
\ao}{(x-1)^4}+16 \ao-\frac{5}{2 (x-1)}+\frac{5}{3 (x-1)^2}-\frac{5}{3 \
(x-1)^3}+\frac{5}{2 (x-1)^4}+\frac{37}{6 (x-1)^5}-\frac{13}{6}\Big) \
H(0;\ao)+\Big(\frac{37}{6}+\frac{5}{2 (x-1)}-\frac{5}{3 \
(x-1)^2}+\frac{5}{3 (x-1)^3}-\frac{5}{2 (x-1)^4}-\frac{37}{6 (x-1)^5}\
\Big) H(0;x)+\Big(-\frac{d_1 \ao^4}{2}-\frac{d_1 \ao^4}{2 \
(x-1)}+\frac{8 d_1 \ao^3}{3}+\frac{2 d_1 \ao^3}{x-1}-\frac{2 d_1 \
\ao^3}{3 (x-1)^2}-6 d_1 \ao^2-\frac{3 d_1 \ao^2}{x-1}+\frac{2 d_1 \
\ao^2}{(x-1)^2}-\frac{d_1 \ao^2}{(x-1)^3}+8 d_1 \ao+\frac{2 d_1 \
\ao}{x-1}-\frac{2 d_1 \ao}{(x-1)^2}+\frac{2 d_1 \ao}{(x-1)^3}-\frac{2 \
d_1 \ao}{(x-1)^4}-\frac{25 d_1}{6}-\frac{d_1}{2 (x-1)}+\frac{2 d_1}{3 \
(x-1)^2}-\frac{d_1}{(x-1)^3}+\frac{2 d_1}{(x-1)^4}\Big) \
H(1;\ao)+\Big(\frac{2 d_1}{(x-1)^5}-\frac{2}{(x-1)^5}+2\Big) H(0;\ao) \
H(1;x)+\Big(-\frac{\ao^4}{2 (x-1)}-\frac{\ao^4}{2}+\frac{2 \
\ao^3}{x-1}-\frac{2 \ao^3}{3 (x-1)^2}+\frac{8 \ao^3}{3}-\frac{3 \
\ao^2}{x-1}+\frac{2 \ao^2}{(x-1)^2}-\frac{\ao^2}{(x-1)^3}-6 \
\ao^2+\frac{2 \ao}{x-1}-\frac{2 \ao}{(x-1)^2}+\frac{2 \
\ao}{(x-1)^3}-\frac{2 \ao}{(x-1)^4}+8 \ao-\frac{4 H(0;\ao)}{(x-1)^5}-\
\frac{2 d_1 H(1;\ao)}{(x-1)^5}-\frac{5}{2 (x-1)}+\frac{5}{3 (x-1)^2}-\
\frac{5}{3 (x-1)^3}+\frac{5}{2 (x-1)^4}+\frac{37}{6 \
(x-1)^5}-\frac{25}{6}\Big) \
H(c_1(\ao);x)+\Big(-4-\frac{4}{(x-1)^5}\Big) \
H(0,0;\ao)+\Big(\frac{4}{(x-1)^5}-4\Big) H(0,0;x)+\Big(-\frac{2 \
d_1}{(x-1)^5}-2 d_1\Big) H(0,1;\ao)+\Big(2-\frac{2}{(x-1)^5}\Big) \
H(0,c_1(\ao);x)+\Big(-\frac{2 d_1}{(x-1)^5}+\frac{2}{(x-1)^5}-2\Big) \
H(1,0;x)+\Big(\frac{2 d_1}{(x-1)^5}-\frac{2}{(x-1)^5}+2\Big) H(1,c_1(\
\ao);x)-\frac{2 H(c_1(\ao),c_1(\ao);x)}{(x-1)^5}+\frac{\pi ^2}{3 \
(x-1)^5}-\frac{\pi ^2}{3}-2,
\erp
% ep^2
\brp
b_2^{(1,0)}=\frac{d_1^2 \ao^4}{16}-\frac{d_1 \ao^4}{2}+\frac{d_1^2 \ao^4}{16 \
(x-1)}-\frac{d_1 \ao^4}{2 (x-1)}-\frac{\pi ^2 \ao^4}{24 \
(x-1)}+\frac{7 \ao^4}{4 (x-1)}-\frac{\pi ^2 \ao^4}{24}+\frac{7 \
\ao^4}{4}-\frac{43 d_1^2 \ao^3}{108}+\frac{349 d_1 \
\ao^3}{108}-\frac{d_1^2 \ao^3}{4 (x-1)}+\frac{2 d_1 \
\ao^3}{x-1}+\frac{\pi ^2 \ao^3}{6 (x-1)}-\frac{7 \ao^3}{x-1}+\frac{4 \
d_1^2 \ao^3}{27 (x-1)^2}-\frac{133 d_1 \ao^3}{108 (x-1)^2}-\frac{\pi \
^2 \ao^3}{18 (x-1)^2}+\frac{191 \ao^3}{54 (x-1)^2}+\frac{2 \pi ^2 \
\ao^3}{9}-\frac{569 \ao^3}{54}+\frac{95 d_1^2 \ao^2}{72}-\frac{85 d_1 \
\ao^2}{8}+\frac{3 d_1^2 \ao^2}{8 (x-1)}-\frac{203 d_1 \ao^2}{72 \
(x-1)}-\frac{\pi ^2 \ao^2}{4 (x-1)}+\frac{353 \ao^2}{36 \
(x-1)}-\frac{4 d_1^2 \ao^2}{9 (x-1)^2}+\frac{31 d_1 \ao^2}{8 \
(x-1)^2}+\frac{\pi ^2 \ao^2}{6 (x-1)^2}-\frac{407 \ao^2}{36 (x-1)^2}+\
\frac{d_1^2 \ao^2}{2 (x-1)^3}-\frac{283 d_1 \ao^2}{72 (x-1)^3}-\frac{\
\pi ^2 \ao^2}{12 (x-1)^3}+\frac{331 \ao^2}{36 (x-1)^3}-\frac{\pi ^2 \
\ao^2}{2}+\frac{1091 \ao^2}{36}-\frac{205 d_1^2 \ao}{36}+\frac{475 \
d_1 \ao}{12}+\frac{2 \ao}{3 (x-2)}-\frac{d_1^2 \ao}{4 \
(x-1)}-\frac{d_1 \ao}{9 (x-1)}+\frac{\pi ^2 \ao}{6 (x-1)}-\frac{13 \
\ao}{36 (x-1)}+\frac{4 d_1^2 \ao}{9 (x-1)^2}-\frac{73 d_1 \ao}{18 \
(x-1)^2}-\frac{\pi ^2 \ao}{6 (x-1)^2}+\frac{35 \ao}{3 \
(x-1)^2}-\frac{d_1^2 \ao}{(x-1)^3}+\frac{173 d_1 \ao}{18 \
(x-1)^3}+\frac{\pi ^2 \ao}{6 (x-1)^3}-\frac{875 \ao}{36 \
(x-1)^3}+\frac{4 d_1^2 \ao}{(x-1)^4}-\frac{937 d_1 \ao}{36 \
(x-1)^4}-\frac{\pi ^2 \ao}{6 (x-1)^4}+\frac{413 \ao}{9 \
(x-1)^4}+\frac{2 \pi ^2 \ao}{3}-\frac{737 \ao}{9}+\Big(\frac{d_1 \
\ao^4}{2}+\frac{d_1 \ao^4}{2 (x-1)}-\frac{3 \ao^4}{x-1}-3 \
\ao^4-\frac{26 d_1 \ao^3}{9}-\frac{2 d_1 \ao^3}{x-1}+\frac{12 \
\ao^3}{x-1}+\frac{8 d_1 \ao^3}{9 (x-1)^2}-\frac{46 \ao^3}{9 (x-1)^2}+\
\frac{154 \ao^3}{9}+\frac{23 d_1 \ao^2}{3}+\frac{3 d_1 \
\ao^2}{x-1}-\frac{53 \ao^2}{3 (x-1)}-\frac{8 d_1 \ao^2}{3 \
(x-1)^2}+\frac{47 \ao^2}{3 (x-1)^2}+\frac{2 d_1 \
\ao^2}{(x-1)^3}-\frac{31 \ao^2}{3 (x-1)^3}-\frac{131 \
\ao^2}{3}-\frac{50 d_1 \ao}{3}-\frac{2 d_1 \ao}{x-1}+\frac{28 \ao}{3 \
(x-1)}+\frac{8 d_1 \ao}{3 (x-1)^2}-\frac{16 \ao}{(x-1)^2}-\frac{4 d_1 \
\ao}{(x-1)^3}+\frac{68 \ao}{3 (x-1)^3}+\frac{8 d_1 \
\ao}{(x-1)^4}-\frac{98 \ao}{3 (x-1)^4}+\frac{242 \ao}{3}+\frac{205 \
d_1}{36}-\frac{4}{x-2}+\frac{17 d_1}{4 (x-1)}-\frac{53}{4 \
(x-1)}+\frac{8}{3 (x-2)^2}-\frac{13 d_1}{9 (x-1)^2}+\frac{317}{36 \
(x-1)^2}+\frac{13 d_1}{9 (x-1)^3}-\frac{371}{36 (x-1)^3}-\frac{17 \
d_1}{4 (x-1)^4}+\frac{193}{12 (x-1)^4}-\frac{205 d_1}{36 \
(x-1)^5}-\frac{\pi ^2}{6 (x-1)^5}+\frac{266}{9 (x-1)^5}-\frac{\pi \
^2}{6}-\frac{194}{9}\Big) H(0;\ao)+\Big(-\frac{17 d_1}{4 \
(x-1)}+\frac{13 d_1}{9 (x-1)^2}-\frac{13 d_1}{9 (x-1)^3}+\frac{17 \
d_1}{4 (x-1)^4}+\frac{205 d_1}{36 (x-1)^5}-\frac{205 \
d_1}{36}+\frac{4}{x-2}+\frac{53}{4 (x-1)}-\frac{8}{3 \
(x-2)^2}-\frac{317}{36 (x-1)^2}+\frac{371}{36 (x-1)^3}-\frac{193}{12 \
(x-1)^4}-\frac{7 \pi ^2}{6 (x-1)^5}-\frac{266}{9 (x-1)^5}+\frac{3 \pi \
^2}{2}+\frac{266}{9}\Big) H(0;x)+\Big(\frac{d_1^2 \ao^4}{4}-\frac{3 \
d_1 \ao^4}{2}+\frac{d_1^2 \ao^4}{4 (x-1)}-\frac{3 d_1 \ao^4}{2 \
(x-1)}-\frac{13 d_1^2 \ao^3}{9}+\frac{77 d_1 \ao^3}{9}-\frac{d_1^2 \
\ao^3}{x-1}+\frac{6 d_1 \ao^3}{x-1}+\frac{4 d_1^2 \ao^3}{9 \
(x-1)^2}-\frac{23 d_1 \ao^3}{9 (x-1)^2}+\frac{23 d_1^2 \
\ao^2}{6}-\frac{131 d_1 \ao^2}{6}+\frac{3 d_1^2 \ao^2}{2 \
(x-1)}-\frac{53 d_1 \ao^2}{6 (x-1)}-\frac{4 d_1^2 \ao^2}{3 \
(x-1)^2}+\frac{47 d_1 \ao^2}{6 (x-1)^2}+\frac{d_1^2 \
\ao^2}{(x-1)^3}-\frac{31 d_1 \ao^2}{6 (x-1)^3}-\frac{25 d_1^2 \
\ao}{3}+\frac{121 d_1 \ao}{3}-\frac{d_1^2 \ao}{x-1}+\frac{14 d_1 \
\ao}{3 (x-1)}+\frac{4 d_1^2 \ao}{3 (x-1)^2}-\frac{8 d_1 \
\ao}{(x-1)^2}-\frac{2 d_1^2 \ao}{(x-1)^3}+\frac{34 d_1 \ao}{3 \
(x-1)^3}+\frac{4 d_1^2 \ao}{(x-1)^4}-\frac{49 d_1 \ao}{3 \
(x-1)^4}+\frac{205 d_1^2}{36}-\frac{230 d_1}{9}+\frac{d_1^2}{4 \
(x-1)}-\frac{d_1}{3 (x-1)}-\frac{4 d_1^2}{9 (x-1)^2}+\frac{49 d_1}{18 \
(x-1)^2}+\frac{d_1^2}{(x-1)^3}-\frac{37 d_1}{6 (x-1)^3}-\frac{4 \
d_1^2}{(x-1)^4}+\frac{49 d_1}{3 (x-1)^4}\Big) H(1;\ao)+\Big(\frac{\pi \
^2}{2 (x-1)^5}-\frac{\pi ^2}{2}\Big) H(2;x)+\Big(\frac{4 \
\ao^4}{x-1}+4 \ao^4-\frac{16 \ao^3}{x-1}+\frac{16 \ao^3}{3 \
(x-1)^2}-\frac{64 \ao^3}{3}+\frac{24 \ao^2}{x-1}-\frac{16 \
\ao^2}{(x-1)^2}+\frac{8 \ao^2}{(x-1)^3}+48 \ao^2-\frac{16 \
\ao}{x-1}+\frac{16 \ao}{(x-1)^2}-\frac{16 \ao}{(x-1)^3}+\frac{16 \
\ao}{(x-1)^4}-64 \ao+\frac{10}{x-1}-\frac{20}{3 (x-1)^2}+\frac{20}{3 \
(x-1)^3}-\frac{10}{(x-1)^4}-\frac{74}{3 (x-1)^5}+\frac{26}{3}\Big) \
H(0,0;\ao)+\Big(-\frac{74}{3}-\frac{10}{x-1}+\frac{20}{3 \
(x-1)^2}-\frac{20}{3 (x-1)^3}+\frac{10}{(x-1)^4}+\frac{74}{3 (x-1)^5}\
\Big) H(0,0;x)+\Big(2 d_1 \ao^4+\frac{2 d_1 \ao^4}{x-1}-\frac{32 d_1 \
\ao^3}{3}-\frac{8 d_1 \ao^3}{x-1}+\frac{8 d_1 \ao^3}{3 (x-1)^2}+24 \
d_1 \ao^2+\frac{12 d_1 \ao^2}{x-1}-\frac{8 d_1 \
\ao^2}{(x-1)^2}+\frac{4 d_1 \ao^2}{(x-1)^3}-32 d_1 \ao-\frac{8 d_1 \
\ao}{x-1}+\frac{8 d_1 \ao}{(x-1)^2}-\frac{8 d_1 \ao}{(x-1)^3}+\frac{8 \
d_1 \ao}{(x-1)^4}+\frac{13 d_1}{3}+\frac{5 d_1}{x-1}-\frac{10 d_1}{3 \
(x-1)^2}+\frac{10 d_1}{3 (x-1)^3}-\frac{5 d_1}{(x-1)^4}-\frac{37 \
d_1}{3 (x-1)^5}\Big) H(0,1;\ao)+H(1;x) \Big(\frac{2 \pi ^2 d_1}{3 \
(x-1)^5}+\Big(-\frac{4 d_1}{x-1}+\frac{2 d_1}{(x-1)^2}-\frac{4 d_1}{3 \
(x-1)^3}+\frac{d_1}{(x-1)^4}+\frac{37 d_1}{3 \
(x-1)^5}+\frac{4}{x-2}+\frac{5}{2 (x-1)}-\frac{8}{3 \
(x-2)^2}-\frac{3}{(x-1)^2}+\frac{8}{3 (x-2)^3}+\frac{5}{3 \
(x-1)^3}-\frac{13}{2 (x-1)^4}-\frac{37}{3 (x-1)^5}+\frac{37}{3}\Big) \
H(0;\ao)+\Big(-\frac{8 d_1}{(x-1)^5}+\frac{8}{(x-1)^5}-8\Big) \
H(0,0;\ao)+\Big(-\frac{4 d_1^2}{(x-1)^5}+\frac{4 d_1}{(x-1)^5}-4 \
d_1\Big) H(0,1;\ao)-\frac{2 \pi ^2}{3 (x-1)^5}+\frac{\pi ^2}{3}\Big)+\
\Big(-\frac{4 d_1}{(x-1)^5}+4 d_1+\frac{8}{(x-1)^5}-8\Big) H(0;\ao) \
H(0,1;x)+\Big(\Big(\frac{8}{(x-1)^5}-8\Big) H(0;\ao)+\Big(\frac{4 \
d_1}{(x-1)^5}-4 d_1\Big) H(1;\ao)+\frac{4}{x-2}+\frac{5}{2 \
(x-1)}-\frac{8}{3 (x-2)^2}-\frac{3}{(x-1)^2}+\frac{8}{3 \
(x-2)^3}+\frac{5}{3 (x-1)^3}-\frac{13}{2 (x-1)^4}-\frac{37}{3 \
(x-1)^5}+\frac{37}{3}\Big) H(0,c_1(\ao);x)+\Big(2 d_1 \ao^4+\frac{2 \
d_1 \ao^4}{x-1}-\frac{32 d_1 \ao^3}{3}-\frac{8 d_1 \
\ao^3}{x-1}+\frac{8 d_1 \ao^3}{3 (x-1)^2}+24 d_1 \ao^2+\frac{12 d_1 \
\ao^2}{x-1}-\frac{8 d_1 \ao^2}{(x-1)^2}+\frac{4 d_1 \
\ao^2}{(x-1)^3}-32 d_1 \ao-\frac{8 d_1 \ao}{x-1}+\frac{8 d_1 \
\ao}{(x-1)^2}-\frac{8 d_1 \ao}{(x-1)^3}+\frac{8 d_1 \
\ao}{(x-1)^4}+\frac{50 d_1}{3}+\frac{2 d_1}{x-1}-\frac{8 d_1}{3 \
(x-1)^2}+\frac{4 d_1}{(x-1)^3}-\frac{8 d_1}{(x-1)^4}\Big) H(1,0;\ao)+\
\Big(\frac{4 d_1}{x-1}-\frac{2 d_1}{(x-1)^2}+\frac{4 d_1}{3 (x-1)^3}-\
\frac{d_1}{(x-1)^4}-\frac{37 d_1}{3 (x-1)^5}-\frac{4}{x-2}-\frac{5}{2 \
(x-1)}+\frac{8}{3 (x-2)^2}+\frac{3}{(x-1)^2}-\frac{8}{3 \
(x-2)^3}-\frac{5}{3 (x-1)^3}+\frac{13}{2 (x-1)^4}+\frac{37}{3 \
(x-1)^5}-\frac{37}{3}\Big) H(1,0;x)+\Big(d_1^2 \ao^4+\frac{d_1^2 \
\ao^4}{x-1}-\frac{16 d_1^2 \ao^3}{3}-\frac{4 d_1^2 \
\ao^3}{x-1}+\frac{4 d_1^2 \ao^3}{3 (x-1)^2}+12 d_1^2 \ao^2+\frac{6 \
d_1^2 \ao^2}{x-1}-\frac{4 d_1^2 \ao^2}{(x-1)^2}+\frac{2 d_1^2 \
\ao^2}{(x-1)^3}-16 d_1^2 \ao-\frac{4 d_1^2 \ao}{x-1}+\frac{4 d_1^2 \
\ao}{(x-1)^2}-\frac{4 d_1^2 \ao}{(x-1)^3}+\frac{4 d_1^2 \
\ao}{(x-1)^4}+\frac{25 d_1^2}{3}+\frac{d_1^2}{x-1}-\frac{4 d_1^2}{3 \
(x-1)^2}+\frac{2 d_1^2}{(x-1)^3}-\frac{4 d_1^2}{(x-1)^4}\Big) \
H(1,1;\ao)+H(c_1(\ao);x) \Big(\frac{d_1 \ao^4}{4}+\frac{d_1 \ao^4}{4 \
(x-1)}-\frac{3 \ao^4}{2 (x-1)}-\frac{3 \ao^4}{2}-\frac{13 d_1 \
\ao^3}{9}-\frac{d_1 \ao^3}{x-1}+\frac{6 \ao^3}{x-1}+\frac{4 d_1 \
\ao^3}{9 (x-1)^2}-\frac{23 \ao^3}{9 (x-1)^2}+\frac{77 \
\ao^3}{9}+\frac{23 d_1 \ao^2}{6}-\frac{\ao^2}{3 (x-2)}+\frac{3 d_1 \
\ao^2}{2 (x-1)}-\frac{103 \ao^2}{12 (x-1)}-\frac{4 d_1 \ao^2}{3 \
(x-1)^2}+\frac{95 \ao^2}{12 (x-1)^2}+\frac{d_1 \
\ao^2}{(x-1)^3}-\frac{31 \ao^2}{6 (x-1)^3}-\frac{131 \
\ao^2}{6}-\frac{25 d_1 \ao}{3}+\frac{2 \ao}{x-2}-\frac{d_1 \ao}{x-1}+\
\frac{10 \ao}{3 (x-1)}-\frac{4 \ao}{3 (x-2)^2}+\frac{4 d_1 \ao}{3 \
(x-1)^2}-\frac{15 \ao}{2 (x-1)^2}-\frac{2 d_1 \ao}{(x-1)^3}+\frac{71 \
\ao}{6 (x-1)^3}+\frac{4 d_1 \ao}{(x-1)^4}-\frac{49 \ao}{3 \
(x-1)^4}+\frac{121 \ao}{3}+\frac{205 d_1}{36}+\Big(\frac{2 \
\ao^4}{x-1}+2 \ao^4-\frac{8 \ao^3}{x-1}+\frac{8 \ao^3}{3 \
(x-1)^2}-\frac{32 \ao^3}{3}+\frac{12 \ao^2}{x-1}-\frac{8 \
\ao^2}{(x-1)^2}+\frac{4 \ao^2}{(x-1)^3}+24 \ao^2-\frac{8 \
\ao}{x-1}+\frac{8 \ao}{(x-1)^2}-\frac{8 \ao}{(x-1)^3}+\frac{8 \
\ao}{(x-1)^4}-32 \ao+\frac{10}{x-1}-\frac{20}{3 (x-1)^2}+\frac{20}{3 \
(x-1)^3}-\frac{10}{(x-1)^4}-\frac{74}{3 (x-1)^5}+\frac{50}{3}\Big) \
H(0;\ao)+\Big(d_1 \ao^4+\frac{d_1 \ao^4}{x-1}-\frac{16 d_1 \ao^3}{3}-\
\frac{4 d_1 \ao^3}{x-1}+\frac{4 d_1 \ao^3}{3 (x-1)^2}+12 d_1 \
\ao^2+\frac{6 d_1 \ao^2}{x-1}-\frac{4 d_1 \ao^2}{(x-1)^2}+\frac{2 d_1 \
\ao^2}{(x-1)^3}-16 d_1 \ao-\frac{4 d_1 \ao}{x-1}+\frac{4 d_1 \
\ao}{(x-1)^2}-\frac{4 d_1 \ao}{(x-1)^3}+\frac{4 d_1 \
\ao}{(x-1)^4}+\frac{25 d_1}{3}+\frac{5 d_1}{x-1}-\frac{10 d_1}{3 \
(x-1)^2}+\frac{10 d_1}{3 (x-1)^3}-\frac{5 d_1}{(x-1)^4}-\frac{37 \
d_1}{3 (x-1)^5}\Big) H(1;\ao)+\frac{16 H(0,0;\ao)}{(x-1)^5}+\frac{8 \
d_1 H(0,1;\ao)}{(x-1)^5}+\frac{8 d_1 H(1,0;\ao)}{(x-1)^5}+\frac{4 \
d_1^2 H(1,1;\ao)}{(x-1)^5}-\frac{4}{x-2}+\frac{17 d_1}{4 \
(x-1)}-\frac{53}{4 (x-1)}+\frac{8}{3 (x-2)^2}-\frac{13 d_1}{9 \
(x-1)^2}+\frac{317}{36 (x-1)^2}+\frac{13 d_1}{9 \
(x-1)^3}-\frac{371}{36 (x-1)^3}-\frac{17 d_1}{4 \
(x-1)^4}+\frac{193}{12 (x-1)^4}-\frac{205 d_1}{36 (x-1)^5}-\frac{\pi \
^2}{6 (x-1)^5}+\frac{266}{9 (x-1)^5}-\frac{230}{9}\Big)+\Big(\frac{4 \
d_1^2}{(x-1)^5}-\frac{8 d_1}{(x-1)^5}+4 d_1+\frac{4}{(x-1)^5}-2\Big) \
H(0;\ao) H(1,1;x)+\Big(-\frac{4 d_1}{x-1}+\frac{2 \
d_1}{(x-1)^2}-\frac{4 d_1}{3 (x-1)^3}+\frac{d_1}{(x-1)^4}+\frac{37 \
d_1}{3 (x-1)^5}+\Big(-\frac{8 d_1}{(x-1)^5}+\frac{8}{(x-1)^5}-8\Big) \
H(0;\ao)+\Big(-\frac{4 d_1^2}{(x-1)^5}+\frac{4 d_1}{(x-1)^5}-4 \
d_1\Big) H(1;\ao)+\frac{4}{x-2}+\frac{5}{2 (x-1)}-\frac{8}{3 \
(x-2)^2}-\frac{3}{(x-1)^2}+\frac{8}{3 (x-2)^3}+\frac{5}{3 \
(x-1)^3}-\frac{13}{2 (x-1)^4}-\frac{37}{3 (x-1)^5}+\frac{37}{3}\Big) \
H(1,c_1(\ao);x)+\Big(2-\frac{2}{(x-1)^5}\Big) H(0;\ao) H(2,1;x)+\Big(\
\frac{\ao^4}{x-1}+\frac{3 \ao^4}{2}-\frac{4 \ao^3}{x-1}+\frac{4 \
\ao^3}{3 (x-1)^2}-8 \ao^3+\frac{6 \ao^2}{x-1}-\frac{4 \
\ao^2}{(x-1)^2}+\frac{2 \ao^2}{(x-1)^3}+18 \ao^2-\frac{4 \
\ao}{x-1}+\frac{4 \ao}{(x-1)^2}-\frac{4 \ao}{(x-1)^3}+\frac{4 \
\ao}{(x-1)^4}-24 \ao+\frac{8 H(0;\ao)}{(x-1)^5}+\frac{4 d_1 \
H(1;\ao)}{(x-1)^5}+\frac{7}{x-1}-\frac{13}{3 \
(x-1)^2}+\frac{4}{(x-1)^3}-\frac{11}{2 (x-1)^4}-\frac{37}{3 (x-1)^5}+\
\frac{25}{2}\Big) H(c_1(\ao),c_1(\ao);x)+\Big(\frac{\ao^4}{2 \
(x-1)}-\frac{\ao^4}{2}-\frac{2 \ao^3}{x-1}+\frac{2 \ao^3}{3 (x-1)^2}+\
\frac{8 \ao^3}{3}+\frac{3 \ao^2}{x-1}-\frac{2 \
\ao^2}{(x-1)^2}+\frac{\ao^2}{(x-1)^3}-6 \ao^2-\frac{2 \
\ao}{x-1}+\frac{2 \ao}{(x-1)^2}-\frac{2 \ao}{(x-1)^3}+\frac{2 \
\ao}{(x-1)^4}+8 \ao-\frac{4}{x-2}+\frac{1}{2 (x-1)}+\frac{8}{3 \
(x-2)^2}+\frac{2}{3 (x-1)^2}-\frac{8}{3 \
(x-2)^3}+\frac{1}{(x-1)^3}+\frac{2}{(x-1)^4}-\frac{25}{6}\Big) H(c_2(\
\ao),c_1(\ao);x)+\Big(16+\frac{16}{(x-1)^5}\Big) \
H(0,0,0;\ao)+\Big(16-\frac{16}{(x-1)^5}\Big) H(0,0,0;x)+\Big(\frac{8 \
d_1}{(x-1)^5}+8 d_1\Big) H(0,0,1;\ao)+\Big(\frac{8}{(x-1)^5}-8\Big) \
H(0,0,c_1(\ao);x)+\Big(\frac{8 d_1}{(x-1)^5}+8 d_1\Big) H(0,1,0;\ao)+\
\Big(\frac{4 d_1}{(x-1)^5}-4 d_1-\frac{8}{(x-1)^5}+8\Big) H(0,1,0;x)+\
\Big(\frac{4 d_1^2}{(x-1)^5}+4 d_1^2\Big) H(0,1,1;\ao)+\Big(-\frac{4 \
d_1}{(x-1)^5}+4 d_1+\frac{8}{(x-1)^5}-8\Big) \
H(0,1,c_1(\ao);x)+\Big(\frac{4}{(x-1)^5}-6\Big) \
H(0,c_1(\ao),c_1(\ao);x)+\Big(2-\frac{2}{(x-1)^5}\Big) \
H(0,c_2(\ao),c_1(\ao);x)+\Big(\frac{8 \
d_1}{(x-1)^5}-\frac{8}{(x-1)^5}+8\Big) H(1,0,0;x)+\Big(-\frac{4 \
d_1}{(x-1)^5}+\frac{4}{(x-1)^5}-2\Big) \
H(1,0,c_1(\ao);x)+\Big(-\frac{4 d_1^2}{(x-1)^5}+\frac{8 \
d_1}{(x-1)^5}-4 d_1-\frac{4}{(x-1)^5}+2\Big) H(1,1,0;x)+\Big(\frac{4 \
d_1^2}{(x-1)^5}-\frac{8 d_1}{(x-1)^5}+4 d_1+\frac{4}{(x-1)^5}-2\Big) \
H(1,1,c_1(\ao);x)+\Big(-\frac{4 \
d_1}{(x-1)^5}+\frac{4}{(x-1)^5}-6\Big) \
H(1,c_1(\ao),c_1(\ao);x)+\Big(2-\frac{2}{(x-1)^5}\Big) \
H(2,0,c_1(\ao);x)+\Big(\frac{2}{(x-1)^5}-2\Big) \
H(2,1,0;x)+\Big(2-\frac{2}{(x-1)^5}\Big) \
H(2,1,c_1(\ao);x)+\Big(\frac{2}{(x-1)^5}-2\Big) \
H(2,c_2(\ao),c_1(\ao);x)+\frac{4 \
H(c_1(\ao),c_1(\ao),c_1(\ao);x)}{(x-1)^5}+\frac{2 \
H(c_1(\ao),c_2(\ao),c_1(\ao);x)}{(x-1)^5}-\frac{\pi ^2}{x-2}-\frac{3 \
\pi ^2}{8 (x-1)}+\frac{2 \pi ^2}{3 (x-2)^2}+\frac{5 \pi ^2}{9 \
(x-1)^2}-\frac{2 \pi ^2}{3 (x-2)^3}-\frac{7 \pi ^2}{36 \
(x-1)^3}+\frac{5 \pi ^2}{4 (x-1)^4}+\frac{37 \pi ^2}{18 \
(x-1)^5}-\frac{21 \zeta_3}{4 (x-1)^5}+\frac{17 \zeta_3}{4}+\frac{\pi \
^2 \ln 2\, }{2 (x-1)^5}-\frac{1}{2} \pi ^2 \ln 2\, -\frac{173 \pi \
^2}{72}-4.
\erp

%
% The B integral for k=1 and delta=1
%

\subsection{The $\cB$ integral for $k=1$ and $\delta=1$}
%
% This file contains the TeX output produced by Mathematica for the integral B1, for delta = -1
%
The $\eps$ expansion for this integral reads
\beq
\bsp
\begin{cal}I\end{cal}(x,\eps;\ao,3+d_1\eps;1,1,1,g_B) &= x\,\bint(\eps,x;3+d_1\eps;1,1)\\
&=\frac{1}{\eps}b_{-1}^{(1,1)}+b_0^{(1,1)}+\eps b_1^{(1,1)}+\eps^2b_2^{(1,1)} +\ocal\left(\eps^3\right),
\esp
\eeq
where
%1/ep piece
\brp
b_{-1}^{(1,1)}=-\frac{1}{4},
\erp
% ep^0
\brp
b_0^{(1,1)}=\frac{\ao^4}{8 (x-1)}+\frac{\ao^4}{8}-\frac{\ao^3}{2 \
(x-1)}+\frac{\ao^3}{6 (x-1)^2}-\frac{2 \ao^3}{3}+\frac{3 \ao^2}{4 \
(x-1)}-\frac{\ao^2}{2 (x-1)^2}+\frac{\ao^2}{4 (x-1)^3}+\frac{3 \
\ao^2}{2}-\frac{\ao}{2 (x-1)}+\frac{\ao}{2 (x-1)^2}-\frac{\ao}{2 \
(x-1)^3}+\frac{\ao}{2 (x-1)^4}-2 \ao+\Big(\frac{1}{2}+\frac{1}{2 \
(x-1)^5}\Big) H(0;\ao)+\Big(\frac{1}{2}-\frac{1}{2 (x-1)^5}\Big) \
H(0;x)+\frac{H(c_1(\ao);x)}{2 (x-1)^5}-\frac{3}{4},
\erp
% ep^1
\brp
b_1^{(1,1)}=-\frac{d_1 \ao^4}{16}-\frac{d_1 \ao^4}{16 (x-1)}+\frac{\ao^4}{2 \
(x-1)}+\frac{\ao^4}{2}+\frac{13 d_1 \ao^3}{36}-\frac{\ao^3}{6 (x-2)}+\
\frac{d_1 \ao^3}{4 (x-1)}-\frac{23 \ao^3}{12 (x-1)}-\frac{d_1 \
\ao^3}{9 (x-1)^2}+\frac{29 \ao^3}{36 (x-1)^2}-\frac{101 \
\ao^3}{36}-\frac{23 d_1 \ao^2}{24}+\frac{5 \ao^2}{6 (x-2)}-\frac{3 \
d_1 \ao^2}{8 (x-1)}+\frac{5 \ao^2}{2 (x-1)}-\frac{2 \ao^2}{3 \
(x-2)^2}+\frac{d_1 \ao^2}{3 (x-1)^2}-\frac{7 \ao^2}{3 \
(x-1)^2}-\frac{d_1 \ao^2}{4 (x-1)^3}+\frac{37 \ao^2}{24 \
(x-1)^3}+\frac{167 \ao^2}{24}+\frac{25 d_1 \ao}{12}-\frac{7 \ao}{3 \
(x-2)}+\frac{d_1 \ao}{4 (x-1)}-\frac{\ao}{12 (x-1)}+\frac{10 \ao}{3 \
(x-2)^2}-\frac{d_1 \ao}{3 (x-1)^2}+\frac{11 \ao}{6 (x-1)^2}-\frac{4 \
\ao}{(x-2)^3}+\frac{d_1 \ao}{2 (x-1)^3}-\frac{37 \ao}{12 \
(x-1)^3}-\frac{d_1 \ao}{(x-1)^4}+\frac{55 \ao}{12 (x-1)^4}-\frac{145 \
\ao}{12}+\Big(-\frac{\ao^4}{2 (x-1)}-\frac{\ao^4}{2}+\frac{2 \
\ao^3}{x-1}-\frac{2 \ao^3}{3 (x-1)^2}+\frac{8 \ao^3}{3}-\frac{3 \
\ao^2}{x-1}+\frac{2 \ao^2}{(x-1)^2}-\frac{\ao^2}{(x-1)^3}-6 \
\ao^2+\frac{2 \ao}{x-1}-\frac{2 \ao}{(x-1)^2}+\frac{2 \
\ao}{(x-1)^3}-\frac{2 \ao}{(x-1)^4}+8 \
\ao+\frac{4}{x-2}-\frac{5}{x-1}-\frac{16}{3 (x-2)^2}+\frac{7}{4 \
(x-1)^2}+\frac{8}{(x-2)^3}-\frac{5}{4 \
(x-1)^3}-\frac{16}{(x-2)^4}+\frac{3}{2 (x-1)^4}+\frac{43}{12 \
(x-1)^5}-\frac{7}{12}\Big) H(0;\ao)+\Big(\frac{5}{x-1}-\frac{7}{4 \
(x-1)^2}+\frac{5}{4 (x-1)^3}-\frac{3}{2 (x-1)^4}-\frac{43}{12 \
(x-1)^5}+\frac{43}{12}-\frac{4}{x-2}+\frac{16}{3 \
(x-2)^2}-\frac{8}{(x-2)^3}+\frac{16}{(x-2)^4}\Big) \
H(0;x)+\Big(-\frac{d_1 \ao^4}{4}-\frac{d_1 \ao^4}{4 (x-1)}+\frac{4 \
d_1 \ao^3}{3}+\frac{d_1 \ao^3}{x-1}-\frac{d_1 \ao^3}{3 (x-1)^2}-3 d_1 \
\ao^2-\frac{3 d_1 \ao^2}{2 (x-1)}+\frac{d_1 \ao^2}{(x-1)^2}-\frac{d_1 \
\ao^2}{2 (x-1)^3}+4 d_1 \ao+\frac{d_1 \ao}{x-1}-\frac{d_1 \
\ao}{(x-1)^2}+\frac{d_1 \ao}{(x-1)^3}-\frac{d_1 \
\ao}{(x-1)^4}-\frac{25 d_1}{12}-\frac{d_1}{4 (x-1)}+\frac{d_1}{3 \
(x-1)^2}-\frac{d_1}{2 (x-1)^3}+\frac{d_1}{(x-1)^4}\Big) \
H(1;\ao)+\Big(\frac{d_1}{(x-1)^5}-\frac{16}{(x-2)^5}-\frac{1}{(x-1)^5}\
+1\Big) H(0;\ao) H(1;x)+\Big(\frac{\ao^4}{4 (x-2)}-\frac{\ao^4}{4 \
(x-1)}-\frac{\ao^4}{4}-\frac{\ao^3}{x-2}+\frac{\ao^3}{x-1}+\frac{2 \
\ao^3}{3 (x-2)^2}-\frac{\ao^3}{3 (x-1)^2}+\frac{4 \ao^3}{3}+\frac{3 \
\ao^2}{2 (x-2)}-\frac{3 \ao^2}{2 (x-1)}-\frac{2 \
\ao^2}{(x-2)^2}+\frac{\ao^2}{(x-1)^2}+\frac{2 \
\ao^2}{(x-2)^3}-\frac{\ao^2}{2 (x-1)^3}-3 \
\ao^2-\frac{\ao}{x-2}+\frac{\ao}{x-1}+\frac{2 \
\ao}{(x-2)^2}-\frac{\ao}{(x-1)^2}-\frac{4 \
\ao}{(x-2)^3}+\frac{\ao}{(x-1)^3}+\frac{8 \
\ao}{(x-2)^4}-\frac{\ao}{(x-1)^4}+4 \ao-\frac{2 \
H(0;\ao)}{(x-1)^5}-\frac{d_1 \
H(1;\ao)}{(x-1)^5}+\frac{4}{x-2}-\frac{5}{x-1}-\frac{16}{3 \
(x-2)^2}+\frac{7}{4 (x-1)^2}+\frac{8}{(x-2)^3}-\frac{5}{4 \
(x-1)^3}-\frac{16}{(x-2)^4}+\frac{3}{2 (x-1)^4}+\frac{43}{12 \
(x-1)^5}-\frac{25}{12}\Big) \
H(c_1(\ao);x)+\Big(-2-\frac{2}{(x-1)^5}\Big) \
H(0,0;\ao)+\Big(\frac{2}{(x-1)^5}-2\Big) \
H(0,0;x)+\Big(-\frac{d_1}{(x-1)^5}-d_1\Big) \
H(0,1;\ao)+\Big(-\frac{1}{(x-1)^5}+1-\frac{16}{(x-2)^5}\Big) H(0,c_1(\
\ao);x)+\Big(-\frac{d_1}{(x-1)^5}+\frac{16}{(x-2)^5}+\frac{1}{(x-1)^5}\
-1\Big) H(1,0;x)+\Big(\frac{d_1}{(x-1)^5}-\frac{16}{(x-2)^5}-\frac{1}{\
(x-1)^5}+1\Big) \
H(1,c_1(\ao);x)-\frac{H(c_1(\ao),c_1(\ao);x)}{(x-1)^5}+\frac{16 \
H(c_2(\ao),c_1(\ao);x)}{(x-2)^5}+\frac{4 \pi ^2}{(x-2)^5}+\frac{\pi \
^2}{6 (x-1)^5}-\frac{\pi ^2}{6}-\frac{7}{4},
\erp
% ep^2
\brp
b_2^{(1,1)}=\frac{d_1^2 \ao^4}{32}-\frac{5 d_1 \ao^4}{16}+\frac{d_1^2 \ao^4}{32 \
(x-1)}-\frac{5 d_1 \ao^4}{16 (x-1)}-\frac{\pi ^2 \ao^4}{48 \
(x-1)}+\frac{11 \ao^4}{8 (x-1)}-\frac{\pi ^2 \ao^4}{48}+\frac{11 \
\ao^4}{8}-\frac{43 d_1^2 \ao^3}{216}+\frac{427 d_1 \
\ao^3}{216}+\frac{7 d_1 \ao^3}{36 (x-2)}-\frac{5 \ao^3}{9 \
(x-2)}-\frac{d_1^2 \ao^3}{8 (x-1)}+\frac{83 d_1 \ao^3}{72 \
(x-1)}+\frac{\pi ^2 \ao^3}{12 (x-1)}-\frac{47 \ao^3}{9 (x-1)}+\frac{2 \
d_1^2 \ao^3}{27 (x-1)^2}-\frac{157 d_1 \ao^3}{216 (x-1)^2}-\frac{\pi \
^2 \ao^3}{36 (x-1)^2}+\frac{139 \ao^3}{54 (x-1)^2}+\frac{\pi ^2 \
\ao^3}{9}-\frac{218 \ao^3}{27}+\frac{95 d_1^2 \ao^2}{144}-\frac{301 \
d_1 \ao^2}{48}-\frac{41 d_1 \ao^2}{36 (x-2)}+\frac{28 \ao^2}{9 \
(x-2)}+\frac{3 d_1^2 \ao^2}{16 (x-1)}-\frac{9 d_1 \ao^2}{8 \
(x-1)}-\frac{\pi ^2 \ao^2}{8 (x-1)}+\frac{151 \ao^2}{24 \
(x-1)}+\frac{10 d_1 \ao^2}{9 (x-2)^2}-\frac{26 \ao^2}{9 \
(x-2)^2}-\frac{2 d_1^2 \ao^2}{9 (x-1)^2}+\frac{25 d_1 \ao^2}{12 \
(x-1)^2}+\frac{\pi ^2 \ao^2}{12 (x-1)^2}-\frac{533 \ao^2}{72 \
(x-1)^2}+\frac{d_1^2 \ao^2}{4 (x-1)^3}-\frac{319 d_1 \ao^2}{144 \
(x-1)^3}-\frac{\pi ^2 \ao^2}{24 (x-1)^3}+\frac{221 \ao^2}{36 \
(x-1)^3}-\frac{\pi ^2 \ao^2}{4}+\frac{199 \ao^2}{9}-\frac{205 d_1^2 \
\ao}{72}+\frac{175 d_1 \ao}{8}+\frac{97 d_1 \ao}{18 (x-2)}-\frac{209 \
\ao}{18 (x-2)}-\frac{d_1^2 \ao}{8 (x-1)}-\frac{281 d_1 \ao}{72 \
(x-1)}+\frac{\pi ^2 \ao}{12 (x-1)}+\frac{191 \ao}{36 (x-1)}-\frac{74 \
d_1 \ao}{9 (x-2)^2}+\frac{178 \ao}{9 (x-2)^2}+\frac{2 d_1^2 \ao}{9 \
(x-1)^2}-\frac{5 d_1 \ao}{6 (x-1)^2}-\frac{\pi ^2 \ao}{12 \
(x-1)^2}+\frac{65 \ao}{18 (x-1)^2}+\frac{12 d_1 \
\ao}{(x-2)^3}-\frac{28 \ao}{(x-2)^3}-\frac{d_1^2 \ao}{2 \
(x-1)^3}+\frac{337 d_1 \ao}{72 (x-1)^3}+\frac{\pi ^2 \ao}{12 \
(x-1)^3}-\frac{233 \ao}{18 (x-1)^3}+\frac{2 d_1^2 \
\ao}{(x-1)^4}-\frac{1009 d_1 \ao}{72 (x-1)^4}-\frac{\pi ^2 \ao}{12 \
(x-1)^4}+\frac{991 \ao}{36 (x-1)^4}+\frac{\pi ^2 \ao}{3}-\frac{1909 \
\ao}{36}+\Big(\frac{d_1 \ao^4}{4}+\frac{d_1 \ao^4}{4 (x-1)}-\frac{2 \
\ao^4}{x-1}-2 \ao^4-\frac{13 d_1 \ao^3}{9}+\frac{2 \ao^3}{3 \
(x-2)}-\frac{d_1 \ao^3}{x-1}+\frac{23 \ao^3}{3 (x-1)}+\frac{4 d_1 \
\ao^3}{9 (x-1)^2}-\frac{29 \ao^3}{9 (x-1)^2}+\frac{101 \
\ao^3}{9}+\frac{23 d_1 \ao^2}{6}-\frac{10 \ao^2}{3 (x-2)}+\frac{3 d_1 \
\ao^2}{2 (x-1)}-\frac{10 \ao^2}{x-1}+\frac{8 \ao^2}{3 \
(x-2)^2}-\frac{4 d_1 \ao^2}{3 (x-1)^2}+\frac{28 \ao^2}{3 \
(x-1)^2}+\frac{d_1 \ao^2}{(x-1)^3}-\frac{37 \ao^2}{6 \
(x-1)^3}-\frac{167 \ao^2}{6}-\frac{25 d_1 \ao}{3}+\frac{28 \ao}{3 \
(x-2)}-\frac{d_1 \ao}{x-1}+\frac{\ao}{3 (x-1)}-\frac{40 \ao}{3 \
(x-2)^2}+\frac{4 d_1 \ao}{3 (x-1)^2}-\frac{22 \ao}{3 \
(x-1)^2}+\frac{16 \ao}{(x-2)^3}-\frac{2 d_1 \ao}{(x-1)^3}+\frac{37 \
\ao}{3 (x-1)^3}+\frac{4 d_1 \ao}{(x-1)^4}-\frac{55 \ao}{3 \
(x-1)^4}+\frac{145 \ao}{3}+\frac{205 d_1}{72}-\frac{34 d_1}{3 (x-2)}+\
\frac{4}{x-2}+\frac{40 d_1}{3 (x-1)}-\frac{155}{12 (x-1)}+\frac{116 \
d_1}{9 (x-2)^2}-\frac{88}{9 (x-2)^2}-\frac{47 d_1}{24 \
(x-1)^2}+\frac{103}{12 (x-1)^2}-\frac{16 \
d_1}{(x-2)^3}+\frac{24}{(x-2)^3}+\frac{25 d_1}{24 \
(x-1)^3}-\frac{103}{12 (x-1)^3}+\frac{32 \
d_1}{(x-2)^4}-\frac{80}{(x-2)^4}-\frac{9 d_1}{4 \
(x-1)^4}+\frac{145}{12 (x-1)^4}-\frac{205 d_1}{72 (x-1)^5}-\frac{\pi \
^2}{12 (x-1)^5}+\frac{661}{36 (x-1)^5}-\frac{\pi \
^2}{12}-\frac{409}{36}\Big) H(0;\ao)+\Big(\frac{34 d_1}{3 \
(x-2)}-\frac{40 d_1}{3 (x-1)}-\frac{116 d_1}{9 (x-2)^2}+\frac{47 \
d_1}{24 (x-1)^2}+\frac{16 d_1}{(x-2)^3}-\frac{25 d_1}{24 \
(x-1)^3}-\frac{32 d_1}{(x-2)^4}+\frac{9 d_1}{4 (x-1)^4}+\frac{205 \
d_1}{72 (x-1)^5}-\frac{205 d_1}{72}-\frac{4}{x-2}+\frac{155}{12 \
(x-1)}+\frac{88}{9 (x-2)^2}-\frac{103}{12 \
(x-1)^2}-\frac{24}{(x-2)^3}+\frac{103}{12 \
(x-1)^3}+\frac{80}{(x-2)^4}-\frac{145}{12 (x-1)^4}-\frac{16 \pi \
^2}{(x-2)^5}-\frac{7 \pi ^2}{12 (x-1)^5}-\frac{661}{36 \
(x-1)^5}+\frac{3 \pi ^2}{4}+\frac{661}{36}\Big) \
H(0;x)+\Big(\frac{d_1^2 \ao^4}{8}-d_1 \ao^4+\frac{d_1^2 \ao^4}{8 \
(x-1)}-\frac{d_1 \ao^4}{x-1}-\frac{13 d_1^2 \ao^3}{18}+\frac{101 d_1 \
\ao^3}{18}+\frac{d_1 \ao^3}{3 (x-2)}-\frac{d_1^2 \ao^3}{2 \
(x-1)}+\frac{23 d_1 \ao^3}{6 (x-1)}+\frac{2 d_1^2 \ao^3}{9 \
(x-1)^2}-\frac{29 d_1 \ao^3}{18 (x-1)^2}+\frac{23 d_1^2 \
\ao^2}{12}-\frac{167 d_1 \ao^2}{12}-\frac{5 d_1 \ao^2}{3 \
(x-2)}+\frac{3 d_1^2 \ao^2}{4 (x-1)}-\frac{5 d_1 \ao^2}{x-1}+\frac{4 \
d_1 \ao^2}{3 (x-2)^2}-\frac{2 d_1^2 \ao^2}{3 (x-1)^2}+\frac{14 d_1 \
\ao^2}{3 (x-1)^2}+\frac{d_1^2 \ao^2}{2 (x-1)^3}-\frac{37 d_1 \
\ao^2}{12 (x-1)^3}-\frac{25 d_1^2 \ao}{6}+\frac{145 d_1 \
\ao}{6}+\frac{14 d_1 \ao}{3 (x-2)}-\frac{d_1^2 \ao}{2 \
(x-1)}+\frac{d_1 \ao}{6 (x-1)}-\frac{20 d_1 \ao}{3 (x-2)^2}+\frac{2 \
d_1^2 \ao}{3 (x-1)^2}-\frac{11 d_1 \ao}{3 (x-1)^2}+\frac{8 d_1 \
\ao}{(x-2)^3}-\frac{d_1^2 \ao}{(x-1)^3}+\frac{37 d_1 \ao}{6 (x-1)^3}+\
\frac{2 d_1^2 \ao}{(x-1)^4}-\frac{55 d_1 \ao}{6 (x-1)^4}+\frac{205 \
d_1^2}{72}-\frac{535 d_1}{36}-\frac{10 d_1}{3 (x-2)}+\frac{d_1^2}{8 \
(x-1)}+\frac{2 d_1}{x-1}+\frac{16 d_1}{3 (x-2)^2}-\frac{2 d_1^2}{9 \
(x-1)^2}+\frac{11 d_1}{18 (x-1)^2}-\frac{8 \
d_1}{(x-2)^3}+\frac{d_1^2}{2 (x-1)^3}-\frac{37 d_1}{12 \
(x-1)^3}-\frac{2 d_1^2}{(x-1)^4}+\frac{55 d_1}{6 (x-1)^4}\Big) \
H(1;\ao)+\Big(-\frac{8 \pi ^2 d_1}{(x-2)^5}+\frac{16 \pi \
^2}{(x-2)^5}+\frac{\pi ^2}{4 (x-1)^5}-\frac{\pi ^2}{4}\Big) \
H(2;x)+\Big(\frac{2 \ao^4}{x-1}+2 \ao^4-\frac{8 \ao^3}{x-1}+\frac{8 \
\ao^3}{3 (x-1)^2}-\frac{32 \ao^3}{3}+\frac{12 \ao^2}{x-1}-\frac{8 \
\ao^2}{(x-1)^2}+\frac{4 \ao^2}{(x-1)^3}+24 \ao^2-\frac{8 \
\ao}{x-1}+\frac{8 \ao}{(x-1)^2}-\frac{8 \ao}{(x-1)^3}+\frac{8 \
\ao}{(x-1)^4}-32 \ao-\frac{16}{x-2}+\frac{20}{x-1}+\frac{64}{3 \
(x-2)^2}-\frac{7}{(x-1)^2}-\frac{32}{(x-2)^3}+\frac{5}{(x-1)^3}+\frac{\
64}{(x-2)^4}-\frac{6}{(x-1)^4}-\frac{43}{3 (x-1)^5}+\frac{7}{3}\Big) \
H(0,0;\ao)+\Big(-\frac{20}{x-1}+\frac{7}{(x-1)^2}-\frac{5}{(x-1)^3}+\frac{6}{(x-1)^4}+\frac{43}{3 \
(x-1)^5}-\frac{43}{3}+\frac{16}{x-2}-\frac{64}{3 \
(x-2)^2}+\frac{32}{(x-2)^3}-\frac{64}{(x-2)^4}\Big) H(0,0;x)+\Big(d_1 \
\ao^4+\frac{d_1 \ao^4}{x-1}-\frac{16 d_1 \ao^3}{3}-\frac{4 d_1 \
\ao^3}{x-1}+\frac{4 d_1 \ao^3}{3 (x-1)^2}+12 d_1 \ao^2+\frac{6 d_1 \
\ao^2}{x-1}-\frac{4 d_1 \ao^2}{(x-1)^2}+\frac{2 d_1 \
\ao^2}{(x-1)^3}-16 d_1 \ao-\frac{4 d_1 \ao}{x-1}+\frac{4 d_1 \
\ao}{(x-1)^2}-\frac{4 d_1 \ao}{(x-1)^3}+\frac{4 d_1 \
\ao}{(x-1)^4}+\frac{7 d_1}{6}-\frac{8 d_1}{x-2}+\frac{10 \
d_1}{x-1}+\frac{32 d_1}{3 (x-2)^2}-\frac{7 d_1}{2 (x-1)^2}-\frac{16 \
d_1}{(x-2)^3}+\frac{5 d_1}{2 (x-1)^3}+\frac{32 d_1}{(x-2)^4}-\frac{3 \
d_1}{(x-1)^4}-\frac{43 d_1}{6 (x-1)^5}\Big) H(0,1;\ao)+H(1;x) \
\Big(\frac{\pi ^2 d_1}{3 (x-1)^5}+\Big(\frac{15 d_1}{2 \
(x-2)}-\frac{19 d_1}{2 (x-1)}-\frac{28 d_1}{3 (x-2)^2}+\frac{17 \
d_1}{6 (x-1)^2}+\frac{12 d_1}{(x-2)^3}-\frac{3 d_1}{2 \
(x-1)^3}-\frac{16 d_1}{(x-2)^4}+\frac{d_1}{(x-1)^4}+\frac{43 d_1}{6 \
(x-1)^5}+\frac{11}{2 (x-2)}-\frac{5}{2 (x-1)}-\frac{16}{3 \
(x-2)^2}-\frac{7}{12 (x-1)^2}+\frac{16}{3 (x-2)^3}+\frac{5}{12 \
(x-1)^3}-\frac{3}{(x-1)^4}-\frac{16}{(x-2)^5}-\frac{43}{6 \
(x-1)^5}+\frac{43}{6}\Big) H(0;\ao)+\Big(-\frac{4 \
d_1}{(x-1)^5}+\frac{64}{(x-2)^5}+\frac{4}{(x-1)^5}-4\Big) H(0,0;\ao)+\
\Big(-\frac{2 d_1^2}{(x-1)^5}+\frac{32 d_1}{(x-2)^5}+\frac{2 \
d_1}{(x-1)^5}-2 d_1\Big) H(0,1;\ao)+\frac{8 \pi ^2}{3 (x-2)^5}-\frac{\
\pi ^2}{3 (x-1)^5}+\frac{\pi ^2}{6}\Big)+\Big(-\frac{32 \
d_1}{(x-2)^5}-\frac{2 d_1}{(x-1)^5}+2 \
d_1+\frac{32}{(x-2)^5}+\frac{4}{(x-1)^5}-4\Big) H(0;\ao) \
H(0,1;x)+\Big(\frac{\ao^4}{2 (x-2)}-\frac{2 \ao^3}{x-2}+\frac{4 \
\ao^3}{3 (x-2)^2}+\frac{3 \ao^2}{x-2}-\frac{4 \ao^2}{(x-2)^2}+\frac{4 \
\ao^2}{(x-2)^3}-\frac{2 \ao}{x-2}+\frac{4 \ao}{(x-2)^2}-\frac{8 \
\ao}{(x-2)^3}+\frac{16 \
\ao}{(x-2)^4}+\Big(\frac{4}{(x-1)^5}-4+\frac{64}{(x-2)^5}\Big) \
H(0;\ao)+\Big(\frac{32 d_1}{(x-2)^5}+\frac{2 d_1}{(x-1)^5}-2 d_1\Big) \
H(1;\ao)+\frac{6}{x-2}-\frac{5}{2 (x-1)}-\frac{20}{3 \
(x-2)^2}-\frac{7}{12 (x-1)^2}+\frac{28}{3 (x-2)^3}+\frac{5}{12 \
(x-1)^3}-\frac{16}{(x-2)^4}-\frac{3}{(x-1)^4}-\frac{16}{(x-2)^5}-\frac{43}{6 (x-1)^5}+\frac{43}{6}\Big) H(0,c_1(\ao);x)+\Big(d_1 \
\ao^4+\frac{d_1 \ao^4}{x-1}-\frac{16 d_1 \ao^3}{3}-\frac{4 d_1 \
\ao^3}{x-1}+\frac{4 d_1 \ao^3}{3 (x-1)^2}+12 d_1 \ao^2+\frac{6 d_1 \
\ao^2}{x-1}-\frac{4 d_1 \ao^2}{(x-1)^2}+\frac{2 d_1 \
\ao^2}{(x-1)^3}-16 d_1 \ao-\frac{4 d_1 \ao}{x-1}+\frac{4 d_1 \
\ao}{(x-1)^2}-\frac{4 d_1 \ao}{(x-1)^3}+\frac{4 d_1 \
\ao}{(x-1)^4}+\frac{25 d_1}{3}+\frac{d_1}{x-1}-\frac{4 d_1}{3 \
(x-1)^2}+\frac{2 d_1}{(x-1)^3}-\frac{4 d_1}{(x-1)^4}\Big) H(1,0;\ao)+\
\Big(-\frac{15 d_1}{2 (x-2)}+\frac{19 d_1}{2 (x-1)}+\frac{28 d_1}{3 \
(x-2)^2}-\frac{17 d_1}{6 (x-1)^2}-\frac{12 d_1}{(x-2)^3}+\frac{3 \
d_1}{2 (x-1)^3}+\frac{16 d_1}{(x-2)^4}-\frac{d_1}{(x-1)^4}-\frac{43 \
d_1}{6 (x-1)^5}-\frac{11}{2 (x-2)}+\frac{5}{2 (x-1)}+\frac{16}{3 \
(x-2)^2}+\frac{7}{12 (x-1)^2}-\frac{16}{3 (x-2)^3}-\frac{5}{12 \
(x-1)^3}+\frac{3}{(x-1)^4}+\frac{16}{(x-2)^5}+\frac{43}{6 \
(x-1)^5}-\frac{43}{6}\Big) H(1,0;x)+\Big(\frac{d_1^2 \
\ao^4}{2}+\frac{d_1^2 \ao^4}{2 (x-1)}-\frac{8 d_1^2 \ao^3}{3}-\frac{2 \
d_1^2 \ao^3}{x-1}+\frac{2 d_1^2 \ao^3}{3 (x-1)^2}+6 d_1^2 \
\ao^2+\frac{3 d_1^2 \ao^2}{x-1}-\frac{2 d_1^2 \
\ao^2}{(x-1)^2}+\frac{d_1^2 \ao^2}{(x-1)^3}-8 d_1^2 \ao-\frac{2 d_1^2 \
\ao}{x-1}+\frac{2 d_1^2 \ao}{(x-1)^2}-\frac{2 d_1^2 \
\ao}{(x-1)^3}+\frac{2 d_1^2 \ao}{(x-1)^4}+\frac{25 \
d_1^2}{6}+\frac{d_1^2}{2 (x-1)}-\frac{2 d_1^2}{3 \
(x-1)^2}+\frac{d_1^2}{(x-1)^3}-\frac{2 d_1^2}{(x-1)^4}\Big) \
H(1,1;\ao)+H(c_1(\ao);x) \Big(\frac{d_1 \ao^4}{8}-\frac{d_1 \ao^4}{8 \
(x-2)}+\frac{\ao^4}{2 (x-2)}+\frac{d_1 \ao^4}{8 \
(x-1)}-\frac{\ao^4}{x-1}-\ao^4-\frac{13 d_1 \ao^3}{18}+\frac{d_1 \
\ao^3}{2 (x-2)}-\frac{5 \ao^3}{3 (x-2)}-\frac{d_1 \ao^3}{2 \
(x-1)}+\frac{43 \ao^3}{12 (x-1)}-\frac{4 d_1 \ao^3}{9 \
(x-2)^2}+\frac{14 \ao^3}{9 (x-2)^2}+\frac{2 d_1 \ao^3}{9 \
(x-1)^2}-\frac{29 \ao^3}{18 (x-1)^2}+\frac{101 \ao^3}{18}+\frac{23 \
d_1 \ao^2}{12}-\frac{3 d_1 \ao^2}{4 (x-2)}+\frac{11 \ao^2}{12 (x-2)}+\
\frac{3 d_1 \ao^2}{4 (x-1)}-\frac{7 \ao^2}{2 (x-1)}+\frac{4 d_1 \
\ao^2}{3 (x-2)^2}-\frac{10 \ao^2}{3 (x-2)^2}-\frac{2 d_1 \ao^2}{3 \
(x-1)^2}+\frac{13 \ao^2}{3 (x-1)^2}-\frac{2 d_1 \
\ao^2}{(x-2)^3}+\frac{6 \ao^2}{(x-2)^3}+\frac{d_1 \ao^2}{2 \
(x-1)^3}-\frac{37 \ao^2}{12 (x-1)^3}-\frac{167 \ao^2}{12}-\frac{25 \
d_1 \ao}{6}+\frac{d_1 \ao}{2 (x-2)}+\frac{37 \ao}{6 (x-2)}-\frac{d_1 \
\ao}{2 (x-1)}-\frac{21 \ao}{4 (x-1)}-\frac{4 d_1 \ao}{3 \
(x-2)^2}-\frac{14 \ao}{3 (x-2)^2}+\frac{2 d_1 \ao}{3 \
(x-1)^2}-\frac{17 \ao}{12 (x-1)^2}+\frac{4 d_1 \ao}{(x-2)^3}-\frac{4 \
\ao}{(x-2)^3}-\frac{d_1 \ao}{(x-1)^3}+\frac{17 \ao}{3 \
(x-1)^3}-\frac{16 d_1 \ao}{(x-2)^4}+\frac{40 \ao}{(x-2)^4}+\frac{2 \
d_1 \ao}{(x-1)^4}-\frac{55 \ao}{6 (x-1)^4}+\frac{145 \
\ao}{6}+\frac{205 d_1}{72}+\Big(-\frac{\ao^4}{x-2}+\frac{\ao^4}{x-1}+\
\ao^4+\frac{4 \ao^3}{x-2}-\frac{4 \ao^3}{x-1}-\frac{8 \ao^3}{3 \
(x-2)^2}+\frac{4 \ao^3}{3 (x-1)^2}-\frac{16 \ao^3}{3}-\frac{6 \
\ao^2}{x-2}+\frac{6 \ao^2}{x-1}+\frac{8 \ao^2}{(x-2)^2}-\frac{4 \
\ao^2}{(x-1)^2}-\frac{8 \ao^2}{(x-2)^3}+\frac{2 \ao^2}{(x-1)^3}+12 \
\ao^2+\frac{4 \ao}{x-2}-\frac{4 \ao}{x-1}-\frac{8 \
\ao}{(x-2)^2}+\frac{4 \ao}{(x-1)^2}+\frac{16 \ao}{(x-2)^3}-\frac{4 \
\ao}{(x-1)^3}-\frac{32 \ao}{(x-2)^4}+\frac{4 \ao}{(x-1)^4}-16 \
\ao-\frac{16}{x-2}+\frac{20}{x-1}+\frac{64}{3 \
(x-2)^2}-\frac{7}{(x-1)^2}-\frac{32}{(x-2)^3}+\frac{5}{(x-1)^3}+\frac{\
64}{(x-2)^4}-\frac{6}{(x-1)^4}-\frac{43}{3 (x-1)^5}+\frac{25}{3}\Big) \
H(0;\ao)+\Big(\frac{d_1 \ao^4}{2}-\frac{d_1 \ao^4}{2 (x-2)}+\frac{d_1 \
\ao^4}{2 (x-1)}-\frac{8 d_1 \ao^3}{3}+\frac{2 d_1 \ao^3}{x-2}-\frac{2 \
d_1 \ao^3}{x-1}-\frac{4 d_1 \ao^3}{3 (x-2)^2}+\frac{2 d_1 \ao^3}{3 \
(x-1)^2}+6 d_1 \ao^2-\frac{3 d_1 \ao^2}{x-2}+\frac{3 d_1 \ao^2}{x-1}+\
\frac{4 d_1 \ao^2}{(x-2)^2}-\frac{2 d_1 \ao^2}{(x-1)^2}-\frac{4 d_1 \
\ao^2}{(x-2)^3}+\frac{d_1 \ao^2}{(x-1)^3}-8 d_1 \ao+\frac{2 d_1 \
\ao}{x-2}-\frac{2 d_1 \ao}{x-1}-\frac{4 d_1 \ao}{(x-2)^2}+\frac{2 d_1 \
\ao}{(x-1)^2}+\frac{8 d_1 \ao}{(x-2)^3}-\frac{2 d_1 \
\ao}{(x-1)^3}-\frac{16 d_1 \ao}{(x-2)^4}+\frac{2 d_1 \
\ao}{(x-1)^4}+\frac{25 d_1}{6}-\frac{8 d_1}{x-2}+\frac{10 \
d_1}{x-1}+\frac{32 d_1}{3 (x-2)^2}-\frac{7 d_1}{2 (x-1)^2}-\frac{16 \
d_1}{(x-2)^3}+\frac{5 d_1}{2 (x-1)^3}+\frac{32 d_1}{(x-2)^4}-\frac{3 \
d_1}{(x-1)^4}-\frac{43 d_1}{6 (x-1)^5}\Big) H(1;\ao)+\frac{8 \
H(0,0;\ao)}{(x-1)^5}+\frac{4 d_1 H(0,1;\ao)}{(x-1)^5}+\frac{4 d_1 \
H(1,0;\ao)}{(x-1)^5}+\frac{2 d_1^2 H(1,1;\ao)}{(x-1)^5}-\frac{34 \
d_1}{3 (x-2)}+\frac{4}{x-2}+\frac{40 d_1}{3 (x-1)}-\frac{155}{12 \
(x-1)}+\frac{116 d_1}{9 (x-2)^2}-\frac{88}{9 (x-2)^2}-\frac{47 \
d_1}{24 (x-1)^2}+\frac{103}{12 (x-1)^2}-\frac{16 \
d_1}{(x-2)^3}+\frac{24}{(x-2)^3}+\frac{25 d_1}{24 \
(x-1)^3}-\frac{103}{12 (x-1)^3}+\frac{32 \
d_1}{(x-2)^4}-\frac{80}{(x-2)^4}-\frac{9 d_1}{4 \
(x-1)^4}+\frac{145}{12 (x-1)^4}-\frac{205 d_1}{72 (x-1)^5}-\frac{\pi \
^2}{12 (x-1)^5}+\frac{661}{36 \
(x-1)^5}-\frac{535}{36}\Big)+\Big(\frac{2 d_1^2}{(x-1)^5}-\frac{32 \
d_1}{(x-2)^5}-\frac{4 d_1}{(x-1)^5}+2 \
d_1-\frac{16}{(x-2)^5}+\frac{2}{(x-1)^5}-1\Big) H(0;\ao) \
H(1,1;x)+\Big(\frac{15 d_1}{2 (x-2)}-\frac{19 d_1}{2 (x-1)}-\frac{28 \
d_1}{3 (x-2)^2}+\frac{17 d_1}{6 (x-1)^2}+\frac{12 \
d_1}{(x-2)^3}-\frac{3 d_1}{2 (x-1)^3}-\frac{16 \
d_1}{(x-2)^4}+\frac{d_1}{(x-1)^4}+\frac{43 d_1}{6 \
(x-1)^5}+\Big(-\frac{4 \
d_1}{(x-1)^5}+\frac{64}{(x-2)^5}+\frac{4}{(x-1)^5}-4\Big) \
H(0;\ao)+\Big(-\frac{2 d_1^2}{(x-1)^5}+\frac{32 d_1}{(x-2)^5}+\frac{2 \
d_1}{(x-1)^5}-2 d_1\Big) H(1;\ao)+\frac{11}{2 (x-2)}-\frac{5}{2 \
(x-1)}-\frac{16}{3 (x-2)^2}-\frac{7}{12 (x-1)^2}+\frac{16}{3 \
(x-2)^3}+\frac{5}{12 \
(x-1)^3}-\frac{3}{(x-1)^4}-\frac{16}{(x-2)^5}-\frac{43}{6 \
(x-1)^5}+\frac{43}{6}\Big) H(1,c_1(\ao);x)+\Big(\frac{32 \
d_1}{(x-2)^5}-\frac{64}{(x-2)^5}-\frac{1}{(x-1)^5}+1\Big) H(0;\ao) \
H(2,1;x)+\Big(-\frac{5 \ao^4}{4 (x-2)}+\frac{\ao^4}{2 (x-1)}+\frac{3 \
\ao^4}{4}+\frac{5 \ao^3}{x-2}-\frac{2 \ao^3}{x-1}-\frac{10 \ao^3}{3 \
(x-2)^2}+\frac{2 \ao^3}{3 (x-1)^2}-4 \ao^3-\frac{15 \ao^2}{2 \
(x-2)}+\frac{3 \ao^2}{x-1}+\frac{10 \ao^2}{(x-2)^2}-\frac{2 \
\ao^2}{(x-1)^2}-\frac{10 \ao^2}{(x-2)^3}+\frac{\ao^2}{(x-1)^3}+9 \
\ao^2+\frac{5 \ao}{x-2}-\frac{2 \ao}{x-1}-\frac{10 \
\ao}{(x-2)^2}+\frac{2 \ao}{(x-1)^2}+\frac{20 \ao}{(x-2)^3}-\frac{2 \
\ao}{(x-1)^3}-\frac{40 \ao}{(x-2)^4}+\frac{2 \ao}{(x-1)^4}-12 \
\ao+\frac{4 H(0;\ao)}{(x-1)^5}+\frac{2 d_1 \
H(1;\ao)}{(x-1)^5}-\frac{20}{x-2}+\frac{89}{4 (x-1)}+\frac{80}{3 \
(x-2)^2}-\frac{27}{4 (x-1)^2}-\frac{40}{(x-2)^3}+\frac{49}{12 \
(x-1)^3}+\frac{80}{(x-2)^4}-\frac{4}{(x-1)^4}-\frac{43}{6 \
(x-1)^5}+\frac{25}{4}\Big) H(c_1(\ao),c_1(\ao);x)+\Big(\frac{\ao^4}{4 \
(x-1)}-\frac{\ao^4}{4}-\frac{\ao^3}{x-1}+\frac{\ao^3}{3 \
(x-1)^2}+\frac{4 \ao^3}{3}+\frac{3 \ao^2}{2 \
(x-1)}-\frac{\ao^2}{(x-1)^2}+\frac{\ao^2}{2 (x-1)^3}-3 \
\ao^2-\frac{\ao}{x-1}+\frac{\ao}{(x-1)^2}-\frac{\ao}{(x-1)^3}+\frac{\ao}{(x-1)^4}+4 \ao-\frac{64 H(0;\ao)}{(x-2)^5}-\frac{32 d_1 \
H(1;\ao)}{(x-2)^5}-\frac{2}{x-2}+\frac{1}{4 (x-1)}+\frac{4}{3 \
(x-2)^2}+\frac{1}{3 (x-1)^2}-\frac{4}{3 (x-2)^3}+\frac{1}{2 (x-1)^3}+\
\frac{1}{(x-1)^4}+\frac{16}{(x-2)^5}-\frac{25}{12}\Big) \
H(c_2(\ao),c_1(\ao);x)+\Big(8+\frac{8}{(x-1)^5}\Big) \
H(0,0,0;\ao)+\Big(8-\frac{8}{(x-1)^5}\Big) H(0,0,0;x)+\Big(\frac{4 \
d_1}{(x-1)^5}+4 d_1\Big) \
H(0,0,1;\ao)+\Big(\frac{4}{(x-1)^5}-4+\frac{32}{(x-2)^5}\Big) \
H(0,0,c_1(\ao);x)+\Big(\frac{4 d_1}{(x-1)^5}+4 d_1\Big) H(0,1,0;\ao)+\
\Big(\frac{32 d_1}{(x-2)^5}+\frac{2 d_1}{(x-1)^5}-2 \
d_1-\frac{32}{(x-2)^5}-\frac{4}{(x-1)^5}+4\Big) \
H(0,1,0;x)+\Big(\frac{2 d_1^2}{(x-1)^5}+2 d_1^2\Big) \
H(0,1,1;\ao)+\Big(-\frac{32 d_1}{(x-2)^5}-\frac{2 d_1}{(x-1)^5}+2 \
d_1+\frac{32}{(x-2)^5}+\frac{4}{(x-1)^5}-4\Big) \
H(0,1,c_1(\ao);x)+\Big(\frac{2}{(x-1)^5}-3+\frac{80}{(x-2)^5}\Big) \
H(0,c_1(\ao),c_1(\ao);x)+\Big(-\frac{1}{(x-1)^5}+1-\frac{64}{(x-2)^5}\
\Big) H(0,c_2(\ao),c_1(\ao);x)+\Big(\frac{4 \
d_1}{(x-1)^5}-\frac{64}{(x-2)^5}-\frac{4}{(x-1)^5}+4\Big) H(1,0,0;x)+\
\Big(-\frac{2 \
d_1}{(x-1)^5}-\frac{16}{(x-2)^5}+\frac{2}{(x-1)^5}-1\Big) \
H(1,0,c_1(\ao);x)+\Big(-\frac{2 d_1^2}{(x-1)^5}+\frac{32 \
d_1}{(x-2)^5}+\frac{4 d_1}{(x-1)^5}-2 \
d_1+\frac{16}{(x-2)^5}-\frac{2}{(x-1)^5}+1\Big) \
H(1,1,0;x)+\Big(\frac{2 d_1^2}{(x-1)^5}-\frac{32 \
d_1}{(x-2)^5}-\frac{4 d_1}{(x-1)^5}+2 \
d_1-\frac{16}{(x-2)^5}+\frac{2}{(x-1)^5}-1\Big) \
H(1,1,c_1(\ao);x)+\Big(-\frac{2 \
d_1}{(x-1)^5}+\frac{80}{(x-2)^5}+\frac{2}{(x-1)^5}-3\Big) \
H(1,c_1(\ao),c_1(\ao);x)+\Big(\frac{32 \
d_1}{(x-2)^5}-\frac{64}{(x-2)^5}-\frac{1}{(x-1)^5}+1\Big) \
H(2,0,c_1(\ao);x)+\Big(-\frac{32 \
d_1}{(x-2)^5}+\frac{64}{(x-2)^5}+\frac{1}{(x-1)^5}-1\Big) H(2,1,0;x)+\
\Big(\frac{32 \
d_1}{(x-2)^5}-\frac{64}{(x-2)^5}-\frac{1}{(x-1)^5}+1\Big) \
H(2,1,c_1(\ao);x)+\Big(-\frac{32 \
d_1}{(x-2)^5}+\frac{64}{(x-2)^5}+\frac{1}{(x-1)^5}-1\Big) \
H(2,c_2(\ao),c_1(\ao);x)+\frac{2 \
H(c_1(\ao),c_1(\ao),c_1(\ao);x)}{(x-1)^5}+\frac{H(c_1(\ao),c_2(\ao),c_1(\ao);x)}{(x-1)^5}+\frac{32 \
H(c_2(\ao),0,c_1(\ao);x)}{(x-2)^5}-\frac{80 \
H(c_2(\ao),c_1(\ao),c_1(\ao);x)}{(x-2)^5}-\frac{7 \pi ^2}{6 \
(x-2)}+\frac{7 \pi ^2}{16 (x-1)}+\frac{11 \pi ^2}{9 \
(x-2)^2}+\frac{\pi ^2}{8 (x-1)^2}-\frac{5 \pi ^2}{3 \
(x-2)^3}-\frac{\pi ^2}{36 (x-1)^3}+\frac{8 \pi ^2}{3 (x-2)^4}+\frac{7 \
\pi ^2}{12 (x-1)^4}+\frac{4 \pi ^2}{(x-2)^5}+\frac{43 \pi ^2}{36 \
(x-1)^5}-\frac{28 \zeta_3}{(x-2)^5}-\frac{21 \zeta_3}{8 \
(x-1)^5}+\frac{17 \zeta_3}{8}+\frac{24 \pi ^2 \ln 2\, \
}{(x-2)^5}+\frac{\pi ^2 \ln 2\, }{4 (x-1)^5}-\frac{1}{4} \pi ^2 \ln 2\
\, -\frac{197 \pi ^2}{144}-\frac{15}{4}
\erp

%
% The B integral for k=2 and delta=1
%

\subsection{The $\cB$ integral for $k=2$ and $\delta=1$ and $d_1=-3$}
%
% This file contains the TeX output produced by Mathematica for the integral B1, for delta = -1
%
The $\eps$ expansion for this integral reads
\beq
\bsp
\begin{cal}I\end{cal}(x,\eps;\ao,3+d_1\eps;1,2,1,g_B) &= x\,\bint(\eps,x;3+d_1\eps;1,2)\\
&=\frac{1}{\eps}b_{-1}^{(1,2)}+b_0^{(1,2)}+\eps b_1^{(1,2)}+\eps^2b_2^{(1,2)} +\ocal\left(\eps^3\right),
\esp
\eeq
where
%1/ep piece
\brp
b_{-1}^{(1,2)}=-\frac{1}{6},
\erp
% ep^0
\brp
b_0^{(1,2)}=-\frac{\ao^6}{3 (x \ao-2 \ao-x)}+\frac{\ao^5}{3 (x-2)}+\frac{5 \
\ao^5}{3 (x \ao-2 \ao-x)}-\frac{5 \ao^4}{4 (x-2)}+\frac{\ao^4}{12 \
(x-1)}-\frac{10 \ao^4}{3 (x \ao-2 \ao-x)}+\frac{5 \ao^4}{6 \
(x-2)^2}+\frac{\ao^4}{12}+\frac{5 \ao^3}{3 (x-2)}-\frac{\ao^3}{3 \
(x-1)}+\frac{10 \ao^3}{3 (x \ao-2 \ao-x)}-\frac{20 \ao^3}{9 (x-2)^2}+\
\frac{\ao^3}{9 (x-1)^2}+\frac{20 \ao^3}{9 (x-2)^3}-\frac{4 \ao^3}{9}-\
\frac{5 \ao^2}{6 (x-2)}+\frac{\ao^2}{2 (x-1)}-\frac{5 \ao^2}{3 (x \
\ao-2 \ao-x)}+\frac{5 \ao^2}{3 (x-2)^2}-\frac{\ao^2}{3 \
(x-1)^2}-\frac{10 \ao^2}{3 (x-2)^3}+\frac{\ao^2}{6 (x-1)^3}+\frac{20 \
\ao^2}{3 (x-2)^4}+\ao^2-\frac{\ao}{3 (x-1)}+\frac{\ao}{3 (x \ao-2 \
\ao-x)}+\frac{\ao}{3 (x-1)^2}-\frac{\ao}{3 (x-1)^3}+\frac{\ao}{3 \
(x-1)^4}+\frac{80 \ao}{3 (x-2)^5}-\frac{4 \ao}{3}+\Big(\frac{1}{3 \
(x-1)^5}+\frac{1}{3}+\frac{80}{3 (x-2)^5}+\frac{160}{3 (x-2)^6}\Big) \
H(0;\ao)+\Big(-\frac{1}{3 (x-1)^5}+\frac{1}{3}-\frac{80}{3 \
(x-2)^5}-\frac{160}{3 (x-2)^6}\Big) H(0;x)+\frac{H(c_1(\ao);x)}{3 \
(x-1)^5}+\Big(\frac{80}{3 (x-2)^5}+\frac{160}{3 (x-2)^6}\Big) \
H(c_2(\ao);x)+\frac{80 \ln 2\, }{3 (x-2)^5}+\frac{160 \ln 2\, }{3 \
(x-2)^6}-\frac{11}{18},
\erp
% ep^1
\brp
b_1^{(1,2)}=\frac{1}{\ao x-x-2\ao}\Big\{\frac{37 x \ao^5}{72}+\frac{31 \ao^5}{36 (x-2)}-\frac{37 \ao^5}{72 \
(x-1)}-\frac{\ao^5}{12}-\frac{245 x \ao^4}{72}-\frac{5 \ao^4}{2 \
(x-2)}+\frac{161 \ao^4}{72 (x-1)}+\frac{65 \ao^4}{18 (x-2)^2}-\frac{5 \
\ao^4}{6 (x-1)^2}+\frac{11 \ao^4}{12}+\frac{91 x \ao^3}{9}+\frac{13 \
\ao^3}{18 (x-2)}-\frac{125 \ao^3}{36 (x-1)}-\frac{40 \ao^3}{9 \
(x-2)^2}+\frac{25 \ao^3}{8 (x-1)^2}+\frac{20 \ao^3}{(x-2)^3}-\frac{59 \
\ao^3}{36 (x-1)^3}-\frac{307 \ao^3}{72}-\frac{61 x \ao^2}{3}+\frac{65 \
\ao^2}{9 (x-2)}-\frac{\ao^2}{6 (x-1)}-\frac{43 \ao^2}{3 \
(x-2)^2}-\frac{265 \ao^2}{72 (x-1)^2}+\frac{400 \ao^2}{9 \
(x-2)^3}+\frac{20 \ao^2}{3 (x-1)^3}+\frac{1880 \ao^2}{9 \
(x-2)^4}-\frac{95 \ao^2}{18 (x-1)^4}+\frac{1091 \
\ao^2}{72}-\frac{1}{9} \pi ^2 x \ao+\frac{623 x \ao}{54}+\frac{14 \
\ao}{9 (x-2)}-\frac{35 \ao}{12 (x-1)}+\frac{2 \ao}{3 \
(x-2)^2}+\frac{61 \ao}{36 (x-1)^2}-\frac{248 \ao}{9 (x-2)^3}-\frac{9 \
\ao}{4 (x-1)^3}+\frac{76 \pi ^2 \ao}{9 (x-2)^4}-\frac{3760 \ao}{9 \
(x-2)^4}+\frac{\pi ^2 \ao}{9 (x-1)^4}-\frac{95 \ao}{18 \
(x-1)^4}+\frac{80 \pi ^2 \ao}{9 (x-2)^5}-\frac{5120 \ao}{9 \
(x-2)^5}-\frac{\pi ^2 \ao}{9 (x-1)^5}+\frac{160 \ln ^22\,  \ao}{3 \
(x-2)^4}+\frac{320 \ln ^22\,  \ao}{3 (x-2)^5}+\frac{160 \ln 2\,  \
\ao}{9 (x-2)^4}+\frac{320 \ln 2\,  \ao}{9 (x-2)^5}+\frac{2 \pi ^2 \
\ao}{9}+\frac{859 \ao}{108}+\frac{\pi ^2 x}{9}+\frac{85 x}{54}+\Big(-\
\frac{x \ao^5}{3}-\frac{2 \ao^5}{3 (x-2)}+\frac{\ao^5}{3 \
(x-1)}+\frac{19 x \ao^4}{9}+\frac{20 \ao^4}{9 (x-2)}-\frac{13 \
\ao^4}{9 (x-1)}-\frac{20 \ao^4}{9 (x-2)^2}+\frac{4 \ao^4}{9 (x-1)^2}-\
\frac{2 \ao^4}{9}-\frac{52 x \ao^3}{9}-\frac{20 \ao^3}{9 \
(x-2)}+\frac{22 \ao^3}{9 (x-1)}+\frac{40 \ao^3}{9 (x-2)^2}-\frac{14 \
\ao^3}{9 (x-1)^2}-\frac{80 \ao^3}{9 (x-2)^3}+\frac{2 \ao^3}{3 \
(x-1)^3}+\frac{4 \ao^3}{3}+\frac{28 x \ao^2}{3}-\frac{2 \
\ao^2}{x-1}+\frac{2 \ao^2}{(x-1)^2}-\frac{2 \ao^2}{(x-1)^3}-\frac{160 \
\ao^2}{3 (x-2)^4}+\frac{4 \ao^2}{3 (x-1)^4}-4 \ao^2-\frac{11 x \
\ao}{2}-\frac{128 \ao}{9 (x-2)}+\frac{31 \ao}{2 (x-1)}+\frac{152 \
\ao}{9 (x-2)^2}-\frac{37 \ao}{9 (x-1)^2}-\frac{16 \
\ao}{(x-2)^3}+\frac{47 \ao}{18 (x-1)^3}+\frac{2080 \ao}{9 \
(x-2)^4}+\frac{97 \ao}{36 (x-1)^4}+\frac{2240 \ao}{9 \
(x-2)^5}-\frac{47 \ao}{18 (x-1)^5}-\frac{37 \
\ao}{12}+\frac{x}{6}-\frac{88}{9 (x-2)}+\frac{10}{x-1}+\frac{104}{9 \
(x-2)^2}-\frac{25}{18 (x-1)^2}-\frac{160}{9 (x-2)^3}+\frac{1}{9 \
(x-1)^3}-\frac{832}{9 (x-2)^4}-\frac{139}{36 (x-1)^4}-\frac{2560}{9 \
(x-2)^5}-\frac{47}{18 (x-1)^5}-\frac{640}{9 (x-2)^6}+\frac{3}{4}\Big) \
H(0;\ao)+\Big(\frac{x \ao^5}{2}+\frac{\ao^5}{x-2}-\frac{\ao^5}{2 \
(x-1)}-\frac{19 x \ao^4}{6}-\frac{10 \ao^4}{3 (x-2)}+\frac{13 \
\ao^4}{6 (x-1)}+\frac{10 \ao^4}{3 (x-2)^2}-\frac{2 \ao^4}{3 (x-1)^2}+\
\frac{\ao^4}{3}+\frac{26 x \ao^3}{3}+\frac{10 \ao^3}{3 \
(x-2)}-\frac{11 \ao^3}{3 (x-1)}-\frac{20 \ao^3}{3 (x-2)^2}+\frac{7 \
\ao^3}{3 (x-1)^2}+\frac{40 \ao^3}{3 (x-2)^3}-\frac{\ao^3}{(x-1)^3}-2 \
\ao^3-14 x \ao^2+\frac{3 \ao^2}{x-1}-\frac{3 \ao^2}{(x-1)^2}+\frac{3 \
\ao^2}{(x-1)^3}+\frac{80 \ao^2}{(x-2)^4}-\frac{2 \ao^2}{(x-1)^4}+6 \
\ao^2+\frac{73 x \ao}{6}-\frac{5 \ao}{3 (x-2)}-\frac{7 \ao}{6 (x-1)}+\
\frac{20 \ao}{3 (x-2)^2}+\frac{5 \ao}{3 (x-1)^2}-\frac{40 \
\ao}{(x-2)^3}-\frac{3 \ao}{(x-1)^3}-\frac{320 \ao}{(x-2)^4}-\frac{320 \
\ao}{(x-2)^5}-\frac{10 \ao}{3}-\frac{25 x}{6}+\frac{2}{3 \
(x-2)}+\frac{1}{6 (x-1)}-\frac{10}{3 (x-2)^2}-\frac{1}{3 \
(x-1)^2}+\frac{80}{3 \
(x-2)^3}+\frac{1}{(x-1)^3}+\frac{240}{(x-2)^4}+\frac{2}{(x-1)^4}+\frac{320}{(x-2)^5}-1\Big) H(1;\ao)+\Big(\frac{2 x \ao}{3}+\frac{112 \
\ao}{3 (x-2)^4}-\frac{8 \ao}{3 (x-1)^4}+\frac{320 \ao}{3 \
(x-2)^5}+\frac{8 \ao}{3 (x-1)^5}-\frac{4 \ao}{3}-\frac{2 \
x}{3}-\frac{112}{3 (x-2)^4}+\frac{8}{3 (x-1)^4}-\frac{544}{3 \
(x-2)^5}+\frac{8}{3 (x-1)^5}-\frac{640}{3 (x-2)^6}\Big) H(0;\ao) \
H(1;x)+\Big(-\frac{x \ao^5}{6}-\frac{\ao^5}{3 (x-2)}+\frac{\ao^5}{6 \
(x-1)}+\frac{\ao^5}{4}+\frac{19 x \ao^4}{18}+\frac{23 \ao^4}{18 \
(x-2)}-\frac{13 \ao^4}{18 (x-1)}-\frac{10 \ao^4}{9 (x-2)^2}+\frac{2 \
\ao^4}{9 (x-1)^2}-\frac{49 \ao^4}{36}-\frac{26 x \ao^3}{9}-\frac{16 \
\ao^3}{9 (x-2)}+\frac{11 \ao^3}{9 (x-1)}+\frac{26 \ao^3}{9 \
(x-2)^2}-\frac{7 \ao^3}{9 (x-1)^2}-\frac{40 \ao^3}{9 \
(x-2)^3}+\frac{\ao^3}{3 (x-1)^3}+\frac{19 \ao^3}{6}+\frac{14 x \
\ao^2}{3}+\frac{\ao^2}{x-2}-\frac{\ao^2}{x-1}-\frac{2 \
\ao^2}{(x-2)^2}+\frac{\ao^2}{(x-1)^2}+\frac{4 \
\ao^2}{(x-2)^3}-\frac{\ao^2}{(x-1)^3}-\frac{80 \ao^2}{3 \
(x-2)^4}+\frac{2 \ao^2}{3 (x-1)^4}-\frac{9 \ao^2}{2}-\frac{73 x \
\ao}{18}-\frac{128 \ao}{9 (x-2)}+\frac{31 \ao}{2 (x-1)}+\frac{152 \
\ao}{9 (x-2)^2}-\frac{37 \ao}{9 (x-1)^2}-\frac{16 \
\ao}{(x-2)^3}+\frac{47 \ao}{18 (x-1)^3}+\frac{144 \
\ao}{(x-2)^4}+\frac{73 \ao}{36 (x-1)^4}+\frac{320 \ao}{3 \
(x-2)^5}-\frac{47 \ao}{18 (x-1)^5}+\frac{61 \ao}{36}+\frac{25 x}{18}+\
\Big(-\frac{4 \ao}{3 (x-1)^4}+\frac{4 \ao}{3 (x-1)^5}+\frac{4}{3 \
(x-1)^4}+\frac{4}{3 (x-1)^5}\Big) H(0;\ao)+\Big(\frac{2 \
\ao}{(x-1)^4}-\frac{2 \
\ao}{(x-1)^5}-\frac{2}{(x-1)^4}-\frac{2}{(x-1)^5}\Big) \
H(1;\ao)-\frac{88}{9 (x-2)}+\frac{10}{x-1}+\frac{104}{9 \
(x-2)^2}-\frac{25}{18 (x-1)^2}-\frac{160}{9 (x-2)^3}+\frac{1}{9 \
(x-1)^3}-\frac{224}{3 (x-2)^4}-\frac{139}{36 (x-1)^4}-\frac{640}{3 \
(x-2)^5}-\frac{47}{18 (x-1)^5}+\frac{3}{4}\Big) \
H(c_1(\ao);x)+\Big(\frac{160 \ao}{9 (x-2)^4}+\frac{320 \ao}{9 \
(x-2)^5}+\Big(-\frac{320 \ao}{3 (x-2)^4}-\frac{640 \ao}{3 \
(x-2)^5}+\frac{320}{3 (x-2)^4}+\frac{1280}{3 (x-2)^5}+\frac{1280}{3 \
(x-2)^6}\Big) H(0;\ao)+\Big(\frac{160 \ao}{(x-2)^4}+\frac{320 \
\ao}{(x-2)^5}-\frac{160}{(x-2)^4}-\frac{640}{(x-2)^5}-\frac{640}{(x-2)\
^6}\Big) H(1;\ao)-\frac{160}{9 (x-2)^4}-\frac{640}{9 \
(x-2)^5}-\frac{640}{9 (x-2)^6}\Big) H(c_2(\ao);x)+\Big(-\frac{4 x \
\ao}{3}-\frac{320 \ao}{3 (x-2)^4}-\frac{4 \ao}{3 (x-1)^4}-\frac{640 \
\ao}{3 (x-2)^5}+\frac{4 \ao}{3 (x-1)^5}+\frac{8 \ao}{3}+\frac{4 \
x}{3}+\frac{320}{3 (x-2)^4}+\frac{4}{3 (x-1)^4}+\frac{1280}{3 \
(x-2)^5}+\frac{4}{3 (x-1)^5}+\frac{1280}{3 (x-2)^6}\Big) \
H(0,0;\ao)+\Big(-\frac{4 x \ao}{3}+\frac{320 \ao}{3 (x-2)^4}+\frac{4 \
\ao}{3 (x-1)^4}+\frac{640 \ao}{3 (x-2)^5}-\frac{4 \ao}{3 \
(x-1)^5}+\frac{8 \ao}{3}+\frac{4 x}{3}-\frac{320}{3 \
(x-2)^4}-\frac{4}{3 (x-1)^4}-\frac{1280}{3 (x-2)^5}-\frac{4}{3 \
(x-1)^5}-\frac{1280}{3 (x-2)^6}\Big) H(0,0;x)+\Big(2 x \ao+\frac{160 \
\ao}{(x-2)^4}+\frac{2 \ao}{(x-1)^4}+\frac{320 \ao}{(x-2)^5}-\frac{2 \
\ao}{(x-1)^5}-4 \ao-2 \
x-\frac{160}{(x-2)^4}-\frac{2}{(x-1)^4}-\frac{640}{(x-2)^5}-\frac{2}{(\
x-1)^5}-\frac{640}{(x-2)^6}\Big) H(0,1;\ao)+\Big(\frac{2 x \
\ao}{3}+\frac{112 \ao}{3 (x-2)^4}-\frac{2 \ao}{3 (x-1)^4}+\frac{320 \
\ao}{3 (x-2)^5}+\frac{2 \ao}{3 (x-1)^5}-\frac{4 \ao}{3}-\frac{2 \
x}{3}-\frac{112}{3 (x-2)^4}+\frac{2}{3 (x-1)^4}-\frac{544}{3 \
(x-2)^5}+\frac{2}{3 (x-1)^5}-\frac{640}{3 (x-2)^6}\Big) \
H(0,c_1(\ao);x)+\Big(-\frac{320 \ao}{3 (x-2)^4}-\frac{640 \ao}{3 \
(x-2)^5}+\frac{320}{3 (x-2)^4}+\frac{1280}{3 (x-2)^5}+\frac{1280}{3 \
(x-2)^6}\Big) H(0,c_2(\ao);x)+\Big(-\frac{2 x \ao}{3}-\frac{112 \
\ao}{3 (x-2)^4}+\frac{8 \ao}{3 (x-1)^4}-\frac{320 \ao}{3 \
(x-2)^5}-\frac{8 \ao}{3 (x-1)^5}+\frac{4 \ao}{3}+\frac{2 \
x}{3}+\frac{112}{3 (x-2)^4}-\frac{8}{3 (x-1)^4}+\frac{544}{3 \
(x-2)^5}-\frac{8}{3 (x-1)^5}+\frac{640}{3 (x-2)^6}\Big) \
H(1,0;x)+\Big(\frac{2 x \ao}{3}+\frac{112 \ao}{3 (x-2)^4}-\frac{8 \
\ao}{3 (x-1)^4}+\frac{320 \ao}{3 (x-2)^5}+\frac{8 \ao}{3 \
(x-1)^5}-\frac{4 \ao}{3}-\frac{2 x}{3}-\frac{112}{3 \
(x-2)^4}+\frac{8}{3 (x-1)^4}-\frac{544}{3 (x-2)^5}+\frac{8}{3 \
(x-1)^5}-\frac{640}{3 (x-2)^6}\Big) H(1,c_1(\ao);x)+\Big(-\frac{800 \
\ao}{3 (x-2)^4}-\frac{1600 \ao}{3 (x-2)^5}+\frac{800}{3 \
(x-2)^4}+\frac{3200}{3 (x-2)^5}+\frac{3200}{3 (x-2)^6}\Big) H(2,0;x)+\
\Big(\frac{800 \ao}{3 (x-2)^4}+\frac{1600 \ao}{3 \
(x-2)^5}-\frac{800}{3 (x-2)^4}-\frac{3200}{3 (x-2)^5}-\frac{3200}{3 \
(x-2)^6}\Big) H(2,c_2(\ao);x)+\Big(-\frac{2 \ao}{3 (x-1)^4}+\frac{2 \
\ao}{3 (x-1)^5}+\frac{2}{3 (x-1)^4}+\frac{2}{3 (x-1)^5}\Big) \
H(c_1(\ao),c_1(\ao);x)+\Big(-\frac{112 \ao}{3 (x-2)^4}-\frac{320 \
\ao}{3 (x-2)^5}+\frac{112}{3 (x-2)^4}+\frac{544}{3 \
(x-2)^5}+\frac{640}{3 (x-2)^6}\Big) H(c_2(\ao),c_1(\ao);x)+H(0;x) \
\Big(\frac{47 x \ao}{18}+\frac{128 \ao}{9 (x-2)}-\frac{31 \ao}{2 \
(x-1)}-\frac{152 \ao}{9 (x-2)^2}+\frac{37 \ao}{9 (x-1)^2}+\frac{16 \
\ao}{(x-2)^3}-\frac{47 \ao}{18 (x-1)^3}-\frac{1120 \ao}{9 \
(x-2)^4}-\frac{49 \ao}{36 (x-1)^4}-\frac{320 \ao}{9 (x-2)^5}+\frac{47 \
\ao}{18 (x-1)^5}-\frac{320 \ln 2\,  \ao}{3 (x-2)^4}-\frac{640 \ln 2\, \
 \ao}{3 (x-2)^5}-\frac{161 \ao}{36}-\frac{47 x}{18}+\frac{88}{9 \
(x-2)}-\frac{10}{x-1}-\frac{104}{9 (x-2)^2}+\frac{25}{18 \
(x-1)^2}+\frac{160}{9 (x-2)^3}-\frac{1}{9 (x-1)^3}+\frac{832}{9 \
(x-2)^4}+\frac{139}{36 (x-1)^4}+\frac{2560}{9 (x-2)^5}+\frac{47}{18 \
(x-1)^5}+\frac{640}{9 (x-2)^6}+\frac{320 \ln 2\, }{3 \
(x-2)^4}+\frac{1280 \ln 2\, }{3 (x-2)^5}+\frac{1280 \ln 2\, }{3 \
(x-2)^6}-\frac{3}{4}\Big)+H(2;x) \Big(\frac{800 \ln 2\,  \ao}{3 \
(x-2)^4}+\frac{1600 \ln 2\,  \ao}{3 (x-2)^5}+\Big(\frac{800 \ao}{3 \
(x-2)^4}+\frac{1600 \ao}{3 (x-2)^5}-\frac{800}{3 \
(x-2)^4}-\frac{3200}{3 (x-2)^5}-\frac{3200}{3 (x-2)^6}\Big) H(0;\ao)-\
\frac{800 \ln 2\, }{3 (x-2)^4}-\frac{3200 \ln 2\, }{3 \
(x-2)^5}-\frac{3200 \ln 2\, }{3 (x-2)^6}\Big)-\frac{76 \pi ^2}{9 \
(x-2)^4}-\frac{\pi ^2}{9 (x-1)^4}-\frac{232 \pi ^2}{9 (x-2)^5}-\frac{\
\pi ^2}{9 (x-1)^5}-\frac{160 \pi ^2}{9 (x-2)^6}-\frac{160 \ln ^22\, \
}{3 (x-2)^4}-\frac{640 \ln ^22\, }{3 (x-2)^5}-\frac{640 \ln ^22\, }{3 \
(x-2)^6}-\frac{160 \ln 2\, }{9 (x-2)^4}-\frac{640 \ln 2\, }{9 \
(x-2)^5}-\frac{640 \ln 2\, }{9 (x-2)^6}\},
\erp
% ep^2
\brp
b_2^{(1,2)}=\frac{1}{\ao x-x-2\ao}\Big\{-\frac{1}{72} \pi ^2 x \ao^5+\frac{895 x \ao^5}{432}-\frac{\pi ^2 \
\ao^5}{36 (x-2)}+\frac{673 \ao^5}{216 (x-2)}+\frac{\pi ^2 \ao^5}{72 \
(x-1)}-\frac{895 \ao^5}{432 (x-1)}-\frac{37 \ao^5}{72}+\frac{19}{216} \
\pi ^2 x \ao^4-\frac{18829 x \ao^4}{1296}+\frac{5 \pi ^2 \ao^4}{54 \
(x-2)}-\frac{2345 \ao^4}{324 (x-2)}-\frac{13 \pi ^2 \ao^4}{216 \
(x-1)}+\frac{12073 \ao^4}{1296 (x-1)}-\frac{5 \pi ^2 \ao^4}{54 \
(x-2)^2}+\frac{5405 \ao^4}{324 (x-2)^2}+\frac{\pi ^2 \ao^4}{54 \
(x-1)^2}-\frac{1351 \ao^4}{324 (x-1)^2}-\frac{\pi ^2 \
\ao^4}{108}+\frac{3691 \ao^4}{648}-\frac{13}{54} \pi ^2 x \
\ao^3+\frac{7859 x \ao^3}{162}-\frac{5 \pi ^2 \ao^3}{54 \
(x-2)}-\frac{3451 \ao^3}{324 (x-2)}+\frac{11 \pi ^2 \ao^3}{108 \
(x-1)}-\frac{7985 \ao^3}{648 (x-1)}+\frac{5 \pi ^2 \ao^3}{27 \
(x-2)^2}-\frac{115 \ao^3}{81 (x-2)^2}-\frac{7 \pi ^2 \ao^3}{108 \
(x-1)^2}+\frac{22919 \ao^3}{1296 (x-1)^2}-\frac{10 \pi ^2 \ao^3}{27 \
(x-2)^3}+\frac{10580 \ao^3}{81 (x-2)^3}+\frac{\pi ^2 \ao^3}{36 \
(x-1)^3}-\frac{2401 \ao^3}{216 (x-1)^3}+\frac{\pi ^2 \
\ao^3}{18}-\frac{12859 \ao^3}{432}+\frac{7}{18} \pi ^2 x \
\ao^2-\frac{7613 x \ao^2}{54}+\frac{211 \ao^2}{2 (x-2)}-\frac{\pi ^2 \
\ao^2}{12 (x-1)}-\frac{823 \ao^2}{18 (x-1)}-\frac{3931 \ao^2}{18 \
(x-2)^2}+\frac{\pi ^2 \ao^2}{12 (x-1)^2}-\frac{685 \ao^2}{48 \
(x-1)^2}+\frac{7600 \ao^2}{9 (x-2)^3}-\frac{\pi ^2 \ao^2}{12 \
(x-1)^3}+\frac{2507 \ao^2}{36 (x-1)^3}-\frac{20 \pi ^2 \ao^2}{9 \
(x-2)^4}+\frac{66760 \ao^2}{27 (x-2)^4}+\frac{\pi ^2 \ao^2}{18 \
(x-1)^4}-\frac{6685 \ao^2}{108 (x-1)^4}-\frac{\pi ^2 \
\ao^2}{6}+\frac{22817 \ao^2}{144}-\frac{29}{24} \pi ^2 x \
\ao+\frac{16423 x \ao}{162}-\frac{10 \pi ^2 \ao}{9 (x-2)}+\frac{65 \
\ao}{3 (x-2)}+\frac{409 \pi ^2 \ao}{216 (x-1)}-\frac{3017 \ao}{72 \
(x-1)}+\frac{2 \pi ^2 \ao}{3 (x-2)^2}+\frac{13 \ao}{(x-2)^2}-\frac{4 \
\pi ^2 \ao}{9 (x-1)^2}+\frac{1715 \ao}{72 (x-1)^2}+\frac{8 \pi ^2 \
\ao}{3 (x-2)^3}-\frac{4936 \ao}{9 (x-2)^3}+\frac{55 \pi ^2 \ao}{108 \
(x-1)^3}-\frac{1033 \ao}{24 (x-1)^3}+\frac{1136 \pi ^2 \ao}{27 \
(x-2)^4}-\frac{133520 \ao}{27 (x-2)^4}+\frac{55 \pi ^2 \ao}{108 \
(x-1)^4}-\frac{6685 \ao}{108 (x-1)^4}+\frac{400 \pi ^2 \ao}{27 \
(x-2)^5}-\frac{154240 \ao}{27 (x-2)^5}-\frac{47 \pi ^2 \ao}{54 \
(x-1)^5}+\frac{17}{12} x \zeta_3 \ao+\frac{56 \zeta_3 \ao}{3 \
(x-2)^4}-\frac{7 \zeta_3 \ao}{4 (x-1)^4}+\frac{280 \zeta_3 \ao}{3 \
(x-2)^5}+\frac{7 \zeta_3 \ao}{4 (x-1)^5}-\frac{17}{6} \zeta_3 \
\ao+\frac{640 \ln ^32\,  \ao}{9 (x-2)^4}+\frac{1280 \ln ^32\,  \ao}{9 \
(x-2)^5}+\frac{320 \ln ^22\,  \ao}{9 (x-2)^4}+\frac{640 \ln ^22\,  \
\ao}{9 (x-2)^5}-\frac{1}{6} \pi ^2 x \ln 2\,  \ao+\frac{112 \pi ^2 \
\ln 2\,  \ao}{3 (x-2)^4}+\frac{320 \ln 2\,  \ao}{27 \
(x-2)^4}+\frac{\pi ^2 \ln 2\,  \ao}{6 (x-1)^4}+\frac{80 \pi ^2 \ln \
2\,  \ao}{3 (x-2)^5}+\frac{640 \ln 2\,  \ao}{27 (x-2)^5}-\frac{\pi ^2 \
\ln 2\,  \ao}{6 (x-1)^5}+\frac{1}{3} \pi ^2 \ln 2\,  \ao+\frac{95 \pi \
^2 \ao}{72}+\frac{20569 \ao}{648}+\frac{71 \pi ^2 x}{72}+\frac{575 \
x}{162}+\Big(-\frac{37 x \ao^5}{18}-\frac{31 \ao^5}{9 (x-2)}+\frac{37 \
\ao^5}{18 (x-1)}+\frac{\ao^5}{3}+\frac{245 x \ao^4}{18}+\frac{10 \
\ao^4}{x-2}-\frac{161 \ao^4}{18 (x-1)}-\frac{130 \ao^4}{9 \
(x-2)^2}+\frac{10 \ao^4}{3 (x-1)^2}-\frac{11 \ao^4}{3}-\frac{364 x \
\ao^3}{9}-\frac{26 \ao^3}{9 (x-2)}+\frac{125 \ao^3}{9 \
(x-1)}+\frac{160 \ao^3}{9 (x-2)^2}-\frac{25 \ao^3}{2 \
(x-1)^2}-\frac{80 \ao^3}{(x-2)^3}+\frac{59 \ao^3}{9 \
(x-1)^3}+\frac{307 \ao^3}{18}+\frac{244 x \ao^2}{3}-\frac{260 \
\ao^2}{9 (x-2)}+\frac{2 \ao^2}{3 (x-1)}+\frac{172 \ao^2}{3 \
(x-2)^2}+\frac{265 \ao^2}{18 (x-1)^2}-\frac{1600 \ao^2}{9 \
(x-2)^3}-\frac{80 \ao^2}{3 (x-1)^3}-\frac{7520 \ao^2}{9 \
(x-2)^4}+\frac{190 \ao^2}{9 (x-1)^4}-\frac{1091 \
\ao^2}{18}-\frac{1}{18} \pi ^2 x \ao-\frac{7109 x \
\ao}{108}-\frac{1804 \ao}{9 (x-2)}+\frac{2002 \ao}{9 \
(x-1)}+\frac{1936 \ao}{9 (x-2)^2}-\frac{413 \ao}{9 (x-1)^2}-\frac{496 \
\ao}{9 (x-2)^3}+\frac{227 \ao}{6 (x-1)^3}-\frac{40 \pi ^2 \ao}{9 \
(x-2)^4}+\frac{76160 \ao}{27 (x-2)^4}-\frac{\pi ^2 \ao}{18 \
(x-1)^4}+\frac{5381 \ao}{216 (x-1)^4}-\frac{80 \pi ^2 \ao}{9 \
(x-2)^5}+\frac{62080 \ao}{27 (x-2)^5}+\frac{\pi ^2 \ao}{18 \
(x-1)^5}-\frac{2125 \ao}{108 (x-1)^5}+\frac{\pi ^2 \ao}{9}-\frac{811 \
\ao}{216}+\frac{\pi ^2 x}{18}+\frac{1445 x}{108}-\frac{1372}{9 \
(x-2)}+\frac{5705}{36 (x-1)}+\frac{512}{3 (x-2)^2}-\frac{475}{36 \
(x-1)^2}-\frac{2432}{9 (x-2)^3}-\frac{35}{12 (x-1)^3}+\frac{40 \pi \
^2}{9 (x-2)^4}-\frac{22112}{27 (x-2)^4}+\frac{\pi ^2}{18 \
(x-1)^4}-\frac{7679}{216 (x-1)^4}+\frac{160 \pi ^2}{9 \
(x-2)^5}-\frac{62720}{27 (x-2)^5}+\frac{\pi ^2}{18 \
(x-1)^5}-\frac{2125}{108 (x-1)^5}+\frac{160 \pi ^2}{9 \
(x-2)^6}-\frac{1280}{27 (x-2)^6}+\frac{271}{24}\Big) \
H(0;\ao)+\Big(\frac{37 x \ao^5}{12}+\frac{31 \ao^5}{6 (x-2)}-\frac{37 \
\ao^5}{12 (x-1)}-\frac{\ao^5}{2}-\frac{245 x \ao^4}{12}-\frac{15 \
\ao^4}{x-2}+\frac{161 \ao^4}{12 (x-1)}+\frac{65 \ao^4}{3 \
(x-2)^2}-\frac{5 \ao^4}{(x-1)^2}+\frac{11 \ao^4}{2}+\frac{182 x \
\ao^3}{3}+\frac{13 \ao^3}{3 (x-2)}-\frac{125 \ao^3}{6 (x-1)}-\frac{80 \
\ao^3}{3 (x-2)^2}+\frac{75 \ao^3}{4 (x-1)^2}+\frac{120 \
\ao^3}{(x-2)^3}-\frac{59 \ao^3}{6 (x-1)^3}-\frac{307 \ao^3}{12}-122 x \
\ao^2+\frac{130 \ao^2}{3 (x-2)}-\frac{\ao^2}{x-1}-\frac{86 \
\ao^2}{(x-2)^2}-\frac{265 \ao^2}{12 (x-1)^2}+\frac{800 \ao^2}{3 \
(x-2)^3}+\frac{40 \ao^2}{(x-1)^3}+\frac{3760 \ao^2}{3 \
(x-2)^4}-\frac{95 \ao^2}{3 (x-1)^4}+\frac{1091 \ao^2}{12}+\frac{513 x \
\ao}{4}-\frac{205 \ao}{6 (x-2)}-\frac{25 \ao}{12 (x-1)}+\frac{344 \
\ao}{3 (x-2)^2}+\frac{211 \ao}{12 (x-1)^2}-\frac{792 \
\ao}{(x-2)^3}-\frac{107 \ao}{2 (x-1)^3}-\frac{12640 \ao}{3 \
(x-2)^4}-\frac{10240 \ao}{3 (x-2)^5}-\frac{245 \ao}{4}-\frac{595 \
x}{12}-\frac{11}{3 (x-2)}+\frac{163}{12 (x-1)}-\frac{71}{3 \
(x-2)^2}-\frac{37}{4 (x-1)^2}+\frac{1216}{3 (x-2)^3}+\frac{70}{3 \
(x-1)^3}+\frac{2960}{(x-2)^4}+\frac{95}{3 (x-1)^4}+\frac{10240}{3 \
(x-2)^5}-\frac{109}{12}\Big) H(1;\ao)+\Big(\frac{4 x \
\ao^5}{3}+\frac{8 \ao^5}{3 (x-2)}-\frac{4 \ao^5}{3 (x-1)}-\frac{76 x \
\ao^4}{9}-\frac{80 \ao^4}{9 (x-2)}+\frac{52 \ao^4}{9 (x-1)}+\frac{80 \
\ao^4}{9 (x-2)^2}-\frac{16 \ao^4}{9 (x-1)^2}+\frac{8 \
\ao^4}{9}+\frac{208 x \ao^3}{9}+\frac{80 \ao^3}{9 (x-2)}-\frac{88 \
\ao^3}{9 (x-1)}-\frac{160 \ao^3}{9 (x-2)^2}+\frac{56 \ao^3}{9 \
(x-1)^2}+\frac{320 \ao^3}{9 (x-2)^3}-\frac{8 \ao^3}{3 \
(x-1)^3}-\frac{16 \ao^3}{3}-\frac{112 x \ao^2}{3}+\frac{8 \
\ao^2}{x-1}-\frac{8 \ao^2}{(x-1)^2}+\frac{8 \ao^2}{(x-1)^3}+\frac{640 \
\ao^2}{3 (x-2)^4}-\frac{16 \ao^2}{3 (x-1)^4}+16 \ao^2+22 x \
\ao+\frac{512 \ao}{9 (x-2)}-\frac{62 \ao}{x-1}-\frac{608 \ao}{9 \
(x-2)^2}+\frac{148 \ao}{9 (x-1)^2}+\frac{64 \ao}{(x-2)^3}-\frac{94 \
\ao}{9 (x-1)^3}-\frac{8320 \ao}{9 (x-2)^4}-\frac{97 \ao}{9 \
(x-1)^4}-\frac{8960 \ao}{9 (x-2)^5}+\frac{94 \ao}{9 (x-1)^5}+\frac{37 \
\ao}{3}-\frac{2 x}{3}+\frac{352}{9 (x-2)}-\frac{40}{x-1}-\frac{416}{9 \
(x-2)^2}+\frac{50}{9 (x-1)^2}+\frac{640}{9 (x-2)^3}-\frac{4}{9 \
(x-1)^3}+\frac{3328}{9 (x-2)^4}+\frac{139}{9 (x-1)^4}+\frac{10240}{9 \
(x-2)^5}+\frac{94}{9 (x-1)^5}+\frac{2560}{9 (x-2)^6}-3\Big) \
H(0,0;\ao)+\Big(-2 x \ao^5-\frac{4 \ao^5}{x-2}+\frac{2 \
\ao^5}{x-1}+\frac{38 x \ao^4}{3}+\frac{40 \ao^4}{3 (x-2)}-\frac{26 \
\ao^4}{3 (x-1)}-\frac{40 \ao^4}{3 (x-2)^2}+\frac{8 \ao^4}{3 (x-1)^2}-\
\frac{4 \ao^4}{3}-\frac{104 x \ao^3}{3}-\frac{40 \ao^3}{3 \
(x-2)}+\frac{44 \ao^3}{3 (x-1)}+\frac{80 \ao^3}{3 (x-2)^2}-\frac{28 \
\ao^3}{3 (x-1)^2}-\frac{160 \ao^3}{3 (x-2)^3}+\frac{4 \
\ao^3}{(x-1)^3}+8 \ao^3+56 x \ao^2-\frac{12 \ao^2}{x-1}+\frac{12 \
\ao^2}{(x-1)^2}-\frac{12 \ao^2}{(x-1)^3}-\frac{320 \
\ao^2}{(x-2)^4}+\frac{8 \ao^2}{(x-1)^4}-24 \ao^2-33 x \ao-\frac{256 \
\ao}{3 (x-2)}+\frac{93 \ao}{x-1}+\frac{304 \ao}{3 (x-2)^2}-\frac{74 \
\ao}{3 (x-1)^2}-\frac{96 \ao}{(x-2)^3}+\frac{47 \ao}{3 \
(x-1)^3}+\frac{4160 \ao}{3 (x-2)^4}+\frac{97 \ao}{6 \
(x-1)^4}+\frac{4480 \ao}{3 (x-2)^5}-\frac{47 \ao}{3 (x-1)^5}-\frac{37 \
\ao}{2}+x-\frac{176}{3 (x-2)}+\frac{60}{x-1}+\frac{208}{3 \
(x-2)^2}-\frac{25}{3 (x-1)^2}-\frac{320}{3 (x-2)^3}+\frac{2}{3 \
(x-1)^3}-\frac{1664}{3 (x-2)^4}-\frac{139}{6 (x-1)^4}-\frac{5120}{3 \
(x-2)^5}-\frac{47}{3 (x-1)^5}-\frac{1280}{3 (x-2)^6}+\frac{9}{2}\Big) \
H(0,1;\ao)+H(1;x) \Big(\frac{1}{9} \pi ^2 x \ao+\frac{16 \pi ^2 \
\ao}{(x-2)^4}-\frac{8 \pi ^2 \ao}{9 (x-1)^4}+\frac{80 \pi ^2 \ao}{3 \
(x-2)^5}+\frac{8 \pi ^2 \ao}{9 (x-1)^5}-\frac{2 \pi ^2 \ao}{9}-\frac{\
\pi ^2 x}{9}+\Big(\frac{47 x \ao}{9}+\frac{838 \ao}{9 \
(x-2)}-\frac{102 \ao}{x-1}-\frac{1000 \ao}{9 (x-2)^2}+\frac{433 \
\ao}{18 (x-1)^2}+\frac{128 \ao}{(x-2)^3}-\frac{113 \ao}{9 \
(x-1)^3}-\frac{4288 \ao}{9 (x-2)^4}-\frac{137 \ao}{9 \
(x-1)^4}+\frac{640 \ao}{9 (x-2)^5}+\frac{188 \ao}{9 (x-1)^5}-\frac{77 \
\ao}{18}-\frac{47 x}{9}+\frac{578}{9 (x-2)}-\frac{203}{3 \
(x-1)}-\frac{676}{9 (x-2)^2}+\frac{185}{18 (x-1)^2}+\frac{848}{9 \
(x-2)^3}-\frac{11}{9 (x-1)^3}+\frac{1984}{9 (x-2)^4}+\frac{239}{9 \
(x-1)^4}+\frac{7936}{9 (x-2)^5}+\frac{188}{9 (x-1)^5}-\frac{1280}{9 \
(x-2)^6}-\frac{37}{6}\Big) H(0;\ao)+\Big(-\frac{8 x \ao}{3}-\frac{448 \
\ao}{3 (x-2)^4}+\frac{32 \ao}{3 (x-1)^4}-\frac{1280 \ao}{3 \
(x-2)^5}-\frac{32 \ao}{3 (x-1)^5}+\frac{16 \ao}{3}+\frac{8 \
x}{3}+\frac{448}{3 (x-2)^4}-\frac{32}{3 (x-1)^4}+\frac{2176}{3 \
(x-2)^5}-\frac{32}{3 (x-1)^5}+\frac{2560}{3 (x-2)^6}\Big) H(0,0;\ao)+\
\Big(4 x \ao+\frac{224 \ao}{(x-2)^4}-\frac{16 \ao}{(x-1)^4}+\frac{640 \
\ao}{(x-2)^5}+\frac{16 \ao}{(x-1)^5}-8 \ao-4 \
x-\frac{224}{(x-2)^4}+\frac{16}{(x-1)^4}-\frac{1088}{(x-2)^5}+\frac{\
16}{(x-1)^5}-\frac{1280}{(x-2)^6}\Big) H(0,1;\ao)-\frac{16 \pi \
^2}{(x-2)^4}+\frac{8 \pi ^2}{9 (x-1)^4}-\frac{176 \pi ^2}{3 (x-2)^5}+\
\frac{8 \pi ^2}{9 (x-1)^5}-\frac{160 \pi ^2}{3 \
(x-2)^6}\Big)+\Big(-\frac{20 x \ao}{3}-\frac{1216 \ao}{3 \
(x-2)^4}+\frac{20 \ao}{3 (x-1)^4}-\frac{3200 \ao}{3 (x-2)^5}-\frac{20 \
\ao}{3 (x-1)^5}+\frac{40 \ao}{3}+\frac{20 x}{3}+\frac{1216}{3 \
(x-2)^4}-\frac{20}{3 (x-1)^4}+\frac{5632}{3 (x-2)^5}-\frac{20}{3 \
(x-1)^5}+\frac{6400}{3 (x-2)^6}\Big) H(0;\ao) \
H(0,1;x)+\Big(\frac{\ao^5}{2}+\frac{\ao^4}{3 (x-2)}-\frac{5 \
\ao^4}{2}-\frac{4 \ao^3}{3 (x-2)}+\frac{4 \ao^3}{3 (x-2)^2}+5 \
\ao^3+\frac{2 \ao^2}{x-2}-\frac{4 \ao^2}{(x-2)^2}+\frac{8 \
\ao^2}{(x-2)^3}-5 \ao^2+\frac{47 x \ao}{9}+\frac{64 \ao}{9 \
(x-2)}-\frac{34 \ao}{3 (x-1)}-\frac{40 \ao}{9 (x-2)^2}+\frac{49 \
\ao}{18 (x-1)^2}-\frac{16 \ao}{(x-2)^3}-\frac{26 \ao}{9 \
(x-1)^3}-\frac{1696 \ao}{9 (x-2)^4}-\frac{55 \ao}{18 \
(x-1)^4}+\frac{640 \ao}{9 (x-2)^5}+\frac{47 \ao}{9 (x-1)^5}-\frac{52 \
\ao}{9}-\frac{47 x}{9}+\Big(-\frac{8 x \ao}{3}-\frac{448 \ao}{3 \
(x-2)^4}+\frac{8 \ao}{3 (x-1)^4}-\frac{1280 \ao}{3 (x-2)^5}-\frac{8 \
\ao}{3 (x-1)^5}+\frac{16 \ao}{3}+\frac{8 x}{3}+\frac{448}{3 (x-2)^4}-\
\frac{8}{3 (x-1)^4}+\frac{2176}{3 (x-2)^5}-\frac{8}{3 \
(x-1)^5}+\frac{2560}{3 (x-2)^6}\Big) H(0;\ao)+\Big(4 x \ao+\frac{224 \
\ao}{(x-2)^4}-\frac{4 \ao}{(x-1)^4}+\frac{640 \ao}{(x-2)^5}+\frac{4 \
\ao}{(x-1)^5}-8 \ao-4 \
x-\frac{224}{(x-2)^4}+\frac{4}{(x-1)^4}-\frac{1088}{(x-2)^5}+\frac{4}{\
(x-1)^5}-\frac{1280}{(x-2)^6}\Big) H(1;\ao)+\frac{56}{9 \
(x-2)}-\frac{22}{3 (x-1)}-\frac{88}{9 (x-2)^2}+\frac{23}{18 (x-1)^2}+\
\frac{224}{9 (x-2)^3}+\frac{13}{9 (x-1)^3}+\frac{1696}{9 \
(x-2)^4}+\frac{133}{18 (x-1)^4}+\frac{2176}{9 (x-2)^5}+\frac{47}{9 \
(x-1)^5}-\frac{1280}{9 (x-2)^6}-\frac{8}{3}\Big) \
H(0,c_1(\ao);x)+\Big(-\frac{640 \ao}{9 (x-2)^4}-\frac{1280 \ao}{9 \
(x-2)^5}+\Big(\frac{1280 \ao}{3 (x-2)^4}+\frac{2560 \ao}{3 \
(x-2)^5}-\frac{1280}{3 (x-2)^4}-\frac{5120}{3 (x-2)^5}-\frac{5120}{3 \
(x-2)^6}\Big) H(0;\ao)+\Big(-\frac{640 \ao}{(x-2)^4}-\frac{1280 \
\ao}{(x-2)^5}+\frac{640}{(x-2)^4}+\frac{2560}{(x-2)^5}+\frac{2560}{(x-\
2)^6}\Big) H(1;\ao)+\frac{640}{9 (x-2)^4}+\frac{2560}{9 \
(x-2)^5}+\frac{2560}{9 (x-2)^6}\Big) H(0,c_2(\ao);x)+\Big(-2 x \ao^5-\
\frac{4 \ao^5}{x-2}+\frac{2 \ao^5}{x-1}+\frac{38 x \ao^4}{3}+\frac{40 \
\ao^4}{3 (x-2)}-\frac{26 \ao^4}{3 (x-1)}-\frac{40 \ao^4}{3 \
(x-2)^2}+\frac{8 \ao^4}{3 (x-1)^2}-\frac{4 \ao^4}{3}-\frac{104 x \
\ao^3}{3}-\frac{40 \ao^3}{3 (x-2)}+\frac{44 \ao^3}{3 (x-1)}+\frac{80 \
\ao^3}{3 (x-2)^2}-\frac{28 \ao^3}{3 (x-1)^2}-\frac{160 \ao^3}{3 \
(x-2)^3}+\frac{4 \ao^3}{(x-1)^3}+8 \ao^3+56 x \ao^2-\frac{12 \
\ao^2}{x-1}+\frac{12 \ao^2}{(x-1)^2}-\frac{12 \
\ao^2}{(x-1)^3}-\frac{320 \ao^2}{(x-2)^4}+\frac{8 \ao^2}{(x-1)^4}-24 \
\ao^2-\frac{146 x \ao}{3}+\frac{20 \ao}{3 (x-2)}+\frac{14 \ao}{3 \
(x-1)}-\frac{80 \ao}{3 (x-2)^2}-\frac{20 \ao}{3 (x-1)^2}+\frac{160 \
\ao}{(x-2)^3}+\frac{12 \ao}{(x-1)^3}+\frac{1280 \
\ao}{(x-2)^4}+\frac{1280 \ao}{(x-2)^5}+\frac{40 \ao}{3}+\frac{50 \
x}{3}-\frac{8}{3 (x-2)}-\frac{2}{3 (x-1)}+\frac{40}{3 \
(x-2)^2}+\frac{4}{3 (x-1)^2}-\frac{320}{3 (x-2)^3}-\frac{4}{(x-1)^3}-\
\frac{960}{(x-2)^4}-\frac{8}{(x-1)^4}-\frac{1280}{(x-2)^5}+4\Big) \
H(1,0;\ao)+\Big(-\frac{47 x \ao}{9}-\frac{838 \ao}{9 (x-2)}+\frac{102 \
\ao}{x-1}+\frac{1000 \ao}{9 (x-2)^2}-\frac{433 \ao}{18 \
(x-1)^2}-\frac{128 \ao}{(x-2)^3}+\frac{113 \ao}{9 (x-1)^3}+\frac{4288 \
\ao}{9 (x-2)^4}+\frac{137 \ao}{9 (x-1)^4}-\frac{640 \ao}{9 \
(x-2)^5}-\frac{188 \ao}{9 (x-1)^5}+\frac{77 \ao}{18}+\frac{47 \
x}{9}-\frac{578}{9 (x-2)}+\frac{203}{3 (x-1)}+\frac{676}{9 \
(x-2)^2}-\frac{185}{18 (x-1)^2}-\frac{848}{9 (x-2)^3}+\frac{11}{9 \
(x-1)^3}-\frac{1984}{9 (x-2)^4}-\frac{239}{9 (x-1)^4}-\frac{7936}{9 \
(x-2)^5}-\frac{188}{9 (x-1)^5}+\frac{1280}{9 \
(x-2)^6}+\frac{37}{6}\Big) H(1,0;x)+\Big(3 x \ao^5+\frac{6 \
\ao^5}{x-2}-\frac{3 \ao^5}{x-1}-19 x \ao^4-\frac{20 \
\ao^4}{x-2}+\frac{13 \ao^4}{x-1}+\frac{20 \ao^4}{(x-2)^2}-\frac{4 \
\ao^4}{(x-1)^2}+2 \ao^4+52 x \ao^3+\frac{20 \ao^3}{x-2}-\frac{22 \
\ao^3}{x-1}-\frac{40 \ao^3}{(x-2)^2}+\frac{14 \
\ao^3}{(x-1)^2}+\frac{80 \ao^3}{(x-2)^3}-\frac{6 \ao^3}{(x-1)^3}-12 \
\ao^3-84 x \ao^2+\frac{18 \ao^2}{x-1}-\frac{18 \
\ao^2}{(x-1)^2}+\frac{18 \ao^2}{(x-1)^3}+\frac{480 \
\ao^2}{(x-2)^4}-\frac{12 \ao^2}{(x-1)^4}+36 \ao^2+73 x \ao-\frac{10 \
\ao}{x-2}-\frac{7 \ao}{x-1}+\frac{40 \ao}{(x-2)^2}+\frac{10 \
\ao}{(x-1)^2}-\frac{240 \ao}{(x-2)^3}-\frac{18 \
\ao}{(x-1)^3}-\frac{1920 \ao}{(x-2)^4}-\frac{1920 \ao}{(x-2)^5}-20 \
\ao-25 x+\frac{4}{x-2}+\frac{1}{x-1}-\frac{20}{(x-2)^2}-\frac{2}{(x-1)\
^2}+\frac{160}{(x-2)^3}+\frac{6}{(x-1)^3}+\frac{1440}{(x-2)^4}+\frac{\
12}{(x-1)^4}+\frac{1920}{(x-2)^5}-6\Big) H(1,1;\ao)+H(c_2(\ao);x) \
\Big(-\frac{40 \pi ^2 \ao}{9 (x-2)^4}+\frac{320 \ao}{27 \
(x-2)^4}-\frac{80 \pi ^2 \ao}{9 (x-2)^5}+\frac{640 \ao}{27 \
(x-2)^5}+\Big(-\frac{640 \ao}{9 (x-2)^4}-\frac{1280 \ao}{9 \
(x-2)^5}+\frac{640}{9 (x-2)^4}+\frac{2560}{9 (x-2)^5}+\frac{2560}{9 \
(x-2)^6}\Big) H(0;\ao)+\Big(\frac{320 \ao}{3 (x-2)^4}+\frac{640 \
\ao}{3 (x-2)^5}-\frac{320}{3 (x-2)^4}-\frac{1280}{3 \
(x-2)^5}-\frac{1280}{3 (x-2)^6}\Big) H(1;\ao)+\Big(\frac{1280 \ao}{3 \
(x-2)^4}+\frac{2560 \ao}{3 (x-2)^5}-\frac{1280}{3 \
(x-2)^4}-\frac{5120}{3 (x-2)^5}-\frac{5120}{3 (x-2)^6}\Big) \
H(0,0;\ao)+\Big(-\frac{640 \ao}{(x-2)^4}-\frac{1280 \
\ao}{(x-2)^5}+\frac{640}{(x-2)^4}+\frac{2560}{(x-2)^5}+\frac{2560}{(x-\
2)^6}\Big) H(0,1;\ao)+\Big(-\frac{640 \ao}{(x-2)^4}-\frac{1280 \
\ao}{(x-2)^5}+\frac{640}{(x-2)^4}+\frac{2560}{(x-2)^5}+\frac{2560}{(x-\
2)^6}\Big) H(1,0;\ao)+\Big(\frac{960 \ao}{(x-2)^4}+\frac{1920 \
\ao}{(x-2)^5}-\frac{960}{(x-2)^4}-\frac{3840}{(x-2)^5}-\frac{3840}{(x-\
2)^6}\Big) H(1,1;\ao)+\frac{40 \pi ^2}{9 (x-2)^4}-\frac{320}{27 \
(x-2)^4}+\frac{160 \pi ^2}{9 (x-2)^5}-\frac{1280}{27 \
(x-2)^5}+\frac{160 \pi ^2}{9 (x-2)^6}-\frac{1280}{27 \
(x-2)^6}\Big)+H(c_1(\ao);x) \Big(-\frac{37 x \ao^5}{36}-\frac{31 \
\ao^5}{18 (x-2)}+\frac{37 \ao^5}{36 (x-1)}+\frac{9 \
\ao^5}{8}+\frac{245 x \ao^4}{36}+\frac{73 \ao^4}{12 (x-2)}-\frac{151 \
\ao^4}{36 (x-1)}-\frac{65 \ao^4}{9 (x-2)^2}+\frac{5 \ao^4}{3 \
(x-1)^2}-\frac{497 \ao^4}{72}-\frac{182 x \ao^3}{9}-\frac{52 \ao^3}{9 \
(x-2)}+\frac{187 \ao^3}{36 (x-1)}+\frac{44 \ao^3}{3 \
(x-2)^2}-\frac{211 \ao^3}{36 (x-1)^2}-\frac{40 \
\ao^3}{(x-2)^3}+\frac{59 \ao^3}{18 (x-1)^3}+\frac{353 \
\ao^3}{18}+\frac{122 x \ao^2}{3}-\frac{91 \ao^2}{9 (x-2)}+\frac{275 \
\ao^2}{36 (x-1)}+\frac{34 \ao^2}{3 (x-2)^2}+\frac{167 \ao^2}{36 \
(x-1)^2}-\frac{268 \ao^2}{9 (x-2)^3}-\frac{38 \ao^2}{3 \
(x-1)^3}-\frac{3760 \ao^2}{9 (x-2)^4}+\frac{95 \ao^2}{9 \
(x-1)^4}-\frac{781 \ao^2}{18}-\frac{171 x \ao}{4}-\frac{610 \ao}{3 \
(x-2)}+\frac{3965 \ao}{18 (x-1)}+\frac{228 \ao}{(x-2)^2}-\frac{797 \
\ao}{18 (x-1)^2}-\frac{1312 \ao}{9 (x-2)^3}+\frac{34 \
\ao}{(x-1)^3}+\frac{1808 \ao}{(x-2)^4}-\frac{\pi ^2 \ao}{18 (x-1)^4}+\
\frac{3101 \ao}{216 (x-1)^4}+\frac{10240 \ao}{9 (x-2)^5}+\frac{\pi ^2 \
\ao}{18 (x-1)^5}-\frac{2125 \ao}{108 (x-1)^5}+\frac{1315 \
\ao}{72}+\frac{595 x}{36}+\Big(\frac{2 x \ao^5}{3}+\frac{4 \ao^5}{3 \
(x-2)}-\frac{2 \ao^5}{3 (x-1)}-\ao^5-\frac{38 x \ao^4}{9}-\frac{46 \
\ao^4}{9 (x-2)}+\frac{26 \ao^4}{9 (x-1)}+\frac{40 \ao^4}{9 \
(x-2)^2}-\frac{8 \ao^4}{9 (x-1)^2}+\frac{49 \ao^4}{9}+\frac{104 x \
\ao^3}{9}+\frac{64 \ao^3}{9 (x-2)}-\frac{44 \ao^3}{9 (x-1)}-\frac{104 \
\ao^3}{9 (x-2)^2}+\frac{28 \ao^3}{9 (x-1)^2}+\frac{160 \ao^3}{9 \
(x-2)^3}-\frac{4 \ao^3}{3 (x-1)^3}-\frac{38 \ao^3}{3}-\frac{56 x \
\ao^2}{3}-\frac{4 \ao^2}{x-2}+\frac{4 \ao^2}{x-1}+\frac{8 \
\ao^2}{(x-2)^2}-\frac{4 \ao^2}{(x-1)^2}-\frac{16 \
\ao^2}{(x-2)^3}+\frac{4 \ao^2}{(x-1)^3}+\frac{320 \ao^2}{3 \
(x-2)^4}-\frac{8 \ao^2}{3 (x-1)^4}+18 \ao^2+\frac{146 x \
\ao}{9}+\frac{512 \ao}{9 (x-2)}-\frac{62 \ao}{x-1}-\frac{608 \ao}{9 \
(x-2)^2}+\frac{148 \ao}{9 (x-1)^2}+\frac{64 \ao}{(x-2)^3}-\frac{94 \
\ao}{9 (x-1)^3}-\frac{576 \ao}{(x-2)^4}-\frac{73 \ao}{9 \
(x-1)^4}-\frac{1280 \ao}{3 (x-2)^5}+\frac{94 \ao}{9 (x-1)^5}-\frac{61 \
\ao}{9}-\frac{50 x}{9}+\frac{352}{9 \
(x-2)}-\frac{40}{x-1}-\frac{416}{9 (x-2)^2}+\frac{50}{9 \
(x-1)^2}+\frac{640}{9 (x-2)^3}-\frac{4}{9 (x-1)^3}+\frac{896}{3 \
(x-2)^4}+\frac{139}{9 (x-1)^4}+\frac{2560}{3 (x-2)^5}+\frac{94}{9 \
(x-1)^5}-3\Big) H(0;\ao)+\Big(-x \ao^5-\frac{2 \
\ao^5}{x-2}+\frac{\ao^5}{x-1}+\frac{3 \ao^5}{2}+\frac{19 x \ao^4}{3}+\
\frac{23 \ao^4}{3 (x-2)}-\frac{13 \ao^4}{3 (x-1)}-\frac{20 \ao^4}{3 \
(x-2)^2}+\frac{4 \ao^4}{3 (x-1)^2}-\frac{49 \ao^4}{6}-\frac{52 x \
\ao^3}{3}-\frac{32 \ao^3}{3 (x-2)}+\frac{22 \ao^3}{3 (x-1)}+\frac{52 \
\ao^3}{3 (x-2)^2}-\frac{14 \ao^3}{3 (x-1)^2}-\frac{80 \ao^3}{3 \
(x-2)^3}+\frac{2 \ao^3}{(x-1)^3}+19 \ao^3+28 x \ao^2+\frac{6 \
\ao^2}{x-2}-\frac{6 \ao^2}{x-1}-\frac{12 \ao^2}{(x-2)^2}+\frac{6 \
\ao^2}{(x-1)^2}+\frac{24 \ao^2}{(x-2)^3}-\frac{6 \
\ao^2}{(x-1)^3}-\frac{160 \ao^2}{(x-2)^4}+\frac{4 \ao^2}{(x-1)^4}-27 \
\ao^2-\frac{73 x \ao}{3}-\frac{256 \ao}{3 (x-2)}+\frac{93 \
\ao}{x-1}+\frac{304 \ao}{3 (x-2)^2}-\frac{74 \ao}{3 (x-1)^2}-\frac{96 \
\ao}{(x-2)^3}+\frac{47 \ao}{3 (x-1)^3}+\frac{864 \
\ao}{(x-2)^4}+\frac{73 \ao}{6 (x-1)^4}+\frac{640 \
\ao}{(x-2)^5}-\frac{47 \ao}{3 (x-1)^5}+\frac{61 \ao}{6}+\frac{25 \
x}{3}-\frac{176}{3 (x-2)}+\frac{60}{x-1}+\frac{208}{3 \
(x-2)^2}-\frac{25}{3 (x-1)^2}-\frac{320}{3 (x-2)^3}+\frac{2}{3 \
(x-1)^3}-\frac{448}{(x-2)^4}-\frac{139}{6 \
(x-1)^4}-\frac{1280}{(x-2)^5}-\frac{47}{3 (x-1)^5}+\frac{9}{2}\Big) \
H(1;\ao)+\Big(\frac{16 \ao}{3 (x-1)^4}-\frac{16 \ao}{3 \
(x-1)^5}-\frac{16}{3 (x-1)^4}-\frac{16}{3 (x-1)^5}\Big) \
H(0,0;\ao)+\Big(-\frac{8 \ao}{(x-1)^4}+\frac{8 \
\ao}{(x-1)^5}+\frac{8}{(x-1)^4}+\frac{8}{(x-1)^5}\Big) \
H(0,1;\ao)+\Big(-\frac{8 \ao}{(x-1)^4}+\frac{8 \
\ao}{(x-1)^5}+\frac{8}{(x-1)^4}+\frac{8}{(x-1)^5}\Big) \
H(1,0;\ao)+\Big(\frac{12 \ao}{(x-1)^4}-\frac{12 \
\ao}{(x-1)^5}-\frac{12}{(x-1)^4}-\frac{12}{(x-1)^5}\Big) \
H(1,1;\ao)-\frac{1372}{9 (x-2)}+\frac{5705}{36 (x-1)}+\frac{512}{3 \
(x-2)^2}-\frac{475}{36 (x-1)^2}-\frac{2432}{9 (x-2)^3}-\frac{35}{12 \
(x-1)^3}-\frac{7264}{9 (x-2)^4}+\frac{\pi ^2}{18 \
(x-1)^4}-\frac{7679}{216 (x-1)^4}-\frac{20480}{9 (x-2)^5}+\frac{\pi \
^2}{18 (x-1)^5}-\frac{2125}{108 \
(x-1)^5}+\frac{271}{24}\Big)+\Big(-\frac{14 x \ao}{3}-\frac{320 \
\ao}{(x-2)^4}+\frac{64 \ao}{3 (x-1)^4}-\frac{800 \
\ao}{(x-2)^5}-\frac{64 \ao}{3 (x-1)^5}+\frac{28 \ao}{3}+\frac{14 \
x}{3}+\frac{320}{(x-2)^4}-\frac{64}{3 \
(x-1)^4}+\frac{1440}{(x-2)^5}-\frac{64}{3 \
(x-1)^5}+\frac{1600}{(x-2)^6}\Big) H(0;\ao) H(1,1;x)+\Big(\frac{47 x \
\ao}{9}+\frac{838 \ao}{9 (x-2)}-\frac{102 \ao}{x-1}-\frac{1000 \ao}{9 \
(x-2)^2}+\frac{433 \ao}{18 (x-1)^2}+\frac{128 \ao}{(x-2)^3}-\frac{113 \
\ao}{9 (x-1)^3}-\frac{4288 \ao}{9 (x-2)^4}-\frac{137 \ao}{9 (x-1)^4}+\
\frac{640 \ao}{9 (x-2)^5}+\frac{188 \ao}{9 (x-1)^5}-\frac{77 \
\ao}{18}-\frac{47 x}{9}+\Big(-\frac{8 x \ao}{3}-\frac{448 \ao}{3 \
(x-2)^4}+\frac{32 \ao}{3 (x-1)^4}-\frac{1280 \ao}{3 (x-2)^5}-\frac{32 \
\ao}{3 (x-1)^5}+\frac{16 \ao}{3}+\frac{8 x}{3}+\frac{448}{3 (x-2)^4}-\
\frac{32}{3 (x-1)^4}+\frac{2176}{3 (x-2)^5}-\frac{32}{3 \
(x-1)^5}+\frac{2560}{3 (x-2)^6}\Big) H(0;\ao)+\Big(4 x \ao+\frac{224 \
\ao}{(x-2)^4}-\frac{16 \ao}{(x-1)^4}+\frac{640 \ao}{(x-2)^5}+\frac{16 \
\ao}{(x-1)^5}-8 \ao-4 \
x-\frac{224}{(x-2)^4}+\frac{16}{(x-1)^4}-\frac{1088}{(x-2)^5}+\frac{\
16}{(x-1)^5}-\frac{1280}{(x-2)^6}\Big) H(1;\ao)+\frac{578}{9 \
(x-2)}-\frac{203}{3 (x-1)}-\frac{676}{9 (x-2)^2}+\frac{185}{18 \
(x-1)^2}+\frac{848}{9 (x-2)^3}-\frac{11}{9 (x-1)^3}+\frac{1984}{9 \
(x-2)^4}+\frac{239}{9 (x-1)^4}+\frac{7936}{9 (x-2)^5}+\frac{188}{9 \
(x-1)^5}-\frac{1280}{9 (x-2)^6}-\frac{37}{6}\Big) \
H(1,c_1(\ao);x)+\Big(\frac{2 x \ao}{3}+\frac{1120 \ao}{3 \
(x-2)^4}-\frac{2 \ao}{3 (x-1)^4}+\frac{3200 \ao}{3 (x-2)^5}+\frac{2 \
\ao}{3 (x-1)^5}-\frac{4 \ao}{3}-\frac{2 x}{3}-\frac{1120}{3 (x-2)^4}+\
\frac{2}{3 (x-1)^4}-\frac{5440}{3 (x-2)^5}+\frac{2}{3 \
(x-1)^5}-\frac{6400}{3 (x-2)^6}\Big) H(0;\ao) \
H(2,1;x)+\Big(\frac{1600 \ao}{9 (x-2)^4}+\frac{3200 \ao}{9 \
(x-2)^5}+\Big(-\frac{3200 \ao}{3 (x-2)^4}-\frac{6400 \ao}{3 (x-2)^5}+\
\frac{3200}{3 (x-2)^4}+\frac{12800}{3 (x-2)^5}+\frac{12800}{3 \
(x-2)^6}\Big) H(0;\ao)+\Big(\frac{1600 \ao}{(x-2)^4}+\frac{3200 \
\ao}{(x-2)^5}-\frac{1600}{(x-2)^4}-\frac{6400}{(x-2)^5}-\frac{6400}{(\
x-2)^6}\Big) H(1;\ao)-\frac{1600}{9 (x-2)^4}-\frac{6400}{9 \
(x-2)^5}-\frac{6400}{9 (x-2)^6}\Big) H(2,c_2(\ao);x)+\Big(\frac{x \
\ao^5}{2}+\frac{5 \ao^5}{6 (x-2)}-\frac{\ao^5}{3 (x-1)}-\frac{3 \
\ao^5}{2}-\frac{19 x \ao^4}{6}-\frac{65 \ao^4}{18 (x-2)}+\frac{13 \
\ao^4}{9 (x-1)}+\frac{25 \ao^4}{9 (x-2)^2}-\frac{4 \ao^4}{9 (x-1)^2}+\
\frac{47 \ao^4}{6}+\frac{26 x \ao^3}{3}+\frac{55 \ao^3}{9 \
(x-2)}-\frac{22 \ao^3}{9 (x-1)}-\frac{80 \ao^3}{9 (x-2)^2}+\frac{14 \
\ao^3}{9 (x-1)^2}+\frac{100 \ao^3}{9 (x-2)^3}-\frac{2 \ao^3}{3 \
(x-1)^3}-17 \ao^3-14 x \ao^2-\frac{5 \ao^2}{x-2}+\frac{2 \ao^2}{x-1}+\
\frac{10 \ao^2}{(x-2)^2}-\frac{2 \ao^2}{(x-1)^2}-\frac{20 \
\ao^2}{(x-2)^3}+\frac{2 \ao^2}{(x-1)^3}+\frac{200 \ao^2}{3 \
(x-2)^4}-\frac{4 \ao^2}{3 (x-1)^4}+21 \ao^2+\frac{73 x \
\ao}{6}+\frac{440 \ao}{9 (x-2)}-\frac{913 \ao}{18 (x-1)}-\frac{560 \
\ao}{9 (x-2)^2}+\frac{245 \ao}{18 (x-1)^2}+\frac{80 \
\ao}{(x-2)^3}-\frac{71 \ao}{9 (x-1)^3}-\frac{320 \
\ao}{(x-2)^4}-\frac{55 \ao}{18 (x-1)^4}-\frac{800 \ao}{3 \
(x-2)^5}+\frac{47 \ao}{9 (x-1)^5}-\frac{53 \ao}{6}-\frac{25 \
x}{6}+\Big(\frac{8 \ao}{3 (x-1)^4}-\frac{8 \ao}{3 (x-1)^5}-\frac{8}{3 \
(x-1)^4}-\frac{8}{3 (x-1)^5}\Big) H(0;\ao)+\Big(-\frac{4 \
\ao}{(x-1)^4}+\frac{4 \
\ao}{(x-1)^5}+\frac{4}{(x-1)^4}+\frac{4}{(x-1)^5}\Big) \
H(1;\ao)+\frac{280}{9 (x-2)}-\frac{581}{18 (x-1)}-\frac{320}{9 \
(x-2)^2}+\frac{29}{6 (x-1)^2}+\frac{400}{9 (x-2)^3}-\frac{8}{9 \
(x-1)^3}+\frac{320}{3 (x-2)^4}+\frac{157}{18 (x-1)^4}+\frac{1600}{3 \
(x-2)^5}+\frac{47}{9 (x-1)^5}-\frac{3}{2}\Big) \
H(c_1(\ao),c_1(\ao);x)+\Big(-\frac{x \ao^5}{6}-\frac{\ao^5}{6 (x-1)}+\
\frac{\ao^5}{2}+\frac{19 x \ao^4}{18}+\frac{13 \ao^4}{18 \
(x-1)}-\frac{2 \ao^4}{9 (x-1)^2}-\frac{47 \ao^4}{18}-\frac{26 x \
\ao^3}{9}-\frac{11 \ao^3}{9 (x-1)}+\frac{7 \ao^3}{9 \
(x-1)^2}-\frac{\ao^3}{3 (x-1)^3}+\frac{17 \ao^3}{3}+\frac{14 x \
\ao^2}{3}+\frac{\ao^2}{x-1}-\frac{\ao^2}{(x-1)^2}+\frac{\ao^2}{(x-1)^3}-\frac{2 \ao^2}{3 (x-1)^4}-7 \ao^2-\frac{73 x \ao}{18}+\frac{8 \
\ao}{9 (x-2)}+\frac{\ao}{18 (x-1)}-\frac{8 \ao}{9 \
(x-2)^2}+\frac{\ao}{9 (x-1)^2}+\frac{\ao}{3 (x-1)^3}-\frac{32 \ao}{9 \
(x-2)^4}-\frac{4 \ao}{3 (x-1)^4}-\frac{640 \ao}{9 (x-2)^5}+\frac{41 \
\ao}{18}+\frac{25 x}{18}+\Big(\frac{448 \ao}{3 (x-2)^4}+\frac{1280 \
\ao}{3 (x-2)^5}-\frac{448}{3 (x-2)^4}-\frac{2176}{3 \
(x-2)^5}-\frac{2560}{3 (x-2)^6}\Big) H(0;\ao)+\Big(-\frac{224 \
\ao}{(x-2)^4}-\frac{640 \
\ao}{(x-2)^5}+\frac{224}{(x-2)^4}+\frac{1088}{(x-2)^5}+\frac{1280}{(x-\
2)^6}\Big) H(1;\ao)+\frac{16}{9 (x-2)}-\frac{7}{18 (x-1)}-\frac{8}{9 \
(x-2)^2}-\frac{5}{9 (x-1)^2}+\frac{16}{9 \
(x-2)^3}-\frac{1}{(x-1)^3}+\frac{32}{9 (x-2)^4}-\frac{2}{3 \
(x-1)^4}+\frac{704}{9 (x-2)^5}+\frac{1280}{9 \
(x-2)^6}+\frac{7}{6}\Big) H(c_2(\ao),c_1(\ao);x)+\Big(\frac{16 x \
\ao}{3}+\frac{1280 \ao}{3 (x-2)^4}+\frac{16 \ao}{3 \
(x-1)^4}+\frac{2560 \ao}{3 (x-2)^5}-\frac{16 \ao}{3 (x-1)^5}-\frac{32 \
\ao}{3}-\frac{16 x}{3}-\frac{1280}{3 (x-2)^4}-\frac{16}{3 \
(x-1)^4}-\frac{5120}{3 (x-2)^5}-\frac{16}{3 (x-1)^5}-\frac{5120}{3 \
(x-2)^6}\Big) H(0,0,0;\ao)+\Big(\frac{16 x \ao}{3}-\frac{1280 \ao}{3 \
(x-2)^4}-\frac{16 \ao}{3 (x-1)^4}-\frac{2560 \ao}{3 (x-2)^5}+\frac{16 \
\ao}{3 (x-1)^5}-\frac{32 \ao}{3}-\frac{16 x}{3}+\frac{1280}{3 \
(x-2)^4}+\frac{16}{3 (x-1)^4}+\frac{5120}{3 (x-2)^5}+\frac{16}{3 \
(x-1)^5}+\frac{5120}{3 (x-2)^6}\Big) H(0,0,0;x)+\Big(-8 x \
\ao-\frac{640 \ao}{(x-2)^4}-\frac{8 \ao}{(x-1)^4}-\frac{1280 \
\ao}{(x-2)^5}+\frac{8 \ao}{(x-1)^5}+16 \ao+8 \
x+\frac{640}{(x-2)^4}+\frac{8}{(x-1)^4}+\frac{2560}{(x-2)^5}+\frac{8}{\
(x-1)^5}+\frac{2560}{(x-2)^6}\Big) H(0,0,1;\ao)+\Big(-\frac{8 x \
\ao}{3}-\frac{544 \ao}{3 (x-2)^4}+\frac{8 \ao}{3 (x-1)^4}-\frac{1280 \
\ao}{3 (x-2)^5}-\frac{8 \ao}{3 (x-1)^5}+\frac{16 \ao}{3}+\frac{8 \
x}{3}+\frac{544}{3 (x-2)^4}-\frac{8}{3 (x-1)^4}+\frac{2368}{3 \
(x-2)^5}-\frac{8}{3 (x-1)^5}+\frac{2560}{3 (x-2)^6}\Big) \
H(0,0,c_1(\ao);x)+\Big(\frac{1280 \ao}{3 (x-2)^4}+\frac{2560 \ao}{3 \
(x-2)^5}-\frac{1280}{3 (x-2)^4}-\frac{5120}{3 (x-2)^5}-\frac{5120}{3 \
(x-2)^6}\Big) H(0,0,c_2(\ao);x)+\Big(-8 x \ao-\frac{640 \
\ao}{(x-2)^4}-\frac{8 \ao}{(x-1)^4}-\frac{1280 \ao}{(x-2)^5}+\frac{8 \
\ao}{(x-1)^5}+16 \ao+8 \
x+\frac{640}{(x-2)^4}+\frac{8}{(x-1)^4}+\frac{2560}{(x-2)^5}+\frac{8}{\
(x-1)^5}+\frac{2560}{(x-2)^6}\Big) H(0,1,0;\ao)+\Big(\frac{20 x \
\ao}{3}+\frac{1216 \ao}{3 (x-2)^4}-\frac{20 \ao}{3 \
(x-1)^4}+\frac{3200 \ao}{3 (x-2)^5}+\frac{20 \ao}{3 (x-1)^5}-\frac{40 \
\ao}{3}-\frac{20 x}{3}-\frac{1216}{3 (x-2)^4}+\frac{20}{3 \
(x-1)^4}-\frac{5632}{3 (x-2)^5}+\frac{20}{3 (x-1)^5}-\frac{6400}{3 \
(x-2)^6}\Big) H(0,1,0;x)+\Big(12 x \ao+\frac{960 \
\ao}{(x-2)^4}+\frac{12 \ao}{(x-1)^4}+\frac{1920 \
\ao}{(x-2)^5}-\frac{12 \ao}{(x-1)^5}-24 \ao-12 x-\frac{960}{(x-2)^4}-\
\frac{12}{(x-1)^4}-\frac{3840}{(x-2)^5}-\frac{12}{(x-1)^5}-\frac{3840}\
{(x-2)^6}\Big) H(0,1,1;\ao)+\Big(-\frac{20 x \ao}{3}-\frac{1216 \
\ao}{3 (x-2)^4}+\frac{20 \ao}{3 (x-1)^4}-\frac{3200 \ao}{3 \
(x-2)^5}-\frac{20 \ao}{3 (x-1)^5}+\frac{40 \ao}{3}+\frac{20 \
x}{3}+\frac{1216}{3 (x-2)^4}-\frac{20}{3 (x-1)^4}+\frac{5632}{3 \
(x-2)^5}-\frac{20}{3 (x-1)^5}+\frac{6400}{3 (x-2)^6}\Big) \
H(0,1,c_1(\ao);x)+\Big(\frac{3200 \ao}{3 (x-2)^4}+\frac{6400 \ao}{3 \
(x-2)^5}-\frac{3200}{3 (x-2)^4}-\frac{12800}{3 \
(x-2)^5}-\frac{12800}{3 (x-2)^6}\Big) H(0,2,0;x)+\Big(-\frac{3200 \
\ao}{3 (x-2)^4}-\frac{6400 \ao}{3 (x-2)^5}+\frac{3200}{3 \
(x-2)^4}+\frac{12800}{3 (x-2)^5}+\frac{12800}{3 (x-2)^6}\Big) \
H(0,2,c_2(\ao);x)+\Big(-2 x \ao-\frac{160 \ao}{3 (x-2)^4}+\frac{4 \
\ao}{3 (x-1)^4}-\frac{800 \ao}{3 (x-2)^5}-\frac{4 \ao}{3 (x-1)^5}+4 \
\ao+2 x+\frac{160}{3 (x-2)^4}-\frac{4}{3 (x-1)^4}+\frac{1120}{3 \
(x-2)^5}-\frac{4}{3 (x-1)^5}+\frac{1600}{3 (x-2)^6}\Big) \
H(0,c_1(\ao),c_1(\ao);x)+\Big(\frac{2 x \ao}{3}+\frac{448 \ao}{3 \
(x-2)^4}-\frac{2 \ao}{3 (x-1)^4}+\frac{1280 \ao}{3 (x-2)^5}+\frac{2 \
\ao}{3 (x-1)^5}-\frac{4 \ao}{3}-\frac{2 x}{3}-\frac{448}{3 \
(x-2)^4}+\frac{2}{3 (x-1)^4}-\frac{2176}{3 (x-2)^5}+\frac{2}{3 \
(x-1)^5}-\frac{2560}{3 (x-2)^6}\Big) \
H(0,c_2(\ao),c_1(\ao);x)+\Big(\frac{8 x \ao}{3}+\frac{448 \ao}{3 \
(x-2)^4}-\frac{32 \ao}{3 (x-1)^4}+\frac{1280 \ao}{3 (x-2)^5}+\frac{32 \
\ao}{3 (x-1)^5}-\frac{16 \ao}{3}-\frac{8 x}{3}-\frac{448}{3 (x-2)^4}+\
\frac{32}{3 (x-1)^4}-\frac{2176}{3 (x-2)^5}+\frac{32}{3 \
(x-1)^5}-\frac{2560}{3 (x-2)^6}\Big) H(1,0,0;x)+\Big(-\frac{2 x \
\ao}{3}-\frac{96 \ao}{(x-2)^4}+\frac{16 \ao}{3 (x-1)^4}-\frac{160 \
\ao}{(x-2)^5}-\frac{16 \ao}{3 (x-1)^5}+\frac{4 \ao}{3}+\frac{2 x}{3}+\
\frac{96}{(x-2)^4}-\frac{16}{3 \
(x-1)^4}+\frac{352}{(x-2)^5}-\frac{16}{3 (x-1)^5}+\frac{320}{(x-2)^6}\
\Big) H(1,0,c_1(\ao);x)+\Big(\frac{14 x \ao}{3}+\frac{320 \
\ao}{(x-2)^4}-\frac{64 \ao}{3 (x-1)^4}+\frac{800 \
\ao}{(x-2)^5}+\frac{64 \ao}{3 (x-1)^5}-\frac{28 \ao}{3}-\frac{14 \
x}{3}-\frac{320}{(x-2)^4}+\frac{64}{3 \
(x-1)^4}-\frac{1440}{(x-2)^5}+\frac{64}{3 \
(x-1)^5}-\frac{1600}{(x-2)^6}\Big) H(1,1,0;x)+\Big(-\frac{14 x \
\ao}{3}-\frac{320 \ao}{(x-2)^4}+\frac{64 \ao}{3 (x-1)^4}-\frac{800 \
\ao}{(x-2)^5}-\frac{64 \ao}{3 (x-1)^5}+\frac{28 \ao}{3}+\frac{14 \
x}{3}+\frac{320}{(x-2)^4}-\frac{64}{3 \
(x-1)^4}+\frac{1440}{(x-2)^5}-\frac{64}{3 \
(x-1)^5}+\frac{1600}{(x-2)^6}\Big) H(1,1,c_1(\ao);x)+\Big(-2 x \
\ao-\frac{160 \ao}{3 (x-2)^4}+\frac{16 \ao}{3 (x-1)^4}-\frac{800 \
\ao}{3 (x-2)^5}-\frac{16 \ao}{3 (x-1)^5}+4 \ao+2 x+\frac{160}{3 \
(x-2)^4}-\frac{16}{3 (x-1)^4}+\frac{1120}{3 (x-2)^5}-\frac{16}{3 \
(x-1)^5}+\frac{1600}{3 (x-2)^6}\Big) \
H(1,c_1(\ao),c_1(\ao);x)+\Big(\frac{3200 \ao}{3 (x-2)^4}+\frac{6400 \
\ao}{3 (x-2)^5}-\frac{3200}{3 (x-2)^4}-\frac{12800}{3 \
(x-2)^5}-\frac{12800}{3 (x-2)^6}\Big) H(2,0,0;x)+\Big(\frac{2 x \
\ao}{3}+\frac{1120 \ao}{3 (x-2)^4}-\frac{2 \ao}{3 (x-1)^4}+\frac{3200 \
\ao}{3 (x-2)^5}+\frac{2 \ao}{3 (x-1)^5}-\frac{4 \ao}{3}-\frac{2 \
x}{3}-\frac{1120}{3 (x-2)^4}+\frac{2}{3 (x-1)^4}-\frac{5440}{3 \
(x-2)^5}+\frac{2}{3 (x-1)^5}-\frac{6400}{3 (x-2)^6}\Big) \
H(2,0,c_1(\ao);x)+\Big(-\frac{3200 \ao}{3 (x-2)^4}-\frac{6400 \ao}{3 \
(x-2)^5}+\frac{3200}{3 (x-2)^4}+\frac{12800}{3 \
(x-2)^5}+\frac{12800}{3 (x-2)^6}\Big) H(2,0,c_2(\ao);x)+\Big(-\frac{2 \
x \ao}{3}-\frac{1120 \ao}{3 (x-2)^4}+\frac{2 \ao}{3 \
(x-1)^4}-\frac{3200 \ao}{3 (x-2)^5}-\frac{2 \ao}{3 (x-1)^5}+\frac{4 \
\ao}{3}+\frac{2 x}{3}+\frac{1120}{3 (x-2)^4}-\frac{2}{3 \
(x-1)^4}+\frac{5440}{3 (x-2)^5}-\frac{2}{3 (x-1)^5}+\frac{6400}{3 \
(x-2)^6}\Big) H(2,1,0;x)+\Big(\frac{2 x \ao}{3}+\frac{1120 \ao}{3 \
(x-2)^4}-\frac{2 \ao}{3 (x-1)^4}+\frac{3200 \ao}{3 (x-2)^5}+\frac{2 \
\ao}{3 (x-1)^5}-\frac{4 \ao}{3}-\frac{2 x}{3}-\frac{1120}{3 (x-2)^4}+\
\frac{2}{3 (x-1)^4}-\frac{5440}{3 (x-2)^5}+\frac{2}{3 \
(x-1)^5}-\frac{6400}{3 (x-2)^6}\Big) \
H(2,1,c_1(\ao);x)+\Big(-\frac{8000 \ao}{3 (x-2)^4}-\frac{16000 \ao}{3 \
(x-2)^5}+\frac{8000}{3 (x-2)^4}+\frac{32000}{3 \
(x-2)^5}+\frac{32000}{3 (x-2)^6}\Big) H(2,2,0;x)+\Big(\frac{8000 \
\ao}{3 (x-2)^4}+\frac{16000 \ao}{3 (x-2)^5}-\frac{8000}{3 \
(x-2)^4}-\frac{32000}{3 (x-2)^5}-\frac{32000}{3 (x-2)^6}\Big) \
H(2,2,c_2(\ao);x)+\Big(-\frac{2 x \ao}{3}-\frac{1120 \ao}{3 (x-2)^4}+\
\frac{2 \ao}{3 (x-1)^4}-\frac{3200 \ao}{3 (x-2)^5}-\frac{2 \ao}{3 \
(x-1)^5}+\frac{4 \ao}{3}+\frac{2 x}{3}+\frac{1120}{3 \
(x-2)^4}-\frac{2}{3 (x-1)^4}+\frac{5440}{3 (x-2)^5}-\frac{2}{3 \
(x-1)^5}+\frac{6400}{3 (x-2)^6}\Big) \
H(2,c_2(\ao),c_1(\ao);x)+\Big(\frac{4 \ao}{3 (x-1)^4}-\frac{4 \ao}{3 \
(x-1)^5}-\frac{4}{3 (x-1)^4}-\frac{4}{3 (x-1)^5}\Big) H(c_1(\ao),c_1(\
\ao),c_1(\ao);x)+\Big(\frac{2 \ao}{3 (x-1)^4}-\frac{2 \ao}{3 \
(x-1)^5}-\frac{2}{3 (x-1)^4}-\frac{2}{3 (x-1)^5}\Big) H(c_1(\ao),c_2(\
\ao),c_1(\ao);x)+\Big(\frac{32 \
\ao}{(x-2)^4}-\frac{32}{(x-2)^4}-\frac{64}{(x-2)^5}\Big) \
H(c_2(\ao),0,c_1(\ao);x)+\Big(\frac{160 \ao}{3 (x-2)^4}+\frac{800 \
\ao}{3 (x-2)^5}-\frac{160}{3 (x-2)^4}-\frac{1120}{3 \
(x-2)^5}-\frac{1600}{3 (x-2)^6}\Big) \
H(c_2(\ao),c_1(\ao),c_1(\ao);x)+H(2,0;x) \Big(-\frac{1600 \ao}{9 \
(x-2)^4}-\frac{3200 \ao}{9 (x-2)^5}-\frac{3200 \ln 2\,  \ao}{3 \
(x-2)^4}-\frac{6400 \ln 2\,  \ao}{3 (x-2)^5}+\frac{1600}{9 \
(x-2)^4}+\frac{6400}{9 (x-2)^5}+\frac{6400}{9 (x-2)^6}+\frac{3200 \ln \
2\, }{3 (x-2)^4}+\frac{12800 \ln 2\, }{3 (x-2)^5}+\frac{12800 \ln 2\, \
}{3 (x-2)^6}\Big)+H(0,2;x) \Big(-\frac{3200 \ln 2\,  \ao}{3 (x-2)^4}-\
\frac{6400 \ln 2\,  \ao}{3 (x-2)^5}+\Big(-\frac{3200 \ao}{3 (x-2)^4}-\
\frac{6400 \ao}{3 (x-2)^5}+\frac{3200}{3 (x-2)^4}+\frac{12800}{3 \
(x-2)^5}+\frac{12800}{3 (x-2)^6}\Big) H(0;\ao)+\frac{3200 \ln 2\, }{3 \
(x-2)^4}+\frac{12800 \ln 2\, }{3 (x-2)^5}+\frac{12800 \ln 2\, }{3 \
(x-2)^6}\Big)+H(0,0;x) \Big(-\frac{94 x \ao}{9}-\frac{512 \ao}{9 \
(x-2)}+\frac{62 \ao}{x-1}+\frac{608 \ao}{9 (x-2)^2}-\frac{148 \ao}{9 \
(x-1)^2}-\frac{64 \ao}{(x-2)^3}+\frac{94 \ao}{9 (x-1)^3}+\frac{4480 \
\ao}{9 (x-2)^4}+\frac{49 \ao}{9 (x-1)^4}+\frac{1280 \ao}{9 \
(x-2)^5}-\frac{94 \ao}{9 (x-1)^5}+\frac{1280 \ln 2\,  \ao}{3 \
(x-2)^4}+\frac{2560 \ln 2\,  \ao}{3 (x-2)^5}+\frac{161 \
\ao}{9}+\frac{94 x}{9}-\frac{352}{9 \
(x-2)}+\frac{40}{x-1}+\frac{416}{9 (x-2)^2}-\frac{50}{9 \
(x-1)^2}-\frac{640}{9 (x-2)^3}+\frac{4}{9 (x-1)^3}-\frac{3328}{9 \
(x-2)^4}-\frac{139}{9 (x-1)^4}-\frac{10240}{9 (x-2)^5}-\frac{94}{9 \
(x-1)^5}-\frac{2560}{9 (x-2)^6}-\frac{1280 \ln 2\, }{3 \
(x-2)^4}-\frac{5120 \ln 2\, }{3 (x-2)^5}-\frac{5120 \ln 2\, }{3 \
(x-2)^6}+3\Big)+H(2,2;x) \Big(\frac{8000 \ln 2\,  \ao}{3 \
(x-2)^4}+\frac{16000 \ln 2\,  \ao}{3 (x-2)^5}+\Big(\frac{8000 \ao}{3 \
(x-2)^4}+\frac{16000 \ao}{3 (x-2)^5}-\frac{8000}{3 \
(x-2)^4}-\frac{32000}{3 (x-2)^5}-\frac{32000}{3 (x-2)^6}\Big) \
H(0;\ao)-\frac{8000 \ln 2\, }{3 (x-2)^4}-\frac{32000 \ln 2\, }{3 \
(x-2)^5}-\frac{32000 \ln 2\, }{3 (x-2)^6}\Big)+H(0;x) \
\Big(\frac{1}{2} \pi ^2 x \ao+\frac{2125 x \ao}{108}+\frac{1748 \
\ao}{9 (x-2)}-\frac{1897 \ao}{9 (x-1)}-\frac{1960 \ao}{9 \
(x-2)^2}+\frac{352 \ao}{9 (x-1)^2}+\frac{496 \ao}{3 \
(x-2)^3}-\frac{173 \ao}{6 (x-1)^3}-\frac{88 \pi ^2 \ao}{3 \
(x-2)^4}-\frac{31040 \ao}{27 (x-2)^4}-\frac{7 \pi ^2 \ao}{18 \
(x-1)^4}-\frac{821 \ao}{216 (x-1)^4}-\frac{80 \pi ^2 \ao}{3 (x-2)^5}-\
\frac{640 \ao}{27 (x-2)^5}+\frac{7 \pi ^2 \ao}{18 (x-1)^5}+\frac{2125 \
\ao}{108 (x-1)^5}-\frac{640 \ln ^22\,  \ao}{3 (x-2)^4}-\frac{1280 \ln \
^22\,  \ao}{3 (x-2)^5}-\frac{640 \ln 2\,  \ao}{9 (x-2)^4}-\frac{1280 \
\ln 2\,  \ao}{9 (x-2)^5}-\pi ^2 \ao-\frac{6061 \ao}{216}-\frac{\pi ^2 \
x}{2}-\frac{2125 x}{108}+\frac{1372}{9 (x-2)}-\frac{5705}{36 \
(x-1)}-\frac{512}{3 (x-2)^2}+\frac{475}{36 (x-1)^2}+\frac{2432}{9 \
(x-2)^3}+\frac{35}{12 (x-1)^3}+\frac{88 \pi ^2}{3 \
(x-2)^4}+\frac{22112}{27 (x-2)^4}+\frac{7 \pi ^2}{18 \
(x-1)^4}+\frac{7679}{216 (x-1)^4}+\frac{256 \pi ^2}{3 \
(x-2)^5}+\frac{62720}{27 (x-2)^5}+\frac{7 \pi ^2}{18 \
(x-1)^5}+\frac{2125}{108 (x-1)^5}+\frac{160 \pi ^2}{3 \
(x-2)^6}+\frac{1280}{27 (x-2)^6}+\frac{640 \ln ^22\, }{3 \
(x-2)^4}+\frac{2560 \ln ^22\, }{3 (x-2)^5}+\frac{2560 \ln ^22\, }{3 \
(x-2)^6}+\frac{640 \ln 2\, }{9 (x-2)^4}+\frac{2560 \ln 2\, }{9 \
(x-2)^5}+\frac{2560 \ln 2\, }{9 (x-2)^6}-\frac{271}{24}\Big)+H(2;x) \
\Big(-\frac{1}{6} \pi ^2 x \ao+\frac{760 \pi ^2 \ao}{9 \
(x-2)^4}+\frac{\pi ^2 \ao}{6 (x-1)^4}+\frac{800 \pi ^2 \ao}{9 \
(x-2)^5}-\frac{\pi ^2 \ao}{6 (x-1)^5}+\frac{1600 \ln ^22\,  \ao}{3 \
(x-2)^4}+\frac{3200 \ln ^22\,  \ao}{3 (x-2)^5}+\frac{1600 \ln 2\,  \
\ao}{9 (x-2)^4}+\frac{3200 \ln 2\,  \ao}{9 (x-2)^5}+\frac{\pi ^2 \
\ao}{3}+\frac{\pi ^2 x}{6}+\Big(\frac{1600 \ao}{9 (x-2)^4}+\frac{3200 \
\ao}{9 (x-2)^5}-\frac{1600}{9 (x-2)^4}-\frac{6400}{9 \
(x-2)^5}-\frac{6400}{9 (x-2)^6}\Big) H(0;\ao)+\Big(-\frac{3200 \ao}{3 \
(x-2)^4}-\frac{6400 \ao}{3 (x-2)^5}+\frac{3200}{3 \
(x-2)^4}+\frac{12800}{3 (x-2)^5}+\frac{12800}{3 (x-2)^6}\Big) \
H(0,0;\ao)+\Big(\frac{1600 \ao}{(x-2)^4}+\frac{3200 \
\ao}{(x-2)^5}-\frac{1600}{(x-2)^4}-\frac{6400}{(x-2)^5}-\frac{6400}{(\
x-2)^6}\Big) H(0,1;\ao)-\frac{760 \pi ^2}{9 (x-2)^4}-\frac{\pi ^2}{6 \
(x-1)^4}-\frac{2320 \pi ^2}{9 (x-2)^5}-\frac{\pi ^2}{6 \
(x-1)^5}-\frac{1600 \pi ^2}{9 (x-2)^6}-\frac{1600 \ln ^22\, }{3 \
(x-2)^4}-\frac{6400 \ln ^22\, }{3 (x-2)^5}-\frac{6400 \ln ^22\, }{3 \
(x-2)^6}-\frac{1600 \ln 2\, }{9 (x-2)^4}-\frac{6400 \ln 2\, }{9 \
(x-2)^5}-\frac{6400 \ln 2\, }{9 (x-2)^6}\Big)-\frac{8 \pi ^2}{9 \
(x-2)}+\frac{257 \pi ^2}{216 (x-1)}+\frac{14 \pi ^2}{9 \
(x-2)^2}-\frac{7 \pi ^2}{27 (x-1)^2}-\frac{4 \pi \
^2}{(x-2)^3}-\frac{35 \pi ^2}{108 (x-1)^3}-\frac{1160 \pi ^2}{27 \
(x-2)^4}-\frac{139 \pi ^2}{108 (x-1)^4}-\frac{2192 \pi ^2}{27 \
(x-2)^5}-\frac{47 \pi ^2}{54 (x-1)^5}-\frac{320 \pi ^2}{27 \
(x-2)^6}-\frac{17}{12} x \zeta_3-\frac{56 \zeta_3}{3 (x-2)^4}+\frac{7 \
\zeta_3}{4 (x-1)^4}-\frac{392 \zeta_3}{3 (x-2)^5}+\frac{7 \zeta_3}{4 \
(x-1)^5}-\frac{560 \zeta_3}{3 (x-2)^6}-\frac{640 \ln ^32\, }{9 \
(x-2)^4}-\frac{2560 \ln ^32\, }{9 (x-2)^5}-\frac{2560 \ln ^32\, }{9 \
(x-2)^6}-\frac{320 \ln ^22\, }{9 (x-2)^4}-\frac{1280 \ln ^22\, }{9 \
(x-2)^5}-\frac{1280 \ln ^22\, }{9 (x-2)^6}+\frac{1}{6} \pi ^2 x \ln 2\
\, -\frac{112 \pi ^2 \ln 2\, }{3 (x-2)^4}-\frac{320 \ln 2\, }{27 \
(x-2)^4}-\frac{\pi ^2 \ln 2\, }{6 (x-1)^4}-\frac{304 \pi ^2 \ln 2\, \
}{3 (x-2)^5}-\frac{1280 \ln 2\, }{27 (x-2)^5}-\frac{\pi ^2 \ln 2\, \
}{6 (x-1)^5}-\frac{160 \pi ^2 \ln 2\, }{3 (x-2)^6}-\frac{1280 \ln 2\, \
}{27 (x-2)^6}+\frac{13 \pi ^2}{24}\}.
\erp

%
% The B integral for k=-1 and delta=1
%

\subsection{The $\cB$ integral for $k=-1$ and $\delta=1$}
%
% This file contains the TeX output produced by Mathematica for the integral Bm1, for delta = -1
%
The $\eps$ expansion for this integral reads
\beq
\bsp
\begin{cal}I\end{cal}(x,\eps;\ao,3+d_1\eps&;1,-1,1,g_B) = x\,\bint(\eps,x;3+d_1\eps;1,-1)\\
&=\frac{1}{\eps^2}b_{-2}^{(1,-1)}+\frac{1}{\eps}b_{-1}^{(1,-1)}+b_0^{(1,-1)}+\eps b_1^{(1,-1)}+\eps^2b_2^{(1,-1)} +\ocal\left(\eps^3\right),
\esp
\eeq
where
%1/ep piece
\brp
b_{-2}^{(1,-1)}=\frac{1}{8},
\erp
\brp
b_{-1}^{(1,-1)}=- H(0;x),
\erp
% ep^0
\brp
b_0^{(1,-1)}=\frac{\ao^3}{12 (x-1)^2}-\frac{\ao^3}{12}+\frac{\ao^2}{24 \
(x-1)}-\frac{5 \ao^2}{24 (x-1)^2}+\frac{7 \ao^2}{24 (x-1)^3}+\frac{13 \
\ao^2}{24}-\frac{\ao}{3 (x-1)}+\frac{\ao}{6 (x-1)^2}-\frac{\ao}{3 \
(x-1)^3}+\frac{13 \ao}{12 (x-1)^4}-\frac{23 \
\ao}{12}+\Big(\frac{25}{12}+\frac{3}{4 (x-1)}-\frac{1}{6 \
(x-1)^2}-\frac{1}{6 (x-1)^3}+\frac{3}{4 (x-1)^4}+\frac{25}{12 \
(x-1)^5}\Big) H(0;\ao)+\Big(-\frac{25}{12}-\frac{3}{4 \
(x-1)}+\frac{1}{6 (x-1)^2}+\frac{1}{6 (x-1)^3}-\frac{3}{4 \
(x-1)^4}-\frac{25}{12 (x-1)^5}\Big) \
H(0;x)+\Big(\frac{1}{(x-1)^5}-1\Big) H(0;\ao) \
H(1;x)+\Big(-\frac{\ao^4}{4 (x-1)}+\frac{\ao^4}{4}+\frac{\ao^3}{x-1}-\
\frac{\ao^3}{3 (x-1)^2}-\frac{4 \ao^3}{3}-\frac{3 \ao^2}{2 \
(x-1)}+\frac{\ao^2}{(x-1)^2}-\frac{\ao^2}{2 (x-1)^3}+3 \
\ao^2+\frac{\ao}{x-1}-\frac{\ao}{(x-1)^2}+\frac{\ao}{(x-1)^3}-\frac{\ao}{(x-1)^4}-4 \ao+\frac{3}{4 (x-1)}-\frac{1}{6 (x-1)^2}-\frac{1}{6 \
(x-1)^3}+\frac{3}{4 (x-1)^4}+\frac{25}{12 (x-1)^5}+\frac{25}{12}\Big) \
H(c_1(\ao);x)+4 H(0,0;x)+\Big(\frac{1}{(x-1)^5}-1\Big) \
H(0,c_1(\ao);x)+\Big(1-\frac{1}{(x-1)^5}\Big) \
H(1,0;x)+\Big(\frac{1}{(x-1)^5}-1\Big) \
H(1,c_1(\ao);x)-\frac{H(c_1(\ao),c_1(\ao);x)}{(x-1)^5}-\frac{\pi \
^2}{6 (x-1)^5}-\frac{\pi ^2}{12},
\erp
% ep^1
\brp
b_1^{(1,-1)}=\frac{7 d_1 \ao^3}{72}-\frac{7 d_1 \ao^3}{72 (x-1)^2}+\frac{7 \
\ao^3}{36 (x-1)^2}-\frac{7 \ao^3}{36}-\frac{109 d_1 \
\ao^2}{144}-\frac{13 d_1 \ao^2}{144 (x-1)}-\frac{11 \ao^2}{72 (x-1)}+\
\frac{29 d_1 \ao^2}{144 (x-1)^2}-\frac{53 \ao^2}{72 (x-1)^2}-\frac{67 \
d_1 \ao^2}{144 (x-1)^3}+\frac{79 \ao^2}{72 (x-1)^3}+\frac{121 \
\ao^2}{72}+\frac{305 d_1 \ao}{72}+\frac{2 \ao}{3 (x-2)}+\frac{19 d_1 \
\ao}{18 (x-1)}-\frac{5 \ao}{18 (x-1)}-\frac{d_1 \ao}{9 \
(x-1)^2}+\frac{5 \ao}{9 (x-1)^2}+\frac{d_1 \ao}{18 (x-1)^3}-\frac{41 \
\ao}{18 (x-1)^3}-\frac{217 d_1 \ao}{72 (x-1)^4}+\frac{271 \ao}{36 \
(x-1)^4}-\frac{371 \ao}{36}+\Big(-\frac{\ao^3}{3 \
(x-1)^2}+\frac{\ao^3}{3}-\frac{\ao^2}{6 (x-1)}+\frac{5 \ao^2}{6 \
(x-1)^2}-\frac{7 \ao^2}{6 (x-1)^3}-\frac{13 \ao^2}{6}+\frac{4 \ao}{3 \
(x-1)}-\frac{2 \ao}{3 (x-1)^2}+\frac{4 \ao}{3 (x-1)^3}-\frac{13 \
\ao}{3 (x-1)^4}+\frac{23 \ao}{3}-\frac{205 \
d_1}{72}-\frac{4}{x-2}-\frac{15 d_1}{8 (x-1)}+\frac{7}{4 \
(x-1)}+\frac{8}{3 (x-2)^2}+\frac{5 d_1}{18 (x-1)^2}-\frac{1}{18 \
(x-1)^2}+\frac{5 d_1}{18 (x-1)^3}-\frac{31}{18 (x-1)^3}-\frac{15 \
d_1}{8 (x-1)^4}+\frac{13}{4 (x-1)^4}-\frac{205 d_1}{72 \
(x-1)^5}+\frac{415}{36 (x-1)^5}+\frac{205}{36}\Big) \
H(0;\ao)+\Big(\frac{15 d_1}{8 (x-1)}-\frac{5 d_1}{18 (x-1)^2}-\frac{5 \
d_1}{18 (x-1)^3}+\frac{15 d_1}{8 (x-1)^4}+\frac{205 d_1}{72 (x-1)^5}+\
\frac{205 d_1}{72}+\frac{4}{x-2}-\frac{7}{4 (x-1)}-\frac{8}{3 \
(x-2)^2}+\frac{1}{18 (x-1)^2}+\frac{31}{18 (x-1)^3}-\frac{13}{4 \
(x-1)^4}+\frac{2 \pi ^2}{3 (x-1)^5}-\frac{415}{36 (x-1)^5}+\frac{\pi \
^2}{3}-\frac{205}{36}\Big) H(0;x)+\Big(\frac{d_1 \ao^3}{6}-\frac{d_1 \
\ao^3}{6 (x-1)^2}-\frac{13 d_1 \ao^2}{12}-\frac{d_1 \ao^2}{12 (x-1)}+\
\frac{5 d_1 \ao^2}{12 (x-1)^2}-\frac{7 d_1 \ao^2}{12 \
(x-1)^3}+\frac{23 d_1 \ao}{6}+\frac{2 d_1 \ao}{3 (x-1)}-\frac{d_1 \
\ao}{3 (x-1)^2}+\frac{2 d_1 \ao}{3 (x-1)^3}-\frac{13 d_1 \ao}{6 \
(x-1)^4}-\frac{35 d_1}{12}-\frac{7 d_1}{12 (x-1)}+\frac{d_1}{12 \
(x-1)^2}-\frac{d_1}{12 (x-1)^3}+\frac{13 d_1}{6 (x-1)^4}\Big) \
H(1;\ao)+\Big(\frac{\pi ^2}{2 (x-1)^5}-\frac{\pi ^2}{2}\Big) \
H(2;x)+\Big(-\frac{d_1 \ao^4}{8}+\frac{d_1 \ao^4}{8 \
(x-1)}-\frac{\ao^4}{4 (x-1)}+\frac{\ao^4}{4}+\frac{13 d_1 \ao^3}{18}-\
\frac{d_1 \ao^3}{2 (x-1)}+\frac{5 \ao^3}{3 (x-1)}+\frac{2 d_1 \
\ao^3}{9 (x-1)^2}-\frac{7 \ao^3}{9 (x-1)^2}-\frac{13 \
\ao^3}{9}-\frac{23 d_1 \ao^2}{12}-\frac{\ao^2}{3 (x-2)}+\frac{3 d_1 \
\ao^2}{4 (x-1)}-\frac{9 \ao^2}{2 (x-1)}-\frac{2 d_1 \ao^2}{3 \
(x-1)^2}+\frac{7 \ao^2}{2 (x-1)^2}+\frac{d_1 \ao^2}{2 \
(x-1)^3}-\frac{13 \ao^2}{6 (x-1)^3}+\frac{23 \ao^2}{6}+\frac{25 d_1 \
\ao}{6}+\frac{2 \ao}{x-2}-\frac{d_1 \ao}{2 (x-1)}+\frac{23 \ao}{3 \
(x-1)}-\frac{4 \ao}{3 (x-2)^2}+\frac{2 d_1 \ao}{3 (x-1)^2}-\frac{20 \
\ao}{3 (x-1)^2}-\frac{d_1 \ao}{(x-1)^3}+\frac{22 \ao}{3 \
(x-1)^3}+\frac{2 d_1 \ao}{(x-1)^4}-\frac{25 \ao}{3 (x-1)^4}-\frac{25 \
\ao}{3}-\frac{205 d_1}{72}+\Big(\frac{\ao^4}{x-1}-\ao^4-\frac{4 \
\ao^3}{x-1}+\frac{4 \ao^3}{3 (x-1)^2}+\frac{16 \ao^3}{3}+\frac{6 \
\ao^2}{x-1}-\frac{4 \ao^2}{(x-1)^2}+\frac{2 \ao^2}{(x-1)^3}-12 \ao^2-\
\frac{4 \ao}{x-1}+\frac{4 \ao}{(x-1)^2}-\frac{4 \ao}{(x-1)^3}+\frac{4 \
\ao}{(x-1)^4}+16 \ao-\frac{3}{x-1}+\frac{2}{3 (x-1)^2}+\frac{2}{3 \
(x-1)^3}-\frac{3}{(x-1)^4}-\frac{25}{3 (x-1)^5}-\frac{25}{3}\Big) \
H(0;\ao)+\Big(-\frac{d_1 \ao^4}{2}+\frac{d_1 \ao^4}{2 (x-1)}+\frac{8 \
d_1 \ao^3}{3}-\frac{2 d_1 \ao^3}{x-1}+\frac{2 d_1 \ao^3}{3 (x-1)^2}-6 \
d_1 \ao^2+\frac{3 d_1 \ao^2}{x-1}-\frac{2 d_1 \
\ao^2}{(x-1)^2}+\frac{d_1 \ao^2}{(x-1)^3}+8 d_1 \ao-\frac{2 d_1 \
\ao}{x-1}+\frac{2 d_1 \ao}{(x-1)^2}-\frac{2 d_1 \ao}{(x-1)^3}+\frac{2 \
d_1 \ao}{(x-1)^4}-\frac{25 d_1}{6}-\frac{3 d_1}{2 (x-1)}+\frac{d_1}{3 \
(x-1)^2}+\frac{d_1}{3 (x-1)^3}-\frac{3 d_1}{2 (x-1)^4}-\frac{25 \
d_1}{6 (x-1)^5}\Big) H(1;\ao)-\frac{4}{x-2}-\frac{15 d_1}{8 \
(x-1)}+\frac{7}{4 (x-1)}+\frac{8}{3 (x-2)^2}+\frac{5 d_1}{18 \
(x-1)^2}-\frac{1}{18 (x-1)^2}+\frac{5 d_1}{18 (x-1)^3}-\frac{31}{18 \
(x-1)^3}-\frac{15 d_1}{8 (x-1)^4}+\frac{13}{4 (x-1)^4}-\frac{205 \
d_1}{72 (x-1)^5}+\frac{415}{36 (x-1)^5}+\frac{205}{36}\Big) \
H(c_1(\ao);x)+\Big(-\frac{25}{3}-\frac{3}{x-1}+\frac{2}{3 \
(x-1)^2}+\frac{2}{3 (x-1)^3}-\frac{3}{(x-1)^4}-\frac{25}{3 \
(x-1)^5}\Big) H(0,0;\ao)+\Big(\frac{25}{3}+\frac{3}{x-1}-\frac{2}{3 \
(x-1)^2}-\frac{2}{3 (x-1)^3}+\frac{3}{(x-1)^4}+\frac{25}{3 \
(x-1)^5}\Big) H(0,0;x)+\Big(-\frac{3 d_1}{2 (x-1)}+\frac{d_1}{3 \
(x-1)^2}+\frac{d_1}{3 (x-1)^3}-\frac{3 d_1}{2 (x-1)^4}-\frac{25 \
d_1}{6 (x-1)^5}-\frac{25 d_1}{6}\Big) H(0,1;\ao)+H(1;x) \
\Big(-\frac{\pi ^2 d_1}{3 (x-1)^5}+\Big(\frac{2 \
d_1}{x-1}-\frac{d_1}{(x-1)^2}+\frac{2 d_1}{3 (x-1)^3}-\frac{d_1}{2 \
(x-1)^4}+\frac{25 d_1}{6 (x-1)^5}+\frac{4}{x-2}-\frac{8}{3 \
(x-2)^2}-\frac{4}{3 (x-1)^2}+\frac{8}{3 \
(x-2)^3}-\frac{4}{(x-1)^4}\Big) \
H(0;\ao)+\Big(4-\frac{4}{(x-1)^5}\Big) H(0,0;\ao)+\Big(2 d_1-\frac{2 \
d_1}{(x-1)^5}\Big) H(0,1;\ao)+\frac{2 \pi ^2}{3 \
(x-1)^5}\Big)+\Big(\frac{2 d_1}{(x-1)^5}-2 \
d_1-\frac{2}{(x-1)^5}+2\Big) H(0;\ao) H(0,1;x)+\Big(-\frac{\ao^4}{2 \
(x-1)}+\frac{\ao^4}{2}+\frac{2 \ao^3}{x-1}-\frac{2 \ao^3}{3 (x-1)^2}-\
\frac{8 \ao^3}{3}-\frac{3 \ao^2}{x-1}+\frac{2 \
\ao^2}{(x-1)^2}-\frac{\ao^2}{(x-1)^3}+6 \ao^2+\frac{2 \
\ao}{x-1}-\frac{2 \ao}{(x-1)^2}+\frac{2 \ao}{(x-1)^3}-\frac{2 \
\ao}{(x-1)^4}-8 \ao+\Big(4-\frac{4}{(x-1)^5}\Big) H(0;\ao)+\Big(2 \
d_1-\frac{2 d_1}{(x-1)^5}\Big) H(1;\ao)+\frac{4}{x-2}-\frac{1}{2 \
(x-1)}-\frac{8}{3 (x-2)^2}-\frac{2}{3 (x-1)^2}+\frac{8}{3 \
(x-2)^3}-\frac{1}{(x-1)^3}-\frac{2}{(x-1)^4}+\frac{25}{6}\Big) \
H(0,c_1(\ao);x)+\Big(-\frac{2 d_1}{x-1}+\frac{d_1}{(x-1)^2}-\frac{2 \
d_1}{3 (x-1)^3}+\frac{d_1}{2 (x-1)^4}-\frac{25 d_1}{6 \
(x-1)^5}-\frac{4}{x-2}+\frac{8}{3 (x-2)^2}+\frac{4}{3 \
(x-1)^2}-\frac{8}{3 (x-2)^3}+\frac{4}{(x-1)^4}\Big) \
H(1,0;x)+\Big(\frac{4 d_1}{(x-1)^5}-2 d_1-\frac{4}{(x-1)^5}\Big) H(0;\
\ao) H(1,1;x)+\Big(\frac{2 d_1}{x-1}-\frac{d_1}{(x-1)^2}+\frac{2 \
d_1}{3 (x-1)^3}-\frac{d_1}{2 (x-1)^4}+\frac{25 d_1}{6 \
(x-1)^5}+\Big(4-\frac{4}{(x-1)^5}\Big) H(0;\ao)+\Big(2 d_1-\frac{2 \
d_1}{(x-1)^5}\Big) H(1;\ao)+\frac{4}{x-2}-\frac{8}{3 \
(x-2)^2}-\frac{4}{3 (x-1)^2}+\frac{8}{3 \
(x-2)^3}-\frac{4}{(x-1)^4}\Big) \
H(1,c_1(\ao);x)+\Big(2-\frac{2}{(x-1)^5}\Big) H(0;\ao) H(2,1;x)+\Big(\
\frac{\ao^4}{x-1}-\ao^4-\frac{4 \ao^3}{x-1}+\frac{4 \ao^3}{3 \
(x-1)^2}+\frac{16 \ao^3}{3}+\frac{6 \ao^2}{x-1}-\frac{4 \
\ao^2}{(x-1)^2}+\frac{2 \ao^2}{(x-1)^3}-12 \ao^2-\frac{4 \
\ao}{x-1}+\frac{4 \ao}{(x-1)^2}-\frac{4 \ao}{(x-1)^3}+\frac{4 \
\ao}{(x-1)^4}+16 \ao+\frac{4 H(0;\ao)}{(x-1)^5}+\frac{2 d_1 \
H(1;\ao)}{(x-1)^5}-\frac{3}{x-1}+\frac{2}{3 (x-1)^2}+\frac{2}{3 \
(x-1)^3}-\frac{3}{(x-1)^4}-\frac{25}{3 (x-1)^5}-\frac{25}{3}\Big) \
H(c_1(\ao),c_1(\ao);x)+\Big(\frac{\ao^4}{2 \
(x-1)}-\frac{\ao^4}{2}-\frac{2 \ao^3}{x-1}+\frac{2 \ao^3}{3 (x-1)^2}+\
\frac{8 \ao^3}{3}+\frac{3 \ao^2}{x-1}-\frac{2 \
\ao^2}{(x-1)^2}+\frac{\ao^2}{(x-1)^3}-6 \ao^2-\frac{2 \
\ao}{x-1}+\frac{2 \ao}{(x-1)^2}-\frac{2 \ao}{(x-1)^3}+\frac{2 \
\ao}{(x-1)^4}+8 \ao-\frac{4}{x-2}+\frac{1}{2 (x-1)}+\frac{8}{3 \
(x-2)^2}+\frac{2}{3 (x-1)^2}-\frac{8}{3 \
(x-2)^3}+\frac{1}{(x-1)^3}+\frac{2}{(x-1)^4}-\frac{25}{6}\Big) H(c_2(\
\ao),c_1(\ao);x)-16 H(0,0,0;x)+\Big(2-\frac{2}{(x-1)^5}\Big) \
H(0,0,c_1(\ao);x)+\Big(-\frac{2 d_1}{(x-1)^5}+2 \
d_1+\frac{2}{(x-1)^5}-2\Big) H(0,1,0;x)+\Big(\frac{2 d_1}{(x-1)^5}-2 \
d_1-\frac{2}{(x-1)^5}+2\Big) H(0,1,c_1(\ao);x)+4 \
H(0,c_1(\ao),c_1(\ao);x)+\Big(2-\frac{2}{(x-1)^5}\Big) \
H(0,c_2(\ao),c_1(\ao);x)+\Big(\frac{4}{(x-1)^5}-4\Big) \
H(1,0,0;x)+\Big(\frac{2 d_1}{(x-1)^5}-\frac{4}{(x-1)^5}\Big) \
H(1,0,c_1(\ao);x)+\Big(-\frac{4 d_1}{(x-1)^5}+2 d_1+\frac{4}{(x-1)^5}\
\Big) H(1,1,0;x)+\Big(\frac{4 d_1}{(x-1)^5}-2 \
d_1-\frac{4}{(x-1)^5}\Big) H(1,1,c_1(\ao);x)+\Big(4-\frac{2 \
d_1}{(x-1)^5}\Big) \
H(1,c_1(\ao),c_1(\ao);x)+\Big(2-\frac{2}{(x-1)^5}\Big) \
H(2,0,c_1(\ao);x)+\Big(\frac{2}{(x-1)^5}-2\Big) \
H(2,1,0;x)+\Big(2-\frac{2}{(x-1)^5}\Big) \
H(2,1,c_1(\ao);x)+\Big(\frac{2}{(x-1)^5}-2\Big) \
H(2,c_2(\ao),c_1(\ao);x)-\frac{2 \
H(c_1(\ao),0,c_1(\ao);x)}{(x-1)^5}+\frac{4 \
H(c_1(\ao),c_1(\ao),c_1(\ao);x)}{(x-1)^5}+\frac{2 \
H(c_1(\ao),c_2(\ao),c_1(\ao);x)}{(x-1)^5}-\frac{\pi \
^2}{x-2}+\frac{\pi ^2}{8 (x-1)}+\frac{2 \pi ^2}{3 (x-2)^2}+\frac{\pi \
^2}{6 (x-1)^2}-\frac{2 \pi ^2}{3 (x-2)^3}+\frac{\pi ^2}{4 \
(x-1)^3}+\frac{\pi ^2}{2 (x-1)^4}-\frac{21 \zeta_3}{4 \
(x-1)^5}-\frac{33 \zeta_3}{4}+\frac{\pi ^2 \ln 2\, }{2 \
(x-1)^5}-\frac{1}{2} \pi ^2 \ln 2\, -\frac{25 \pi ^2}{24},
\erp
% ep^2
\brp
b_2^{(1,-1)}=-\frac{37}{432} d_1^2 \ao^3+\frac{37 d_1 \ao^3}{108}+\frac{37 d_1^2 \
\ao^3}{432 (x-1)^2}-\frac{37 d_1 \ao^3}{108 (x-1)^2}-\frac{\pi ^2 \
\ao^3}{72 (x-1)^2}+\frac{37 \ao^3}{108 (x-1)^2}+\frac{\pi ^2 \
\ao^3}{72}-\frac{37 \ao^3}{108}+\frac{715 d_1^2 \ao^2}{864}-\frac{793 \
d_1 \ao^2}{216}+\frac{115 d_1^2 \ao^2}{864 (x-1)}+\frac{41 d_1 \
\ao^2}{216 (x-1)}-\frac{\pi ^2 \ao^2}{144 (x-1)}-\frac{197 \ao^2}{216 \
(x-1)}-\frac{107 d_1^2 \ao^2}{864 (x-1)^2}+\frac{263 d_1 \ao^2}{216 \
(x-1)^2}+\frac{5 \pi ^2 \ao^2}{144 (x-1)^2}-\frac{419 \ao^2}{216 \
(x-1)^2}+\frac{493 d_1^2 \ao^2}{864 (x-1)^3}-\frac{571 d_1 \ao^2}{216 \
(x-1)^3}-\frac{7 \pi ^2 \ao^2}{144 (x-1)^3}+\frac{649 \ao^2}{216 \
(x-1)^3}-\frac{13 \pi ^2 \ao^2}{144}+\frac{871 \ao^2}{216}-\frac{3515 \
d_1^2 \ao}{432}+\frac{1040 d_1 \ao}{27}-\frac{25 d_1 \ao}{9 \
(x-2)}+\frac{50 \ao}{9 (x-2)}-\frac{265 d_1^2 \ao}{108 \
(x-1)}+\frac{523 d_1 \ao}{108 (x-1)}+\frac{\pi ^2 \ao}{18 \
(x-1)}+\frac{62 \ao}{27 (x-1)}-\frac{d_1^2 \ao}{108 \
(x-1)^2}-\frac{d_1 \ao}{54 (x-1)^2}-\frac{\pi ^2 \ao}{36 \
(x-1)^2}+\frac{2 \ao}{27 (x-1)^2}+\frac{113 d_1^2 \ao}{108 \
(x-1)^3}+\frac{307 d_1 \ao}{108 (x-1)^3}+\frac{\pi ^2 \ao}{18 \
(x-1)^3}-\frac{325 \ao}{27 (x-1)^3}+\frac{2911 d_1^2 \ao}{432 \
(x-1)^4}-\frac{1739 d_1 \ao}{54 (x-1)^4}-\frac{13 \pi ^2 \ao}{72 \
(x-1)^4}+\frac{4099 \ao}{108 (x-1)^4}+\frac{23 \pi ^2 \
\ao}{72}-\frac{4859 \ao}{108}+\Big(-\frac{7 d_1 \ao^3}{18}+\frac{7 \
d_1 \ao^3}{18 (x-1)^2}-\frac{7 \ao^3}{9 (x-1)^2}+\frac{7 \
\ao^3}{9}+\frac{109 d_1 \ao^2}{36}+\frac{13 d_1 \ao^2}{36 \
(x-1)}+\frac{11 \ao^2}{18 (x-1)}-\frac{29 d_1 \ao^2}{36 \
(x-1)^2}+\frac{53 \ao^2}{18 (x-1)^2}+\frac{67 d_1 \ao^2}{36 (x-1)^3}-\
\frac{79 \ao^2}{18 (x-1)^3}-\frac{121 \ao^2}{18}-\frac{305 d_1 \
\ao}{18}-\frac{8 \ao}{3 (x-2)}-\frac{38 d_1 \ao}{9 (x-1)}+\frac{10 \
\ao}{9 (x-1)}+\frac{4 d_1 \ao}{9 (x-1)^2}-\frac{20 \ao}{9 \
(x-1)^2}-\frac{2 d_1 \ao}{9 (x-1)^3}+\frac{82 \ao}{9 \
(x-1)^3}+\frac{217 d_1 \ao}{18 (x-1)^4}-\frac{271 \ao}{9 \
(x-1)^4}+\frac{371 \ao}{9}+\frac{2035 d_1^2}{432}-\frac{2035 \
d_1}{108}+\frac{38 d_1}{3 (x-2)}-\frac{68}{3 (x-2)}+\frac{63 \
d_1^2}{16 (x-1)}-\frac{161 d_1}{12 (x-1)}-\frac{\pi ^2}{8 \
(x-1)}+\frac{3}{2 (x-1)}-\frac{76 d_1}{9 (x-2)^2}+\frac{152}{9 \
(x-2)^2}-\frac{19 d_1^2}{54 (x-1)^2}+\frac{323 d_1}{108 \
(x-1)^2}+\frac{\pi ^2}{36 (x-1)^2}-\frac{215}{108 (x-1)^2}-\frac{19 \
d_1^2}{54 (x-1)^3}+\frac{605 d_1}{108 (x-1)^3}+\frac{\pi ^2}{36 \
(x-1)^3}-\frac{1643}{108 (x-1)^3}+\frac{63 d_1^2}{16 \
(x-1)^4}-\frac{50 d_1}{3 (x-1)^4}-\frac{\pi ^2}{8 \
(x-1)^4}+\frac{8}{(x-1)^4}+\frac{2035 d_1^2}{432 (x-1)^5}-\frac{895 \
d_1}{27 (x-1)^5}-\frac{25 \pi ^2}{72 (x-1)^5}+\frac{5665}{108 \
(x-1)^5}-\frac{25 \pi ^2}{72}+\frac{1855}{108}\Big) \
H(0;\ao)+\Big(-\frac{7}{36} d_1^2 \ao^3+\frac{7 d_1 \
\ao^3}{18}+\frac{7 d_1^2 \ao^3}{36 (x-1)^2}-\frac{7 d_1 \ao^3}{18 \
(x-1)^2}+\frac{109 d_1^2 \ao^2}{72}-\frac{121 d_1 \ao^2}{36}+\frac{13 \
d_1^2 \ao^2}{72 (x-1)}+\frac{11 d_1 \ao^2}{36 (x-1)}-\frac{29 d_1^2 \
\ao^2}{72 (x-1)^2}+\frac{53 d_1 \ao^2}{36 (x-1)^2}+\frac{67 d_1^2 \
\ao^2}{72 (x-1)^3}-\frac{79 d_1 \ao^2}{36 (x-1)^3}-\frac{305 d_1^2 \
\ao}{36}+\frac{371 d_1 \ao}{18}-\frac{4 d_1 \ao}{3 (x-2)}-\frac{19 \
d_1^2 \ao}{9 (x-1)}+\frac{5 d_1 \ao}{9 (x-1)}+\frac{2 d_1^2 \ao}{9 \
(x-1)^2}-\frac{10 d_1 \ao}{9 (x-1)^2}-\frac{d_1^2 \ao}{9 \
(x-1)^3}+\frac{41 d_1 \ao}{9 (x-1)^3}+\frac{217 d_1^2 \ao}{36 \
(x-1)^4}-\frac{271 d_1 \ao}{18 (x-1)^4}+\frac{515 \
d_1^2}{72}-\frac{635 d_1}{36}+\frac{4 d_1}{3 (x-2)}+\frac{139 \
d_1^2}{72 (x-1)}-\frac{31 d_1}{36 (x-1)}-\frac{d_1^2}{72 \
(x-1)^2}+\frac{d_1}{36 (x-1)^2}-\frac{59 d_1^2}{72 (x-1)^3}-\frac{85 \
d_1}{36 (x-1)^3}-\frac{217 d_1^2}{36 (x-1)^4}+\frac{271 d_1}{18 \
(x-1)^4}\Big) H(1;\ao)+\Big(\frac{4 \ao^3}{3 (x-1)^2}-\frac{4 \
\ao^3}{3}+\frac{2 \ao^2}{3 (x-1)}-\frac{10 \ao^2}{3 (x-1)^2}+\frac{14 \
\ao^2}{3 (x-1)^3}+\frac{26 \ao^2}{3}-\frac{16 \ao}{3 (x-1)}+\frac{8 \
\ao}{3 (x-1)^2}-\frac{16 \ao}{3 (x-1)^3}+\frac{52 \ao}{3 \
(x-1)^4}-\frac{92 \ao}{3}+\frac{205 d_1}{18}+\frac{16}{x-2}+\frac{15 \
d_1}{2 (x-1)}-\frac{7}{x-1}-\frac{32}{3 (x-2)^2}-\frac{10 d_1}{9 \
(x-1)^2}+\frac{2}{9 (x-1)^2}-\frac{10 d_1}{9 (x-1)^3}+\frac{62}{9 \
(x-1)^3}+\frac{15 d_1}{2 (x-1)^4}-\frac{13}{(x-1)^4}+\frac{205 \
d_1}{18 (x-1)^5}-\frac{415}{9 (x-1)^5}-\frac{205}{9}\Big) H(0,0;\ao)+\
\Big(-\frac{15 d_1}{2 (x-1)}+\frac{10 d_1}{9 (x-1)^2}+\frac{10 d_1}{9 \
(x-1)^3}-\frac{15 d_1}{2 (x-1)^4}-\frac{205 d_1}{18 \
(x-1)^5}-\frac{205 d_1}{18}-\frac{16}{x-2}+\frac{7}{x-1}+\frac{32}{3 \
(x-2)^2}-\frac{2}{9 (x-1)^2}-\frac{62}{9 (x-1)^3}+\frac{13}{(x-1)^4}-\
\frac{8 \pi ^2}{3 (x-1)^5}+\frac{415}{9 (x-1)^5}-\frac{4 \pi \
^2}{3}+\frac{205}{9}\Big) H(0,0;x)+\Big(-\frac{2 d_1 \
\ao^3}{3}+\frac{2 d_1 \ao^3}{3 (x-1)^2}+\frac{13 d_1 \
\ao^2}{3}+\frac{d_1 \ao^2}{3 (x-1)}-\frac{5 d_1 \ao^2}{3 \
(x-1)^2}+\frac{7 d_1 \ao^2}{3 (x-1)^3}-\frac{46 d_1 \ao}{3}-\frac{8 \
d_1 \ao}{3 (x-1)}+\frac{4 d_1 \ao}{3 (x-1)^2}-\frac{8 d_1 \ao}{3 \
(x-1)^3}+\frac{26 d_1 \ao}{3 (x-1)^4}+\frac{205 d_1^2}{36}-\frac{205 \
d_1}{18}+\frac{8 d_1}{x-2}+\frac{15 d_1^2}{4 (x-1)}-\frac{7 d_1}{2 \
(x-1)}-\frac{16 d_1}{3 (x-2)^2}-\frac{5 d_1^2}{9 \
(x-1)^2}+\frac{d_1}{9 (x-1)^2}-\frac{5 d_1^2}{9 (x-1)^3}+\frac{31 \
d_1}{9 (x-1)^3}+\frac{15 d_1^2}{4 (x-1)^4}-\frac{13 d_1}{2 \
(x-1)^4}+\frac{205 d_1^2}{36 (x-1)^5}-\frac{415 d_1}{18 (x-1)^5}\Big) \
H(0,1;\ao)+\Big(\frac{4 \pi ^2 d_1}{3 (x-1)^5}+\Big(\frac{8 \
d_1}{x-2}-\frac{16 d_1}{3 (x-2)^2}-\frac{8 d_1}{3 (x-1)^2}+\frac{16 \
d_1}{3 (x-2)^3}-\frac{8 \
d_1}{(x-1)^4}-\frac{8}{x-2}+\frac{5}{x-1}+\frac{16}{3 \
(x-2)^2}-\frac{2}{3 (x-1)^2}-\frac{16}{3 (x-2)^3}+\frac{10}{3 \
(x-1)^3}+\frac{3}{(x-1)^4}+\frac{25}{3 (x-1)^5}-\frac{25}{3}\Big) \
H(0;\ao)+\Big(-\frac{8 d_1}{(x-1)^5}+8 d_1+\frac{8}{(x-1)^5}-8\Big) \
H(0,0;\ao)+\Big(-\frac{4 d_1^2}{(x-1)^5}+4 d_1^2+\frac{4 \
d_1}{(x-1)^5}-4 d_1\Big) H(0,1;\ao)-\frac{8 \pi ^2}{3 (x-1)^5}\Big) \
H(0,1;x)+\Big(\frac{\pi ^2 d_1}{(x-1)^5}-\pi ^2 d_1-\frac{\pi \
^2}{(x-1)^5}+\pi ^2\Big) H(0,2;x)+\Big(-\frac{2 d_1 \ao^3}{3}+\frac{2 \
d_1 \ao^3}{3 (x-1)^2}+\frac{13 d_1 \ao^2}{3}+\frac{d_1 \ao^2}{3 \
(x-1)}-\frac{5 d_1 \ao^2}{3 (x-1)^2}+\frac{7 d_1 \ao^2}{3 \
(x-1)^3}-\frac{46 d_1 \ao}{3}-\frac{8 d_1 \ao}{3 (x-1)}+\frac{4 d_1 \
\ao}{3 (x-1)^2}-\frac{8 d_1 \ao}{3 (x-1)^3}+\frac{26 d_1 \ao}{3 \
(x-1)^4}+\frac{35 d_1}{3}+\frac{7 d_1}{3 (x-1)}-\frac{d_1}{3 \
(x-1)^2}+\frac{d_1}{3 (x-1)^3}-\frac{26 d_1}{3 (x-1)^4}\Big) \
H(1,0;\ao)+\Big(\frac{4 d_1^2}{x-1}-\frac{d_1^2}{(x-1)^2}+\frac{4 \
d_1^2}{9 (x-1)^3}-\frac{d_1^2}{4 (x-1)^4}+\frac{205 d_1^2}{36 \
(x-1)^5}+\frac{46 d_1}{3 (x-2)}-\frac{46 d_1}{3 (x-1)}-\frac{52 \
d_1}{9 (x-2)^2}+\frac{40 d_1}{9 (x-1)^2}+\frac{28 d_1}{9 \
(x-2)^3}-\frac{98 d_1}{9 (x-1)^3}+\frac{13 d_1}{6 (x-1)^4}+\frac{4 \
\pi ^2 d_1}{3 (x-1)^5}-\frac{415 d_1}{18 (x-1)^5}-\frac{80}{3 (x-2)}-\
\frac{43}{6 (x-1)}+\frac{56}{9 (x-2)^2}+\frac{299}{18 \
(x-1)^2}-\frac{56}{9 (x-2)^3}+\frac{23}{6 (x-1)^3}+\frac{203}{6 \
(x-1)^4}-\frac{5 \pi ^2}{2 (x-1)^5}+\frac{35}{6 (x-1)^5}-\frac{\pi \
^2}{6}-\frac{35}{6}\Big) H(1,0;x)+\Big(-\frac{1}{3} d_1^2 \
\ao^3+\frac{d_1^2 \ao^3}{3 (x-1)^2}+\frac{13 d_1^2 \
\ao^2}{6}+\frac{d_1^2 \ao^2}{6 (x-1)}-\frac{5 d_1^2 \ao^2}{6 \
(x-1)^2}+\frac{7 d_1^2 \ao^2}{6 (x-1)^3}-\frac{23 d_1^2 \
\ao}{3}-\frac{4 d_1^2 \ao}{3 (x-1)}+\frac{2 d_1^2 \ao}{3 \
(x-1)^2}-\frac{4 d_1^2 \ao}{3 (x-1)^3}+\frac{13 d_1^2 \ao}{3 \
(x-1)^4}+\frac{35 d_1^2}{6}+\frac{7 d_1^2}{6 (x-1)}-\frac{d_1^2}{6 \
(x-1)^2}+\frac{d_1^2}{6 (x-1)^3}-\frac{13 d_1^2}{3 (x-1)^4}\Big) \
H(1,1;\ao)+H(0,c_1(\ao);x) \Big(-\frac{d_1 \ao^4}{4}+\frac{d_1 \
\ao^4}{4 (x-1)}-\frac{\ao^4}{2 (x-1)}+\frac{\ao^4}{2}+\frac{13 d_1 \
\ao^3}{9}-\frac{d_1 \ao^3}{x-1}+\frac{10 \ao^3}{3 (x-1)}+\frac{4 d_1 \
\ao^3}{9 (x-1)^2}-\frac{11 \ao^3}{9 (x-1)^2}-\frac{29 \
\ao^3}{9}-\frac{23 d_1 \ao^2}{6}-\frac{2 \ao^2}{3 (x-2)}+\frac{3 d_1 \
\ao^2}{2 (x-1)}-\frac{53 \ao^2}{6 (x-1)}-\frac{4 d_1 \ao^2}{3 \
(x-1)^2}+\frac{37 \ao^2}{6 (x-1)^2}+\frac{d_1 \
\ao^2}{(x-1)^3}-\frac{19 \ao^2}{6 (x-1)^3}+\frac{59 \
\ao^2}{6}+\frac{25 d_1 \ao}{3}+\frac{4 \ao}{x-2}-\frac{d_1 \ao}{x-1}+\
\frac{14 \ao}{x-1}-\frac{8 \ao}{3 (x-2)^2}+\frac{4 d_1 \ao}{3 \
(x-1)^2}-\frac{38 \ao}{3 (x-1)^2}-\frac{2 d_1 \ao}{(x-1)^3}+\frac{40 \
\ao}{3 (x-1)^3}+\frac{4 d_1 \ao}{(x-1)^4}-\frac{37 \ao}{3 \
(x-1)^4}-\frac{73 \ao}{3}-\frac{205 d_1}{36}+\Big(\frac{2 \
\ao^4}{x-1}-2 \ao^4-\frac{8 \ao^3}{x-1}+\frac{8 \ao^3}{3 \
(x-1)^2}+\frac{32 \ao^3}{3}+\frac{12 \ao^2}{x-1}-\frac{8 \
\ao^2}{(x-1)^2}+\frac{4 \ao^2}{(x-1)^3}-24 \ao^2-\frac{8 \
\ao}{x-1}+\frac{8 \ao}{(x-1)^2}-\frac{8 \ao}{(x-1)^3}+\frac{8 \
\ao}{(x-1)^4}+32 \ao-\frac{16}{x-2}+\frac{2}{x-1}+\frac{32}{3 \
(x-2)^2}+\frac{8}{3 (x-1)^2}-\frac{32}{3 \
(x-2)^3}+\frac{4}{(x-1)^3}+\frac{8}{(x-1)^4}-\frac{50}{3}\Big) \
H(0;\ao)+\Big(-d_1 \ao^4+\frac{d_1 \ao^4}{x-1}+\frac{16 d_1 \
\ao^3}{3}-\frac{4 d_1 \ao^3}{x-1}+\frac{4 d_1 \ao^3}{3 (x-1)^2}-12 \
d_1 \ao^2+\frac{6 d_1 \ao^2}{x-1}-\frac{4 d_1 \ao^2}{(x-1)^2}+\frac{2 \
d_1 \ao^2}{(x-1)^3}+16 d_1 \ao-\frac{4 d_1 \ao}{x-1}+\frac{4 d_1 \
\ao}{(x-1)^2}-\frac{4 d_1 \ao}{(x-1)^3}+\frac{4 d_1 \
\ao}{(x-1)^4}-\frac{25 d_1}{3}-\frac{8 \
d_1}{x-2}+\frac{d_1}{x-1}+\frac{16 d_1}{3 (x-2)^2}+\frac{4 d_1}{3 \
(x-1)^2}-\frac{16 d_1}{3 (x-2)^3}+\frac{2 d_1}{(x-1)^3}+\frac{4 \
d_1}{(x-1)^4}\Big) H(1;\ao)+\Big(\frac{16}{(x-1)^5}-16\Big) \
H(0,0;\ao)+\Big(\frac{8 d_1}{(x-1)^5}-8 d_1\Big) \
H(0,1;\ao)+\Big(\frac{8 d_1}{(x-1)^5}-8 d_1\Big) \
H(1,0;\ao)+\Big(\frac{4 d_1^2}{(x-1)^5}-4 d_1^2\Big) \
H(1,1;\ao)-\frac{32 d_1}{3 (x-2)}+\frac{70}{3 (x-2)}+\frac{35 d_1}{12 \
(x-1)}-\frac{5}{6 (x-1)}+\frac{28 d_1}{9 (x-2)^2}-\frac{32}{9 \
(x-2)^2}+\frac{28 d_1}{9 (x-1)^2}-\frac{80}{9 (x-1)^2}-\frac{28 \
d_1}{9 (x-2)^3}+\frac{56}{9 (x-2)^3}+\frac{5 \
d_1}{(x-1)^3}-\frac{14}{(x-1)^3}+\frac{4 d_1}{(x-1)^4}-\frac{43}{2 \
(x-1)^4}-\frac{\pi ^2}{6 (x-1)^5}-\frac{35}{6 (x-1)^5}+\frac{\pi \
^2}{6}+\frac{415}{18}\Big)+H(c_1(\ao);x) \Big(\frac{d_1^2 \ao^4}{16}-\
\frac{d_1 \ao^4}{4}-\frac{d_1^2 \ao^4}{16 (x-1)}+\frac{d_1 \ao^4}{4 \
(x-1)}+\frac{\pi ^2 \ao^4}{24 (x-1)}-\frac{\ao^4}{4 (x-1)}-\frac{\pi \
^2 \ao^4}{24}+\frac{\ao^4}{4}-\frac{43 d_1^2 \ao^3}{108}+\frac{43 d_1 \
\ao^3}{27}+\frac{d_1^2 \ao^3}{4 (x-1)}-\frac{16 d_1 \ao^3}{9 \
(x-1)}-\frac{\pi ^2 \ao^3}{6 (x-1)}+\frac{23 \ao^3}{9 (x-1)}-\frac{4 \
d_1^2 \ao^3}{27 (x-1)^2}+\frac{53 d_1 \ao^3}{54 (x-1)^2}+\frac{\pi ^2 \
\ao^3}{18 (x-1)^2}-\frac{37 \ao^3}{27 (x-1)^2}+\frac{2 \pi ^2 \
\ao^3}{9}-\frac{43 \ao^3}{27}+\frac{95 d_1^2 \ao^2}{72}-\frac{95 d_1 \
\ao^2}{18}+\frac{13 d_1 \ao^2}{18 (x-2)}-\frac{13 \ao^2}{9 \
(x-2)}-\frac{3 d_1^2 \ao^2}{8 (x-1)}+\frac{16 d_1 \ao^2}{3 \
(x-1)}+\frac{\pi ^2 \ao^2}{4 (x-1)}-\frac{59 \ao^2}{6 (x-1)}+\frac{4 \
d_1^2 \ao^2}{9 (x-1)^2}-\frac{173 d_1 \ao^2}{36 (x-1)^2}-\frac{\pi ^2 \
\ao^2}{6 (x-1)^2}+\frac{59 \ao^2}{6 (x-1)^2}-\frac{d_1^2 \ao^2}{2 \
(x-1)^3}+\frac{139 d_1 \ao^2}{36 (x-1)^3}+\frac{\pi ^2 \ao^2}{12 \
(x-1)^3}-\frac{56 \ao^2}{9 (x-1)^3}-\frac{\pi ^2 \ao^2}{2}+\frac{46 \
\ao^2}{9}-\frac{205 d_1^2 \ao}{36}+\frac{205 d_1 \ao}{9}-\frac{17 d_1 \
\ao}{3 (x-2)}+\frac{10 \ao}{x-2}+\frac{d_1^2 \ao}{4 (x-1)}-\frac{127 \
d_1 \ao}{9 (x-1)}-\frac{\pi ^2 \ao}{6 (x-1)}+\frac{1247 \ao}{36 \
(x-1)}+\frac{38 d_1 \ao}{9 (x-2)^2}-\frac{76 \ao}{9 (x-2)^2}-\frac{4 \
d_1^2 \ao}{9 (x-1)^2}+\frac{34 d_1 \ao}{3 (x-1)^2}+\frac{\pi ^2 \
\ao}{6 (x-1)^2}-\frac{565 \ao}{18 (x-1)^2}+\frac{d_1^2 \ao}{(x-1)^3}-\
\frac{146 d_1 \ao}{9 (x-1)^3}-\frac{\pi ^2 \ao}{6 (x-1)^3}+\frac{1495 \
\ao}{36 (x-1)^3}-\frac{4 d_1^2 \ao}{(x-1)^4}+\frac{505 d_1 \ao}{18 \
(x-1)^4}+\frac{\pi ^2 \ao}{6 (x-1)^4}-\frac{803 \ao}{18 \
(x-1)^4}+\frac{2 \pi ^2 \ao}{3}-\frac{377 \ao}{18}+\frac{2035 \
d_1^2}{432}-\frac{2035 d_1}{108}+\Big(\frac{d_1 \ao^4}{2}-\frac{d_1 \
\ao^4}{2 (x-1)}+\frac{\ao^4}{x-1}-\ao^4-\frac{26 d_1 \
\ao^3}{9}+\frac{2 d_1 \ao^3}{x-1}-\frac{20 \ao^3}{3 (x-1)}-\frac{8 \
d_1 \ao^3}{9 (x-1)^2}+\frac{28 \ao^3}{9 (x-1)^2}+\frac{52 \
\ao^3}{9}+\frac{23 d_1 \ao^2}{3}+\frac{4 \ao^2}{3 (x-2)}-\frac{3 d_1 \
\ao^2}{x-1}+\frac{18 \ao^2}{x-1}+\frac{8 d_1 \ao^2}{3 \
(x-1)^2}-\frac{14 \ao^2}{(x-1)^2}-\frac{2 d_1 \
\ao^2}{(x-1)^3}+\frac{26 \ao^2}{3 (x-1)^3}-\frac{46 \
\ao^2}{3}-\frac{50 d_1 \ao}{3}-\frac{8 \ao}{x-2}+\frac{2 d_1 \
\ao}{x-1}-\frac{92 \ao}{3 (x-1)}+\frac{16 \ao}{3 (x-2)^2}-\frac{8 d_1 \
\ao}{3 (x-1)^2}+\frac{80 \ao}{3 (x-1)^2}+\frac{4 d_1 \
\ao}{(x-1)^3}-\frac{88 \ao}{3 (x-1)^3}-\frac{8 d_1 \
\ao}{(x-1)^4}+\frac{100 \ao}{3 (x-1)^4}+\frac{100 \ao}{3}+\frac{205 \
d_1}{18}+\frac{16}{x-2}+\frac{15 d_1}{2 \
(x-1)}-\frac{7}{x-1}-\frac{32}{3 (x-2)^2}-\frac{10 d_1}{9 \
(x-1)^2}+\frac{2}{9 (x-1)^2}-\frac{10 d_1}{9 (x-1)^3}+\frac{62}{9 \
(x-1)^3}+\frac{15 d_1}{2 (x-1)^4}-\frac{13}{(x-1)^4}+\frac{205 \
d_1}{18 (x-1)^5}-\frac{415}{9 (x-1)^5}-\frac{205}{9}\Big) \
H(0;\ao)+\Big(\frac{d_1^2 \ao^4}{4}-\frac{d_1 \ao^4}{2}-\frac{d_1^2 \
\ao^4}{4 (x-1)}+\frac{d_1 \ao^4}{2 (x-1)}-\frac{13 d_1^2 \
\ao^3}{9}+\frac{26 d_1 \ao^3}{9}+\frac{d_1^2 \ao^3}{x-1}-\frac{10 d_1 \
\ao^3}{3 (x-1)}-\frac{4 d_1^2 \ao^3}{9 (x-1)^2}+\frac{14 d_1 \ao^3}{9 \
(x-1)^2}+\frac{23 d_1^2 \ao^2}{6}-\frac{23 d_1 \ao^2}{3}+\frac{2 d_1 \
\ao^2}{3 (x-2)}-\frac{3 d_1^2 \ao^2}{2 (x-1)}+\frac{9 d_1 \
\ao^2}{x-1}+\frac{4 d_1^2 \ao^2}{3 (x-1)^2}-\frac{7 d_1 \
\ao^2}{(x-1)^2}-\frac{d_1^2 \ao^2}{(x-1)^3}+\frac{13 d_1 \ao^2}{3 \
(x-1)^3}-\frac{25 d_1^2 \ao}{3}+\frac{50 d_1 \ao}{3}-\frac{4 d_1 \
\ao}{x-2}+\frac{d_1^2 \ao}{x-1}-\frac{46 d_1 \ao}{3 (x-1)}+\frac{8 \
d_1 \ao}{3 (x-2)^2}-\frac{4 d_1^2 \ao}{3 (x-1)^2}+\frac{40 d_1 \ao}{3 \
(x-1)^2}+\frac{2 d_1^2 \ao}{(x-1)^3}-\frac{44 d_1 \ao}{3 \
(x-1)^3}-\frac{4 d_1^2 \ao}{(x-1)^4}+\frac{50 d_1 \ao}{3 \
(x-1)^4}+\frac{205 d_1^2}{36}-\frac{205 d_1}{18}+\frac{8 \
d_1}{x-2}+\frac{15 d_1^2}{4 (x-1)}-\frac{7 d_1}{2 (x-1)}-\frac{16 \
d_1}{3 (x-2)^2}-\frac{5 d_1^2}{9 (x-1)^2}+\frac{d_1}{9 \
(x-1)^2}-\frac{5 d_1^2}{9 (x-1)^3}+\frac{31 d_1}{9 (x-1)^3}+\frac{15 \
d_1^2}{4 (x-1)^4}-\frac{13 d_1}{2 (x-1)^4}+\frac{205 d_1^2}{36 \
(x-1)^5}-\frac{415 d_1}{18 (x-1)^5}\Big) H(1;\ao)+\Big(-\frac{4 \
\ao^4}{x-1}+4 \ao^4+\frac{16 \ao^3}{x-1}-\frac{16 \ao^3}{3 \
(x-1)^2}-\frac{64 \ao^3}{3}-\frac{24 \ao^2}{x-1}+\frac{16 \
\ao^2}{(x-1)^2}-\frac{8 \ao^2}{(x-1)^3}+48 \ao^2+\frac{16 \
\ao}{x-1}-\frac{16 \ao}{(x-1)^2}+\frac{16 \ao}{(x-1)^3}-\frac{16 \
\ao}{(x-1)^4}-64 \ao+\frac{12}{x-1}-\frac{8}{3 (x-1)^2}-\frac{8}{3 \
(x-1)^3}+\frac{12}{(x-1)^4}+\frac{100}{3 (x-1)^5}+\frac{100}{3}\Big) \
H(0,0;\ao)+\Big(2 d_1 \ao^4-\frac{2 d_1 \ao^4}{x-1}-\frac{32 d_1 \
\ao^3}{3}+\frac{8 d_1 \ao^3}{x-1}-\frac{8 d_1 \ao^3}{3 (x-1)^2}+24 \
d_1 \ao^2-\frac{12 d_1 \ao^2}{x-1}+\frac{8 d_1 \
\ao^2}{(x-1)^2}-\frac{4 d_1 \ao^2}{(x-1)^3}-32 d_1 \ao+\frac{8 d_1 \
\ao}{x-1}-\frac{8 d_1 \ao}{(x-1)^2}+\frac{8 d_1 \ao}{(x-1)^3}-\frac{8 \
d_1 \ao}{(x-1)^4}+\frac{50 d_1}{3}+\frac{6 d_1}{x-1}-\frac{4 d_1}{3 \
(x-1)^2}-\frac{4 d_1}{3 (x-1)^3}+\frac{6 d_1}{(x-1)^4}+\frac{50 \
d_1}{3 (x-1)^5}\Big) H(0,1;\ao)+\Big(2 d_1 \ao^4-\frac{2 d_1 \
\ao^4}{x-1}-\frac{32 d_1 \ao^3}{3}+\frac{8 d_1 \ao^3}{x-1}-\frac{8 \
d_1 \ao^3}{3 (x-1)^2}+24 d_1 \ao^2-\frac{12 d_1 \ao^2}{x-1}+\frac{8 \
d_1 \ao^2}{(x-1)^2}-\frac{4 d_1 \ao^2}{(x-1)^3}-32 d_1 \ao+\frac{8 \
d_1 \ao}{x-1}-\frac{8 d_1 \ao}{(x-1)^2}+\frac{8 d_1 \
\ao}{(x-1)^3}-\frac{8 d_1 \ao}{(x-1)^4}+\frac{50 d_1}{3}+\frac{6 \
d_1}{x-1}-\frac{4 d_1}{3 (x-1)^2}-\frac{4 d_1}{3 (x-1)^3}+\frac{6 \
d_1}{(x-1)^4}+\frac{50 d_1}{3 (x-1)^5}\Big) H(1,0;\ao)+\Big(d_1^2 \
\ao^4-\frac{d_1^2 \ao^4}{x-1}-\frac{16 d_1^2 \ao^3}{3}+\frac{4 d_1^2 \
\ao^3}{x-1}-\frac{4 d_1^2 \ao^3}{3 (x-1)^2}+12 d_1^2 \ao^2-\frac{6 \
d_1^2 \ao^2}{x-1}+\frac{4 d_1^2 \ao^2}{(x-1)^2}-\frac{2 d_1^2 \
\ao^2}{(x-1)^3}-16 d_1^2 \ao+\frac{4 d_1^2 \ao}{x-1}-\frac{4 d_1^2 \
\ao}{(x-1)^2}+\frac{4 d_1^2 \ao}{(x-1)^3}-\frac{4 d_1^2 \
\ao}{(x-1)^4}+\frac{25 d_1^2}{3}+\frac{3 d_1^2}{x-1}-\frac{2 d_1^2}{3 \
(x-1)^2}-\frac{2 d_1^2}{3 (x-1)^3}+\frac{3 d_1^2}{(x-1)^4}+\frac{25 \
d_1^2}{3 (x-1)^5}\Big) H(1,1;\ao)+\frac{38 d_1}{3 (x-2)}-\frac{68}{3 \
(x-2)}+\frac{63 d_1^2}{16 (x-1)}-\frac{161 d_1}{12 (x-1)}-\frac{\pi \
^2}{8 (x-1)}+\frac{3}{2 (x-1)}-\frac{76 d_1}{9 (x-2)^2}+\frac{152}{9 \
(x-2)^2}-\frac{19 d_1^2}{54 (x-1)^2}+\frac{323 d_1}{108 \
(x-1)^2}+\frac{\pi ^2}{36 (x-1)^2}-\frac{215}{108 (x-1)^2}-\frac{19 \
d_1^2}{54 (x-1)^3}+\frac{605 d_1}{108 (x-1)^3}+\frac{\pi ^2}{36 \
(x-1)^3}-\frac{1643}{108 (x-1)^3}+\frac{63 d_1^2}{16 \
(x-1)^4}-\frac{50 d_1}{3 (x-1)^4}-\frac{\pi ^2}{8 \
(x-1)^4}+\frac{8}{(x-1)^4}+\frac{2035 d_1^2}{432 (x-1)^5}-\frac{895 \
d_1}{27 (x-1)^5}-\frac{25 \pi ^2}{72 (x-1)^5}+\frac{5665}{108 \
(x-1)^5}-\frac{25 \pi ^2}{72}+\frac{1855}{108}\Big)+\Big(-\frac{2 \pi \
^2 d_1^2}{3 (x-1)^5}+\frac{8 \pi ^2 d_1}{3 (x-1)^5}+\Big(\frac{4 \
d_1^2}{x-1}-\frac{2 d_1^2}{(x-1)^2}+\frac{4 d_1^2}{3 \
(x-1)^3}-\frac{d_1^2}{(x-1)^4}+\frac{25 d_1^2}{3 (x-1)^5}+\frac{8 \
d_1}{x-2}-\frac{16 d_1}{3 (x-2)^2}-\frac{8 d_1}{3 (x-1)^2}+\frac{16 \
d_1}{3 (x-2)^3}-\frac{8 d_1}{(x-1)^4}+\frac{4}{x-2}-\frac{7}{2 \
(x-1)}-\frac{8}{3 (x-2)^2}+\frac{8}{3 (x-2)^3}-\frac{1}{3 \
(x-1)^3}-\frac{5}{(x-1)^4}-\frac{25}{3 (x-1)^5}-\frac{25}{6}\Big) \
H(0;\ao)+\Big(-\frac{16 d_1}{(x-1)^5}+8 d_1+\frac{16}{(x-1)^5}\Big) \
H(0,0;\ao)+\Big(-\frac{8 d_1^2}{(x-1)^5}+4 d_1^2+\frac{8 \
d_1}{(x-1)^5}\Big) H(0,1;\ao)-\frac{2 \pi ^2}{(x-1)^5}-\frac{2 \pi \
^2}{3}\Big) H(1,1;x)+\Big(\frac{\pi ^2 d_1}{(x-1)^5}-\frac{\pi ^2}{2 \
(x-1)^5}-\frac{3 \pi ^2}{2}\Big) H(1,2;x)+\Big(-\frac{4 \
d_1^2}{x-1}+\frac{d_1^2}{(x-1)^2}-\frac{4 d_1^2}{9 \
(x-1)^3}+\frac{d_1^2}{4 (x-1)^4}-\frac{205 d_1^2}{36 \
(x-1)^5}-\frac{46 d_1}{3 (x-2)}+\frac{46 d_1}{3 (x-1)}+\frac{52 \
d_1}{9 (x-2)^2}-\frac{40 d_1}{9 (x-1)^2}-\frac{28 d_1}{9 \
(x-2)^3}+\frac{98 d_1}{9 (x-1)^3}-\frac{13 d_1}{6 (x-1)^4}+\frac{415 \
d_1}{18 (x-1)^5}+\Big(-\frac{8 d_1}{x-1}+\frac{4 \
d_1}{(x-1)^2}-\frac{8 d_1}{3 (x-1)^3}+\frac{2 d_1}{(x-1)^4}-\frac{50 \
d_1}{3 (x-1)^5}-\frac{16}{x-2}+\frac{32}{3 (x-2)^2}+\frac{16}{3 \
(x-1)^2}-\frac{32}{3 (x-2)^3}+\frac{16}{(x-1)^4}\Big) H(0;\ao)+\Big(-\
\frac{4 d_1^2}{x-1}+\frac{2 d_1^2}{(x-1)^2}-\frac{4 d_1^2}{3 \
(x-1)^3}+\frac{d_1^2}{(x-1)^4}-\frac{25 d_1^2}{3 (x-1)^5}-\frac{8 \
d_1}{x-2}+\frac{16 d_1}{3 (x-2)^2}+\frac{8 d_1}{3 (x-1)^2}-\frac{16 \
d_1}{3 (x-2)^3}+\frac{8 d_1}{(x-1)^4}\Big) \
H(1;\ao)+\Big(\frac{16}{(x-1)^5}-16\Big) H(0,0;\ao)+\Big(\frac{8 \
d_1}{(x-1)^5}-8 d_1\Big) H(0,1;\ao)+\Big(\frac{8 d_1}{(x-1)^5}-8 \
d_1\Big) H(1,0;\ao)+\Big(\frac{4 d_1^2}{(x-1)^5}-4 d_1^2\Big) \
H(1,1;\ao)+\frac{80}{3 (x-2)}+\frac{43}{6 (x-1)}-\frac{56}{9 \
(x-2)^2}-\frac{299}{18 (x-1)^2}+\frac{56}{9 (x-2)^3}-\frac{23}{6 \
(x-1)^3}-\frac{203}{6 (x-1)^4}-\frac{\pi ^2}{6 (x-1)^5}-\frac{35}{6 \
(x-1)^5}+\frac{\pi ^2}{6}+\frac{35}{6}\Big) H(1,c_1(\ao);x)+\Big(2 \
\pi ^2-\frac{2 \pi ^2}{(x-1)^5}\Big) H(2,0;x)+\Big(\Big(-\frac{8 \
d_1}{x-2}+\frac{16 d_1}{3 (x-2)^2}+\frac{8 d_1}{3 (x-1)^2}-\frac{16 \
d_1}{3 (x-2)^3}+\frac{8 d_1}{(x-1)^4}+\frac{16}{x-2}-\frac{15}{2 \
(x-1)}-\frac{32}{3 (x-2)^2}-\frac{1}{3 (x-1)^2}+\frac{32}{3 (x-2)^3}-\
\frac{5}{(x-1)^3}-\frac{17}{2 (x-1)^4}-\frac{25}{2 \
(x-1)^5}+\frac{25}{2}\Big) H(0;\ao)+\Big(\frac{8}{(x-1)^5}-8\Big) \
H(0,0;\ao)+\Big(\frac{4 d_1}{(x-1)^5}-4 d_1\Big) H(0,1;\ao)+\frac{\pi \
^2}{3 (x-1)^5}-\frac{\pi ^2}{3}\Big) H(2,1;x)+\Big(-\frac{\pi ^2 \
d_1}{(x-1)^5}+\pi ^2 d_1+\frac{2 \pi ^2}{(x-1)^5}-2 \pi ^2\Big) \
H(2,2;x)+\Big(\frac{d_1 \ao^4}{2}-\frac{d_1 \ao^4}{2 \
(x-1)}+\frac{\ao^4}{x-1}-\ao^4-\frac{26 d_1 \ao^3}{9}+\frac{2 d_1 \
\ao^3}{x-1}-\frac{41 \ao^3}{6 (x-1)}-\frac{8 d_1 \ao^3}{9 \
(x-1)^2}+\frac{25 \ao^3}{9 (x-1)^2}+\frac{107 \ao^3}{18}+\frac{23 d_1 \
\ao^2}{3}+\frac{5 \ao^2}{3 (x-2)}-\frac{3 d_1 \ao^2}{x-1}+\frac{221 \
\ao^2}{12 (x-1)}+\frac{8 d_1 \ao^2}{3 (x-1)^2}-\frac{27 \ao^2}{2 \
(x-1)^2}-\frac{2 d_1 \ao^2}{(x-1)^3}+\frac{15 \ao^2}{2 \
(x-1)^3}-\frac{197 \ao^2}{12}-\frac{50 d_1 \ao}{3}-\frac{10 \
\ao}{x-2}+\frac{2 d_1 \ao}{x-1}-\frac{181 \ao}{6 (x-1)}+\frac{20 \
\ao}{3 (x-2)^2}-\frac{8 d_1 \ao}{3 (x-1)^2}+\frac{80 \ao}{3 (x-1)^2}+\
\frac{4 d_1 \ao}{(x-1)^3}-\frac{29 \ao}{(x-1)^3}-\frac{8 d_1 \
\ao}{(x-1)^4}+\frac{29 \ao}{(x-1)^4}+\frac{223 \ao}{6}+\frac{205 \
d_1}{18}+\Big(-\frac{4 \ao^4}{x-1}+4 \ao^4+\frac{16 \
\ao^3}{x-1}-\frac{16 \ao^3}{3 (x-1)^2}-\frac{64 \ao^3}{3}-\frac{24 \
\ao^2}{x-1}+\frac{16 \ao^2}{(x-1)^2}-\frac{8 \ao^2}{(x-1)^3}+48 \
\ao^2+\frac{16 \ao}{x-1}-\frac{16 \ao}{(x-1)^2}+\frac{16 \
\ao}{(x-1)^3}-\frac{16 \ao}{(x-1)^4}-64 \ao+\frac{12}{x-1}-\frac{8}{3 \
(x-1)^2}-\frac{8}{3 (x-1)^3}+\frac{12}{(x-1)^4}+\frac{100}{3 \
(x-1)^5}+\frac{100}{3}\Big) H(0;\ao)+\Big(2 d_1 \ao^4-\frac{2 d_1 \
\ao^4}{x-1}-\frac{32 d_1 \ao^3}{3}+\frac{8 d_1 \ao^3}{x-1}-\frac{8 \
d_1 \ao^3}{3 (x-1)^2}+24 d_1 \ao^2-\frac{12 d_1 \ao^2}{x-1}+\frac{8 \
d_1 \ao^2}{(x-1)^2}-\frac{4 d_1 \ao^2}{(x-1)^3}-32 d_1 \ao+\frac{8 \
d_1 \ao}{x-1}-\frac{8 d_1 \ao}{(x-1)^2}+\frac{8 d_1 \
\ao}{(x-1)^3}-\frac{8 d_1 \ao}{(x-1)^4}+\frac{50 d_1}{3}+\frac{6 \
d_1}{x-1}-\frac{4 d_1}{3 (x-1)^2}-\frac{4 d_1}{3 (x-1)^3}+\frac{6 \
d_1}{(x-1)^4}+\frac{50 d_1}{3 (x-1)^5}\Big) H(1;\ao)-\frac{16 \
H(0,0;\ao)}{(x-1)^5}-\frac{8 d_1 H(0,1;\ao)}{(x-1)^5}-\frac{8 d_1 \
H(1,0;\ao)}{(x-1)^5}-\frac{4 d_1^2 \
H(1,1;\ao)}{(x-1)^5}+\frac{20}{x-2}+\frac{15 d_1}{2 \
(x-1)}-\frac{49}{4 (x-1)}-\frac{40}{3 (x-2)^2}-\frac{10 d_1}{9 \
(x-1)^2}+\frac{5}{36 (x-1)^2}-\frac{10 d_1}{9 (x-1)^3}+\frac{275}{36 \
(x-1)^3}+\frac{15 d_1}{2 (x-1)^4}-\frac{6}{(x-1)^4}+\frac{205 d_1}{18 \
(x-1)^5}+\frac{\pi ^2}{6 (x-1)^5}-\frac{725}{18 \
(x-1)^5}-\frac{925}{36}\Big) H(c_1(\ao),c_1(\ao);x)+\Big(\frac{d_1 \
\ao^4}{4}-\frac{d_1 \ao^4}{4 (x-1)}+\frac{\ao^4}{2 \
(x-1)}-\frac{\ao^4}{2}-\frac{13 d_1 \ao^3}{9}+\frac{d_1 \
\ao^3}{x-1}-\frac{3 \ao^3}{x-1}-\frac{4 d_1 \ao^3}{9 \
(x-1)^2}+\frac{25 \ao^3}{18 (x-1)^2}+\frac{61 \ao^3}{18}+\frac{23 d_1 \
\ao^2}{6}-\frac{3 d_1 \ao^2}{2 (x-1)}+\frac{31 \ao^2}{4 \
(x-1)}+\frac{4 d_1 \ao^2}{3 (x-1)^2}-\frac{71 \ao^2}{12 \
(x-1)^2}-\frac{d_1 \ao^2}{(x-1)^3}+\frac{15 \ao^2}{4 \
(x-1)^3}-\frac{131 \ao^2}{12}-\frac{25 d_1 \ao}{3}+\frac{d_1 \
\ao}{x-1}-\frac{13 \ao}{x-1}-\frac{4 d_1 \ao}{3 (x-1)^2}+\frac{35 \
\ao}{3 (x-1)^2}+\frac{2 d_1 \ao}{(x-1)^3}-\frac{12 \
\ao}{(x-1)^3}-\frac{4 d_1 \ao}{(x-1)^4}+\frac{29 \ao}{2 \
(x-1)^4}+\frac{169 \ao}{6}+\frac{205 d_1}{36}+\Big(-\frac{2 \
\ao^4}{x-1}+2 \ao^4+\frac{8 \ao^3}{x-1}-\frac{8 \ao^3}{3 \
(x-1)^2}-\frac{32 \ao^3}{3}-\frac{12 \ao^2}{x-1}+\frac{8 \
\ao^2}{(x-1)^2}-\frac{4 \ao^2}{(x-1)^3}+24 \ao^2+\frac{8 \
\ao}{x-1}-\frac{8 \ao}{(x-1)^2}+\frac{8 \ao}{(x-1)^3}-\frac{8 \
\ao}{(x-1)^4}-32 \ao+\frac{16}{x-2}-\frac{2}{x-1}-\frac{32}{3 \
(x-2)^2}-\frac{8}{3 (x-1)^2}+\frac{32}{3 \
(x-2)^3}-\frac{4}{(x-1)^3}-\frac{8}{(x-1)^4}+\frac{50}{3}\Big) \
H(0;\ao)+\Big(d_1 \ao^4-\frac{d_1 \ao^4}{x-1}-\frac{16 d_1 \ao^3}{3}+\
\frac{4 d_1 \ao^3}{x-1}-\frac{4 d_1 \ao^3}{3 (x-1)^2}+12 d_1 \
\ao^2-\frac{6 d_1 \ao^2}{x-1}+\frac{4 d_1 \ao^2}{(x-1)^2}-\frac{2 d_1 \
\ao^2}{(x-1)^3}-16 d_1 \ao+\frac{4 d_1 \ao}{x-1}-\frac{4 d_1 \
\ao}{(x-1)^2}+\frac{4 d_1 \ao}{(x-1)^3}-\frac{4 d_1 \
\ao}{(x-1)^4}+\frac{25 d_1}{3}+\frac{8 \
d_1}{x-2}-\frac{d_1}{x-1}-\frac{16 d_1}{3 (x-2)^2}-\frac{4 d_1}{3 \
(x-1)^2}+\frac{16 d_1}{3 (x-2)^3}-\frac{2 d_1}{(x-1)^3}-\frac{4 \
d_1}{(x-1)^4}\Big) H(1;\ao)+\frac{32 d_1}{3 (x-2)}-\frac{82}{3 \
(x-2)}-\frac{35 d_1}{12 (x-1)}+\frac{73}{12 (x-1)}-\frac{28 d_1}{9 \
(x-2)^2}+\frac{56}{9 (x-2)^2}-\frac{28 d_1}{9 (x-1)^2}+\frac{323}{36 \
(x-1)^2}+\frac{28 d_1}{9 (x-2)^3}-\frac{56}{9 (x-2)^3}-\frac{5 \
d_1}{(x-1)^3}+\frac{53}{4 (x-1)^3}-\frac{4 d_1}{(x-1)^4}+\frac{29}{2 \
(x-1)^4}-\frac{725}{36}\Big) \
H(c_2(\ao),c_1(\ao);x)+\Big(\frac{100}{3}+\frac{12}{x-1}-\frac{8}{3 \
(x-1)^2}-\frac{8}{3 (x-1)^3}+\frac{12}{(x-1)^4}+\frac{100}{3 (x-1)^5}\
\Big) H(0,0,0;\ao)+\Big(-\frac{100}{3}-\frac{12}{x-1}+\frac{8}{3 \
(x-1)^2}+\frac{8}{3 (x-1)^3}-\frac{12}{(x-1)^4}-\frac{100}{3 (x-1)^5}\
\Big) H(0,0,0;x)+\Big(\frac{6 d_1}{x-1}-\frac{4 d_1}{3 \
(x-1)^2}-\frac{4 d_1}{3 (x-1)^3}+\frac{6 d_1}{(x-1)^4}+\frac{50 \
d_1}{3 (x-1)^5}+\frac{50 d_1}{3}\Big) H(0,0,1;\ao)+\Big(-\frac{4 \
d_1}{(x-1)^5}+4 d_1+\frac{12}{(x-1)^5}-12\Big) H(0;\ao) \
H(0,0,1;x)+\Big(-\frac{\ao^4}{x-1}+\ao^4+\frac{4 \ao^3}{x-1}-\frac{4 \
\ao^3}{3 (x-1)^2}-\frac{16 \ao^3}{3}-\frac{6 \ao^2}{x-1}+\frac{4 \
\ao^2}{(x-1)^2}-\frac{2 \ao^2}{(x-1)^3}+12 \ao^2+\frac{4 \
\ao}{x-1}-\frac{4 \ao}{(x-1)^2}+\frac{4 \ao}{(x-1)^3}-\frac{4 \
\ao}{(x-1)^4}-16 \ao+\Big(\frac{8}{(x-1)^5}-8\Big) \
H(0;\ao)+\Big(\frac{4 d_1}{(x-1)^5}-4 d_1\Big) \
H(1;\ao)-\frac{8}{x-2}+\frac{4}{x-1}+\frac{16}{3 (x-2)^2}+\frac{2}{3 \
(x-1)^2}-\frac{16}{3 (x-2)^3}+\frac{4}{3 \
(x-1)^3}+\frac{7}{(x-1)^4}+\frac{25}{3 (x-1)^5}\Big) \
H(0,0,c_1(\ao);x)+\Big(\frac{6 d_1}{x-1}-\frac{4 d_1}{3 \
(x-1)^2}-\frac{4 d_1}{3 (x-1)^3}+\frac{6 d_1}{(x-1)^4}+\frac{50 \
d_1}{3 (x-1)^5}+\frac{50 d_1}{3}\Big) H(0,1,0;\ao)+\Big(-\frac{8 \
d_1}{x-2}+\frac{16 d_1}{3 (x-2)^2}+\frac{8 d_1}{3 (x-1)^2}-\frac{16 \
d_1}{3 (x-2)^3}+\frac{8 \
d_1}{(x-1)^4}+\frac{8}{x-2}-\frac{5}{x-1}-\frac{16}{3 \
(x-2)^2}+\frac{2}{3 (x-1)^2}+\frac{16}{3 (x-2)^3}-\frac{10}{3 \
(x-1)^3}-\frac{3}{(x-1)^4}-\frac{25}{3 (x-1)^5}+\frac{25}{3}\Big) \
H(0,1,0;x)+\Big(\frac{3 d_1^2}{x-1}-\frac{2 d_1^2}{3 (x-1)^2}-\frac{2 \
d_1^2}{3 (x-1)^3}+\frac{3 d_1^2}{(x-1)^4}+\frac{25 d_1^2}{3 (x-1)^5}+\
\frac{25 d_1^2}{3}\Big) H(0,1,1;\ao)+\Big(\frac{4 d_1^2}{(x-1)^5}-4 \
d_1^2-\frac{12 d_1}{(x-1)^5}+4 d_1+\frac{16}{(x-1)^5}\Big) H(0;\ao) \
H(0,1,1;x)+\Big(\frac{8 d_1}{x-2}-\frac{16 d_1}{3 (x-2)^2}-\frac{8 \
d_1}{3 (x-1)^2}+\frac{16 d_1}{3 (x-2)^3}-\frac{8 d_1}{(x-1)^4}+\Big(-\
\frac{8 d_1}{(x-1)^5}+8 d_1+\frac{8}{(x-1)^5}-8\Big) \
H(0;\ao)+\Big(-\frac{4 d_1^2}{(x-1)^5}+4 d_1^2+\frac{4 \
d_1}{(x-1)^5}-4 d_1\Big) \
H(1;\ao)-\frac{8}{x-2}+\frac{5}{x-1}+\frac{16}{3 (x-2)^2}-\frac{2}{3 \
(x-1)^2}-\frac{16}{3 (x-2)^3}+\frac{10}{3 (x-1)^3}+\frac{3}{(x-1)^4}+\
\frac{25}{3 (x-1)^5}-\frac{25}{3}\Big) \
H(0,1,c_1(\ao);x)+\Big(-\frac{4 d_1}{(x-1)^5}+4 \
d_1+\frac{4}{(x-1)^5}-4\Big) H(0;\ao) H(0,2,1;x)+\Big(\frac{2 \
\ao^4}{x-1}-2 \ao^4-\frac{8 \ao^3}{x-1}+\frac{8 \ao^3}{3 \
(x-1)^2}+\frac{32 \ao^3}{3}+\frac{12 \ao^2}{x-1}-\frac{8 \
\ao^2}{(x-1)^2}+\frac{4 \ao^2}{(x-1)^3}-24 \ao^2-\frac{8 \
\ao}{x-1}+\frac{8 \ao}{(x-1)^2}-\frac{8 \ao}{(x-1)^3}+\frac{8 \
\ao}{(x-1)^4}+32 \ao-16 H(0;\ao)-8 d_1 \
H(1;\ao)-\frac{20}{x-2}+\frac{11}{2 (x-1)}+\frac{40}{3 \
(x-2)^2}+\frac{8}{3 (x-1)^2}-\frac{40}{3 (x-2)^3}+\frac{13}{3 \
(x-1)^3}+\frac{13}{(x-1)^4}+\frac{25}{3 (x-1)^5}-\frac{25}{2}\Big) \
H(0,c_1(\ao),c_1(\ao);x)+\Big(\frac{\ao^4}{x-1}-\ao^4-\frac{4 \
\ao^3}{x-1}+\frac{4 \ao^3}{3 (x-1)^2}+\frac{16 \ao^3}{3}+\frac{6 \
\ao^2}{x-1}-\frac{4 \ao^2}{(x-1)^2}+\frac{2 \ao^2}{(x-1)^3}-12 \ao^2-\
\frac{4 \ao}{x-1}+\frac{4 \ao}{(x-1)^2}-\frac{4 \ao}{(x-1)^3}+\frac{4 \
\ao}{(x-1)^4}+16 \ao+\Big(\frac{8}{(x-1)^5}-8\Big) \
H(0;\ao)+\Big(\frac{4 d_1}{(x-1)^5}-4 d_1\Big) \
H(1;\ao)+\frac{16}{x-2}-\frac{13}{2 (x-1)}-\frac{32}{3 \
(x-2)^2}-\frac{5}{3 (x-1)^2}+\frac{32}{3 \
(x-2)^3}-\frac{3}{(x-1)^3}-\frac{25}{2 (x-1)^4}-\frac{25}{2 (x-1)^5}+\
\frac{25}{6}\Big) H(0,c_2(\ao),c_1(\ao);x)+\Big(\frac{8 \
d_1}{x-1}-\frac{4 d_1}{(x-1)^2}+\frac{8 d_1}{3 (x-1)^3}-\frac{2 \
d_1}{(x-1)^4}+\frac{50 d_1}{3 (x-1)^5}+\frac{16}{x-2}-\frac{32}{3 \
(x-2)^2}-\frac{16}{3 (x-1)^2}+\frac{32}{3 (x-2)^3}-\frac{16}{(x-1)^4}\
\Big) H(1,0,0;x)+\Big(\frac{4 d_1^2}{(x-1)^5}-\frac{12 d_1}{(x-1)^5}+\
\frac{12}{(x-1)^5}-4\Big) H(0;\ao) \
H(1,0,1;x)+\Big(\Big(\frac{16}{(x-1)^5}-\frac{8 d_1}{(x-1)^5}\Big) \
H(0;\ao)+\Big(\frac{8 d_1}{(x-1)^5}-\frac{4 d_1^2}{(x-1)^5}\Big) H(1;\
\ao)+\frac{4}{x-2}-\frac{7}{2 (x-1)}-\frac{8}{3 (x-2)^2}+\frac{8}{3 \
(x-2)^3}-\frac{1}{3 (x-1)^3}-\frac{5}{(x-1)^4}-\frac{25}{3 \
(x-1)^5}-\frac{25}{6}\Big) H(1,0,c_1(\ao);x)+\Big(-\frac{4 \
d_1^2}{x-1}+\frac{2 d_1^2}{(x-1)^2}-\frac{4 d_1^2}{3 \
(x-1)^3}+\frac{d_1^2}{(x-1)^4}-\frac{25 d_1^2}{3 (x-1)^5}-\frac{8 \
d_1}{x-2}+\frac{16 d_1}{3 (x-2)^2}+\frac{8 d_1}{3 (x-1)^2}-\frac{16 \
d_1}{3 (x-2)^3}+\frac{8 d_1}{(x-1)^4}-\frac{4}{x-2}+\frac{7}{2 \
(x-1)}+\frac{8}{3 (x-2)^2}-\frac{8}{3 (x-2)^3}+\frac{1}{3 \
(x-1)^3}+\frac{5}{(x-1)^4}+\frac{25}{3 (x-1)^5}+\frac{25}{6}\Big) \
H(1,1,0;x)+\Big(\frac{12 d_1^2}{(x-1)^5}-4 d_1^2-\frac{24 \
d_1}{(x-1)^5}+\frac{12}{(x-1)^5}+4\Big) H(0;\ao) \
H(1,1,1;x)+\Big(\frac{4 d_1^2}{x-1}-\frac{2 d_1^2}{(x-1)^2}+\frac{4 \
d_1^2}{3 (x-1)^3}-\frac{d_1^2}{(x-1)^4}+\frac{25 d_1^2}{3 \
(x-1)^5}+\frac{8 d_1}{x-2}-\frac{16 d_1}{3 (x-2)^2}-\frac{8 d_1}{3 \
(x-1)^2}+\frac{16 d_1}{3 (x-2)^3}-\frac{8 \
d_1}{(x-1)^4}+\Big(-\frac{16 d_1}{(x-1)^5}+8 \
d_1+\frac{16}{(x-1)^5}\Big) H(0;\ao)+\Big(-\frac{8 d_1^2}{(x-1)^5}+4 \
d_1^2+\frac{8 d_1}{(x-1)^5}\Big) H(1;\ao)+\frac{4}{x-2}-\frac{7}{2 \
(x-1)}-\frac{8}{3 (x-2)^2}+\frac{8}{3 (x-2)^3}-\frac{1}{3 \
(x-1)^3}-\frac{5}{(x-1)^4}-\frac{25}{3 (x-1)^5}-\frac{25}{6}\Big) \
H(1,1,c_1(\ao);x)+\Big(-\frac{4 \
d_1}{(x-1)^5}+\frac{2}{(x-1)^5}+6\Big) H(0;\ao) \
H(1,2,1;x)+\Big(-\frac{8 d_1}{x-1}+\frac{4 d_1}{(x-1)^2}-\frac{8 \
d_1}{3 (x-1)^3}+\frac{2 d_1}{(x-1)^4}-\frac{50 d_1}{3 \
(x-1)^5}+\Big(\frac{8 d_1}{(x-1)^5}-16\Big) H(0;\ao)+\Big(\frac{4 \
d_1^2}{(x-1)^5}-8 d_1\Big) H(1;\ao)-\frac{20}{x-2}+\frac{7}{2 (x-1)}+\
\frac{40}{3 (x-2)^2}+\frac{16}{3 (x-1)^2}-\frac{40}{3 \
(x-2)^3}+\frac{1}{3 (x-1)^3}+\frac{21}{(x-1)^4}+\frac{25}{3 (x-1)^5}+\
\frac{25}{6}\Big) H(1,c_1(\ao),c_1(\ao);x)+\Big(-\frac{4 \
d_1}{(x-1)^5}+4 d_1+\frac{4}{(x-1)^5}-4\Big) H(0;\ao) \
H(2,0,1;x)+\Big(-\frac{8 d_1}{x-2}+\frac{16 d_1}{3 (x-2)^2}+\frac{8 \
d_1}{3 (x-1)^2}-\frac{16 d_1}{3 (x-2)^3}+\frac{8 \
d_1}{(x-1)^4}+\Big(\frac{8}{(x-1)^5}-8\Big) H(0;\ao)+\Big(\frac{4 \
d_1}{(x-1)^5}-4 d_1\Big) H(1;\ao)+\frac{16}{x-2}-\frac{15}{2 \
(x-1)}-\frac{32}{3 (x-2)^2}-\frac{1}{3 (x-1)^2}+\frac{32}{3 (x-2)^3}-\
\frac{5}{(x-1)^3}-\frac{17}{2 (x-1)^4}-\frac{25}{2 \
(x-1)^5}+\frac{25}{2}\Big) H(2,0,c_1(\ao);x)+\Big(\frac{8 \
d_1}{x-2}-\frac{16 d_1}{3 (x-2)^2}-\frac{8 d_1}{3 (x-1)^2}+\frac{16 \
d_1}{3 (x-2)^3}-\frac{8 d_1}{(x-1)^4}-\frac{16}{x-2}+\frac{15}{2 \
(x-1)}+\frac{32}{3 (x-2)^2}+\frac{1}{3 (x-1)^2}-\frac{32}{3 (x-2)^3}+\
\frac{5}{(x-1)^3}+\frac{17}{2 (x-1)^4}+\frac{25}{2 \
(x-1)^5}-\frac{25}{2}\Big) H(2,1,0;x)+\Big(-\frac{4 d_1}{(x-1)^5}+4 \
d_1-\frac{2}{(x-1)^5}+2\Big) H(0;\ao) H(2,1,1;x)+\Big(-\frac{8 \
d_1}{x-2}+\frac{16 d_1}{3 (x-2)^2}+\frac{8 d_1}{3 (x-1)^2}-\frac{16 \
d_1}{3 (x-2)^3}+\frac{8 d_1}{(x-1)^4}+\Big(\frac{8}{(x-1)^5}-8\Big) \
H(0;\ao)+\Big(\frac{4 d_1}{(x-1)^5}-4 d_1\Big) \
H(1;\ao)+\frac{16}{x-2}-\frac{15}{2 (x-1)}-\frac{32}{3 \
(x-2)^2}-\frac{1}{3 (x-1)^2}+\frac{32}{3 \
(x-2)^3}-\frac{5}{(x-1)^3}-\frac{17}{2 (x-1)^4}-\frac{25}{2 (x-1)^5}+\
\frac{25}{2}\Big) H(2,1,c_1(\ao);x)+\Big(\frac{4 d_1}{(x-1)^5}-4 d_1-\
\frac{8}{(x-1)^5}+8\Big) H(0;\ao) H(2,2,1;x)+\Big(\frac{8 \
d_1}{x-2}-\frac{16 d_1}{3 (x-2)^2}-\frac{8 d_1}{3 (x-1)^2}+\frac{16 \
d_1}{3 (x-2)^3}-\frac{8 d_1}{(x-1)^4}+\Big(8-\frac{8}{(x-1)^5}\Big) \
H(0;\ao)+\Big(4 d_1-\frac{4 d_1}{(x-1)^5}\Big) \
H(1;\ao)-\frac{16}{x-2}+\frac{15}{2 (x-1)}+\frac{32}{3 \
(x-2)^2}+\frac{1}{3 (x-1)^2}-\frac{32}{3 \
(x-2)^3}+\frac{5}{(x-1)^3}+\frac{17}{2 (x-1)^4}+\frac{25}{2 (x-1)^5}-\
\frac{25}{2}\Big) \
H(2,c_2(\ao),c_1(\ao);x)+\Big(\frac{\ao^4}{x-1}-\ao^4-\frac{4 \
\ao^3}{x-1}+\frac{4 \ao^3}{3 (x-1)^2}+\frac{16 \ao^3}{3}+\frac{6 \
\ao^2}{x-1}-\frac{4 \ao^2}{(x-1)^2}+\frac{2 \ao^2}{(x-1)^3}-12 \ao^2-\
\frac{4 \ao}{x-1}+\frac{4 \ao}{(x-1)^2}-\frac{4 \ao}{(x-1)^3}+\frac{4 \
\ao}{(x-1)^4}+16 \ao+\frac{8 H(0;\ao)}{(x-1)^5}+\frac{4 d_1 \
H(1;\ao)}{(x-1)^5}-\frac{3}{x-1}+\frac{2}{3 (x-1)^2}+\frac{2}{3 \
(x-1)^3}-\frac{3}{(x-1)^4}-\frac{25}{3 (x-1)^5}-\frac{25}{3}\Big) \
H(c_1(\ao),0,c_1(\ao);x)+\Big(-\frac{3 \ao^4}{x-1}+3 \ao^4+\frac{12 \
\ao^3}{x-1}-\frac{4 \ao^3}{(x-1)^2}-16 \ao^3-\frac{18 \
\ao^2}{x-1}+\frac{12 \ao^2}{(x-1)^2}-\frac{6 \ao^2}{(x-1)^3}+36 \
\ao^2+\frac{12 \ao}{x-1}-\frac{12 \ao}{(x-1)^2}+\frac{12 \
\ao}{(x-1)^3}-\frac{12 \ao}{(x-1)^4}-48 \ao-\frac{16 \
H(0;\ao)}{(x-1)^5}-\frac{8 d_1 \
H(1;\ao)}{(x-1)^5}+\frac{9}{x-1}-\frac{2}{(x-1)^2}-\frac{2}{(x-1)^3}+\
\frac{9}{(x-1)^4}+\frac{25}{(x-1)^5}+25\Big) H(c_1(\ao),c_1(\ao),c_1(\
\ao);x)+\Big(-\frac{3 \ao^4}{2 (x-1)}+\frac{3 \ao^4}{2}+\frac{6 \
\ao^3}{x-1}-\frac{2 \ao^3}{(x-1)^2}-8 \ao^3-\frac{9 \
\ao^2}{x-1}+\frac{6 \ao^2}{(x-1)^2}-\frac{3 \ao^2}{(x-1)^3}+18 \ao^2+\
\frac{6 \ao}{x-1}-\frac{6 \ao}{(x-1)^2}+\frac{6 \ao}{(x-1)^3}-\frac{6 \
\ao}{(x-1)^4}-24 \ao-\frac{8 H(0;\ao)}{(x-1)^5}-\frac{4 d_1 \
H(1;\ao)}{(x-1)^5}+\frac{9}{2 \
(x-1)}-\frac{1}{(x-1)^2}-\frac{1}{(x-1)^3}+\frac{9}{2 \
(x-1)^4}+\frac{25}{2 (x-1)^5}+\frac{25}{2}\Big) \
H(c_1(\ao),c_2(\ao),c_1(\ao);x)+\Big(\frac{\ao^4}{x-1}-\ao^4-\frac{4 \
\ao^3}{x-1}+\frac{4 \ao^3}{3 (x-1)^2}+\frac{16 \ao^3}{3}+\frac{6 \
\ao^2}{x-1}-\frac{4 \ao^2}{(x-1)^2}+\frac{2 \ao^2}{(x-1)^3}-12 \ao^2-\
\frac{4 \ao}{x-1}+\frac{4 \ao}{(x-1)^2}-\frac{4 \ao}{(x-1)^3}+\frac{4 \
\ao}{(x-1)^4}+16 \ao-\frac{8}{x-2}+\frac{1}{x-1}+\frac{16}{3 \
(x-2)^2}+\frac{4}{3 (x-1)^2}-\frac{16}{3 \
(x-2)^3}+\frac{2}{(x-1)^3}+\frac{4}{(x-1)^4}-\frac{25}{3}\Big) H(c_2(\
\ao),0,c_1(\ao);x)+\Big(-\frac{5 \ao^4}{2 (x-1)}+\frac{5 \
\ao^4}{2}+\frac{10 \ao^3}{x-1}-\frac{10 \ao^3}{3 (x-1)^2}-\frac{40 \
\ao^3}{3}-\frac{15 \ao^2}{x-1}+\frac{10 \ao^2}{(x-1)^2}-\frac{5 \
\ao^2}{(x-1)^3}+30 \ao^2+\frac{10 \ao}{x-1}-\frac{10 \
\ao}{(x-1)^2}+\frac{10 \ao}{(x-1)^3}-\frac{10 \ao}{(x-1)^4}-40 \
\ao+\frac{20}{x-2}-\frac{5}{2 (x-1)}-\frac{40}{3 (x-2)^2}-\frac{10}{3 \
(x-1)^2}+\frac{40}{3 \
(x-2)^3}-\frac{5}{(x-1)^3}-\frac{10}{(x-1)^4}+\frac{125}{6}\Big) \
H(c_2(\ao),c_1(\ao),c_1(\ao);x)+64 \
H(0,0,0,0;x)+\Big(\frac{12}{(x-1)^5}-12\Big) \
H(0,0,0,c_1(\ao);x)+\Big(\frac{4 d_1}{(x-1)^5}-4 \
d_1-\frac{12}{(x-1)^5}+12\Big) H(0,0,1,0;x)+\Big(-\frac{4 \
d_1}{(x-1)^5}+4 d_1+\frac{12}{(x-1)^5}-12\Big) \
H(0,0,1,c_1(\ao);x)+\Big(-8-\frac{8}{(x-1)^5}\Big) \
H(0,0,c_1(\ao),c_1(\ao);x)+\Big(\frac{4}{(x-1)^5}-4\Big) \
H(0,0,c_2(\ao),c_1(\ao);x)+\Big(\frac{8 d_1}{(x-1)^5}-8 \
d_1-\frac{8}{(x-1)^5}+8\Big) \
H(0,1,0,0;x)+\Big(\frac{16}{(x-1)^5}-\frac{8 d_1}{(x-1)^5}\Big) \
H(0,1,0,c_1(\ao);x)+\Big(-\frac{4 d_1^2}{(x-1)^5}+4 d_1^2+\frac{12 \
d_1}{(x-1)^5}-4 d_1-\frac{16}{(x-1)^5}\Big) H(0,1,1,0;x)+\Big(\frac{4 \
d_1^2}{(x-1)^5}-4 d_1^2-\frac{12 d_1}{(x-1)^5}+4 \
d_1+\frac{16}{(x-1)^5}\Big) H(0,1,1,c_1(\ao);x)+\Big(8 \
d_1-\frac{8}{(x-1)^5}-8\Big) H(0,1,c_1(\ao),c_1(\ao);x)+\Big(-\frac{4 \
d_1}{(x-1)^5}+4 d_1+\frac{4}{(x-1)^5}-4\Big) \
H(0,2,0,c_1(\ao);x)+\Big(\frac{4 d_1}{(x-1)^5}-4 \
d_1-\frac{4}{(x-1)^5}+4\Big) H(0,2,1,0;x)+\Big(-\frac{4 \
d_1}{(x-1)^5}+4 d_1+\frac{4}{(x-1)^5}-4\Big) \
H(0,2,1,c_1(\ao);x)+\Big(\frac{4 d_1}{(x-1)^5}-4 \
d_1-\frac{4}{(x-1)^5}+4\Big) \
H(0,2,c_2(\ao),c_1(\ao);x)+\Big(4+\frac{4}{(x-1)^5}\Big) \
H(0,c_1(\ao),0,c_1(\ao);x)+\Big(-12-\frac{4}{(x-1)^5}\Big) \
H(0,c_1(\ao),c_1(\ao),c_1(\ao);x)+\Big(-6-\frac{2}{(x-1)^5}\Big) \
H(0,c_1(\ao),c_2(\ao),c_1(\ao);x)+\Big(4-\frac{4}{(x-1)^5}\Big) \
H(0,c_2(\ao),0,c_1(\ao);x)+\Big(\frac{10}{(x-1)^5}-10\Big) \
H(0,c_2(\ao),c_1(\ao),c_1(\ao);x)+\Big(16-\frac{16}{(x-1)^5}\Big) \
H(1,0,0,0;x)+\Big(-\frac{4 d_1}{(x-1)^5}+\frac{12}{(x-1)^5}-4\Big) \
H(1,0,0,c_1(\ao);x)+\Big(-\frac{4 d_1^2}{(x-1)^5}+\frac{12 \
d_1}{(x-1)^5}-\frac{12}{(x-1)^5}+4\Big) H(1,0,1,0;x)+\Big(\frac{4 \
d_1^2}{(x-1)^5}-\frac{12 d_1}{(x-1)^5}+\frac{12}{(x-1)^5}-4\Big) \
H(1,0,1,c_1(\ao);x)+\Big(\frac{4}{(x-1)^5}-4\Big) H(1,0,c_1(\ao),c_1(\
\ao);x)+\Big(-\frac{4 d_1}{(x-1)^5}+\frac{2}{(x-1)^5}+6\Big) \
H(1,0,c_2(\ao),c_1(\ao);x)+\Big(\frac{16 d_1}{(x-1)^5}-8 \
d_1-\frac{16}{(x-1)^5}\Big) H(1,1,0,0;x)+\Big(\frac{4 \
d_1^2}{(x-1)^5}-\frac{16 d_1}{(x-1)^5}+\frac{12}{(x-1)^5}+4\Big) \
H(1,1,0,c_1(\ao);x)+\Big(-\frac{12 d_1^2}{(x-1)^5}+4 d_1^2+\frac{24 \
d_1}{(x-1)^5}-\frac{12}{(x-1)^5}-4\Big) H(1,1,1,0;x)+\Big(\frac{12 \
d_1^2}{(x-1)^5}-4 d_1^2-\frac{24 \
d_1}{(x-1)^5}+\frac{12}{(x-1)^5}+4\Big) \
H(1,1,1,c_1(\ao);x)+\Big(-\frac{4 d_1^2}{(x-1)^5}+8 \
d_1+\frac{4}{(x-1)^5}-4\Big) H(1,1,c_1(\ao),c_1(\ao);x)+\Big(-\frac{4 \
d_1}{(x-1)^5}+\frac{2}{(x-1)^5}+6\Big) \
H(1,2,0,c_1(\ao);x)+\Big(\frac{4 \
d_1}{(x-1)^5}-\frac{2}{(x-1)^5}-6\Big) H(1,2,1,0;x)+\Big(-\frac{4 \
d_1}{(x-1)^5}+\frac{2}{(x-1)^5}+6\Big) \
H(1,2,1,c_1(\ao);x)+\Big(\frac{4 \
d_1}{(x-1)^5}-\frac{2}{(x-1)^5}-6\Big) \
H(1,2,c_2(\ao),c_1(\ao);x)+\Big(-\frac{4 \
d_1}{(x-1)^5}+\frac{4}{(x-1)^5}+4\Big) \
H(1,c_1(\ao),0,c_1(\ao);x)+\Big(\frac{8 \
d_1}{(x-1)^5}-\frac{4}{(x-1)^5}-12\Big) \
H(1,c_1(\ao),c_1(\ao),c_1(\ao);x)+\Big(\frac{4 \
d_1}{(x-1)^5}-\frac{2}{(x-1)^5}-6\Big) \
H(1,c_1(\ao),c_2(\ao),c_1(\ao);x)+\Big(\frac{4}{(x-1)^5}-4\Big) \
H(2,0,0,c_1(\ao);x)+\Big(\frac{4 d_1}{(x-1)^5}-4 \
d_1-\frac{4}{(x-1)^5}+4\Big) H(2,0,1,0;x)+\Big(-\frac{4 \
d_1}{(x-1)^5}+4 d_1+\frac{4}{(x-1)^5}-4\Big) \
H(2,0,1,c_1(\ao);x)+\Big(\frac{10}{(x-1)^5}-10\Big) \
H(2,0,c_1(\ao),c_1(\ao);x)+\Big(8-\frac{8}{(x-1)^5}\Big) \
H(2,0,c_2(\ao),c_1(\ao);x)+\Big(8-\frac{8}{(x-1)^5}\Big) \
H(2,1,0,0;x)+\Big(2-\frac{2}{(x-1)^5}\Big) \
H(2,1,0,c_1(\ao);x)+\Big(\frac{4 d_1}{(x-1)^5}-4 \
d_1+\frac{2}{(x-1)^5}-2\Big) H(2,1,1,0;x)+\Big(-\frac{4 \
d_1}{(x-1)^5}+4 d_1-\frac{2}{(x-1)^5}+2\Big) \
H(2,1,1,c_1(\ao);x)+\Big(\frac{10}{(x-1)^5}-10\Big) \
H(2,1,c_1(\ao),c_1(\ao);x)+\Big(\frac{4 d_1}{(x-1)^5}-4 \
d_1-\frac{8}{(x-1)^5}+8\Big) H(2,2,0,c_1(\ao);x)+\Big(-\frac{4 \
d_1}{(x-1)^5}+4 d_1+\frac{8}{(x-1)^5}-8\Big) \
H(2,2,1,0;x)+\Big(\frac{4 d_1}{(x-1)^5}-4 \
d_1-\frac{8}{(x-1)^5}+8\Big) H(2,2,1,c_1(\ao);x)+\Big(-\frac{4 \
d_1}{(x-1)^5}+4 d_1+\frac{8}{(x-1)^5}-8\Big) \
H(2,2,c_2(\ao),c_1(\ao);x)+\Big(\frac{4}{(x-1)^5}-4\Big) \
H(2,c_2(\ao),0,c_1(\ao);x)+\Big(10-\frac{10}{(x-1)^5}\Big) \
H(2,c_2(\ao),c_1(\ao),c_1(\ao);x)-\frac{4 \
H(c_1(\ao),0,0,c_1(\ao);x)}{(x-1)^5}+\frac{8 \
H(c_1(\ao),0,c_1(\ao),c_1(\ao);x)}{(x-1)^5}+\frac{4 H(c_1(\ao),0,c_2(\
\ao),c_1(\ao);x)}{(x-1)^5}+\frac{4 \
H(c_1(\ao),c_1(\ao),0,c_1(\ao);x)}{(x-1)^5}-\frac{12 \
H(c_1(\ao),c_1(\ao),c_1(\ao),c_1(\ao);x)}{(x-1)^5}-\frac{6 \
H(c_1(\ao),c_1(\ao),c_2(\ao),c_1(\ao);x)}{(x-1)^5}+\frac{4 \
H(c_1(\ao),c_2(\ao),0,c_1(\ao);x)}{(x-1)^5}-\frac{10 \
H(c_1(\ao),c_2(\ao),c_1(\ao),c_1(\ao);x)}{(x-1)^5}+H(2;x) \
\Big(\frac{2 \pi ^2 d_1}{x-2}-\frac{4 \pi ^2 d_1}{3 (x-2)^2}-\frac{2 \
\pi ^2 d_1}{3 (x-1)^2}+\frac{4 \pi ^2 d_1}{3 (x-2)^3}-\frac{2 \pi ^2 \
d_1}{(x-1)^4}-\frac{4 \pi ^2}{x-2}+\frac{15 \pi ^2}{8 (x-1)}+\frac{8 \
\pi ^2}{3 (x-2)^2}+\frac{\pi ^2}{12 (x-1)^2}-\frac{8 \pi ^2}{3 \
(x-2)^3}+\frac{5 \pi ^2}{4 (x-1)^3}+\frac{17 \pi ^2}{8 \
(x-1)^4}+\frac{25 \pi ^2}{8 (x-1)^5}-\frac{7 \zeta_3}{2 \
(x-1)^5}+\frac{7 \zeta_3}{2}+\frac{3 \pi ^2 \ln 2\, }{(x-1)^5}-3 \pi \
^2 \ln 2\, -\frac{25 \pi ^2}{8}\Big)+H(0;x) \Big(-\frac{63 d_1^2}{16 \
(x-1)}+\frac{19 d_1^2}{54 (x-1)^2}+\frac{19 d_1^2}{54 \
(x-1)^3}-\frac{63 d_1^2}{16 (x-1)^4}-\frac{2035 d_1^2}{432 \
(x-1)^5}-\frac{2035 d_1^2}{432}-\frac{38 d_1}{3 (x-2)}+\frac{161 \
d_1}{12 (x-1)}+\frac{76 d_1}{9 (x-2)^2}-\frac{323 d_1}{108 \
(x-1)^2}-\frac{605 d_1}{108 (x-1)^3}+\frac{50 d_1}{3 \
(x-1)^4}+\frac{895 d_1}{27 (x-1)^5}+\frac{2035 d_1}{108}+\frac{4 \pi \
^2}{x-2}+\frac{68}{3 (x-2)}-\frac{3 \pi ^2}{8 (x-1)}-\frac{3}{2 \
(x-1)}-\frac{8 \pi ^2}{3 (x-2)^2}-\frac{152}{9 (x-2)^2}-\frac{25 \pi \
^2}{36 (x-1)^2}+\frac{215}{108 (x-1)^2}+\frac{8 \pi ^2}{3 \
(x-2)^3}-\frac{37 \pi ^2}{36 (x-1)^3}+\frac{1643}{108 \
(x-1)^3}-\frac{15 \pi ^2}{8 (x-1)^4}-\frac{8}{(x-1)^4}+\frac{25 \pi \
^2}{72 (x-1)^5}-\frac{5665}{108 (x-1)^5}+\frac{21 \
\zeta_3}{(x-1)^5}+33 \zeta_3-\frac{2 \pi ^2 \ln 2\, }{(x-1)^5}+2 \pi \
^2 \ln 2\, +\frac{325 \pi ^2}{72}-\frac{1855}{108}\Big)+H(1;x) \Big(-\
\frac{21 \zeta_3 d_1}{2 (x-1)^5}+\frac{\pi ^2 \ln 2\,  d_1}{(x-1)^5}+\
\Big(-\frac{4 d_1^2}{x-1}+\frac{d_1^2}{(x-1)^2}-\frac{4 d_1^2}{9 \
(x-1)^3}+\frac{d_1^2}{4 (x-1)^4}-\frac{205 d_1^2}{36 \
(x-1)^5}-\frac{46 d_1}{3 (x-2)}+\frac{46 d_1}{3 (x-1)}+\frac{52 \
d_1}{9 (x-2)^2}-\frac{40 d_1}{9 (x-1)^2}-\frac{28 d_1}{9 \
(x-2)^3}+\frac{98 d_1}{9 (x-1)^3}-\frac{13 d_1}{6 (x-1)^4}+\frac{415 \
d_1}{18 (x-1)^5}+\frac{80}{3 (x-2)}+\frac{43}{6 (x-1)}-\frac{56}{9 \
(x-2)^2}-\frac{299}{18 (x-1)^2}+\frac{56}{9 (x-2)^3}-\frac{23}{6 \
(x-1)^3}-\frac{203}{6 (x-1)^4}-\frac{\pi ^2}{6 (x-1)^5}-\frac{35}{6 \
(x-1)^5}+\frac{\pi ^2}{6}+\frac{35}{6}\Big) H(0;\ao)+\Big(-\frac{8 \
d_1}{x-1}+\frac{4 d_1}{(x-1)^2}-\frac{8 d_1}{3 (x-1)^3}+\frac{2 \
d_1}{(x-1)^4}-\frac{50 d_1}{3 (x-1)^5}-\frac{16}{x-2}+\frac{32}{3 \
(x-2)^2}+\frac{16}{3 (x-1)^2}-\frac{32}{3 (x-2)^3}+\frac{16}{(x-1)^4}\
\Big) H(0,0;\ao)+\Big(-\frac{4 d_1^2}{x-1}+\frac{2 \
d_1^2}{(x-1)^2}-\frac{4 d_1^2}{3 \
(x-1)^3}+\frac{d_1^2}{(x-1)^4}-\frac{25 d_1^2}{3 (x-1)^5}-\frac{8 \
d_1}{x-2}+\frac{16 d_1}{3 (x-2)^2}+\frac{8 d_1}{3 (x-1)^2}-\frac{16 \
d_1}{3 (x-2)^3}+\frac{8 d_1}{(x-1)^4}\Big) \
H(0,1;\ao)+\Big(\frac{16}{(x-1)^5}-16\Big) H(0,0,0;\ao)+\Big(\frac{8 \
d_1}{(x-1)^5}-8 d_1\Big) H(0,0,1;\ao)+\Big(\frac{8 d_1}{(x-1)^5}-8 \
d_1\Big) H(0,1,0;\ao)+\Big(\frac{4 d_1^2}{(x-1)^5}-4 d_1^2\Big) \
H(0,1,1;\ao)-\frac{2 \pi ^2}{3 (x-2)}+\frac{7 \pi ^2}{12 \
(x-1)}+\frac{4 \pi ^2}{9 (x-2)^2}-\frac{4 \pi ^2}{9 \
(x-2)^3}+\frac{\pi ^2}{18 (x-1)^3}+\frac{5 \pi ^2}{6 \
(x-1)^4}+\frac{25 \pi ^2}{18 (x-1)^5}+\frac{21 \zeta_3}{4 \
(x-1)^5}+\frac{63 \zeta_3}{4}-\frac{\pi ^2 \ln 2\, }{2 \
(x-1)^5}-\frac{3}{2} \pi ^2 \ln 2\, +\frac{25 \pi \
^2}{36}\Big)+\frac{8 d_1 \pi ^2}{3 (x-2)}-\frac{37 \pi ^2}{6 \
(x-2)}-\frac{35 d_1 \pi ^2}{48 (x-1)}+\frac{31 \pi ^2}{48 \
(x-1)}-\frac{7 d_1 \pi ^2}{9 (x-2)^2}+\frac{10 \pi ^2}{9 \
(x-2)^2}-\frac{7 d_1 \pi ^2}{9 (x-1)^2}+\frac{107 \pi ^2}{48 \
(x-1)^2}+\frac{7 d_1 \pi ^2}{9 (x-2)^3}-\frac{14 \pi ^2}{9 \
(x-2)^3}-\frac{5 d_1 \pi ^2}{4 (x-1)^3}+\frac{55 \pi ^2}{16 (x-1)^3}-\
\frac{d_1 \pi ^2}{(x-1)^4}+\frac{115 \pi ^2}{24 (x-1)^4}-\frac{263 \
\pi ^4}{720 (x-1)^5}+\frac{35 \pi ^2}{36 (x-1)^5}+\frac{7 \
\zeta_3}{x-2}-\frac{203 \zeta_3}{16 (x-1)}-\frac{14 \zeta_3}{3 \
(x-2)^2}+\frac{35 \zeta_3}{24 (x-1)^2}+\frac{14 \zeta_3}{3 \
(x-2)^3}+\frac{7 \zeta_3}{8 (x-1)^3}-\frac{245 \zeta_3}{16 \
(x-1)^4}-\frac{525 \zeta_3}{16 (x-1)^5}-\frac{1225 \
\zeta_3}{48}+\frac{4 \text{Li}_4\Big(\frac{1}{2}\Big)}{(x-1)^5}-4 \text{Li}_4\frac{1}{2}+\frac{\ln ^42\, }{6 (x-1)^5}-\frac{\ln \
^42\, }{6}+\frac{4 \pi ^2 \ln ^22\, }{3 (x-1)^5}-\frac{4}{3} \pi ^2 \
\ln ^22\, -\frac{6 \pi ^2 \ln 2\, }{x-2}+\frac{15 \pi ^2 \ln 2\, }{8 \
(x-1)}+\frac{4 \pi ^2 \ln 2\, }{(x-2)^2}+\frac{3 \pi ^2 \ln 2\, }{4 \
(x-1)^2}-\frac{4 \pi ^2 \ln 2\, }{(x-2)^3}+\frac{5 \pi ^2 \ln 2\, }{4 \
(x-1)^3}+\frac{33 \pi ^2 \ln 2\, }{8 (x-1)^4}+\frac{25 \pi ^2 \ln 2\, \
}{8 (x-1)^5}-\frac{25}{8} \pi ^2 \ln 2\, -\frac{21 \pi \
^4}{80}+\frac{205 d_1 \pi ^2}{144}-\frac{265 \pi ^2}{48}.
\erp

%%%
%%% The J*I integrals
%%%

\section{The $\cJI$-type integrals}
\label{app:JIIntegrals}

%
% The J*I integral for k=0
%

\subsection{The $\cJI$ integral for $k=0$}
%
% This file contains the TeX output produced by Mathematica for the integral JI for arbitrary kap=0  and D0 = 3 +d'1 ep
%
The $\eps$ expansion for this integral reads
\beq
\cJI(Y,\ep;y_{0},d'_{0},\alpha_{0},d_{0};0)=\frac{1}{\eps^3}\cji_{-3}^{(0)}+\frac{1}{\eps^2}\cji_{-2}^{(0)}+\frac{1}{\eps}\cji_{-1}^{(0)}+\cji_0^{(0)}+\ocal\left(\eps\right),
\eeq
where
%% 1/ep^3
\brp
\cji_{-3}^{(0)}=\frac{1}{2},
\erp
%% 1/ep^2
\brp
\cji_{-2}^{(0)}=\frac{2 \yo^3}{3}-3 \yo^2+6 \yo-\frac{1}{2} H(0;Y)-2 H(0;\yo)+1,
\erp
%% 1/ep
\brp
\cji_{-1}^{(0)}=-\frac{1}{6} \yo^3 \ao^4+\frac{\yo^2 \ao^4}{2}-\frac{\yo \
\ao^4}{2}+\frac{8 \yo^3 \ao^3}{9}-3 \yo^2 \ao^3+\frac{10 \yo \
\ao^3}{3}-2 \yo^3 \ao^2+\frac{15 \yo^2 \ao^2}{2}-10 \yo \ao^2+\frac{8 \
\yo^3 \ao}{3}-11 \yo^2 \ao+18 \yo \ao-\frac{2 d_1' \yo^3}{9}+2 \yo^3+\
\frac{7 d_1' \yo^2}{6}-\frac{21 \yo^2}{2}-\frac{11 d_1' \yo}{3}+30 \
\yo+\Big(-\frac{2 \yo^3}{3}+3 \yo^2-6 \yo+\frac{2}{\yo-1}+2\Big) H(0;\
\ao)-\frac{2}{3} \yo^3 H(0;Y)+3 \yo^2 H(0;Y)-6 \yo \
H(0;Y)-H(0;Y)+\Big(-2 \yo^3+9 \yo^2-18 \yo+2 H(0;Y)-\frac{2}{\yo-1}-6\
\Big) H(0;\yo)+\Big(-\frac{2 d_1' \yo^3}{3}+3 d_1' \yo^2-6 d_1' \
\yo+\frac{11 d_1'}{3}-2 H(0;\ao)\Big) H(1;\yo)+\Big(-2 \
\ao+\frac{2}{\yo-1}+2\Big) H(c_1(\ao);\yo)+\frac{1}{2} H(0,0;Y)+8 \
H(0,0;\yo)+2 d_1' H(0,1;\yo)-2 H(0,c_1(\ao);\yo)+\frac{1}{2} \
H(1,0;Y)+2 H(1,0;\yo)-2 H(1,c_1(\ao);\yo)+\frac{\pi ^2}{12}+2,
\erp
%% ep^0
\brp
\cji_0^{(0)} =\frac{1}{12} d_1 \yo^3 \ao^4+\frac{1}{18} d_1' \yo^3 \ao^4-\frac{7 \
\yo^3 \ao^4}{12}-\frac{1}{4} d_1 \yo^2 \ao^4-\frac{1}{6} d_1' \yo^2 \
\ao^4+\frac{23 \yo^2 \ao^4}{12}+\frac{1}{4} d_1 \yo \ao^4+\frac{1}{6} \
d_1' \yo \ao^4-\frac{29 \yo \ao^4}{12}+\frac{1}{6} \yo^3 H(0;Y) \
\ao^4-\frac{1}{2} \yo^2 H(0;Y) \ao^4+\frac{1}{2} \yo H(0;Y) \
\ao^4-\frac{13}{27} d_1 \yo^3 \ao^3-\frac{8}{27} d_1' \yo^3 \
\ao^3+\frac{173 \yo^3 \ao^3}{54}+\frac{5}{3} d_1 \yo^2 \
\ao^3+\frac{19}{18} d_1' \yo^2 \ao^3-\frac{217 \yo^2 \
\ao^3}{18}-\frac{17}{9} d_1 \yo \ao^3-\frac{11}{9} d_1' \yo \
\ao^3+\frac{305 \yo \ao^3}{18}-\frac{8}{9} \yo^3 H(0;Y) \ao^3+3 \yo^2 \
H(0;Y) \ao^3-\frac{10}{3} \yo H(0;Y) \ao^3+\frac{23}{18} d_1 \yo^3 \
\ao^2+\frac{2}{3} d_1' \yo^3 \ao^2-\frac{275 \yo^3 \ao^2}{36}-5 d_1 \
\yo^2 \ao^2-\frac{11}{4} d_1' \yo^2 \ao^2+\frac{581 \yo^2 \ao^2}{18}+\
\frac{43}{6} d_1 \yo \ao^2+\frac{9}{2} d_1' \yo \ao^2-\frac{1003 \yo \
\ao^2}{18}+2 \yo^3 H(0;Y) \ao^2-\frac{15}{2} \yo^2 H(0;Y) \ao^2+10 \
\yo H(0;Y) \ao^2-\frac{25}{9} d_1 \yo^3 \ao-\frac{8}{9} d_1' \yo^3 \
\ao+\frac{217 \yo^3 \ao}{18}+12 d_1 \yo^2 \ao+\frac{25}{6} d_1' \yo^2 \
\ao-\frac{503 \yo^2 \ao}{9}-\frac{65 d_1 \yo \ao}{3}-\frac{29 d_1' \
\yo \ao}{3}+\frac{237 \yo \ao}{2}-\frac{8}{3} \yo^3 H(0;Y) \ao+11 \
\yo^2 H(0;Y) \ao-18 \yo H(0;Y) \ao+\frac{2 d_1'^2 \yo^3}{27}-\frac{8 \
d_1' \yo^3}{9}+\frac{14 \yo^3}{3}-\frac{17 d_1'^2 \yo^2}{36}+6 d_1' \
\yo^2-\frac{111 \yo^2}{4}+\frac{49 d_1'^2 \yo}{18}-\frac{65 d_1' \
\yo}{2}+114 \yo+\frac{2}{9} d_1' \yo^3 H(0;Y)-2 \yo^3 \
H(0;Y)-\frac{7}{6} d_1' \yo^2 H(0;Y)+\frac{21}{2} \yo^2 \
H(0;Y)+\frac{11}{3} d_1' \yo H(0;Y)-30 \yo H(0;Y)-\frac{1}{12} \pi ^2 \
H(0;Y)-2 H(0;Y)+H(0;\ao) \Big(\frac{\yo^3 \ao^4}{3}-\yo^2 \ao^4+\yo \
\ao^4-\frac{16 \yo^3 \ao^3}{9}+6 \yo^2 \ao^3-\frac{20 \yo \ao^3}{3}+4 \
\yo^3 \ao^2-15 \yo^2 \ao^2+20 \yo \ao^2-\frac{16 \yo^3 \ao}{3}+22 \
\yo^2 \ao-36 \yo \ao+\frac{2 d_1' \yo^3}{9}-\frac{11 \
\yo^3}{18}-\frac{7 d_1' \yo^2}{6}+\frac{31 \yo^2}{6}-4 d_1+2 \
d_1'+\frac{11 d_1' \yo}{3}-\frac{131 \yo}{6}+\frac{2}{3} \yo^3 \
H(0;Y)-3 \yo^2 H(0;Y)+6 \yo H(0;Y)-\frac{2 H(0;Y)}{\yo-1}-2 \
H(0;Y)-\frac{4 d_1}{\yo-1}+\frac{2 d_1'}{\yo-1}+\frac{61}{6 (\yo-1)}+\
\frac{61}{6}\Big)+\Big(\frac{1}{3} d_1 \yo^3 \ao^4-d_1 \yo^2 \
\ao^4+d_1 \yo \ao^4-\frac{16}{9} d_1 \yo^3 \ao^3+6 d_1 \yo^2 \
\ao^3-\frac{20}{3} d_1 \yo \ao^3+4 d_1 \yo^3 \ao^2-15 d_1 \yo^2 \
\ao^2+20 d_1 \yo \ao^2-\frac{16}{3} d_1 \yo^3 \ao+22 d_1 \yo^2 \ao-36 \
d_1 \yo \ao+\frac{25 d_1 \yo^3}{9}-12 d_1 \yo^2+\frac{65 d_1 \
\yo}{3}\Big) H(1;\ao)-\frac{1}{12} \pi ^2 H(1;Y)+\Big(\frac{\yo^3 \
\ao^4}{6}-\frac{\yo^2 \ao^4}{2}+\frac{\yo \
\ao^4}{2}-\frac{\ao^4}{6}-\frac{8 \yo^3 \ao^3}{9}+3 \yo^2 \
\ao^3-\frac{10 \yo \ao^3}{3}+\frac{11 \ao^3}{9}+2 \yo^3 \
\ao^2-\frac{15 \yo^2 \ao^2}{2}+10 \yo \ao^2-\frac{9 \ao^2}{2}-\frac{8 \
\yo^3 \ao}{3}+11 \yo^2 \ao+4 d_1 \ao-2 d_1' \ao-18 \yo \ao+2 H(0;Y) \
\ao-2 \ao+\frac{25 \yo^3}{18}-\frac{16 \yo^2}{3}-4 d_1+2 \
d_1'+\frac{49 \yo}{6}+\Big(4 \ao-\frac{4}{\yo-1}-4\Big) \
H(0;\ao)-\frac{2 H(0;Y)}{\yo-1}-2 H(0;Y)+\Big(4 \ao d_1-\frac{4 d_1}{\
\yo-1}-4 d_1\Big) H(1;\ao)-\frac{4 d_1}{\yo-1}+\frac{2 \
d_1'}{\yo-1}+\frac{61}{6 (\yo-1)}+\frac{61}{6}\Big) \
H(c_1(\ao);\yo)+\Big(\frac{4 \yo^3}{3}-6 \yo^2+12 \
\yo-\frac{4}{\yo-1}-4\Big) H(0,0;\ao)+\frac{2}{3} \yo^3 H(0,0;Y)-3 \
\yo^2 H(0,0;Y)+6 \yo H(0,0;Y)+H(0,0;Y)+\Big(\frac{20 \yo^3}{3}-30 \
\yo^2+60 \yo-8 H(0;Y)+\frac{12}{\yo-1}+28\Big) \
H(0,0;\yo)+\Big(\frac{4 d_1 \yo^3}{3}-6 d_1 \yo^2+12 d_1 \yo-4 \
d_1-\frac{4 d_1}{\yo-1}\Big) H(0,1;\ao)+H(1;\yo) \Big(\frac{1}{6} \
d_1' \yo^3 \ao^4-\frac{1}{2} d_1' \yo^2 \ao^4-\frac{d_1' \
\ao^4}{6}+\frac{1}{2} d_1' \yo \ao^4-\frac{8}{9} d_1' \yo^3 \ao^3+3 \
d_1' \yo^2 \ao^3+\frac{11 d_1' \ao^3}{9}-\frac{10}{3} d_1' \yo \
\ao^3+2 d_1' \yo^3 \ao^2-\frac{15}{2} d_1' \yo^2 \ao^2-\frac{9 d_1' \
\ao^2}{2}+10 d_1' \yo \ao^2-\frac{8}{3} d_1' \yo^3 \ao+11 d_1' \yo^2 \
\ao+\frac{23 d_1' \ao}{3}-18 d_1' \yo \ao+\frac{2 d_1'^2 \yo^3}{9}-2 \
d_1' \yo^3-\frac{49 d_1'^2}{18}-\frac{7 d_1'^2 \yo^2}{6}+\frac{21 \
d_1' \yo^2}{2}+\frac{43 d_1'}{2}+\frac{11 d_1'^2 \yo}{3}-30 d_1' \yo+\
\frac{2}{3} d_1' \yo^3 H(0;Y)-3 d_1' \yo^2 H(0;Y)-\frac{11}{3} d_1' \
H(0;Y)+6 d_1' \yo H(0;Y)+H(0;\ao) \Big(\frac{2 d_1' \yo^3}{3}-\frac{2 \
\yo^3}{3}-3 d_1' \yo^2+3 \yo^2+6 d_1' \yo-6 \yo-\frac{11 d_1'}{3}+2 \
H(0;Y)+\frac{4 d_1}{\yo-1}-\frac{2 \
d_1'}{\yo-1}-\frac{2}{\yo-1}-10\Big)+4 H(0,0;\ao)+4 d_1 \
H(0,1;\ao)-\frac{\pi ^2}{3}\Big)+\Big(2 d_1' \yo^3-9 d_1' \yo^2+18 \
d_1' \yo+6 d_1'+(4-4 d_1) H(0;\ao)-2 d_1' H(0;Y)+\frac{2 d_1'}{\yo-1}\
\Big) H(0,1;\yo)+\Big(-\frac{2 \yo^3}{3}+3 \yo^2-6 \yo+4 H(0;\ao)+2 \
H(0;Y)+4 d_1 H(1;\ao)-\frac{6}{\yo-1}-10\Big) \
H(0,c_1(\ao);\yo)+H(0;\yo) \Big(\frac{\yo^3 \ao^4}{3}-\yo^2 \ao^4+\yo \
\ao^4-\frac{16 \yo^3 \ao^3}{9}+6 \yo^2 \ao^3-\frac{20 \yo \ao^3}{3}+4 \
\yo^3 \ao^2-15 \yo^2 \ao^2+20 \yo \ao^2-\frac{16 \yo^3 \ao}{3}+22 \
\yo^2 \ao-36 \yo \ao+\frac{2 d_1' \yo^3}{3}-\frac{133 \
\yo^3}{18}-\frac{7 d_1' \yo^2}{2}+\frac{221 \yo^2}{6}+4 d_1-2 d_1'+11 \
d_1' \yo-\frac{589 \yo}{6}+\Big(\frac{4 \yo^3}{3}-6 \yo^2+12 \
\yo-\frac{4}{\yo-1}-4\Big) H(0;\ao)+2 \yo^3 H(0;Y)-9 \yo^2 H(0;Y)+18 \
\yo H(0;Y)+\frac{2 H(0;Y)}{\yo-1}+6 H(0;Y)-2 H(0,0;Y)-2 \
H(1,0;Y)+\frac{4 d_1}{\yo-1}-\frac{2 d_1'}{\yo-1}-\frac{61}{6 \
(\yo-1)}-\frac{\pi ^2}{3}-\frac{109}{6}\Big)+\frac{2}{3} \yo^3 \
H(1,0;Y)-3 \yo^2 H(1,0;Y)+6 \yo H(1,0;Y)+H(1,0;Y)+\Big(2 d_1' \
\yo^3+\frac{2 \yo^3}{3}-9 d_1' \yo^2-3 \yo^2+18 d_1' \yo+6 \yo-11 \
d_1'+4 H(0;\ao)-2 H(0;Y)-\frac{4 d_1}{\yo-1}+\frac{2 \
d_1'}{\yo-1}+\frac{2}{\yo-1}+10\Big) H(1,0;\yo)+\Big(\frac{2 d_1'^2 \
\yo^3}{3}-3 d_1'^2 \yo^2+6 d_1'^2 \yo-\frac{11 d_1'^2}{3}+(-4 d_1+2 \
d_1'+2) H(0;\ao)\Big) H(1,1;\yo)+\Big(-\frac{2 \yo^3}{3}+3 \yo^2-6 \
\yo+4 H(0;\ao)+2 H(0;Y)+4 d_1 H(1;\ao)+\frac{4 d_1}{\yo-1}-\frac{2 \
d_1'}{\yo-1}-\frac{2}{\yo-1}-10\Big) H(1,c_1(\ao);\yo)+\Big(4 \
\ao-\frac{4}{\yo-1}-4\Big) H(c_1(\ao),0;\yo)+\Big(2 \ao d_1'-\frac{2 \
d_1'}{\yo-1}-2 d_1'\Big) H(c_1(\ao),1;\yo)+\Big(2 \
\ao-\frac{2}{\yo-1}-2\Big) H(c_1(\ao),c_1(\ao);\yo)-\frac{1}{2} \
H(0,0,0;Y)-32 H(0,0,0;\yo)-8 d_1' H(0,0,1;\yo)+8 H(0,0,c_1(\ao);\yo)-\
\frac{1}{2} H(0,1,0;Y)+(4 d_1-8 d_1'-4) H(0,1,0;\yo)-2 d_1'^2 \
H(0,1,1;\yo)+(-4 d_1+2 d_1'+4) H(0,1,c_1(\ao);\yo)+4 \
H(0,c_1(\ao),0;\yo)+2 d_1' H(0,c_1(\ao),1;\yo)+2 \
H(0,c_1(\ao),c_1(\ao);\yo)-\frac{1}{2} H(1,0,0;Y)-12 H(1,0,0;\yo)-2 \
d_1' H(1,0,1;\yo)+6 H(1,0,c_1(\ao);\yo)-\frac{1}{2} H(1,1,0;Y)+(4 \
d_1-2 d_1'-2) H(1,1,0;\yo)+(-4 d_1+2 d_1'+2) H(1,1,c_1(\ao);\yo)+4 \
H(1,c_1(\ao),0;\yo)+2 d_1' H(1,c_1(\ao),1;\yo)+2 \
H(1,c_1(\ao),c_1(\ao);\yo)+\frac{\pi ^2}{3 (\yo-1)}-3 \
\zeta_3+\frac{\pi ^2}{2}+4.
\erp

%
% The J*I integral for k=1
%

\subsection{The $\cJI$ integral for $k=1$}
%
% This file contains the TeX output produced by Mathematica for the integral JI for arbitrary kap=0  and D0 = 3 +d'1 ep
%
The $\eps$ expansion for this integral reads
\beq
\cJI(Y,\ep;y_{0},d'_{0},\alpha_{0},d_{0};1)=\frac{1}{\eps^3}\cji_{-3}^{(1)}+\frac{1}{\eps^2}\cji_{-2}^{(1)}+\frac{1}{\eps}\cji_{-1}^{(1)}+\cji_0^{(1)}+\ocal\left(\eps\right),
\eeq
where

%% 1/ep^3
\brp
\cji_{-3}^{(1)}=\frac{1}{4},
\erp
%% 1/ep^2
\brp
\cji_{-2}^{(1)}=\frac{\yo^3}{3}-\frac{3 \yo^2}{2}+3 \yo-\frac{1}{4} \
H(0;Y)-H(0;\yo)+\frac{1}{2},
\erp
%% 1/ep
\brp
\cji_{-1}^{(1)}=-\frac{1}{12} \yo^3 \ao^4+\frac{\yo^2 \ao^4}{4}-\frac{\yo \
\ao^4}{4}+\frac{4 \yo^3 \ao^3}{9}-\frac{3 \yo^2 \ao^3}{2}+\frac{5 \yo \
\ao^3}{3}-\yo^3 \ao^2+\frac{15 \yo^2 \ao^2}{4}-5 \yo \ao^2+\frac{4 \
\yo^3 \ao}{3}-\frac{11 \yo^2 \ao}{2}+9 \yo \ao-\frac{d_1' \
\yo^3}{9}+\yo^3+\frac{7 d_1' \yo^2}{12}-\frac{21 \yo^2}{4}-\frac{11 \
d_1' \yo}{6}+15 \yo+\Big(-\frac{\yo^3}{3}+\frac{3 \yo^2}{2}-3 \
\yo+\frac{1}{\yo-1}+1\Big) H(0;\ao)-\frac{1}{3} \yo^3 \
H(0;Y)+\frac{3}{2} \yo^2 H(0;Y)-3 \yo H(0;Y)-\frac{1}{2} \
H(0;Y)+\Big(-\yo^3+\frac{9 \yo^2}{2}-9 \
\yo+H(0;Y)-\frac{1}{\yo-1}-3\Big) H(0;\yo)+\Big(-\frac{d_1' \
\yo^3}{3}+\frac{3 d_1' \yo^2}{2}-3 d_1' \yo+\frac{11 \
d_1'}{6}-H(0;\ao)\Big) H(1;\yo)+\Big(-\ao+\frac{1}{\yo-1}+1\Big) \
H(c_1(\ao);\yo)+\frac{1}{4} H(0,0;Y)+4 H(0,0;\yo)+d_1' \
H(0,1;\yo)-H(0,c_1(\ao);\yo)+\frac{1}{4} H(1,0;Y)+H(1,0;\yo)-H(1,c_1(\
\ao);\yo)+\frac{\pi ^2}{24}+1,
\erp
%% ep^0
\brp
\cji_0^{(1)} =\frac{1}{24} d_1 \yo^3 \ao^4+\frac{1}{36} d_1' \yo^3 \ao^4-\frac{7 \
\yo^3 \ao^4}{24}-\frac{1}{8} d_1 \yo^2 \ao^4-\frac{1}{12} d_1' \yo^2 \
\ao^4+\frac{23 \yo^2 \ao^4}{24}+\frac{1}{8} d_1 \yo \
\ao^4+\frac{1}{12} d_1' \yo \ao^4-\frac{29 \yo \
\ao^4}{24}+\frac{1}{12} \yo^3 H(0;Y) \ao^4-\frac{1}{4} \yo^2 H(0;Y) \
\ao^4+\frac{1}{4} \yo H(0;Y) \ao^4-\frac{13}{54} d_1 \yo^3 \
\ao^3-\frac{4}{27} d_1' \yo^3 \ao^3+\frac{173 \yo^3 \
\ao^3}{108}+\frac{5}{6} d_1 \yo^2 \ao^3+\frac{19}{36} d_1' \yo^2 \
\ao^3-\frac{217 \yo^2 \ao^3}{36}-\frac{17}{18} d_1 \yo \
\ao^3-\frac{11}{18} d_1' \yo \ao^3+\frac{305 \yo \
\ao^3}{36}-\frac{4}{9} \yo^3 H(0;Y) \ao^3+\frac{3}{2} \yo^2 H(0;Y) \
\ao^3-\frac{5}{3} \yo H(0;Y) \ao^3+\frac{23}{36} d_1 \yo^3 \
\ao^2+\frac{1}{3} d_1' \yo^3 \ao^2-\frac{275 \yo^3 \
\ao^2}{72}-\frac{5}{2} d_1 \yo^2 \ao^2-\frac{11}{8} d_1' \yo^2 \ao^2+\
\frac{581 \yo^2 \ao^2}{36}+\frac{43}{12} d_1 \yo \ao^2+\frac{9}{4} \
d_1' \yo \ao^2-\frac{1003 \yo \ao^2}{36}+\yo^3 H(0;Y) \
\ao^2-\frac{15}{4} \yo^2 H(0;Y) \ao^2+5 \yo H(0;Y) \
\ao^2-\frac{25}{18} d_1 \yo^3 \ao-\frac{4}{9} d_1' \yo^3 \
\ao+\frac{217 \yo^3 \ao}{36}+6 d_1 \yo^2 \ao+\frac{25}{12} d_1' \yo^2 \
\ao-\frac{503 \yo^2 \ao}{18}-\frac{65 d_1 \yo \ao}{6}-\frac{29 d_1' \
\yo \ao}{6}+\frac{237 \yo \ao}{4}-\frac{4}{3} \yo^3 H(0;Y) \
\ao+\frac{11}{2} \yo^2 H(0;Y) \ao-9 \yo H(0;Y) \ao+\frac{d_1'^2 \
\yo^3}{27}-\frac{4 d_1' \yo^3}{9}+\frac{7 \yo^3}{3}-\frac{17 d_1'^2 \
\yo^2}{72}+3 d_1' \yo^2-\frac{111 \yo^2}{8}+\frac{49 d_1'^2 \yo}{36}-\
\frac{65 d_1' \yo}{4}+57 \yo+\frac{1}{9} d_1' \yo^3 H(0;Y)-\yo^3 \
H(0;Y)-\frac{7}{12} d_1' \yo^2 H(0;Y)+\frac{21}{4} \yo^2 \
H(0;Y)+\frac{11}{6} d_1' \yo H(0;Y)-15 \yo H(0;Y)-\frac{1}{24} \pi ^2 \
H(0;Y)-H(0;Y)+H(0;\ao) \Big(\frac{\yo^3 \ao^4}{6}-\frac{\yo^2 \
\ao^4}{2}+\frac{\yo \ao^4}{2}-\frac{8 \yo^3 \ao^3}{9}+3 \yo^2 \
\ao^3-\frac{10 \yo \ao^3}{3}+2 \yo^3 \ao^2-\frac{15 \yo^2 \
\ao^2}{2}+10 \yo \ao^2-\frac{8 \yo^3 \ao}{3}+11 \yo^2 \ao-18 \yo \ao+\
\frac{d_1' \yo^3}{9}-\frac{11 \yo^3}{36}-\frac{7 d_1' \
\yo^2}{12}+\frac{31 \yo^2}{12}-2 d_1+d_1'+\frac{11 d_1' \
\yo}{6}-\frac{131 \yo}{12}+\frac{1}{3} \yo^3 H(0;Y)-\frac{3}{2} \yo^2 \
H(0;Y)+3 \yo H(0;Y)-\frac{H(0;Y)}{\yo-1}-H(0;Y)-\frac{2 \
d_1}{\yo-1}+\frac{d_1'}{\yo-1}+\frac{61}{12 \
(\yo-1)}+\frac{61}{12}\Big)+\Big(\frac{1}{6} d_1 \yo^3 \
\ao^4-\frac{1}{2} d_1 \yo^2 \ao^4+\frac{1}{2} d_1 \yo \
\ao^4-\frac{8}{9} d_1 \yo^3 \ao^3+3 d_1 \yo^2 \ao^3-\frac{10}{3} d_1 \
\yo \ao^3+2 d_1 \yo^3 \ao^2-\frac{15}{2} d_1 \yo^2 \ao^2+10 d_1 \yo \
\ao^2-\frac{8}{3} d_1 \yo^3 \ao+11 d_1 \yo^2 \ao-18 d_1 \yo \
\ao+\frac{25 d_1 \yo^3}{18}-6 d_1 \yo^2+\frac{65 d_1 \yo}{6}\Big) \
H(1;\ao)-\frac{1}{24} \pi ^2 H(1;Y)+\Big(\frac{\yo^3 \
\ao^4}{12}-\frac{\yo^2 \ao^4}{4}+\frac{\yo \
\ao^4}{4}-\frac{\ao^4}{12}-\frac{4 \yo^3 \ao^3}{9}+\frac{3 \yo^2 \
\ao^3}{2}-\frac{5 \yo \ao^3}{3}+\frac{11 \ao^3}{18}+\yo^3 \
\ao^2-\frac{15 \yo^2 \ao^2}{4}+5 \yo \ao^2-\frac{9 \ao^2}{4}-\frac{4 \
\yo^3 \ao}{3}+\frac{11 \yo^2 \ao}{2}+2 d_1 \ao-d_1' \ao-9 \yo \
\ao+H(0;Y) \ao-\ao+\frac{25 \yo^3}{36}-\frac{8 \yo^2}{3}-2 \
d_1+d_1'+\frac{49 \yo}{12}+\Big(2 \ao-\frac{2}{\yo-1}-2\Big) \
H(0;\ao)-\frac{H(0;Y)}{\yo-1}-H(0;Y)+\Big(2 \ao d_1-\frac{2 \
d_1}{\yo-1}-2 d_1\Big) H(1;\ao)-\frac{2 \
d_1}{\yo-1}+\frac{d_1'}{\yo-1}+\frac{61}{12 \
(\yo-1)}+\frac{61}{12}\Big) H(c_1(\ao);\yo)+\Big(\frac{2 \yo^3}{3}-3 \
\yo^2+6 \yo-\frac{2}{\yo-1}-2\Big) H(0,0;\ao)+\frac{1}{3} \yo^3 \
H(0,0;Y)-\frac{3}{2} \yo^2 H(0,0;Y)+3 \yo H(0,0;Y)+\frac{1}{2} \
H(0,0;Y)+\Big(\frac{10 \yo^3}{3}-15 \yo^2+30 \yo-4 \
H(0;Y)+\frac{6}{\yo-1}+14\Big) H(0,0;\yo)+\Big(\frac{2 d_1 \
\yo^3}{3}-3 d_1 \yo^2+6 d_1 \yo-2 d_1-\frac{2 d_1}{\yo-1}\Big) H(0,1;\
\ao)+H(1;\yo) \Big(\frac{1}{12} d_1' \yo^3 \ao^4-\frac{1}{4} d_1' \
\yo^2 \ao^4-\frac{d_1' \ao^4}{12}+\frac{1}{4} d_1' \yo \
\ao^4-\frac{4}{9} d_1' \yo^3 \ao^3+\frac{3}{2} d_1' \yo^2 \
\ao^3+\frac{11 d_1' \ao^3}{18}-\frac{5}{3} d_1' \yo \ao^3+d_1' \yo^3 \
\ao^2-\frac{15}{4} d_1' \yo^2 \ao^2-\frac{9 d_1' \ao^2}{4}+5 d_1' \yo \
\ao^2-\frac{4}{3} d_1' \yo^3 \ao+\frac{11}{2} d_1' \yo^2 \ao+\frac{23 \
d_1' \ao}{6}-9 d_1' \yo \ao+\frac{d_1'^2 \yo^3}{9}-d_1' \
\yo^3-\frac{49 d_1'^2}{36}-\frac{7 d_1'^2 \yo^2}{12}+\frac{21 d_1' \
\yo^2}{4}+\frac{43 d_1'}{4}+\frac{11 d_1'^2 \yo}{6}-15 d_1' \
\yo+\frac{1}{3} d_1' \yo^3 H(0;Y)-\frac{3}{2} d_1' \yo^2 \
H(0;Y)-\frac{11}{6} d_1' H(0;Y)+3 d_1' \yo H(0;Y)+H(0;\ao) \
\Big(\frac{d_1' \yo^3}{3}-\frac{\yo^3}{3}-\frac{3 d_1' \
\yo^2}{2}+\frac{3 \yo^2}{2}+3 d_1' \yo-3 \yo-\frac{11 \
d_1'}{6}+H(0;Y)+\frac{2 \
d_1}{\yo-1}-\frac{d_1'}{\yo-1}-\frac{1}{\yo-1}-5\Big)+2 H(0,0;\ao)+2 \
d_1 H(0,1;\ao)-\frac{\pi ^2}{6}\Big)+\Big(d_1' \yo^3-\frac{9 d_1' \
\yo^2}{2}+9 d_1' \yo+3 d_1'+(2-2 d_1) H(0;\ao)-d_1' \
H(0;Y)+\frac{d_1'}{\yo-1}\Big) \
H(0,1;\yo)+\Big(-\frac{\yo^3}{3}+\frac{3 \yo^2}{2}-3 \yo+2 \
H(0;\ao)+H(0;Y)+2 d_1 H(1;\ao)-\frac{3}{\yo-1}-5\Big) \
H(0,c_1(\ao);\yo)+H(0;\yo) \Big(\frac{\yo^3 \ao^4}{6}-\frac{\yo^2 \
\ao^4}{2}+\frac{\yo \ao^4}{2}-\frac{8 \yo^3 \ao^3}{9}+3 \yo^2 \
\ao^3-\frac{10 \yo \ao^3}{3}+2 \yo^3 \ao^2-\frac{15 \yo^2 \
\ao^2}{2}+10 \yo \ao^2-\frac{8 \yo^3 \ao}{3}+11 \yo^2 \ao-18 \yo \ao+\
\frac{d_1' \yo^3}{3}-\frac{133 \yo^3}{36}-\frac{7 d_1' \
\yo^2}{4}+\frac{221 \yo^2}{12}+2 d_1-d_1'+\frac{11 d_1' \
\yo}{2}-\frac{589 \yo}{12}+\Big(\frac{2 \yo^3}{3}-3 \yo^2+6 \
\yo-\frac{2}{\yo-1}-2\Big) H(0;\ao)+\yo^3 H(0;Y)-\frac{9}{2} \yo^2 \
H(0;Y)+9 \yo H(0;Y)+\frac{H(0;Y)}{\yo-1}+3 \
H(0;Y)-H(0,0;Y)-H(1,0;Y)+\frac{2 \
d_1}{\yo-1}-\frac{d_1'}{\yo-1}-\frac{61}{12 (\yo-1)}-\frac{\pi \
^2}{6}-\frac{109}{12}\Big)+\frac{1}{3} \yo^3 H(1,0;Y)-\frac{3}{2} \
\yo^2 H(1,0;Y)+3 \yo H(1,0;Y)+\frac{1}{2} H(1,0;Y)+\Big(d_1' \
\yo^3+\frac{\yo^3}{3}-\frac{9 d_1' \yo^2}{2}-\frac{3 \yo^2}{2}+9 d_1' \
\yo+3 \yo-\frac{11 d_1'}{2}+2 H(0;\ao)-H(0;Y)-\frac{2 \
d_1}{\yo-1}+\frac{d_1'}{\yo-1}+\frac{1}{\yo-1}+5\Big) \
H(1,0;\yo)+\Big(\frac{d_1'^2 \yo^3}{3}-\frac{3 d_1'^2 \yo^2}{2}+3 \
d_1'^2 \yo-\frac{11 d_1'^2}{6}+(-2 d_1+d_1'+1) H(0;\ao)\Big) \
H(1,1;\yo)+\Big(-\frac{\yo^3}{3}+\frac{3 \yo^2}{2}-3 \yo+2 \
H(0;\ao)+H(0;Y)+2 d_1 H(1;\ao)+\frac{2 \
d_1}{\yo-1}-\frac{d_1'}{\yo-1}-\frac{1}{\yo-1}-5\Big) \
H(1,c_1(\ao);\yo)+\Big(2 \ao-\frac{2}{\yo-1}-2\Big) \
H(c_1(\ao),0;\yo)+\Big(\ao d_1'-\frac{d_1'}{\yo-1}-d_1'\Big) \
H(c_1(\ao),1;\yo)+\Big(\ao-\frac{1}{\yo-1}-1\Big) \
H(c_1(\ao),c_1(\ao);\yo)-\frac{1}{4} H(0,0,0;Y)-16 H(0,0,0;\yo)-4 \
d_1' H(0,0,1;\yo)+4 H(0,0,c_1(\ao);\yo)-\frac{1}{4} H(0,1,0;Y)+(2 \
d_1-4 d_1'-2) H(0,1,0;\yo)-d_1'^2 H(0,1,1;\yo)+(-2 d_1+d_1'+2) \
H(0,1,c_1(\ao);\yo)+2 H(0,c_1(\ao),0;\yo)+d_1' \
H(0,c_1(\ao),1;\yo)+H(0,c_1(\ao),c_1(\ao);\yo)-\frac{1}{4} \
H(1,0,0;Y)-6 H(1,0,0;\yo)-d_1' H(1,0,1;\yo)+3 \
H(1,0,c_1(\ao);\yo)-\frac{1}{4} H(1,1,0;Y)+(2 d_1-d_1'-1) \
H(1,1,0;\yo)+(-2 d_1+d_1'+1) H(1,1,c_1(\ao);\yo)+2 \
H(1,c_1(\ao),0;\yo)+d_1' \
H(1,c_1(\ao),1;\yo)+H(1,c_1(\ao),c_1(\ao);\yo)+\frac{\pi ^2}{6 \
(\yo-1)}-\frac{3 \zeta_3}{2}+\frac{\pi ^2}{4}+2.
\erp

%
% The J*I integral for k=2
%

\subsection{The $\cJI$ integral for $k=2$}
%
% This file contains the TeX output produced by Mathematica for the integral JI for arbitrary kap=0  and D0 = 3 +d'1 ep
%
\sloppy
The $\eps$ expansion for this integral reads
\beq
\cJI(Y,\ep;y_{0},d'_{0},\alpha_{0},d_{0};2)=\frac{1}{\eps^3}\cji_{-3}^{(2)}+\frac{1}{\eps^2}\cji_{-2}^{(2)}+\frac{1}{\eps}\cji_{-1}^{(2)}+\cji_0^{(2)}+\ocal\left(\eps\right),
\eeq
where
%% 1/ep^3
\brp
\cji_{-3}^{(2)}=\frac{1}{6},
\erp
%% 1/ep^2
\brp
\cji_{-2}^{(2)}=\frac{2 \yo^3}{9}-\yo^2+2 \yo-\frac{1}{6} H(0;Y)-\frac{2}{3} \
H(0;\yo)+\frac{4}{9},
\erp
%% 1/ep
\brp
\cji_{-1}^{(2)}=-\frac{1}{18} \yo^3 \ao^4+\frac{\yo^2 \ao^4}{12}-\frac{\yo \ao^4}{3}+\
\frac{\ao^4}{6 (\yo-2)}+\frac{\ao^4}{12}+\frac{8 \yo^3 \
\ao^3}{27}-\frac{2 \yo^2 \ao^3}{3}+\frac{8 \yo \
\ao^3}{9}+\frac{\ao^3}{\yo-2}+\frac{4 \ao^3}{9 (\yo-2)^2}+\frac{7 \
\ao^3}{18}-\frac{2 \yo^3 \ao^2}{3}+2 \yo^2 \ao^2-\frac{5 \yo \
\ao^2}{3}+\frac{8 \ao^2}{3 (\yo-2)}+\frac{11 \ao^2}{3 \
(\yo-2)^2}+\frac{4 \ao^2}{3 (\yo-2)^3}+\frac{7 \ao^2}{12}+\frac{8 \
\yo^3 \ao}{9}-\frac{10 \yo^2 \ao}{3}+4 \yo \ao+\frac{13 \ao}{3 \
(\yo-2)}+\frac{18 \ao}{(\yo-2)^2}+\frac{52 \ao}{3 (\yo-2)^3}+\frac{16 \
\ao}{3 (\yo-2)^4}-\frac{\ao}{2}-\frac{2 d_1' \yo^3}{27}+\frac{19 \
\yo^3}{27}+\frac{7 d_1' \yo^2}{18}-\frac{11 \yo^2}{3}-\frac{11 d_1' \
\yo}{9}+\frac{31 \yo}{3}+\Big(-\frac{2 \yo^3}{9}+\yo^2-2 \
\yo+\frac{2}{3 (\yo-1)}+\frac{80}{3 (\yo-2)^2}+\frac{160}{3 \
(\yo-2)^3}+\frac{40}{(\yo-2)^4}+\frac{32}{3 \
(\yo-2)^5}-\frac{3}{2}\Big) H(0;\ao)-\frac{2}{9} \yo^3 H(0;Y)+\yo^2 \
H(0;Y)-2 \yo H(0;Y)-\frac{4}{9} H(0;Y)+\Big(-\frac{2 \yo^3}{3}+3 \
\yo^2-6 \yo+\frac{2}{3} H(0;Y)-\frac{2}{3 (\yo-1)}-\frac{80}{3 \
(\yo-2)^2}-\frac{160}{3 (\yo-2)^3}-\frac{40}{(\yo-2)^4}-\frac{32}{3 (\
\yo-2)^5}-\frac{5}{18}\Big) H(0;\yo)+\Big(-\frac{2 d_1' \
\yo^3}{9}+d_1' \yo^2-2 d_1' \yo+\frac{11 d_1'}{9}-\frac{2}{3} \
H(0;\ao)\Big) H(1;\yo)+\Big(-\frac{2 \ao}{3}+\frac{2}{3 \
(\yo-1)}+\frac{2}{3}\Big) \
H(c_1(\ao);\yo)+\Big(-\frac{\ao^4}{2}-\frac{2 \ao^3}{3}+\ao^2+2 \
\ao+\frac{80}{3 (\yo-2)^2}+\frac{160}{3 \
(\yo-2)^3}+\frac{40}{(\yo-2)^4}+\frac{32}{3 \
(\yo-2)^5}-\frac{11}{6}\Big) H(c_2(\ao);\yo)+\frac{1}{6} \
H(0,0;Y)+\frac{8}{3} H(0,0;\yo)+\frac{2}{3} d_1' \
H(0,1;\yo)-\frac{2}{3} H(0,c_1(\ao);\yo)+\frac{1}{6} \
H(1,0;Y)+\frac{2}{3} H(1,0;\yo)-\frac{2}{3} \
H(1,c_1(\ao);\yo)+\frac{80 \ln 2\, }{3 (\yo-2)^2}+\frac{160 \ln 2\, \
}{3 (\yo-2)^3}+\frac{40 \ln 2\, }{(\yo-2)^4}+\frac{32 \ln 2\, }{3 \
(\yo-2)^5}-\frac{13 \ln 2\, }{6}+\frac{\pi ^2}{36}+\frac{26}{27},
\erp
%% ep^0
\brp
\cji_0^{(2)} =\frac{1}{36} d_1 \yo^3 \ao^4+\frac{1}{54} d_1' \yo^3 \ao^4-\frac{11 \
\yo^3 \ao^4}{54}-\frac{1}{24} d_1 \yo^2 \ao^4-\frac{1}{72} d_1' \yo^2 \
\ao^4+\frac{5 \yo^2 \ao^4}{18}-\frac{d_1 \ao^4}{24}+\frac{1}{6} d_1 \
\yo \ao^4+\frac{11}{36} d_1' \yo \ao^4-\frac{35 \yo \
\ao^4}{18}+\frac{1}{18} \yo^3 H(0;Y) \ao^4-\frac{1}{12} \yo^2 H(0;Y) \
\ao^4+\frac{1}{3} \yo H(0;Y) \ao^4-\frac{H(0;Y) \ao^4}{6 \
(\yo-2)}-\frac{1}{12} H(0;Y) \ao^4-\frac{d_1 \ao^4}{12 \
(\yo-2)}+\frac{19 \ao^4}{36 (\yo-2)}+\frac{19 \
\ao^4}{72}-\frac{13}{81} d_1 \yo^3 \ao^3-\frac{8}{81} d_1' \yo^3 \
\ao^3+\frac{181 \yo^3 \ao^3}{162}+\frac{7}{18} d_1 \yo^2 \
\ao^3+\frac{5}{27} d_1' \yo^2 \ao^3-\frac{281 \yo^2 \
\ao^3}{108}-\frac{43 d_1 \ao^3}{108}+\frac{d_1' \
\ao^3}{18}-\frac{10}{27} d_1 \yo \ao^3-\frac{14}{27} d_1' \yo \
\ao^3+\frac{493 \yo \ao^3}{108}-\frac{8}{27} \yo^3 H(0;Y) \
\ao^3+\frac{2}{3} \yo^2 H(0;Y) \ao^3-\frac{8}{9} \yo H(0;Y) \
\ao^3-\frac{H(0;Y) \ao^3}{\yo-2}-\frac{4 H(0;Y) \ao^3}{9 \
(\yo-2)^2}-\frac{7}{18} H(0;Y) \ao^3-\frac{17 d_1 \ao^3}{18 (\yo-2)}+\
\frac{d_1' \ao^3}{9 (\yo-2)}+\frac{61 \ao^3}{18 (\yo-2)}-\frac{8 d_1 \
\ao^3}{27 (\yo-2)^2}+\frac{40 \ao^3}{27 (\yo-2)^2}+\frac{143 \
\ao^3}{108}+\frac{23}{54} d_1 \yo^3 \ao^2+\frac{2}{9} d_1' \yo^3 \
\ao^2-\frac{287 \yo^3 \ao^2}{108}-\frac{17}{12} d_1 \yo^2 \
\ao^2-\frac{2}{3} d_1' \yo^2 \ao^2+\frac{1883 \yo^2 \
\ao^2}{216}-\frac{113 d_1 \ao^2}{72}+\frac{13 d_1' \
\ao^2}{36}+\frac{10}{9} d_1 \yo \ao^2+\frac{1}{3} d_1' \yo \
\ao^2-\frac{224 \yo \ao^2}{27}+\frac{2}{3} \yo^3 H(0;Y) \ao^2-2 \yo^2 \
H(0;Y) \ao^2+\frac{5}{3} \yo H(0;Y) \ao^2-\frac{8 H(0;Y) \ao^2}{3 \
(\yo-2)}-\frac{11 H(0;Y) \ao^2}{3 (\yo-2)^2}-\frac{4 H(0;Y) \ao^2}{3 \
(\yo-2)^3}-\frac{7}{12} H(0;Y) \ao^2-\frac{46 d_1 \ao^2}{9 \
(\yo-2)}+\frac{8 d_1' \ao^2}{9 (\yo-2)}+\frac{179 \ao^2}{18 (\yo-2)}-\
\frac{83 d_1 \ao^2}{18 (\yo-2)^2}+\frac{d_1' \ao^2}{3 \
(\yo-2)^2}+\frac{247 \ao^2}{18 (\yo-2)^2}-\frac{4 d_1 \ao^2}{3 \
(\yo-2)^3}+\frac{44 \ao^2}{9 (\yo-2)^3}+\frac{155 \
\ao^2}{72}-\frac{25}{27} d_1 \yo^3 \ao-\frac{8}{27} d_1' \yo^3 \
\ao+\frac{25 \yo^3 \ao}{6}+\frac{23}{6} d_1 \yo^2 \ao+\frac{11}{9} \
d_1' \yo^2 \ao-\frac{1883 \yo^2 \ao}{108}-\frac{83 d_1 \
\ao}{36}+\frac{19 d_1' \ao}{18}-\frac{52 d_1 \yo \ao}{9}-\frac{14 \
d_1' \yo \ao}{9}+\frac{949 \yo \ao}{36}-\frac{8}{9} \yo^3 H(0;Y) \ao+\
\frac{10}{3} \yo^2 H(0;Y) \ao-4 \yo H(0;Y) \ao-\frac{13 H(0;Y) \ao}{3 \
(\yo-2)}-\frac{18 H(0;Y) \ao}{(\yo-2)^2}-\frac{52 H(0;Y) \ao}{3 \
(\yo-2)^3}-\frac{16 H(0;Y) \ao}{3 (\yo-2)^4}+\frac{1}{2} H(0;Y) \
\ao-\frac{383 d_1 \ao}{18 (\yo-2)}+\frac{35 d_1' \ao}{9 \
(\yo-2)}+\frac{113 \ao}{6 (\yo-2)}-\frac{151 d_1 \ao}{3 \
(\yo-2)^2}+\frac{38 d_1' \ao}{9 (\yo-2)^2}+\frac{263 \ao}{3 \
(\yo-2)^2}-\frac{118 d_1 \ao}{3 (\yo-2)^3}+\frac{4 d_1' \ao}{3 \
(\yo-2)^3}+\frac{746 \ao}{9 (\yo-2)^3}-\frac{32 d_1 \ao}{3 \
(\yo-2)^4}+\frac{224 \ao}{9 (\yo-2)^4}-\frac{133 \ao}{36}+\frac{2 \
d_1'^2 \yo^3}{81}-\frac{25 d_1' \yo^3}{81}+\frac{137 \
\yo^3}{81}-\frac{17 d_1'^2 \yo^2}{108}+\frac{223 d_1' \
\yo^2}{108}-\frac{179 \yo^2}{18}+\frac{49 d_1'^2 \yo}{54}-\frac{298 \
d_1' \yo}{27}+\frac{359 \yo}{9}+\frac{2}{27} d_1' \yo^3 \
H(0;Y)-\frac{19}{27} \yo^3 H(0;Y)-\frac{7}{18} d_1' \yo^2 \
H(0;Y)+\frac{11}{3} \yo^2 H(0;Y)+\frac{11}{9} d_1' \yo \
H(0;Y)-\frac{31}{3} \yo H(0;Y)-\frac{1}{36} \pi ^2 \
H(0;Y)-\frac{26}{27} H(0;Y)+H(0;\ao) \Big(\frac{\yo^3 \
\ao^4}{9}-\frac{\yo^2 \ao^4}{6}+\frac{2 \yo \ao^4}{3}-\frac{\ao^4}{3 \
(\yo-2)}-\frac{\ao^4}{6}-\frac{16 \yo^3 \ao^3}{27}+\frac{4 \yo^2 \
\ao^3}{3}-\frac{16 \yo \ao^3}{9}-\frac{2 \ao^3}{\yo-2}-\frac{8 \
\ao^3}{9 (\yo-2)^2}-\frac{7 \ao^3}{9}+\frac{4 \yo^3 \ao^2}{3}-4 \yo^2 \
\ao^2+\frac{10 \yo \ao^2}{3}-\frac{16 \ao^2}{3 (\yo-2)}-\frac{22 \
\ao^2}{3 (\yo-2)^2}-\frac{8 \ao^2}{3 (\yo-2)^3}-\frac{7 \
\ao^2}{6}-\frac{16 \yo^3 \ao}{9}+\frac{20 \yo^2 \ao}{3}-8 \yo \
\ao-\frac{26 \ao}{3 (\yo-2)}-\frac{36 \ao}{(\yo-2)^2}-\frac{104 \
\ao}{3 (\yo-2)^3}-\frac{32 \ao}{3 (\yo-2)^4}+\ao+\frac{2 d_1' \
\yo^3}{27}-\frac{13 \yo^3}{54}-\frac{7 d_1' \yo^2}{18}+\frac{71 \
\yo^2}{36}+\frac{41 d_1}{36}+\frac{d_1'}{18}+\frac{11 d_1' \
\yo}{9}-\frac{25 \yo}{3}+\frac{2}{9} \yo^3 H(0;Y)-\yo^2 H(0;Y)+2 \yo \
H(0;Y)-\frac{2 H(0;Y)}{3 (\yo-1)}-\frac{80 H(0;Y)}{3 \
(\yo-2)^2}-\frac{160 H(0;Y)}{3 (\yo-2)^3}-\frac{40 \
H(0;Y)}{(\yo-2)^4}-\frac{32 H(0;Y)}{3 (\yo-2)^5}+\frac{3}{2} \
H(0;Y)+\frac{8 d_1'}{3 (\yo-2)}-\frac{20}{3 (\yo-2)}-\frac{4 d_1}{3 (\
\yo-1)}+\frac{2 d_1'}{3 (\yo-1)}+\frac{7}{2 (\yo-1)}-\frac{16 \
d_1}{(\yo-2)^2}+\frac{12 \
d_1'}{(\yo-2)^2}+\frac{88}{(\yo-2)^2}-\frac{128 d_1}{9 \
(\yo-2)^3}+\frac{88 d_1'}{9 (\yo-2)^3}+\frac{168}{(\yo-2)^3}-\frac{4 \
d_1}{(\yo-2)^4}+\frac{8 d_1'}{3 \
(\yo-2)^4}+\frac{116}{(\yo-2)^4}+\frac{256}{9 \
(\yo-2)^5}-\frac{259}{36}\Big)+\Big(\frac{1}{9} d_1 \yo^3 \
\ao^4-\frac{1}{6} d_1 \yo^2 \ao^4-\frac{d_1 \ao^4}{6}+\frac{2}{3} d_1 \
\yo \ao^4-\frac{d_1 \ao^4}{3 (\yo-2)}-\frac{16}{27} d_1 \yo^3 \
\ao^3+\frac{4}{3} d_1 \yo^2 \ao^3-\frac{7 d_1 \ao^3}{9}-\frac{16}{9} \
d_1 \yo \ao^3-\frac{2 d_1 \ao^3}{\yo-2}-\frac{8 d_1 \ao^3}{9 \
(\yo-2)^2}+\frac{4}{3} d_1 \yo^3 \ao^2-4 d_1 \yo^2 \ao^2-\frac{7 d_1 \
\ao^2}{6}+\frac{10}{3} d_1 \yo \ao^2-\frac{16 d_1 \ao^2}{3 \
(\yo-2)}-\frac{22 d_1 \ao^2}{3 (\yo-2)^2}-\frac{8 d_1 \ao^2}{3 \
(\yo-2)^3}-\frac{16}{9} d_1 \yo^3 \ao+\frac{20}{3} d_1 \yo^2 \ao+d_1 \
\ao-8 d_1 \yo \ao-\frac{26 d_1 \ao}{3 (\yo-2)}-\frac{36 d_1 \
\ao}{(\yo-2)^2}-\frac{104 d_1 \ao}{3 (\yo-2)^3}-\frac{32 d_1 \ao}{3 (\
\yo-2)^4}+\frac{25 d_1 \yo^3}{27}-\frac{23 d_1 \yo^2}{6}+\frac{10 \
d_1}{9}+\frac{52 d_1 \yo}{9}+\frac{49 d_1}{3 (\yo-2)}+\frac{398 \
d_1}{9 (\yo-2)^2}+\frac{112 d_1}{3 (\yo-2)^3}+\frac{32 d_1}{3 \
(\yo-2)^4}\Big) H(1;\ao)-\frac{1}{36} \pi ^2 H(1;Y)+\Big(\frac{\yo^3 \
\ao^4}{18}-\frac{\yo^2 \ao^4}{12}+\frac{\yo \ao^4}{3}-\frac{\ao^4}{6 \
(\yo-2)}-\frac{17 \ao^4}{36}-\frac{8 \yo^3 \ao^3}{27}+\frac{2 \yo^2 \
\ao^3}{3}-\frac{8 \yo \ao^3}{9}-\frac{\ao^3}{\yo-2}-\frac{4 \ao^3}{9 \
(\yo-2)^2}-\frac{29 \ao^3}{54}+\frac{2 \yo^3 \ao^2}{3}-2 \yo^2 \ao^2+\
\frac{5 \yo \ao^2}{3}-\frac{8 \ao^2}{3 (\yo-2)}-\frac{11 \ao^2}{3 \
(\yo-2)^2}-\frac{4 \ao^2}{3 (\yo-2)^3}-\frac{13 \ao^2}{12}-\frac{8 \
\yo^3 \ao}{9}+\frac{10 \yo^2 \ao}{3}+\frac{4 d_1 \ao}{3}-\frac{2 d_1' \
\ao}{3}-4 \yo \ao+\frac{2}{3} H(0;Y) \ao-\frac{13 \ao}{3 \
(\yo-2)}-\frac{18 \ao}{(\yo-2)^2}-\frac{52 \ao}{3 (\yo-2)^3}-\frac{16 \
\ao}{3 (\yo-2)^4}+\frac{7 \ao}{18}+\frac{25 \yo^3}{54}-\frac{61 \
\yo^2}{36}-\frac{4 d_1}{3}+\frac{2 d_1'}{3}+2 \yo+\Big(\frac{4 \
\ao}{3}-\frac{4}{3 (\yo-1)}-\frac{4}{3}\Big) H(0;\ao)-\frac{2 \
H(0;Y)}{3 (\yo-1)}-\frac{2}{3} H(0;Y)+\Big(\frac{4 \ao \
d_1}{3}-\frac{4 d_1}{3 (\yo-1)}-\frac{4 d_1}{3}\Big) \
H(1;\ao)+\frac{32}{3 (\yo-2)}-\frac{4 d_1}{3 (\yo-1)}+\frac{2 d_1'}{3 \
(\yo-1)}+\frac{7}{2 (\yo-1)}+\frac{332}{9 (\yo-2)^2}+\frac{104}{3 \
(\yo-2)^3}+\frac{32}{3 (\yo-2)^4}+\frac{59}{18}\Big) H(c_1(\ao);\yo)+\
\Big(\frac{d_1 \ao^4}{4}-\frac{d_1' \ao^4}{6}+\frac{1}{2} H(0;Y) \
\ao^4-\frac{5 \ao^4}{4}+\frac{7 d_1 \ao^3}{9}-\frac{5 d_1' \ao^3}{9}+\
\frac{2}{3} H(0;Y) \ao^3-\ao^3+\frac{d_1 \ao^2}{6}-\frac{d_1' \
\ao^2}{3}-H(0;Y) \ao^2+\frac{29 \ao^2}{6}-\frac{11 d_1 \
\ao}{3}+\frac{5 d_1' \ao}{3}-2 H(0;Y) \ao+7 \ao+\frac{89 \
d_1}{36}-\frac{11 d_1'}{18}+\Big(\ao^4+\frac{4 \ao^3}{3}-2 \ao^2-4 \
\ao-\frac{160}{3 (\yo-2)^2}-\frac{320}{3 \
(\yo-2)^3}-\frac{80}{(\yo-2)^4}-\frac{64}{3 \
(\yo-2)^5}+\frac{11}{3}\Big) H(0;\ao)-\frac{80 H(0;Y)}{3 \
(\yo-2)^2}-\frac{160 H(0;Y)}{3 (\yo-2)^3}-\frac{40 \
H(0;Y)}{(\yo-2)^4}-\frac{32 H(0;Y)}{3 (\yo-2)^5}+\frac{11}{6} H(0;Y)+\
\Big(d_1 \ao^4+\frac{4 d_1 \ao^3}{3}-2 d_1 \ao^2-4 d_1 \ao+\frac{11 \
d_1}{3}-\frac{160 d_1}{3 (\yo-2)^2}-\frac{320 d_1}{3 \
(\yo-2)^3}-\frac{80 d_1}{(\yo-2)^4}-\frac{64 d_1}{3 (\yo-2)^5}\Big) \
H(1;\ao)+\frac{8 d_1'}{3 (\yo-2)}-\frac{52}{3 (\yo-2)}-\frac{16 \
d_1}{(\yo-2)^2}+\frac{12 d_1'}{(\yo-2)^2}+\frac{460}{9 \
(\yo-2)^2}-\frac{128 d_1}{9 (\yo-2)^3}+\frac{88 d_1'}{9 \
(\yo-2)^3}+\frac{400}{3 (\yo-2)^3}-\frac{4 d_1}{(\yo-2)^4}+\frac{8 \
d_1'}{3 (\yo-2)^4}+\frac{316}{3 (\yo-2)^4}+\frac{256}{9 \
(\yo-2)^5}-\frac{115}{12}\Big) H(c_2(\ao);\yo)+\Big(\frac{4 \
\yo^3}{9}-2 \yo^2+4 \yo-\frac{4}{3 (\yo-1)}-\frac{160}{3 \
(\yo-2)^2}-\frac{320}{3 (\yo-2)^3}-\frac{80}{(\yo-2)^4}-\frac{64}{3 (\
\yo-2)^5}+3\Big) H(0,0;\ao)+\frac{2}{9} \yo^3 H(0,0;Y)-\yo^2 \
H(0,0;Y)+2 \yo H(0,0;Y)+\frac{4}{9} H(0,0;Y)+\Big(\frac{20 \
\yo^3}{9}-10 \yo^2+20 \yo-\frac{8}{3} \
H(0;Y)+\frac{4}{\yo-1}+\frac{160}{(\yo-2)^2}+\frac{320}{(\yo-2)^3}+\frac{240}{(\yo-2)^4}+\frac{64}{(\yo-2)^5}-\frac{17}{9}\Big) \
H(0,0;\yo)+\Big(\frac{4 d_1 \yo^3}{9}-2 d_1 \yo^2+4 d_1 \yo+3 \
d_1-\frac{4 d_1}{3 (\yo-1)}-\frac{160 d_1}{3 (\yo-2)^2}-\frac{320 \
d_1}{3 (\yo-2)^3}-\frac{80 d_1}{(\yo-2)^4}-\frac{64 d_1}{3 (\yo-2)^5}\
\Big) H(0,1;\ao)+\Big(\frac{2 d_1' \yo^3}{3}-3 d_1' \yo^2+6 d_1' \yo+\
\frac{5 d_1'}{18}+\Big(\frac{4}{3}-\frac{4 d_1}{3}\Big) \
H(0;\ao)-\frac{2}{3} d_1' H(0;Y)+\frac{2 d_1'}{3 (\yo-1)}+\frac{80 \
d_1'}{3 (\yo-2)^2}+\frac{160 d_1'}{3 (\yo-2)^3}+\frac{40 \
d_1'}{(\yo-2)^4}+\frac{32 d_1'}{3 (\yo-2)^5}\Big) \
H(0,1;\yo)+\Big(-\frac{2 \yo^3}{9}+\yo^2-2 \yo+\frac{4}{3} \
H(0;\ao)+\frac{2}{3} H(0;Y)+\frac{4}{3} d_1 H(1;\ao)-\frac{2}{\yo-1}+\
\frac{80}{3 (\yo-2)^2}+\frac{160}{3 \
(\yo-2)^3}+\frac{40}{(\yo-2)^4}+\frac{32}{3 (\yo-2)^5}-\frac{107}{18}\
\Big) H(0,c_1(\ao);\yo)+\Big(\frac{26}{3}-\frac{320}{3 \
(\yo-2)^2}-\frac{640}{3 (\yo-2)^3}-\frac{160}{(\yo-2)^4}-\frac{128}{3 \
(\yo-2)^5}\Big) H(0,c_2(\ao);\yo)+\frac{2}{9} \yo^3 H(1,0;Y)-\yo^2 \
H(1,0;Y)+2 \yo H(1,0;Y)+\frac{4}{9} H(1,0;Y)+\Big(\frac{2 d_1' \
\yo^3}{3}+\frac{2 \yo^3}{9}-3 d_1' \yo^2-\yo^2+6 d_1' \yo+2 \
\yo-\frac{19 d_1'}{3}+\frac{4}{3} H(0;\ao)-\frac{2}{3} H(0;Y)-\frac{4 \
d_1}{3 (\yo-1)}+\frac{2 d_1'}{3 (\yo-1)}+\frac{2}{3 (\yo-1)}+\frac{80 \
d_1'}{3 (\yo-2)^2}-\frac{80}{3 (\yo-2)^2}+\frac{160 d_1'}{3 \
(\yo-2)^3}-\frac{160}{3 (\yo-2)^3}+\frac{40 \
d_1'}{(\yo-2)^4}-\frac{40}{(\yo-2)^4}+\frac{32 d_1'}{3 \
(\yo-2)^5}-\frac{32}{3 (\yo-2)^5}+\frac{107}{18}\Big) \
H(1,0;\yo)+\Big(\frac{2 d_1'^2 \yo^3}{9}-d_1'^2 \yo^2+2 d_1'^2 \
\yo-\frac{11 d_1'^2}{9}+\Big(-\frac{4 d_1}{3}+\frac{2 \
d_1'}{3}+\frac{2}{3}\Big) H(0;\ao)\Big) H(1,1;\yo)+\Big(-\frac{2 \
\yo^3}{9}+\yo^2-2 \yo+\frac{4}{3} H(0;\ao)+\frac{2}{3} \
H(0;Y)+\frac{4}{3} d_1 H(1;\ao)+\frac{4 d_1}{3 (\yo-1)}-\frac{2 \
d_1'}{3 (\yo-1)}-\frac{2}{3 (\yo-1)}+\frac{80}{3 \
(\yo-2)^2}+\frac{160}{3 (\yo-2)^3}+\frac{40}{(\yo-2)^4}+\frac{32}{3 (\
\yo-2)^5}-\frac{107}{18}\Big) H(1,c_1(\ao);\yo)+\Big(-\frac{80 \
d_1'}{3 (\yo-2)^2}-\frac{160 d_1'}{3 (\yo-2)^3}-\frac{40 \
d_1'}{(\yo-2)^4}-\frac{32 d_1'}{3 (\yo-2)^5}+\frac{8 d_1'}{3}\Big) \
H(1,c_2(\ao);\yo)+\Big(\frac{160 d_1}{3 (\yo-2)^2}+\frac{320 d_1}{3 (\
\yo-2)^3}+\frac{80 d_1}{(\yo-2)^4}+\frac{64 d_1}{3 (\yo-2)^5}-\frac{8 \
d_1'}{3}-\frac{160}{3 (\yo-2)^2}-\frac{320}{3 \
(\yo-2)^3}-\frac{80}{(\yo-2)^4}-\frac{64}{3 \
(\yo-2)^5}+\frac{26}{3}\Big) H(2,0;\yo)+\Big(-\frac{160 d_1}{3 \
(\yo-2)^2}-\frac{320 d_1}{3 (\yo-2)^3}-\frac{80 \
d_1}{(\yo-2)^4}-\frac{64 d_1}{3 (\yo-2)^5}+\frac{8 \
d_1'}{3}+\frac{160}{3 (\yo-2)^2}+\frac{320}{3 \
(\yo-2)^3}+\frac{80}{(\yo-2)^4}+\frac{64}{3 \
(\yo-2)^5}-\frac{26}{3}\Big) H(2,c_2(\ao);\yo)+\Big(\frac{4 \
\ao}{3}-\frac{4}{3 (\yo-1)}-\frac{4}{3}\Big) \
H(c_1(\ao),0;\yo)+\Big(\frac{2 \ao d_1'}{3}-\frac{2 d_1'}{3 (\yo-1)}-\
\frac{2 d_1'}{3}\Big) H(c_1(\ao),1;\yo)+\Big(\frac{2 \
\ao}{3}-\frac{2}{3 (\yo-1)}-\frac{2}{3}\Big) \
H(c_1(\ao),c_1(\ao);\yo)+\Big(\ao^4+\frac{4 \ao^3}{3}-2 \ao^2-4 \
\ao-\frac{160}{3 (\yo-2)^2}-\frac{320}{3 \
(\yo-2)^3}-\frac{80}{(\yo-2)^4}-\frac{64}{3 \
(\yo-2)^5}+\frac{11}{3}\Big) H(c_2(\ao),0;\yo)+\Big(\frac{d_1' \
\ao^4}{2}+\frac{2 d_1' \ao^3}{3}-d_1' \ao^2-2 d_1' \ao+\frac{11 \
d_1'}{6}-\frac{80 d_1'}{3 (\yo-2)^2}-\frac{160 d_1'}{3 \
(\yo-2)^3}-\frac{40 d_1'}{(\yo-2)^4}-\frac{32 d_1'}{3 (\yo-2)^5}\Big) \
H(c_2(\ao),1;\yo)+\Big(\frac{\ao^4}{2}+\frac{2 \ao^3}{3}-\ao^2-2 \ao-\
\frac{80}{3 (\yo-2)^2}-\frac{160}{3 \
(\yo-2)^3}-\frac{40}{(\yo-2)^4}-\frac{32}{3 \
(\yo-2)^5}+\frac{11}{6}\Big) H(c_2(\ao),c_1(\ao);\yo)-\frac{1}{6} \
H(0,0,0;Y)-\frac{32}{3} H(0,0,0;\yo)-\frac{8}{3} d_1' \
H(0,0,1;\yo)+\frac{8}{3} H(0,0,c_1(\ao);\yo)-\frac{1}{6} \
H(0,1,0;Y)+\Big(\frac{4 d_1}{3}-\frac{8 d_1'}{3}-\frac{4}{3}\Big) \
H(0,1,0;\yo)-\frac{2}{3} d_1'^2 H(0,1,1;\yo)+\Big(-\frac{4 \
d_1}{3}+\frac{2 d_1'}{3}+\frac{4}{3}\Big) \
H(0,1,c_1(\ao);\yo)+\frac{4}{3} H(0,c_1(\ao),0;\yo)+\frac{2}{3} d_1' \
H(0,c_1(\ao),1;\yo)+\frac{2}{3} \
H(0,c_1(\ao),c_1(\ao);\yo)-\frac{1}{6} H(1,0,0;Y)-4 \
H(1,0,0;\yo)-\frac{2}{3} d_1' H(1,0,1;\yo)+2 \
H(1,0,c_1(\ao);\yo)-\frac{1}{6} H(1,1,0;Y)+\Big(\frac{4 \
d_1}{3}-\frac{2 d_1'}{3}-\frac{2}{3}\Big) H(1,1,0;\yo)+\Big(-\frac{4 \
d_1}{3}+\frac{2 d_1'}{3}+\frac{2}{3}\Big) \
H(1,1,c_1(\ao);\yo)+\frac{4}{3} H(1,c_1(\ao),0;\yo)+\frac{2}{3} d_1' \
H(1,c_1(\ao),1;\yo)+\frac{2}{3} H(1,c_1(\ao),c_1(\ao);\yo)+H(0;\yo) \
\Big(\frac{\yo^3 \ao^4}{9}-\frac{\yo^2 \ao^4}{6}+\frac{2 \yo \
\ao^4}{3}-\frac{\ao^4}{3 (\yo-2)}-\frac{\ao^4}{6}-\frac{16 \yo^3 \
\ao^3}{27}+\frac{4 \yo^2 \ao^3}{3}-\frac{16 \yo \ao^3}{9}-\frac{2 \
\ao^3}{\yo-2}-\frac{8 \ao^3}{9 (\yo-2)^2}-\frac{7 \ao^3}{9}+\frac{4 \
\yo^3 \ao^2}{3}-4 \yo^2 \ao^2+\frac{10 \yo \ao^2}{3}-\frac{16 \
\ao^2}{3 (\yo-2)}-\frac{22 \ao^2}{3 (\yo-2)^2}-\frac{8 \ao^2}{3 \
(\yo-2)^3}-\frac{7 \ao^2}{6}-\frac{16 \yo^3 \ao}{9}+\frac{20 \yo^2 \
\ao}{3}-8 \yo \ao-\frac{26 \ao}{3 (\yo-2)}-\frac{36 \
\ao}{(\yo-2)^2}-\frac{104 \ao}{3 (\yo-2)^3}-\frac{32 \ao}{3 \
(\yo-2)^4}+\ao+\frac{2 d_1' \yo^3}{9}-\frac{139 \yo^3}{54}-\frac{7 \
d_1' \yo^2}{6}+\frac{457 \yo^2}{36}-\frac{41 \
d_1}{36}-\frac{d_1'}{18}+\frac{11 d_1' \yo}{3}-33 \yo+\Big(\frac{4 \
\yo^3}{9}-2 \yo^2+4 \yo-\frac{4}{3 (\yo-1)}-\frac{160}{3 \
(\yo-2)^2}-\frac{320}{3 (\yo-2)^3}-\frac{80}{(\yo-2)^4}-\frac{64}{3 (\
\yo-2)^5}+3\Big) H(0;\ao)+\frac{2}{3} \yo^3 H(0;Y)-3 \yo^2 H(0;Y)+6 \
\yo H(0;Y)+\frac{2 H(0;Y)}{3 (\yo-1)}+\frac{80 H(0;Y)}{3 \
(\yo-2)^2}+\frac{160 H(0;Y)}{3 (\yo-2)^3}+\frac{40 \
H(0;Y)}{(\yo-2)^4}+\frac{32 H(0;Y)}{3 (\yo-2)^5}+\frac{5}{18} H(0;Y)-\
\frac{2}{3} H(0,0;Y)-\frac{2}{3} H(1,0;Y)-\frac{8 d_1'}{3 \
(\yo-2)}+\frac{20}{3 (\yo-2)}+\frac{4 d_1}{3 (\yo-1)}-\frac{2 d_1'}{3 \
(\yo-1)}-\frac{7}{2 (\yo-1)}+\frac{16 d_1}{(\yo-2)^2}-\frac{12 \
d_1'}{(\yo-2)^2}-\frac{88}{(\yo-2)^2}+\frac{128 d_1}{9 \
(\yo-2)^3}-\frac{88 d_1'}{9 (\yo-2)^3}-\frac{168}{(\yo-2)^3}+\frac{4 \
d_1}{(\yo-2)^4}-\frac{8 d_1'}{3 \
(\yo-2)^4}-\frac{116}{(\yo-2)^4}-\frac{256}{9 (\yo-2)^5}-\frac{320 \
\ln 2\, }{3 (\yo-2)^2}-\frac{640 \ln 2\, }{3 (\yo-2)^3}-\frac{160 \ln \
2\, }{(\yo-2)^4}-\frac{128 \ln 2\, }{3 (\yo-2)^5}+\frac{26 \ln 2\, \
}{3}-\frac{\pi ^2}{9}+\frac{361}{108}\Big)+H(2;\yo) \Big(-\frac{160 \
\ln 2\,  d_1}{3 (\yo-2)^2}-\frac{320 \ln 2\,  d_1}{3 \
(\yo-2)^3}-\frac{80 \ln 2\,  d_1}{(\yo-2)^4}-\frac{64 \ln 2\,  d_1}{3 \
(\yo-2)^5}+\Big(-\frac{160 d_1}{3 (\yo-2)^2}-\frac{320 d_1}{3 \
(\yo-2)^3}-\frac{80 d_1}{(\yo-2)^4}-\frac{64 d_1}{3 \
(\yo-2)^5}+\frac{8 d_1'}{3}+\frac{160}{3 (\yo-2)^2}+\frac{320}{3 \
(\yo-2)^3}+\frac{80}{(\yo-2)^4}+\frac{64}{3 \
(\yo-2)^5}-\frac{26}{3}\Big) H(0;\ao)+\frac{8}{3} d_1' \ln 2\, \
+\frac{160 \ln 2\, }{3 (\yo-2)^2}+\frac{320 \ln 2\, }{3 \
(\yo-2)^3}+\frac{80 \ln 2\, }{(\yo-2)^4}+\frac{64 \ln 2\, }{3 \
(\yo-2)^5}-\frac{26 \ln 2\, }{3}\Big)+H(1;\yo) \Big(\frac{1}{18} d_1' \
\yo^3 \ao^4-\frac{1}{12} d_1' \yo^2 \ao^4-\frac{17 d_1' \
\ao^4}{36}+\frac{1}{3} d_1' \yo \ao^4-\frac{d_1' \ao^4}{6 \
(\yo-2)}-\frac{8}{27} d_1' \yo^3 \ao^3+\frac{2}{3} d_1' \yo^2 \
\ao^3-\frac{d_1' \ao^3}{27}-\frac{8}{9} d_1' \yo \ao^3-\frac{d_1' \
\ao^3}{\yo-2}-\frac{4 d_1' \ao^3}{9 (\yo-2)^2}+\frac{2}{3} d_1' \yo^3 \
\ao^2-2 d_1' \yo^2 \ao^2-\frac{2 d_1' \ao^2}{3}+\frac{5}{3} d_1' \yo \
\ao^2-\frac{8 d_1' \ao^2}{3 (\yo-2)}-\frac{11 d_1' \ao^2}{3 \
(\yo-2)^2}-\frac{4 d_1' \ao^2}{3 (\yo-2)^3}-\frac{8}{9} d_1' \yo^3 \
\ao+\frac{10}{3} d_1' \yo^2 \ao+\frac{23 d_1' \ao}{9}-4 d_1' \yo \ao-\
\frac{13 d_1' \ao}{3 (\yo-2)}-\frac{18 d_1' \ao}{(\yo-2)^2}-\frac{52 \
d_1' \ao}{3 (\yo-2)^3}-\frac{16 d_1' \ao}{3 (\yo-2)^4}+\frac{2 d_1'^2 \
\yo^3}{27}-\frac{19 d_1' \yo^3}{27}-\frac{49 d_1'^2}{54}-\frac{7 \
d_1'^2 \yo^2}{18}+\frac{11 d_1' \yo^2}{3}+\frac{199 \
d_1'}{27}+\frac{11 d_1'^2 \yo}{9}-\frac{31 d_1' \yo}{3}+H(0;\ao) \
\Big(\frac{2 d_1' \yo^3}{9}-\frac{2 \yo^3}{9}-d_1' \yo^2+\yo^2+2 d_1' \
\yo-2 \yo+\frac{13 d_1'}{9}+\frac{2}{3} H(0;Y)+\frac{4 d_1}{3 \
(\yo-1)}-\frac{2 d_1'}{3 (\yo-1)}-\frac{2}{3 (\yo-1)}-\frac{80 \
d_1'}{3 (\yo-2)^2}+\frac{80}{3 (\yo-2)^2}-\frac{160 d_1'}{3 \
(\yo-2)^3}+\frac{160}{3 (\yo-2)^3}-\frac{40 \
d_1'}{(\yo-2)^4}+\frac{40}{(\yo-2)^4}-\frac{32 d_1'}{3 \
(\yo-2)^5}+\frac{32}{3 (\yo-2)^5}-\frac{107}{18}\Big)+\frac{2}{9} \
d_1' \yo^3 H(0;Y)-d_1' \yo^2 H(0;Y)-\frac{11}{9} d_1' H(0;Y)+2 d_1' \
\yo H(0;Y)+\frac{4}{3} H(0,0;\ao)+\frac{4}{3} d_1 \
H(0,1;\ao)+\frac{8}{3} d_1' \ln 2\, -\frac{80 d_1' \ln 2\, }{3 \
(\yo-2)^2}-\frac{160 d_1' \ln 2\, }{3 (\yo-2)^3}-\frac{40 d_1' \ln \
2\, }{(\yo-2)^4}-\frac{32 d_1' \ln 2\, }{3 (\yo-2)^5}-\frac{\pi \
^2}{9}\Big)+\frac{\pi ^2}{9 (\yo-1)}+\frac{20 \pi ^2}{9 \
(\yo-2)^2}+\frac{40 \pi ^2}{9 (\yo-2)^3}+\frac{10 \pi ^2}{3 \
(\yo-2)^4}+\frac{8 \pi ^2}{9 (\yo-2)^5}-\zeta_3+\frac{80 \ln ^22\, \
}{3 (\yo-2)^2}+\frac{160 \ln ^22\, }{3 (\yo-2)^3}+\frac{40 \ln ^22\, \
}{(\yo-2)^4}+\frac{32 \ln ^22\, }{3 (\yo-2)^5}-\frac{13 \ln ^22\, \
}{6}+\frac{89}{36} d_1 \ln 2\, -\frac{11}{18} d_1' \ln 2\, -\frac{80 \
H(0;Y) \ln 2\, }{3 (\yo-2)^2}-\frac{160 H(0;Y) \ln 2\, }{3 \
(\yo-2)^3}-\frac{40 H(0;Y) \ln 2\, }{(\yo-2)^4}-\frac{32 H(0;Y) \ln 2\
\, }{3 (\yo-2)^5}+\frac{13}{6} H(0;Y) \ln 2\, +\frac{8 d_1' \ln 2\, \
}{3 (\yo-2)}-\frac{52 \ln 2\, }{3 (\yo-2)}-\frac{16 d_1 \ln 2\, \
}{(\yo-2)^2}+\frac{12 d_1' \ln 2\, }{(\yo-2)^2}+\frac{460 \ln 2\, }{9 \
(\yo-2)^2}-\frac{128 d_1 \ln 2\, }{9 (\yo-2)^3}+\frac{88 d_1' \ln 2\, \
}{9 (\yo-2)^3}+\frac{400 \ln 2\, }{3 (\yo-2)^3}-\frac{4 d_1 \ln 2\, \
}{(\yo-2)^4}+\frac{8 d_1' \ln 2\, }{3 (\yo-2)^4}+\frac{316 \ln 2\, \
}{3 (\yo-2)^4}+\frac{256 \ln 2\, }{9 (\yo-2)^5}-\frac{377 \ln 2\, \
}{36}+\frac{\pi ^2}{216}+\frac{160}{81}.
\erp

%
% The J*I integral for k=-1
%

\subsection{The $\cJI$ integral for $k=-1$}
%
% This file contains the TeX output produced by Mathematica for the integral JI for arbitrary kap=0  and D0 = 3 +d'1 ep
%
The $\eps$ expansion for this integral reads
\beq
\cJI(Y,\ep;y_{0},d'_{0},\alpha_{0},d_{0};-1)=\frac{1}{\eps^4}\cji_{-4}^{(-1)}+\frac{1}{\eps^3}\cji_{-3}^{(-1)}+\frac{1}{\eps^2}\cji_{-2}^{(-1)}+\frac{1}{\eps}\cji_{-1}^{(-1)}+\cji_0^{(-1)}+\ocal\left(\eps\right),
\eeq
where
%% 1/ep^4
\brp
\cji_{-4}^{(-1)}=-\frac{1}{4},
\erp
%% 1/ep^3
\brp
\cji_{-3}^{(-1)}=-\frac{\yo^3}{3}+\frac{3 \yo^2}{2}-3 \yo+\frac{1}{4} H(0;Y)+H(0;\yo),
\erp
%% 1/ep^2
\brp
\cji_{-2}^{(-1)}=\frac{d_1' \yo^3}{9}+\frac{1}{3} H(0;Y) \yo^3-\frac{4 \
\yo^3}{9}-\frac{7 d_1' \yo^2}{12}-\frac{3}{2} H(0;Y) \yo^2+3 \
\yo^2+\frac{11 d_1' \yo}{6}+3 H(0;Y) \yo-12 \yo+\Big(\frac{4 \
\yo^3}{3}-6 \yo^2+12 \yo-H(0;Y)\Big) H(0;\yo)+\Big(\frac{d_1' \
\yo^3}{3}-\frac{3 d_1' \yo^2}{2}+3 d_1' \yo-\frac{11 d_1'}{6}\Big) \
H(1;\yo)-\frac{1}{4} H(0,0;Y)-4 H(0,0;\yo)-d_1' \
H(0,1;\yo)-\frac{1}{4} H(1,0;Y)-\frac{\pi ^2}{24},
\erp
%% 1/ep
\brp
\cji_{-1}^{(-1)} =\frac{\yo^3 \ao^4}{18}-\frac{5 \yo^2 \ao^4}{12}+\frac{5 \yo \
\ao^4}{3}-\frac{13 \yo^3 \ao^3}{54}+\frac{17 \yo^2 \ao^3}{9}-\frac{80 \
\yo \ao^3}{9}+\frac{11 \yo^3 \ao^2}{36}-\frac{109 \yo^2 \
\ao^2}{36}+\frac{158 \yo \ao^2}{9}+\frac{7 \yo^3 \ao}{18}+\frac{7 \
\yo^2 \ao}{18}-\frac{43 \yo \ao}{3}-\frac{d_1'^2 \yo^3}{27}+\frac{8 \
d_1' \yo^3}{27}+\frac{\pi ^2 \yo^3}{18}-\frac{14 \yo^3}{27}+\frac{17 \
d_1'^2 \yo^2}{72}-\frac{22 d_1' \yo^2}{9}-\frac{\pi ^2 \
\yo^2}{4}+\frac{21 \yo^2}{4}-\frac{49 d_1'^2 \yo}{36}+\frac{151 d_1' \
\yo}{9}+\frac{\pi ^2 \yo}{2}-42 \yo+\Big(-\frac{7 \yo^3}{6}+\frac{13 \
\yo^2}{3}-\frac{11 \yo}{2}+\frac{13}{6 (\yo-1)}+\frac{13}{6}\Big) \
H(0;\ao)-\frac{1}{9} d_1' \yo^3 H(0;Y)+\frac{4}{9} \yo^3 \
H(0;Y)+\frac{7}{12} d_1' \yo^2 H(0;Y)-3 \yo^2 H(0;Y)-\frac{11}{6} \
d_1' \yo H(0;Y)+12 \yo H(0;Y)+\frac{1}{24} \pi ^2 H(0;Y)+\frac{1}{24} \
\pi ^2 H(1;Y)+\Big(-\frac{1}{9} d_1'^2 \yo^3+\frac{4 d_1' \
\yo^3}{9}-\frac{1}{3} d_1' H(0;Y) \yo^3+\frac{7 d_1'^2 \yo^2}{12}-3 \
d_1' \yo^2+\frac{3}{2} d_1' H(0;Y) \yo^2-\frac{11 d_1'^2 \yo}{6}+12 \
d_1' \yo-3 d_1' H(0;Y) \yo+\frac{49 d_1'^2}{36}-\frac{85 \
d_1'}{9}+\Big(\frac{2 \yo^3}{3}-3 \yo^2+6 \
\yo+\frac{2}{\yo-1}-\frac{19}{3}\Big) H(0;\ao)+\frac{11}{6} d_1' \
H(0;Y)+\frac{\pi ^2}{3}\Big) H(1;\yo)+\Big(-\frac{1}{6} \yo^3 \
\ao^4+\yo^2 \ao^4-\frac{5 \yo \ao^4}{2}+\frac{5 \ao^4}{3}+\frac{8 \
\yo^3 \ao^3}{9}-5 \yo^2 \ao^3+\frac{38 \yo \ao^3}{3}-\frac{68 \
\ao^3}{9}-2 \yo^3 \ao^2+\frac{21 \yo^2 \ao^2}{2}-26 \yo \
\ao^2+\frac{38 \ao^2}{3}+\frac{8 \yo^3 \ao}{3}-13 \yo^2 \ao+30 \yo \
\ao-\frac{41 \ao}{3}-\frac{25 \yo^3}{18}+\frac{35 \yo^2}{6}-\frac{23 \
\yo}{2}+\frac{13}{6 (\yo-1)}+\frac{13}{6}\Big) \
H(c_1(\ao);\yo)-\frac{1}{3} \yo^3 H(0,0;Y)+\frac{3}{2} \yo^2 \
H(0,0;Y)-3 \yo H(0,0;Y)+\Big(-\frac{16 \yo^3}{3}+24 \yo^2-48 \yo+4 \
H(0;Y)\Big) H(0,0;\yo)+\Big(-\frac{4 d_1' \yo^3}{3}+6 d_1' \yo^2-12 \
d_1' \yo-4 H(0;\ao)+d_1' H(0;Y)\Big) H(0,1;\yo)+\Big(\ao^4-\frac{16 \
\ao^3}{3}+12 \ao^2-16 \ao+\frac{2 \yo^3}{3}-3 \yo^2+6 \
\yo+\frac{2}{\yo-1}+2\Big) H(0,c_1(\ao);\yo)-\frac{1}{3} \yo^3 \
H(1,0;Y)+\frac{3}{2} \yo^2 H(1,0;Y)-3 \yo H(1,0;Y)+H(0;\yo) \
\Big(-\frac{4 d_1' \yo^3}{9}-\frac{4}{3} H(0;Y) \yo^3+\frac{53 \
\yo^3}{18}+\frac{7 d_1' \yo^2}{3}+6 H(0;Y) \yo^2-\frac{49 \
\yo^2}{3}-\frac{22 d_1' \yo}{3}-12 H(0;Y) \yo+\frac{107 \
\yo}{2}+H(0,0;Y)+H(1,0;Y)-\frac{13}{6 (\yo-1)}+\frac{\pi \
^2}{6}-\frac{13}{6}\Big)+\Big(-\frac{4 d_1' \yo^3}{3}-\frac{2 \
\yo^3}{3}+6 d_1' \yo^2+3 \yo^2-12 d_1' \yo-6 \yo+\frac{22 \
d_1'}{3}-\frac{2}{\yo-1}+\frac{19}{3}\Big) \
H(1,0;\yo)+\Big(-\frac{1}{3} d_1'^2 \yo^3+\frac{3 d_1'^2 \yo^2}{2}-3 \
d_1'^2 \yo+\frac{11 d_1'^2}{6}-2 H(0;\ao)\Big) \
H(1,1;\yo)+\Big(\frac{2 \yo^3}{3}-3 \yo^2+6 \
\yo+\frac{2}{\yo-1}-\frac{19}{3}\Big) H(1,c_1(\ao);\yo)+\Big(2 \
\ao-\frac{2}{\yo-1}-2\Big) H(c_1(\ao),c_1(\ao);\yo)+\frac{1}{4} \
H(0,0,0;Y)+16 H(0,0,0;\yo)+4 d_1' H(0,0,1;\yo)-4 H(0,0,c_1(\ao);\yo)+\
\frac{1}{4} H(0,1,0;Y)+(4 d_1'+4) H(0,1,0;\yo)+d_1'^2 H(0,1,1;\yo)-4 \
H(0,1,c_1(\ao);\yo)+2 H(0,c_1(\ao),c_1(\ao);\yo)+\frac{1}{4} \
H(1,0,0;Y)-2 H(1,0,c_1(\ao);\yo)+\frac{1}{4} H(1,1,0;Y)+2 \
H(1,1,0;\yo)-2 H(1,1,c_1(\ao);\yo)+2 \
H(1,c_1(\ao),c_1(\ao);\yo)-\frac{\pi ^2}{3 (\yo-1)}+\frac{3 \
\zeta_3}{2}-\frac{\pi ^2}{3},
\erp
%% ep^0
\brp
\cji_{0}^{(-1)} =-\frac{1}{36} d_1 \yo^3 \ao^4-\frac{1}{27} d_1' \yo^3 \ao^4+\frac{13 \
\yo^3 \ao^4}{108}+\frac{5}{24} d_1 \yo^2 \ao^4+\frac{13}{36} d_1' \
\yo^2 \ao^4-\frac{7 \yo^2 \ao^4}{6}-\frac{5}{6} d_1 \yo \
\ao^4-\frac{47}{18} d_1' \yo \ao^4+\frac{97 \yo \
\ao^4}{12}-\frac{1}{18} \yo^3 H(0;Y) \ao^4+\frac{5}{12} \yo^2 H(0;Y) \
\ao^4-\frac{5}{3} \yo H(0;Y) \ao^4+\frac{31}{324} d_1 \yo^3 \
\ao^3+\frac{29}{162} d_1' \yo^3 \ao^3-\frac{185 \yo^3 \
\ao^3}{324}-\frac{91}{108} d_1 \yo^2 \ao^3-\frac{187}{108} d_1' \yo^2 \
\ao^3+\frac{631 \yo^2 \ao^3}{108}+\frac{245}{54} d_1 \yo \
\ao^3+\frac{1549}{108} d_1' \yo \ao^3-\frac{5095 \yo \
\ao^3}{108}+\frac{13}{54} \yo^3 H(0;Y) \ao^3-\frac{17}{9} \yo^2 \
H(0;Y) \ao^3+\frac{80}{9} \yo H(0;Y) \ao^3+\frac{17}{216} d_1 \yo^3 \
\ao^2-\frac{35}{108} d_1' \yo^3 \ao^2+\frac{179 \yo^3 \
\ao^2}{216}+\frac{17}{27} d_1 \yo^2 \ao^2+\frac{707}{216} d_1' \yo^2 \
\ao^2-\frac{589 \yo^2 \ao^2}{54}-\frac{895}{108} d_1 \yo \
\ao^2-\frac{809}{27} d_1' \yo \ao^2+\frac{2836 \yo \
\ao^2}{27}-\frac{11}{36} \yo^3 H(0;Y) \ao^2+\frac{109}{36} \yo^2 \
H(0;Y) \ao^2-\frac{158}{9} \yo H(0;Y) \ao^2-\frac{205}{108} d_1 \yo^3 \
\ao+\frac{1}{6} d_1' \yo^3 \ao+\frac{95 \yo^3 \
\ao}{54}+\frac{659}{108} d_1 \yo^2 \ao-\frac{74}{27} d_1' \yo^2 \
\ao+\frac{187 \yo^2 \ao}{108}+\frac{d_1 \yo \ao}{36}+\frac{278 d_1' \
\yo \ao}{9}-\frac{219 \yo \ao}{2}-\frac{7}{18} \yo^3 H(0;Y) \
\ao-\frac{7}{18} \yo^2 H(0;Y) \ao+\frac{43}{3} \yo H(0;Y) \
\ao+\frac{d_1'^3 \yo^3}{81}-\frac{4 d_1'^2 \yo^3}{27}+\frac{14 d_1' \
\yo^3}{27}-\frac{1}{54} d_1' \pi ^2 \yo^3+\frac{29 \pi ^2 \
\yo^3}{108}-\frac{46 \yo^3}{81}-\frac{43 d_1'^3 \yo^2}{432}+\frac{167 \
d_1'^2 \yo^2}{108}-\frac{1435 d_1' \yo^2}{216}+\frac{7}{72} d_1' \pi \
^2 \yo^2-\frac{11 \pi ^2 \yo^2}{9}+\frac{69 \yo^2}{8}+\frac{251 \
d_1'^3 \yo}{216}-\frac{542 d_1'^2 \yo}{27}+\frac{10367 d_1' \
\yo}{108}-\frac{11}{36} d_1' \pi ^2 \yo+\frac{35 \pi ^2 \yo}{12}-138 \
\yo+\frac{1}{27} d_1'^2 \yo^3 H(0;Y)-\frac{8}{27} d_1' \yo^3 \
H(0;Y)-\frac{1}{18} \pi ^2 \yo^3 H(0;Y)+\frac{14}{27} \yo^3 \
H(0;Y)-\frac{17}{72} d_1'^2 \yo^2 H(0;Y)+\frac{22}{9} d_1' \yo^2 \
H(0;Y)+\frac{1}{4} \pi ^2 \yo^2 H(0;Y)-\frac{21}{4} \yo^2 \
H(0;Y)+\frac{49}{36} d_1'^2 \yo H(0;Y)-\frac{151}{9} d_1' \yo H(0;Y)-\
\frac{1}{2} \pi ^2 \yo H(0;Y)+42 \yo H(0;Y)+\frac{\pi ^2 H(0;Y)}{3 \
(\yo-1)}+\frac{1}{3} \pi ^2 H(0;Y)+H(0;\ao) \Big(-\frac{1}{9} \yo^3 \
\ao^4+\frac{5 \yo^2 \ao^4}{6}-\frac{10 \yo \ao^4}{3}+\frac{13 \yo^3 \
\ao^3}{27}-\frac{34 \yo^2 \ao^3}{9}+\frac{160 \yo \ao^3}{9}-\frac{11 \
\yo^3 \ao^2}{18}+\frac{109 \yo^2 \ao^2}{18}-\frac{316 \yo \
\ao^2}{9}-\frac{7 \yo^3 \ao}{9}-\frac{7 \yo^2 \ao}{9}+\frac{86 \yo \
\ao}{3}+\frac{205 d_1 \yo^3}{108}+\frac{17 d_1' \yo^3}{54}-\frac{35 \
\yo^3}{9}-\frac{22 d_1 \yo^2}{3}-\frac{10 d_1' \yo^2}{9}+\frac{649 \
\yo^2}{36}-\frac{217 d_1}{36}+\frac{d_1'}{6}+\frac{469 d_1 \
\yo}{36}-\frac{7 d_1' \yo}{18}-\frac{299 \yo}{9}+\frac{7}{6} \yo^3 \
H(0;Y)-\frac{13}{3} \yo^2 H(0;Y)+\frac{11}{2} \yo H(0;Y)-\frac{13 \
H(0;Y)}{6 (\yo-1)}-\frac{13}{6} H(0;Y)-\frac{217 d_1}{36 \
(\yo-1)}+\frac{d_1'}{6 (\yo-1)}+\frac{149}{18 (\yo-1)}+\frac{149}{18}\
\Big)+\Big(-\frac{1}{9} d_1 \yo^3 \ao^4+\frac{5}{6} d_1 \yo^2 \
\ao^4-\frac{10}{3} d_1 \yo \ao^4+\frac{13}{27} d_1 \yo^3 \
\ao^3-\frac{34}{9} d_1 \yo^2 \ao^3+\frac{160}{9} d_1 \yo \
\ao^3-\frac{11}{18} d_1 \yo^3 \ao^2+\frac{109}{18} d_1 \yo^2 \
\ao^2-\frac{316}{9} d_1 \yo \ao^2-\frac{7}{9} d_1 \yo^3 \
\ao-\frac{7}{9} d_1 \yo^2 \ao+\frac{86 d_1 \yo \ao}{3}+\frac{55 d_1 \
\yo^3}{54}-\frac{7 d_1 \yo^2}{3}-8 d_1 \yo\Big) H(1;\ao)+\frac{1}{18} \
\pi ^2 \yo^3 H(1;Y)-\frac{1}{4} \pi ^2 \yo^2 H(1;Y)+\frac{1}{2} \pi \
^2 \yo H(1;Y)+\Big(\frac{1}{12} d_1 \yo^3 \ao^4+\frac{1}{18} d_1' \
\yo^3 \ao^4-\frac{\yo^3 \ao^4}{4}-\frac{1}{2} d_1 \yo^2 \
\ao^4-\frac{5}{12} d_1' \yo^2 \ao^4+\frac{7 \yo^2 \ao^4}{4}-\frac{5 \
d_1 \ao^4}{6}-\frac{47 d_1' \ao^4}{36}+\frac{5}{4} d_1 \yo \
\ao^4+\frac{5}{3} d_1' \yo \ao^4-\frac{25 \yo \ao^4}{4}+\frac{1}{6} \
\yo^3 H(0;Y) \ao^4-\yo^2 H(0;Y) \ao^4+\frac{5}{2} \yo H(0;Y) \
\ao^4-\frac{5}{3} H(0;Y) \ao^4+\frac{19 \ao^4}{4}-\frac{13}{27} d_1 \
\yo^3 \ao^3-\frac{8}{27} d_1' \yo^3 \ao^3+\frac{77 \yo^3 \
\ao^3}{54}+\frac{8}{3} d_1 \yo^2 \ao^3+\frac{37}{18} d_1' \yo^2 \
\ao^3-\frac{29 \yo^2 \ao^3}{3}+\frac{221 d_1 \ao^3}{54}+\frac{313 \
d_1' \ao^3}{54}-\frac{61}{9} d_1 \yo \ao^3-\frac{77}{9} d_1' \yo \
\ao^3+\frac{331 \yo \ao^3}{9}-\frac{8}{9} \yo^3 H(0;Y) \ao^3+5 \yo^2 \
H(0;Y) \ao^3-\frac{38}{3} \yo H(0;Y) \ao^3+\frac{68}{9} H(0;Y) \ao^3-\
\frac{676 \ao^3}{27}+\frac{23}{18} d_1 \yo^3 \ao^2+\frac{2}{3} d_1' \
\yo^3 \ao^2-\frac{131 \yo^3 \ao^2}{36}-\frac{13}{2} d_1 \yo^2 \
\ao^2-\frac{17}{4} d_1' \yo^2 \ao^2+\frac{283 \yo^2 \
\ao^2}{12}-\frac{287 d_1 \ao^2}{36}-\frac{28 d_1' \
\ao^2}{3}+\frac{95}{6} d_1 \yo \ao^2+\frac{35}{2} d_1' \yo \
\ao^2-\frac{533 \yo \ao^2}{6}+2 \yo^3 H(0;Y) \ao^2-\frac{21}{2} \yo^2 \
H(0;Y) \ao^2+26 \yo H(0;Y) \ao^2-\frac{38}{3} H(0;Y) \ao^2+\frac{149 \
\ao^2}{3}-\frac{25}{9} d_1 \yo^3 \ao-\frac{8}{9} d_1' \yo^3 \
\ao+\frac{121 \yo^3 \ao}{18}+13 d_1 \yo^2 \ao+\frac{31}{6} d_1' \yo^2 \
\ao-\frac{241 \yo^2 \ao}{6}+\frac{343 d_1 \ao}{18}+\frac{47 d_1' \
\ao}{6}-\frac{85 d_1 \yo \ao}{3}-\frac{59 d_1' \yo \ao}{3}+\frac{404 \
\yo \ao}{3}-\frac{8}{3} \yo^3 H(0;Y) \ao+13 \yo^2 H(0;Y) \ao-30 \yo \
H(0;Y) \ao+\frac{41}{3} H(0;Y) \ao-\frac{143 \ao}{2}+\frac{205 d_1 \
\yo^3}{108}+\frac{25 d_1' \yo^3}{54}-\frac{115 \yo^3}{27}-\frac{22 \
d_1 \yo^2}{3}-\frac{7 d_1' \yo^2}{3}+\frac{196 \yo^2}{9}-\frac{217 \
d_1}{36}+\frac{d_1'}{6}+\frac{469 d_1 \yo}{36}+8 d_1' \yo-\frac{569 \
\yo}{9}+\Big(\frac{\yo^3 \ao^4}{3}-2 \yo^2 \ao^4+5 \yo \ao^4-\frac{10 \
\ao^4}{3}-\frac{16 \yo^3 \ao^3}{9}+10 \yo^2 \ao^3-\frac{76 \yo \
\ao^3}{3}+\frac{136 \ao^3}{9}+4 \yo^3 \ao^2-21 \yo^2 \ao^2+52 \yo \
\ao^2-\frac{76 \ao^2}{3}-\frac{16 \yo^3 \ao}{3}+26 \yo^2 \ao-60 \yo \
\ao+\frac{82 \ao}{3}+\frac{25 \yo^3}{9}-\frac{35 \yo^2}{3}+23 \
\yo-\frac{13}{3 (\yo-1)}-\frac{13}{3}\Big) H(0;\ao)+\frac{25}{18} \
\yo^3 H(0;Y)-\frac{35}{6} \yo^2 H(0;Y)+\frac{23}{2} \yo \
H(0;Y)-\frac{13 H(0;Y)}{6 (\yo-1)}-\frac{13}{6} \
H(0;Y)+\Big(\frac{1}{3} d_1 \yo^3 \ao^4-2 d_1 \yo^2 \ao^4-\frac{10 \
d_1 \ao^4}{3}+5 d_1 \yo \ao^4-\frac{16}{9} d_1 \yo^3 \ao^3+10 d_1 \
\yo^2 \ao^3+\frac{136 d_1 \ao^3}{9}-\frac{76}{3} d_1 \yo \ao^3+4 d_1 \
\yo^3 \ao^2-21 d_1 \yo^2 \ao^2-\frac{76 d_1 \ao^2}{3}+52 d_1 \yo \
\ao^2-\frac{16}{3} d_1 \yo^3 \ao+26 d_1 \yo^2 \ao+\frac{82 d_1 \
\ao}{3}-60 d_1 \yo \ao+\frac{25 d_1 \yo^3}{9}-\frac{35 d_1 \yo^2}{3}-\
\frac{13 d_1}{3}+23 d_1 \yo-\frac{13 d_1}{3 (\yo-1)}\Big) \
H(1;\ao)-\frac{217 d_1}{36 (\yo-1)}+\frac{d_1'}{6 \
(\yo-1)}+\frac{149}{18 (\yo-1)}+\frac{149}{18}\Big) \
H(c_1(\ao);\yo)+\Big(\frac{7 \yo^3}{3}-\frac{26 \yo^2}{3}+11 \
\yo-\frac{13}{3 (\yo-1)}-\frac{13}{3}\Big) H(0,0;\ao)+\frac{1}{9} \
d_1' \yo^3 H(0,0;Y)-\frac{4}{9} \yo^3 H(0,0;Y)-\frac{7}{12} d_1' \
\yo^2 H(0,0;Y)+3 \yo^2 H(0,0;Y)+\frac{11}{6} d_1' \yo H(0,0;Y)-12 \yo \
H(0,0;Y)-\frac{1}{24} \pi ^2 H(0,0;Y)+\Big(\frac{7 d_1 \
\yo^3}{3}-\frac{26 d_1 \yo^2}{3}+11 d_1 \yo-\frac{13 d_1}{3}-\frac{13 \
d_1}{3 (\yo-1)}\Big) H(0,1;\ao)-\frac{1}{24} \pi ^2 \
H(0,1;Y)+\Big(-\frac{d_1 \ao^4}{2}-H(0;Y) \
\ao^4+\frac{\ao^4}{2}+\frac{26 d_1 \ao^3}{9}+\frac{16}{3} H(0;Y) \
\ao^3-\frac{38 \ao^3}{9}-\frac{23 d_1 \ao^2}{3}-12 H(0;Y) \
\ao^2+\frac{49 \ao^2}{3}+\frac{50 d_1 \ao}{3}+16 H(0;Y) \ao-\frac{142 \
\ao}{3}-\frac{2 d_1' \yo^3}{9}+\frac{37 \yo^3}{18}+\frac{7 d_1' \
\yo^2}{6}-\frac{31 \yo^2}{3}-4 d_1+2 d_1'-\frac{11 d_1' \
\yo}{3}+\frac{59 \yo}{2}+\Big(-2 \ao^4+\frac{32 \ao^3}{3}-24 \ao^2+32 \
\ao-\frac{4 \yo^3}{3}+6 \yo^2-12 \yo-\frac{4}{\yo-1}-4\Big) H(0;\ao)-\
\frac{2}{3} \yo^3 H(0;Y)+3 \yo^2 H(0;Y)-6 \yo H(0;Y)-\frac{2 H(0;Y)}{\
\yo-1}-2 H(0;Y)+\Big(-2 d_1 \ao^4+\frac{32 d_1 \ao^3}{3}-24 d_1 \
\ao^2+32 d_1 \ao-\frac{4 d_1 \yo^3}{3}+6 d_1 \yo^2-4 d_1-12 d_1 \
\yo-\frac{4 d_1}{\yo-1}\Big) H(1;\ao)-\frac{4 d_1}{\yo-1}+\frac{2 \
d_1'}{\yo-1}+\frac{11}{6 (\yo-1)}+\frac{11}{6}\Big) \
H(0,c_1(\ao);\yo)+H(0,0;\yo) \Big(\frac{16 d_1' \
\yo^3}{9}+\frac{16}{3} H(0;Y) \yo^3-\frac{127 \yo^3}{9}-\frac{28 d_1' \
\yo^2}{3}-24 H(0;Y) \yo^2+74 \yo^2+\frac{88 d_1' \yo}{3}+48 H(0;Y) \
\yo-225 \yo-4 H(0,0;Y)-4 H(1,0;Y)+\frac{13}{\yo-1}-\frac{2 \pi \
^2}{3}+13\Big)+\frac{1}{9} d_1' \yo^3 H(1,0;Y)-\frac{4}{9} \yo^3 \
H(1,0;Y)-\frac{7}{12} d_1' \yo^2 H(1,0;Y)+3 \yo^2 \
H(1,0;Y)+\frac{11}{6} d_1' \yo H(1,0;Y)-12 \yo H(1,0;Y)-\frac{1}{24} \
\pi ^2 H(1,0;Y)+H(0,1;\yo) \Big(\frac{4 d_1'^2 \yo^3}{9}-\frac{53 \
d_1' \yo^3}{18}+\frac{4}{3} d_1' H(0;Y) \yo^3-\frac{7 d_1'^2 \
\yo^2}{3}+\frac{49 d_1' \yo^2}{3}-6 d_1' H(0;Y) \yo^2+\frac{22 d_1'^2 \
\yo}{3}-\frac{107 d_1' \yo}{2}+12 d_1' H(0;Y) \yo+\frac{13 \
d_1'}{6}+H(0;\ao) \Big(\frac{4 d_1 \yo^3}{3}-\frac{4 \yo^3}{3}-6 d_1 \
\yo^2+6 \yo^2+12 d_1 \yo-12 \yo-\frac{38 d_1}{3}+4 H(0;Y)+\frac{4 \
d_1}{\yo-1}-\frac{4}{\yo-1}-\frac{62}{3}\Big)+8 H(0,0;\ao)-d_1' \
H(0,0;Y)+8 d_1 H(0,1;\ao)-d_1' H(1,0;Y)+\frac{13 d_1'}{6 \
(\yo-1)}+\frac{2 d_1 \pi ^2}{3}-\frac{d_1' \pi \
^2}{6}\Big)+\Big(\frac{4 d_1'^2 \yo^3}{9}-\frac{49 d_1' \
\yo^3}{18}+\frac{4}{3} d_1' H(0;Y) \yo^3+\frac{2}{3} H(0;Y) \
\yo^3-\frac{37 \yo^3}{18}-\frac{7 d_1'^2 \yo^2}{3}+\frac{4 d_1 \
\yo^2}{3}+\frac{91 d_1' \yo^2}{6}-6 d_1' H(0;Y) \yo^2-3 H(0;Y) \yo^2+\
\frac{35 \yo^2}{3}+\frac{22 d_1'^2 \yo}{3}-\frac{16 d_1 \
\yo}{3}-\frac{299 d_1' \yo}{6}+12 d_1' H(0;Y) \yo+6 H(0;Y) \
\yo-\frac{209 \yo}{6}-\frac{49 d_1'^2}{9}+\frac{37 d_1}{18}+\frac{673 \
d_1'}{18}+\Big(-\frac{4 \yo^3}{3}+6 \yo^2-12 \
\yo-\frac{4}{\yo-1}+\frac{38}{3}\Big) H(0;\ao)-\frac{22}{3} d_1' \
H(0;Y)+\frac{2 H(0;Y)}{\yo-1}-\frac{19}{3} H(0;Y)-\frac{d_1}{3 \
(\yo-1)}+\frac{d_1'}{6 (\yo-1)}-\frac{37}{6 (\yo-1)}-\frac{4 \pi \
^2}{3}+\frac{127}{3}\Big) H(1,0;\yo)-\frac{1}{24} \pi ^2 \
H(1,1;Y)+\Big(\frac{\yo^3 d_1'^3}{9}-\frac{7 \yo^2 \
d_1'^3}{12}+\frac{11 \yo d_1'^3}{6}-\frac{49 d_1'^3}{36}-\frac{4 \
\yo^3 d_1'^2}{9}+3 \yo^2 d_1'^2-12 \yo d_1'^2+\frac{1}{3} \yo^3 \
H(0;Y) d_1'^2-\frac{3}{2} \yo^2 H(0;Y) d_1'^2+3 \yo H(0;Y) \
d_1'^2-\frac{11}{6} H(0;Y) d_1'^2+\frac{85 d_1'^2}{9}-\frac{\pi ^2 \
d_1'}{3}+H(0;\ao) \Big(\frac{4 d_1 \yo^3}{3}-\frac{4 d_1' \
\yo^3}{3}+\frac{2 \yo^3}{3}-6 d_1 \yo^2+6 d_1' \yo^2-3 \yo^2+12 d_1 \
\yo-12 d_1' \yo+6 \yo-\frac{38 d_1}{3}+10 d_1'+2 H(0;Y)+\frac{8 d_1}{\
\yo-1}-\frac{4 d_1'}{\yo-1}-\frac{2}{\yo-1}-\frac{43}{3}\Big)+4 \
H(0,0;\ao)+4 d_1 H(0,1;\ao)+\frac{2 d_1 \pi ^2}{3}-\frac{\pi \
^2}{3}\Big) H(1,1;\yo)+\Big(\frac{1}{6} d_1' \yo^3 \ao^4-d_1' \yo^2 \
\ao^4-\frac{5 d_1' \ao^4}{3}+\frac{5}{2} d_1' \yo \ao^4-\frac{8}{9} \
d_1' \yo^3 \ao^3+5 d_1' \yo^2 \ao^3+\frac{77 d_1' \
\ao^3}{9}-\frac{38}{3} d_1' \yo \ao^3+2 d_1' \yo^3 \ao^2-\frac{21}{2} \
d_1' \yo^2 \ao^2-\frac{35 d_1' \ao^2}{2}+26 d_1' \yo \
\ao^2-\frac{8}{3} d_1' \yo^3 \ao+13 d_1' \yo^2 \ao+\frac{65 d_1' \
\ao}{3}-30 d_1' \yo \ao+\frac{7 d_1' \yo^3}{6}+\frac{37 \
\yo^3}{18}-\frac{4 d_1 \yo^2}{3}-\frac{14 d_1' \yo^2}{3}-\frac{35 \
\yo^2}{3}-\frac{37 d_1}{18}-\frac{13 d_1'}{3}+\frac{16 d_1 \
\yo}{3}+\frac{47 d_1' \yo}{6}+\frac{209 \yo}{6}+\Big(-\frac{4 \
\yo^3}{3}+6 \yo^2-12 \yo-\frac{4}{\yo-1}+\frac{38}{3}\Big) \
H(0;\ao)-\frac{2}{3} \yo^3 H(0;Y)+3 \yo^2 H(0;Y)-6 \yo H(0;Y)-\frac{2 \
H(0;Y)}{\yo-1}+\frac{19}{3} H(0;Y)+\Big(-\frac{4 d_1 \yo^3}{3}+6 d_1 \
\yo^2-12 d_1 \yo+\frac{38 d_1}{3}-\frac{4 d_1}{\yo-1}\Big) \
H(1;\ao)+\frac{d_1}{3 (\yo-1)}-\frac{d_1'}{6 (\yo-1)}+\frac{37}{6 \
(\yo-1)}-\frac{127}{3}\Big) H(1,c_1(\ao);\yo)+\Big(\frac{\yo^3 \
\ao^4}{3}-2 \yo^2 \ao^4+5 \yo \ao^4-\frac{10 \ao^4}{3}-\frac{16 \yo^3 \
\ao^3}{9}+10 \yo^2 \ao^3-\frac{76 \yo \ao^3}{3}+\frac{136 \ao^3}{9}+4 \
\yo^3 \ao^2-21 \yo^2 \ao^2+52 \yo \ao^2-\frac{76 \ao^2}{3}-\frac{16 \
\yo^3 \ao}{3}+26 \yo^2 \ao-60 \yo \ao+\frac{82 \ao}{3}+\frac{25 \
\yo^3}{9}-\frac{35 \yo^2}{3}+23 \yo-\frac{13}{3 (\yo-1)}-\frac{13}{3}\
\Big) H(c_1(\ao),0;\yo)+\Big(\frac{1}{6} d_1' \yo^3 \ao^4-d_1' \yo^2 \
\ao^4-\frac{5 d_1' \ao^4}{3}+\frac{5}{2} d_1' \yo \ao^4-\frac{8}{9} \
d_1' \yo^3 \ao^3+5 d_1' \yo^2 \ao^3+\frac{68 d_1' \
\ao^3}{9}-\frac{38}{3} d_1' \yo \ao^3+2 d_1' \yo^3 \ao^2-\frac{21}{2} \
d_1' \yo^2 \ao^2-\frac{38 d_1' \ao^2}{3}+26 d_1' \yo \
\ao^2-\frac{8}{3} d_1' \yo^3 \ao+13 d_1' \yo^2 \ao+\frac{41 d_1' \
\ao}{3}-30 d_1' \yo \ao+\frac{25 d_1' \yo^3}{18}-\frac{35 d_1' \
\yo^2}{6}-\frac{13 d_1'}{6}+\frac{23 d_1' \yo}{2}-\frac{13 d_1'}{6 \
(\yo-1)}\Big) H(c_1(\ao),1;\yo)+\Big(\frac{\yo^3 \ao^4}{2}-3 \yo^2 \
\ao^4+\frac{15 \yo \ao^4}{2}-5 \ao^4-\frac{8 \yo^3 \ao^3}{3}+15 \yo^2 \
\ao^3-38 \yo \ao^3+\frac{68 \ao^3}{3}+6 \yo^3 \ao^2-\frac{63 \yo^2 \
\ao^2}{2}+78 \yo \ao^2-38 \ao^2-8 \yo^3 \ao+39 \yo^2 \ao-4 d_1 \ao+2 \
d_1' \ao-90 \yo \ao-2 H(0;Y) \ao+45 \ao+\frac{25 \yo^3}{6}-\frac{35 \
\yo^2}{2}+4 d_1-2 d_1'+\frac{69 \yo}{2}+\Big(-4 \ao+\frac{4}{\yo-1}+4\
\Big) H(0;\ao)+\frac{2 H(0;Y)}{\yo-1}+2 H(0;Y)+\Big(-4 \ao \
d_1+\frac{4 d_1}{\yo-1}+4 d_1\Big) H(1;\ao)+\frac{4 \
d_1}{\yo-1}-\frac{2 d_1'}{\yo-1}-\frac{21}{2 \
(\yo-1)}-\frac{21}{2}\Big) H(c_1(\ao),c_1(\ao);\yo)+\frac{1}{3} \yo^3 \
H(0,0,0;Y)-\frac{3}{2} \yo^2 H(0,0,0;Y)+3 \yo \
H(0,0,0;Y)+\Big(\frac{64 \yo^3}{3}-96 \yo^2+192 \yo-16 H(0;Y)\Big) \
H(0,0,0;\yo)+\Big(\frac{16 d_1' \yo^3}{3}-24 d_1' \yo^2+48 d_1' \
\yo+(8-8 d_1) H(0;\ao)-4 d_1' H(0;Y)\Big) H(0,0,1;\yo)+\Big(-\frac{8 \
\yo^3}{3}+12 \yo^2-24 \yo+8 H(0;\ao)+4 H(0;Y)+8 d_1 \
H(1;\ao)-\frac{8}{\yo-1}-8\Big) H(0,0,c_1(\ao);\yo)+\frac{1}{3} \yo^3 \
H(0,1,0;Y)-\frac{3}{2} \yo^2 H(0,1,0;Y)+3 \yo \
H(0,1,0;Y)+\Big(-\frac{4 d_1 \yo^3}{3}+\frac{16 d_1' \
\yo^3}{3}+\frac{4 \yo^3}{3}+6 d_1 \yo^2-24 d_1' \yo^2-6 \yo^2-12 d_1 \
\yo+48 d_1' \yo+12 \yo+\frac{38 d_1}{3}+8 H(0;\ao)-4 d_1' H(0;Y)-4 \
H(0;Y)-\frac{4 d_1}{\yo-1}+\frac{4}{\yo-1}+\frac{62}{3}\Big) H(0,1,0;\
\yo)+\Big(\frac{4 d_1'^2 \yo^3}{3}-6 d_1'^2 \yo^2+12 d_1'^2 \yo+(8 \
d_1'-12 d_1) H(0;\ao)-d_1'^2 H(0;Y)\Big) H(0,1,1;\yo)+\Big(-d_1' \
\ao^4+\frac{16 d_1' \ao^3}{3}-12 d_1' \ao^2+16 d_1' \ao+\frac{4 d_1 \
\yo^3}{3}-\frac{2 d_1' \yo^3}{3}-\frac{4 \yo^3}{3}-6 d_1 \yo^2+3 d_1' \
\yo^2+6 \yo^2-\frac{38 d_1}{3}-2 d_1'+12 d_1 \yo-6 d_1' \yo-12 \yo+8 \
H(0;\ao)+4 H(0;Y)+8 d_1 H(1;\ao)+\frac{4 d_1}{\yo-1}-\frac{2 \
d_1'}{\yo-1}-\frac{4}{\yo-1}-\frac{62}{3}\Big) \
H(0,1,c_1(\ao);\yo)+\Big(-2 \ao^4+\frac{32 \ao^3}{3}-24 \ao^2+32 \ao-\
\frac{4 \yo^3}{3}+6 \yo^2-12 \yo-\frac{4}{\yo-1}-4\Big) \
H(0,c_1(\ao),0;\yo)+\Big(-d_1' \ao^4+\frac{16 d_1' \ao^3}{3}-12 d_1' \
\ao^2+16 d_1' \ao-\frac{2 d_1' \yo^3}{3}+3 d_1' \yo^2-2 d_1'-6 d_1' \
\yo-\frac{2 d_1'}{\yo-1}\Big) H(0,c_1(\ao),1;\yo)+\Big(-3 \ao^4+16 \
\ao^3-36 \ao^2+48 \ao-2 \yo^3+9 \yo^2-18 \yo-4 H(0;\ao)-2 H(0;Y)-4 \
d_1 H(1;\ao)+\frac{2}{\yo-1}+2\Big) \
H(0,c_1(\ao),c_1(\ao);\yo)+\frac{1}{3} \yo^3 H(1,0,0;Y)-\frac{3}{2} \
\yo^2 H(1,0,0;Y)+3 \yo H(1,0,0;Y)+\Big(\frac{16 d_1' \yo^3}{3}+4 \
\yo^3-24 d_1' \yo^2-18 \yo^2+48 d_1' \yo+36 \yo-\frac{88 \
d_1'}{3}+\frac{12}{\yo-1}-38\Big) H(1,0,0;\yo)+\Big(\frac{4 d_1'^2 \
\yo^3}{3}+\frac{2 d_1' \yo^3}{3}-6 d_1'^2 \yo^2-3 d_1' \yo^2+12 \
d_1'^2 \yo+6 d_1' \yo-\frac{22 d_1'^2}{3}-\frac{19 d_1'}{3}+(4-4 d_1) \
H(0;\ao)+\frac{2 d_1'}{\yo-1}\Big) H(1,0,1;\yo)+\Big(-\frac{2 d_1' \
\yo^3}{3}-\frac{2 \yo^3}{3}+3 d_1' \yo^2+3 \yo^2-6 d_1' \yo-6 \
\yo+\frac{11 d_1'}{3}+4 H(0;\ao)+2 H(0;Y)+4 d_1 H(1;\ao)+\frac{4 \
d_1}{\yo-1}-\frac{2 d_1'}{\yo-1}-\frac{6}{\yo-1}-\frac{5}{3}\Big) \
H(1,0,c_1(\ao);\yo)+\frac{1}{3} \yo^3 H(1,1,0;Y)-\frac{3}{2} \yo^2 \
H(1,1,0;Y)+3 \yo H(1,1,0;Y)+\Big(\frac{4 d_1'^2 \yo^3}{3}-\frac{4 d_1 \
\yo^3}{3}+\frac{4 d_1' \yo^3}{3}-\frac{2 \yo^3}{3}-6 d_1'^2 \yo^2+6 \
d_1 \yo^2-6 d_1' \yo^2+3 \yo^2+12 d_1'^2 \yo-12 d_1 \yo+12 d_1' \yo-6 \
\yo-\frac{22 d_1'^2}{3}+\frac{38 d_1}{3}-10 d_1'+4 H(0;\ao)-2 H(0;Y)-\
\frac{8 d_1}{\yo-1}+\frac{4 d_1'}{\yo-1}+\frac{2}{\yo-1}+\frac{43}{3}\
\Big) H(1,1,0;\yo)+\Big(\frac{\yo^3 d_1'^3}{3}-\frac{3 \yo^2 \
d_1'^3}{2}+3 \yo d_1'^3-\frac{11 d_1'^3}{6}+(-8 d_1+4 d_1'+2) \
H(0;\ao)\Big) H(1,1,1;\yo)+\Big(\frac{4 d_1 \yo^3}{3}-\frac{4 d_1' \
\yo^3}{3}+\frac{2 \yo^3}{3}-6 d_1 \yo^2+6 d_1' \yo^2-3 \yo^2+12 d_1 \
\yo-12 d_1' \yo+6 \yo-\frac{38 d_1}{3}+10 d_1'+4 H(0;\ao)+2 H(0;Y)+4 \
d_1 H(1;\ao)+\frac{8 d_1}{\yo-1}-\frac{4 \
d_1'}{\yo-1}-\frac{2}{\yo-1}-\frac{43}{3}\Big) \
H(1,1,c_1(\ao);\yo)+\Big(-\frac{4 \yo^3}{3}+6 \yo^2-12 \
\yo-\frac{4}{\yo-1}+\frac{38}{3}\Big) \
H(1,c_1(\ao),0;\yo)+\Big(-\frac{2 d_1' \yo^3}{3}+3 d_1' \yo^2-6 d_1' \
\yo+\frac{19 d_1'}{3}-\frac{2 d_1'}{\yo-1}\Big) \
H(1,c_1(\ao),1;\yo)+\Big(-2 \yo^3+9 \yo^2-18 \yo-4 H(0;\ao)-2 \
H(0;Y)-4 d_1 H(1;\ao)-\frac{4 d_1}{\yo-1}+\frac{2 \
d_1'}{\yo-1}-\frac{2}{\yo-1}+27\Big) \
H(1,c_1(\ao),c_1(\ao);\yo)+\Big(-2 \ao d_1'+\frac{2 d_1'}{\yo-1}+2 \
d_1'\Big) H(c_1(\ao),1,c_1(\ao);\yo)+\Big(-4 \
\ao+\frac{4}{\yo-1}+4\Big) H(c_1(\ao),c_1(\ao),0;\yo)+\Big(-2 \ao \
d_1'+\frac{2 d_1'}{\yo-1}+2 d_1'\Big) \
H(c_1(\ao),c_1(\ao),1;\yo)+\Big(-6 \ao+\frac{6}{\yo-1}+6\Big) \
H(c_1(\ao),c_1(\ao),c_1(\ao);\yo)-\frac{1}{4} H(0,0,0,0;Y)-64 \
H(0,0,0,0;\yo)-16 d_1' H(0,0,0,1;\yo)+16 \
H(0,0,0,c_1(\ao);\yo)-\frac{1}{4} H(0,0,1,0;Y)+(8 d_1-16 d_1'-8) \
H(0,0,1,0;\yo)-4 d_1'^2 H(0,0,1,1;\yo)+(-8 d_1+4 d_1'+8) H(0,0,1,c_1(\
\ao);\yo)+8 H(0,0,c_1(\ao),0;\yo)+4 d_1' H(0,0,c_1(\ao),1;\yo)+4 \
H(0,0,c_1(\ao),c_1(\ao);\yo)-\frac{1}{4} H(0,1,0,0;Y)+(-16 d_1'-24) \
H(0,1,0,0;\yo)+\Big(-4 d_1'^2-4 d_1'\Big) H(0,1,0,1;\yo)+(-4 d_1+4 \
d_1'+8) H(0,1,0,c_1(\ao);\yo)-\frac{1}{4} H(0,1,1,0;Y)+\Big(-4 \
d_1'^2-8 d_1'+12 d_1\Big) H(0,1,1,0;\yo)-d_1'^3 H(0,1,1,1;\yo)+(8 \
d_1'-12 d_1) H(0,1,1,c_1(\ao);\yo)+8 H(0,1,c_1(\ao),0;\yo)+4 d_1' \
H(0,1,c_1(\ao),1;\yo)+(4 d_1-2 d_1'+8) H(0,1,c_1(\ao),c_1(\ao);\yo)-2 \
d_1' H(0,c_1(\ao),1,c_1(\ao);\yo)-4 H(0,c_1(\ao),c_1(\ao),0;\yo)-2 \
d_1' H(0,c_1(\ao),c_1(\ao),1;\yo)-6 \
H(0,c_1(\ao),c_1(\ao),c_1(\ao);\yo)-\frac{1}{4} H(1,0,0,0;Y)+8 \
H(1,0,0,c_1(\ao);\yo)-\frac{1}{4} H(1,0,1,0;Y)+(4 d_1-4) \
H(1,0,1,0;\yo)+(-4 d_1+2 d_1'+4) H(1,0,1,c_1(\ao);\yo)+4 \
H(1,0,c_1(\ao),0;\yo)+2 d_1' H(1,0,c_1(\ao),1;\yo)-2 \
H(1,0,c_1(\ao),c_1(\ao);\yo)-\frac{1}{4} H(1,1,0,0;Y)-12 \
H(1,1,0,0;\yo)-2 d_1' H(1,1,0,1;\yo)+(-4 d_1+2 d_1'+6) \
H(1,1,0,c_1(\ao);\yo)-\frac{1}{4} H(1,1,1,0;Y)+(8 d_1-4 d_1'-2) \
H(1,1,1,0;\yo)+(-8 d_1+4 d_1'+2) H(1,1,1,c_1(\ao);\yo)+4 \
H(1,1,c_1(\ao),0;\yo)+2 d_1' H(1,1,c_1(\ao),1;\yo)+(4 d_1-2 d_1'+2) \
H(1,1,c_1(\ao),c_1(\ao);\yo)-2 d_1' H(1,c_1(\ao),1,c_1(\ao);\yo)-4 \
H(1,c_1(\ao),c_1(\ao),0;\yo)-2 d_1' H(1,c_1(\ao),c_1(\ao),1;\yo)-6 \
H(1,c_1(\ao),c_1(\ao),c_1(\ao);\yo)+H(0;\yo) \Big(-\frac{1}{9} \yo^3 \
\ao^4+\frac{5 \yo^2 \ao^4}{6}-\frac{10 \yo \ao^4}{3}+\frac{13 \yo^3 \
\ao^3}{27}-\frac{34 \yo^2 \ao^3}{9}+\frac{160 \yo \ao^3}{9}-\frac{11 \
\yo^3 \ao^2}{18}+\frac{109 \yo^2 \ao^2}{18}-\frac{316 \yo \
\ao^2}{9}-\frac{7 \yo^3 \ao}{9}-\frac{7 \yo^2 \ao}{9}+\frac{86 \yo \
\ao}{3}+\frac{4 d_1'^2 \yo^3}{27}-\frac{205 d_1 \yo^3}{108}-\frac{3 \
d_1' \yo^3}{2}-\frac{2 \pi ^2 \yo^3}{9}+\frac{161 \yo^3}{27}-\frac{17 \
d_1'^2 \yo^2}{18}+\frac{22 d_1 \yo^2}{3}+\frac{98 d_1' \yo^2}{9}+\pi \
^2 \yo^2-\frac{1405 \yo^2}{36}+\frac{217 \
d_1}{36}-\frac{d_1'}{6}+\frac{49 d_1'^2 \yo}{9}-\frac{469 d_1 \
\yo}{36}-\frac{1201 d_1' \yo}{18}-2 \pi ^2 \yo+\frac{1811 \
\yo}{9}+\Big(\frac{7 \yo^3}{3}-\frac{26 \yo^2}{3}+11 \yo-\frac{13}{3 \
(\yo-1)}-\frac{13}{3}\Big) H(0;\ao)+\frac{4}{9} d_1' \yo^3 \
H(0;Y)-\frac{53}{18} \yo^3 H(0;Y)-\frac{7}{3} d_1' \yo^2 \
H(0;Y)+\frac{49}{3} \yo^2 H(0;Y)+\frac{22}{3} d_1' \yo \
H(0;Y)-\frac{107}{2} \yo H(0;Y)+\frac{13 H(0;Y)}{6 \
(\yo-1)}-\frac{1}{6} \pi ^2 H(0;Y)+\frac{13}{6} H(0;Y)-\frac{1}{6} \
\pi ^2 H(1;Y)+\frac{4}{3} \yo^3 H(0,0;Y)-6 \yo^2 H(0,0;Y)+12 \yo \
H(0,0;Y)+\frac{4}{3} \yo^3 H(1,0;Y)-6 \yo^2 H(1,0;Y)+12 \yo \
H(1,0;Y)-H(0,0,0;Y)-H(0,1,0;Y)-H(1,0,0;Y)-H(1,1,0;Y)+\frac{217 \
d_1}{36 (\yo-1)}-\frac{d_1'}{6 (\yo-1)}+\frac{4 \pi ^2}{3 \
(\yo-1)}-\frac{149}{18 (\yo-1)}-6 \zeta_3+\frac{4 \pi \
^2}{3}-\frac{149}{18}\Big)+H(1;\yo) \Big(-\frac{1}{18} d_1' \yo^3 \
\ao^4+\frac{5}{12} d_1' \yo^2 \ao^4+\frac{47 d_1' \
\ao^4}{36}-\frac{5}{3} d_1' \yo \ao^4+\frac{13}{54} d_1' \yo^3 \ao^3-\
\frac{17}{9} d_1' \yo^2 \ao^3-\frac{391 d_1' \ao^3}{54}+\frac{80}{9} \
d_1' \yo \ao^3-\frac{11}{36} d_1' \yo^3 \ao^2+\frac{109}{36} d_1' \
\yo^2 \ao^2+\frac{89 d_1' \ao^2}{6}-\frac{158}{9} d_1' \yo \
\ao^2-\frac{7}{18} d_1' \yo^3 \ao-\frac{7}{18} d_1' \yo^2 \
\ao-\frac{247 d_1' \ao}{18}+\frac{43 d_1' \yo \ao}{3}-\frac{251 \
d_1'^3}{216}+\frac{d_1'^3 \yo^3}{27}-\frac{8 d_1'^2 \
\yo^3}{27}+\frac{14 d_1' \yo^3}{27}-\frac{1}{18} d_1' \pi ^2 \
\yo^3-\frac{\pi ^2 \yo^3}{9}+\frac{395 d_1'^2}{27}-\frac{17 d_1'^3 \
\yo^2}{72}+\frac{22 d_1'^2 \yo^2}{9}-\frac{21 d_1' \
\yo^2}{4}+\frac{1}{4} d_1' \pi ^2 \yo^2+\frac{\pi ^2 \
\yo^2}{2}-\frac{4025 d_1'}{108}+\frac{49 d_1'^3 \yo}{36}-\frac{151 \
d_1'^2 \yo}{9}+42 d_1' \yo-\frac{1}{2} d_1' \pi ^2 \yo-\pi ^2 \
\yo+\frac{1}{9} d_1'^2 \yo^3 H(0;Y)-\frac{4}{9} d_1' \yo^3 \
H(0;Y)-\frac{49}{36} d_1'^2 H(0;Y)-\frac{7}{12} d_1'^2 \yo^2 H(0;Y)+3 \
d_1' \yo^2 H(0;Y)+\frac{85}{9} d_1' H(0;Y)+\frac{11}{6} d_1'^2 \yo \
H(0;Y)-12 d_1' \yo H(0;Y)-\frac{1}{3} \pi ^2 H(0;Y)+H(0;\ao) \
\Big(\frac{17 d_1' \yo^3}{18}-\frac{2}{3} H(0;Y) \yo^3+\frac{37 \
\yo^3}{18}-\frac{4 d_1 \yo^2}{3}-\frac{19 d_1' \yo^2}{6}+3 H(0;Y) \
\yo^2-\frac{35 \yo^2}{3}+\frac{16 d_1 \yo}{3}+\frac{11 d_1' \yo}{6}-6 \
H(0;Y) \yo+\frac{209 \yo}{6}-\frac{37 d_1}{18}+\frac{7 \
d_1'}{18}-\frac{2 H(0;Y)}{\yo-1}+\frac{19}{3} H(0;Y)+\frac{d_1}{3 \
(\yo-1)}-\frac{d_1'}{6 (\yo-1)}+\frac{37}{6 \
(\yo-1)}-\frac{127}{3}\Big)+\Big(-\frac{4 \yo^3}{3}+6 \yo^2-12 \
\yo-\frac{4}{\yo-1}+\frac{38}{3}\Big) H(0,0;\ao)+\frac{1}{3} d_1' \
\yo^3 H(0,0;Y)-\frac{3}{2} d_1' \yo^2 H(0,0;Y)-\frac{11}{6} d_1' \
H(0,0;Y)+3 d_1' \yo H(0,0;Y)+\Big(-\frac{4 d_1 \yo^3}{3}+6 d_1 \
\yo^2-12 d_1 \yo+\frac{38 d_1}{3}-\frac{4 d_1}{\yo-1}\Big) \
H(0,1;\ao)+\frac{1}{3} d_1' \yo^3 H(1,0;Y)-\frac{3}{2} d_1' \yo^2 \
H(1,0;Y)-\frac{11}{6} d_1' H(1,0;Y)+3 d_1' \yo H(1,0;Y)-\frac{2 d_1 \
\pi ^2}{3 (\yo-1)}+\frac{d_1' \pi ^2}{3 (\yo-1)}+\frac{\pi ^2}{3 \
(\yo-1)}+6 \zeta_3+\frac{11 d_1' \pi ^2}{36}+\frac{43 \pi \
^2}{18}\Big)+\frac{2 d_1 \pi ^2}{3 (\yo-1)}-\frac{d_1' \pi ^2}{3 \
(\yo-1)}-\frac{37 \pi ^2}{36 (\yo-1)}+4 \yo^3 \zeta_3-18 \yo^2 \
\zeta_3+36 \yo \zeta_3-\frac{3}{2} H(0;Y) \zeta_3-\frac{6 \
\zeta_3}{\yo-1}-6 \zeta_3+\frac{\pi ^4}{288}+\frac{2 d_1 \pi \
^2}{3}-\frac{d_1' \pi ^2}{3}-\frac{37 \pi ^2}{36}.
\erp

%%%
%%% The K*I integrals
%%%

\section{The $\cKI$-type integrals}
\label{app:KIIntegrals}

%
% The K*I integral for k=0
%

\subsection{The $\cKI$ integral for $k=0$}
%
% This file contains the TeX output produced by Mathematica for the integral JI for arbitrary kap=0  and D0 = 3 +d'1 ep
%
The $\eps$ expansion for this integral reads
\beq
\cKI(\ep;y_{0},d'_{0},\alpha_{0},d_{0};0)=\frac{1}{\eps^3}\cki_{-3}^{(0)}+\frac{1}{\eps^2}\cki_{-2}^{(0)}+\frac{1}{\eps}\cki_{-1}^{(0)}+\cki_0^{(0)}+\ocal\left(\eps\right),
\eeq
where
%% 1/ep^3
\brp
\cki_{-3}^{(0)}=-\frac{1}{2},
\erp
%% 1/ep^2
\brp
\cki_{-2}^{(0)}=-\frac{2 \yo^3}{3}+3 \yo^2-6 \yo+2 H(0;\yo)-1,
\erp
%% 1/ep
\brp
\cki_{-1}^{(0)}=\frac{\yo^3 \ao^4}{6}-\frac{\yo^2 \ao^4}{2}+\frac{\yo \
\ao^4}{2}-\frac{8 \yo^3 \ao^3}{9}+3 \yo^2 \ao^3-\frac{10 \yo \
\ao^3}{3}+2 \yo^3 \ao^2-\frac{15 \yo^2 \ao^2}{2}+10 \yo \ao^2-\frac{8 \
\yo^3 \ao}{3}+11 \yo^2 \ao-18 \yo \ao+\frac{2 d_1' \yo^3}{9}-\frac{8 \
\yo^3}{3}-\frac{7 d_1' \yo^2}{6}+\frac{25 \yo^2}{2}+\frac{11 d_1' \
\yo}{3}-32 \yo+\Big(\frac{2 \yo^3}{3}-3 \yo^2+6 \yo-\frac{2}{\yo-1}-2\
\Big) H(0;\ao)+\Big(2 \yo^3-9 \yo^2+18 \yo+\frac{2}{\yo-1}+6\Big) \
H(0;\yo)+\Big(\frac{2 d_1' \yo^3}{3}-3 d_1' \yo^2+6 d_1' \yo-\frac{11 \
d_1'}{3}+2 H(0;\ao)\Big) H(1;\yo)+\Big(2 \ao-\frac{2}{\yo-1}-2\Big) \
H(c_1(\ao);\yo)-8 H(0,0;\yo)-2 d_1' H(0,1;\yo)+2 H(0,c_1(\ao);\yo)-2 \
H(1,0;\yo)+2 H(1,c_1(\ao);\yo)-2,
\erp
%% ep^0
\brp
\cki_0^{(0)} =-\frac{1}{12} d_1 \yo^3 \ao^4-\frac{1}{18} d_1' \yo^3 \ao^4+\frac{3 \
\yo^3 \ao^4}{4}+\frac{1}{4} d_1 \yo^2 \ao^4+\frac{1}{6} d_1' \yo^2 \
\ao^4-\frac{13 \yo^2 \ao^4}{6}-\frac{1}{4} d_1 \yo \ao^4-\frac{1}{6} \
d_1' \yo \ao^4+\frac{29 \yo \ao^4}{12}+\frac{13}{27} d_1 \yo^3 \ao^3+\
\frac{8}{27} d_1' \yo^3 \ao^3-\frac{221 \yo^3 \ao^3}{54}-\frac{5}{3} \
d_1 \yo^2 \ao^3-\frac{19}{18} d_1' \yo^2 \ao^3+\frac{247 \yo^2 \
\ao^3}{18}+\frac{17}{9} d_1 \yo \ao^3+\frac{11}{9} d_1' \yo \
\ao^3-\frac{305 \yo \ao^3}{18}-\frac{23}{18} d_1 \yo^3 \
\ao^2-\frac{2}{3} d_1' \yo^3 \ao^2+\frac{347 \yo^3 \ao^2}{36}+5 d_1 \
\yo^2 \ao^2+\frac{11}{4} d_1' \yo^2 \ao^2-\frac{331 \yo^2 \
\ao^2}{9}-\frac{43}{6} d_1 \yo \ao^2-\frac{9}{2} d_1' \yo \
\ao^2+\frac{1021 \yo \ao^2}{18}+\frac{25}{9} d_1 \yo^3 \
\ao+\frac{8}{9} d_1' \yo^3 \ao-\frac{265 \yo^3 \ao}{18}-12 d_1 \yo^2 \
\ao-\frac{25}{6} d_1' \yo^2 \ao+\frac{566 \yo^2 \ao}{9}+\frac{65 d_1 \
\yo \ao}{3}+\frac{29 d_1' \yo \ao}{3}-\frac{245 \yo \
\ao}{2}-\frac{\ao}{\yo-1}-\ao-\frac{2 d_1'^2 \yo^3}{27}+\frac{10 d_1' \
\yo^3}{9}+\frac{\pi ^2 \yo^3}{9}-8 \yo^3+\frac{17 d_1'^2 \
\yo^2}{36}-\frac{20 d_1' \yo^2}{3}-\frac{\pi ^2 \yo^2}{2}+\frac{155 \
\yo^2}{4}-\frac{49 d_1'^2 \yo}{18}+\frac{199 d_1' \yo}{6}+\pi ^2 \
\yo-128 \yo+\Big(-\frac{1}{3} \yo^3 \ao^4+\yo^2 \ao^4-\yo \
\ao^4+\frac{16 \yo^3 \ao^3}{9}-6 \yo^2 \ao^3+\frac{20 \yo \ao^3}{3}-4 \
\yo^3 \ao^2+15 \yo^2 \ao^2-20 \yo \ao^2+\frac{16 \yo^3 \ao}{3}-22 \
\yo^2 \ao+36 \yo \ao-\frac{2 d_1' \yo^3}{9}+\frac{23 \
\yo^3}{18}+\frac{7 d_1' \yo^2}{6}-\frac{43 \yo^2}{6}+4 d_1-2 \
d_1'-\frac{11 d_1' \yo}{3}+\frac{143 \yo}{6}+\frac{4 \
d_1}{\yo-1}-\frac{2 d_1'}{\yo-1}-\frac{61}{6 \
(\yo-1)}-\frac{1}{(\yo-1)^2}-\frac{55}{6}\Big) \
H(0;\ao)+\Big(-\frac{1}{3} \yo^3 \ao^4+\yo^2 \ao^4-\yo \ao^4+\frac{16 \
\yo^3 \ao^3}{9}-6 \yo^2 \ao^3+\frac{20 \yo \ao^3}{3}-4 \yo^3 \ao^2+15 \
\yo^2 \ao^2-20 \yo \ao^2+\frac{16 \yo^3 \ao}{3}-22 \yo^2 \ao+36 \yo \
\ao-\frac{2 d_1' \yo^3}{3}+\frac{169 \yo^3}{18}+\frac{7 d_1' \
\yo^2}{2}-\frac{257 \yo^2}{6}-4 d_1+2 d_1'-11 d_1' \yo+\frac{625 \
\yo}{6}+\Big(-\frac{4 \yo^3}{3}+6 \yo^2-12 \yo+\frac{4}{\yo-1}+4\Big) \
H(0;\ao)-\frac{4 d_1}{\yo-1}+\frac{2 d_1'}{\yo-1}+\frac{61}{6 \
(\yo-1)}+\frac{1}{(\yo-1)^2}+\frac{103}{6}\Big) \
H(0;\yo)+\Big(-\frac{1}{3} d_1 \yo^3 \ao^4+d_1 \yo^2 \ao^4-d_1 \yo \
\ao^4+\frac{16}{9} d_1 \yo^3 \ao^3-6 d_1 \yo^2 \ao^3+\frac{20}{3} d_1 \
\yo \ao^3-4 d_1 \yo^3 \ao^2+15 d_1 \yo^2 \ao^2-20 d_1 \yo \
\ao^2+\frac{16}{3} d_1 \yo^3 \ao-22 d_1 \yo^2 \ao+36 d_1 \yo \
\ao-\frac{25 d_1 \yo^3}{9}+12 d_1 \yo^2-\frac{65 d_1 \yo}{3}\Big) \
H(1;\ao)+\Big(-\frac{1}{6} \yo^3 \ao^4+\frac{\yo^2 \
\ao^4}{2}-\frac{\yo \ao^4}{2}+\frac{\ao^4}{6}+\frac{8 \yo^3 \
\ao^3}{9}-3 \yo^2 \ao^3+\frac{10 \yo \ao^3}{3}-\frac{11 \ao^3}{9}-2 \
\yo^3 \ao^2+\frac{15 \yo^2 \ao^2}{2}-10 \yo \ao^2+\frac{11 \ao^2}{2}+\
\frac{8 \yo^3 \ao}{3}-11 \yo^2 \ao-4 d_1 \ao+2 d_1' \ao+18 \yo \
\ao-\frac{25 \yo^3}{18}+\frac{16 \yo^2}{3}+4 d_1-2 d_1'-\frac{49 \
\yo}{6}+\Big(-4 \ao+\frac{4}{\yo-1}+4\Big) H(0;\ao)+\Big(-4 \ao \
d_1+\frac{4 d_1}{\yo-1}+4 d_1\Big) H(1;\ao)+\frac{4 \
d_1}{\yo-1}-\frac{2 d_1'}{\yo-1}-\frac{61}{6 \
(\yo-1)}-\frac{1}{(\yo-1)^2}-\frac{55}{6}\Big) H(c_1(\ao);\yo)+\Big(-\
\frac{4 \yo^3}{3}+6 \yo^2-12 \yo+\frac{4}{\yo-1}+4\Big) \
H(0,0;\ao)+\Big(-\frac{20 \yo^3}{3}+30 \yo^2-60 \
\yo-\frac{12}{\yo-1}-28\Big) H(0,0;\yo)+\Big(-\frac{4 d_1 \yo^3}{3}+6 \
d_1 \yo^2-12 d_1 \yo+4 d_1+\frac{4 d_1}{\yo-1}\Big) \
H(0,1;\ao)+H(1;\yo) \Big(-\frac{1}{6} d_1' \yo^3 \ao^4+\frac{1}{2} \
d_1' \yo^2 \ao^4+\frac{d_1' \ao^4}{6}-\frac{1}{2} d_1' \yo \
\ao^4+\frac{8}{9} d_1' \yo^3 \ao^3-3 d_1' \yo^2 \ao^3-\frac{11 d_1' \
\ao^3}{9}+\frac{10}{3} d_1' \yo \ao^3-2 d_1' \yo^3 \ao^2+\frac{15}{2} \
d_1' \yo^2 \ao^2+\frac{9 d_1' \ao^2}{2}-10 d_1' \yo \ao^2+\frac{8}{3} \
d_1' \yo^3 \ao-11 d_1' \yo^2 \ao-\frac{23 d_1' \ao}{3}+18 d_1' \yo \
\ao-\frac{2 d_1'^2 \yo^3}{9}+\frac{8 d_1' \yo^3}{3}+\frac{49 \
d_1'^2}{18}+\frac{7 d_1'^2 \yo^2}{6}-\frac{25 d_1' \
\yo^2}{2}-\frac{133 d_1'}{6}-\frac{11 d_1'^2 \yo}{3}+32 d_1' \
\yo+\Big(-\frac{2 d_1' \yo^3}{3}+\frac{2 \yo^3}{3}+3 d_1' \yo^2-3 \
\yo^2-6 d_1' \yo+6 \yo+\frac{11 d_1'}{3}-\frac{4 d_1}{\yo-1}+\frac{2 \
d_1'}{\yo-1}+\frac{2}{\yo-1}+10\Big) H(0;\ao)-4 H(0,0;\ao)-4 d_1 \
H(0,1;\ao)+\frac{\pi ^2}{3}\Big)+\Big(-2 d_1' \yo^3+9 d_1' \yo^2-18 \
d_1' \yo-6 d_1'+(4 d_1-4) H(0;\ao)-\frac{2 d_1'}{\yo-1}\Big) \
H(0,1;\yo)+\Big(\frac{2 \yo^3}{3}-3 \yo^2+6 \yo-4 H(0;\ao)-4 d_1 H(1;\
\ao)+\frac{6}{\yo-1}+10\Big) H(0,c_1(\ao);\yo)+\Big(-2 d_1' \
\yo^3-\frac{2 \yo^3}{3}+9 d_1' \yo^2+3 \yo^2-18 d_1' \yo-6 \yo+11 \
d_1'-4 H(0;\ao)+\frac{4 d_1}{\yo-1}-\frac{2 \
d_1'}{\yo-1}-\frac{2}{\yo-1}-10\Big) H(1,0;\yo)+\Big(-\frac{2}{3} \
d_1'^2 \yo^3+3 d_1'^2 \yo^2-6 d_1'^2 \yo+\frac{11 d_1'^2}{3}+(4 d_1-2 \
d_1'-2) H(0;\ao)\Big) H(1,1;\yo)+\Big(\frac{2 \yo^3}{3}-3 \yo^2+6 \
\yo-4 H(0;\ao)-4 d_1 H(1;\ao)-\frac{4 d_1}{\yo-1}+\frac{2 \
d_1'}{\yo-1}+\frac{2}{\yo-1}+10\Big) H(1,c_1(\ao);\yo)+\Big(-4 \
\ao+\frac{4}{\yo-1}+4\Big) H(c_1(\ao),0;\yo)+\Big(-2 \ao d_1'+\frac{2 \
d_1'}{\yo-1}+2 d_1'\Big) H(c_1(\ao),1;\yo)+\Big(-2 \
\ao+\frac{2}{\yo-1}+2\Big) H(c_1(\ao),c_1(\ao);\yo)+32 H(0,0,0;\yo)+8 \
d_1' H(0,0,1;\yo)-8 H(0,0,c_1(\ao);\yo)+(-4 d_1+8 d_1'+4) \
H(0,1,0;\yo)+2 d_1'^2 H(0,1,1;\yo)+(4 d_1-2 d_1'-4) \
H(0,1,c_1(\ao);\yo)-4 H(0,c_1(\ao),0;\yo)-2 d_1' \
H(0,c_1(\ao),1;\yo)-2 H(0,c_1(\ao),c_1(\ao);\yo)+12 H(1,0,0;\yo)+2 \
d_1' H(1,0,1;\yo)-6 H(1,0,c_1(\ao);\yo)+(-4 d_1+2 d_1'+2) \
H(1,1,0;\yo)+(4 d_1-2 d_1'-2) H(1,1,c_1(\ao);\yo)-4 \
H(1,c_1(\ao),0;\yo)-2 d_1' H(1,c_1(\ao),1;\yo)-2 \
H(1,c_1(\ao),c_1(\ao);\yo)-\frac{\pi ^2}{3 (\yo-1)}+3 \
\zeta_3-\frac{\pi ^2}{3}-4.
\erp

%
% The K*I integral for k=1
%

\subsection{The $\cKI$ integral for $k=1$}
%
% This file contains the TeX output produced by Mathematica for the integral JI for arbitrary kap=0  and D0 = 3 +d'1 ep
%
The $\eps$ expansion for this integral reads
\beq
\cKI(\ep;y_{0},d'_{0},\alpha_{0},d_{0};1)=\frac{1}{\eps^3}\cki_{-3}^{(1)}+\frac{1}{\eps^2}\cki_{-2}^{(1)}+\frac{1}{\eps}\cki_{-1}^{(1)}+\cki_0^{(1)}+\ocal\left(\eps\right),
\eeq
where
%% 1/ep^3
\brp
\cki_{-3}^{(1)}=-\frac{1}{4},
\erp
%% 1/ep^2
\brp
\cki_{-2}^{(1)}=-\frac{\yo^3}{3}+\frac{3 \yo^2}{2}-3 \yo+H(0;\yo)-\frac{1}{2},
\erp
%% 1/ep
\brp
\cki_{-1}^{(1)}=\frac{\yo^3 \ao^4}{12}-\frac{\yo^2 \ao^4}{4}+\frac{\yo \
\ao^4}{4}-\frac{4 \yo^3 \ao^3}{9}+\frac{3 \yo^2 \ao^3}{2}-\frac{5 \yo \
\ao^3}{3}+\yo^3 \ao^2-\frac{15 \yo^2 \ao^2}{4}+5 \yo \ao^2-\frac{4 \
\yo^3 \ao}{3}+\frac{11 \yo^2 \ao}{2}-9 \yo \ao+\frac{d_1' \
\yo^3}{9}-\frac{4 \yo^3}{3}-\frac{7 d_1' \yo^2}{12}+\frac{25 \
\yo^2}{4}+\frac{11 d_1' \yo}{6}-16 \yo+\Big(\frac{\yo^3}{3}-\frac{3 \
\yo^2}{2}+3 \yo-\frac{1}{\yo-1}-1\Big) H(0;\ao)+\Big(\yo^3-\frac{9 \
\yo^2}{2}+9 \yo+\frac{1}{\yo-1}+3\Big) H(0;\yo)+\Big(\frac{d_1' \
\yo^3}{3}-\frac{3 d_1' \yo^2}{2}+3 d_1' \yo-\frac{11 \
d_1'}{6}+H(0;\ao)\Big) H(1;\yo)+\Big(\ao-\frac{1}{\yo-1}-1\Big) \
H(c_1(\ao);\yo)-4 H(0,0;\yo)-d_1' H(0,1;\yo)+H(0,c_1(\ao);\yo)-H(1,0;\
\yo)+H(1,c_1(\ao);\yo)-1,
\erp
%% ep^0
\brp
\cki_0^{(1)} =-\frac{1}{24} d_1 \yo^3 \ao^4-\frac{1}{36} d_1' \yo^3 \ao^4+\frac{3 \
\yo^3 \ao^4}{8}+\frac{1}{8} d_1 \yo^2 \ao^4+\frac{1}{12} d_1' \yo^2 \
\ao^4-\frac{13 \yo^2 \ao^4}{12}-\frac{1}{8} d_1 \yo \
\ao^4-\frac{1}{12} d_1' \yo \ao^4+\frac{29 \yo \
\ao^4}{24}+\frac{13}{54} d_1 \yo^3 \ao^3+\frac{4}{27} d_1' \yo^3 \
\ao^3-\frac{221 \yo^3 \ao^3}{108}-\frac{5}{6} d_1 \yo^2 \
\ao^3-\frac{19}{36} d_1' \yo^2 \ao^3+\frac{247 \yo^2 \
\ao^3}{36}+\frac{17}{18} d_1 \yo \ao^3+\frac{11}{18} d_1' \yo \
\ao^3-\frac{305 \yo \ao^3}{36}-\frac{23}{36} d_1 \yo^3 \
\ao^2-\frac{1}{3} d_1' \yo^3 \ao^2+\frac{347 \yo^3 \
\ao^2}{72}+\frac{5}{2} d_1 \yo^2 \ao^2+\frac{11}{8} d_1' \yo^2 \ao^2-\
\frac{331 \yo^2 \ao^2}{18}-\frac{43}{12} d_1 \yo \ao^2-\frac{9}{4} \
d_1' \yo \ao^2+\frac{1021 \yo \ao^2}{36}+\frac{25}{18} d_1 \yo^3 \ao+\
\frac{4}{9} d_1' \yo^3 \ao-\frac{265 \yo^3 \ao}{36}-6 d_1 \yo^2 \
\ao-\frac{25}{12} d_1' \yo^2 \ao+\frac{283 \yo^2 \ao}{9}+\frac{65 d_1 \
\yo \ao}{6}+\frac{29 d_1' \yo \ao}{6}-\frac{245 \yo \
\ao}{4}-\frac{\ao}{2 (\yo-1)}-\frac{\ao}{2}-\frac{d_1'^2 \
\yo^3}{27}+\frac{5 d_1' \yo^3}{9}+\frac{\pi ^2 \yo^3}{18}-4 \
\yo^3+\frac{17 d_1'^2 \yo^2}{72}-\frac{10 d_1' \yo^2}{3}-\frac{\pi ^2 \
\yo^2}{4}+\frac{155 \yo^2}{8}-\frac{49 d_1'^2 \yo}{36}+\frac{199 d_1' \
\yo}{12}+\frac{\pi ^2 \yo}{2}-64 \yo+\Big(-\frac{1}{6} \yo^3 \
\ao^4+\frac{\yo^2 \ao^4}{2}-\frac{\yo \ao^4}{2}+\frac{8 \yo^3 \
\ao^3}{9}-3 \yo^2 \ao^3+\frac{10 \yo \ao^3}{3}-2 \yo^3 \ao^2+\frac{15 \
\yo^2 \ao^2}{2}-10 \yo \ao^2+\frac{8 \yo^3 \ao}{3}-11 \yo^2 \ao+18 \
\yo \ao-\frac{d_1' \yo^3}{9}+\frac{23 \yo^3}{36}+\frac{7 d_1' \
\yo^2}{12}-\frac{43 \yo^2}{12}+2 d_1-d_1'-\frac{11 d_1' \
\yo}{6}+\frac{143 \yo}{12}+\frac{2 \
d_1}{\yo-1}-\frac{d_1'}{\yo-1}-\frac{61}{12 (\yo-1)}-\frac{1}{2 \
(\yo-1)^2}-\frac{55}{12}\Big) H(0;\ao)+\Big(-\frac{1}{6} \yo^3 \ao^4+\
\frac{\yo^2 \ao^4}{2}-\frac{\yo \ao^4}{2}+\frac{8 \yo^3 \ao^3}{9}-3 \
\yo^2 \ao^3+\frac{10 \yo \ao^3}{3}-2 \yo^3 \ao^2+\frac{15 \yo^2 \
\ao^2}{2}-10 \yo \ao^2+\frac{8 \yo^3 \ao}{3}-11 \yo^2 \ao+18 \yo \ao-\
\frac{d_1' \yo^3}{3}+\frac{169 \yo^3}{36}+\frac{7 d_1' \
\yo^2}{4}-\frac{257 \yo^2}{12}-2 d_1+d_1'-\frac{11 d_1' \
\yo}{2}+\frac{625 \yo}{12}+\Big(-\frac{2 \yo^3}{3}+3 \yo^2-6 \
\yo+\frac{2}{\yo-1}+2\Big) H(0;\ao)-\frac{2 \
d_1}{\yo-1}+\frac{d_1'}{\yo-1}+\frac{61}{12 (\yo-1)}+\frac{1}{2 \
(\yo-1)^2}+\frac{103}{12}\Big) H(0;\yo)+\Big(-\frac{1}{6} d_1 \yo^3 \
\ao^4+\frac{1}{2} d_1 \yo^2 \ao^4-\frac{1}{2} d_1 \yo \
\ao^4+\frac{8}{9} d_1 \yo^3 \ao^3-3 d_1 \yo^2 \ao^3+\frac{10}{3} d_1 \
\yo \ao^3-2 d_1 \yo^3 \ao^2+\frac{15}{2} d_1 \yo^2 \ao^2-10 d_1 \yo \
\ao^2+\frac{8}{3} d_1 \yo^3 \ao-11 d_1 \yo^2 \ao+18 d_1 \yo \
\ao-\frac{25 d_1 \yo^3}{18}+6 d_1 \yo^2-\frac{65 d_1 \yo}{6}\Big) \
H(1;\ao)+\Big(-\frac{1}{12} \yo^3 \ao^4+\frac{\yo^2 \
\ao^4}{4}-\frac{\yo \ao^4}{4}+\frac{\ao^4}{12}+\frac{4 \yo^3 \
\ao^3}{9}-\frac{3 \yo^2 \ao^3}{2}+\frac{5 \yo \ao^3}{3}-\frac{11 \
\ao^3}{18}-\yo^3 \ao^2+\frac{15 \yo^2 \ao^2}{4}-5 \yo \ao^2+\frac{11 \
\ao^2}{4}+\frac{4 \yo^3 \ao}{3}-\frac{11 \yo^2 \ao}{2}-2 d_1 \ao+d_1' \
\ao+9 \yo \ao-\frac{25 \yo^3}{36}+\frac{8 \yo^2}{3}+2 \
d_1-d_1'-\frac{49 \yo}{12}+\Big(-2 \ao+\frac{2}{\yo-1}+2\Big) \
H(0;\ao)+\Big(-2 \ao d_1+\frac{2 d_1}{\yo-1}+2 d_1\Big) \
H(1;\ao)+\frac{2 d_1}{\yo-1}-\frac{d_1'}{\yo-1}-\frac{61}{12 \
(\yo-1)}-\frac{1}{2 (\yo-1)^2}-\frac{55}{12}\Big) \
H(c_1(\ao);\yo)+\Big(-\frac{2 \yo^3}{3}+3 \yo^2-6 \
\yo+\frac{2}{\yo-1}+2\Big) H(0,0;\ao)+\Big(-\frac{10 \yo^3}{3}+15 \
\yo^2-30 \yo-\frac{6}{\yo-1}-14\Big) H(0,0;\yo)+\Big(-\frac{2 d_1 \
\yo^3}{3}+3 d_1 \yo^2-6 d_1 \yo+2 d_1+\frac{2 d_1}{\yo-1}\Big) H(0,1;\
\ao)+H(1;\yo) \Big(-\frac{1}{12} d_1' \yo^3 \ao^4+\frac{1}{4} d_1' \
\yo^2 \ao^4+\frac{d_1' \ao^4}{12}-\frac{1}{4} d_1' \yo \
\ao^4+\frac{4}{9} d_1' \yo^3 \ao^3-\frac{3}{2} d_1' \yo^2 \
\ao^3-\frac{11 d_1' \ao^3}{18}+\frac{5}{3} d_1' \yo \ao^3-d_1' \yo^3 \
\ao^2+\frac{15}{4} d_1' \yo^2 \ao^2+\frac{9 d_1' \ao^2}{4}-5 d_1' \yo \
\ao^2+\frac{4}{3} d_1' \yo^3 \ao-\frac{11}{2} d_1' \yo^2 \ao-\frac{23 \
d_1' \ao}{6}+9 d_1' \yo \ao-\frac{d_1'^2 \yo^3}{9}+\frac{4 d_1' \
\yo^3}{3}+\frac{49 d_1'^2}{36}+\frac{7 d_1'^2 \yo^2}{12}-\frac{25 \
d_1' \yo^2}{4}-\frac{133 d_1'}{12}-\frac{11 d_1'^2 \yo}{6}+16 d_1' \
\yo+\Big(-\frac{d_1' \yo^3}{3}+\frac{\yo^3}{3}+\frac{3 d_1' \
\yo^2}{2}-\frac{3 \yo^2}{2}-3 d_1' \yo+3 \yo+\frac{11 \
d_1'}{6}-\frac{2 \
d_1}{\yo-1}+\frac{d_1'}{\yo-1}+\frac{1}{\yo-1}+5\Big) H(0;\ao)-2 \
H(0,0;\ao)-2 d_1 H(0,1;\ao)+\frac{\pi ^2}{6}\Big)+\Big(-d_1' \
\yo^3+\frac{9 d_1' \yo^2}{2}-9 d_1' \yo-3 d_1'+(2 d_1-2) \
H(0;\ao)-\frac{d_1'}{\yo-1}\Big) \
H(0,1;\yo)+\Big(\frac{\yo^3}{3}-\frac{3 \yo^2}{2}+3 \yo-2 H(0;\ao)-2 \
d_1 H(1;\ao)+\frac{3}{\yo-1}+5\Big) H(0,c_1(\ao);\yo)+\Big(-d_1' \
\yo^3-\frac{\yo^3}{3}+\frac{9 d_1' \yo^2}{2}+\frac{3 \yo^2}{2}-9 d_1' \
\yo-3 \yo+\frac{11 d_1'}{2}-2 H(0;\ao)+\frac{2 \
d_1}{\yo-1}-\frac{d_1'}{\yo-1}-\frac{1}{\yo-1}-5\Big) \
H(1,0;\yo)+\Big(-\frac{1}{3} d_1'^2 \yo^3+\frac{3 d_1'^2 \yo^2}{2}-3 \
d_1'^2 \yo+\frac{11 d_1'^2}{6}+(2 d_1-d_1'-1) H(0;\ao)\Big) \
H(1,1;\yo)+\Big(\frac{\yo^3}{3}-\frac{3 \yo^2}{2}+3 \yo-2 H(0;\ao)-2 \
d_1 H(1;\ao)-\frac{2 d_1}{\yo-1}+\frac{d_1'}{\yo-1}+\frac{1}{\yo-1}+5\
\Big) H(1,c_1(\ao);\yo)+\Big(-2 \ao+\frac{2}{\yo-1}+2\Big) \
H(c_1(\ao),0;\yo)+\Big(-\ao d_1'+\frac{d_1'}{\yo-1}+d_1'\Big) \
H(c_1(\ao),1;\yo)+\Big(-\ao+\frac{1}{\yo-1}+1\Big) \
H(c_1(\ao),c_1(\ao);\yo)+16 H(0,0,0;\yo)+4 d_1' H(0,0,1;\yo)-4 \
H(0,0,c_1(\ao);\yo)+(-2 d_1+4 d_1'+2) H(0,1,0;\yo)+d_1'^2 \
H(0,1,1;\yo)+(2 d_1-d_1'-2) H(0,1,c_1(\ao);\yo)-2 \
H(0,c_1(\ao),0;\yo)-d_1' \
H(0,c_1(\ao),1;\yo)-H(0,c_1(\ao),c_1(\ao);\yo)+6 H(1,0,0;\yo)+d_1' \
H(1,0,1;\yo)-3 H(1,0,c_1(\ao);\yo)+(-2 d_1+d_1'+1) H(1,1,0;\yo)+(2 \
d_1-d_1'-1) H(1,1,c_1(\ao);\yo)-2 H(1,c_1(\ao),0;\yo)-d_1' \
H(1,c_1(\ao),1;\yo)-H(1,c_1(\ao),c_1(\ao);\yo)-\frac{\pi ^2}{6 \
(\yo-1)}+\frac{3 \zeta_3}{2}-\frac{\pi ^2}{6}-2.
\erp

%
% The K*I integral for k=2
%

\subsection{The $\cKI$ integral for $k=2$}
%
% This file contains the TeX output produced by Mathematica for the integral JI for arbitrary kap=0  and D0 = 3 +d'1 ep
%
The $\eps$ expansion for this integral reads
\beq
\cKI(\ep;y_{0},d'_{0},\alpha_{0},d_{0};1)=\frac{1}{\eps^3}\cki_{-3}^{(2)}+\frac{1}{\eps^2}\cki_{-2}^{(2)}+\frac{1}{\eps}\cki_{-1}^{(2)}+\cki_0^{(2)}+\ocal\left(\eps\right),
\eeq
where
%% 1/ep^3
\brp
\cki_{-3}^{(2)}=-\frac{1}{6},
\erp
%% 1/ep^2
\brp
\cki_{-2}^{(2)}=-\frac{2 \yo^3}{9}+\yo^2-2 \yo+\frac{2}{3} H(0;\yo)-\frac{4}{9},
\erp
%% 1/ep
\brp
\cki_{-1}^{(2)}=\frac{\yo^3 \ao^4}{18}-\frac{\yo^2 \ao^4}{12}+\frac{\yo \
\ao^4}{3}-\frac{\ao^4}{6 (\yo-2)}-\frac{\ao^4}{12}-\frac{8 \yo^3 \
\ao^3}{27}+\frac{2 \yo^2 \ao^3}{3}-\frac{8 \yo \
\ao^3}{9}-\frac{\ao^3}{\yo-2}-\frac{4 \ao^3}{9 (\yo-2)^2}-\frac{7 \
\ao^3}{18}+\frac{2 \yo^3 \ao^2}{3}-2 \yo^2 \ao^2+\frac{5 \yo \
\ao^2}{3}-\frac{8 \ao^2}{3 (\yo-2)}-\frac{11 \ao^2}{3 \
(\yo-2)^2}-\frac{4 \ao^2}{3 (\yo-2)^3}-\frac{7 \ao^2}{12}-\frac{8 \
\yo^3 \ao}{9}+\frac{10 \yo^2 \ao}{3}-4 \yo \ao-\frac{13 \ao}{3 \
(\yo-2)}-\frac{18 \ao}{(\yo-2)^2}-\frac{52 \ao}{3 (\yo-2)^3}-\frac{16 \
\ao}{3 (\yo-2)^4}+\frac{\ao}{2}+\frac{2 d_1' \yo^3}{27}-\frac{25 \
\yo^3}{27}-\frac{7 d_1' \yo^2}{18}+\frac{13 \yo^2}{3}+\frac{11 d_1' \
\yo}{9}-11 \yo+\Big(\frac{2 \yo^3}{9}-\yo^2+2 \yo-\frac{2}{3 \
(\yo-1)}-\frac{80}{3 (\yo-2)^2}-\frac{160}{3 \
(\yo-2)^3}-\frac{40}{(\yo-2)^4}-\frac{32}{3 \
(\yo-2)^5}+\frac{3}{2}\Big) H(0;\ao)+\Big(\frac{2 \yo^3}{3}-3 \yo^2+6 \
\yo+\frac{2}{3 (\yo-1)}+\frac{80}{3 (\yo-2)^2}+\frac{160}{3 \
(\yo-2)^3}+\frac{40}{(\yo-2)^4}+\frac{32}{3 \
(\yo-2)^5}+\frac{5}{18}\Big) H(0;\yo)+\Big(\frac{2 d_1' \
\yo^3}{9}-d_1' \yo^2+2 d_1' \yo-\frac{11 d_1'}{9}+\frac{2}{3} \
H(0;\ao)\Big) H(1;\yo)+\Big(\frac{2 \ao}{3}-\frac{2}{3 \
(\yo-1)}-\frac{2}{3}\Big) \
H(c_1(\ao);\yo)+\Big(\frac{\ao^4}{2}+\frac{2 \ao^3}{3}-\ao^2-2 \
\ao-\frac{80}{3 (\yo-2)^2}-\frac{160}{3 \
(\yo-2)^3}-\frac{40}{(\yo-2)^4}-\frac{32}{3 \
(\yo-2)^5}+\frac{11}{6}\Big) H(c_2(\ao);\yo)-\frac{8}{3} \
H(0,0;\yo)-\frac{2}{3} d_1' H(0,1;\yo)+\frac{2}{3} H(0,c_1(\ao);\yo)-\
\frac{2}{3} H(1,0;\yo)+\frac{2}{3} H(1,c_1(\ao);\yo)-\frac{80 \ln 2\, \
}{3 (\yo-2)^2}-\frac{160 \ln 2\, }{3 (\yo-2)^3}-\frac{40 \ln 2\, \
}{(\yo-2)^4}-\frac{32 \ln 2\, }{3 (\yo-2)^5}+\frac{13 \ln 2\, \
}{6}-\frac{26}{27},
\erp
%% ep^0
\brp
\cki_0^{(2)} =-\frac{1}{36} d_1 \yo^3 \ao^4-\frac{1}{54} d_1' \yo^3 \ao^4+\frac{7 \
\yo^3 \ao^4}{27}+\frac{1}{24} d_1 \yo^2 \ao^4+\frac{1}{72} d_1' \yo^2 \
\ao^4-\frac{5 \yo^2 \ao^4}{18}+\frac{d_1 \ao^4}{24}-\frac{1}{6} d_1 \
\yo \ao^4-\frac{11}{36} d_1' \yo \ao^4+\frac{41 \yo \
\ao^4}{18}+\frac{d_1 \ao^4}{12 (\yo-2)}-\frac{31 \ao^4}{36 \
(\yo-2)}-\frac{31 \ao^4}{72}+\frac{13}{81} d_1 \yo^3 \
\ao^3+\frac{8}{81} d_1' \yo^3 \ao^3-\frac{229 \yo^3 \
\ao^3}{162}-\frac{7}{18} d_1 \yo^2 \ao^3-\frac{5}{27} d_1' \yo^2 \
\ao^3+\frac{305 \yo^2 \ao^3}{108}+\frac{43 d_1 \ao^3}{108}-\frac{d_1' \
\ao^3}{18}+\frac{10}{27} d_1 \yo \ao^3+\frac{14}{27} d_1' \yo \
\ao^3-\frac{541 \yo \ao^3}{108}+\frac{17 d_1 \ao^3}{18 \
(\yo-2)}-\frac{d_1' \ao^3}{9 (\yo-2)}-\frac{29 \ao^3}{6 \
(\yo-2)}+\frac{8 d_1 \ao^3}{27 (\yo-2)^2}-\frac{64 \ao^3}{27 \
(\yo-2)^2}-\frac{197 \ao^3}{108}-\frac{23}{54} d_1 \yo^3 \
\ao^2-\frac{2}{9} d_1' \yo^3 \ao^2+\frac{359 \yo^3 \
\ao^2}{108}+\frac{17}{12} d_1 \yo^2 \ao^2+\frac{2}{3} d_1' \yo^2 \
\ao^2-\frac{2099 \yo^2 \ao^2}{216}+\frac{113 d_1 \ao^2}{72}-\frac{13 \
d_1' \ao^2}{36}-\frac{10}{9} d_1 \yo \ao^2-\frac{1}{3} d_1' \yo \
\ao^2+\frac{215 \yo \ao^2}{27}+\frac{46 d_1 \ao^2}{9 (\yo-2)}-\frac{8 \
d_1' \ao^2}{9 (\yo-2)}-\frac{25 \ao^2}{2 (\yo-2)}+\frac{83 d_1 \
\ao^2}{18 (\yo-2)^2}-\frac{d_1' \ao^2}{3 (\yo-2)^2}-\frac{349 \
\ao^2}{18 (\yo-2)^2}+\frac{4 d_1 \ao^2}{3 (\yo-2)^3}-\frac{68 \
\ao^2}{9 (\yo-2)^3}-\frac{169 \ao^2}{72}+\frac{25}{27} d_1 \yo^3 \ao+\
\frac{8}{27} d_1' \yo^3 \ao-\frac{91 \yo^3 \ao}{18}-\frac{23}{6} d_1 \
\yo^2 \ao-\frac{11}{9} d_1' \yo^2 \ao+\frac{2099 \yo^2 \
\ao}{108}+\frac{83 d_1 \ao}{36}-\frac{19 d_1' \ao}{18}+\frac{52 d_1 \
\yo \ao}{9}+\frac{14 d_1' \yo \ao}{9}-\frac{949 \yo \
\ao}{36}+\frac{383 d_1 \ao}{18 (\yo-2)}-\frac{35 d_1' \ao}{9 \
(\yo-2)}-\frac{385 \ao}{18 (\yo-2)}-\frac{\ao}{3 (\yo-1)}+\frac{151 \
d_1 \ao}{3 (\yo-2)^2}-\frac{38 d_1' \ao}{9 (\yo-2)^2}-\frac{983 \
\ao}{9 (\yo-2)^2}+\frac{118 d_1 \ao}{3 (\yo-2)^3}-\frac{4 d_1' \ao}{3 \
(\yo-2)^3}-\frac{998 \ao}{9 (\yo-2)^3}+\frac{32 d_1 \ao}{3 \
(\yo-2)^4}-\frac{320 \ao}{9 (\yo-2)^4}+\frac{167 \ao}{36}-\frac{2 \
d_1'^2 \yo^3}{81}+\frac{31 d_1' \yo^3}{81}+\frac{\pi ^2 \
\yo^3}{27}-\frac{230 \yo^3}{81}+\frac{17 d_1'^2 \yo^2}{108}-\frac{247 \
d_1' \yo^2}{108}-\frac{\pi ^2 \yo^2}{6}+\frac{247 \yo^2}{18}-\frac{49 \
d_1'^2 \yo}{54}+\frac{304 d_1' \yo}{27}+\frac{\pi ^2 \
\yo}{3}-\frac{134 \yo}{3}+\Big(-\frac{1}{9} \yo^3 \ao^4+\frac{\yo^2 \
\ao^4}{6}-\frac{2 \yo \ao^4}{3}+\frac{\ao^4}{3 \
(\yo-2)}+\frac{\ao^4}{6}+\frac{16 \yo^3 \ao^3}{27}-\frac{4 \yo^2 \
\ao^3}{3}+\frac{16 \yo \ao^3}{9}+\frac{2 \ao^3}{\yo-2}+\frac{8 \
\ao^3}{9 (\yo-2)^2}+\frac{7 \ao^3}{9}-\frac{4 \yo^3 \ao^2}{3}+4 \yo^2 \
\ao^2-\frac{10 \yo \ao^2}{3}+\frac{16 \ao^2}{3 (\yo-2)}+\frac{22 \
\ao^2}{3 (\yo-2)^2}+\frac{8 \ao^2}{3 (\yo-2)^3}+\frac{7 \
\ao^2}{6}+\frac{16 \yo^3 \ao}{9}-\frac{20 \yo^2 \ao}{3}+8 \yo \
\ao+\frac{26 \ao}{3 (\yo-2)}+\frac{36 \ao}{(\yo-2)^2}+\frac{104 \
\ao}{3 (\yo-2)^3}+\frac{32 \ao}{3 (\yo-2)^4}-\ao-\frac{2 d_1' \
\yo^3}{27}+\frac{25 \yo^3}{54}+\frac{7 d_1' \yo^2}{18}-\frac{95 \
\yo^2}{36}-\frac{41 d_1}{36}-\frac{d_1'}{18}-\frac{11 d_1' \yo}{9}+9 \
\yo-\frac{8 d_1'}{3 (\yo-2)}+\frac{20}{3 (\yo-2)}+\frac{4 d_1}{3 \
(\yo-1)}-\frac{2 d_1'}{3 (\yo-1)}-\frac{7}{2 (\yo-1)}+\frac{16 d_1}{(\
\yo-2)^2}-\frac{12 d_1'}{(\yo-2)^2}-\frac{344}{3 \
(\yo-2)^2}-\frac{1}{3 (\yo-1)^2}+\frac{128 d_1}{9 (\yo-2)^3}-\frac{88 \
d_1'}{9 (\yo-2)^3}-\frac{2152}{9 (\yo-2)^3}+\frac{4 \
d_1}{(\yo-2)^4}-\frac{8 d_1'}{3 (\yo-2)^4}-\frac{548}{3 \
(\yo-2)^4}-\frac{448}{9 (\yo-2)^5}+\frac{317}{36}\Big) \
H(0;\ao)+\Big(-\frac{1}{9} d_1 \yo^3 \ao^4+\frac{1}{6} d_1 \yo^2 \
\ao^4+\frac{d_1 \ao^4}{6}-\frac{2}{3} d_1 \yo \ao^4+\frac{d_1 \
\ao^4}{3 (\yo-2)}+\frac{16}{27} d_1 \yo^3 \ao^3-\frac{4}{3} d_1 \yo^2 \
\ao^3+\frac{7 d_1 \ao^3}{9}+\frac{16}{9} d_1 \yo \ao^3+\frac{2 d_1 \
\ao^3}{\yo-2}+\frac{8 d_1 \ao^3}{9 (\yo-2)^2}-\frac{4}{3} d_1 \yo^3 \
\ao^2+4 d_1 \yo^2 \ao^2+\frac{7 d_1 \ao^2}{6}-\frac{10}{3} d_1 \yo \
\ao^2+\frac{16 d_1 \ao^2}{3 (\yo-2)}+\frac{22 d_1 \ao^2}{3 \
(\yo-2)^2}+\frac{8 d_1 \ao^2}{3 (\yo-2)^3}+\frac{16}{9} d_1 \yo^3 \
\ao-\frac{20}{3} d_1 \yo^2 \ao-d_1 \ao+8 d_1 \yo \ao+\frac{26 d_1 \
\ao}{3 (\yo-2)}+\frac{36 d_1 \ao}{(\yo-2)^2}+\frac{104 d_1 \ao}{3 \
(\yo-2)^3}+\frac{32 d_1 \ao}{3 (\yo-2)^4}-\frac{25 d_1 \
\yo^3}{27}+\frac{23 d_1 \yo^2}{6}-\frac{10 d_1}{9}-\frac{52 d_1 \
\yo}{9}-\frac{49 d_1}{3 (\yo-2)}-\frac{398 d_1}{9 \
(\yo-2)^2}-\frac{112 d_1}{3 (\yo-2)^3}-\frac{32 d_1}{3 \
(\yo-2)^4}\Big) H(1;\ao)+\Big(-\frac{1}{18} \yo^3 \ao^4+\frac{\yo^2 \
\ao^4}{12}-\frac{\yo \ao^4}{3}+\frac{\ao^4}{6 (\yo-2)}+\frac{17 \
\ao^4}{36}+\frac{8 \yo^3 \ao^3}{27}-\frac{2 \yo^2 \ao^3}{3}+\frac{8 \
\yo \ao^3}{9}+\frac{\ao^3}{\yo-2}+\frac{4 \ao^3}{9 \
(\yo-2)^2}+\frac{29 \ao^3}{54}-\frac{2 \yo^3 \ao^2}{3}+2 \yo^2 \ao^2-\
\frac{5 \yo \ao^2}{3}+\frac{8 \ao^2}{3 (\yo-2)}+\frac{11 \ao^2}{3 \
(\yo-2)^2}+\frac{4 \ao^2}{3 (\yo-2)^3}+\frac{17 \ao^2}{12}+\frac{8 \
\yo^3 \ao}{9}-\frac{10 \yo^2 \ao}{3}-\frac{4 d_1 \ao}{3}+\frac{2 d_1' \
\ao}{3}+4 \yo \ao+\frac{13 \ao}{3 (\yo-2)}+\frac{18 \
\ao}{(\yo-2)^2}+\frac{52 \ao}{3 (\yo-2)^3}+\frac{16 \ao}{3 \
(\yo-2)^4}-\frac{19 \ao}{18}-\frac{25 \yo^3}{54}+\frac{61 \yo^2}{36}+\
\frac{4 d_1}{3}-\frac{2 d_1'}{3}-2 \yo+\Big(-\frac{4 \
\ao}{3}+\frac{4}{3 (\yo-1)}+\frac{4}{3}\Big) H(0;\ao)+\Big(-\frac{4 \
\ao d_1}{3}+\frac{4 d_1}{3 (\yo-1)}+\frac{4 d_1}{3}\Big) \
H(1;\ao)-\frac{32}{3 (\yo-2)}+\frac{4 d_1}{3 (\yo-1)}-\frac{2 d_1'}{3 \
(\yo-1)}-\frac{7}{2 (\yo-1)}-\frac{332}{9 (\yo-2)^2}-\frac{1}{3 \
(\yo-1)^2}-\frac{104}{3 (\yo-2)^3}-\frac{32}{3 \
(\yo-2)^4}-\frac{53}{18}\Big) H(c_1(\ao);\yo)+\Big(-\frac{d_1 \
\ao^4}{4}+\frac{d_1' \ao^4}{6}+\frac{25 \ao^4}{12}-\frac{7 d_1 \
\ao^3}{9}+\frac{5 d_1' \ao^3}{9}+\frac{11 \ao^3}{9}-\frac{d_1 \
\ao^2}{6}+\frac{d_1' \ao^2}{3}-\frac{13 \ao^2}{2}+\frac{11 d_1 \
\ao}{3}-\frac{5 d_1' \ao}{3}-\frac{23 \ao}{3}-\frac{89 \
d_1}{36}+\frac{11 d_1'}{18}+\Big(-\ao^4-\frac{4 \ao^3}{3}+2 \ao^2+4 \
\ao+\frac{160}{3 (\yo-2)^2}+\frac{320}{3 \
(\yo-2)^3}+\frac{80}{(\yo-2)^4}+\frac{64}{3 \
(\yo-2)^5}-\frac{11}{3}\Big) H(0;\ao)+\Big(-d_1 \ao^4-\frac{4 d_1 \
\ao^3}{3}+2 d_1 \ao^2+4 d_1 \ao-\frac{11 d_1}{3}+\frac{160 d_1}{3 \
(\yo-2)^2}+\frac{320 d_1}{3 (\yo-2)^3}+\frac{80 \
d_1}{(\yo-2)^4}+\frac{64 d_1}{3 (\yo-2)^5}\Big) H(1;\ao)-\frac{8 \
d_1'}{3 (\yo-2)}+\frac{52}{3 (\yo-2)}+\frac{16 \
d_1}{(\yo-2)^2}-\frac{12 d_1'}{(\yo-2)^2}-\frac{700}{9 \
(\yo-2)^2}+\frac{128 d_1}{9 (\yo-2)^3}-\frac{88 d_1'}{9 \
(\yo-2)^3}-\frac{1840}{9 (\yo-2)^3}+\frac{4 d_1}{(\yo-2)^4}-\frac{8 \
d_1'}{3 (\yo-2)^4}-\frac{172}{(\yo-2)^4}-\frac{448}{9 \
(\yo-2)^5}+\frac{391}{36}\Big) H(c_2(\ao);\yo)+\Big(-\frac{4 \
\yo^3}{9}+2 \yo^2-4 \yo+\frac{4}{3 (\yo-1)}+\frac{160}{3 \
(\yo-2)^2}+\frac{320}{3 (\yo-2)^3}+\frac{80}{(\yo-2)^4}+\frac{64}{3 (\
\yo-2)^5}-3\Big) H(0,0;\ao)+\Big(-\frac{20 \yo^3}{9}+10 \yo^2-20 \yo-\
\frac{4}{\yo-1}-\frac{160}{(\yo-2)^2}-\frac{320}{(\yo-2)^3}-\frac{240}\
{(\yo-2)^4}-\frac{64}{(\yo-2)^5}+\frac{17}{9}\Big) \
H(0,0;\yo)+\Big(-\frac{4 d_1 \yo^3}{9}+2 d_1 \yo^2-4 d_1 \yo-3 \
d_1+\frac{4 d_1}{3 (\yo-1)}+\frac{160 d_1}{3 (\yo-2)^2}+\frac{320 \
d_1}{3 (\yo-2)^3}+\frac{80 d_1}{(\yo-2)^4}+\frac{64 d_1}{3 (\yo-2)^5}\
\Big) H(0,1;\ao)+\Big(-\frac{2 d_1' \yo^3}{3}+3 d_1' \yo^2-6 d_1' \
\yo-\frac{5 d_1'}{18}+\Big(\frac{4 d_1}{3}-\frac{4}{3}\Big) H(0;\ao)-\
\frac{2 d_1'}{3 (\yo-1)}-\frac{80 d_1'}{3 (\yo-2)^2}-\frac{160 \
d_1'}{3 (\yo-2)^3}-\frac{40 d_1'}{(\yo-2)^4}-\frac{32 d_1'}{3 \
(\yo-2)^5}\Big) H(0,1;\yo)+\Big(\frac{2 \yo^3}{9}-\yo^2+2 \
\yo-\frac{4}{3} H(0;\ao)-\frac{4}{3} d_1 \
H(1;\ao)+\frac{2}{\yo-1}-\frac{80}{3 (\yo-2)^2}-\frac{160}{3 \
(\yo-2)^3}-\frac{40}{(\yo-2)^4}-\frac{32}{3 (\yo-2)^5}+\frac{107}{18}\
\Big) H(0,c_1(\ao);\yo)+\Big(-\frac{26}{3}+\frac{320}{3 \
(\yo-2)^2}+\frac{640}{3 (\yo-2)^3}+\frac{160}{(\yo-2)^4}+\frac{128}{3 \
(\yo-2)^5}\Big) H(0,c_2(\ao);\yo)+\Big(-\frac{2 d_1' \
\yo^3}{3}-\frac{2 \yo^3}{9}+3 d_1' \yo^2+\yo^2-6 d_1' \yo-2 \
\yo+\frac{19 d_1'}{3}-\frac{4}{3} H(0;\ao)+\frac{4 d_1}{3 \
(\yo-1)}-\frac{2 d_1'}{3 (\yo-1)}-\frac{2}{3 (\yo-1)}-\frac{80 \
d_1'}{3 (\yo-2)^2}+\frac{80}{3 (\yo-2)^2}-\frac{160 d_1'}{3 \
(\yo-2)^3}+\frac{160}{3 (\yo-2)^3}-\frac{40 \
d_1'}{(\yo-2)^4}+\frac{40}{(\yo-2)^4}-\frac{32 d_1'}{3 \
(\yo-2)^5}+\frac{32}{3 (\yo-2)^5}-\frac{107}{18}\Big) \
H(1,0;\yo)+\Big(-\frac{2}{9} d_1'^2 \yo^3+d_1'^2 \yo^2-2 d_1'^2 \
\yo+\frac{11 d_1'^2}{9}+\Big(\frac{4 d_1}{3}-\frac{2 \
d_1'}{3}-\frac{2}{3}\Big) H(0;\ao)\Big) H(1,1;\yo)+\Big(\frac{2 \
\yo^3}{9}-\yo^2+2 \yo-\frac{4}{3} H(0;\ao)-\frac{4}{3} d_1 \
H(1;\ao)-\frac{4 d_1}{3 (\yo-1)}+\frac{2 d_1'}{3 (\yo-1)}+\frac{2}{3 \
(\yo-1)}-\frac{80}{3 (\yo-2)^2}-\frac{160}{3 \
(\yo-2)^3}-\frac{40}{(\yo-2)^4}-\frac{32}{3 (\yo-2)^5}+\frac{107}{18}\
\Big) H(1,c_1(\ao);\yo)+\Big(\frac{80 d_1'}{3 (\yo-2)^2}+\frac{160 \
d_1'}{3 (\yo-2)^3}+\frac{40 d_1'}{(\yo-2)^4}+\frac{32 d_1'}{3 \
(\yo-2)^5}-\frac{8 d_1'}{3}\Big) H(1,c_2(\ao);\yo)+\Big(-\frac{160 \
d_1}{3 (\yo-2)^2}-\frac{320 d_1}{3 (\yo-2)^3}-\frac{80 \
d_1}{(\yo-2)^4}-\frac{64 d_1}{3 (\yo-2)^5}+\frac{8 \
d_1'}{3}+\frac{160}{3 (\yo-2)^2}+\frac{320}{3 \
(\yo-2)^3}+\frac{80}{(\yo-2)^4}+\frac{64}{3 \
(\yo-2)^5}-\frac{26}{3}\Big) H(2,0;\yo)+\Big(\frac{160 d_1}{3 \
(\yo-2)^2}+\frac{320 d_1}{3 (\yo-2)^3}+\frac{80 \
d_1}{(\yo-2)^4}+\frac{64 d_1}{3 (\yo-2)^5}-\frac{8 \
d_1'}{3}-\frac{160}{3 (\yo-2)^2}-\frac{320}{3 \
(\yo-2)^3}-\frac{80}{(\yo-2)^4}-\frac{64}{3 \
(\yo-2)^5}+\frac{26}{3}\Big) H(2,c_2(\ao);\yo)+\Big(-\frac{4 \ao}{3}+\
\frac{4}{3 (\yo-1)}+\frac{4}{3}\Big) H(c_1(\ao),0;\yo)+\Big(-\frac{2 \
\ao d_1'}{3}+\frac{2 d_1'}{3 (\yo-1)}+\frac{2 d_1'}{3}\Big) \
H(c_1(\ao),1;\yo)+\Big(-\frac{2 \ao}{3}+\frac{2}{3 \
(\yo-1)}+\frac{2}{3}\Big) \
H(c_1(\ao),c_1(\ao);\yo)+\Big(-\ao^4-\frac{4 \ao^3}{3}+2 \ao^2+4 \ao+\
\frac{160}{3 (\yo-2)^2}+\frac{320}{3 (\yo-2)^3}+\frac{80}{(\yo-2)^4}+\
\frac{64}{3 (\yo-2)^5}-\frac{11}{3}\Big) \
H(c_2(\ao),0;\yo)+\Big(-\frac{d_1' \ao^4}{2}-\frac{2 d_1' \
\ao^3}{3}+d_1' \ao^2+2 d_1' \ao-\frac{11 d_1'}{6}+\frac{80 d_1'}{3 \
(\yo-2)^2}+\frac{160 d_1'}{3 (\yo-2)^3}+\frac{40 \
d_1'}{(\yo-2)^4}+\frac{32 d_1'}{3 (\yo-2)^5}\Big) \
H(c_2(\ao),1;\yo)+\Big(-\frac{\ao^4}{2}-\frac{2 \ao^3}{3}+\ao^2+2 \
\ao+\frac{80}{3 (\yo-2)^2}+\frac{160}{3 \
(\yo-2)^3}+\frac{40}{(\yo-2)^4}+\frac{32}{3 \
(\yo-2)^5}-\frac{11}{6}\Big) H(c_2(\ao),c_1(\ao);\yo)+\frac{32}{3} \
H(0,0,0;\yo)+\frac{8}{3} d_1' H(0,0,1;\yo)-\frac{8}{3} \
H(0,0,c_1(\ao);\yo)+\Big(-\frac{4 d_1}{3}+\frac{8 \
d_1'}{3}+\frac{4}{3}\Big) H(0,1,0;\yo)+\frac{2}{3} d_1'^2 \
H(0,1,1;\yo)+\Big(\frac{4 d_1}{3}-\frac{2 d_1'}{3}-\frac{4}{3}\Big) \
H(0,1,c_1(\ao);\yo)-\frac{4}{3} H(0,c_1(\ao),0;\yo)-\frac{2}{3} d_1' \
H(0,c_1(\ao),1;\yo)-\frac{2}{3} H(0,c_1(\ao),c_1(\ao);\yo)+4 H(1,0,0;\
\yo)+\frac{2}{3} d_1' H(1,0,1;\yo)-2 \
H(1,0,c_1(\ao);\yo)+\Big(-\frac{4 d_1}{3}+\frac{2 \
d_1'}{3}+\frac{2}{3}\Big) H(1,1,0;\yo)+\Big(\frac{4 d_1}{3}-\frac{2 \
d_1'}{3}-\frac{2}{3}\Big) H(1,1,c_1(\ao);\yo)-\frac{4}{3} \
H(1,c_1(\ao),0;\yo)-\frac{2}{3} d_1' H(1,c_1(\ao),1;\yo)-\frac{2}{3} \
H(1,c_1(\ao),c_1(\ao);\yo)+H(0;\yo) \Big(-\frac{1}{9} \yo^3 \
\ao^4+\frac{\yo^2 \ao^4}{6}-\frac{2 \yo \ao^4}{3}+\frac{\ao^4}{3 \
(\yo-2)}+\frac{\ao^4}{6}+\frac{16 \yo^3 \ao^3}{27}-\frac{4 \yo^2 \
\ao^3}{3}+\frac{16 \yo \ao^3}{9}+\frac{2 \ao^3}{\yo-2}+\frac{8 \
\ao^3}{9 (\yo-2)^2}+\frac{7 \ao^3}{9}-\frac{4 \yo^3 \ao^2}{3}+4 \yo^2 \
\ao^2-\frac{10 \yo \ao^2}{3}+\frac{16 \ao^2}{3 (\yo-2)}+\frac{22 \
\ao^2}{3 (\yo-2)^2}+\frac{8 \ao^2}{3 (\yo-2)^3}+\frac{7 \
\ao^2}{6}+\frac{16 \yo^3 \ao}{9}-\frac{20 \yo^2 \ao}{3}+8 \yo \
\ao+\frac{26 \ao}{3 (\yo-2)}+\frac{36 \ao}{(\yo-2)^2}+\frac{104 \
\ao}{3 (\yo-2)^3}+\frac{32 \ao}{3 (\yo-2)^4}-\ao-\frac{2 d_1' \
\yo^3}{9}+\frac{175 \yo^3}{54}+\frac{7 d_1' \yo^2}{6}-\frac{529 \
\yo^2}{36}+\frac{41 d_1}{36}+\frac{d_1'}{18}-\frac{11 d_1' \yo}{3}+35 \
\yo+\Big(-\frac{4 \yo^3}{9}+2 \yo^2-4 \yo+\frac{4}{3 \
(\yo-1)}+\frac{160}{3 (\yo-2)^2}+\frac{320}{3 \
(\yo-2)^3}+\frac{80}{(\yo-2)^4}+\frac{64}{3 (\yo-2)^5}-3\Big) \
H(0;\ao)+\frac{8 d_1'}{3 (\yo-2)}-\frac{20}{3 (\yo-2)}-\frac{4 d_1}{3 \
(\yo-1)}+\frac{2 d_1'}{3 (\yo-1)}+\frac{7}{2 (\yo-1)}-\frac{16 d_1}{(\
\yo-2)^2}+\frac{12 d_1'}{(\yo-2)^2}+\frac{344}{3 \
(\yo-2)^2}+\frac{1}{3 (\yo-1)^2}-\frac{128 d_1}{9 (\yo-2)^3}+\frac{88 \
d_1'}{9 (\yo-2)^3}+\frac{2152}{9 (\yo-2)^3}-\frac{4 \
d_1}{(\yo-2)^4}+\frac{8 d_1'}{3 (\yo-2)^4}+\frac{548}{3 \
(\yo-2)^4}+\frac{448}{9 (\yo-2)^5}+\frac{320 \ln 2\, }{3 \
(\yo-2)^2}+\frac{640 \ln 2\, }{3 (\yo-2)^3}+\frac{160 \ln 2\, \
}{(\yo-2)^4}+\frac{128 \ln 2\, }{3 (\yo-2)^5}-\frac{26 \ln 2\, \
}{3}-\frac{535}{108}\Big)+H(2;\yo) \Big(\frac{160 \ln 2\,  d_1}{3 \
(\yo-2)^2}+\frac{320 \ln 2\,  d_1}{3 (\yo-2)^3}+\frac{80 \ln 2\,  \
d_1}{(\yo-2)^4}+\frac{64 \ln 2\,  d_1}{3 (\yo-2)^5}+\Big(\frac{160 \
d_1}{3 (\yo-2)^2}+\frac{320 d_1}{3 (\yo-2)^3}+\frac{80 \
d_1}{(\yo-2)^4}+\frac{64 d_1}{3 (\yo-2)^5}-\frac{8 \
d_1'}{3}-\frac{160}{3 (\yo-2)^2}-\frac{320}{3 \
(\yo-2)^3}-\frac{80}{(\yo-2)^4}-\frac{64}{3 \
(\yo-2)^5}+\frac{26}{3}\Big) H(0;\ao)-\frac{8}{3} d_1' \ln 2\, \
-\frac{160 \ln 2\, }{3 (\yo-2)^2}-\frac{320 \ln 2\, }{3 \
(\yo-2)^3}-\frac{80 \ln 2\, }{(\yo-2)^4}-\frac{64 \ln 2\, }{3 \
(\yo-2)^5}+\frac{26 \ln 2\, }{3}\Big)+H(1;\yo) \Big(-\frac{1}{18} \
d_1' \yo^3 \ao^4+\frac{1}{12} d_1' \yo^2 \ao^4+\frac{17 d_1' \
\ao^4}{36}-\frac{1}{3} d_1' \yo \ao^4+\frac{d_1' \ao^4}{6 \
(\yo-2)}+\frac{8}{27} d_1' \yo^3 \ao^3-\frac{2}{3} d_1' \yo^2 \
\ao^3+\frac{d_1' \ao^3}{27}+\frac{8}{9} d_1' \yo \ao^3+\frac{d_1' \
\ao^3}{\yo-2}+\frac{4 d_1' \ao^3}{9 (\yo-2)^2}-\frac{2}{3} d_1' \yo^3 \
\ao^2+2 d_1' \yo^2 \ao^2+\frac{2 d_1' \ao^2}{3}-\frac{5}{3} d_1' \yo \
\ao^2+\frac{8 d_1' \ao^2}{3 (\yo-2)}+\frac{11 d_1' \ao^2}{3 \
(\yo-2)^2}+\frac{4 d_1' \ao^2}{3 (\yo-2)^3}+\frac{8}{9} d_1' \yo^3 \
\ao-\frac{10}{3} d_1' \yo^2 \ao-\frac{23 d_1' \ao}{9}+4 d_1' \yo \ao+\
\frac{13 d_1' \ao}{3 (\yo-2)}+\frac{18 d_1' \ao}{(\yo-2)^2}+\frac{52 \
d_1' \ao}{3 (\yo-2)^3}+\frac{16 d_1' \ao}{3 (\yo-2)^4}-\frac{2 d_1'^2 \
\yo^3}{27}+\frac{25 d_1' \yo^3}{27}+\frac{49 d_1'^2}{54}+\frac{7 \
d_1'^2 \yo^2}{18}-\frac{13 d_1' \yo^2}{3}-\frac{205 \
d_1'}{27}-\frac{11 d_1'^2 \yo}{9}+11 d_1' \yo+\Big(-\frac{2 d_1' \
\yo^3}{9}+\frac{2 \yo^3}{9}+d_1' \yo^2-\yo^2-2 d_1' \yo+2 \
\yo-\frac{13 d_1'}{9}-\frac{4 d_1}{3 (\yo-1)}+\frac{2 d_1'}{3 \
(\yo-1)}+\frac{2}{3 (\yo-1)}+\frac{80 d_1'}{3 (\yo-2)^2}-\frac{80}{3 \
(\yo-2)^2}+\frac{160 d_1'}{3 (\yo-2)^3}-\frac{160}{3 \
(\yo-2)^3}+\frac{40 d_1'}{(\yo-2)^4}-\frac{40}{(\yo-2)^4}+\frac{32 \
d_1'}{3 (\yo-2)^5}-\frac{32}{3 (\yo-2)^5}+\frac{107}{18}\Big) \
H(0;\ao)-\frac{4}{3} H(0,0;\ao)-\frac{4}{3} d_1 \
H(0,1;\ao)-\frac{8}{3} d_1' \ln 2\, +\frac{80 d_1' \ln 2\, }{3 \
(\yo-2)^2}+\frac{160 d_1' \ln 2\, }{3 (\yo-2)^3}+\frac{40 d_1' \ln \
2\, }{(\yo-2)^4}+\frac{32 d_1' \ln 2\, }{3 (\yo-2)^5}+\frac{\pi \
^2}{9}\Big)-\frac{\pi ^2}{9 (\yo-1)}-\frac{20 \pi ^2}{9 \
(\yo-2)^2}-\frac{40 \pi ^2}{9 (\yo-2)^3}-\frac{10 \pi ^2}{3 \
(\yo-2)^4}-\frac{8 \pi ^2}{9 (\yo-2)^5}+\zeta_3-\frac{80 \ln ^22\, \
}{3 (\yo-2)^2}-\frac{160 \ln ^22\, }{3 (\yo-2)^3}-\frac{40 \ln ^22\, \
}{(\yo-2)^4}-\frac{32 \ln ^22\, }{3 (\yo-2)^5}+\frac{13 \ln ^22\, \
}{6}-\frac{89}{36} d_1 \ln 2\, +\frac{11}{18} d_1' \ln 2\, -\frac{8 \
d_1' \ln 2\, }{3 (\yo-2)}+\frac{52 \ln 2\, }{3 (\yo-2)}+\frac{16 d_1 \
\ln 2\, }{(\yo-2)^2}-\frac{12 d_1' \ln 2\, }{(\yo-2)^2}-\frac{700 \ln \
2\, }{9 (\yo-2)^2}+\frac{128 d_1 \ln 2\, }{9 (\yo-2)^3}-\frac{88 d_1' \
\ln 2\, }{9 (\yo-2)^3}-\frac{1840 \ln 2\, }{9 (\yo-2)^3}+\frac{4 d_1 \
\ln 2\, }{(\yo-2)^4}-\frac{8 d_1' \ln 2\, }{3 (\yo-2)^4}-\frac{172 \
\ln 2\, }{(\yo-2)^4}-\frac{448 \ln 2\, }{9 (\yo-2)^5}+\frac{47 \ln \
2\, }{4}+\frac{5 \pi ^2}{72}-\frac{160}{81}.
\erp

%
% The K*I integral for k=-1
%

\subsection{The $\cKI$ integral for $k=-1$}
%
% This file contains the TeX output produced by Mathematica for the integral JI for arbitrary kap=0  and D0 = 3 +d'1 ep
%
The $\eps$ expansion for this integral reads
\beq
\cKI(\ep;y_{0},d'_{0},\alpha_{0},d_{0};1)=\frac{1}{\eps^4}\cki_{-4}^{(-1)}+\frac{1}{\eps^3}\cki_{-3}^{(-1)}+\frac{1}{\eps^2}\cki_{-2}^{(-1)}+\frac{1}{\eps}\cki_{-1}^{(-1)}+\cki_0^{(-1)}+\ocal\left(\eps\right),
\eeq
where
%% 1/ep^4
\brp
\cki_{-4}^{(-1)}=\frac{1}{4},
\erp
%% 1/ep^3
\brp
\cki_{-3}^{(-1)}=\frac{\yo^3}{3}-\frac{3 \yo^2}{2}+3 \yo-H(0;\yo),
\erp
%% 1/ep^2
\brp
\cki_{-2}^{(-1)}=-\frac{d_1' \yo^3}{9}+\frac{7 \yo^3}{9}+\frac{7 d_1' \yo^2}{12}-4 \
\yo^2-\frac{11 d_1' \yo}{6}+13 \yo+\Big(-\frac{4 \yo^3}{3}+6 \yo^2-12 \
\yo\Big) H(0;\yo)+\Big(-\frac{d_1' \yo^3}{3}+\frac{3 d_1' \yo^2}{2}-3 \
d_1' \yo+\frac{11 d_1'}{6}\Big) H(1;\yo)+4 H(0,0;\yo)+d_1' H(0,1;\yo),
\erp
%% 1/ep
\brp
\cki_{-1}^{(-1)} =-\frac{1}{18} \yo^3 \ao^4+\frac{5 \yo^2 \ao^4}{12}-\frac{5 \yo \
\ao^4}{3}+\frac{13 \yo^3 \ao^3}{54}-\frac{17 \yo^2 \ao^3}{9}+\frac{80 \
\yo \ao^3}{9}-\frac{11 \yo^3 \ao^2}{36}+\frac{109 \yo^2 \
\ao^2}{36}-\frac{158 \yo \ao^2}{9}-\frac{7 \yo^3 \ao}{18}-\frac{7 \
\yo^2 \ao}{18}+\frac{43 \yo \ao}{3}+\frac{d_1'^2 \yo^3}{27}-\frac{11 \
d_1' \yo^3}{27}-\frac{\pi ^2 \yo^3}{9}+\frac{44 \yo^3}{27}-\frac{17 \
d_1'^2 \yo^2}{72}+\frac{25 d_1' \yo^2}{9}+\frac{\pi ^2 \
\yo^2}{2}-\frac{37 \yo^2}{4}+\frac{49 d_1'^2 \yo}{36}-\frac{154 d_1' \
\yo}{9}-\pi ^2 \yo+48 \yo+\Big(\frac{7 \yo^3}{6}-\frac{13 \
\yo^2}{3}+\frac{11 \yo}{2}-\frac{13}{6 (\yo-1)}-\frac{13}{6}\Big) \
H(0;\ao)+\Big(\frac{4 d_1' \yo^3}{9}-\frac{77 \yo^3}{18}-\frac{7 d_1' \
\yo^2}{3}+\frac{61 \yo^2}{3}+\frac{22 d_1' \yo}{3}-\frac{115 \yo}{2}+\
\frac{13}{6 (\yo-1)}+\frac{13}{6}\Big) H(0;\yo)+\Big(\frac{d_1'^2 \
\yo^3}{9}-\frac{7 d_1' \yo^3}{9}-\frac{7 d_1'^2 \yo^2}{12}+4 d_1' \
\yo^2+\frac{11 d_1'^2 \yo}{6}-13 d_1' \yo-\frac{49 \
d_1'^2}{36}+\frac{88 d_1'}{9}+\Big(-\frac{2 \yo^3}{3}+3 \yo^2-6 \
\yo-\frac{2}{\yo-1}+\frac{19}{3}\Big) H(0;\ao)-\frac{\pi ^2}{3}\Big) \
H(1;\yo)+\Big(\frac{\yo^3 \ao^4}{6}-\yo^2 \ao^4+\frac{5 \yo \
\ao^4}{2}-\frac{5 \ao^4}{3}-\frac{8 \yo^3 \ao^3}{9}+5 \yo^2 \
\ao^3-\frac{38 \yo \ao^3}{3}+\frac{68 \ao^3}{9}+2 \yo^3 \
\ao^2-\frac{21 \yo^2 \ao^2}{2}+26 \yo \ao^2-\frac{38 \
\ao^2}{3}-\frac{8 \yo^3 \ao}{3}+13 \yo^2 \ao-30 \yo \ao+\frac{41 \
\ao}{3}+\frac{25 \yo^3}{18}-\frac{35 \yo^2}{6}+\frac{23 \
\yo}{2}-\frac{13}{6 (\yo-1)}-\frac{13}{6}\Big) \
H(c_1(\ao);\yo)+\Big(\frac{16 \yo^3}{3}-24 \yo^2+48 \yo\Big) \
H(0,0;\yo)+\Big(\frac{4 d_1' \yo^3}{3}-6 d_1' \yo^2+12 d_1' \yo+4 \
H(0;\ao)\Big) H(0,1;\yo)+\Big(-\ao^4+\frac{16 \ao^3}{3}-12 \ao^2+16 \
\ao-\frac{2 \yo^3}{3}+3 \yo^2-6 \yo-\frac{2}{\yo-1}-2\Big) \
H(0,c_1(\ao);\yo)+\Big(\frac{4 d_1' \yo^3}{3}+\frac{2 \yo^3}{3}-6 \
d_1' \yo^2-3 \yo^2+12 d_1' \yo+6 \yo-\frac{22 \
d_1'}{3}+\frac{2}{\yo-1}-\frac{19}{3}\Big) \
H(1,0;\yo)+\Big(\frac{d_1'^2 \yo^3}{3}-\frac{3 d_1'^2 \yo^2}{2}+3 \
d_1'^2 \yo-\frac{11 d_1'^2}{6}+2 H(0;\ao)\Big) \
H(1,1;\yo)+\Big(-\frac{2 \yo^3}{3}+3 \yo^2-6 \
\yo-\frac{2}{\yo-1}+\frac{19}{3}\Big) H(1,c_1(\ao);\yo)+\Big(-2 \
\ao+\frac{2}{\yo-1}+2\Big) H(c_1(\ao),c_1(\ao);\yo)-16 H(0,0,0;\yo)-4 \
d_1' H(0,0,1;\yo)+4 H(0,0,c_1(\ao);\yo)+(-4 d_1'-4) \
H(0,1,0;\yo)-d_1'^2 H(0,1,1;\yo)+4 H(0,1,c_1(\ao);\yo)-2 \
H(0,c_1(\ao),c_1(\ao);\yo)+2 H(1,0,c_1(\ao);\yo)-2 H(1,1,0;\yo)+2 \
H(1,1,c_1(\ao);\yo)-2 H(1,c_1(\ao),c_1(\ao);\yo)+\frac{\pi ^2}{3 \
(\yo-1)}-\frac{3 \zeta_3}{2}+\frac{\pi ^2}{3},
\erp
% ep^0
\brp
\cki_0^{(-1)} = \frac{1}{36} d_1 \yo^3 \ao^4+\frac{1}{27} d_1' \yo^3 \ao^4-\frac{19 \
\yo^3 \ao^4}{108}-\frac{5}{24} d_1 \yo^2 \ao^4-\frac{13}{36} d_1' \
\yo^2 \ao^4+\frac{35 \yo^2 \ao^4}{24}+\frac{5}{6} d_1 \yo \
\ao^4+\frac{47}{18} d_1' \yo \ao^4-\frac{17 \yo \
\ao^4}{2}-\frac{31}{324} d_1 \yo^3 \ao^3-\frac{29}{162} d_1' \yo^3 \
\ao^3+\frac{263 \yo^3 \ao^3}{324}+\frac{91}{108} d_1 \yo^2 \
\ao^3+\frac{187}{108} d_1' \yo^2 \ao^3-\frac{193 \yo^2 \
\ao^3}{27}-\frac{245}{54} d_1 \yo \ao^3-\frac{1549}{108} d_1' \yo \
\ao^3+\frac{1354 \yo \ao^3}{27}-\frac{17}{216} d_1 \yo^3 \
\ao^2+\frac{35}{108} d_1' \yo^3 \ao^2-\frac{245 \yo^3 \
\ao^2}{216}-\frac{17}{27} d_1 \yo^2 \ao^2-\frac{707}{216} d_1' \yo^2 \
\ao^2+\frac{2803 \yo^2 \ao^2}{216}+\frac{895}{108} d_1 \yo \
\ao^2+\frac{809}{27} d_1' \yo \ao^2-\frac{3013 \yo \
\ao^2}{27}+\frac{205}{108} d_1 \yo^3 \ao-\frac{1}{6} d_1' \yo^3 \
\ao-\frac{58 \yo^3 \ao}{27}-\frac{659}{108} d_1 \yo^2 \
\ao+\frac{74}{27} d_1' \yo^2 \ao-\frac{55 \yo^2 \ao}{27}-\frac{d_1 \
\yo \ao}{36}-\frac{278 d_1' \yo \ao}{9}+\frac{1379 \yo \
\ao}{12}-\frac{7 \ao}{12 (\yo-1)}-\frac{7 \ao}{12}-\frac{d_1'^3 \
\yo^3}{81}+\frac{5 d_1'^2 \yo^3}{27}-\frac{28 d_1' \
\yo^3}{27}+\frac{1}{27} d_1' \pi ^2 \yo^3-\frac{49 \pi ^2 \
\yo^3}{108}+\frac{268 \yo^3}{81}+\frac{43 d_1'^3 \
\yo^2}{432}-\frac{179 d_1'^2 \yo^2}{108}+\frac{1891 d_1' \yo^2}{216}-\
\frac{7}{36} d_1' \pi ^2 \yo^2+\frac{37 \pi ^2 \yo^2}{18}-\frac{161 \
\yo^2}{8}-\frac{251 d_1'^3 \yo}{216}+\frac{545 d_1'^2 \
\yo}{27}-\frac{10847 d_1' \yo}{108}+\frac{11}{18} d_1' \pi ^2 \
\yo-\frac{21 \pi ^2 \yo}{4}+164 \yo+\Big(\frac{\yo^3 \
\ao^4}{9}-\frac{5 \yo^2 \ao^4}{6}+\frac{10 \yo \ao^4}{3}-\frac{13 \
\yo^3 \ao^3}{27}+\frac{34 \yo^2 \ao^3}{9}-\frac{160 \yo \
\ao^3}{9}+\frac{11 \yo^3 \ao^2}{18}-\frac{109 \yo^2 \
\ao^2}{18}+\frac{316 \yo \ao^2}{9}+\frac{7 \yo^3 \ao}{9}+\frac{7 \
\yo^2 \ao}{9}-\frac{86 \yo \ao}{3}-\frac{205 d_1 \yo^3}{108}-\frac{17 \
d_1' \yo^3}{54}+\frac{91 \yo^3}{18}+\frac{22 d_1 \yo^2}{3}+\frac{10 \
d_1' \yo^2}{9}-\frac{187 \yo^2}{9}+\frac{217 d_1}{36}-\frac{d_1'}{6}-\
\frac{469 d_1 \yo}{36}+\frac{7 d_1' \yo}{18}+\frac{314 \
\yo}{9}+\frac{217 d_1}{36 (\yo-1)}-\frac{d_1'}{6 (\yo-1)}-\frac{88}{9 \
(\yo-1)}-\frac{19}{12 (\yo-1)^2}-\frac{295}{36}\Big) \
H(0;\ao)+\Big(\frac{1}{9} d_1 \yo^3 \ao^4-\frac{5}{6} d_1 \yo^2 \
\ao^4+\frac{10}{3} d_1 \yo \ao^4-\frac{13}{27} d_1 \yo^3 \
\ao^3+\frac{34}{9} d_1 \yo^2 \ao^3-\frac{160}{9} d_1 \yo \
\ao^3+\frac{11}{18} d_1 \yo^3 \ao^2-\frac{109}{18} d_1 \yo^2 \
\ao^2+\frac{316}{9} d_1 \yo \ao^2+\frac{7}{9} d_1 \yo^3 \
\ao+\frac{7}{9} d_1 \yo^2 \ao-\frac{86 d_1 \yo \ao}{3}-\frac{55 d_1 \
\yo^3}{54}+\frac{7 d_1 \yo^2}{3}+8 d_1 \yo\Big) \
H(1;\ao)+\Big(-\frac{1}{12} d_1 \yo^3 \ao^4-\frac{1}{18} d_1' \yo^3 \
\ao^4+\frac{5 \yo^3 \ao^4}{12}+\frac{1}{2} d_1 \yo^2 \
\ao^4+\frac{5}{12} d_1' \yo^2 \ao^4-\frac{5 \yo^2 \ao^4}{2}+\frac{5 \
d_1 \ao^4}{6}+\frac{47 d_1' \ao^4}{36}-\frac{5}{4} d_1 \yo \
\ao^4-\frac{5}{3} d_1' \yo \ao^4+\frac{29 \yo \ao^4}{4}-\frac{31 \
\ao^4}{6}+\frac{13}{27} d_1 \yo^3 \ao^3+\frac{8}{27} d_1' \yo^3 \
\ao^3-\frac{125 \yo^3 \ao^3}{54}-\frac{8}{3} d_1 \yo^2 \
\ao^3-\frac{37}{18} d_1' \yo^2 \ao^3+\frac{40 \yo^2 \
\ao^3}{3}-\frac{221 d_1 \ao^3}{54}-\frac{313 d_1' \
\ao^3}{54}+\frac{61}{9} d_1 \yo \ao^3+\frac{77}{9} d_1' \yo \
\ao^3-\frac{379 \yo \ao^3}{9}+\frac{745 \ao^3}{27}-\frac{23}{18} d_1 \
\yo^3 \ao^2-\frac{2}{3} d_1' \yo^3 \ao^2+\frac{203 \yo^3 \
\ao^2}{36}+\frac{13}{2} d_1 \yo^2 \ao^2+\frac{17}{4} d_1' \yo^2 \
\ao^2-\frac{373 \yo^2 \ao^2}{12}+\frac{287 d_1 \ao^2}{36}+\frac{28 \
d_1' \ao^2}{3}-\frac{95}{6} d_1 \yo \ao^2-\frac{35}{2} d_1' \yo \
\ao^2+\frac{599 \yo \ao^2}{6}-\frac{161 \ao^2}{3}+\frac{25}{9} d_1 \
\yo^3 \ao+\frac{8}{9} d_1' \yo^3 \ao-\frac{169 \yo^3 \ao}{18}-13 d_1 \
\yo^2 \ao-\frac{31}{6} d_1' \yo^2 \ao+\frac{295 \yo^2 \
\ao}{6}-\frac{343 d_1 \ao}{18}-\frac{47 d_1' \ao}{6}+\frac{85 d_1 \yo \
\ao}{3}+\frac{59 d_1' \yo \ao}{3}-\frac{440 \yo \
\ao}{3}+\frac{\ao}{\yo-1}+\frac{149 \ao}{2}-\frac{205 d_1 \
\yo^3}{108}-\frac{25 d_1' \yo^3}{54}+\frac{305 \yo^3}{54}+\frac{22 \
d_1 \yo^2}{3}+\frac{7 d_1' \yo^2}{3}-\frac{919 \yo^2}{36}+\frac{217 \
d_1}{36}-\frac{d_1'}{6}-\frac{469 d_1 \yo}{36}-8 d_1' \yo+\frac{602 \
\yo}{9}+\Big(-\frac{1}{3} \yo^3 \ao^4+2 \yo^2 \ao^4-5 \yo \
\ao^4+\frac{10 \ao^4}{3}+\frac{16 \yo^3 \ao^3}{9}-10 \yo^2 \
\ao^3+\frac{76 \yo \ao^3}{3}-\frac{136 \ao^3}{9}-4 \yo^3 \ao^2+21 \
\yo^2 \ao^2-52 \yo \ao^2+\frac{76 \ao^2}{3}+\frac{16 \yo^3 \ao}{3}-26 \
\yo^2 \ao+60 \yo \ao-\frac{82 \ao}{3}-\frac{25 \yo^3}{9}+\frac{35 \
\yo^2}{3}-23 \yo+\frac{13}{3 (\yo-1)}+\frac{13}{3}\Big) \
H(0;\ao)+\Big(-\frac{1}{3} d_1 \yo^3 \ao^4+2 d_1 \yo^2 \ao^4+\frac{10 \
d_1 \ao^4}{3}-5 d_1 \yo \ao^4+\frac{16}{9} d_1 \yo^3 \ao^3-10 d_1 \
\yo^2 \ao^3-\frac{136 d_1 \ao^3}{9}+\frac{76}{3} d_1 \yo \ao^3-4 d_1 \
\yo^3 \ao^2+21 d_1 \yo^2 \ao^2+\frac{76 d_1 \ao^2}{3}-52 d_1 \yo \
\ao^2+\frac{16}{3} d_1 \yo^3 \ao-26 d_1 \yo^2 \ao-\frac{82 d_1 \
\ao}{3}+60 d_1 \yo \ao-\frac{25 d_1 \yo^3}{9}+\frac{35 d_1 \yo^2}{3}+\
\frac{13 d_1}{3}-23 d_1 \yo+\frac{13 d_1}{3 (\yo-1)}\Big) \
H(1;\ao)+\frac{217 d_1}{36 (\yo-1)}-\frac{d_1'}{6 \
(\yo-1)}-\frac{88}{9 (\yo-1)}-\frac{19}{12 \
(\yo-1)^2}-\frac{295}{36}\Big) H(c_1(\ao);\yo)+\Big(-\frac{7 \
\yo^3}{3}+\frac{26 \yo^2}{3}-11 \yo+\frac{13}{3 (\yo-1)}+\frac{13}{3}\
\Big) H(0,0;\ao)+\Big(-\frac{16 d_1' \yo^3}{9}+\frac{175 \
\yo^3}{9}+\frac{28 d_1' \yo^2}{3}-90 \yo^2-\frac{88 d_1' \yo}{3}+241 \
\yo-\frac{13}{\yo-1}-13\Big) H(0,0;\yo)+\Big(-\frac{7 d_1 \
\yo^3}{3}+\frac{26 d_1 \yo^2}{3}-11 d_1 \yo+\frac{13 d_1}{3}+\frac{13 \
d_1}{3 (\yo-1)}\Big) H(0,1;\ao)+\Big(-\frac{4}{9} d_1'^2 \
\yo^3+\frac{77 d_1' \yo^3}{18}+\frac{7 d_1'^2 \yo^2}{3}-\frac{61 d_1' \
\yo^2}{3}-\frac{22 d_1'^2 \yo}{3}+\frac{115 d_1' \yo}{2}-\frac{13 \
d_1'}{6}+\Big(-\frac{4 d_1 \yo^3}{3}+\frac{4 \yo^3}{3}+6 d_1 \yo^2-6 \
\yo^2-12 d_1 \yo+12 \yo+\frac{38 d_1}{3}-\frac{4 \
d_1}{\yo-1}+\frac{4}{\yo-1}+\frac{62}{3}\Big) H(0;\ao)-8 H(0,0;\ao)-8 \
d_1 H(0,1;\ao)-\frac{13 d_1'}{6 (\yo-1)}-\frac{2 d_1 \pi ^2}{3}\Big) \
H(0,1;\yo)+\Big(\frac{d_1 \ao^4}{2}-\frac{\ao^4}{2}-\frac{26 d_1 \
\ao^3}{9}+\frac{38 \ao^3}{9}+\frac{23 d_1 \ao^2}{3}-\frac{49 \
\ao^2}{3}-\frac{50 d_1 \ao}{3}+\frac{142 \ao}{3}+\frac{2 d_1' \
\yo^3}{9}-\frac{49 \yo^3}{18}-\frac{7 d_1' \yo^2}{6}+\frac{37 \
\yo^2}{3}+4 d_1-2 d_1'+\frac{11 d_1' \yo}{3}-\frac{63 \yo}{2}+\Big(2 \
\ao^4-\frac{32 \ao^3}{3}+24 \ao^2-32 \ao+\frac{4 \yo^3}{3}-6 \yo^2+12 \
\yo+\frac{4}{\yo-1}+4\Big) H(0;\ao)+\Big(2 d_1 \ao^4-\frac{32 d_1 \
\ao^3}{3}+24 d_1 \ao^2-32 d_1 \ao+\frac{4 d_1 \yo^3}{3}-6 d_1 \yo^2+4 \
d_1+12 d_1 \yo+\frac{4 d_1}{\yo-1}\Big) H(1;\ao)+\frac{4 d_1}{\yo-1}-\
\frac{2 d_1'}{\yo-1}-\frac{11}{6 \
(\yo-1)}-\frac{1}{(\yo-1)^2}-\frac{5}{6}\Big) \
H(0,c_1(\ao);\yo)+\Big(-\frac{4}{9} d_1'^2 \yo^3+\frac{73 d_1' \
\yo^3}{18}+\frac{49 \yo^3}{18}+\frac{7 d_1'^2 \yo^2}{3}-\frac{4 d_1 \
\yo^2}{3}-\frac{115 d_1' \yo^2}{6}-\frac{41 \yo^2}{3}-\frac{22 d_1'^2 \
\yo}{3}+\frac{16 d_1 \yo}{3}+\frac{323 d_1' \yo}{6}+\frac{221 \
\yo}{6}+\frac{49 d_1'^2}{9}-\frac{37 d_1}{18}-\frac{697 \
d_1'}{18}+\Big(\frac{4 \yo^3}{3}-6 \yo^2+12 \
\yo+\frac{4}{\yo-1}-\frac{38}{3}\Big) H(0;\ao)+\frac{d_1}{3 (\yo-1)}-\
\frac{d_1'}{6 (\yo-1)}+\frac{37}{6 \
(\yo-1)}+\frac{1}{(\yo-1)^2}+\frac{4 \pi ^2}{3}-\frac{130}{3}\Big) \
H(1,0;\yo)+\Big(-\frac{1}{9} \yo^3 d_1'^3+\frac{7 \yo^2 \
d_1'^3}{12}-\frac{11 \yo d_1'^3}{6}+\frac{49 d_1'^3}{36}+\frac{7 \
\yo^3 d_1'^2}{9}-4 \yo^2 d_1'^2+13 \yo d_1'^2-\frac{88 \
d_1'^2}{9}+\frac{\pi ^2 d_1'}{3}+\Big(-\frac{4 d_1 \yo^3}{3}+\frac{4 \
d_1' \yo^3}{3}-\frac{2 \yo^3}{3}+6 d_1 \yo^2-6 d_1' \yo^2+3 \yo^2-12 \
d_1 \yo+12 d_1' \yo-6 \yo+\frac{38 d_1}{3}-10 d_1'-\frac{8 \
d_1}{\yo-1}+\frac{4 d_1'}{\yo-1}+\frac{2}{\yo-1}+\frac{43}{3}\Big) \
H(0;\ao)-4 H(0,0;\ao)-4 d_1 H(0,1;\ao)-\frac{2 d_1 \pi \
^2}{3}+\frac{\pi ^2}{3}\Big) H(1,1;\yo)+\Big(-\frac{1}{6} d_1' \yo^3 \
\ao^4+d_1' \yo^2 \ao^4+\frac{5 d_1' \ao^4}{3}-\frac{5}{2} d_1' \yo \
\ao^4+\frac{8}{9} d_1' \yo^3 \ao^3-5 d_1' \yo^2 \ao^3-\frac{77 d_1' \
\ao^3}{9}+\frac{38}{3} d_1' \yo \ao^3-2 d_1' \yo^3 \ao^2+\frac{21}{2} \
d_1' \yo^2 \ao^2+\frac{35 d_1' \ao^2}{2}-26 d_1' \yo \
\ao^2+\frac{8}{3} d_1' \yo^3 \ao-13 d_1' \yo^2 \ao-\frac{65 d_1' \
\ao}{3}+30 d_1' \yo \ao-\frac{7 d_1' \yo^3}{6}-\frac{49 \
\yo^3}{18}+\frac{4 d_1 \yo^2}{3}+\frac{14 d_1' \yo^2}{3}+\frac{41 \
\yo^2}{3}+\frac{37 d_1}{18}+\frac{13 d_1'}{3}-\frac{16 d_1 \
\yo}{3}-\frac{47 d_1' \yo}{6}-\frac{221 \yo}{6}+\Big(\frac{4 \
\yo^3}{3}-6 \yo^2+12 \yo+\frac{4}{\yo-1}-\frac{38}{3}\Big) \
H(0;\ao)+\Big(\frac{4 d_1 \yo^3}{3}-6 d_1 \yo^2+12 d_1 \yo-\frac{38 \
d_1}{3}+\frac{4 d_1}{\yo-1}\Big) H(1;\ao)-\frac{d_1}{3 \
(\yo-1)}+\frac{d_1'}{6 (\yo-1)}-\frac{37}{6 \
(\yo-1)}-\frac{1}{(\yo-1)^2}+\frac{130}{3}\Big) \
H(1,c_1(\ao);\yo)+\Big(-\frac{1}{3} \yo^3 \ao^4+2 \yo^2 \ao^4-5 \yo \
\ao^4+\frac{10 \ao^4}{3}+\frac{16 \yo^3 \ao^3}{9}-10 \yo^2 \
\ao^3+\frac{76 \yo \ao^3}{3}-\frac{136 \ao^3}{9}-4 \yo^3 \ao^2+21 \
\yo^2 \ao^2-52 \yo \ao^2+\frac{76 \ao^2}{3}+\frac{16 \yo^3 \ao}{3}-26 \
\yo^2 \ao+60 \yo \ao-\frac{82 \ao}{3}-\frac{25 \yo^3}{9}+\frac{35 \
\yo^2}{3}-23 \yo+\frac{13}{3 (\yo-1)}+\frac{13}{3}\Big) H(c_1(\ao),0;\
\yo)+\Big(-\frac{1}{6} d_1' \yo^3 \ao^4+d_1' \yo^2 \ao^4+\frac{5 d_1' \
\ao^4}{3}-\frac{5}{2} d_1' \yo \ao^4+\frac{8}{9} d_1' \yo^3 \ao^3-5 \
d_1' \yo^2 \ao^3-\frac{68 d_1' \ao^3}{9}+\frac{38}{3} d_1' \yo \
\ao^3-2 d_1' \yo^3 \ao^2+\frac{21}{2} d_1' \yo^2 \ao^2+\frac{38 d_1' \
\ao^2}{3}-26 d_1' \yo \ao^2+\frac{8}{3} d_1' \yo^3 \ao-13 d_1' \yo^2 \
\ao-\frac{41 d_1' \ao}{3}+30 d_1' \yo \ao-\frac{25 d_1' \
\yo^3}{18}+\frac{35 d_1' \yo^2}{6}+\frac{13 d_1'}{6}-\frac{23 d_1' \
\yo}{2}+\frac{13 d_1'}{6 (\yo-1)}\Big) \
H(c_1(\ao),1;\yo)+\Big(-\frac{1}{2} \yo^3 \ao^4+3 \yo^2 \
\ao^4-\frac{15 \yo \ao^4}{2}+5 \ao^4+\frac{8 \yo^3 \ao^3}{3}-15 \yo^2 \
\ao^3+38 \yo \ao^3-\frac{68 \ao^3}{3}-6 \yo^3 \ao^2+\frac{63 \yo^2 \
\ao^2}{2}-78 \yo \ao^2+37 \ao^2+8 \yo^3 \ao-39 \yo^2 \ao+4 d_1 \ao-2 \
d_1' \ao+90 \yo \ao-43 \ao-\frac{25 \yo^3}{6}+\frac{35 \yo^2}{2}-4 \
d_1+2 d_1'-\frac{69 \yo}{2}+\Big(4 \ao-\frac{4}{\yo-1}-4\Big) \
H(0;\ao)+\Big(4 \ao d_1-\frac{4 d_1}{\yo-1}-4 d_1\Big) \
H(1;\ao)-\frac{4 d_1}{\yo-1}+\frac{2 d_1'}{\yo-1}+\frac{21}{2 \
(\yo-1)}+\frac{1}{(\yo-1)^2}+\frac{19}{2}\Big) \
H(c_1(\ao),c_1(\ao);\yo)+\Big(-\frac{64 \yo^3}{3}+96 \yo^2-192 \
\yo\Big) H(0,0,0;\yo)+\Big(-\frac{16 d_1' \yo^3}{3}+24 d_1' \yo^2-48 \
d_1' \yo+(8 d_1-8) H(0;\ao)\Big) H(0,0,1;\yo)+\Big(\frac{8 \
\yo^3}{3}-12 \yo^2+24 \yo-8 H(0;\ao)-8 d_1 H(1;\ao)+\frac{8}{\yo-1}+8\
\Big) H(0,0,c_1(\ao);\yo)+\Big(\frac{4 d_1 \yo^3}{3}-\frac{16 d_1' \
\yo^3}{3}-\frac{4 \yo^3}{3}-6 d_1 \yo^2+24 d_1' \yo^2+6 \yo^2+12 d_1 \
\yo-48 d_1' \yo-12 \yo-\frac{38 d_1}{3}-8 H(0;\ao)+\frac{4 \
d_1}{\yo-1}-\frac{4}{\yo-1}-\frac{62}{3}\Big) \
H(0,1,0;\yo)+\Big(-\frac{4}{3} d_1'^2 \yo^3+6 d_1'^2 \yo^2-12 d_1'^2 \
\yo+(12 d_1-8 d_1') H(0;\ao)\Big) H(0,1,1;\yo)+\Big(d_1' \
\ao^4-\frac{16 d_1' \ao^3}{3}+12 d_1' \ao^2-16 d_1' \ao-\frac{4 d_1 \
\yo^3}{3}+\frac{2 d_1' \yo^3}{3}+\frac{4 \yo^3}{3}+6 d_1 \yo^2-3 d_1' \
\yo^2-6 \yo^2+\frac{38 d_1}{3}+2 d_1'-12 d_1 \yo+6 d_1' \yo+12 \yo-8 \
H(0;\ao)-8 d_1 H(1;\ao)-\frac{4 d_1}{\yo-1}+\frac{2 \
d_1'}{\yo-1}+\frac{4}{\yo-1}+\frac{62}{3}\Big) \
H(0,1,c_1(\ao);\yo)+\Big(2 \ao^4-\frac{32 \ao^3}{3}+24 \ao^2-32 \
\ao+\frac{4 \yo^3}{3}-6 \yo^2+12 \yo+\frac{4}{\yo-1}+4\Big) \
H(0,c_1(\ao),0;\yo)+\Big(d_1' \ao^4-\frac{16 d_1' \ao^3}{3}+12 d_1' \
\ao^2-16 d_1' \ao+\frac{2 d_1' \yo^3}{3}-3 d_1' \yo^2+2 d_1'+6 d_1' \
\yo+\frac{2 d_1'}{\yo-1}\Big) H(0,c_1(\ao),1;\yo)+\Big(3 \ao^4-16 \
\ao^3+36 \ao^2-48 \ao+2 \yo^3-9 \yo^2+18 \yo+4 H(0;\ao)+4 d_1 \
H(1;\ao)-\frac{2}{\yo-1}-2\Big) \
H(0,c_1(\ao),c_1(\ao);\yo)+\Big(-\frac{16 d_1' \yo^3}{3}-4 \yo^3+24 \
d_1' \yo^2+18 \yo^2-48 d_1' \yo-36 \yo+\frac{88 \
d_1'}{3}-\frac{12}{\yo-1}+38\Big) H(1,0,0;\yo)+\Big(-\frac{4}{3} \
d_1'^2 \yo^3-\frac{2 d_1' \yo^3}{3}+6 d_1'^2 \yo^2+3 d_1' \yo^2-12 \
d_1'^2 \yo-6 d_1' \yo+\frac{22 d_1'^2}{3}+\frac{19 d_1'}{3}+(4 d_1-4) \
H(0;\ao)-\frac{2 d_1'}{\yo-1}\Big) H(1,0,1;\yo)+\Big(\frac{2 d_1' \
\yo^3}{3}+\frac{2 \yo^3}{3}-3 d_1' \yo^2-3 \yo^2+6 d_1' \yo+6 \
\yo-\frac{11 d_1'}{3}-4 H(0;\ao)-4 d_1 H(1;\ao)-\frac{4 \
d_1}{\yo-1}+\frac{2 d_1'}{\yo-1}+\frac{6}{\yo-1}+\frac{5}{3}\Big) \
H(1,0,c_1(\ao);\yo)+\Big(-\frac{4}{3} d_1'^2 \yo^3+\frac{4 d_1 \
\yo^3}{3}-\frac{4 d_1' \yo^3}{3}+\frac{2 \yo^3}{3}+6 d_1'^2 \yo^2-6 \
d_1 \yo^2+6 d_1' \yo^2-3 \yo^2-12 d_1'^2 \yo+12 d_1 \yo-12 d_1' \yo+6 \
\yo+\frac{22 d_1'^2}{3}-\frac{38 d_1}{3}+10 d_1'-4 H(0;\ao)+\frac{8 \
d_1}{\yo-1}-\frac{4 d_1'}{\yo-1}-\frac{2}{\yo-1}-\frac{43}{3}\Big) \
H(1,1,0;\yo)+\Big(-\frac{1}{3} \yo^3 d_1'^3+\frac{3 \yo^2 \
d_1'^3}{2}-3 \yo d_1'^3+\frac{11 d_1'^3}{6}+(8 d_1-4 d_1'-2) H(0;\ao)\
\Big) H(1,1,1;\yo)+\Big(-\frac{4 d_1 \yo^3}{3}+\frac{4 d_1' \
\yo^3}{3}-\frac{2 \yo^3}{3}+6 d_1 \yo^2-6 d_1' \yo^2+3 \yo^2-12 d_1 \
\yo+12 d_1' \yo-6 \yo+\frac{38 d_1}{3}-10 d_1'-4 H(0;\ao)-4 d_1 \
H(1;\ao)-\frac{8 d_1}{\yo-1}+\frac{4 \
d_1'}{\yo-1}+\frac{2}{\yo-1}+\frac{43}{3}\Big) \
H(1,1,c_1(\ao);\yo)+\Big(\frac{4 \yo^3}{3}-6 \yo^2+12 \
\yo+\frac{4}{\yo-1}-\frac{38}{3}\Big) \
H(1,c_1(\ao),0;\yo)+\Big(\frac{2 d_1' \yo^3}{3}-3 d_1' \yo^2+6 d_1' \
\yo-\frac{19 d_1'}{3}+\frac{2 d_1'}{\yo-1}\Big) \
H(1,c_1(\ao),1;\yo)+\Big(2 \yo^3-9 \yo^2+18 \yo+4 H(0;\ao)+4 d_1 H(1;\
\ao)+\frac{4 d_1}{\yo-1}-\frac{2 d_1'}{\yo-1}+\frac{2}{\yo-1}-27\Big) \
H(1,c_1(\ao),c_1(\ao);\yo)+\Big(2 \ao d_1'-\frac{2 d_1'}{\yo-1}-2 \
d_1'\Big) H(c_1(\ao),1,c_1(\ao);\yo)+\Big(4 \
\ao-\frac{4}{\yo-1}-4\Big) H(c_1(\ao),c_1(\ao),0;\yo)+\Big(2 \ao \
d_1'-\frac{2 d_1'}{\yo-1}-2 d_1'\Big) \
H(c_1(\ao),c_1(\ao),1;\yo)+\Big(6 \ao-\frac{6}{\yo-1}-6\Big) \
H(c_1(\ao),c_1(\ao),c_1(\ao);\yo)+64 H(0,0,0,0;\yo)+16 d_1' \
H(0,0,0,1;\yo)-16 H(0,0,0,c_1(\ao);\yo)+(-8 d_1+16 d_1'+8) H(0,0,1,0;\
\yo)+4 d_1'^2 H(0,0,1,1;\yo)+(8 d_1-4 d_1'-8) H(0,0,1,c_1(\ao);\yo)-8 \
H(0,0,c_1(\ao),0;\yo)-4 d_1' H(0,0,c_1(\ao),1;\yo)-4 \
H(0,0,c_1(\ao),c_1(\ao);\yo)+(16 d_1'+24) H(0,1,0,0;\yo)+\Big(4 \
d_1'^2+4 d_1'\Big) H(0,1,0,1;\yo)+(4 d_1-4 d_1'-8) \
H(0,1,0,c_1(\ao);\yo)+\Big(4 d_1'^2+8 d_1'-12 d_1\Big) \
H(0,1,1,0;\yo)+d_1'^3 H(0,1,1,1;\yo)+(12 d_1-8 d_1') \
H(0,1,1,c_1(\ao);\yo)-8 H(0,1,c_1(\ao),0;\yo)-4 d_1' \
H(0,1,c_1(\ao),1;\yo)+(-4 d_1+2 d_1'-8) \
H(0,1,c_1(\ao),c_1(\ao);\yo)+2 d_1' H(0,c_1(\ao),1,c_1(\ao);\yo)+4 \
H(0,c_1(\ao),c_1(\ao),0;\yo)+2 d_1' H(0,c_1(\ao),c_1(\ao),1;\yo)+6 \
H(0,c_1(\ao),c_1(\ao),c_1(\ao);\yo)-8 H(1,0,0,c_1(\ao);\yo)+(4-4 d_1) \
H(1,0,1,0;\yo)+(4 d_1-2 d_1'-4) H(1,0,1,c_1(\ao);\yo)-4 \
H(1,0,c_1(\ao),0;\yo)-2 d_1' H(1,0,c_1(\ao),1;\yo)+2 \
H(1,0,c_1(\ao),c_1(\ao);\yo)+12 H(1,1,0,0;\yo)+2 d_1' \
H(1,1,0,1;\yo)+(4 d_1-2 d_1'-6) H(1,1,0,c_1(\ao);\yo)+(-8 d_1+4 \
d_1'+2) H(1,1,1,0;\yo)+(8 d_1-4 d_1'-2) H(1,1,1,c_1(\ao);\yo)-4 \
H(1,1,c_1(\ao),0;\yo)-2 d_1' H(1,1,c_1(\ao),1;\yo)+(-4 d_1+2 d_1'-2) \
H(1,1,c_1(\ao),c_1(\ao);\yo)+2 d_1' H(1,c_1(\ao),1,c_1(\ao);\yo)+4 \
H(1,c_1(\ao),c_1(\ao),0;\yo)+2 d_1' H(1,c_1(\ao),c_1(\ao),1;\yo)+6 \
H(1,c_1(\ao),c_1(\ao),c_1(\ao);\yo)+H(1;\yo) \Big(\frac{1}{18} d_1' \
\yo^3 \ao^4-\frac{5}{12} d_1' \yo^2 \ao^4-\frac{47 d_1' \
\ao^4}{36}+\frac{5}{3} d_1' \yo \ao^4-\frac{13}{54} d_1' \yo^3 \ao^3+\
\frac{17}{9} d_1' \yo^2 \ao^3+\frac{391 d_1' \ao^3}{54}-\frac{80}{9} \
d_1' \yo \ao^3+\frac{11}{36} d_1' \yo^3 \ao^2-\frac{109}{36} d_1' \
\yo^2 \ao^2-\frac{89 d_1' \ao^2}{6}+\frac{158}{9} d_1' \yo \
\ao^2+\frac{7}{18} d_1' \yo^3 \ao+\frac{7}{18} d_1' \yo^2 \
\ao+\frac{247 d_1' \ao}{18}-\frac{43 d_1' \yo \ao}{3}+\frac{251 \
d_1'^3}{216}-\frac{d_1'^3 \yo^3}{27}+\frac{11 d_1'^2 \
\yo^3}{27}-\frac{44 d_1' \yo^3}{27}+\frac{1}{9} d_1' \pi ^2 \
\yo^3+\frac{\pi ^2 \yo^3}{9}-\frac{398 d_1'^2}{27}+\frac{17 d_1'^3 \
\yo^2}{72}-\frac{25 d_1'^2 \yo^2}{9}+\frac{37 d_1' \
\yo^2}{4}-\frac{1}{2} d_1' \pi ^2 \yo^2-\frac{\pi ^2 \
\yo^2}{2}+\frac{4361 d_1'}{108}-\frac{49 d_1'^3 \yo}{36}+\frac{154 \
d_1'^2 \yo}{9}-48 d_1' \yo+d_1' \pi ^2 \yo+\pi ^2 \yo+\Big(-\frac{17 \
d_1' \yo^3}{18}-\frac{49 \yo^3}{18}+\frac{4 d_1 \yo^2}{3}+\frac{19 \
d_1' \yo^2}{6}+\frac{41 \yo^2}{3}-\frac{16 d_1 \yo}{3}-\frac{11 d_1' \
\yo}{6}-\frac{221 \yo}{6}+\frac{37 d_1}{18}-\frac{7 \
d_1'}{18}-\frac{d_1}{3 (\yo-1)}+\frac{d_1'}{6 (\yo-1)}-\frac{37}{6 \
(\yo-1)}-\frac{1}{(\yo-1)^2}+\frac{130}{3}\Big) H(0;\ao)+\Big(\frac{4 \
\yo^3}{3}-6 \yo^2+12 \yo+\frac{4}{\yo-1}-\frac{38}{3}\Big) \
H(0,0;\ao)+\Big(\frac{4 d_1 \yo^3}{3}-6 d_1 \yo^2+12 d_1 \yo-\frac{38 \
d_1}{3}+\frac{4 d_1}{\yo-1}\Big) H(0,1;\ao)+\frac{2 d_1 \pi ^2}{3 \
(\yo-1)}-\frac{d_1' \pi ^2}{3 (\yo-1)}-\frac{\pi ^2}{3 (\yo-1)}-6 \
\zeta_3-\frac{11 d_1' \pi ^2}{18}-\frac{43 \pi ^2}{18}\Big)+H(0;\yo) \
\Big(\frac{\yo^3 \ao^4}{9}-\frac{5 \yo^2 \ao^4}{6}+\frac{10 \yo \
\ao^4}{3}-\frac{13 \yo^3 \ao^3}{27}+\frac{34 \yo^2 \
\ao^3}{9}-\frac{160 \yo \ao^3}{9}+\frac{11 \yo^3 \ao^2}{18}-\frac{109 \
\yo^2 \ao^2}{18}+\frac{316 \yo \ao^2}{9}+\frac{7 \yo^3 \
\ao}{9}+\frac{7 \yo^2 \ao}{9}-\frac{86 \yo \ao}{3}-\frac{4 d_1'^2 \
\yo^3}{27}+\frac{205 d_1 \yo^3}{108}+\frac{35 d_1' \yo^3}{18}+\frac{4 \
\pi ^2 \yo^3}{9}-\frac{625 \yo^3}{54}+\frac{17 d_1'^2 \
\yo^2}{18}-\frac{22 d_1 \yo^2}{3}-\frac{110 d_1' \yo^2}{9}-2 \pi ^2 \
\yo^2+\frac{520 \yo^2}{9}-\frac{217 d_1}{36}+\frac{d_1'}{6}-\frac{49 \
d_1'^2 \yo}{9}+\frac{469 d_1 \yo}{36}+\frac{1225 d_1' \yo}{18}+4 \pi \
^2 \yo-\frac{2042 \yo}{9}+\Big(-\frac{7 \yo^3}{3}+\frac{26 \
\yo^2}{3}-11 \yo+\frac{13}{3 (\yo-1)}+\frac{13}{3}\Big) \
H(0;\ao)-\frac{217 d_1}{36 (\yo-1)}+\frac{d_1'}{6 (\yo-1)}-\frac{4 \
\pi ^2}{3 (\yo-1)}+\frac{88}{9 (\yo-1)}+\frac{19}{12 (\yo-1)^2}+6 \
\zeta_3-\frac{4 \pi ^2}{3}+\frac{295}{36}\Big)-\frac{2 d_1 \pi ^2}{3 \
(\yo-1)}+\frac{d_1' \pi ^2}{3 (\yo-1)}+\frac{37 \pi ^2}{36 \
(\yo-1)}+\frac{\pi ^2}{6 (\yo-1)^2}-4 \yo^3 \zeta_3+18 \yo^2 \
\zeta_3-36 \yo \zeta_3+\frac{6 \zeta_3}{\yo-1}+6 \zeta_3-\frac{\pi \
^4}{120}-\frac{2 d_1 \pi ^2}{3}+\frac{d_1' \pi ^2}{3}+\frac{31 \pi \
^2}{36}.
\erp

\end{document}